\documentclass[11pt,twoside,a4paper]{cernrep} 
\pdfoutput=1
\usepackage{rep_common,mathrsfs}
\usepackage{tabulary}
\usepackage[colorlinks, urlcolor=blue, linkcolor=blue, citecolor=blue, plainpages=false]{hyperref}
\usepackage{chngcntr}
\def\bibfiles{\main/bib/chapter}
\providecommand{\biblio}{\nocite{article-minimal}\bibliographystyle{report}\clearpage\bibliography{\bibfiles}}  % *Modification: added `\main/` to specify relative file location.
\newcommand{\minipageheightsp}{6.6cm}
\newcommand{\minipagewidthsp}{\textwidth}
\newcommand{\figuremaxwidthsp}{9cm}
\newcommand{\figuremaxheightsp}{6.5cm}
\newcommand{\minipageheightdp}{5.5cm}
\newcommand{\minipagewidthdp}{0.49\textwidth}
\newcommand{\figuremaxwidthdp}{7.8cm}
\newcommand{\figuremaxheightdp}{6.5cm}
\newcommand{\minipageheighttp}{3.67cm}
\newcommand{\minipagewidthtp}{0.327\textwidth}
\newcommand{\figuremaxwidthtp}{5.6cm}
\newcommand{\figuremaxheighttp}{6.5cm}

\counterwithin{figure}{subsection}

\counterwithin{table}{subsection}

\counterwithin{equation}{subsection}

\makeatletter
\newcommand{\getCurrentSectionNumber}{%
  \ifnum\c@section=0 %
  \thechapter
  \else
  \ifnum\c@subsection=0 %
  \thesection
  \else
  \thesubsection
  \fi
  \fi
}
\makeatother
\newcounter{resetdummycounter}
\newcommand{\resetcounters}{\stepcounter{resetdummycounter}}

\newcommand{\KeV}{\ensuremath{\rm~KeV}\xspace}
\newcommand{\MeV}{\ensuremath{\rm~MeV}\xspace}
\newcommand{\GeV}{\ensuremath{\rm~GeV}\xspace}
\newcommand{\TeV}{\ensuremath{\rm~TeV}\xspace}

\newcommand{\tp}{T_p}
\newcommand{\tpbar}{\bar{T}_p}
\newcommand{\stp}{\tilde{t}}

\newcommand{\go}{\tilde{g}}
\newcommand{\Q}{\mathcal{Q}}
\newcommand{\Qbar}{\overline{\newResonance}}
\newcommand{\QQbar}{\Q\Qbar}

\newcommand{\newResonance}{\Q}

\newcommand{\dm}{X}
\newcommand{\jet}{j}
\newcommand{\q}{q}
\newcommand{\qbar}{\bar{\q}}
\newcommand{\qqbar}{\q\qbar}

\newcommand{\smtop}{t}
\newcommand{\ttbar}{\smtop\bar{\smtop}}

\newcommand{\tptp}{\tp\tpbar}

\newcommand{\ppQQbarj}{pp\rightarrow\QQbar+\jet}
\newcommand{\pptptp}{pp\rightarrow\tptp}

\newcommand{\pt}{\ensuremath{p_{T}\xspace}}

\newcommand{\ptjcut}{p_{T,\mathrm{cut}}^{j_1}}
\newcommand{\met}{E_{\mathrm{T}}^{\mathrm{miss}}}

\newcommand{\ptmiss}{\ensuremath{\textbf{p}_{\mathrm{T}}^{\mathrm{miss}}}\xspace}
\newcommand{\mtop}{\ensuremath{m_{\mathrm{top}}}\xspace}
\newcommand{\Dmstop}{\ensuremath{\Delta m(\tone,\neut)}\xspace}
\newcommand{\largeDM}{large $\Delta m$\xspace}
\newcommand{\diagonal}{diagonal\xspace}
\newcommand{\antikt}{\ensuremath{\textrm{anti-}k_{t}}\xspace}
\newcommand{\akt}[1]{\ensuremath{\textrm{anti-}k_{t}^{{#1}}}\xspace}
\newcommand{\nbjet}{\ensuremath{N_{b-\mathrm{jet}}}\xspace}
\newcommand{\aktm}[2]{\ensuremath{m_{#2}^{\akt{{#1}}}}\xspace}

\newcommand{\RISR}{\ensuremath{R_{\mathrm{ISR}}}\xspace}
\newcommand{\Ci}{\ensuremath{\widetilde{\chi}_{i}^{\pm}}\xspace} % C
\newcommand{\Nj}{\ensuremath{\widetilde{\chi}^{0}_{j}}\xspace} % N
\newcommand{\COne}{\ensuremath{\widetilde{\chi}_{1}^{\pm}}\xspace} % C1
\newcommand{\NTwo}{\ensuremath{\widetilde{\chi}^{0}_{2}}\xspace} % N2
\newcommand{\NOne}{\ensuremath{\widetilde{\chi}^{0}_{1}}\xspace} % N1
\newcommand{\Wstar}{\ensuremath{\PW^{*}}\xspace}
\newcommand{\Zstar}{\ensuremath{\PZ^{*}}\xspace}
\newcommand{\jisr}{\ensuremath{j_{\text{ISR}}}\xspace}
\newcommand{\ptisr}{\ensuremath{p_{\text{T}}(\jisr)}\xspace}
\newcommand{\dphijisr}{\ensuremath{\Delta\phi(\ptmiss,  \ptisr)}\xspace}
\newcommand{\DRll}{\ensuremath{\Delta R(\ell_1\ell_2)}\xspace}
\newcommand{\Njet}{\ensuremath{N_{\text{jet}}}\xspace}
\newcommand{\Nlep}{\ensuremath{N_{\ell}}\xspace}
\newcommand{\Nbjet}{\ensuremath{N_{\text{b-jet}}}\xspace}
\newcommand{\Nisr}{\ensuremath{N_{\text{ISR}}}\xspace}
\newcommand{\mll}{\ensuremath{m_{\ell_1, \ell_2}}\xspace}
\newcommand{\pttwo}{\ensuremath{p_{\text{T}}({\ell_2})}\xspace}
\newcommand{\BR}{\ensuremath{\mathcal{B}}\xspace}
\newcommand{\mleadJ}{\ensuremath{m_{\mathrm{J}}^{\mathrm{lead}}}\xspace}
\newcommand{\msublJ}{\ensuremath{m_{\mathrm{J}}^{\mathrm{subl}}}\xspace}
\def\kMPl{\ensuremath{k/\overline{M}_{\mathrm{Pl}}}\xspace}
\newcommand{\kt}{\ensuremath{k_{\mathrm{T}}}\xspace}
\newcommand{\tauTT}{\ensuremath{\tau_3/\tau_2}\xspace}
\newcommand{\HT}{\ensuremath{H_{\text{T}}\xspace}}
\newcommand{\abs}[1]{\ensuremath{\lvert #1 \rvert}}
\newcommand*{\zptt}{\ensuremath{Z^{\prime} \rightarrow \ttbar}}
\newcommand*{\qjj}{\ensuremath{Q^{\prime} \rightarrow jj}}
\newcommand*{\rsg}{\ensuremath{G_{RS} \rightarrow WW}}
\newcommand*{\tauh}{\ensuremath{\tau_h}}

\newcommand*{\ZpSSM}{\ensuremath{Z^{\prime}_{\mathrm{SSM}}}}
\newcommand*{\metvec}{\vec{\pt}^{\textrm{miss}}}
\newcommand{\meters}{\mbox{\ensuremath{\rm~m}}\xspace}
\newcommand{\cmeters}{\mbox{\ensuremath{\rm~cm}}\xspace}
\newcommand{\mumeters}{\mbox{\ensuremath{~\mu{\rm m}}}\xspace}
\newcommand{\yr}{\mbox{\ensuremath{\rm~yr}}\xspace}
\newcommand{\psec}{\mbox{\ensuremath{\rm~ps}}\xspace}
\newcommand{\nsec}{\mbox{\ensuremath{\rm~ns}}\xspace}
\newcommand{\fb}{\mbox{\ensuremath{\rm~fb}}\xspace}
\newcommand{\pb}{\mbox{\ensuremath{\rm~pb}}\xspace}
\newcommand{\ab}{\mbox{\ensuremath{\rm~ab}}\xspace}
\newcommand{\fbinv}{\mbox{\ensuremath{{\rm~fb}^{-1}}}\xspace}
\newcommand{\abinv}{\mbox{\ensuremath{{\rm~ab}^{-1}}}\xspace}
\newcommand{\intLumi}{\ensuremath{3\abinv}}
\newcommand{\intlumihelhc}{\ensuremath{15\abinv}}

\newcommand{\mx}{\ensuremath{m_{\PX}}\xspace}
\newcommand{\bbbar}{\ensuremath{\Pqb\Paqb}\xspace}
\newcommand{\Mpl}{\ensuremath{{M_\mathrm{Pl}}}\xspace}

\newcommand{\LambdaR}{\ensuremath{\Lambda_{\text{R}}}\xspace}

\newcommand{\nsub}{\ensuremath{\tau_{21}}\xspace}
\newcommand*{\Z}{\ensuremath{Z}}
\newcommand*{\Zp}{\ensuremath{Z^{\prime}}}
\newcommand*{\Zptata}{\ensuremath{Z^{\prime}\rightarrow \tau\tau}}
\newcommand*{\Zpee}{\ensuremath{Z^{\prime}\rightarrow e e}}
\newcommand*{\Zpmumu}{\ensuremath{Z^{\prime}\rightarrow \mu\mu}}
\newcommand*{\Zpll}{\ensuremath{Z^{\prime}\rightarrow \ell\ell}}

\newcommand*{\ptZp}{\ensuremath{p_{\text{T}}^{\ensuremath{Z^{\prime}}}}}
\newcommand*{\Wp}{\ensuremath{W^{\prime}}}
\newcommand\T{\rule{0pt}{2.6ex}}
\newcommand\B{\rule[-1.2ex]{0pt}{0pt}}
\newcommand\lstar{\ensuremath{\ell^*}\xspace }
\newcommand\estar{\ensuremath{\Pe^*}\xspace }
\newcommand\mustar{\ensuremath{\Pgm^*}\xspace }
\newcommand\mlstar{\ensuremath{M_{\ell^*}}\xspace }
\providecommand\Mmin{\ensuremath{M_{\ell \gamma}^\text{min}}\xspace }
\providecommand\Mmax{\ensuremath{M_{\ell \gamma}^\text{max}}\xspace }
\newcommand{\bone}   {\tilde{b}^{}_{1}}
\newcommand{\tone}   {\tilde{t}^{}_{1}}
\newcommand{\neut}  {\tilde{\chi}^{0}_{1}}
\def\tg{\tilde g}
\def\tw{\widetilde\chi^{\pm}}
\def\tz{\widetilde\chi^0}
\def\tst{\tilde t}

\newcommand{\contributors}[1]{{\small \it \underline{Contributors}: #1\normalsize}}

\newcommand{\cms}[1]{#1}
\newcommand{\atlas}[1]{#1}
\newcommand{\lhcb}[1]{#1}
\newcommand{\theory}[1]{#1}
\newcommand{\cmsC}[1]{#1}
\newcommand{\atlasC}[1]{#1}
\newcommand{\lhcbC}[1]{#1}
\newcommand{\theoryC}[1]{#1}
\usepackage{marginnote}
\newcommand{\sherpa}[1]{\textsc{SHERPA#1}}
\newcommand{\pythia}[1]{\textsc{Pythia#1}}
\newcommand{\powhegbox}[1]{\textsc{Powheg-Box#1}}
\newcommand{\powheg}[1]{\textsc{Powheg#1}}
\newcommand{\herwig}[1]{\textsc{Herwig#1}}
\newcommand{\madgraph}[1]{\textsc{MadGraph#1}}
\newcommand{\madanalysis}[1]{\textsc{MadAnalysis#1}}
\newcommand{\geant}[1]{\textsc{Geant#1}}
\newcommand{\fastjet}[1]{\textsc{FastJet#1}}
\newcommand{\micromegas}[1]{\textsc{micrOMEGAs#1}}
\newcommand{\spheno}[1]{\textsc{SPheno#1}}
\newcommand{\delphes}[1]{\textsc{Delphes#1}}
\newcommand{\alpgen}[1]{\textsc{ALPGEN#1}}
\newcommand{\calchep}[1]{\textsc{CalcHep#1}}
\newcommand{\cp}{CP~}%$\mathcal{CP}$\ }
\newcommand{\RESUMMINO}{\textsc{Resummino}\xspace}

\newcommand{\sarah}{{\textsc{sarah}}\xspace}
\newcommand{\PYTHIA} {{\textsc{pythia}}\xspace}
\newcommand{\MADGRAPH} {\textsc{MadGraph}\xspace}
\newcommand{\feynrules} {\textsc{FeynRules}\xspace}
\newcommand{\MCATNLO} {\textsc{mc@nlo}\xspace}
\newcommand{\maddm} {\textsc{MadDM}\xspace}
\newcommand{\MGvATNLO}{\MADGRAPH{}5\_a\MCATNLO}
\newcommand{\GEANTfour} {{\textsc{Geant4}}\xspace}
\newcommand{\fig}[1]{Fig.~\ref{#1}}
\newcommand{\figs}[1]{Fig.s~\ref{#1}}
\newcommand{\citeref}[1]{Ref.~\cite{#1}}
\newcommand{\citerefs}[1]{Ref.s~\cite{#1}}
\newcommand{\tabb}[1]{Table~\ref{#1}}
\newcommand{\tabbs}[1]{Tables~\ref{#1}}
\newcommand{\eq}[1]{Eq.~\eqref{#1}}
\newcommand{\eqs}[1]{Eq.s~\eqref{#1}}
\newcommand{\secc}[1]{Section~\ref{#1}}
\newcommand{\seccs}[1]{Sections~\ref{#1}}
\newcommand{\com}{c.o.m.}
\newcommand{\cpeven}{CP-even~}
\newcommand{\cpodd}{CP-odd~}
\newcommand{\cl}{\ensuremath{\rm~C.L.}\xspace}
\newcommand{\vev}{\textsc{vev}~}
\newcommand{\vevs}{\textsc{vevs}~}
\newcommand{\iee}{{\it i.e.}}
\newcommand{\egg}{{\it e.g.}}
\newcommand{\etcc}{{\it etc.}}

\def\gsim{\mathrel{\lower2.5pt\vbox{\lineskip=0pt\baselineskip=0pt
           \hbox{$>$}\hbox{$\sim$}}}}
\def\lsim{\mathrel{\lower2.5pt\vbox{\lineskip=0pt\baselineskip=0pt
           \hbox{$<$}\hbox{$\sim$}}}}

\begin{document}
\newcommand{\main}{.}
\def\biblio{}
% Title --------------------------------------------------
\title{{\normalfont\bfseries\boldmath\huge
\begin{center}
%  HL/HE-LHC Physics Workshop Report\\
%  WG3 : Beyond the Standard Model Physics
Beyond the Standard Model Physics\\
at the HL-LHC and HE-LHC\\
\begin{normalsize} 
\href{http://lpcc.web.cern.ch/hlhe-lhc-physics-workshop}{Report from Working Group 3 on the Physics of the HL-LHC, and Perspectives at the HE-LHC} 
\end{normalsize}
\end{center}\vspace*{0.2cm}
}}

% Authors -------------------------------------------------
\author{Convenors: \\ 
\href{http://inspirehep.net/record/1070808}{X.~Cid Vidal}$^{1}$,\, 
\href{http://inspirehep.net/record/1028663}{M.~D'Onofrio}$^{2}$,\, 
\href{http://inspirehep.net/record/1009609}{P.~J.~Fox}$^{3}$,\, 
\href{http://inspirehep.net/record/1064514}{R.~Torre}$^{4, 5}$,\, 
\href{http://inspirehep.net/record/1028311}{K.~A.~Ulmer}$^{6}$,\, 
\\ \vspace*{4mm} 
 Contributors: 
 \\ 
\href{http://inspirehep.net/record/1272238}{A.~Aboubrahim}$^{7}$,\, 
\href{http://inspirehep.net/record/1467342}{A.~Albert}$^{8}$,\, 
\href{http://inspirehep.net/record/1068305}{J.~Alimena}$^{9}$,\, 
\href{http://inspirehep.net/record/1018633}{B.~C.~Allanach}$^{10}$,\, 
\href{http://inspirehep.net/record/1228960}{C.~Alpigiani}$^{11}$
% ...
\vspace*{0.5cm} 
 }
\institute{\small 
$^{1}$ \href{http://inspirehep.net/record/910711}{U. Santiago de Compostela, IGFAE}, $^{2}$ \href{http://inspirehep.net/record/902964}{U. Liverpool, Dept. Phys.}, $^{3}$ \href{http://inspirehep.net/record/902796}{Fermilab}, $^{4}$ \href{http://inspirehep.net/record/902725}{CERN, Geneva}, $^{5}$ \href{http://inspirehep.net/record/902881}{INFN, Genoa}, $^{6}$ \href{http://inspirehep.net/record/902748}{U. Colorado, Boulder, Dept. Phys.}, $^{7}$ \href{http://inspirehep.net/record/946076}{Northeastern U.}, $^{8}$ \href{http://inspirehep.net/record/910724}{RWTH Aachen}, $^{9}$ \href{http://inspirehep.net/record/1118764}{Ohio State U., Columbus}, $^{10}$ \href{http://inspirehep.net/record/907623}{U. Cambridge, DAMTP}, $^{11}$ \href{http://inspirehep.net/record/903338}{U. Washington, Seattle, Dept. Phys.}, $^{12}$ \href{http://inspirehep.net/record/902828}{LPSC, Grenoble}, $^{13}$ \href{http://inspirehep.net/record/902666}{DESY, Zeuthen}, $^{14}$ \href{http://inspirehep.net/record/908594}{U. Bern, LHEP}, $^{15}$ \href{http://inspirehep.net/record/902752}{Concordia U., Montreal, Dept. Phys.}, $^{16}$ \href{http://inspirehep.net/record/902974}{IPNL, Lyon}, $^{17}$ \href{http://inspirehep.net/record/902884}{INFN, Padua}, $^{18}$ \href{http://inspirehep.net/record/902770}{DESY, Hamburg}, $^{19}$ \href{http://inspirehep.net/record/1273509}{U. Oklahoma, Norman}, $^{20}$ \href{http://inspirehep.net/record/903370}{U. Zurich, Phys. Inst.}, $^{21}$ \href{http://inspirehep.net/record/904416}{SISSA, Trieste}, $^{22}$ \href{http://inspirehep.net/record/1189714}{U. Wisconsin, Madison}, $^{23}$ \href{http://inspirehep.net/record/902990}{U. Maryland, College Park, Dept. Phys.}, $^{24}$ \href{http://inspirehep.net/record/907907}{IFIC, Valencia}, $^{25}$ \href{http://inspirehep.net/record/1218068}{UC, Santa Cruz, SCIPP}, $^{26}$ \href{http://inspirehep.net/record/903523}{EPFL, Lausanne, LPHE}, $^{27}$ \href{http://inspirehep.net/record/1120892}{TIFR, Mumbai, DHEP}, $^{28}$ \href{http://inspirehep.net/record/911897}{RKMVU, West Bengal}, $^{29}$ \href{http://inspirehep.net/record/911366}{IPHC, Strasbourg}, $^{30}$ \href{http://inspirehep.net/record/907904}{U. Barcelona, IFAE}, $^{31}$ \href{http://inspirehep.net/record/903294}{U. Tubingen, Dept. Phys.}, $^{32}$ \href{http://inspirehep.net/record/903110}{U. Oslo, Dept. Phys.}, $^{33}$ \href{http://inspirehep.net/record/902886}{INFN, Pisa}, $^{34}$ \href{http://inspirehep.net/record/902804}{U. Florida, Gainesville, Dept. Phys.}, $^{35}$ \href{http://inspirehep.net/record/902989}{CPPM, Marseille}, $^{36}$ \href{http://inspirehep.net/record/912344}{U. Fed. Rio Grande do Norte, Intl. Inst. Phys.}, $^{37}$ \href{http://inspirehep.net/record/902882}{INFN, Milan}, $^{38}$ \href{http://inspirehep.net/record/903009}{U. Milan, Dept. Phys.}, $^{39}$ \href{http://inspirehep.net/record/903802}{U. Andes, Bogota, Dept. Phys.}, $^{40}$ \href{http://inspirehep.net/record/905096}{NICPB, Tallinn}, $^{41}$ \href{http://inspirehep.net/record/902689}{Brookhaven Natl. Lab., Dept. Phys.}, $^{42}$ \href{http://inspirehep.net/record/904144}{U. Perugia, Dept. Phys.}, $^{43}$ \href{http://inspirehep.net/record/904996}{INFN, Perugia}, $^{44}$ \href{http://inspirehep.net/record/903038}{LMU Munich, Dept. Phys.}, $^{45}$ \href{http://inspirehep.net/record/903239}{U. Sussex, Brighton, Dept. Phys. Astron.}, $^{46}$ \href{http://inspirehep.net/record/902868}{Imperial Coll., London, Dept. Phys.}, $^{47}$ \href{http://inspirehep.net/record/908399}{Durham U., IPPP}, $^{48}$ \href{http://inspirehep.net/record/902916}{KEK, Tsukuba}, $^{49}$ \href{http://inspirehep.net/record/902645}{Argonne Natl. Lab., HEP Div.}, $^{50}$ \href{http://inspirehep.net/record/903123}{CAS, IHEP, Beijing}, $^{51}$ \href{http://inspirehep.net/record/908583}{LPTHE, Paris}, $^{52}$ \href{http://inspirehep.net/record/903307}{UC, Santa Barbara, Dept. Phys.}, $^{53}$ \href{http://inspirehep.net/record/902807}{INFN, LNF, Frascati}, $^{54}$ \href{http://inspirehep.net/record/902887}{INFN, Rome 1}, $^{55}$ \href{http://inspirehep.net/record/903168}{U. Rome 1, La Sapienza, Dept. Phys.}, $^{56}$ \href{http://inspirehep.net/record/902889}{INFN, Turin}, $^{57}$ \href{http://inspirehep.net/record/922848}{U. Turin, Dept. Exp. Phys.}, $^{58}$ \href{http://inspirehep.net/record/903537}{U. Louisville, Dept. Phys.}, $^{59}$ \href{http://inspirehep.net/record/911834}{U. Lyon 1}, $^{60}$ \href{http://inspirehep.net/record/902880}{INFN, Florence}, $^{61}$ \href{http://inspirehep.net/record/903006}{Michigan State U., East Lansing, Dept. Phys. Astron.}, $^{62}$ \href{http://inspirehep.net/record/902801}{U. Florence, Dept. Phys. Astron.}, $^{63}$ \href{http://inspirehep.net/record/945357}{U. Rome 1, La Sapienza}, $^{64}$ \href{http://inspirehep.net/record/903357}{Yale U., Dept. Phys.}, $^{65}$ \href{http://inspirehep.net/record/902874}{Indiana U., Bloomington, Dept. Phys.}, $^{66}$ \href{http://inspirehep.net/record/1279835}{CCNU, IOPP, Wuhan}, $^{67}$ \href{http://inspirehep.net/record/902740}{LPC, Clermont-Ferrand}, $^{68}$ \href{http://inspirehep.net/record/902823}{U. Glasgow, Sch. Phys. Astron.}, $^{69}$ \href{http://inspirehep.net/record/902825}{Chalmers U. Technol., Gothenburg}, $^{70}$ \href{http://inspirehep.net/record/906950}{U. Munster, Inst. Theor. Phys.}, $^{71}$ \href{http://inspirehep.net/record/911469}{KIT, Karlsruhe}, $^{72}$ \href{http://inspirehep.net/record/1297569}{IBS, Daejeon}, $^{73}$ \href{http://inspirehep.net/record/902953}{LBNL, Berkeley, Div. Phys.}, $^{74}$ \href{http://inspirehep.net/record/903308}{UCSC, Santa Cruz, Dept. Phys.}, $^{75}$ \href{http://inspirehep.net/record/903599}{U. Oklahoma, Norman, Dept. Phys. Astron.}, $^{76}$ \href{http://inspirehep.net/record/903342}{Weizmann Inst. Sci., Rehovot, Fac. Phys.}, $^{77}$ \href{http://inspirehep.net/record/908245}{Inst. U. de France}, $^{78}$ \href{http://inspirehep.net/record/903287}{U. Trieste, Dept. Phys.}, $^{79}$ \href{http://inspirehep.net/record/945007}{U. Hawaii}, $^{80}$ \href{http://inspirehep.net/record/903828}{Hanyang U., Seoul, Dept. Phys.}, $^{81}$ \href{http://inspirehep.net/record/907960}{U. Milan Bicocca, Dept. Phys.}, $^{82}$ \href{http://inspirehep.net/record/909939}{INFN, Milan Bicocca}, $^{83}$ \href{http://inspirehep.net/record/903321}{Vanderbilt U., Dept. Phys. Astron.}, $^{84}$ \href{http://inspirehep.net/record/1625414}{IRFU, Saclay, DPP}, $^{85}$ \href{http://inspirehep.net/record/904557}{Korea U., Seoul}, $^{86}$ \href{http://inspirehep.net/record/903036}{MPI Phys., Munich}, $^{87}$ \href{http://inspirehep.net/record/902832}{U. Hamburg, Inst. Exp. Phys.}, $^{88}$ \href{http://inspirehep.net/record/1272953}{U. Pittsburgh}, $^{89}$ \href{http://inspirehep.net/record/902843}{U. Helsinki, Dept. Phys.}, $^{90}$ \href{http://inspirehep.net/record/902692}{Brown U., Dept. Phys.}, $^{91}$ \href{http://inspirehep.net/record/1241132}{Bethel Coll.}, $^{92}$ \href{http://inspirehep.net/record/907484}{Helsinki Inst. Phys.}, $^{93}$ \href{http://inspirehep.net/record/902671}{U. Birmingham, Sch. Phys. Astron.}, $^{94}$ \href{http://inspirehep.net/record/903130}{U. Pittsburgh, Dept. Phys. Astron.}, $^{95}$ \href{http://inspirehep.net/record/902883}{INFN, Naples}, $^{96}$ \href{http://inspirehep.net/record/904413}{Acad. Sin., Inst. Phys., Taipei}, $^{97}$ \href{http://inspirehep.net/record/903097}{U. Oregon, Eugene, Dept. Phys.}, $^{98}$ \href{http://inspirehep.net/record/903139}{Princeton U., Dept. Phys.}, $^{99}$ \href{http://inspirehep.net/record/903738}{Hungarian Acad. Sci., Debrecen, Inst. Nucl. Res.}, $^{100}$ \href{http://inspirehep.net/record/908200}{KLTE-ATOMKI}, $^{101}$ \href{http://inspirehep.net/record/902663}{Ben-Gurion U., Beer-Sheva, Dept. Phys.}, $^{102}$ \href{http://inspirehep.net/record/903094}{OKState, Stillwater, Dept. Phys.}, $^{103}$ \href{http://inspirehep.net/record/904756}{Kyungpook Natl. U., Daegu, Dept. Phys.}, $^{104}$ \href{http://inspirehep.net/record/919235}{Natl. U. Educ., Gwangju}, $^{105}$ \href{http://inspirehep.net/record/909080}{Cornell U., LEPP}, $^{106}$ \href{http://inspirehep.net/record/902658}{Indian Inst. Sci., Bangalore}, $^{107}$ \href{http://inspirehep.net/record/902624}{RWTH, Aachen, Phys. Inst.}, $^{108}$ \href{http://inspirehep.net/record/902992}{UMass, Amherst, Dept. Phys.}, $^{109}$ \href{http://inspirehep.net/record/1441143}{OEAW, Vienna}, $^{110}$ \href{http://inspirehep.net/record/903328}{U. Virginia, Charlottesville, Dept. Phys.}, $^{111}$ \href{http://inspirehep.net/record/902665}{U. Bergen, Dept. Phys. Technol.}, $^{112}$ \href{http://inspirehep.net/record/902835}{Harvard U.}, $^{113}$ \href{http://inspirehep.net/record/903282}{U. Toronto, Dept. Phys.}, $^{114}$ \href{http://inspirehep.net/record/902912}{U. Kansas, Lawrence, Dept. Phys. Astron.}, $^{115}$ \href{http://inspirehep.net/record/903603}{Peking U., Beijing, Sch. Phys.}, $^{116}$ \href{http://inspirehep.net/record/906082}{Nankai U., Tianjin}, $^{117}$ \href{http://inspirehep.net/record/903349}{U. Wisconsin, Madison,  Dept. Phys.}, $^{118}$ \href{http://inspirehep.net/record/903101}{LPT, Orsay}, $^{119}$ \href{http://inspirehep.net/record/907933}{Vrije U. Brussels, Dept. Phys. Astrophys.}, $^{120}$ \href{http://inspirehep.net/record/905888}{U. Laguna, Tenerife, Dept. Phys.}, $^{121}$ \href{http://inspirehep.net/record/910482}{IAC, La Laguna}, $^{122}$ \href{http://inspirehep.net/record/1255221}{ARC, CoEPP, Melbourne}, $^{123}$ \href{http://inspirehep.net/record/903002}{CINVESTAV, Mexico}, $^{124}$ \href{http://inspirehep.net/record/1280366}{U. Mainz, PRISMA}, $^{125}$ \href{http://inspirehep.net/record/903174}{RAL, Didcot}, $^{126}$ \href{http://inspirehep.net/record/903212}{U. Southampton, Phys. Astron.}, $^{127}$ \href{http://inspirehep.net/record/903883}{IACS, Kolkata, Dept. Theor. Phys.}, $^{128}$ \href{http://inspirehep.net/record/903299}{UC, Berkeley, Dept. Phys.}, $^{129}$ \href{http://inspirehep.net/record/1236475}{Seoul Natl. U.}, $^{130}$ \href{http://inspirehep.net/record/903336}{Waseda U., Tokyo, Dept. Phys.}, $^{131}$ \href{http://inspirehep.net/record/902712}{U. Cambridge, Cavendish Lab.}, $^{132}$ \href{http://inspirehep.net/record/911254}{U. Tokyo, KIPMU}, $^{133}$ \href{http://inspirehep.net/record/912351}{Sokendai, Tsukuba}, $^{134}$ \href{http://inspirehep.net/record/904336}{Pontificia U. Catol. Chile, Santiago, Dept. Phys.}, $^{135}$ \href{http://inspirehep.net/record/902814}{U. Genoa, Dept. Phys.}, $^{136}$ \href{http://inspirehep.net/record/906106}{U. Oviedo, Dept. Phys.}, $^{137}$ \href{http://inspirehep.net/record/903314}{Uppsala U., Dept. Phys. Astron.}, $^{138}$ \href{http://inspirehep.net/record/903113}{U. Padua, Dept. Phys.}, $^{139}$ \href{http://inspirehep.net/record/903100}{LAL, Orsay}, $^{140}$ \href{http://inspirehep.net/record/903369}{ETH, Zurich, Dept. Phys.}, $^{141}$ \href{http://inspirehep.net/record/903007}{U. Michigan, Ann Arbor, Dept. Phys.}, $^{142}$ \href{http://inspirehep.net/record/902885}{INFN, Pavia}, $^{143}$ \href{http://inspirehep.net/record/904307}{U. Fed. Rio Grande do Norte}, $^{144}$ \href{http://inspirehep.net/record/903138}{IAS, Princeton}, $^{145}$ \href{http://inspirehep.net/record/903259}{Tel-Aviv U., Dept. Part. Phys.}, $^{146}$ \href{http://inspirehep.net/record/903206}{SLAC}, $^{147}$ \href{http://inspirehep.net/record/910783}{Cathol. U. Louvain, CP3}, $^{148}$ \href{http://inspirehep.net/record/903906}{Antioquia U.}, $^{149}$ \href{http://inspirehep.net/record/1224033}{Karamanoglu Mehmetbey U., Karaman}, $^{150}$ \href{http://inspirehep.net/record/903275}{U. Tokyo, ICEPP}, $^{151}$ \href{http://inspirehep.net/record/908554}{Lund U., THEP}, $^{152}$ \href{http://inspirehep.net/record/902730}{Chicago U., EFI}, $^{153}$ \href{http://inspirehep.net/record/903829}{Yonsei U., Dept. Phys.}, $^{154}$ \href{http://inspirehep.net/record/902782}{Durham U., Dept. Phys.}, $^{155}$ \href{http://inspirehep.net/record/902803}{Florida State U., Tallahassee, Dept. Phys.}, $^{156}$ \href{http://inspirehep.net/record/906319}{U. Cyprus, Nicosia, Dept. Phys.}, $^{157}$ \href{http://inspirehep.net/record/902888}{INFN, Trieste}, $^{158}$ \href{http://inspirehep.net/record/903832}{Nikhef, Amsterdam}, $^{159}$ \href{http://inspirehep.net/record/903416}{Inst. Phys., Belgrade}, $^{160}$ \href{http://inspirehep.net/record/907455}{MIT, Cambridge}, $^{161}$ \href{http://inspirehep.net/record/903192}{Seikei U.}, $^{162}$ \href{http://inspirehep.net/record/913314}{U. Cambridge}, $^{163}$ \href{http://inspirehep.net/record/902982}{U. Mainz, Inst. Phys.}, $^{164}$ \href{http://inspirehep.net/record/903305}{UC, San Diego, Dept. Phys.}, $^{165}$ \href{http://inspirehep.net/record/902734}{U. Cincinnati, Dept. Phys.}, $^{166}$ \href{http://inspirehep.net/record/907966}{Karlsruhe U., TTP}.  }\normalsize

\begin{titlepage}

% Header ---------------------------------------------------
\vspace*{-1.8cm}

\noindent
\begin{tabular*}{\linewidth}{lc@{\extracolsep{\fill}}r@{\extracolsep{0pt}}}
\vspace*{-1.2cm}\mbox{\!\!\!\includegraphics[width=.14\textwidth]{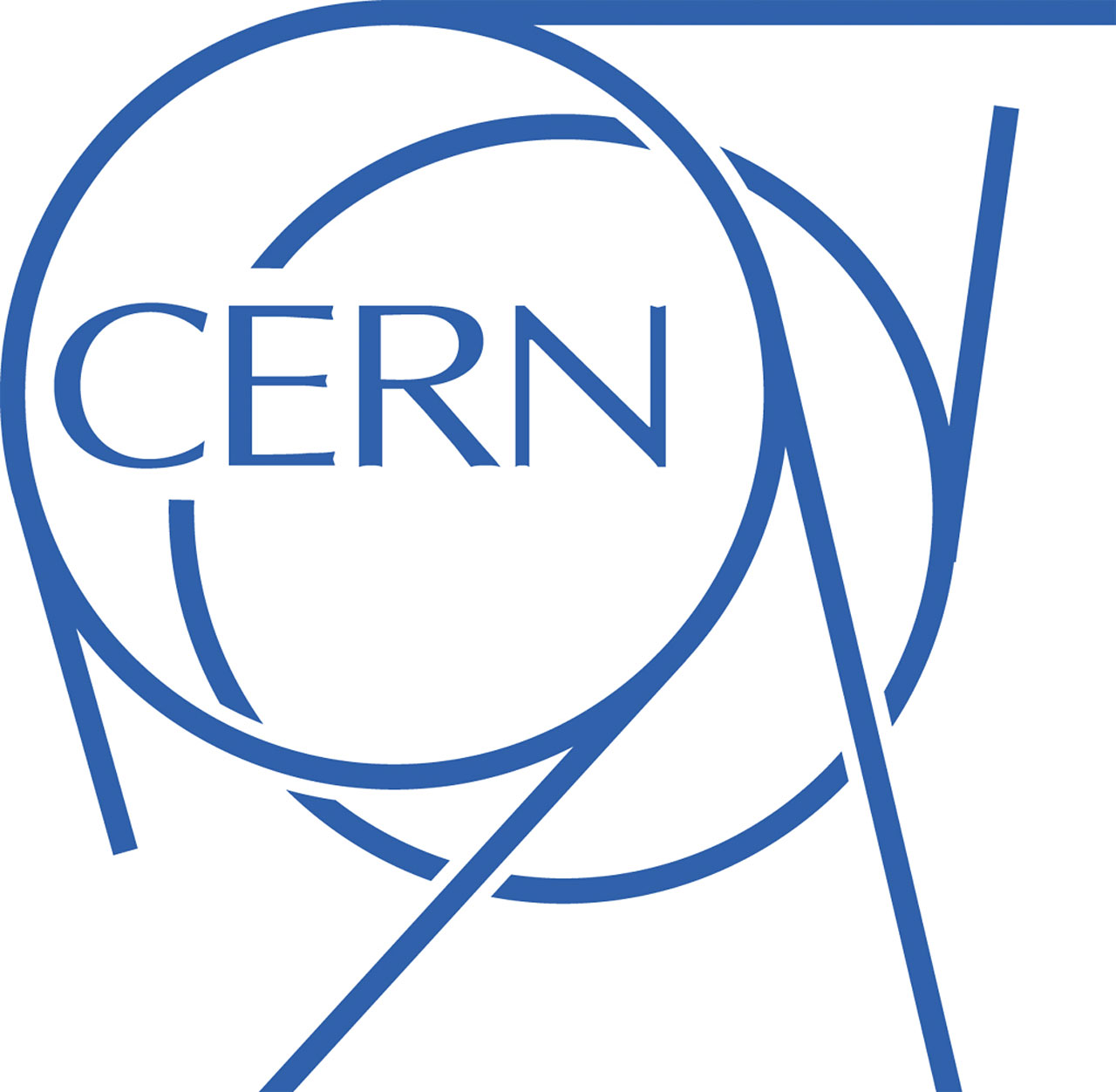}} & & \\
 & & CERN-LPCC-2018-05 \\  % ID 
 & & \today \\ % Date - Can also hardwire e.g.: 23 March 2010
 & & \\
\hline
\end{tabular*}

\vspace*{0.3cm}
%
%\vspace*{2.0cm}
%
\maketitle
\vspace{\fill}

\newpage
% Abstract ------
\begin{center}
    \begin{abstract}
  \noindent
  
\end{abstract}
\end{center}
This is the third out of five chapters of the final report \cite{YRHLHE} of the Workshop on {\it Physics at HL-LHC, and perspectives on HE-LHC} \cite{WorkshopHLHE}. It is devoted to the study of the potential, in the search for Beyond the Standard Model (BSM) physics, of the High Luminosity (HL) phase of the LHC, defined as $3\abinv$ of data taken at a centre-of-mass energy of $14\TeV$, and of a possible future upgrade, the High Energy (HE) LHC, defined as $15\abinv$ of data at a centre-of-mass energy of $27\TeV$. We consider a large variety of new physics models, both in a simplified model fashion and in a more model-dependent one. A long list of contributions from the theory and experimental (ATLAS, CMS, LHCb) communities have been collected and merged together to give a complete, wide, and consistent view of future prospects for BSM physics at the considered colliders. On top of the usual {\it standard candles}, such as supersymmetric simplified models and resonances, considered for the evaluation of future collider potentials, this report contains results on dark matter and dark sectors, long lived particles, leptoquarks, sterile neutrinos, axion-like particles, heavy scalars, vector-like quarks, and more. Particular attention is placed, especially in the study of the HL-LHC prospects, to the detector upgrades, the assessment of the future systematic uncertainties, and new experimental techniques. The general conclusion is that the HL-LHC, on top of allowing to extend the present LHC mass and coupling reach by $20-50\%$ on most new physics scenarios, will also be able to constrain, and potentially discover, new physics that is presently unconstrained. Moreover, compared to the HL-LHC, the reach in most observables will generally more than double at the HE-LHC, which may represent a good candidate future facility for a final test of TeV-scale new physics.

\vspace*{2.0cm}
\vspace{\fill}

\end{titlepage}
%\title{HL/HE-LHC Physics WG3 Report\\[5mm]
%  BSM}
%\input{authors_sorted}

%\maketitle
%\begin{enumerate}
%\item 
%\begin{itemize}
%\item 
%\begin{enumerate}
%\item 
%\begin{itemize}
%\item 
%\begin{enumerate}
%\item 
%\end{enumerate}
%\end{itemize}
%\end{enumerate}
%\end{itemize}
%\end{enumerate}

%\begin{abstract}
%This is the abstract
%\end{abstract}

% -- Set the level of the TOC: 2 means including subsection, 1 means
% including section 
\setcounter{tocdepth}{3}
{
  \hypersetup{linkcolor=black}
  \tableofcontents
}
\newpage

% -- List of sections
%\noindent {\bf Some global fixes still missing:}
%\begin{itemize}
%\item All numbers in dollars\rt{almost impossible to do}
%\end{itemize}

\resetcounters

\section{Introduction and overview}

The LHC physics program represents one of the most successful experimental programs in Science, and has been rewarded as such with the discovery, in 2012, of the Higgs boson \cite{Aad:2012tfa,Chatrchyan:2012ufa}. However, this discovery was only one of the targets of the LHC, which aims at constraining, and possibly discovering, an incredible variety of new physics (NP) scenarios with imprints at the TeV scale. In order to fully profit from the LHC potential, an upgrade of its luminosity \cite{HLLHC,ApollinariG.:2017ojx}, together with consistent upgrades of the major experiments \cite{ATLAS_PERF_Note,Collaboration:2650976}, has already been approved by the CERN Council \cite{HLLHCapproval}. The High Luminosity LHC  (HL-LHC) upgrade will eventually collect an integrated luminosity of $3\abinv$ of data in $pp$ collisions at a centre-of-mass (\com) energy of $14\TeV$, which should maximise the LHC potential to uncover new phenomena. 

The lack of indications for the presence of NP so far may imply that either NP is not where we expect it, or that it is elusive. The first case should not be seen as a negative result. Indeed the theoretical and phenomenological arguments suggesting NP close to the electroweak (EW) scale are so compelling, that a null result should be considered itself as a great discovery. This would shake our grounds, falsifying some of the paradigms that guided research in fundamental physics so far. In the second case, while these paradigms would be vindicated, Nature may have been clever in protecting its secrets.  It may be hiding the NP at slightly higher masses or lower couplings than we expected or, perhaps, in more compressed spectra and involved signatures, making it extremely difficult to address experimentally.  Both cases would lead to a discovery happening at the edge of the LHC potential, with little space left for identifying the new particles, or the new paradigms. 

These considerations drove, in the last few years, intense activity worldwide to assess the future of collider experiments beyond the HL-LHC. Several proposals and studies have been performed, also in the view of the forthcoming update of the European Strategy for Particle Physics (ESPP), that will take place in 2019-2020. Several options for future colliders have been and are being considered, such as future lepton colliders, either linear $e^{+}e^{-}$ machines like ILC \cite{Behnke:2013xla,Baer:2013cma,Adolphsen:2013jya,Adolphsen:2013kya,Behnke:2013lya} and CLIC \cite{Linssen:2012hp,Aicheler:2012bya}, or circular $e^{+}e^{-}$ ones like FCC-ee/TLEP \cite{FCCee} and CepC \cite{CEPC-SPPCStudyGroup:2015csa,CEPC-SPPCStudyGroup:2015esa} and $\mu^{+}\mu^{-}$ accelerators like MAP \cite{MAP} and LEMMA \cite{Antonelli:2015nla,Boscolo:2018tlu}, or hadron $pp$ colliders such as a $27\TeV$ \com~upgraded HE-LHC \cite{FCCCDRHELHC}, a $50-100\TeV$ SppC \cite{CEPC-SPPCStudyGroup:2015csa,CEPC-SPPCStudyGroup:2015esa}, and a $100\TeV$ FCC-hh \cite{FCCCDR,Mangano:2017tke,Mangano:2016jyj,Contino:2016spe,Golling:2016gvc}. Comparing the physics potentials, the needed technology and prospects for its availability, and the cost to benefit ratio of such machines is extremely challenging, but also very timely. The proposal for an $e^-p$ collider, the LHeC~\cite{LHeC}, is also being considered to further upgrade the HL-LHC with a $60\GeV$ energy, high current electron beam by using novel Energy Recovery Linear Accelerator (ERL) techniques. 
%Simultaneous operations of the LHeC and HL-LHC would improve the potential to explore the TeV energy scale and to measure Higgs properties. 
The same facility could be hosted at the FCC~\cite{FCCCDR}.
 
A crucial ingredient to allow a comparison of proposed future machines is the assessment of our understanding of physics at the end of the HL-LHC program. Knowing which scenarios remain open at the end of the approved HL-LHC allows one to set standard benchmarks for all the interesting phenomena to study, that could be used to infer the potential of different future machines. Moreover, in the perspective of pushing the LHC program even further, one may wonder if the LHC tunnel and the whole CERN infrastructure, together with future magnet technology, could be exploited to push the energy up into an unexplored region with the HE-LHC, that could collect an integrated luminosity of $15\abinv$.

These two points are the foundations of the Workshop on {\it Physics at HL-LHC, and perspectives on HE-LHC} \cite{WorkshopHLHE}, that has been devoted, between 2017 and 2018, to the study of the physics potential of the HL- and HE-LHC. This document is the third out of five chapters of the final report \cite{YRHLHE} of the Workshop. In this chapter, the attention is focused on beyond the SM (BSM) phenomena, one of the key reasons to continue to pursue an hadron collider physics program.

Naturalness, also often referred to as the Hierarchy Problem (HP), is the main motivation to expect new physics close to the EW scale. 
This theoretical puzzle can be understood in different ways: from a more technical perspective, it refers to radiative corrections to the Higgs mass parameter, which can receive contributions from new physics present up at ultraviolet (UV) scales. We have at least one important example: the scale at which gravity becomes strongly coupled, usually identified with the Planck scale, $\Mpl$.  From a more conceptual point of view it can be phrased as the question why is the Fermi constant $G_{F}\approx 1.2\cdot 10^{-5}\GeV^{-2}$ (EW scale $v=246\GeV$) so much bigger (smaller) than the Newton constant $G_{N}\approx 6.7\cdot 10^{-39}\GeV^{-2}$ (Planck scale $\Mpl=1.22\cdot 10^{19}\GeV$).

Despite the different ways of phrasing and understanding the HP, its importance is intimately related to our reductionist approach to physics and our understanding of effective field theory. We do not expect the infrared (IR) physics, \iee, for instance, at the energies that we are able to probe at colliders, to be strongly affected by the details of the UV theory. Therefore, unless Naturalness is only an apparent problem and has an anthropic explanation, or it is just the outcome of the dynamical evolution of our universe, all of its solutions are based on mechanisms that screen the effects of UV physics from the IR, by effectively reducing the UV cut-off to the TeV scale. 

Such mechanisms can be dynamical, similarly to what occurs for the QCD scale, or can instead arise from extended space-time symmetries, such as in Supersymmetry (SUSY) or in Extra Dimensions (ED). All of these solutions share the prediction of new degrees of freedom close to the EW scale. How close is determined by where we are willing to push the UV scale, still accepting IR parameters to strongly dependent on it. In other words, it depends on the level of cancellation between different UV parameters that we are willing to accept to reproduce the observed IR parameters. Nature gives few examples of such large cancellations, which could be a few percent accidents, but are never far below the percent level. The LHC is a machine designed to test such cancellations at the percent level in most of the common solutions to the HP. There are some exceptions, as for instance in models where the so-called top partners are neutral under the SM colour group, where the LHC can only probe the \sloppy\mbox{few-to-$10\%$} region. Obviously, tests of our understanding of Naturalness pass through three main approaches, addressed in the first four chapters of this report. The first is the precise test of the SM observables, both in the EW and QCD sectors, discussed in the first chapter \cite{Nason:2650160}, and in the flavour sector, discussed in the fourth chapter \cite{MartinCamalich:2650175}; the second is the study of the properties of the Higgs boson, presented in the second chapter \cite{Gori:2650162}; the third is the direct search for new physics, which is the topic of this chapter.

Since the top quark is the particle that contributes the most to the radiative correction to the Higgs mass, the main prediction of the majority of models addressing the HP is the existence of coloured particles ``related'' to the top quark, that can generally be called ``top-partners''. These may be scalars, like the top squarks (stops) in SUSY, or (vector-like) fermions, like in models of Higgs compositeness. These particles have to be light for Naturalness to be properly addressed and, due to their strong production cross section, they are among  the primary signatures of Naturalness at hadron colliders. To address the HP other particles have to be light too, such as for instance the gluinos in SUSY, that in turn affect the stops masses, and the EW partners of the Higgs boson. However, while the gluino profits from a strong production cross section at hadron colliders, the EW sector remains much more difficult to test, due to the smaller cross sections. All these signatures, together with others, less tightly related to Naturalness, are studied in details in this report.

Dark Matter (DM) is one of the big puzzles of fundamental physics. While there is stunning evidence for its existence, in the form of non-baryonic contribution to the matter abundance in the Universe, there are no particular indications on what it actually is. This is due to the fact that, so far, we have only probed it through its gravitational interactions, which tell us about its abundance, but do not tell us anything about its form. It could be made of particles, but this is not the only option. However, if DM is made of weakly interacting massive particles (WIMPs), then the observed abundance can only be reproduced for a relatively small window in its mass/coupling parameter space, which turns out to lie roughly in the ten GeV to ten TeV range, making it relevant for collider experiments. 

Several theoretical constructions addressing the HP also naturally predict a WIMP DM candidate. The most notable is SUSY, where EW neutral fermionic partners of the Higgs and the SM gauge bosons, the neutralinos, could be, in proper regions of the parameter space, good WIMPs. Another compelling paradigm for DM that may be relevant for collider experiments is that of the so-called Minimal Dark Matter (MDM), that corresponds to neutral particles belonging to EW multiplets that remains stable due to accidental symmetries. The simplest examples are just the wino and higgsino DM candidates arising in SUSY, but larger multiplets are also allowed. In this case the DM mass required to provide the observed abundance grows with the dimension of the EW group representation (multiplet) and usually lies between one to ten TeV. Therefore, a coverage of the whole MDM window provides a good benchmark for future hadron colliders, such as the HE-LHC or the FCC-hh (see \citeref{Golling:2016gvc} for prospect studies of MDM at a $100$ TeV collider). 

Finally, the third big mystery of the SM is flavour. Why are there such big hierarchies among fermion masses, and how do neutrino masses arise? These are two of the most compelling questions of fundamental physics. The generation of the flavour structure of the SM (the Yukawa couplings) and of the neutrino masses may be tied to a scale much above the EW scale.  Thus, precision flavour observables are the most sensitive window to high-scale UV physics. Indeed, the ability of LHC experiments, with a leading role of LHCb in this context, to observe extremely rare flavour transitions, allows one to set constraints on new physics corresponding to scales of hundreds, or even thousands of TeV, completely inaccessible to direct searches. 

Flavour transitions indirectly constraining NP at the TeV scale and above, have a crucial interplay with direct searches for the particles that may induce such transitions. A clear example of this interplay is given by the recent flavour anomalies in neutral and charged current $B$ decays ($R_{K}$-$R_{K*}$, $R_{D}$-$R_{D*}$, \etcc, which are discussed at length in the fourth chapter of this report \cite{MartinCamalich:2650175}. Due to the relevance of such anomalies at the time of writing this report, prospect studies on high transverse momentum particles, as vector resonances or lepto-quarks (LQ), that could explain them, are presented by both working groups.
%both from the BSM and from the flavour points of view, we decided, together with the flavour working group, to replicate the material on projections for high transverse momentum particles that could explain such anomalies, such as vector resonances or leptoquarks. While these anomalies could still go away, this represents a good proxy study addressing the question of what are, in the case of some evidence, the prospects for discovery at the HL- and HE-LHC, and for probing the relevant parameter space.

Concerning neutrino masses, the seesaw mechanism predicts the existence of heavy (sterile) neutrinos that can provide, in particular regions of the parameter space, peculiar signatures with several leptons in the final state. These neutral particles, coupled to leptons, can also arise in cascade decays of heavy right-handed charged gauge bosons.  Whether produced directly, or in decays, the HL- and HE-LHC will be able to significantly reduce the parameter space of models predicting heavy neutrinos.
%allowing\mdo{maybe re-write? suggestion commented} to shrink the allowed parameter space of models of neutrino masses that can be tested at colliders.\pf{do we need this paragraph?}
%MDO: These neutral particles, coupled to leptons, can also arise in cascade decays of heavy right-handed charged gauge bosons. Studies at the HL-LHC will allow to significantly reduce the parameter space of models predicting heavy neutrinos. 

The report is not structured based on a separation of the HL-LHC from the HE-LHC studies, since several analyses were done for both options, and showing them together allows for a clearer understanding. However, when summarising our results in \secc{sec:conclusions}, we present conclusions separately for HL- and HE-LHC. The report is organised as follows. The introductory part includes a brief discussion of the future detector performances in analysis methods and objects identification and of the projected systematic uncertainties. Section \ref{sec:SUSY} is devoted to the study of SUSY prospects. Section \ref{sec:DM} shows projections for DM and Dark Sectors. Section \ref{sec:LLP} contains studies relevant for Long Lived Particles (LLPs). Section \ref{sec:flavor} presents prospects for high-$p_{T}$ signatures in the context of flavour physics.
Section \ref{sec:otherBSM} is devoted to resonances, either singly or doubly produced, and to other BSM signatures. Finally, in \secc{sec:conclusions} we present our conclusions, with a separate executive summary of the HL- and HE-LHC potentials. 

%\subfile{\main/section1IntroductionAndOverview/sub11}
\subsection{Analysis methods and objects definitions}

Different approaches have been used by the experiments and in theoretical prospect studies, hereafter named projections, to assess the sensitivity in searching for new physics at the HL-LHC and HE-LHC.
For some of the projections, a mix of the approaches described below is used, in order to deliver the most realistic result.
The total integrated luminosity for the HL-LHC dataset is assumed to be $3\abinv$ at a \com~energy of $14\TeV$. For HE-LHC studies the dataset is assumed to be $15\abinv$ at a \com~of $27\TeV$.
The effect of systematic uncertainties is taken into account based on the studies performed for the existing analyses and using common guidelines for projecting the expected improvements that are foreseen thanks to the large dataset and upgraded detectors, as described in Section~\ref{sec:methods:syst}.

{\bf Detailed-simulations} are used to assess the performance of reconstructed objects in the upgraded detectors and HL-LHC conditions, as described in Sections~\ref{sec:methods:perf} and~\ref{sec:methods:perf_LHCb}.
For some of the projections, such simulations are directly interfaced to different event generators, parton showering (PS) and hadronisation generators. Monte Carlo (MC) generated events are used for SM and BSM processes, and are employed in the various projections to estimate the expected contributions of each process.

{\bf Extrapolations} of existing results rely on the existent statistical frameworks to estimate the expected sensitivity for the HL-LHC dataset.
The increased \com~energy and the performance of the upgraded detectors are taken into account for most of the extrapolations using scale factors on the individual processes contributing to the signal regions. Such scale factors are derived from the expected cross sections and from detailed simulation studies.

{\bf Fast-simulations} are employed for some of the projections in order to produce a large number of Monte Carlo events and estimate the reconstruction efficiency for the upgraded detectors. The upgraded CMS detector performance is taken into account encoding the expected performance of the upgraded detector in \delphes{3}~\cite{deFavereau:2013fsa}, including the effects of pile-up interactions. Theoretical contributions use \delphes with the commonly accepted HL-LHC card corresponding to the upgraded ATLAS and CMS detectors.

{\bf Parametric-simulations} are used for some of the projections to allow a full re-optimisation of the analysis selection criteria that benefit from the larger available datasets.
Particle-level definitions are used for electrons, photons, muons, taus, jets and missing transverse momentum. These are constructed from stable particles from the MC event record with a lifetime larger than $0.3 \times 10^{-10}$~s within the observable pseudorapidity range. Jets are reconstructed using the anti-$k_t$ algorithm~\cite{Cacciari:2008gp} implemented in the Fastjet~\cite{Cacciari:2011ma} library, with a radius parameter of 0.4. All stable final-state particles are used to reconstruct the jets, except the neutrinos, leptons and photons associated to $W$ or $Z$ boson or $\tau$ lepton decays. The effects of an upgraded ATLAS detector are taken into account by applying energy smearing, efficiencies and fake rates to generator level quantities, following parameterisations based on detector performance studies with the detailed simulations. The effect of the high pileup at the HL-LHC is incorporated by overlaying pileup jets onto the hard-scatter events. Jets from pileup are randomly selected as jets to be considered for analysis with $\sim 2\%$ efficiency, based on studies of pile-up jet rejection and experience from Run-2 of the LHC.

\subsubsection{ATLAS and CMS performance}
\label{sec:methods:perf}

The expected performance of the upgraded ATLAS and CMS detectors has been studied in detail in the context of the Technical Design Reports
and subsequent studies; the assumptions used for this report and a more detailed description are available in~\citerefs{ATLAS_PERF_Note,Collaboration:2650976}. For CMS, the object performance in the central region assumes a barrel calorimeter ageing conditions corresponding to an integrated luminosity of $1\abinv$.

The triggering system for both experiments will be replaced and its impact on the triggering abilities of each experiment assessed;
new capabilities will be added, and, despite the more challenging conditions, most of the trigger thresholds for common objects are expected
to either remain similar to the current ones or even to decrease~\cite{ATLAS_TDAQ_TDR,CMSL1interim}.

The inner detector is expected to be completely replaced by both experiments, notably extending its coverage to $|\eta|<4.0$.
The performance for reconstructing charged particles has been studied in detail in~\citerefs{ATLAS_Pixel_TDR,ATLAS_Strip_TDR,CMS_Tracker_TDR}.
Electrons and photons are reconstructed from energy deposits in the electromagnetic calorimeter and information from the inner tracker\cite{ATLAS_LAr_TDR,CMS_Barrel_TDR,CMS_HGCAL_TDR,CMS_MTD_TP}.
Several identification working points have been studied and are employed by the projection studies as most appropriate.
Muons are reconstructed combining muon spectrometer and inner tracker information~\cite{ATLAS_Muon_TDR,CMS_Muon_TDR}.

Jets are reconstructed by clustering energy deposits in the electromagnetic and hadronic calorimeters\cite{ATLAS_Tile_TDR,ATLAS_LAr_TDR,CMS_Barrel_TDR} using the anti-$k_{T}$ algorithm\cite{Cacciari:2008gp}.
B-jets are identified via $b$-tagging algorithms. B-tagging is performed if the jet is within the tracker acceptance ($|\eta|<4.0$).
Multivariate techniques are employed in order to identify $b-$jets and $c-$jets, and were fully re-optimised for the upgraded detectors~\cite{ATLAS_Pixel_TDR,CMS_Tracker_TDR}.
A working point with $70\%$ efficiency for $b-$jet identification is used, unless otherwise noted.
High $p_T$ boosted jets are reconstructed using large-radius anti-$k_{T}$ jets with a distance parameter of $0.8$. Various jet substructure variables are employed to identify boosted $W/Z/H$ boson and top quark jets with good discrimination against generic QCD jets. 

Missing transverse momentum (its modulus referred to as $\met$) is reconstructed following similar algorithms as employed in the Run-2 data taking. Its performance has been evaluated for standard processes, such as top-quark pair production~\cite{ATLAS_Pixel_TDR,CMSCollaboration:2015zni}.

The addition of new precise-timing detectors and its effect on object reconstruction has also been studied in \citerefs{ATLAS_TP_HGTD,CMS_MTD_TP}, although its results are only taken into account in a small subset of the projections in this report.

\subsubsection{LHCb performance}
\label{sec:methods:perf_LHCb}
The LHCb upgrades are shifted with respect to those of ATLAS and CMS. A first upgrade will happen at the end of Run-2 of the LHC, to run at a luminosity five times larger  ($2\times 10^{33}\text{cm}^{-2}\text{s}^{-1}$) in LHC Run-3 compared to those in Run-1 and Run-2, while maintaining or improving the current detector performance. This first upgrade (named \mbox{Upgrade~I}) will be followed by by the so-called \mbox{Upgrade~II} (planned at the end of Run-4) to run at a luminosity of $\sim 2\times 10^{34}\text{cm}^{-2}\text{s}^{-1}$.

The LHCb MC simulation used in this document mainly relies on the \pythia{ 8} generator~\cite{Sjostrand:2007gs} with a specific LHCb configuration~\cite{Belyaev:2011zza}, using the CTEQ6 leading-order set of parton density functions~\cite{Pumplin:2002vw}. The interaction of the generated particles with the detector, and its response, are implemented using the \geant{} toolkit~\cite{Allison:2006ve,Agostinelli:2002hh} as described in \citeref{Clemencic:2011zza}. 

The reconstruction of jets is done using a particle flow algorithm, with the output of this clustered using
the anti-kT algorithm as implemented in \fastjet{}, with a distance parameter of
$0.5$. Requirements are placed on the candidate jet in order to reduce the background
formed by particles which are either incorrectly reconstructed or produced in additional pp interactions in the same event.
Different assumptions are made regarding the increased pile-up, though in general the effect is assumed to be similar to that in Run-2.
\subsection{Treatment of systematic uncertainties}
\label{sec:methods:syst}
It is a significant challenge to predict the expected systematic uncertainties of physics results at the end of HL-LHC running.
It is reasonable to anticipate improvements to techniques of determining systematic uncertainties over an additional decade of data-taking.
To estimate the expected performance, experts in the various physics objects and detector systems from ATLAS and CMS have looked at current limitations to systematic uncertainties in detail to determine which contributions are limited by statistics and where there are more fundamental limitations.
Predictions were made taking into account the increased integrated luminosity and expected potential gains in technique.
These recommendations were then harmonised between the experiments to take advantage of a wider array of expert opinions and to allow the experiments to make sensitivity predictions on equal footing~\cite{ATLAS_PERF_Note,Collaboration:2650976}. For theorists' contributions, a simplified approach is often adopted, loosely inspired by the improvements predicted by experiments. 

General guide-lining principles were defined in assessing the expected systematic uncertainties.
Theoretical uncertainties are assumed to be reduced by a factor of two with respect to the current knowledge, thanks to both
higher-order calculation as well as reduced PDF uncertainties~\cite{Khalek:2018mdn}.
All the uncertainties related to the limited number of simulated events are neglected, under the assumption that sufficiently large simulation samples will be available by the time the HL-LHC becomes operational. For all scenarios, the intrinsic statistical uncertainty in the measurement is reduced by a factor $1/\sqrt{\mathcal{L}}$, where $\mathcal{L}$ is the projection integrated luminosity divided by that of the reference Run-2 analysis.
Systematics driven by intrinsic detector limitations are left unchanged, or revised according to detailed simulation studies of the upgraded detector.
Uncertainties on methods are kept at the same value as in the latest public results available, assuming that the harsher HL-LHC conditions will be compensated by improvements to the experimental methods.

The uncertainty in the integrated luminosity of the data sample is expected to be reduced down to $1\%$ by better understanding of the calibration methods, improved stability in applying those methods, and making use of the new capabilities of the upgraded detectors~\cite{Nason:2650160}.

In addition to the above scenario (often referred to as ``YR18 systematics uncertainties'' scenario), results are often
compared to the case where the current level of understanding of systematic uncertainties is assumed (``Run-2 systematic uncertainties'') or to the case of statistical-only uncertainties.

\newpage
\resetcounters

\section{Supersymmetry}\label{sec:SUSY}

One of the main goal of collider physics is to uncover the nature of EW symmetry breaking (EWSB). Supersymmetry can resolve the hierarchy problem, as well as provide gauge coupling unification and a dark matter candidate. 
SUSY might be realised in nature in various ways and superpartners of the SM particles could be produced at colliders leading to many different possible detector signatures. 
Coloured superpartners such as squarks ($\tilde{q}$) and gluinos ($\tilde{g}$) are strongly produced and have the highest cross sections. 
Scalar partners of the left-handed and right-handed chiral components of the bottom 
quark ($\tilde{b}_{\mathrm{L},\mathrm{R}}$) or top quark ($\tilde{t}_{\mathrm{L},\mathrm{R}}$) mix to form 
mass eigenstates for which the bottom squark ($\bone$) and top squark ($\tone$) are defined as the lighter of the two. The lightest bottom and top squark  mass eigenstates might be significantly lighter than the other squarks and the gluinos. As a consequence, $\bone$ and $\tone$ could be pair-produced with relatively large cross-sections at the HL- and HE-LHC. 
In the EW sector, SUSY partners of the Higgs, photon, $Z$, and $W$ bosons are the spin-$1/2$ higgsinos, photino, zino, and winos that further mix in neutralino ($\tilde{\chi}_{1,2,3,4}^{0}$) and chargino ($\tilde{\chi}_{1,2}^{\pm}$) states, also called the electroweakinos. Their production rate is a few order of magnitudes lower than that of coloured superpartners. Superpartners of charged leptons, the sleptons ($\tilde{\ell}$), can also have sizeable production rates and are searched for at hadron colliders.     
Provided that R-parity conservation is assumed, SUSY particles typically decay to final states involving SM particles in addition to significant momentum imbalance due to a collider-stable lightest supersymmetric particle (LSP). 

\begin{figure}[t]
\begin{center}
\centering
\begin{minipage}[t][\minipageheightdp][t]{\minipagewidthdp}
\centering
\includegraphics[width=\figuremaxwidthdp-0.3cm,height=\figuremaxheightdp,keepaspectratio]{\main/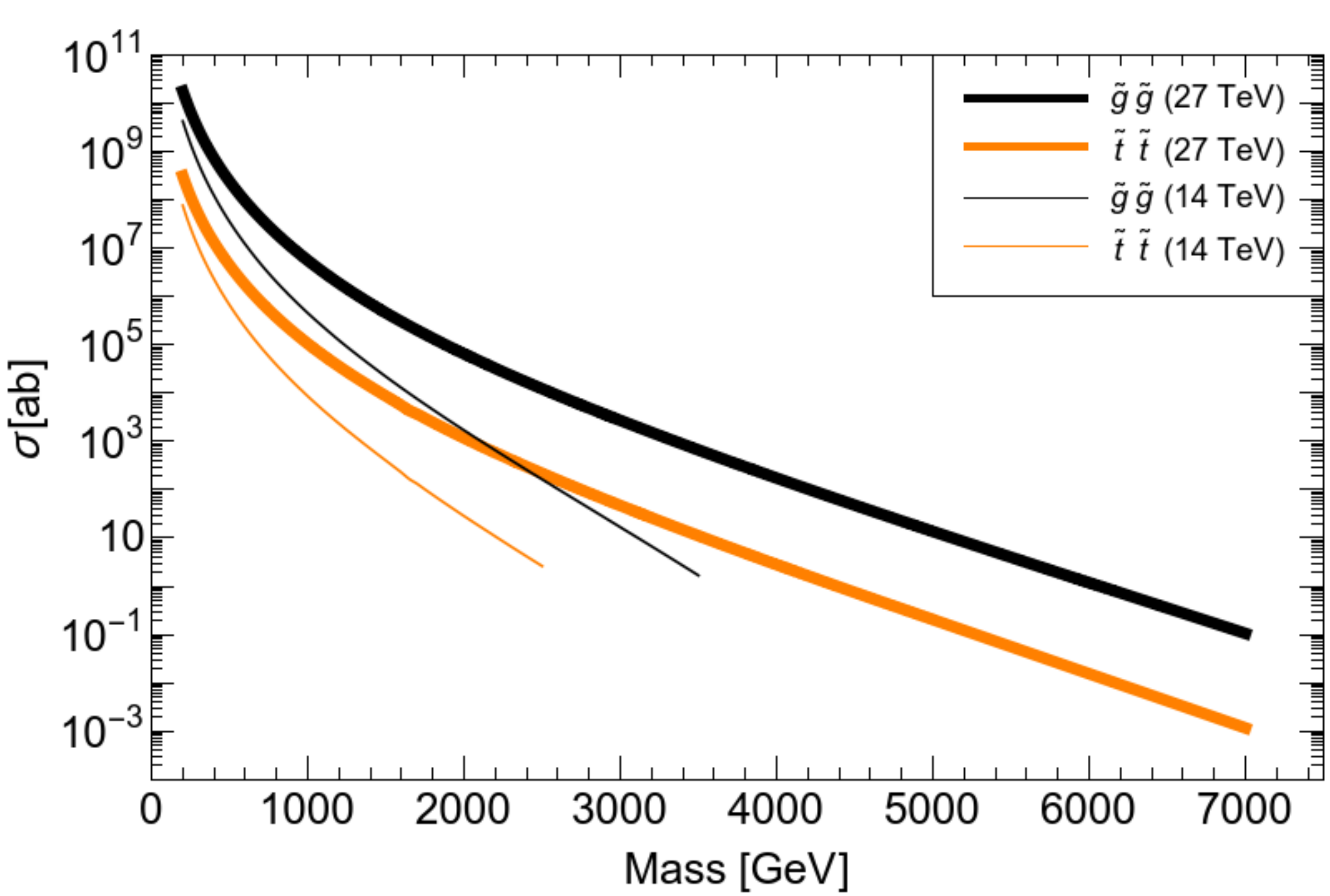}
\end{minipage}
\begin{minipage}[t][5.3cm][t]{\minipagewidthdp}
\centering
\includegraphics[width=\figuremaxwidthdp,height=\figuremaxheightdp,keepaspectratio]{\main/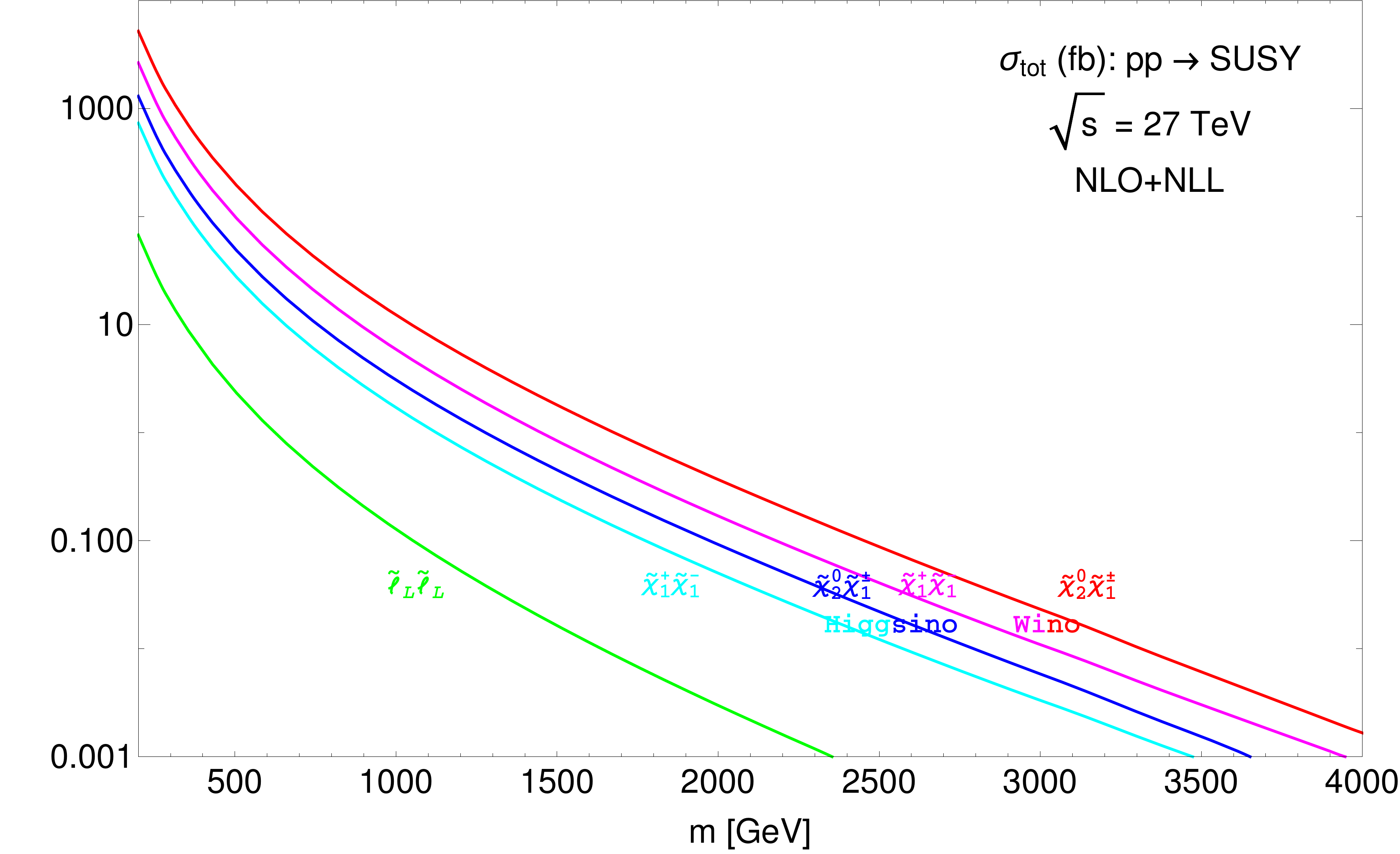}
\end{minipage}
\caption{Left: NLL+NLO predictions~\cite{Borschensky:2014cia} 
of $\sigma (pp\to \tg\tg X)$ and $\sigma (pp\to\tst_1\tst_1^* X)$ production processes at the LHC for 
$\sqrt{s}=14$ and $27\TeV$ \com~energy (\emph{Contribution from C. Borschensky, M. Kramer, A. Kulesza}). 
\sloppy\mbox{Right: NLO} predictions~\cite{Fuks:2012qx,Fuks:2013lya,Fuks:2013vua} for electroweakinos and sleptons pair production for $27\TeV$ \com~energy (\emph{Contribution from J. Fiaschi, M. Klasen, M. Sunder}). 
\label{fig:sigma}}
\end{center}
\end{figure}

%The extension of the kinematic reach for supersymmetry searches at the HL-LHC is reflected foremost in the sensitivity to EW states, including sleptons, as well as for gluinos and squarks.  
Searches of SUSY particles are presented in the following targeting HL- and HE-LHC, under various theoretical hypotheses such as the Minimal Supersymmetric Standard Model (MSSM)~\cite{Han:2018xxx}, phenomenological MSSM, light higgsinos models and more. R-parity conservation and prompt particle decays are generally assumed, whilst dedicated searches for long-lived particles are depicted in \secc{sec:LLP}. Simplified models are also used to optimise the searches and interpret the results. 
The cross-sections used to evaluate the signal yields at $14\TeV$ are calculated to next-to-leading order in the strong coupling constant, adding the resummation of soft gluon emission at next-to-leading-logarithmic accuracy (NLO+NLL), see \citeref{Borschensky:2014cia} for squarks and gluinos, and \citerefs{Fuks:2012qx,Fuks:2013vua} for electroweakinos. 
The nominal cross sections and the uncertainties are taken from an envelope of cross section predictions using different PDF sets and factorisation and renormalisation scales.  
%More details can be found in the SUSYXCWG page \url{https://twiki.cern.ch/twiki/bin/view/LHCPhysics/SUSYCrossSections}. 
PDF uncertainties are dominant for strongly-produced particles. In particular, PDF uncertainties on gluino pair production ranges between $30\%$ and $60\%$ depending on the gluino mass, in a mass range between $1$ and $4$ TeV. Expected improvements due to precision SM measurements in the jet and top sector are expected at HL-LHC. In this report, nominal predictions are considered and, unless stated otherwise, no attempt to evaluate the impact of theoretical uncertainties on the reach of the searches is made. 
For $27\TeV$ \com~energy, cross sections are also evaluated at NLO+NLL as shown in \fig{fig:sigma} for gluinos and top squarks pair production and for electroweakinos and sleptons pair production. For the latter, the NLO set from PDF4LHC is used and cross sections are presented for wino and higgsino hypotheses. %Calculations for electroweakinos and slepton pair production processes are not yet publicly available.    

Prospects for exclusion and discovery of gluinos and top squarks are reported in \secc{sec:susy221}. We show that HL-LHC will probe gluino masses up to $3.2\TeV$, about $0.8-1\TeV$ above the Run-2 $\tilde{g}$ mass reach for $80\fbinv$. Top squarks can be discovered (excluded) up to masses of $1.25$ ($1.7$)\TeV. This extends by about 700~GeV the reach of Run-2 for $80\fbinv$. 
Charginos and neutralinos studies are presented in \secc{sec:susy23}, considering electroweakino decays via $W$, $Z$ (also off-shell) and Higgs boson and various hypotheses for sparticles mass hierarchy. 
As an example, masses up to $850$ ($680$)\GeV can be excluded (discovered) for charginos decaying as $\tilde{\chi}_{1}^{\pm} \rightarrow W^{(*)} \neut$: the results extend by about $500\GeV$ the mass reach obtained with $80\fbinv$ of $13\TeV$ $pp$ collisions, and extend beyond the LEP limit by almost an order of magnitude.  HL-LHC searches for low momentum leptons will be sensitive to $\tilde{\chi}^\pm$ masses up to $350\GeV$ for $\Delta m(\tilde{\chi}_{1}^{\pm}, \neut) \approx 5\GeV$, and to mass splittings between $2$ and $50\GeV$, thus bringing significant new reach to Higgsino models. In~\secc{sec:sleptons}, dedicated searches for sleptons are presented, and in particular for the \emph{currently unconstrained} pair production of staus exploiting hadronically decaying tau leptons.    
Finally, identification of benchmark models and probing of various natural scenarios at HL- and HE-LHC are presented in~\secc{sec:gluinos}. 
For gluinos and stops HE-LHC will further increase the reach, above that of HL-LHC, by about a factor of two, and several benchmark MSSM and pMSSM models will be discoverable. 
%width=0.331\textheight

\subsection{Searches for gluinos and third generation squarks}
\label{sec:susy221}
%\subsubsection{\atlas{SUSY strong - improved searches for squark/gluinos}}
%{\bf Author(s): ATLAS}
%\emph{Status Authors are presenting internally in ATLAS this week. Status to be checked, and update expected next week.} 

%Scalar partners of the left-handed and right-handed chiral components of the bottom  quark ($\tilde{b}_{\mathrm{L},\mathrm{R}}$) or top quark ($\tilde{t}_{\mathrm{L},\mathrm{R}}$) mix to form mass eigenstates for which the bottom squark ($\bone$) and top squark ($\tone$) are defined as the lighter of the two. 
Naturalness considerations suggest that the supersymmetric partners of the third-generation SM quarks are the lightest coloured supersymmetric particles and gluinos are also within a range of few TeV. 
%This results in the lightest bottom and top squark  mass eigenstates ($\bone$ and $\tone$) being significantly lighter than the other squarks, and gluinos being accessible thanks to their relatively large cross section at the hadron colliders. 
Several prospect studies have been presented by ATLAS and CMS for gluinos, bottom and top squarks (see, for example, \citerefs{ATL-PHYS-PUB-2014-010,ATL-PHYS-PUB-2016-022}). New studies and further considerations on the HL- and HE-LHC potential for gluinos and top squarks are presented in the following sections. 
\subsubsection{%\theoryC{Probing SUSY at HL- and HE-LHC via gluino pair-production}
\theoryC{Gluino pair production at HL- and HE-LHC}
}
\label{sec:susy212}
\contributors{T. Han, A. Ismail, B. Shams Es Haghi}
%{\bf Author(s): T. Han\footnote{than@pitt.edu}, A. Ismail\footnote{aismail@pitt.edu}, B. Shams Es Haghi\footnote{bas143@pitt.edu}}
%\begin{center}
%PITT-PACC, Department of Physics and Astronomy, University of Pittsburgh, Pittsburgh, PA 15260, USA
%\end{center}

%A main goal of collider physics is to uncover the nature of EWSB. Supersymmetry can resolve the hierarchy problem, as well as provide for gauge coupling unification and a dark matter candidate. The superpartners of the SM particles should be produced at colliders, with many different possible detector signatures. Generally speaking, coloured superpartners such as squarks and gluinos are easiest to produce. Provided that R-parity conservation is assumed, such particles typically decay to final states involving SM particles in addition to significant missing energy in the form of a collider-stable lightest supersymmetric particle (LSP).

The potential of the HL- and HE-LHC to discover supersymmetry is presented in this section focusing on searches for gluinos within MSSM scenarios. Gluino pair production has relatively large cross section and naturalness considerations indicate that gluino masses should not exceed few TeV and lie not too far above the EW scale. Hence they are certainly among the first particles that could be discovered at HL-LHC. 

In the following we assume that a simplified topology dominates the gluino decay chain, culminating in jets plus missing energy in the form of a bino-like LSP $\tilde{\chi}^0$. We evaluate the sensitivity of future proton colliders to gluino pair production with gluinos decaying exclusively to $q \bar{q} \tilde{\chi}^0$ through off-shell first and second generation squarks, using a standard jets + $\met$ search. Currently, the reach for this simplified model with $36\fbinv$ of $13\TeV$ data is roughly $2\TeV$ gluinos, for a massless LSP~\cite{Aaboud:2017vwy,Sirunyan:2018vjp}. A single search region requiring four jets and missing transverse momentum is optimised. In the compressed region where the gluino and LSP masses are similar, a search region with fewer jets is expected to be more effective (see, for example, \citerefs{Cohen:2013xda,ATL-PHYS-PUB-2014-010}) but is not considered in this study. 

The main SM backgrounds contributing to the final states considered are $Z (\to \nu \nu)$ + jets, \sloppy\mbox{$W (\to \ell \nu)$ + jets}, and $t\bar{t}$ production. Other SM background sources such as dibosons and multi-jet are considered negligible. Signal and background samples are generated with MLM matching using \madgraph{ 5}~\cite{Alwall:2014hca}, \pythia{ 8.2}~\cite{Sjostrand:2014zea}. Detector performance are simulated using \delphes{ 3}~\cite{deFavereau:2013fsa}, which employs FastJet~\cite{Cacciari:2011ma} to cluster jets and uses the commonly accepted HL-LHC card corresponding to the upgraded ATLAS and CMS detectors prescribing anti-$k_T$ jets~\cite{Cacciari:2008gp} with radius $0.4$. Effects due to high pile-up are not taken into account, as we expect it to have a negligible impact on our results~\cite{Cohen:2013xda}. An overall systematic uncertainty of $20\%$ is assumed on the SM background contributions covering, among others, jet energy scale and resolution uncertainties.   
A generic $10\%$ uncertainty is assumed on the signal. This does not take into account PDF-related uncertainty which might be as large as $50\%$ for gluinos around $3\TeV$, although the impact of an uncertainty of this kind is presented below for a massless LSP scenario.  

%%%%%%%%%%%%%%% FIGURE %%%%%%%%%%%%%%%%%%%%%
\begin{figure}[t]
\centering
\begin{minipage}[t][\minipageheightdp][t]{\minipagewidthdp}
\centering
\includegraphics[width=\figuremaxwidthdp,height=\figuremaxheightdp,keepaspectratio]{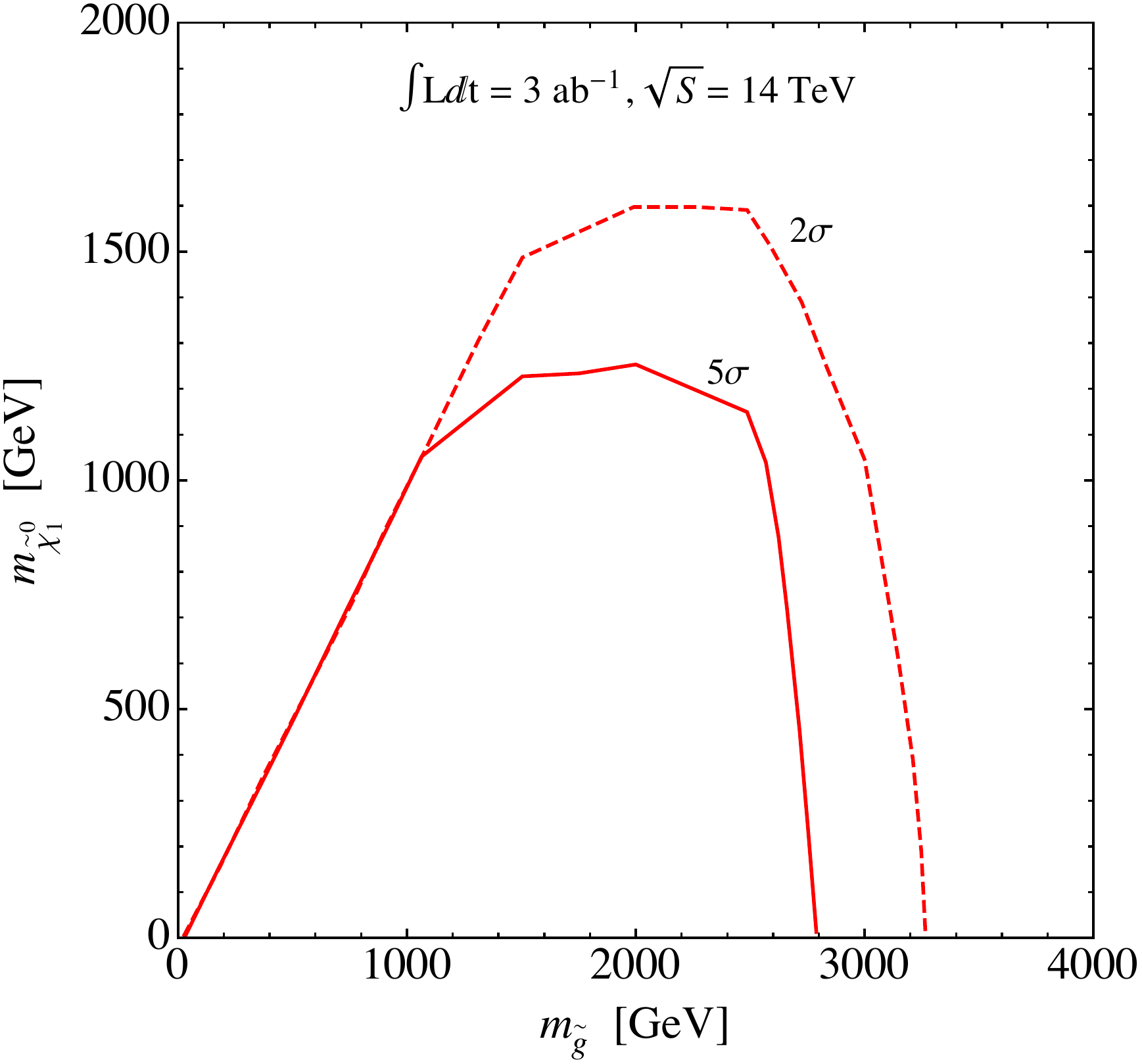}
\end{minipage}
\begin{minipage}[t][5.2cm][t]{\minipagewidthdp}
\centering
\includegraphics[width=\figuremaxwidthdp,height=\figuremaxheightdp,keepaspectratio]{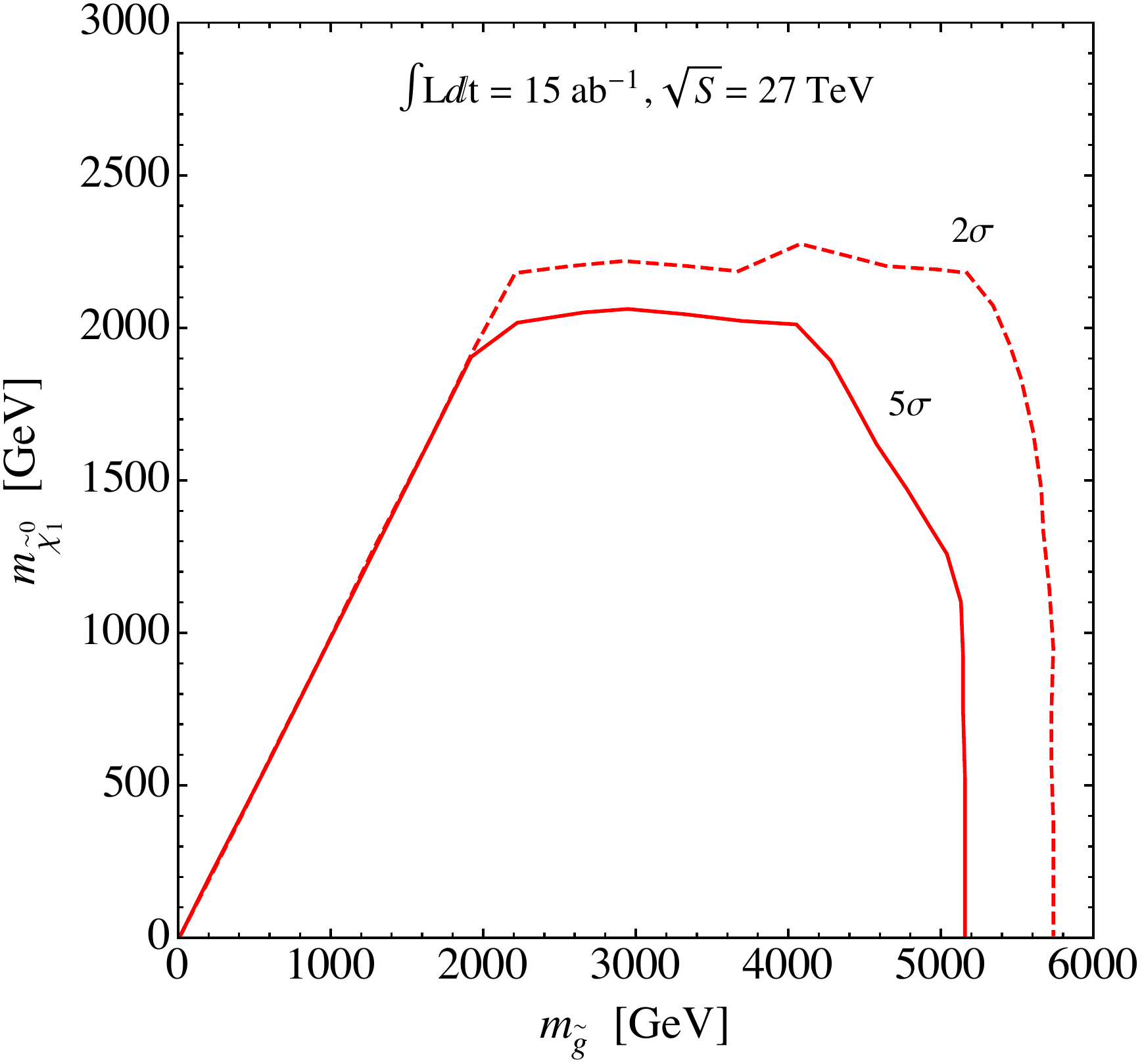}
\end{minipage}
\caption{\footnotesize Expected reach of HL- and HE-LHC in probing gluinos, in the gluino-LSP mass plane. The left (right) plots show the gluino mass reach in $14$ ($27$)\TeV $pp$ collisions with $\intLumi$ ($\intlumihelhc$) of data. The decay $\tilde{g} \to q \bar{q} \tilde{\chi}_1^0$ is assumed to occur with $100\%$ branching fraction, with a bino-like LSP. Both $2\sigma$ exclusion (dashed) and $5\sigma$ discovery contours are shown.}
\label{fig:gluino}
\end{figure}

Following previous works~\cite{Cohen:2013xda,ATL-PHYS-PUB-2014-010,Aaboud:2016zdn}, we apply a set of baseline selections at both $14$ and $27\TeV$. We require that signal events contain no electrons (muons) with $p_T$ above $10$ ($10$)\GeV and $|\eta|$ below $2.47$ ($2.4$). Events are also required to contain a leading jet with $p_T > 160\GeV$ and three additional jets with $p_T > 60\GeV$. In addition, a minimum missing transverse momentum of $160\GeV$ is required to fulfil trigger-based requirements. We reject events with $\Delta \phi(j, \met) > 0.4$ for any of the first three jets to avoid contamination from multi-jet background with mis-measured jets. To further reduce SM contributions, we demand $\met / \sqrt{H_T}>10\GeV^{1/2}$ and $p_T(\mathrm{j_4})/H_T > 0.1$ where ${\mathrm{j_4}}$ indicates the fourth leading jet and $H_T$ is the sum of the transverse momentum of the jets considered in the analysis. After this baseline selection, a two dimensional optimisation over selections on $\met$ and $H_T$ is performed to obtain the maximum significance. 
For the HL-LHC (HE-LHC), we vary $\met$ in steps of $0.5$ ($0.5$)\TeV from $0.5$ ($0.5$) up to $3.0$ ($7.0$)\TeV and $H_T$ in steps of $0.5$ ($0.5$)\TeV from $0.5$ ($0.5$) up to $5.0$ ($7.0$)\TeV. The optimisation aims to maximise the signal significance, defined as $S / \sqrt{(B + (sysB)^2 B^2 + (sysS)^2 S^2)}$, where $S$ indicates the number of signal events, $B$ the total SM background events, and $sysB = 0.2$ and $sysS = 0.1$ are the systematic uncertainties on background and signal, respectively.
Thanks to the optimisation procedure used in this study, the results present an improvement with respect to the existing ATLAS HL-LHC study~\cite{ATL-PHYS-PUB-2014-010}, although the impact related to different assumptions on systematic uncertainties and pile-up conditions might play a non-negligible role.

Exclusion and discovery contours are shown in \fig{fig:gluino} as $2\sigma$ and $5\sigma$ contours of the significance previously defined. For a massless LSP, a gluino of approximately $3.2\TeV$ can be probed by the \sloppy\mbox{HL-LHC} with $3\abinv$ of integrated luminosity, with a discovery potential up to $2.9\TeV$. At $27\TeV$ with \intlumihelhc of integrated luminosity, the exclusion (discovery) reach is roughly $5.7$ ($5.2$)\TeV for massless LSP. With the signal varied within a $50\%$ band, mimicking current PDF uncertainties for high mass gluinos, the HL-LHC (HE-LHC) exclusion reach will decrease by about $200$ ($400$)\GeV and become approximately $3$ ($5.3$)\TeV.

\subsubsection{%\atlas{Prospects for third generation squarks at HL-LHC}
\atlasC{Third generation squarks at HL-LHC}}
\label{sec:susy223}
\contributors{I. Vivarelli, ATLAS}
%{\bf Author(s): I. Vivarelli et al. ATLAS}

The expected ATLAS sensitivity to stop pair production at the HL-LHC is investigated, based upon \citeref{ATL-PHYS-PUB-2018-021}. The Run-2 analysis described in \citeref{Aaboud:2017ayj} is taken as reference and an event selection yielding optimal sensitivity to stop pair production with \intLumi of $pp$ collisions is developed. 
The $\tone$ decaying in $t^{(*)} \neut$ mode is considered, 
where the star indicates that the top quark can possibly be off mass-shell, depending on the mass difference 
between the stop and the neutralino  masses, $\Dmstop$. The final state analysed is that where both top quarks 
decay hadronically hence characterised by the presence of many jets and $b$-jets, and by missing transverse 
momentum $\ptmiss$ (whose magnitude will be indicated by $\met$ in the following) stemming from the presence 
of the two $\neut$. Two kinematic regimes are considered, referred to as ``\largeDM'' and ``\diagonal'' in the following.  
The \largeDM regime is %that of signal models 
where the difference between the stop and neutralino masses is 
large with respect to the top quark mass $\Dmstop \gg \mtop$.  
The top quarks emitted in the stop decay are produced on-shell, and they have a boost in the laboratory frame proportional to $\Dmstop$. 
The final state is hence characterised by high $\pt$ jets and $b$-jets, and large $\met$. Typical analyses in this kinematic regimes 
have large acceptance, and the sensitivity is limited by the signal cross section that decreases steeply with increasing $m\left(\tone\right)$. 
%The sensitivity in this  kinematic regime was already studied in \citeref{ATL-PHYS-PUB-2013-011}. 
If $\Dmstop \sim \mtop$, hence the \diagonal regime, the extraction of the signal from the SM background stemming from mainly $\ttbar$ 
production requires a focus on events where the stop pair system recoils against substantial initial-state hadronic activity (ISR).
%The upgrade sensitivity to this scenario has never been investigated before by ATLAS in final states with no leptons.  
%(see \citeref{ATL-PHYS-PUB-2016-022} for a study in final states with two leptons). 

The analysis is performed on datasets of SM background processes and supersymmetric signals simulated through different 
event generators. The event selection is based on variables constructed from the kinematics of particle-level objects, 
selected according to reconstruction-level quantities obtained from the emulation of the detector response for HL-LHC. 
Particularly relevant for this analysis, jets arising from the fragmentation of $b$-hadrons which are tagged with a nominal 
efficiency of $70\%$, computed on a $\ttbar$ sample simulated assuming $\langle \mu \rangle = 200$. 
Reclustered jets are created by applying the $\antikt$ algorithm with distance parameters $\Delta R = 0.8$ and $\Delta R = 1.2$ 
on signal jets, indicated in the following as $\akt{0.8}$ and $\akt{1.2}$ jet collections. 
A trimming procedure is applied that removes $R = 0.4$ jets from the reclustered jets if their $\pt$ is less 
than $5\%$ of the $\pt$ of the \akt{0.8} or \akt{1.2} jet $\pt$.

\begin{figure}[t]
\centering
\begin{minipage}[t][\minipageheightdp][t]{\minipagewidthdp}
\centering
\includegraphics[width=\figuremaxwidthdp,height=\figuremaxheightdp,keepaspectratio]{\main/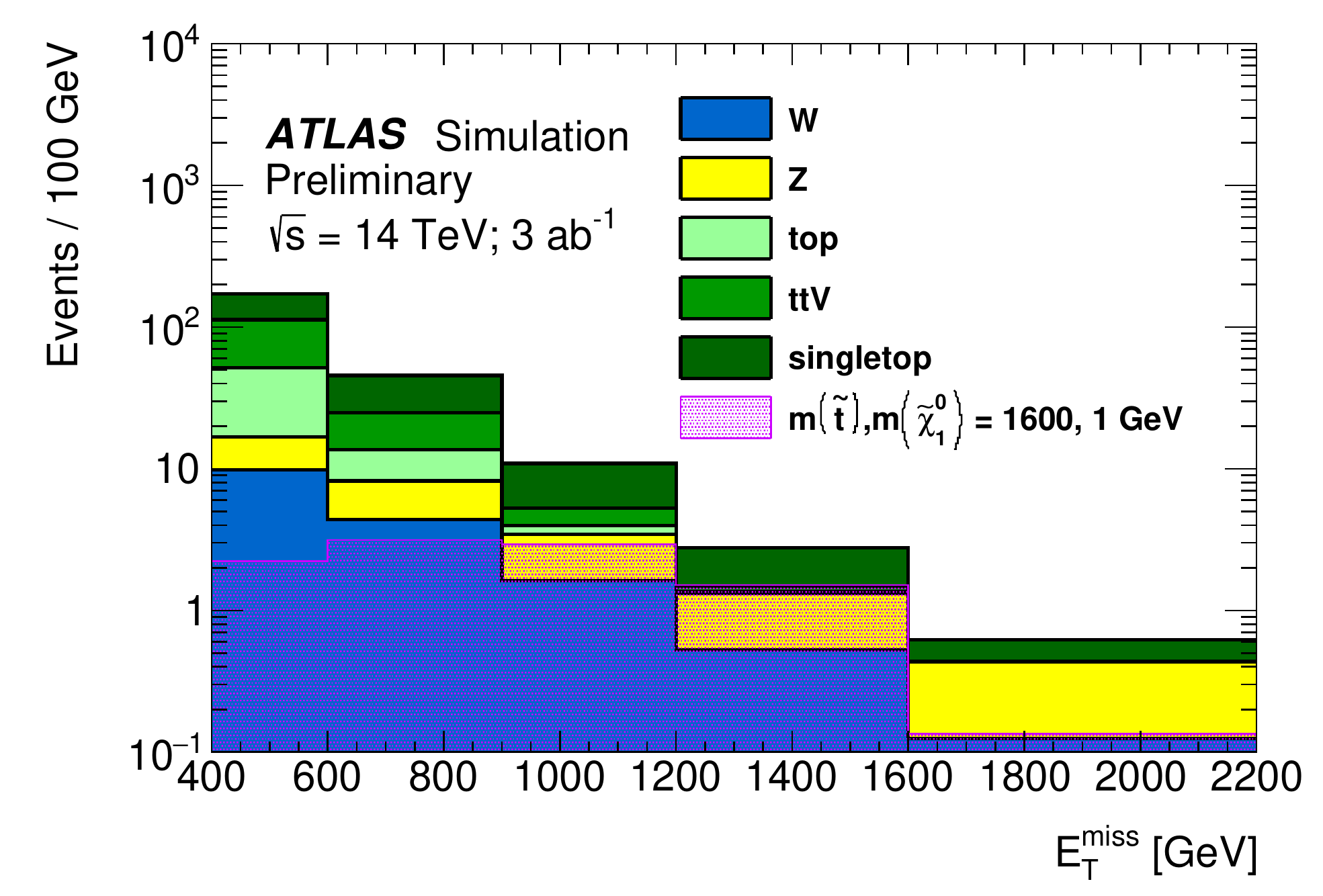}
\end{minipage}
\begin{minipage}[t][\minipageheightdp][t]{\minipagewidthdp}
\centering
\includegraphics[width=\figuremaxwidthdp,height=\figuremaxheightdp,keepaspectratio]{\main/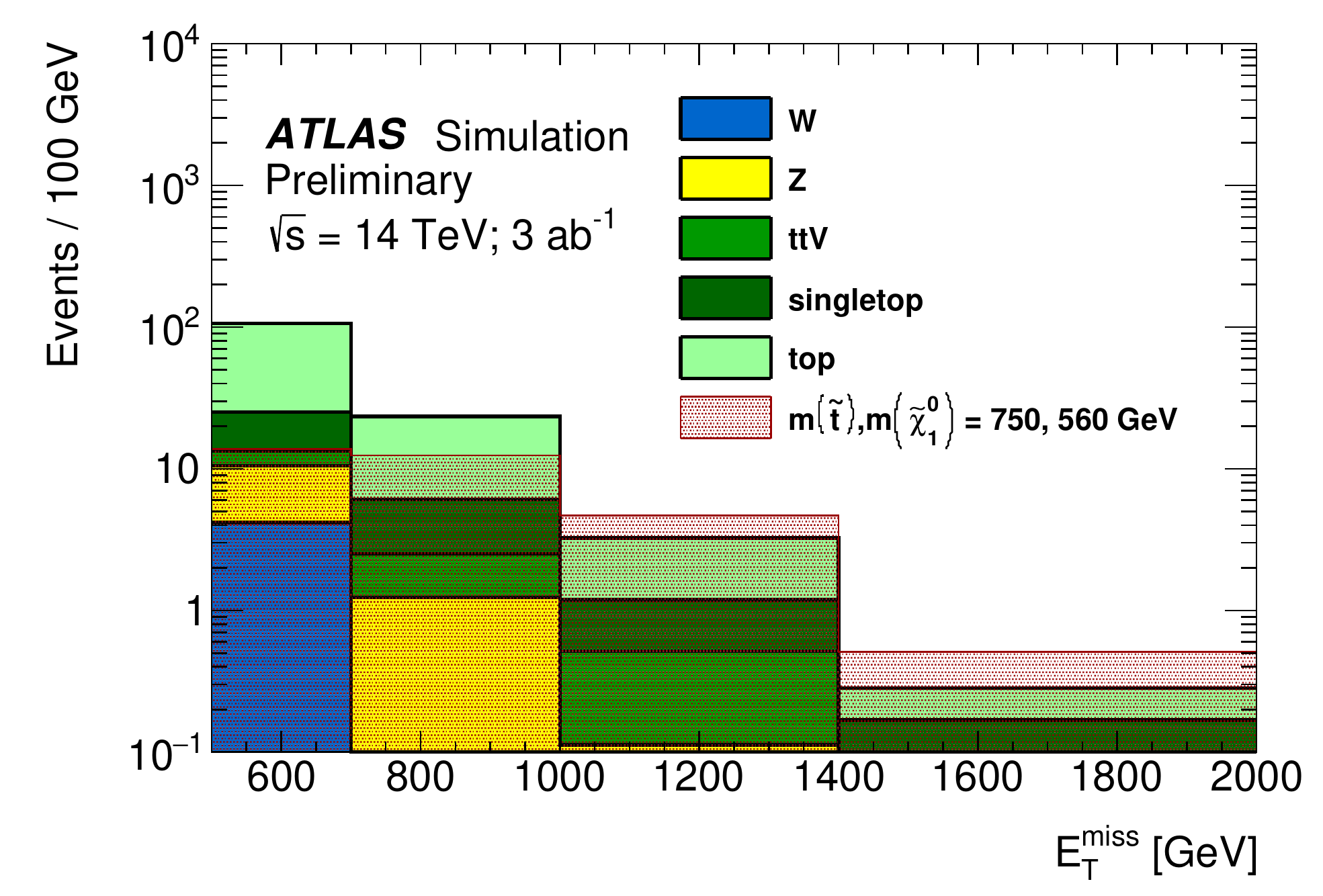}
\end{minipage}
    \caption{ $\met$ distribution for the $\aktm{1.2}{2} > 120\ \GeV$, $\nbjet \ge 2$ bin of the \largeDM analysis (left) and $\RISR > 0.65$ bin of the \diagonal analysis (right). 
The last bin includes overflow events.}
    \label{fig:atlas_stop_final_distributions}
\end{figure}  

Several variables are used for the event selection in the signal regions targeting the large $\Dmstop$ regime. 
The selection on $\met$ exploits the presence of the non-interacting neutralinos in the final state whilst 
the selections on the \akt{1.2} and \akt{0.8} jet masses exploit the potential presence of boosted top quarks 
and $W$-bosons in the final state. %The selection on \mtbmin is effective in suppressing events from SM $\ttbar$~production. 
For the evaluation of the final exclusion sensitivity, a set of mutually exclusive signal regions is defined. 
%The background after preselection is dominated by $\ttbar$ and single top $Wt$ production. For both of these processes, 
%the largest contribution comes from events where one of the two $W$ bosons decays hadronically and the other decays 
%leptonically (there including $W\rightarrow \tau \nu$). %The dominant background processes hence feature at most one hadronic top and/or $W$ decay, while the signal features two of them. 
The events are classified in 30 different signal regions according to the number of identified $b$-jets, the value of the mass of the second (ordering done in mass) reclustered jet reconstructed with distance parameter $R=1.2$, $\aktm{1.2}{2}$ mass, and the value of the $\met$. 
For the evaluation of the discovery sensitivity, a set of single bin cut-and-count signal regions is defined, which apply the full preselection, and then require $\nbjet \ge 2$, $\aktm{1.2}{2} > 120\ \GeV$. Four different thresholds in $\met$ are then defined to achieve optimal sensitivity for a $5\sigma$ discovery: $\met >  400, 600, 800, 1000\GeV$. 
For each model considered, the signal region giving the lowest $p$-value against the background-only hypothesis in presence of the signal is used. 
The basic idea of the \diagonal analysis arises from the fact that, given the mass relation between the stop and the neutralino, 
the stop decay products (the top quark and the neutralino) are produced nearly at rest in the stop reference frame. 
When looked at from the lab reference frame, the transverse momentum acquired by the decay products will be proportional to their mass. 
If $\pt^{\mathrm{ISR}}$ is the transverse momentum of everything that recoils against the stop pair, it can be shown that 
\begin{equation}
\RISR \equiv \frac{\met}{\pt^{\mathrm{ISR}}} \sim \frac{m\left(\neut\right)}{m\left(\tone\right)}. 
\label{eq:RISR}
\end{equation}
Following this considerations, a recursive jigsaw reconstruction is performed, which makes assumptions that allow the definition 
of a set of variables in different reference frames. %In this specific case, it first defines the \com~of the primary proton-proton collision, or \textbf{CM} frame. In the \textbf{CM} frame, the sparticles frame \textbf{S} and the ISR system (\textbf{ISR}) are back-to-back to each other. One can then define the Visible (\textbf{V}) and Invisible (\textbf{I}) reference frames, composed respectively by the visible particles produced in the stop pair decay and the invisible particles produced in the stop pair decay.   New variables are defined to exploit the relation suggested by \eq{eq:RISR}. 
The final strategy for the assessment of exclusion sensitivity for the \diagonal analysis is thus to use a set of mutually 
exclusive signal region defined in bins of $\RISR$ and $\met$. 
For the evaluation of the discovery sensitivity, four cut-and-count signal regions are defined, which apply the full preselection, 
and then require $\RISR > 0.7$ and $\met >  500, 700, 900, 1100\ \GeV$. For each model considered, the signal region giving the lowest $p$-value against 
the SM hypothesis in presence of signal is used. 

\begin{figure}[t]
\centering
\begin{minipage}[t][\minipageheightsp][t]{\minipagewidthsp}
\centering
\includegraphics[width=\figuremaxwidthsp,height=\figuremaxheightsp,keepaspectratio]{\main/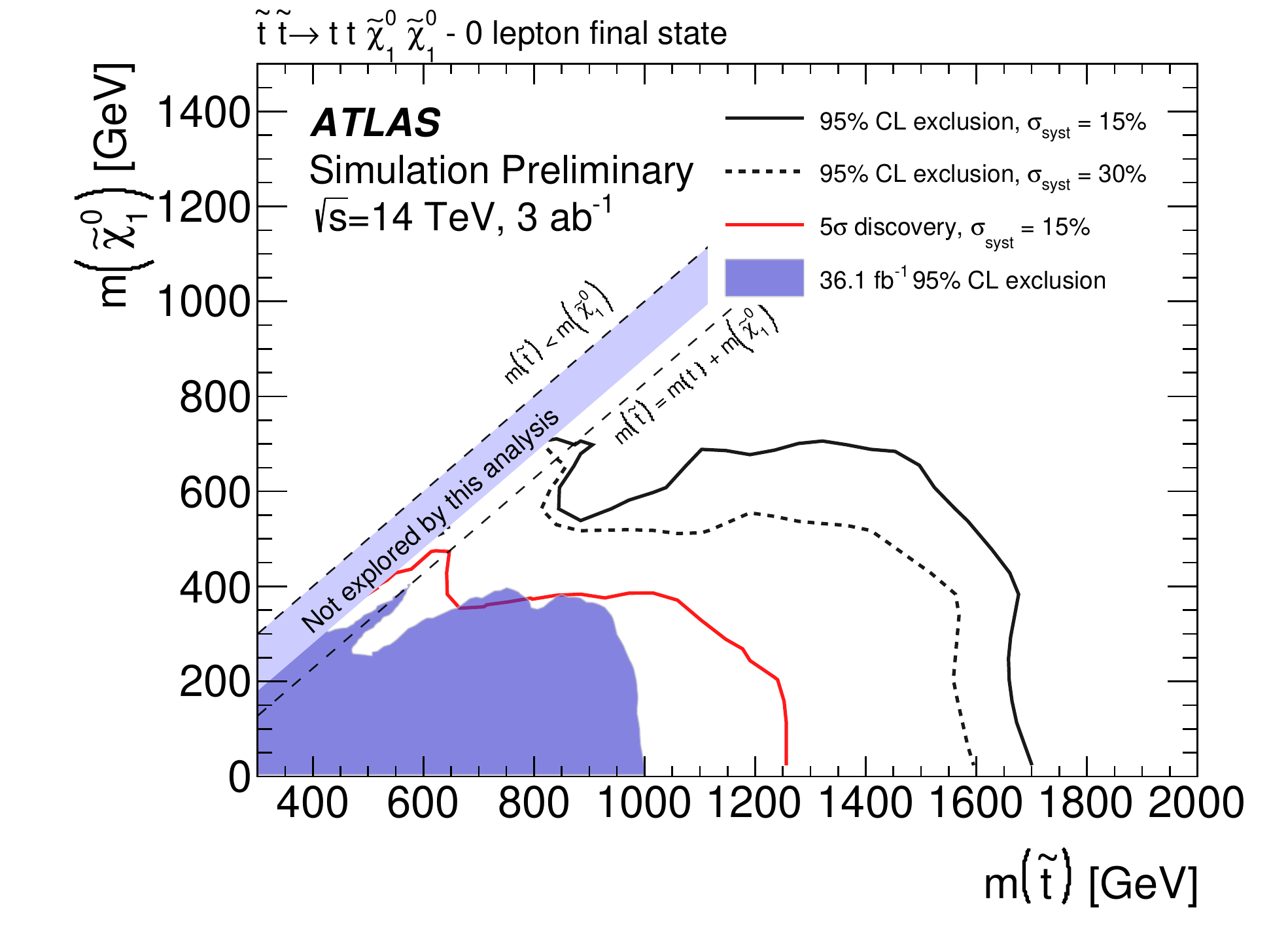}
\end{minipage}
\caption{Final $95\%\cl$ exclusion reach and $5\sigma$ discovery contour corresponding to \intLumi of proton-proton collisions collected by ATLAS at the HL-LHC.}
\label{fig:atlas_stop_final_result}
\end{figure}  

The final $\met$ distribution in the bins with $\aktm{1.2}{2} > 120\ \GeV$, $\nbjet \ge 2$ (for the \largeDM analysis) and $\RISR > 0.65$ 
(for the \diagonal analysis) are shown in \fig{fig:atlas_stop_final_distributions}. In all cases, the main background process is $\ttbar$, 
with significant contribution of $W$+jets events for the \largeDM analysis. 
A $15\%$ uncertainty is retained as a baseline value of the expected uncertainty for both analyses to determine both 
the $5\sigma$ and the $95\%\cl$ exclusion reach of the analysis. For the case of the estimation of the $95\%\cl$ exclusion sensitivity, a further scenario 
with doubled uncertainty ($30\%$) is also evaluated. 

The final exclusion sensitivity evaluation is done by performing a profile-likelihood fit to a set of pseudo-data providing bin-by-bin yields corresponding 
to the background expectations. For each of the two analyses the likelihood is built as the product of Poissonian terms, 
one for each of the considered bins. 
Systematic uncertainties are accounted for by introducing one independent nuisance parameter for each of the bins considered. 
%The likelihood is modified introducing gaussian terms representing the assumed uncertainty. $95\%\cl$ exclusion contours on the masses of the supersymmetric particles are extracted using the CLs method. 
For each mass of the stop and the neutralino, the analysis yielding the smallest CLs among the \largeDM and the \diagonal is used. 
The discovery sensitivity is obtained similarly from each of the single cut-and-count regions independently. For each signal point, 
the profile likelihood ratio fit is performed on pseudo-data corresponding to the sum of the expected background and the signal. 
The discovery contour corresponds to points expected to give a $5\sigma$ $p$-value against the background-only hypothesis. 
For each signal point, the discovery signal region yielding the smallest $p$-value is considered. 
The final sensitivity of the analysis is summarised in \fig{fig:atlas_stop_final_result} assuming a $15\%$ uncertainty 
for the $5\sigma$ discovery and $95\%\cl$ exclusion contour, and also assuming $30\%$ uncertainty for the $95\%\cl$ exclusion contour.

Top squarks can be discovered (excluded) up to masses of $1.25$ ($1.7$)\TeV for $m\left(\neut\right) \sim 0$ under realistic uncertainty assumptions. The 
 reach in stop mass degrades for larger neutralino masses. If $\Dmstop \sim \mtop$, then the discovery (exclusion) 
reach is $650$ ($850$)\GeV.

\subsubsection{\cmsC{Gluinos and top squarks at HL-LHC in hadronic boosted signatures}}
\contributors{J. Karancsi, S. W. Lee, S. Sekmen, R. Ye, CMS}

This section presents the projection of a CMS search for new physics with boosted W bosons or top quarks using the razor kinematic variables to the HL-LHC conditions.  The original search performed on the Run-2 2016 dataset is part of a larger inclusive new physics search with razor variables that includes an extensive set of hadronic and leptonic search regions documented in~\cite{Sirunyan:2018ell}.

\begin{figure}[t]
\centering
\begin{minipage}[t][\minipageheighttp][t]{\minipagewidthtp-4mm}
\centering
\includegraphics[width=\figuremaxwidthtp-7mm,height=\figuremaxheighttp-7mm,keepaspectratio]{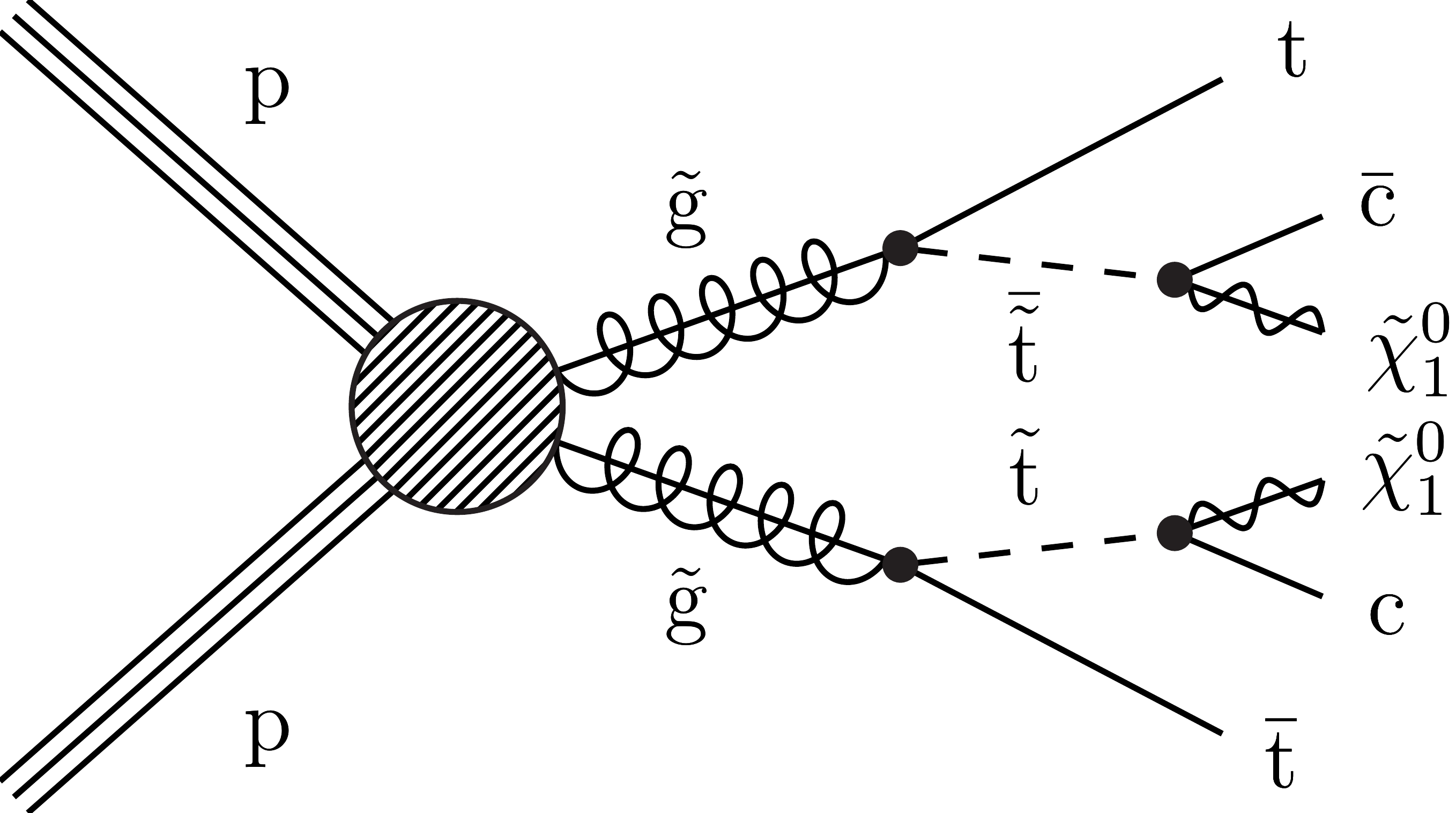} 
\end{minipage}
\begin{minipage}[t][\minipageheighttp][t]{\minipagewidthtp+2mm}
\centering
\includegraphics[width=\figuremaxwidthtp-7mm,height=\figuremaxheighttp-7mm,keepaspectratio]{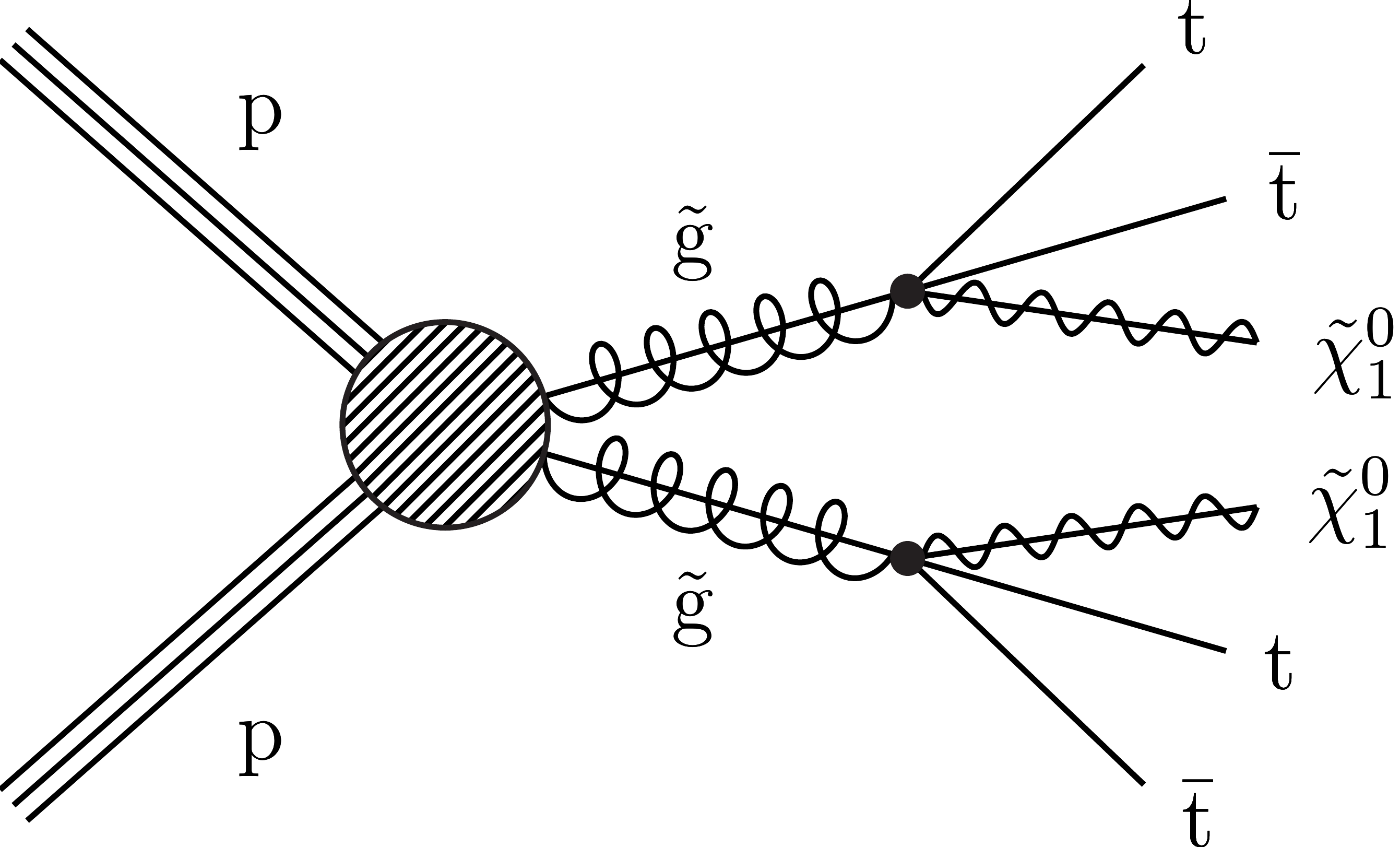} 
\end{minipage}
\begin{minipage}[t][\minipageheighttp][t]{\minipagewidthtp+2mm}
\centering
\includegraphics[width=\figuremaxwidthtp-7mm,height=\figuremaxheighttp-7mm,keepaspectratio]{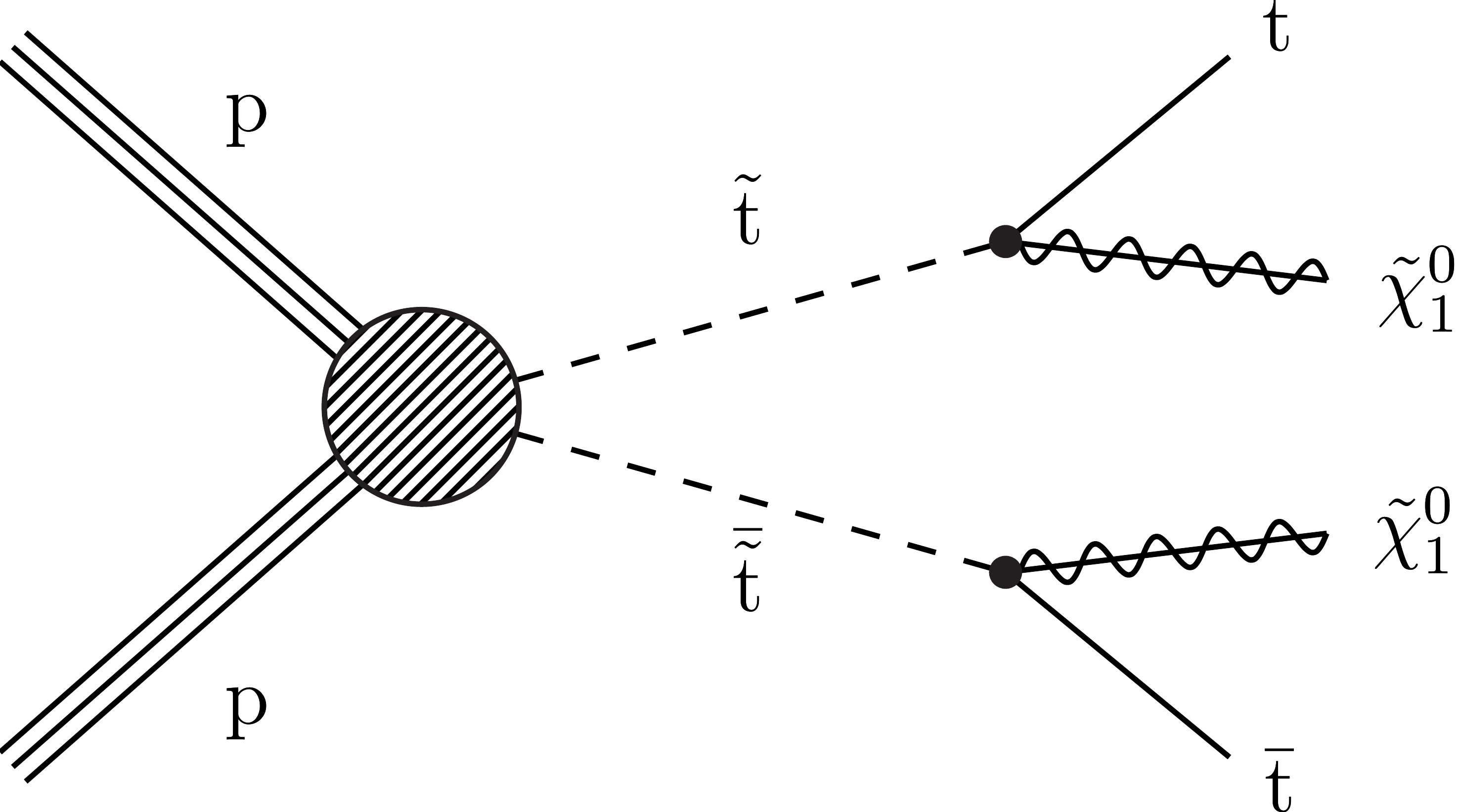}
\end{minipage}
\caption{Signal models considered in this analysis: T5ttcc (left), T1tttt (middle), and T2tt (right).
\label{fig:diagrams}}
\end{figure}

The analysis targets final states consistent with natural SUSY. %SUSY, and in particular, with Natural SUSY, which requires the existence of a light top squark, $\tilde{t}$, and a somewhat light gluino, $\tilde{g}$, that stabilize the Higgs field mass-squared term without excessive fine tuning. 
The primary model of interest is gluino-pair production, where the gluino decays to a top squark and a top quark, with a mass gap between the gluino and the top squark being large enough to give the top quark from the gluino decay a significant boost.  The top squark is light, and decays to $c\tilde{\chi}^0_1$ for small mass differences with respect to the neutralino.  In this simplified model, referred to as \emph{T5ttcc} in the following, decay products of the top squark have very low transverse momentum and thus are very hard to detect. Therefore, the boosted top quark from the gluino decay is used as a handle for enhancing sensitivity. In addition, we also consider scenarios with gluinos directly decaying to $t\bar{t} \tilde{\chi}^0_1$, called \emph{T1tttt}, and with direct production of top squark pairs, where each top squark decays to a top quark and a neutralino, referred to as \emph{T2tt} in the following. The stop model here is equivalent to the model considered in~\secc{sec:susy223}. All models are illustrated by the diagrams in \fig{fig:diagrams}.

Boosted objects, which have high \pt, are characterised by merged decay products separated by ${\Delta R \sim 2m/\pt}$, where $m$ is the mass of the decaying massive particle. A top quark or $W$ boson can be identified via boosted objects within a jet of size $0.8$ if it has a momentum of $\gtrsim 430\GeV$ or $\gtrsim 200\GeV$, respectively.  Boosted objects are more accessible at  increased centre-of-mass energies, and thus will be produced more frequently at the HL-LHC and especially the HE-LHC.  The search is performed in hadronic final states with at least one boosted $W$-jet and one $b$-jet or at least one boosted top-jet, using razor kinematic variables $M_R$ and $R^2$~\cite{Rogan:2010kb}, which are powerful tools that can discriminate between SM processes and production of heavy new particles decaying to massive invisible particles and massless visible particles. The analysis searches for an excess in events with high values of $M_R$ and $R^2$. 

The projection study, explained in detail in~\cite{CMS-PAS-FTR-18-037}, uses the same data and MC events as in the 2016 analysis.  It also follows the same object selection, event selection, background estimation, systematic uncertainty calculation, and limit setting procedures as used in the 2016 analysis. Boosted $W$ bosons and top quarks are identified using the jet mass, the n-subjettiness variables $\tau_2/\tau_1$ and \tauTT~\cite{Thaler:2010tr}, and subjet b-tagging.  Events in all signal, control, and validation regions in the analysis are required to have at least four selected anti-$k_{T}$ jets with radius parameter $0.4$ (AK4 jets), at least one anti-$k_{T}$ jet with radius parameter $0.8$ (AK8 jets) and $\pt > 200\GeV$ defining the boosted phase space, and $M_R > 800\GeV$ and $R^2 > 0.08$. In addition, the signal regions are required to have zero leptons and an azimuthal distance between the two megajets, two partitioned sets of jets in the event used for computing the razor variables~\cite{Rogan:2010kb}, $\Delta\phi_{\text{megajets}}$, to be greater than $2.8$. Three event categories are defined based on boosted object and jet multiplicities: i) W 4-5 jet: ${\geq}1$ reconstructed AK8 $W$-jet, ${\geq}1$ AK4 $b$-jet, $4 \le n_{\text{jet}} \le 5$; ii) W 6 jet: ${\geq}1$ reconstructed AK8 $W$-jet, ${\geq}1$ AK4 b jet, $n_{\text{jet}} \ge 6$; and iii) Top category: ${\geq}1$ reconstructed AK8 top-jet.

The dominant SM backgrounds in the signal regions originate from $t\bar{t}+$jets, single top quark production, multijet events that have jets produced through the strong interaction, and the $W+$jets and $Z+$jets processes.  Data-driven methods are employed to estimate the background contributions to the signal regions. Control regions are used to isolate a process to be estimated, defined by modifying one or more signal selection criteria.  After applying the signal and control region selections, resulting data and MC event distributions are scaled to the HL-LHC cross sections and integrated luminosities.  For data, a procedure is  designed to mimic both the statistical precision and potential modifications in shape due to different levels of cross section scaling in the various contributing processes.  After scaling all distributions, background estimates in the signal regions are obtained by multiplying the observed data yields, binned in $M_R$ and $R^2$, by the simulation transfer factors computed as the ratios of the yields of background MC simulation events in the signal regions to the yields in control regions.  Other SM processes that contribute less significantly, such as diboson, triboson, and $\mathrm{t\bar{t}V}$, are estimated directly from the simulation.  The simulated events used for obtaining both the transfer factors and the direct estimates were corrected using various data-to-simulation correction factors and event weights. The uncertainties in these correction factors and weights were taken into account as systematic uncertainties.  Three different scenarios for systematic uncertainties are considered, in which the systematic uncertainties are i) taken as they are in the original analysis, (Run-2) ii) scaled down according to the expected improvements in the detector and theory calculations (YR18), and iii) neglected in order to test a case with statistical uncertainties only (stat-only). Statistical uncertainties are scaled down by $1/\sqrt{\mathcal{L}_{HL-LHC}/\mathcal{L}_{2016}}$.  

\begin{figure}[!htb] 
\centering
\begin{minipage}[t][\minipageheighttp][t]{\minipagewidthtp}
\centering
\includegraphics[width=\figuremaxwidthtp-2mm,height=\figuremaxheighttp,keepaspectratio]{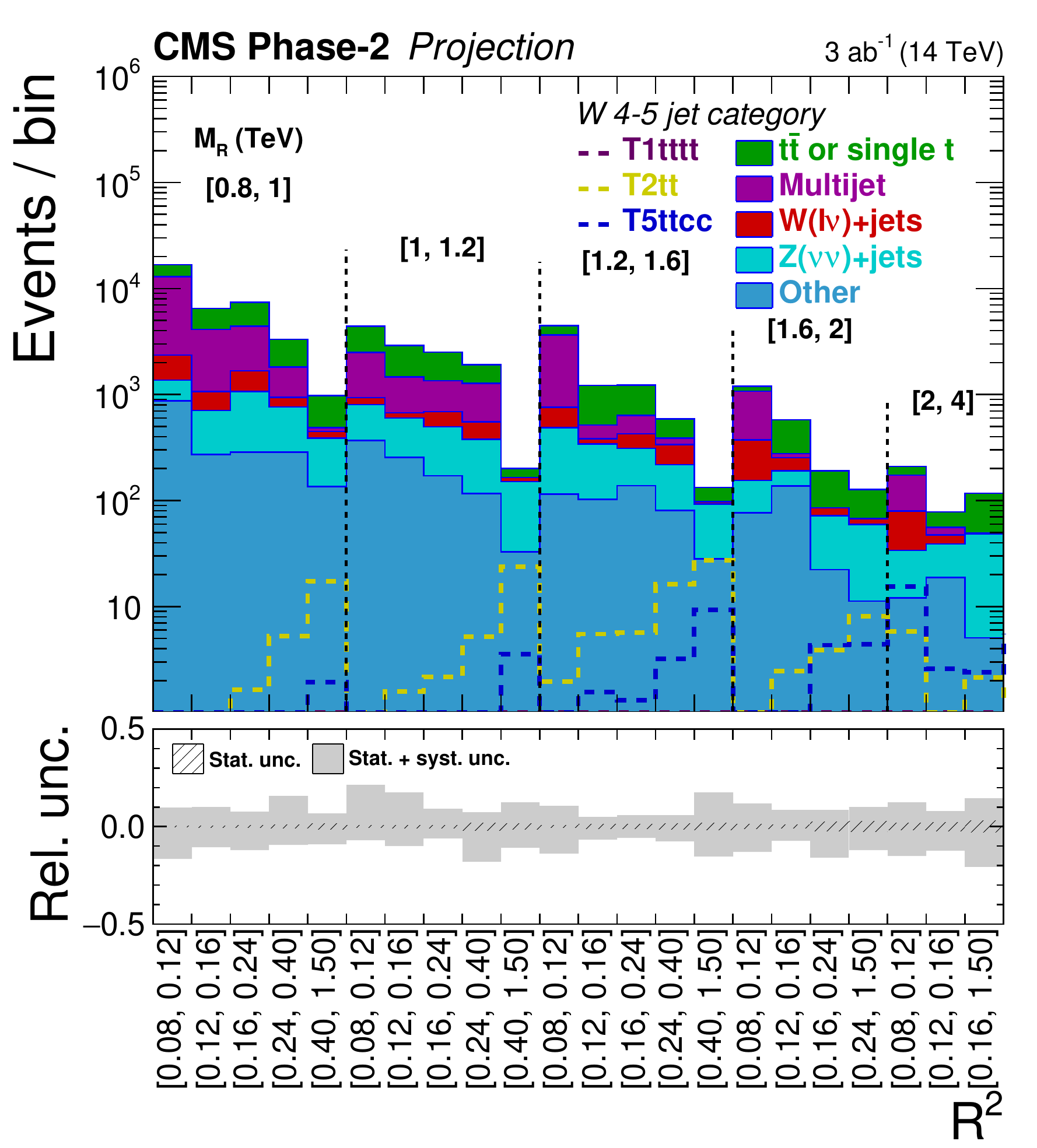}
\end{minipage}
\begin{minipage}[t][\minipageheighttp][t]{\minipagewidthtp}
\centering
\includegraphics[width=\figuremaxwidthtp-2mm,height=\figuremaxheighttp,keepaspectratio]{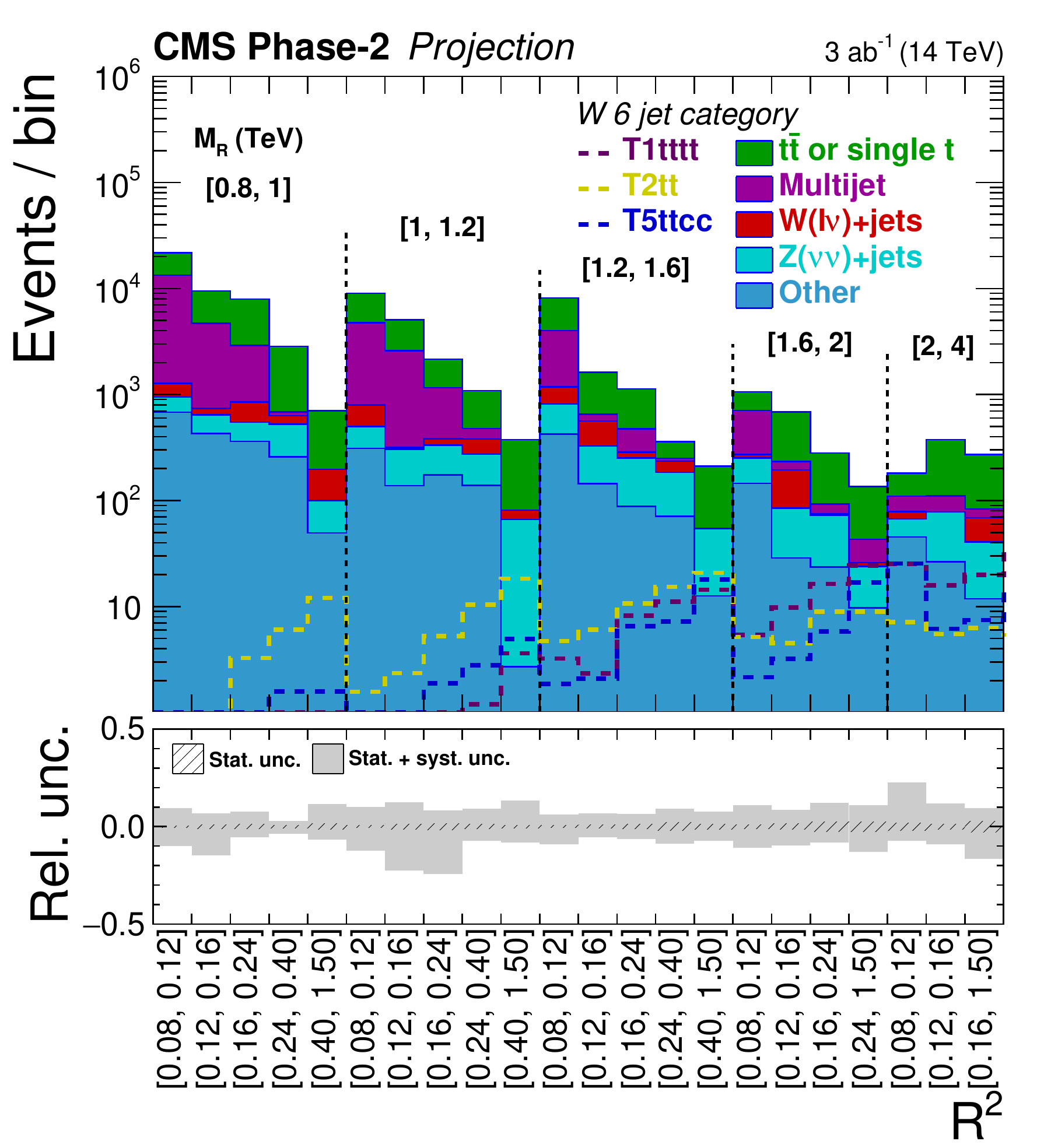} 
\end{minipage}
\begin{minipage}[t][\minipageheighttp][t]{\minipagewidthtp}
\centering
\includegraphics[width=\figuremaxwidthtp-2mm,height=\figuremaxheighttp,keepaspectratio]{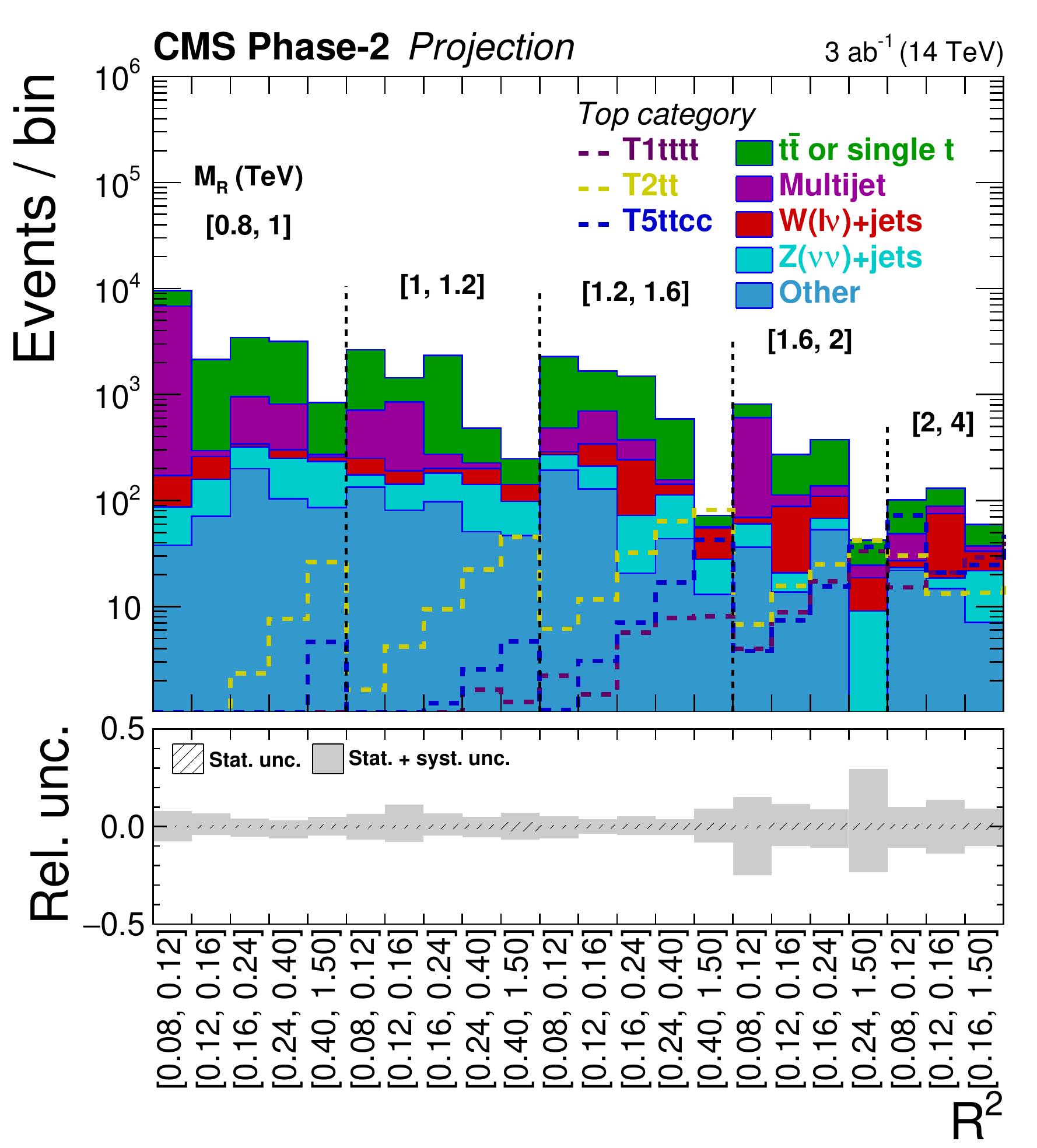}
\end{minipage}
\caption{$M_R$-$R^2$ distributions shown in a one-dimensional representation for background predictions obtained for the W 4-5 jet (upper left), W 6 jet (upper right), and Top (lower) categories for the HL-LHC. Statistical and systematic uncertainties for the YR18 scenario are shown with the hatched and shaded error bars, respectively. Also shown are the signal benchmark models T5ttcc with $m_{\tilde{g}} = 2\TeV$, $m_{\tilde{t}} = 320\GeV$ and $m_{\tilde{\chi}^0_1} = 300\GeV$; T1tttt with $m_{\tilde{g}} = 2\TeV$ and $m_{\tilde{\chi}^0_1} = 300\GeV$; and T2tt with 
$m_{\tilde{t}} = 1.2\TeV$ and $m_{\tilde{\chi}^0_1} = 100\GeV$.
\label{fig:results_hl}}
\end{figure}

The overall background estimation for the W 4-5 jet, W 6 jet, and Top categories along with distributions for several signal benchmark scenarios versus a one-dimensional representation of the bins in $M_R$ and $R^2$ are shown in Figure~\ref{fig:results_hl}, considering the YR18 systematic scenario.  The most dominant systematic uncertainties on the total background estimate come from $W$/top tagging ($\sim 1-4 \%$), \sloppy\mbox{b-tagging} ($\sim3\%$), jet energy scale (JES) ($\sim3\%$), and pile-up ($\sim1-3\%$) variations along with QCD multijet background shape uncertainties ($\sim3-7\%$). For the simulated signal event yields, the largest contributions come from $W$/top tagging ($\sim8\%$), jet energy scale (JES) ($\sim3\%$) and b-tagging ($\sim2\%$) variations.  

\begin{figure}[htpb]
\centering
\begin{minipage}[t][\minipageheighttp][t]{\minipagewidthtp}
\centering
\includegraphics[width=\figuremaxwidthtp-2mm,height=\figuremaxheighttp,keepaspectratio]{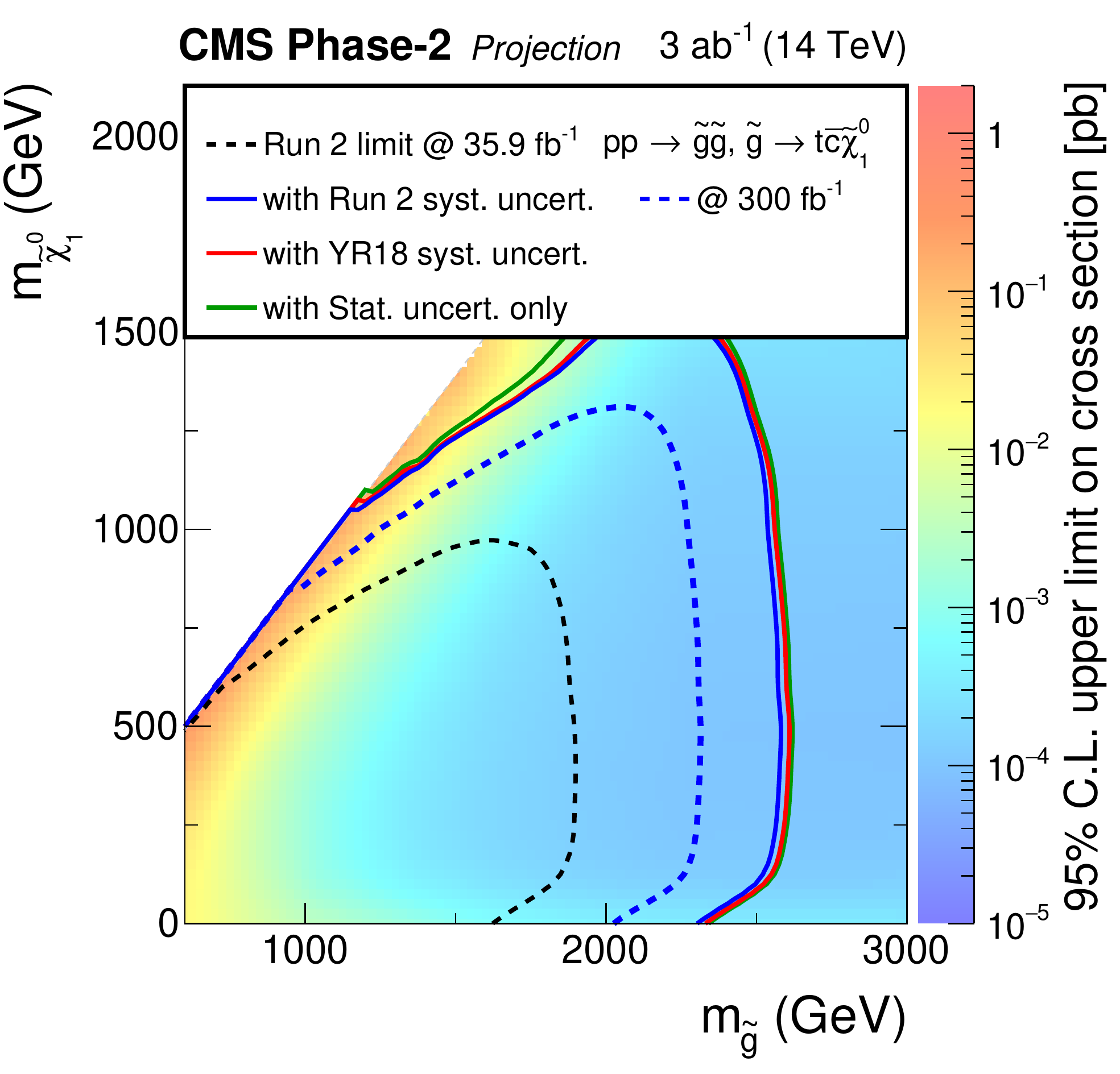} 
\end{minipage}
\begin{minipage}[t][\minipageheighttp][t]{\minipagewidthtp}
\centering
\includegraphics[width=\figuremaxwidthtp-2mm,height=\figuremaxheighttp,keepaspectratio]{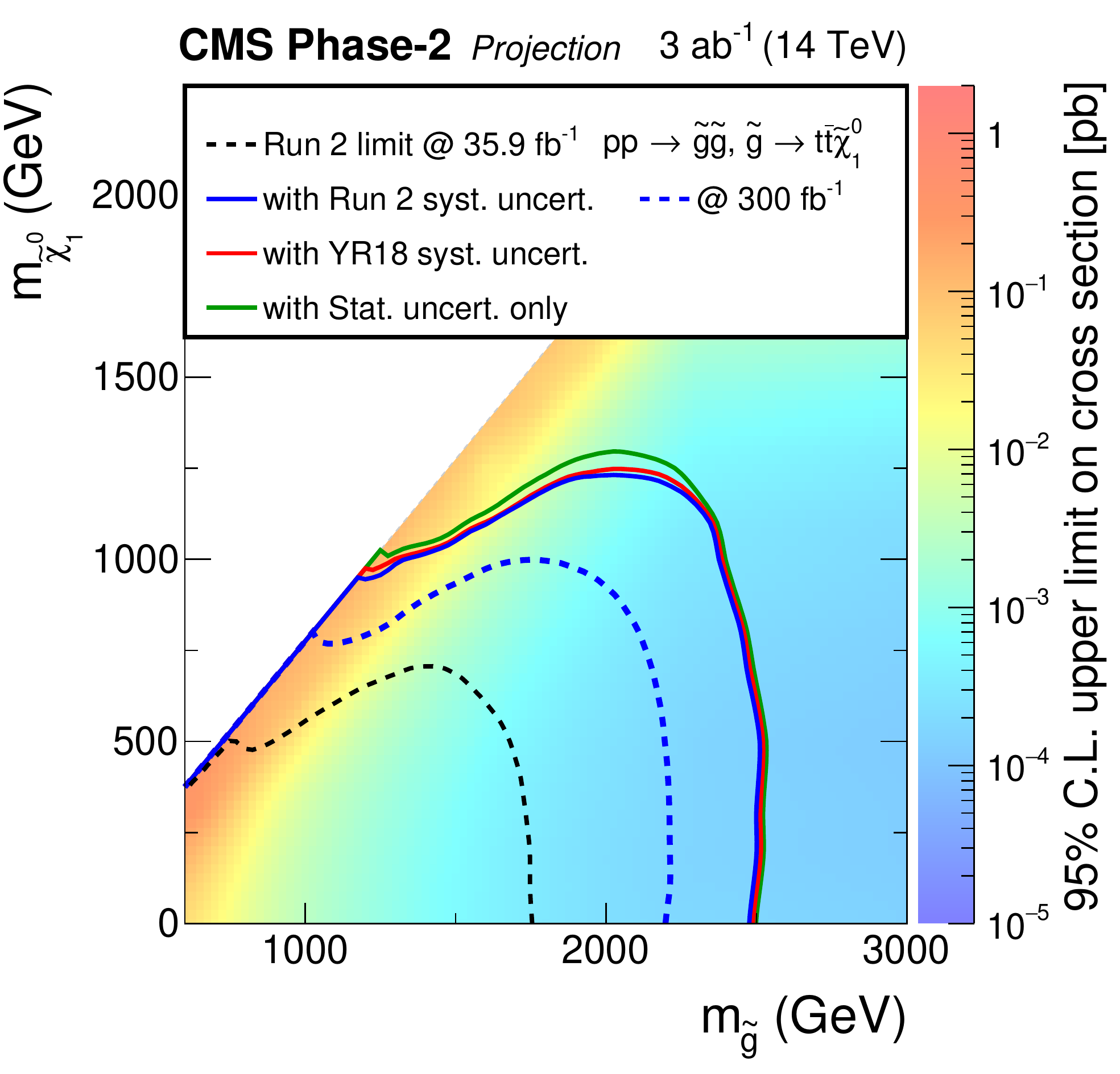} 
\end{minipage}
\begin{minipage}[t][\minipageheighttp][t]{\minipagewidthtp}
\centering
\includegraphics[width=\figuremaxwidthtp-2mm,height=\figuremaxheighttp,keepaspectratio]{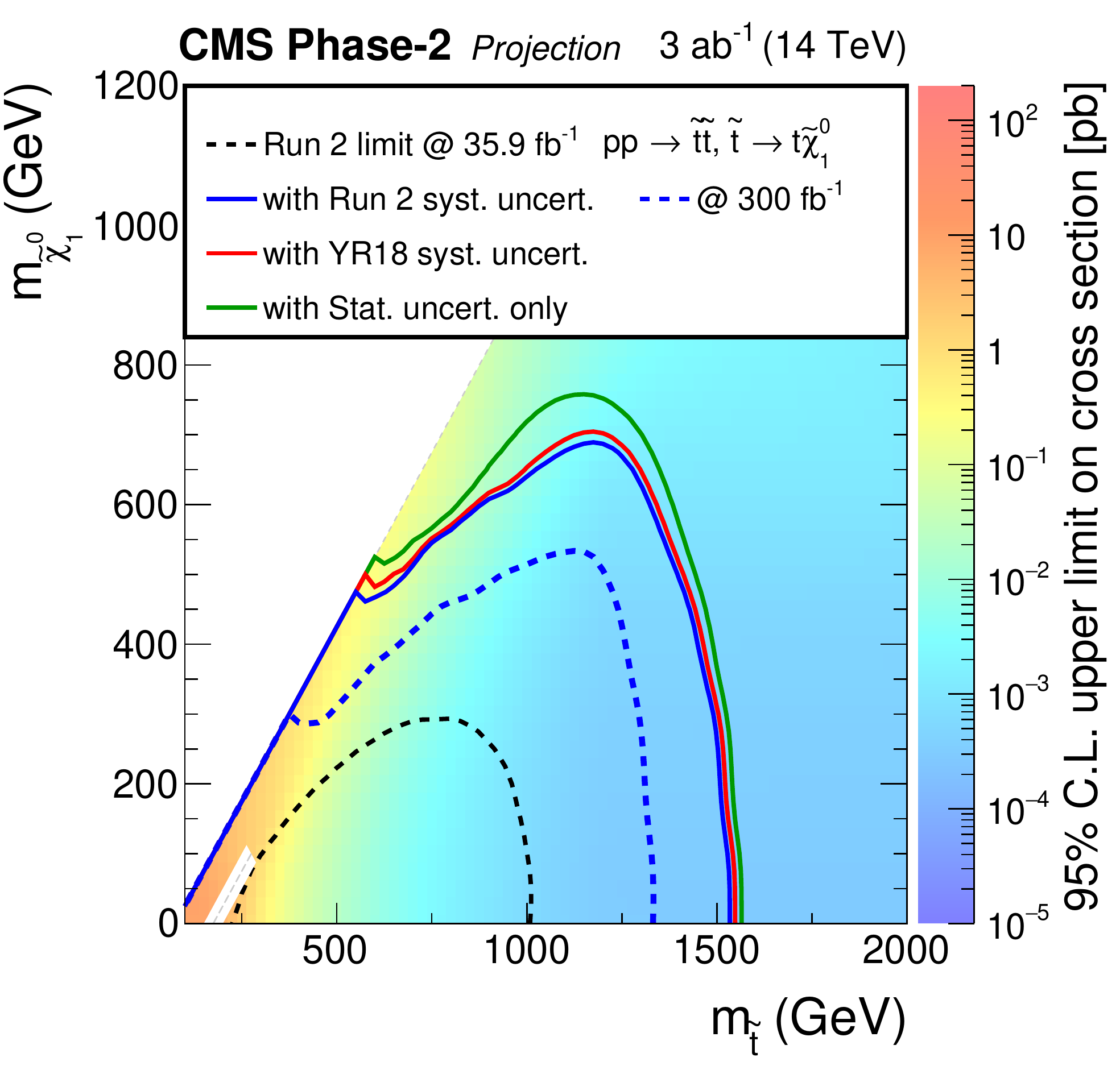} \end{minipage}
\caption{Projected expected upper limits on the signal cross sections for the HL-LHC using the asymptotic CLs method versus gluino/top squark and neutralino masses for the T5ttcc (top left), T1tttt (top right), and T2tt (bottom) models for the combined W 4-5 jet, W 6 jet, and Top categories for the YR18 scenario.  The contours show the expected lower limits on the gluino/top squark and neutralino masses based on the Run-2 systematic uncertainties, YR18 systematic uncertainties, and statistical-only scenarios, along with the 2016 razor boost limit and the $300\fbinv$ limit for comparison.  
\label{fig:excllimitshl}}
\end{figure}

\begin{figure}[htpb]
\centering
\begin{minipage}[t][\minipageheighttp][t]{\minipagewidthtp}
\centering
\includegraphics[width=\figuremaxwidthtp-2mm,height=\figuremaxheighttp,keepaspectratio]{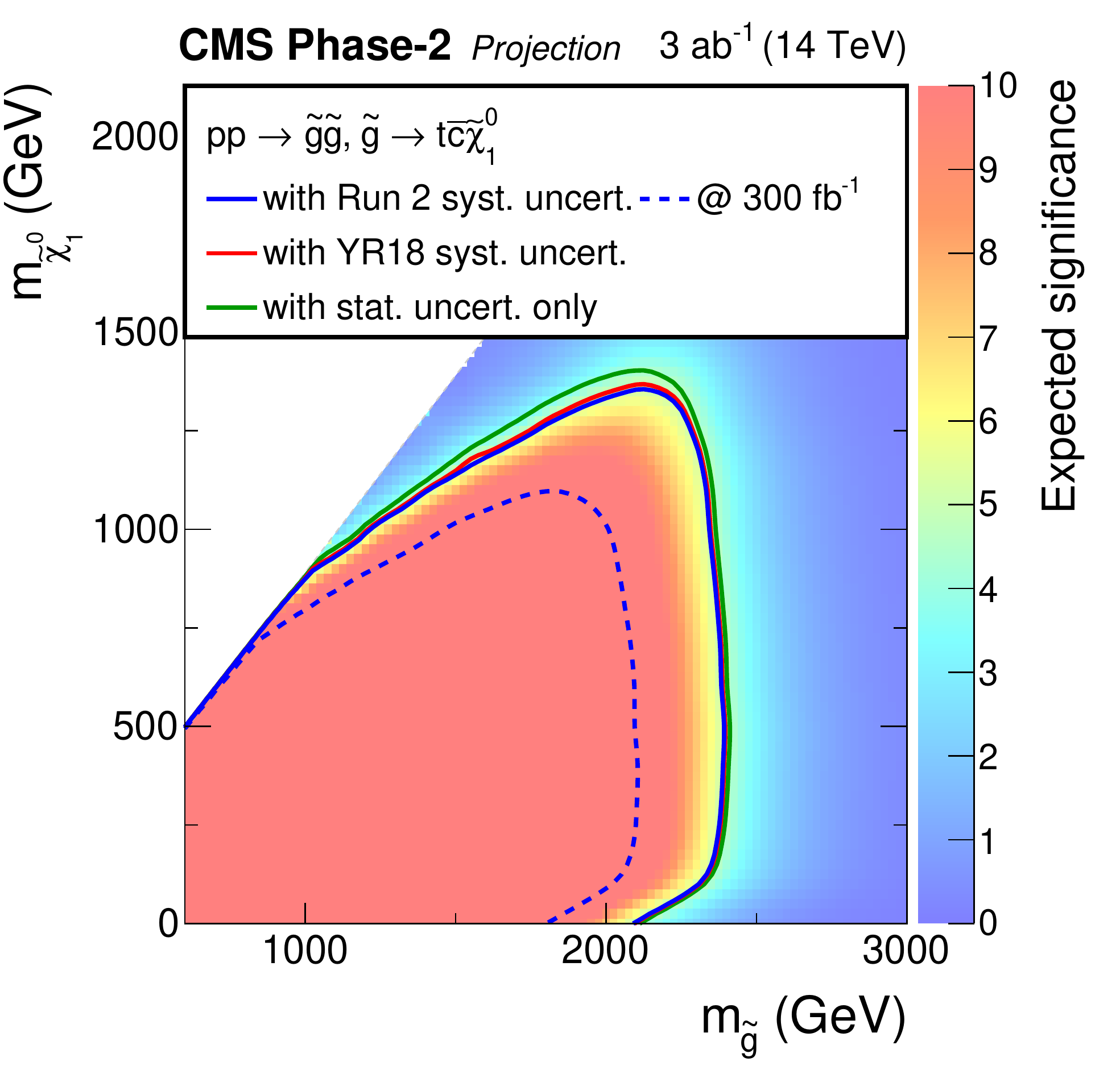} 
\end{minipage}
\begin{minipage}[t][\minipageheighttp][t]{\minipagewidthtp}
\centering
\includegraphics[width=\figuremaxwidthtp-2mm,height=\figuremaxheighttp,keepaspectratio]{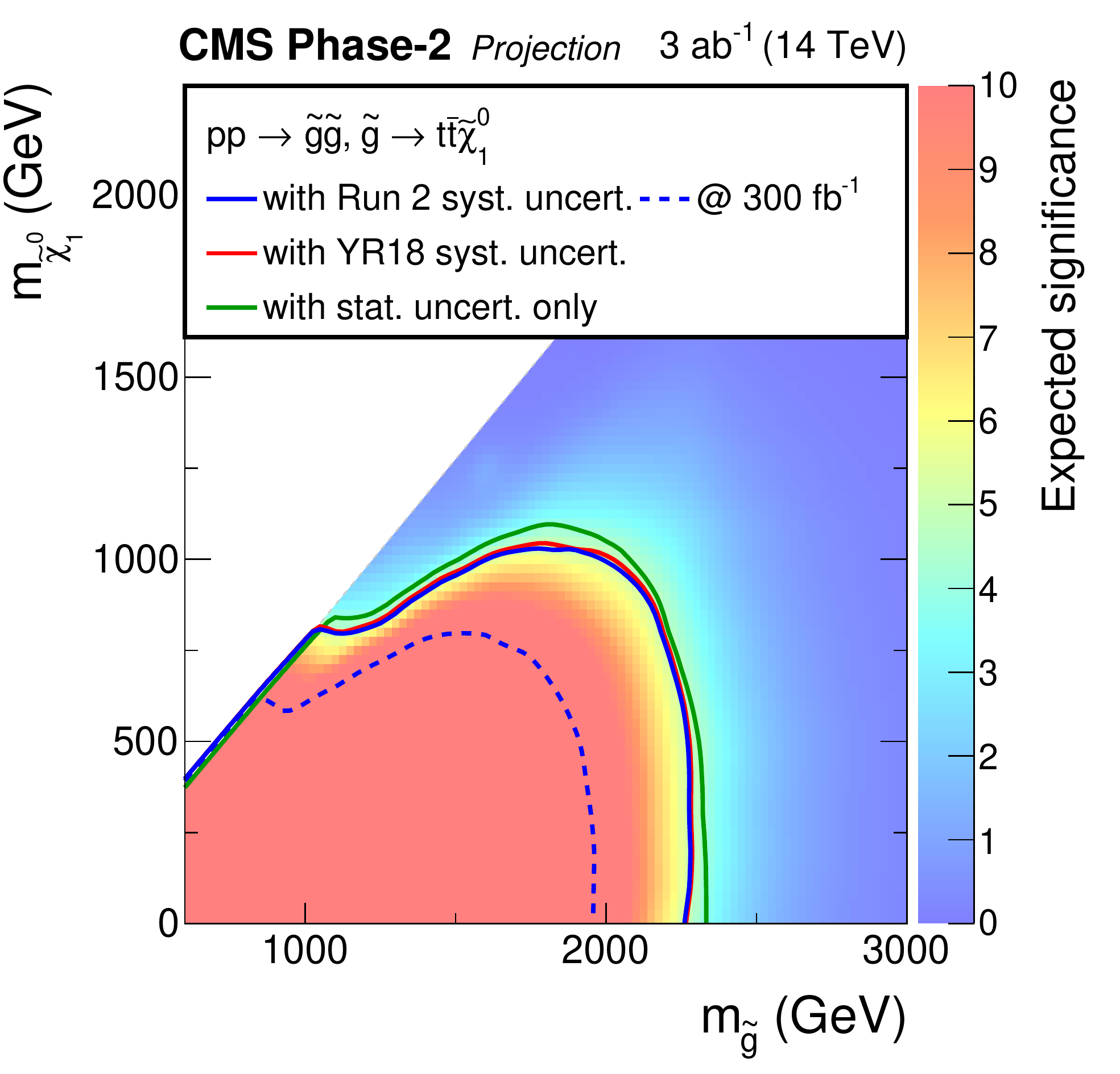} 
\end{minipage}
\begin{minipage}[t][\minipageheighttp][t]{\minipagewidthtp}
\centering
\includegraphics[width=\figuremaxwidthtp-2mm,height=\figuremaxheighttp,keepaspectratio]{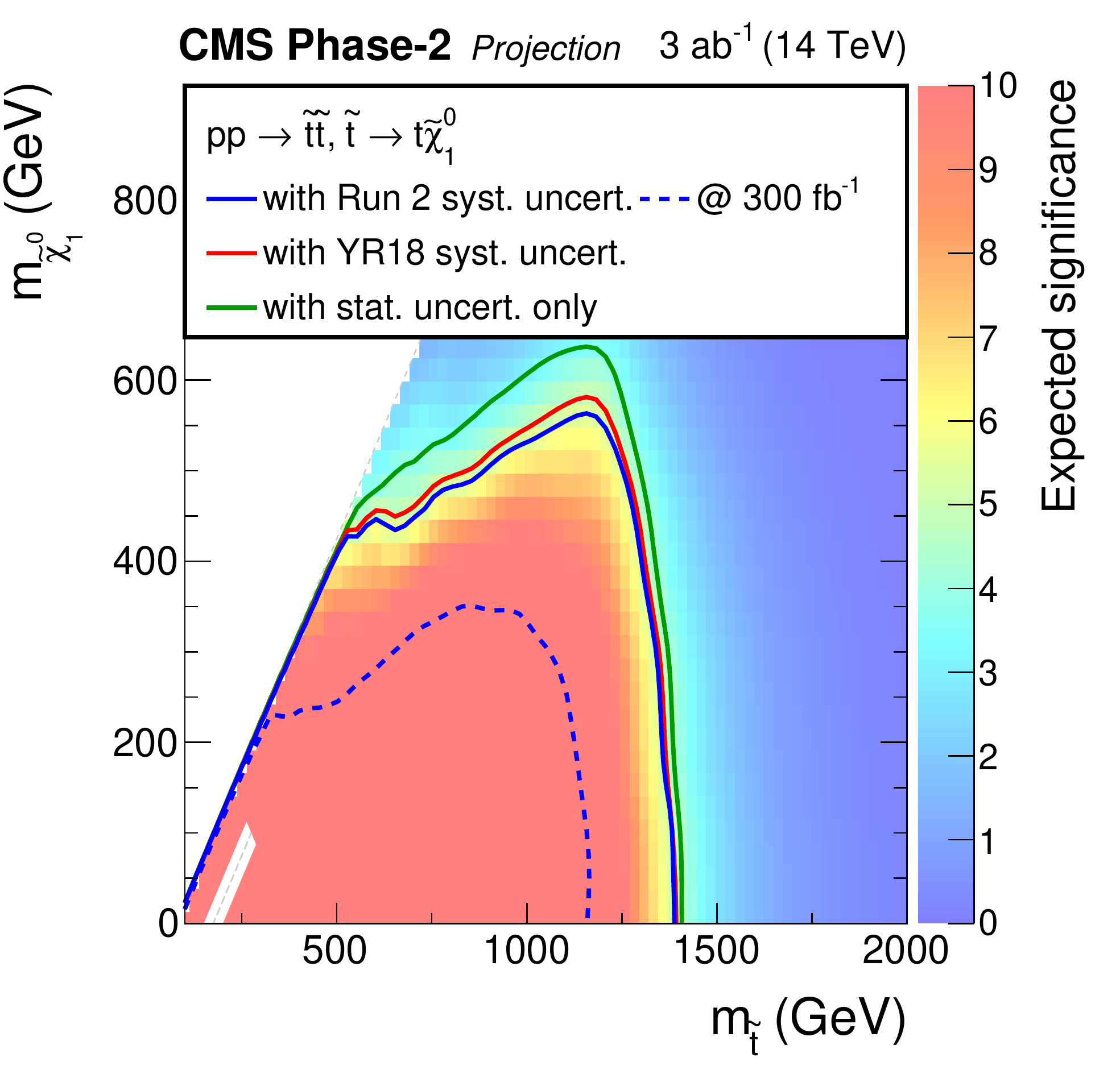} 
\end{minipage}
\caption{Projected expected significance for the HL-LHC versus gluino/stop and neutralino masses for the T5ttcc (top left), T2tttt (top right), and T2tt (bottom) models for the combined W 4-5 jet, W 6 jet, and Top categories for the YR18 scenario.  The contours show the expected discovery bounds on the gluino/top squark and neutralino masses based on the Run-2 systematic uncertainties, YR18 systematic uncertainties, and statistical-only scenarios.
\label{fig:disclimitshl}}
\end{figure}

The results are used to set expected upper limits on the production cross sections of various SUSY simplified models.
Figure~\ref{fig:excllimitshl} shows the expected upper limits on the signal cross sections for the T5ttcc, T1tttt and T2tt simplified models for the combination of the W 4-5 jet, W 6 jet, and Top categories for the HL-LHC projection based on the YR18 scenario.  Additionally, lower limits on gluino/top squark versus neutralino masses are shown for the cases of Run-2 systematic uncertainties, YR18 systematic uncertainties, and statistical-only scenarios for the HL-LHC case. Furthermore, projections of expected discovery sensitivity in the presence of a signal were computed.  The $p$-values for the signal plus background and background-only hypotheses were used to obtain the expected significances in terms of number of standard deviations.  Figure~\ref{fig:disclimitshl} shows the projected expected significance for the T5ttcc, T1tttt, and T2tt models based on the YR18 systematic uncertainties, along with the discovery upper bounds on the gluino/top squark versus neutralino masses for the three uncertainty scenarios for the HL-LHC.

The projection results show that HL-LHC would improve the gluino mass exclusion limits via top-quark by around $750\GeV$, while making discovery possible for gluinos up to masses of $2.4\TeV$. For top squark pair production, the discovery reach is up to $1.4\TeV$, consistent with the ATLAS prospect studies in~\secc{sec:susy223}.
\subsubsection{\theory{Implications of a stop sector signal at the HL-LHC}}
\contributors{A. Pierce, B. Shakya}
%{\bf Author(s): A. Pierce, B. Shakya}

The stop sector is intricately tied to the mass of the Higgs boson. A stop sector discovery therefore provides an opportunity to test the Higgs mass relation, as well as predict subsequent signals at the LHC. This section is devoted to illustrations of such scenarios at the HL-LHC, based on the studies in \citeref{Pierce:2016nwg}. 

The Higgs boson mass at one-loop in the MSSM is~\cite{Martin:1997ns}
\begin{eqnarray}
m_h^2&=&m_Z^2\cos^2(2\beta)+\frac{3 \sin^2\beta\, y_t^2}{4\pi^2}\, [\, m_t^2\ln\left(\frac{m_{\tilde{t}_1}m_{\tilde{t}_2}}{m_t^2}\right)+c_t^2s_t^2(m_{\tilde{t}_2}^2-m_{\tilde{t}_1}^2)\ln\left(\frac{m_{\tilde{t}_2}^2}{m_{\tilde{t}_1}^2}\right)\nonumber\\
&+&c_t^4s_t^4\left\{(m_{\tilde{t}_2}^2-m_{\tilde{t}_1}^2)^2-\frac{1}{2}(m_{\tilde{t}_2}^4-m_{\tilde{t}_1}^4)\ln\left(\frac{m_{\tilde{t}_2}^2}{m_{\tilde{t}_1}^2}\right)\right\}/m_t^2 \,],
\label{eq:higgsmass}
\end{eqnarray}
where $\tilde{t}_1$ and $\tilde{t}_2$ are the stop mass eigenstates, $\theta_t$ is the stop mixing angle, with $\tilde{t}_1=\cos\theta_t\, \tilde{t}_L + \sin\theta_t \,\tilde{t}_R$, $s_t (c_t)=\sin\theta_t (\cos\theta_t)$, and $y_t$ is the top Yukawa coupling. As the mass splitting between stops increases, the latter two terms in the loop correction grow stronger; in particular, the final term switches sign and becomes negative for $m_{\tilde{t}_2}\,\gtrsim\, 2.7 m_{\tilde{t}_1}$, and can dominate for non-vanishing stop mixing and $m_{\tilde{t}_2}\gg m_{\tilde{t}_1}$. Consequently, there exists an $\textit{upper}$ limit on $m_{\tilde{t}_2}$ (as a function of $m_{\tilde{t}_1}$ and $\theta_t$), beyond which it is impossible to accommodate $m_h=125$ GeV in the MSSM. In other words, a measurement of $m_{\tilde{t}_1}$ and some knowledge of $\theta_t$ allows for an upper limit on $m_{\tilde{t}_2}$, and ruling out this window rules out the MSSM. In the following, we discuss some scenarios where such ideas can be implemented at the HL-LHC. 

%\vskip 0.2cm
\noindent\textit{A Sbottom Signal in Multileptons: } 

\begin{figure}[t]
\centering
\begin{minipage}[t][\minipageheightsp][t]{\minipagewidthsp}
\centering
\includegraphics[width=\figuremaxwidthsp,height=\figuremaxheightsp,keepaspectratio]{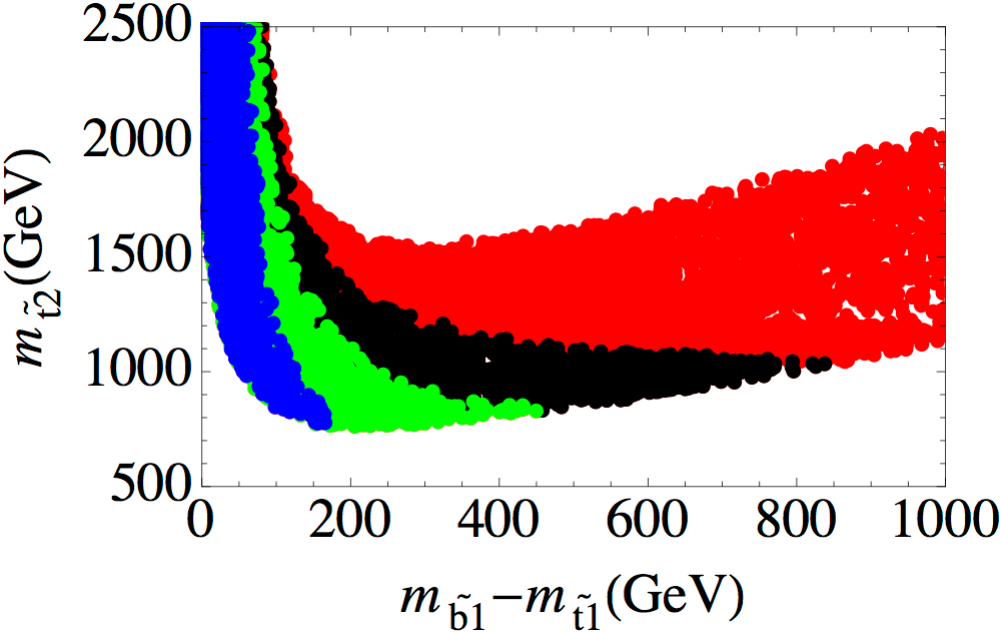}
\end{minipage}
\caption{$m_{\tilde{t}_2}$ as a function of the mass splitting $m_{\tilde{b}_1}-m_{\tilde{t}_1}$  for points with $120< m_h<130\GeV$  and $m_{\tilde{t}_1}<1\TeV$ in the MSSM. Red, black, green, and blue points correspond to $m_{\tilde{b}_1}>1000\GeV$, \sloppy\mbox{$750<m_{\tilde{b}_1}<1000\GeV$}, $500<m_{\tilde{b}_1}<750\GeV$, and $m_{\tilde{b}_1}<500\GeV$ respectively.}
\label{fig:stop2splitting}
\end{figure}

Consider a spectrum such that $\tilde{b}_1\to\tilde{t}_1 W$ (which requires both $\tilde{t}_1$ and $\tilde{b}_1$ to be somewhat left-handed) and $\tilde{t}_1\to t\chi_0$, the mass splitting $\Delta m_{\tilde{b}_1\tilde{t}_1}$ is sufficiently large that sbottom decays give visible multilepton signals, but direct $\tilde{t}_1$ discovery is elusive because of a squeezed spectrum. In the MSSM, $m_{\tilde{t}_2}$ is correlated with $\Delta m_{\tilde{b}_1\tilde{t}_1}$, as shown in \fig{fig:stop2splitting}. A large splitting involves a large stop mixing angle (otherwise, $\tilde{b}_1$ is approximately degenerate with $\tilde{t}_1$); in this case, as discussed above, consistency with the Higgs mass enforces an upper limit on $m_{\tilde{t}_2}$, as seen in \fig{fig:stop2splitting}. Alternatively, the splitting can be raised by making $\tilde{t}_1$ mostly right-handed; however, in this case, $m_{\tilde{t}_2}\approx m_{\tilde{b}_1}$ and again faces an upper limit. For a sub-TeV sbottom, such limits are particularly sharp, as compatibility with the Higgs mass forces $m_{\tilde{t}_2}$ to lie within a narrow wedge shaped region, as seen in the figure. 

Detailed analysis of the multilepton excess can shed further light on the properties of $\tilde{t}_2$: inferring $\Delta m_{\tilde{b}_1\tilde{t}_1}$ (from, \egg, the lepton $p_T$ distribution) and the sbottom mass (from, \egg, the signal rate) not only narrows the allowed range of $m_{\tilde{t}_2}$ (\fig{fig:stop2splitting}) but also constrains the stop mixing angle. It is therefore possible to not only predict a relatively narrow mass window for $\tilde{t}_2$, but also get a profile of its decay channels.  Such a $\tilde{t}_2$  may well be within reach of the LHC, and ruling out such a $\tilde{t}_2$ is sufficient to rule out the MSSM. This concept was implemented in \citeref{Pierce:2016nwg} for a benchmark point with $m_{\tilde{t}_2}=1022\GeV$, $m_{\tilde{b}_1}=885\GeV$, $m_{\tilde{t}_1}=646\GeV$, and $m_{\chi_0}=445\GeV$. Using existing CMS search strategies for same-sign dileptons~\cite{Khachatryan:2016kod,CMS:2016vfu} as well as multileptons~\cite{CMS:2016kuv, CMS:2016ahn}, it was shown that the sbottom decay signal could be identified at the HL-LHC at $3-5\sigma$ significance with \intLumi of data. As discussed above, the discovery of such a sbottom signal with energetic multileptons imposes an upper limit on $m_{\tilde{t}_2}$ in the MSSM. For this benchmark point, it was also shown that modifications of the aforementioned multilepton search strategy would also allow for a $3-5\sigma$ significance discovery of the heavier stop decay $\tilde{t}_2\to\tilde{t}_1 Z$ with \intLumi of data, illustrating how using the MSSM Higgs mass relation in conjunction with a sbottom signal discovery can lead to predictions and discovery of $\tilde{t}_2$ at the HL-LHC. For details of the analysis, the interested reader is referred to~\cite{Pierce:2016nwg}.

%\vskip 0.2cm
\noindent\textit{Using Multiple Decay Channels for the Heavier Stop:} 

The HL-LHC could enable measurements in multiple channels with significant statistics. In particular, the heavier stop $\tilde{t}_2$ could be observed in multiple channels $\tilde{t}_1 Z,\,\tilde{t}_1 h, \,\tilde{b}_1 W$, and $t \chi_0$, with branching ratios (BRs) determined by the stop masses and mixing angle. We focus on the two decays $\tilde{t}_2\to\tilde{t}_1 Z$ and $\tilde{t}_2\to\tilde{t}_1 h$, which give rise to boosted dibosons if the mass splitting between the two stop mass eigenstates is large. In the decoupling limit in the Higgs sector, the ratio of the decay widths into these two channels is~\cite{Baer:2006rs}:
 \begin{equation}
R_{hZ} \equiv\frac{\Gamma(\tilde{t}_2\rightarrow\tilde{t}_1\,h)}{\Gamma(\tilde{t}_2\rightarrow\tilde{t}_1\,Z)}=\left[\left(1-\frac{m_{\tilde{t}_1}^2}{m_{\tilde{t}_2}^2}\right) \cos{2\theta_{\tilde{t}}} +\frac{m_W^2}{m_{\tilde{t}_2}^2}\left(1-\frac{5}{3}\tan^2\theta_W\right)\right]^2\approx \left(1-\frac{m_{\tilde{t}_1}^2}{m_{\tilde{t}_2}^2}\right)^2 \cos^2{2\theta_{\tilde{t}}}.
\label{eq:Rdef}
\end{equation}
Phase space factors as well as many experimental uncertainties cancel in this ratio, offering a clean dependence on the stop mixing angle if the two stop masses are known from other measurements, enabling a check of the MSSM Higgs mass relation. 

\begin{figure}[t]
\centering
\begin{minipage}[t][\minipageheightsp][t]{\minipagewidthsp}
\centering
\includegraphics[width=\figuremaxwidthsp,height=\figuremaxheightsp,keepaspectratio]{\main/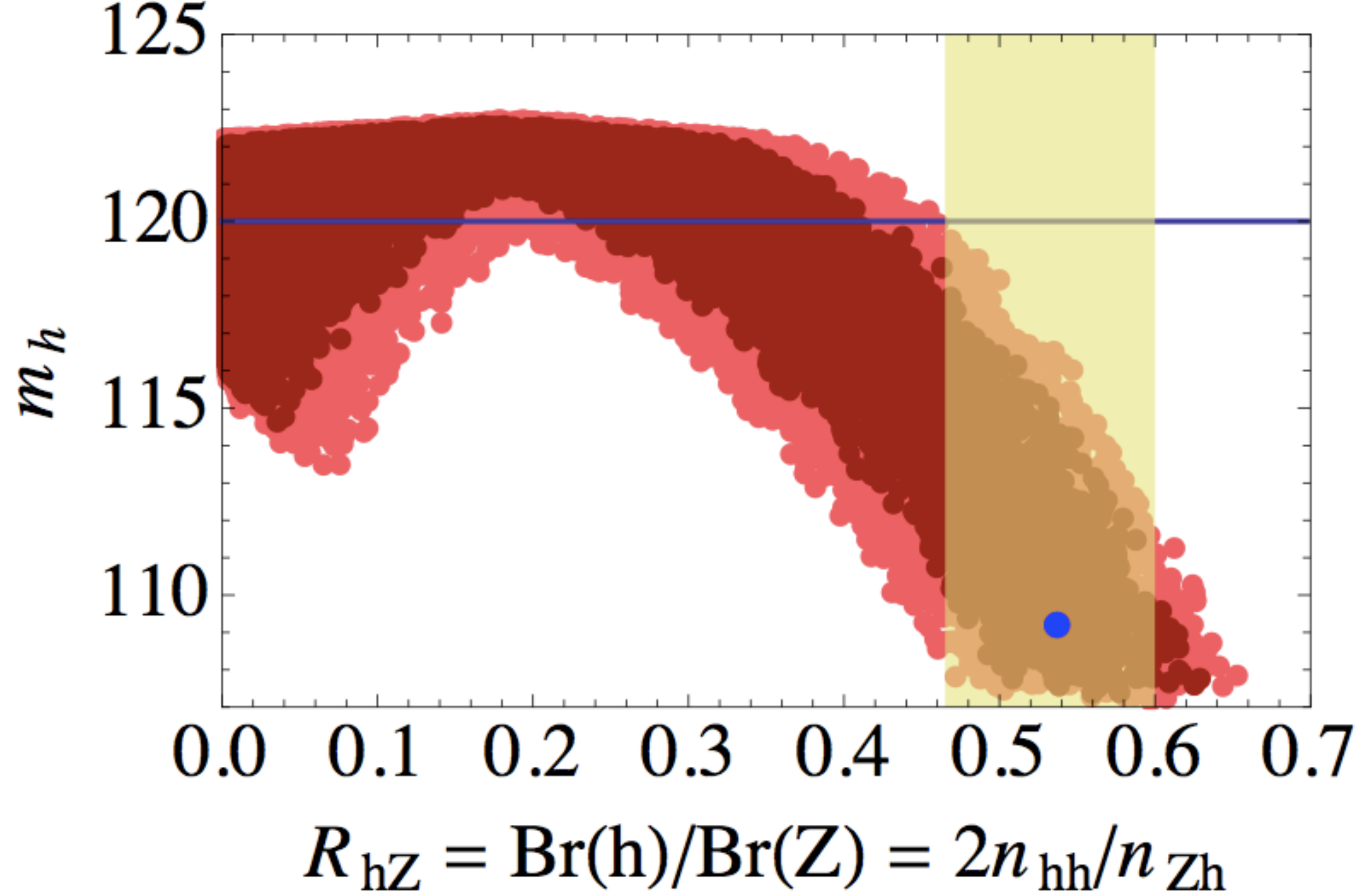}
\end{minipage}
\caption{MSSM Higgs mass as a function of $R_{hZ}$. The horizontal blue line denotes $m_h=120$ GeV, the cutoff below which the Higgs mass is assumed to be inconsistent with the MSSM (see~\cite{Pierce:2016nwg} for details). Light (dark) red bands correspond to $440\leq m_{\tilde{t}_1}\leq 520\GeV$ and $930\leq m_{\tilde{t}_2}\leq 1030\GeV$ ($450\leq m_{\tilde{t}_1}\leq  510\GeV$ and $945\leq m_{\tilde{t}_2}\leq 1015\GeV$). The blue dot denotes the benchmark point in our analysis, which is consistent with all constraints but does not produce the correct Higgs mass in the MSSM. The golden band shows the uncertainty in the calculated value of $R_{hZ}$ with \intLumi of data at the HL-LHC.}
\label{fig:hZratio}
\end{figure}

The viability of such a strategy was explored in \citeref{Pierce:2016nwg} for a benchmark scenario with \sloppy\mbox{$m_{\tilde{t}_2}=994\GeV$}, $m_{\tilde{b}_1}=977\GeV$, $m_{\tilde{t}_1}=486\GeV$, and $m_{\chi_0}=406\GeV$, with $\tilde{t}_2$ decay BRs of $52\%$ and $28\%$ into $\tilde{t}_1\,Z$ and $\tilde{t}_1\,h$ respectively, using the strategy from \citeref{Ghosh:2013qga} to reconstruct boosted dibosons via fat jets. Note that measuring the ratio $R_{hZ}$ with reasonable precision requires high statistics, motivating searches for the boosted $Z$ and $h$ bosons in their dominant (hadronic) decay channels rather than the cleaner decays into leptons or photons. Assuming that $m_{\tilde{t}_1}$ has been measured to lie in the range $486\pm 40\GeV$ from monojet or charm-tagged events, while $m_{\tilde{t}_2}$ is known to fall in the $994.2\pm 50\GeV$ range from various measurements (such as by combining the knowledge of $m_{\tilde{t}_1}$ with information on $p_T(Z)$  in $\tilde{t}_2\to \tilde{t}_1 Z$ events), the MSSM Higgs mass can be calculated as a function of $\theta_t$ using \eq{eq:higgsmass}, which can be converted to a function of the ratio $R_{hZ}$ using \eq{eq:Rdef}; this relation is plotted in \fig{fig:hZratio} in the broad red band for the above stop mass windows. The narrower, darker red band corresponds to narrower windows for the two stop masses, reflecting the improvement in the Higgs mass uncertainty with better knowledge of the stop masses.

The Higgs mass is small for vanishing stop mixing $\theta_t\to 0,\,\pi/2$, which corresponds to \sloppy\mbox{$R_{hZ}\sim \cos^2{2\theta_t}$} approaching $1$. On the other hand, achieving the correct Higgs mass with sub-TeV stops requires large stop mixing, which correlates with a smaller value of $R_{hZ}$. As seen in \fig{fig:hZratio}, an inferred value of $R_{hZ}$ above some cutoff value $R_0$ ($\approx 0.45$ in this case) is incompatible with the MSSM Higgs mass relation. Such an observation would rule out the MSSM, pointing to the need for additional contributions to the Higgs mass (as is the case for the chosen benchmark point). The golden band encodes the uncertainty in the calculated value of $R_{hZ}$ for the benchmark point that can be achieved at the HL-LHC with \intLumi of data (see~\cite{Pierce:2016nwg} for details). This benchmark point study illustrates that measurements of the two decay channels can be used as a consistency check of the Higgs mass and possibly rule out the MSSM at the HL-LHC. 

%\subfile{\main/section2SUSY/sub21_5}
%%%%%%%%%%%%

\subsection{Searches for charginos and neutralinos}
\label{sec:susy23}
The direct production of charginos and neutralinos through EW interactions may dominate the SUSY production at the LHC if the masses of the gluinos and squarks are beyond $3-4\TeV$. In this section, the sensitivity at the end of HL-LHC to the direct production of various SUSY partners in the EW sector under the assumption of R-parity conservation is presented. Charginos and heavier neutralinos production processes are considered, assuming they decay into the LSP via on-shell or off-shell $W$ and $Z$ or Higgs bosons. Final state events characterised by the presence of charged leptons, missing transverse momentum and possibly jets and $b$-jets are studied and prospects are presented. Dedicated searches for higgsino-like, 'compressed' SUSY models, characterised by small mass splittings between electroweakinos, are also reported, and possible complements with new facilities are illustrated.  % for HL-LHC.   
\subsubsection{%\atlas{Prospects for chargino pair production at HL- and HE-LHC}
\atlasC{Chargino pair production at HL-LHC}}
\contributors{S. Carra, T. Lari, D. L. Noel, C. Potter, ATLAS}
%{\bf Author(s): S. Carra (Milano), T. Larri (Milano), C. Potter (Cambridge)}

%The charged wino or higgsino states might be light and decay via SM gauge bosons. In this search, chargino pair production is searched for considering dilepton final states with large missing transverse momentum.   \emph{Status: in ATLAS approval.}

The charged wino or higgsino states might be light and decay via SM gauge bosons. 
In this search~\cite{ATL-PHYS-PUB-2018-048}, the direct production of $\tilde{\chi}^+_1\tilde{\chi}^-_1$ is studied. The $\tilde{\chi}^\pm_1$ is assumed to be pure wino, while the $\tilde{\chi}^0_1$ is the LSP and is assumed to be pure bino and stable. The $\tilde{\chi}^\pm_1$ decays with $100\%$ branching fraction to $W^\pm$ and $\tilde{\chi}^0_1$. 
Only the leptonic decays of the $W$ are considered, resulting in final states with two opposite electric charge (OS) leptons and missing transverse energy from the two undetected $\tilde{\chi}^0_1$.

The selection closely follows the strategies adopted in the $8\TeV$~\cite{Aad:2014vma} and $13\TeV$~\cite{ATLAS:2018gfq} searches.
Events are required to contain exactly two leptons (electrons or muons) with $p_{\mathrm{T}} > 20 $ GeV and $|\eta|<2.5$ ($2.47$ for electrons). 
The lepton pair must satisfy $m_{\ell\ell} > 25$ GeV to remove contributions from low mass resonances. 
The two leptons must be OS, pass ``tight'' identification criteria, and be isolated (the scalar sum of the transverse momenta of charged particles with $p_{\mathrm{T}}>1$ GeV within a cone of $\Delta R = 0.3$ around the lepton candidate, excluding the lepton candidate track itself, must be less than $15\%$ of the lepton $p_{\mathrm{T}}$).
All leptons are required to be separated from each other and from candidate jets defined with $p_{\mathrm{T}} > 30$ GeV and $|\eta|<2.5$. 
The latter requirement is imposed to suppress the background from semi-leptonic decays of heavy-flavour quarks, which is further suppressed by vetoing events having one or more jets tagged as originating from $b$-decays, ``$b$-tagged jets''. 
The chosen working point of the $b$-tagging algorithm correctly identifies $b$-quark jets in simulated $t\bar{t}$ samples with an average efficiency of $85\%$, with a light-flavour jet misidentification probability of a few percent (parametrised as a function of jet $p_{\mathrm{T}}$ and $\eta$).

\begin{figure*}[t]
\centering
\begin{minipage}[t][\minipageheightdp][t]{\minipagewidthdp}
\centering
\includegraphics[width=\figuremaxwidthdp,height=\figuremaxheightdp,keepaspectratio]{\main/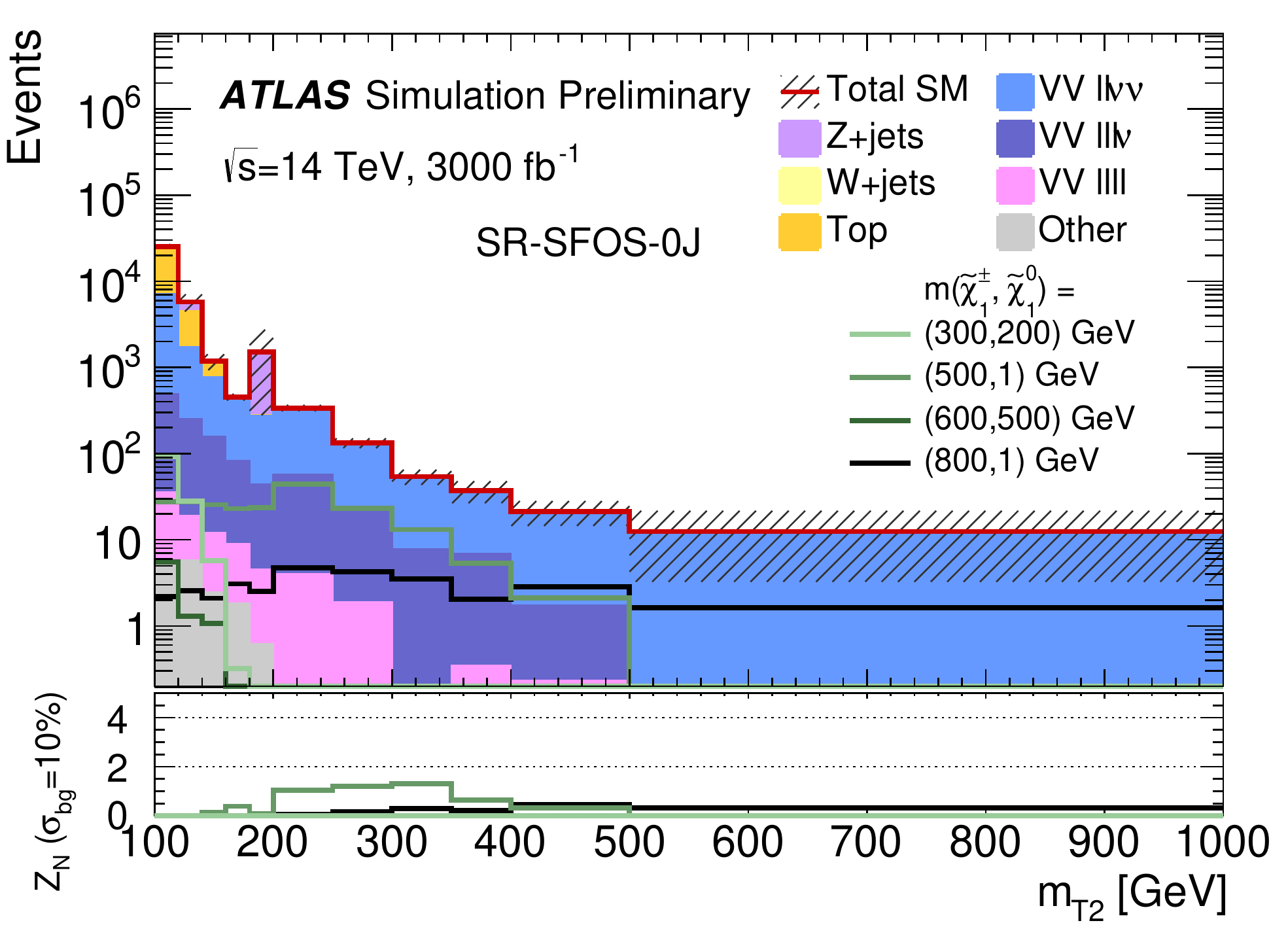}
\end{minipage}
\begin{minipage}[t][\minipageheightdp][t]{\minipagewidthdp}
\centering
\includegraphics[width=\figuremaxwidthdp,height=\figuremaxheightdp,keepaspectratio]{\main/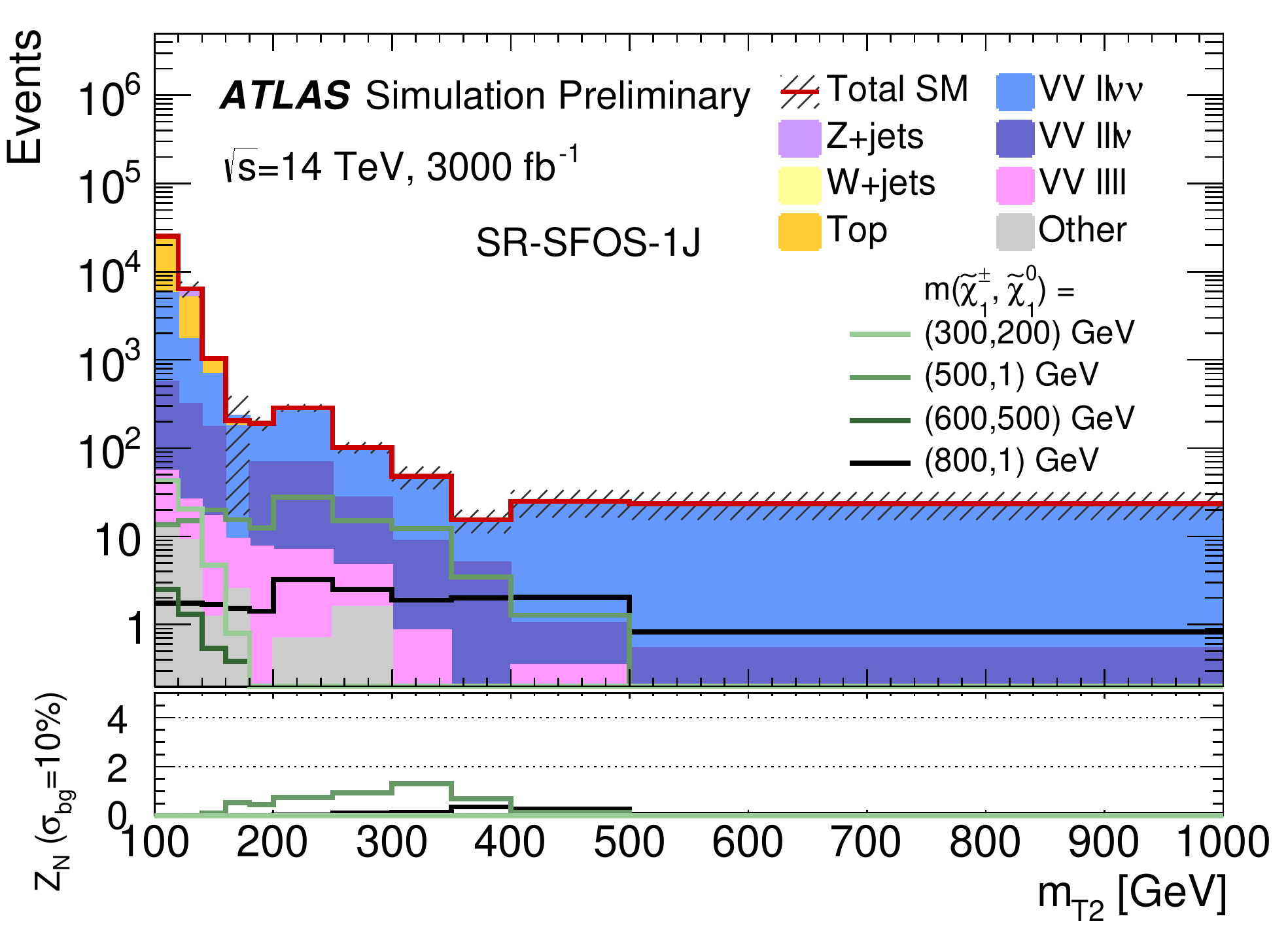}
\end{minipage}\\
\begin{minipage}[t][\minipageheightdp][t]{\minipagewidthdp}
\centering
\includegraphics[width=\figuremaxwidthdp,height=\figuremaxheightdp,keepaspectratio]{\main/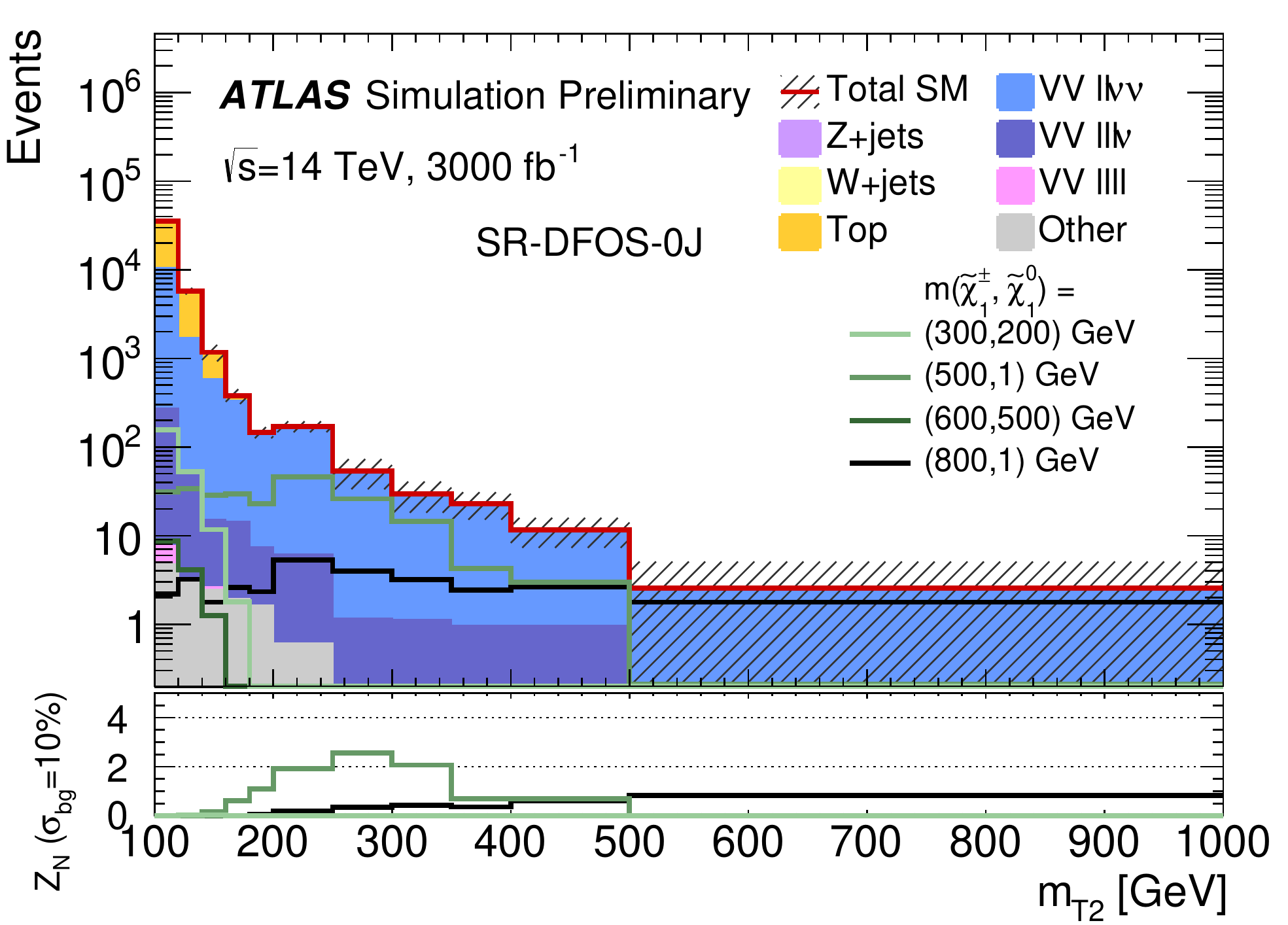}
\end{minipage}
\begin{minipage}[t][\minipageheightdp][t]{\minipagewidthdp}
\centering
\includegraphics[width=\figuremaxwidthdp,height=\figuremaxheightdp,keepaspectratio]{\main/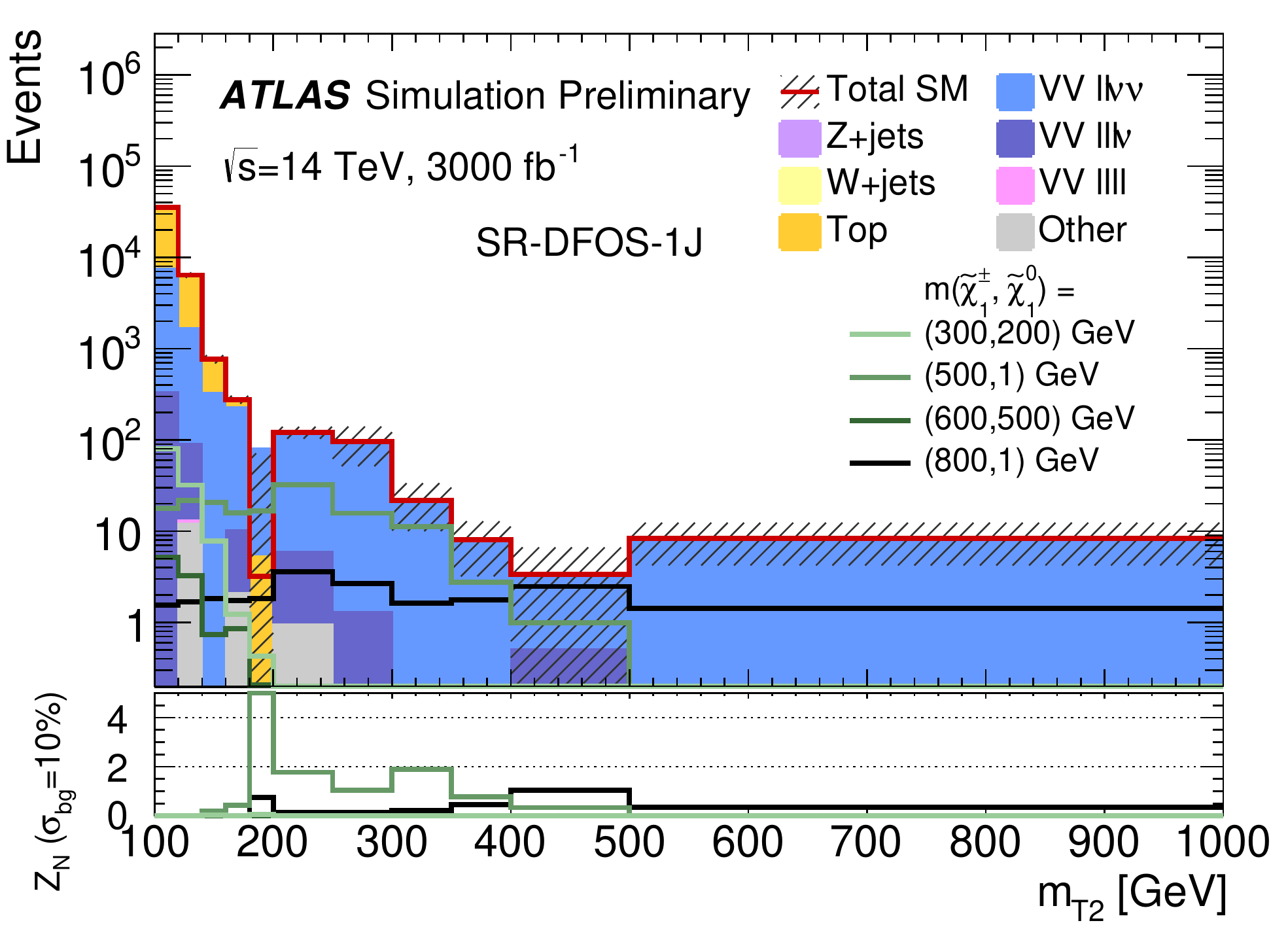}
\end{minipage}
\caption{\label{fig:charginos:yields}                                                                            
    Expected number of events from SM and SUSY processes in the signal regions optimised for $\tilde{\chi}^+_1\tilde{\chi}^-_1$ production, for the HL-LHC. Uncertainties shown are the MC statistical uncertainties only. 
   The lower pad in each plot shows the significance, $Z_N$ using a background uncertainty of $10\%$, for a selection of SUSY scenarios in each $m_\mathrm{T2}$ interval. 
   }
\end{figure*}

The signal region is divided into two disjoint regions with a Same Flavour Opposite Sign (SFOS: $e^+e^-$, $\mu^+\mu^-$) or Different Flavour Opposite Sign (DFOS: $e^{\pm}\mu^{\mp}$) lepton pair to take advantage of the differing SM background composition for each flavour combination. 
The SFOS and DFOS regions are divided again into events with exactly zero jets or one jet, which target scenarios with large or small $\tilde{\chi}^\pm_1-\tilde{\chi}^0_1$ mass splittings, respectively.
One lepton must have $p_{\mathrm{T}}>40\GeV$ to suppress the SM background, and with $p_{\mathrm{T}}^{\ell 1}>40\GeV$ and $p_{\mathrm{T}}^{\ell 2}>20\GeV$, either the single or double lepton triggers may be used to accept the event at the HL-LHC. 
Events with SFOS lepton pairs with an invariant mass within $30\GeV$ of the $Z$ boson mass are rejected to suppress the large $Z \rightarrow \ell\ell$ SM background.
Events with $E_{\mathrm{T}}^{\mathrm{miss}}$ larger than $110\GeV$ and $E_{\mathrm{T}}^{\mathrm{miss}}$ significance (defined as $E_{\mathrm{T}}^{\mathrm{miss}}/\sqrt{\sum \vec{p}_{\mathrm T}^{\mathrm{~leptons,\, jets}}}$) larger than $10\GeV^{1/2}$  are selected in to suppress $Z$+jets events with poorly measured leptons.

The stransverse mass $m_\mathrm{T2}$ is calculated using the two leptons and $E_{\mathrm{T}}^{\mathrm{miss}}$, and used as the main discriminator in the SR selection to suppress the SM background. 
For $t\bar{t}$ or $WW$ decays, assuming an ideal detector with perfect momentum resolution, $m_\mathrm{T2}(\ell,\ell,E_{\mathrm{T}}^{\mathrm{miss}})$ has a kinematic endpoint at the mass of the $W$ boson. 
Signal models with sufficient mass splittings between the $\tilde{\chi}^\pm_1$ and the $\tilde{\chi}^0_1$ feature $m_\mathrm{T2}$ distributions that extend beyond this kinematic endpoint expected for the dominant SM backgrounds. 
Therefore, events in this search are required to have high $m_\mathrm{T2}$ values. 
A set of disjoint signal regions ``binned'' in $m_\mathrm{T2}$ are used to maximise model-dependent exclusion sensitivity.
Each SR is identified by the lepton flavour combination (SFOS or DFOS), number of jets (-0J or -1J) and the range of the $m_\mathrm{T2}$ interval. 
Ten high $m_\mathrm{T2}$ intervals: $[120,140]$, $[140,160]$,  $[160,180]$, $[180,200]$, $[200,250]$, $[250,300]$, $[300,350]$, $[350,400]$, $[400,500]$ and $[500,\infty]$, are used to maximise the sensitivity to $\tilde{\chi}^+_1\tilde{\chi}^-_1$ production. 
After the application of the full selection criteria, no $Z$+jets or $W$+jets events remain.
The diboson process $WW$ is seen to dominate the total SM background across all signal regions, due to its similarity with the SUSY signal. 
The stransverse mass $m_\mathrm{T2}$ of SM and SUSY events in the signal regions is shown in \fig{fig:charginos:yields}, for events passing $m_\mathrm{T2}>100\GeV$. 
The SM background drops off at lower $m_\mathrm{T2}$ values (around the $W$ mass), while the SUSY signal is seen to have long tails to high $m_\mathrm{T2}$ values. 
The $2\ell$ diboson SM processes show long $m_\mathrm{T2}$ tails, which is mostly from $ZZ \rightarrow \ell^+\ell^- \nu\bar{\nu}$ production; a small contribution from $WW$ will be present due to the imperfect measurement of the leptons and $E_{\mathrm{T}}^{\mathrm{miss}}$. 
\begin{figure*}[t]
\centering
\begin{minipage}[t][\minipageheightsp][t]{\minipagewidthsp}
\centering
\includegraphics[width=\figuremaxwidthsp,height=\figuremaxheightsp,keepaspectratio]{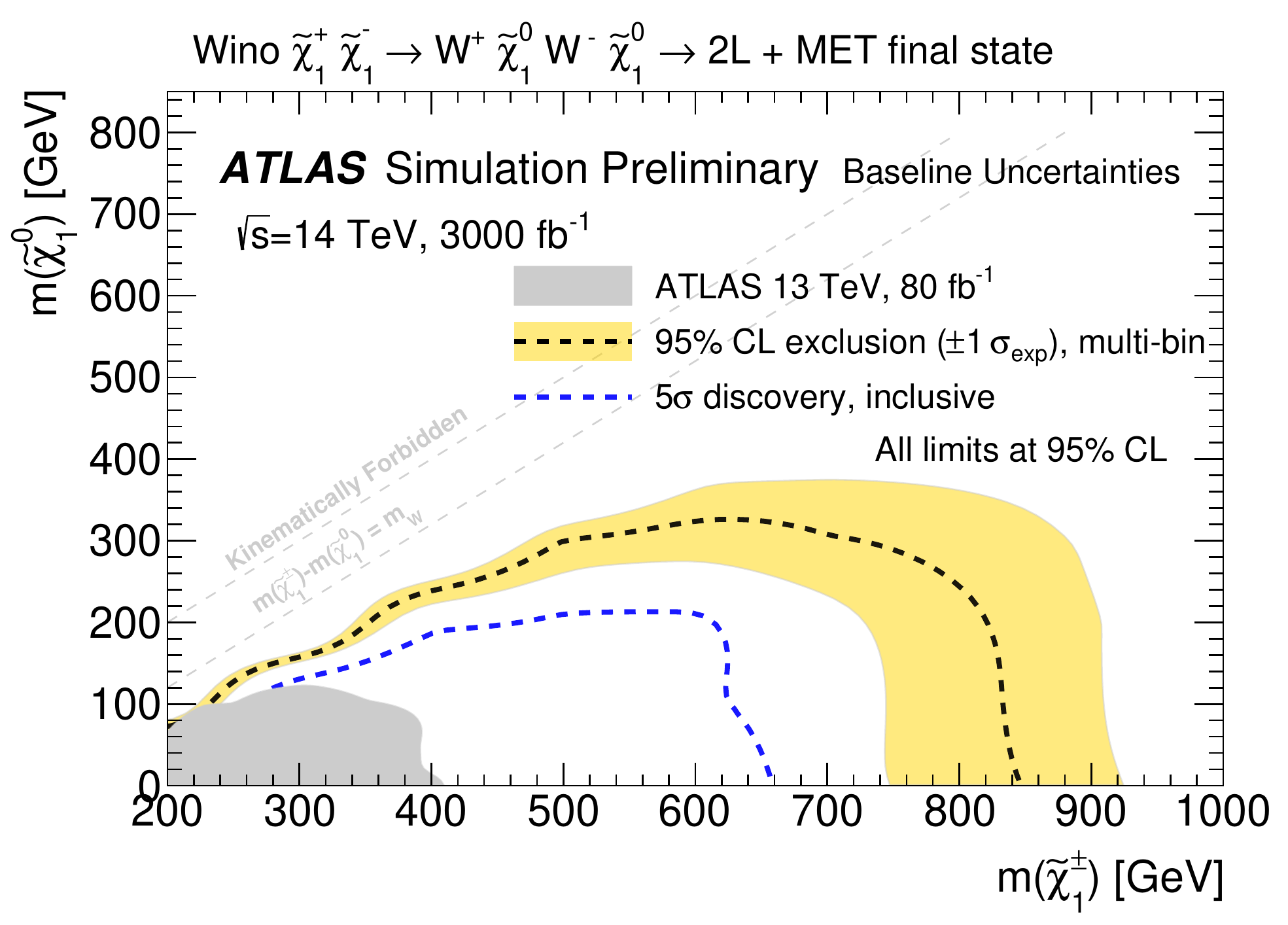}
\end{minipage}
 \caption{\label{fig:charginos:limit}                                                                            
    The $95\%\cl$ exclusion and discovery potential for $\tilde{\chi}^+_1\tilde{\chi}^-_1$ production at the HL-LHC (\intLumi at $\sqrt{s}=14\TeV$), assuming $\tilde{\chi}^\pm_1 \rightarrow W \tilde{\chi}^0_1$ with a BR of $100\%$, for an uncertainty on the modelling of the SM background of $5\%$ (baseline uncertainty). 
    The observed limits from the analyses of $13\TeV$ data~\cite{ATLAS:2018gfq} are also shown. }
\end{figure*}	

To calculate the expected sensitivity to $\tilde{\chi}^+_1\tilde{\chi}^-_1$ production and decay via $W$ bosons, the uncertainties from the normalisation of the $WW$ background are assumed to scale inversely with the increase in luminosity, and thus decrease to $\sim1\%$, while a better understanding of $WW$ could halve the theoretical uncertainties to $\sim 2.5-5\%$.
It is assumed that the experimental uncertainties will be understood to the same level, or better, than the $13\TeV$ analysis~\cite{ATLAS:2018gfq}. 
Two scenarios are considered for $\tilde{\chi}^+_1\tilde{\chi}^-_1$ production and decay via $W$ bosons at the HL-LHC: the so-called Run-2 scenario, with $5\%$ experimental uncertainty on the signal and SM background, a $10\%$ theoretical uncertainty on the signal, and a $10\%$ modelling uncertainty on the SM background, and the so-called baseline scenario, where the $WW$ theoretical uncertainty can be understood to a better level and modelling uncertainty on SM background halves to $5\%$.

The statistical combination of all disjoint signal regions is used to set model-dependent exclusion limits. 
For each of the three uncertainties considered, half of the value is treated as correlated across signal regions, and the other half as uncorrelated. 
The exclusion potentials for $\tilde{\chi}^+_1\tilde{\chi}^-_1$ production and decay via $W$ bosons at the HL-LHC are shown in \fig{fig:charginos:limit} for the baseline scenario.  
In the absence of an excess, $\tilde{\chi}^+_1\tilde{\chi}^-_1$ production may be excluded up to $840\GeV$ in $\tilde{\chi}^\pm_1$ mass. 
For the Run-2, where the modelling uncertainty on the SM background are raised from $5\%$ to $10\%$, the expected
exclusion potential decreases by $10\GeV$ in $\tilde{\chi}^\pm_1$ mass and $30\GeV$ in $\tilde{\chi}^0_1$ mass. 
To calculate the discovery potential, eleven inclusive signal regions are defined with $m_\mathrm{T2}$ larger than the lower bound of each $m_\mathrm{T2}$ interval, and the inclusive signal region with the best expected sensitivity is used. 
At the HL-LHC, the discovery potential reaches up to $660\GeV$ in $\tilde{\chi}^\pm_1$ mass with the baseline scenario assumption for the background modelling uncertainty, and it decreases by $30\GeV$ in $\tilde{\chi}^\pm_1$ mass and $60\GeV$ in $\tilde{\chi}^0_1$ mass if uncertainties doubled.

\subsubsection{%\atlas{Prospects for Chargino-Neutralino searches in multileptons at HL-LHC}
\atlasC{Chargino-Neutralino searches in multileptons at HL-LHC}}
\contributors{A. De Santo, B. Safarzadeh Samani, F. Trovato, ATLAS}
%{\bf Author(s): A. de Santo, DE SANTO, Antonella (Sussex), B. Safarzadeh Samani (Sussex), F. Trovato (Sussex)}

\begin{table}[!t]
\centering
   \small{
     \begin{tabular}{ c |cc|cc  }
  \hline
  \multicolumn{5}{c}{Common}\\
  \hline
lepton flavour/sign & \multicolumn{4}{|c}{$e^+e^-\ell^\pm$ or $\mu^+\mu^-\ell^\pm$} \\
%$p_{\mathrm T}^{\textrm lep 1}, p_{\mathrm T}^{\textrm lep 2}, p_{\mathrm T}^{\textrm lep 3} > $ [GeV] &   \multicolumn{4}{|c}{$25,25,20$} \\
%$|m_\mathrm{SFOS} - m_Z| < $ [GeV] & \multicolumn{4}{|c}{10} \\
%$m_{\ell\ell\ell} > $ [GeV] & \multicolumn{4}{|c}{$20$} \\
%number of $b$ jets &  \multicolumn{4}{|c}{$=0$} \\
\hline
  & \multicolumn{2}{|c}{SR-0J} & \multicolumn{2}{|c}{SR-1J} \\
\hline
number of jets & \multicolumn{2}{|c}{$=0$} & \multicolumn{2}{|c}{$\geq 1$} \\
\hline
Binned SR & $m_{\mathrm T}^{\mathrm{min}}~[\mathrm{GeV}]$  & $E_{\mathrm{T}}^{\mathrm{miss}}~[\mathrm{GeV}]$ & $m_{\mathrm T}^{\mathrm{min}}~[\mathrm{GeV}]$ & $E_{\mathrm{T}}^{\mathrm{miss}}~[\mathrm{GeV}]$\\
\hline
    & $\in [150,250]$ &  $\in [200,250]$ & $\in [150,250]$ &  $\in [200,250]$ \\
    &                 &  $\in [250,350]$    &                 &  $\in [250,350]$ \\
    &                 &  $\in [350,450]$    &                 &  $\in [350,450]$ \\
    &                 &  $\in [450,\infty]$ &                 &  $\in [450,600]$ \\
    &                 &                     &                 &  $\in [600,\infty]$ \\
\hline
    & $\in [250,400]$ &  $\in [150,250]$    & $\in [250,400]$ &  $\in [150,250]$ \\
    &                 &  $\in [250,350]$    &                 &  $\in [250,350]$ \\
    &                 &  $\in [350,500]$    &                 &  $\in [350,500]$ \\
    &                 &  $\in [500,\infty]$ &                 &  $\in [500,\infty]$ \\
\hline
    & $\in [400,\infty]$ &  $\in [150,350]$    & $\in [400,\infty]$ &  $\in [150,350]$ \\
    &                    &  $\in [350,450]$    &                    &  $\in [350,450]$ \\
    &                    &  $\in [450,600]$    &                    &  $\in [450,600]$ \\
    &                    &  $\in [600,\infty]$ &                    &  $\in [600,\infty]$ \\
\hline
\end{tabular}}\caption{ Signal regions for the chargino/next-to-lightest neutralino production analysis.  \label{tab:c1n23L:SRdefs}}
\end{table}

Charginos and next-to-lightest neutralinos decaying via $W$ and $Z$ or Higgs bosons and LSP are searched for using three-lepton signatures characterised by large missing transverse momentum~\cite{ATL-PHYS-PUB-2018-048}. %Sensitivity to higgs decays from $WW$ and $\tau\tau$ decay modes is exploited.
A simplified model describing the direct production of $\tilde{\chi}^\pm_1\tilde{\chi}^0_2$ is studied here, where the $\tilde{\chi}^\pm_1$ and $\tilde{\chi}^0_2$ are assumed to be pure wino and equal mass, while the $\tilde{\chi}^0_1$ is the LSP and is assumed to be pure bino and stable.
The selection for $\tilde{\chi}^\pm_1\tilde{\chi}^0_2 \rightarrow W\tilde{\chi}^0_1 Z \tilde{\chi}^0_1$ at the HL-LHC follows the strategy used in the $13\TeV$ search~\cite{Aaboud:2018jiw}. 
Events are selected with exactly three leptons (electrons or muons) with $p_{\mathrm T} > 20\GeV$ and $|\eta|<2.5$, two of which must form an SFOS pair consistent with a $Z$ boson decay and have \sloppy\mbox{$|m_{\ell\ell}-m_Z|<10\GeV$}.
To resolve ambiguities when multiple SFOS pairings are present, the transverse mass $m_{\mathrm T}$ is calculated using the unpaired lepton for each possible SFOS pairing, and the combination that minimises the transverse mass, $m_{\mathrm T}^{\mathrm{min}}$, is chosen. 
The two leading leptons must have $p_{\mathrm T}>25\GeV$, and $m_{\ell\ell\ell}$ must be larger than $20\GeV$ to reject low mass SM decays. 
To suppress the $t\bar{t}$ background, events are vetoed if they contain $b$-tagged jets with $p_{\mathrm T} > 30\GeV$, while the $Z$+jets background is suppressed by
requiring $E_{\mathrm{T}}^{\mathrm{miss}}> 50\GeV$.

\begin{figure*}[t]
\centering
\begin{minipage}[t][\minipageheightdp][t]{\minipagewidthdp}
\centering
\includegraphics[width=\figuremaxwidthdp,height=\figuremaxheightdp,keepaspectratio]{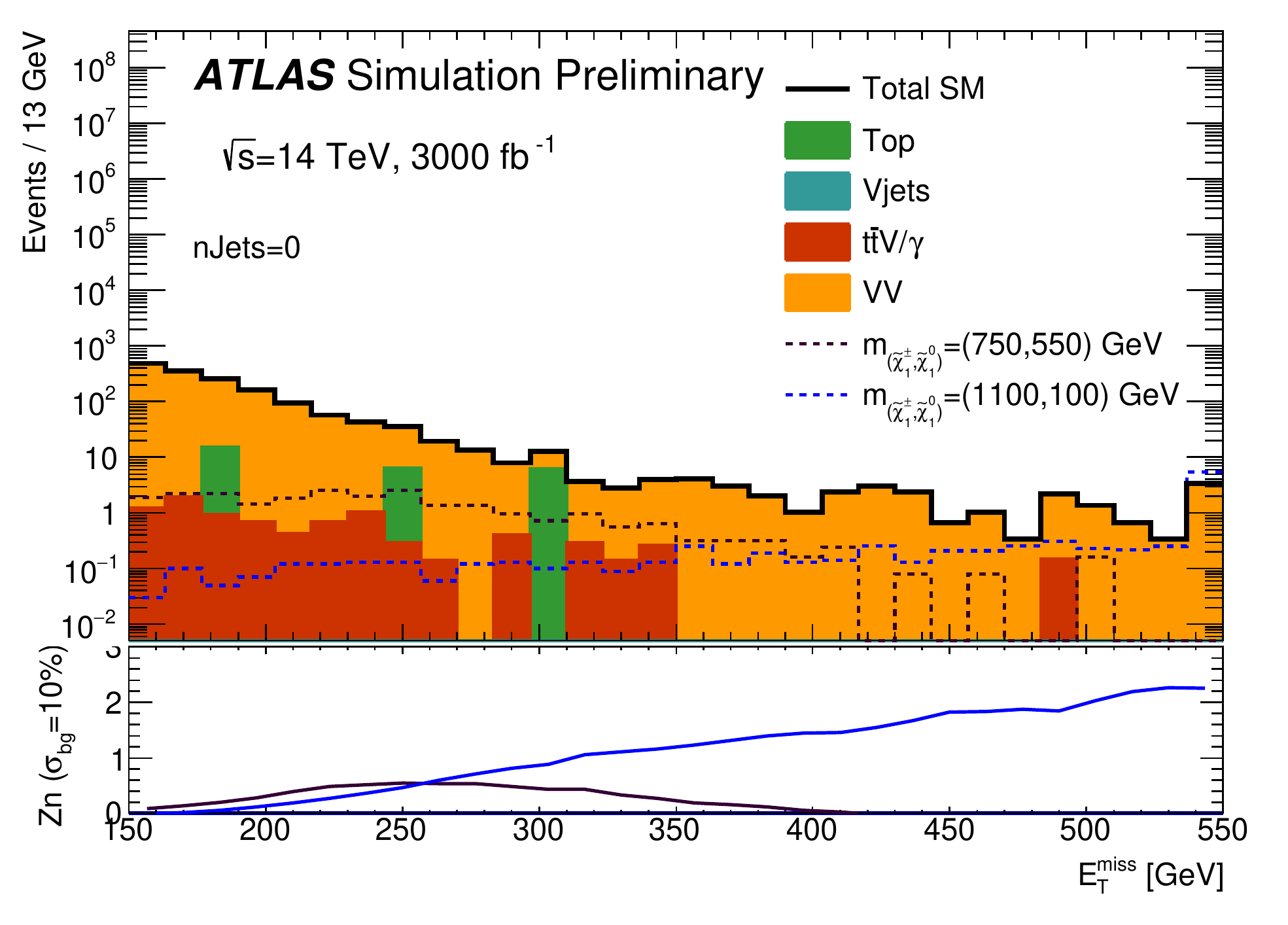}
\end{minipage}
\begin{minipage}[t][\minipageheightdp][t]{\minipagewidthdp}
\centering
\includegraphics[width=\figuremaxwidthdp,height=\figuremaxheightdp,keepaspectratio]{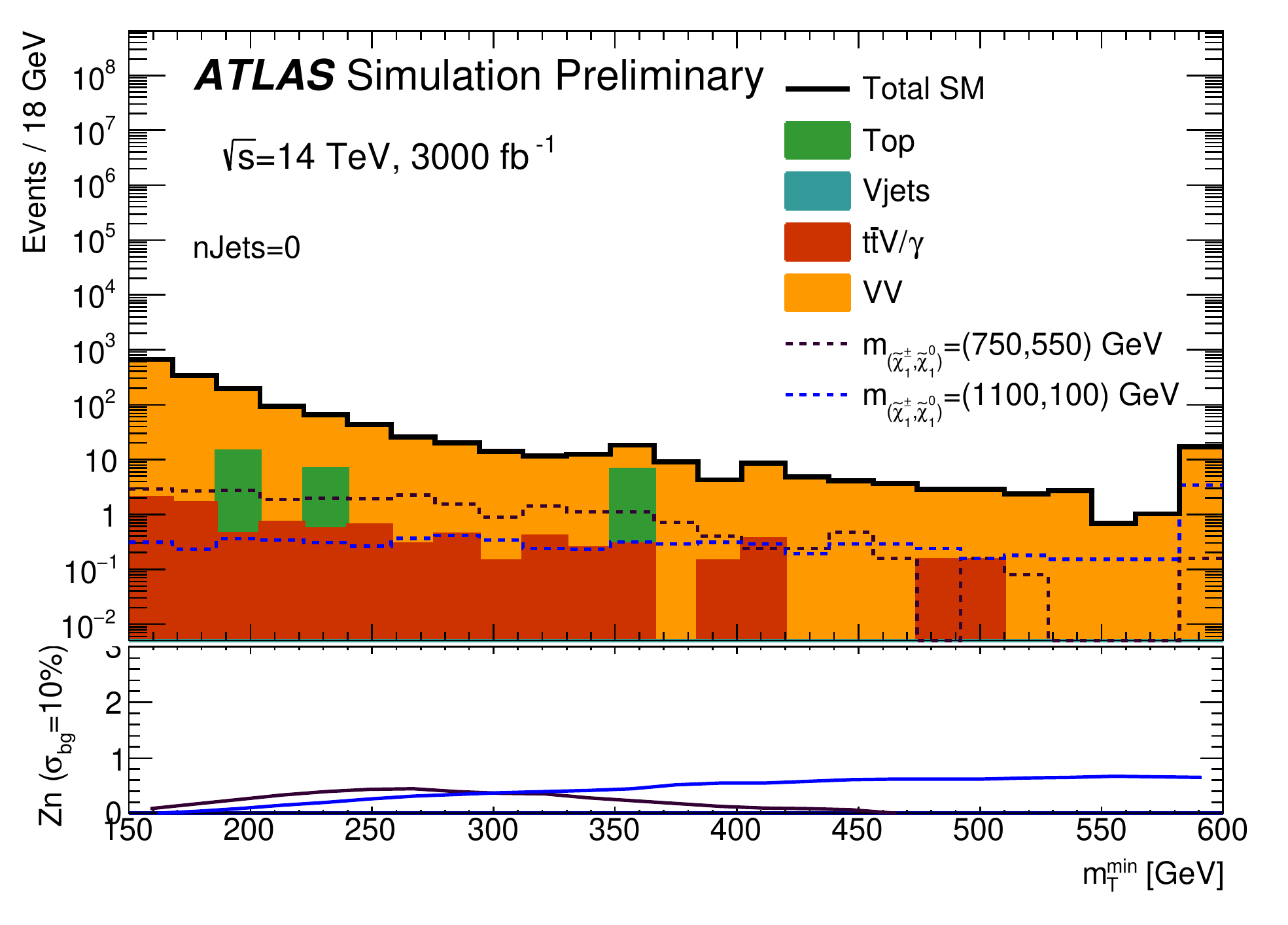}
\end{minipage}\\
\begin{minipage}[t][\minipageheightdp][t]{\minipagewidthdp}
\centering
\includegraphics[width=\figuremaxwidthdp,height=\figuremaxheightdp,keepaspectratio]{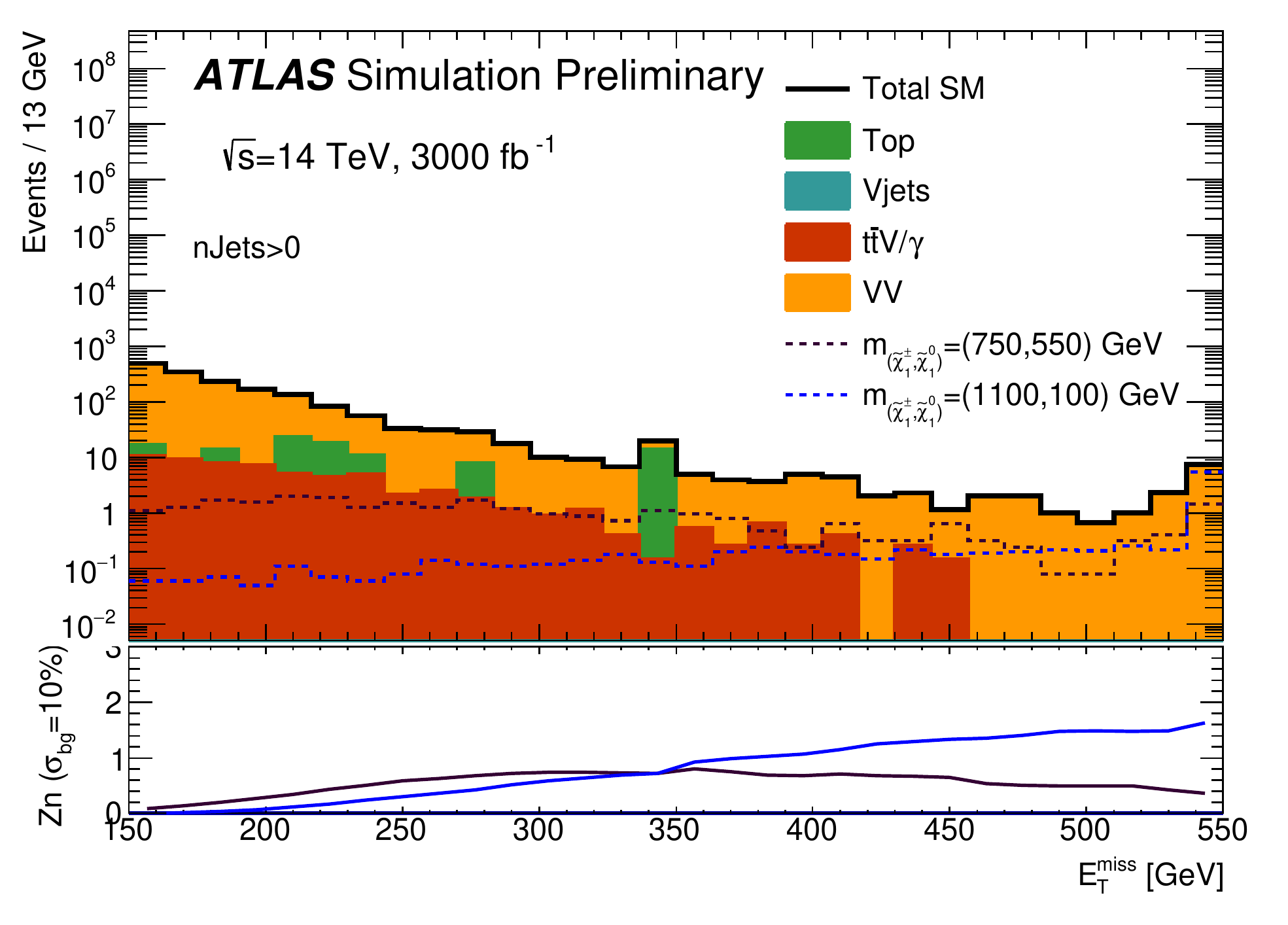}
\end{minipage}
\begin{minipage}[t][\minipageheightdp][t]{\minipagewidthdp}
\centering
\includegraphics[width=\figuremaxwidthdp,height=\figuremaxheightdp,keepaspectratio]{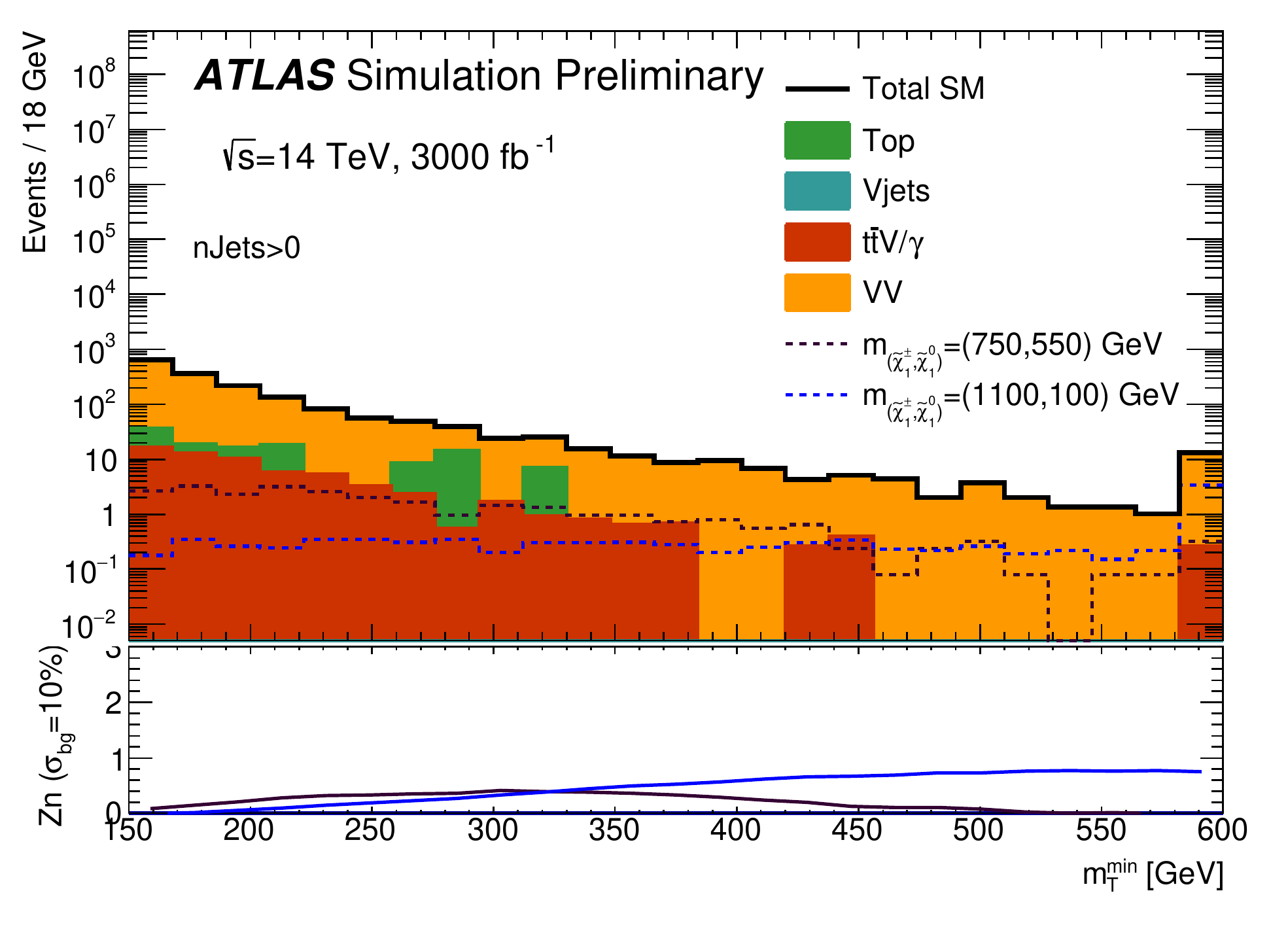}
\end{minipage}
%  \centering
%\includegraphics[width=0.48\textwidth]{}
%\includegraphics[width=0.48\textwidth]{}\\
%\includegraphics[width=0.48\textwidth]{}
%\includegraphics[width=0.48\textwidth]{}
 \caption{Distribution of $E_{\mathrm{T}}^{\mathrm{miss}}$ and $m_{\mathrm T}^{\mathrm{min}} $ in the events with zero jets (top) and the events with at least one jet (bottom). All baseline requirements along with the $E_{\mathrm{T}}^{\mathrm{miss}}$ and  $m_{\mathrm T}^{\mathrm{min}} $ selections of $150\GeV$ are applied. The lower pad in each plot shows the significance, $Z_N$ using a background uncertainty of $10\%$, for the SUSY reference points.}
  \label{fig:nJ}
\end{figure*}

A set of disjoint signal regions binned in $m_{\mathrm T}^{\mathrm{min}}$ and $E_{\mathrm{T}}^{\mathrm{miss}}$ are used to maximise model-dependent exclusion sensitivity. 
Each SR is identified by the number of jets (-0J or -1J), the range of the $E_{\mathrm{T}}^{\mathrm{miss}}$ interval and the range of the $m_{\mathrm T}^{\mathrm{min}}$ interval, as seen in \tabb{tab:c1n23L:SRdefs}. 
The SRs with at least one jet are defined to extend the sensitivity for the signal benchmark points in which the mass differences between the $\tilde{\chi}^+_1$ and $ \tilde{\chi}^0_1$ is small. In such scenarios higher $E_{\mathrm{T}}^{\mathrm{miss}}$in the event is expected when the $\tilde{\chi}^\pm_1\tilde{\chi}^0_2$ system recoils against the initial-state-radiation jets.   
The distribution of $E_{\mathrm{T}}^{\mathrm{miss}}$ and $m_{\mathrm T}^{\mathrm{min}}$ in the 0-jet and 1-jet categories are shown in \fig{fig:nJ} for events with $E_{\mathrm{T}}^{\mathrm{miss}}> 150$ GeV and $m_{\mathrm T}^{\mathrm{min}}>150\GeV$.

\begin{figure*}[t]
\centering
\begin{minipage}[t][\minipageheightsp][t]{\minipagewidthsp}
\centering
\includegraphics[width=\figuremaxwidthsp,height=\figuremaxheightsp,keepaspectratio]{\main/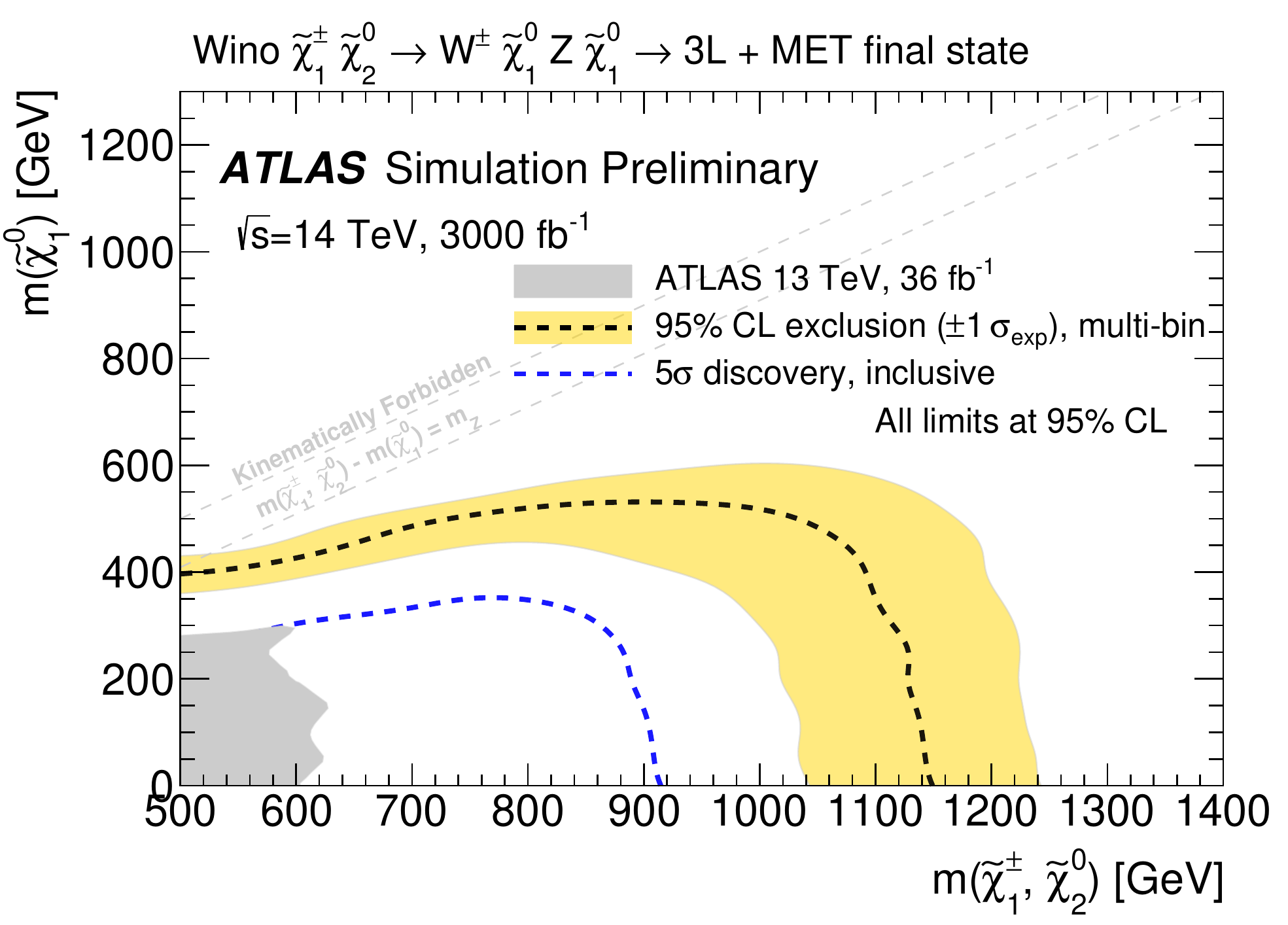}
\end{minipage}
 \caption{Expected exclusion limit and discovery potential on SUSY simplified models for $(\tilde{\chi}^\pm_1/\tilde{\chi}^0_2)$ production with decays via W/Z bosons, assuming $15\%$ uncertainty on the modelling of the SM backgrounds.  \label{fig:c1n23L_limit}
}

\end{figure*}

To calculate the expected sensitivity to $(\tilde{\chi}^\pm_1/\tilde{\chi}^0_2)$ production, a $5\%$ experimental uncertainty on the SM background and signal, a $10\%$ theoretical uncertainty on the signal, and a $10\%$ modelling uncertainty on the SM are assumed. 
With these uncertainty assumptions, \fig{fig:c1n23L_limit} shows the expected exclusion for $\tilde{\chi}^\pm_1\tilde{\chi}^0_2 \rightarrow W\tilde{\chi}^0_1 Z\tilde{\chi}^0_1$. 
In the absence of an excess, chargino and neutralino masses up to $1150\GeV$ may be excluded. 
The discovery potential is also shown in \fig{fig:c1n23L_limit}, which reaches up to $920\GeV$ in chargino and neutralino masses. 

\subsubsection{%\atlas{Search for Chargino-Neutralino production in the $Wh\rightarrow \ell b\bar{b}$ channel}
\atlasC{Chargino-Neutralino production in the $Wh\rightarrow \ell \nu \, b\bar{b}$ channel at HL-LHC}
}
\contributors{D. Bogavac, M. D'Onofrio, Y. Gao, M. Sullivan,  H. Teagle, ATLAS}
%{\bf Author(s): M. Sullivan (Liverpool), M. D'Onofrio (Liverpool), Y. Gao (Liverpool), D. Bogavac (Munich LMU)}

%Chargino and next-to-lightest neutralinos can be searched for in one lepton plus b-jets final state events if the next-to-lightest neutralino decays into a SM-like Higgs boson and the LSP. The Higgs decay mode into two b-quarks is exploited. \emph{Status: in ATLAS approval.}

Chargino and next-to-lightest neutralinos can be searched for in one lepton plus $b$-jets final state events if the next-to-lightest neutralino decays into a SM-like Higgs boson and the LSP~\cite{ATL-PHYS-PUB-2018-048}. The Higgs decay mode into two $b$-quarks is exploited. 
Signal models with $\tilde{\chi}^\pm_1$ and $\tilde{\chi}^0_2$ masses
up to $1.5\TeV$ are considered in this search. 
The analysis is performed separately in three signal regions
targeting signal models with different values of
mass difference $\Delta m = m(\tilde{\chi}^\pm_1/\tilde{\chi}^0_2) - m(\tilde{\chi}^0_1)$: low ($\Delta m < 300\GeV$),  medium ($\Delta m\in [300, 600]\GeV$) and high ($\Delta m > 600\GeV$).

The expected SM background is dominated by  
top quark pair-production $t\bar{t}$ and single top production,
with smaller contributions from vector boson production W$+$jets,
associated production of $t\bar{t}$ and a vector boson $t\bar{t}V$ and dibosons. 

The event selection follows a similar strategy as in
the previous studies documented in \citeref{ATL-PHYS-PUB-2015-032}.
Candidate leptons (electrons or muons) are required to have
$p_\mathrm{T}>25\GeV$ and $\left| \eta \right| < 2.47 \; (2.7)$,
and pass ``tight" and ``medium" identification criteria
for electrons and muons respectively. 
Candidate jets are reconstructed
using the anti-$k_t$ algorithm with $R = 0.4$, are required to have $p_\mathrm{T}$
greater than $25\GeV$ and $\left| \eta \right| < 2.5$.
The jets tagged as originating from $b$-decays are required to pass the jet requirements described previously,
and pass the MV2c10 tagging algorithm operating at $77\%$ $b$-jet tagging efficiency.
Candidate jets and electrons are required to satisfy $\Delta R(e, \textrm{jet}) \textgreater 0.2$.
Any leptons within $\Delta R = 0.4$ of the remaining jet are removed.
The $\met$ at generator level is calculated as the vectorial
sum of the momenta of neutral weakly-interacting particles, in this case neutrinos and neutralinos. 

Events containing exactly one lepton, and two or three jets passing the 
above object definitions are selected.
Two of the jets are required to be $b$-tagged with the criteria defined above.
Four key variables are further used to discriminate signal from background: the invariant mass of the two $b$-tagged jets, $m_{bb}$, the transverse momentum imbalance in the event, $\met$, the transverse mass constructed using the leading lepton $p_\mathrm{T}$ and the $\met$, $m_{\mathrm{T}}$, and the contransverse mass constructed using the two $b$-tagged jets, $m_{\mathrm{CT}}$. 
The $m_{bb}$ is used to select events which have dijet masses within a window of the Higgs boson mass.
The transverse mass variable $m_{\mathrm{T}}$, defined from the $\met$ and the leading lepton in the event,
is effective at suppressing SM backgrounds
containing W bosons due to the expected kinematic endpoint
around the W boson mass assuming an ideal detector with perfect momentum resolution.
The contransverse mass variable $m_{\mathrm{CT}}$ is defined for the $b\bar{b}$ system as 
$m_{\mathrm{CT}} = 2 p_{\mathrm{T}}^{b_1} p_{\mathrm{T}}^{b_2} (1+\cos\Delta\phi_{bb})$, 
where $p_{\mathrm{T}}^{b_1}$ and $p_{\mathrm{T}}^{b_2}$ are transverse momenta
of the two leading $b-$jets and $\Delta\phi_{bb}$ is the azimuthal angle between them.
It is an effective variable to select Higgs boson decays into $b-$quarks and
to suppress the $t\bar{t}$ backgrounds. 

%%%%%%%%%%%%%%%%%%%%%%%
\begin{figure*}[t]
\centering
\begin{minipage}[t][\minipageheighttp][t]{\minipagewidthtp}
\centering
\includegraphics[width=\figuremaxwidthtp,height=\figuremaxheighttp,keepaspectratio]{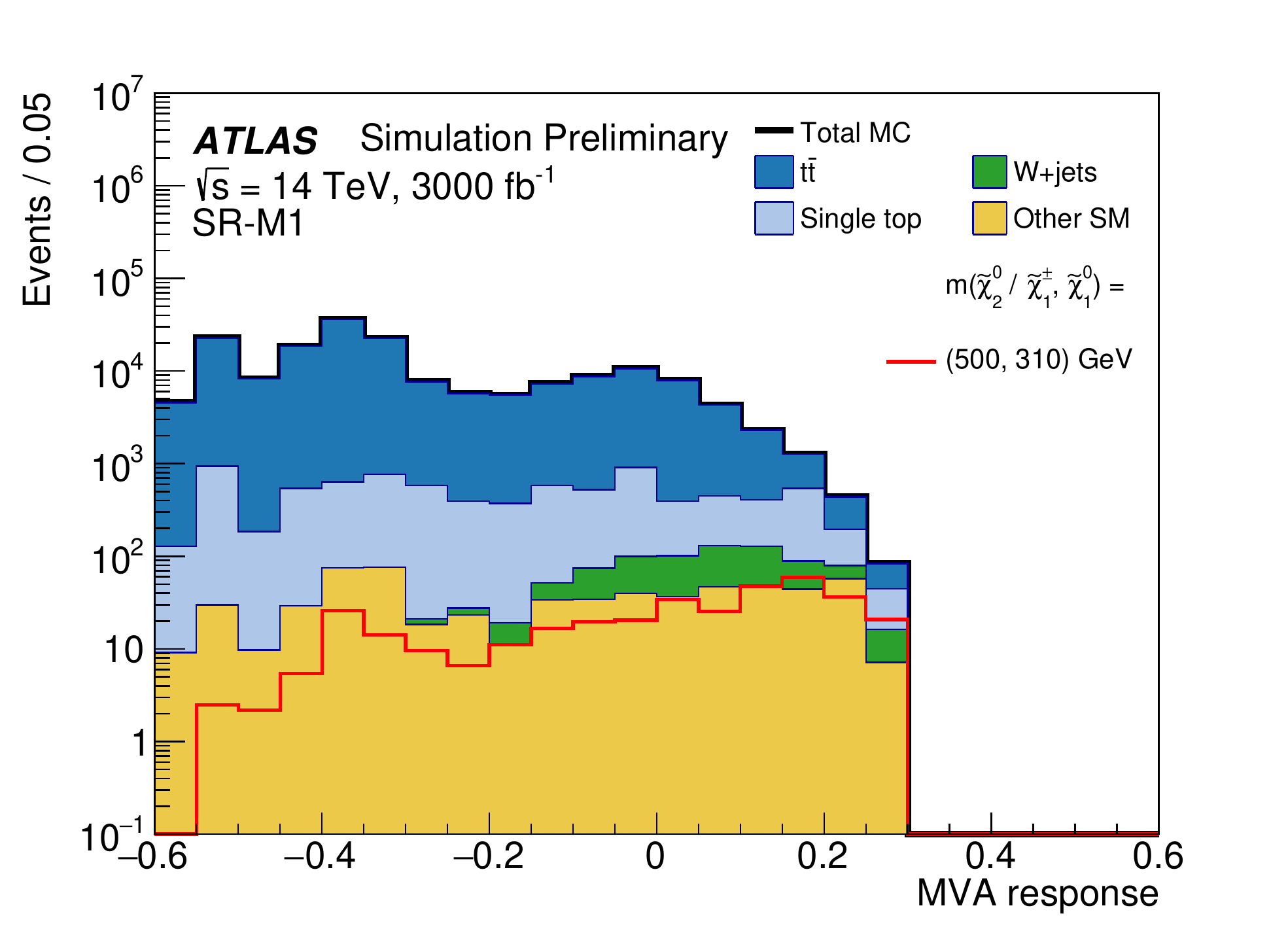}
\end{minipage}
\begin{minipage}[t][\minipageheighttp][t]{\minipagewidthtp}
\centering
\includegraphics[width=\figuremaxwidthtp,height=\figuremaxheighttp,keepaspectratio]{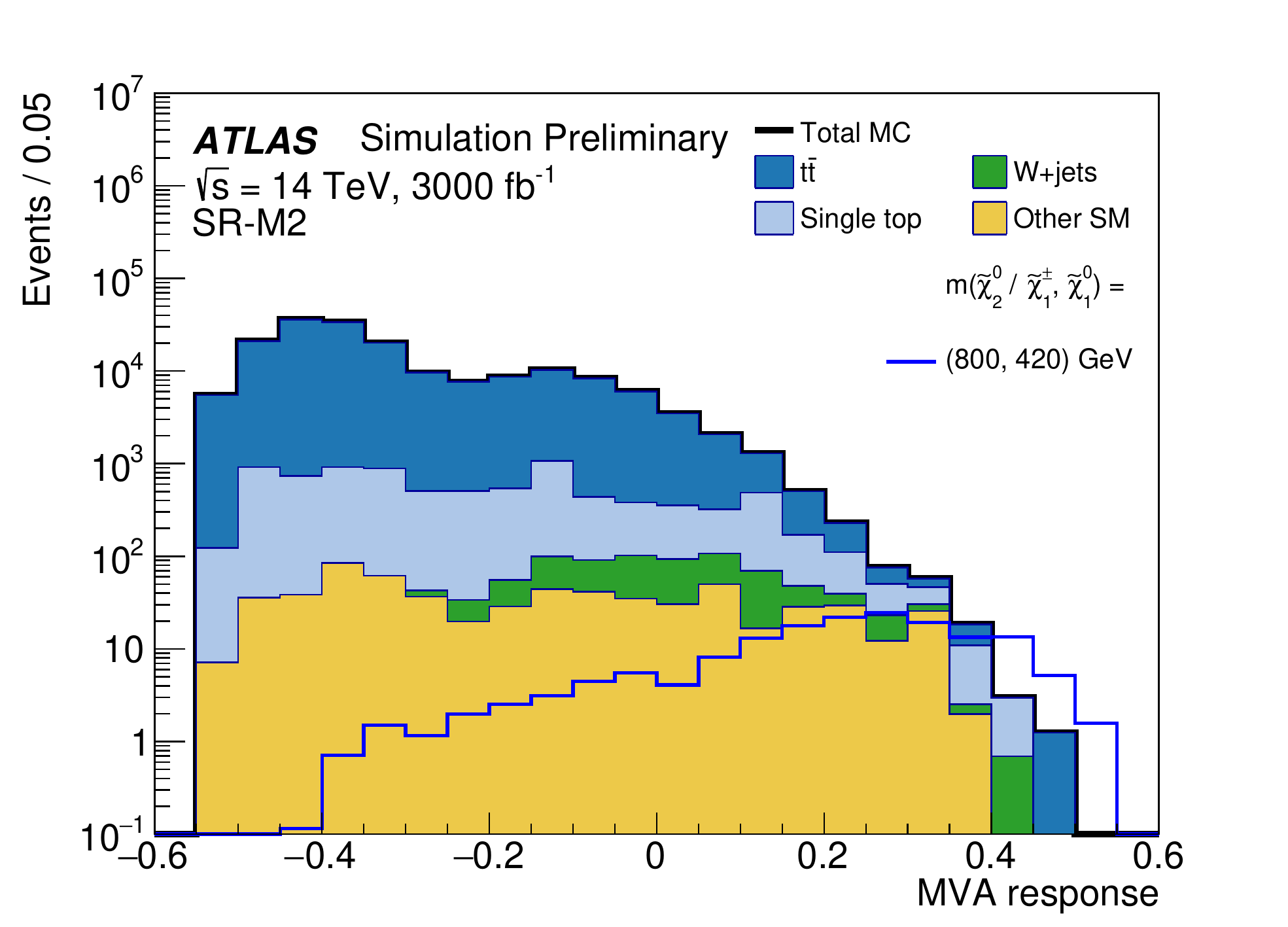}
\end{minipage}
\begin{minipage}[t][\minipageheighttp][t]{\minipagewidthtp}
\centering
\includegraphics[width=\figuremaxwidthtp,height=\figuremaxheighttp,keepaspectratio]{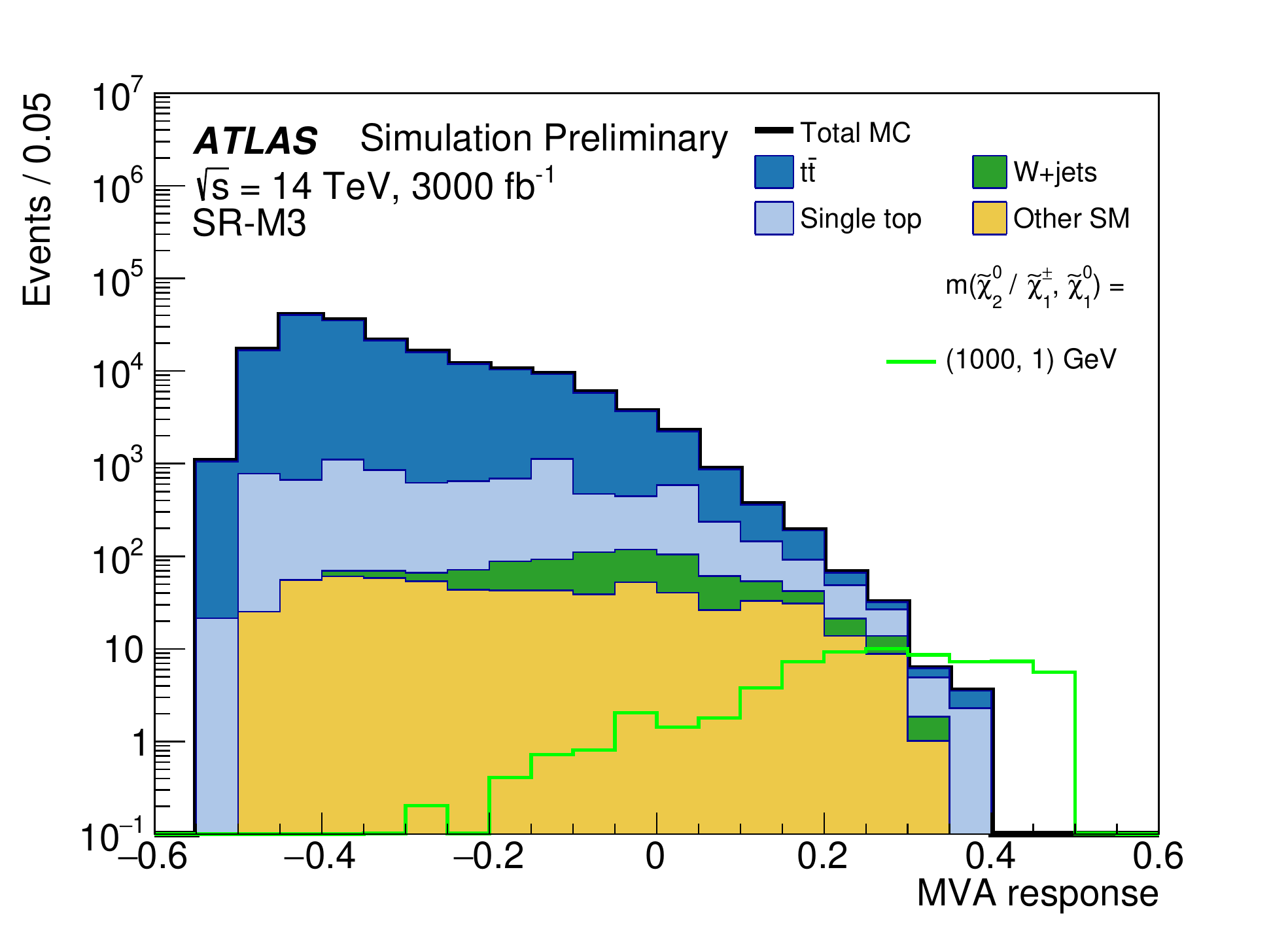}
\end{minipage}
%	\centering
%    \includegraphics[width=0.329\textwidth]{\main/section2SUSY/img/BDT1}
%    \includegraphics[width=0.329\textwidth]{\main/section2SUSY/img/BDT2}
%    \includegraphics[width=0.329\textwidth]{\main/section2SUSY/img/BDT3}
\caption{
Distributions of the BDT responses in the three signal regions for
the events that pass the preselection and are within $m_{bb}$ mass window of $[105, 135]\GeV$.
The contributions from all SM background are shown as stacked, and
the expected distribution from the benchmark signal models are overlaid. 
}
	\label{fig:C1N2_1Lbb_BDT_output}
\end{figure*}

A set of common loose requirements, referred to as preselection,
are applied first to suppress the fully hadronic multijet and $W$+jets backgrounds: 
$m_{\mathrm{T}}>40\GeV$, $m_{bb}>50\GeV$, $\met>200\GeV$.
%Figure~\ref{fig:variables} shows the distributions of the key
%discriminating variables after this selection, comparing
%the three benchmark signal models with the expected SM background.
%%%%%%%%%%%%%%%%%%%%%%%%%%%%%%
%\begin{figure}[h!]
%	\begin{center}	
%
%			\includegraphics[width=0.45\textwidth]{\main/section2SUSY/img/mbb_Preselection}
%			\includegraphics[width=0.45\textwidth]{\main/section2SUSY/img/mt_Preselection} \\
%			\includegraphics[width=0.45\textwidth]{\main/section2SUSY/img/mct_Preselection}	
%			\includegraphics[width=0.45\textwidth]{\main/section2SUSY/img/met_Preselection}
%	\end{center}
%	\caption{
%	Distributions of the key discriminating variables at the preselection level.
%  The contributions from the SM backgrounds are shown as stacked.
%  The distributions from three benchmark signal models as described in the text
%  are shown with overall normalisation comparable to the expected background for illustration.
%}
%	\label{fig:variables}
%\end{figure}
%%%%%%%%%%%%%%%%%%%%%%%%%%%%%%
%To further distinguish between signal and background processes,
%a set of rectangular selections based on these kinematic observables
%is first studied to evaluate possible optimal selections and 
%residual SM background. 
A multivariate method based on boosted decision trees (BDT) is then chosen for the optimal sensitivity.
In this approach, three independent BDTs (referred to as M1, M2 and M3),
are trained separately in each signal region
for events passing the preselection and within the $m_{bb}$ mass window of $[105, 135]\GeV$.  
In all regions, the following seven variables are used as inputs:
$\met$, $m_{\mathrm{T}}$, $m_{\mathrm{CT}}$, the leading lepton $p_\mathrm{T}$, the leading
and sub-leading b-jet $p_\mathrm{T}$, as well as the angular separation of the two b-jets $\Delta R(b_1, b_2)$.
The BDT output distributions are then used to optimise signal regions
maximising the expected significance $Z_N$ of the benchmark signal model.
Examples of the BDT output distributions are shown in \fig{fig:C1N2_1Lbb_BDT_output}.
The resulting signal region regions targeting models with low (SR-M1), medium (SR-M3) and high (SR-M3) $\Delta m$, are defined by requiring the BDT ranged larger than $0.25$, $0.35$ and $0.30$, respectively.

%\tabb{tab:yields_1Lbb_BDT} shows the expected number of events for the 
%SM background and three benchmark SUSY scenarios respectively.
The SM background is dominated by the top backgrounds, including
both the $t\bar{t}$ and single top processes.  The largest systematic uncertainties arise from the theoretical modelling
of the irreducible backgrounds of $t\bar{t}$ and single top,
mainly from the generator difference,
renormalisation and factorisation scale variations
and the interference between
the $t\bar{t}$ and single top background.
The total theoretical uncertainty is estimated to be about $7\%$.
Experimental uncertainties are dominated by the jet energy scale (JES) and
jet energy resolution (JER), on the order of $6\%$.
\begin{figure*}[t]
\centering
\begin{minipage}[t][\minipageheightsp][t]{\minipagewidthsp}
\centering
\includegraphics[width=\figuremaxwidthsp,height=\figuremaxheightsp,keepaspectratio]{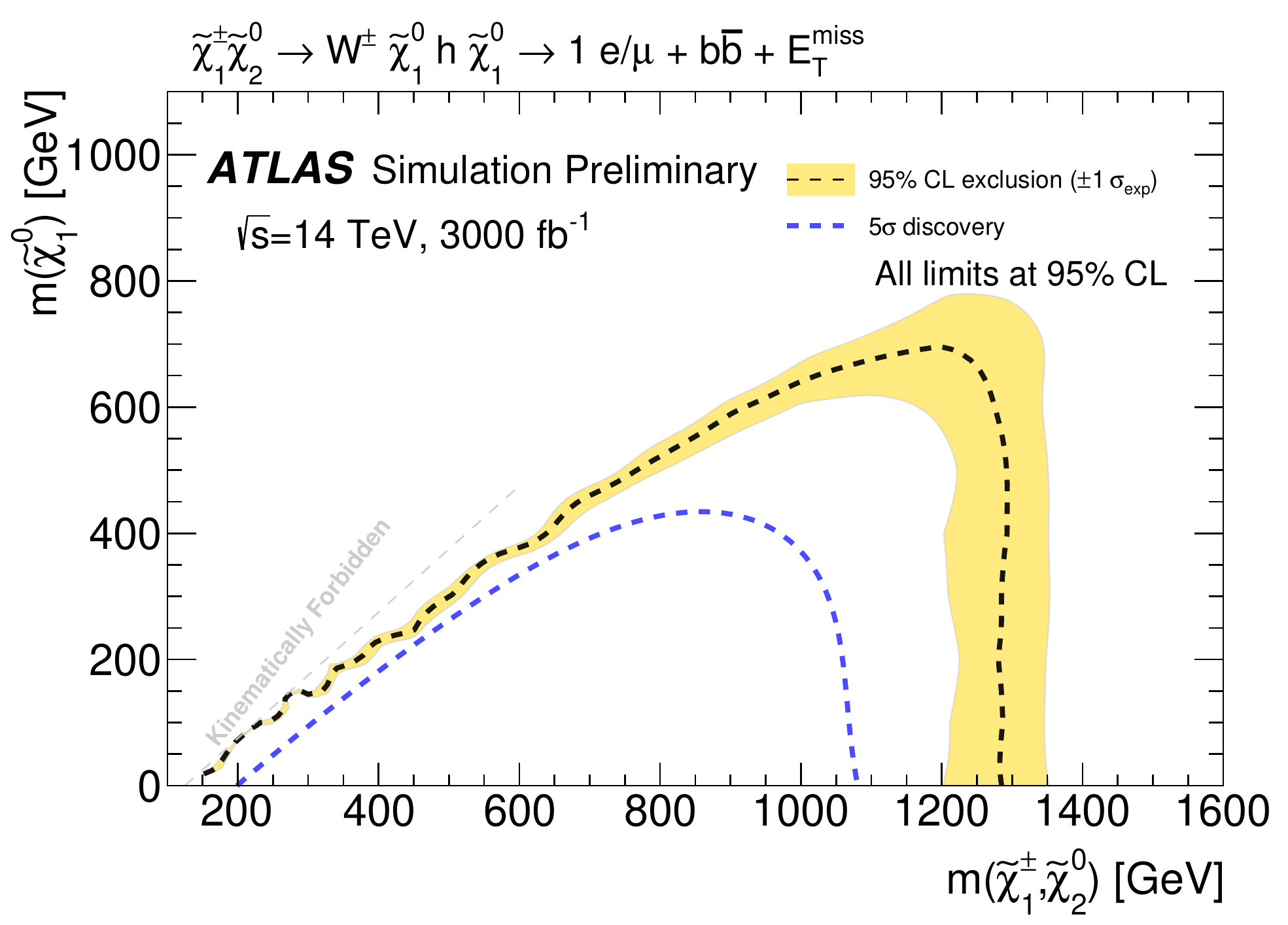}
\end{minipage}
	\caption{ 
  Expected $95\%\cl$ exclusion and $5\sigma$ discovery contours in the $m(\tilde{\chi}^0_1), m(\tilde{\chi}^\pm_1/\tilde{\chi}^0_2)$ masses plane
  for the $Wh$-mediated simplified model.  }
	\label{fig:C1N2_1Lbb_results}
\end{figure*}
Figure~\ref{fig:C1N2_1Lbb_results} shows the expected 
$95\%\cl$ exclusion and $5\sigma$ discovery contours 
for the simplified models described earlier.
In this model, masses of $\tilde{\chi}^\pm_1/\tilde{\chi}^0_2$ up to about
$1280\GeV$ are excluded at $95\%\cl$ for a massless $\tilde{\chi}^0_1$. 
The discovery potential at $5\sigma$ can be extended up to
$1080\GeV$ for a massless $\tilde{\chi}^0_1$. 

%\begin{table}[h!]
%	\centering
%	\caption{
%	Expected number of events for the SM background and 
%	the benchmark signal models in the signal regions SR-M1, SR-M2, and SR-M3.
%  The uncertainties are statistical only. 
%}
%\begin{tabular}{cccc}
%  \hline
%  Processes & SR-M1 & SR-M2 & SR-M3 \\
%  \hline 
% $t\bar t$        &  38.9 $\pm$ 8.4  & 8.7  $\pm$ 3.3 & 2.5 $\pm$ 1.8 \\
% single top       &  28.3 $\pm$ 4.8  & 10.7 $\pm$ 3.2 & 5.4 $\pm$ 2.5 \\
% W$+$jets         &  22.2 $\pm$ 5.4  &  3.0 $\pm$ 2.0 & 2.0 $\pm$ 1.8\\
% $ttV$            &  5.1  $\pm$ 2.4  &  2.0 $\pm$ 1.4 & 1.0 $\pm$ 1.0 \\
% Diboson          &  2.0  $\pm$ 2.0  &  -              & - \\
% \hline
% total background &  97 $\pm$ 12 &  24.4$\pm$ 5.2 & 10.9 $\pm$ 3.4 \\
% \hline
% $m(\tilde{\chi}^\pm_1/\tilde{\chi}^0_2, \tilde{\chi}^0_1) = (500, 310)$ GeV & 20.7 $\pm$ 4.8 & 4.6 $\pm$ 2.3 & 1.0 $\pm$ 1.0 \\
% $m(\tilde{\chi}^\pm_1/\tilde{\chi}^0_2, \tilde{\chi}^0_1) = (800, 420)$ GeV & 44.3 $\pm$ 2.3 & 33.6 $\pm$ 2.0 & 21.2 $\pm$ 1.6 \\
% $m(\tilde{\chi}^\pm_1/\tilde{\chi}^0_2, \tilde{\chi}^0_1) = (1000, 1)$  GeV & 32.2 $\pm$ 1.8 & 31.9 $\pm$ 1.8 & 28.9 $\pm$ 1.7 \\
%\hline
%	\end{tabular}
%	\label{tab:yields_1Lbb_BDT}
%\end{table}

\subsubsection{%\cms{Prospects for Chargino-Neutralino search with same-charge dilepton final states}
\cms{Chargino-Neutralino searches with same-charge dilepton final states at HL-LHC}}
\contributors{G. Zevi Della Porta, A. Canepa, CMS}
%{\bf Author(s): G. Zevi Della Porta, A. Canepa, CMS}

% Text based on the CMS endcap calorimeter TDR (CMS-TDR-17-007)

\begin{figure}[t]
\begin{center}
\includegraphics[width=0.3\textwidth]{\main/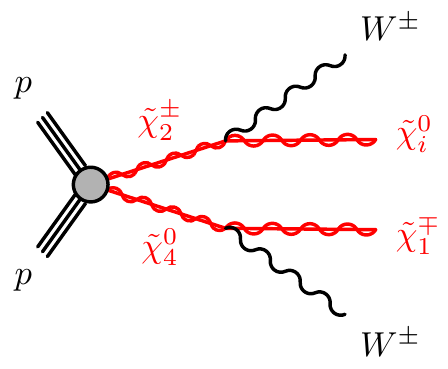}
\caption{Diagram for wino-like $\tw_2 \tz_4$ pair-production and decay into a final 
state with two same charge $W$ bosons.}
\label{fig:smsSS_TDR}\end{center}
\end{figure}

This section presents a search from CMS for the pair production of $\tw_2$, $\tz_4$ in the final
states with two same charge leptons, large $\met$ and modest jet activity.
The search is motivated by radiatively-driven natural 
supersymmetry (RNS) models, such as those presented in \secc{sec:BaerSection}. 
In these models, the mass spectra of the supersymmetric
partners of the gauge and Higgs bosons are characterised by low-mass higgsino-like $\tz_1$, 
$\tz_2$, $\tw_1$, and heavier bino-like $\tz_3$ along with mass-degenerate wino-like $\tw_2$, $\tz_4$. 
Two complementary analyses are designed to probe the wino and higgsino sectors of this model.
The final states resulting from higgsino production, discussed in \secc{sec:LightHiggsino},
are characterised by very low $\pt$ SM particles, due to the small mass difference between the 
low mass states and the $\tz_1$.
The final states resulting from wino production, discussed in this section, are expected to have a significant contribution
 of events (around $25\%$ of the total BR) where $\tw_2\tz_4$ decay into the higgsino sector 
emitting same-charge W bosons as in \fig{fig:smsSS_TDR}~\cite{Baer:2013xua, Baer:2016usl}.
This analysis is based on~\citeref{CMS-TDR-17-007}.

Estimates of signal and background yields are based on Monte Carlo samples followed by a \delphes{}
simulation~\cite{deFavereau:2013fsa} of the CMS Phase-2 detector. 
The signal samples are generated by \MGvATNLO (v2.3.3)~\cite{Alwall:2014hca} with up to two additional
jets at leading order precision. The  supersymmetric particles are then decayed by the 
\pythia{ 8.2}~\cite{Sjostrand:2014zea} package also providing showering and hadronisation. 
The cross-sections  for SUSY production have been  calculated for $\sqrt{s}=14\TeV$ at NLO-NLL
using the resumming code from \citeref{Fuks:2012qx,Fuks:2013vua} with CTEQ6.6 and  
MSTW2008nlo90cl PDFs. 
The background samples are generated with \madgraph{ 5} at LO, followed by parton showering and hadronisation with 
\pythia{ 6}~\cite{Sjostrand:2006za}. 
The \delphes{}-based yields of processes containing prompt leptons are corrected by the lepton reconstruction,
identification and isolation efficiencies measured in Run-2 collision data. For example, the reconstruction
efficiency for centrally produced electrons ranges from $60$ to $86\%$ for $\pt$ values between $20$ and $200\GeV$.
The \delphes{}-based yields of processes containing non-prompt leptons are increased by $25\%$, based on
\citeref{Sirunyan:2017lae}, to account for events with misidentified leptons from light flavour quarks, 
which are not included by \delphes{}~\cite{Sirunyan:2017lae}.

Candidate signal events are selected if they contain two high quality and isolated
leptons with $\pt\geq 20 \GeV$, $|\eta|\leq1.6$, and the same charge. 
Discrimination from the background processes is achieved 
by selecting events with no additional leptons with $\pt\geq 5\GeV$ 
and $|\eta| \leq 4.0$ (to suppress multi-boson production), and 
no  $\pt \geq 30 \GeV$ jets (to suppress events with top quarks).  
The remaining background processes include the pair 
production of W and Z/Z$^*$ bosons, as well as the W+jets and $t\bar{t}$ processes in 
association with a non-prompt or misidentified lepton. These are suppressed by 
imposing a tight selection on the $m_{\rm T, min}$ based on $\met$ and the $\pt$
of the leptons and defined as 
\begin{equation} 
m_{T,min} = \min[ m_T(\pt^{lep1}, \met), m_T(\pt^{lep2}, \met) ]. 
\end{equation} 

\begin{figure*}[t]
\centering
\begin{minipage}[t][\minipageheightsp][t]{\minipagewidthsp}
\centering
\includegraphics[width=\figuremaxwidthsp,height=\figuremaxheightsp,keepaspectratio]{\main/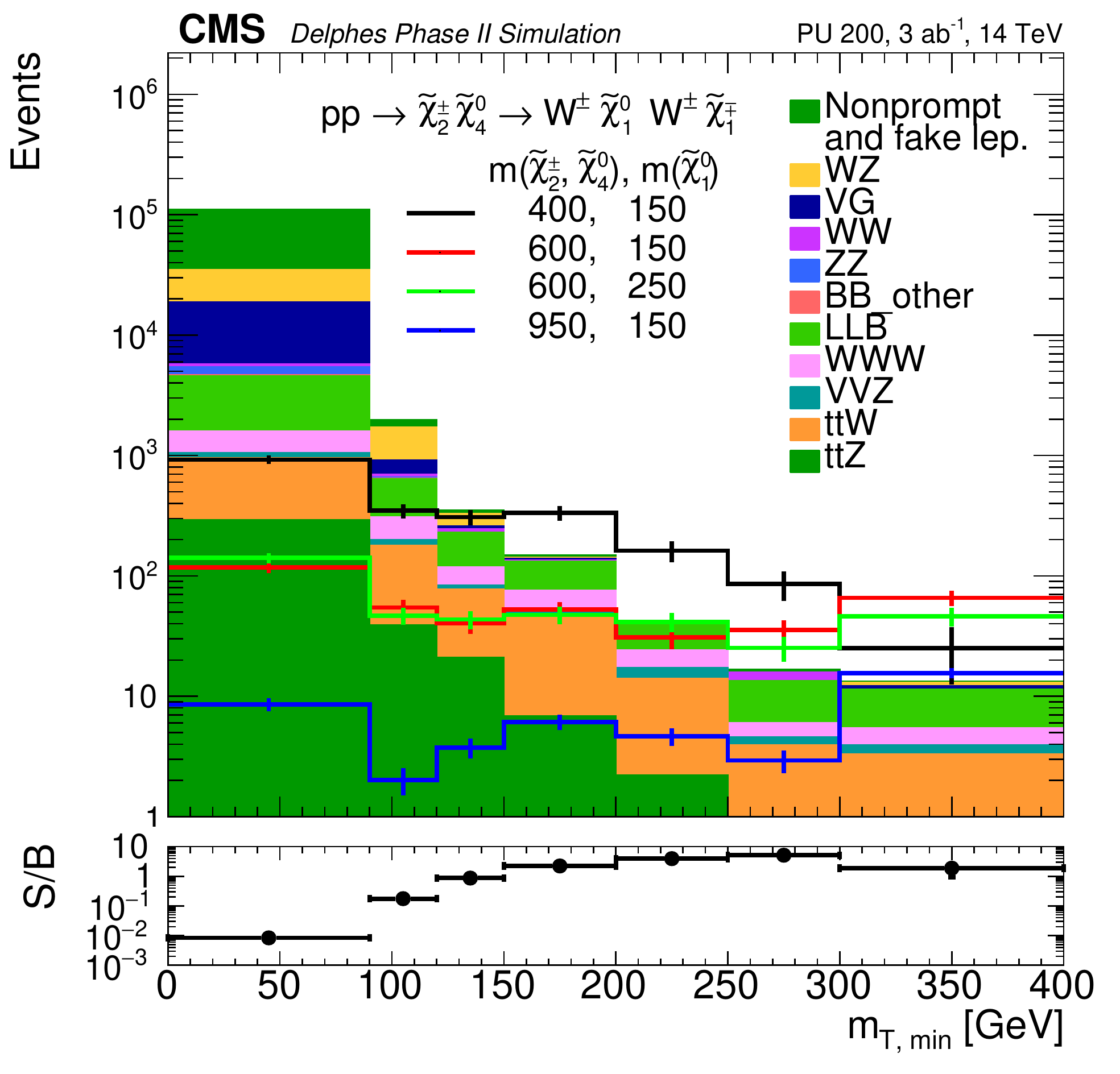}
\end{minipage}
\caption{Distribution of $m_{\rm T, min}$ in candidate events satisfying the signal region selection.
For coloured lines displaying the signal yields, the first number in the legend refers to the $\tw_2 \tz_4$ mass
while the second number refers to the  value of the $\tz_1$ mass.}
\label{fig:SS_limit_TDR_SS}
\end{figure*}

Figure~\ref{fig:SS_limit_TDR_SS} shows the distribution of the $m_{\rm T, min}$ 
observable in events satisfying the signal region selection
described above. To maximise the sensitivity, seven signal regions are then defined with 
 $m_{\rm T, min}$  in the ranges $[0, 90)$, $[90, 120)$, $[120, 150)$, $[150; 200)$, $[200; 250)$, 
 $[250; 300)$, and $[300; \infty)\GeV$. 

The search sensitivity is calculated using a modified frequentist approach with the 
CL$_\mathrm S$ criterion and asymptotic results for the test statistic~\cite{Junk:1999kv,Read:2002hq}. 
The systematic uncertainty on the prompt (fake, signal) yields is assumed to be 
$20\%$ ($50\%$, $20\%$) based on the estimates computed in the corresponding  search carried out 
in Run-2 collision data~\cite{Sirunyan:2017lae}.  

\begin{figure*}[t]
\centering
\begin{minipage}[t][\minipageheightsp][t]{\minipagewidthsp}
\centering
\includegraphics[width=\figuremaxwidthsp,height=\figuremaxheightsp,keepaspectratio]{\main/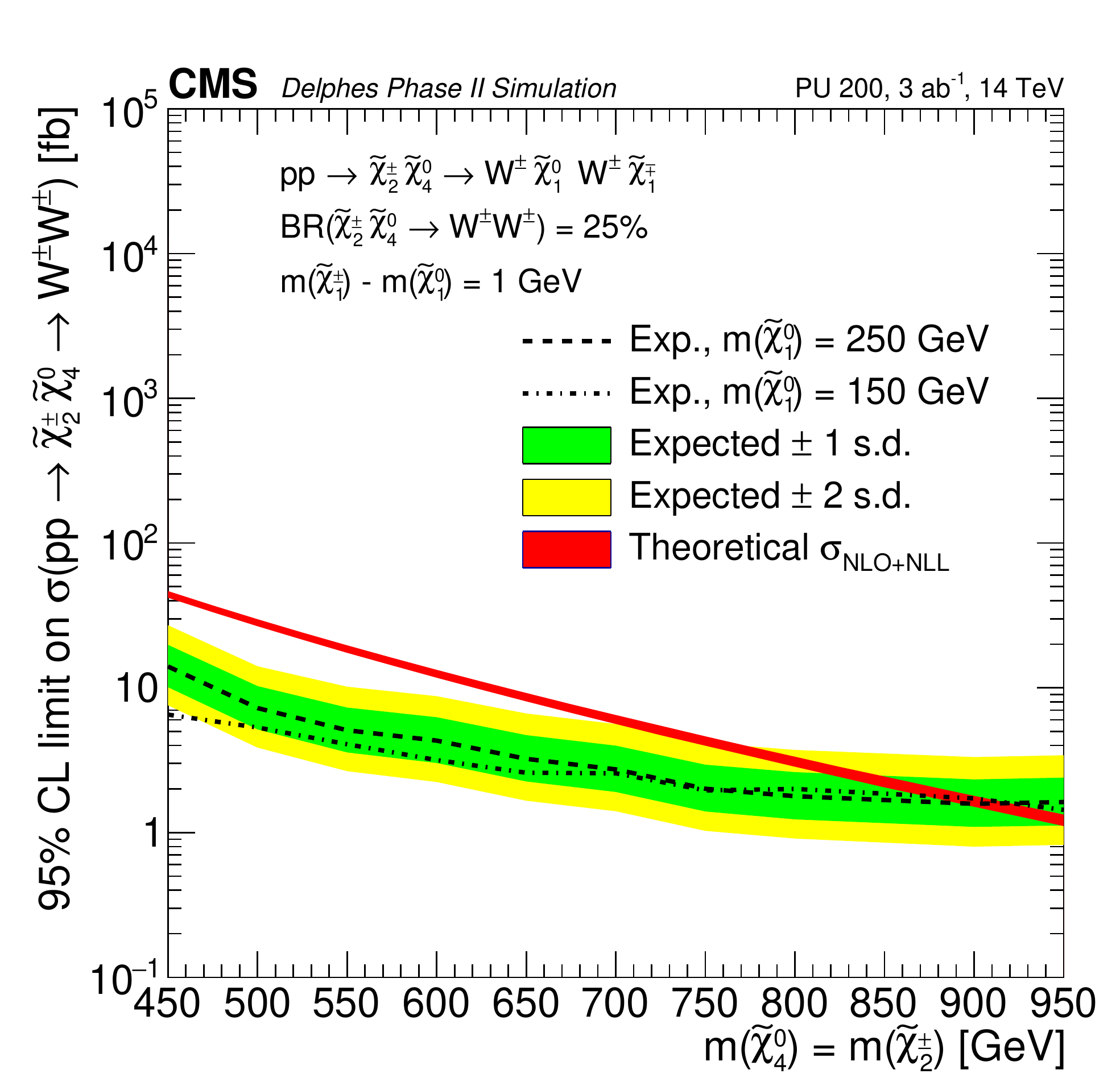}
\end{minipage}
\caption{Upper limit on the production cross-section of pair produced $\tw_2 \tz_4$ decaying 
into a final state with two same charge W boson with a BR of $25\%$ for two assumptions
on the $\tz_1$ mass.}
\label{fig:SS_limit_TDR_SS_mu250}
\end{figure*}

The upper limit on the production cross-section of pair produced $\tw_2 \tz_4$ decaying into a 
final state with two same charge W bosons with a BR of $25\%$ is shown in 
\fig{fig:SS_limit_TDR_SS_mu250} for two $\mu$ scenarios (where $\mu \sim m_{\tw_1}, m_{\tz_2}, m_{\tz_1}$). 
The value $\tz_1=150\GeV$ is representative of the region of  parameter space outside the reach of the Run-2
search for direct production of higgsinos in the final states with two same flavour opposite sign 
 leptons~\cite{Sirunyan:2018iwl}, while $\tz_1=250\GeV$ is close to the sensitivity reach 
 of the same search when extrapolated to the HL-LHC (\secc{sec:LightHiggsino}). 
 As expected, the sensitivity depends 
 only mildly on the value of $\tz_1$ at large $\tw_2 \tz_4$ mass values, while as the $\tw_2 \tz_4$ mass
 approaches $\tz_1$ the dependence is more significant.
Wino-like mass degenerate $\tw_2 \tz_4$ are excluded  at $95\%\cl$  for masses up to  $900\GeV$ 
in  both the $\tz_1=150\GeV$ and  $250\GeV$ scenarios.   
This demonstrates that the \sloppy\mbox{HL-LHC}  has the potential to probe most of the natural SUSY  parameter 
space with EW naturalness measure $\Delta_{\mathrm{EW}}\leq30$~\cite{Baer:2017gzf}. 

\subsubsection{Searches for SUSY models with compressed electroweakino mass spectra}
\label{sec:susy236}
In several SUSY scenarios, higgsinos could be light, with masses below $1\TeV$,  and the absolute value of the higgsino mass parameter $\mu$ is expected to be near the weak scale, while the magnitude of the bino and wino mass parameters, $M_1$ and $M_2$ can be significantly larger, \iee~$|\mu|\ll |M_1|,|M_2|$. This results in the three lightest electroweakino states being dominated by the higgsino component. 
In this scenario, their masses are separated by hundreds of MeV to tens of GeV depending on the composition of these mass eigenstates, which is determined by the specific values of $M_1$ 
and $M_2$. 
Investigating either of these scenarios, with very small mass splitting between the lightest electroweakinos, 
is particularly challenging at hadron colliders, both due to the small cross-sections and the small transverse momenta of the final state particles. As of writing the ATLAS and CMS collaborations have searched for higgsinos in up to $36\fbinv$ of proton-proton collision
data~\cite{Aaboud:2017leg, Sirunyan:2018iwl} and  just started  probing the parameter space beyond the LEP experiments' limits~\cite{Heister:2002mn, Akeroyd:2018axd}. By providing \intLumi of proton-proton collision data at
a \com~energy of $14\TeV$, the HL-LHC has the potential
to significantly extend the sensitivity to higgsinos and thus to natural SUSY. This is depicted also in \secc{sec:BaerSection} of this report. 

\begin{figure*}[t]
\begin{center}
\includegraphics[width=0.3\textwidth]{\main/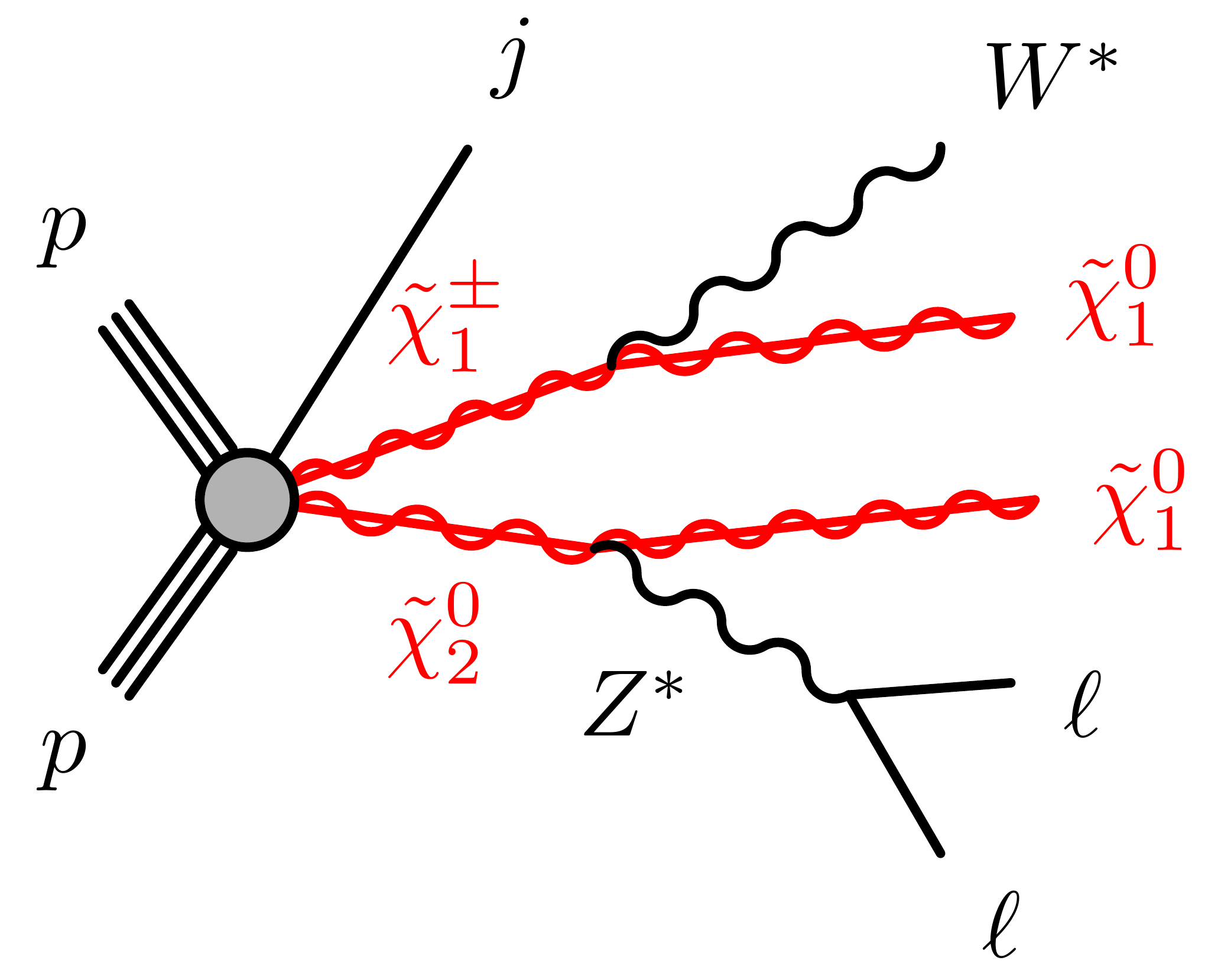}\hspace{1.5cm}
\includegraphics[width=0.3\textwidth]{\main/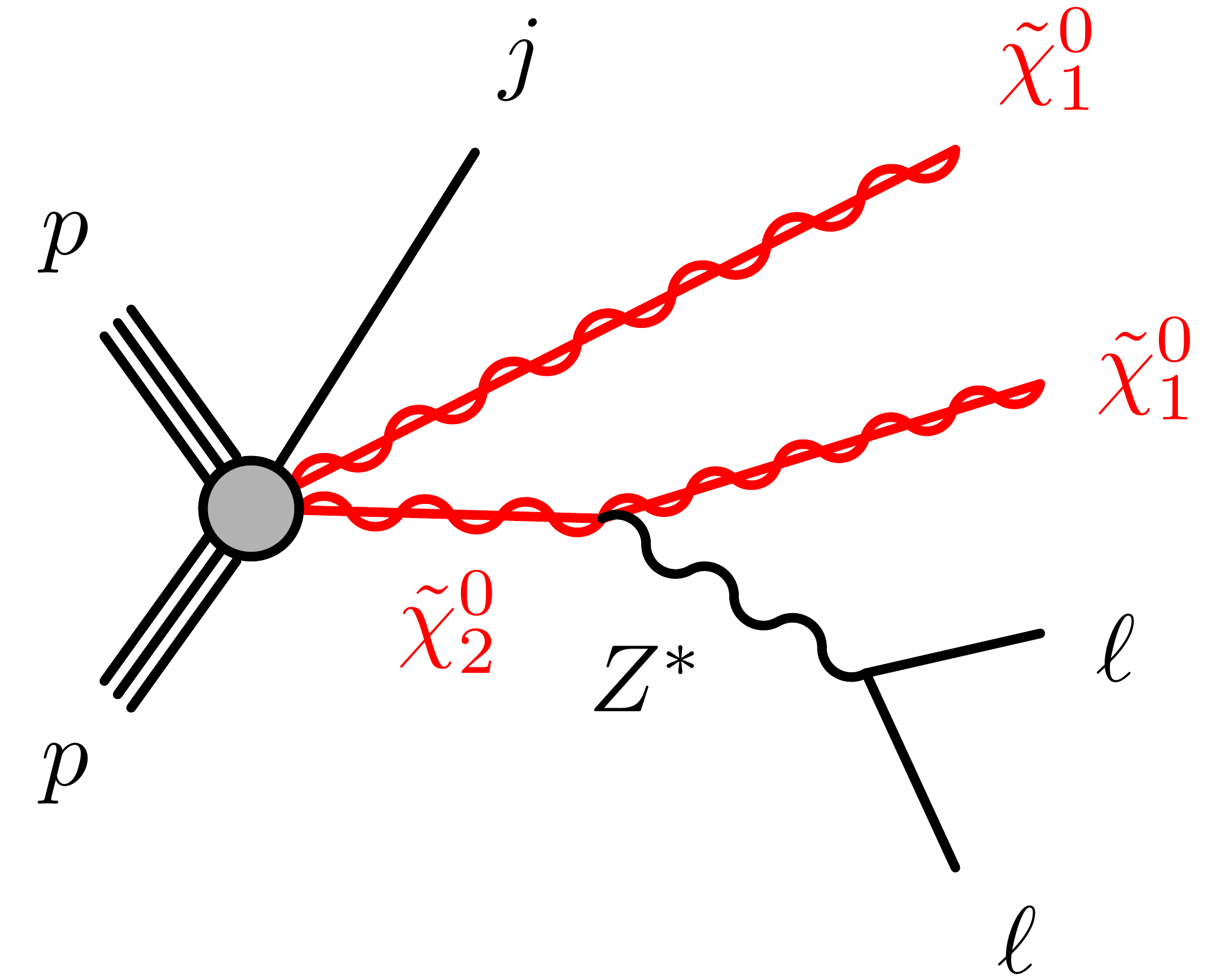}
\caption{Example Feynman diagrams for
\COne{}\NTwo (left) and \NTwo{}\NOne (right) $s$-channel pair production, followed by the leptonic decay of the \NTwo.}
\label{fig:C1N2_FD_AN_ISR_YR}\end{center}
\end{figure*}

The model used for the development of the searches for higgsino-like \Ci and \Nj by ATLAS and CMS is a SUSY simplified model where the higgsino-like \COne and \NTwo are assumed to be quasi mass-degenerate and produced in pairs. The model contains 
both the  \COne{}\NTwo and the \NTwo{}\NOne production, 
where \COne decays into $\Wstar\NOne$ and \NTwo into $\Zstar\NOne$, respectively, with a branching
fraction of $100\%$ (\fig{fig:C1N2_FD_AN_ISR_YR}). 

Both ATLAS and CMS analyses presented in the following 
exploit the presence of charged leptons with low transverse momenta 
arising from the off-shell $W$ and $Z$ bosons in the $\tilde{\chi}_{1}^{\pm} \rightarrow W^{*} \tilde{\chi}_{1}^{0}$ 
and $\tilde{\chi}_{2}^{0} \rightarrow Z^{*} \tilde{\chi}_{1}^{0}$ decays, and 
large missing transverse momentum due to the presence of an ISR jet. 
%The cross sections   are calculated for $\sqrt{s}=14\TeV$
%at next-to-leading-order (NLO) plus next-to-leading-logarithmic (NLL) precision with the CTEQ~6.6 and MSTW~2008nlo90cl parton distribution functions (PDFs) using the \RESUMMINO code~\cite{Fuks:2012qx, Fuks:2013vua}.
%\vspace{1cm}

\subsubsubsection{\cms{Higgsino search prospects at HL- and HE-LHC at CMS}}
\label{sec:LightHiggsino}
\contributors{A. Canepa, J. Hogan, S. Kulkarni, B. Schneider, CMS}
%{\bf Author(s): A. Canepa, J. Hogan, S. Kulkarni, B. Schneider, CMS}

The results presented here are from~\citeref{CMS:2018qsc} from the CMS Collaboration.
If the \COne, \NTwo, and \NOne are higgsino-like, the mass splitting
is just driven by radiative corrections and acquires values up to a few GeV.
As a result, pair-produced \COne{}\NTwo or pair-produced
\NTwo{}\NOne can decay promptly into \NOne only via off-shell
{\PW} and {\PZ} bosons,
leading to events with low transverse momentum ($p_T$) SM particles. 
 In leptonic decays of the {\PZ} boson, 
the events will contain   
one same-flavour, opposite-charge lepton pair,
the invariant mass of which has a kinematic endpoint at
${\Delta M(\NTwo,\NOne)=m(\NTwo)-m(\NOne)}$.
Sensitivity to the signal is achieved by requiring at least
one jet from initial-state radiation (ISR) that recoils against the two \NOne
and produces significant missing transverse momentum (\ptmiss) in
the event. 

In the analysis muons (electrons) are  selected with
${5 \leq \pt \leq 30\GeV}$ and ${|\eta|\leq 2.4\,(1.6)}$.
 Dedicated lepton identification criteria are then applied, providing 40\% to 90\% efficiency for
 muons and $20\%$ to $80\%$ efficiency for electrons. Finally, identified
 leptons are considered candidate leptons if they are isolated. 
The anti-$k_t$ algorithm with a size parameter of $0.4$  is adopted to
reconstruct jets. Candidate jets  are reconstructed jets with
$\pt > 40\GeV$ and ${|\eta|\leq 4.0}$ and are referred to as ISR jets if $\pt > 200\GeV$ and ${|\eta|\leq 2.4}$  (\jisr).
Candidate jets consistent with the decay and
hadronisation of a B hadron are tagged as b jets with
an efficiency of $74\%$.  Spacial separation is imposed
between each candidate lepton and jet. 

To be considered for this analysis, events are requested to contain at least two low-$p_T$, same-flavour,
opposite-charge candidate leptons,  $\ptmiss \geq 250\GeV$ 3and  at least one \jisr.
To further exploit the boosted topology of
the signal, events are accepted only if the \ptmiss and the ISR candidate jet $p_T$
satisfy $\dphijisr\geq 2.0$ and the angular
separation between the two candidate leptons satisfies  $\DRll\leq 2.0$. 
Since minor hadronic activity is expected from the
EW production of \COne and \NTwo,
an upper bound of 4 is placed on the number of candidate jets \Njet.

\begin{figure*}[t]
\centering
\begin{minipage}[t][\minipageheightdp][t]{\minipagewidthdp}
\centering
\includegraphics[width=\figuremaxwidthdp,height=\figuremaxheightdp,keepaspectratio]{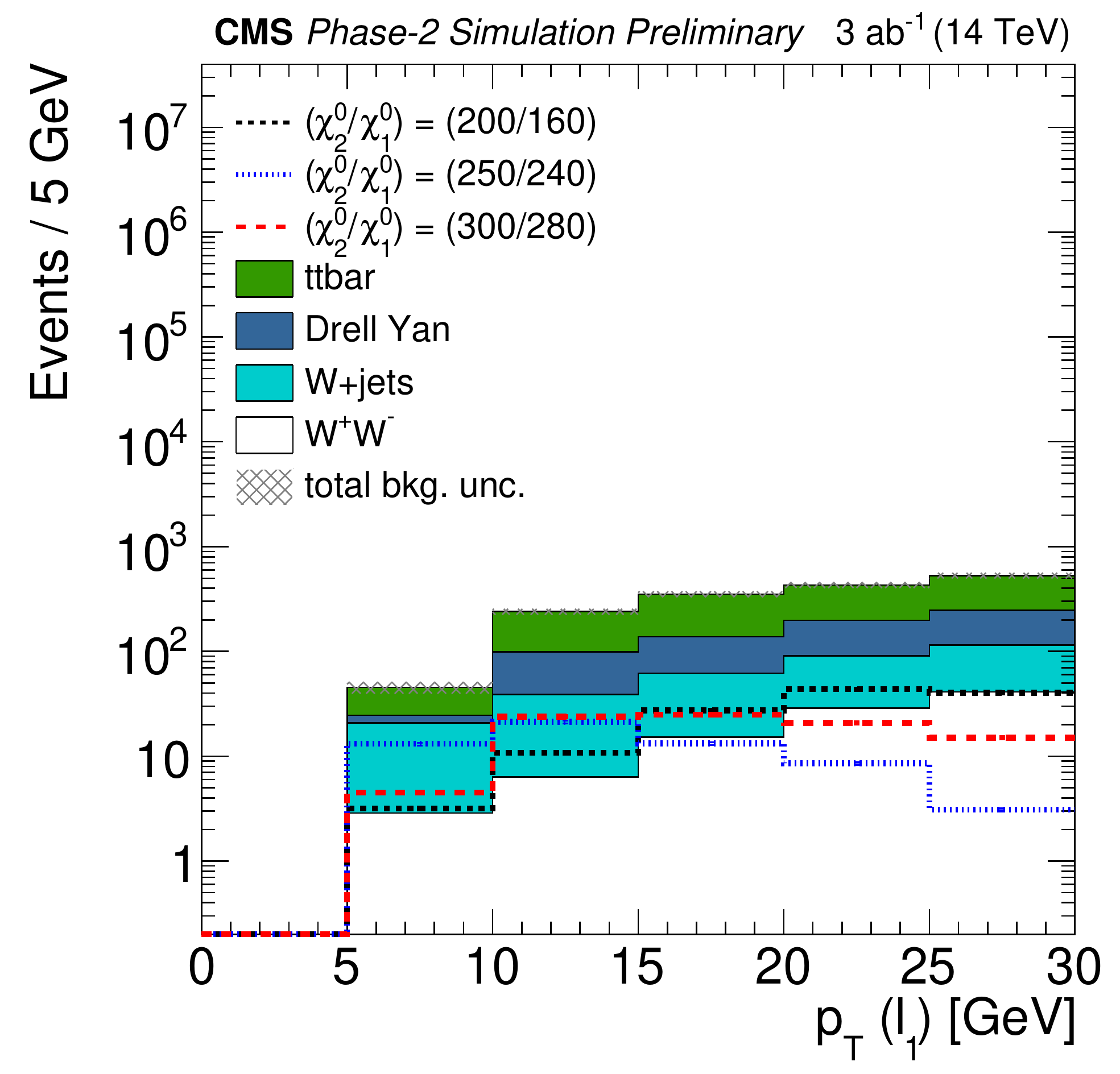}
\end{minipage}
\begin{minipage}[t][\minipageheightdp][t]{\minipagewidthdp}
\centering
\includegraphics[width=\figuremaxwidthdp,height=\figuremaxheightdp,keepaspectratio]{\main/section2SUSY/img/lep1_pt_log}
\end{minipage}
    \caption{Distributions of the $p_T$ of the candidate lepton with the highest $p_T$ (left) and the second-highest $p_T$ (right)
      for background and signal events in the baseline signal region. Three selected ${\COne{}\NTwo+\NTwo{}\NOne}$ signal models are shown,
      where the first number corresponds to the mass of the \NTwo (and \COne) and the second one to the mass of the \NOne. The uncertainty band represents
      systematical uncertainties.
    \label{fig:inclusiveSR1_SR_PAS_YR}
    }
\end{figure*}

\begin{figure*}[t]
\centering
\begin{minipage}[t][\minipageheightdp][t]{\minipagewidthdp}
\centering
\includegraphics[width=\figuremaxwidthdp,height=\figuremaxheightdp,keepaspectratio]{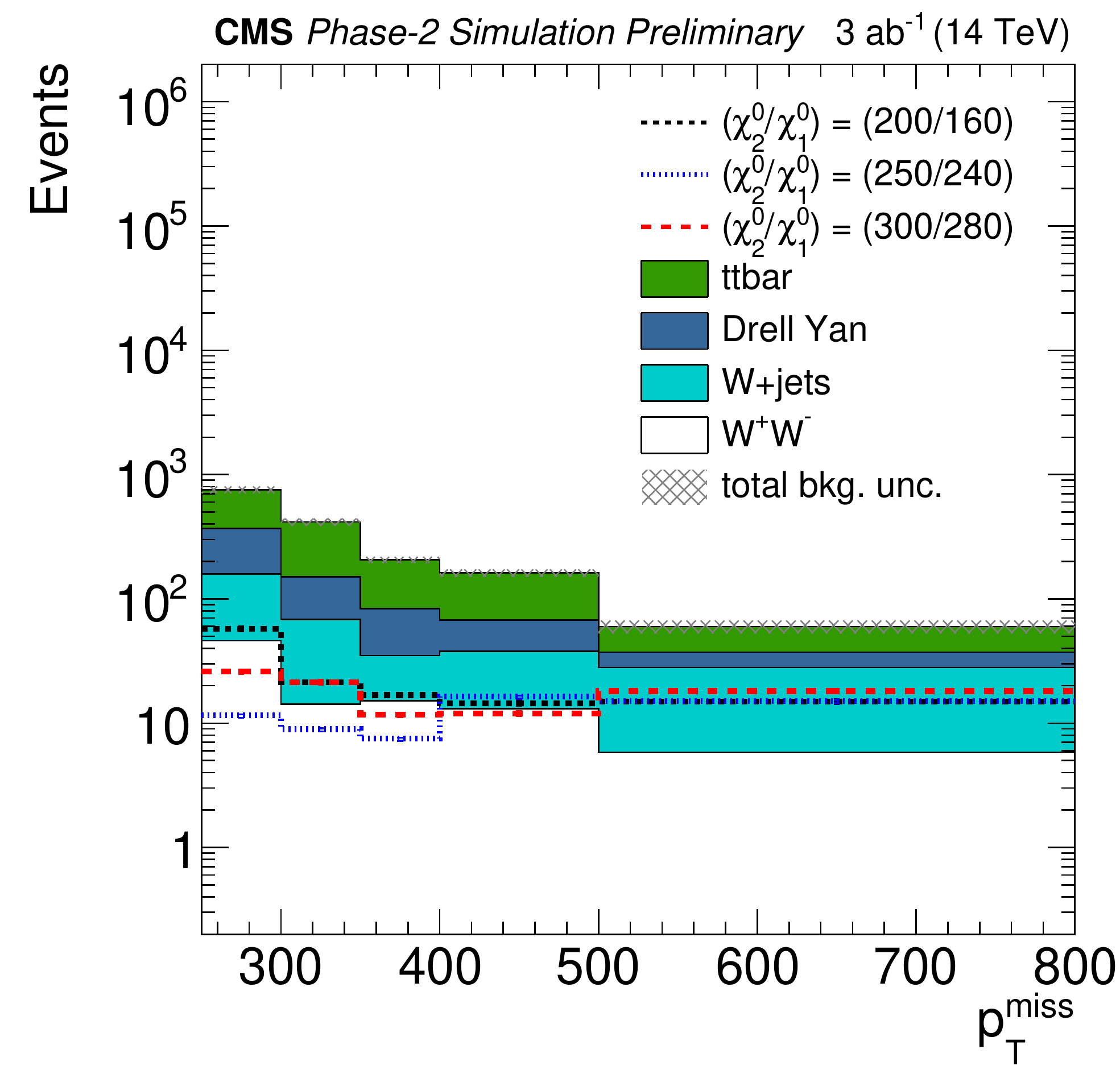}
\end{minipage}
\begin{minipage}[t][\minipageheightdp][t]{\minipagewidthdp}
\centering
\includegraphics[width=\figuremaxwidthdp,height=\figuremaxheightdp,keepaspectratio]{\main/section2SUSY/img/met_log_bins}
\end{minipage}
    \caption{Distributions of the \ptmiss (left) and \mll (right)
      for background and signal events in the baseline signal region. Three selected ${\COne{}\NTwo+\NTwo{}\NOne}$ signal models are shown,
      where the first number corresponds to the mass of \NTwo (and \COne) and the second one to the mass of \NOne. The uncertainty band represents
      systematical uncertainties.
    \label{fig:inclusiveSR2_SR_PAS_YR}
    }
\end{figure*}

Several SM processes exhibit a signature similar to that of the signal. One background category consists of
prompt processes, where both candidate leptons originate from the prompt
decay of {\PW} and {\PZ} bosons. Another category is misclassified processes, where at least one
of the two candidate leptons originates from a semi-leptonic decay of a B hadron,
a photon conversion, a decay in flight, or a misidentified quark or gluon. The prompt
background is dominated by Drell-Yan (DY), diboson, and $t\bar{t}$ production where
both W bosons decay leptonically. The DY contribution is suppressed by requiring significant
\ptmiss, while rejecting events with at least one $b$ jet reduces
the $t\bar{t}$ background. The dominant misclassified processes are
{\PW} and $t\bar{t}$ production where one candidate lepton originates from the {\PW} boson decay and an additional misclassified lepton
is selected in the event.
Rejecting events with at least one $b$ jet reduces both contributions.
Events satisfying the criteria described above, which are summarised in \tabb{tab:inclusiveSR_SR_PAS_YR},
form the baseline signal region for which relevant distributions are presented in \fig{fig:inclusiveSR1_SR_PAS_YR} and \ref{fig:inclusiveSR2_SR_PAS_YR}. 

\begin{table*}[t]
  \begin{center}
    \small\begin{tabular}{ll}
      Observable   & Requirement   \\ \hline
      \Nlep        &  $= 2$ (same flavour, opposite charge)  \\
      \DRll 	   &  $\leq 2.0$ \\
      \Nbjet       &  $= 0$                         \\
      \Njet        &  $\leq 4$                          \\
      \Nisr        &  $\geq 1$                      \\
      \ptmiss      &  $\geq 250\GeV$  \\
      \dphijisr    &  $\geq 2.0$ \\
      \mll         & [5,40]\GeV \\ \hline
       \end{tabular}\normalsize
    \caption{Definition of the baseline signal region. In the table,
      \Nlep is the number of  candidate leptons;
      $\Delta R(\ell_1,\ell_2)$
      is the angular separation between the two candidate
      leptons in the $\phi, \eta$ space;
      \Nbjet is the number of $b$ jets;       \Njet is the number of candidate jets (including any ISR jet reconstructed in the event);
      \Nisr is the number of ISR jets;
      $\Delta\phi(\ptmiss, \ptisr)$ is the azimuthal distance between the \ptmiss vector and
    the \jisr $p_T$ vector; and \mll is the invariant mass of the two candidate leptons.    \label{tab:inclusiveSR_SR_PAS_YR}}
\end{center}
\end{table*}

The missing transverse momentum, the invariant mass of the two candidate
leptons, and the sub-leading lepton \pttwo observables are found to provide
the best discrimination between signal and background. Events in the baseline signal
region are therefore classified in 60 categories with \ptmiss values in
$[250, 300, 350, 400, 500, \infty]\GeV$, \mll values in  $[5, 10, 20, 30, 40]\GeV$,
and \pttwo in \sloppy\mbox{$[5, 13, 21, 30]\GeV$}.

Several systematic uncertainties affect the yields of both the background and the
signal processes. The dominant experimental uncertainties are those originating from the jet
energy corrections ($1-2.5\%$), b-tagging efficiency ($1\%$), lepton identification efficiency and isolation ($0.5\%$, $2.5\%$ for
muons and electrons, respectively), and integrated luminosity ($1\%$). 
An additional systematic uncertainty of $30\%$ on the yield of the misclassified background is also assumed  
based on the estimate in \citeref{Sirunyan:2018iwl}. It is assumed that the yields are not affected by the statistical uncertainty deriving from the limited number of generated events.
Theoretical uncertainties in the signal cross sections and in the acceptance from the choice
of parton distribution functions are considered negligible and are not included. However,
a systematic uncertainty of $10\%$ in the signal acceptance, similar to the value from \citeref{Sirunyan:2018iwl},
is included to account for the modelling of the ISR jet.

\begin{figure*}[t]
\centering
\begin{minipage}[t][\minipageheightsp][t]{\minipagewidthsp}
\centering
\includegraphics[width=\figuremaxwidthsp,height=\figuremaxheightsp,keepaspectratio]{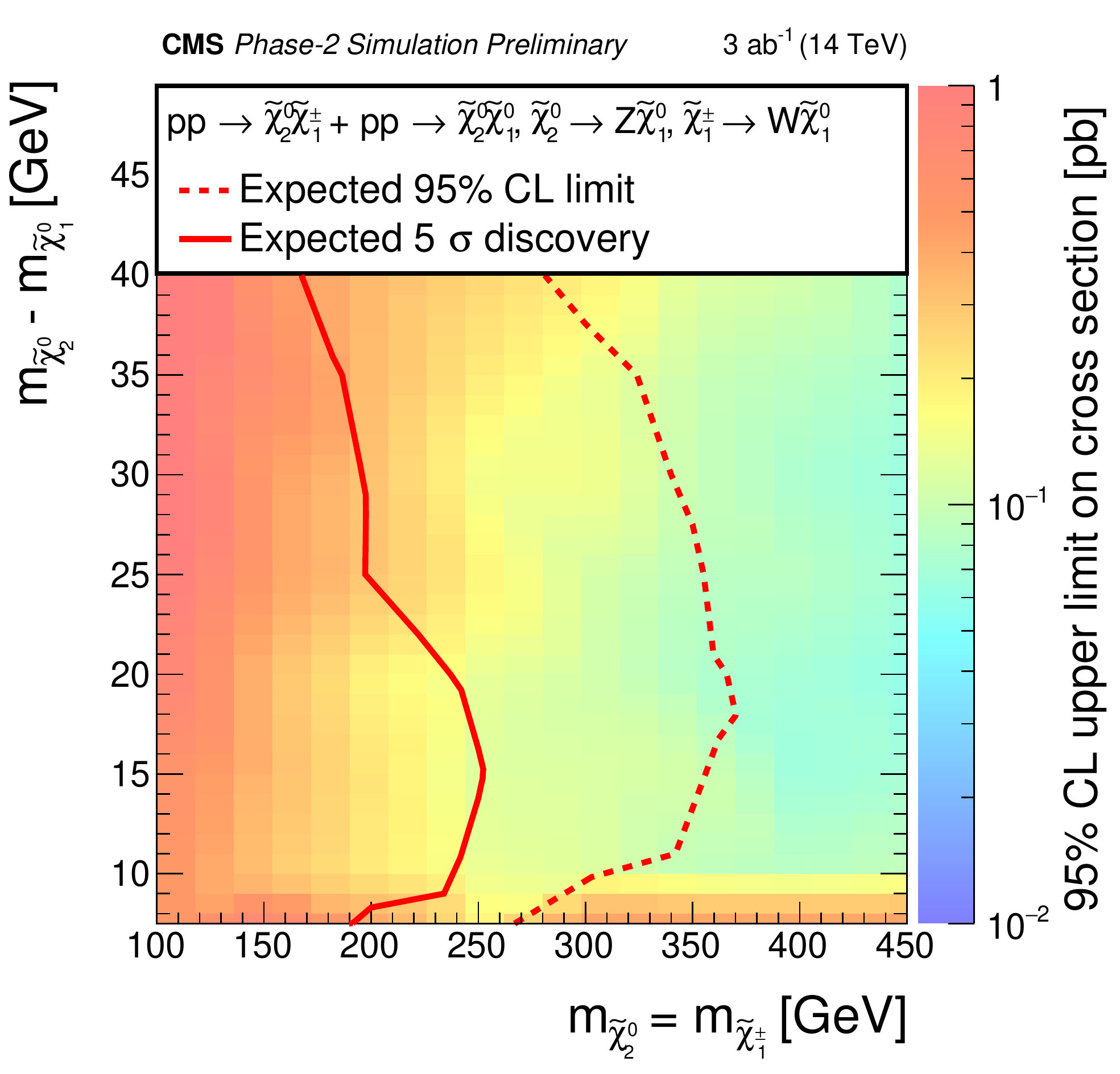}
\includegraphics[width=\figuremaxwidthsp,height=\figuremaxheightsp,keepaspectratio]{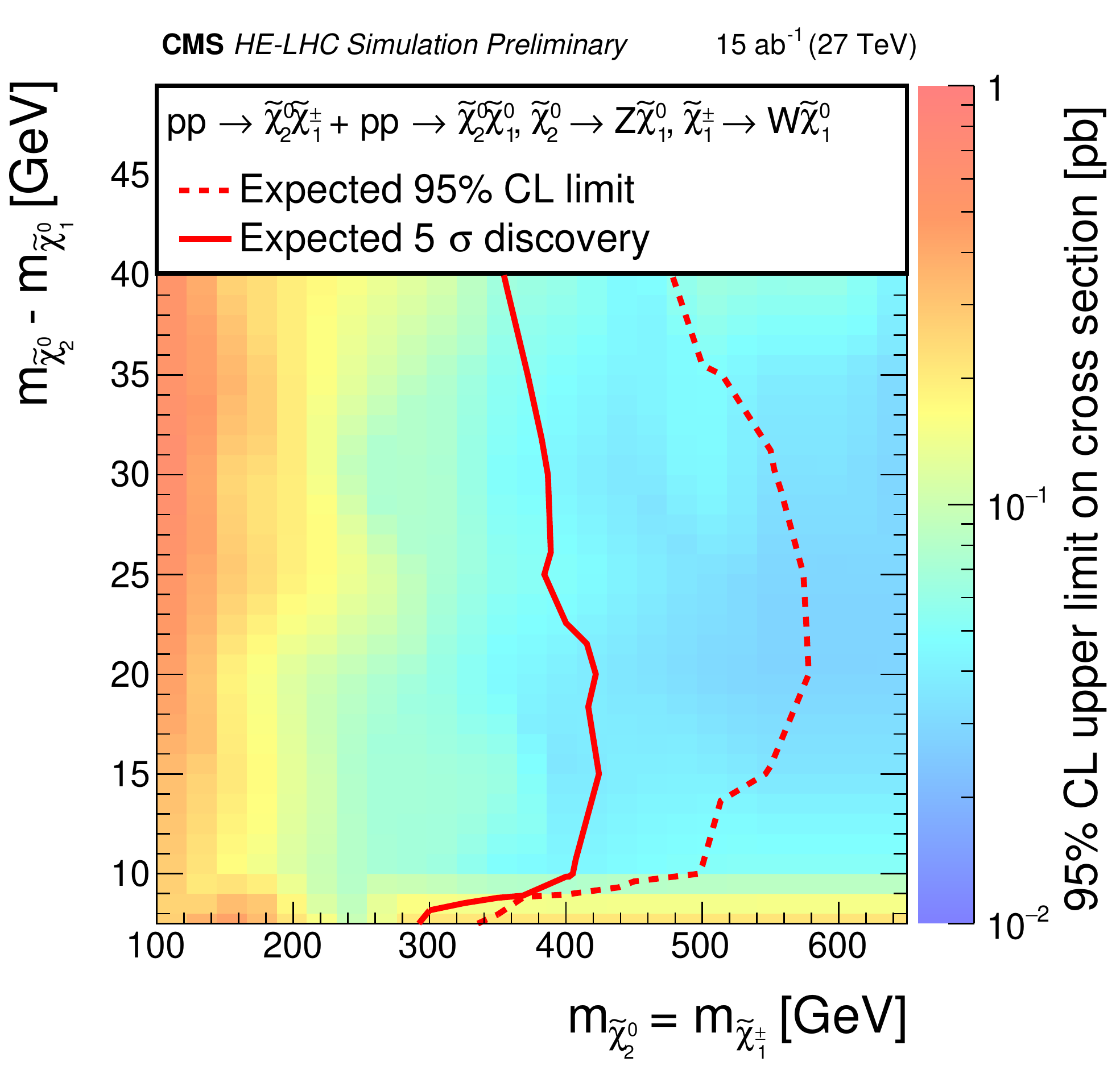}

\end{minipage}
  \caption{$5\sigma$ discovery contours and expected $95\%\cl$ exclusion contours
  for the combined \COne\NTwo and \NTwo\NOne production (left). Projection of the HL-LHC $5\sigma$ discovery contours and expected $95\%\cl$ exclusion contours for the combined \COne\NTwo and \NTwo\NOne production for a centre-of-mass energy of $27\TeV$ and an integrated luminosity of $15\abinv$ (HE-LHC). Except for the cross sections and the integrated luminosity, the HL-LHC analysis was not modified (right). Results are presented
for $\Delta M(\NTwo, \NOne) > 7.5\GeV$.}
\label{fig:ISR_limit_here_YR}
\end{figure*}

The upper limit on the cross sections is
computed at $95\%\cl$ and shown in
\fig{fig:ISR_limit_here_YR}. 
Higgsino-like mass-degenerate \COne and \NTwo are excluded
for masses up to $360\GeV$ if the mass difference
with respect to the lightest neutralino \NOne is $15\GeV$, extending the sensitivity achieved in \citeref{Sirunyan:2018iwl} by  ${\approx}210\GeV$.
Figure~\ref{fig:ISR_limit_here_YR} also shows the $5\sigma$ discovery contour,
computed using all signal regions without taking the
look-elsewhere-effect into account.
Under this assumption \COne and \NTwo can be discovered for masses as large as $250\GeV$.
These results demonstrate that the HL-LHC can significantly improve the sensitivity
to natural SUSY.

Figure~\ref{fig:ISR_limit_here_YR} also shows the $5\sigma$ discovery contours and expected $95\%\cl$ exclusion contours for the
combined \COne\NTwo and \NTwo\NOne production for the HE-LHC. The main gain in sensitivity comes from the increased luminosity, since the cross section increase for signal is the same order as that for background.
Except for the cross sections and the integrated luminosity, the HL-LHC analysis was not modified for this HE-LHC projection.

\subsubsubsection{\atlas{Higgsino search prospects at HL-LHC at ATLAS}}
\contributors{S. Amoroso, J. K. Anders, F. Meloni, C. Merlassino, B. Petersen, J. A. Sabater Iglesias, M. Saito, R. Sawada, P. Tornambe, M. Weber, ATLAS}

The presented dilepton search~\cite{ATL-PHYS-PUB-2018-031} investigates final states containing two soft muons and a large transverse momentum imbalance, 
which arise in scenarios where $\tilde{\chi}_{2}^{0}$ and $\tilde{\chi}_{1}^{\pm}$ are produced and decay via an off-shell $Z$ and $W$ boson, as depicted in \fig{fig:C1N2_FD_AN_ISR_YR}. 
Considering the $Z \rightarrow ee$ decay is beyond the scope of this prospect study, but could further improve the sensitivity to these scenarios. 
Due to the very small mass splitting of the electroweakinos in this scenario, a jet arising from initial-state radiation (ISR) is required, to boost the sparticle system. 
First constraints surpassing the LEP limits have recently been set by the ATLAS experiment~\cite{Aaboud:2017leg}, excluding mass splittings down 
to $2.5\GeV$ for $m(\tilde{\chi}_{1}^{0}) = 100\GeV$. 

The search targets scenarios that contain low $\pt$ muons selected with $\pt$ $>3\GeV$ and \sloppy\mbox{$|\eta|<2.5$}. 
Muons that originate from pile up interactions or from heavy flavour decays, referred as fake or non-prompt muons, are rejected by applying an 
isolation to the muon candidates. The main source of these fake muons are decays from heavy flavour mesons and baryons created in the quark hadronisation process. 
The signal region (SR) optimisation is performed by scanning a set of variables which are expected to provide discrimination between the signal scenario under 
consideration and the expected SM background processes. 
Only events with two opposite-sign muons are used in the final selection, as the muon reconstruction rate is not expected to fall dramatically and the muon 
fake rate is not expected to grow largely with increased pile-up.
Additional requirements are applied on the leading jet of $\pt(\mathrm{jet}_1) > 100\GeV$, 
and on the azimuthal separation $\Delta\phi(\mathrm{jet}_1,\met) > 2.0$. In order to discriminate the signal from SM background processes, kinematic variables are used such as the total number of muons in the event, the total number of jets and $b$-jets with $p_{\mathrm{T}} > 30\GeV$, the $\met$, the invariant mass of the dilepton system ($m_{\ell\ell}$), the angular separation between the leptons (${\Delta R(\ell,\ell)}$) and more.  

\begin{figure}[t]
\centering
\begin{minipage}[t][\minipageheightdp][t]{\minipagewidthdp}
\centering
\includegraphics[width=\figuremaxwidthdp,height=\figuremaxheightdp,keepaspectratio]{\main/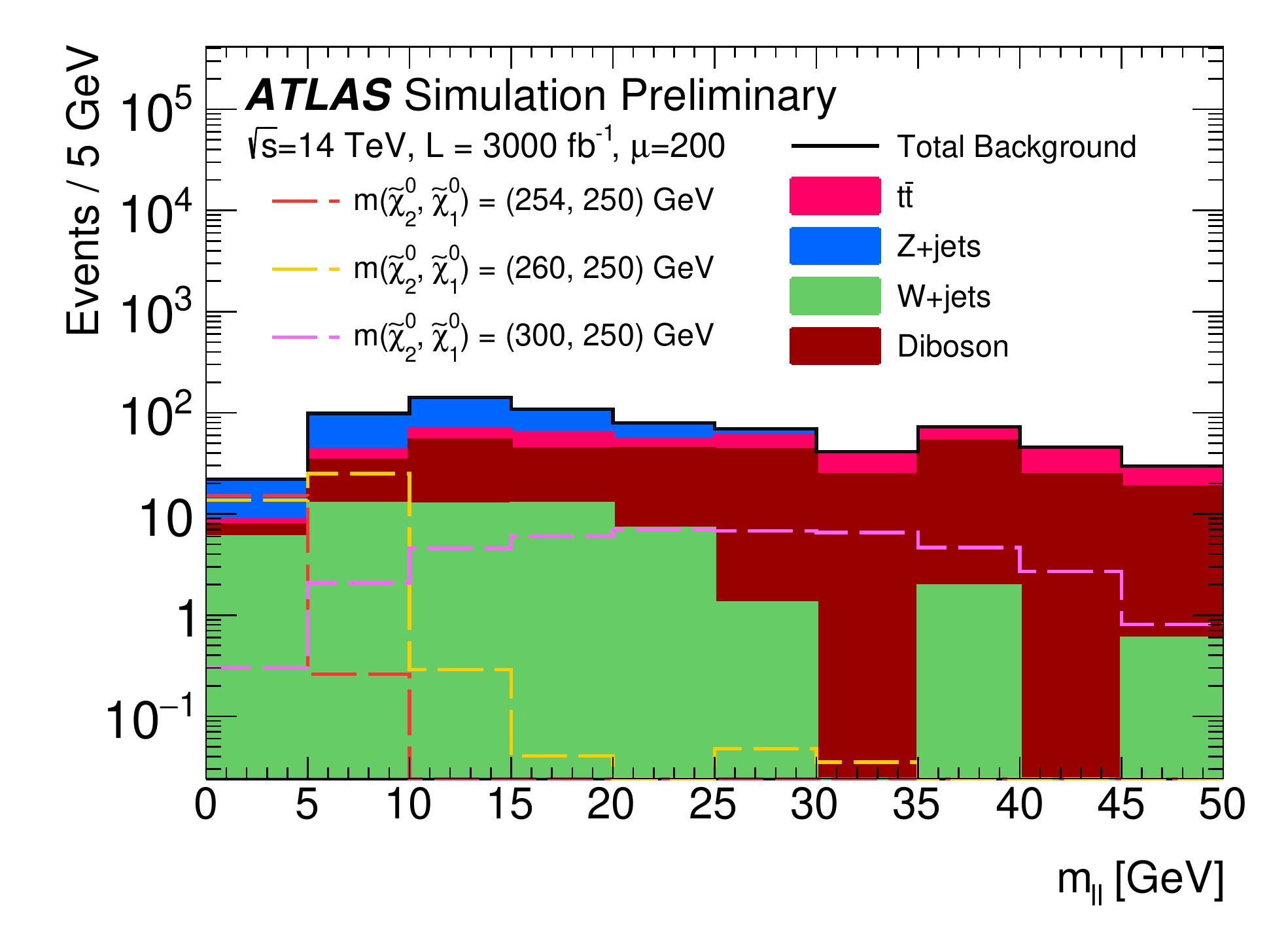}
\end{minipage}
\begin{minipage}[t][\minipageheightdp][t]{\minipagewidthdp}
\centering
\includegraphics[width=\figuremaxwidthdp,height=\figuremaxheightdp,keepaspectratio]{\main/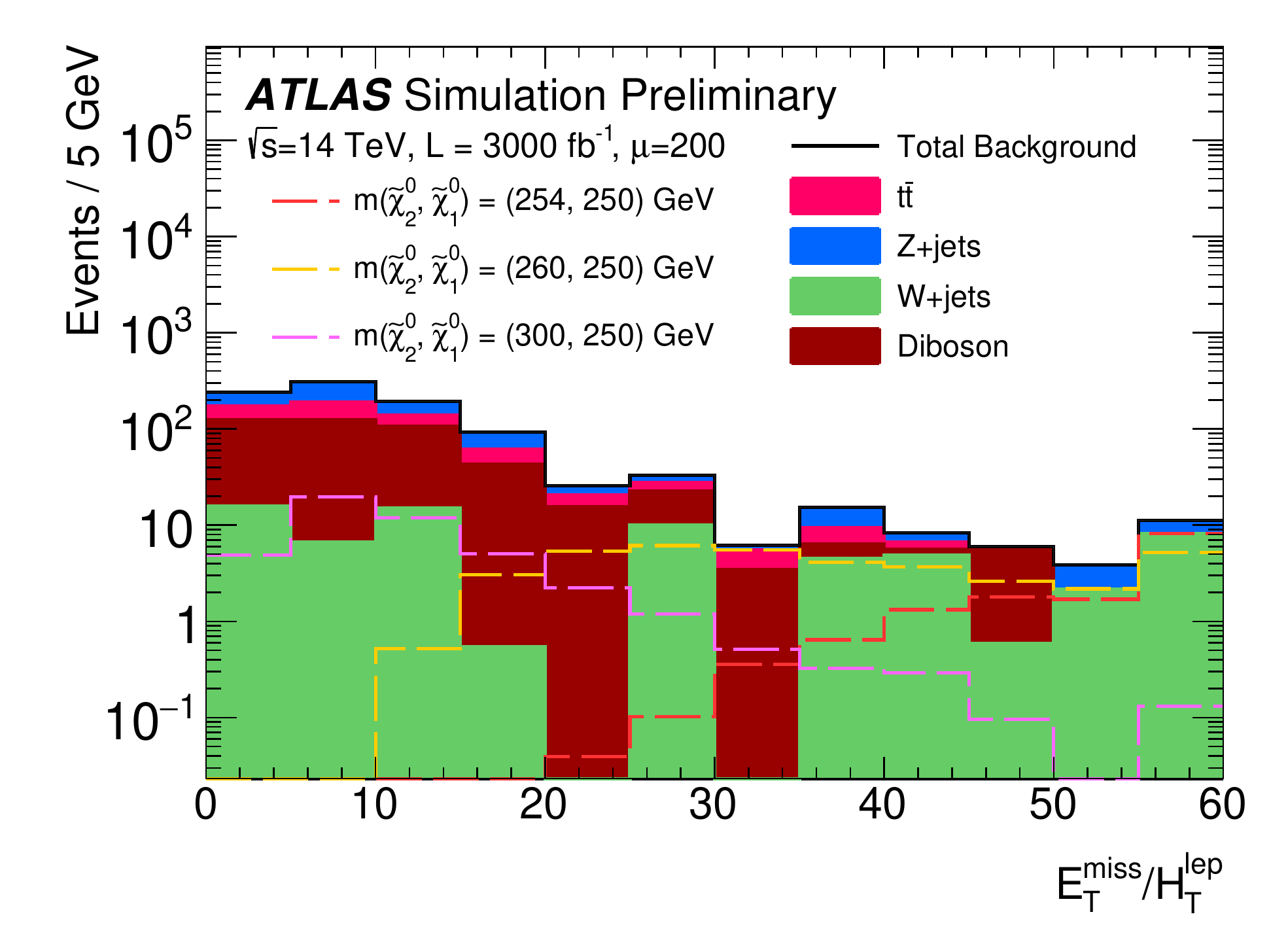}
\end{minipage}
\caption{\label{fig:2L_Presel_plots}Distributions of a selection of kinematic variables used for the SR optimisation in the dilepton search. The variables are presented with the full SR selections implemented aside from the selection on the variable shown. %Left, $m_{\ell\ell}$; Right, $\met/H_{\mathrm{T}}^{\mathrm{lep}}$. 
Three signal models with $m(\chi_{1}^0) = 250\GeV$ and different mass splittings ($\Delta m(\chi_{2}^0, \chi_{1}^0) = 4,\,10,\, \mathrm{ and }\,5\GeV$) are overlaid.}
\end{figure}

Figure \ref{fig:2L_Presel_plots} presents a selection of kinematic distributions after the full SR selection is applied, minus the selection on the variable under consideration. The final SR definitions split the $m_{\ell\ell}$ into six non-overlapping SRs, with $m_{\ell\ell}$ selections of $[1, 3]$, $[3.2, 5]$, $[5, 10]$, $[10, 20]$, $[20, 30]$ and $[30, 50]\GeV$. 

The leading sources of background in the SR are from $\ttbar$, single-top, $WW$ + jets, and $Z/\gamma^*(\to\tau\tau)$ + jets. The dominant source of reducible background arises from processes where one or more leptons are fake or non-prompt, such as in $W$+jets production. The fake/non-prompt lepton background arises from jets misidentified as leptons, photon conversions, or semileptonic decays of heavy-flavour hadrons. The total uncertainty for the dilepton search is extrapolated to be $30\%$ and are dominated 
by the modelling of the fake and non-prompt lepton backgrounds, followed by the experimental uncertainties related to the jet energy scale and flavour tagging. 
The experimental uncertainty is assumed to be fully correlated between the background and the signal. 

\begin{figure}[t]
\begin{center}
\includegraphics[width=.55\textwidth]{\main/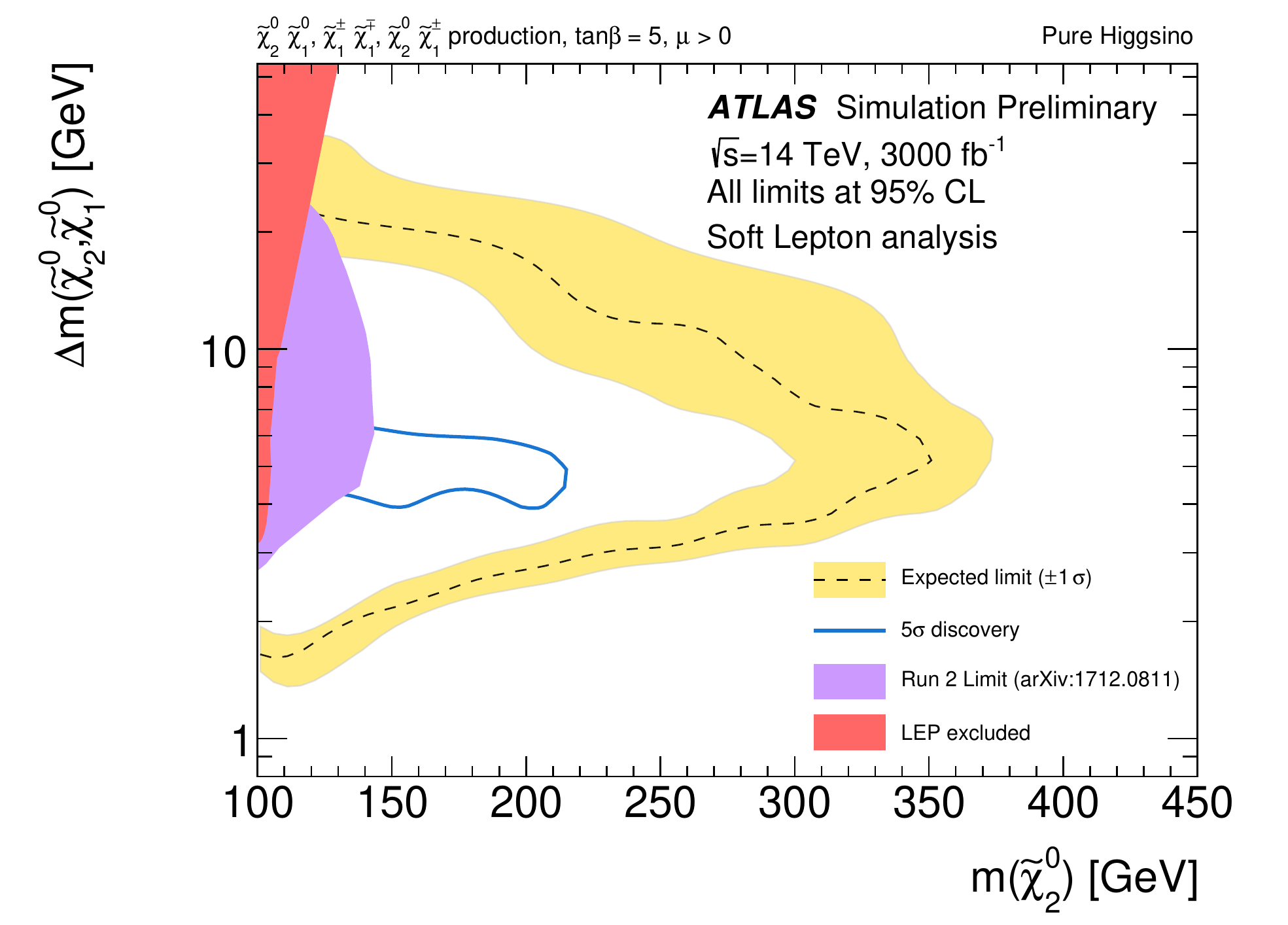}
\end{center}
\caption{Expected exclusion limit (dashed line) in the $\Delta m(\tilde{\chi}_{2}^{0}, \tilde{\chi}_{1}^{0})$, $m(\tilde{\chi}_{2}^{0})$ mass plane, 
at $95\%\cl$ from the dilepton analysis with \intLumi of $14\TeV$, proton-proton collision data in the context of a pure Higgsino LSP with $\pm 1 \sigma$ (yellow band) from the associated systematic uncertainties. The blue curve presents the $5\sigma$ discovery potential of the search. The purple contour is the observed exclusion limit from the Run-2 analysis. The figure also presents the limits on chargino production from LEP. The relationship between the masses of the chargino and the two lightest neutralinos in this scenario is $m(\tilde{\chi}_{1}^{\pm}) = \frac{1}{2}(m(\tilde{\chi}_{1}^{0}) + m(\tilde{\chi}_{2}^{0}))$.}
\label{fig:limit_compress_atlas}
\end{figure}

Figure~\ref{fig:limit_compress_atlas} shows the $95\%\cl$ exclusion limits in the $m(\tilde{\chi}_{2}^{0})$, $\Delta m (\tilde{\chi}_{2}^{0}, \tilde{\chi}_{1}^{0})$ plane. 
With \intLumi, $\tilde{\chi}_{2}^{0}$ masses up to $350\GeV$ could be excluded, as well as $\Delta m (\tilde{\chi}_{2}^{0}, \tilde{\chi}_{1}^{0})$ between $2$ and $25\GeV$ 
for \sloppy\mbox{$m(\tilde{\chi}_{2}^{0}) = 100\GeV$}. In the figure the blue curve presents the $5\sigma$ discovery potential of the search. To calculate the discovery potential a single-bin discovery test is performed by removing the lower bound on $m_{\ell\ell}$ in the SRs previously defined. 

\subsubsection{%\theoryC{Multilepton signals from resonant electroweakino production in left-right supersymmetry}
\theoryC{Multileptons from resonant electroweakinos in left-right SUSY at HL- and HE-LHC
%Multileptons from resonant electroweakino production in left-right SUSY at HL- and HE-LHC
}
}
\contributors{M. Frank, B. Fuks, K. Huitu, S. Mondal, S. Kumar Rai, H. Waltari}

%{\bf Authors:} Mariana Frank$^{a}$, Benjamin Fuks$^{b,c}$, Katri Huitu$^{d}$, Subhadeep Mondal$^{d}$, Santosh Kumar Rai$^{e}$ and Harri Waltari$^{d,f}$
%
%\noindent {\bf Affiliations:} \\
%a) Department of Physics, Concordia University, 7141 Sherbrooke St.West, Montreal, QC, Canada H4B 1R6\\
%b) Sorbonne Universit\'e, CNRS, Laboratoire de Physique Th\'eorique et Hautes Energies, LPTHE, F-75005 Paris, France\\
%c) Institut Universitaire de France, 103 boulevard Saint-Michel, 75005 Paris, France\\
%d) Department of Physics and Helsinki Institute of Physics, University of Helsinki, P.O. Box 64 (Gustaf H\"allstr\"omin katu 2), FI-00014 University of Helsinki, Finland\\
%e) Regional Centre for Accelerator-based Particle Physics, Harish-Chandra Research Institute, HBNI, Chhatnag Road, Jhusi, Allahabad 211019, India.\\
%f) Department of Physics and Astronomy, University of Southampton, Highfield, Southampton SO17 1BJ, United Kingdom

%\paragraph*{Left-right supersymmetry, dark matter and the LHC}

Left-right supersymmetric (LRSUSY) models, based on the gauge symmetry 
$SU(3)_{C}\times SU(2)_{L}\times SU(2)_{R}\times
U(1)_{\mathrm{B-L}}$, inherit the attractive features of the left-right (LR)
symmetry~\cite{Pati:1974yy,Mohapatra:1974hk}, whereas they forbid any $R$-parity
violating operators thanks to the gauged $B-L$ symmetry.
To naturally describe the small magnitude of the neutrino masses and preserve
$R$-parity, the model superfield content includes both $SU(2)_{L}$ and
$SU(2)_{R}$ triplets of Higgs supermultiplets. The neutral component of the
$SU(2)_{R}$ Higgs scalar field then acquires a large vacuum expectation
value $v_{R}$, which breaks the LR symmetry and makes the $SU(2)_{R}$ gauge
sector heavy. In order to prevent the tree-level vacuum from being a
charge-breaking one, we can either rely on spontaneous $R$-parity violation~\cite{Kuchimanchi:1993jg}, one-loop corrections~\cite{Babu:2008ep},
higher-dimensional operators~\cite{Mohapatra:1995xd} or additional $B-L=0$
triplets~\cite{Aulakh:1997ba}. Whereas the first two options restrict $v_{R}$ to
be of at most about $10\TeV$, the latter ones enforce $v_{R}$ to lie above
$10^{10}\GeV$. In this work, we rely on radiative corrections to stabilise the
vacuum, so that the lightest supersymmetric particle (LSP) is stable and can act
as a dark matter candidate.

%%%%%%%%%%%%%%%%%%%%%%%%%%%%%%%n{figure}
\begin{figure}[t]
\centering
\begin{minipage}[t][\minipageheightsp][t]{\minipagewidthsp}
\centering
\includegraphics[width=\figuremaxwidthsp,height=\figuremaxheightsp,keepaspectratio]{\main/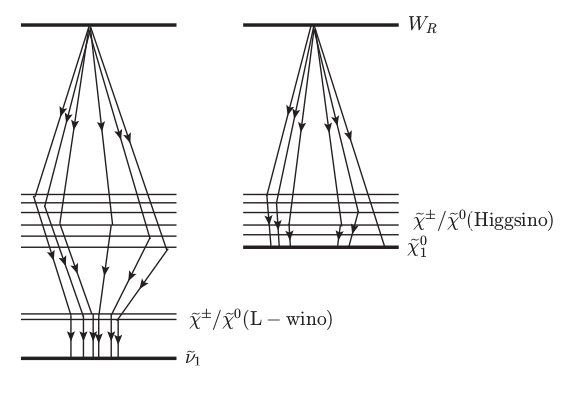}
\end{minipage}
\caption{The two LRSUSY spectra chosen for our study. Left: sneutrino LSP spectrum, where the $W_R$ boson decays into multiple higgsino-like
chargino-neutralino pairs, each electroweakino subsequently decaying into the
sneutrino LSP either directly or through an intermediate $SU(2)_{L}$
wino-like chargino-neutralino pair. Right: neutralino LSP spectrum, where
the $W_R$ bosons decays into higgsino-like chargino-neutralino pairs, the
electroweakinos subsequently decaying into the lightest (higgsino-like)
neutralino.}
\label{fig:spect}
\end{figure}
%%%%%%%%%%%%%%%%%%%%%%%%%%%%%%%n{figure}

Two viable LSP options emerge from LRSUSY, neutralinos and right
sneutrinos. Out of the 12 neutralinos, gauginos and LR bidoublet,
higgsinos can generally be lighter than $1\TeV$. The correct relic density
can be accommodated with dominantly-bino LSPs with a mass close to
$m_{h}/2$~\cite{Frank:2017tsm}, whilst in the bidoublet higgsinos case
(featuring four neutralinos and two charginos that are nearly-degenerate), 
co-annihilations play a crucial role and impose higgsino masses close to
$700\GeV$. In this setup, the rest of the spectrum is always heavier, so that
SUSY could be challenging to discover. Right sneutrino LSP
annihilate via the exchange of an $s$-channel Higgs boson through gauge
interactions stemming from the
$D$-terms~\cite{Frank:2017tsm}. Without options for co-annihilating, the LSP
sneutrino mass must lie between $250$ and $300\GeV$. However, potential
co-annihilations with neutralinos enhance the effective annihilation cross
section so that the relic density constraints can be satisfied with heavier
sneutrinos. The fully degenerate sneutrinos and higgsinos case impose an upper
limit on the sneutrino mass of $700\GeV$. Additionally, right neutrinos can also
be part of the dark sector, together with the LSP~\cite{Bhattacharya:2013nya}.

Direct detection constraints imposed by the XENON1T~\cite{Aprile:2018dbl} and
PANDA~\cite{Cui:2017nnn} collaborations put light DM scenarios under severe
scrutiny. Hence, in LRSUSY, in order to account for the relic density and direct
detection constraints simultaneously, we  need to focus on various
co-annihilation options. In this work, we consider one right sneutrino and one
higgsino LSP scenario and highlight the corresponding implications for $W_R$
searches at the LHC.
A robust signal of left-right symmetry consists in the discovery of a right
gauge boson $W_R$, possibly together with a right neutrino $N_R$. Both the ATLAS
and CMS collaborations have  looked for a $W_R$ signal in the $\ell
\ell jj$ channel, excluding $W_R$ masses up to about $4.5\TeV$ when at least one
right-handed neutrino is lighter than the $W_R$-boson~\cite{Sirunyan:2018pom}.
These exclusion limits nevertheless strongly depend on the spectrum and could be
weakened or even evaded, for instance when $M_{N_R}\simeq M_{W_R}$ or for
$M_{N_R}\lesssim 150\GeV$ and $M_{W_R} > 3\TeV$. In addition, dijet resonance
searches yield $M_{W_R}\gtrsim 3.5\TeV$~\cite{Aaboud:2017yvp,Sirunyan:2018xlo},
even if these bounds can once again be relaxed by virtue of the supersymmetric
$W_R$ decay modes. On different grounds, dark matter considerations lead to favoured LRSUSY scenarios in which several neutralinos and charginos are light (so that they could co-annihilate). 
This motivates the investigation of a new $W_R$ search channel where decays into pairs of electroweakinos are considered.
In many LRSUSY setups, the corresponding combined BR can be as large as $25\%$, so that the production of multileptonic systems featuring a large amount of missing transverse momentum is enhanced. Whilst such a multilepton signal with $\met$ is a characteristic SUSY signal, it also provides an additional search channel for $W_R$-bosons at the LHC. Moreover, the resonant production mode offers the opportunity to reconstruct the $W_R$-boson mass through kinematic thresholds featured by various transverse observables.  

%\paragraph*{Resonant chargino and neutralino production}

In order to illustrate the above features, we perform an analysis in
the context of two LRSUSY scenarios respectively featuring a sneutrino and a neutralino LSP. The results are presented for both the high \com~energy ($\sqrt{s}$) and high luminosity ($\mathcal L$)
cases, $\sqrt{s}=14\TeV$ with ${\mathcal L}= \intLumi$ and $\sqrt{s} = 27\TeV$ with ${\mathcal L}= \intlumihelhc$
%($3~{\rm ab}^{-1}$) and high-energy ($\sqrt{s}$ = 27 TeV) 
options for the future run of the LHC.
For the higgsino-like neutralino LSP case, we  kept the
bidoublet higgsino masses in the $700-750\GeV$ region. In contrast, for the
sneutrino DM case, the LSP mass can be much lower and has been fixed to about
$400\GeV$, with the second lightest superpartner being an $SU(2)_{L}$ wino
lying about $30\GeV$ above and the higgsinos being again in the $700-750\GeV$
regime. These two mass spectra are illustrated in \fig{fig:spect}. 
The sneutrino DM option is expected to be reachable with a lower luminosity due to the harder charged leptons arising from the cascade decays.

%%%%%%%%%%%%%%%%%%%%%%%%%%%%%%%%
\begin{table}[t]
\begin{center}
\small\begin{tabular}{c||c} 
Signal Regions & Requirements \\
\hline\hline
{\bf SRA44}  & $N_{\ell}=3$, $N_{\rm OSSF}\ge 1$, $N_{\tau}=0$,
  $M_T > 160\GeV$, $\slashed{E}_T \ge 200\GeV$, $M_{\ell\ell}\ge 105\GeV$   \\
\hline
{\bf SRD16}  & $N_{\ell}=2$, $N_{\rm OS}=1$, $N_{\rm SF}=0$, $N_{\tau}=1$,
  $M_{T2} > 100\GeV$, $\slashed{E}_T \ge 200\GeV$ \\
\end{tabular}\normalsize
\caption{The two signal regions of the analysis of \citeref{Sirunyan:2017lae}
that are the most suitable for discovering our considered LRSUSY $W_R$-boson
signal. Here, $N_{\rm OSSF}$ stands for the number of opposite-sign same-flavour
lepton pairs, $N_{\rm OS}$ for the number of opposite-sign lepton pair and
$N_{\rm SF}$ for the number of same-flavour lepton pair. Moreover, $\ell\equiv
e, \mu$.}
\label{tab:def_sr}
\end{center}
\end{table} 
%%%%%%%%%%%%%%%%%%%%%%%%%%%%%%%

For our study, we  followed the CMS search of multileptonic new physics
signals as could emerge from electroweakino production~\cite{Sirunyan:2017lae}.
We  tested several signal regions introduced in this CMS search, all
featuring different lepton multiplicities and selections on transverse kinematic
variables like the missing transverse energy, the transverse momenta of various
systems, the transverse mass $M_T$ of systems made of one lepton and the missing
momentum, the stransverse mass $M_{T2}$ or the dilepton invariant mass
$M_{\ell\ell}$. The two signal regions that are most suitable for the
considered types of
LRSUSY spectra, are listed in \tabb{tab:def_sr}. In the {\bf SRA44} region,
one requires the presence of three charged leptons (electrons or muons), with at
least two of them forming a pair of opposite-sign same-flavour (OSSF) leptons.
One further constrains the invariant mass of this OSSF lepton system
$M_{\ell\ell}$,
relying on the pair that is the most compatible with a $Z$-boson if several
combinations are possible. The transverse mass $M_T$ of the system constructed
from the third lepton and the missing momentum is finally constrained, together
with the missing transverse momentum. In the {\bf SRD16} region,
one instead asks for two opposite-sign leptons (electrons and muons) and one
tau-lepton. The stransverse mass $M_{T2}$ originating from the lepton-pair is
then constrained, together with the missing transverse energy.

For our simulations, we used the \sarah implementation of the LRSUSY
model~\cite{Staub:2013tta,Basso:2015pka} and generated the particle spectrum by
means of \spheno~\cite{Porod:2003um}. DM calculations were performed using
\maddm~\cite{Ambrogi:2018jqj} and LHC simulations were performed at the
parton-level using \madgraph{5 (v2.5.5)} \cite{Alwall:2014hca} with the
UFO~\cite{Degrande:2011ua} model obtained from SARAH. We used the leading
order set of NNPDF parton distribution functions~\cite{Ball:2014uwa}. Showering
and hadronisation were performed using \pythia{ 8}~\cite{Sjostrand:2014zea},
and we have used \madanalysis{ 5 (v1.6.40)} \cite{Conte:2012fm,Conte:2014zja,Dumont:2014tja} to handle the simulation of the response of the CMS detector
(through its interface to \delphes{ 3 (v3.4.1)}~\cite{deFavereau:2013fsa} and
\fastjet{ (v3.3.0)}~\cite{Cacciari:2011ma}) and to recast the CMS analysis of
\citeref{Sirunyan:2017lae}, available from the MadAnalysis 5 Public Analysis
Database~\cite{1676304}.

%%%%%%%%%%%%%%%%%%%%%%%%%%%%%%%n{figure}
\begin{figure}[t]
\centering
\begin{minipage}[t][\minipageheightdp][t]{\minipagewidthdp}
\centering
\includegraphics[width=\figuremaxwidthdp,height=\figuremaxheightdp,keepaspectratio]{\main/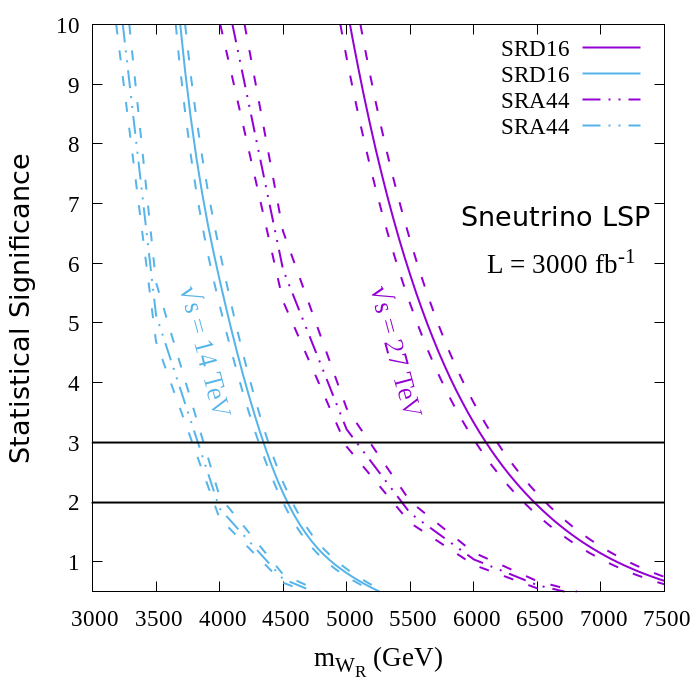}
\end{minipage}
\begin{minipage}[t][\minipageheightdp][t]{\minipagewidthdp}
\centering
\includegraphics[width=\figuremaxwidthdp,height=\figuremaxheightdp,keepaspectratio]{\main/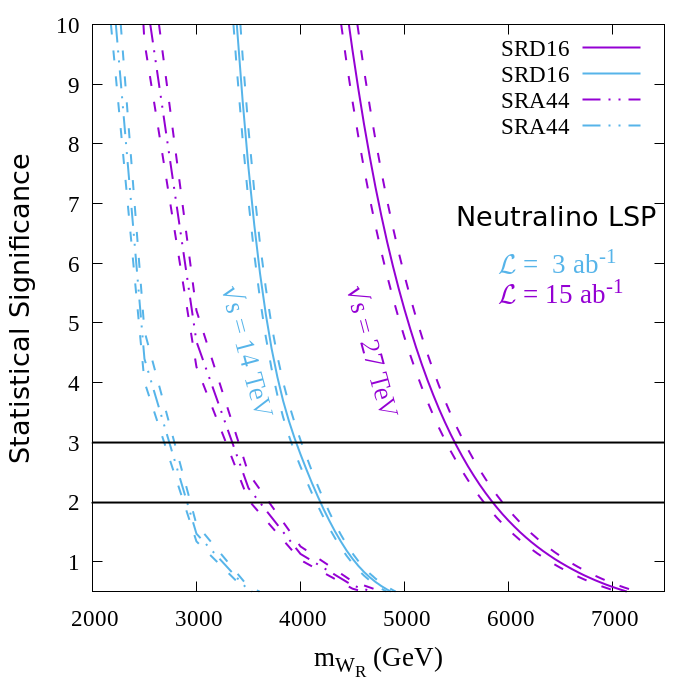}
\end{minipage}
\caption{Statistical significance of the two signal regions of the analysis of
\citeref{Sirunyan:2017lae} to a LRSUSY scenarios in the case of a sneutrino
(left) and a neutralino (right) LSP setup. We consider \com~energies
of $\sqrt{s}=14$ and $27\TeV$, and an integrated luminosities of $3$ and $15\abinv$.}
\label{fig:sig3000}
\end{figure}
%%%%%%%%%%%%%%%%%%%%%%%%%%%%%%%

The results are presented in \fig{fig:sig3000} for
\com~energies of $14$ and $27\TeV$. The two figures depict the reach in
the $W_R$-boson mass $M_{W_R}$ for the two signal regions of
\tabb{tab:def_sr} for the sneutrino LSP (left) and neutralino LSP (right)
scenarios. The two horizontal black lines represent the $2\sigma$ (mostly
equivalent to a $95\%\cl$ exclusion) and $3\sigma$ statistical significance. For the
$14 TeV$ analysis, we considered the same
SM background as in \citeref{Sirunyan:2017lae}, appropriately scaled to the
required luminosity and assuming relative errors similar to the
$35.9\fbinv$ case. For the $27\TeV$ analysis, we scaled all
background contributions by a factor of 3, since the increase in the background
cross-sections (for all the dominant channels) is approximately of 3 compared
with $\sqrt{s}=14\TeV$. Whilst this approximation is crude due to lack of
information regarding the background contributions from non-prompt leptons and
conversions at $\sqrt{s}=27\TeV$, it allows us to get back-of-the-envelope
estimations. As the targeted $W_R$ masses are way larger than the electroweakino
masses, the overall cut efficiencies do not change by more than $10\%$, so that
uniform signal selection efficiencies could be considered throughout the entire
$W_R$ mass range of $[2, 7.5]\TeV$. We however include in our estimations
the effect of a $10\%$ variation on the background uncertainties, as depicted by
the dashed lines. 

%\paragraph*{Discussion}

The {\bf SRD16} region proves to be the more favoured channel for both benchmark scenarios mainly because it features an almost background-free environment. It would even be more significant for a tau sneutrino LSP, as this leads to chargino decays into tau leptons. Our calculations however only consider cases where the sneutrino LSP is the electron sneutrino, so that they could be taken as conservative. The reach to sneutrino LSP scenarios is however better,
as could be expected from the potentially substantial mass gap featured by the particle spectrum. Multiple hard leptons can indeed arise from the cascade decays, in contrast to the higgsino LSP scenarios where the decay products have softer momenta as the spectrum is more compressed. We observe that for sneutrino and
neutralino LSP scenarios, $W_R$-boson masses up to respectively about $4.5\TeV$ and $4.2\TeV$ can be reached while providing enough events for electroweakinos signal 
sensitivity, when considering \intLumi of proton-proton
collisions at $\sqrt{s}= 14\TeV$. With $\sqrt{s} = 27\TeV$ and ${\mathcal L} = \intlumihelhc$,
the reach extends to about $6.5\TeV$ and $5.7\TeV$ respectively. All the limits are obtained from the sole {\bf SRD16} signal region, so that the {\bf SRA44} region could be used as a confirmatory channel if some excess would be observed.  

In conclusion, within the LRSUSY framework, $W_R$-boson-induced neutralino and chargino production could be used as a probe for dark-matter motivated scenarios. The HL and HE phases of the LHC could hence push the limits on the sensitivity to electroweakino searches as well as on the $W_R$-boson mass, relying on multilepton production in association with missing transverse momentum.
%%%%%%%%%%%%
\subsection{Searches for Sleptons: stau pair production at HE- and HL-LHC}\label{sec:sleptons}

Slepton pair production cross sections are less than $1\fbinv$ for sparticles above $400\GeV$ at $14\TeV$ \com~energy, hence searches for these processes will benefit considerably of the large datasets to be collected at the HL-LHC. 
In many SUSY scenarios with large $\tan \beta$, the stau ($\tilde{\tau}$) is lighter than the selectron and smuon, resulting in tau-rich final states. Co-annihilation processes favour a light stau that has a small mass splitting with a bino LSP, as it can set the relic density to the observed value. Searches for $\tilde{\tau}$ pair production are presented in this section using final state events with at least one hadronically decaying $\tau$ lepton as performed by ATLAS and CMS. 

\begin{figure}[t]
\begin{center}
  \includegraphics[width=0.3\textwidth]{\main/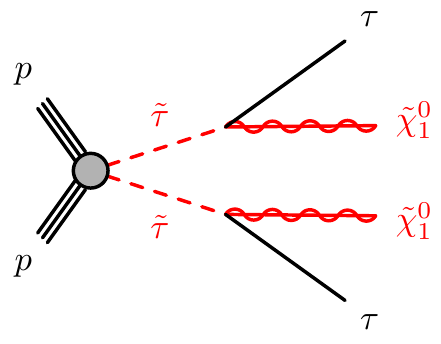}
\caption{Diagram for the $\tilde{\tau}$ pair-production.}
\label{fig:sms_stau_YR}
\end{center}
\end{figure}

The simplified model used for the optimisation of the searches and the
interpretation of the results is shown in \fig{fig:sms_stau_YR}. Assumptions on the mixture of left- and right-handed $\tau$ leptons as considered by the experiments are detailed where relevant.  

%\subsection{Prospects for direct stau production at the upgraded ATLAS and CMS}
%\label{sec:susy237}
%While searches for strong production of SUSY particles have already placed stringent limits of about 2\TeV on the strongly interacting SUSY particles, the EW production
%is more challenging because of lower production cross sections for given sparticle masses.

\subsubsection{\atlasC{Searches for $\tilde{\tau}$ pair production in the hadronic channel ($\tau_h$$\tau_h$) at ATLAS at the HL-LHC}}
\contributors{H. Cheng, D. Xu, C. Zhu, X. Zhuang, ATLAS}
%{\bf Author(s): H. Cheng (Beijing IHEP), C. Zhu (Beijing IHEP), X. Zhuang (Beijing IHEP)}

%In the ATLAS search, two simplified models describing the direct production of stau are used: one considers stau partners of the
%left-handed $\tau$ lepton ($\tilde{\tau_L}$), and a second considers stau partners of the right-handed $\tau$ lepton ($\tilde{\tau_R}$). In both models, the stau decays with a branching fraction of 100\% to the SM tau-lepton and the LSP.  \emph{Status: in ATLAS approval.}

In the ATLAS search~\cite{ATL-PHYS-PUB-2018-048}, two models describing the direct production of stau are employed: one considers stau partners of the
left-handed $\tau$ lepton ($\tilde{\tau_L}$), and a second considers stau partners of the right-handed $\tau$ lepton ($\tilde{\tau_R}$). In both models, the stau decays with a branching fraction of $100\%$ to the SM tau-lepton and the LSP. A search for stau production is presented here, which uses a final state with two hadronically decaying $\tau$ leptons, low jet activity, and large missing transverse energy ($E_{\mathrm{T}}^{\mathrm{miss}}$) from the $\tilde{\chi}^0_1$ and neutrinos. 
The SM background is dominated by $W$+jets, multi-boson production and top pair production.

The event pre-selection is based on that of the previous $8\TeV$ analysis~\cite{Aad:2015eda} and $13\TeV$ analysis~\cite{Aaboud:2017nhr}. 
Hadronically decaying taus are selected with $p_{\mathrm{T}} > 20\GeV$ and $|\eta| < 4$, while electrons and muons are selected with $p_{\mathrm{T}} > 10\GeV$ and $|\eta|< 2.47$ ($|\eta|<2.5$ for muons). 
Jets are reconstructed with the anti-$k_t$ algorithm with a radius parameter of $0.4$, with $p_{\mathrm{T}}>20\GeV$ and $|\eta|<4$.
To remove close-by objects from one another, an overlap removal based on $\Delta R$ is applied. 
In processes where jets may be misidentified as hadronically decaying taus, each jet is assigned a weight corresponding to the tau fake rate in the HL-LHC detector performance parameterisation. 

\begin{table}[t]
\centering
\renewcommand\arraystretch{1.2}
\small{
\begin{tabular}{ ccccc }
\hline
%\multicolumn{5}{c}{Common Selection} \\
%\hline
%\multicolumn{5}{c}{exactly two tight taus with opposite sign} \\
%\multicolumn{5}{c}{$Z$-veto}\\
%\multicolumn{5}{c}{$E_{\mathrm{T}}^{\mathrm{miss}} >200$ GeV} \\
%\multicolumn{5}{c}{$p_{\mathrm{T}\tau2} > 75 $ GeV} \\
%\multicolumn{5}{c}{$\Delta R (\tau1,\tau2)<3$} \\
%\multicolumn{5}{c}{$\Delta \phi (\tau1,\tau2)$  $>2$} \\
\hline
Selection & SR-low $[\mathrm{GeV}]$  & SR-med $[\mathrm{GeV}]$ & SR-high $[\mathrm{GeV}]$ & SR-exclHigh $[\mathrm{GeV}]$\\
\hline
$p_{\mathrm{T}\,\textrm{jet}} >$  & $ 40$  &$ 40$  & $20 $ & -\\
$p_{\mathrm{T}\tau1} >$ & $150 $  & $200 $  & $200 $  & $200 $ \\
$m_\mathrm{T\tau1}+m_\mathrm{T\tau2}>$ & $500$    & $700$   & $800$ & $800$\\
\hline
$m_\mathrm{T2} (\tau1,\tau2) $ & $\in [80, \infty]$ & $\in [130, \infty]$ & $\in [130, \infty]$  &  $\in [80, 130]$ \\
& &  &   & $\in [130,180]$ \\
& &  &   & $\in [180,230]$  \\
& &  &   & $\in [230, \infty]$\\
\hline
\end{tabular}\caption{Summary of selection requirements for the direct stau signal regions.
\label{tab:final_2tau_SR_text}}
}
\end{table}

Events are selected with exactly two tightly identified hadronic taus
with $|\eta|<2.5$, and the two taus must have opposite electric charge (OS).
The tight tau algorithm correctly identifies one-prong (three-prong) taus with an efficiency of $60\%$ ($45\%$), with a light-flavour jet misidentification probability of $0.06\%$ ($0.02\%$). 
Events with electrons, muons, $b$-jets or forward jets ($|\eta| > 2.5$) are vetoed. 
The effect of a di-tau trigger is considered by requiring that
the leading tau $p_{\mathrm{T}}$ is larger than $50\GeV$ and the sub-leading tau $p_{\mathrm{T}}$ is larger than $40\GeV$, with an assumed trigger efficiency of $64\%$. 
To suppress the SM background, a loose jet veto is applied that
rejects events containing jets with $|\eta| < 2.5$ and $p_{\mathrm{T}}>100\GeV$.

\begin{figure}[t]
\centering
\begin{minipage}[t][\minipageheightdp][t]{\minipagewidthdp}
\centering
\includegraphics[width=\figuremaxwidthdp,height=\figuremaxheightdp,keepaspectratio]{\main/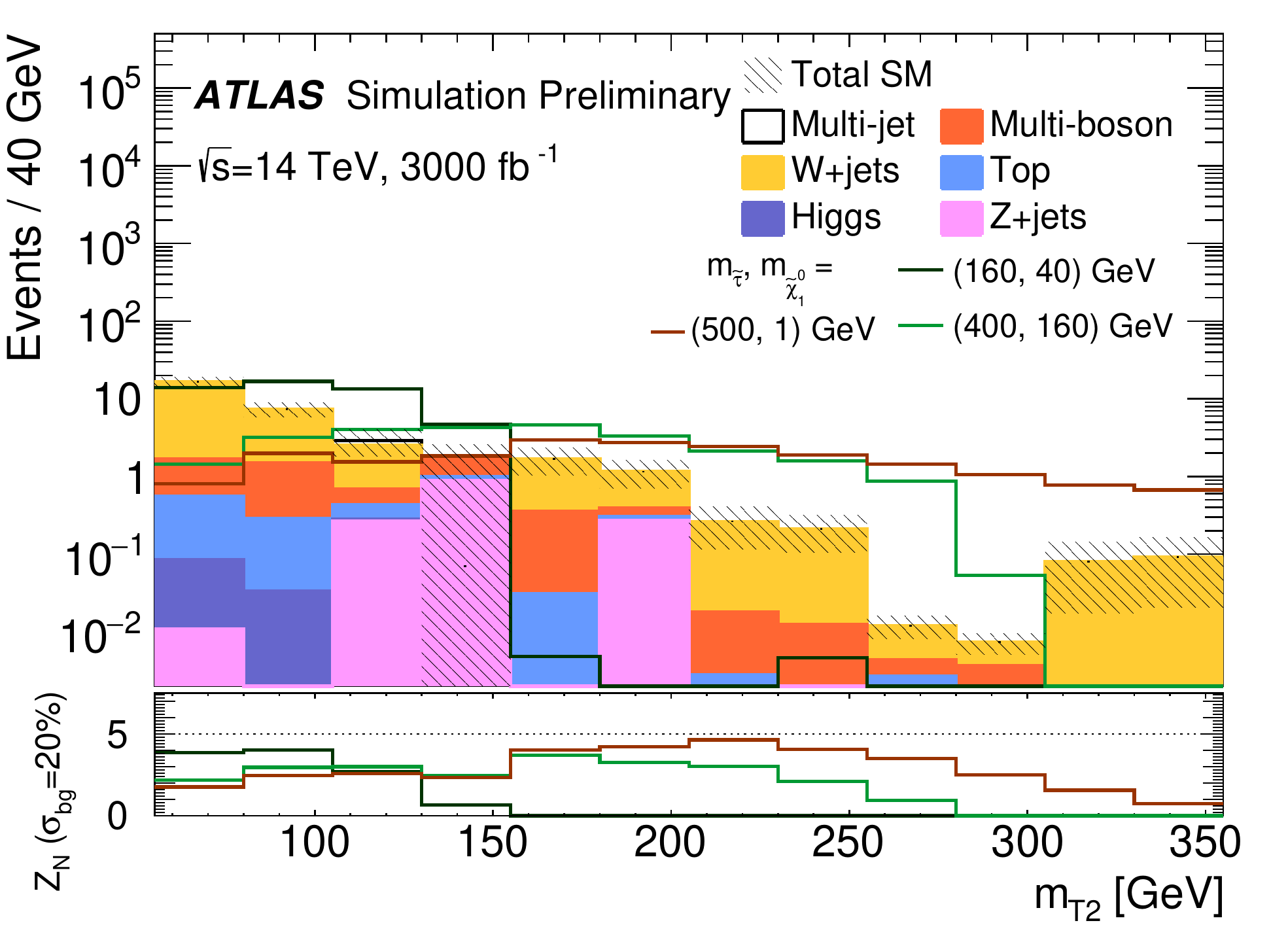}
\end{minipage}
\begin{minipage}[t][\minipageheightdp][t]{\minipagewidthdp}
\centering
\includegraphics[width=\figuremaxwidthdp,height=\figuremaxheightdp,keepaspectratio]{\main/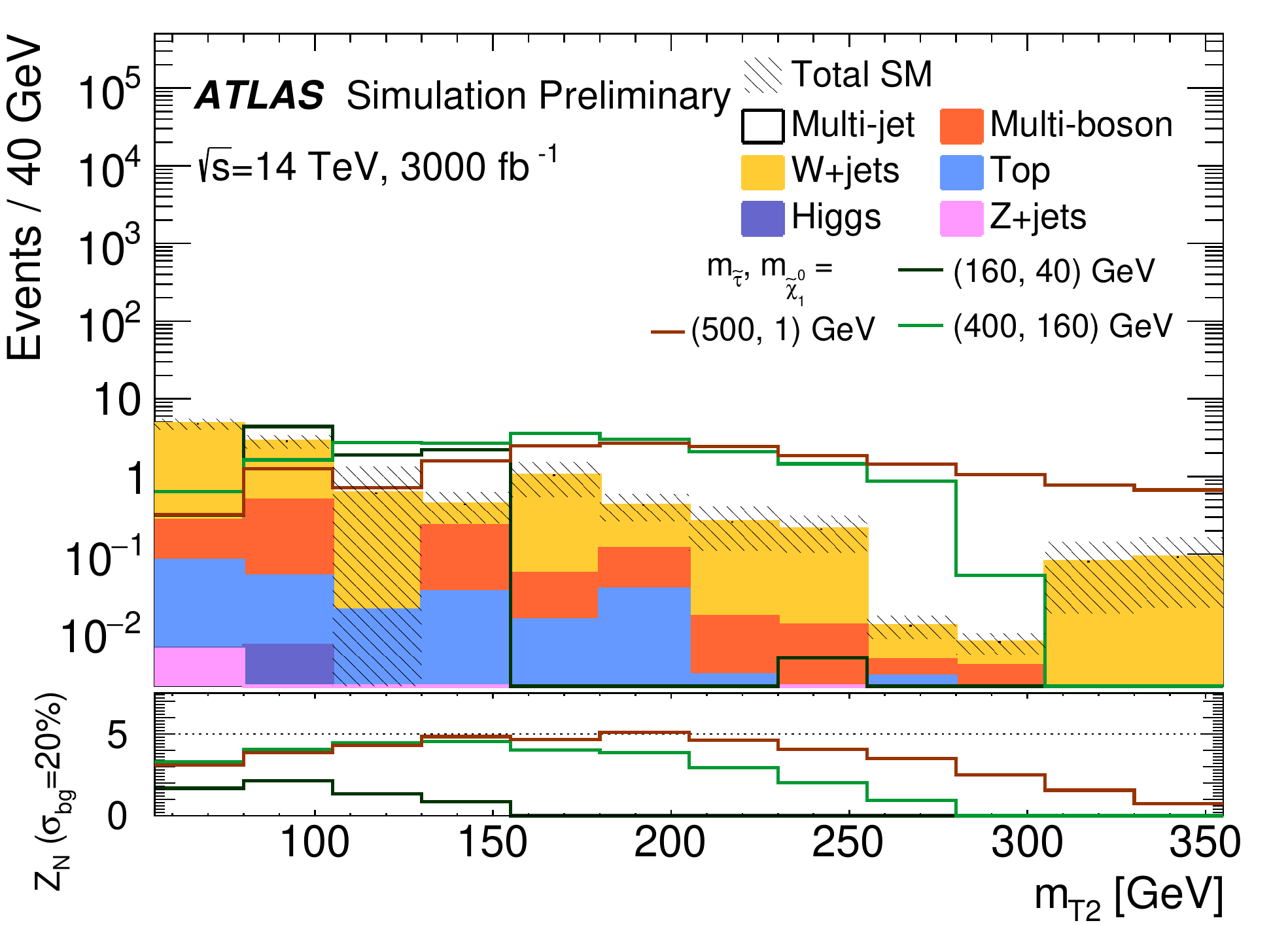}
\end{minipage}\\
\begin{minipage}[t][\minipageheightdp][t]{\minipagewidthdp}
\centering
\includegraphics[width=\figuremaxwidthdp,height=\figuremaxheightdp,keepaspectratio]{\main/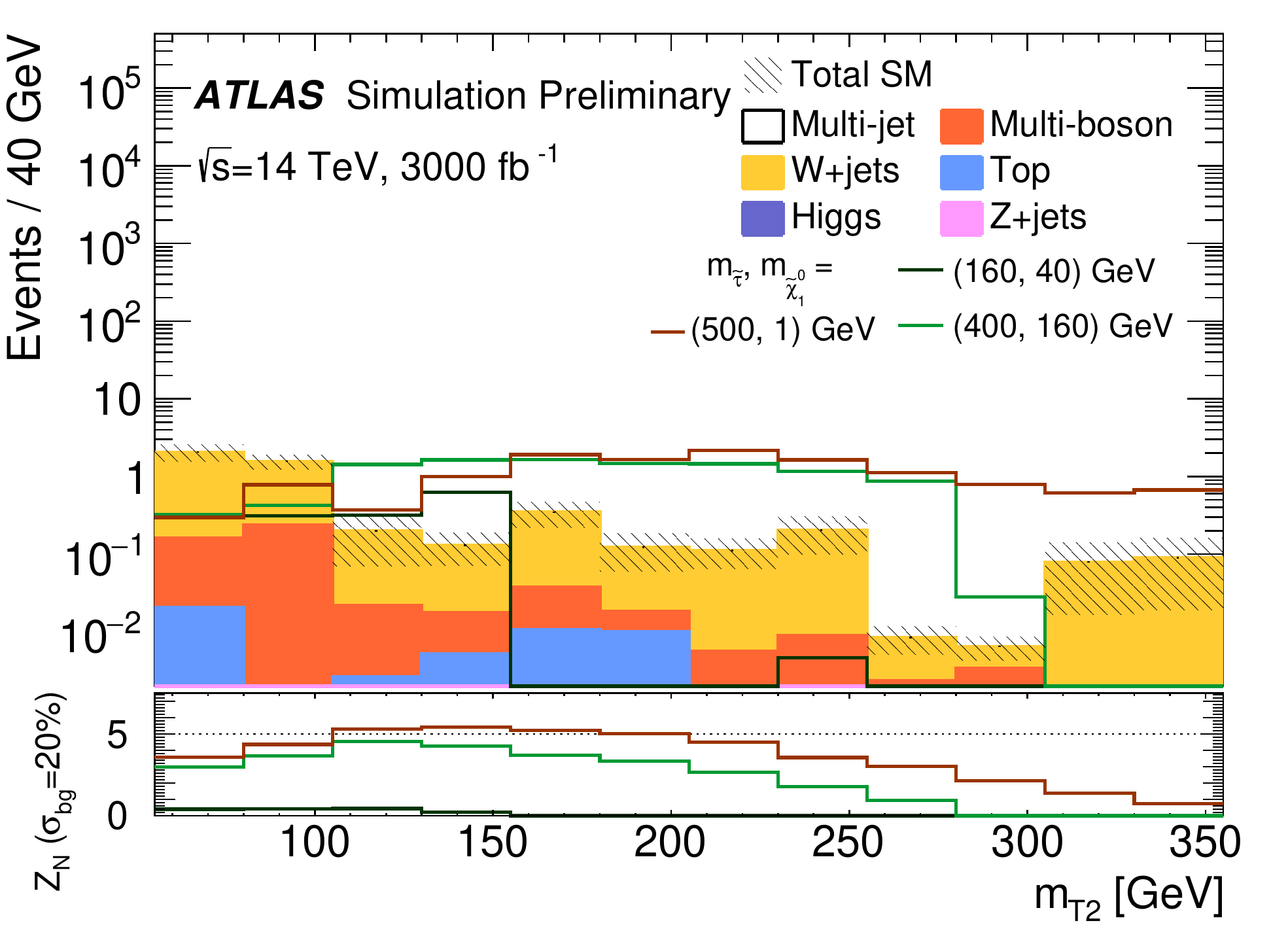}
\end{minipage}
\begin{minipage}[t][\minipageheightdp][t]{\minipagewidthdp}
\centering
\includegraphics[width=\figuremaxwidthdp,height=\figuremaxheightdp,keepaspectratio]{\main/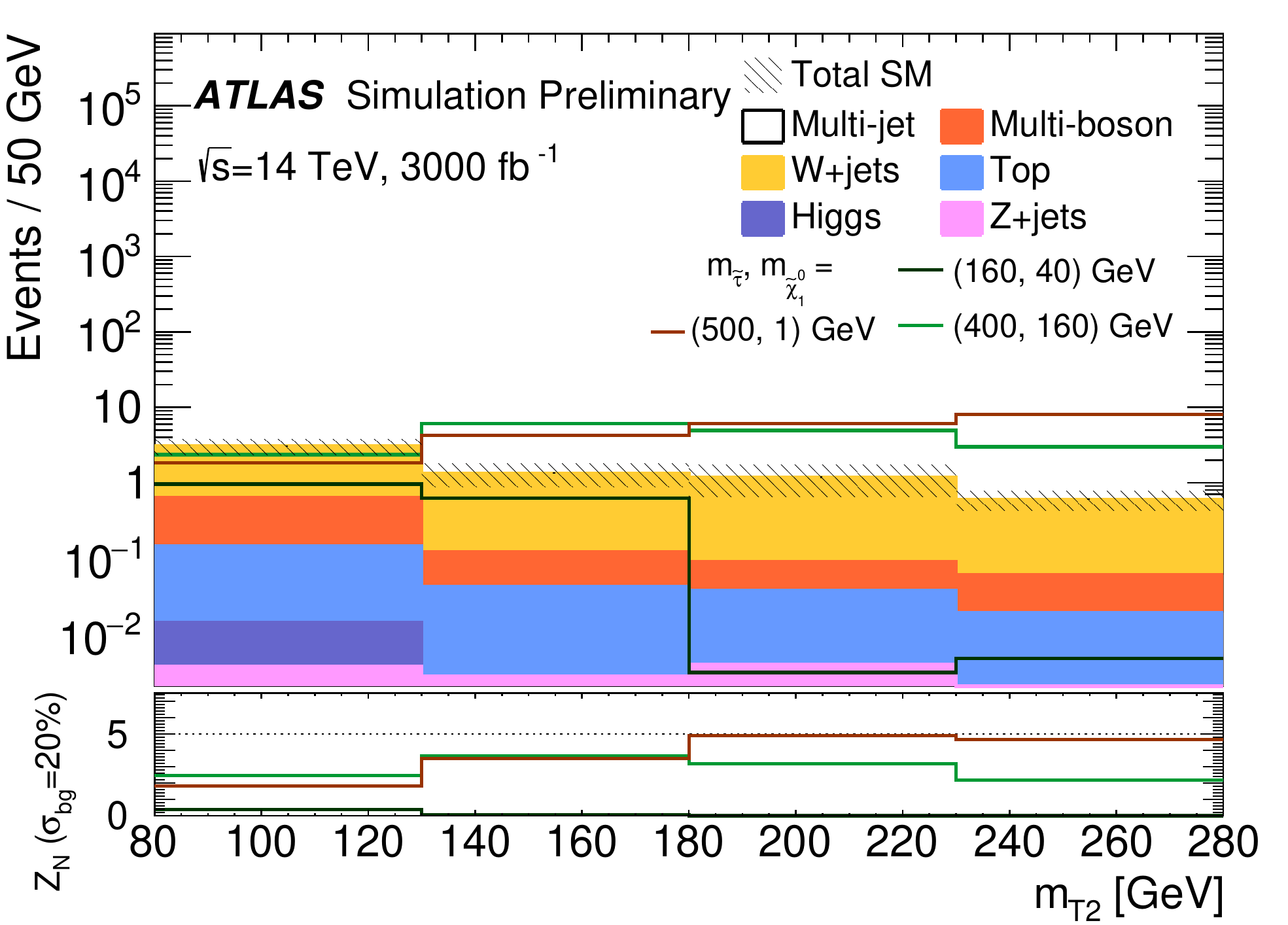}
\end{minipage}
\caption{
Distributions of each $m_\mathrm{T2}$ variable in the SR-low, SR-med, SR-high and SR-exclHigh regions, 
applying all selections as \tabb{tab:final_2tau_SR_text} with the exception of $m_\mathrm{T2}$ itself. 
The stacked histograms show the expected SM backgrounds normalised to \intLumi.
The hatched bands represent the statistical uncertainties on the total SM background.
For illustration, the distributions of the SUSY reference points are also shown as dashed lines. 
The lower pad in each plot shows the significance, $Z_N$ using a background uncertainty of $20\%$, for the SUSY reference points.
In the SR-exclHigh plot, the sensitivity distribution is the distribution for each $m_\mathrm{T2}$ bin.
\label{fig:2tau_mttwo}}
\end{figure}

Since the SUSY signal involves two undetected $\tilde{\chi}^0_1$, the resulting $E_{\mathrm{T}}^{\mathrm{miss}}$ spectrum tends to be harder than that for the the major SM backgrounds, thus $E_{\mathrm{T}}^{\mathrm{miss}} > 200\GeV$ is required to reject the multi-jet background. 
A $Z$ veto is imposed, where the invariant mass of the two taus, $m_{\tau\tau}$, is required to be larger than $100\GeV$ to suppress contributions from
$Z/\gamma^{*}+\mathrm{jets}$ production. 
To suppress the top quark and multi-jet backgrounds, the sum of the two-tau transverse mass defined using the transverse momentum of the leading (next-to-leading) tau and $E_{\mathrm{T}}^{\mathrm{miss}}$, must be larger than $450\GeV$.
The transverse mass requirement of $m_\mathrm{T2} >35\GeV$ is used to to further suppress the top, $W$+jets and $Z/\gamma^{*}+\mathrm{jets}$ backgrounds. 

In order to increase the discrimination power between signal and SM backgrounds several kinematic variables are further applied:
the $p_\mathrm{T}$ of the next-to-leading tau, $p_{\mathrm{T}\tau2} > 75\GeV$, and    
the angular separation between the leading and next-to-leading tau is required to be $\Delta \phi(\tau1,\tau2) > 2$ and $\Delta R(\tau1,\tau2) < 3$.  

Following these preselection requirements, three signal regions are defined to maximise model-independent discovery sensitivity targeting scenarios 
with low (SR-low), medium (SR-med) and high (SR-high) mass differences between the $\tilde{\tau}$ and $\tilde{\chi}^0_1$. 
A set of disjoint signal regions binned in $m_\mathrm{T2}$ are also defined to maximise model-dependent exclusion sensitivity based on the previous SR-high signal region
with the jet veto threshold cut removed. Each SR is identified by the range of the $m_\mathrm{T2}$. 
All signal regions are shown in \tabb{tab:final_2tau_SR_text}.

Figure~\ref{fig:2tau_mttwo} shows the distributions of $m_\mathrm{T2}$ 
in these signal regions, applying all SR selections with the exception of $m_\mathrm{T2}$ itself.

The systematic uncertainties are evaluated based on the SR-high systematic uncertainty in \sloppy\mbox{\citeref{Aaboud:2017nhr}}.  
A few of the experimental uncertainties are expected to be smaller at the HL-LHC compared to the $13\TeV$ studies.
In particular, the tau energy scale in-situ uncertainty is scaled by a factor of $0.6$ and the tau ID efficiency uncertainty is scaled by a factor of $0.45 - 0.9$. 
The multi-jet uncertainties scale with the increased integrated luminosity, and the background theoretical uncertainties are halved. 
The theoretical cross-section uncertainty for direct stau production is taken as $10\%$, while the MC/data related systematics are considered negligible. 
All other uncertainties are assumed to be the same as in the $13\TeV$ studies. 
In this assumption, the total background experimental uncertainty is $\sim 19\%$, with theoretical uncertainties on the Top, $Z/\gamma^{*}+\mathrm{jets}$ and Higgs backgrounds of $13\%$, theoretical uncertainties on the $W$+jets and multi-jet backgrounds of $10\%$, and uncertainties on the multi-boson background of $8\%$. This is referred to as the "Baseline uncertainty" scenario. 
The total uncertainty on the SUSY signal is $\sim 14\%$. 
Another scenario is also considered, where the expected uncertainties at the HL-LHC do not improve upon the $13\TeV$ studies for the SM background and signal. 
This results in a total background uncertainty of $\sim 38\%$ and a signal uncertainty of $\sim 21\%$ and is referred to as "Run-2 scenario". 

\begin{figure}[t]
\centering
\begin{minipage}[t][\minipageheightsp][t]{\minipagewidthsp}
\centering
\includegraphics[width=\figuremaxwidthsp,height=\figuremaxheightsp,keepaspectratio]{\main/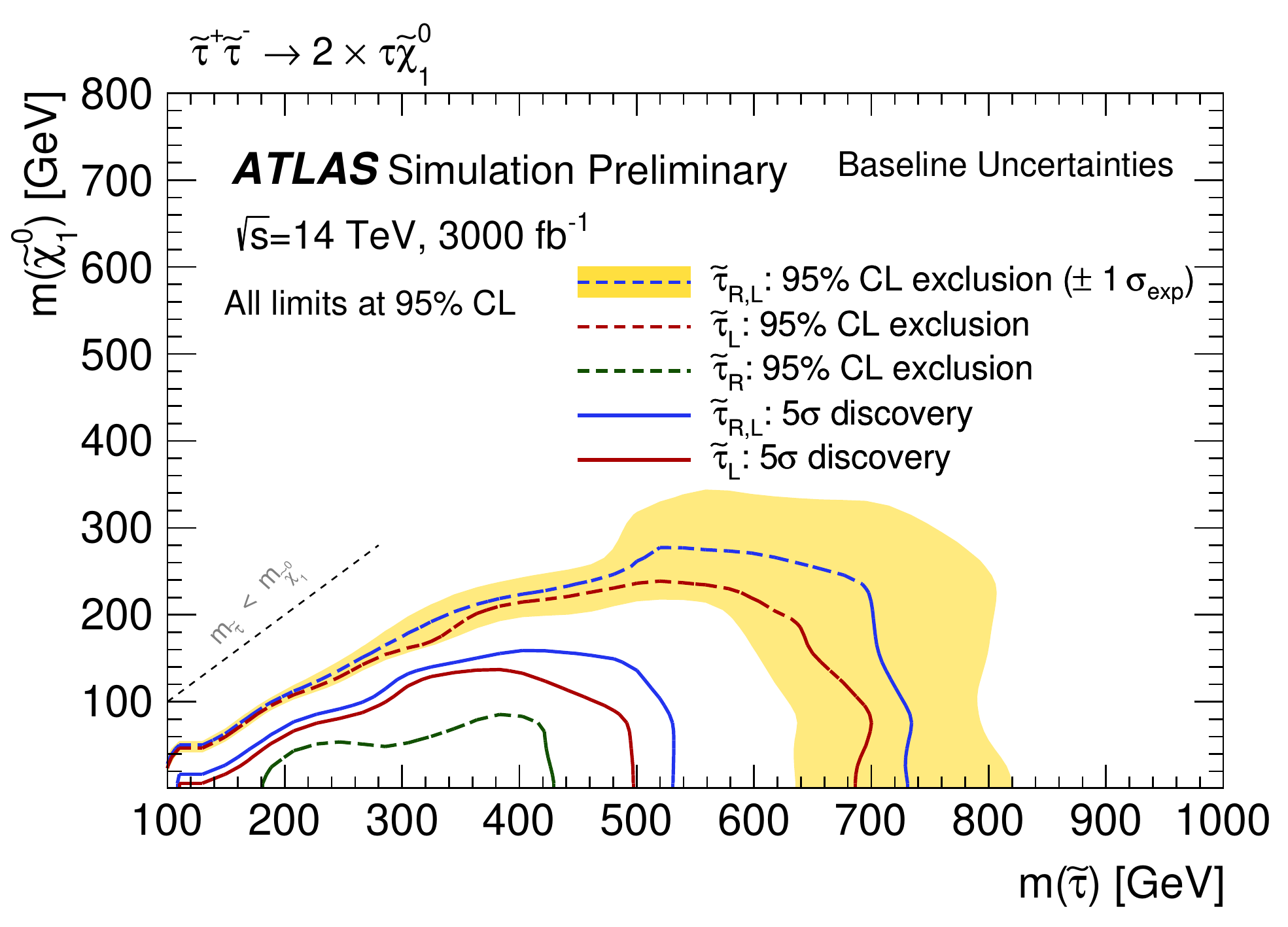}
\end{minipage}
\caption{$95\%\cl$ exclusion limits and $5\sigma$ discovery contours for \intLumi luminosity on the pure 
$\tilde{\tau_L}\tilde{\tau_L}$ or $\tilde{\tau_R}\tilde{\tau_R}$ and combined $\tilde{\tau_L}\tilde{\tau_L}$, $\tilde{\tau_R}\tilde{\tau_R}$ production in HL-LHC under the baseline systematic uncertainty assumptions. %The systematic uncertainties on signal and SM backgrounds are taken into account.
\label{fig:2tau_limit}}
\end{figure}

To calculate the discovery potential, SR-low, SR-med and SR-High defined in \tabb{tab:final_2tau_SR_text} are used, 
while for the final exclusion limit, the best expected exclusion resulting from these and one additional region, SR-exclHigh, are used. 
The $95\%\cl$ exclusion limits and $5\sigma$ discovery contours on the
combined $\tilde{\tau_L}\tilde{\tau_L}$ and $\tilde{\tau_R}\tilde{\tau_R}$ production, and separate $\tilde{\tau_L}\tilde{\tau_L}$ and $\tilde{\tau_R}\tilde{\tau_R}$ productions 
under baseline systematic uncertainty assumptions are shown in \fig{fig:2tau_limit}. 
The exclusion limit reaches $730\GeV$ in $\tilde{\tau}$ mass for the
combined $\tilde{\tau_L}\tilde{\tau_L}$ and $\tilde{\tau_R}\tilde{\tau_R}$ production, and $690\GeV$ ($430\GeV$)
for pure $\tilde{\tau_L}\tilde{\tau_L}$ (pure $\tilde{\tau_R}\tilde{\tau_R}$) production with a massless $\tilde{\chi}^0_1$.  
The discovery sensitivity reaches $110-530\GeV$ ($110-500\GeV$) in $\tilde{\tau}$ mass for the combined $\tilde{\tau_L}\tilde{\tau_L}$ and $\tilde{\tau_R}\tilde{\tau_R}$ (pure $\tilde{\tau_L}\tilde{\tau_L}$) production with a massless $\tilde{\chi}^0_1$.
No discovery sensitivity is found for pure $\tilde{\tau_R}\tilde{\tau_R}$ production as the production cross section is very small although a further reduction of the systematic uncertainties might open a window for discovery in the $100-200\GeV$ mass range. In general, sensitivity is achieved for scenarios with large mass difference between the stau and neutralino, \iee~$\Delta m(\tilde{\tau}, \tilde{\chi}^0_1) > 100\GeV$.

Under the assumption where the expected uncertainties at the HL-LHC do not improve upon the $13\TeV$ (Run-2 scenario), the exclusion limit is reduced slightly, which down to $720\GeV$ in $\tilde{\tau}$ mass for the combined $\tilde{\tau_L}\tilde{\tau_L}$  and $\tilde{\tau_R}\tilde{\tau_R}$ production and $670\GeV$ ($400\GeV$) for pure $\tilde{\tau_L}\tilde{\tau_L}$ (pure $\tilde{\tau_R}\tilde{\tau_R}$) production with a massless $\tilde{\chi}^0_1$.
The discovery sensitivity is also slightly reduced by about $20-50\GeV$.%, reaching $200-500$ GeV ($210-460$ GeV) in $\tilde{\tau}$  mass for the combined $\tilde{\tau_L}\tilde{\tau_L}$ and $\tilde{\tau_R}\tilde{\tau_R}$ (pure $\tilde{\tau_L}\tilde{\tau_L}$) production with a massless $\tilde{\chi}^0_1$.

\subsubsection{\cmsC{Searches for $\tilde{\tau}$ pair production in the $\tau_h \tau_h$ and $\tau_{\ell} \tau_h$ channels at CMS at the HL-LHC}}
\label{sec:cms-stau}
\contributors{I. Babounikau, A. Canepa, O. Colegrove, V. Dutta, I. Melzer-Pellmann, CMS}

CMS investigates the expected reach for direct stau ($\tilde{\tau}$) pair production, where the $\tilde{\tau}$ decays to a $\tau$ and the lightest
SUSY particle, the neutralino (\PSGczDo)~\cite{CMS:2018imu}.  
Final states with either two hadronically
decaying tau leptons ($\tau_h$) or one $\tau_h$ and one electron or muon, referred to in the following as the $\tau\tau$ 
and $\ell\tau$ channels, respectively, are considered. In both cases we expect missing transverse momentum from the two LSPs.

The search assumes $\tilde{\tau}$ pair production in the mass-degenerate scenario. The cross-sections have been
computed for $\sqrt{s} = 14\TeV$ at NLO using the Prospino code~\cite{Beenakker:1996ed}. Final values are
calculated using the
PDF4LHC recommendations for the two sets of cross sections following the prescriptions of the
LHC SUSY Cross Section Working Group~\cite{Fuks:2013lya}.

The event selection for each final state requires the presence of exactly two reconstructed leptons with
opposite charges, corresponding to the $\tau\tau$ or $\ell\tau$ final states.
In order to pass the selection, electrons (muons) are required to have transverse momentum $\pt > 30\GeV$ and pseudorapidity $|\eta| < 1.6 (2.4)$ and a minimum azimuthal angle between each other of $1.5$. 
Dedicated lepton identification criteria are applied, providing $50\%$ to $90\%$ efficiency for muons and $25\%$ to $80\%$ efficiency for electrons, depending on $p_T$ and $\eta$. Both muons and electrons are required to be isolated.

The momentum of the $\tau_h$ candidates is required to be above $40\GeV$ in the $\ell\tau$ final state, while we require
$\pt>50\GeV$ for the $\tau_h$ in the $\tau\tau$ final state. For both final states, the $\tau_h$ is required to be within $|\eta|<2.3$. A tight working point is chosen for the $\tau_h$ identification in order to
obtain a small rate of jets being misidentified as $\tau_h$. The $\tau_h$ reconstruction efficiency for this working
point is about $30\%$, with a fake rate of about $0.08\%$ assuming an MVA optimisation.
Overlaps between the two reconstructed leptons in the $\ell\tau$ final state are avoided by requiring them
to have a minimum separation of $\Delta R > 0.3$.

In the $\ell\tau$ final state, all events with at least one jet are rejected.
In the $\tau\tau$ channel, in order to suppress backgrounds with top quarks, we veto events containing any \PQb-tagged jet
with $p_T > 40\GeV$ identified with the loose CSV working point
in both final states, which corresponds to an identification efficiency of about $60-65\%$.

\begin{figure}[t]
\centering
\begin{minipage}[t][\minipageheightdp][t]{\minipagewidthdp}
\centering
\includegraphics[width=\figuremaxwidthdp,height=\figuremaxheightdp,keepaspectratio]{\main/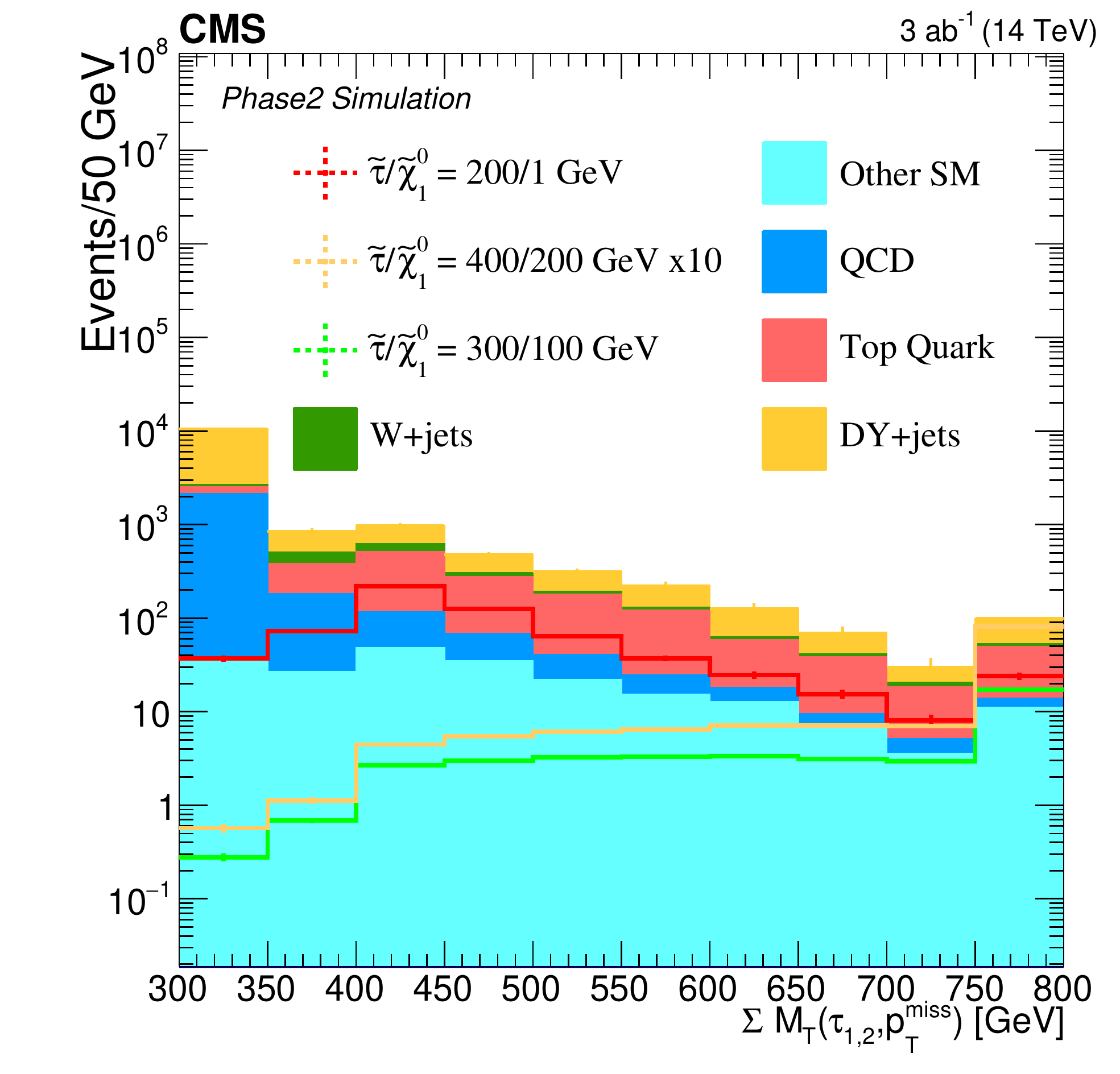}
\end{minipage}
\begin{minipage}[t][\minipageheightdp][t]{\minipagewidthdp}
\centering
\includegraphics[width=\figuremaxwidthdp,height=\figuremaxheightdp,keepaspectratio]{\main/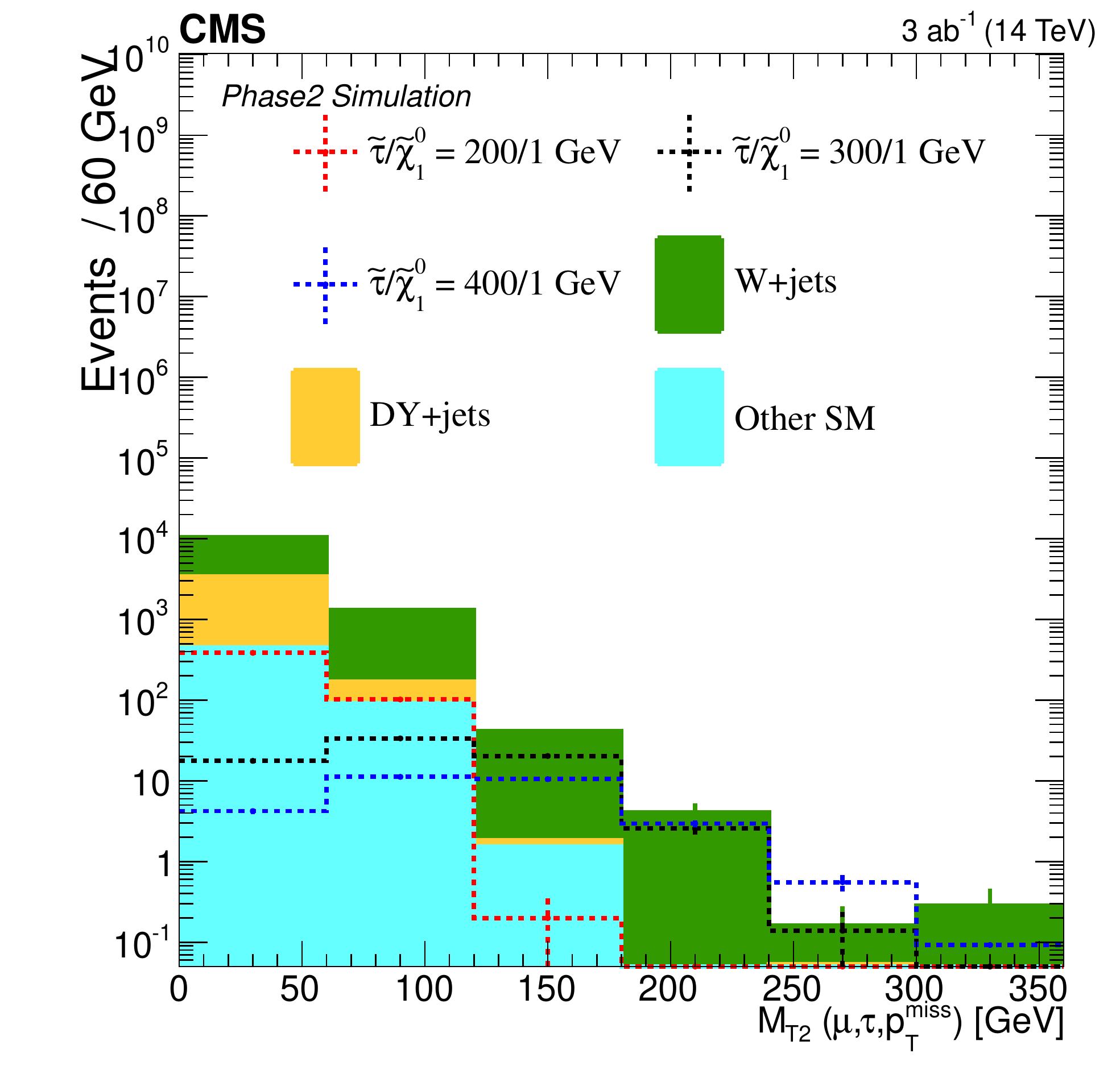}
\end{minipage}
\caption{\label{fig:SR_var} Example plots for the main search variables: $\Sigma M_T$ for the $\tau\tau$ analysis (left), and $M_{\rm T2}$ for the $\mu\tau$ analysis (right), both after the baseline selection.}
\end{figure}

The main background for the $\tau\tau$ final state after this selection consists of quantum chromodynamics
(QCD) multijet events, W+jets, DY+jets, and top quark events. Separating the background into 
prompt $\tau_h$ events, where both reconstructed taus are matched to a generator $\tau_h$, and misidentified events, 
where one or more non-generator matched jets has been misidentified as prompt $\tau_h$, we find that the misidentified background dominates our search regions.

In the $\ell\tau$ final state, all events with at least one jet are rejected. 
Due to kinematical constrains in the signal, we reduce the
background from QCD multijet events by requiring a maximum separation of the two leptons in $\Delta R$ of $3.5$.

In order to further improve discrimination against the SM background, we take advantage 
of the expected presence of two \PSGczDo in the final state for signal events, which would lead to missing transverse momentum, $\vec{p_T}_{miss}$, defined as the projection onto the
plane perpendicular to the beam axis of the negative vector sum of the momenta of all reconstructed objects in an event. Its magnitude, referred to as \ptmiss, is an important discriminator between signal and SM background.

Events are then further selected using discriminating kinematic variables for each of the two final states to improve the sensitivity of the search to a range of sparticle masses, such as the transverse mass, $M_T(\ell, \vec{p_T}_miss) \equiv \sqrt{2 p_{\ell} \ptmiss (1 - \cos \Delta\phi(\vec{p}_{\ell}, \vec{p_T}_{miss}))}$, where $\ell$ represents the lepton. In addition, the scalar sum of the $M_T$ calculated with the highest \pt\ ($\ell_1$) and second highest \pt\ ($\ell_2$) lepton and the missing transverse momentum is used to further reduce the background events: \sloppy\mbox{$\sum M_T = M_T(\ell_1, \vec{p_T}_{miss}) + M_T(\ell_2, \vec{p_T}_{miss})$}.
Finally the stransverse mass 
$M_{\rm T2}$~\cite{Lester:1999tx,Barr:2003rg} is used to discriminate the signal from the background.

The main variables that are used to define the search regions in the $\tau\tau$ final state are $\Sigma M_T$ and $M_{\rm T2}$, where the former is shown for the baseline selection in \fig{fig:SR_var}~(left). While we apply a stringent requirement of at least $400\GeV$ for $\Sigma M_T$, we require $M_{\rm T2}$ to be above $50\GeV$.
The $\tau\tau$ search regions are then binned in $M_{\rm T2}, \Sigma M_T$, and the number of jets $n_{\rm jet}$.

In the $\ell\tau$ final state, we require $M_T(\mu, \vec{p_T}_{miss})>120\GeV$, which reduces the W+jets background significantly.
To further suppress the SM background in the leptonic final states, \ptmiss has to be above $150\GeV$,
which mainly reduces QCD multijets and Drell Yan events.
Additional binning in $M_{\rm T2}$ and the \pt\ of the $\tau_h$ is applied to define the search regions in the $\ell\tau$ selection. Figure~\ref{fig:SR_var}~(right) shows the $M_{\rm T2}$ distribution after the baseline selection.

\begin{figure}[t]
\centering
\begin{minipage}[t][\minipageheightsp][t]{\minipagewidthsp}
\centering
\includegraphics[width=\figuremaxwidthsp,height=\figuremaxheightsp,keepaspectratio]{\main/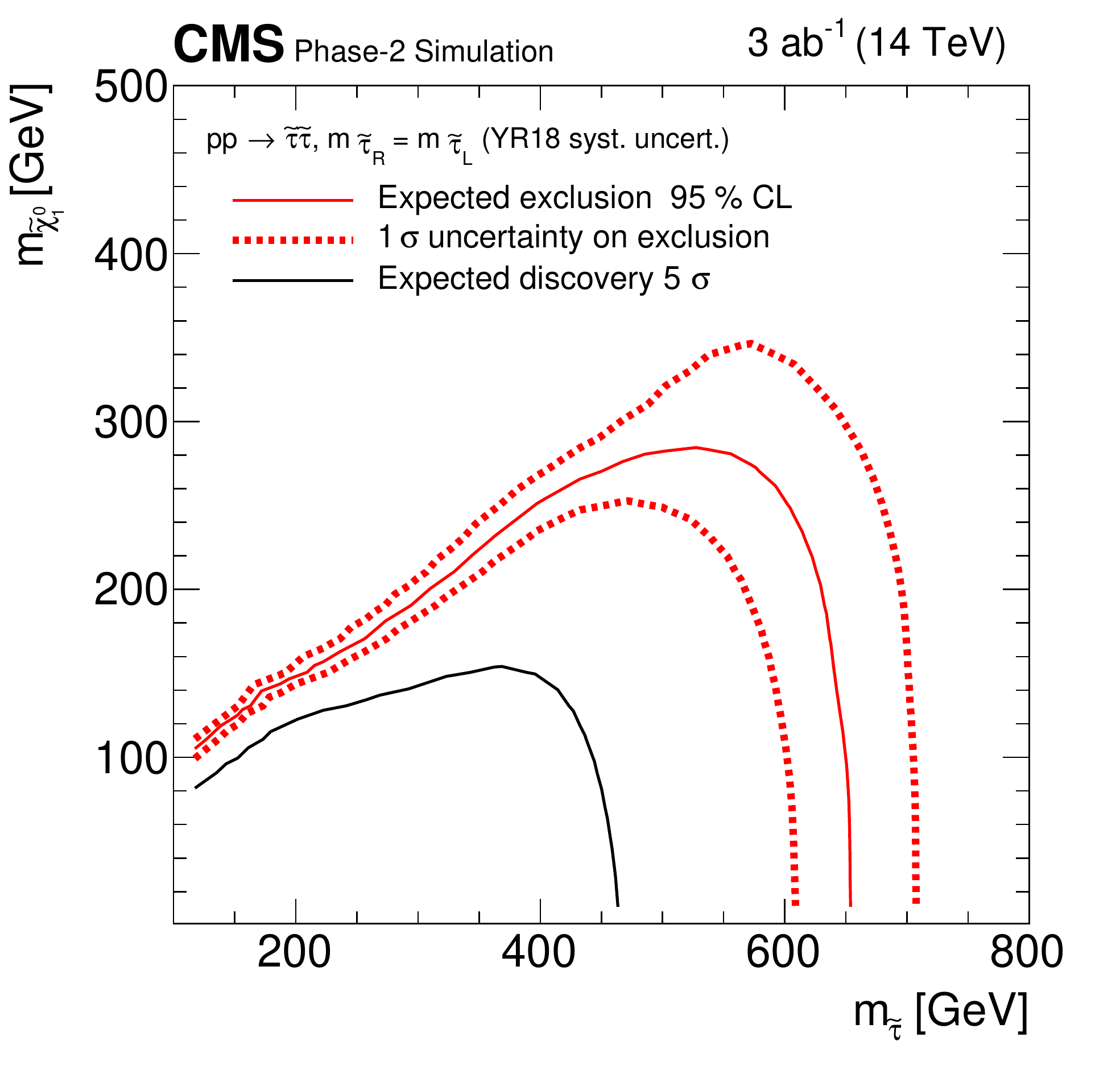}
\end{minipage}
\caption{Expected upper limits at the $95\%\cl$ (dashed line) and the $5 \sigma$ discovery potential (full line) for the combination of the results of the $\tau\tau$ and $\ell\tau$ channels. }
\label{fig:limits_YR}
\end{figure}

The dominant experimental uncertainties are those originating from jets being misidentified as $\tau_h$ ($15\%$), $\tau_h$ identification efficiency ($2.5\%$), the muon identification efficiency ($0.5\%$), the electron identification efficiency ($1\%$), the jet energy scale ($1-3.5\%$) and resolution ($3-5\%$), b-tagging efficiency ($1\%$) and the integrated luminosity ($1\%$). These systematic uncertainties are correlated between the signal and the irreducible background yields.

The expected upper limits and the discovery potential are given in \fig{fig:limits_YR}.
In mass-degenerate scenarios, degenerate production of $\tau$ sleptons are excluded up to $650\GeV$ with the discovery contour reaching up to $470\GeV$ for a massless neutralino.
The $\tau\tau$ analysis has been found to drive the sensitivity, but adding the $\ell\tau$ channel enlarges the exclusion bounds by about $60-80\GeV$.

\subsubsection{Remarks on stau pair production searches at HL-LHC} 

Prospects for stau pair production presented by ATLAS and CMS in the previous sections generally cover a similar region of the stau-neutralino mass plane. 
Stau masses up to $730\GeV$ are excluded by ATLAS for scenarios with large mass difference between stau and neutralino, \iee~$\Delta m(\tilde{\tau}, \tilde{\chi}^0_1) > 100\GeV$. CMS contours reach up to about $650\GeV$ covering a similar region in the parameter space. 
Differences in the reaches are small but noticeable, and are briefly highlighted in the following.   
The main difference between the ATLAS and CMS searches is the definition of the tau object. ATLAS has optimised the so-called working point (WP), \iee~the combination of selection requirements leading to a certain level of identification efficiency and jet-rejection rate, and chosen a WP leading to $45\%$ to $60\%$ efficiency as a function of $\pt$ and an average jet-rejection rate of $0.6\%$ ($0.02\%$) for 1-prong (3-prong) taus. The CMS analysis considers a tighter WP, resulting in an almost negligible level of misidentified taus but with lower efficiency ($\sim 30\%$). This leads to a small difference in terms of acceptance $\times$ efficiency which translates to $80$ ($50$)\GeV differences in the exclusion (discovery) contours. %Other differences arise from the parameterisation of the detectors response and the methods used for reconstructing missing transverse momentum and jets used for vetos. 

Finally, we underline that the sensitivity to more compressed scenarios, as predicted in theoretically favoured co-annihilation scenarios, might be partially recovered exploiting the presence of a high $\pt$ ISR jet, similarly to studies presented in~\secc{sec:susy236}. For this, identification of tau objects at low $\pt$ will be crucial.

%\subsubsection{Constraining slepton and chargino through compressed top squark search}
%
%{\bf Author(s): P. Konar et al.}
%
%EMPTY? 
\subsubsection{\cmsC{Searches for $\tilde{\tau}$ pair production in the $\tau_h \tau_h$ and $\tau_{\ell} \tau_h$ channels at CMS at HE-LHC}}
\contributors{I. Babounikau, A. Canepa, O. Colegrove, V. Dutta, I. Melzer-Pellmann, CMS}

On top of the CMS HL-LHC analysis, we also study the influence of the increased cross section for $27\TeV$ and the increased luminosity of $15\abinv$ expected to be achieved in HE-LHC~\cite{CMS:2018imu}. For this study the cross sections of all backgrounds and signal contributions are recalculated for $\sqrt{s} = 27\TeV$ at NLO using {\sc Prospino}. The signal region definition and kinematic distributions are the same as described in \secc{sec:cms-stau} for the HL-LHC study, but are scaled with the new cross sections and luminosity. The main gain in sensitivity comes from the increased luminosity, since the cross section increase for signal is the same order as that for background. The applied uncertainties are the same as for HL-LHC study described in \secc{sec:cms-stau}.

\begin{figure}[t]
\centering
\begin{minipage}[t][\minipageheightsp][t]{\minipagewidthsp}
\centering
\includegraphics[width=\figuremaxwidthsp,height=\figuremaxheightsp,keepaspectratio]{\main/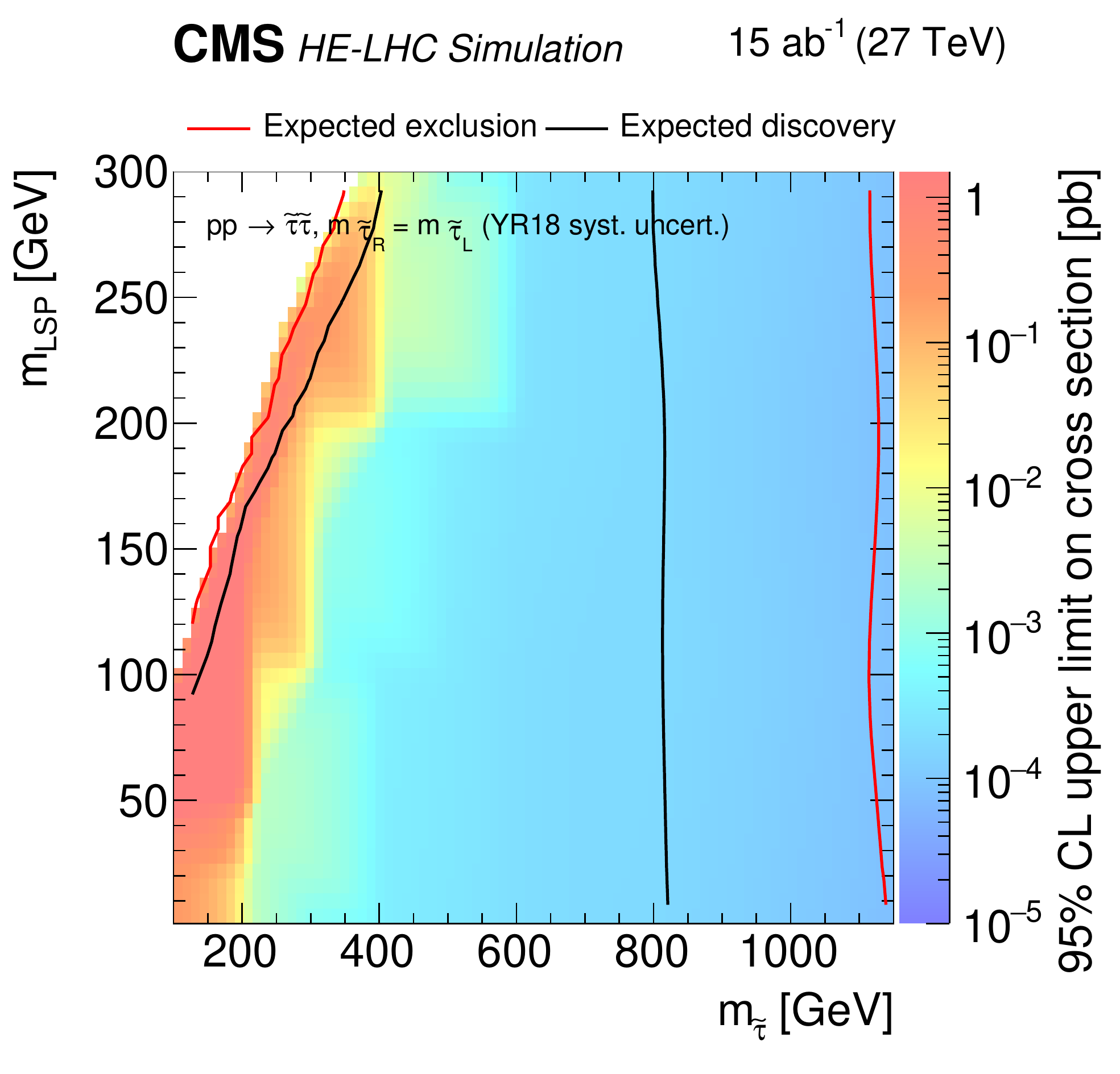}
\end{minipage}
\caption{Expected upper limits at the $95\%\cl$ (red line) and the $5 \sigma$ discovery potential (black line) for the combination of the results of the $\tau\tau$ and $\ell\tau$ channels for HE-LHC.}
\label{fig:limits_YR_HELHC}
\end{figure}

The expected upper limits and the discovery potential are given in~\fig{fig:limits_YR_HELHC}.
In the mass-degenerate scenario, $\tau$ slepton production is excluded up to $1150\GeV$ with the discovery contour reaching up to $810\GeV$ for a massless neutralino. Signal events were generated up to neutralino mass of $300\GeV$, at which point the discovery (exclusion) potential ranges from $400 - 800\GeV$ ($350 - 1100\GeV$).

\subsection{Other SUSY signatures and implications on SUSY models}\label{sec:gluinos}
Supersymmetry might manifest in different ways at hadron colliders. Simplified models help in setting the search strategy and illustrate the reach for individual processes, as shown in the prospects presented in previous sections. In this section, analyses of the discovery potential of HL- and HE-LHC are reported considering benchmark points in supergravity grand unified models, light higgsino scenarios, pMSSM and U(1)$^{\prime}$-extended  MSSM models. 

\subsubsection{%\theory{HE-LHC vs HL-LHC potential for SUSY discovery}
\theoryC{SUSY discovery potential at HL- and HE-LHC}}
\contributors{A. Aboubrahim, P. Nath}
%
%{\bf Author(s): A. Aboubrahim\footnote{Email: a.abouibrahim@northeastern.edu}, P. Nath\footnote{Email: p.nath@northeastern.edu}}
%\begin{center}
%Department of Physics, Northeastern University, Boston, MA 02115-5000, USA
%\end{center}

We give an analysis of the  discovery potential of HE-LHC with respect to the HL-LHC for supersymmetry, based on studies presented in ~\citeref{Aboubrahim:2018bil}. Specifically, a set of benchmark points which are discoverable both at HE-LHC and HL-LHC are presented. In addition, we also report on a set of benchmarks which are beyond the reach of HL-LHC but are discoverable at HE-LHC. The  models we consider are supergravity grand unified models~\cite{Chamseddine:1982jx,Nath:1983aw,Hall:1983iz,Arnowitt:1992aq} with non-universalities in the gaugino sector. Thus, the models are described by the set of parameters $m_0, m_1, m_2, m_3, A_0, \tan\beta, {\rm sgn}(\mu)$ where 
$m_0$ is the universal scalar mass (which can be large consistent with naturalness~\cite{Chan:1997bi}),
 $m_1,m_2, m_3$ are the $U(1), SU(2)_L, SU(3)_C$ gaugino masses, $A_0$ is the universal trilinear coupling, and 
$\tan\beta= \langle H_2\rangle/\langle H_1\rangle$  is the ratio of the Higgs \vevs. The analysis is done under the constraints of Higgs boson mass and the relic density constraints which requires coannihilation~\cite{Aboubrahim:2017wjl,Aboubrahim:2017aen,Nath:2016kfp}. The analysis
uses  signatures involving a single charged lepton and jets, two charged leptons and jets and three charged leptons and jets, resulting from the decay of a gluino pair (points (a)-(f) of \tabb{tab:hmass}) and the decay of $\tilde\chi^0_2\tilde\chi^{\pm}_1$ (points (g)-(j) of \tabb{tab:hmass}). It is found that most often
the dominant signature is the single lepton and jets signature, indicated as SR-1$\ell$-B or C in the figures, depending on the specific selections applied. Twelve different kinematic variables are used to discriminate the signal from the background. These consist of 
\begin{align}
N_{\rm jets}, \,\, E_T^{\rm miss}, \,\, H_T, \,\, m_{\rm eff}, \,\, R, \,\, H_{20}, \,\, p_T(j_n), \,\, m^{\ell}_T, \,\, m^{\rm min}_T(j_{1-2},E^{\rm miss}_T)
\end{align}
 where $N_{\rm jets}$ is the number of jets, $E_T^{\rm miss}$ is the missing transverse energy, $H_T$ is the sum of the jets' transverse momenta, $m_{\rm eff}$ is the effective mass, $R=E_T^{\rm miss}/(E_T^{\rm miss}+p^{\ell}_T)$, $H_{20}$ is the second Fox-Wolfram moment, $p_T(j_n)$ is the $n^{\rm th}$ jet transverse momentum, $m^{\ell}_T$ is the leading lepton transverse mass and $m^{\rm min}_T(j_{1-2},E^{\rm miss}_T)$ is the minimum of the transverse masses of the first and second leading jets. Finally, $p^{\ell}_T$ denotes the transverse momentum of the leading lepton. 
  
 \begin{figure}[t]
\centering
\begin{minipage}[t][\minipageheightdp][t]{\minipagewidthdp}
\centering
\includegraphics[width=\figuremaxwidthdp,height=\figuremaxheightdp,keepaspectratio]{\main/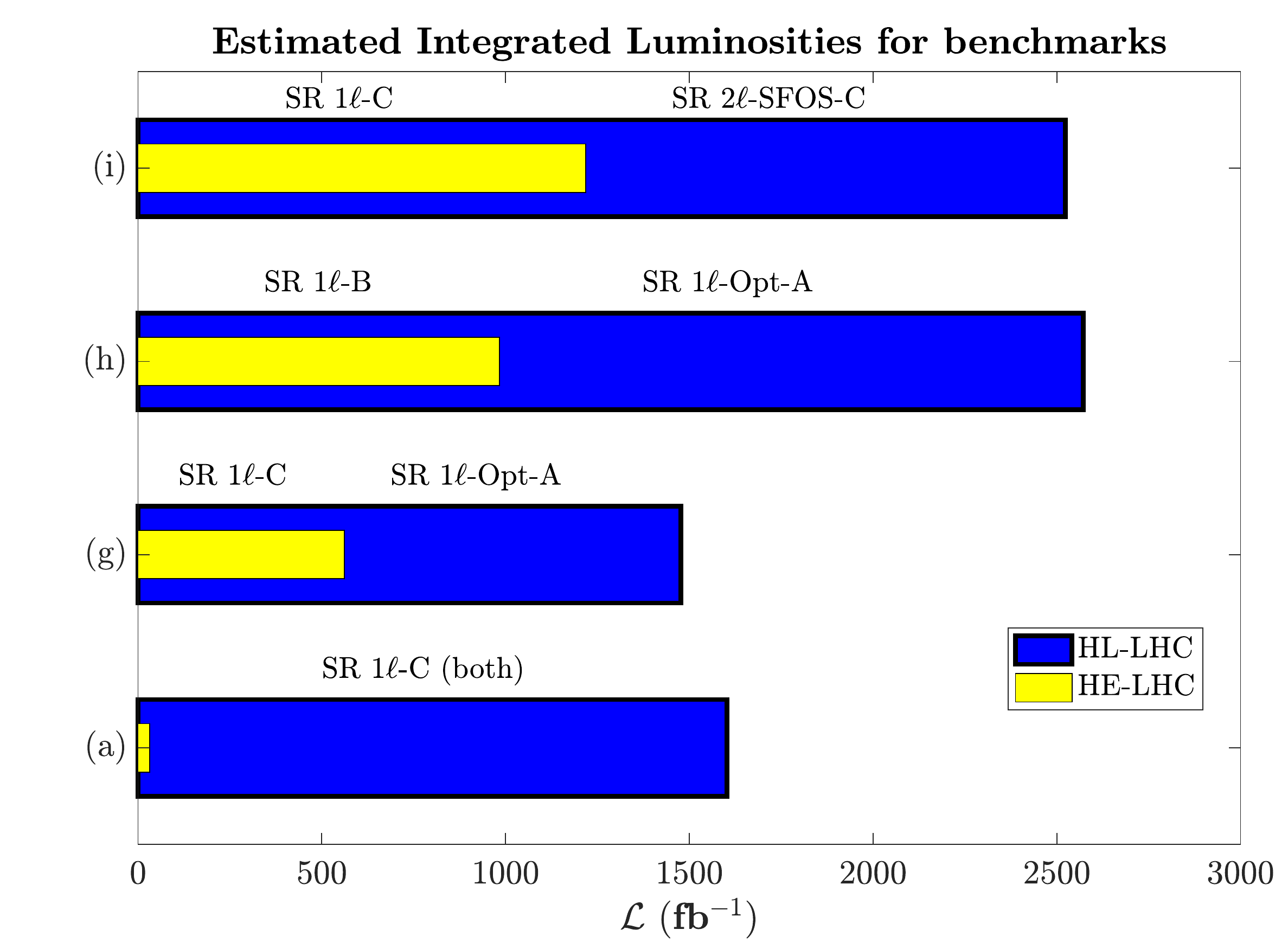}
\end{minipage}
\begin{minipage}[t][\minipageheightdp][t]{\minipagewidthdp}
\centering
\includegraphics[width=\figuremaxwidthdp,height=\figuremaxheightdp,keepaspectratio]{\main/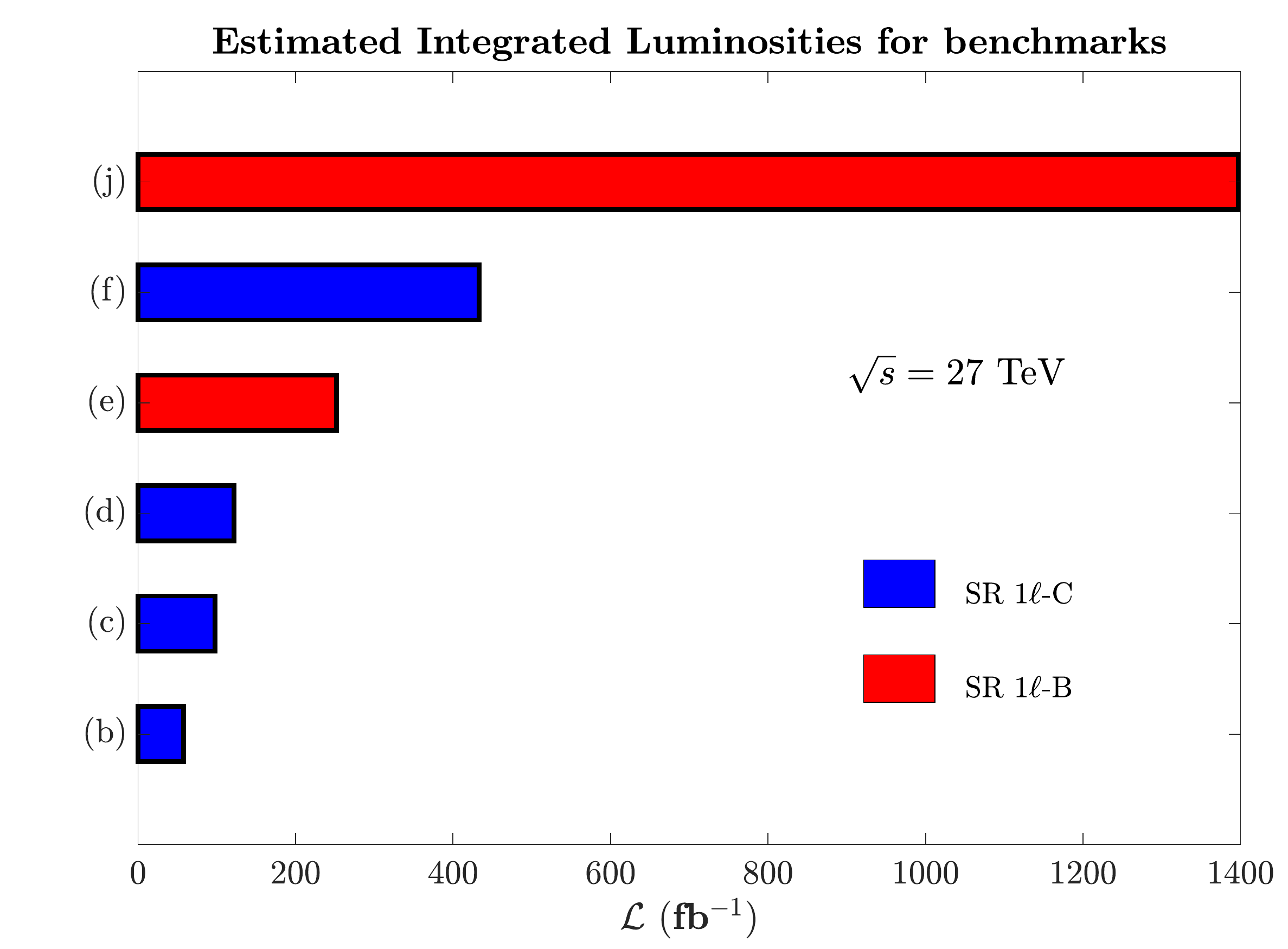}
\end{minipage}
    \caption{Estimated integrated luminosities, $\mathcal{L}$, for a $5\sigma$ discovery of the benchmark points of \tabb{tab:hmass}. Left: comparison between $\mathcal{L}$ at HL-LHC and HE-LHC for points (a), (g), (h) and (i). Right: HE-LHC analysis for points (b), (c), (d), (e), (f) and (j). The SRs that appear in the panels are as defined in \citeref{Aboubrahim:2018bil}.}
	\label{fig:lum1-lum2}
\end{figure}
 
  The left panel of \fig{fig:lum1-lum2} exhibits four parameter points which are discoverable both at  HE-LHC and at HL-LHC. 
   Here one finds that the integrated luminosities needed for discovery at HL-LHC (blue bars) are 2-50 times larger than what is needed at 
   HE-LHC (yellow bars). For these model points one finds that discovery would require an HL-LHC run between 5-8 years while the same  parameter points can be discovered in a period of few weeks to $\sim 1.5\yr$ at HE-LHC running at its optimal luminosity of $2.5\times 10^{35}\cmeters^{-2} \mathrm{s}^{-1}$.  The right panel of 
   \fig{fig:lum1-lum2} gives a set of benchmarks which are not accessible at HL-LHC but will be visible at  HE-LHC. We note that half of the benchmarks in the right panel of
    \fig{fig:lum1-lum2} can be discovered with less than $200\fbinv$ of integrated luminosity at HE-LHC with few months of running at its optimal luminosity. Considering points that are just beyond the HL-LHC reach, point (j) requires a run of $\sim 1.6~\mathrm{yr}$ while point (b) $\sim 3.5~\mathrm{weeks}$ for discovery at the HE-LHC.
         In summary, the analysis above indicates that a transition from HL-LHC to HE-LHC can aid in the discovery of supersymmetry for part of the parameter space accessible to both. Further to that, HE-LHC can explore significantly beyond the realm of the parameter space accessible to HL-LHC.

\begin{table}[t]
\begin{center}
\small\begin{tabulary}{1.1\textwidth}{l|CCCCCCC}
\hline\hline\rule{0pt}{3ex}
Model  & $h^0~[\mathrm{GeV}]$ & $\mu~[\mathrm{TeV}]$ & $\tilde\chi_1^0~[\times 10^2~\mathrm{GeV}]$ & $\tilde\chi_1^\pm~[\times 10^2~\mathrm{GeV}]$ & $\tilde t~[\mathrm{TeV}]$ & $\tilde g~[\mathrm{TeV}]$ & $\Omega^{\rm th}_{\tilde{\chi}^0_1} h^2$ \\
\hline\rule{0pt}{3ex} 
\!\!(a) & 124 & 8.02 & 9.73 & 10.6 & 4.73 & 1.36 & 0.039 \\
(b) & 125 & 6.29 & 10.2 & 10.3 & 2.08 & 1.40 & 0.035 \\
(c) & 123 & 5.59 & 11.1 & 11.9 & 2.88 & 1.51 & 0.048 \\
(d) & 124 & 15.5 & 11.9 & 12.7 & 10.0 & 1.75 & 0.048 \\
(e) & 124 & 11.7 & 9.48 & 9.48 & 6.78 & 1.33 & 0.020 \\
(f) & 124 & 13.7 & 12.4 & 13.5 & 6.98 & 1.62 & 0.112 \\
(g) & 124 & 10.4 & 1.34 & 1.51 & 5.27 & 3.93 & 0.121 \\
(h) & 124 & 26.1 & 1.54 & 1.76 & 18.6 & 5.88 & 0.105  \\
(i) & 124 & 1.15 & 1.65 & 1.89 & 4.17 & 6.71 & 0.114 \\
(j) & 125 & 29.7 & 1.62 & 1.87 & 10.4 & 15.6 & 0.105 \\
\hline
\end{tabulary}\normalsize
\caption{The Higgs boson ($h^0$) mass, the $\mu$ parameter and 
some relevant sparticle masses, and the relic density for the benchmark points used in this analysis~\cite{Aboubrahim:2018bil}. }
\label{tab:hmass}
\end{center}
\end{table}

\vspace{1cm}

%\textbf{ACKNOWLEDGMENTS:} The research of  Amin Aboubrahim  and Pran Nath was supported in part by  NSF Grant PHY-1620575.

\subsubsection{%\theoryC{Prospects for natural SUSY at HL- and HE-LHC}
\theoryC{Natural SUSY at HL- and HE-LHC}
}
\label{sec:BaerSection}
% - COMPLETED and updated 13/9/2018
\contributors{H. Baer, V. Barger, J. Gainer, H. Serce, D. Sengupta, X. Tata}
%{\bf Author(s): H. Baer, V. Barger, J. Gainer, H. Serce, D. Sengupta and X. Tata}

We present HL- and HE-LHC reach calculations for
supersymmetry in models with light higgsinos.  The light higgsino
scenario is inspired by the requirement of naturalness in that if the
superpotential (higgsino) mass parameter $\mu$ is much beyond the weak
scale, then the weak scale soft term $m_{H_u}^2$ will have to be
fine-tuned in order to maintain $m_{W,Z,h}$ at their measured mass
values.

\paragraph*{HL/HE-LHC reach for gluino pair production}
In \citeref{Baer:2016wkz} we evaluated the reach of the HL-LHC for
gluino pair production, assuming that $\tg\rightarrow
t\tst_1$ and $\tst_1 \rightarrow b\tilde{\chi}_{1}^+$ or $t\tz_{1,2}$
and that the decay products of the higgsinos $\tilde{\chi}_{1}^\pm$ and
$\tz_2$ are essentially invisible.%: see supplemental \fig{fig:FDgluino}
In \citeref{Baer:2017pba} we computed the reach of HE-LHC  
for both gluinos and top squarks in the light higgsino scenario
(with $\sqrt{s}=33\TeV$). 
These results have been updated for HE-LHC with $\sqrt{s}=27\TeV$ and \intlumihelhc of integrated luminosity 
in \citeref{Baer:2018hpb} where more details can be found.
We use \madgraph{}~\cite{Alwall:2014hca} to generate gluino pair production events and
SM backgrounds.
We interface \madgraph{} with \pythia{}~\cite{Sjostrand:2014zea} for initial/final state 
showering, hadronisation and underlying event simulation. 
The \delphes{} detector simulation~\cite{deFavereau:2013fsa} is used with specifications
as listed in \citeref{Baer:2017pba}.
SM backgrounds include $t\bar{t}$, $t\bar{t}b\bar{b}$, $t\bar{t}t\bar{t}$, 
$t\bar{t}Z$, $t\bar{t}h$, $b\bar{b}Z$ and single top production. 
We require at least four high $\pt$ jets, with two or more identified as $b$-jets,
no isolated leptons and large $\met$ selections.

\begin{figure}[t]
\centering
\begin{minipage}[t][\minipageheightdp][t]{\minipagewidthdp}
\centering
\includegraphics[width=\figuremaxwidthdp-1mm,height=\figuremaxheightdp,keepaspectratio]{\main/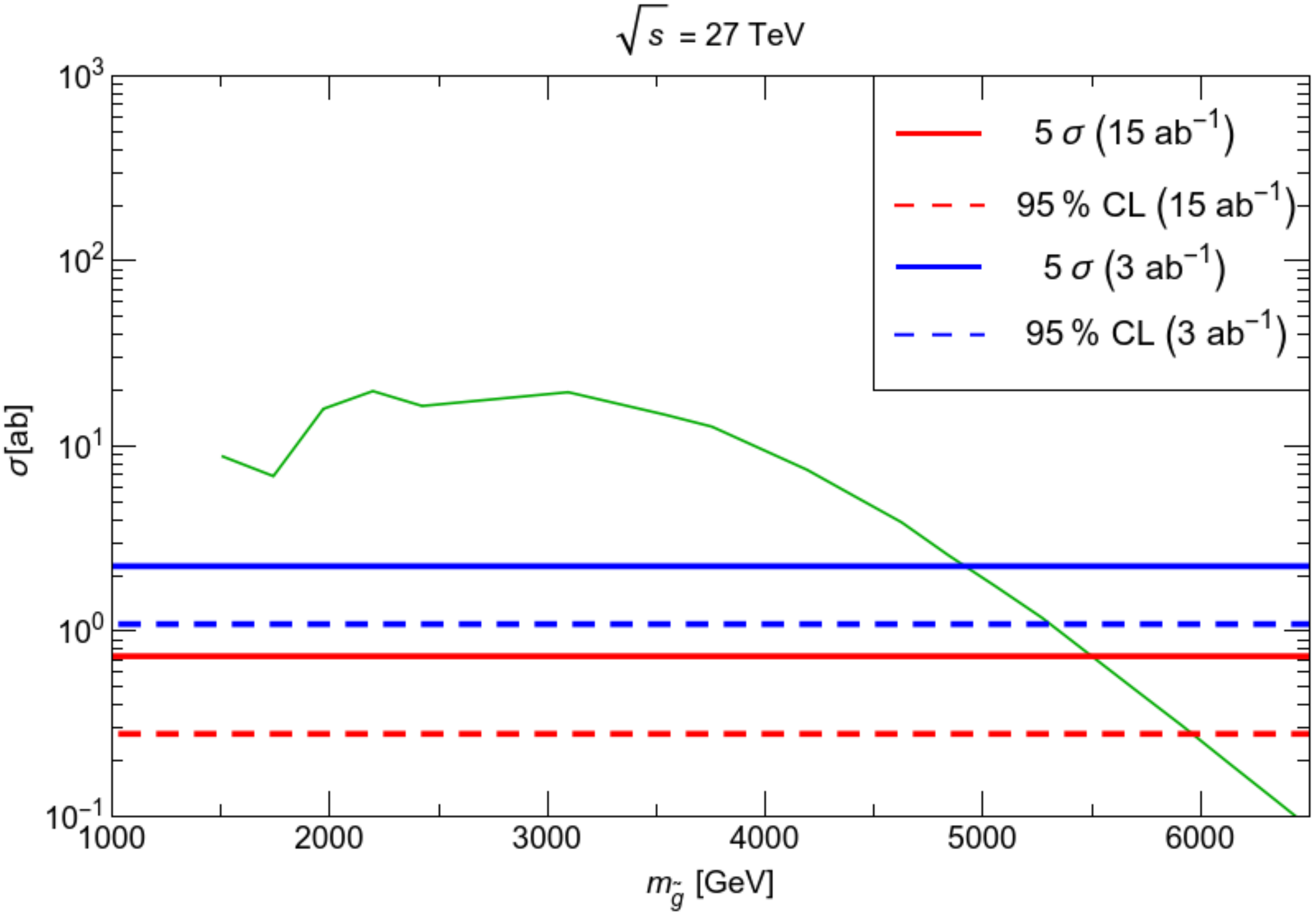}
\end{minipage}
\begin{minipage}[t][\minipageheightdp][t]{\minipagewidthdp}
\centering
\includegraphics[width=\figuremaxwidthdp,height=\figuremaxheightdp,keepaspectratio]{\main/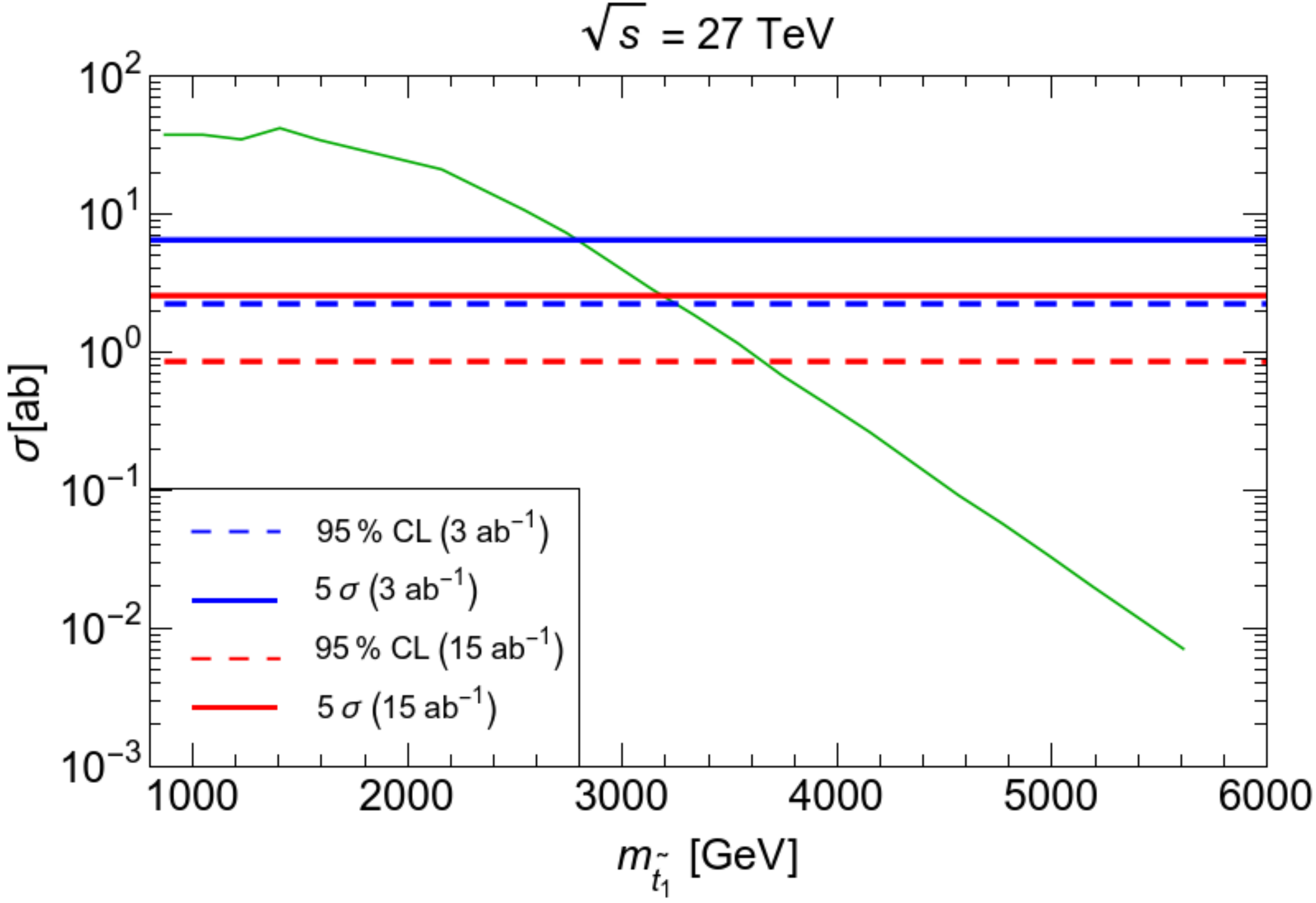}
\end{minipage}
\caption{Left: gluino pair production cross section vs. $m_{\tg}$ 
after selections at HE-LHC with $\sqrt{s}=27\TeV$ (green curve). 
Right: top-squark pair production cross section vs. $m_{\tst_1}$ 
after selections at HE-LHC with $\sqrt{s}=27\TeV$ (green curve). 
Both panels also show the $5\sigma$ reach and $95\%\cl$ exclusion lines
assuming $3$ and $\intlumihelhc$ of integrated luminosity.
\label{fig:mgl}\label{fig:mt1}}
\end{figure}

Results are shown in the left panel of \fig{fig:mgl} where we report the gluino pair
production signal versus $m_{\tg}$ for a natural NUHM2 model line with 
parameter choice $m_0=5m_{1/2}$, $A_0=-1.6 m_0$, $m_A=m_{1/2}$,
$\tan\beta =10$ and $\mu =150\GeV$ with varying $m_{1/2}$. 
The results are not expected to be sensitive to this precise choice of parameters as long as first generation squarks are much heavier than gluinos.
From the figure, we see that the $5\sigma$ discovery reach of HE-LHC
extends to $m_{\tg}=4.9\TeV$ for $3\abinv$ and to $m_{\tg}=5.5\TeV$
for \intlumihelhc of integrated luminosity. The corresponding $95\%\cl$ exclusion
reaches extend to $m_{\tg}=5.3\TeV$ for $3\abinv$ and to $m_{\tg}=5.9\TeV$ for \intlumihelhc of integrated luminosity. The impact of the theoretical uncertainties related to the total production rate of gluinos is not taken into account. 
For comparison, the $5\sigma$ discovery reach of LHC14 is ($2.4$) $2.8\TeV$ for an integrated luminosity of ($300\fbinv$) \intLumi~\cite{Baer:2016wkz}. %: see \fig{fig:figreach}  

\paragraph*{Top-squark pair production}
In \citeref{Baer:2017pba}, the reach of a $33\TeV$ LHC upgrade for 
top-squark pair production was investigated. Here, we repeat the analysis but for updated LHC energy upgrade $\sqrt{s}=27\TeV$. 
We use \madgraph{}~\cite{Alwall:2014hca} to generate top-squark pair production events within a
simplified model where $\tst_1\to b\tw_1$ at 50\%, and
$\tst_1\to t\tz_{1,2}$ each at 25\% branching fraction, which are typical
of most SUSY models~\cite{Baer:2016bwh} with light higgsinos. 
The higgsino-like
electroweakino masses are $m_{\tz_{1,2},\tw_1}\simeq 150\GeV$.
We also used \madgraph{}-\pythia{}-\delphes{} for the same SM background processes as listed above for the gluino pair production case. We required at least two high $\pt$ $b$-jets, no isolated leptons and large $\met$, see \citeref{Baer:2018hpb} for details.

Using these background rates for LHC at $\sqrt{s}=27\TeV$, we compute the $5\sigma$ reach and $95\%\cl$ exclusion of HE-LHC for $3$ and $\intlumihelhc$ of integrated luminosity using Poisson statistics. 
Our results are shown in the right panel of \fig{fig:mt1} along with the
top-squark pair production cross section after cuts versus $m_{\tst_1}$.
From the figure, we see the $5\sigma$ discovery reach of HE-LHC
extends to $m_{\tst_1}=2.8\TeV$ for $3\abinv$ and to $3.16\TeV$ for
\intlumihelhc. The $95\%\cl$ exclusion limits extend to $m_{\tst_1}=3.25\TeV$
for $3\abinv$ and to $m_{\tst_1}=3.65\TeV$ for \intlumihelhc. 
We checked that $S/B$ exceeds $0.8$ whenever we deem the signal to be 
observable~\cite{Baer:2018hpb}. 
For comparison, the Atlas projected $95\%\cl$ LHC14 reach~\cite{ATL-PHYS-PUB-2013-011} for $3\abinv$ extends to $m_{\tst_1} \simeq  1.7\TeV$ (see \secc{sec:susy221} for details) assuming $\tst_1\to t\tz_{1}$ decays.  
%
%\begin{figure}[t]
%\begin{center}
%%\includegraphics[height=0.35\textheight]{section2SUSY/img/mt_mz_2.png}
%\includegraphics[width=0.5\textwidth]{\main/section2SUSY/img/stop.pdf}
%\caption{Top-squark pair production cross section vs. $m_{\tst_1}$ 
%after selections at HE-LHC with $\sqrt{s}=27$ TeV (green curve). 
%Also shown are the $5\sigma$ reach  and $95\%\cl$ exclusion lines,
%assuming  3 and \intlumihelhc of integrated luminosity.% (for a single detector).
%\label{fig:mt1}}
%\end{center}
%\end{figure}
%

\begin{figure}[t]
\centering
\begin{minipage}[t][\minipageheightdp][t]{\minipagewidthdp}
\centering
\includegraphics[width=\figuremaxwidthdp,height=\figuremaxheightdp,keepaspectratio]{\main/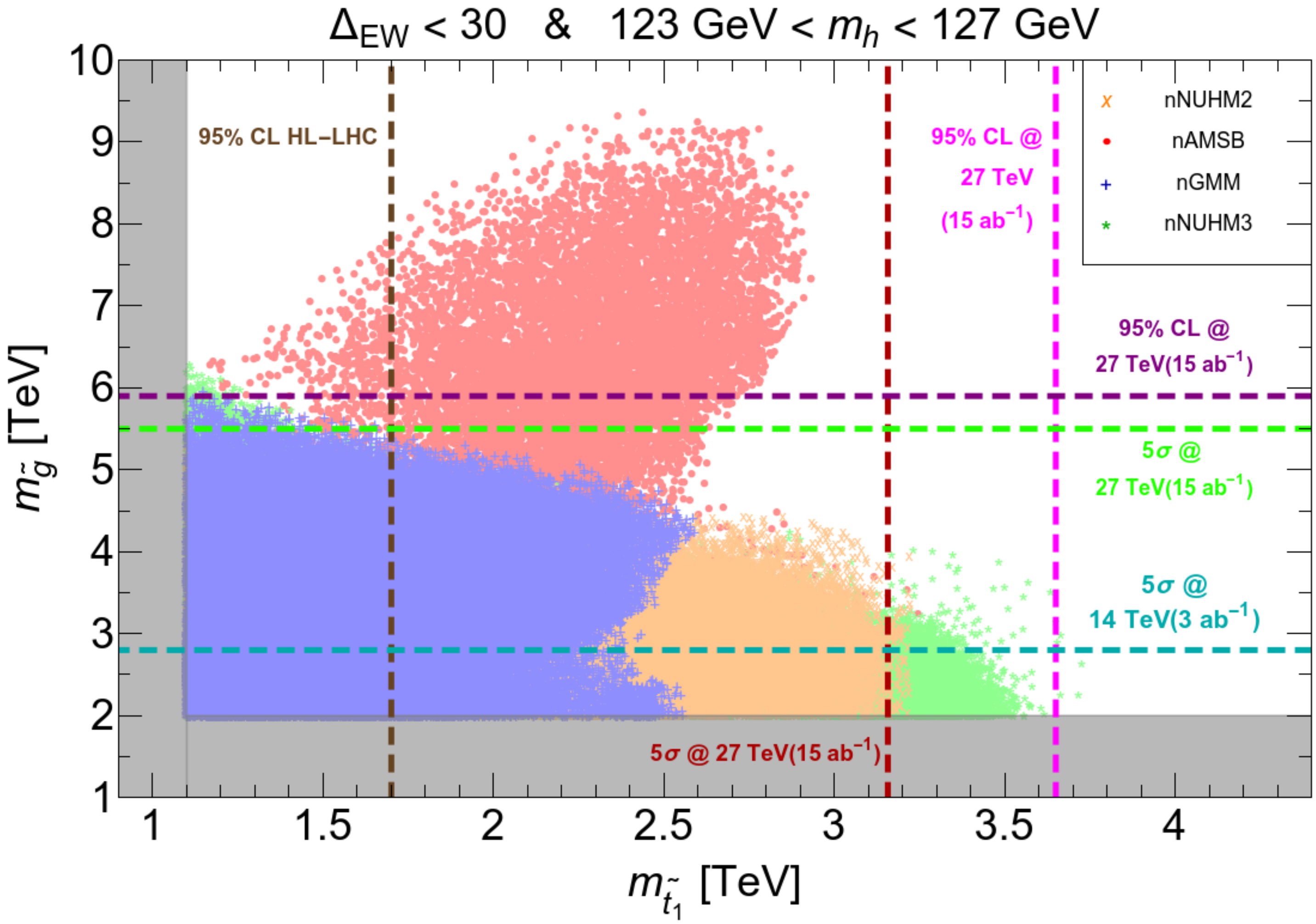}
\end{minipage}
\begin{minipage}[t][\minipageheightdp][t]{\minipagewidthdp}
\centering
\includegraphics[width=\figuremaxwidthdp,height=\figuremaxheightdp,keepaspectratio]{\main/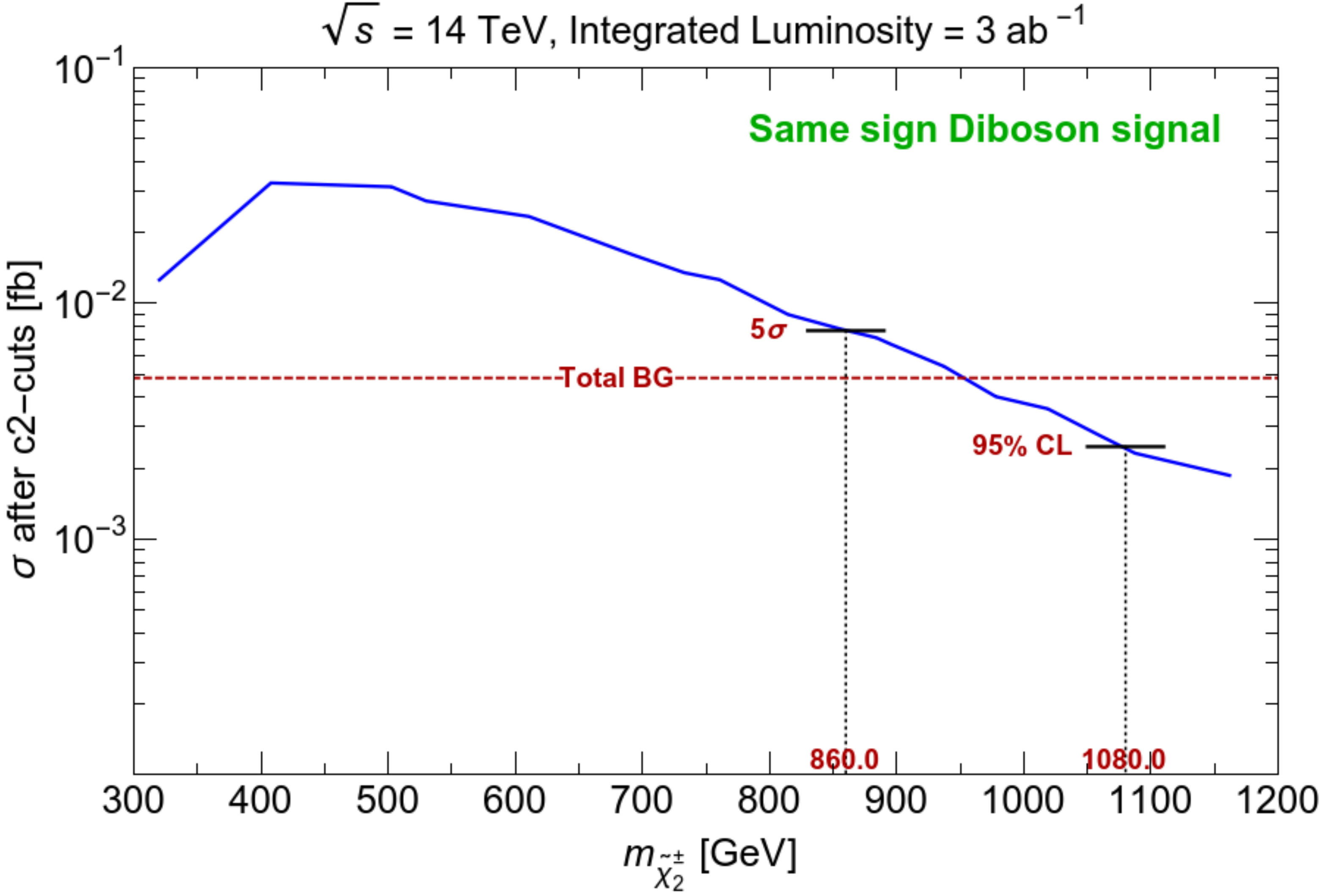}
\end{minipage}
\caption{Left: points in the $m_{\tst_1}$ vs. $m_{\tg}$ plane from a
  scan over nNUHM2, nNUHM3, nGMM and nAMSB model parameter space.  We
  compare to recent search limits from the ATLAS/CMS experiments (shaded regions) and show
  the projected reach of HL- and HE-LHC. Right: cross section for SSdB production 
after cuts versus wino mass at the LHC with 
$\sqrt{s}=14\TeV$.  We show the $5\sigma$ and $95\%\cl$ reach 
assuming a HL-LHC integrated luminosity of $3\abinv$.
\label{fig:mt1mgl}\label{fig:ssdb}}
\end{figure}

\paragraph*{Combined reach for stops and gluinos}
In \fig{fig:mt1mgl} we exhibit the gluino and top-squark reach
values in the $m_{\tst_1}$ vs. $m_{\tg}$ plane.  We compare the reach of HL- and HE-LHC to values of gluino and stop masses (shown by the dots) in a variety of natural SUSY models defined to have $\Delta_{\rm EW} < 30$~\cite{Baer:2012up,Baer:2012cf},
\footnote{
%We  define natural SUSY models as those with EW fine-tuning
 % parameter $\Delta_{\rm EW}<30$.  
The onset of fine-tuning for larger
  values of $\Delta_{\rm EW}$ is visually displayed in
  \citeref{Baer:2015rja}.} including the two- and three-extra parameter
non-universal Higgs models~\cite{Baer:2005bu} (nNUHM2 and nNUHM3),
natural generalised mirage mediation~\cite{Baer:2016hfa} (nGMM) and
natural anomaly-mediation~\cite{Baer:2018hwa} (nAMSB).
These models all allow
for input of the SUSY $\mu$ parameter at values $\mu\sim 100-350$ GeV which
is a necessary (though not sufficient) condition for naturalness in the
MSSM. 

The highlight of this figure is that at least one of the gluino or the
stop should be discoverable at the HE-LHC.  We also see that in natural
SUSY models (with the exception of nAMSB), the highest values of
$m_{\tg}$ coincide with the lowest values of $m_{\tst_1}$ while the
highest top squark masses occur at the lowest gluino masses.  Thus, a
marginal signal in one channel (due to the sparticle mass being near
their upper limit) should be accompanied by a robust signal in the other
channel.  Over most of the parameter range of weak scale natural SUSY
there should be a $5\sigma$ signal in {\it both} the top-squark and
gluino pair production channels at HE-LHC.

\paragraph*{Same-sign diboson signature from wino pair production}
The wino pair production reaction $pp\to\tilde{\chi}_2^\pm\tilde{\chi}_4^0$ can occur at observable rates
for SUSY models with light higgsinos.  The decays
$\tilde{\chi}_2^\pm\rightarrow W^\pm\tilde{\chi}_{1,2}^0$ and
$\tilde{\chi}_4^0\rightarrow W^\pm\tilde{\chi}_1^\mp$ lead to final
state dibosons which half the time give a relatively jet-free same-sign
diboson signature (SSdB). This has only tiny SM
backgrounds~\cite{Baer:2013yha,Baer:2013xua,Baer:2017gzf} and excellent prospects for discovery.%: see supplemental \fig{fig:FDssdb}.

%%
%\begin{figure}[t]
%\begin{center}
%%\includegraphics[height=0.35\textheight]{section2SUSY/img/mu_mwt_2.png}
%
%\caption{Cross section for SSdB production 
%after cuts versus wino mass at the LHC with 
%$\sqrt{s}=14$ TeV.  We show the $5\sigma$ and $95\%\cl$ reach 
%assuming a HL-LHC integrated luminosity of 3\abinv.
%\label{fig:ssdb}}
%\end{center}
%\end{figure}
%

We have computed the reach of HL-LHC for the SSdB signature 
in \fig{fig:ssdb} including $t\bar{t}$, $WZ$, $t\bar{t}W$,
$t\bar{t}Z$, $t\bar{t}t\bar{t}$, $WWW$ and $WWjj$ backgrounds. 
For LHC14 with $3\abinv$ of integrated luminosity,
the $5\sigma$ reach extends to $m(\textrm{wino})\sim 860$ GeV while
the $95\%\cl$ exclusion extends to $m(\textrm{wino})\sim 1080\GeV$.
In models with unified gaugino masses, these would correspond to 
a reach in terms of $m_{\tg}$ of $2.4$ ($3$)\TeV, respectively.
These values are comparable to what LHC14 can achieve via gluino pair
searches with $3\abinv$. The SSdB signature is distinctive for the case of SUSY models with light higgsinos.

While \fig{fig:ssdb} presents the HL-LHC reach for SUSY in the
SSdB channel, the corresponding reach of HE-LHC has not yet 
been computed.
 The SSdB signal arises via  
EW production, and  
the signal rates are expected to rise by a factor of a few 
by moving from $\sqrt{s}=14\TeV$ to $\sqrt{s}=27\TeV$. In contrast,
some of the strongly-produced SM backgrounds like $t\bar{t}$
production will rise by much larger factors.
Thus, it is not yet clear whether the reach for SUSY in the SSdB 
channel will be increased by moving from HL-LHC to HE-LHC.
We note though that other signals channels from wino decays to higgsinos
plus a $W$, $Z$ and Higgs boson may offer further SUSY detection
possibilities.

\paragraph*{Higgsino pair production at LHC upgrades}
The four higgsino-like charginos $\tw_1$ and neutralinos $\tz_{1,2}$
are the only SUSY particles required by naturalness to lie near to the
weak scale at $m_{weak}\sim 100\GeV$.  In spite of their lightness, they
are very challenging to detect at LHC.  The lightest neutralino
evidently comprises just a portion of dark matter~\cite{Baer:2018rhs},
and if produced at LHC via $pp\to\tz_1\tz_1,\ \tilde{\chi}_1^\pm
\tilde{\chi}_1^\mp\ {\rm and}\ \tilde{\chi}_1^\pm\tilde{\chi}_{1,2}^0$
could escape detection. This is because the decay products of $\tilde{\chi}_2^0$
and $\tilde{\chi}_{1}^\pm$ are expected to be very soft, causing the
signal to be well below SM processes like $WW$ and $t\bar{t}$
production.  The monojet signal arising from initial state radiation (ISR) 
$pp\to\tz_1\tz_1 j$, $\tilde{\chi}_1^\pm\tilde{\chi}_1^\mp j$ and
$\tilde{\chi}_1^\pm\tilde{\chi}_{1,2}^0 j$ has been evaluated in
\citeref{Baer:2014cua} and was found to have similar shape
distributions to the dominant $pp\to Zj$ background but with background
levels about 100 times larger than signal.  However, at HE-LHC harder monojet-like 
selections may be possible~\cite{Han:2018wus}, and generic prospects studies are presented in \secc{sec:DMjet} of this report.

\begin{figure}[t]
\centering
\begin{minipage}[t][\minipageheightsp][t]{\minipagewidthsp}
\centering
\includegraphics[width=\figuremaxwidthsp,height=\figuremaxheightsp,keepaspectratio]{\main/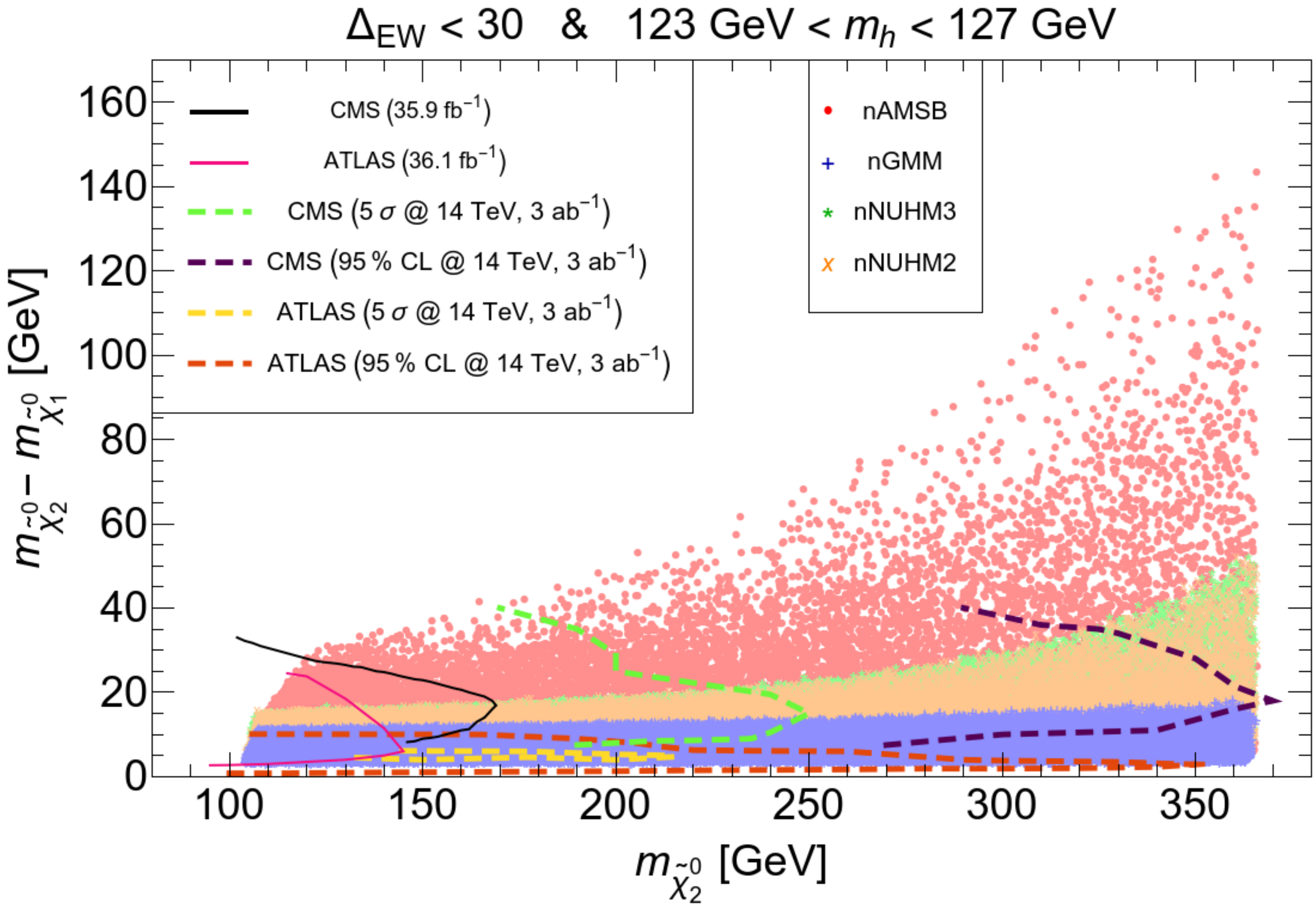}
\end{minipage}
\caption{Points in the $m_{\tz_2}$ vs. $m_{\tz_2}-m_{\tz_1}$ plane 
from a scan over nNUHM2, nNUHM3, nGMM and nAMSB model parameter space.
We compare to recent search limits from the ATLAS/CMS experiments
and to future reach contours for HL-LHC.
\label{fig:mz2mz1}}
\end{figure}

A way forward has been proposed via the $pp\to\tz_1\tz_2 j$ channel
where $\tz_2\to\ell^+\ell^-\tz_1$: a soft opposite-sign dilepton pair  recoils against a hard initial state jet radiation which serves as a 
trigger~\cite{Baer:2014kya}. Experimental prospect searches 
presented in \secc{sec:susy236} by ATLAS and CMS exploit this kind of 
signature.  
%: see supplemental \fig{fig:FDsoftOSdilepton}. 
%Recent searches in this $\ell^+\ell^-j+MET$ 
%channel have been performed by CMS~\cite{CMS:2017fij} 
%and by ATLAS~\cite{Aaboud:2017leg}.
The projected reach for $5\sigma$ and $95\%\cl$ reach at the HL-LHC with $3\abinv$ in $\ell^+\ell^+\met$ final state events are shown in the $m_{\tz_2}$ vs. $m_{\tz_2}-m_{\tz_1}$ plane in~\fig{fig:mz2mz1}. 
The ATLAS and CMS experiments contours are shown as the yellow, green, purple and red dashed contours. 
We see that these contours can probe considerably more parameter space 
although some of natural SUSY parameter space (shown by dots for the
same set of models as in \fig{fig:mt1mgl}) might lie beyond these
projected reaches. So far, reach contours for HE-LHC in this search 
channel have not been computed but it is again anticipated that HE-LHC will not be greatly beneficial here since $pp\to\tz_1\tz_2$ is an EW production process so the signal cross section will increase only marginally while SM background processes like $t\bar{t}$ production will increase substantially. 
%We also show via (faded) dots the locus of various natural SUSY models.

It is imperative that future search channels try to squeeze
their reach to the lowest $m_{\tz_2}-m_{\tz_1}$ mass gaps which are 
favoured to lie in the $3-5\GeV$ region for string landscape 
projections~\cite{Baer:2017uvn} of SUSY mass spectra. 
The Atlas red-dashed contour appears to go a long way in this regard, 
though the corresponding $5\sigma$ reach is considerably smaller.

\paragraph*{Summary} 
We have delineated the reach of the HE-LHC and compared it to the
corresponding reach of the \sloppy\mbox{HL-LHC} for SUSY models with light higgsinos, expected in a variety of natural SUSY models. While the HL-LHC increases the SUSY search range, it appears that the HE-LHC will definitively probe natural SUSY models with $\Delta_{\rm EW} < 30$ via a $5\sigma$ discovery of at least one of the top squark or the gluino (likely even both), possibly also with signals in other channels.

\subsubsection{%\theoryC{Prospects for Higgs and SUSY searches in pMSSM}
\theoryC{The pMSSM at HL- and HE-LHC}
}

\contributors{A. Arbey, M. Battaglia, F. Mahmoudi}
%{\bf Author(s): A. Arbey$^{a,b,1}$, M. Battaglia$^{c,b,2}$, F. Mahmoudi$^{a,b,3}$}\\
%%
%{\it \small $^a$ Univ. Lyon, Univ. Lyon 1, CNRS/IN2P3, Institut de Physique Nucl\'eaire de Lyon,\\ UMR5822, F-69622 Villeurbanne, France}\\
%{\it \small $^b$ CERN, CH-1211 Geneva 23, Switzerland}\\
%{\it \small $^c$ University of California at Santa Cruz, Santa Cruz Institute of Particle Physics, CA 95064, USA}\\
%{\small $^1$ alexandre.arbey@ens-lyon.fr, $^2$ marco.battaglia@ucsc.edu $^3$ nazila@cern.ch}\\

The phenomenological MSSM (pMSSM)~\cite{Djouadi:1998di}, contains 20 free parameters, and is the most general \cp and R-parity conserving MSSM scenario with minimal flavour violation. It was introduced in order to reduce the theoretical prejudices of the constrained MSSM scenarios. In the following, we consider the case where the lightest neutralino is the LSP and can constitute part or all of the dark matter. Technical details concerning the pMSSM scans and software required for the presented analyses can be found in \citeref{Arbey:2011un,Arbey:2011aa}.  

\paragraph*{SUSY and Higgs searches:} 

The direct SUSY searches at the LHC are extremely powerful in probing the strongly interacting sector of the MSSM. Nevertheless, scenarios with compressed spectra or with long decay chains can escape the current searches and remain challenging. In the pMSSM, such cases are not rare, and thus the complementary information from other sectors can be of interest. In particular the Higgs sector, namely the measurement of the couplings of the lightest Higgs boson as well as searches for heavier states can unveil additional MSSM phase space~\cite{Gori:2650162}, especially during HL and HE runs of the LHC.

In the extended Higgs sector of the MSSM, the couplings of the lightest Higgs boson to up- and down-type quarks are modified by terms inversely proportional to the \cpodd $A$ boson mass as $2 M_Z^2 / M_A^2 \tan^2 \beta$ and $2 M_Z^2 / M_A^2$ respectively, providing an indirect sensitivity to the scale of $M_A$, if deviations in the branching fractions to up- and down-type quarks are detected, or a lower bound on $M_A$, if the coupling properties agree with the SM predictions. At present, the direct sensitivity to the $A$ (and $H$) boson at the LHC comes mostly from the $pp \rightarrow A/H \rightarrow \tau^+ \tau^-$ process.
On the other hand, the $bbH$ associate production and gluon fusion processes decrease the total cross section with $\tan \beta$ up to the point where the $b$-quark loops take over and the cross section increases. For $\tan \beta < 10$, the decay branching fraction is proportional to $\tan \beta$ .
Thus, the bounds from the $\tau \tau$ final state become particularly strong for large values of $\tan \beta$ but quite unconstrained for $\tan \beta \simeq 10$.

The modification of the Higgs couplings to fermions induced by loops of strongly interacting SUSY particles, namely the $\Delta_b$ correction in the Higgs coupling to $b\bar{b}$ is of special importance, as the SUSY contribution scales with $\mu \tan \beta M_{\tilde{g}} / M^2_{\tilde{g}, \tilde{b}, \tilde{t}}$. 
Since the value of $\mu \tan \beta$ can be much larger than the mass of the SUSY particles in the denominator, the SUSY strongly interacting sector does not decouple. Therefore, the study of the Higgs branching fractions, or the Higgs signal strengths, can unravel SUSY scenarios with strongly interacting particles at masses well beyond the kinematic reach of the LHC~\cite{Arbey:2015aca}.

\begin{figure}[t]
\centering
\begin{minipage}[t][\minipageheightdp][t]{\minipagewidthdp}
\centering
\includegraphics[width=\figuremaxwidthdp,height=\figuremaxheightdp,keepaspectratio]{\main/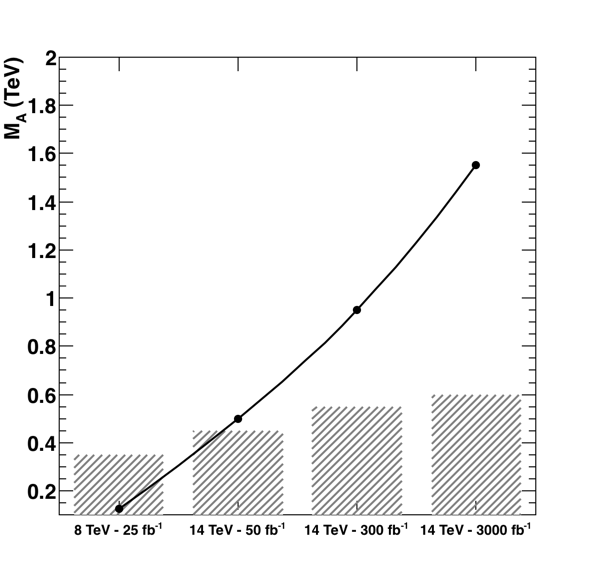}
\end{minipage}
\begin{minipage}[t][\minipageheightdp][t]{\minipagewidthdp}
\centering
\includegraphics[width=\figuremaxwidthdp,height=\figuremaxheightdp,keepaspectratio]{\main/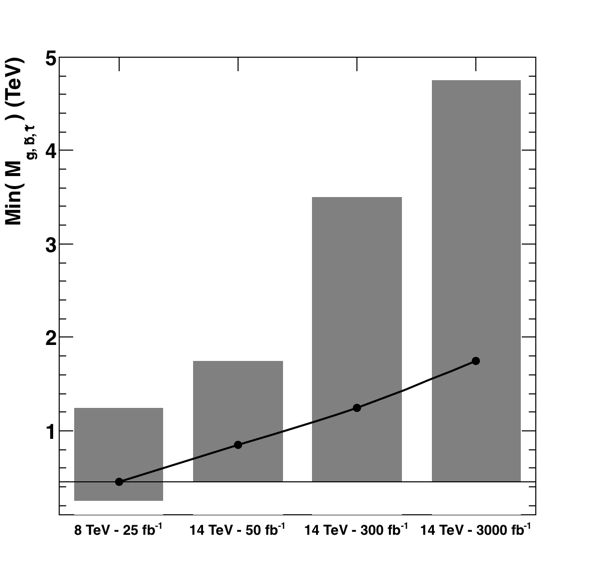}
\end{minipage}
  \caption{\label{fig:mass-dir-indir-LHC} 
    Sensitivity to the mass of the \cpodd $A$ boson (left) and the lightest state among $\tilde{g}$, $\tilde{b}$ and $\tilde{t}$ (right) as a function of the energy and luminosity of the LHC, in the pMSSM. The 95\%~\cl\ exclusion range, when the MSSM parameters are varied, are given for the direct search by the continuous line and for the indirect constraints from the $h$ decay properties by the filled bars (from~\citeref{Arbey:2015aca}).}
\end{figure}
The sensitivity to SUSY mass scales is summarised in \fig{fig:mass-dir-indir-LHC}, which gives a comparison of the direct and indirect sensitivity to $M_A$ with the mass of the gluino and squarks of the third generation, as a function of the different energies and integrated luminosities of the LHC.
Direct searches are accounted for by implementing LHC Run-1 searches in jets + $\met$~\cite{Aad:2014wea,Aad:2015pfx}, jets + leptons + $\met$~\cite{Aad:2014qaa,Aad:2015mia,Aad:2015jqa}, leptons + $\met$~\cite{Aad:2014vma,Aad:2014nua} and monojets~\cite{Aad:2014nra,Khachatryan:2014rra}. Signal selection cuts corresponding to each of the analyses are applied to these simulated signal events. The number of SM background events in the signal regions are taken from the estimates reported by the experiments. Results are projected to $14\TeV$ for $300\fbinv$ and $3\abinv$ of integrated luminosity, by generating events at $14\TeV$ and rescaling the $8\TeV$ backgrounds by the corresponding increase in cross section and signal cut acceptance at the higher energy~\cite{Arvey:2015nra}. The use of $7+8\TeV$ analyses at the higher energy ensures a constant scaling for the various energy and integrated luminosity conditions considered here. In addition, the constraints from the Higgs signal strengths for the $\gamma \gamma$, $WW$, $ZZ$, $\tau \tau$ and $b \bar b$ channels have been added. Here we assume SM-like central values and evolve the statistical uncertainties according to the increase of signal events with energy and integrated luminosity~\cite{Dawson:2013bba}
following the same procedures as in \citeref{Arbey:2015aca}.
\begin{table}[t]
\centering
 \small \begin{tabular}{|c|c|c|c|c|}
      \hline
    & LHC8 -- $25\fbinv$  & LHC14 -- $50\fbinv$ & LHC14 -- $300\fbinv$ & HL-LHC -- $3\abinv$\\
    \hline\hline
    js+$\ell$s+MET   & $0.145 $     & $ 0.570$ &$ 0.698$      & $0.820 $ \\
    \hline
    +h$^0$ $\mu$s    & $0.317 $     & $ 0.622 $& $0.793  $    &$ 0.920$ \\
    \hline
  \end{tabular}\normalsize
  \caption{Fractions of pMSSM points excluded by the combination of LHC MET searches, and the LHC Higgs data.
  \label{tab:methiggs}}
\end{table}
Table \ref{tab:methiggs} summarises the fraction of pMSSM points with SUSY masses up to $5\TeV$ excluded by the LHC searches based on $\met$+jets signatures, and by the addition of the Higgs data.

\paragraph*{Monojets}

Monojet searches remain a powerful tool for discovery at $pp$ colliders of increasing energy and luminosity and specific prospects for WIMP searches using this signature are presented in this report (Section 3). %~\cite{Arbey:2015hca}.
Beyond those scenarios, the monojet signature can be sensitive to specific MSSM scenarios such as decays with two gluinos or scalar tops and an ISR hard jet, when the scalar top decays are soft enough for the event to be classified as monojet-like. Expanding on the work of \cite{Arbey:2015hca}, we consider here these monojet-like signatures at $\sqrt{s}$ energies of $8$, $13$, $14$ and $27\TeV$ for two different pMSSM scenarios featuring a light gluino and a light bino neutralino with a mass splitting of $10\GeV$, and a light stop and bino-wino neutralino and chargino with a mass splitting slightly smaller than the top quark mass so that the stop decays into three soft jets and the lightest neutralino.

\begin{figure*}[t]
\centering
\begin{minipage}[t][\minipageheightdp][t]{\minipagewidthdp}
\centering
\includegraphics[width=\figuremaxwidthdp,height=\figuremaxheightdp,keepaspectratio]{\main/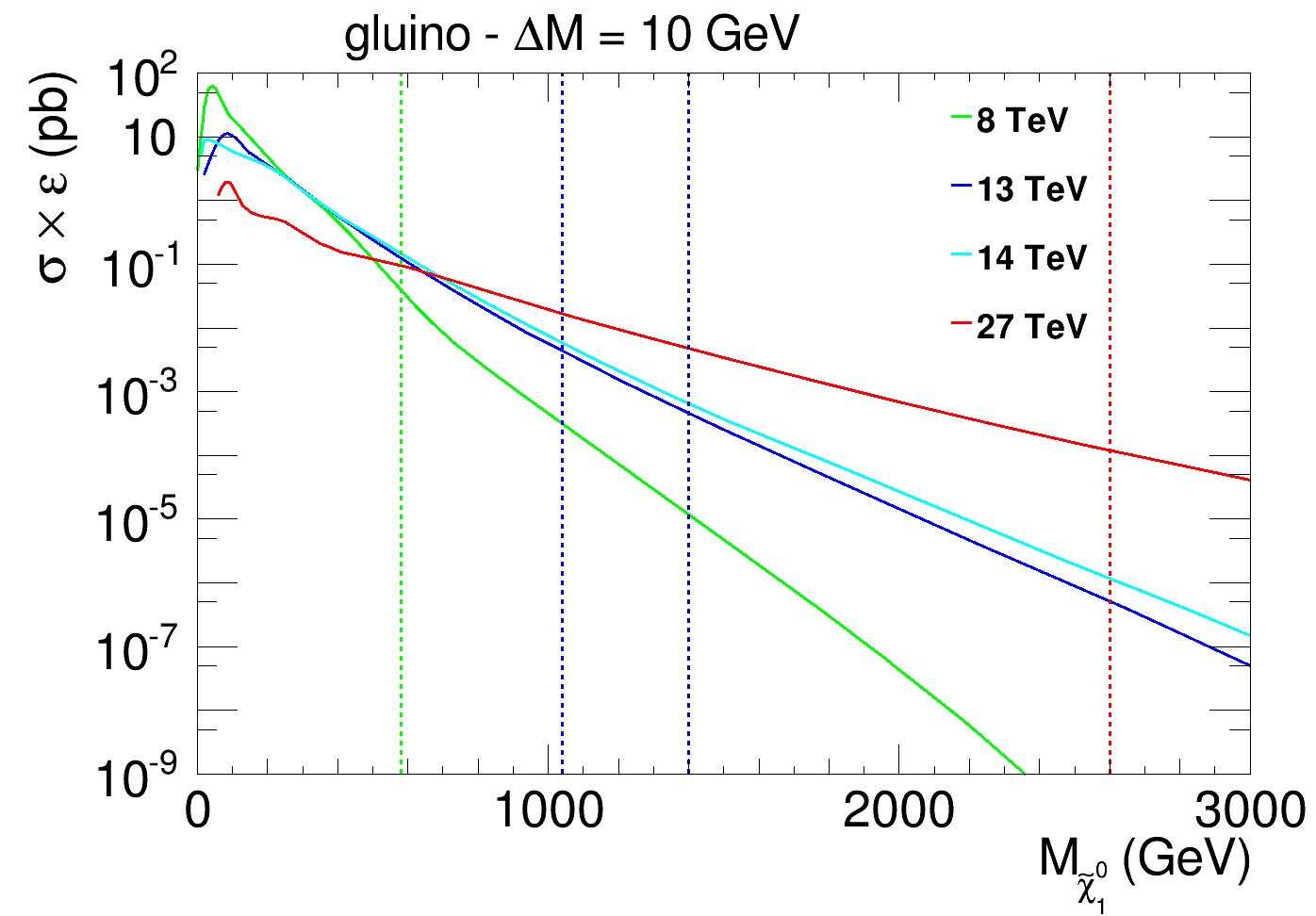}
\end{minipage}
\begin{minipage}[t][\minipageheightdp][t]{\minipagewidthdp}
\centering
\includegraphics[width=\figuremaxwidthdp,height=\figuremaxheightdp,keepaspectratio]{\main/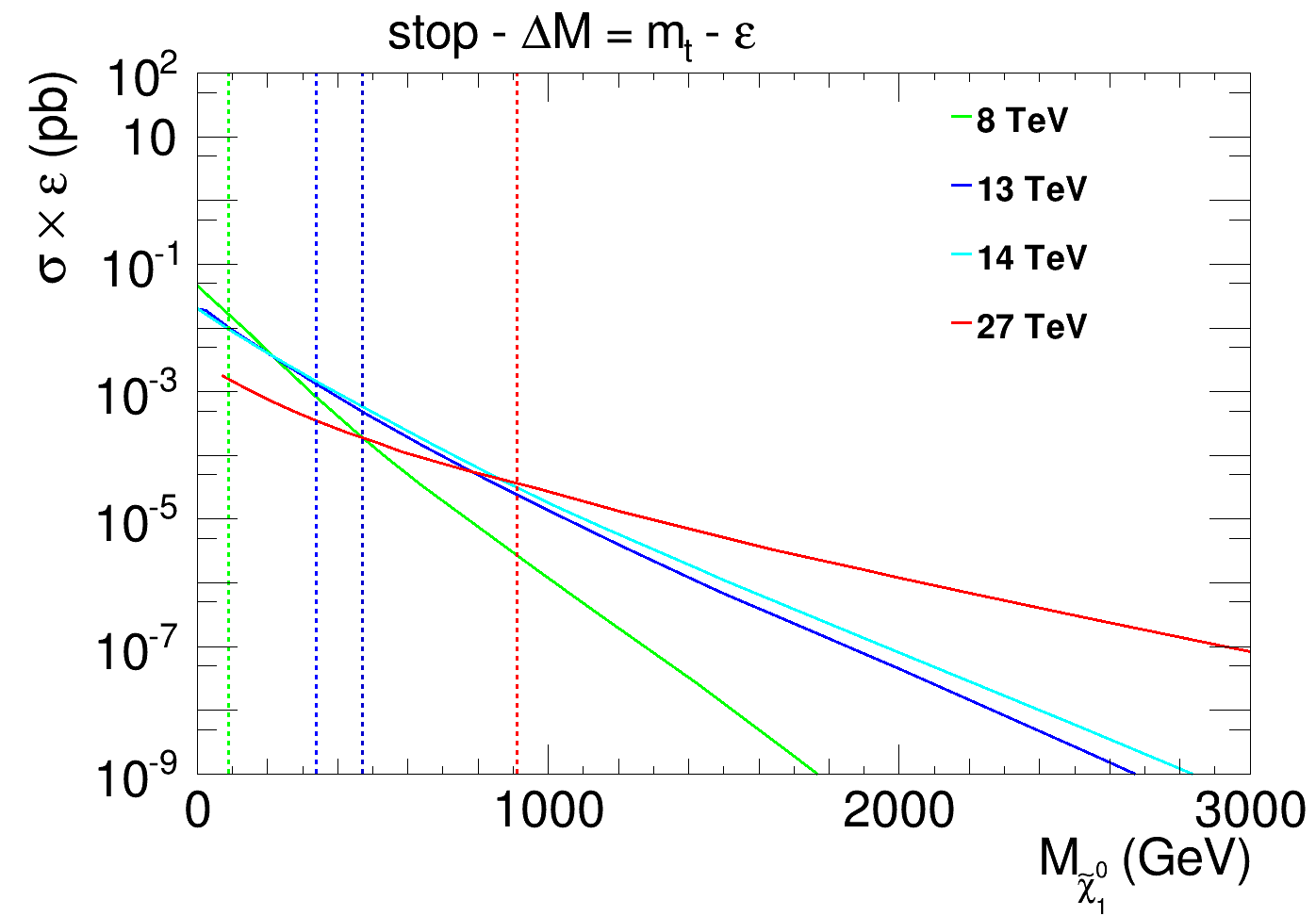}
\end{minipage}
\caption{Monojet production cross section times acceptance and efficiency as a function of the neutralino mass, for scenarios with a gluino (left) and a stop (right) with small mass splittings with the neutralino LSP. The different curves correspond to $\sqrt{s} = 8$, $13$, $14$ and $27\TeV$ LHC \com~energies. The green vertical dashed line corresponds to an indicative exclusion limit by the LHC Run-1, the light (dark) blue line to a prospective limit for the LHC $14\TeV$ run with $300\fbinv$ ($3\abinv$) of data, and the red line to a potential limit at $27\TeV$ with \intlumihelhc of data.\label{fig:mssm:energy_dep}}
\end{figure*}

The mass splittings have been chosen to maximise the number of monojet events, and also to ensure the consistency with the dark matter relic density constraint which requires small mass splittings for co-annihilations. It is instructive to consider the scaling of the product of the monojet production cross section times efficiency with respect to the neutralino mass and the collider energy. The acceptance is defined by $\sqrt{s}$-dependent 
lower cuts on the jet $p_T$ and missing energy ($\met$), scaled from early LHC monojet and
monojet-like analyses~\cite{Aad:2014nra,Khachatryan:2014rra} as
discussed in details in \citeref{Arbey:2015hca}. The results are shown in \fig{fig:mssm:energy_dep}.
Although the change in cross section times efficiency from $8$ to $14\TeV$ as a function of the mass has been relatively small, the increase in mass coverage afforded by $14\TeV$ is very significant. This motivates a possible further increase of the energy up to $27\TeV$, and beyond.

\subsubsection{%\theoryC{\Zp~bosons in supersymmetric and leptophobic scenarios}
\theoryC{\Zp~bosons in supersymmetric and leptophobic scenarios at HL- and HE-LHC}}
\label{sec:susy216} 
\contributors{J. Y. Araz, G. Corcella, M. Frank, B. Fuks}
%
%{\bf Author(s): Jack Y. Araz$^{a,1}$, Gennaro Corcella$^{b,2}$, Mariana Frank$^{a,3}$, Benjamin Fuks$^{c,d,4}$ }\\
%
%{\it \small $^a$ Department of Physics,
% Concordia University, 7141 Sherbrooke St.West, Montreal, QC, Canada H4B 1R6}\\ 
% {\it \small $^b$ INFN, Laboratori Nazionali di Frascati, Via E. Fermi 40, I-00044, Frascati (RM), Italy}\\
%{\it \small $^c$ Sorbonne Universit\'e, CNRS, Laboratoire de Physique Th\'eorique et Hautes \'Energies, LPTHE, F-75005 Paris, France}\\
%{\it\small $^d$ Institut Universitaire de France, 103 boulevard Saint-Michel,
%  75005 Paris, France}
  
%  \subsubsection{Introduction}
%\paragraph*{Introduction}
Searching for heavy neutral vector bosons \Zp~is one of the challenging objectives of the LHC.
Such heavy bosons are  predicted by $U(1)^{\prime}$ models inspired by Grand Unification Theories (GUT) as well as by the Sequential Standard Model (SSM), one of the simplest extensions of the SM wherein \Zp~and possible $W'$ bosons have the same couplings as the SM $Z$ and $W$. The LHC experiments have searched for \Zp~signals by exploring high-mass dilepton and dijet systems and have set exclusion limits of a few TeV on the \Zp~mass. For studies of \Zp-bosons at HL/HE-LHC see \secc{sec:otherBSM}. 
%More specifically, by using high-mass dilepton data at 13 TeV, the ATLAS collaboration~\cite{Aaboud:2017buh} set $M_{Z'}>4.5$~TeV in the SSM and $M_{Z'}>3.8$-4.1~TeV in various U(1)$'$ models, whereas CMS obtained $M_{Z'}>4.0$~TeV (SSM) and   $M_{Z'}>3.5$~TeV (GUT models)~\cite{CMS:2016abv}.For what concern dijet probes, the limits are less stringent and are given by $M_{Z'}>2.1$-2.9~TeV (ATLAS)~\cite{Aaboud:2017yvp} $M_{Z'}>2.7$~TeV (CMS)~\cite{Sirunyan:2016iap}.  

While such analyses have assumed that the \Zp-boson can only decay into SM channels,
recent investigations (see, \egg, \citeref{Corcella:2012dw,Corcella:2014lha}) have considered the possibility that the \Zp-boson could decay according to modes BSM, like for instance in supersymmetric final states in the so-called UMSSM framework.  From the  MSSM viewpoint,
extending it via a $U(1)^{\prime}$ group has the advantage that the extra symmetry
forbids a too quick proton decay and allows to stabilise all particle
masses with respect to quantum corrections. Regarding the searches,
assuming BSM decays lowers the rates into lepton and quark pairs, and
therefore milder exclusion limits are to be expected.

Unlike direct sparticle production in $pp$ collisions, 
the \Zp~mass sets one further kinematic constrain on the invariant masses of the produced
supersymmetric particles. Furthermore, as will be discussed in the following, there are realisations
of the $U(1)^{\prime}$ symmetry wherein, due to the kinetic mixing with the 
SM $U(1)$ group, the \Zp~is leptophobic. Leptonic final states can therefore occur,
in the UMSSM, only through supersymmetric cascades.
Such scenarios avoid the present dilepton bounds and may well be worth to be
investigated, especially in the high-luminosity LHC phase.
%In Subsection \ref{par:zpmodels} we shall review the theoretical framework of our exploration; in \ref{par:zpresults} and \ref{par:zpsign}, we will present some phenomenological results at HL-LHC; in \ref{zpconclusions} some final remarks will be given.

In what follows we shall review the theoretical framework of our exploration, present some phenomenological results at the HL-LHC and then some final remarks will be given.

%\paragraph*{\Zp~bosons in the UMSSM}\label{par:zpmodels}

Grand-unified theories are based on a rank-6 group $E_6$, 
where the symmetry-breaking scheme proceeds via multiple steps:
 \begin{equation}
E_6 \to SO(10) \otimes U(1)_\psi
\to SU(5) \otimes U(1)_\chi \otimes U(1)_\psi
\to SU(3)_C\otimes SU(2)_L \otimes U(1)_Y  \otimes U(1)' \ .
\end{equation}
The $U(1)^{\prime}$ symmetry surviving at the EW scale can be expressed as a
combination of $U(1)_\chi$ and $U(1)_\psi$,
%\begin{equation}
$U(1)' =U(1)_\psi\cos \theta  - U(1)_\chi\sin\theta $,
%\label{eq:u1psichi} 
%\end{equation}
where $\theta$ is the $E_6$ mixing angle. The neutral
vector bosons associated with $U(1)_\psi$ and $U(1)_\chi$ are
called $Z'_\psi$ and $Z'_\chi$, while a generic
\Zp~is given by their mixing.

We investigate possible \Zp~supersymmetric decays in the UMSSM. 
As for the particle content of the UMSSM after EWSB,
one is left in the Higgs sector with two charged $H^\pm$ and four neutral
scalar bosons, namely one pseudoscalar $A$ and three
neutral scalars $h$, $H$ and $H^\prime$, where $h$ and $H$ are MSSM-like,
with $h$ roughly corresponding to the SM Higgs, and $H'$ is a new singlet-like 
Higgs boson related to the extra $U(1)^{\prime}$.
In the gaugino sector, one has two extra neutralinos with respect to the MSSM,
related to the supersymmetric partners of \Zp~and $H'$,
for a total of six $\tilde\chi^0_1, \dots, \tilde\chi^0_6$ neutralinos.
The chargino sector is unchanged, since the \Zp~is electrically neutral.

It was found in \citeref{Araz:2017wbp} that the very inclusion of supersymmetric modes
lowers the exclusion limits on $M_{Z'}$ obtained from the analysis of the dilepton 
channels by about $200-300\GeV$, depending on the $U(1)^{\prime}$ model, while searches 
relying on the dijet mode, which already exhibit milder limits, are basically 
unconstraining once BSM channels are accounted for.
\begin{table}[t]
 \renewcommand{\arraystretch}{1.4}
 \begin{center}
  \small\begin{tabular}{c||c|c|c|c|c|c}
   Parameter &
     $\theta$ & $\tan\beta$ & $\mu_{\rm eff}~[\mathrm{GeV}]$ &
     $M_{Z'}~[\mathrm{TeV}]$ & $M_0~[\mathrm{TeV}]$ &  $M_1~[\mathrm{GeV}]$\\ \hline \hline
   {\bf BM I} &
     $-0.79~\pi$ & 9.11 & 218.9  & 2.5 & 2.6 & 106.5 \\ \hline
   {\bf BM II} &
     $0.2~\pi$  & 16.08 & 345.3 & 2.5 & 1.9 & 186.7\\[.1cm]
  \multicolumn{7}{c}{}\\
   Parameter &
     $M_2~[\mathrm{GeV}]$ & $M_3~[\mathrm{TeV}]$ & $M_1^\prime~[\mathrm{GeV}]$ & $A_0~[\mathrm{TeV}]$ & 
     $A_\lambda~[\mathrm{TeV}]$& $\sin\chi$ \\ \hline \hline
   {\bf BM I} &
     230.0 & 3.6 & 198.9 & 2 & 5.9 & $-0.35$ \\ \hline
   {\bf BM II} &
     545.5 & 5.5 & 551.7 & 1.5 & 5.1 & 0.33
  \end{tabular}\normalsize
  \caption{\label{tab:parabm}
   UMSSM parameters for the reference points
   {\bf BM I} and {\bf BM II}.}\end{center}
\end{table}

In the present investigation, the mixing between the new $U(1)^{\prime}$ and the SM
groups plays a crucial role. First, there could be some mass mixing 
between the $Z$ and $\Zp$ eigenstates parametrised by
a mixing angle $\alpha_{Z\Zp}$, which is nevertheless
constrained by the EW precision tests (EWPT) to be very small~\cite{Erler:2002pr}.
Then, the $Z$ and \Zp~can kinematically mix through an angle $\chi$, which
modifies the interaction term between the \Zp~and the fermions.
In detail, after accounting for the kinetic mixing, the interaction of the \Zp~with a fermion 
$\psi_i$ having charges $Y_i$ and $Q^\prime_i$ under $U(1)_Y$ and $U(1)^{\prime}$ is given by the Lagrangian
\begin{equation}
{\cal L}_{\rm int}=-g^\prime\bar\psi_i\gamma^\mu\bar
Q_i \Zp_\mu\psi_i\ , 
\end{equation}
where 
$
\bar{Q}_i=\frac{Q'_i}{\cos\chi}-\frac{g_1}{g'}Y_i\tan\chi
$
is the  modified fermion $U(1)^{\prime}$ charge after kinetic mixing and $g^\prime$ 
is the $U(1)^{\prime}$ coupling constant.
Leptophobic scenarios can hence be obtained by
requiring $\bar{Q}_i=0$ for both left- and right-handed leptons, \iee,
$\bar Q_L=\bar Q_E =0$~\cite{Babu:1996vt}.
As discussed in \citeref{Araz:2017wbp}, the leptophobic condition can be naturally
achieved for the model labelled as $U(1)'_\eta$, corresponding to an $E_6$ mixing angle
$\theta=\arccos\sqrt{5/8}$. Furthermore, using the typical GUT-inspired
proportionality relation between the coupling constants of $U(1)$ and
$U(1)^{\prime}$ 
$
g^\prime=\sqrt{\frac{5}{3}}~g_1,
$ the leptophobic condition is 
realised for \sloppy\mbox{$\sin\chi\approx-0.3$}.
As pointed out in \citeref{Araz:2017wbp}, this relation can be defined either at the $\Zp$
mass scale, \iee~{\cal O}(TeV), or at the GUT scale. Imposing unification at the $\Zp$ scale clearly yields
a higher value of $g^\prime$ and hence a larger $\Zp$ production cross section at the LHC.
In the following, we assume unification at the TeV scale.
\begin{figure}[t]
\centering
\begin{minipage}[t][\minipageheightdp][t]{\minipagewidthdp}
\centering
\includegraphics[width=\figuremaxwidthdp,height=\figuremaxheightdp,keepaspectratio]{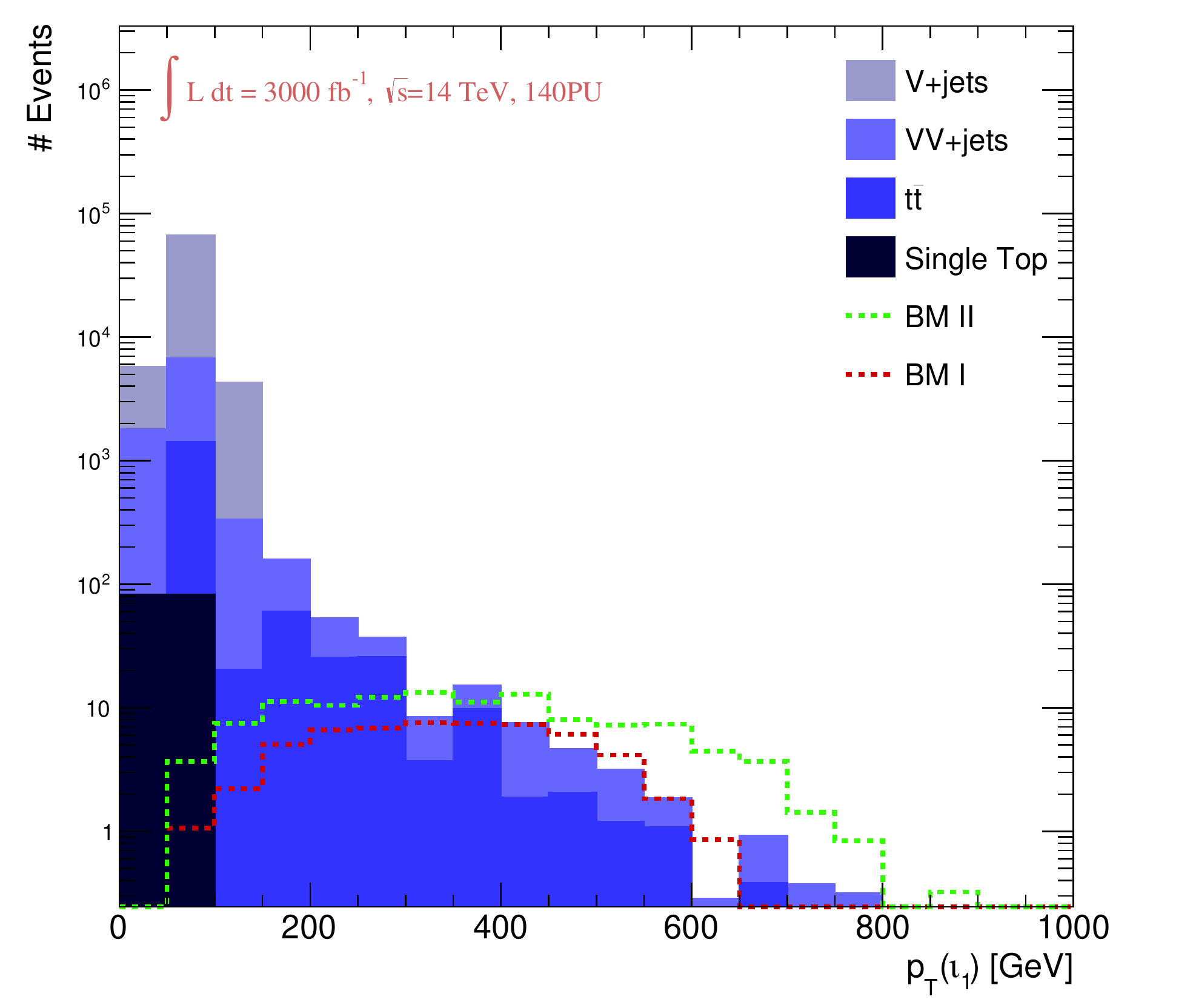}
\end{minipage}
\begin{minipage}[t][\minipageheightdp][t]{\minipagewidthdp}
\centering
\includegraphics[width=\figuremaxwidthdp,height=\figuremaxheightdp,keepaspectratio]{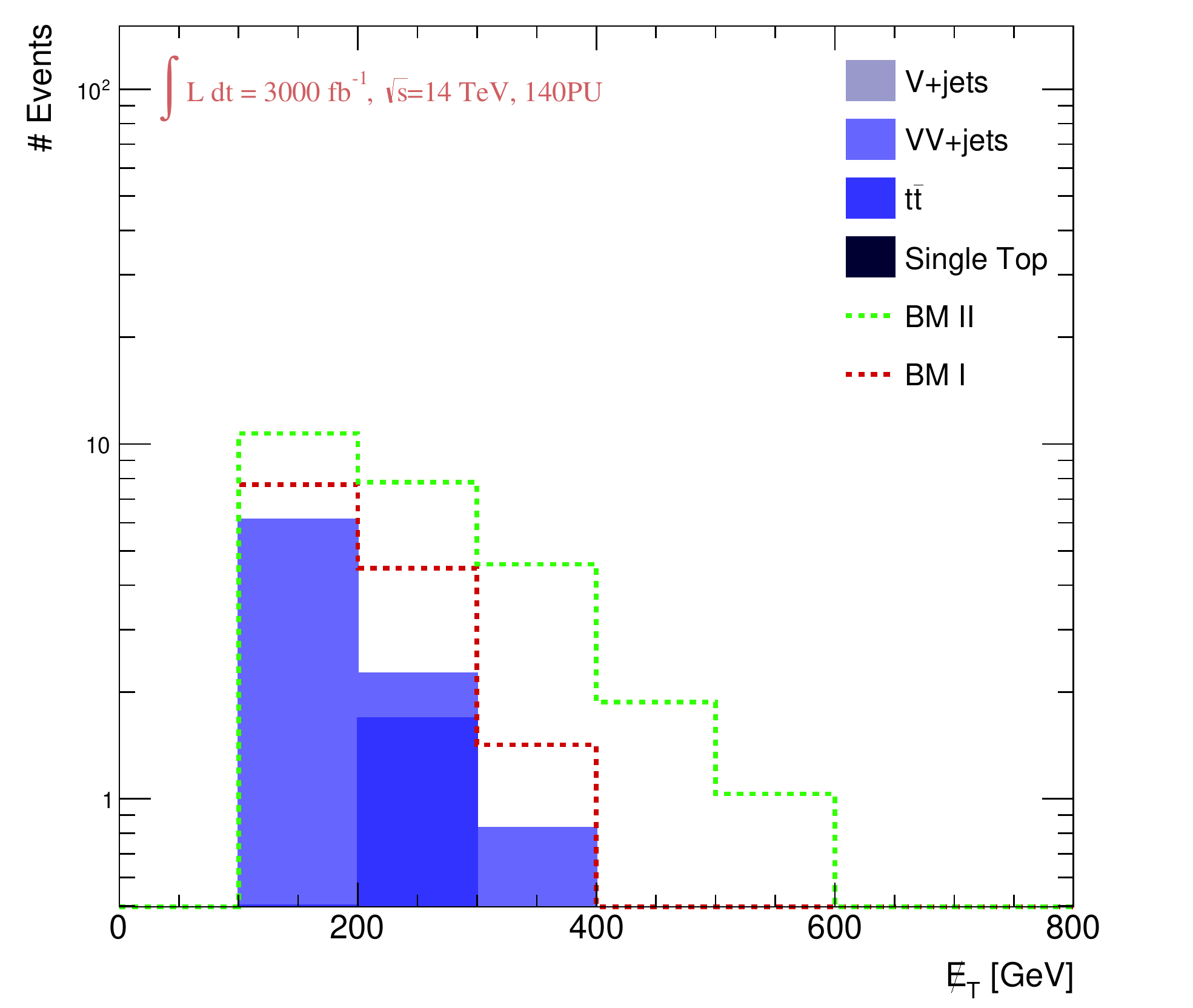}
\end{minipage}
 \caption{Transverse momentum distribution of the leading muon $l_1$ after
  applying the first 6 cuts (left) and missing transverse-energy
  spectrum after all cuts (right) for both leptophobic UMSSM benchmark signals and backgrounds.}
 \label{fig:distri}
\end{figure}
%\paragraph*{Results at HL-LHC}\label{par:zpresults}

Following~\cite{Araz:2017wbp}, two UMSSM benchmark points are considered for this study,  
denoted by {\bf BM I} and {\bf BM II}, consistent with the current experimental data and 
featuring a leptophobic $\Zp$. In both cases the $\Zp$ mass is set to $M_{\Zp}=2.5\TeV$.
In \tabb{tab:parabm}, the relevant parameters for these reference points are reported:
the particle mass spectrum can be calculated by using 
the \sarah{} code~\cite{Staub:2013tta} and its interface with \spheno{}~\cite{Porod:2011nf}.
The particle masses and the decay tables of {\bf BMI} and {\bf BMII} have been given in \citeref{Araz:2017wbp}
and we do not quote them here for the sake of brevity.
The $\Zp$ BRs into BSM final states are 
of about $12\%$ for {\bf BM I} and $15\%$ for {\bf BM II}. In particular, 
the BR of the $\Zp$ into chargino pairs
$\tilde\chi^+_1\tilde\chi^-_1$  is about $2\%$ in {\bf BM I} and $6\%$ in {\bf BM II}. 

Since in {\bf BM I} and {\bf BM II} the 
mass splitting between the lightest charginos and neutralinos is
larger than the $W$ mass ($M_{\tilde\chi^\pm_1}\simeq 177\GeV$ and $M_{\tilde\chi^0_1}\simeq 95\GeV$ in {\bf BM I}, 
$M_{\tilde\chi^\pm_1}\simeq 344\GeV$ and $ M_{\tilde\chi^0_1}\simeq 178\GeV$ in
{\bf BM II}), then $ \tilde\chi^\pm_1 $ can undergo the
transition $\tilde\chi^\pm\to W^\pm \tilde\chi^0_1$ with real $W$-bosons.
As a case study of a leptophobic $\Zp$ in the UMSSM at the HL-LHC, we then
explore the decay chain
\begin{equation}\label{eq:signal}
pp\to Z'\to \tilde\chi^+_1\tilde\chi^-_1\to l^+l^-\slashed{E}_T,
\end{equation}
where we have assumed that both $W$-bosons decay leptonically and $\slashed{E}_T$
is the missing transverse energy carried away by final-state neutrinos and neutralinos.
The amplitudes of the process \eqref{eq:signal} have been computed at the NLO accuracy
by means of \MGvATNLO~\cite{Alwall:2014hca}, yielding a cross section of about 
$\sigma(pp\to \Zp)\simeq 120\pb$.
In our phenomenological study, parton showers and hadronisation are provided by \pythia{ 8}~\cite{Sjostrand:2014zea}, with the
response of a typical LHC detector modelled 
according to the \delphes{ 3} package~\cite{deFavereau:2013fsa} (version 3.3.2), the detector
parameterisation being the one provided by Snowmass~\cite{Anderson:2013kxz,%
Avetisyan:2013onh}. While the default mean number of pile-up events is in this
case of $140$ and thus a bit low, our analysis essentially relies on very hard
isolated leptons (with transverse momenta larger than $200\GeV$) and a large
amount of missing energy (greater than $100\GeV$) which are expected to only be
slightly affected by the differences.
Jets are clustered
following the anti-$k_T$ algorithm~\cite{Cacciari:2008gp}
with a radius parameter $R=0.6$, as implemented in
the \fastjet program (version 3.1.3)~\cite{Cacciari:2011ma}.

As backgrounds to our signal, we consider single
vector-boson ($V$) and vector-boson pair ($VV$) production, possibly accompanied by jets,
top-quark pairs and single-top events.
Our results concern $pp$ collisions at $14\TeV$
and an integrated luminosity of ${\cal L}=3\abinv$, which corresponds to the HL-LHC.
We require two charged muons in
the final state at an invariant opening angle $\Delta R$ larger than $2.5$ and take into account only isolated muons, with an activity of at most $15\%$ of the muon transverse momentum lying in a cone of radius $0.4$ centred on the muon; also, we set cuts of $300\GeV$ and $200\GeV$ on the hardest and next-to-hardest lepton and force the missing transverse momentum to be above $100\GeV$. 
Setting such cuts, we are able to separate the signal from the background with
significances $s$ and $Z_A$, defined by
\begin{equation}
  s  = \frac{S}{\sqrt{B+\sigma_B^2}}\,,\quad
  Z_A= \sqrt{ 2\left( (S+B)\ln\left[\frac{(S+B)(S+\sigma^2_B)}{B^2+(S+B)\sigma^2_B}\right] - \frac{B^2}{\sigma^2_B}\ln\left[1+\frac{\sigma^2_BS}{B(B+\sigma^2_B)}\right]\right)} \ ,
\end{equation}
varying between $3\sigma$ and $7\sigma$.
Besides the total number of events, one can explore differential distributions, such as the
leading-lepton transverse momentum or the missing transverse energy presented in \fig{fig:distri}.
As for the $p_T(l_1)$ spectrum (left), all four considered backgrounds
contribute at small $p_T$, while above $100\GeV$ only $VV$ and $t\bar t$ events survive.
The signal spectra are broad and below the
backgrounds at low transverse momentum, whereas, for
$p_T(l_1)>300\GeV$, both {\bf BM I} and {\bf BM II} signals are
competitive with the background.
For even larger $p_T$, say $p_T(l_1)>500\GeV$,
muons coming from supersymmetric decays of a leptophobic $\Zp$
become dominant, especially for the reference point {\bf BM II}.
In \fig{fig:distri} (right) we present the missing transverse energy,
due to the lightest neutralinos $\tilde\chi^0_1$ in the signal and
to neutrinos in the backgrounds, after all cuts are imposed.
Our  $\slashed{E}_T$ signal spectra are 
well above the backgrounds due to $VV$ and
$t\bar t$ production.
The {\bf BM II} scenario, in particular, is capable of
yielding a few events up to $\slashed{E}_T\simeq 600\GeV$,
while all backgrounds are negligible for
$\slashed{E}_T> 400\GeV$.

\begin{figure}[t]
\centering
\begin{minipage}[t][\minipageheightdp][t]{\minipagewidthdp}
\centering
\includegraphics[width=\figuremaxwidthdp,height=\figuremaxheightdp,keepaspectratio]{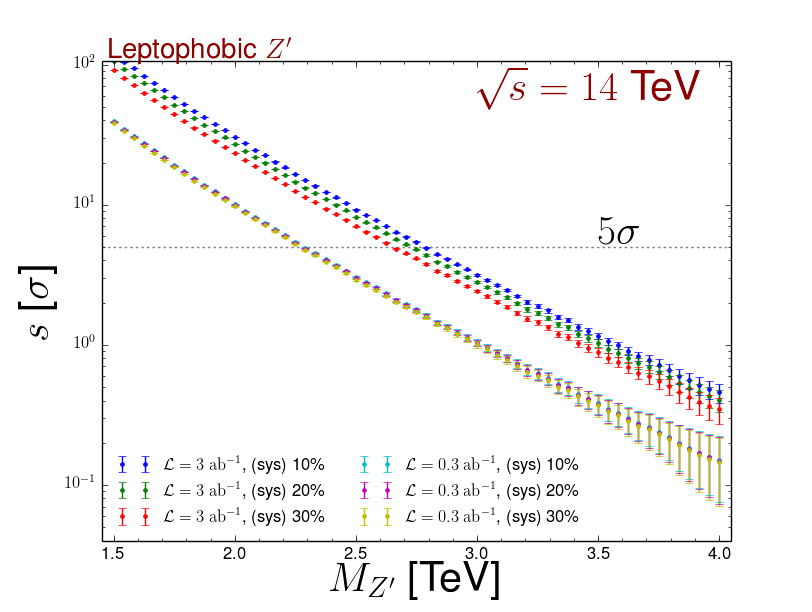}
\end{minipage}
\begin{minipage}[t][\minipageheightdp][t]{\minipagewidthdp}
\centering
\includegraphics[width=\figuremaxwidthdp,height=\figuremaxheightdp,keepaspectratio]{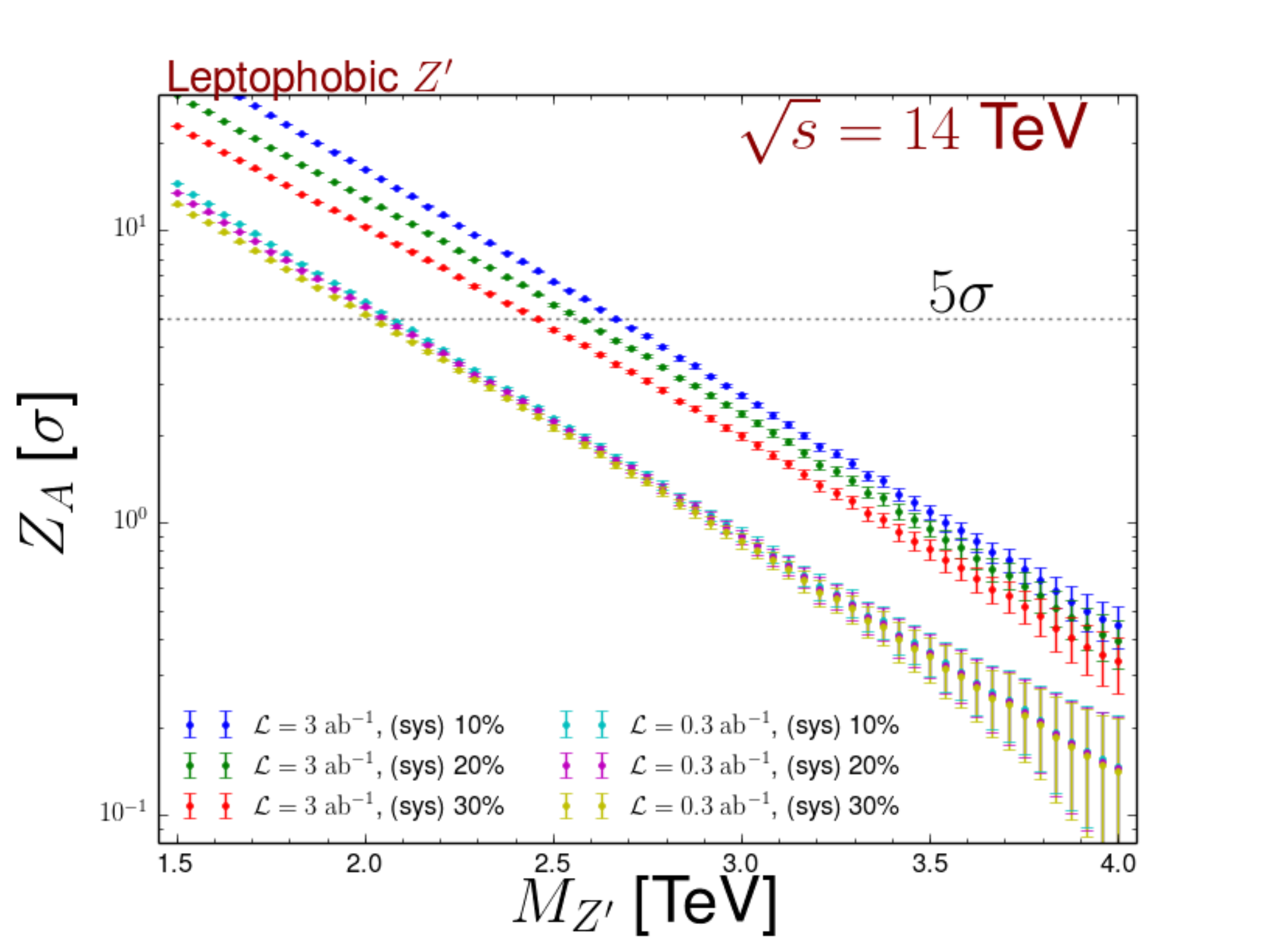}
\end{minipage}\\
\begin{minipage}[t][\minipageheightdp][t]{\minipagewidthdp}
\centering
\includegraphics[width=\figuremaxwidthdp,height=\figuremaxheightdp,keepaspectratio]{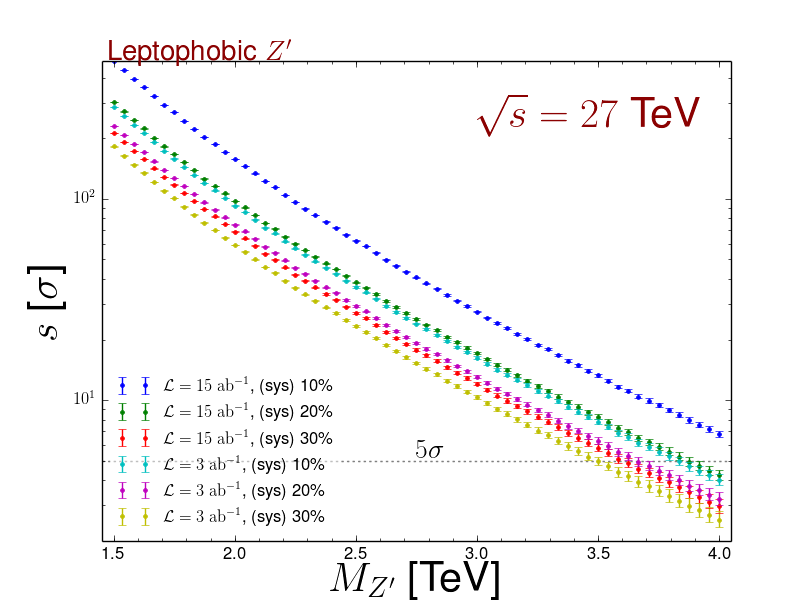}
\end{minipage}
\begin{minipage}[t][\minipageheightdp][t]{\minipagewidthdp}
\centering
\includegraphics[width=\figuremaxwidthdp,height=\figuremaxheightdp,keepaspectratio]{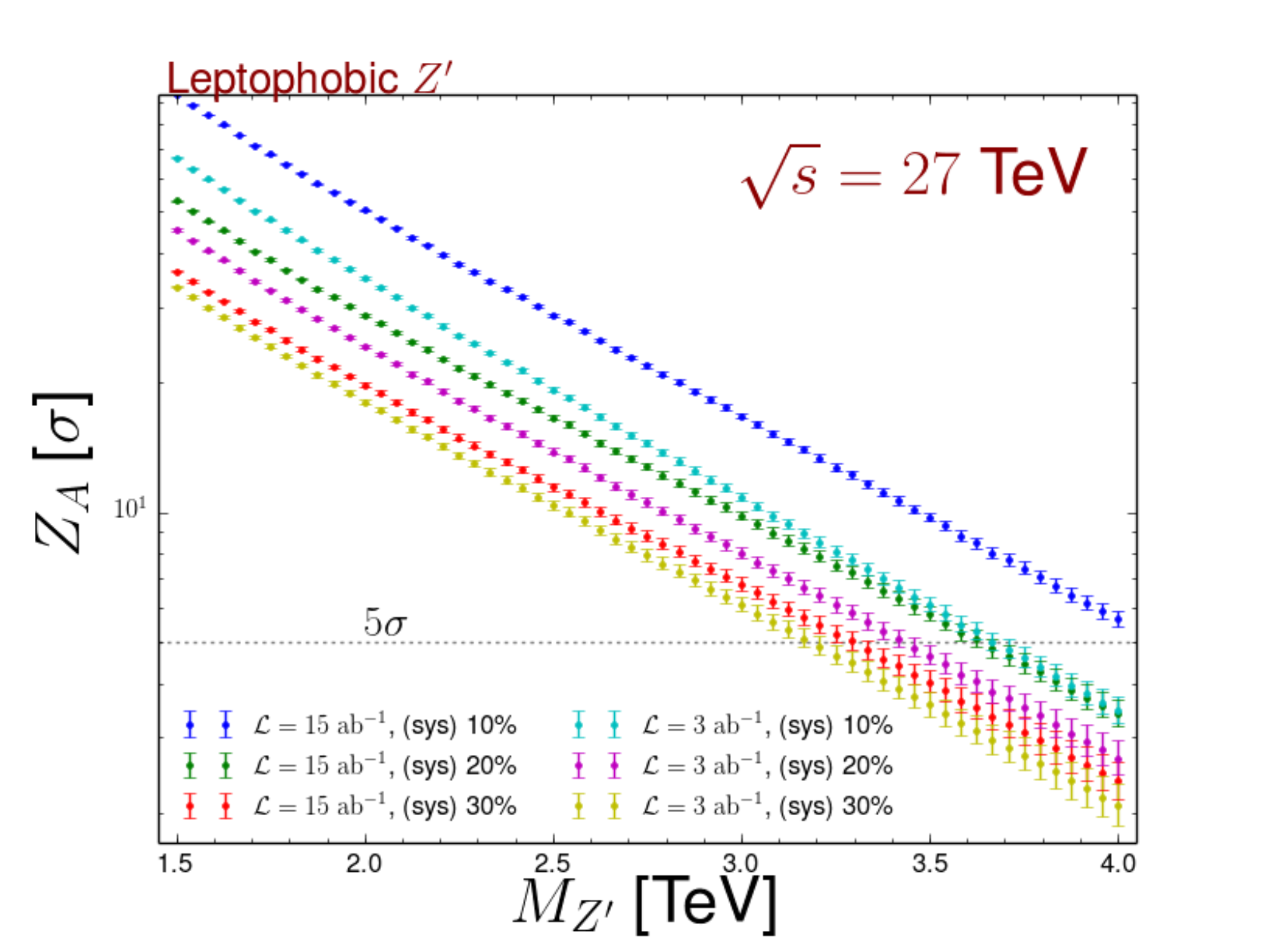}
\end{minipage}
 \caption{Significances $s$ (right) and $Z_A$ (left) to discover a leptophobic \Zp~boson decaying into charginos as a
 function of its mass $M_{Z'}$, for a few values of luminosity and systematic error on the SM background. All results are obtained for
 a \com~energy $\sqrt{s}=14\TeV$ (top) and $\sqrt{s}=27\TeV$ (bottom).}
 \label{fig:sz14}
\end{figure}

We present projections for proton-proton collisions at $14\TeV$
and $27\TeV$ as functions of the luminosity, $M_{Z'}$ and different assumptions
for the systematic uncertainties on the background. We consider a leptophobic
\Zp~decaying into charginos and refer to the BM II reference point as it turns
out to be the more promising setup for a possible discovery of a \Zp~boson in
supersymmetric and leptophobic scenarios. In \fig{fig:sz14} we show the significances $s$ (left) and $Z_A$ (right) at $14\TeV$ (up)
and $27\TeV$ (down), for a \Zp~mass in the
$1.5~{\rm TeV}<M_{Z'}<4~{\rm TeV}$ mass window. In each figure, we consider two
luminosity options chosen to be $300\fbinv$ and $\intLumi$ for a collisions at a
\com~energy of $14\TeV$, and \intLumi and \intlumihelhc for $27\TeV$ collisions.
We moreover allow the systematics $\sigma_B$ on the SM background to vary from
$10\%$ to $30\%$ of the background itself.

It turns out that with $300\fbinv$ of $14\TeV$ collisions, leptophobic \Zp
with masses ranging up to $2.3\TeV$ could be discovered (\iee~$s \geq 5
\sigma$) regardless of the assumption on the systematic errors (\fig{fig:sz14},
top left). In the high-luminosity phase of the LHC, with a luminosity of
$3\abinv$, the lower bound on the \Zp~mass increases to $2.6\TeV$ (for
$\sigma_B=0.3 B$) or $2.8\TeV$ (for more optimistic systematics of
$\sigma_B=0.1 B$).
In terms of $Z_A$, the discovery reach is reduced to $M_{Z'}=2\TeV$ for
$300\fbinv$, again independently of any assumption on the
background systematics $\sigma_B$, and to $2.5~{\rm TeV}<M_{Z'}<2.7~{\rm TeV}$
for $3\abinv$ (\fig{fig:sz14}, top right).

In the high-energy phase of the LHC at $27\TeV$ (\fig{fig:sz14} bottom), a visible \Zp
signal can be obviously obtained even for much higher masses. In detail, a
significance of $s=5\sigma$ can be reached for \Zp~masses ranging up to
$3.8-4\TeV$ at $27\TeV$ and for a luminosity of \intlumihelhc. The limits are
slightly worse for a reduced luminosity of \intLumi and then reach
$3.5-3.7\TeV$. Using in contrast a more conservative definition of the
significance $Z_A$, we obtain a reach of $3.7-4\TeV$ for \intlumihelhc and only
of $3.2-3.5\TeV$ for a luminosity of \intLumi.

%\paragraph*{Conclusions}\label{par:zpconclusions}
We explored possible loop-holes in GUT-inspired $\Zp$ searches at the LHC by investigating 
supersymmetric and leptophobic models, finding that the inclusion of 
BSM decay modes lowers the exclusion limits on $M_{\Zp}$ 
in dilepton final states by a few hundred GeV and that the limits from 
dijets can be evaded as well. In leptophobic models,
only supersymmetric $\Zp$ decay chains can give rise to charged leptons.
As a case study, we considered the decay of a leptophobic \Zp~with mass $M_{Z'}=2.5\TeV$
into charginos, leading to final states with leptons and missing energy.
We chose two benchmark points in the UMSSM parameter space 
and found that both yield a substantial signal at LHC, which one can
separate from the background with a sensitivity between $3\sigma$ and $7\sigma$
at $14\TeV$ and \intLumi. We finally investigated the reach of the high-luminosity
and high-energy LHC runs in terms of the \Zp~mass and systematic uncertainty on the background.
We found that at $14\TeV$ and \intLumi a leptophobic \Zp~can be discovered with a significance about
$5\sigma$ for a mass $M_{Z'}<2.8\TeV$, while at $27\TeV$ and \intlumihelhc one can
explore leptophobic and supersymmetric \Zp~models up to about $M_{Z'}\simeq 4\TeV$.
These result make therefore the investigation of such
scenarios both feasible and worthwhile at HL- and HE-LHC.

%%%%%%%%%%%%

\clearpage
\resetcounters

\section{Dark Matter and Dark Sectors Searches}\label{sec:DM}
%The global consistency of the SM predictions against the first years of data at $7,8$, and $13$ TeV at the LHC is dramatically changing our perspective on the new-physics origin and characteristic energy scale.
%Most beyond-SM scenarios, introduced so far either to solve or ameliorate the hierarchy problem in the SM Higgs-boson sector, are predicting new observable phenomena at the TeV scale. The lack of any experimental evidence for these scenarios has in turn set very strong constraints on their free-parameter space, eventually questioning the foundations behind the very idea of naturalness in the Higgs sector.
There is now overwhelming evidence for the existence of a new matter component of the universe, dark matter (DM).  Precision measurements of the cosmic microwave background and gravitational lensing measurements confirm the presence of this non-luminous matter. Through its gravitational interactions we know that dark matter makes up $\sim 26\%$ of the matter-energy budget of the universe, and that the stars of the Milky Way move inside a far larger, approximately spherical, dark matter halo.  
However, the nature and properties of DM remain largely unknown. Searches for DM particles are performed using multiple, complementary, approaches: the measurement of elastic scattering of DM with nuclei and electrons in a detector (direct detection)~\cite{Goodman:1984dc}, the detection of SM particles produced in the annihilations or decays of DM in the universe (indirect detection)~\cite{Gunn:1978gr,Stecker:1978du,Silk:1984zy,Krauss:1985ks,Silk:1985ax}, the study of the effect of DM self interactions on astrophysical systems~\cite{Spergel:1999mh}, and the production of DM particles at colliders~\cite{Buchmueller:2017qhf,Penning:2017tmb}.  The latter process is the focus of this section.  

These various approaches are very complementary in nature.  For instance, the first three techniques all require relic DM to interact and thus suffer from uncertainties related to our knowledge of DM's distribution whereas the production of DM at colliders does not, but is instead limited by the kinematic reach of the machine.  By combining the results of all search techniques we gain a deeper understanding of the nature of dark matter.

Dark Matter production by itself does not lead to an observable signal at hadronic machines, where the total \com~energy of the collision is not known. Instead if the DM system recoils against visible activity it can be searched for as missing transverse energy and momentum.  We catalogue the search strategies for DM composed of a  by what this visible activity is, and report on prospective DM studies in \secc{sec:DMjet} to \secc{sec:DMWZ}, while \secc{sec:DS} is focused on searches for light vector bosons associated with forces in the dark sector.  In particular, if this dark sector force is abelian then the associated ``dark photon" can kinetically mix with the $U(1)$ in the SM, leading to lepton pairs, possibly with displacement, or missing energy signals.  

DM production in association with a high $\pt$ jet is presented in \secc{sec:DMjet}, with sensitivities depending on the careful assessment of systematics.  This channel is a useful probe of DM production  through the exchange of a neutral mediator that couples to the SM. It may also be produced in the decay of an exotic coloured state.  Furthermore, compressed SUSY scenarios, such as higgsino or wino DM, can be probed through the monojet signature. 

DM production in association with heavy flavour quarks is presented in \secc{sec:DMHF}.  The HL-LHC will improve the sensitivity to mediator masses by a factor of 3-8 relative to the Run-2 searches with $36$\fbinv, while HE-LHC will extend the coverage to otherwise inaccessible regions of the parameter space.  The case of 2HDMa models is complemented by 4-top final states at HL-LHC, searched in events with two same-charge leptons, or with at least three leptons. While searches using $36\fbinv$ Run-2 data have limited sensitivity considering the most favourable signal scenarios (\egg~$\tan \beta = 0.5$), HL-LHC will probe possible evidence of a signal with $\tan \beta = 1$, $m_{H}=600\GeV$ and mixing angle of $\sin \theta=0.35$, assuming $m_{a}$ masses between $400\GeV$ and $1\TeV$, and will allow exclusion for all $200\GeV<m_{a}<1\TeV$.
%For DM produced in association with bottom or top quarks, where a (pseudo)scalar mediator decays to a DM pair, the HL-LHC will improve the sensitivity to mediator masses by a factor of 3-8 relative to the Run-2 searches with 36\fbinv.

Prospect studies where DM is produced in association with, or through interactions with, EW gauge bosons are reported in~\secc{sec:DMWZ}. Compressed SUSY scenarios, as well as other DM models, can be targeted using signatures such as mono-photon and vector-boson-fusion (VBF) production, in addition to the classic monojet channel. We show that mono-photon and VBF events allow targeting an EW fermionic triplet (minimal DM), equivalent to a wino-like signature in SUSY, for which there is no sensitivity in Run-2 searches with $36\fbinv$. Masses of the $\neut$ up to $310\GeV$ ($130\GeV$) can be excluded by the mono-photon (VBF) channel, with improvements being possible, reducing the theoretical uncertainties. Projections for searches for a mono-$Z$ signature
with $Z\to\ell^+\ell^{-}$ recoiling against missing $E_T$, have been interpreted in terms of models with a spin-1 mediator and 2HDMa models. %models with two Higgs doublets and an additional pseudoscalar mediator $a$ coupling to DM (2HDMa). 
The exclusion is expected for mediator masses up to $1.5\TeV$, and for DM and pseudoscalar masses up to $600\GeV$,  a factor of $\sim 3$ better than the $36\fbinv$ Run-2 constraints. The potential to target Higgs portal models 
and prospects for the HL- and HE-LHC to probe viable multi-TeV dark matter are also presented. 

Simple DM models consist of a DM particle and a mediator that couples it to the SM.  The DM may, however, sit in a larger {\it hidden} (or {\it dark}) sector with additional new states and new interactions.  If these interactions include a light $U(1)$ gauge boson, $A'$, this dark photon may mix, through the kinetic-mixing portal, with the SM photon leading to interesting collider signatures.
%A new class of models containing {\it hidden} (or {\it dark}) sectors --made up of matter neutral under the SM gauge interactions-- are also emerging as suitable candidates possibly fitting DM properties.
In \secc{sec:DS}, searches for dark photons under various hypotheses of dark sectors are presented.  Prospects for an inclusive search for dark photons decaying into muon or electron pairs indicate that the HL-LHC could cover a large fraction of the theoretically favoured $\epsilon - m_{A'}$ space, where $\epsilon$ is the size of the kinetic mixing between the photon and the dark photon.

%%%%%%%%%%%%%%%%%%%
\subsection{Dark Matter and Jets}
\label{sec:DMjet}
If DM is produced in association with QCD activity it is typically searched for in the monojet channel.  The DM may be produced through a SM neutral mediator, see \secc{sec:atlasmonojet}, or it may be produced as the decay product of a new heavy coloured state, see \secc{sec:theoryDMqstar}.  Monojet-like signatures can be exploited to search for higgsinos and winos in SUSY, see \secc{sec:theoryewinosmono}.  
If the parent state that decays to DM is charged and relatively long lived it is more efficient to search in the disappearing track topology. Additional discussion of disappearing track analyses, in the context of long lived particle searches, can be found in \secc{sec:disappearingtrackssec4}.

\subsubsection{\atlas{Studies on the sensitivity to Dark Matter of the monojet channel at HL-LHC}}
\label{sec:atlasmonojet}
\contributors{G.~Frattari, V.~Ippolito, G.~Gustavino, J.~Stupak, ATLAS}

The goal of this study, detailed in~\cite{ATL-PHYS-PUB-2018-043}, is to evaluate the impact of different assumed systematic uncertainty scenarios on the expected sensitivity to WIMP Dark Matter in the jet+$\met$ channel, based on the extrapolation to higher luminosity of the limits published by the ATLAS Collaboration with $36.\fbinv$ of $pp$ collisions at a \com~energy $\sqrt{s} = 13\TeV$~\cite{Aaboud:2017phn}. 
The WIMPs escape the detector giving rise to large $\met$ arising if they recoil against a jet from initial state radiation (ISR) off the colliding partons, leading to the so-called monojet topology. The $\met$ in this study is calculated treating electrons and muons as invisible particles.  
The strategy pursued takes signal and background $\met$ distributions from the Run-2 ATLAS data analysis and scales them to $300\fbinv$and \intLumi, exploring various scenarios for the scaling of the systematic uncertainties. 

The dominant backgrounds for the $13\TeV$ data analysis come from $W/Z$+jets processes.  These backgrounds are estimated using MC samples generated with \sherpa{ 2.2.1}. MC samples are re-weighted to account for higher-order QCD and EW corrections following the procedure described in \citeref{Lindert:2017olm} and are normalised in dedicated control regions (CR) described below. Sub-leading backgrounds consist of $t\bar{t}$ and single-top
production, which are generated via \powhegbox{ v2}
%{\sc Powheg-Box v2} 
 and showered with \pythia{ 8},
%{\sc Pythia 8}, 
 and
diboson processes, which are taken from \sherpa{}.
%{\sc SHERPA}. 
The benchmark signal process is generated
for WIMP masses in the range $1\GeV - 1\TeV$ and mediator masses in the range $10\GeV - 10\TeV$ using
\powhegbox{ v2}
%{\sc Powheg-Box v2} 
 with the DMV model~\cite{Haisch:2013ata}, assuming mediator couplings to the quark and WIMP of $g_q=0.25$ and
$g_{\chi}=1$, respectively.  The small multijet and non-collision backgrounds, which are estimated
from data in the $13\TeV$ analysis, are neglected.

The event selection follows that applied in the $13\TeV$ analysis and selects events with \sloppy\mbox{$\met >250\GeV$} in association with at least one high-$p_T$ jet. Up to three additional jets are allowed and all jets are required to be well separated from the missing transverse momentum direction in azimuth. The signal region is required to contain no reconstructed electron or muon. Four additional CRs to isolate the dominant backgrounds are defined based on the
number and type of leptons.  Events with exactly one muon, no other leptons, and transverse mass in the 30--100 GeV range form the $W+$jets ($t\bar{t}$) CR if they have zero (at least one) $b$-tagged jet.
A second $W+$jets CR is built by requiring exactly one electron, no other leptons, and the same transverse mass
requirement. Finally, events with 2 muons with an invariant mass consistent with the $Z$ boson form the $Z$+jets CR. 
A simultaneous, binned likelihood fit of a signal plus background model to the simulated $\met$ distributions 
of the analysis regions is performed. The signal normalisation and two additional normalisation factors,
one which rescales the prediction for processes containing $Z$ and $W$ bosons produced in association with
jets, and one for $t\bar{t}$ and single-top production, are free parameters of the fit. 
Nuisance parameters with Gaussian constraints are used to describe the effect of systematic uncertainties on the signal and background $\met$ distributions. Correlations of systematic uncertainties across $\met$ bins are taken into account.

\begin{figure}[t]
\centering
\begin{minipage}[t][\minipageheightdp][t]{\minipagewidthdp}
\centering
\includegraphics[width=\figuremaxwidthdp,height=\figuremaxheightdp,keepaspectratio]{\main/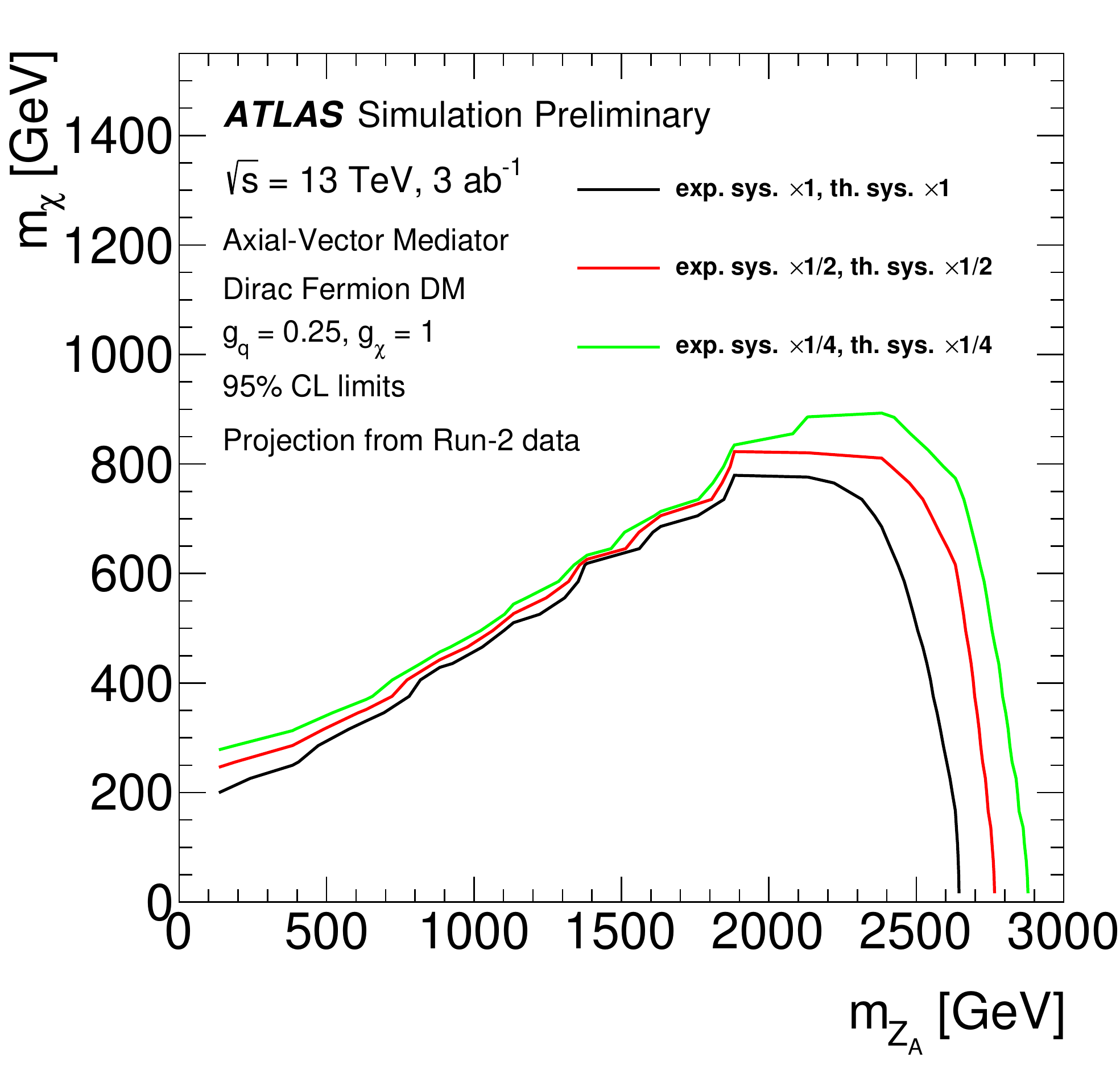}
\end{minipage}
\begin{minipage}[t][\minipageheightdp][t]{\minipagewidthdp}
\centering
\includegraphics[width=\figuremaxwidthdp,height=\figuremaxheightdp,keepaspectratio]{\main/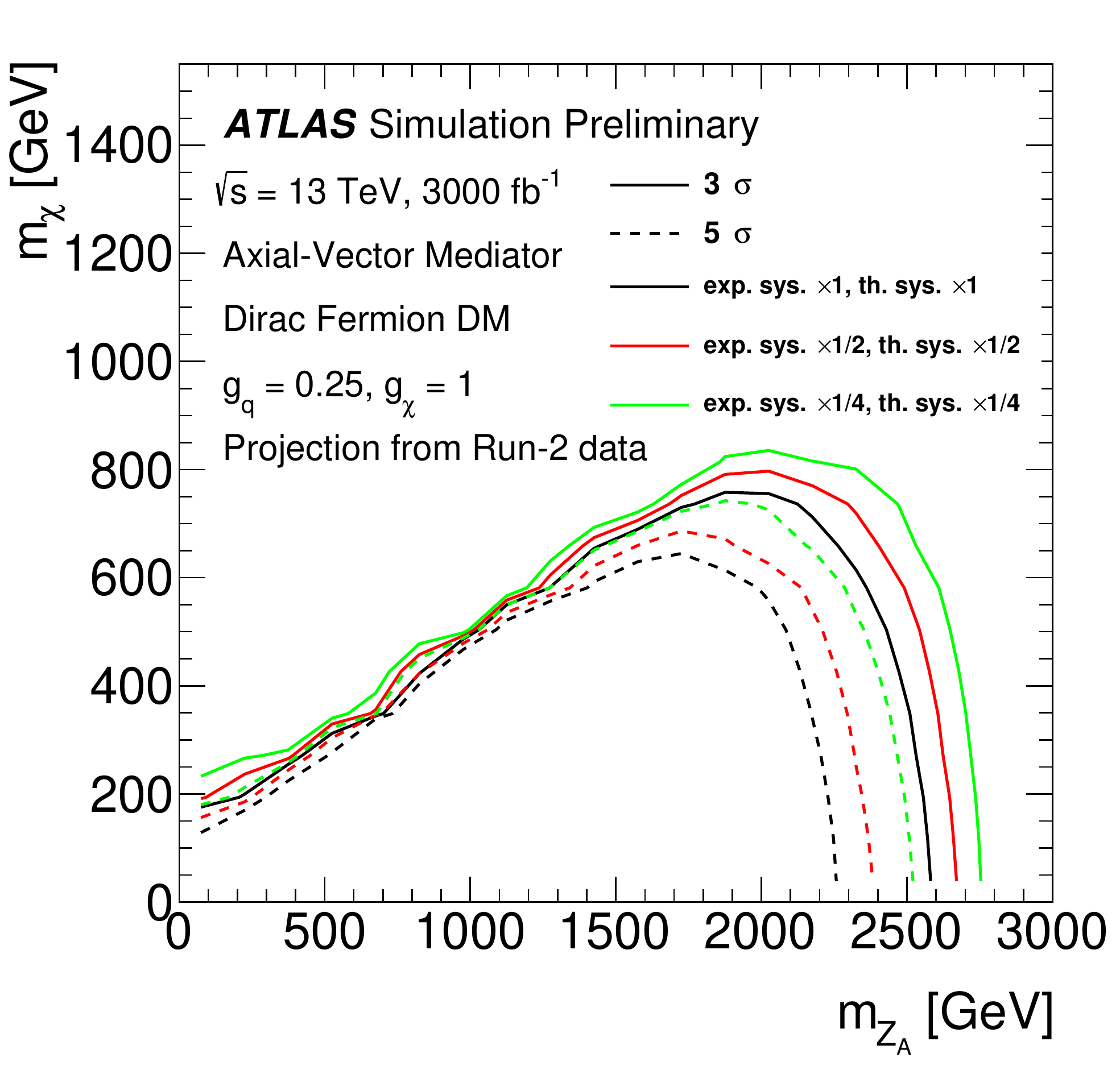}
\end{minipage}
\caption{Left: expected $95\%\cl$ excluded regions in the $(m_{\chi}, m_{Z_A})$ mass plane for the axial-vector simplified model with couplings $g_q=0.25$ and $g_{\chi}=1$, for a luminosity of \intLumi. 
Three contours are shown in each plot, corresponding to the three different systematic uncertainty scenarios: standard (black), 
reduced by a factor 2 (red) and 4 (green). Right: $3\sigma$ and $5\sigma$ discovery contours corresponding to the three different systematic uncertainty scenarios: standard (black), reduced by a factor 2 (red) and 4 (green).} 
\label{fig:ATLAS_monojet_projection}
\end{figure}

The projection for high luminosity proceeds as follows:
\begin{itemize}
\item The $\met$ distributions for signal and the main backgrounds from the $13\TeV$ data analysis in Run-2 are scaled from 36.1\fbinv to \intLumi. The increased statistics
achieved by the higher luminosity allows the discriminant to be binned more finely, increasing the
number of $\met$ bins from 10 in the recent data analysis to 17 for the high-luminosity projection. 
The lower end of the last $\met$ bin, $1.6\TeV$, is chosen in order to keep a similar level of uncertainty 
on signal and control region as for the Run-2 search. 
This turns into an improvement of about $100\GeV$ in the projected mediator mass reach.
\item The background distributions are further scaled up by a factor of $1.27$ ($1.06$) for $Z/W$+jets
($t\bar{t}$ and single top) in order to reflect the observation in the CRs of the Run-2 data
analysis.
\item No correction is made for the increase in the \com~energy to $14\TeV$ at the HL-LHC
since the dedicated NLO QCD and EW corrections for the main $W/Z$+jets background are not available.  
Given the signal cross-section would be expected to increase by $20-40\%$, while the dominant
$W/Z$+jets background cross section will increase by $10-15\%$, this leads to a conservative estimate
of the potential and the actual sensitivity will be slightly better than estimated here.
\end{itemize}

In \citeref{Aaboud:2017phn}, the main background experimental uncertainties for $\met >250\GeV$~($>1\TeV$) are 
related to the leptons (jet and $\met$ scales and resolution), amounting up to $1.7\%$ ($5.3\%$), 
while the main background theoretical uncertainties are related to the $W/Z$ parton shower modelling and PDF 
($W/Z$ QCD and EW corrections), amounting to $0.8\%$ ($2\%$). The signal predictions are mainly affected by 
jet and $\met$ scale and resolution uncertainties on the experimental side, and on the theory side by initial/final-state radiation and PDF uncertainties. In the high-luminosity projection, three different systematic uncertainty scenarios are tested to reflect the possible improvements in detector performance and in the theoretical modelling of signal and background
processes, which could be achieved in the next years thanks to the foreseen detector upgrades and to progress in QCD and EW calculations: 
\begin{itemize}
\item standard: same uncertainties as in \citeref{Aaboud:2017phn};
\item reduced by factor 2: all pre-fit signal and background uncertainties are reduced by a factor two;
\item reduced by factor 4: all pre-fit signal and background uncertainties are reduced by a factor four;
\end{itemize}

The projected exclusion limits with a luminosity of \intLumi for these three scenarios are plotted in the $(m_{\chi}, m_{Z_A})$ mass plane in \fig{fig:ATLAS_monojet_projection} (left), where $\chi$ is the WIMP DM candidate and $Z_A$ the axial-vector mediator. 
The $95\%\cl$ exclusion contour for $m_{\chi}=1\GeV$can be up to $m_{Z_A}=2.65\TeV$, assuming the same uncertainties as in \citeref{Aaboud:2017phn}. 
The excluded region that can be obtained by reducing by a factor two (four) all systematic uncertainties reaches, 
for low $m_{\chi}$, mediator masses of about \sloppy\mbox{$2.77$ ($2.88$)\TeV}. 
Small differences between systematic uncertainty scenarios are observed when approaching the region where the decay of the mediator in two WIMPs is off-shell ($m_{Z^{}_{A}} < 2 m_{\chi}$), due to the decrease of the signal cross-section. 
The discovery contours at $3$ and $5\sigma$ are shown in~\fig{fig:ATLAS_monojet_projection} (right): for $m_{\chi}=1\GeV$, a background incompatibility greater than $5\sigma$ would be reached for $m_{Z_A}=2.25\TeV$, $2.38\TeV$ and $2.52\TeV$ assuming the same uncertainties as in \citeref{Aaboud:2017phn}, 
the scenario obtained by reducing by a factor two, and by a factor four all the systematic uncertainties, respectively. 
The increase in sensitivity of the scenarios with lowered 
systematic uncertainties was checked to be mainly driven by the reduction in the theoretical uncertainties. %This is shown in \fig{fig:ATLAS_monojet_projection} (right). 
Among these, $V$+jets and diboson uncertainties, as well as theoretical uncertainties on signal processes, are similar in size and give the leading contributions. 
%For comparison, the excluded region reaches $m_{Z_A}=2.20$~TeV with $300\fbinv$in the nominal systematic scenario. 

%For comparison, the discovery reach is $m_{Z_A}=1.81$~TeV with $300\fbinv$in the nominal systematic scenario. 

\subsubsection{\theoryC{Monojet Signatures from Heavy Coloured Particles at HL- and HE-LHC}}
\label{sec:theoryDMqstar}
\contributors{A. Chakraborty, S. Kuttimalai, S. H. Lim, M. M. Nojiri, and R. Ruiz}

%%=====================================================================
%At hadron colliders, search strategies for hypothetical coloured particles $\newResonance$ that can decay to dark matter candidates usually involve jets and leptons produced in association  with large missing transverse energy $\met$. In the context of simplified SUSY models, such signatures are now severely constrained by the LHC data~\cite{Sirunyan:2017mrs,Sirunyan:2017xse,Aaboud:2017aeu,Aaboud:2017bac}. It is possible, however, to evade such strong constraints on $\newResonance$ by considering, for example, a compressed mass spectrum scenario~\cite{Martin:2007gf,Fan:2011yu,Murayama:2012jh}. The visible decay products in the  $\newResonance\to$DM+SM process do not have sufficient momenta to be readily distinguished from SM backgrounds because of the small mass splitting between the DM and new coloured particles.  To select such events, one usually relies on hard radiations against which the $\QQbar$ system recoils. This selection leads to significantly smaller selection and acceptance efficiencies, and hence significantly weaker bounds on heavy particle masses. Therefore, such a compressed mass spectra naturally allows relatively light, coloured DM partners in light of present-day LHC data. 

Search strategies for hypothetical coloured particles $\newResonance$ that can decay to dark matter candidates usually involve jets and leptons produced in association  with large missing transverse energy $\met$. In compressed mass spectrum scenarios the visible decay products in the  $\newResonance\to$DM+SM process do not have sufficient momenta to be readily distinguished from SM backgrounds and monojet-like topologies arise. 

%-----------------------------------------
%\begin{figure}
%\begin{center}
%\includegraphics[width=0.325\textwidth]{\main/section3DarkMatterSearches/img/Sec313_tptp_NLO_1j_mVar_varRenoFac_met.pdf}	
%\includegraphics[width=0.325\textwidth]{\main/section3DarkMatterSearches/img/Sec313_gogo_NLO_1j_mVar_varRenoFac_met.pdf}
%\includegraphics[width=0.325\textwidth]{\main/section3DarkMatterSearches/img/Sec313_t1t1_NLO_1j_mVar_varRenoFac_met.pdf}
%\end{center}
%\caption{
%The $\ppQQbarj$ cross section as a function of minimum $\met$ after the experimental selection criteria at 13 TeV,
%for $\Q=\tp,~\go$, and $\stp$, with current $95\%\cl$ limits after $\mathcal{L}=36.1$\fbinv of data at the $13\TeV$ LHC. 
%Also shown is the estimated sensitivity with $\mathcal{L}=3$\abinv, assuming $\delta_{\rm Syst.}=$2.5\% and 1\% systematical errors.
%}\label{fig:comp13TeV}
%\end{figure}
\begin{figure}[t]
% 13 TeV, 36.1\fbinv Plot
\centering
\begin{minipage}[t][\minipageheighttp][t]{\minipagewidthtp}
\centering
\includegraphics[width=\figuremaxwidthtp,height=\figuremaxheighttp,keepaspectratio]{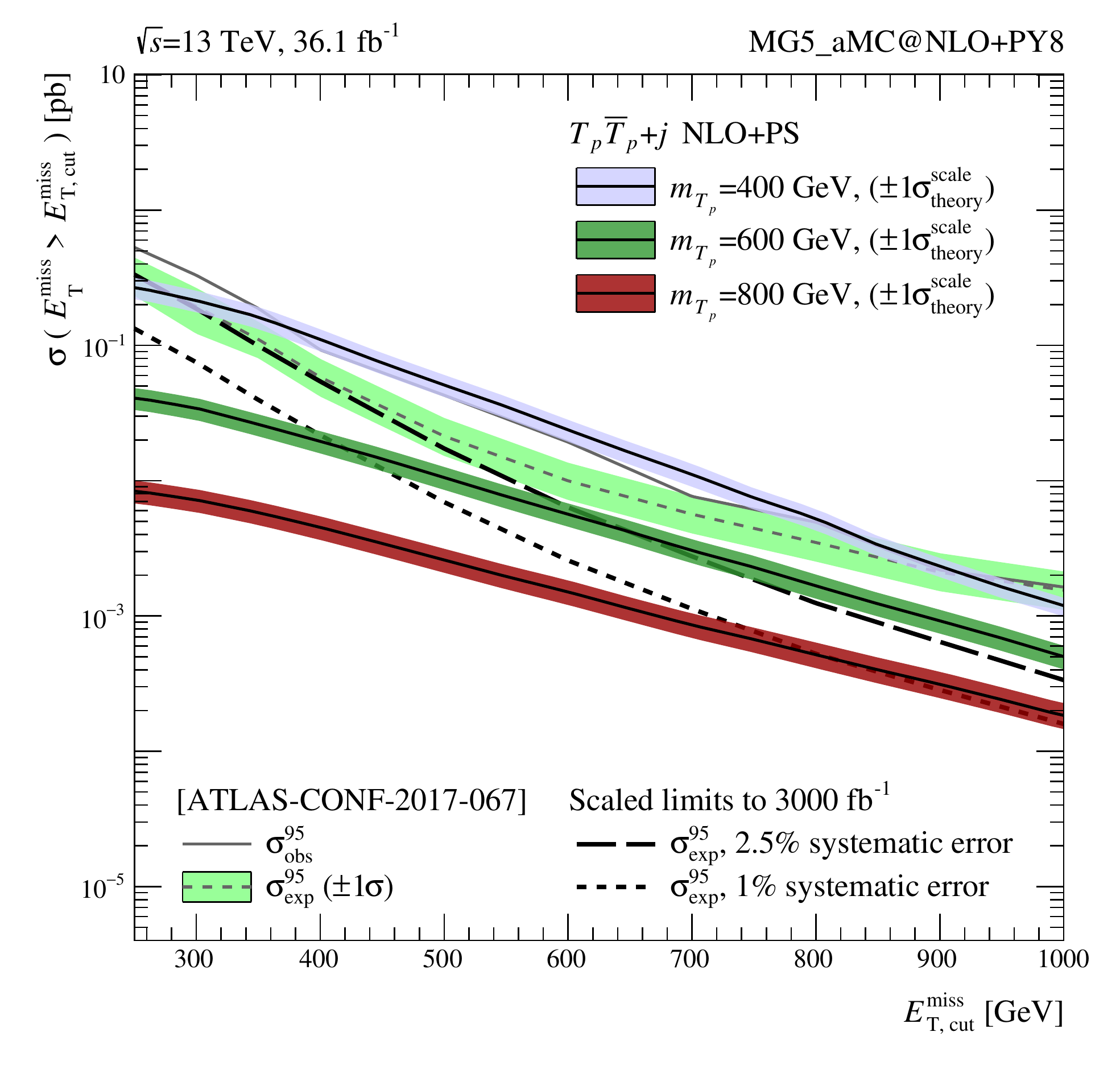}
\end{minipage}
\begin{minipage}[t][\minipageheighttp][t]{\minipagewidthtp}
\centering
\includegraphics[width=\figuremaxwidthtp,height=\figuremaxheighttp,keepaspectratio]{\main/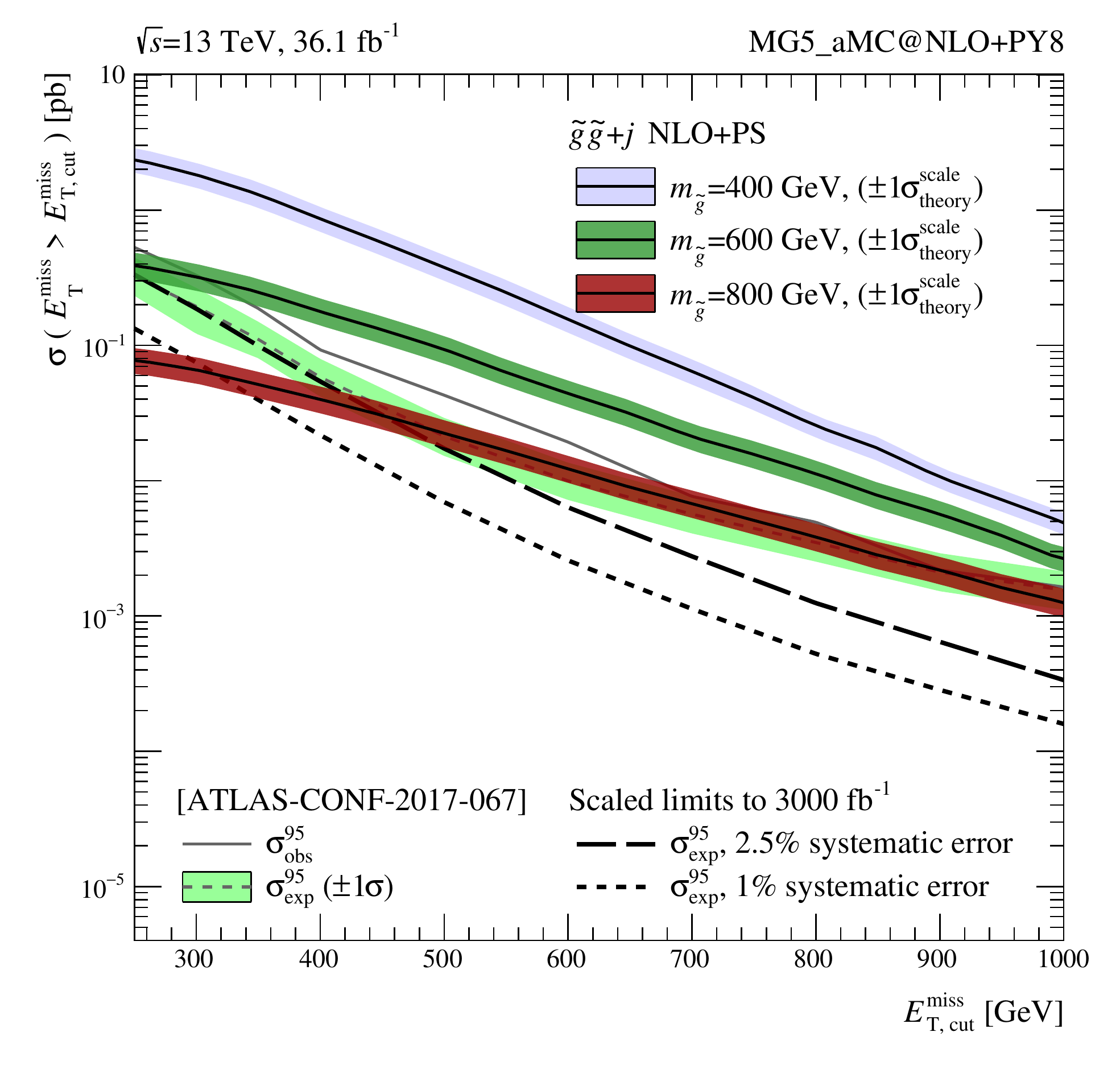}
\end{minipage}
\begin{minipage}[t][\minipageheighttp][t]{\minipagewidthtp}
\centering
\includegraphics[width=\figuremaxwidthtp,height=\figuremaxheighttp,keepaspectratio]{\main/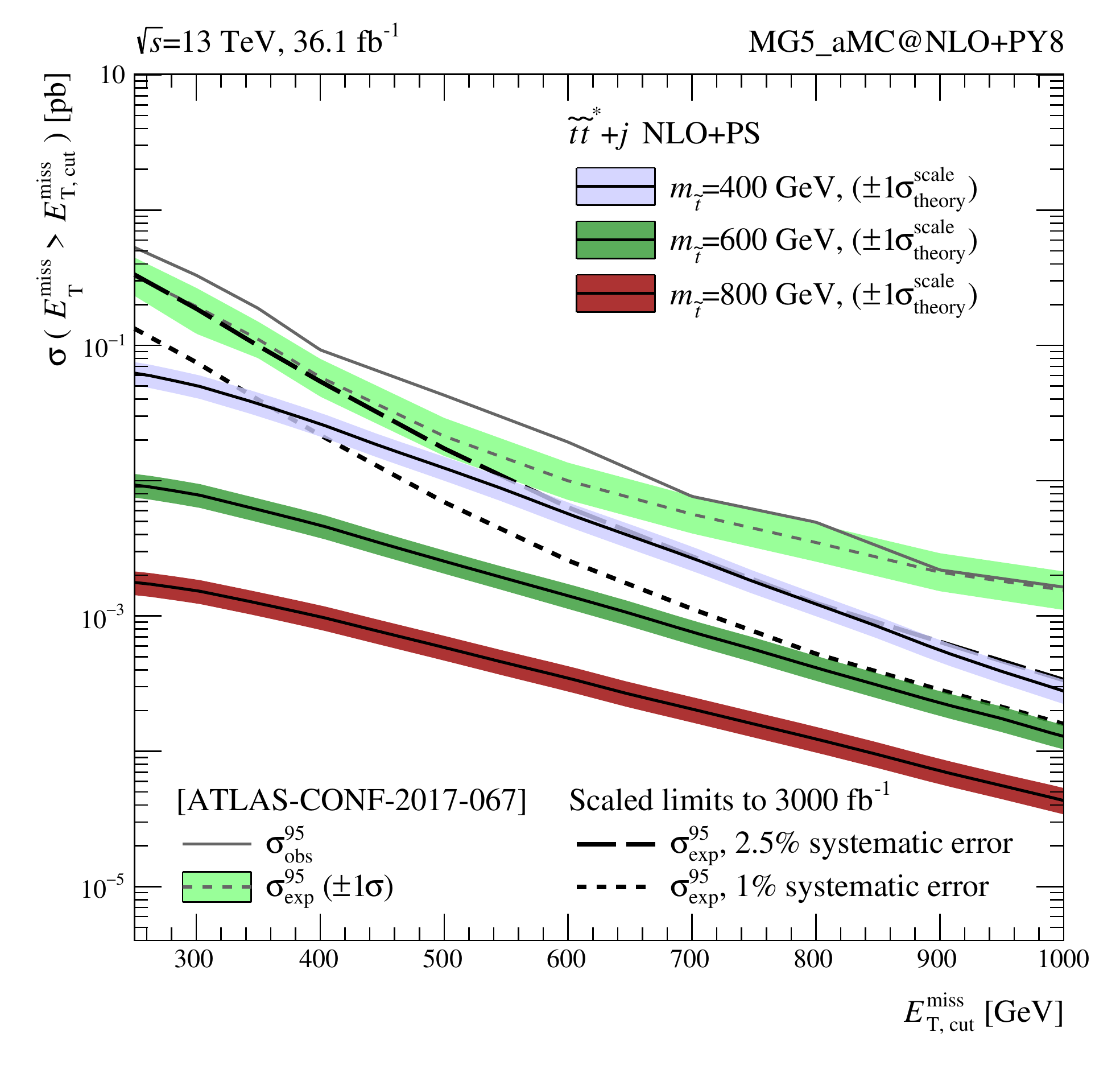}
\end{minipage}\\
\centering
\begin{minipage}[t][\minipageheighttp][t]{\minipagewidthtp}
\centering
\includegraphics[width=\figuremaxwidthtp,height=\figuremaxheighttp,keepaspectratio]{\main/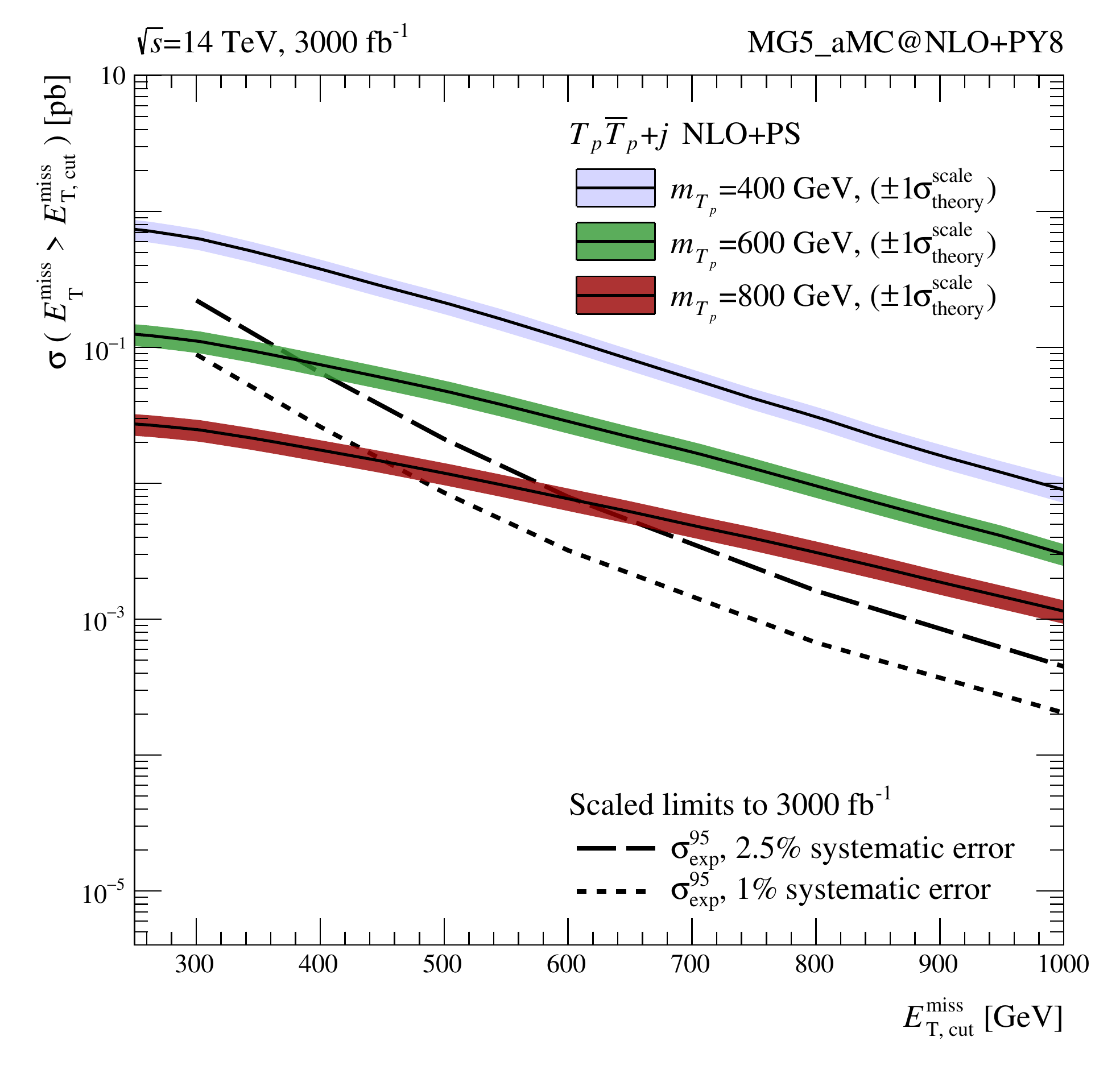}
\end{minipage}
\begin{minipage}[t][\minipageheighttp][t]{\minipagewidthtp}
\centering
\includegraphics[width=\figuremaxwidthtp,height=\figuremaxheighttp,keepaspectratio]{\main/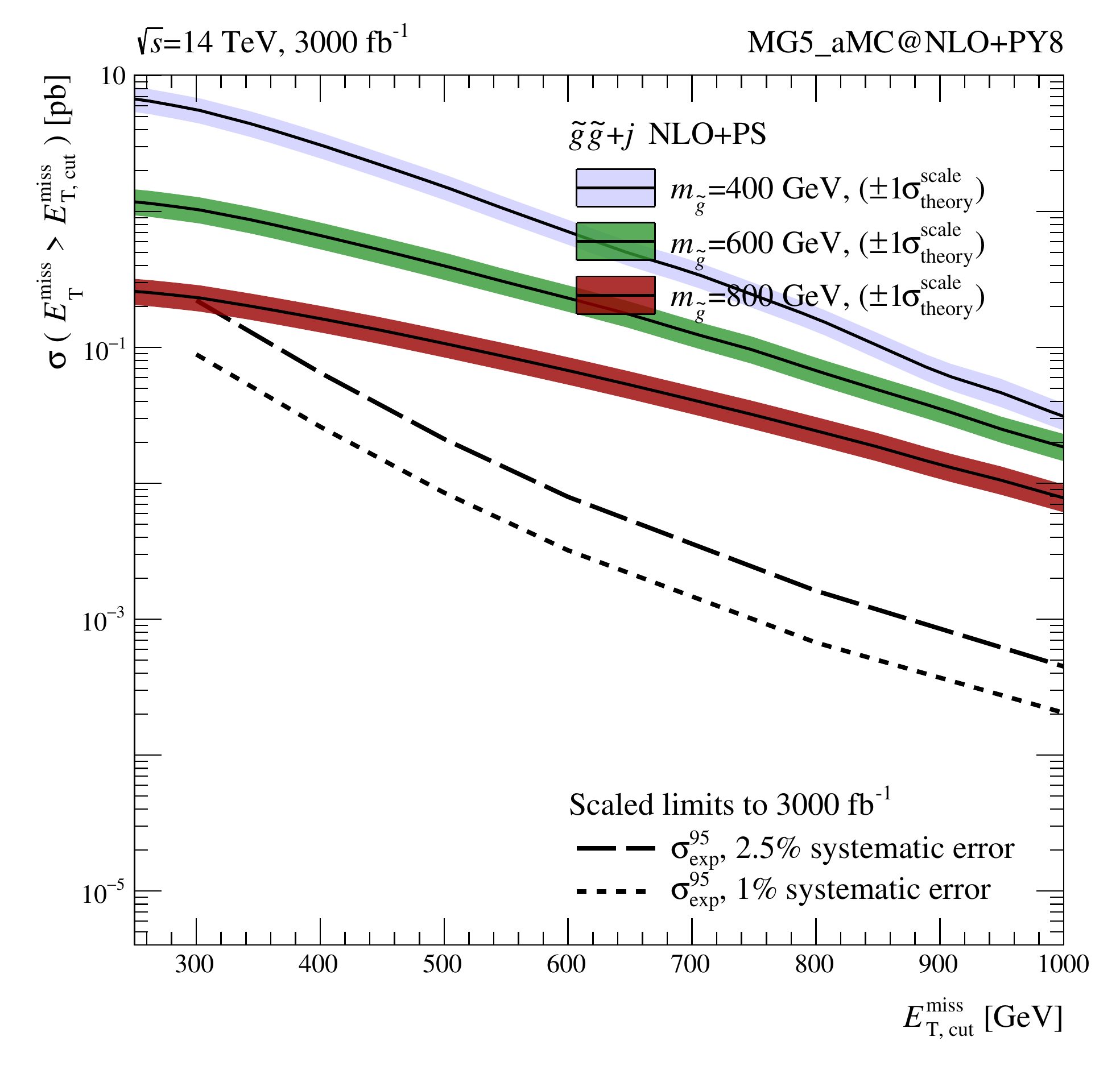}
\end{minipage}
\begin{minipage}[t][\minipageheighttp][t]{\minipagewidthtp}
\centering
\includegraphics[width=\figuremaxwidthtp,height=\figuremaxheighttp,keepaspectratio]{\main/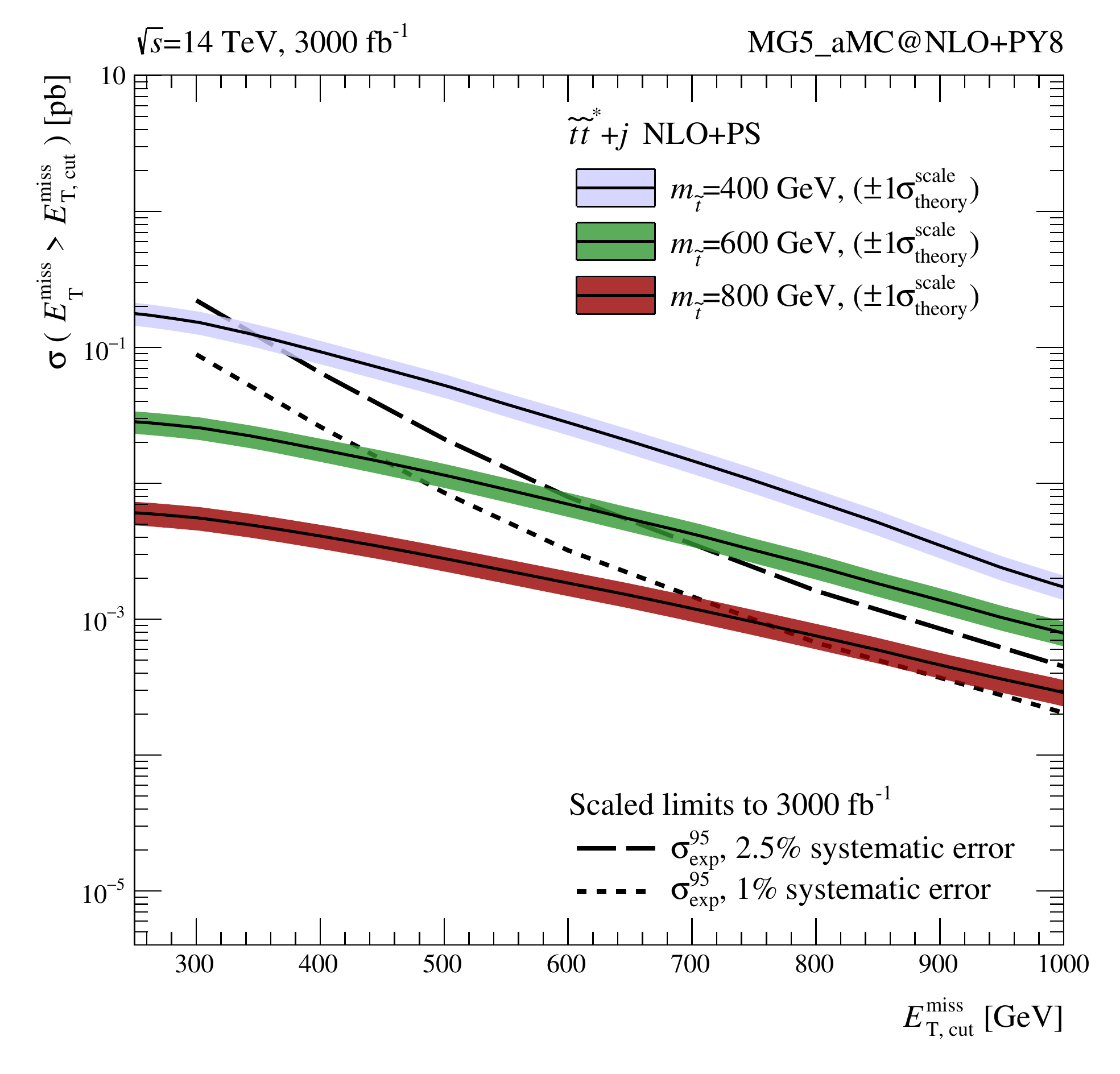}
\end{minipage}
%\begin{center}
%\includegraphics[width=0.327\textwidth]{\main/section3DarkMatterSearches/img/Sec313_tptp_NLO_1j_mVar_varRenoFac_met_13TeV}
%\includegraphics[width=0.327\textwidth]{\main/section3DarkMatterSearches/img/Sec313_gogo_NLO_1j_mVar_varRenoFac_met_13TeV.pdf}
%\includegraphics[width=0.327\textwidth]{\main/section3DarkMatterSearches/img/Sec313_t1t1_NLO_1j_mVar_varRenoFac_met_13TeV.pdf}
%\end{center}
%% 14 TeV, 3\abinv Plot
%\begin{center}
%\includegraphics[width=0.327\textwidth]{\main/section3DarkMatterSearches/img/Sec313_tptp_NLO_1j_mVar_varRenoFac_met_14TeV.pdf}
%\includegraphics[width=0.327\textwidth]{\main/section3DarkMatterSearches/img/Sec313_gogo_NLO_1j_mVar_varRenoFac_met_14TeV.pdf}
%\includegraphics[width=0.327\textwidth]{\main/section3DarkMatterSearches/img/Sec313_t1t1_NLO_1j_mVar_varRenoFac_met_14TeV.pdf}
%\end{center}
\caption{
Top: $\ppQQbarj$ cross section as a function of minimum $\met$ after the experimental selection criteria at $13\TeV$,
for $\Q=\tp,~\go$, and $\stp$, with current $95\%\cl$ limits after $\mathcal{L}=36.1\fbinv$ of data at the $13\TeV$ LHC. 
Bottom: same plots for $14\TeV$ LHC. We also shown the estimated sensitivity with $\mathcal{L}=3\abinv$, assuming $\delta_{\rm Syst.}=2.5\%$ and $1\%$ systematical errors.
}\label{fig:comp13TeV}
\end{figure}
%--------------------------

Were evidence for a new particle $\Q$ established at the LHC, or a successor experiment such as the HE-LHC, it would be crucial to determine the properties of $\Q$, especially its mass, spin, and colour representation,
in order to help understand the nature of DM.
Such a program would typically include investigating various collider observables that can discriminate against possible candidates for $\Q$,
and hence requires that observables are known to sufficiently high precision.
It is the case though that leading order (LO) calculations are poor approximations for QCD processes, even when using sensible scale choices. 
The situation, however, is more hopeful with the advent of general-purpose precision Monte Carlo event generators  
\herwig{}~\cite{Bellm:2015jjp},~\madgraph{5\_aMC@NLO}+\pythia{ 8}~\cite{Alwall:2014hca,Sjostrand:2014zea}, and \sherpa{}~\cite{Gleisberg:2008ta}. 
With automated event generation up to NLO in QCD with parton shower (PS) matching and multijet merging, even for BSM processes~\cite{Degrande:2014vpa}, 
one can now systematically investigate the impact of crucial $\mathcal{O}(\alpha_s)$ corrections on the inclusive monojet process.

We now summarise the key findings of a recent~\cite{Chakraborty:2018kqn} investigation into the prospect for determining the properties 
of a hypothetical heavy resonance $\Q$ associated with DM via the the monojet signature. This includes systematically quantifying theory uncertainties associated with the renormalisation, factorisation, and parton shower scales 
as well as those originating from distribution functions and and multijet merging using state-of-the-art technology.
One finds that in aggregate, the total uncertainties are comparable to differences observed when varying $\Q$ itself. 
However, the precision achievable with next-to-next-to-leading order (NNLO) calculations, where available, can resolve this dilemma.
In light of this, we further emphasise that the HL- and HE-LHC runs possibly resolve different candidates for the new coloured particle. 
For additional details beyond what is provided here, see \citeref{Chakraborty:2018kqn}. \\
%
%{\bf{ \underline{Monojet searches at HL- and HE-LHC}}}: 
The ATLAS and CMS collaborations have studied $\mathcal{L}=36.1\fbinv$ of $\sqrt{s} = 13\TeV$ collision data 
using signatures with significant transverse momentum imbalance and at least one energetic jet~\cite{ATLAS:2017dnw,sirunyan:2017jix}. 
The non-observation of significant deviations from 
SM predictions leads to model-independent $95\%\cl$ upper limits on the production cross section 
of new particles. In \fig{fig:comp13TeV}, we show these experimental limits along with 
NLO+PS-accurate cross section,
and associated scale uncertainty,
for the $\QQbar+j$ process, where we include a hard jet at the matrix-element level, and for $\Q \in \{\stp,\tp,\go\}$, as listed in \tabb{tab:model}. 

\begin{table}[t]
\begin{center}
\hrule
\begin{tabular}{rccllll}
Particle name & 	& Colour Rep.    &   Lorentz Rep. &       Decay  	& UFO Refs. \\ 
\hline
Fermionic Top partner 	& $(\tp)$  	&  $\mathbf{3}$ &   Dirac fermion	&  $\q + \dm$   &~\cite{Fuks:2016ftf,ufo_nlo_database} \\
Top squark  		& $(\stp)$	&  $\mathbf{3}$ &   Complex scalar	&  $t^{*}\dm\to b\qqbar'+\dm$ &~\cite{Degrande:2015vaa,ufo_nlo_database} \\
Gluino 			& $(\go)$       &  $\mathbf{8}$ &   Majorana fermion	&  $\qqbar + \dm$ &~\cite{Degrande:2015vaa,ufo_nlo_database}  \\
 \end{tabular}
 \hrule
\end{center}
\caption{Summary of signal particles, their SU$(3)_c$ and Lorentz representations (Rep.), and decay mode to stable DM candidate $(\dm)$.}
\label{tab:model}
\end{table} 

We find that the lower limits on $\Q$ masses stand at around $m_{\tp}= 400\GeV$ for 
the fermionic top partner and $m_{\go}=600\GeV$ for the gluino,
while no constraint on stop masses is found within the mass range under consideration. It has been observed that 
for high-$p_T$ bins both the systematic and statistical experimental uncertainties play a crucial role. 
Sensitivity is expected to improve at the HL-LHC due to a much larger dataset with better control on uncertainties. 
We calculate the expected upper limits at \sloppy\mbox{HL-LHC} with $\mathcal{L} = 3\abinv$ and $\sqrt{s} = 13\TeV$ by 
rescaling the numbers at $13\TeV$ with two values of total systematic uncertainty, 
namely $\delta_{\rm Sys.}=2.5\%$ and $1\%$. From the scaled limits, we find 
that fermionic top partners with masses $m_{\tp}\lesssim 800\GeV$, gluinos with $m_{\go}\lesssim 1000\GeV$, 
and stops with masses $m_{\stp}\lesssim 600\GeV$ can be excluded at $13\TeV$ with $\mathcal{L}=3\abinv$, 
using the inclusive monojet signature for a compressed mass spectrum. 

\begin{figure}[t]
\centering
\begin{minipage}[t][\minipageheightdp][t]{\minipagewidthdp}
\centering
\includegraphics[width=\figuremaxwidthdp,height=\figuremaxheightdp,keepaspectratio]{\main/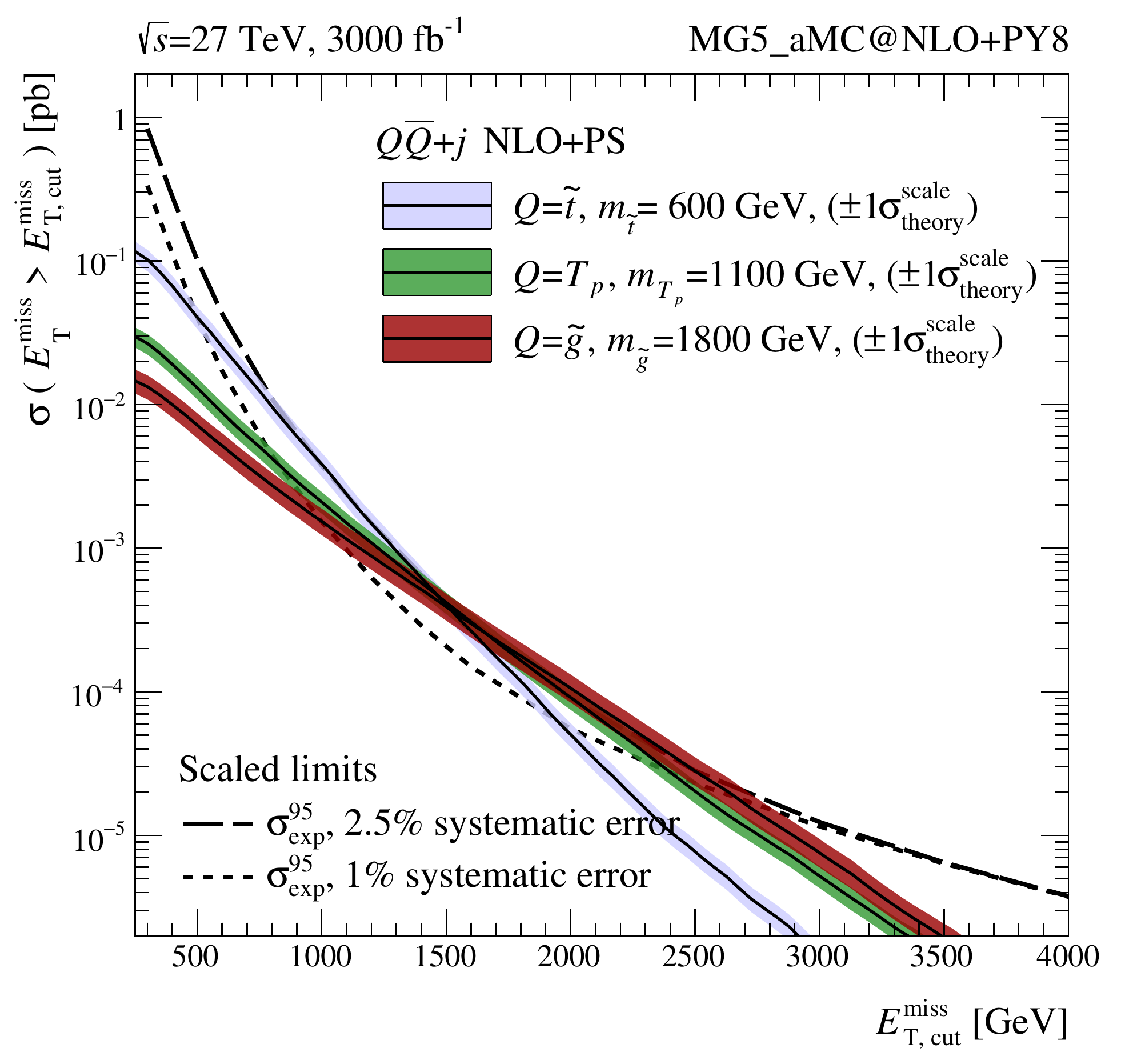}
\end{minipage}
\begin{minipage}[t][\minipageheightdp][t]{\minipagewidthdp}
\centering
\includegraphics[width=\figuremaxwidthdp,height=\figuremaxheightdp,keepaspectratio]{\main/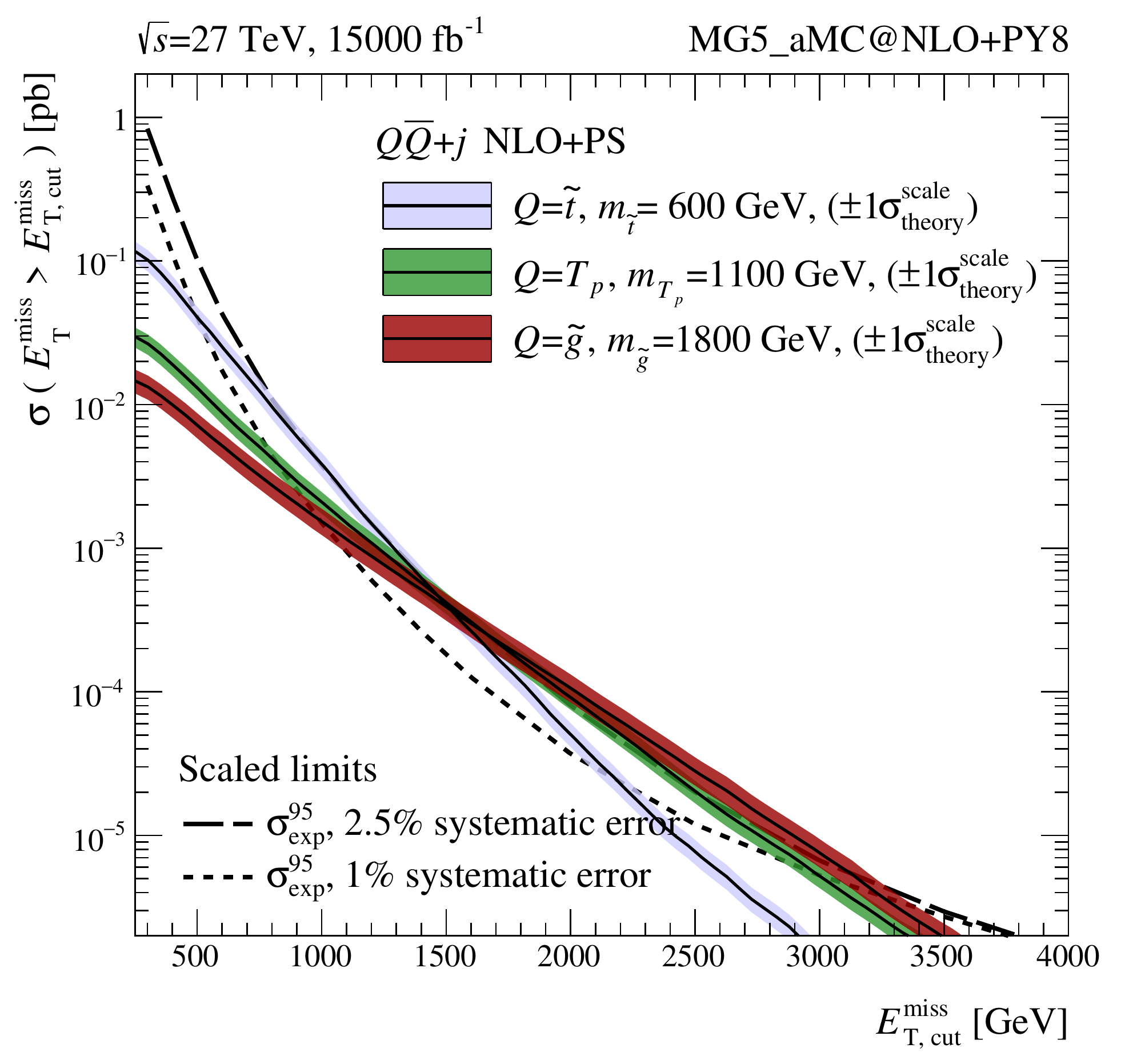}
\end{minipage}
\caption{Same as \fig{fig:comp13TeV} but scaled for $\sqrt{s}=27\TeV$ assuming $\mathcal{L}=3\abinv$ (left) and $15\abinv$ (right).}
\label{fig:comp27TeV}
\end{figure}

Another possibility that can significantly improve the 
sensitivity to heavy coloured particles is increasing the beam energy of the LHC to the proposed 
$\sqrt{s}=27\TeV$~ HE-LHC~\cite{conf27tev}. 
Here, we assume that the SM background is still dominated by $Z+j$ process and then scale the model-independent 
$95\%\cl$ upper limit at $\sqrt{s}=13\TeV$ according to the production cross section ratio. 
In \fig{fig:comp27TeV}, we estimate the expected reach at the $27\TeV$ LHC for 
$p p \rightarrow \Q \Qbar + j$ process. Note, here also we assume that the detector acceptance 
and efficiencies are the same at $13$ and $27\TeV$. 
For comparison, we consider systematic uncertainties of $2.5\%$ and $1\%$, the same considered before. 
We observe that with $\mathcal{L}=3-15\abinv$, 
one can probe the compressed spectra featuring fermionic top partners with masses $m_{\tp}\lesssim1100\GeV$,
gluinos with masses $m_{\go}\lesssim1800\GeV$,
and stops with masses $m_{\stp}\lesssim600\GeV$.

%----------------------------------------------------------------
%--------------------------------------------------------

\begin{figure}[t]
\centering
\begin{minipage}[t][\minipageheightdp][t]{\minipagewidthdp}
\centering
\includegraphics[width=\figuremaxwidthdp,height=\figuremaxheightdp-3mm,keepaspectratio]{\main/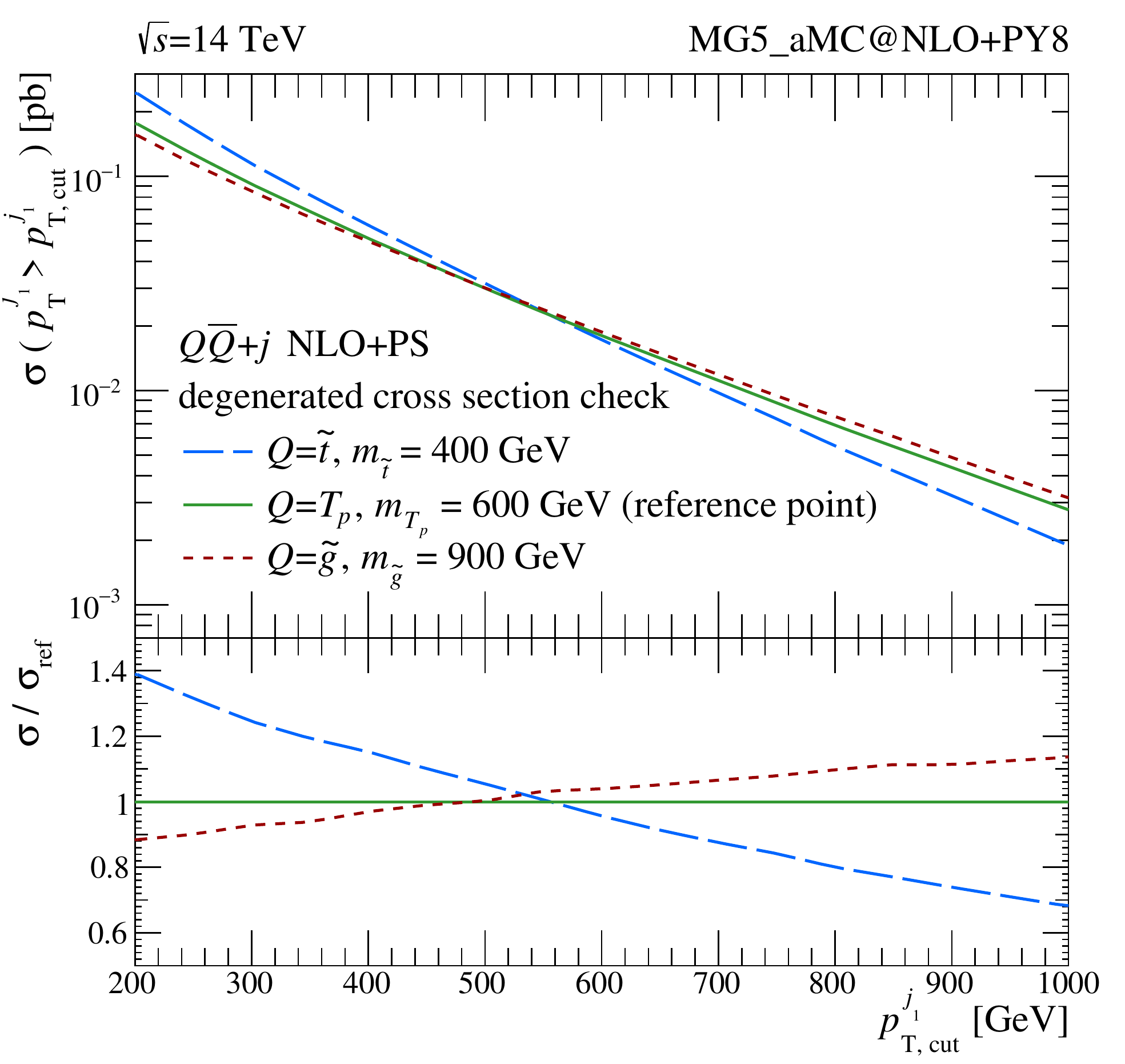}
\end{minipage}
\begin{minipage}[t][\minipageheightdp][t]{\minipagewidthdp}
\centering
\includegraphics[width=\figuremaxwidthdp,height=\figuremaxheightdp,keepaspectratio]{\main/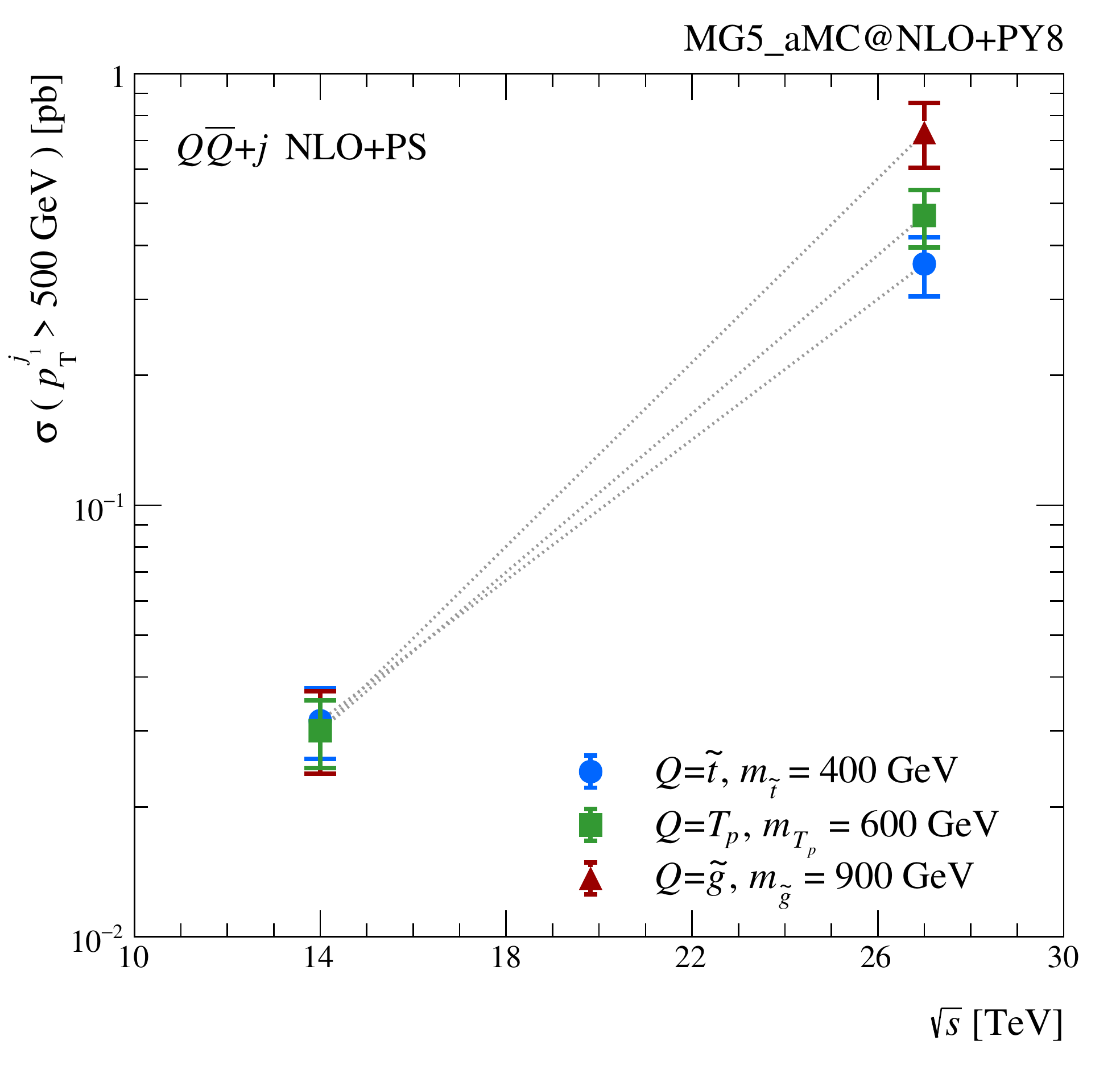}
\end{minipage}
     \caption{
Left: $\ppQQbarj$ cross section at $\sqrt{s}=14\TeV$ 
as a function of jet $\pt$ selection criterion $(\ptjcut)$, for representative $(\Q,m_{\Q})$ combinations.
%Right: The same but normalised to the $(\Q,m_{\Q})=(\tp,600\GeV)$ curve.
Right: $\ppQQbarj$ cross section at $\sqrt{s} =14$ and $27\TeV$, with $\ptjcut = 500\GeV$. 
The error bar reflects the renormalisation and factorisation scale variation.}
\label{fig:ptj-dependence}
\end{figure}

%{\bf{\underline{Theory Uncertainties and Outlook}}}: 
Theoretical uncertainties associated with the monojet signal process are estimated by 
employing the state-of-the-art MC suites \madgraph{5\_aMC@NLO}~and \sherpa{}. 
A single measurement of signal cross section does not constrain the nature of $\Q$ uniquely as different spin and colour hypotheses can lead to identical cross sections if the mass is tuned accordingly. 
For example, a stop of mass $400\GeV$, a fermionic top partner of mass $600\GeV$, 
and a gluino of mass $800\GeV$ have practically the same $\ppQQbarj$ cross section 
for $\ptjcut$ around $500\GeV$ at $\sqrt{s} = 14\TeV$, see \fig{fig:ptj-dependence} (left). The degeneracy, however, can be resolved through additional cross section measurements with larger $\ptjcut$. Cross section measurements at a higher \com~energy can 
also lift this degeneracy, see \fig{fig:ptj-dependence} (right). To distinguish different new physics 
candidates, one, however, needs theoretical uncertainties smaller than $\mathcal{O}(30\%)$ and 
$\mathcal{O}(5\%/100\GeV)$, respectively, on the total cross section normalisations and on 
the change of the cross section for $\sigma(\ppQQbarj)$ as a function of $\ptjcut$. 
Alternatively, one can also consider ratios of cross sections measured at two different $\ptjcut$ and two different energies $\sqrt{s}= 14$ and $27\TeV$. 

%----------------------------------------------------------------
%----------------------------------------------------------------

We find that the monojet signal process under consideration exhibits a residual (factorisation, renormalisation, and shower) scale uncertainty
around $40\%$ at LO and $20\%$ at NLO, 
and is about twice as large as the the inclusive $\pptptp$ cross section due to the presence of additional $\alpha_s$ factors.
Based on available NNLO calculations~\cite{Aliev:2010zk,Czakon:2011xx},
we anticipate such uncertainties can be reduced to the $10\%$ level at NNLO.
As the corrections beyond NLO mainly impact the overall normalisation, cross sections at different $\ptjcut$ possess correlated uncertainties.
Hence, constructing ratios and double-ratios of cross sections at different $\ptjcut$, 
can partially cancel uncertainties and help with the identification of $\Q$.

\subsubsection{\theory{Searching for Electroweakinos in monojet final states at HL- and HE-LHC}}\label{sec:theoryewinosmono}
%%%% NOTE - contribution split in two parts! MD 
%\subsubsection{\theoryC{Searching for Electroweakinos with disappearing tracks}}

\contributors{T. Han, S. Mukhopadhyay, X. Wang}
	
    Among the multitude of possibilities for WIMP dark matter, particles that belong to a multiplet of the SM weak interactions are one of the best representatives. We focus on two representative scenarios for EW DM, namely a wino-like $SU(2)_{L}$ triplet and a Higgsino-like $SU(2)_{L}$ doublet. Such models are often challenging to probe in direct detection experiments due to loop-suppressed scattering cross-sections. Searches at hadron colliders are thus crucial for testing such a scenario, and depending upon the gauge representation, can be complementary to indirect detection probes in different mass windows. Moreover, since the relic abundance of EW DM is uniquely determined by its mass value, $3\TeV$ for wino-like triplet~\cite{Hisano:2003ec, Hisano:2004ds, Hisano:2006nn} and $1\TeV$ for Higgsino-like doublet~\cite{Mizuta:1992qp}, they represent a well-defined target in the  collider search for DM in general. Without large additional corrections from higher-dimensional operators, the mass splitting between the charged and neutral components of the DM ${\rm SU}(2)_{\rm L}$ multiplets is only of the order of a few hundred MeV~\cite{Ibe:2012sx, Thomas:1998wy}. This nearly degenerate spectrum motivates two major search channels at hadron colliders for the EW DM sector, namely, the monojet with missing transverse momentum search and the disappearing charged track search. The first one is reported in this section, the second in \secc{sec:theoryewinosdisappear}.

We present our results~\cite{Han:2018wus} on the future reach of three different scenarios of collider energy and integrated luminosity: HL-LHC, HE-LHC, and FCC-hh/SppC ($100\TeV$, $30\abinv$).
%	\begin{eqnarray}
%	&{\rm HL\text{-}LHC:} \quad &14{\rm~TeV},\quad \intLumi, \nonumber \\
%	&{\rm HE\text{-}LHC:} \quad &27{\rm~TeV},\quad \intlumihelhc, \nonumber \\
%	&{\rm FCC\text{-}hh/SppC:} \quad &100{\rm~TeV},\quad 30\abinv.
%	\end{eqnarray}
We adopt as a definition of significance $S/\sqrt{B+(\Delta_B B)^2 + (\Delta_S S)^2}$ where $S$ and $B$ are the total number of signal and background events, and $\Delta_S, \Delta_B$ refer to the corresponding percentage systematic uncertainties, respectively.
	
\begin{figure}[t]
\centering
\begin{minipage}[t][\minipageheightdp][t]{\minipagewidthdp}
\centering
\includegraphics[width=\figuremaxwidthdp,height=\figuremaxheightdp,keepaspectratio]{\main/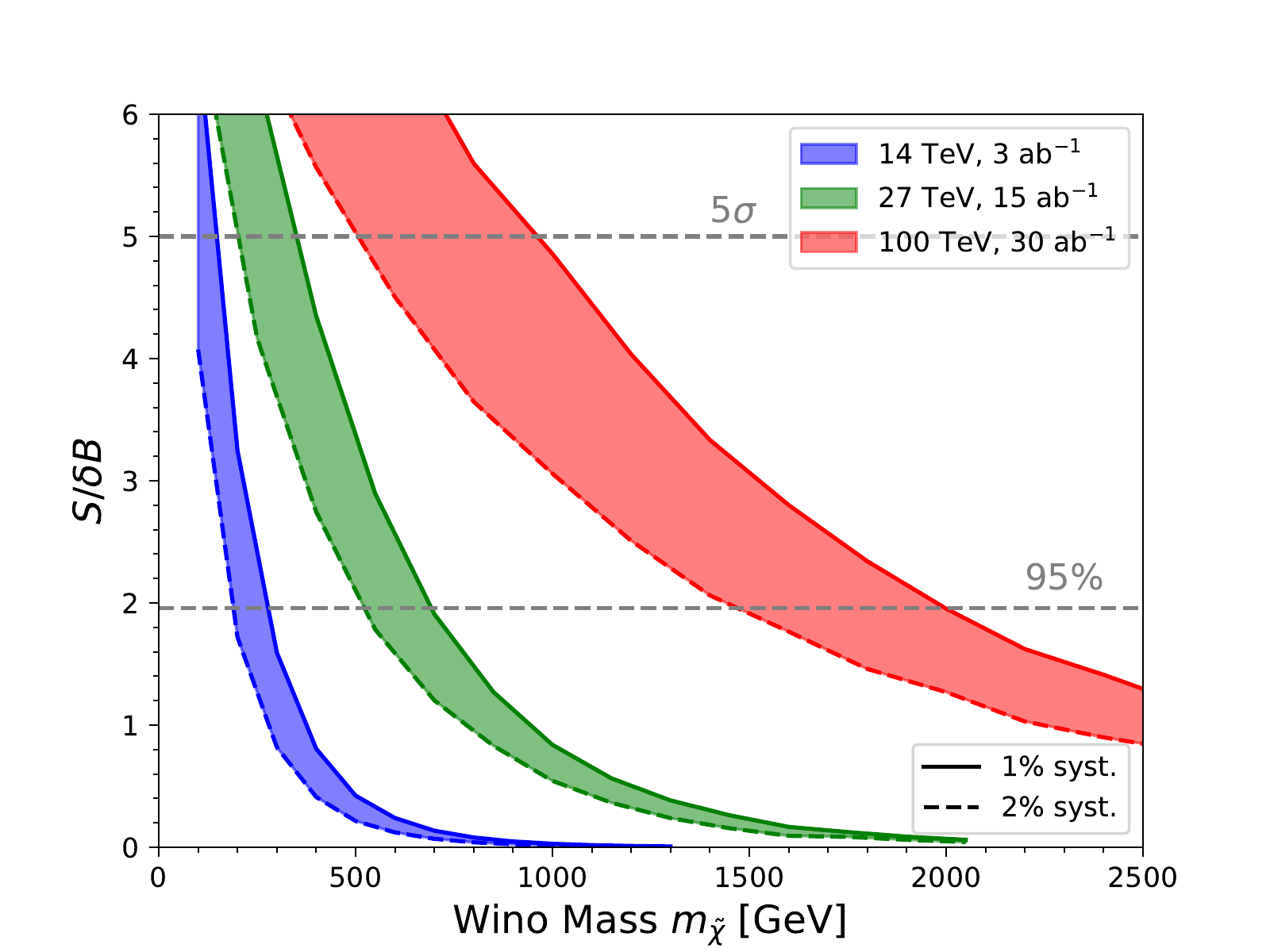}
\end{minipage}
\begin{minipage}[t][\minipageheightdp][t]{\minipagewidthdp}
\centering
\includegraphics[width=\figuremaxwidthdp,height=\figuremaxheightdp,keepaspectratio]{\main/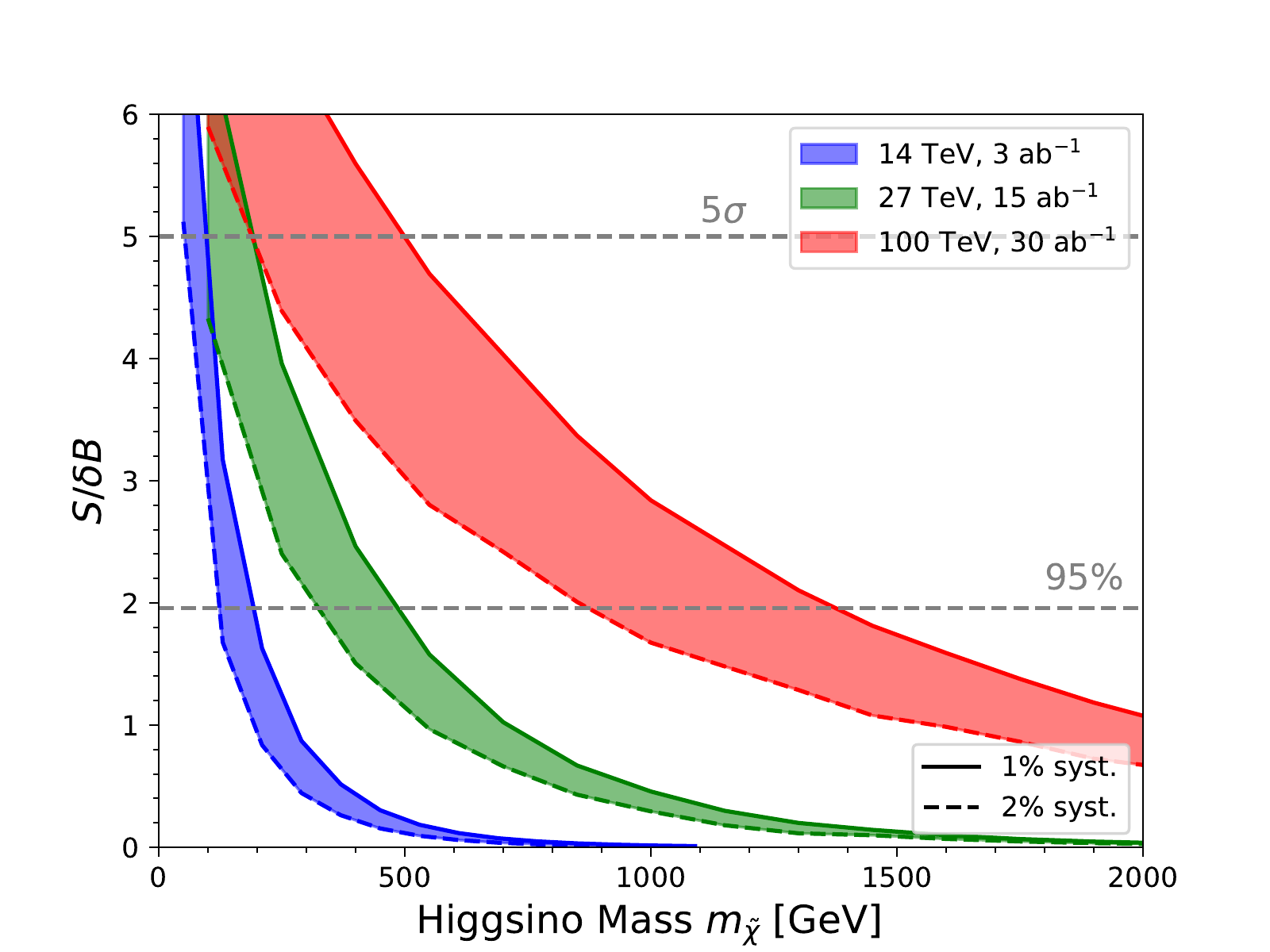}
\end{minipage}
\caption{\small Comparative reach of the HL-LHC, HE-LHC and FCC-hh/SppC options in the monojet channel for wino-like (left) and Higgsino-like (right) DM search. The solid and dashed lines correspond to optimistic values of the systematic uncertainties on the background estimate of $1\%$ and $2\%$ respectively, which might be achievable using data-driven methods with the accumulation of large statistics.}
\label{fig:sec344_reach_monojet}
\end{figure}

The classic monojet and missing transverse momentum search for pair production of a DM particle in association with a hadronic jet originating from initial state radiation is considered. Pair production of both the charged state $\chi^\pm$ and the neutral state $\chi^0$ would contribute to the signal in the monojet search channel, since the charged pions from the charged state $\chi^\pm$ decay are too soft to detect at hadron colliders. Systematic uncertainties $\Delta_B=1-2\%$ and $\Delta_S=10\%$ are assumed. In \fig{fig:sec344_reach_monojet} we compare the reach of the HL-LHC, HE-LHC and FCC-hh/SppC options in the monojet channel for wino-like (left) and Higgsino-like (right) DM search. The solid and dashed lines correspond to systematic uncertainties on the background estimate of $1\%$ and $2\%$ respectively.  
Results are summarised in \tabb{tab:sec344_mono}.
In an optimistic scenario, wino-like DM mass of up to $280, 700$ and $2000\GeV$is expected to be probed at the $95\%\cl$, at the $14,27$ and $100\TeV$ colliders respectively. For the Higgsino-like scenario, these numbers decrease to $200$, $490$ and $1370\GeV$, primarily due to the reduced production cross-section. Clearly, a $27\TeV$ collider can substantially improve the reach by a factor of two or more compared to the HL-LHC, while improvement of another factor of three can be further achieved at the $100\TeV$ collider. 

	\begin{table}[t]
		\def\arraystretch{1.0}\centering
		\begin{tabular}{|c|c|c|}
			\hline
			$95\%\cl$ & Wino  & Higgsino  \\		
			\hline
			$14\TeV$ &  $280\GeV$ & $200\GeV$  \\
			\hline
			$27\TeV$ & $700\GeV$ & $490\GeV$ \\
			\hline
			$100\TeV$ & $2\TeV$ &  $1.4\TeV$ \\
			\hline
		\end{tabular}
		\caption{\small Summary of DM mass reach at $95\%\cl$ for an EW triplet (wino-like) and a doublet (Higgsino-like) representation, at the HL-LHC, HE-LHC and the FCC-hh/SppC colliders, in optimistic scenarios for the background systematics.}
		\label{tab:sec344_mono}
	\end{table}
    
%    \textbf{Acknowledgment:} We would like to thank Matthew Low and Lian-Tao Wang for discussions. This work was supported in part by the U.S.~Department of Energy under grant No.~DE-FG02- 95ER40896 and by the PITT PACC. TH also acknowledges the hospitality of the Aspen Center for Physics, which is supported by National Science Foundation grant PHY-1607611.

%%%%%%%%%%%%%%%
\subsection{Dark Matter and Heavy Flavour}%$t\bar{t}$ / $b\bar{b}$}
\label{sec:DMHF}

When the mediator between the dark sector and the SM is a scalar or pseudoscalar one expects the couplings to the SM to scale with the SM fermion mass.  Thus, a natural place to look for DM production is in association with pairs of top or bottom quarks, see \secc{sec:atlasDMheavypairs} and \secc{sec:theoryDMtop}.  Alternatively, a neutral vector mediator with flavour-changing interactions can produce DM in association with a single top, see \secc{sec:atlasDMsingletop}. Finally, scalar mediators may be searched for directly in four top final states, as shown in \secc{sec:atlas4top}. 

%\subsubsection{\cms{Prospects for associated production of dark matter and top quark pairs using dilepton angular correlations at the HL-LHC}}

%{\bf Author(s): K. Beernaert, A. Grohsjean et al. CMS}
% Note: KU: Not going to make December.

\subsubsection{\atlas{Associated production of dark matter and heavy flavour quarks at HL-LHC}}
\label{sec:atlasDMheavypairs}
\contributors{M. Rimoldi, E. McDonald, F. Meloni, P. Pani, F. Ungaro, ATLAS}

The prospects of a search for dark matter produced in association with heavy flavour (bottom or top) quarks at the HL-LHC are presented in this section~\cite{ATL-PHYS-PUB-2018-036}. The study therefore focuses on two simplified models, defined by either a scalar, $\phi$, or pseudoscalar, $a$, mediator. 
In both cases, the mediating particle is taken to be colour-neutral and the dark matter candidate is assumed to be a weakly interacting Dirac 
fermion, $\chi$, uncharged under the SM. 

The $\chi\bar{\chi}$ production in association with top-quarks is expected to dominate at the HL-LHC. Two signatures featuring 
top quarks in the final state are therefore considered. The first signature, denoted $DM\ensuremath{+t\bar{t}}$, is characterised by two tops decaying 
di-leptonically. The second signature, $DM\ensuremath{+W\bar{t}}$, involves a single top produced in tandem with a $W$-boson, both of which decay leptonically.
Dark matter production in association with $b$-quarks is also considered in this study, as it is relevant if the coupling to up-type 
quarks is suppressed.  The $DM\ensuremath{+b\bar{b}}$ final state is equivalently well motivated as an avenue 
for probing the parameter space of two-Higgs doublet models. In the 2HDM$+a$ model for example, 
the rate for $pp\rightarrow b\bar{b}+a$ is enhanced by the ratio of the Higgs doublet vacuum 
expectation values, $\tan\beta$, if a Yukawa sector of type-II is realised.  
A straightforward recasting of exclusion limits on the simplified pseudoscalar mediator model can then be used to extract constraints on $\tan\beta$.  

A search targeting the $DM\ensuremath{+b\bar{b}}$ and $DM\ensuremath{+t\bar{t}}$ signatures was performed at the LHC using $36.1\fbinv$ of data collected 
in 2015 and 2016 at a centre of mass energy of $13\TeV$. This study presents the prospects for further constraining 
these models with HL-LHC data and is divided into two independent analyses. 

\begin{figure}[t]
\centering
\begin{minipage}[t][\minipageheightdp][t]{\minipagewidthdp}
\centering
\includegraphics[width=\figuremaxwidthdp,height=\figuremaxheightdp,keepaspectratio]{\main/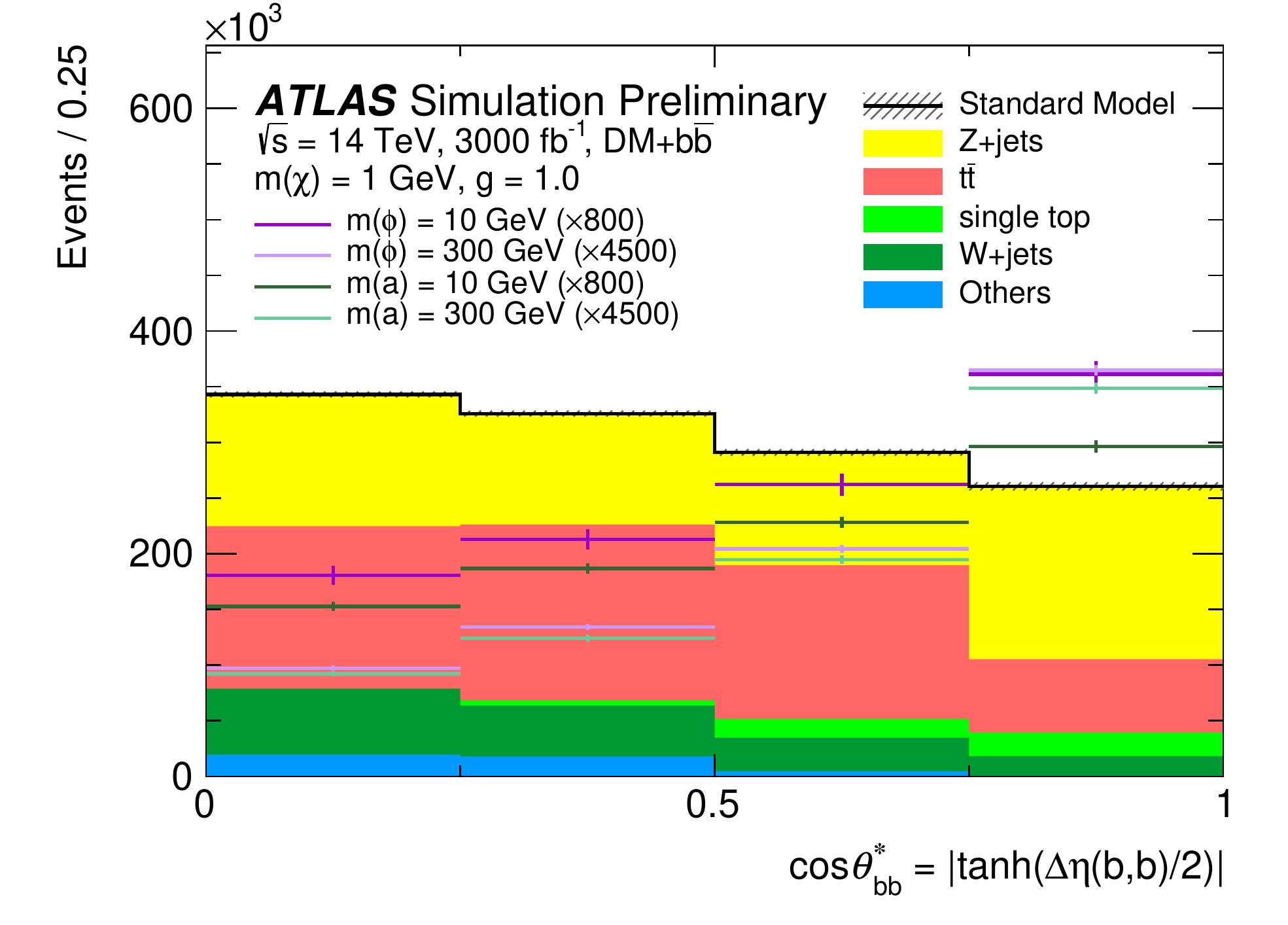}
\end{minipage}
\begin{minipage}[t][\minipageheightdp][t]{\minipagewidthdp}
\centering
\includegraphics[width=\figuremaxwidthdp,height=\figuremaxheightdp-3mm,keepaspectratio]{\main/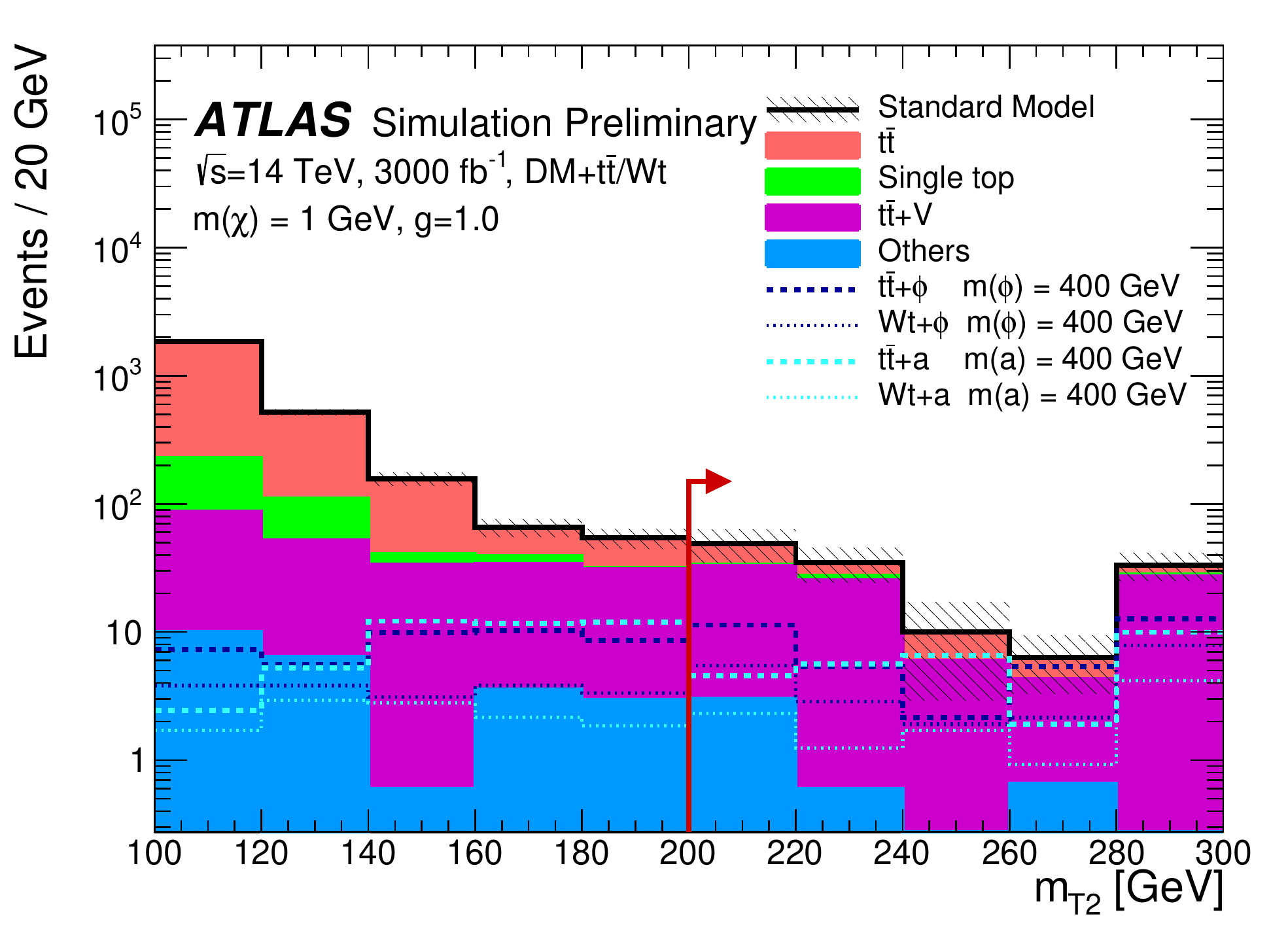}
\end{minipage}
\caption{Distributions of the main discriminants used for the $DM\ensuremath{+b\bar{b}}$ and $DM\ensuremath{+t\bar{t}}$ searches: $\cos\theta_{bb}^{*}$ (left) and $\ensuremath{m_{\mbox{\tiny T2}}}$ (right). For the $\cos\theta_{bb}^{*}$ distribution events are required to have $\met>210\GeV$, no leptons, 2 or 3 jets, and at least two $b$-jets. For the $\ensuremath{m_{\mbox{\tiny T2}}}$ distribution events must satisfy the corresponding signal region criteria except for that on the variable shown.}
\label{fig:distribution_DMHF_atlas}
\end{figure}

\textbf{Signatures with $b$-quarks and $\met$} 
To isolate the event topology of the $DM\ensuremath{+b\bar{b}}$ final state, events are required to have at least two $b$-tagged jets. 
The contribution from SM background processes is suppressed via the application of selection criteria based on that of the $13\TeV$ analysis 
and updated to align with HL-LHC design considerations.
To reduce the contribution from leptonic and semi-leptonic $t\bar{t}$ decays and from leptonic decays of $W$ and $Z$ bosons, 
events containing at least one baseline lepton are vetoed. 
A further requirement of no more than 2 or 3 jets is imposed 
in order to control the large background from hadronic $t\bar{t}$ decays.
The main background from $Z(\rightarrow\nu\bar{\nu})$+jets events is reduced by cutting on variables which exploit the difference in spin between the scalar and pseudoscalar particles and the $Z$ boson. 
These variables make use of the pseudorapidity and azimuthal separations between jets, $b$-jets, and the missing transverse momentum. Among them,  
the hyperbolic tangent of the pseudorapidity separation between the leading and sub-leading $b$-jet, $\Delta\eta(b,b)=\eta(b_{1})-\eta(b_{2})$,  
%\begin{equation*}
%\cos\theta_{bb}^{*} = \left|\tanh\left(\frac{|\Delta\eta_{bb}|}{2}\right)\right|
%\end{equation*}
is used. 
This variable, referred to as $\cos\theta_{bb}^{*}$, is expected to yield a reasonably flat distribution for $b$-jets produced in association with vector particles. For $b$-jets accompanying the production of a heavy scalar or pseudoscalar mediator however, $\cos\theta_{bb}^{*}$ is expected to peak around 1. 
Owing to this shape difference, the $\cos\theta_{bb}^{*}$ variable provides the best discrimination between signal and background events in the $DM\ensuremath{+b\bar{b}}$ channel. The signal region for this search is therefore defined by four equal-width exclusive bins in $\cos\theta_{bb}^{*}$ as shown in~\fig{fig:distribution_DMHF_atlas}, reflecting the configuration used in Run-2. 
Separate selections are derived for \sloppy\mbox{$m(\phi/a)<100\GeV$} 
and $m(\phi/a)\geq100\GeV$ to further enhance the difference in shape, which can depend strongly on the mass of the mediating particle. The resulting signal regions are denoted by $SR\ensuremath{_{b\text{,low}}}$ and $SR\ensuremath{_{b\text{,high}}}$ respectively. For more details, see \citeref{ATL-PHYS-PUB-2018-036}.

\begin{figure}[t]
\centering
\begin{minipage}[t][\minipageheightdp][t]{\minipagewidthdp}
\centering
\includegraphics[width=\figuremaxwidthdp,height=\figuremaxheightdp,keepaspectratio]{\main/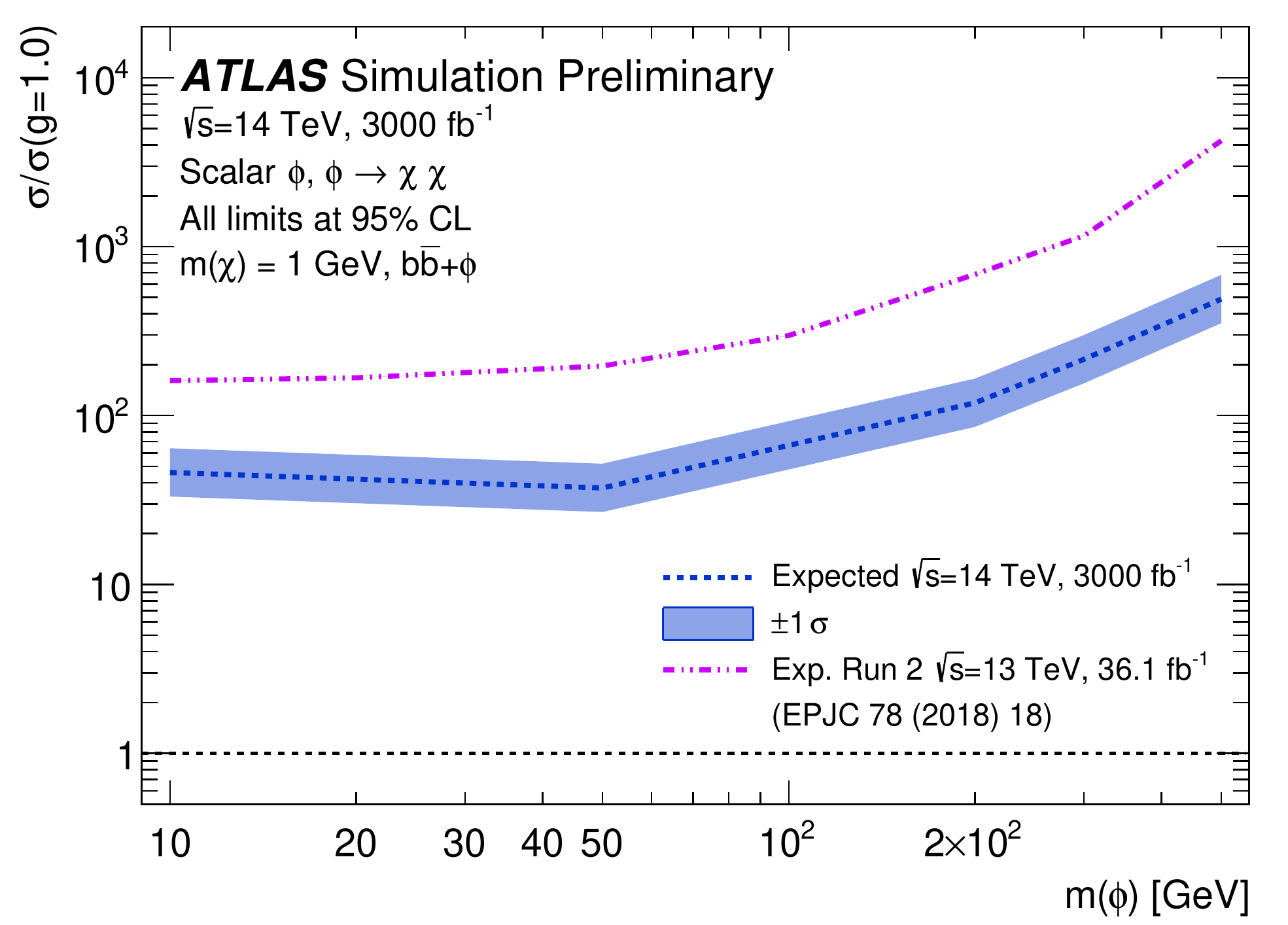}
\end{minipage}
\begin{minipage}[t][\minipageheightdp][t]{\minipagewidthdp}
\centering
\includegraphics[width=\figuremaxwidthdp,height=\figuremaxheightdp,keepaspectratio]{\main/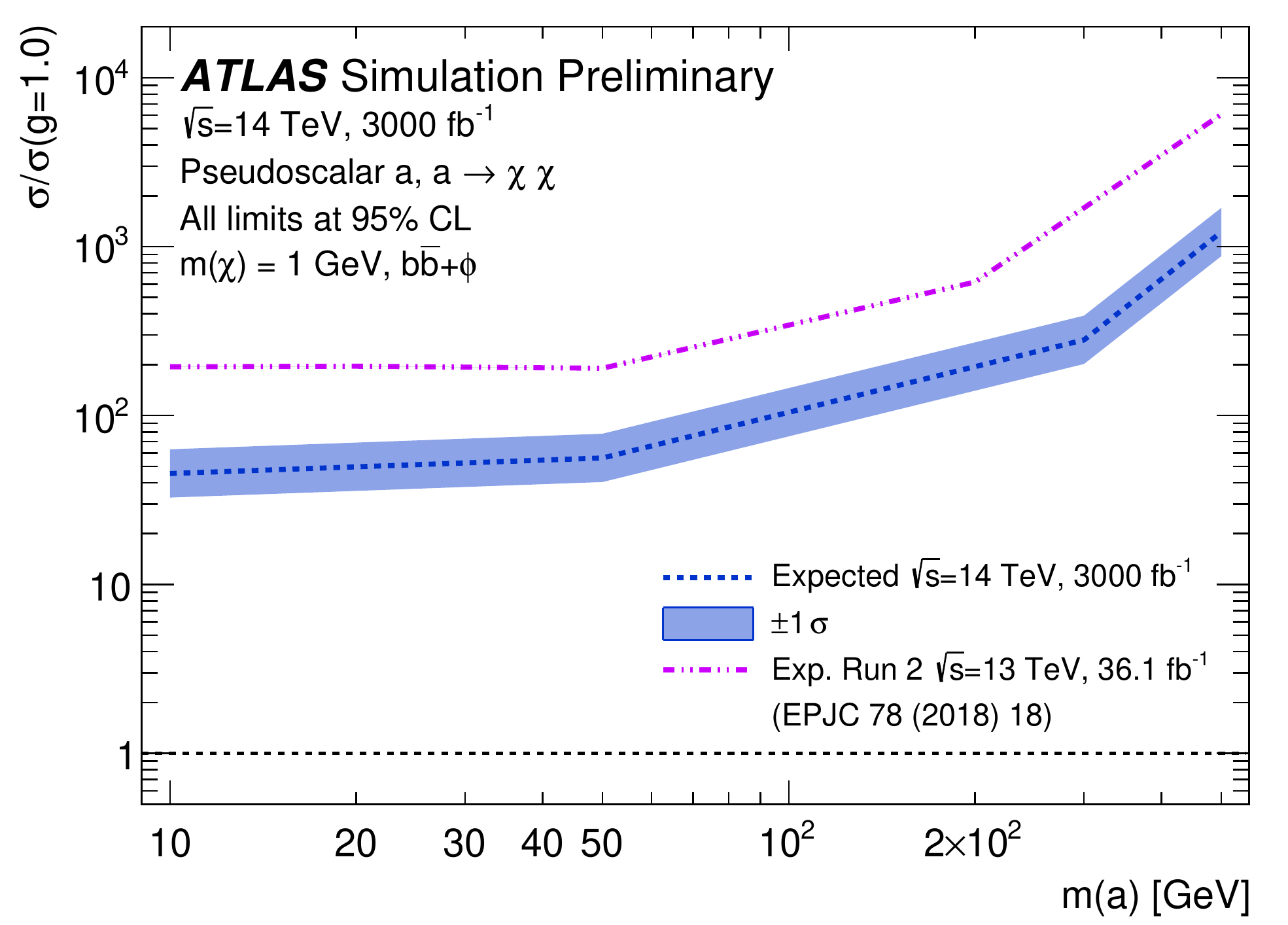}
\end{minipage}\\
\begin{minipage}[t][\minipageheightdp][t]{\minipagewidthdp}
\centering
\includegraphics[width=\figuremaxwidthdp,height=\figuremaxheightdp,keepaspectratio]{\main/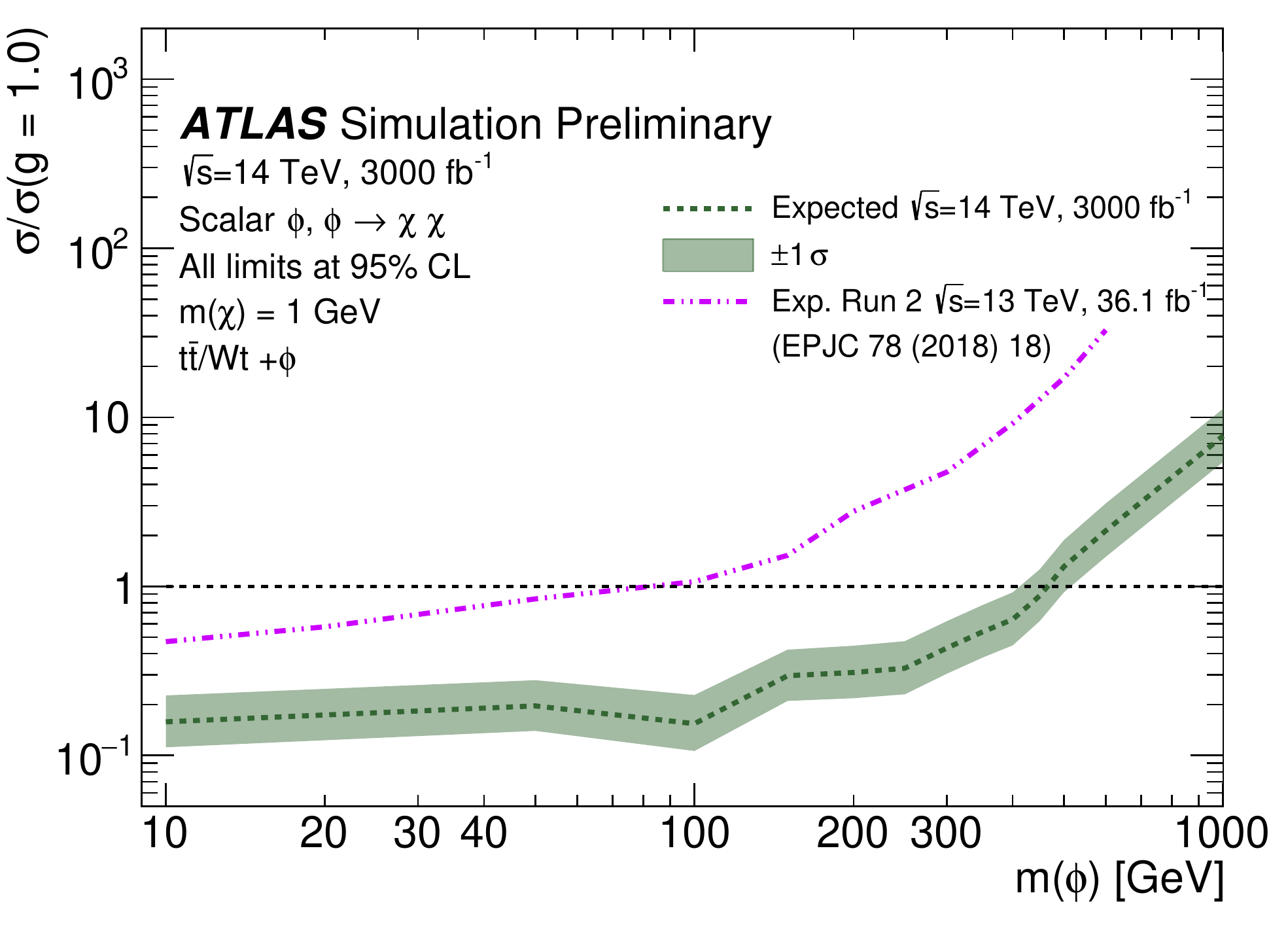}
\end{minipage}
\begin{minipage}[t][\minipageheightdp][t]{\minipagewidthdp}
\centering
\includegraphics[width=\figuremaxwidthdp,height=\figuremaxheightdp,keepaspectratio]{\main/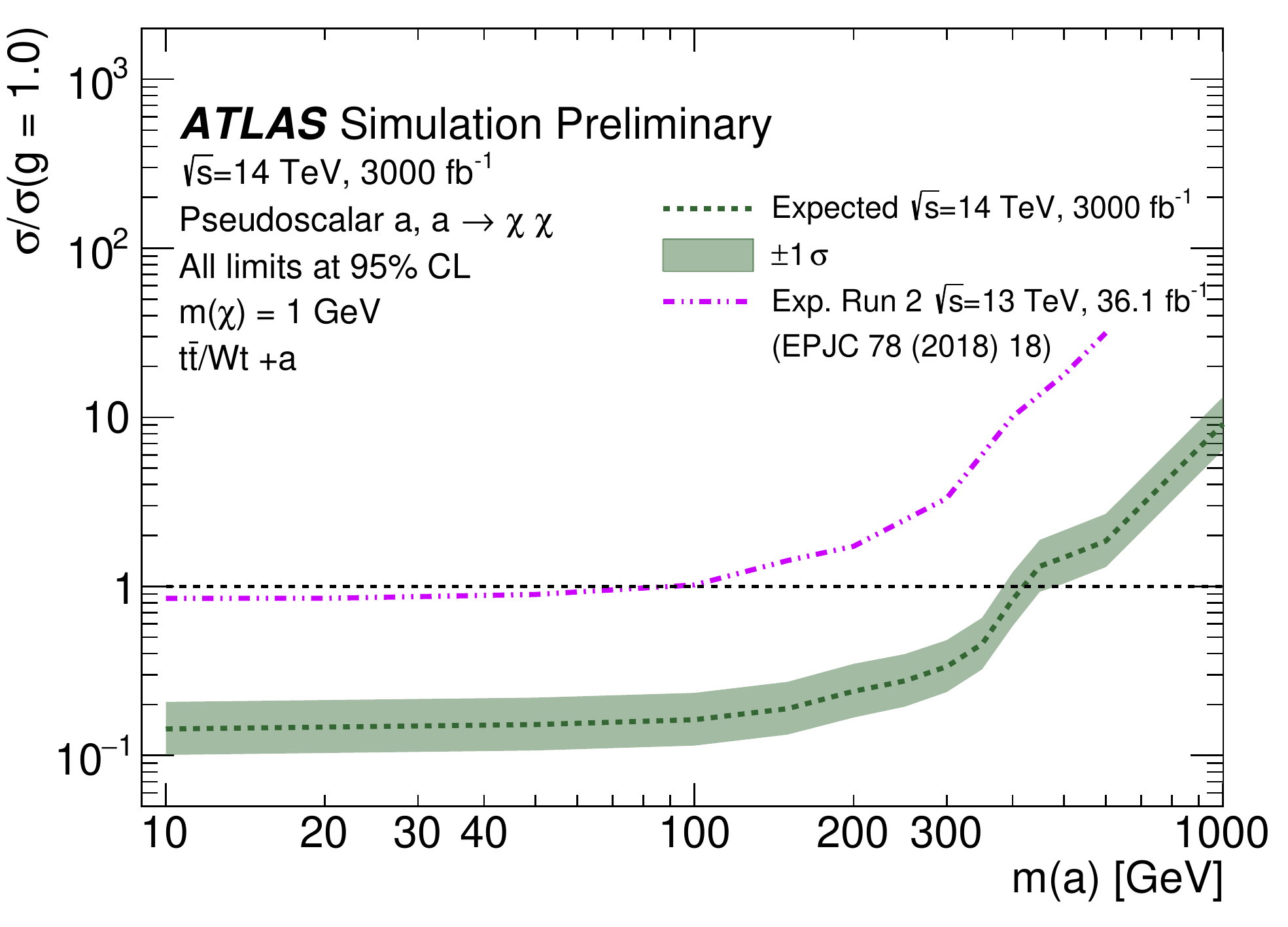}
\end{minipage}
\caption{
Exclusion limits for the production of a colour-neutral mediator in association with bottom quarks (top) or with top quarks (bottom) in case of scalar (left) and pseudoscalar (right) mediator decaying 
to a pair of dark matter particles with mass $1\GeV$. Also shown for comparison is the expected limit from the current analysis~\cite{Aaboud:2017rzf}.
\label{fig:dmbbtt_limits_DMHF_atlas}}
\end{figure}

\textbf{Signatures with top quarks and $\met$}
A single signal region, denoted $SR\ensuremath{_{2\ell}}$, is used for the search targeting DM production in association with one or two top quarks. 
Events are required to have exactly two leptons (electrons or muons), possessing the same or different flavour and opposite electric charge. To reduce the $t\bar{t}$ background, the lepton pair must have an invariant mass larger than $100\GeV$.
Furthermore, candidate signal events are required to have at least one identified $b$-jet.

Different discriminators and kinematic variables are used to further separate the $\ttbar+ \phi/a$ and $Wt+ \phi/a$ signal 
from the SM background. These variables include the lepton-based transverse mass $\ensuremath{m_{\mbox{\tiny T2}}}$, the distribution for which is shown in~\fig{fig:distribution_DMHF_atlas} for events passing all of the SR requirements except that on $\ensuremath{m_{\mbox{\tiny T2}}}$.
For the calculation of exclusion limits, the $\ensuremath{m_{\mbox{\tiny T2}}}$ distribution is divided into five equal-width ($20\GeV$) exclusive bins.  

\textbf{Results}
For both $SR\ensuremath{_{b\text{,low}}}$ and $SR\ensuremath{_{b\text{,high}}}$, the main background consists of Z+jets 
events followed by hadronic decays of $\ttbar$. A significant contribution also comes from single top quark processes 
and events featuring a $W$-boson produced in association with jets. 
In $SR\ensuremath{_{2\ell}}$, the dominant background consists of di-leptonic decays of $\ttbar$~ and $\ttbar Z$ with $Z\rightarrow\nu\nu$. 
Systematic uncertainties include theory modelling and experimental uncertainties related to, for example, the Jet Energy Scale and $b$-jet mis-identification. The total systematic uncertainty on the SM background is $14\%$ for $SR\ensuremath{_{b\text{,low}}}$/$SR\ensuremath{_{b\text{,high}}}$ and $13\%$ for $SR\ensuremath{_{2\ell}}$.

\begin{figure}[!t]
\centering
\begin{minipage}[t][\minipageheightsp][t]{\minipagewidthsp}
\centering
\includegraphics[width=\figuremaxwidthsp,height=\figuremaxheightsp,keepaspectratio]{\main/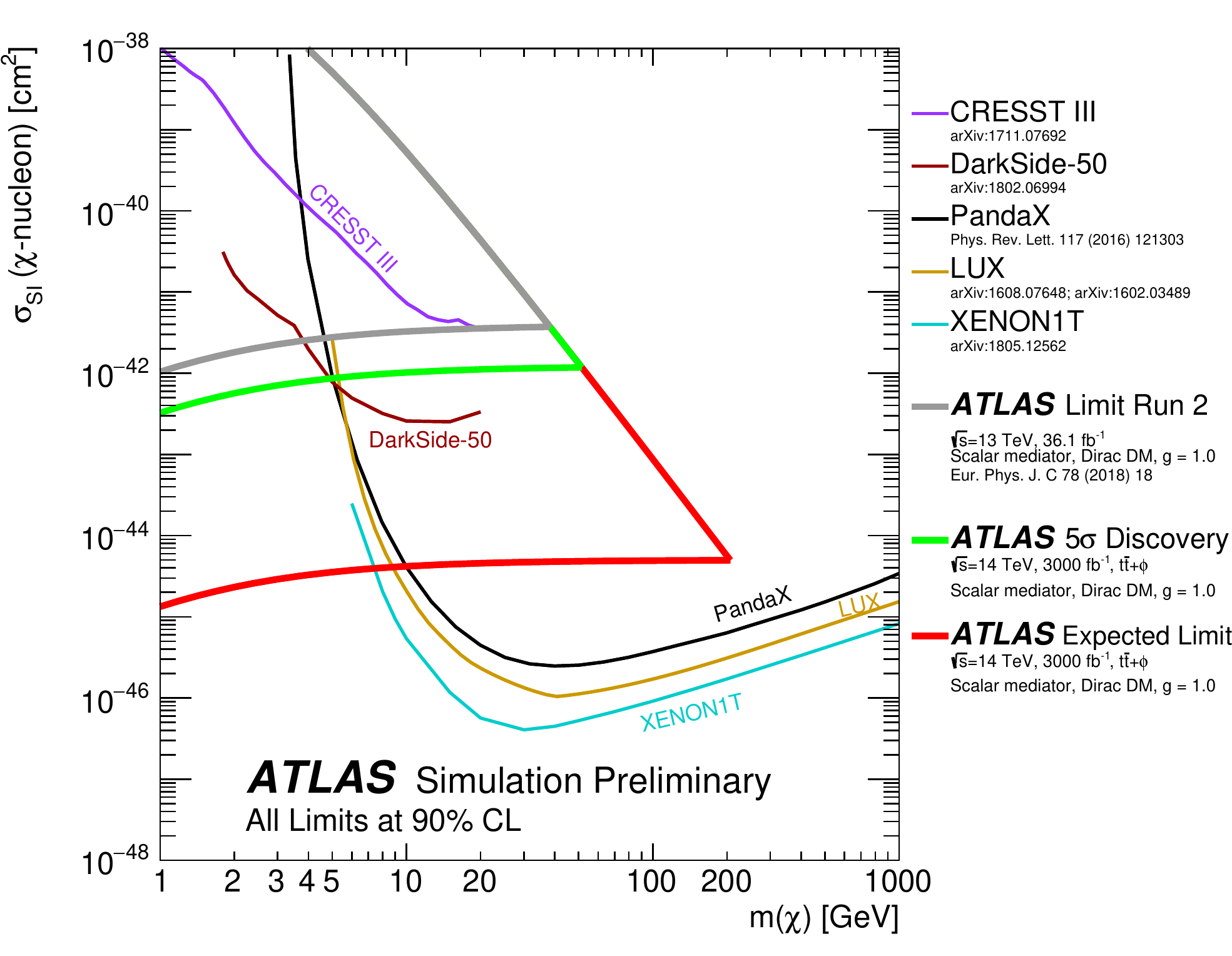}
\end{minipage}
\caption{\label{fig:XsecDMscalarTT} Comparison of the $90\%\cl$ limits on the spin-independent DM-nucleon cross-section as a function of DM mass between these results and the direct-detection experiments, in the context of the colour-neutral simplified model with scalar mediator.
The green contour indicates the $5\sigma$ discovery potential at HL--LHC.
The lower horizontal line of the DM--nucleon scattering cross-section for the red (green) contour corresponds to value of the cross section for $m(\phi)=430\GeV$ ($m(\phi)=105\GeV$).
The grey contour indicates the exclusion derived from the observed limits for Run-2 taken from \citeref{Aaboud:2017rzf}. The results are compared with limits from direct detection experiments.}
\label{fig:DMScatteringxsecPUBNOTE}
\end{figure}

%\pf{I don't understand how the bounds in the text are extracted from the plots.  Also, the results do not seem to correspond to``unitary couplings".  Finally, would be good if scalar and pseudoscalar could be shown for both $b\bar{b}$ and $t\bar{t}$.}

Exclusion limits are derived at $95\%\cl$ for mediator masses in the range $10-500\GeV$ assuming a DM mass of $1\GeV$ and a coupling ($g$) of $1.0$. The limits are shown in \fig{fig:dmbbtt_limits_DMHF_atlas} for $\phi/a\rightarrow\chi\bar{\chi}$ production in association with either bottom quarks or top quarks for $\mathcal{L}=3\abinv$ at $\sqrt{s}=14\TeV$. Also shown for comparison are the corresponding limits at $13\TeV$ with $36.1\fbinv$ taken from the previous Run-2 analysis. 

For $\ensuremath{\phi/a+b\bar{b}}$, the exclusion potential at the HL-LHC is found to improve by a factor of $\sim 3-8.7$ with respect to Run-2. In the context of the 2HDM$+a$ model with $m(A)\gg m(a)$, $\sin\theta=0.35$ and $y_{\chi}=1$, the HL-LHC limits translate to an approximate upper bound on $\tan\beta$ ranging from $\sim 19$ for $m(a) = 10\GeV$ to $\sim 100$ for $m(a) = 500\GeV$, significantly extending the current phase space coverage. In final states with one or two leptonically-decaying top quarks, 
the mass range for which a colour-neutral scalar mediator is excluded extends from $80\GeV$ to $405\GeV$. Similarly, exclusion of pseudoscalar masses up to 385~GeV is expected. In the case of the scalar mediator model, this represents a factor of 5 improvement with respect to the $36\fbinv$ $13\TeV$ results in the same channel. An additional improvement of approximately 3 is possible when considering a statistical combination of all relevant top decay channels~\cite{Sirunyan:2018dub}, which is not explored for the HL-LHC in this work.

For each DM and mediator mass pair, the exclusion limit on the cross-section for producing colour-neutral scalar mediator particles can be converted into a limit on the cross-section for spin-independent DM-nucleon scattering with the procedure described in \citeref{Boveia:2016mrp}. Limits on the $\ttbar+\phi$ model at $90\%\cl$, corresponding to exclusion of mediator masses up to $m(\phi)=430\GeV$, are used for this purpose.
\fig{fig:DMScatteringxsecPUBNOTE} shows the resulting constraints in the plane defined by the DM mass and the scattering cross-section. The maximum value of the DM-nucleon scattering cross-section depicted in the plot corresponds to the value of the cross section for a mediator mass of 10 GeV. The exclusion limits at $90\%\cl$ are shown in red and the $5\sigma$ discovery potential is illustrated in green. The lower horizontal line in the green (red) contour corresponds to the value of the cross section for $m(\phi)=105\GeV$ ($m(\phi)=430\GeV$). Overlaid for comparison are the most stringent limits to date from several DM direct detection experiments.

\subsubsection{\theoryC{Production of dark matter in association with top quarks at HL- and HE-LHC}}
\label{sec:theoryDMtop}
\contributors{U. Haisch, P. Pani, and G. Polesello}

The prospects of the HL-LHC and the HE-LHC to search for DM production in association with top-quark pairs~($t \bar t + E_T^{\rm miss}$) and in single-top quark events~($t X +  E_T^{\rm miss}$) are investigated. Our  sensitivity studies are based on the analysis strategies presented in \citeref{Haisch:2016gry,Pani:2017qyd,Brooijmans:2018xbu}. In the case of the $t \bar t + E_T^{\rm miss}$ signal, the two-lepton final state is considered.  Since the selections employed in \citeref{Haisch:2016gry} turn out to lead to the best performance also at the HL-LHC and the HE-LHC, the selections in our $t \bar t + E_T^{\rm miss}$ sensitivity study were not changed with respect to that used in the earlier analysis. In the case of the $t X + E_T^{\rm miss}$ signature, both the two-lepton and one-lepton final state is studied. Since modifying the selections did not notably increase the sensitivity, the selections of the dilepton search were kept identical to the ones used in \citeref{Pani:2017qyd}. The single-lepton final state selections employed in \citeref{Brooijmans:2018xbu} were instead reoptimised in our sensitivity study to take full advantage of the increased data set expected at future high-luminosity and high-energy LHC runs. 

\begin{figure}[t]
\centering
\begin{minipage}[t][\minipageheightdp][t]{\minipagewidthdp}
\centering
\includegraphics[width=\figuremaxwidthdp,height=\figuremaxheightdp,keepaspectratio]{\main/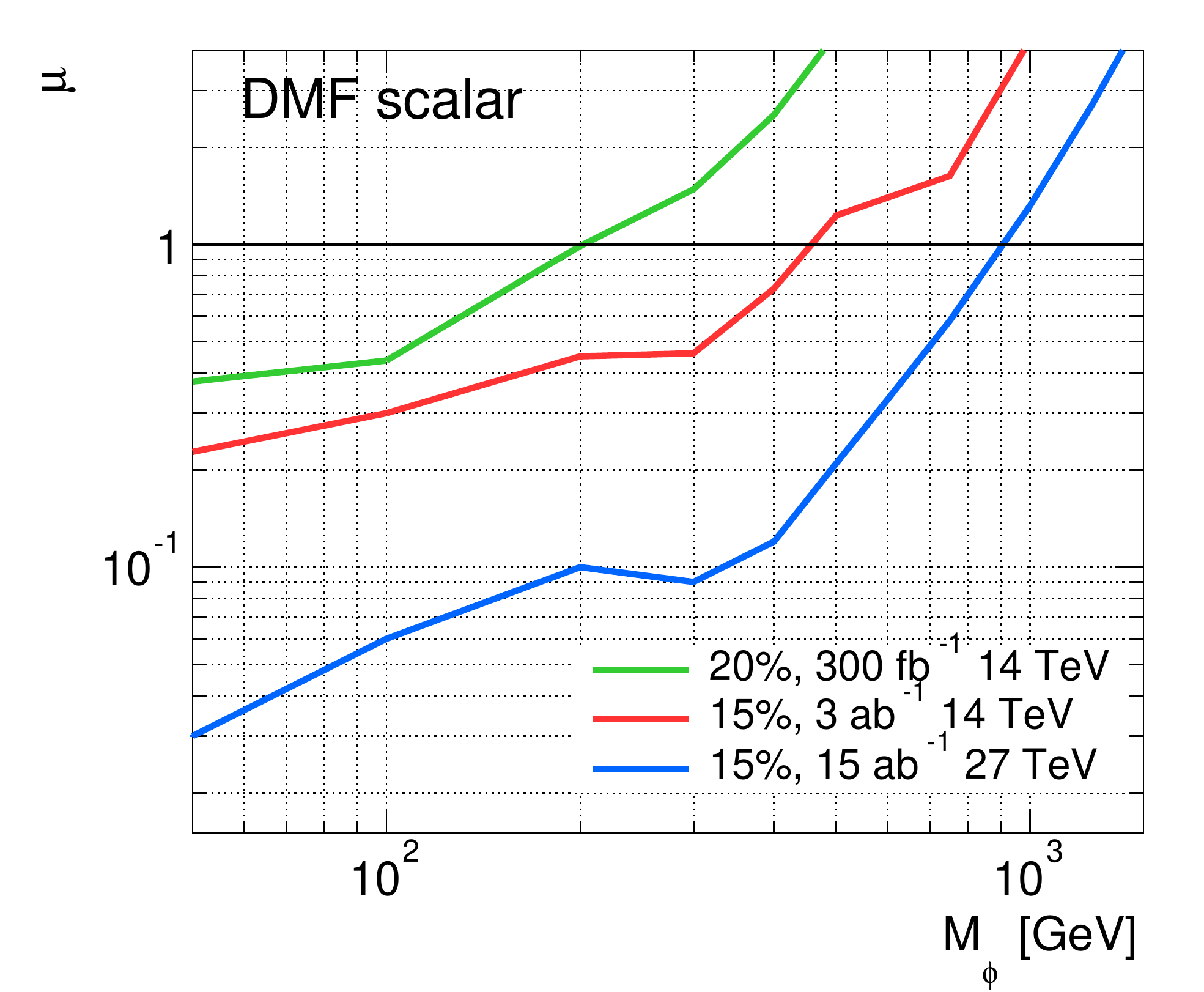}
\end{minipage}
\begin{minipage}[t][\minipageheightdp][t]{\minipagewidthdp}
\centering
\includegraphics[width=\figuremaxwidthdp,height=\figuremaxheightdp,keepaspectratio]{\main/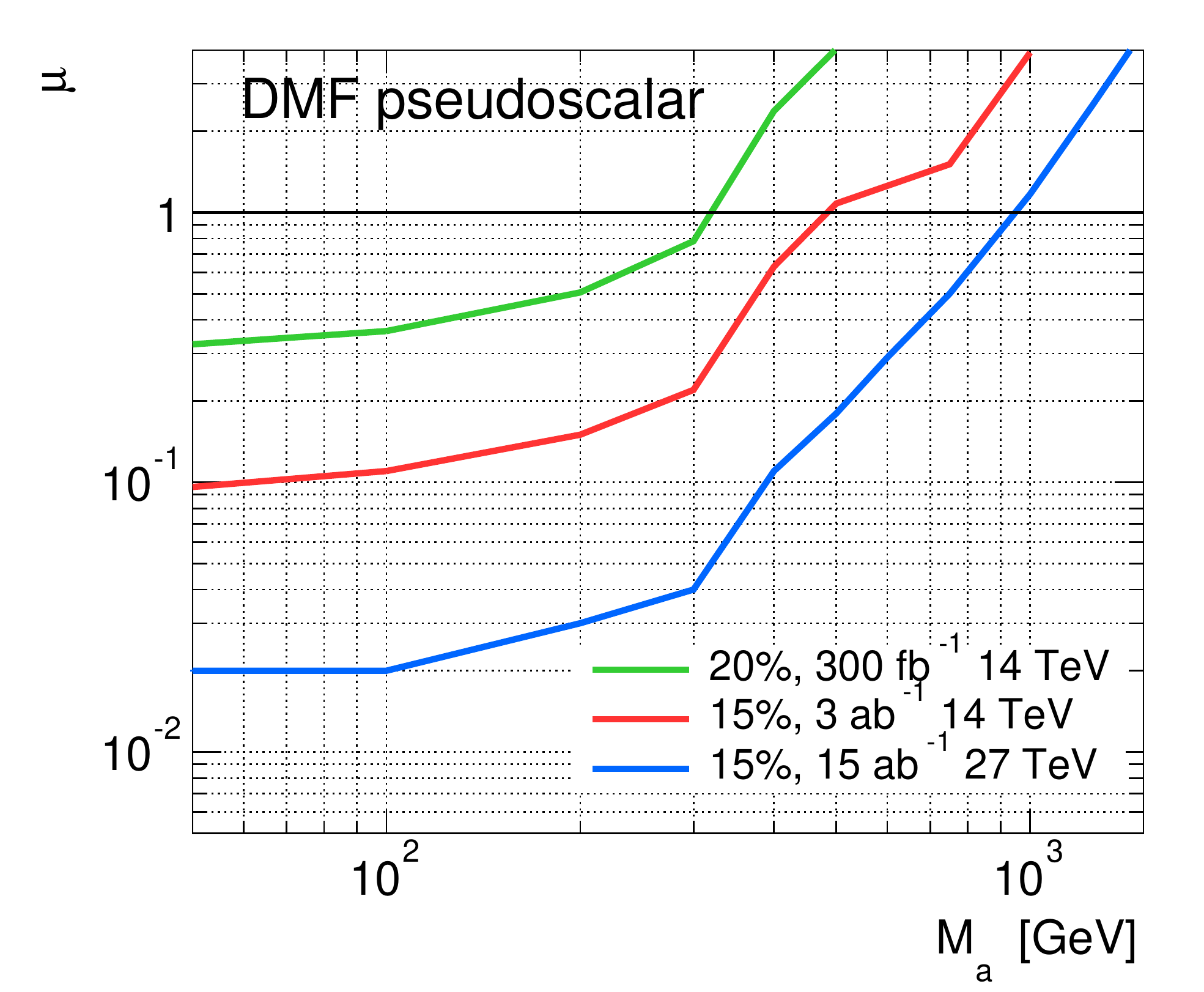}
\end{minipage}
\caption{Values of the signal strength $\mu$ that can be excluded at $95\%\cl$ as a function of the mass for DMF scalar~(left) and pseudoscalar~(right) mediators. The reach with $300\fbinv$ (LHC~Run-3)  and $3\abinv$ (HL-LHC) of $\sqrt{s} = 14\TeV$ data is given for a 5-bin shape fit with 20\%~(green curves) and 15\%~(red curves) errors. A~hypothetical shape-fit scenario based on $15\abinv$ of  $\sqrt{s} = 27\TeV$  of data  (HE-LHC) and 15\% systematics is also shown~(blue curves).}
\label{fig:ttMETDMF}
\end{figure}

{\bf Sensitivity study of the $t \bar t + E_T^{\rm miss}$ signature:} Given the presence of a sizeable irreducible background surviving all the selections, the experimental sensitivity of future $t \bar t + E_T^{\rm miss}$  searches will be largely determined by the systematic uncertainty on the estimate of the SM backgrounds. This uncertainty  has  two main sources: first, uncertainties on the parameters of the detector performance such as the energy scale for hadronic jets and the identification efficiency for leptons, and second, uncertainties plaguing the modelling of SM processes. Depending on the process and on the kinematic selection, the total uncertainty can vary between a few percent and a few tens of percent. The present analysis does not select extreme kinematic configurations for the dominant $t \bar t Z$ background, and it thus should be possible to control the experimental systematics at the $10\%$ to $30\%$ level.  In the following, we will assume a systematic error of either $20\%$ or $15\%$ on both background and signal, fully correlated between the two and across kinematic bins.  We have checked that in the absence of an external measurement (\egg~a background control region) which profiles uncertainties, the use of correlated uncertainties provides the most conservative results. In addition, we consider a $5\%$ uncertainty on the signal only to account for the theoretical uncertainty on the $t \bar t + E_T^{\rm miss}$  signal. 

\begin{figure}[t]
\centering
\begin{minipage}[t][\minipageheightdp][t]{\minipagewidthdp}
\centering
\includegraphics[width=\figuremaxwidthdp,height=\figuremaxheightdp,keepaspectratio]{\main/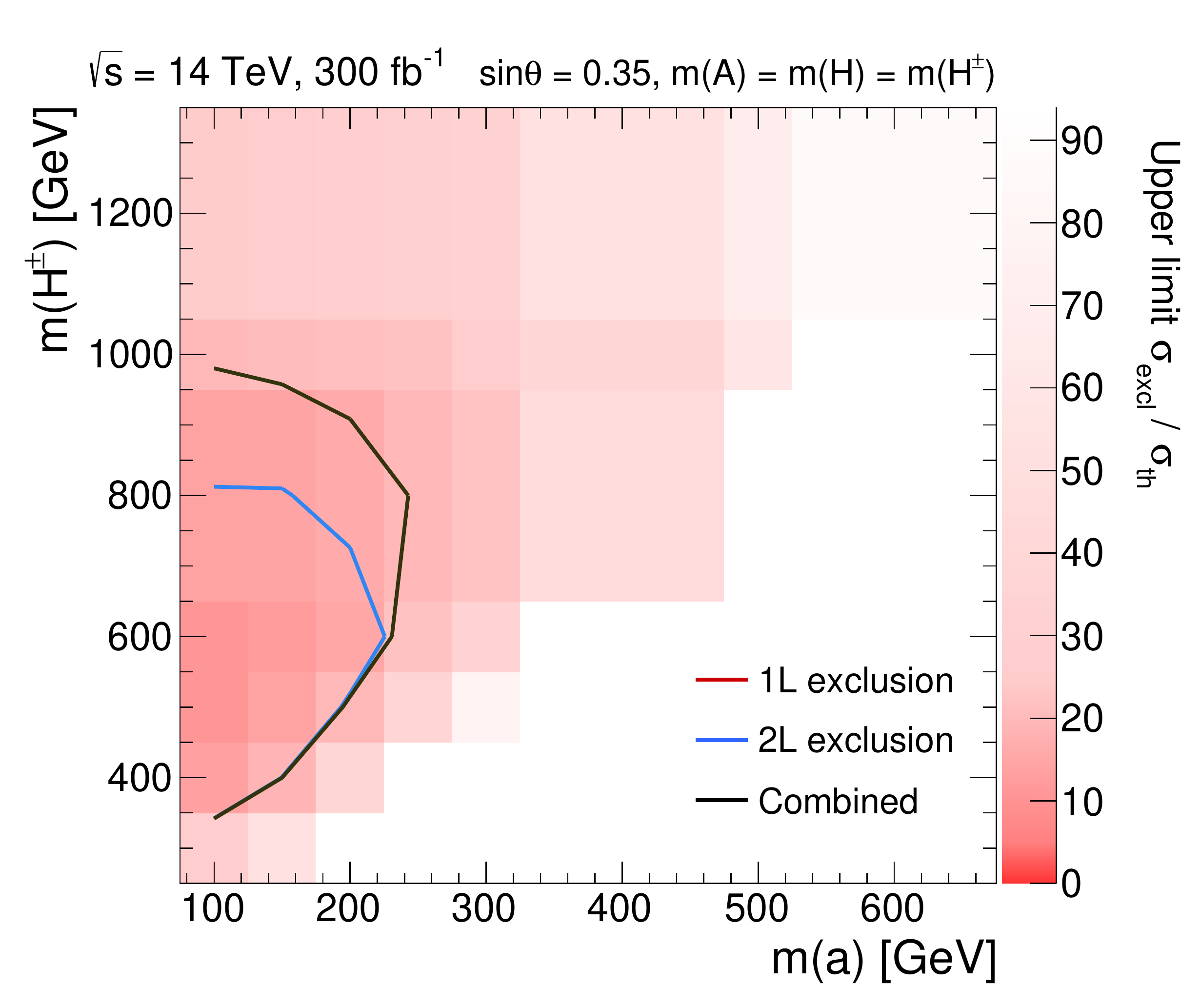}
\end{minipage}
\begin{minipage}[t][\minipageheightdp][t]{\minipagewidthdp}
\centering
\includegraphics[width=\figuremaxwidthdp,height=\figuremaxheightdp,keepaspectratio]{\main/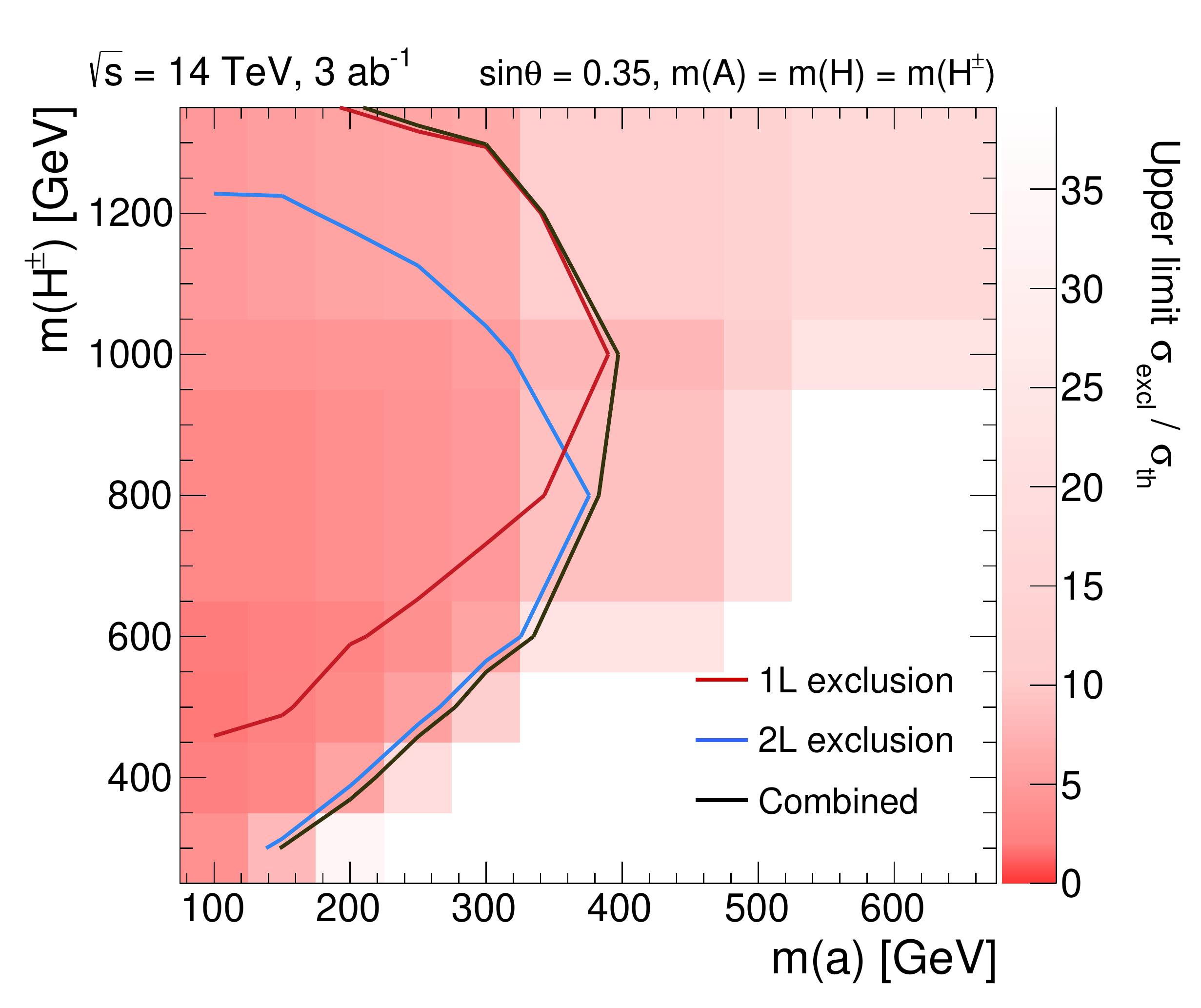}
\end{minipage}\\
\begin{minipage}[t][\minipageheightdp][t]{\minipagewidthdp}
\centering
\includegraphics[width=\figuremaxwidthdp,height=\figuremaxheightdp,keepaspectratio]{\main/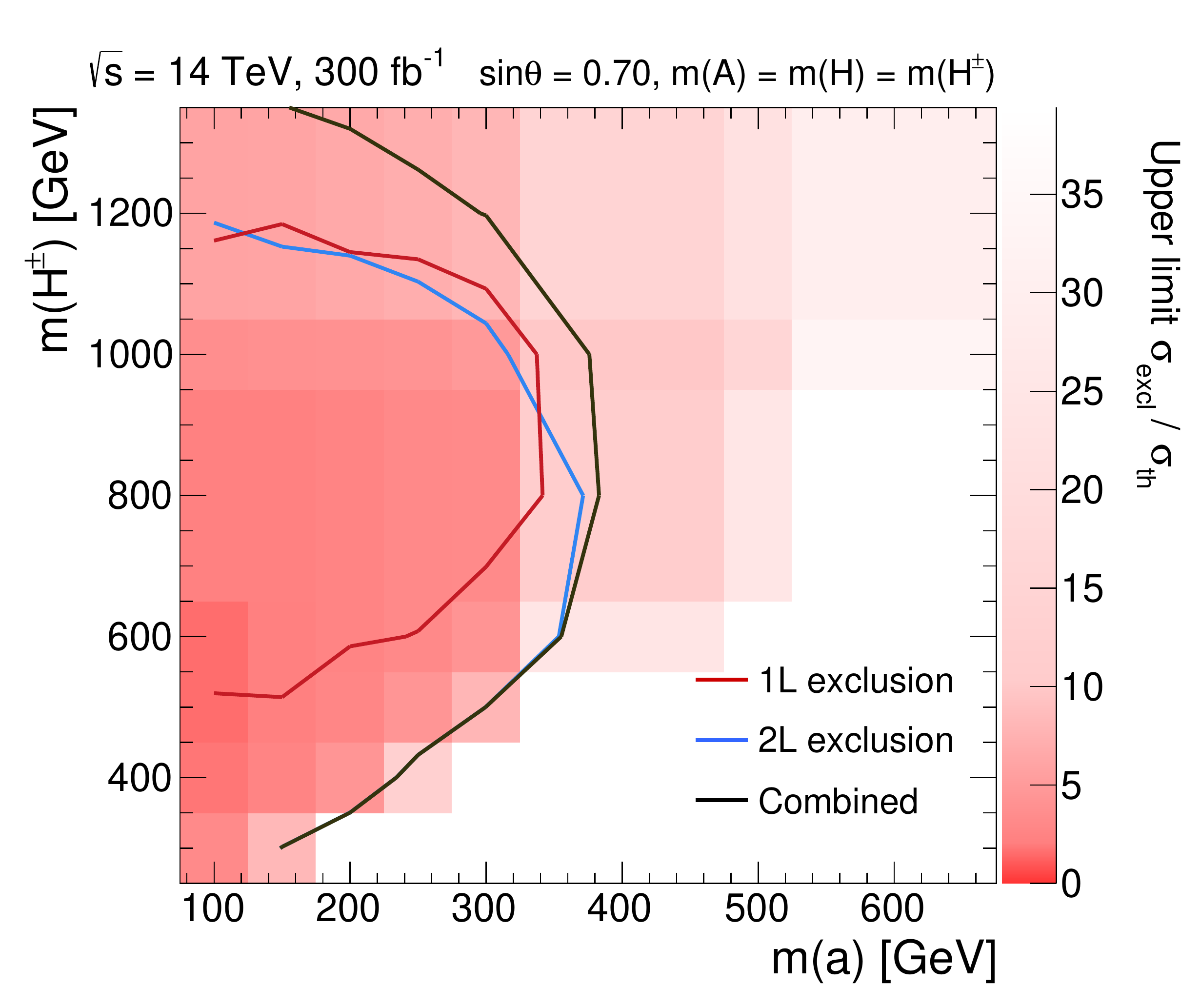}
\end{minipage}
\begin{minipage}[t][\minipageheightdp][t]{\minipagewidthdp}
\centering
\includegraphics[width=\figuremaxwidthdp,height=\figuremaxheightdp,keepaspectratio]{\main/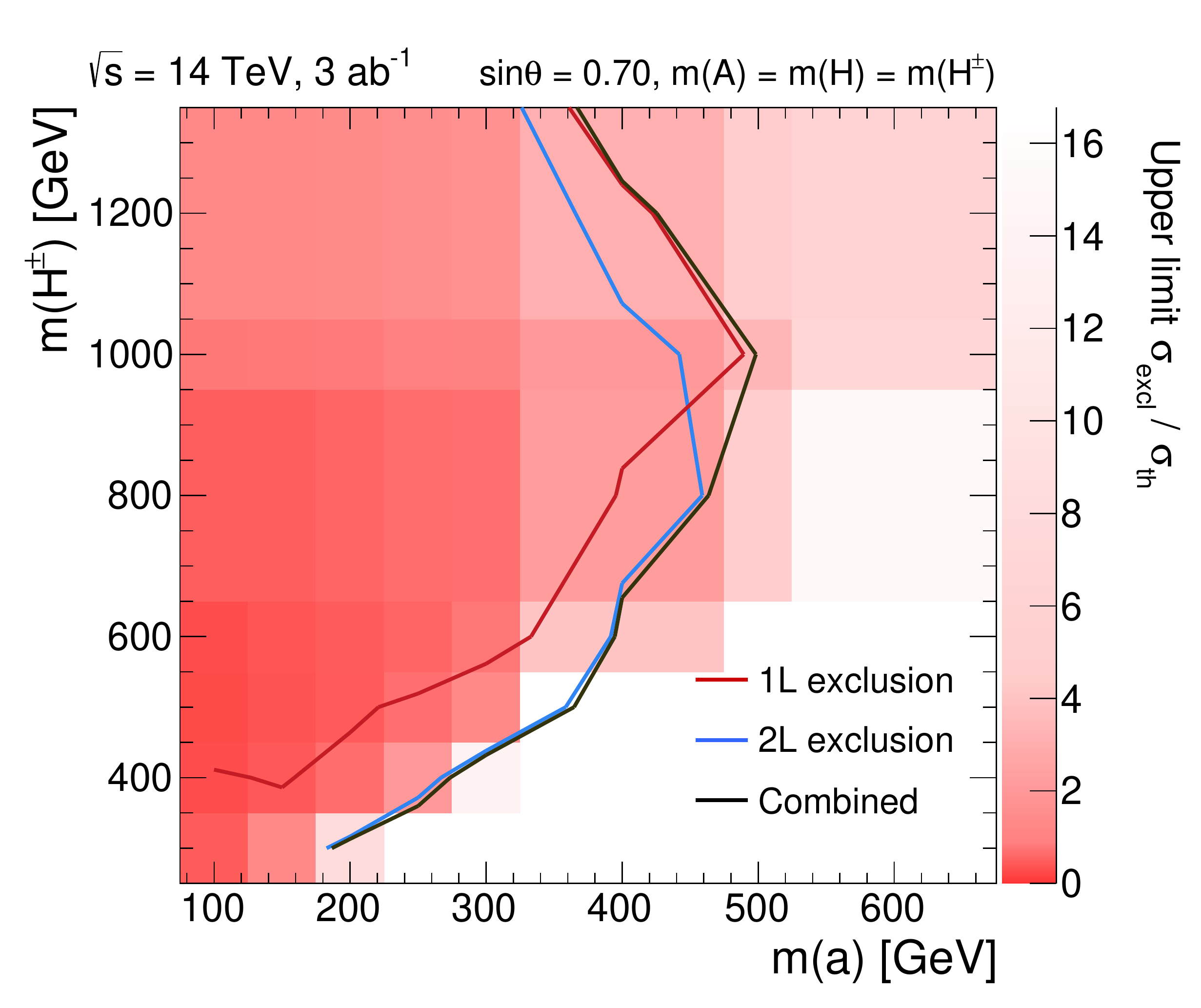}
\end{minipage}
\caption{95\%~\cl exclusion limits in the $m(a)-m(H^\pm)$ plane following from our one-lepton~(1L) and two-lepton~(2L) analysis shown as red and blue lines. The black curves indicate the bounds obtained by a combination of the two search strategies. The used  {\textrm{2HDM+a}} parameters are indicated.  The left~(right) panels correspond to $300\fbinv$ ($3\abinv$) of $\sqrt{s} = 14\TeV$ data.  Systematic uncertainties of $20\%$ ($5\%$) on the SM background (signal) are assumed.}
\label{fig:tXMETprojections1}
\end{figure}

In \fig{fig:ttMETDMF} we present sensitivity estimates at  LHC~Run-3, the HL-LHC and the HE-LHC for scalar (left panel) and pseudoscalar (right panel) simplified DM model mediators. 
The shown results correspond to a DM mass of $m_\chi = 1\GeV$ and the coupling choices $g_q = g_\chi = 1$, as recommended by the ATLAS/CMS DM Forum~(DMF)~\cite{Abercrombie:2015wmb} and the LHC DM Working Group~\cite{Boveia:2016mrp}. 
The estimated $95\%\cl$ exclusion limits are obtained from a  5-bin likelihood fit to the $|\!\cos\theta_{\ell\ell}| = \tanh \left (\Delta \eta_{\ell\ell}/2 \right)$ distribution. The inclusion of shape information is motivated by the observation that the distributions of events as a function of the pseudorapidity difference~$\Delta \eta_{\ell\ell}$  of the  dilepton pair is different for signal and background~\cite{Haisch:2016gry}. At LHC Run-3 it should be possible to exclude DMF scalar (pseudoscalar) models that predict a signal strength of $\mu =1$ for mediator masses up to around $200\GeV$ ($300\GeV$) using the~5-bin likelihood fit employed in our study. It should be possible to improve the maximal mass reach by roughly a factor of 2 when going from LHC~Run-3 to the HL-LHC and from the HL-LHC to the HE-LHC. The corresponding $95\%\cl$ exclusion limits on DMF scalar (pseudoscalar) mediators are thus expected to be around $450\GeV$ ($500\GeV$) and $900\GeV$ ($950\GeV$) at the HL-LHC and the \sloppy\mbox{HE-LHC}, respectively. Another conclusion that can be drawn from our sensitivity study is that the reach of future LHC runs depends strongly on the systematic background uncertainty, and as a result  a good  experimental understanding of $t \bar t Z$ production within the SM will be a key ingredient to  a possible discovery of DM in the~$t \bar t + E_T^{\rm miss}$ channel.

Using the recasting procedure described in \citeref{Abe:2018bpo}, the sensitivity estimates presented in \sloppy\mbox{\fig{fig:ttMETDMF}} for the DMF spin-0 models can be translated into exclusion limits on next-generation spin-0  DM models. In the case of the {\textrm{2HDM+a}} model~\cite{Ipek:2014gua,No:2015xqa,Goncalves:2016iyg,Bauer:2017ota,Tunney:2017yfp} for instance and adopting the benchmark~(4.5)~and~(4.6) introduced in \citeref{Abe:2018bpo}, one finds for $\tan \beta =1$ the $95\%\cl$ bounds \sloppy\mbox{$m(a) \lesssim 150\GeV$}~(HL-LHC) and $m(a) \lesssim 350\GeV$ (HE-LHC) on the mass of the  {\textrm{2HDM+a}}  pseudoscalar mediator $a$. These numbers show again that a HE-LHC is expected to be able to significantly improve upon the HL-LHC reach, in particular for spin-0 DM model realisations that predict small $t \bar t + E_T^{\rm miss}$ signal cross sections. 

{\bf Sensitivity study of the $t X + E_T^{\rm miss}$ signature:}  Following the analyses~\cite{Pani:2017qyd,Brooijmans:2018xbu}, we interpret our $t X + E_T^{\rm miss}$ results in the context of the {\textrm{2HDM+a}} model. The total background in the two-lepton selection is approximately 100 events, dominantly composed of the $t \bar t Z/W$ and $tWZ$ background processes. For charged Higgs masses $m(H^\pm)$ between $300\GeV$ and $700\GeV$, the acceptance  for signal events containing at least two leptons is in the range $[0.1, 0.7]\%$ for $m(a) = 150\GeV$ and $\tan\beta = 1$. The total background in the one-lepton selection is approximately 30 events for the leptonic-$H^\pm$ signal selection and 45 events for the hadronic-$H^\pm$ one. More than $70\%$ of the SM background arises from $t \bar t Z/W$ and $tZ$ processes in both selections and the rest is in equal parts due to  the contributions of top pairs (dileptonic decays) and the single-top $tW$ channel for the hadronic-$H^\pm$ selection, while in the case of the leptonic-$H^\pm$ selection the remaining $30\%$ are dominated by single-top processes. For $m(H^\pm)$  in the range of $600\GeV$  and $1\TeV$, the acceptance for signal events containing at  least one lepton amounts to approximately $[0.2, 0.5]\%$  for $m(a) = 150\GeV$ and $\tan\beta = 1$.

\begin{figure}[t]
\centering
\begin{minipage}[t][\minipageheightsp][t]{\minipagewidthsp}
\centering
\includegraphics[width=\figuremaxwidthsp,height=\figuremaxheightsp,keepaspectratio]{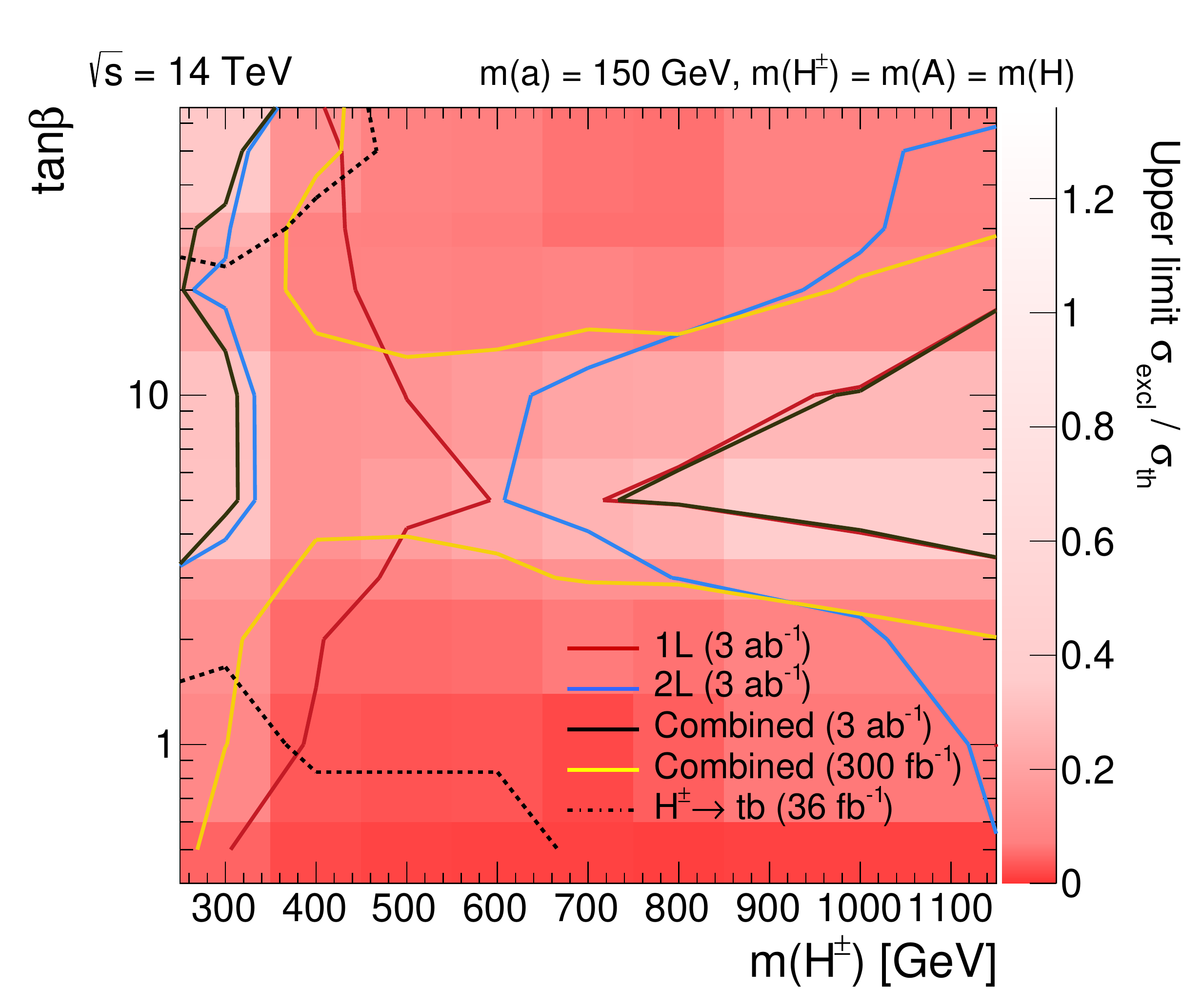}
\end{minipage} 
\caption{Regions in the $m(H^\pm)-\tan \beta$ which can be excluded at $95\%\cl$ through the one-lepton~(1L), two-lepton~(2L) searches and their combination. The used  {\textrm{2HDM+a}} parameters are shown in the headline of the figure. The projections assume $300\fbinv$ or $3\abinv$ of $\sqrt{s} = 14\TeV$ data and a systematic uncertainty of $20\%$ ($5\%$) on the SM background (signal). The shown $H^\pm \to tb$  limits have been obtained by recasting the experimental results presented in \citeref{ATLAS:2016qiq}.}
\label{fig:tXMETprojections2}
\end{figure}

In \fig{fig:tXMETprojections1} we present the results of our $t X + E_T^{\rm miss}$ sensitivity study in the $m(a)-m(H^\pm)$ plane of the {\textrm{2HDM+a}} model employing a DM mass of $m_\chi = 1\GeV$. As indicated by the headlines of the individual panels, two different values for the mixing angle in the pseudoscalar sector ($\sin\theta$) are employed. The parameters not explicitly specified have been set to the benchmark choices (4.5) made in \citeref{Abe:2018bpo}. One observes that at LHC~Run-3 a combination of the one-lepton and two-lepton search should allow one to exclude masses $m(a)$ up to around $250\GeV$ ($375\GeV$) for $\sin \theta = 0.35$ ($\sin \theta = 0.7$). The corresponding HL-LHC limits instead read $400\GeV$ ($500\GeV$), implying that collecting ten times more luminosity is expected to lead to an improvement in the LHC reach by a factor of around 1.5 in the case at hand. Also notice that the one-lepton and two-lepton analyses are complementary because they have different sensitivities on $m(H^\pm)$.

One can also compare the HL-LHC reach in the $m(H^\pm)-\tan \beta$ plane to that derived in \citeref{Pani:2017qyd} for LHC Run-3. Such a comparison is presented in \fig{fig:tXMETprojections2}. Numerically, we find that all values of $\tan\beta$ can be excluded for a charged Higgs mass between $300\GeV$ and $700\GeV$, improving the LHC Run-3 forecast, which showed a coverage in $\tan\beta$ up to $3$ and above $15$ for the same mass range.  For $m(H^\pm) = 1\TeV$ instead, the upper limit in $\tan\beta$ is extended from $2$ to $3$ and the lower limit is extended from around $30$ to $20$. Compared to LHC Run-3 the HL-LHC is thus expected to have a significantly improved coverage in $\tan \beta$, in particular for not too heavy 2HDM spin-0 states. Limits on $m(H^\pm)$ and $\tan \beta$ also derive from $H^\pm$ production followed by the decay of the charged Higgs into SM final states such as $\tau \nu$ or $tb$. As indicated in \fig{fig:tXMETprojections2}, in the {\textrm{2HDM+a}} model the searches for $H^\pm \to tb$ cover an area largely complementary to the results of the  $t X + E_T^{\rm miss}$ searches.

\subsubsection{\atlas{Dark matter production in single-top events at HL-LHC}}
\label{sec:atlasDMsingletop}
\contributors{L. Barranco, F. Castillo, M. J. Costa, C. Escobar, J. Garc\'{i}a-Navarro, D. Madaffari, J. Navarro, ATLAS}

The expected sensitivity of a search for the non-resonant production of an exotic state decaying into a 
pair of invisible DM particle candidates in association with a right-handed top quark is presented~\cite{ATL-PHYS-PUB-2018-024}.
Such final-state events, commonly referred to as ``monotop''
events, are expected to have a reasonably small background contribution
from SM processes. In this analysis only the topologies where the $W$ boson from the top quark
decays into a lepton and a neutrino are considered. 

The non-resonant monotop is produced via a flavour-changing neutral
interaction where a top quark, a light-flavour up-type quark and an
exotic massive vector-like particle $V$ can be parametrised through a general
Lagrangian~\cite{Abercrombie:2015wmb,Andrea:2011ws}:
%%%
\begin{equation}
  \mathcal{L}_{\text{int}} =  a V_{\mu} \bar{u}\gamma^{\mu} \ensuremath{P_{\text{R}}} t + g_{\chi} V_{\mu} \bar{\chi} \gamma^{\mu}\chi + \text{h.c.}\,, 
\end{equation}
where $V$ is coupled to a pair of DM particles (represented by Dirac fermions $\chi\bar{\chi}$) whose strength can be
controlled through a parameter $g_{\chi}$ and where $\ensuremath{P_{\text{R}}}$ represents the right-handed chirality projector. 
The parameter $a$ stands for the coupling constant between the massive invisible vector boson
$V$, and the $t$- and $u$-quarks, and $\gamma^\mu$ are the Dirac matrices. 

The experimental signature of the non-resonant monotop events with $W$
boson decaying leptonically is one lepton from the $W$-boson decay, large $\met$, and one jet identified as likely
to be originated from a $b$-quark. The signal event candidates are selected by requiring exactly one lepton with 
$ \pT > 30\GeV$, exactly one jet with $\pt> 30\GeV$identified as a $b$-jet and $\met > 100\GeV$. 
Since the considered monotop process favours final states with positive leptons,
events with negative lepton charge are rejected. These criteria
define the base selection.%%%%%%

In order to maximise the sensitivity of the study, in addition to the
base selection further discrimination is achieved by applying
additional criteria according to the kinematic properties of the
signal while rejecting background. The transverse mass of the lepton--$\met$~system, 
$\ensuremath{m_{\text{T}}(\ell, \met)}$, is required to be larger than $100\GeV$ in order 
to reduce the background contribution. 
In background events the spectrum of this quantity decreases rapidly for 
values higher than the $W$-boson mass. In signal events instead, the
spectrum has a tail at higher mass values.
When originating from the decay of a top quark, 
the lepton and the $b$-jet are close to each other. Therefore, events are required to have 
an azimuthal difference between the lepton
momentum and the $b$-jet momentum directions ($\ensuremath{\Delta\phi(\ell, $b$-jet)}$) 
of less than $2.0$, which disfavours the $W$+jets and diboson backgrounds. 

Further selection is performed via a BDT algorithm provided by the Toolkit for Multivariate Analysis \cite{Speckmayer:2010zz}.
The BDT is trained to discriminate the monotop signal from the dominant $t\bar{t}$ background.
For the training, since no significant difference is observed for the different mass values, the sample 
with $m_V = 2.5\TeV$ is used.
Half of the events of both signal and background samples are selected randomly and used to train the BDT.
The other half is used to probe the BDT behaviour in order to avoid the presence of overtraining.
The variables entering the BDT are selected from a pool of fundamental quantities,
like $\pt$ of jets and $b$-jets, and angular distances. 
%The transverse mass of the lepton and \met system (\mtW) is also used,
%being powerful in rejecting background, as seen in the search performed at 8~\TeV~\cite{Aad:2014wza}.
The variables selected are the ones showing the best discriminating power. In particular,
$\ensuremath{\Delta\phi(\ell, \met)}$ and $\ensuremath{m_{\text{T}}(\ell, \met)}$ are found to be the most effective variables.
Only events with BDT response $>0.9$ and $\met > 150\GeV$enter in the signal region and are used in 
the extraction of the result. The shape of the $\met$ distribution is used in the
statistical analysis, as it is expected to be the most sensitive
variable to the presence of new physics. The binning of this
distribution is optimised for the sensitivity of the analysis in the signal region
while ensuring the stability of the fit. This results in a
non-equidistant binning which exhibits wider bins in regions with a
large signal contribution, while preserving a sufficiently large number
of background events in each bin. 

\begin{figure}[t]
\centering
\begin{minipage}[t][\minipageheightdp][t]{\minipagewidthdp}
\centering
\includegraphics[width=\figuremaxwidthdp,height=\figuremaxheightdp+2mm,keepaspectratio]{\main/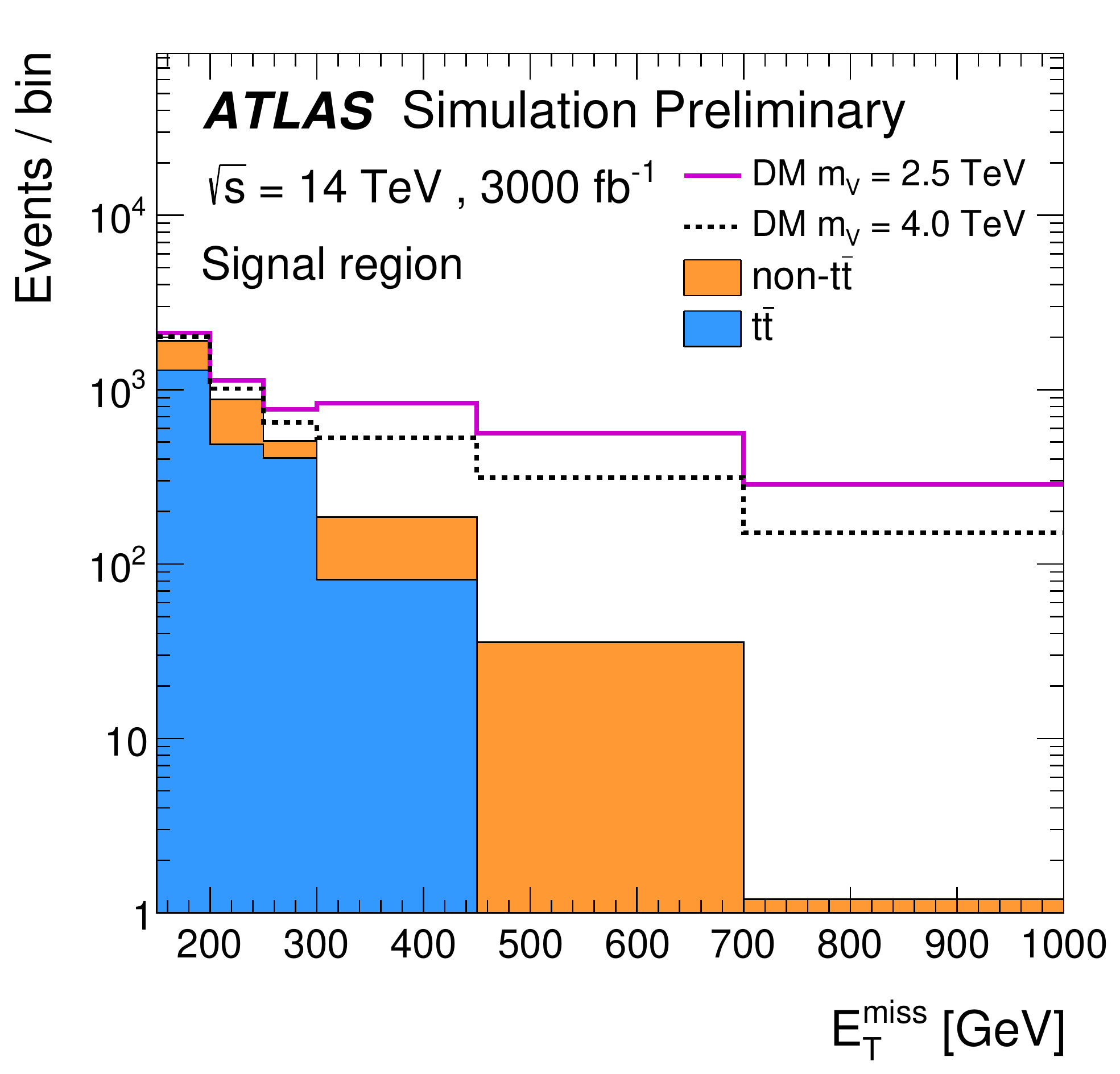}
\end{minipage}
\begin{minipage}[t][\minipageheightdp][t]{\minipagewidthdp}
\centering
\includegraphics[width=\figuremaxwidthdp,height=\figuremaxheightdp,keepaspectratio]{\main/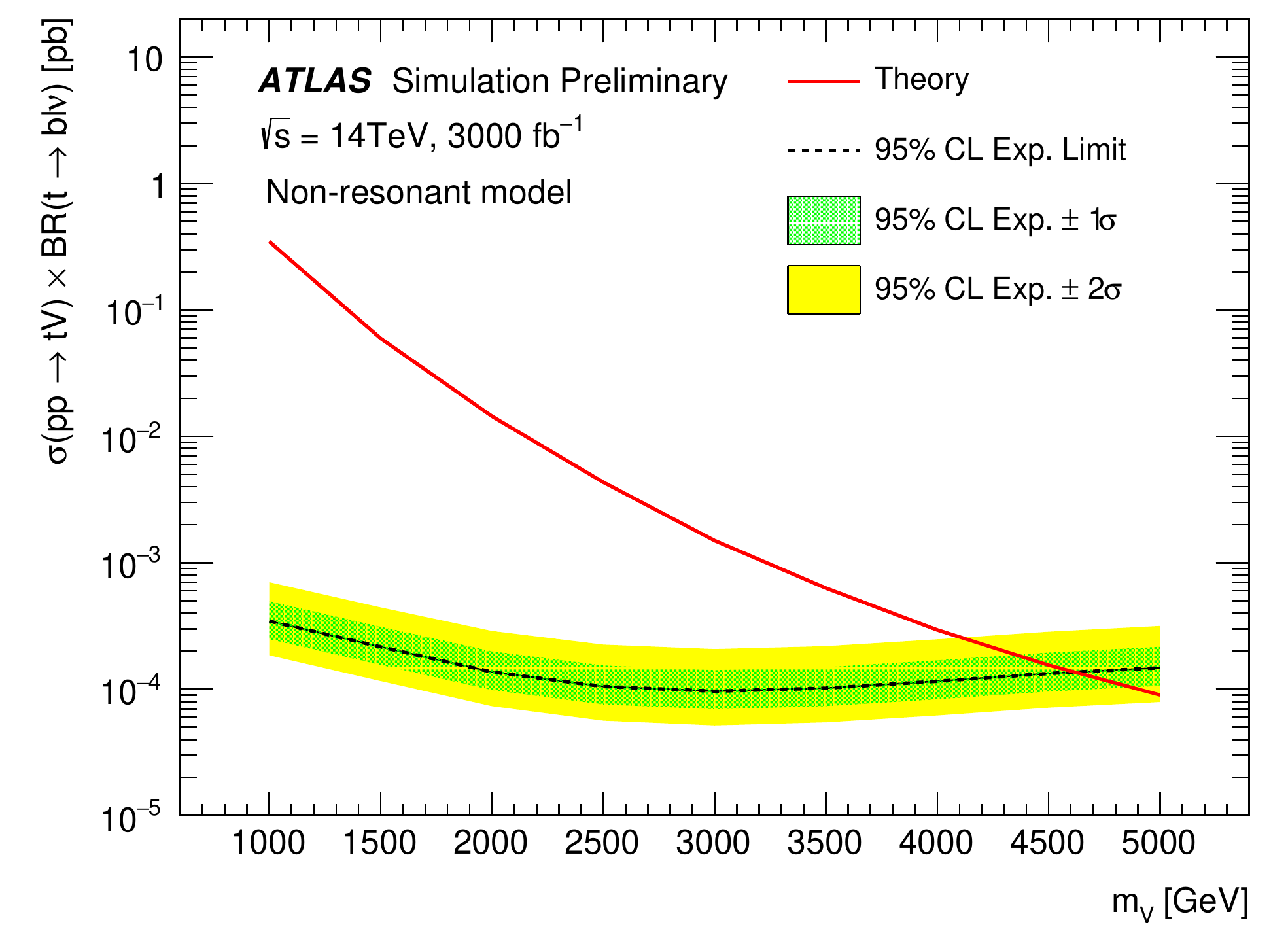}
\end{minipage}
  \caption{Left: expected post-fit $\met$ distribution in the signal region.
    The stack distribution shows the $t\bar{t}$ and non-$t\bar{t}$ background predictions.
    Solid and dashed lines represent the signal corresponding
    to a mediator mass of $2.5$ and $4.0\TeV$, respectively.
    The signal event samples are normalised to the number of
    background events. The binning is the same as the optimised, non-equidistant binning used in the fit.
    Last bin includes overflow events. Right: expected $95\%\cl$ upper limits on the signal cross-section as a function of the mass of
    the mediator for the non-resonant model assuming $m_\chi = 1\GeV$,
    $a = 0.5$ and  $g_\chi = 1$ using a BDT analysis. The MC statistical
  uncertainty is not considered but the full set of systematics,
  extrapolated from the $13\TeV$ analysis, is used.}
  \label{fig:MET_singleTop_atlas}\label{fig:limit_singleTop_atlas}
\end{figure}

Figure~\ref{fig:MET_singleTop_atlas} (left) shows the post-fit $\met$ distribution in the
signal region. The result does not include MC statistical uncertainties but incorporates
effects of systematic uncertainties.
The theoretical modelling of signal and background has the largest prior, $15\%$.
The second largest source of uncertainty is the one relative to the $\met$ reconstruction, with $6\%$ prior.
Jet energy scale (JES) and jet energy resolution (JER) contribute with a total of $5\%$.
The uncertainty on the requirements for pile-up jets rejection is  $5\%$, whilst uncertainties on 
lepton identification, $b$-tagging efficiencies and luminosity are $1.2\%$, $2.5\%$ and $1\%$, respectively.

Figure~\ref{fig:limit_singleTop_atlas} (right) shows the expected $95\%\cl$ upper limits as a function of the mediator mass for the
non-resonant model assuming $m_\chi = 1\GeV$, $a = 0.5$ and  $g_\chi = 1$.
After the fit, the largest impact on the result is coming from the uncertainty
on the $\met$ reconstruction. 
This is expected since the $\met$ is the final discriminant in the analysis.
The second largest contribution is coming from background and signal modelling.
The other contributions are, in order of importance: pile-up jet rejection requirements,
JES and JER, lepton reconstruction efficiency and $b$-tagging efficiency.
The uncertainty on the expected luminosity is found to have the smallest effect.
The expected mass limit at $95\%\cl$ is $4.6\TeV$ while
the discovery reach (based on $5\sigma$ significance) is $4.0\TeV$. 
For the current analysis the effect of
possible improvements in the systematic uncertainties is estimated by
reducing by half the uncertainties. This has the effect of increasing
the exclusion limit (discovery reach) by $80$ ($50$)\GeV.
The expectation for the equivalent of Run-3 integrated luminosity
($300\fbinv$) is checked, obtaining an exclusion limit (discovery
reach) of $3.7\TeV$ ($3.2\TeV$).

\subsubsection{\atlas{Four-top signatures at the HL-LHC}}
\label{sec:atlas4top}
\contributors{P. Pani, F. Meloni, ATLAS}

A class of simplified models for dark matter searches at the LHC involving a two-Higgs-doublet extended sector  
together with an additional pseudoscalar mediator to DM, the 2HDM+$a$, are considered in this study~\cite{ATL-PHYS-PUB-2018-027}.
%In these models, the 2HDM sector consists of  a type-II coupling structure. 
%Furthermore,  the alignment ($\cos(\beta-\alpha) = 0$) and decoupling limit  is assumed, such that the lightest \cpeven
%state of the Higgs-sector, $h$, can be identified with the Standard Model (SM) Higgs boson and the EW \vev, $\nu$, is set to $246\GeV$.
The additional pseudoscalar mediator of the model, $a$, couples the DM particles to the SM and mixes with the
pseudoscalar partner of the SM Higgs boson, $A$. 
%Following the prescriptions in \citeref{2HDMWGproxi}, the masses of the heavy \cpeven Higgs boson, $H$, and charged bosons, $H^\pm$, are
%set equal to the mass of the heavy \cpodd partner $m_{A}$, while the the three quartic couplings between the scalar doublets and the $a$ boson ($\lambda_{P1}, \lambda_{P2}$ and $\lambda_3$) are all set equal to $3$, in order to simplify the phenomenology and evade the
%constraints from EW precision measurements.
%In addition,  to reduce the multiplicity of the parameter space we consider unitary couplings between the $a$ and the DM particle $\chi$ ($y_\chi = 1$) and we fix the DM particle mass to $10\GeV$. The ratio of the \vevs of the
%two Higgs-doublets, $\tan\beta$, is set to unity as well. 
This model is characterised by a rich phenomenology and can produce very different final states according to the production 
and decay modes for the various bosons composing the Higgs sector, which can decay both into dark matter or SM particles. 
The four-top signature is interesting if at least some of the neutral Higgs partners
masses are kept above the $\ttbar$ threshold, since, when kinematically allowed, all
four neutral bosons can contribute to this final state. The total four top-quark production cross-section is dominated by the
light pseudoscalar and the heavy scalar bosons.  In order to highlight this interplay, four
benchmark models will be considered, assuming different choices for the mass of the light
\cpodd and heavy \cpeven bosons and the mixing angle between the two \cpodd weak eigenstates ($\sin\theta$). 
\begin{description}
\item[Scenario 1] $m_{a}$ sensitivity scan assuming:
\begin{enumerate}
\item[a)]  $m_H = 600\GeV$ , $\sin\theta = 0.35$.
\item[b)]  $m_H = 1\TeV$ , $\sin\theta = 0.7$.
\end{enumerate}
  \item[Scenario 2] $\sin\theta$ sensitivity scan assuming:
\begin{enumerate}
\item[a)]  $m_H = 600\GeV$ , $m_{a} = 200\GeV$. 
\item[b)]  $m_H = 1\TeV$ , $m_{a} = 350\GeV$.
\end{enumerate}
\end{description}
%Scenario 1a and Scenario 2 closely follow the DM Working Group recommendations and are therefore ideal benchmarks to compare 
%the four top-quarks signature with other final states that characterise the phenomenology of this model~\cite{2HDMWGproxi}.
%The production cross section of the four benchmark scenarios are presented in \fig{fig:THDMpa_xsec}.
%It is found that for $m_H = 600\GeV$, 
%the kinematic properties of the signal is relatively independent of
%$m_{a}$ (except for masses below the top-quark
%threshold)
%and of the mixing angle. Conversely, when the heavy scalar Higgs boson
%is chosen to be heavier ($m_H = 1\TeV$), the mass of the light
%pseudoscalar and the mixing angle choice play an important 
%role in determining the kinematic properties of the final state.
\begin{figure}[t]
\centering
\begin{minipage}[t][\minipageheightdp][t]{\minipagewidthdp}
\centering
\includegraphics[width=\figuremaxwidthdp,height=\figuremaxheightdp,keepaspectratio]{\main/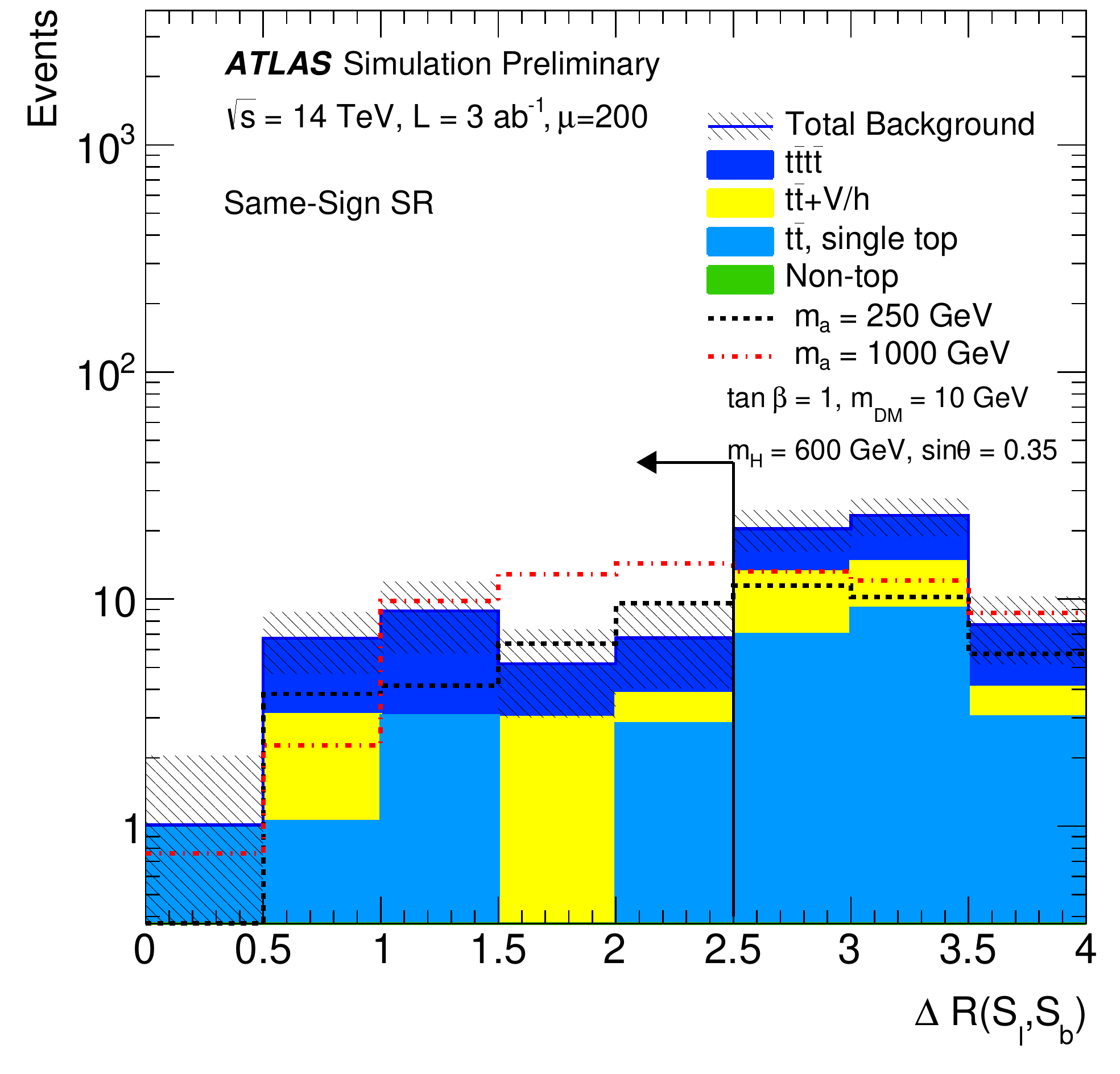}
\end{minipage}
\begin{minipage}[t][\minipageheightdp][t]{\minipagewidthdp}
\centering
\includegraphics[width=\figuremaxwidthdp,height=\figuremaxheightdp,keepaspectratio]{\main/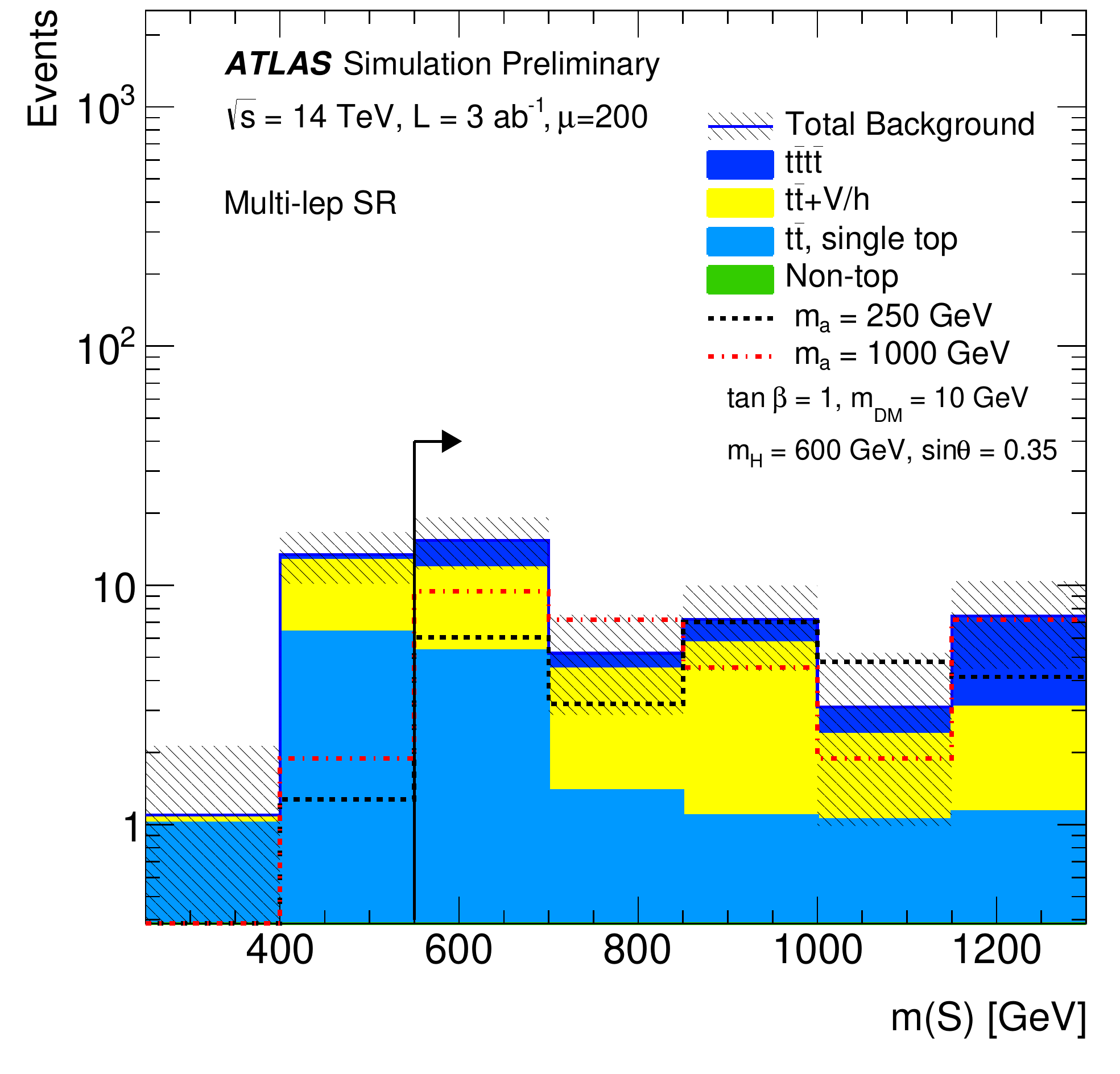}
\end{minipage}
\caption{Key distributions for events passing all Same-Sign (left) or Multi-lep (right) selection requirements except 
that on the distribution itself. The contributions from all SM backgrounds are shown, and the hashed band represents 
the statistical uncertainty on the total SM background prediction. The expected distributions for signal models 
with $m_{a}=250\GeV$ and $m_{a}=1000\GeV$, assuming  $m_{H}=600\GeV$, $\sin \theta=0.35$ 
are also shown as dashed lines for comparison.}
\label{fig:final_N1_4topatlas}
\end{figure}
This prospect study considers four top-quarks final states involving at least two leptons with the same electric charge 
or at least three or more leptons. Final states with high jet multiplicity and one lepton are also very powerful 
to constrain these signatures, but are not considered here. 
Complementary studies of the potential for the measurement of standard model production of the four top final state at CMS~\cite{CMS-PAS-FTR-18-031} and ATLAS~\cite{ATL-PHYS-PUB-2018-047} are also discussed in working group chapter 1~\cite{Nason:2650160}.

Events are accepted if they contain at least two electrons, two muons or one electron and one muon with the same electric 
charge or at least three leptons ($\pt>25\GeV$). 
Furthermore, events are required to contain at least three $b$-jets. The up to four leading leptons and up to four leading 
$b$-jets in the event are grouped respectively in two systems, called $\mathcal{S}_{\ell}$ and $\mathcal{S}_{b}$. 
A signal system $\mathcal{S}$ is defined by $\mathcal{S} = \mathcal{S}_{\ell} \cup \mathcal{S}_{b}$.
Different discriminators and kinematic variables are used in the analysis to separate the signal from the SM background. 
\begin{itemize}
\item[-] $\ensuremath{ \pt (\mathcal{S}_{\ell})}$: the vector sum of the lepton four momenta in $\mathcal{S}_{\ell}$;
\item[-] $\ensuremath{ \Delta R (\mathcal{S}_{\ell}, \mathcal{S}_{b})}$: 
the $\Delta R$ between the vectorial sum of the leptons in $\mathcal{S}_{\ell}$ and the vectorial sum of the $b$-jets in $\mathcal{S}_{b}$;
\item[-] m($\mathcal{S}$): the invariant mass of the signal system $\mathcal{S}$;
\end{itemize}
A common selection is applied to all events, before further categorisations. 
Events are required to have at least two jets with a $\pt>50\GeV$. In events with exactly two (anti-)electrons, the
contribution of SM processes including an on-shell $Z$ boson decaying leptonically with a lepton charge 
misidentification is reduced by vetoing events with $81.2\GeV<\ensuremath{m_{\ell\ell}}<101.2\GeV$. 
Furthermore, low mass resonances are vetoed by requiring $\ensuremath{m_{\ell\ell}}>15\GeV$.  
\begin{figure}[t]
\centering
\begin{minipage}[t][\minipageheightdp][t]{\minipagewidthdp}
\centering
\includegraphics[width=\figuremaxwidthdp,height=\figuremaxheightdp,keepaspectratio]{\main/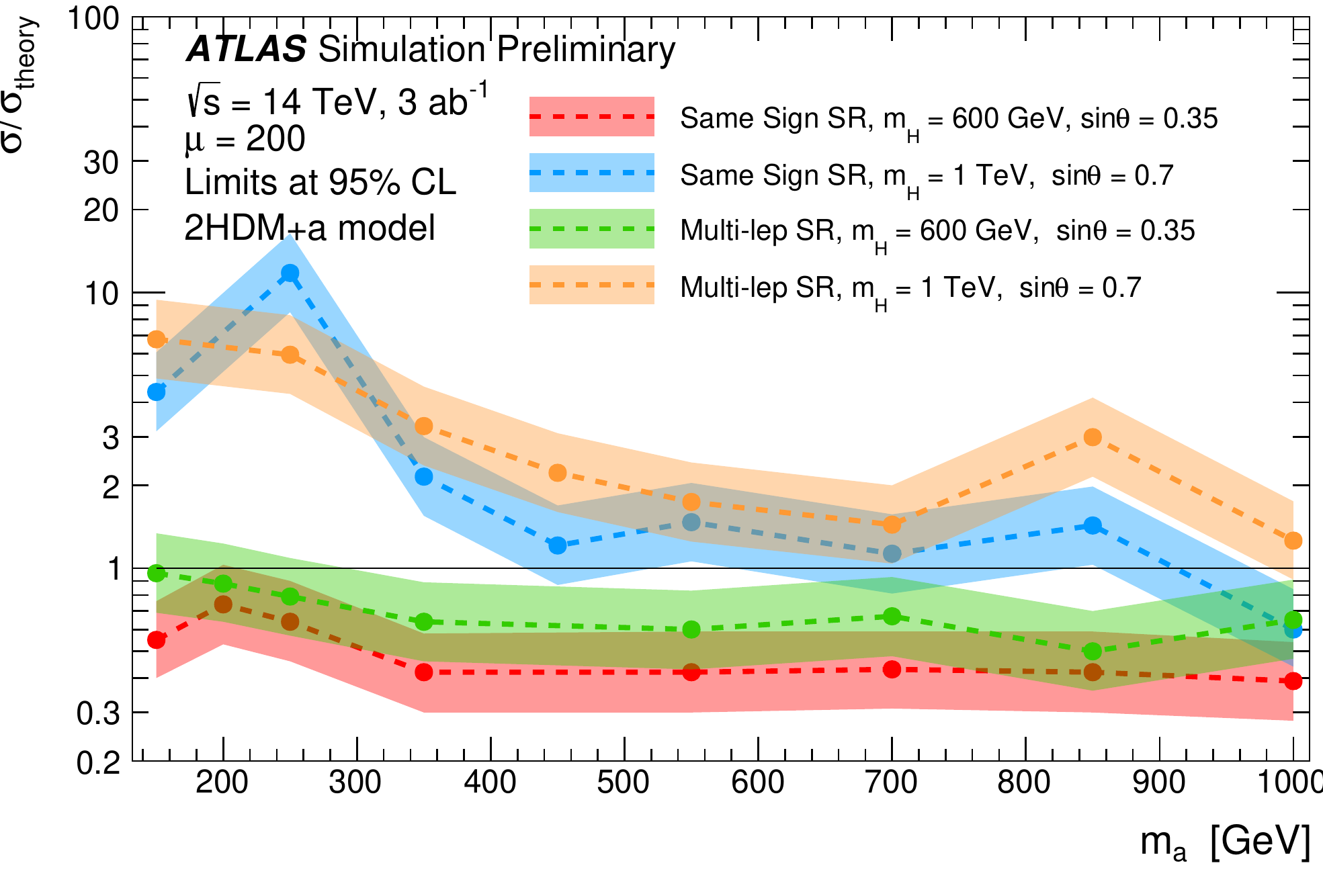}
\end{minipage}
\begin{minipage}[t][\minipageheightdp][t]{\minipagewidthdp}
\centering
\includegraphics[width=\figuremaxwidthdp,height=\figuremaxheightdp,keepaspectratio]{\main/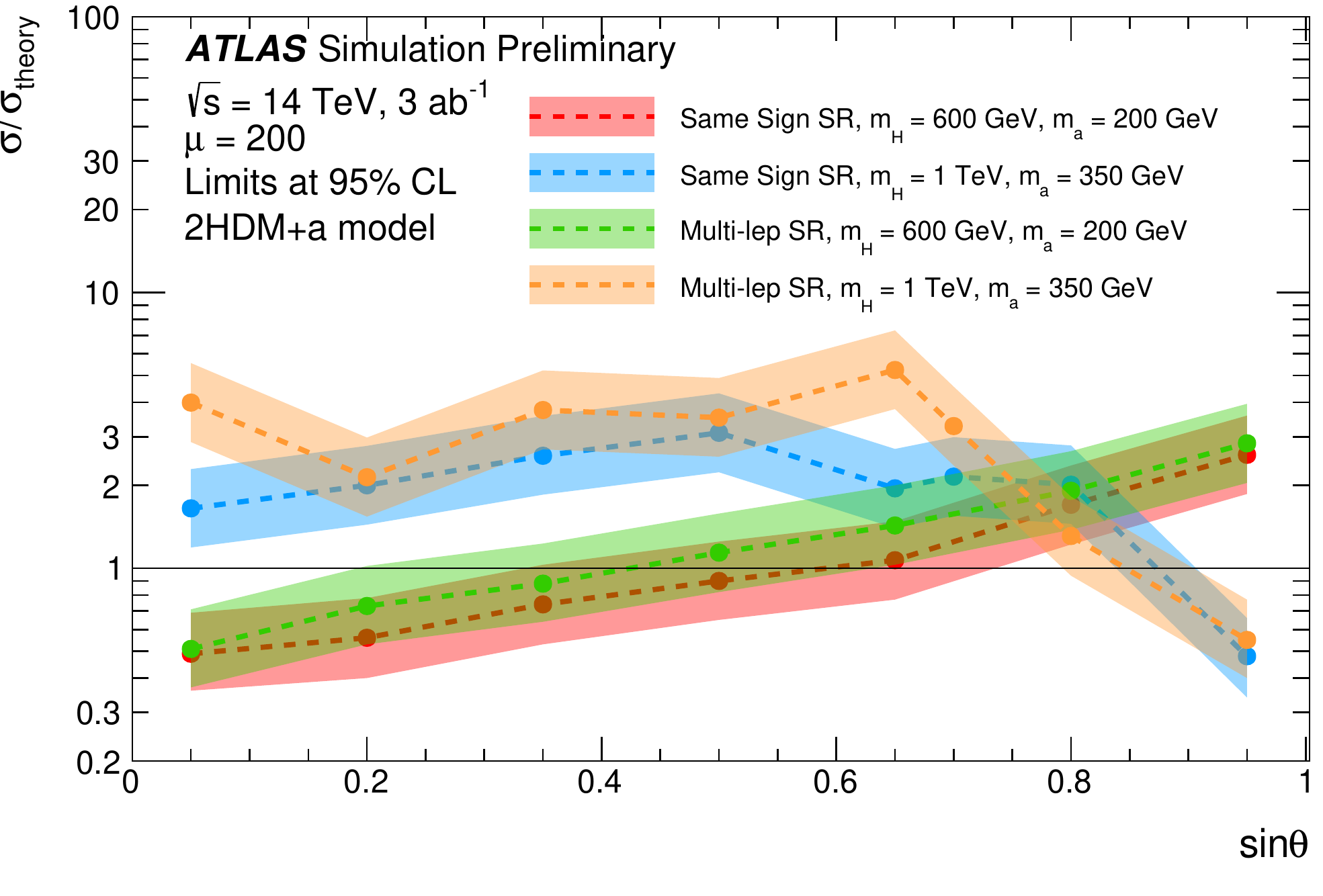}
\end{minipage}
\caption{Exclusion limits at $95\%\cl$ for Same-Sign and Multi-lep SRs
  in terms of excluded cross-section ($\sigma$) over the cross-section
  predicted by the model ($\sigma_{theory}$). Limits are derived from
  the analysis of \intLumi of $14\TeV$ proton-proton collision data
  as a function of $m_{a}$ (left) or as a function of $\sin\theta$
  (right) for each parameter assumptions described and indicated in the legend. The $1\sigma$
  variation of the total uncertainty on the limit is indicated as a
  band around each exclusion line.}
\label{fig:CLs_mascanB_4topatlas} 
\end{figure}
Two signal regions (SRs) are defined selecting events with exactly two charged leptons with the same electric charge 
(denoted Same-Sign) or three or more charged leptons (denoted Multi-lep).
Figure~\ref{fig:final_N1_4topatlas} shows two key distributions ($\ensuremath{ \Delta R (\mathcal{S}_{\ell}, \mathcal{S}_{b}) }$ 
and $m(\mathcal{S})$) for events passing one set of SRs requirements except for the requirement on the shown variable itself. 
The main backgrounds that survive the selections are the irreducible $\ttbar\ttbar$ and $\ttbar$+$V/h$ channels.
The dominant uncertainties are expected to be due to theoretical modelling of the
irreducible backgrounds and, to a lesser extent, to the jet energy scale and 
resolution, and the b-tagging efficiency.  Owing to the
reduced statistical uncertainty and a better understanding of the
physics models, it is expected that JES, JER, \sloppy\mbox{b-tagging} efficiency and
irreducible background modelling uncertainties will all be reduced. 
This leads to an estimate of the total background uncertainty of about $20\%$.   
The resulting experimental uncertainty is assumed to be fully correlated between the background and the signal when setting
$95\%\cl$ exclusion limits. Furthermore, an additional systematic of $5\%$ is considered for the signal, in order to
account for the theoretical systematic uncertainty on the model. 
%For the expected discovery $p$-values values, only the uncertainty on the background is considered. 

Scans of expected exclusion limits at $95\%\cl$ are shown in Figures~\ref{fig:CLs_mascanB_4topatlas} 
as a function of $m_{a}$, for fixed $m_{H}$ and $\sin \theta$ and as a function of $\sin \theta$ 
for fixed $m_{a}$ and $m_{H}$. In all benchmarks, it is assumed that $\tan\beta = 1$ and $m_{\chi} = 10\GeV$.
For light pseudoscalar masses above the $\ttbar$ decay threshold, a significance of about $3\sigma$ is 
expected if $m_{H} = 600\GeV$ and $\sin\theta = 0.35$. The same benchmark is expected to be excluded for all 
light-pseudoscalar masses and for $\sin\theta < 0.35$ if $m_{a} = 200\GeV$. 
Mixing angles such that $\sin\theta > 0.95$ are also expected to be excluded
for $m_{a} = 350\GeV$, $m_{H} = 1\TeV$ and, under the same assumptions, an upper limit of 
about two times the theoretical cross section is set for $\sin\theta < 0.8$. Finally, 
$\sin\theta < 0.4$ is excluded for $m_H = 600\GeV$, $m_{a} = 200\GeV$.
In almost all cases the Same-Sign SR yields the strongest constraints
on the parameter space considered in this work. However, the Multi-lep SR offers a complementary channel whose sensitivity
is of the same order of magnitude. Possibly, exploiting dedicated techniques developed to suppress or better estimate the $\ttbar+V$ background that
affects the Multi-lep SR, this signature can achieve sensitivity comparable to the Same-Sign selection.

%%%%%%%%%%%%%%
\subsection{Dark Matter and Electroweak Bosons}
\label{sec:DMWZ}
DM can be produced in association with, or through interactions with, EW gauge bosons.  The DM may recoil against a (leptonically decaying) $Z$ boson that was produced as ISR or in the decay of a heavy mediator to a lighter mediator and a $Z$, see \secc{sec:cmsDMwithZ}.  It may recoil against a photon or be produced through its couplings to $W,Z$ in VBF, as in \secc{sec:atlasDMmonophoton}.  A standard way to couple to the dark sector is through SM ``portals".  %, which allow gauge invariant operators built out of SM fields to couple to operators in the dark sector.  
UV completing the Higgs portal leads to signals only involving the mediators and not the DM, such as diHiggs or di-mediator production. Prospects are presented in~\secc{sec:theoryhiggsportal}. 
Alternatively, heavier dark sector states with couplings to the $Z$ boson can produce DM in their decays, as shown in~\secc{sec:theorysingletDM}.

\subsubsection{\cms{Dark matter produced in association with a $Z$ boson at HL-LHC}}
\label{sec:cmsDMwithZ}
\contributors{A. Albert, K. Hoepfner, CMS}

Collider searches for DM production critically rely on a visible particle being produced in association with the sought-after invisible DM candidate. One possible choice of an accompanying SM signature is a Z boson reconstructed from an $e^{+}e^{-}$ or $\mu^{+}\mu^{-}$ pair. In the hadronic environment of the LHC, this leptonic signature is well reconstructible and the resonant behaviour of the dilepton mass allows for efficient rejection of non-Z background processes. The presence of a signal is determined from a maximum-likelihood fit of the \ptmiss spectrum of selected events, which would be hardened by the presence of a DM signal relative to the SM backgrounds.

\begin{figure}[t]
\centering
\begin{minipage}[t][\minipageheightsp][t]{\minipagewidthsp}
\centering
\includegraphics[width=\figuremaxwidthsp,height=\figuremaxheightsp,keepaspectratio]{\main/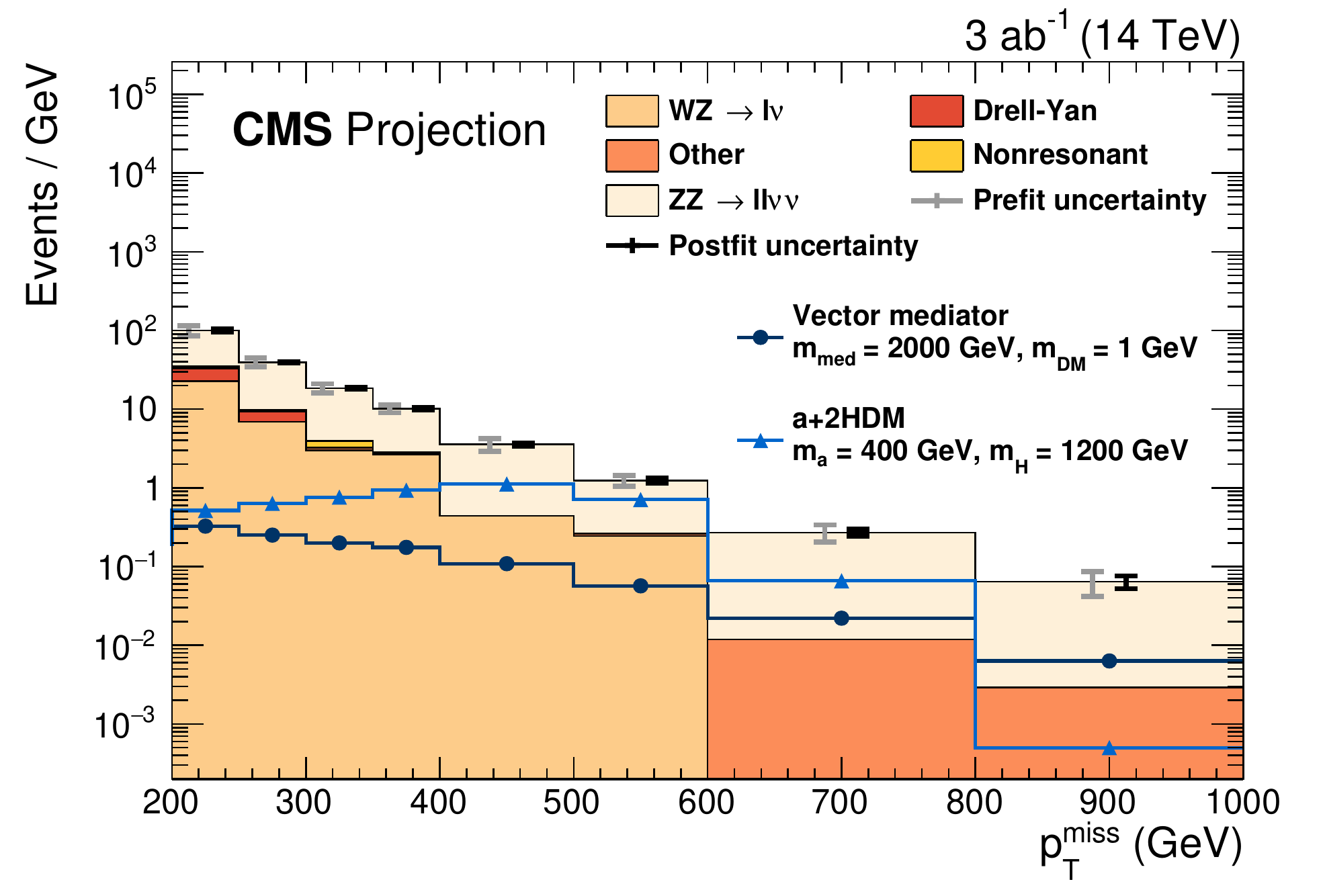}
\end{minipage}
        \caption{Spectrum of \ptmiss in the signal region. Uncertainty bands for the background prediction are shown before and after applying a background-only maximum-likelihood fit to the Asimov dataset in signal and control regions.}
        \label{fig:monoz_results_met_sr}
\end{figure}

This study from CMS is a projection based on the results of \citeref{Sirunyan:2017qfc}. Event-by-event weights are applied to simulated samples to account for the difference in \com~energy and \ptmiss resolution between the Run-2 and HL-LHC scenarios~\cite{CMS:2018fux}.
The \ptmiss spectrum in the signal region is shown in \fig{fig:monoz_results_met_sr}.

The results are interpreted in two simplified models of DM production. In the first model, a minimal scenario is assumed where there is one new mediator boson and one new DM Dirac fermion $\chi$~\cite{Abercrombie:2015wmb}. The mediator is assumed to have vector couplings $g_{q}$ and $g_{DM}$ to quarks and DM, respectively.
In the second model, referred to as ``a+2HDM'', the SM is extended by a second Higgs doublet as well as a light pseudoscalar DM mediator, $a$~\cite{Bauer:2017ota}. By allowing for the light pseudoscalar to mix with the heavier pseudoscalar state from the second Higgs doublet, the mediation mechanism can be realised without violating any of the various existing direct and indirect constraints on the scalar sector~\cite{Bauer:2017ota, Abe:2018bpo}. Importantly, this second model allows for the production of the pseudoscalar mediator and $Z$ boson through the decay of a new heavy scalar H. This production mode provides excellent sensitivity for the $Z+\ptmiss$ search compared to other searches such as jets$+\ptmiss$~\cite{Abe:2018bpo}.

\begin{figure}[t]
\centering
\begin{minipage}[t][\minipageheightdp][t]{\minipagewidthdp}
\centering
\includegraphics[width=\figuremaxwidthdp,height=\figuremaxheightdp,keepaspectratio]{\main/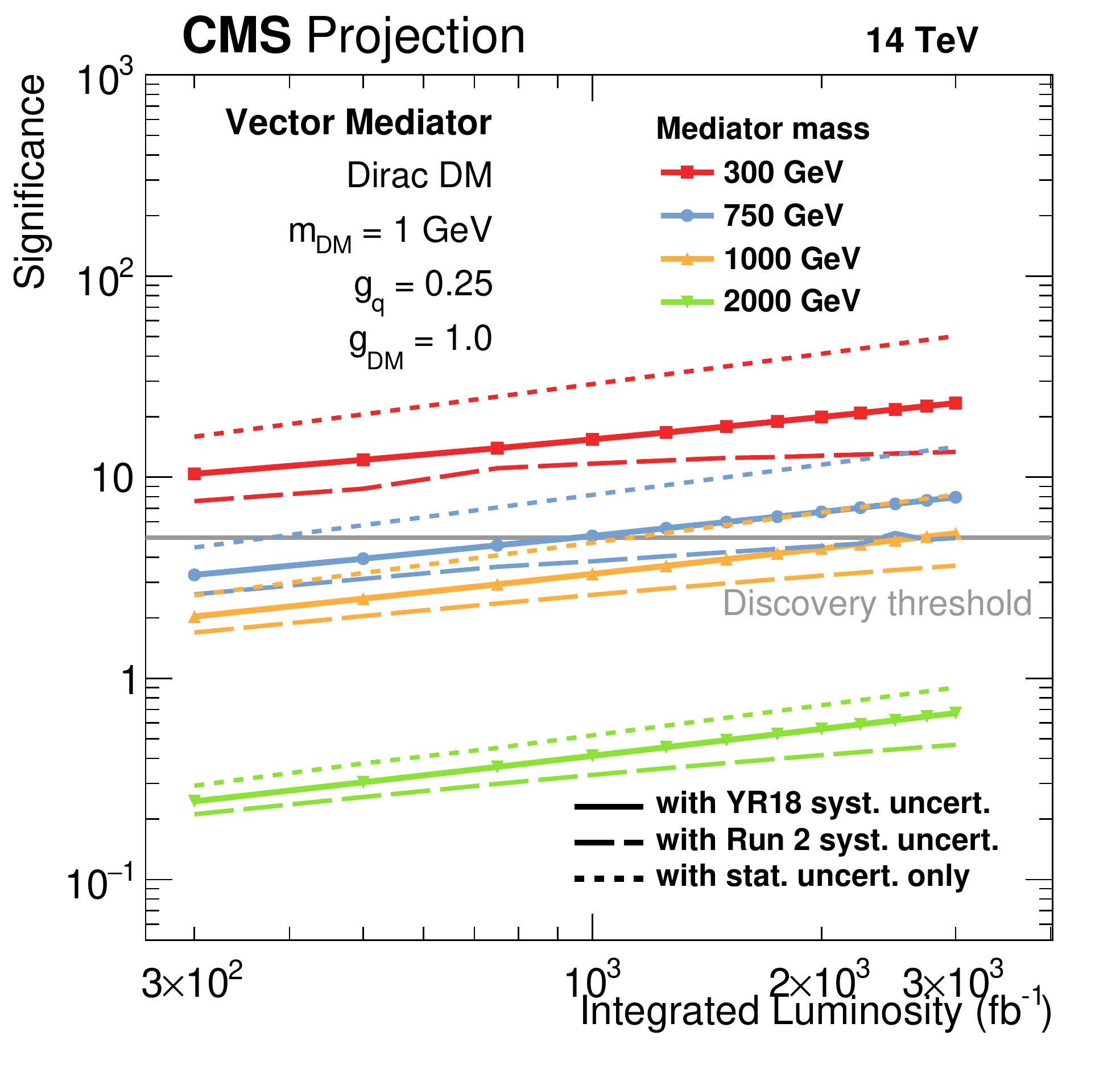}
\end{minipage}
\begin{minipage}[t][\minipageheightdp][t]{\minipagewidthdp}
\centering
\includegraphics[width=\figuremaxwidthdp,height=\figuremaxheightdp,keepaspectratio]{\main/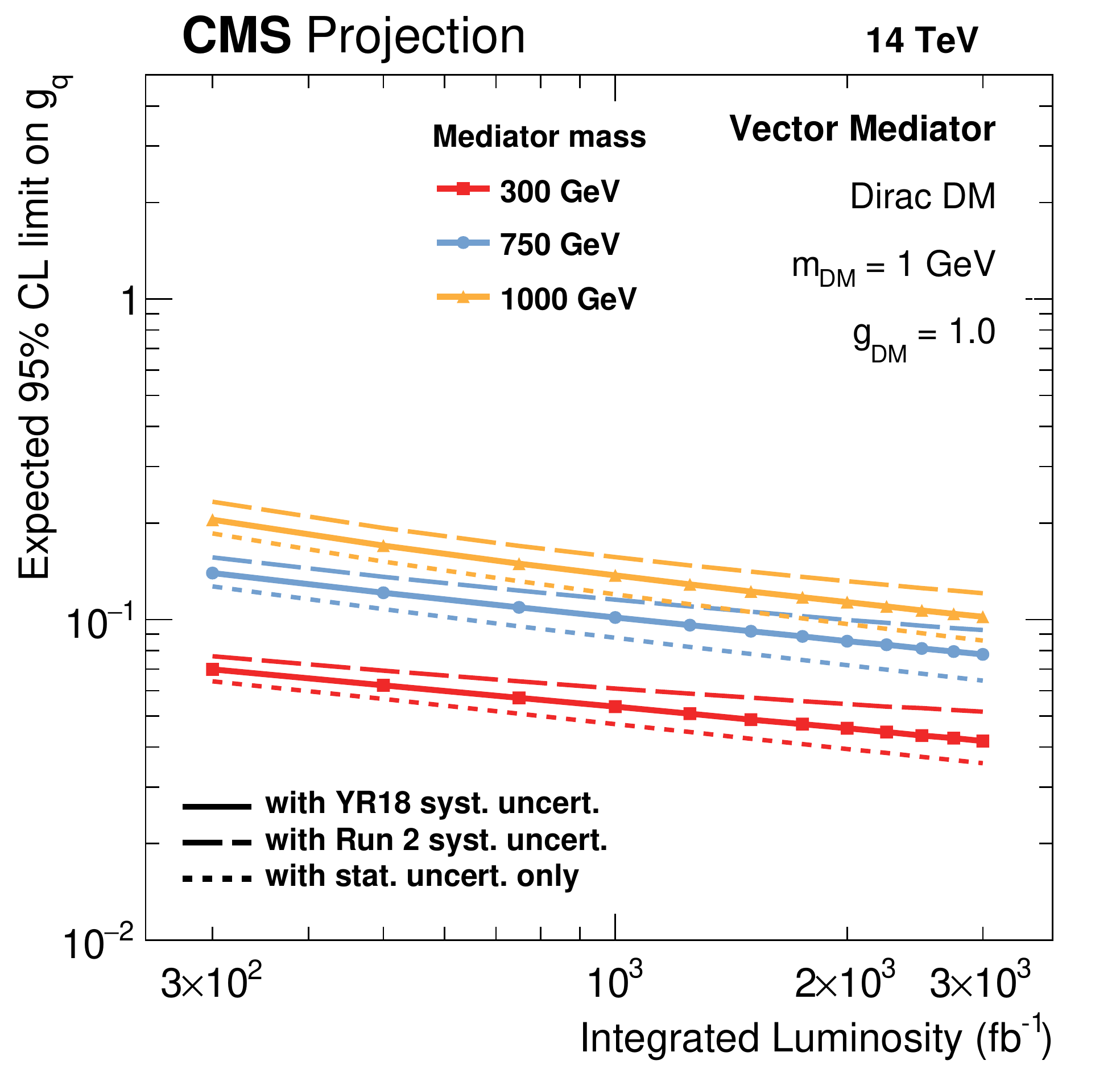}
\end{minipage}
\caption{Expected significance of a vector mediator signal with unity signal strength (left) and  $95\%\cl$ exclusion limits on the coupling $g_{q}$ (right). Both quantities are shown as a function of integrated luminosity for multiple choices of the mediator mass $m_\text{med}$.}
        \label{fig:monoz_results_vector}
\end{figure}

\begin{figure}[t]
\centering
\begin{minipage}[t][\minipageheightdp][t]{\minipagewidthdp}
\centering
\includegraphics[width=\figuremaxwidthdp,height=\figuremaxheightdp,keepaspectratio]{\main/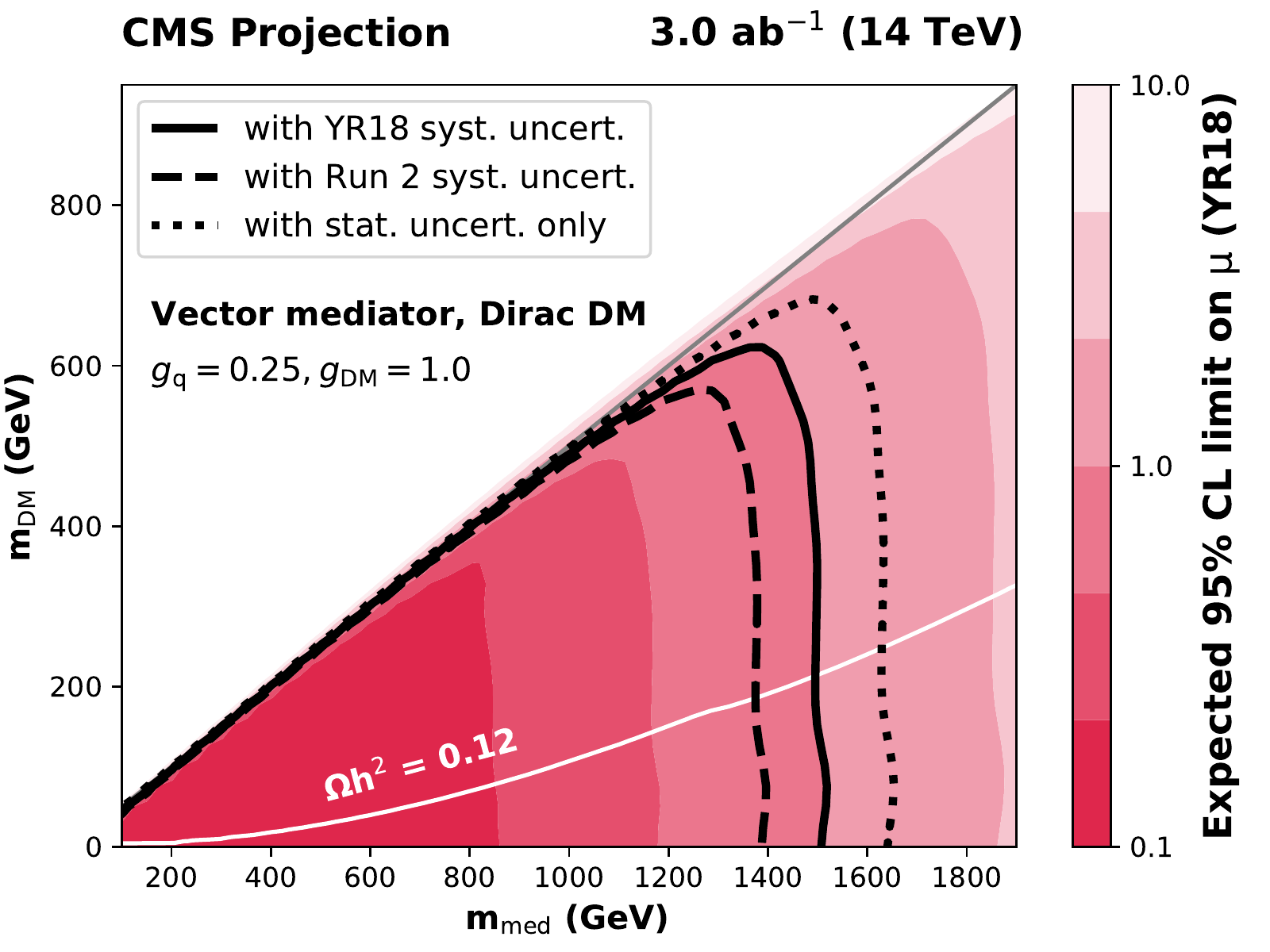}
\end{minipage}
\begin{minipage}[t][\minipageheightdp][t]{\minipagewidthdp}
\centering
\includegraphics[width=\figuremaxwidthdp,height=\figuremaxheightdp,keepaspectratio]{\main/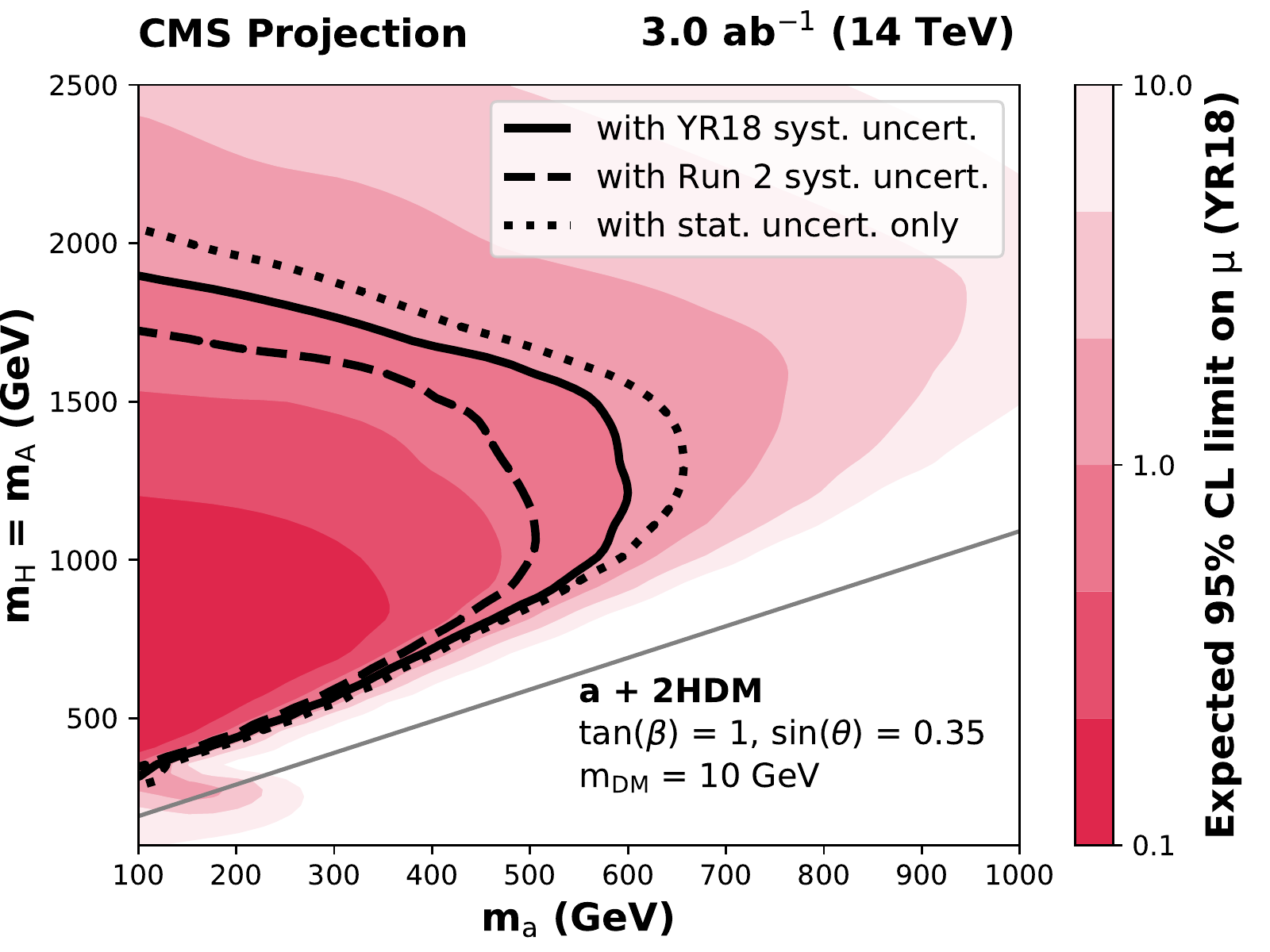}
\end{minipage}
        \caption{Expected $95\%\cl$ exclusion limits on the signal strength in the vector mediator (left) and a+2HDM scenarios (right). In the vector mediator case, the exclusion is presented in the plane of the mediator and dark matter masses, while the result is shown in the plane of the pseudoscalar mediator mass $m_\text{a}$ and the heavy boson masses $m_\text{H}=m_\text{A}$. The grey lines indicate the relevant kinematic boundaries that limit the sensitive regions: $m_\text{med} / 2 = m_\text{DM}$ in the vector mediator case, and $m_\text{H}=m_\text{a}+m_\text{Z}$ in the a+2HDM case. For the vector mediator scenario, the white line indicates the parameter combinations that reproduce the observed DM relic density in the universe~\cite{Ade:2015xua,Albert:2017onk}.}
        \label{fig:monoz_results_2d}
\end{figure}

For the vector mediator scenario, the expected signal significance and expected exclusion limits on $g_{q}$ are shown in \fig{fig:monoz_results_vector}.
A signal with a mediator of mass $m_\text{med}=750\GeV$ could be discovered with $\mathcal{L}_\text{int}\approx1\abinv$, while a heavier mediator with $m_\text{med}=1\TeV$ would require $\mathcal{L}_\text{int}\approx3\abinv$. Especially the latter case highlights the effect of the systematic uncertainty scenarios. Improved handling of systematic uncertainties could reduce the integrated luminosity required for a discovery by a factor three, and thus advance the discovery by years. Framed as an exclusion on the mediator-quark coupling $g_{q}$, values down to 0.04 will be probed for a lighter mediator with $m_\text{med}=300\GeV$, and $g_{q}\approx0.1$ will be testable for $m_\text{med}=1\TeV$. A heavier mediator of mass $m_\text{med}=2\TeV$ will remain out of reach even with the final HL-LHC dataset of $3\abinv$. The two-dimensional exclusion as a function of the relevant particle masses for both models is shown in \fig{fig:monoz_results_2d}. In the case of the vector mediator, mediator masses up to $\sim1.5\TeV$ will be probed, assuming $m_\text{med}/2>m_\text{DM}$. Depending on the choice of systematic uncertainty scenario, the mediator mass exclusion varies by $\approx100\GeV$. In the a+2HDM model, light pseudoscalar masses up to $600\GeV$ and heavy boson masses up to $1.9\TeV$ will be probed. Again, the choice of systematic uncertainty scenarios may influence these values by $\approx100\GeV$ ($m_\text{a}$) and $100-150\GeV$ ($m_\text{H}$). These mass exclusion ranges show an improvement of a factor $\sim 2.5$ in the mediator masses and up to $\sim 3$ for the pseudoscalar mass compared to the Run-2 result with $\mathcal{L}_\text{int}=36\fbinv$~\cite{Sirunyan:2017qfc,Sirunyan:2018gdw}.

\subsubsection{\atlas{Dark matter searches in mono-photon and VBF+$\met$ final states at HL-LHC}}
\label{sec:atlasDMmonophoton}
\contributors{L. Carminati, D. Cavalli, M. Cirelli, C. Guyot, A. Demela, I. Lim, B. Nachman, M. M. Perego, S. Resconi, F. Sala, ATLAS}

%\subsubsubsection*{\bf Overview} 
A prospect study for DM searches with the ATLAS detector is presented in a scenario where the SM is extended by the addition of an EW fermionic triplet with null hypercharge~\cite{ATL-PHYS-PUB-2018-038}. 
The lightest mass state of the triplet constitutes a weakly interacting massive particle DM candidate. 
This model is inspired by SUSY with anomaly-mediated SUSY breaking~\cite{Randall:1998uk,Giudice:1998xp,moroi:1999zb} and by models of Minimal Dark Matter (MDM)~\cite{Cirelli:2005uq,Cirelli:2007xd,Cirelli:2009uv}, 
and provides a benchmark in the spirit of simplified models~\cite{Abercrombie:2015wmb} where the mediator is a SM particle.
Projections for an integrated luminosity of \intLumi are presented for the DM searches in the mono-photon~\cite{Aaboud:2017dor} 
 and VBF$+\met$ ~\cite{Aaboud:2018sfi} final states, based on the Run-2 analyses strategy.
To illustrate the experimental challenges associated to a high pile-up environment due to the high luminosity,  the effect of the pile-up on the VBF invisibly decaying Higgs boson is studied as a benchmark process. 

\subsubsubsection*{\bf{EW fermionic WIMP Dark Matter triplet }}
A fermionic triplet $\chi$ of the $SU(2)_L$ group with null hypercharge ($Y$): 
%\begin{equation}
$\chi = \Bigg( 
\begin{array}{c}
\chi^+ \\
\chi_0 \\
\chi^- \\
\end{array}
\Bigg)$
%\end{equation}
\noindent
is added to the SM with a Lagrangian:
\begin{align*}
\nonumber \mathcal{L}_{MDM} & = \frac{1}{2} \bar{\chi}(i \slashed{D}+M)\chi \\
& = \frac{1}{2} \bar{\chi_0} (i \slashed{\partial} - M_{\chi^{0}}) \chi_0 + \bar{\chi^+} (i \slashed{\partial} - M_{\chi^+}) \chi^+  \\
& + g (\bar{\chi^+} \gamma_{\mu} \chi^+ (\sin\theta_W A_\mu +\cos\theta_W Z_{\mu})) + \bar{\chi^+} \gamma_\mu \chi_0 W^{-}_{\mu} + \bar{\chi_0}\gamma_\mu \chi^+ W^{+}_{\mu} &
\end{align*}
where \textit{g} is the $SU(2)$ gauge coupling; $M$ is the tree-level mass of the particle; $\sin\theta_W$ and $\cos\theta_W$ are the sine and cosine of the Weinberg angle; $A_\mu$, $Z_\mu$, $W_{\mu}$ are the SM boson fields.  The lightest component of the triplet is stable if some extra symmetry is imposed, like lepton number, baryon minus lepton number or a new symmetry under which $\chi$ is charged (\egg~R-parity in SUSY).

At tree level all the $\chi$ components have the same mass, but a mass splitting is induced by the EW corrections given by loops of SM gauge bosons between the charged and neutral components of $\chi$. These corrections make the charged components heavier than the  neutral one ($\chi_0$). %The neutral component ($\chi_0$) is therefore the lightest one and 
Its mass differs by %164 $\div$ 165 MeV 
$\simeq 165\MeV$~\cite{Ibe:2012sx} from the one of the charged components. 
Being neutral and stable, $\chi_0$  constitutes a potential DM candidate. 
If the thermal relic abundance is assumed, the mass of $\chi_0$ is $M_{\chi_0}\simeq 3\TeV$.  
However, if $\chi$ is not the only particle composing dark matter or if it is not thermally produced~\cite{moroi:1999zb}, its mass can be $M_{\chi_0} <3\TeV$. 

This model provides a benchmark of a typical WIMP DM candidate and its phenomenology recreates the one of supersymmetric models where the Wino is the lightest SUSY particle (LSP), for this reason this triplet is referred to as \textit{Wino-like}.
As studied in \citeref{Cirelli:2014dsa},  treating $M$ as a free parameter,  this triplet can be probed at the LHC in different ways.  Once produced, the charged components of the triplet decay into the lightest neutral component $\chi_0$ plus very soft  charged pions. $\chi_0$ is identified as $\met$ in the detector while the pions, because of the small mass splitting between the neutral and charged components,  are so soft that they are lost and not reconstructed.  
Therefore, the production of $\chi$ can be searched for by 
in mono-X events, such as mono-jet~\cite{Low:2014cba} and mono-photon; 
in VBF +$\met$ events as $\chi$ can also be produced via VBF~\cite{Berlin:2015aba};
and also  in events characterised by high $p_{T}$ tracks (caused by $\chi^{\pm}$) which end inside the detector once they have decayed into $\chi_0$ and soft pions, disappearing tracks~\cite{Feng:1999fu}.
The VBF production mode and the mono-photon final state, studied in this contribution, constitute a necessary complement to the mono-jet, discussed in \seccs{sec:atlasmonojet} and \ref{sec:theoryewinosmono}, and disappearing track searches, see \secc{sec:theoryewinosdisappear}, because of the very different dependencies on the model parameters like the EW representation and the value of the mass splitting.  
LEP limits exclude masses below $\sim 90\GeV$~\cite{Heister:2002mn,Abbiendi:2002vz,feng2010searches}, therefore the focus here is on $M_{\chi_0}\geq 90\GeV$. 

Signal events with a pair of $\chi$  produced in the framework of this model~\cite{Cirelli:2014dsa} have been  generated in the $\gamma+\met$  and VBF$+\met$ final state and simulated for different values of $\chi_0$ mass with the official ATLASFAST-II simulation of the current detector~\cite{atlas2010simulation} at $\sqrt{s}=13\TeV$. 
For the VBF+$\met$ analysis, diagrams not properly originating from two vector
bosons (in contrast to pure VBF processes) also contribute to the signal as they produce
a jets+$\met$ signature where the jets have large pseudorapidity separation.
To consider the realistic conditions at the HL-LHC, VBF H (H$\rightarrow ZZ^{*}\rightarrow \nu\bar{\nu}\nu\bar{\nu}$) events 
have been fully simulated, using \geant{ 4}~\cite{Aad:2010ah,Agostinelli:2002hh}, in the upgraded ATLAS detector 
including the upgraded inner tracker (ITk)~\cite{Collaboration:2257755,Collaboration:2285585}, with $\langle\mu\rangle=200$ and at $\sqrt{s}=14\TeV$.  

\subsubsubsection*{\bf{Mono-Photon final state }}
The mono-photon analysis is characterised by a relatively clean final state, containing a photon with a high transverse energy and large $\met$, which can be mimicked by few SM processes.
The search for new phenomena performed in mono-photon events 
 in $pp$ collisions at  $\sqrt{s}=13\TeV$ at the LHC, using data collected by the ATLAS experiment in Run-2 corresponding to an integrated luminosity of $36.1\fbinv$~\cite{Aaboud:2017dor}, has shown no deviations from the SM expectations. 
 The Run-2 mono-photon search is reinterpreted in the context of the WIMP triplet model at HL by keeping the same strategy for background estimates and event selection to exploit the full complexity of the analysis.
The dominant backgrounds consist in processes with a $Z$ or $W$ boson produced in association with a photon, mainly Z$(\rightarrow \nu \nu)$ + $\gamma$. They are estimated by rescaling the MC prediction for those backgrounds with factors obtained from a simultaneous fitting technique, based on control regions (CRs) built by reverting one or more cuts of the signal region such that one type of process becomes dominant in that region. Other backgrounds, like $W$/$Z$ + jet, top and diboson, in which electrons or jets can fake photons are estimated with data-driven techniques.
 
Events passing the lowest unprescaled single photon trigger are selected requiring \sloppy\mbox{$\met > 150\GeV$}.  The leading photon has to satisfy the ``tight" identification criteria and is required to have $p_\text{T}^{\gamma}>150\GeV$,  $|\eta|<2.37$ and to be isolated. The photon and $\met$  are required to be well separated, with $\Delta\phi(\gamma,~\met)  > 0.4$. Finally, events are required to have no electrons or muons and no more than one jet with  $\Delta\phi(\mathrm{jet},~\met) > 0.4$. 
In the  Run-2 analysis the  total background prediction uncertainty is dominated by the statistical uncertainty and the largest systematic uncertainties are due to the uncertainty in the rate of fake photons from jets and to the uncertainty in the jet energy scale. 

\begin{figure}[t]
\centering
\begin{minipage}[t][\minipageheightdp][t]{\minipagewidthdp}
\centering
\includegraphics[width=\figuremaxwidthdp+4mm,height=\figuremaxheightdp,keepaspectratio]{\main/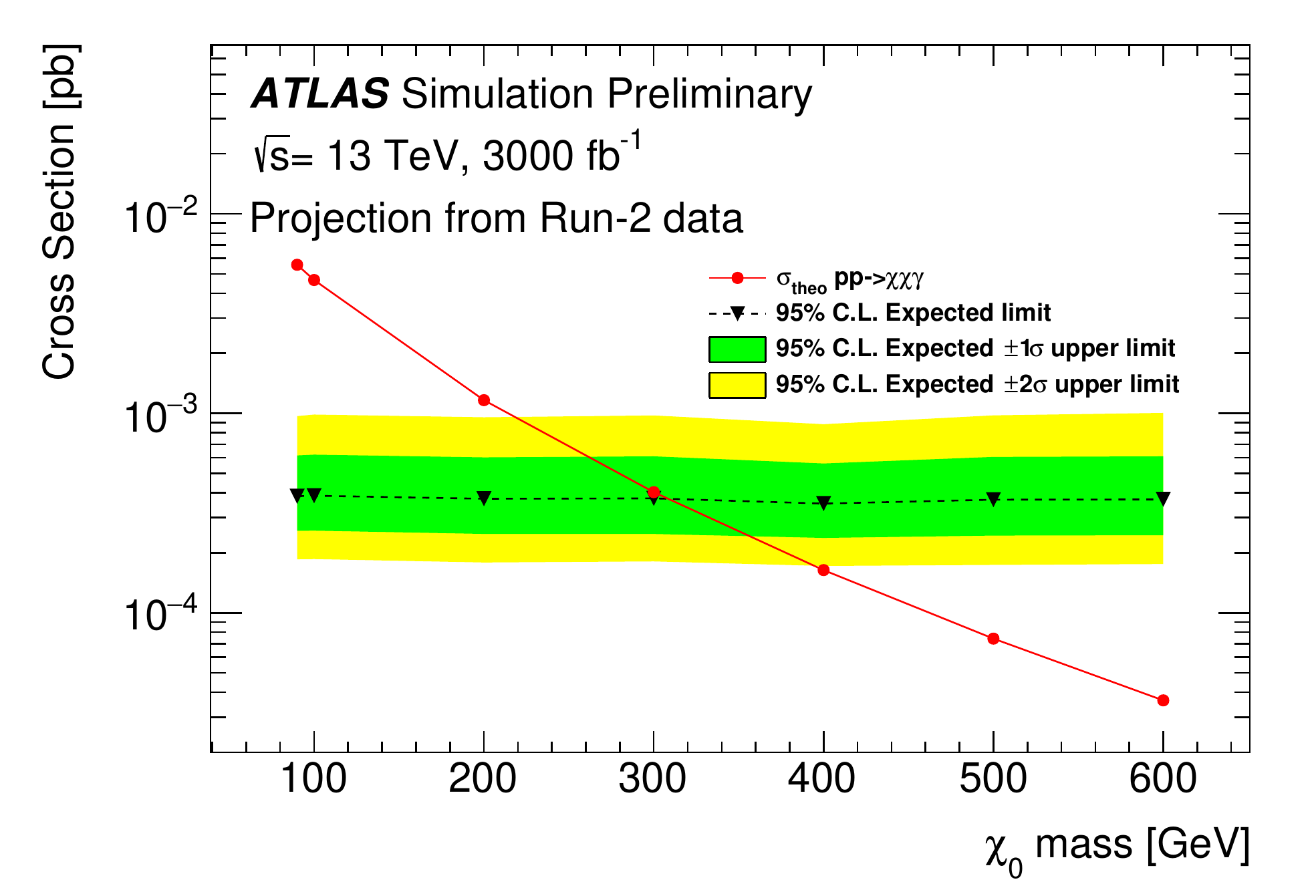}
\end{minipage}
\begin{minipage}[t][\minipageheightdp][t]{\minipagewidthdp}
\centering
\includegraphics[width=\figuremaxwidthdp,height=\figuremaxheightdp,keepaspectratio]{\main/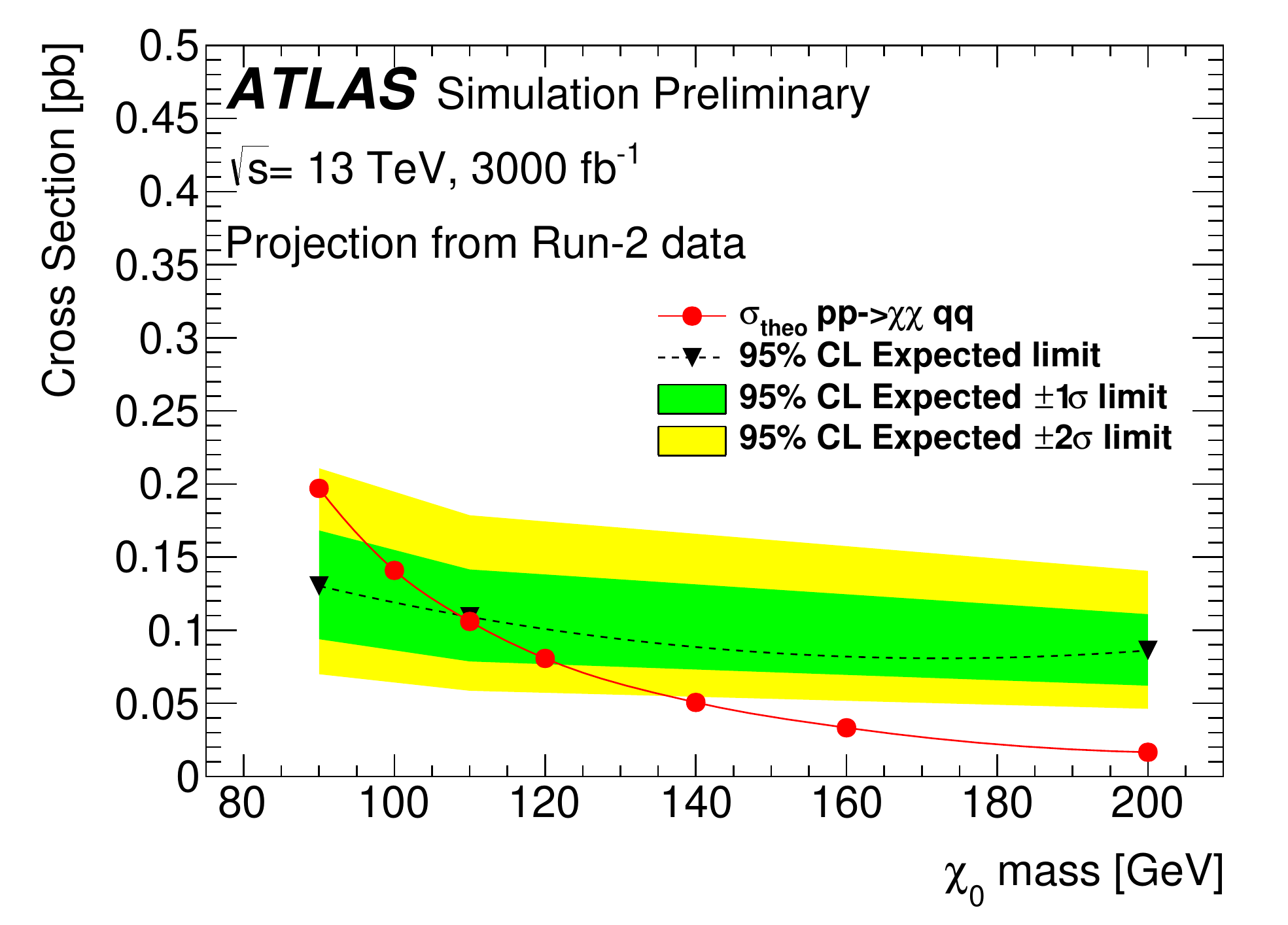}
\end{minipage}
  \caption{Expected upper limits at $95\%\cl$ on the production cross section of $\chi$ as a function of $\chi_0$ mass in 
   (left) mono-photon final state and (right) VBF+$\met$ final state.  Results are shown for an integrated luminosity of \intLumi. 
   The  red line shows the theoretical cross section.  
     \label{fig:Excl_3000fb_all}}
\end{figure}

The reinterpretation of the mono-photon analysis in the WIMP triplet model uses  full simulated MC signal samples  and performs a simultaneous fit on the most inclusive signal region (SR), corresponding to $\met>150\GeV$, that provides the best expected sensitivity.
All backgrounds, including fake photons estimated with data-driven techniques, have been  included in the fit rescaling the Run-2 results to the high luminosity scenario. 
All the systematic uncertainties on the MC background samples have been taken into account to obtain upper limits on the $\chi_0$  production cross section. 
Projections of the expected upper limits on the production cross section of $\chi_0$ at  $95\%\cl$ for an integrated luminosity of \intLumi and $\sqrt{s}=13 \TeV$, are shown in Figure~\ref{fig:Excl_3000fb_all} (left). 
Masses of $\chi_0$ below $310\GeV$ can be excluded at $95\%\cl$ by the analysis assuming the same systematic uncertainties adopted in \citeref{Aaboud:2017dor}. 
The impact of the systematic uncertainty on the sensitivity of the analysis has been checked considering that the analysis will no more be limited by the statistical uncertainty at high luminosity. In a scenario in which the current systematic uncertainties are halved, an exclusion of $\chi_0$ masses up to about 340 GeV could be reached. 
Thanks to the increased statistics, the analysis at high luminosity could be further optimised by performing a multiple-bin fit, thus on more bins in $\met$ improving the overall sensitivity of the analysis.   
This study is done for a \com~energy of $13\TeV$, a  slight improvement in the signal significance is expected from the increase of the \com~energy  to $14\TeV$ foreseen for the HL-LHC.

\subsubsubsection*{\bf VBF plus $\met$  final state }\label{sec:VBFplusMETfinalState}
The VBF+$\met$  topology is characterised by two quark-initiated jets with a large separation in rapidity and $\met$.  
 The sensitivity of the VBF+$\met$ analysis to the WIMP triplet model is presented as a reinterpretation of the Run-2 results for the high luminosity scenario foreseen for the HL-LHC. As pile-up is a key experimental challenge for event reconstruction in the VBF topology at the HL-LHC, a dedicated study of its impact is also shown using VBF $H\to$invisible as benchmark.
\vskip 0.5cm

\noindent
\emph{Projections at high luminosity for DM for EW triplet DM.} \\
\noindent
A search for an invisibly decaying Higgs boson produced via VBF has been performed by ATLAS using a dataset corresponding to an integrated luminosity of 36\fbinv of \textit{pp} collision at $\sqrt{s}=13\TeV$ \cite{Aaboud:2018sfi}.
The final state is defined by the presence of two energetic jets,  largely separated in $\eta$ and with $\mathcal{O}(1)\TeV$ invariant mass, and large
 $\met$. 

This analysis set limits on the BR $\BR$ of the H$\rightarrow\text{invisible}$. 
The main backgrounds arise from  $Z\to\nu\nu$+jets and $W\to\ell\nu$+jets events. 
The contribution of $W/Z$ is estimated from events in CRs enriched in $W\to\ell\nu$ (where the lepton is found) and $Z\to\ell\ell$ (with $\ell$ being electrons or muons) that are used to normalise the MC estimates to data 
through a simultaneous fitting technique and to extrapolate the normalisation  to the SR.   
The multijet background comes from multijet events where large $\met$ is generated mainly by jet mismeasurements.
This is highly reduced by a tight $\met$ cut and is estimated via data-driven methods resulting in less than $1\%$ of the total background. 

The results are interpreted in the context of the WIMP model for an integrated luminosity of \intLumi.
 The same selections and analysis strategy  are used to set limits on the cross section of the WIMP triplet produced via VBF; 
the only selection which has been changed is the request on the separation in pseudorapidity between the two leading jets ($\Delta\eta(j_1,j_2)$) which has been relaxed from $>4.8$ to $>3.5$, thus  increasing the sensitivity to the model as, in addition to the pure VBF Feynman diagrams, also diagrams with strong production contribute to the signal. 
A SR is defined by selecting events passing the lowest unprescaled $\met$ trigger, containing no electron and muon, having exactly two jets with transverse momentum $p_T(j_1)> 80\GeV$ and $p_T(j_2)> 50\GeV$, which are not back to back in the transverse plane ($\Delta\Phi(j_1,j_2)<1.8$) and which are  separated in pseudorapidity ($\Delta\eta(j_1, j_2)>3.5$). Events are required to have large $\met$ ($ >180\GeV$), the two leading jets are separated from the $\met$ ($\Delta\Phi(j_1, ~\met)>1$, $\Delta\Phi(j_2, ~\met)>1$), the vectorial sum of all the jets (including the pile-up ones) is required to be $H_\text{T}^\text{miss} >150\GeV$ and the invariant mass of the dijet system is required to be $M(j_1,j_2)>1\TeV$.
%As done in the Run-2 analysis,  
The events in SR and in CRs are then split into three categories (\textit{bins}) according to the invariant mass of the dijet system; the following $M(j_1,j_2)$ bins are considered: $1-1.5\TeV$, $1.5-2\TeV$ and $>2\TeV$. 

A simultaneous fit in SR and CRs, using the three $M(j_1,j_2)$ bins to increase the signal sensitivity,  is used for the $W/Z +$ jets background estimation and for the limit setting. 
Exclusion limits are set on the production cross section of the model using a one-sided profile likelihood ratio and the CLs technique~\cite{Read:2002hq,Junk:1999kv} with the  asymptotic approximation~\cite{Cowan:2010js}. 
Experimental and theoretical systematic uncertainties  have been taken into account and are included in the likelihood as  Gaussian-distributed nuisance parameters. 
The main experimental systematic uncertainties for the Run-2 VBF+$\met$ analysis come from JES and JER~\cite{Aaboud:2017jcu} and have been rescaled according to the HL expectations which are discussed in \citeref{ATLAS_PERF_Note}. 
The main theoretical sources of uncertainty for the run-2 analysis come from choices on the resummation, renormalisation, factorisation and CKKW matching scale for the $W/Z+$jets backgrounds processes. A significant improvement in these systematic uncertainties is expected; 
therefore, the current run-2 theoretical systematic uncertainties on the $W/Z+$jets backgrounds have been rescaled down to reach the level of few $\%$ ($5\%$ of the run-2 theoretical systematic uncertainties is kept).
Here is assumed that such an improvement in the theoretical systematics for VBF final state will be reached for the HL-LHC phase.
The same correlation scheme that has been used in \citeref{Aaboud:2018sfi} is also used for the projections presented here. 
Uncertainties arising from the finite MC statistics of the samples used are assumed to be negligible. 

The results obtained by rescaling the signals and backgrounds to an integrated luminosity of \intLumi are shown in 
Figure~\ref{fig:Excl_3000fb_all} (right) and indicate that the lowest masses considered ($M_{\chi} \sim 110\GeV$) can be excluded at $95\%~\cl$.
This study is done for a \com~energy of $13\TeV$, a slight improvement in the signal significance is expected from the increase of the 
\com~energy  to $14\TeV$ foreseen for the HL-LHC.
The analysis is very sensitive to the systematic uncertainties and a further optimisation of the selection cuts on this model, together with the increase of  the MC statistics in the VBF phase space, could help to achieve a better reach. 

VBF analyses will probably  benefit from a combination of $\met$ and VBF jet triggers; however, even with $\met$ thresholds raised by $50-100\GeV$ with respect to the current ones, the analysis is still sensitive to this model for the masses considered. 
\vskip 0.5cm
\noindent
\emph{The challenge of pile-up for VBF at HL-LHC.}\\
\noindent
In the study of the pile-up effects for VBF at the HL-LHC,  jets are built from particle flow objects~\cite{Aaboud:2017aca} using the anti-$k_t$ algorithm with radius parameter $R=0.4$ as implemented in \fastjet{}; 
they are only considered if $p_\text{T}>25\GeV$ and $|\eta | <  4.5$.
Charged particle tracks are reconstructed from hits in the Itk.  
Tracks are %ghost 
associated to the jets and required to have $p_\text{T}>0.9\GeV$ and $p_\text{T} < 40\GeV$ (to suppress fake tracks). 
The difference between the primary vertex (this is the vertex with the highest $\sum p_\text{T}^2$) $z$ position and the track $z_0$ (longitudinal impact parameter) must be less than $2\sigma$, where $\sigma$ is the sum in quadrature of the track $z_0$ and the vertex $z$ uncertainties. 

One of the key discriminating observables between pile-up jets and hard-scatter jets is $R_{p_\text{T}}$~\cite{Aad:2015ina}, which is the sum of the $p_\text{T}$ of the tracks associated to the jet normalised by the jet $p_\text{T}$.  Only tracks with $\Delta R <0.3$ are considered in the calculation of $R_{p_\text{T}}$.  
Jets are declared `hard-scatter' if $R_{p_\text{T}}>0.05$ which corresponds to $85\%$ hard-scatter efficiency and $2\%$ pile-up jet efficiency when $|\eta| < 1.2$ and $|z^\text{reco}-z^\text{true}|<0.1$.  
$\met$  is critical to the $H\rightarrow\text{invisible}$ search; as an optimisation for  the $\met$ reconstruction for the upgraded ATLAS detector is not yet available,  the negative sum of the transverse momenta of all reconstructed jets ($E_\text{T,jet}^\text{miss}$) is used in this analysis.  

Due to limitations of MC statistics, a simplified version of the Run-2 VBF $H\rightarrow\text{invisible}$ analysis is used.  In particular, all of the angular requirements with jets are removed and there is no binning in $M(j_1,j_2)$ and $E_\text{T,jet}^\text{miss}$ is required to be $>150\GeV$.
For the Run-2 analysis, the event selection efficiency for $Z\rightarrow \nu\bar{\nu}$ events is about $2\times 10^{-6}$ and about $0.5\%$ for the signal with a \sloppy\mbox{$\BR(H\rightarrow\text{invisible})=100\%$} (which is about $85\%$ from VBF).  
Contrary to the Run-2 analysis, here the ggF $H\rightarrow\text{invisible}$ contribution has been neglected.
The background is nearly half QCD $Z\rightarrow \nu\bar{\nu}$ and half QCD $W$+jets. Since only $Z$+jets are used in this analysis, BR limits are computed by doubling the $Z$+jets background.   It is likely that with the extended coverage of the ITk relative to the current tracker  the lost leptons will be suppressed and thus the $W$+jets background will be less than the $Z$+jets rate so this approximation is conservative.  

A simplified statistical analysis is performed to assess the impact of several scenarios on the \sloppy\mbox{$H\rightarrow\text{invisible}$} BR limit with the full HL-LHC dataset.  A one-bin statistical test with one overall source of systematic uncertainty is performed to determine if a particular signal yield is excluded.  The signal yield is scanned to determine the largest BR that would be not excluded at the $95\%\cl$.  
Table \ref{table:pileup} presents the limits on the $H\rightarrow\text{invisible}$ BR normalised to the one for the run-2 systematic uncertainties and truth-based pile-up tagging to show the relative gains and losses under various pile-up scenarios, 
corresponding to the three rows, 
and with different assumptions on the systematic uncertainties, corresponding to the four columns: $10\%$ (similar to run-2), $5\%$, 
$5\%$ assuming the same signal efficiency as in Run-2 and the background efficiency to be $10\%$ of the signal efficiency; and finally $1\%$.
With a realistic reduction in the systematic uncertainty and tighter selection criteria, it may be possible to significantly improve the sensitivity. 
The limit improves from including a simple $R_{p_\text{T}}$-based pile-up jet rejection, though the gap with the truth-information-based tagger indicates that there is room (and reward) for developing a more sophisticated approach.  
\begin{table}[t]
	\centering
	\bgroup
\def\arraystretch{1.5}
	\small\begin{tabular}{cc|c|c|c|c|} %
	\cline{3-6}
	& & \multicolumn{4}{c|}{Systematic Uncertainties}  \\ \cline{3-6}
 	\multicolumn{2}{c|}{ $\frac{\BR(H\rightarrow \text{invs.})}{\BR_\text{Truth, Nominal}(H\rightarrow \text{invs.})}$} & 10\%& 5\%& 5\% + fixed efficiency & 1\% \\ \cline{1-6}
	 \multicolumn{1}{|c|}{\parbox[t]{3mm}{\multirow{3}{*}{\rotatebox[origin=c]{90}{PU jet rejec}}}}&None & -- & -- & 0.31 & 0.59\\ \cline{2-6}
	\multicolumn{1}{|c|}{}& $R_{p_\text{T}}$  & --& --& 0.28& 0.48\\ \cline{2-6}
	\multicolumn{1}{|c|}{}& Truth & 1.0& 0.48& 0.07& 0.10\\ \hline
	\end{tabular}\normalsize
	\egroup
	   \caption{The limit on the $H\rightarrow\text{invisible}$ BR using the full HL-LHC dataset (\intLumi) normalised to the one for the Run-2 systematic uncertainties and truth-based pile-up tagging to show the relative gains and losses possible under various scenarios.  A `--' indicates a value bigger than 1.  
}   
	   \label{table:pileup}
\end{table}

As an overall  conclusion, with a combination of pile-up robustness studies, analysis optimisation, and theory uncertainty reduction, 
the EW triplet DM searches at the HL-LHC, both in mono-photon and VBF+$\met$  final states, may be significantly improved.

\subsubsection{\theoryC{Search for Higgs portal dark matter models at HL- and HE-LHC}} 
\label{sec:theoryhiggsportal}
\contributors{Y. G. Kim, C. B. Park, S. Shin}
%$^{3,4}$} \\
%$^1$Department of Science Education, Gwangju National
%    University of Education, Gwangju 61204, Korea email:{\tt ygkim@gnue.ac.kr}\\
%$^2$Center for Theoretical Physics of the Universe,
%Institute for Basic Science (IBS), Daejeon 34051, Korea email:{\tt cbpark@ibs.re.kr}\\
%$^3$Enrico Fermi Institute, University of Chicago, Chicago, IL 60637, USA\\
%$^4$Department of Physics \& IPAP, Yonsei University, Seoul 03722, Korea email:{\tt seodongshin@yonsei.ac.kr}
%\paragraph*{Introduction}

In a variety of BSM models, the Higgs boson is often considered as a particle mediating the interactions between the dark matter and the SM particles, dubbed as {\it Higgs portal}.
The Higgs portal models can be categorised by the spin of DM as scalar, fermionic, or vector Higgs portal models.

The scalar Higgs portal models consider a SM singlet scalar DM ($S$) which has interactions with the SM Higgs doublet ($H$) as~\cite{Silveira:1985rk,Burgess:2000yq} $S S^\ast H H^\dagger$, which is a four-dimensional operator. 
This type of models is often considered as a simplest reference DM model so that there exist various complementary searches combining the results from the direct detection, indirect detection, and the LHC.

The second category of Higgs portal models is the fermionic Higgs portal model considering a fermion ($\psi$) as DM.
The interaction term between the DM ($\psi$) and the SM Higgs doublet can be effectively given as~\cite{Kim:2006af}
%--
\begin{align}
\frac{\psi \bar \psi H H^\dagger}{\Lambda}\,,
\label{eq:fermionic}
\end{align}
%--
assuming the interaction is mediated by additional heavy particle(s) with mass scale $\Lambda$ and the DM $\psi$ is a Dirac fermion.
The searches at the LHC, \eg mono-jet with missing energy, provide constraints directly to $\Lambda$ and the mass of DM.
Because this is a five-dimensional operator, a renormalisable simplified model was first introduced in \citeref{Kim:2008pp} by adding a SM real singlet scalar $S$ which mixes
between the SM Higgs doublet and the singlet scalar.
Beyond the minimal set-up in \citeref{Kim:2008pp}, one can also consider the SM scalar field $S$ a complex scalar, pseudo scalar field~\cite{LopezHonorez:2012kv,Fedderke:2014wda,Ghorbani:2014qpa,Kim:2016csm,Kim:2018uov}.
Equipped with the mixing and the existence of additional mediator (singlet-like mass eigenstate), this kind of model has been widely used in consistently explaining various experimental/observational results within the context of DM, such as possible direct detection experimental anomalies for light DM region~\cite{Kim:2009ke}, $\gamma$-ray observation from the Galactic Centre~\cite{Kim:2016csm,Kim:2018uov}, baryon-antibaryon asymmetry~\cite{Chao:2015uoa}, and so on.

The third category of Higgs portal models is the vector DM model with interaction term between the DM ($V^\mu$) and the SM Higgs doublet~\cite{Lebedev:2011iq} $V^\mu V_\mu H H^\dagger$.
The vector dark matter can be, \eg a $U(1)$ vector field which gets a mass term though the Stueckelberg mechanism.

In this report, we focus on the Dirac fermion DM model as a benchmark model and show proper strategies searching for the signals at HL- and HE-LHC.
Note that the results would apply similarly to other kind of models, \iee~scalar or vector DM models.

\textbf{Benchmark model:} As a benchmark model we choose the Singlet Fermionic Dark Matter (SFDM) \sloppy\mbox{model~\cite{Kim:2008pp,Kim:2009ke,Shin:2010gn,Kim:2016csm,Kim:2018uov}} because the analysis methods and results are readily applicable to other type of models.
The SFDM has a dark sector composed of a SM singlet real scalar field $S$ and a singlet Dirac fermion field $\psi$ which is the DM candidate.
The dark sector Lagrangian with most general renormalisable interactions is given by
%--
\begin{align}
  \mathcal{L}^{\rm dark} = \bar\psi (i \partial  \!\!\!\!/
   - m_{\psi_0}) \psi + \frac{1}{2} \partial_\mu S \partial^\mu S  - g_S
  (\cos\theta \,\bar\psi \psi + \sin\theta \,\bar\psi i \gamma^5 \psi) S
  - V_S (S, \, H),
  \label{eq:lagrangian}
\end{align}
where
\begin{align}
  V_S (S, \, H) = \frac{1}{2} m_0^2 S^2 + \lambda_1 H^\dagger H S +
  \lambda_2 H^\dagger H S^2 + \frac{\lambda_3}{3!} S^3 +
  \frac{\lambda_4}{4!} S^4 .
\end{align}
%--
The interactions of the singlet sector to the SM sector arise
only through the Higgs portal $H^\dagger H$ as given above.
Note that the Lagrangian in \eq{eq:lagrangian} generally includes both scalar and pseudoscalar interaction terms in the singlet sector, in contrast to the basic model in \citerefs{Kim:2008pp,Kim:2009ke,Kim:2006af}, following the set-up explaining the galactic $\gamma$-ray signal~\cite{Kim:2016csm,Kim:2018uov}.~\footnote{See also \citeref{Fedderke:2014wda}.}

The SM Higgs potential is given as $V_\mathrm{SM} = -\mu^2 H^\dagger H + \lambda_0 (H^\dagger H)^2$ and the Higgs boson gets a \vev after electroweak symmetry breaking (EWSB), $v_h = \simeq 246\GeV$.
The singlet scalar field generically develops a \vev, $v_s$, and hence we can expand $S = v_s + s$.
There is mixing between the states $h, s$%\mdo{commented out matrix, they should be in references}
%
%\begin{align}
%  \mathcal{L}_\mathrm{mass} = -\frac{1}{2}
%  \begin{pmatrix}
%    h & s
%  \end{pmatrix}
%  \begin{pmatrix}
%    \mu_h^2 & \mu_{hs}^2\\
%    \mu_{hs}^2 & \mu_s^2
%  \end{pmatrix}
%  \begin{pmatrix}
%    h \\ s
%  \end{pmatrix}
%  \label{eq:higgs_mass_matrix}
%\end{align}
%
and the physical mass states are admixtures of
$h$ and $s$, %such that
%%
%\begin{align}
%  \begin{pmatrix}
%    h_1 \\ h_2
%  \end{pmatrix} = \begin{pmatrix}
%    \cos\theta_s & \sin\theta_s\\
%    -\sin\theta_s & \cos\theta_s
%  \end{pmatrix} \begin{pmatrix}
%    h \\ s
%  \end{pmatrix} ,
%\end{align}
%
where the mixing angle is determined by $\tan\theta_s =y/(1 + \sqrt{1 + y^2})$ with $y \equiv 2\mu_{hs}^2 / (\mu_h^2 - \mu_s^2)$.
The expressions of each matrix element in terms of the Lagrangian parameters are given in \citeref{Kim:2016csm,Kim:2018uov}.
Then, the tree-level Higgs boson masses are obtained as
\begin{align}
  m_{h_1,\,h_2}^2 = \frac{1}{2} \left[ (\mu_h^2 + \mu_s^2) \pm (\mu_h^2 -
  \mu_s^2) \sqrt{1 + y^2} \right] ,
\end{align}
where we assume that $h_1$ corresponds to the SM-like Higgs boson.

An interesting feature of this model is that there are extra scalar self-interaction terms.
The cubic self-couplings $c_{ijk}$ for $h_i h_j h_k$ interactions, \iee~$c_{111} h_1^3/3! + c_{112} h_1^2 h_2/2 + c_{122} h_1 h_2^2/2 +  c_{222} h_2^3/3!$, are functions of the scalar couplings, vacuum expectation values, and $\theta_s$, where the exact forms are written in \citeref{Kim:2016csm,Kim:2018uov}.
Note that $c_{112}$ is proportional to $\sin\theta_s$ due to the fact that $\lambda_1 + 2 \lambda_2 v_s$ is also proportional to $\sin\theta_s$, while the other couplings can remain non-vanishing.

%--
\begin{table}[t]
\centering
\scriptsize
\begin{tabular}{|c|c|c|c|}\hline
%\diagbox[width=10em]{$h_i$ production}{Signals}  
&  $E \!\!\!\!/_T^{\rm DM}$ + jet & $E \!\!\!\!/_T^{\rm DM}$ + ($h_1 / h_2 \to$ SM particles) & No $E \!\!\!\!/_T^{\rm DM}$ \\
\hline
%\hline
Single $h_1$  & $p p \to h_1 \to \psi \bar \psi$ & N/A & SM Higgs precision \\
\hline
Single $h_2$ & $p p \to h_2 \to \psi \bar \psi$ & N/A & $p p \to h_2 \to$ SM ($m_{h_2}$) \\
\hline
\multirow{2}{*}{Double $h_1 h_1$: $c_{111}$}  & \multirow{2}{*}{$p p \to h_1 \to h_1 h_1 \to 4\psi$} & \multirow{2}{*}{$p p \to h_1 \to h_1 h_1 \to 2\psi$ + SM}  & $p p \to h_1 \to h_1 h_1$ \\
 & & & (double Higgs production) \\
\hline
\multirow{2}{*}{Double $h_1 h_1$, $h_1 h_2$: $c_{112}$} & $p p \to h_2 \to h_1 h_1 \to 4\psi$ & $p p \to h_2 \to h_1 h_1 \to 2\psi$ + SM & $p p \to h_2 \to h_1 h_1$ ($m_{h_2}$)\\
 & $p p \to h_1 \to h_1 h_2 \to 4\psi$ & $p p \to h_1 \to h_1 h_2 \to 2\psi$ + SM ($m_{h_2}$) & $p p \to h_1 \to h_1 h_2$ ($m_{h_2}$)\\
\hline
\multirow{2}{*}{Double $h_2 h_2$, $h_1 h_2$: $c_{122}$} & $p p \to h_1 \to h_2 h_2 \to 4\psi$ & $p p \to h_1 \to h_2 h_2 \to 2\psi$ + SM ($m_{h_2}$) & $p p \to h_1 \to h_2 h_2$ ($m_{h_2}$)\\
 & $p p \to h_2 \to h_1 h_2 \to 4\psi$ & $p p \to h_2 \to h_1 h_2 \to 2\psi$ + SM ($m_{h_2}$) & $p p \to h_2 \to h_1 h_2$ ($m_{h_2}$)\\
\hline
Double $h_2 h_2$: $c_{222}$ & $p p \to h_2 \to h_2 h_2 \to 4\psi$ & $p p \to h_2 \to h_2 h_2 \to 2\psi$ + SM ($m_{h_2}$) & $p p \to h_2 \to h_2 h_2$ ($m_{h_2}$) \\
\hline
\end{tabular}
\caption{Production channels of $h_1 / h_2$, dubbed as ``$h_i$ production", categorised by the signal types. $E \!\!\!\!/_T^{\rm DM}$ is the missing transverse energy originated from the DM, ``SM particles" means the SM particles produced from the decay of the Higgs bosons ($h_1$ and $h_2$), and $m_{h_2}$ means that we can observe a $h_2$ resonance signal.
}
\label{table:signals}
\end{table}
%--

%%%
\textbf{Signals at the LHC:} The search strategies for SFDM at the LHC rely on the production methods of the SM-like Higgs $h_1$ and the singlet-like Higgs $h_2$. Hence, we first categorise the production channels of $h_1 / h_2$ as following.
%---
\begin{itemize}
\item Single $h_1 / h_2$ production from the Yukawa or gauge interactions
\item Double $h_1 / h_2$ production from the scalar self-interactions
\end{itemize}
%---
The first category implies the conventional single Higgs production mechanisms such as gluon fusion, vector boson fusion, $t \bar t h$, Higgsstrahlung, \etcc.
Hence, for single $h_1$ production, the precise measurements of the SM Higgs production mechanisms, dubbed as {\it Higgs precision}, would provide the indirect hints of the SFDM, unless $h_1$ decays to DM pair.
The second category includes exotic signatures depending on the values of trilinear couplings %in \eq{eq:cijk} 
so we further divide the production channels affected by each coupling.

As a next step, the signal type should be classified for each production channel of $h_1 / h_2$.
Here, we categorise the signal types as
%--
\begin{itemize}
\item $E \!\!\!\!/_T^{\rm DM}$ + jets
\item $E \!\!\!\!/_T^{\rm DM}$ + ($h_1 / h_2 \to$ SM particles)
\item No $E \!\!\!\!/_T^{\rm DM}$
\end{itemize}
%--
where $E \!\!\!\!/_T^{\rm DM}$ is the missing energy from the DM production, defined to separate  from the missing energy from the neutrino production in the SM, \eg $Z \to \nu \bar \nu$.
Such a missing energy signal from the neutrino production belongs to the last category, ``No $E \!\!\!\!/_T^{\rm DM}$".

Table \ref{table:signals} summarises the production channels of $h_1 / h_2$ and the suitable signal type for each channel.
As stated above, the double Higgs production channels are further classified by the trilinear coupling involved in the process.
In many channels, searches for exotic resonance signals by the decay of $h_2$, remarked as ($m_{h_2}$), can be effective methods in probing the Higgs portal DM models.
As long as the DM mass is larger than $m_{h_2}/2$, the BRs of $h_2 \to$ SM particles are the same as those expected for the hypothetical SM Higgs (with an arbitrary mass not restricted to $\sim 125\GeV$).
Note that all the processes can occur altogether so one needs a combined analysis to confirm the scenario.

\begin{figure}[t]
\centering
\begin{minipage}[t][\minipageheightdp][t]{\minipagewidthdp}
\centering
\includegraphics[width=\figuremaxwidthdp,height=\figuremaxheightdp,keepaspectratio]{\main/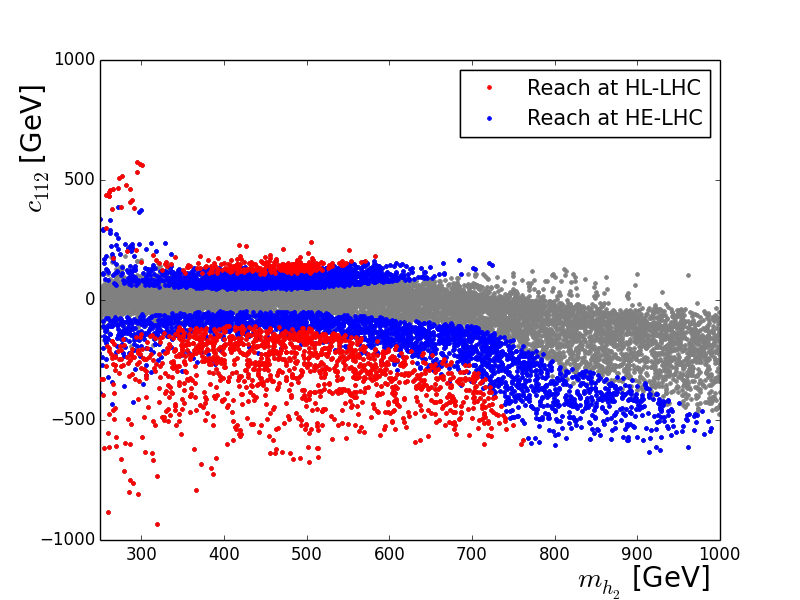}
\end{minipage}
\begin{minipage}[t][\minipageheightdp][t]{\minipagewidthdp}
\centering
\includegraphics[width=\figuremaxwidthdp,height=\figuremaxheightdp,keepaspectratio]{\main/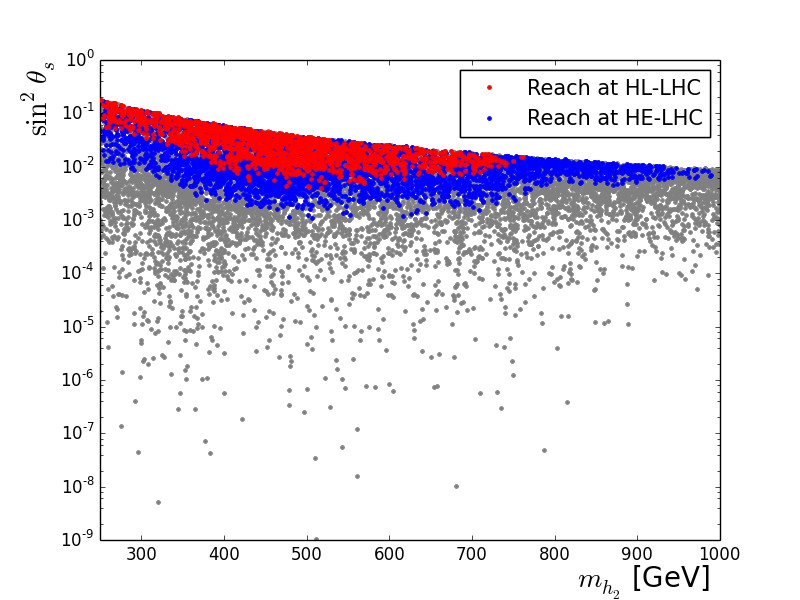}
\end{minipage}
\caption{Expected parameter reach by searching for on-shell $p p \to h_2 \to h_1 h_1$ at HL-LHC (HE-LHC) is shown with red (blue) scatter points in the $m_{h_2} - c_{112}$ (left) and $m_{h_2} - \sin^2\theta_s$ (right) plane.
The gray scatter points are out of the reach of HE-LHC but satisfying the stability of the scalar potential $V_S (S, \, H) + V_{\rm SM}$ and the current constraints from \citeref{Aaboud:2018knk}.
}
\label{fig:c112}
\end{figure}

As preliminary but simple examples, we analyse the expected sensitivities of the on-shell processes $p p \to h_2 \to h_1 h_1$ and $p p \to h_1 \to h_2 h_2$, in the signal type ``No $E \!\!\!\!/_T^{\rm DM}$, assuming the Higgs bosons are produced on-shell.
For the on-shell process $p p \to h_2 \to h_1 h_1$, we apply the search results for heavy scalar into $h_1 h_1 \to b \bar b b \bar b$ in ATLAS~\cite{Aaboud:2018knk} with $36.1\fbinv$ data at $\sqrt{s} = 13\TeV$.
The sensitivity at \sloppy\mbox{HL-LHC} is estimated from rescaling the upper limits in \citeref{Aaboud:2018knk} by $\sqrt{36.1/3000}$ assuming the number of background events at $\sqrt{s} = 13\TeV$ and $\sqrt{s} = 14\TeV$ are similar.
We also assume the signal significance is well approximated by signal/$\sqrt{\rm background}$ where the statistical uncertainty is dominant.
On the other hand, it is non-trivial to obtain the sensitivity at HE-LHC.
For simplicity, we only consider the ratio of the dominant background events (multi-jets~\cite{Aaboud:2018knk}) at $\sqrt{s} = 27\TeV$ and $\sqrt{s} = 13\TeV$, which is $2.9$ from running \madgraph{5NLO}.  %MadGraph5$_{\rm NLO}$.
In \fig{fig:c112}, we show the expected reach at HL-LHC (HE-LHC).
% with red (blue) scatter points in $m_{h_2} - c_{112}$ (left) and $m_{h_2} - \sin^2\theta_s$ plane.
The gray scatter points are out of the reach of HE-LHC.
Note that all the scatter points satisfy the stability of the scalar potential $V_S (S, \, H) + V_{\rm SM}$ and the current constraints from \citeref{Aaboud:2018knk}.
From this result, we conclude that HL-LHC (HE-LHC) can constrain the parameter $|c_{112}|$ up to $100$ ($50$), which correspond to $\sin^2\theta_s \sim 0.004$ ($0.001$).

%%%
%%%

%%%
\begin{figure}[t]
\centering
\begin{minipage}[t][\minipageheightdp][t]{\minipagewidthdp}
\centering
\includegraphics[width=\figuremaxwidthdp,height=\figuremaxheightdp,keepaspectratio]{\main/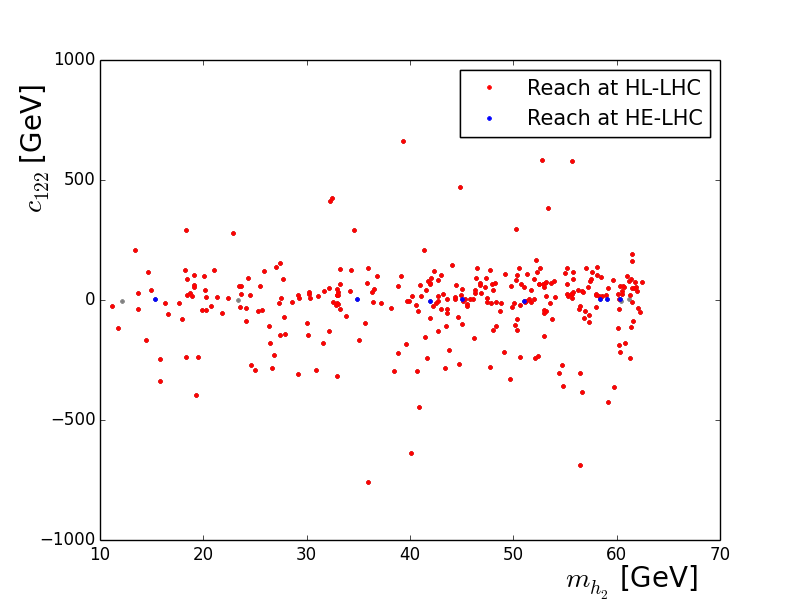}
\end{minipage}
\begin{minipage}[t][\minipageheightdp][t]{\minipagewidthdp}
\centering
\includegraphics[width=\figuremaxwidthdp,height=\figuremaxheightdp,keepaspectratio]{\main/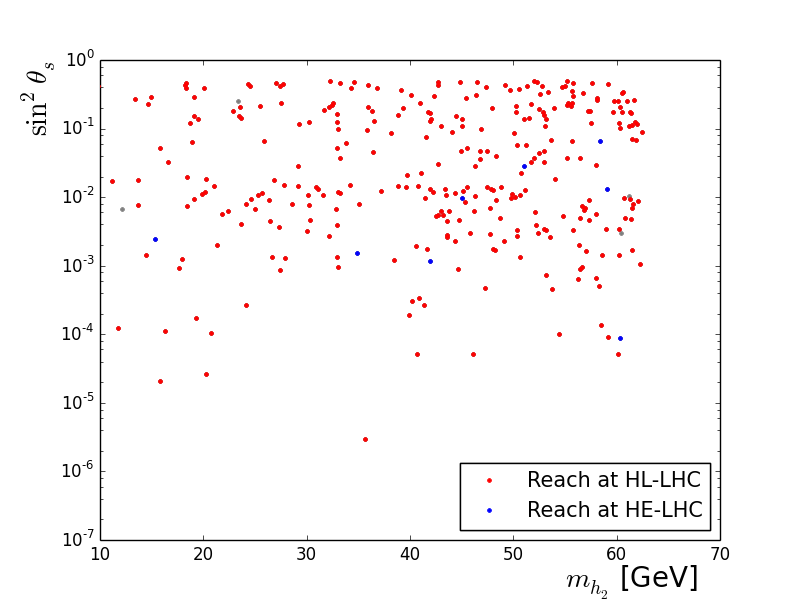}
\end{minipage}
\caption{Expected parameter reach by searching for on-shell $p p \to h_1 \to h_2 h_2$ (exotic Higgs decay) at HL-LHC (HE-LHC) is shown with red (blue) scatter points in the $m_{h_2} - c_{122}$ (left) and $m_{h_2} - \sin^2\theta_s$ (right) plane.
The gray scatter points are out of the reach of HE-LHC but satisfying the stability of the scalar potential $V_S (S, \, H) + V_{\rm SM}$ and the current constraints from \citeref{Aaboud:2018iil}.
}
\label{fig:c122}
\end{figure}
%%%

For the on-shell process $p p \to h_1 \to h_2 h_2$, we apply the search results for \sloppy\mbox{$p p \to W h_1 \to W h_2 h_2 \to \ell \nu_\ell b \bar b b \bar b$} or $p p \to Z h_1 \to Z h_2 h_2 \to \ell \bar \ell b \bar b b \bar b$ in ATLAS~\cite{Aaboud:2018iil} with $36.1\fbinv$ data at $\sqrt{s} = 13\TeV$.
Since the dominant background is $t \bar t$ + light jets, we apply the aforementioned method in obtaining the sensitivities at HL-LHC and HE-LHC for simplicity.
In \fig{fig:c122}, we show the expected reach at HL-LHC (HE-LHC).% with red (blue) scatter points in $m_{h_2} - c_{122}$ (left) and $m_{h_2} - \sin^2\theta_s$ plane.
The gray scatter points are out of the reach of HE-LHC but satisfying the stability of the scalar potential $V_S (S, \, H) + V_{\rm SM}$ and the current constraints from \citeref{Aaboud:2018iil}.
Interestingly, we estimate that HL-LHC can cover most of the parameter space unless $|c_{122}|$ is as small as $\sim 5\GeV$.

In summary, in this section we have explained the SFDM as a reference Higgs portal model. Since the existence of the extra scalar $h_2$ is a key ingredient of the model, we categorise the possible production channels of the two Higgs bosons ($h_1$ and $h_2$) and suitable signal types at the LHC.
As a simple example, we estimate the experimental sensitivities of HL-LHC and HE-LHC for two simple on-shell processes $p p \to h_2 \to h_1 h_1$ and $p p \to h_1 \to h_2 h_2$ with several naive assumptions.

\subsubsection{\theoryC{Singlet dark matter with slepton-like partners at HL- and HE-LHC}}
\label{sec:theorysingletDM}
\contributors{M. J. Baker and A. Thamm}

In this contribution we use a simplified model framework to 
explore the prospects for the HL and HE-LHC to probe 
viable multi-TeV dark matter.  We consider 
a bino motivated (gauge-singlet Dirac or Majorana fermion) dark matter candidate accompanied by 
$n$ dark-sector scalars with unit hypercharge, table \ref{tab:particles}.    
We consider the three possible Yukawa couplings with SM electrons, muons and taus 
individually.
A pure singlet with no other nearby states cannot efficiently annihilate, 
resulting in overclosure of the universe.  However, when dark sector scalars are included, the 
observed relic abundance can be recovered for a relatively wide range of masses. 
In \citeref{Baker:2018uox} we determine the couplings which produce the observed abundance 
of dark matter and calculate the reach of a range of present and future direct detection, 
indirect detection and collider experiments.
In this summary, we will see that there is a large region of viable parameter space for Majorana dark matter 
which only future colliders, such as the HL- and HE-LHC, can probe.

%--------------------------------------------------------------------%
\begin{table}[t]
  \centering
 % \begin{minipage}{13cm}
\small\begin{tabular}{cccccc}
Field & Spin & $su(3)\times su(2)_L \times u(1)_Y$ & $\mathbb{Z}_2$ & Copies & DOF\\
\hline
$\chi$ & 1/2 & (1,1,0) & -1 & 1 & 4\\
$\phi_i$ & 0 & (1,1,-1) & -1 & $n$ & $2n$
\end{tabular}\normalsize
 % \end{minipage}
  \caption{The new particles we introduce with their respective charges, the number of copies we consider and the number of degrees of freedom.}
  \label{tab:particles}
\end{table}
%--------------------------------------------------------------------%

In addition to kinetic and mass terms, the Lagrangian only has one new interaction term (ignoring the scalar quartic, which plays no role in our phenomenology)
\begin{align}
\mathcal{L} \supset \overline{\chi} (i \slashed \partial - m_\chi) \chi + \frac{1}{2} |D_\mu \phi_i|^2 - \frac{1}{2} m_\phi^2 \phi_i^2 + (y_\chi \phi_i \overline{\chi} \ell_R + h.c. ) \,,
\end{align}
where $D_\mu = \partial_\mu - i g' Y B_\mu$ and the coupling is taken to be universal, \iee{} $y_\chi$ is the same for all $\phi_i$. We consider the cases $\ell_R = e_R, \mu_R$ and $\tau_R$, and assume that all $\phi_i$ have the same mass, $m_{\phi_i} = m_\phi$, and that $m_\chi < m_\phi$. We parametrise their mass splitting by $\Delta = (m_\phi-m_\chi)/m_\chi$.
For illustration, we focus on $n \in \{1, 10\}$. In a supersymmetric context, the DM particle $\chi$ would correspond to a bino  and the scalar $\phi_i$ can be identified with a right-handed slepton. In SUSY, the number of degrees of freedom of one right-handed slepton corresponds to $n=1$, all right-handed sleptons corresponds to $n=3$, while all right-and left-handed sleptons correspond to $n=9$. 
We calculate the relic abundances in our models using \micromegas{\,v4.3.5}~\cite{Belanger:2014vza}, 
and then restrict the new Yukawa couplings to lie on the relic surface when considering the reach of various experiments.

%----------------------------------------------------------------------------

\begin{figure}[t]
\centering
\begin{minipage}[t][\minipageheightdp][t]{\minipagewidthdp}
\centering
\raisebox{0.25\height}{\includegraphics[width=0.6\textwidth,height=\figuremaxheightdp,keepaspectratio]{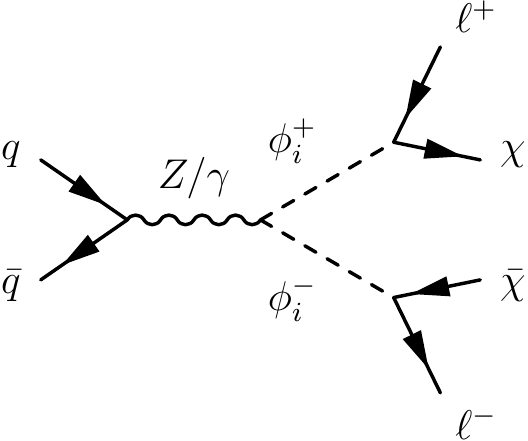}}
\end{minipage}
\begin{minipage}[t][\minipageheightdp][t]{\minipagewidthdp}
\centering
\includegraphics[width=\figuremaxwidthdp,height=\figuremaxheightdp,keepaspectratio]{\main/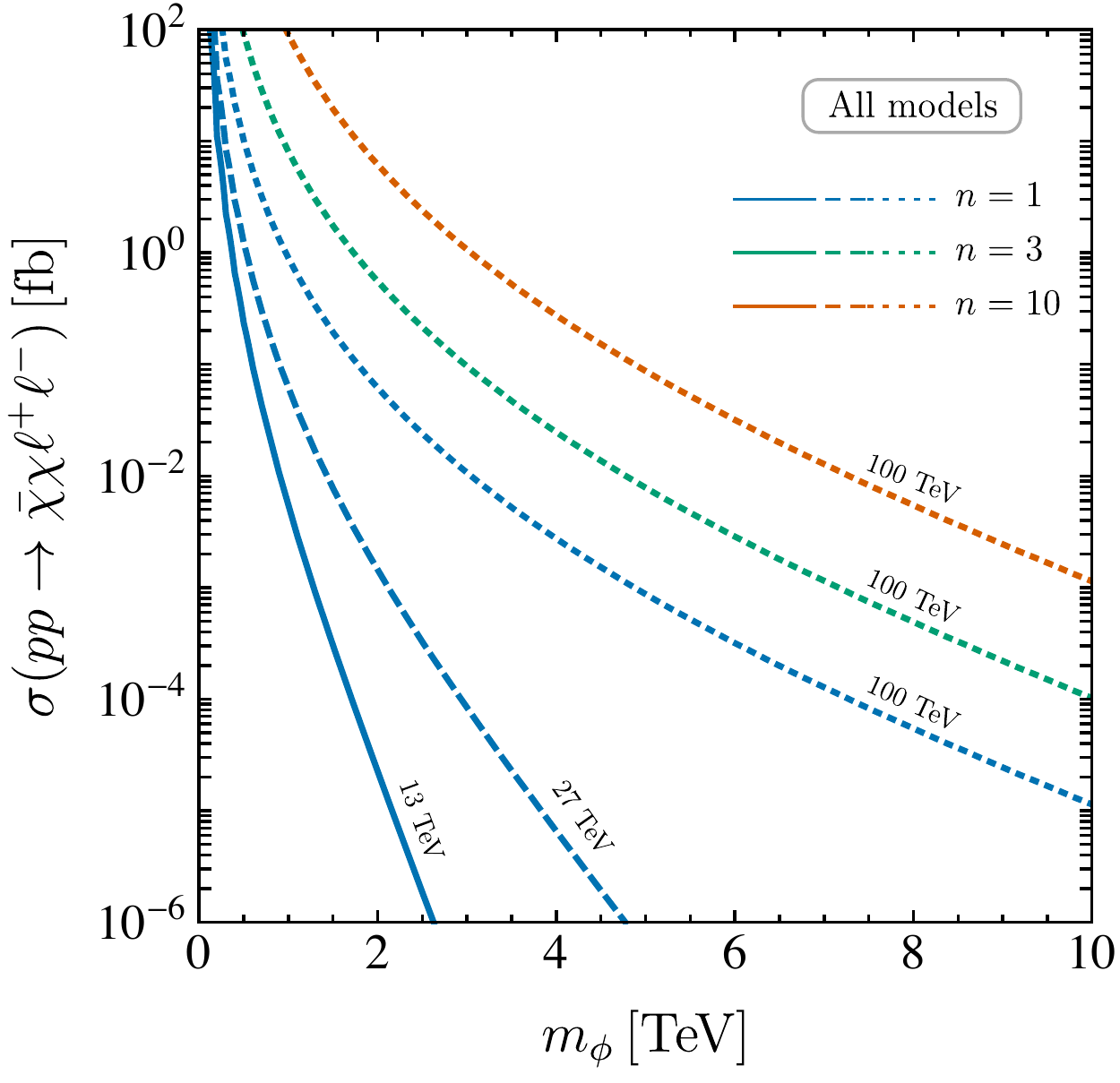}
\end{minipage}
  \caption{Leading order partonic process contributing to $p p \to \phi^+ \phi^- \to \bar \chi \chi \ell^+ \ell^-$ (left) and 
  its cross-section at $13\TeV$, $27\TeV$ and $100\TeV$ including a $K$-factor of 2 (right).
  }
  \label{fig:collider-diagrams}
\end{figure}

%----------------------------------------------------------------------------

It is challenging to search for our dark matter models directly at a hadron collider, since the dark matter is a gauge singlet which only couples to leptons. The coannihilation partner, $\phi_i^\pm$, however is a charged scalar of similar mass.  It will be pair produced in the process $p p \to \phi_i^+ \phi_i^-$ with a subsequent decay to a lepton, $\ell$, and $\chi$, depicted in \fig{fig:collider-diagrams} (left), where $\BR(\phi_i^\pm \to \chi \ell^\pm)=1$. We focus on final states containing two opposite-sign same-flavour leptons and missing energy. As $\tau$ reconstruction at future colliders is particularly challenging to model, we do not provide collider limits for the $\tau$ models. However, it is clear that the collider reach on $\tau$ models will be somewhat worse than the limits on the models involving electrons and muons.

We present sensitivity projections for the HE-LHC with $\sqrt{s} = 27\TeV$ assuming an integrated luminosity of \intlumihelhc~\cite{HELHC} and for the FCC-hh with $\sqrt{s} =100\TeV$ and $20\abinv$~\cite{Hinchliffe:2015qma}.
We estimate the sensitivity of future colliders to our models by adapting the analysis 
used in \citeref{Aad:2014vma} to search for slepton pair production with subsequent decay to neutralinos and leptons.

The signal $pp \to \phi^+ \phi^-$ is simulated using a custom \sarah{\,v4.12.1}~\cite{Staub:2008uz} model, we generate the signal and background parton level events using \madgraph{5\,v2.6.2}~\cite{Alwall:2014hca}, simulate the showering using \pythia{ 6.4.28}~\cite{Sjostrand:2006za} and perform the detector simulation with \delphes{\,v3.3.3}~\cite{deFavereau:2013fsa}.
For our $27\TeV$ simulations, we use the default \delphes{} card.  For the simulations at $100\TeV$ we use the FCC \delphes{} card implementing the configurations proposed by the FCC working group~\cite{FCCDelphes}. For the signal simulation, we adapt the card to treat the DM particle as missing energy. We use the LO partonic production cross-sections and multiply by a generous $K$-factor of 2, as we want to find the exclusion limits in the optimistic case, \fig{fig:collider-diagrams} (right). To validate our analysis, we reproduce the relevant backgrounds in \citeref{Aad:2014vma}.

The main SM backgrounds to our signal are $WW$,  $VV$, $WV$, $t \bar{t}$, $Wt$ and $V$+jets, where $V = Z, \gamma$. While only $WW$ and $VV$ are irreducible backgrounds, $WV$, $t \bar{t}$ and $Wt$ contribute if a lepton or one or two $b$-jets are missed. The $V$+jets background is important at low values of $m_{\text{T}2}$, but is negligible above $m_{\text{T}2} \approx 100\GeV$. In order to isolate the signal, we impose the following cuts.  Two opposite-sign same-flavour light leptons are required with $p_T > 35\GeV$ and $p_T > 20\GeV$ for leading and subleading leptons, respectively. We veto events with any other leptons, which reduces the $WV$ background. Removing events with $m_{\mu\mu} < 20\GeV$ and $|m_{\mu\mu} - m_Z | < 10\GeV$ significantly reduces backgrounds with a $Z$-boson in the final state. Finally we cut on the transverse mass~\cite{Lester:1999tx, Barr:2003rg}, $m_{\text{T}2} > 200\GeV$.  
For a process where two particles each decay to a lepton and missing energy, the $m_{\text{T}2}$ distribution will have an end point at the mass of the heavier particle~\cite{Barr:2009jv}.  Although in \citeref{Aad:2014vma} a cut of $m_{\text{T}2} > 90\GeV$ is used, we increase this to $m_{\text{T}2} > 200\GeV$.  This has a small effect on our signal efficiency, as we are mostly interested in dark matter candidates with mass larger than $200\GeV$, while strongly reducing the background from $t \bar{t}$, $Wt$.  However, even with this large cut, we find a significant background from $WW$, $WV$ and $VV$, where at least one of the vector bosons is extremely off-shell.  
To include this effect in \madgraph{} we simulate $pp \rightarrow \ell^+ \ell^- \nu_\text{all} \bar{\nu_\text{all}}$  
and $pp \rightarrow \ell^+ \ell^- \ell_\text{all} \nu$, where $\nu_\text{all}$ is $\nu_e$, $\nu_\mu$ or $\nu_\tau$ 
and $\ell_\text{all}$ is any charged lepton.
We do not find a similar large contribution from off-shell particles in the $t \bar{t}$ and $Wt$ channels. Even though the cross-section of these gluon initiated channels grows faster than the di-boson processes as the collider energy is increased, they remain a subdominant background as the $t$ is narrower and 
as this background only passes the cuts if a jet is missed, reducing the $m_{\text{T}2}$ endpoint. 
Finally, we checked that the contribution from jets faking muons is negligible.
In table \ref{tab:cut-flow-1} we show the cross-sections at each stage in the analysis 
for the background and for an example signal, $m_\chi = 0.6 \TeV$, $\Delta = 0.34$ ($27\TeV$) and $m_\chi = 0.8 \TeV$, $\Delta = 0.2$ ($100\TeV$), both for the $n=10$ muon type model.

%--------------------------------------------------------------------%
\begin{table}[t]
  \centering
  \begin{minipage}{16cm}
\small\begin{tabular}{ccccccc}
 Channel& \multicolumn{2}{c}{$\mu^+ \mu^- \nu_\text{all} \bar{\nu_\text{all}}$} & 
 \multicolumn{2}{c}{$\mu^+ \mu^- l_\text{all} \nu$}  & 
 \multicolumn{2}{c}{Example Signal}\\
 Energy [TeV]& $27$  & $100$ & $27$ & $100$ & $27$ & $100$\\
\hline
No Cuts & 
$2100$ & $6900$ & $560$ & $1800$ & $17$ & $100$\\
\begin{minipage}{7cm}$p_T^{\mu_1 (\mu_2)} > 35 (20)\GeV$ \& Lepton veto\end{minipage}& 
$1100$ & $620$ & $120$ & $160$ & $12$ & $14$\\
Jet veto & 
$690$ & $530$ & $45$ & $61$ & $3.3$ & $9.4$\\
\begin{minipage}{7cm}$m_{\mu\mu}>20\GeV$ \& $|m_{\mu\mu} - m_Z | > 10\GeV$\end{minipage}&  
$470$ & $370$ & $6.6$ & $13$ & $3.3$ & $8.9$\\
$m_{\text{T}2} > 200\GeV$ & 
$0.26$ & $0.44$ & $0.022$ & $0.076$ & $1.3$ & $2.5$
\end{tabular}\normalsize
  \end{minipage}
  \caption{Cross-sections at each stage in fb.  The example signals are for the 
  muon type model with $n=10$ for the parameter points 
  $m_\chi = 0.6 \TeV$, $\Delta = 0.34$ ($27\TeV$) and $m_\chi = 0.8 \TeV$, $\Delta = 0.2$ ($100\TeV$).}
  \label{tab:cut-flow-1}
\end{table}
%--------------------------------------------------------------------%

\begin{figure}[t]
\centering
\begin{minipage}[t][\minipageheightdp][t]{\minipagewidthdp}
\centering
\includegraphics[width=\figuremaxwidthdp,height=\figuremaxheightdp,keepaspectratio]{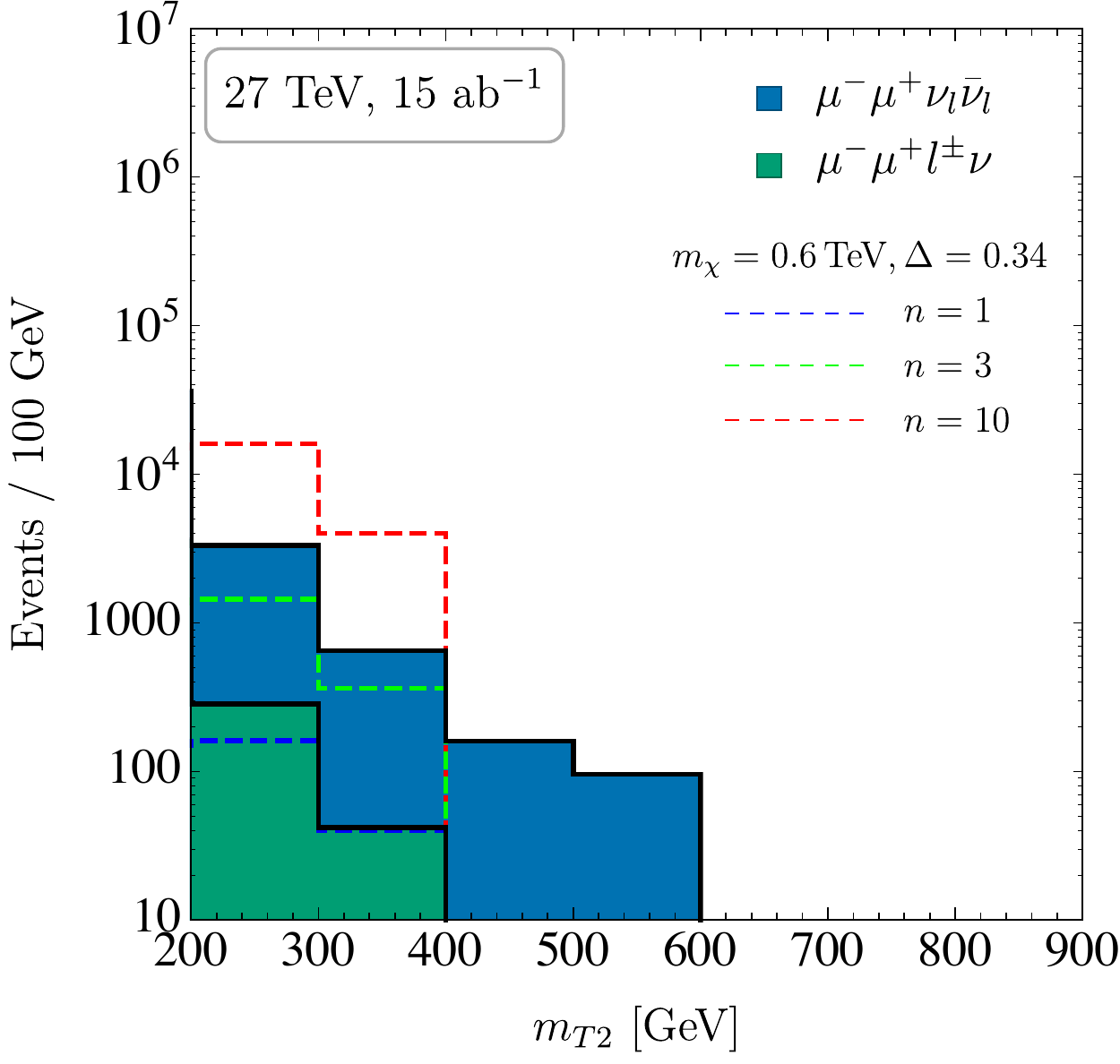}
\end{minipage}
\begin{minipage}[t][\minipageheightdp][t]{\minipagewidthdp}
\centering
\includegraphics[width=\figuremaxwidthdp,height=\figuremaxheightdp,keepaspectratio]{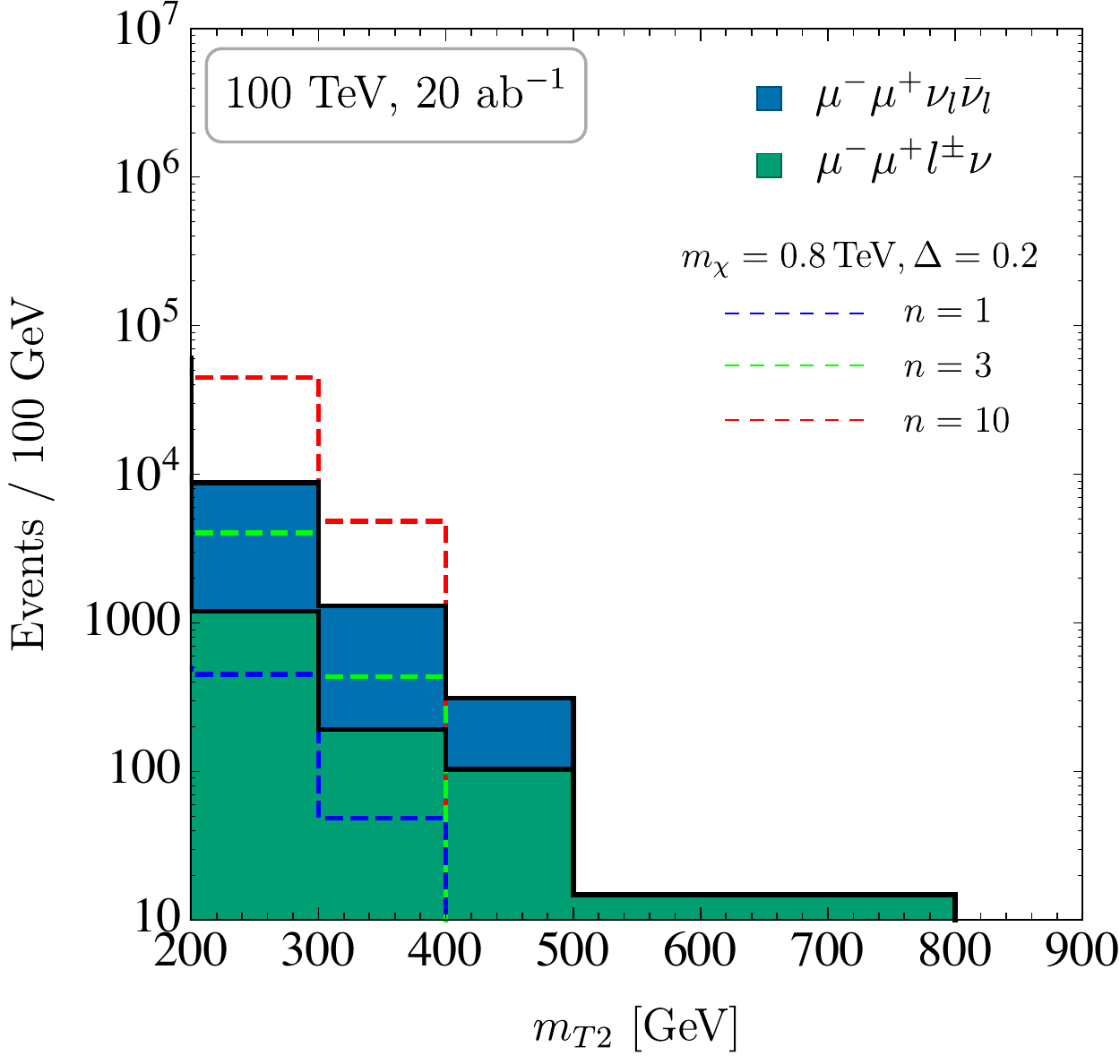}
\end{minipage}
  \caption{$m_{\text{T}2}$ distribution for background events passing all cuts for the muon model, 
  and an example signal for $n=1,3,10$, at $27\TeV$ (left) and $100\TeV$ (right).  We do not use this information 
  in determining the reach, 
  but simply perform a cut-and-count analysis based on these events.}
  \label{fig:muon-mt2}
\end{figure}

In \fig{fig:muon-mt2} we show the differential distribution in $m_{\text{T}2}$ for the muon-type model 
for the events passing all cuts, for the background and example signal.  
We see that $\mu^+ \mu^- \nu_\text{all} \bar{\nu_\text{all}}$ is the dominant background, 
and $\mu^+ \mu^- \ell_\text{all} \nu$ is around an order of magnitude smaller.  This is due to 
both the smaller initial cross-section and the smaller efficiency.  We see that both the background and the example signal falls sharply from $m_{\text{T}2} = 200 \GeV$ to $m_{\text{T}2} \approx 500\GeV$.
However, the signal will continue to higher values of $m_{\text{T}2}$ for other points in our parameter space.
We also see that at $27\TeV$, the $\mu^+ \mu^- \nu_\text{all} \bar{\nu_\text{all}}$ continues out to higher values 
of $m_{\text{T}2}$, while at $100\TeV$ the situation is reversed.

%--------------------------------------------------------------------%

\begin{figure}[t!]
\centering
\begin{minipage}[t][\minipageheightdp][t]{\minipagewidthdp}
\centering
\includegraphics[width=\figuremaxwidthdp,height=\figuremaxheightdp,keepaspectratio]{\main/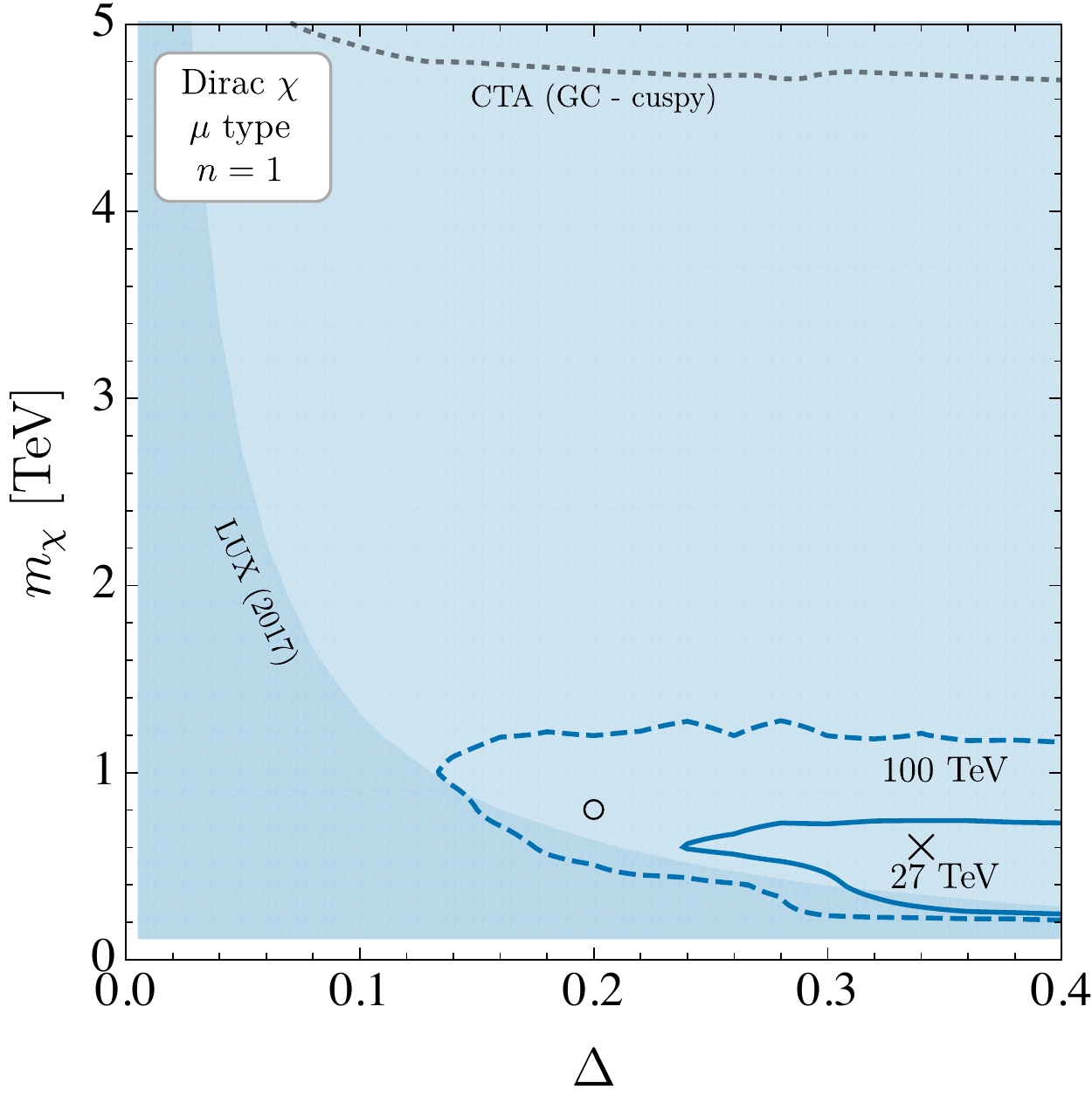}
\end{minipage}
\begin{minipage}[t][\minipageheightdp][t]{\minipagewidthdp}
\centering
\includegraphics[width=\figuremaxwidthdp,height=\figuremaxheightdp,keepaspectratio]{\main/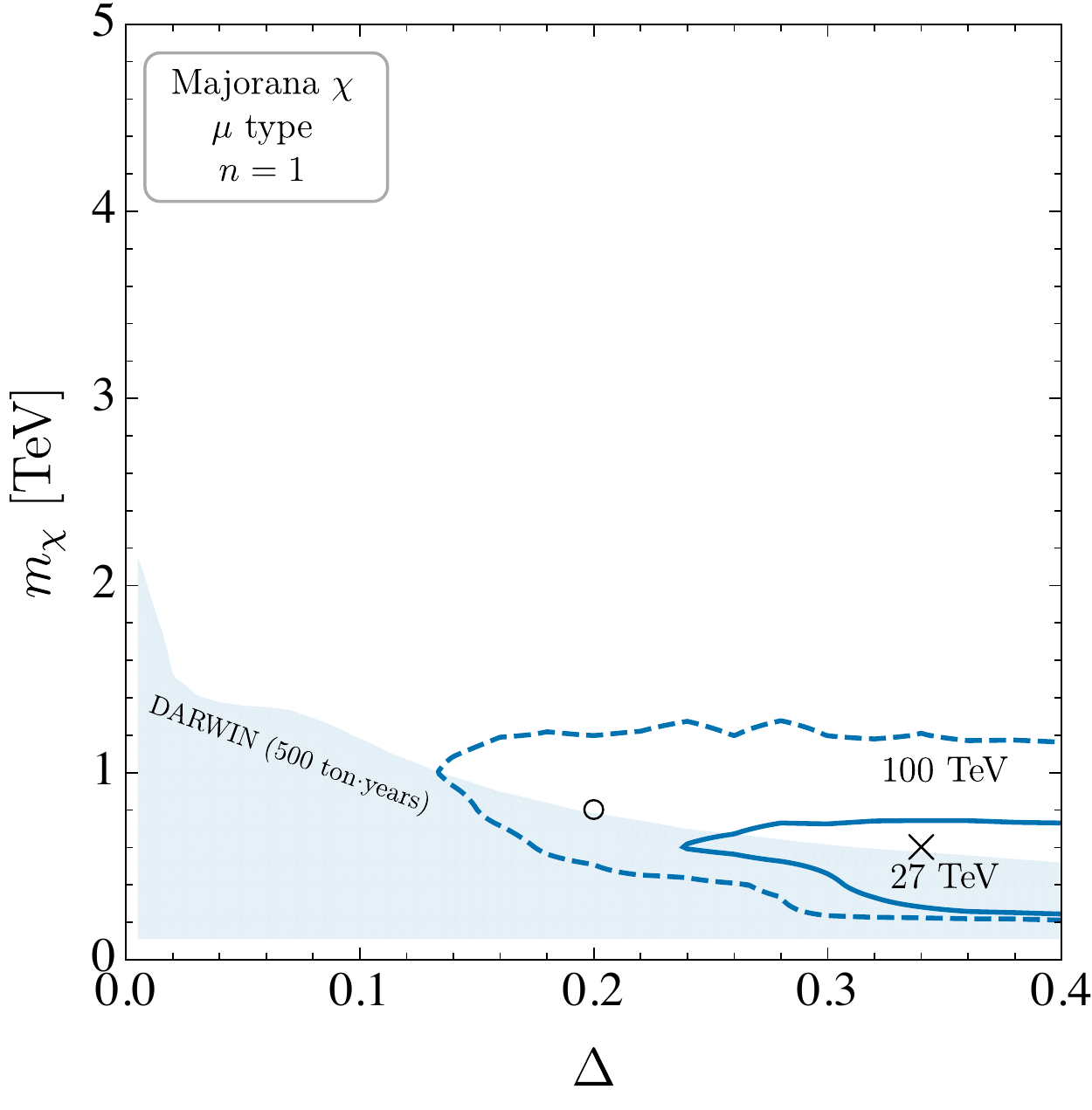}
\end{minipage}\\
\begin{minipage}[t][\minipageheightdp][t]{\minipagewidthdp}
\centering
\includegraphics[width=\figuremaxwidthdp,height=\figuremaxheightdp,keepaspectratio]{\main/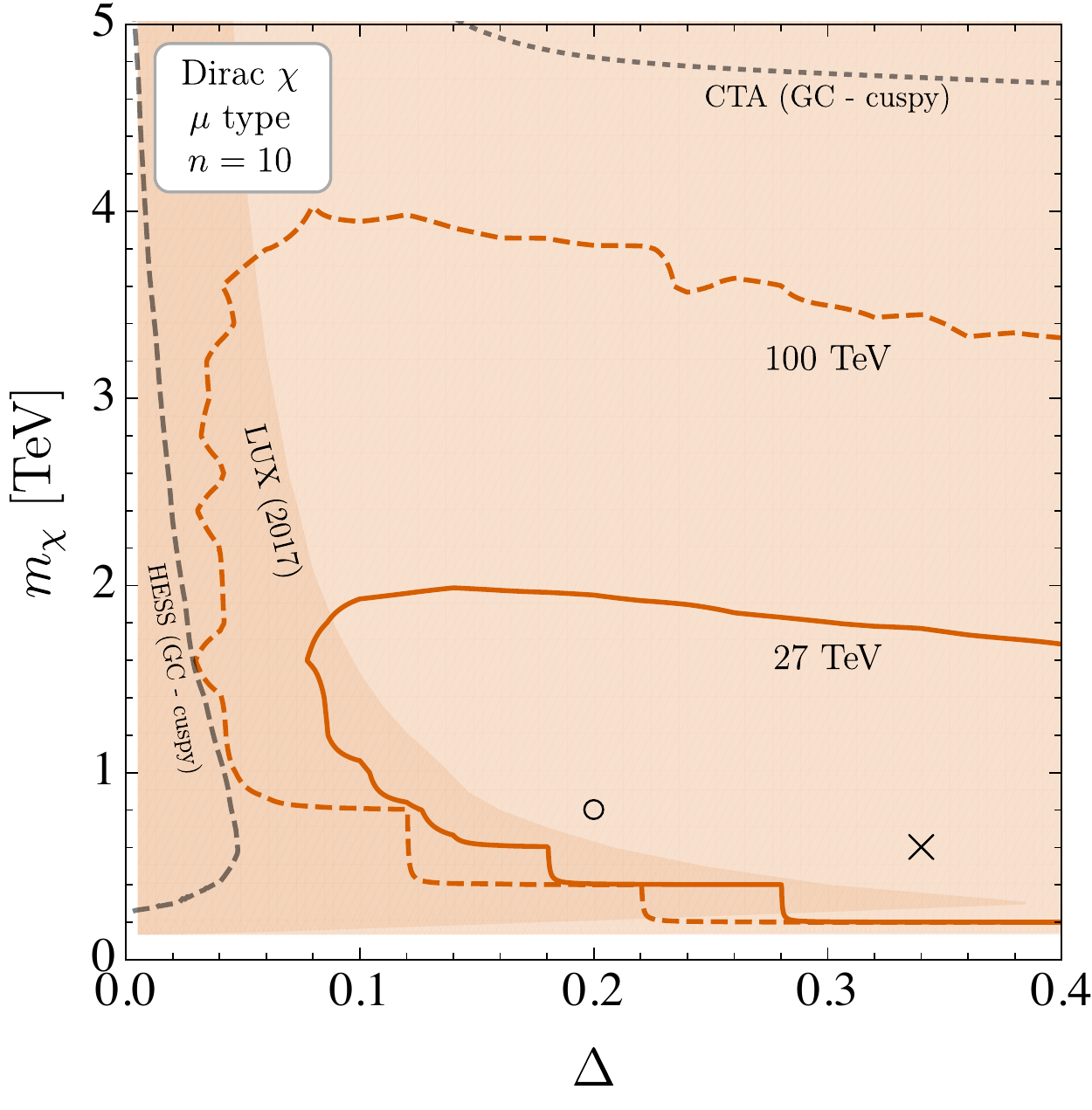}
\end{minipage}
\begin{minipage}[t][\minipageheightdp][t]{\minipagewidthdp}
\centering
\includegraphics[width=\figuremaxwidthdp,height=\figuremaxheightdp,keepaspectratio]{\main/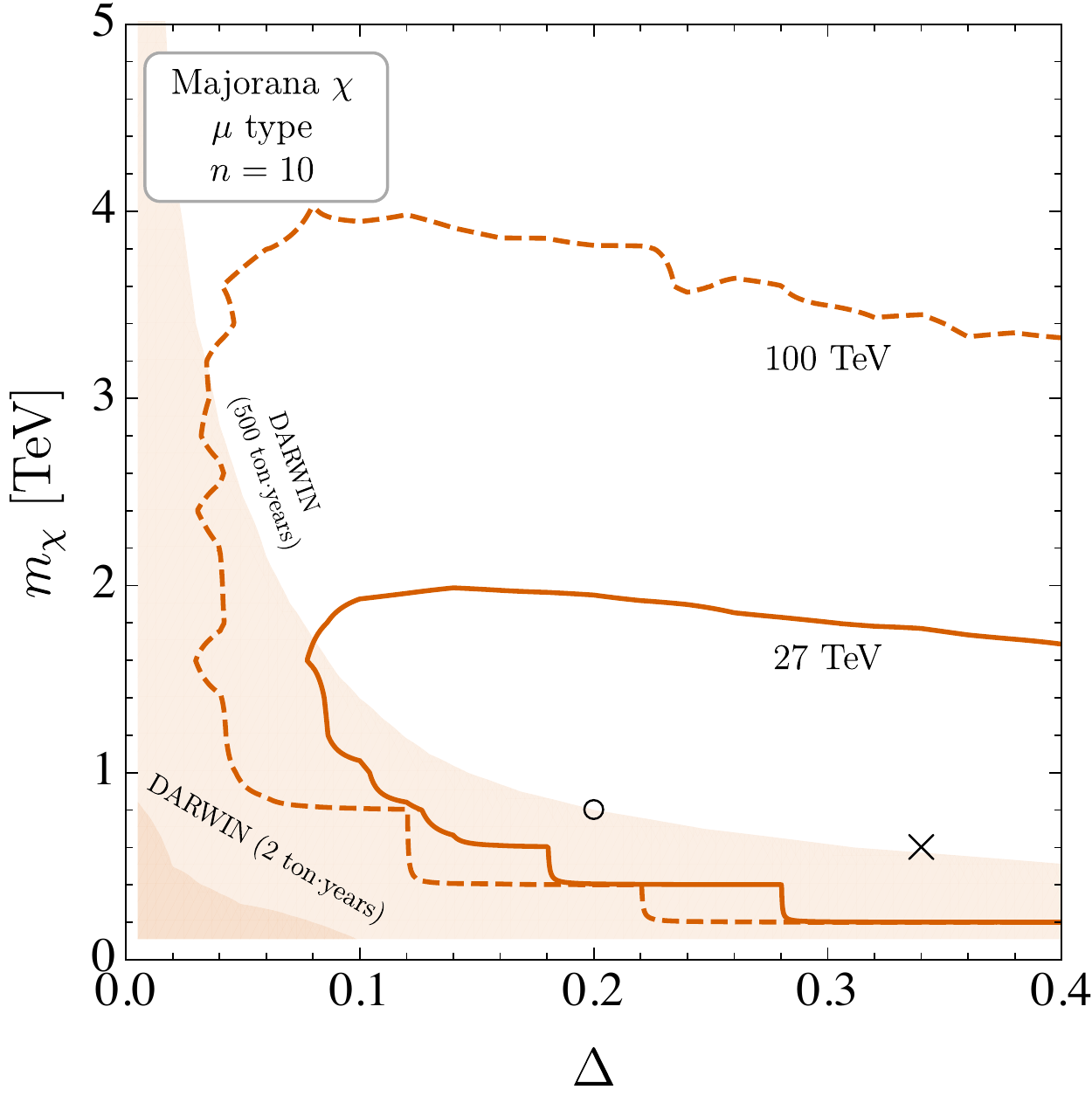}
\end{minipage}
  \caption{Reach of future colliders at $90\%\cl$,  current and future direct detection experiments at $90\%\cl$, 
  and current and future indirect detection experiments at $95\%\cl$ for Dirac (left) and Majorana (right) DM interacting with a muon and one (top) and ten (bottom) coannihilation partners in the $\Delta-m_\chi$ plane. The lightly, moderately and strongly shaded regions correspond to the direct detection limits by the future DARWIN experiment with $500\, \mathrm{ton}\cdot\mathrm{years}$, $2\, \mathrm{ton}\cdot\mathrm{years}$ and the LUX limits, respectively, which are discussed in detail in \citeref{Baker:2018uox}. The circle and cross signify our example signals shown in \fig{fig:muon-mt2}.}
  \label{fig:collider-exclusion}
\end{figure}

%--------------------------------------------------------------------%

To estimate the expected exclusion limit, we use a Poisson counting procedure 
for the signal and background events which pass all the cuts, 
based on a frequentist framework~\cite{Cowan:2010js, Patrignani:2016xqp}.  
In \fig{fig:collider-exclusion} we present the $90\%\cl$ sensitivity for 
the muon type models at a $27\TeV$ and a $100\TeV$ proton-proton collider.  
The parameter space probed is where $m_\chi$ is small and $\Delta$ is relatively large.  
The reach is independent of whether dark matter is Majorana or Dirac, 
since it depends on the $\phi$-pair production cross-section 
and the fact that $\BR(\phi_i^\pm \to \chi \ell^\pm)=1$.
The large $m_\chi$ region is not probed as $m_\phi$ increases with $m_\chi$, 
and the $\phi$-pair production cross-section decreases rapidly as $m_\phi$ increases, \fig{fig:collider-diagrams} (right).  
We see that in both cases the limits are strongest when there are more coannihilation partners. 
This is because the $pp\to \chi \bar{\chi} \ell^+ \ell^-$ cross-section scales as $n^2$. 
For $n=1$, the $27\TeV$ ($100\TeV$) machine can probe $m_\chi < 0.75\TeV \, (1.2\TeV)$, 
for $n=3$ it can probe $m_\chi < 1.3\TeV \, (2.3\TeV)$ while for $n=10$ the limits are 
$m_\chi < 2.0\TeV \, (4.0\TeV)$.
The small $\Delta$ region is not probed as in this region the 
momentum of the leptons is small and they are not efficiently reconstructed.  
This is a well known problem in the coannihilation region.  The gap for 
lower $\Delta$ can be closed, \eg by looking for 
ISR~\cite{Gori:2013ala, Schwaller:2013baa} or for disappearing charged 
tracks~\cite{Evans:2016zau,Khoze:2017ixx,Mahbubani:2017gjh}.

We also overlay the direct and indirect detection bounds 
from~\cite{Baker:2018uox}, to give 
a summary of all the relevant current and future experimental constraints.  
We see that the situation is dramatically different for Dirac and Majorana $\chi$.  
For Dirac $\chi$, small masses and mass splittings have already been excluded by LUX~\cite{Akerib:2015rjg}.  
In the future, DARWIN~\cite{Aalbers:2016jon} will probe the full parameter space, while colliders and indirect detection 
(only under the assumption of a cuspy dark matter halo profile) will be sensitive for relatively low masses and large or small $\Delta$, respectively.
We see that the challenging small $\Delta$ region at colliders is excluded by the 
existing bound from LUX.

For Majorana $\chi$, on the other hand, DARWIN, with the maximum exposure, 
is limited to probing only small masses and small $\Delta$, while there are no constraints from indirect detection.  
This is due to the velocity suppression of both the DM-nucleus and the annihilation cross-sections.  
The collider bounds are the same as in the Dirac case, since the mass term of $\chi$ 
does not enter into the production and decay of $\phi$-pairs.  In this case, future colliders 
are essential for probing the large $\Delta$ region of the parameter space.

Finally, the reach for electron final states is marginally worse than for muon 
final states due to the fact that the electron reconstruction efficiency is slightly worse than for muons. 
Again, we conclude that future colliders are essential for probing 
the large $\Delta$ region of the parameter space.

\subsection{Dark sectors}
\label{sec:DS}
As in our ordinary world, a dark sector could allow for long-range forces among its matter constituents. Evidence from both cosmology and astrophysics may supporting the possibility of long-range interactions among DM constituents (see, for instance, the role of massless dark photons in galaxy formation and dynamics~\cite{Gradwohl:1992ue,Carlson:1992fn,Foot:2004pa,Ackerman:mha,Fan:2013tia,Agrawal:2016quu,Foot:2014uba,Heikinheimo:2015kra}). In the following sections, prospect studies for searches for dark photons at HL- and HE-LHC are presented. 

%\subsection{Dark photons}
%\subfile{\main/section5DarkSectorDarkPhotons/sub55}
%\subfile{\main/section5DarkSectorDarkPhotons/sub52}
%\subfile{\main/section5DarkSectorDarkPhotons/sub53-LLDarkPhotons_1}
\subsubsection{\lhcb{Prospects for dark-photon at the HL-LHCb}}
\label{sec:darkphotonLHCb}
\contributors{P. Ilten, M. Williams and X. Cid Vidal, LHCb}
%{\bf Author(s): P. Ilten, X. Cid et al, LHCb}

\def\ep         {{\ensuremath{\Pe^+}}\xspace}
\def\en         {{\ensuremath{\Pe^-}}\xspace}   % electron negative 
\def\mup        {{\ensuremath{\Pmu^+}}\xspace}
\def\mun        {{\ensuremath{\Pmu^-}}\xspace} % muon negative
\def\upgradeone {\mbox{Upgrade~I}\xspace}
\def\upgradetwo {\mbox{Upgrade~II}\xspace}
\def\PD      {\ensuremath{D}\xspace}
\def\D       {{\ensuremath{\PD}}\xspace}
\def\Dstarz  {{\ensuremath{\D^{*0}}}\xspace}
\def\Dz      {{\ensuremath{\D^0}}\xspace}

A compelling scenario in the search for dark forces and other portals between the visible and dark sectors is that of the dark photon $A'$. 
In this case, a new $U(1)$ dark force, analogous to the electromagnetic (EM) force, can be introduced into the SM, where the dark photon is the corresponding force mediator which couples to dark matter (or matter) carrying dark charge. The $A'$ can kinetically mix with the photon, allowing the $A'$ to be observed in the spectra of final states produced by the EM current. This mixing can be thought of as a low-energy consequence of a loop process, potentially involving very high mass particles, that connect the visible and dark sectors.

\begin{figure}[t]
\centering
\begin{minipage}[t][\minipageheightsp][t]{\minipagewidthsp}
\centering
\includegraphics[height=\figuremaxheightsp,keepaspectratio]{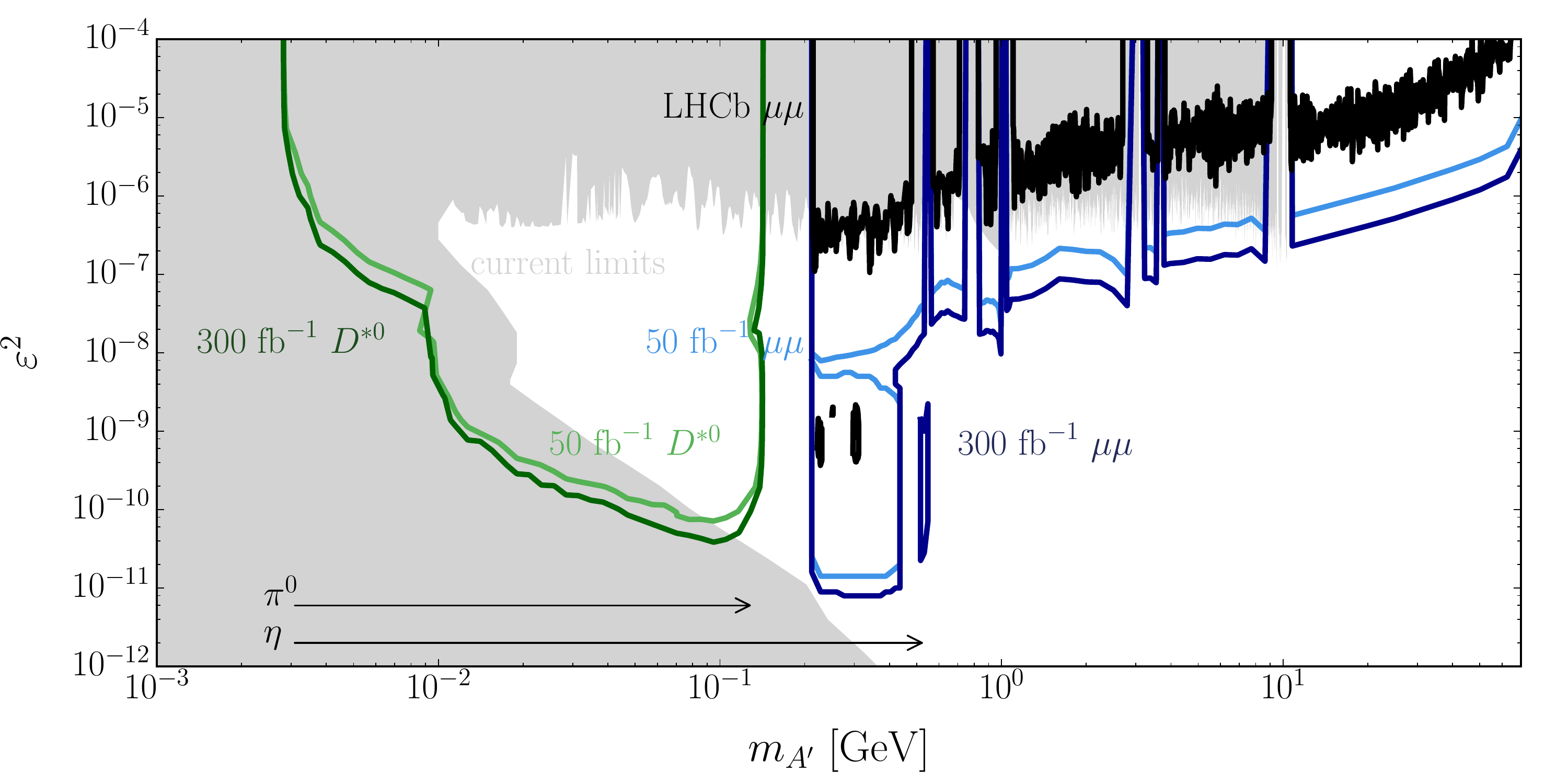}
\end{minipage}
%  \begin{center}
%    \includegraphics[width=\textwidth]{\main/section3DarkMatterSearches/img/DarkPhoton_LHCb}
%  \end{center}
  \caption{Current limits (grey fills), current LHCb limits (black band), and proposed future
    experimental reach (coloured bands) on $A'$ parameter
    space. The arrows indicate the available mass range from light meson decays into $\ep \en \gamma$.\label{fig:DarkPhoton}}
\end{figure}

The study of the $A'$ model is based on two free parameters: the mixing term $\varepsilon^2$ and the invariant mass of the $A'$, $m_{A'}$. The mixing term, $\varepsilon^2$, can be interpreted as the ratio of the dark force strength to the EM force strength. Note that for smaller values of $\varepsilon^2$ the dark photon can be long-lived and fly away from its production vertex. Figure~\ref{fig:DarkPhoton} shows the $\varepsilon^2-m_{A'}$ parameter space with current limits (grey fills, see \citeref{Alexander:2016aln} for details), current LHCb limits (black bands)~\cite{Aaij:2017rft}, and prospects on the LHCb future reach (coloured bands). The light (dark) coloured band corresponds to discovery reach assuming $50$ ($300$)\fbinv datasets. These are the expected integrated luminosities at LHCb at the end of Run-4 and Run-5 of the LHC, respectively. These discovery reaches assume increased pileup within LHCb will not have a significant effect on the dark photon reconstruction.

There are at least two complementary ways for LHCb to explore large portions of unconstrained $A'$ parameter space. They address different regions of this space. The first involves prompt and displaced resonance searches using $\Dstarz \to \Dz \ep \en$ decays~\cite{Ilten:2015hya} (green bands in \fig{fig:DarkPhoton}). The second is an inclusive di-muon search ~\cite{Ilten:2016tkc} (blue bands in \fig{fig:DarkPhoton}) where the di-muon can be prompt or displaced. In both cases the lepton pair is produced from an EM current which kinetically mixes with the $A'$, producing a sharp resonance at the $A'$ mass. In the first case, the $A'$ is detected in its decay to an $\ep \en$ pair and in the second to a $\mup \mun$ pair. The advantage of these approaches is that they do not require the calculation of absolute efficiencies. In both cases, the signal can be normalised to the di-lepton mass sidebands near the $A'$ resonance, where dark photon mixing with the SM virtual photon is negligible. In general, these search strategies depend on three core capabilities of LHCb: excellent secondary vertex resolution, particle identification, and real-time data-analysis. These features are also important for flavour physics, which mainly drives the design of the detector and its upgrades. In particular, the improvement in the impact parameter resolution, expected after the upgrade of the LHCb vertex locator, will be key to tackle the background produced by heavy quark decays.

For the LHCb sensitivities in \fig{fig:DarkPhoton}, the sensitivity calculated using $\Dstarz \to \Dz \ep \en$ decays~\cite{Ilten:2015hya} is based on the normalisation to this channel, which at the same time is the main background for the prompt search. $\Dstarz \to \Dz \ep \en$ decays are generated using \pythia{ 8} \cite{Sjostrand:2007gs}, and the \Dz is reconstructed or partially reconstructed through its decay into at least two charged particles. The selection is designed to maximise the \ep\en mass resolution and to minimise the background. The resolution and efficiencies are obtained using public LHCb information, combined with a simplified simulation of the upgraded vertex locator. 
For the di-muon search~\cite{Ilten:2016tkc}, a fiducial selection is designed so that the reconstruction efficiency is essentially flat across the dark photon parameter space, while minimising the presence of background. The relevant experimental resolutions and efficiencies, including those foreseen after subsequent detector upgrades, are taken from public LHCb documents. The normalisation channel, i.e \mup\mun production originating from electromagnetic processes, and backgrounds are again studied using \pythia{ 8}, corrected with experimental LHC inputs.

\begin{figure}[t]
\centering
\begin{minipage}[t][\minipageheightsp][t]{\minipagewidthsp}
\centering
\includegraphics[height=\figuremaxheightsp,keepaspectratio]{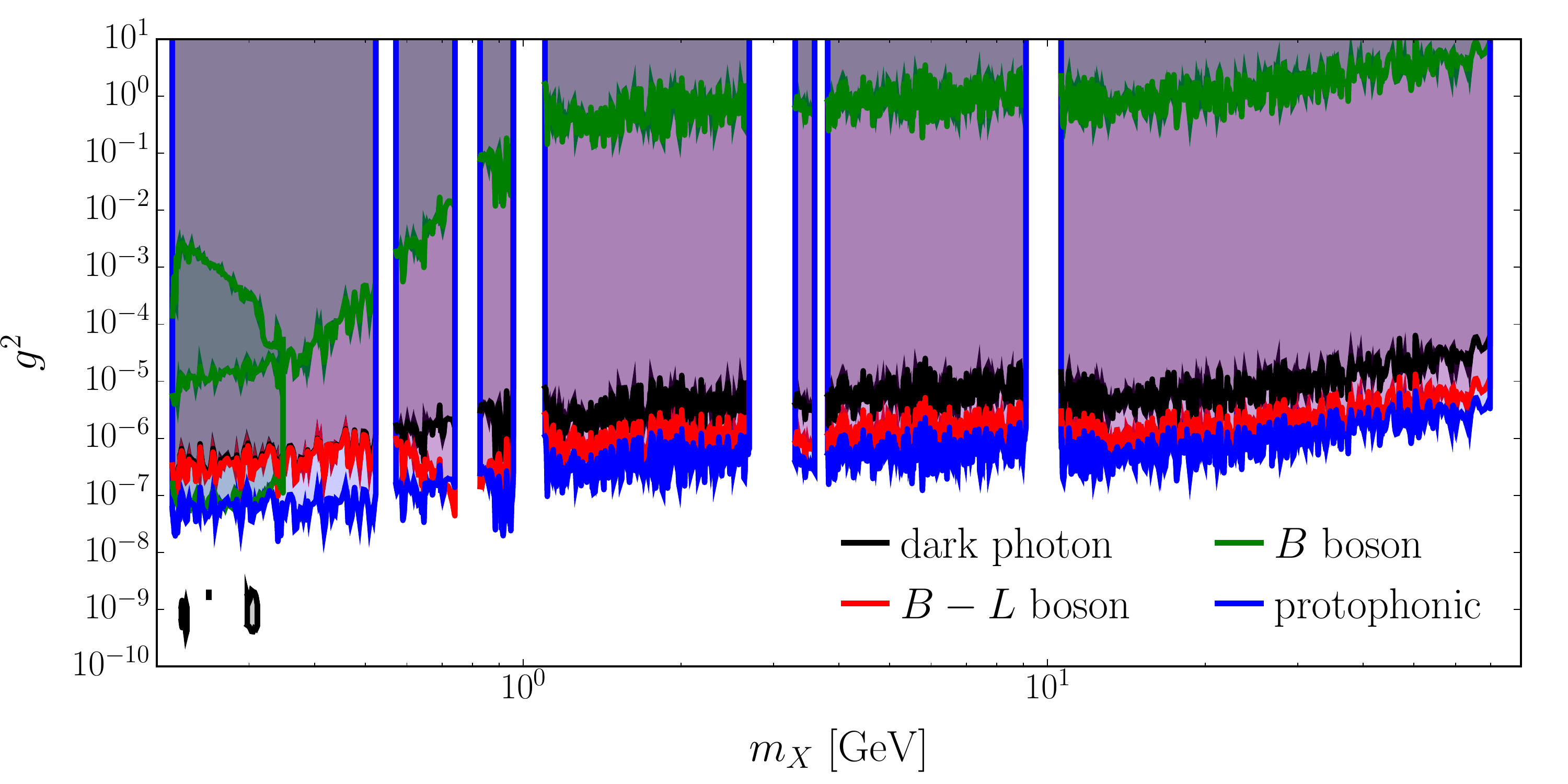}
\end{minipage}
%  \begin{center}
%    \includegraphics[width=\textwidth]{\main/section3DarkMatterSearches/img/Recast_LHCb}
%  \end{center}
  \caption{Recast of the (blue) LHCb dark photon limits into the (red) B-L boson, (green) B boson, and (yellow) protophobic models using the \textsc{Darkcast} tool~\cite{Ilten:2018crw}.
\label{fig:Recast}}
\end{figure}

A first inclusive search for $A'$ bosons decaying into muon pairs was performed by the LHCb collaboration~\cite{Aaij:2017rft} (black band in \fig{fig:DarkPhoton}). This search, an implementation of the second strategy described above, used a data sample corresponding to an integrated luminosity of $1.6\fbinv$ from $pp$ collisions taken at $\sqrt{s}=13\TeV$. Although the data sample used was significantly smaller than the one that will be available at the HL-LHC, this search already produced world-best upper limits in regions of $\varepsilon^2-m_{A'}$ space. This search was limited by the presence of the LHCb hardware trigger, which severely compromised the detection efficiency of low mass dark photons at LHCb. However, this hardware level trigger will be removed from Run-3 of the LHC onwards. At the same time, this was the first simultaneous prompt and displaced $A'$ search. With around $300\fbinv$, LHCb will either confirm or reject the presence of a dark photon for significant portions of the theoretically favoured parameter space. It should be noted that, in non-minimal models, such as those producing dark photons through the Higgs portal, part of this parameter space can be further constrained by other experiments. Examples are given in \seccs{sec:cmsLLdarkphoton} and \ref{sec:dark_photons_th}.

To cover the gap in reach between the two primary search strategies, presented above, production of $A'$ bosons from light meson decays, $\pi^0 \to \ep \en \gamma$ and $\eta \to \ep \en \gamma$, is expected to be used. The parameter space coverage from light meson decays depends on the ability of the LHCb triggerless readout to quickly and efficiently reconstruct low mass and low momentum di-electron pairs. Additionally, the electron momentum resolution, degraded by incomplete bremsstrahlung recovery, dictates the di-electron mass resolution which drives parameter space coverage. More detailed studies are needed to quantify how searches for dark photons from light meson decays will help to constrain the dark photon parameter space.

One of the advantages of the dark photon model is that results can be recast into more complex vector current models, given some knowledge of the dark photon production mechanism and the detector efficiency for displaced dark photon reconstruction. Examples of these models are the $B-L$ boson which couples to the $B-L$ current, the $B$ boson which is leptophobic and couples to baryon number, and a vector boson which mediates a protophobic force. All these models can be fully specified with two parameters: the global coupling $g$ of the vector current for the model with the electromagnetic current, and the mass $m_X$ of the mediating boson. For the dark photon model, this is just $\varepsilon$ and $m_{A^\prime}$, respectively. In \fig{fig:Recast} the initial inclusive di-muon results from LHCb~\cite{Aaij:2017rft} have been recast into these example models using the \textsc{Darkcast} package~\cite{Ilten:2018crw}. Dark photon searches can also be recast to non-vector models, but such a recasting is no longer as straight forward.

\subsubsection{Long-lived dark-photon decays at the HL-LHC}
\label{sec:atlascmsLLdarkphoton}
Among the numerous models predicting dark photons,  one class of models that is particularly interesting for the LHC features the hidden sector communicating with the SM through a Higgs portal. Dark photons are produced through BSM Higgs decays. They couple to SM particles via a small kinetic mixing parameter $\epsilon$, as described in \secc{sec:darkphotonLHCb}. If $\epsilon$ is very weak, the lifetime of the dark photon can range from a few millimetres up to several meters. The lower $\epsilon$ is, the longer the dark photon lifetime will be, which then decays displaced from the primary vertex. The following searches from ATLAS and CMS target complementary scenarios and illustrate possible improvements in trigger and analysis strategies which can be used at HL-LHC.  

\subsubsubsection{\cms{Long-lived dark-photon decays into displaced muons at HL-LHC CMS}}
\label{sec:cmsLLdarkphoton}
\contributors{K.~Hoepfner, H.~Keller, CMS}
%{\bf Author(s): K.~Hoepfner, H.~Keller, CMS}

%%% Introduction and strategy
In the so-called Dark SUSY model~\cite{Baumgart:2009tn,Falkowski:2010cm}, an additional dark $U_D(1)$ symmetry is added as a supersymmetric SM extension. Breaking this symmetry gives rise to an additional massive boson, the dark photon $\gamma_D$, which couples to SM particles via a small kinetic mixing parameter $\epsilon$. A golden channel for such searches is the decay to displaced muons. 

\begin{figure}[t]
\begin{center}
\includegraphics[width=0.3\textwidth]{\main/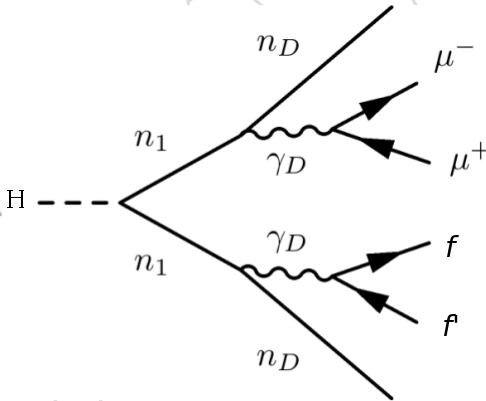}\hspace{1.5cm}
\includegraphics[width=0.3\textwidth]{\main/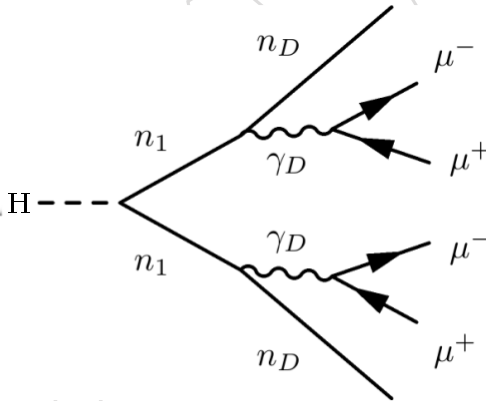}
 \caption{Feynman diagram of the decay of SM Higgs boson to a final state containing two or more muons in Dark SUSY models. Decay chain leading to a final state containing exactly two (left) or four (right) muons.}
\label{fig:Feynman_DarkPhoton}
\end{center}
\end{figure}

%%% Reconstruction
The reconstruction of muons with large displacements is challenging both at trigger level and for the final event reconstruction, especially when the long lived particle decays outside the tracker volume and the precision of the tracker cannot be used for the analysis. To identify displacements of physical objects during reconstruction, the transverse impact parameter $d_{0}$ of the reconstructed track with respect to the primary interaction vertex is used.
This analysis from CMS~\cite{CMS:2018lqx} relies on a dedicated muon reconstruction algorithm that is designed for non-prompt muons leaving hits only in the muon system. This is the displaced standalone (DSA) algorithm, using the same reconstruction techniques as prompt muons, but removing any constraint to the interaction point which is still present in the standard standalone (SA) algorithm. The DSA muon algorithm improves transverse impact parameter and transverse momentum ($p_{T}$) resolutions for displaced muons compared to the SA muon algorithm~\cite{CMS-DP-2015-015}. 

%%% Model Discussion
In the model studied here~\cite{CMS:2018lqx}, dark photons are produced in cascade decays of the SM Higgs boson that would first decay to a pair of MSSM-like lightest neutralinos
($n_1$), each of which can decay further to a dark sector neutralino ($n_D$) and the dark photon, as shown in \fig{fig:Feynman_DarkPhoton}.
For the branching fraction BR$(H \to 2\gamma_D + \mathrm{X})$, where X denotes the particles produced in the decay of the SM Higgs boson apart
from the dark photons, $20\%$ is used. This value is in agreement with recent Run-2 studies~\cite{CMS:2018rdr} and taking into account
the upper limit on invisible/non-conventional decays of the SM Higgs boson~\cite{Khachatryan:2016whc}. We assume neutralino masses $m(n_1) =
50\GeV$ and $m(n_D) = 1\GeV$, and explore the search sensitivity for dark photon masses and lifetimes in the following
ranges: $m(\gamma_D) = (1, 5, 10, 20, 30)\GeV$ and $c\tau = (10, 10^2, 10^3, 5 \times 10^3, 10^4)~\mathrm{mm}$. Final states with two
and four muons are included in the analysis. In the former case, one dark photon decays to a pair of muons while the other dark photon decays
to some other fermions (2-muon final state). In the latter case, both dark photons decay to muon pairs (4-muon final state). Both decay chains
are shown in \fig{fig:Feynman_DarkPhoton}. The assumed Higgs production cross section via gluon-gluon fusion is $49.97\pb$~\cite{Dittmaier:2011ti}.

\begin{figure}[t]
\centering
\begin{minipage}[t][\minipageheightdp][t]{\minipagewidthdp}
\centering
\raisebox{1mm}{\includegraphics[width=\figuremaxwidthdp,height=\figuremaxheightdp-2mm,keepaspectratio]{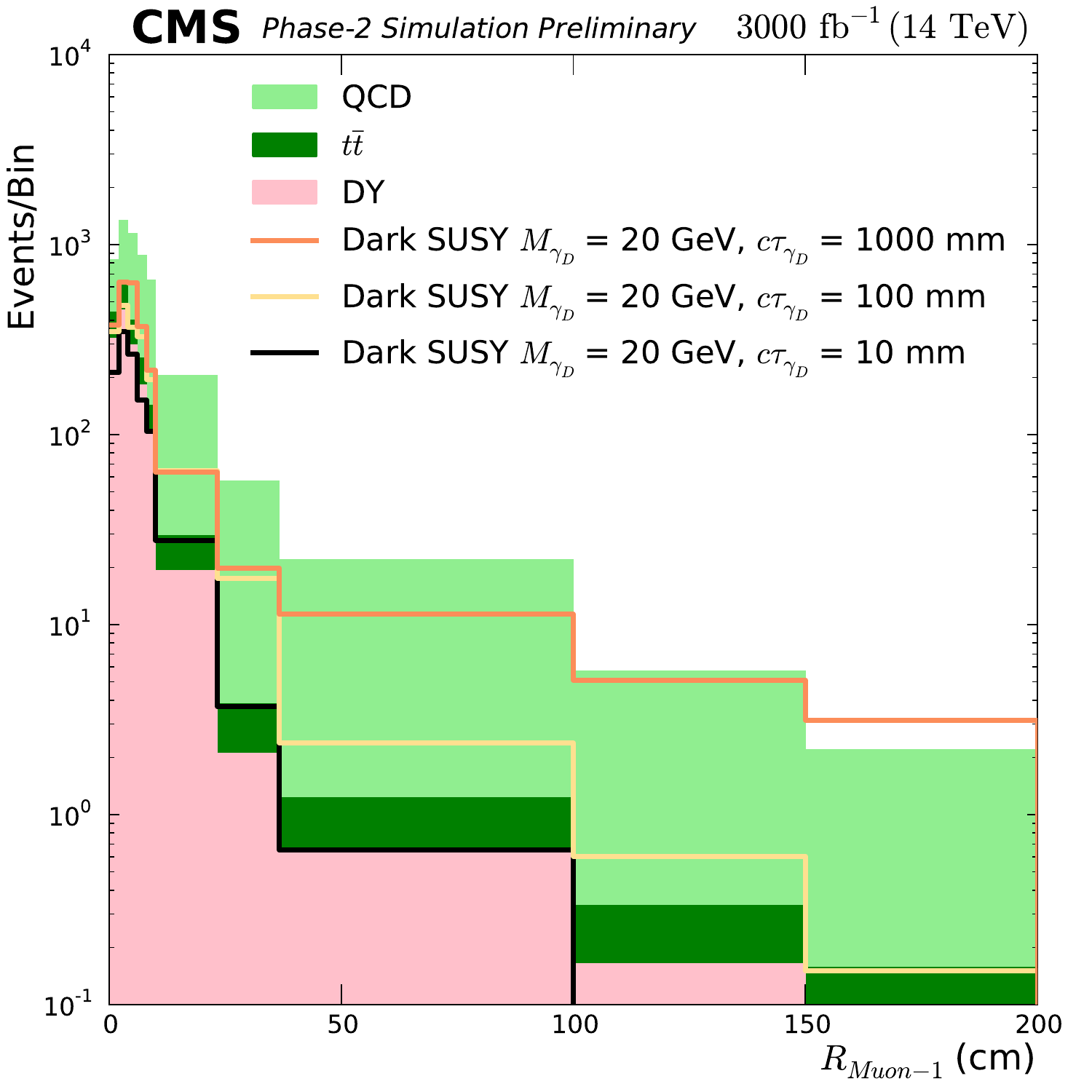}}
\end{minipage}
\begin{minipage}[t][\minipageheightdp][t]{\minipagewidthdp}
\centering
\includegraphics[width=\figuremaxwidthdp,height=\figuremaxheightdp+2mm,keepaspectratio]{\main/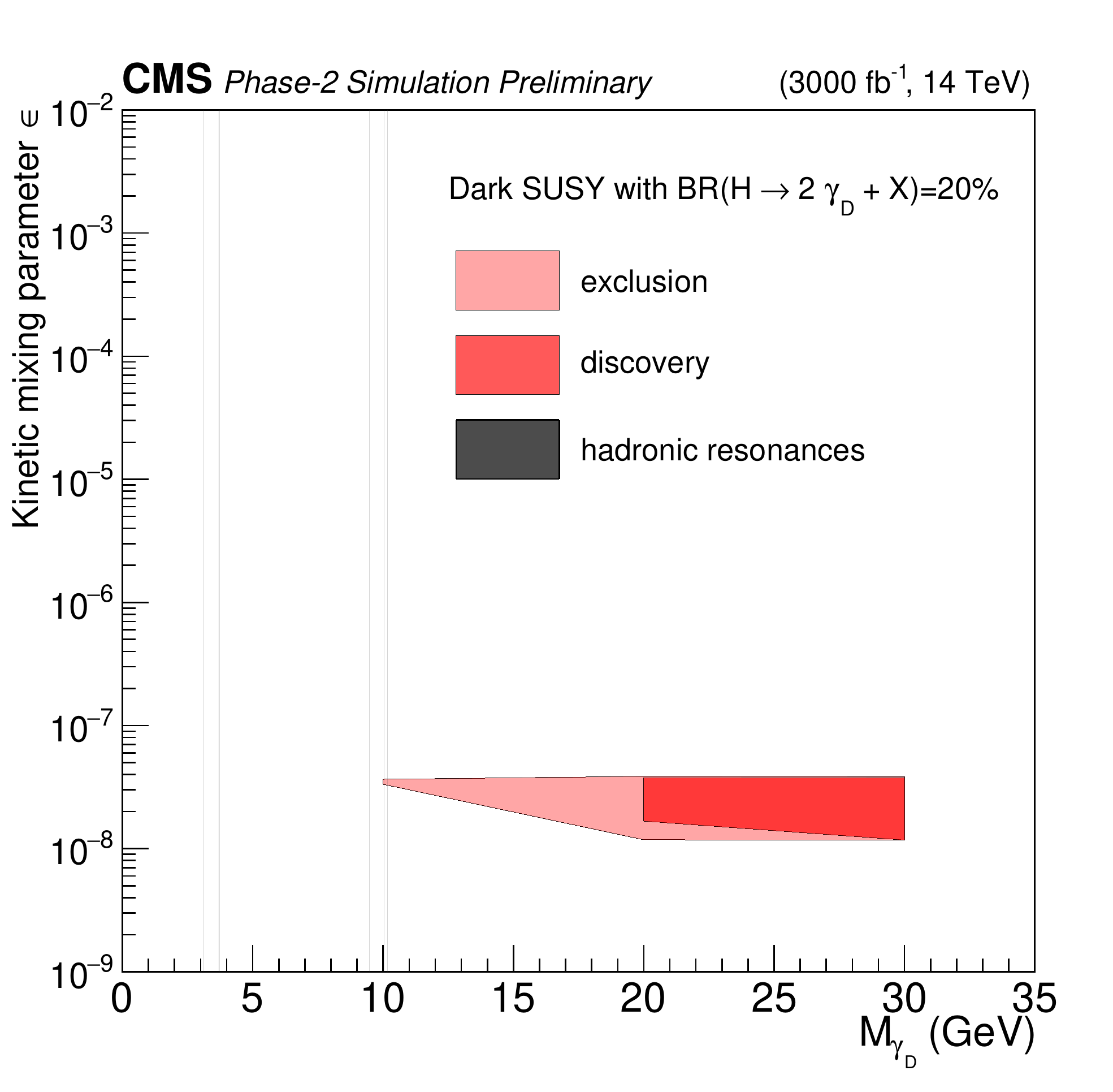}
\end{minipage}
 \caption{Left: distance of the closest approach of the displaced muon track with maximum $p_T$  to the primary interaction vertex, $R_{\rm Muon-1}$, for signal and background after the final event selection. 
Right: parameter scan in the $\epsilon-m_{\gamma_D}$ plane. The gray lines indicate the regions of narrow hadronic resonances where the analysis does not claim any sensitivity.}
\label{fig:DarkPhoton_Distr_Disc}
\end{figure}

%%% Background and Selection
The main background for this search comes from multi-jet production (QCD), $t\bar{t}$~production, 
and Z/DY $\to\ell\ell$ events  where large impact parameters are (mis)reconstructed. Cosmic ray muons can travel through the detector far away from the primary vertex and mimic the signature of displaced muons. 
However, thanks to their striking detector signature, muons from cosmic rays can be suppressed by rejecting back-to-back kinematics. 

For each event, at least two DSA muons are required. If more than two exist, the ones with the
highest $p_T$ are chosen. The two muons must have opposite charge ($q_{\mu,1} \cdot q_{\mu,2} = -1$) and must be separated by $\Delta R = \sqrt{\Delta \phi^2 + \Delta
\eta^2} > 0.05$. The three-dimensional angle between the two displaced muons is required to be less than $\pi - 0.05$ (not back-to-back) in order to suppress cosmic
ray backgrounds. Additionally, $\ptmiss \geq 50\GeV$ is imposed to account for the dark neutralinos escaping the detector without leaving any signal.

In order to discriminate between background and signal, the three-dimensional distance from the primary vertex to the point of closest approach of the extrapolated
displaced muon track, called $R_{\text{Muon}}$, is used. 
The event yield after full event selection of both selected muons as a function of $R_{\rm Muon-1}$ and $R_{\rm Muon-2}$ is used to search for the signal. Figure~\ref{fig:DarkPhoton_Distr_Disc} (left) shows $R_{\rm Muon-1}$ of the first selected muon for signal and background samples.

%%% Sensitivity in exclusion and discovery
The search is performed using a simple counting experiment approach. In the presence of the expected signal, the significance of the corresponding event excess over the expected background is assessed using the likelihood method.
In order to evaluate the discovery sensitivity the same input is used as in the limit calculation, now with the assumption that one would have such a signal in the data. The discovery sensitivity is shown in the two-dimensional $m_{\gamma_D}$-$c\tau$ plane 
in \fig{fig:DarkPhoton_Distr_Disc} (right). This search is sensitive to large decay lengths of the dark photon.

\begin{figure}[t]
\centering
\begin{minipage}[t][\minipageheightdp][t]{\minipagewidthdp}
\centering
\includegraphics[width=\figuremaxwidthdp,height=\figuremaxheightdp,keepaspectratio]{\main/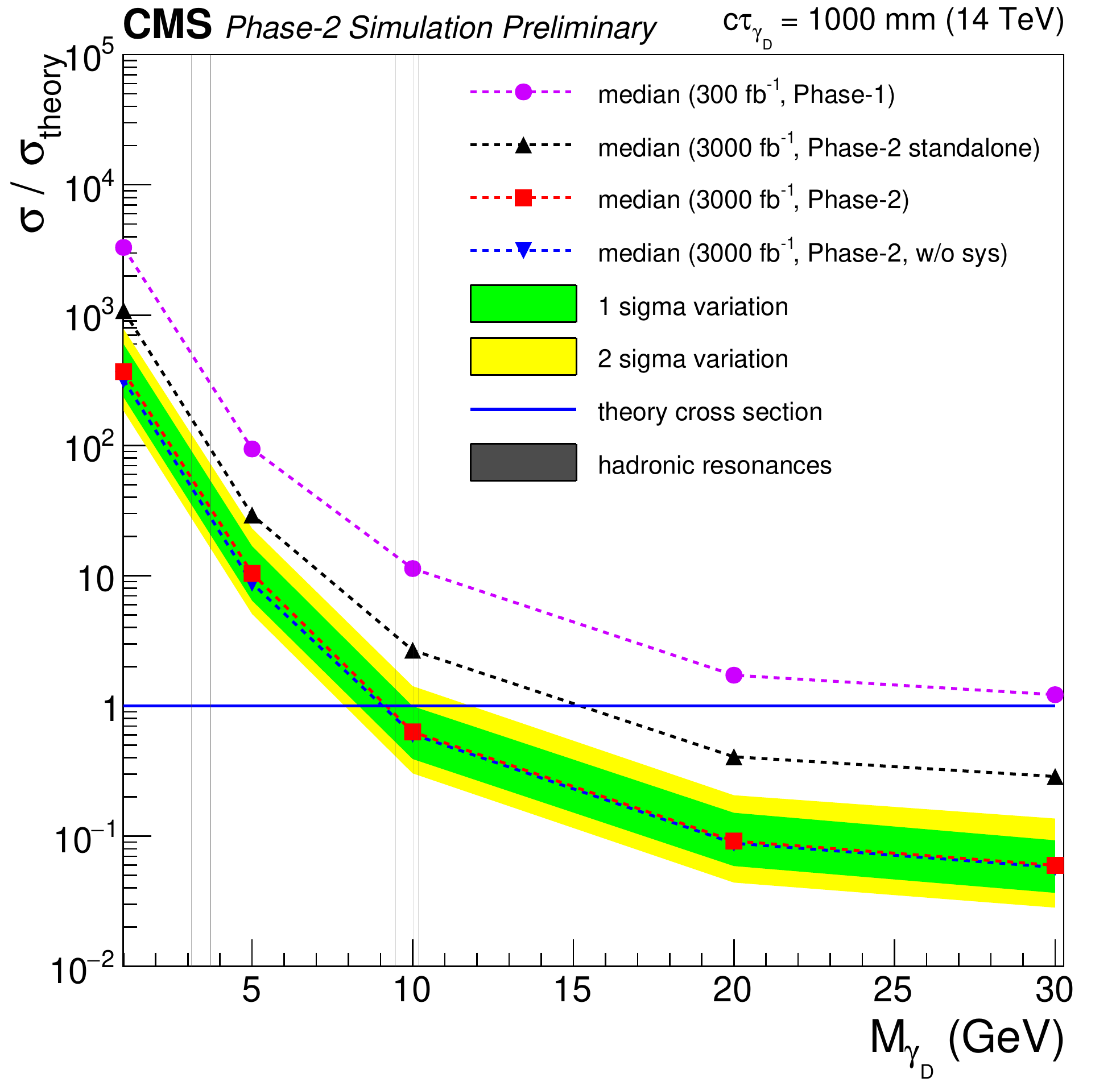}
\end{minipage}
\begin{minipage}[t][\minipageheightdp][t]{\minipagewidthdp}
\centering
\includegraphics[width=\figuremaxwidthdp,height=\figuremaxheightdp,keepaspectratio]{\main/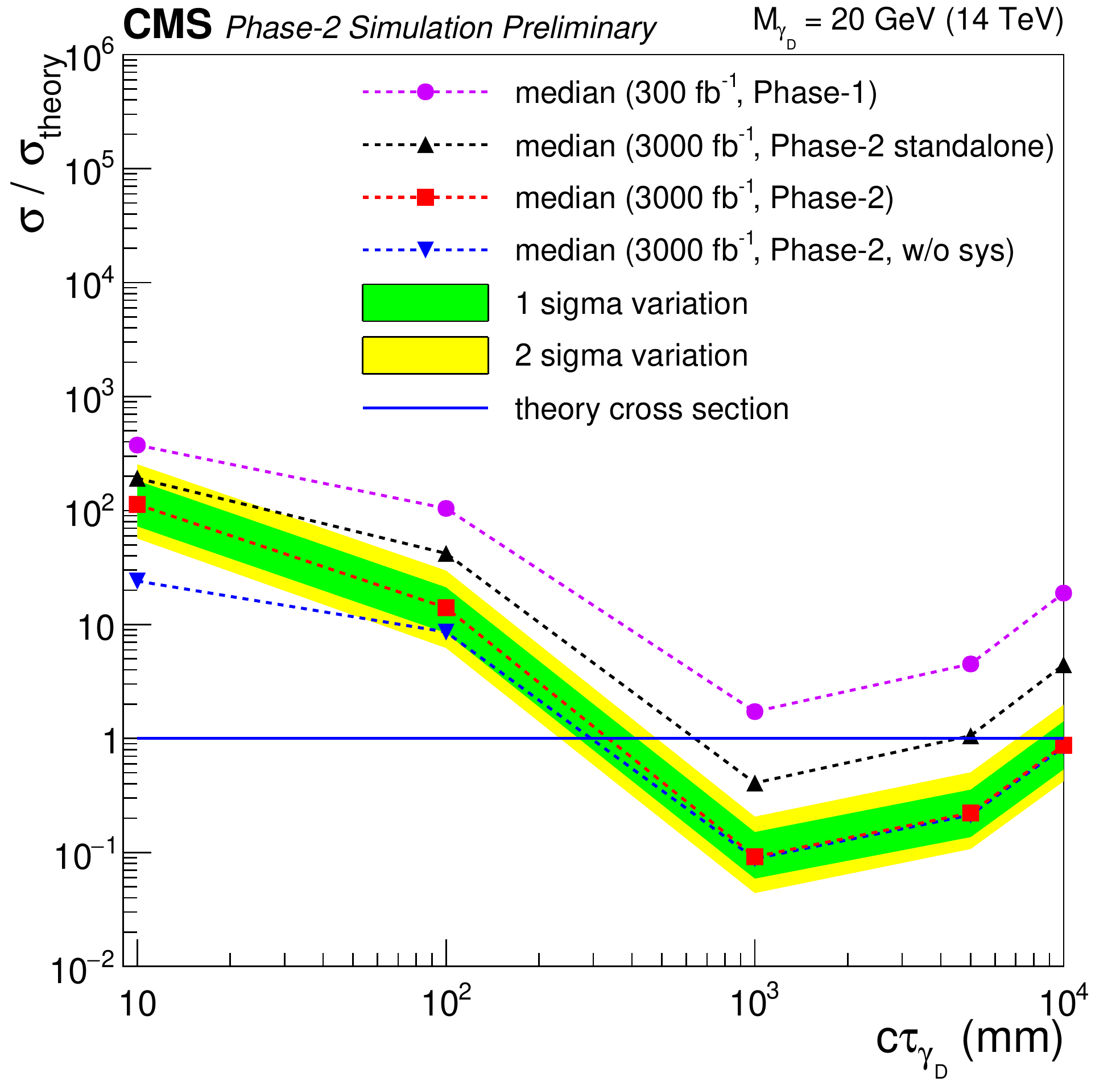}
\end{minipage}
 \caption{$95\%~\cl$ upper limits on production cross section $\sigma / \sigma_{\text{theory}}$ for various dark photon mass hypotheses and a fixed decay length of $c\tau = 1000~\mathrm{mm}$ (left) and a fixed mass of $M_{\gamma_D} = 20\GeV$(right). Green and yellow shaded bands show the one and two sigma range of variation of the expected $95\%\cl$ limits, respectively. The black dashed lines (``Phase-2 standalone'') compare the expected sensitivity of the displaced muon search to the algorithm with a beamspot constraint. The gray lines indicate the regions of narrow hadronic resonances where the analysis does not claim any sensitivity. }
\label{fig:NewCompLimit_DarkPhoton}
\end{figure}

In the absence of a signal, upper limits at $95\%\cl$ are obtained on a signal
event yields with respect to the one expected for the considered model. A Bayesian method with a uniform prior for the signal event rate is used and the nuisance
parameters associated with the systematic uncertainties are modelled with log-normal distributions. 
The resulting limits for the Dark SUSY models are depicted in \fig{fig:NewCompLimit_DarkPhoton}. While the results shown in \fig{fig:NewCompLimit_DarkPhoton} (left) are for a dark photon with a decay length of $1\meters$ as a function of the dark photon mass, \fig{fig:NewCompLimit_DarkPhoton} (right) shows the results for a dark photon mass of $20\GeV$ as a function of the decay length~\cite{CMS:2018lqx}.  The relatively long lifetimes accessible in this search provide complementary sensitivity at lower values of $\epsilon$.

%\vspace{1cm}

\subsubsubsection{\atlas{Searching for dark-photons decays to displaced collimated jets of muons at HL-LHC ATLAS}}
\label{sec:atlasLLdarkphoton}
\contributors{C. Sebastiani, M. Corradi, S. Giagu, A. Policicchio, ATLAS}

\begin{figure}[t!]
\centering
\includegraphics[width=0.3\textwidth]{\main/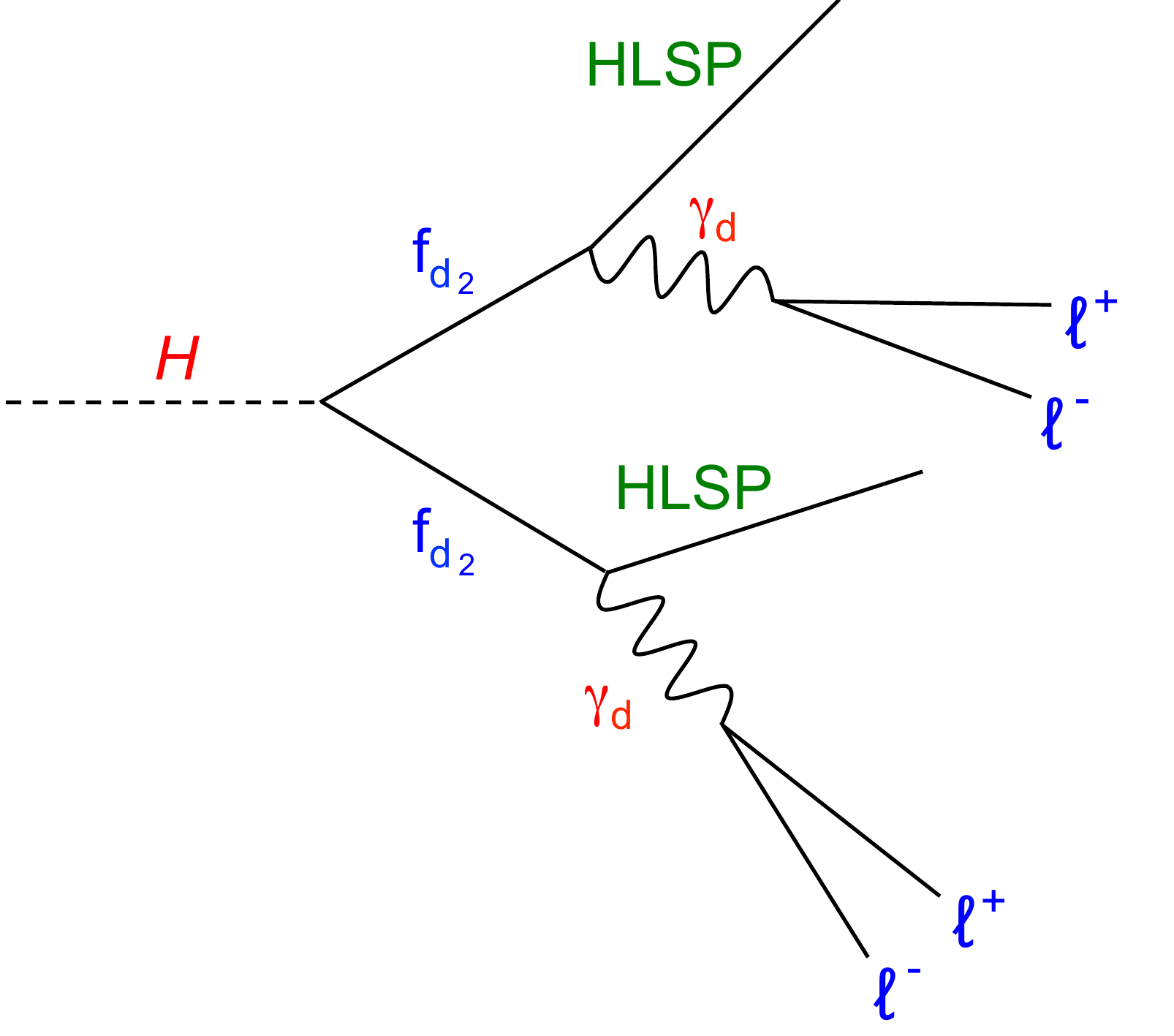}
\caption{The Higgs boson decay to hidden particles according to the FRVZ model.}
\label{fig:modelATLASDP}
\end{figure}

Prospects for searches for Hidden Sectors 
%and in particular for dark photons (here referred to as $\gamma_d$) produced via Higgs boson and subsequently decaying into a pair of displaced muons 
performed by ATLAS are presented in this section~\cite{ATL-PHYS-PUB-2019-002}. 
The benchmark model used in this analysis is the Falkowsky-Ruderman-Volansky-Zupan (FRVZ) vector portal model. In this case, a pair of dark fermions $f_{d2}$ is produced in the Higgs boson decay. As shown in \fig{fig:modelATLASDP}, the dark fermion decays in turn to a $\gamma_d$~and a lighter dark fermion assumed to be the Hidden Lightest Stable Particle (HLSP).  The dark photon, assumed as vector mediator, mixes kinetically with the SM photon and decays to leptons or light hadrons. The branching fractions depend on its mass. At the LHC, these dark photons would typically be produced with large boost, due to their small mass, resulting in collimated structures containing pairs of leptons and/or light hadrons, known as lepton-jets (LJs).  If produced away from the interaction point (IP), they are referred to as "displaced LJs''. The mean lifetime $\tau$ of the dark photon is a free parameter of the model, and is related to the kinetic mixing parameter $\epsilon$ by the relation:
\begin{equation*}
 \beta \gamma c \tau ~ \propto ~ \Biggl(\frac{10^{-4}}{\epsilon}\Biggr)^2 \Biggl(\frac{100~{\mathrm{MeV}}}{m_{\gamma_{d}}}\Biggr)^2 ~\mathrm{s}.
  \label{eq:mixing}
\end{equation*}

Two new muon trigger algorithms are also studied to improve the selection efficiency of displaced muon pairs. 
MC samples have been produced at $13$ and $14\TeV$~\com~energy for the FRVZ model assuming $\mu$ equal to $65$ and $200$, respectively, and various possible $c\tau$. The samples are used to assess the sensitivity of the analysis and study new triggers.  
%\paragraph*{New proposed triggers at the HL-LHC} 

%Additional sensitivity to dark photons might be achievable thanks to newly proposed triggers which could be implemented at HL-LHC.  
The standard ATLAS triggers are designed assuming prompt production of particles at the interaction point and therefore are very inefficient in selecting the products of displaced decays.
The searches for $\gamma_d$ decays are thus based on events selected by specialised triggers dedicated to the selection of events with displaced muon pairs. However these triggers are still far from optimal. If the dark photon is highly boosted, muons are collimated and the trigger efficiency is limited by the finite granularity of the current hardware trigger level. In terms of an interval of the azimuthal angle $\phi$ and pseudorapidity $\eta$, the granularity is $\Delta\eta\times\Delta\phi~\simeq~0.2\times0.2$ (Region of Interest, RoI). If the dark photon is not boosted sufficiently, the out-going muons from a displaced decay are more open and may not point to the IP. The current hardware trigger level has a tight constraint on IP pointing resulting in non-optimal selection efficiency of displaced non-pointing muon tracks.

The new ATLAS detector setup and Trigger \& Data Acquisition system for the  HL-LHC will offer the opportunity to develop new trigger algorithms overcoming the current limitations. Two new trigger selections have been studied in this work: one dedicated to triggering on collimated LJs in boosted scenarios, based on requiring muons in a single RoI and referred to as \emph{L0 multi-muon scan}; a second one dedicated to triggering on non-boosted scenarios, loosening the pointing requirements applied in Run-2 and referred to as \emph{L0 sagitta muon}. A summary of the two triggers is given below, for details see~\cite{ATL-PHYS-PUB-2019-002}.

\textbf{L0 multi-muon scan:} this new approach allows to include in the sector logic multiple trigger candidates in the same RoI and leads to a new trigger selection with lower $\pt$ thresholds resulting in a higher efficiency without increasing sensibly the trigger rate. The new trigger algorithm is designed to analyse hit patterns in the Muon Spectrometer. As a first step, the algorithm searches for the pattern with the highest number of hit points, called best pattern, in the MS to form the primary L0 muon candidate. Then all the other possible hit patterns, not compatible with the best pattern, are searched for in the same RoI to form the secondary L0 muon candidates. A quality cut is applied to reduce the influence of noisy hits, requiring patterns with hits on at least three different RPC layers.
Patterns are requested to not share RPC hits. If at least one secondary pattern is found, an additional L0 muon is assumed to be found in the RoI. The new L0 trigger algorithm is defined by the logical OR of a single muon L0 with $\pt=20\GeV$ threshold and a multi-muon L0 with $\pt=10\GeV$ threshold. Based on signal MC samples, an overall improvement up to $7\%$ is achieved with respect to the baseline $\pt=20\GeV$ selection, in particular for small opening angle between the two muons from the $\gamma_d$ decays. 

\textbf{L0 sagitta muon:} this approach allows to recover for loss of efficiency in case of out-going muons from non-boosted $\gamma_d$ that may not be pointing to the IP. The L1 Run-2 trigger has a tight constraint on selecting only pointing muons resulting in non optimal selection of these exotic signatures. A benchmark FRVZ sample with $10\GeV$ $\gamma_d$ mass is used to for this study.
The sagitta, defined as the vertical distance from the midpoint~\footnote{The midpoint is defined as the middle point of a segment} of the chord~\footnote{The chord of a circle is a line segment that connects two points of the circle itself} to the arc~\footnote{The arc is a portion of the circumference of a circle.} of the muon trajectory itself, can be used to estimate the momentum of a charged particle travelling inside a magnetic field. The sagitta of a muon track can be computed at the L0 trigger level using $\eta-\phi$ measurement points in the various RPC stations. 
The map between the inverse of the sagitta and the muon transverse momentum has been studied using a MC sample of single muons generated according to a uniform transverse momentum distribution. 
The mean value of the inverse of the sagitta for $\pt=20\GeV$ pointing truth muons is $s^{-1}=9 \times 10^{-6}~\mathrm{mm}^{-1}$.
High transverse momentum non-pointing muons can be thus selected using a L0 muon trigger with low \sloppy\mbox{$\pt = 5\GeV$} threshold, computing the inverse of the sagitta and requesting a cut on $s^{-1}\le 9 \times 10^{-6}~\mathrm{mm}^{-1}$.  Overall, a $\sim 20\%$ improvement in efficiency is achieved by adding this new trigger according to MC signal studies. 

\begin{figure}[t]
\centering
\begin{minipage}[t][\minipageheightdp][t]{\minipagewidthdp}
\centering
\includegraphics[width=\figuremaxwidthdp,height=\figuremaxheightdp,keepaspectratio]{\main/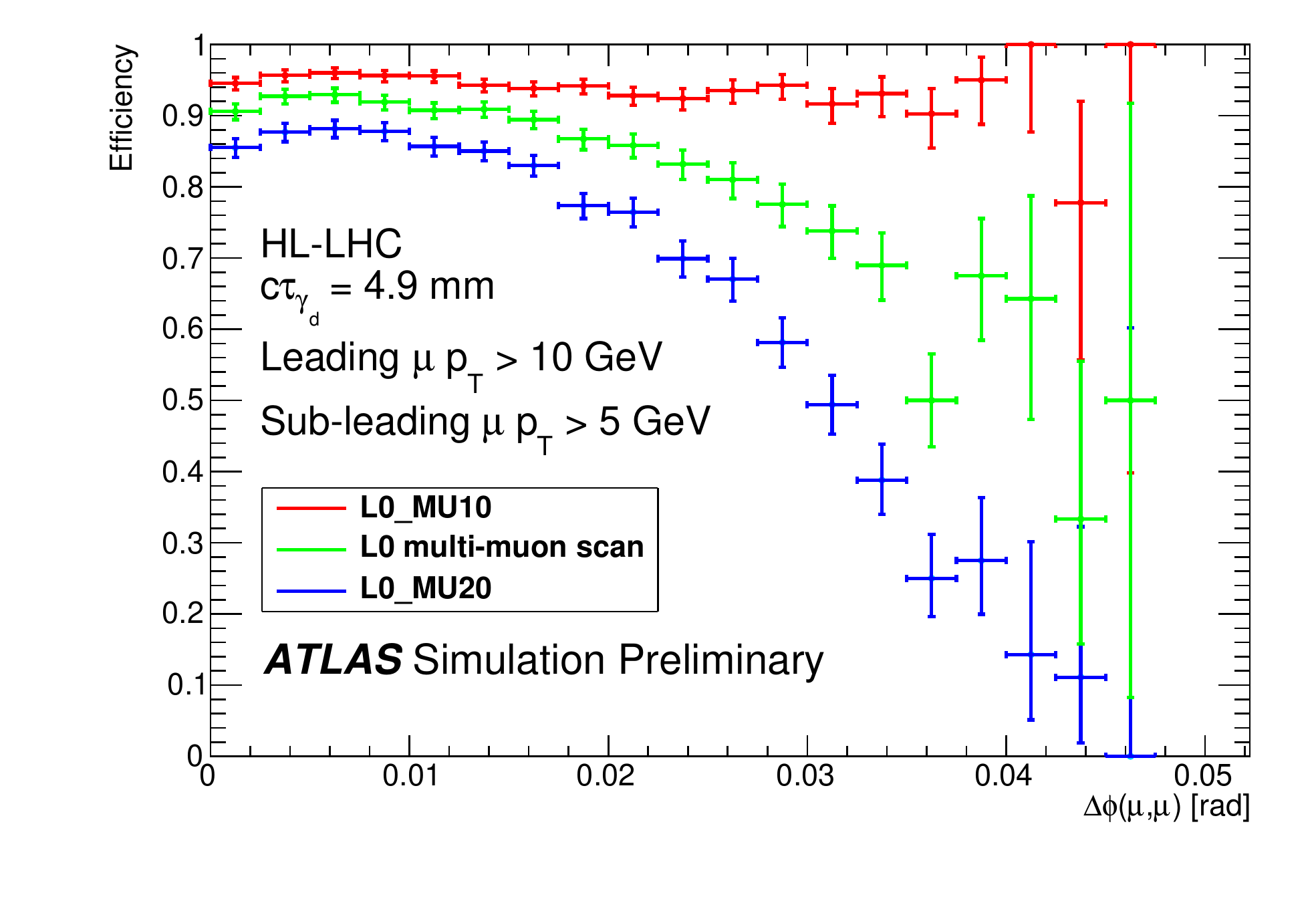}
\end{minipage}
\begin{minipage}[t][\minipageheightdp][t]{\minipagewidthdp}
\centering
\includegraphics[width=\figuremaxwidthdp,height=\figuremaxheightdp,keepaspectratio]{\main/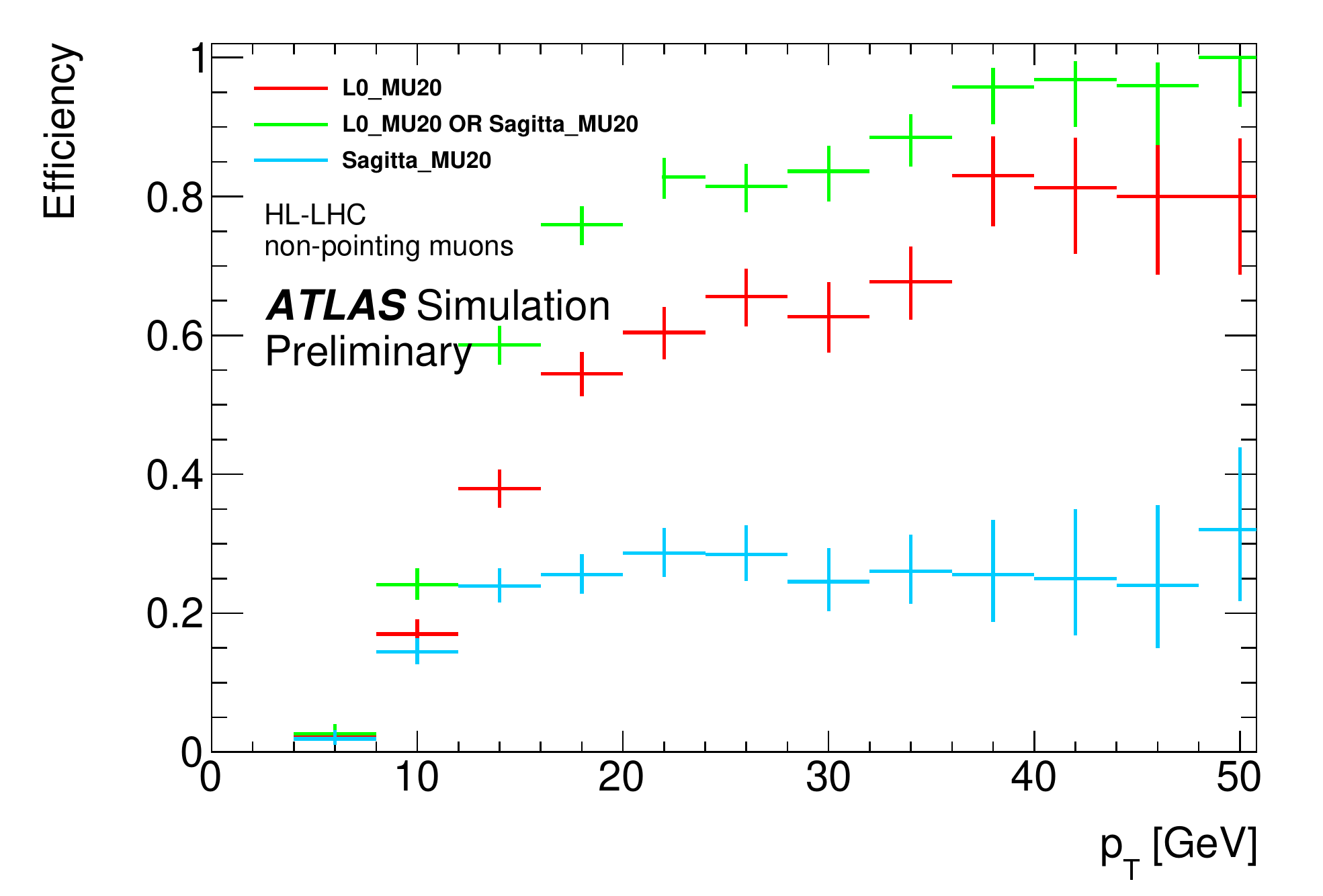}
\end{minipage}
\caption{Left:  Efficiency for different trigger selections as a function of the opening angle of the two muons of the dark photon decay (the new L0 multi-muon scan trigger is shown in green). As a reference, two single-muon selections are shown for $10$ (L1$\_$MU10) and $20$ (L1$\_$MU20) \GeV $\pt$ threshold. Right: Trigger efficiency comparison for a FRVZ sample (the new L0 sagitta muon trigger is shown in green) as a function of the muon transverse momentum. 
\label{fig:triggerDP}}
\end{figure}

Overall, with these new approaches it is possible to choose a lower single muon $\pt$ threshold as compared to the Run-2 configuration, improving the selection efficiency of events with displaced muon pairs without increasing significantly the trigger rate, see~\fig{fig:triggerDP}. \\

The evaluation of the expected sensitivity of the displaced dark photon search after Run-3 and \sloppy\mbox{HL-LHC} operations is based on the 2015+2016 Run-2 ATLAS analysis where multivariate techniques are used for signal discrimination against the backgrounds. The benchmark signal model used in the Run-2 search is a FRVZ model with $400\MeV$ $\gamma_d$ mass and lifetime $c\tau = 49~\mathrm{mm}$. The branching fraction of the $\gamma_d$ decay to muons is $45\%$.  
Only the dominant ggF Higgs production mechanism is considered. One of the main SM backgrounds to the dark photon signal is multijet production. Samples of simulated $14\TeV$ multijet events are used to compute scale factors to rescale the data-driven estimates at $13\TeV$ \com~energy to $14\TeV$. These samples are also used to evaluate the systematic uncertainties. Other sources of background include cosmic rays. This is assumed to scale with duration of data taking, and the cosmic ray background from the Run 2 analysis has been scaled accordingly. 

Uncertainties have been extrapolated from the Run-2 reference analysis. The statistical sources of uncertainties have been scaled with the expected integrated luminosity, for both Run-3 and HL-LHC. The systematic uncertainties for Run-3 have been assumed to be the same as in the Run-2 analysis. For the HL-LHC projection systematic uncertainties have been evaluated according to the specifications of the ATLAS collaboration for upgrade studies. Overall, the dominant uncertainties ($\sim 20\%$) are expected to be arising from pile-up. 

\begin{figure}[t]
\centering
\begin{minipage}[t][\minipageheightsp][t]{\minipagewidthsp}
\centering
\includegraphics[height=\figuremaxheightsp,keepaspectratio]{\main/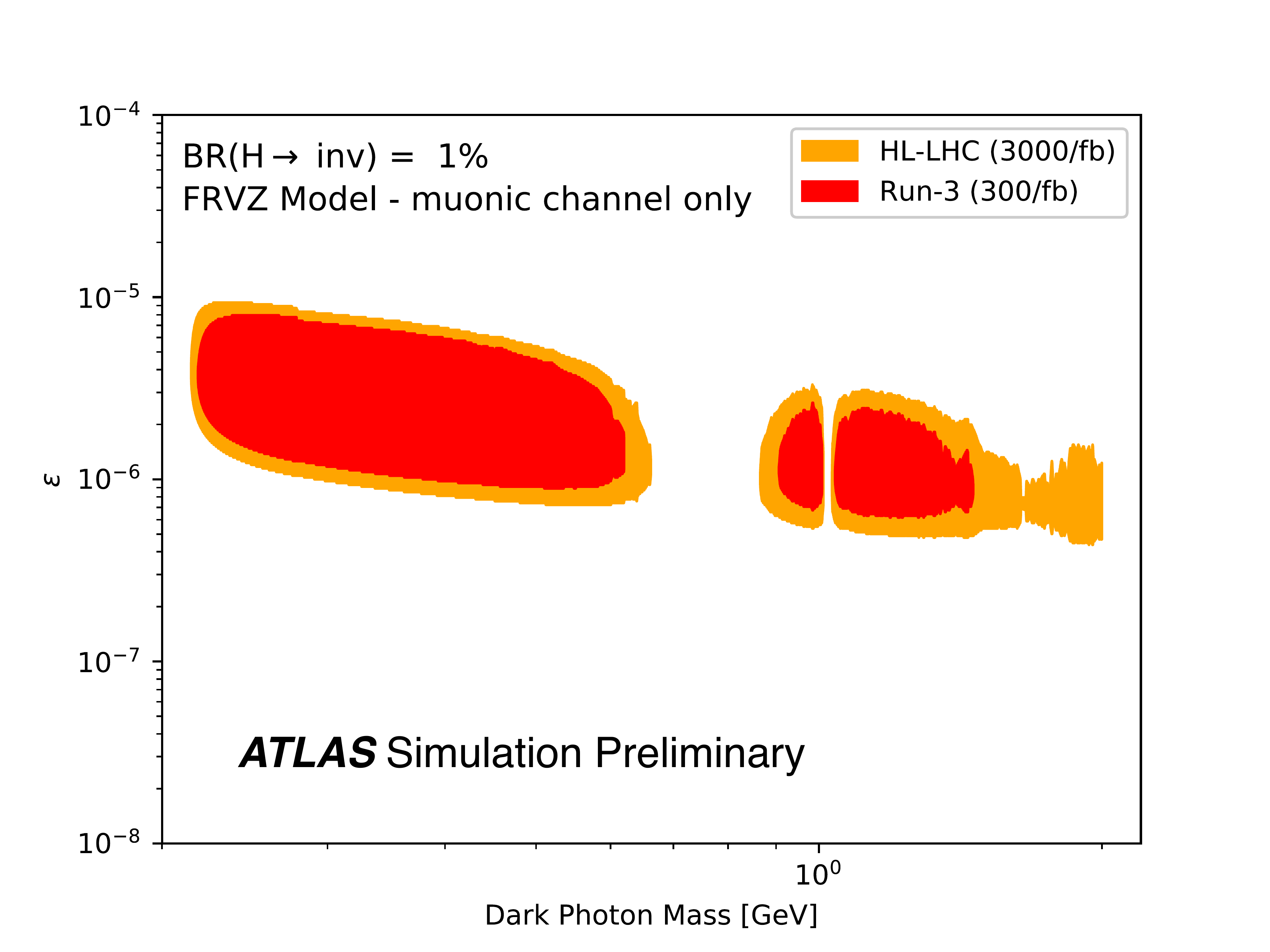}
\end{minipage}
\caption{Exclusion contour plot in the plane defined by the $\gamma_d$ mass and the kinetic mixing parameter $\epsilon$. Two different scenarios are shown assuming a Higgs decay branching fraction to the hidden sector of $1\%$: $300\fbinv$ after Run-3 (red) ad $3\abinv$ after HL-LHC including multi-muon scan trigger improvement (orange).
\label{fig:limitplotDP}}
\end{figure}

Results for the three different scenarios (Run 3, HL-LHC and HL-LHC with trigger improvements) are presented in \tabb{tab:limitsDP} for dark photons with $m^{}_{\gamma_d}=400\MeV$. The excluded $c\tau$ ranges assuming Higgs into dark photons branching ratio of $10\%$ and $1\%$ respectively are shown. The exclusion limits are re-interpreted in the context of the vector portal model. The exclusion contour plot in the plane defined by the dark photon mass and the kinetic mixing parameter $\epsilon$ is presented in~\fig{fig:limitplotDP}, assuming a Higgs decay branching fraction to the hidden sector of $1\%$ and where gaps correspond to hadronic decays not covered by this analysis. 
\begin{table}[t!]
\small
\centering

\begin{tabular}{| c| c| c | c | c| }
\hline
%Excluded $c\tau~[\mathrm{mm} ]$  & Run-2& Run-3 &  HL-LHC\\
Excluded $c\tau~[\mathrm{mm} ]$ & Run-2 & Run-3 &  HL-LHC & HL-LHC\\
 muonic-muonic &  & & &w/ L0 muon-scan \\

\hline
\hline
$\BR(h\rightarrow 2 \gamma_d + X)=10~\%$     &  $2.2 \leq c\tau \leq 111$ &   $1.15 \leq c\tau \leq 435$  &  $0.97 \leq c\tau \leq 553$   &  $0.97 \leq c\tau \leq  597$     \\\hline
$\BR(h\rightarrow 2 \gamma_d + X)=1~\%$     &  - &   $2.76 \leq c\tau \leq 102$  &  $2.18 \leq c\tau \leq 142$   &  $2.13 \leq c\tau \leq  148$     \\
\hline
\end{tabular}
\caption{Ranges of $\gamma_{d}$  $c\tau$ excluded at $95\%\cl$ for $h\rightarrow 2 \gamma_d + X$ assuming  $\BR(h\rightarrow 2 \gamma_d + X) = 10\%$ and $\BR(h\rightarrow 2 \gamma_d + X) = 1\%$ and dark photon mass of $400\MeV$.}
\label{tab:limitsDP}
\end{table}

\subsubsubsection{Summary of sensitivity for dark photons from Higgs decays}

\begin{figure}[h]
\centering
\begin{minipage}[t][\minipageheightsp][t]{\minipagewidthsp}
\centering
\includegraphics[height=\figuremaxheightsp,keepaspectratio]{\main/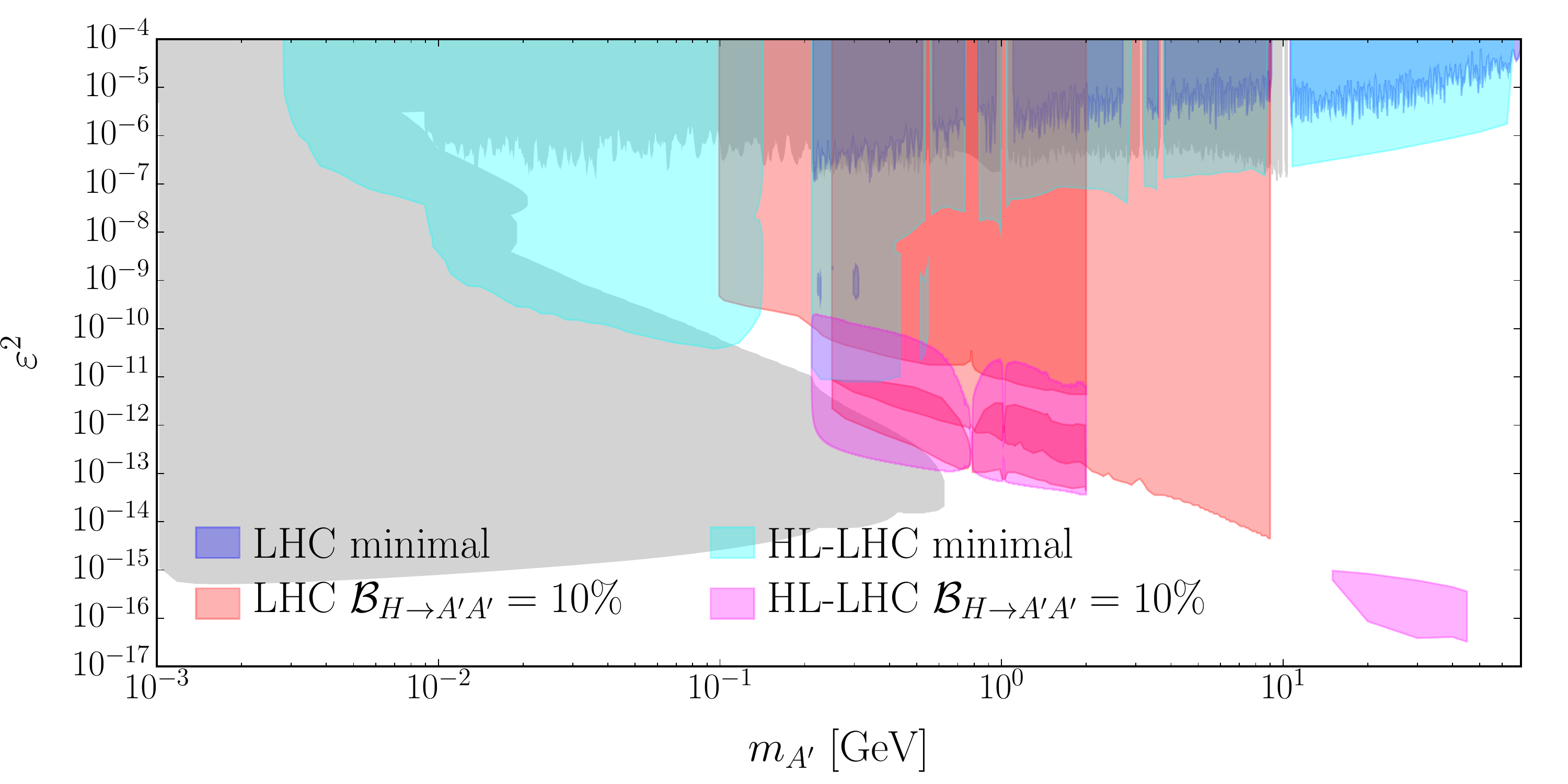}
\end{minipage}
  \caption{Summary of the contour reach of searches for dark photons from Higgs decays. The purple, grey and blue areas are explained in \secc{sec:darkphotonLHCb}, and correspond to the minimal dark photon model, with best sensitivity achieved by LHCb and low-energy experiments. The red and pink areas, explained in \secc{sec:atlasLLdarkphoton} and \secc{sec:cmsLLdarkphoton}, correspond to results from ATLAS and CMS where dark photons are produced through a Higgs boson decay with a branching fraction of $10\%$.
\label{fig:summaryDP}}
\end{figure}

The discovery reach from the ATLAS and CMS searches for dark photons can be compared to that from the generic $\gamma_d$ search results shown in~\fig{fig:DarkPhoton}. This is reported in ~\fig{fig:summaryDP} as a function of the dark photon mass and $\epsilon^2$: the reach of minimal models is shown together with that of models with additional assumptions on the dark photon production mechanism via Higgs decays.  A $10\%$ decay rate of the Higgs boson into dark photons is assumed for the latter. Under these assumptions, the HL-LHC ATLAS search will allow to target a crucial region with dark photon mass between $0.2$ and $10\GeV$ and low $\epsilon^2$, while the CMS search will cover higher $\gamma_d$ masses and even lower mixing parameters. This is complementary to the LHCb and low-energy experiments reach as well as with the coverage of prompt-lepton searches at the LHC.  

\subsubsection{\theoryC{Searching for dark photons via Higgs-boson production at the HL- and HE-LHC}}
\label{sec:dark_photons_th}
\contributors{S. Biswas, E. Gabrielli, M. Heikinheimo, B. Mele}

The dark-sector scenario proposed in \citeref{Gabrielli:2013jka,Gabrielli:2016vbb} is studied in this section. It aims at naturally solving the Flavour hierarchy problem, while providing suitable candidates for DM constituents. In particular, the scenario envisages the existence of stable dark-fermion fields, acting as DM particles, and heavy messenger scalar fields that communicate the interactions between the dark and the SM sectors. Both dark fermions and messenger fields are charged under an unbroken $U(1)$ interaction in the dark sector, whose non-perturbative dynamics are responsible for an exponential hierarchy in the dark fermion spectrum. Consequently, 
 exponentially spread Yukawa couplings are radiatively generated by the dark sector, thus naturally solving the Flavour hierarchy problem.

A crucial aspect of this flavour model is that it  foresees the existence of a massless dark photon. 
Until recently, most attention in collider physics has been given to the search for massive dark photons, whose $U(1)$ gauge field can naturally develop a {\it tree-level} millicharge coupling with ordinary matter fields.
On the contrary, strictly massless dark photons, although very appealing from the theoretical point of view, in general lack  tree-level couplings to SM fields. Indeed, the latter  (even if induced, for instance, by a kinetic mixing with the ordinary photon field) can be rotated away, and reabsorbed in the gauge- and matter-field redefinition~\cite{Holdom:1985ag}. Nevertheless, thanks to the messenger fields, massless dark photons can develop higher-dimensional effective interactions with the SM fields which are  suppressed by the effective scale controlling  the corresponding higher-dimensional-operator coupling. Then, new dedicated search strategies for the massless dark photons are required with respect to the massive case.

The Higgs boson could play a crucial role in the discovery of massless dark photons at the LHC. As discussed in \citeref{Gabrielli:2014oya}, by using as a  benchmark the model in \citeref{Gabrielli:2013jka}, an effective $H\gamma \bar{\gamma}$ interaction can be generated at one loop by the exchange of virtual messenger fields in the 3-point loop function.
This interaction can be parametrised as 
$
{\cal L}_{H\!\gamma \bar{\gamma}} = \frac{1}{\Lambda_{H\!\gamma \bar{\gamma}}} H F^{\mu\nu} \bar{F}_{\mu\nu},
$
where  $F^{\mu\nu}$ and $\bar{F}_{\mu\nu}$ are the field strength of the photon $\gamma$ and the dark photon $\bar{\gamma}$, respectively~\cite{Gabrielli:2014oya}.
While in general the higher-dimensional dark-photon interactions with the SM fields are suppressed, in the present case the Higgs boson can enter a nondecoupling regime in particular model parameter regions, just as happens in the SM for the   Higgs couplings to two photons or gluons   in the large top-quark mass limit. The effective high-energy scale $\Lambda_{H\!\gamma \bar{\gamma}}$  will  then be proportional to the EW Higgs \vev $v$, rather than the characteristic new-physics mass scale. In particular, it  will be given by
\begin{eqnarray}
\Lambda_{H\!\gamma \bar{\gamma}}&=& \frac{6\pi v}{R\sqrt{\alpha\alpha_{D}}}\frac{1-\xi}{\xi^2}
\end{eqnarray}
where  $\xi=\Delta/\bar{m}^2$ is a mixing parameter,
with $\Delta$ the left-right mixing term in the messenger square-mass matrix, $\bar{m}$ is the average messenger mass, $\alpha$ and $\alpha_{D}$ the electromagnetic and the dark $U(1)$ fine structure couplings, respectively, while $R$ is  a product of quantum 
charges~\cite{Biswas:2017lyg}.

This regime can give rise to an exotic signature corresponding to the Higgs decay $$H\to \gamma \bar{\gamma},$$ given by  a monochromatic photon plus {\it massless missing momentum} (both resonating at the Higgs boson mass) with BRs $\BR_{\gamma \bar{\gamma}}$ as large as a few percent.
Below we  report the results of a study of the LHC searches for this decay signature in  gluon-fusion  Higgs production in both the HL- and HE-LHC phases, assuming that $\BR_{\gamma \bar{\gamma}}$ is the only parameter that affects the corresponding production mechanism.

%\subsubsection{Gluon fusion analysis}

The search strategy for the $gg\rightarrow H\rightarrow \gamma\bar{\gamma}$ process was outlined in \citeref{Gabrielli:2014oya} for 8 TeV and in \citeref{Biswas:2016jsh} for 14 TeV, where we also discussed the vector-boson-fusion process.  The final state consists of a single photon and missing transverse momentum, possibly accompanied by one or more jets arising from initial state radiation. The event selection criteria proposed in \citeref{Biswas:2016jsh} were:
\begin{itemize}
\item one isolated ($\Delta R > 0.4$) photon with $p_T^\gamma > 50\GeV$, 
and $|\eta^\gamma|<1.44$;
\item missing transverse momentum satisfying $\slashed{E}_T > 50\GeV$;
\item transverse mass in  the range $100\GeV < M_{T}^{\gamma\bar\gamma} < 130\GeV$;
\item no isolated leptons within $|\eta^\ell|<2.5$,
\end{itemize}
where the transverse-mass variable  is defined as $M_{T}^{\gamma\bar\gamma}=\sqrt{2p_T^\gamma \slashed{E}_T(1-\cos\Delta\phi)}$, and   $\Delta\phi$ is the azimuthal distance between the photon transverse momentum $p_T^\gamma$, and the missing transverse momentum $\slashed{E}_T$.

The most important SM backgrounds are: {\it (i)} $\gamma j$, where missing energy is created from mismeasurement of the jet energy and/or neutrinos from heavy flavour decays, and 
{\it (ii)} $jj$ where in addition to the above, a central jet is misidentified as a photon. In our analysis we assume a probability of $0.1\%$ for mis-tagging a jet as a photon, and a $90\%$ reconstruction efficiency for real photons. In addition to the QCD backgrounds, we identify the following EW backgrounds: $Z\gamma$, where the $Z$ decays into neutrinos; $W\gamma$, where the $W$ decays leptonically (excluding taus) and the charged lepton is outside the acceptance of $|\eta^\ell|<2.5$; and $W\rightarrow e\nu$, where the electron is misidentified as a photon. We also assume a $0.5\%$ probability for the electron to photon mis-tagging.

We have analysed the EW backgrounds at parton level with \madgraph{ 5 v2.3.3}. For the QCD backgrounds we use \madgraph{ 5} interfaced with \pythia{}, and follow the procedure outlined in \citeref{Biswas:2016jsh}. In particular, we have generated event samples at 8 TeV and applied the {\it{SUSY benchmark}} event selection criteria described in the CMS analysis~\cite{Khachatryan:2015vta}, not including  the "$\chi^2$", "$E_{\rm T}^{\rm miss}$ significance" and "$\alpha$" cuts. With these omissions, the event selection criteria is very similar to our selection criteria described above. We then approximate the effect of these further, more sophisticated cuts on the QCD backgrounds, by matching our event samples with the background yield after these cuts reported in \citeref{Khachatryan:2015vta}. This results in a rescaling $k$-factor of $k=0.11$ for the $\gamma j$ background, and $k=0.058$ for the $jj$ background at 8 TeV. Finally, we have generated the signal event samples with \alpgen{} interfaced with \pythia{}, and included the gluon fusion Higgs production processes with zero to one jets.

Assuming the same rescaling factors for the QCD backgrounds at $14$ and $27\TeV$, we obtain the signal and background event yields reported in \tabb{tab:event-yields}, clearly showing a worsening of the signal-to-background ratio at larger energies.

\begin{table}[t]
\begin{center}
\small\begin{tabular}{c|c|c}
 & $\sigma\times A~[14\TeV]$ & $\sigma\times A~[27\TeV]$ \\ \hline
$H\!\to\! \gamma \bar{\gamma}\;\;$\small{ ($\BR_{\gamma \bar{\gamma}}=1\%)$} & $101$ &  $236$ \\ \hline
 $\gamma j$ & $202$ & -- \\
 $jj\rightarrow\gamma j$ & $432$ & $4738$  \\
 $e\rightarrow \gamma$ & $93$ & $169$ \\
 $W(\rightarrow \!\ell\nu) \gamma$ & $123$ & $239$ \\
 $Z(\rightarrow \!\nu\nu) \gamma$ & $283$ & $509$ \\ \hline
\small{total background} & $1133$ & $5655$ \\ \hline
\end{tabular}\normalsize
\caption{Event yields in femtobarn for signal and backgrounds after the cuts $p_T^\gamma > 50\GeV$, $\slashed{E}_T > 50\GeV$, \mbox{$100\GeV < M_{T}^{\gamma\bar\gamma} < 130\GeV$}. The $\gamma j$ and $jj$ backgrounds are obtained via the rescaling $k$-factors described in the text. $A$ is the acceptance described in the text.}
\label{tab:event-yields}
\end{center}
\end{table}

We then tried an alternative strategy to control the QCD background, by analysing the effect of applying a jet veto within $|\eta^j|<4.5$, where a jet is defined as a cluster of hadrons within a cone of size $R=0.4$ and $p_T \ge 20\GeV$, using a simple cone algorithm.
In this case we no longer apply the rescaling $k$-factors obtained from our previous analysis, as now the cut-flow deviates from the CMS analysis presented in \citeref{Khachatryan:2015vta}. The resulting event yields are shown in \tabb{tab:event-yields-jet-veto}.
%\vskip - 2cm
\begin{table}[t]
\small\begin{center}
\begin{tabular}{c|c|c}
 & $\sigma\times A~[14 \TeV]$ & $\sigma\times A~ [27 \TeV]$ \\ \hline
$H\!\to\! \gamma \bar{\gamma}\;\;$\small{ ($\BR_{\gamma \bar{\gamma}}=1\%)$} & $66.6$ & $139.1$  \\ \hline
 $\gamma j$ & -- & -- \\
 $jj\rightarrow\gamma j$ & $886$ & $31235$ \\
 $e\rightarrow \gamma$ & $93$ & $169$ \\
 $W(\rightarrow \!\ell\nu) \gamma$ & $123$ & $239$ \\
 $Z(\rightarrow \!\nu\nu) \gamma$ & $283$ & $509$ \\ \hline
\small{total background} & $1385$ & $32153$ \\ \hline
\end{tabular}\normalsize
\caption{Event yields in femtobarn for signal and backgrounds after the cuts $p_T^\gamma > 50\GeV$, $\slashed{E}_T > 50\GeV$, \mbox{$100\GeV < M_{T}^{\gamma\bar\gamma} < 130\GeV$}, and jet veto within $|\eta^j|<4.5$. $A$ is the acceptance described in the text.}
\label{tab:event-yields-jet-veto}
\end{center}
\end{table}
Based on the event yields reported in \tabbs{tab:event-yields} and \ref{tab:event-yields-jet-veto}, we estimate the reach of the HL-LHC and HE-LHC in terms of the BR of the decay mode $H\rightarrow \gamma\bar{\gamma}$ as shown in \tabb{tab:reach}.
%\vskip - 2cm
\begin{table}[t]
\begin{center}
\small\begin{tabular}{c|c c|c c}
$\BR_{\gamma \bar{\gamma}}(\%)$&\multicolumn{2}{|c|}{$3\abinv$@$14\TeV$}&\multicolumn{2}{|c}{\intlumihelhc@$27\TeV$} \\ \hline
significance & $2\sigma$ & $5\sigma$ & $2\sigma$ & $5\sigma$ \\ \hline
CMS inspired & $0.012$  & $0.030$ & $0.0052$ & $0.013$ \\ \hline
jet veto in $|\eta^j|<4.5$ & $0.020$  & $0.051$ & $0.021$  & $0.053$  \\ \hline
\end{tabular}\normalsize
\caption{Discovery ($5\sigma$) and exclusion ($2\sigma$) reach for the $H\rightarrow \gamma\bar{\gamma}$ BR (in \%) at the HL-LHC and HE-LHC.}
\label{tab:reach}
\end{center}
\end{table}
On the basis of the present analysis, a quite good potential for HL-LHC is expected, that would allow for  a ($5\sigma$) discovery reach  on the corresponding $\BR_{\gamma \bar{\gamma}}$ down to $3\times 10^{-4}$, for  $3\abinv$ of integrated luminosity, provided the CMS inspired analysis of the $jj$ background can be reliably applied in this case. 
Same conclusions hold for the HE-LHC project, where a $1\times 10^{-4}$ ($5\sigma$) discovery reach can be achieved, for  \intlumihelhc of expected luminosity, assuming that  the CMS inspired analysis of the $jj$ background is  still reliable at $27\TeV$. On the other hand, if a jet veto in $|\eta^j<4.5|$ is applied instead, lower sensitivities on $\BR_{\gamma \bar{\gamma}}$ can be obtained, leading to discovery just for $\BR_{\gamma \bar{\gamma}}$ down to $5\times 10^{-4}$ at both HL-LHC and HE-LHC facilities. 

We nevertheless think that a more realistic detector simulation and optimisation strategy would be needed in order to make the present reach estimates more robust.
In \citeref{Biswas:2016jsh}, one can also find a study of the vector-boson-fusion channel sensitivity to $\BR_{\gamma \bar{\gamma}}$ at $14\TeV$.

%\subsubsection*{acknowledgments}
%The work of Matti Heikinheimo has been supported by the Academy of Finland, Grant NO. 310130.

\clearpage
\resetcounters

\section{Long Lived Particles}
\label{sec:LLP}

There are many examples of BSM physics where new particles that can be produced at the LHC will be long lived, on collider timescales, and may travel macroscopic distances before decaying.  Long lifetimes may be due to small couplings, small mass splittings, a high multiplicity of the decay final state, or a combination of these effects.  Details, such as the quantum numbers of the long lived particle (LLP) and the decay products, the typical boost of the LLP and its lifetime, will determine the best search strategy.  In all cases, LLPs present unique challenges for the experiments, both in terms of reconstruction/analysis and triggering, especially in the high pile up environment of the HL-LHC.
A wide variety of signatures can be produced by these later decaying LLPs which depend on their charge, decay position, branching fractions, masses, and other properties, and which traditional analyses are unlikely to be sensitive to. 

If the LLP is charged and decays while still in the tracker to final state particles that are either neutral or too soft to be reconstructed it will appear as a disappearing track: hits in the first few layers of the tracker with no corresponding hits in the outer layers, see \secc{sec:disappearingtrackssec4}.  Such a scenario occurs in models (\eg\ SUSY) with nearly degenerate charged and neutral states, where the charged pion in the decay is too soft to be seen as a track. A  complementary study in the context of disappearing track searches is presented in~\secc{sec:theorydisappearingtracksLHeC}, where the potential of LLP searches at $e^- p$ colliders is presented.  We present studies for disappearing tracks searches using simplified models of $\tilde{\chi}^\pm$ production which lead to exclusions of chargino masses up to $m(\tilde{\chi}_{1}^{\pm}) = 750\GeV$ ($1100\GeV$) for lifetimes of $1\nsec$ for the higgsino (wino) hypothesis. When considering the lifetime predicted by theory, masses up to $300\GeV$ and $830\GeV$ can be excluded in higgsino and wino models, respectively. This improves the $36\fbinv$ Run-2 mass reach by a factor of $2-3$.

Decays of LLPs where the decay products are not missed but instead include multiple tracks will lead to events containing at least one displaced vertex (DV).  Such a signal is sensitive to both charged and neutral LLPs.  If the displacement of the vertex is large, $\gamma c\tau\gsim 1\meters$, then the only available hits are in the muon system, limiting the final states to muons as in \secc{sec:cmsLLdarkphoton} and \secc{sec:cmsdisplacedmuons}.

%If the displacement of the vertex is large, $\gamma c\tau\gsim 1\meters$, then the only available hits are in the muon system, limiting the final states to muons. An example of this kind of signature has been already presented in~\secc{sec:cmsLLdarkphoton} for dedicated dark photon searches. In section \secc{sec:cmsdisplacedmuons} an additional study is presented. 

If the lifetime is shorter the DV can be reconstructed in the tracker.  One such analysis of gluinos decaying to a displaced jet and $\met$ is presented \secc{sec:atlastllpjetmet}.  Searches for long lived dark photons decaying to muons and/or jets are reported in ~\secc{sec:darkphotonLHCb} and  ~\secc{sec:atlascmsLLdarkphoton}.  The signature of long-lived dark photons decaying to displaced muons can be reconstructed with dedicated algorithms and is sensitive to very small coupling $\epsilon^2 \sim 10^{-14}$ for masses of the dark photons between $10$ and $35\GeV$.  Furthermore, LHCb is the only LHC experiment to be fully instrumented in the forward region $2<\eta<5$ and has proved to be sensitive to LLPs. This is particularly true in the low mass (few GeV) and low lifetime (few picoseconds) region of the LLPs. Prospects studies from LHCb on LLPs resulting from Higgs decays are shown in~\seccs{sec:lhcbllpmuons} and \ref{sec:lhcbllpdijets}. 

For displacement of several meters %For lifetimes above $\sim 10$ ns 
LLPs will transit all of the detector before decaying.  Heavy LLPs that are also charged, so called heavy stable charged particles (HSCPs), will behave in a similar fashion to a muon.  However, due to their increased mass it may be possible to distinguish them from muons through their time of flight, \secc{sec:cmsHSCPTOF}, or anomalous energy loss, \secc{sec:cmsHSCPdEdx}.  Finally, two examples of specialised techniques for LLP with jet-like signatures are presented in \secc{sec:techniques}, using timing or EM calorimeter information. %, which may also be used more generally to help tag boosted objects, see \secc{sec:cmsfasttiming},   %and \secc{sec:theoryboostedtiming}.

In addition to searching for LLPs in ATLAS, CMS, and LHCb there are complementary proposals to build new detectors specifically focused on LLP searches, often for light new physics produced in rare meson decays.  A detailed discussion of their capabilities is beyond the scope of this work, and will be discussed elsewhere.  
The Beyond Collider experiments are 
AL3X (A Laboratory for \sloppy\mbox{Long-Lived} eXotics)~\cite{Gligorov:2018vkc}, 
CODEX-b (COmpact Detector for EXotics at LHCb)~\cite{Gligorov:2017nwh}, 
FASER (ForwArd Search ExpeRiment)~\cite{Feng:2017uoz,Feng:2017vli,Kling:2018wct}, 
milliQan~\cite{Haas:2014dda,Ball:2016zrp}, 
MATHUSLA (MAssive Timing Hodoscope for Ultra Stable neutraL pArticles)~\cite{Chou:2016lxi,Curtin:2018mvb},
and SHiP (Search for Hidden Particles)~\cite{Alekhin:2015byh,Anelli:2015pba}.  They use alternative search strategies and often give complementary coverage of the available parameter space. In addition to the afore-mentioned study on disappearing tracks, complementary studies on LLPs \egg~from higgs decays have been performed in the context of a future $e^- p$ collider, resulting in good sensitivity for a wide range in $c\tau$ and mass~\cite{Curtin:2017bxr}.   

%%%%%%%%%%%%%%%%%%%%%%5
\subsection{Disappearing Tracks} 
\label{sec:disappearingtrackssec4}
A disappearing track occurs when the decay products of a charged particle, like a supersymmetric chargino, are not detected (disappear) because they either interact only weakly or have soft momenta and hence are not reconstructed. In the following, prospect studies for HL-, HE- and new proposed $e^- p$ collider are presented, illustrating the potential of this signature as well as its experimental challenges. 

\subsubsection{\atlas{Prospects for disappearing track analysis at HL-LHC}}
\label{sec:atlasdisappearingtracks}
\contributors{S. Amoroso, J. K. Anders, F. Meloni, C. Merlassino, B. Petersen, J. A. Sabater Iglesias, M. Saito, R. Sawada, P. Tornambe, M. Weber, ATLAS}
%{\bf Author(s):Michele Weber, Brian Petersen, Federico Meloni, Simone Amoroso, John Kenneth Anders, Masahiko Saito, Peter Tornambe, Claudia Merlassino, Ryu Sawada, Jorge Andres Sabater Iglesias et al. ATLAS}

\begin{figure}[t]
\centering
\begin{minipage}[t][\minipageheightdp-1.3cm][t]{\minipagewidthdp}
\centering
\includegraphics[width=0.6\textwidth,keepaspectratio]{\main/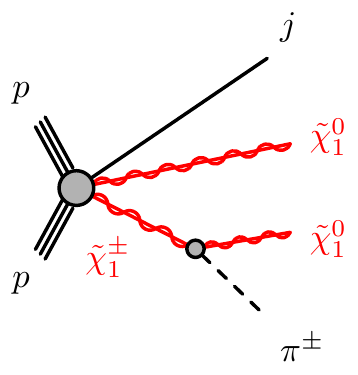}
\end{minipage}
\begin{minipage}[t][\minipageheightdp-1.3cm][t]{\minipagewidthdp}
\centering
\raisebox{1cm}{\includegraphics[width=\figuremaxwidthdp,height=\figuremaxheightdp,keepaspectratio]{\main/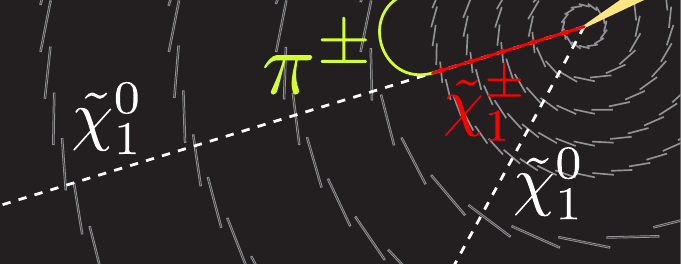}}
\end{minipage}
\caption{Diagram depicting $\tilde{\chi}_{1}^{\pm}\tilde{\chi}_{1}^{0}$ production (left), and schematic illustration of a $pp\rightarrow \tilde{\chi}_{1}^{\pm}\tilde{\chi}_{1}^{0} +~\mathrm{jet}$ event in the HL-LHC ATLAS detector, with a long-lived chargino (right). Particles produced in pile-up $pp$ interactions are not shown. The $\tilde{\chi}_{1}^{\pm}$ decays into a low-momentum pion and a $\tilde{\chi}_{1}^{0}$ after leaving hits in the pixel layers.}
\label{fig:DT_sketch}
\end{figure}

The disappearing track search~\cite{ATL-PHYS-PUB-2018-031} investigates scenarios where the $\tilde{\chi}_{1}^{\pm}$, and $\tilde{\chi}_{1}^{0}$ are almost mass degenerate, leading to a long lifetime for the $\tilde{\chi}_{1}^{\pm}$ which decays after the first few layers of the inner detector, leaving a track in the innermost layers of the detector. 
The chargino decays as $\tilde{\chi}_{1}^{\pm} \rightarrow \pi^{\pm}\tilde{\chi}_{1}^{0}$. The $\tilde{\chi}_{1}^{0}$ escapes the detector and the pion has a very low energy and is not reconstructed, leading to the disappearing track signature. Diagram and schematic illustration of production and decay process are shown in in \fig{fig:DT_sketch}.
%The latest ATLAS results for this scenario~\cite{SUSY-2016-06} using the 2015--2016 dataset of the LHC $pp$ run at a \com~energy of $\sqrt{s}=13$~TeV, excluded wino masses below 430~GeV with a chargino lifetime, $\tau(\tilde{\chi}_{1}^{\pm})$, of 0.2~ns.
The main signature of the search is a short ``tracklet'' which is reconstructed in the inner layers of the detector and subsequently disappears. 
The tracklet reconstruction efficiency for signal charginos is estimated using fully simulated samples of 
$\tilde{\chi}_{1}^{\pm}$ pair production with $m(\tilde{\chi}_{1}^{\pm}) = 600\GeV$. 
Tracklet reconstruction is performed in two stages. Firstly ``standard'' tracks, hereafter referred to as tracks are reconstructed. Afterwards the track 
reconstruction is then rerun with looser criteria, requiring at least four pixel-detector hits. This second reconstruction uses only input hits which are not 
associated with tracks, referred to as ``tracklets''. The tracklets are then extrapolated to the strip detectors, and any compatible hits are assigned 
to the tracklet candidate. Tracklets are required to have $\pt>5\GeV$and $|\eta| < 2.2$. 
Candidate leptons, which are used only to veto events, are selected with $\pt>20\GeV$ and $|\eta|<2.47$ ($2.7$) for electrons (muons).

The signal region (SR) optimisation is performed by scanning a set of variables which are expected to provide discrimination between the signal scenario under 
consideration and the expected SM background processes. The final state contains zero leptons, large $\met$ and at least one tracklet, 
and events are reweighted by the expected efficiencies of tracklet reconstruction. The small mass splitting between the $\tilde{\chi}_{1}^{\pm}$ and $\tilde{\chi}_{1}^{0}$ implies they are generally produced back to back with similar transverse momentum. Hence it is necessary to select events where the system is boosted by the recoil of at least one energetic ISR jet. The minimum azimuthal angular distance between the first four jets (ordered in $p_{\mathrm{T}}$) and the $\met$ is required to be greater than 1, in order to reject events with mis-measured $\met$.

There are two main background contributions: SM particles that are reconstructed as tracklets, and events which contain fake tracklets. The SM particles reconstructed as tracklets are typically hadrons scattering in the detector material or electrons undergoing bremsstrahlung. The probability of an isolated electron or hadron leaving a disappearing track is calculated using samples of single electrons or pions passing through the current ATLAS detector layout, and is then scaled to take into account the ratio of material in the current ATLAS inner detector and the upgraded inner tracker. The second background contribution arises from events which contain ``fake'' tracklets. These events arise from $Z \rightarrow \nu\nu$ or $W \rightarrow \ell\nu$ (where the lepton is not reconstructed) and are scaled by the expected fake tracklet probability:
\begin{equation}
p_{\mathrm {fake,tight}}^{\mathrm {ITk}} = p_{\mathrm {fake,tight}}^{\mathrm {ATLAS}} \times \frac{R_{\mathrm {fake,loose}}^{\mathrm {ITk}}}{R_{\mathrm {fake,loose}}^{\mathrm {ATLAS}}} \times \frac{\epsilon_{z_{0}}^{\mathrm {ITk}}}{\epsilon_{z_{0}}^{\mathrm {ATLAS}}}.
\end{equation}
In this equation, $p_{\mathrm {fake,tight}}^{\mathrm {ATLAS}}$ is the fake rate of the current Run-2 analysis~\cite{ATLAS:2017bna}, computed using a $d_{0}$ sideband for the track reconstruction, $R_{\mathrm {fake,loose}}^{\mathrm {ITk}}$ is the fake rate in the same $d_{0}$ sideband for ITk computed with a neutrino particle gun sample, such that all tracks are purely a result of pile-up interactions, $R_{\mathrm {fake,loose}}^{\mathrm {ATLAS}}$ is the fake rate in the $d_{0}$ sideband for ATLAS computed on data, $\epsilon_{z_{0}}^{\mathrm {ITk}}$ is the selection efficiency of the tracklet $z_{0}$ selection in ITk, and $\epsilon_{z_{0}}^{\mathrm {ATLAS}}$ is the selection efficiency of the tracklet $z_{0}$ selection in ATLAS.

Systematic uncertainty projections for both searches have been determined starting from the systematic uncertainties studied in Run-2 
and evolving them to a level which the ATLAS and CMS collaborations have agreed to consider as a sensible extrapolation to HL--LHC. 
Hence, the theory modelling uncertainties are expected to halve while the recommendations for detector-level and experimental uncertainties are dependent 
upon the systematic uncertainty under consideration and are scaled appropriately from the Run-2 analysis. 
When setting exclusion limits, an additional systematic uncertainty of $20\%$ is set to account for the theoretical systematic uncertainty on the models under consideration.
The dominant uncertainties in the disappearing track analysis arise from the modelling of the fake tracklet component, and the total uncertainty on the background yield is extrapolated to be $30\%$.

\begin{table}[t]
\small\begin{center}
\setlength{\tabcolsep}{0.0pc}
{\small
\begin{tabular*}{\textwidth}{@{\extracolsep{\fill}}lr}
\noalign{\smallskip}\hline\noalign{\smallskip}
           & {\textbf {SR}}             \\[-0.05cm]
\noalign{\smallskip}\hline\noalign{\smallskip}
Total SM         & $4.6 \pm 1.3$                \\
\noalign{\smallskip}\hline\noalign{\smallskip}
        $V$+jets events         & $0.17 \pm 0.05$              \\
        $t\bar{t}$ events          & $0.02 \pm 0.01$              \\
        Fake tracklets       & $4.4 \pm 1.3$\\
        
 \noalign{\smallskip}\hline\noalign{\smallskip}
\end{tabular*}\normalsize
%%%
}
\end{center}
\caption{Yields are presented for the disappearing track SR selection with an integrated luminosity of \intLumi at $\sqrt{s} = 14\TeV$. The errors shown are the total statistical and systematic uncertainty.}
\label{tab:disTrk_Fit_Results}
\end{table}

\begin{figure}[t]
\centering
\begin{minipage}[t][\minipageheightdp][t]{\minipagewidthdp}
\centering
\includegraphics[width=\figuremaxwidthdp,height=\figuremaxheightdp,keepaspectratio]{\main/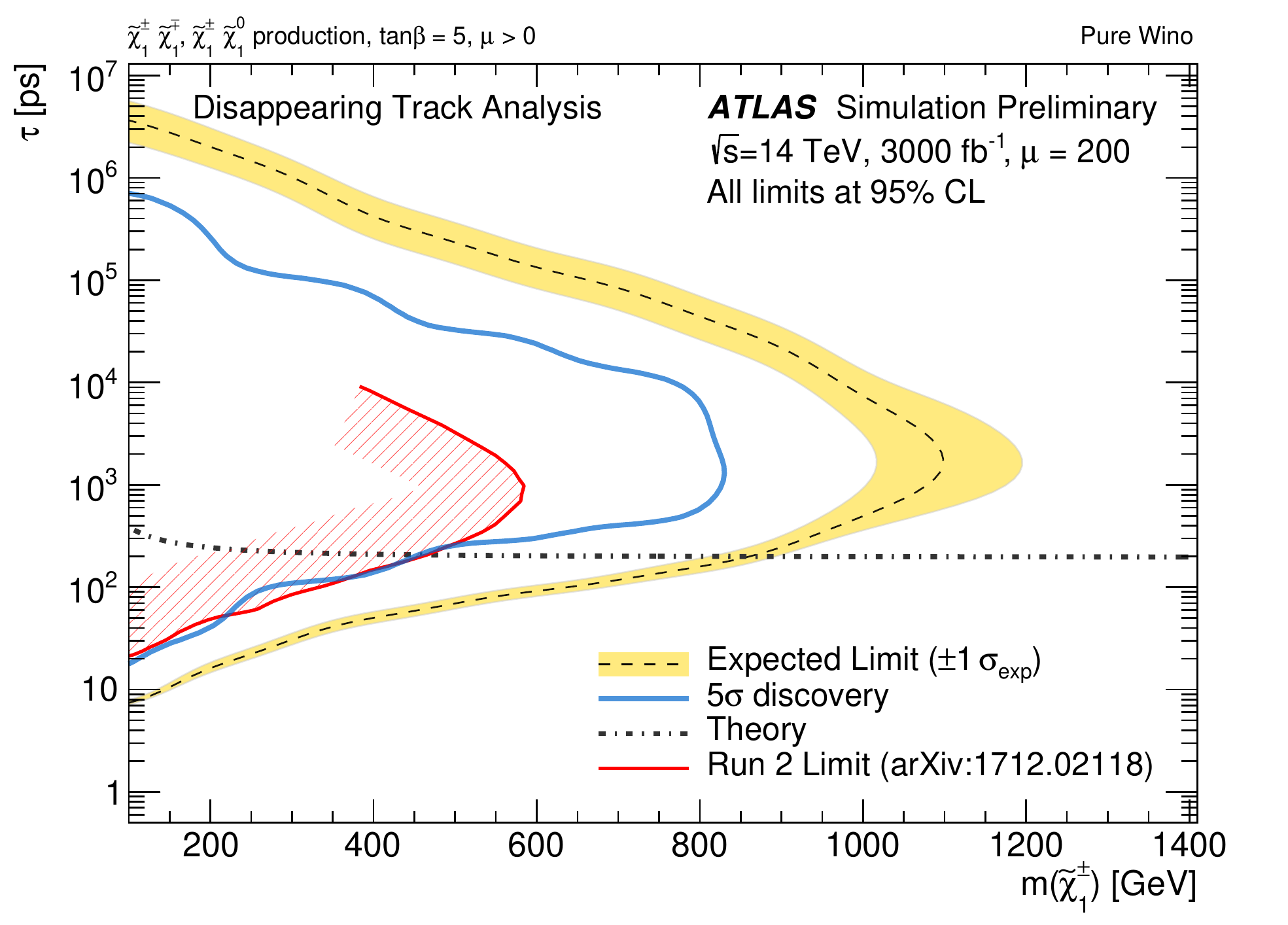}
\end{minipage}
\begin{minipage}[t][\minipageheightdp][t]{\minipagewidthdp}
\centering
\includegraphics[width=\figuremaxwidthdp,height=\figuremaxheightdp,keepaspectratio]{\main/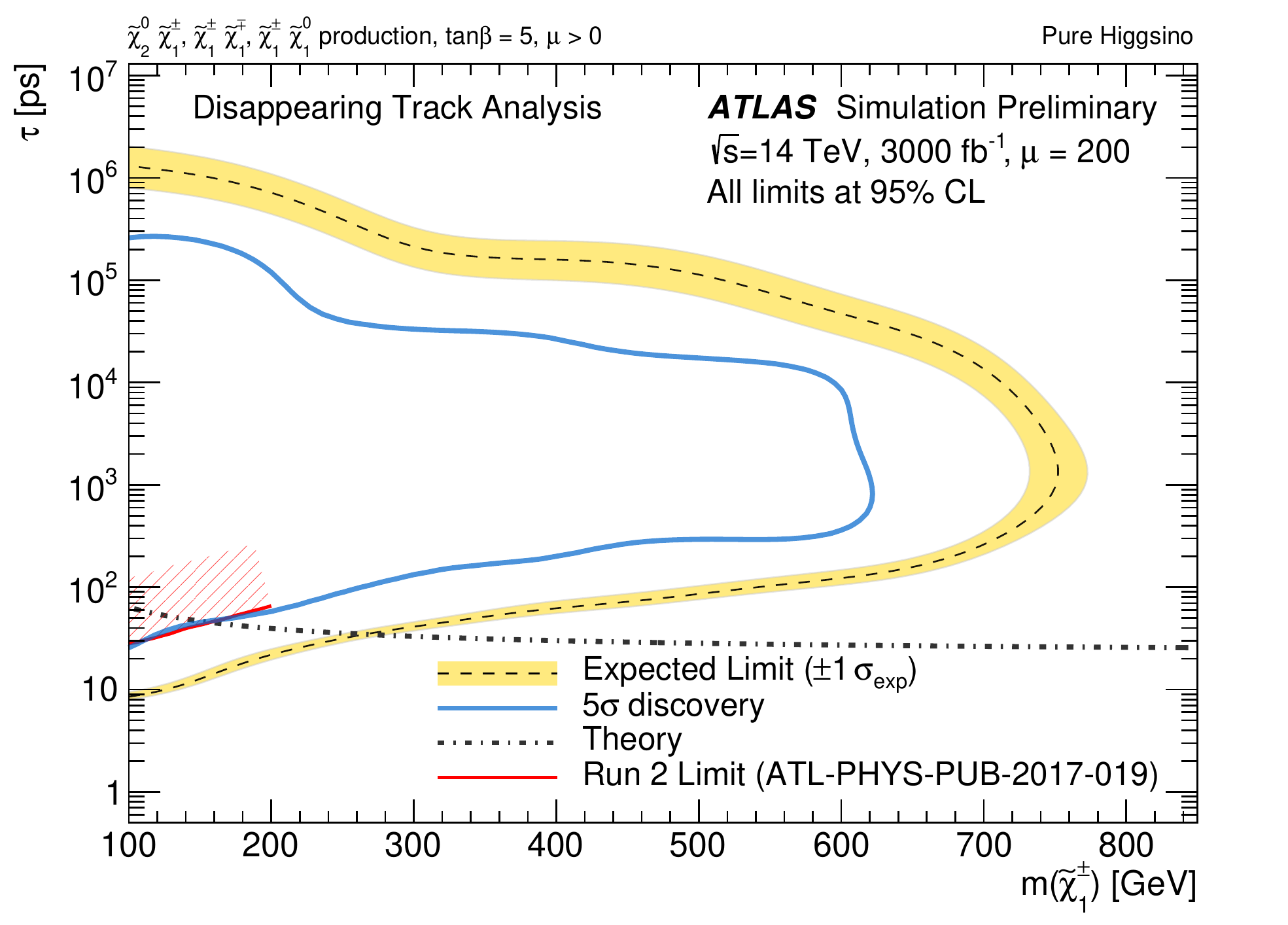}
\end{minipage}
\caption{Expected exclusion limits at $95\%\cl$ from the disappearing track search using of \intLumi of $14\TeV$ proton-proton collision data 
as a function of the $\tilde{\chi}_{1}^{\pm}$ mass and lifetime. Simplified models including both chargino pair production and associated production $\tilde{\chi}_{1}^{\pm} \tilde{\chi}_{1}^{0}$  are considered assuming pure-wino production cross sections (left) and pure-higgsino production cross sections (right). The yellow band shows the $1\sigma$ region of the distribution of the expected limits. The median of the expected limits is shown by a dashed line. The red line presents the current limits from the Run-2 analysis and the hashed region is used to show the direction of the exclusion. The expected limits with the upgraded ATLAS detector would extend these limits significantly. The chargino lifetime as a function of the chargino mass is shown in the almost pure wino LSP scenario (light grey) calculated at one loop level. The relationship between the masses of the chargino and the two lightest neutralinos in this scenario is $m(\tilde{\chi}_{1}^{\pm}) = (m(\tilde{\chi}_{1}^{0}) + m(\tilde{\chi}_{2}^{0}))/2$. The theory curve is a prediction from a pure higgsino scenario.}
\label{fig:limits_wino_DT_atlas}
\end{figure}

Table~\ref{tab:disTrk_Fit_Results} presents the expected yields in the SR for the disappearing track search for each background source, corresponding to an integrated luminosity of \intLumi. As seen in the table the dominant background source corresponds to events with a ``fake'' tracklet, arising predominantly from $Z\rightarrow\nu\nu$ events with an ISR jet and high $E_{\mathrm{T}}^{\mathrm{miss}}$, which contain spurious hits that are reconstructed as a tracklet.

Limits at $95\%\cl$ on the chargino lifetime are shown in \fig{fig:limits_wino_DT_atlas} as a function of the $\tilde{\chi}_{1}^{\pm}$ mass. The simplified models of chargino production considered include chargino pair production and chargino-neutralino production (both $\tilde{\chi}_{1}^{\pm} \tilde{\chi}_{1}^{0}$ and $\tilde{\chi}_{1}^{\pm} \tilde{\chi}_{2}^{0}$. The potential for the full HL-LHC dataset is expected to exclude at the $95\%\cl$ chargino lifetimes, assuming a wino-like (higgsino-like) LSP, of between 7\,ps (10\,ps) and 4\,$\mu$s (1.5\,$\mu$s) for light charginos with a mass of 100\,GeV. Heavier wino-like (higgsino-like) charginos are excluded up to $m(\tilde{\chi}_{1}^{\pm}) = 1100\GeV$ ($750\GeV$) for lifetimes of $1\nsec$. The discovery potential of the analysis would allow for the discovery of wino-like (higgsino-like) charginos of mass $100\GeV$ with lifetimes between $20\psec$ and $700\nsec$ ($30\psec$ and $250\nsec$), or for a lifetime of $1\nsec$ would allow the discovery of wino-like (higgsino-like) charginos of mass up to $800\GeV$ ($600\GeV$). 
%omparing the results to the theoretical prediction in SECTION XX of this report would allow for the exclusion at $95\%\cl$ of the theory with masses up to 850\,GeV for the pure wino scenario and 250\,GeV for the pure higgsino scenario. The discovery potential would be up to 450\,GeV for the pure wino scenario and 150\,GeV for the pure higgsino scenario. 

\begin{figure}[t]
\centering
\begin{minipage}[t][\minipageheightsp][t]{\minipagewidthsp}
\centering
\includegraphics[width=\figuremaxwidthsp,height=\figuremaxheightsp,keepaspectratio]{\main/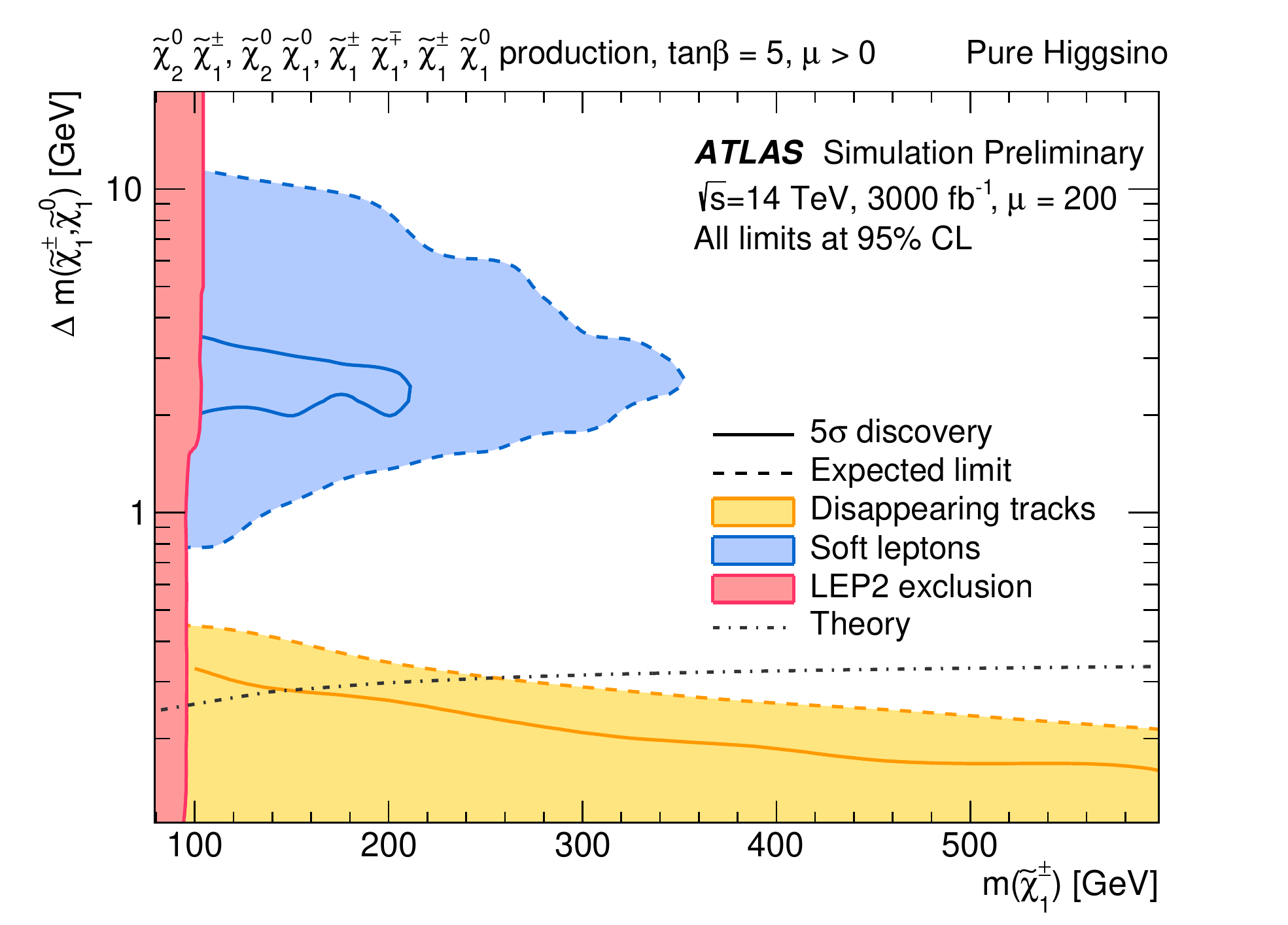}
\end{minipage}
\caption{Expected exclusion at the $95\%\cl$ from the disappearing track and dilepton searches in the $\Delta m(\tilde{\chi}_{1}^{\pm}, \tilde{\chi}_{1}^{0})$, $m(\tilde{\chi}_{1}^{\pm})$ mass plane. The blue curve presents the exclusion limits from the dilepton search. The yellow contour presents the exclusion limit from the disappearing track search. The figure also presents the limits on chargino production from LEP. The relationship between the masses of the chargino and the two lightest neutralinos in this scenario is $m(\tilde{\chi}_{1}^{\pm}) = \frac{1}{2}(m(\tilde{\chi}_{1}^{0}) + m(\tilde{\chi}_{2}^{0}))$. The theory curve is a prediction from a pure higgsino scenario taken from \citeref{Thomas:1998wy}.}
\label{fig:disap_dilept_final}
\end{figure}

Finally, \fig{fig:disap_dilept_final} presents the $95\%\cl$ expected exclusion limits in the $\tilde{\chi}_{1}^{0},\Delta m(\tilde{\chi}_{1}^{\pm},\tilde{\chi}_{1}^{0})$ mass plane, from both the disappearing track and dilepton searches. The yellow contour shows the expected exclusion limit from the disappearing track search, with the possibility to exclude $m(\tilde{\chi}_{1}^{\pm})$ up to $600\GeV$ for $\Delta m(\tilde{\chi}_{1}^{\pm}, \tilde{\chi}_{1}^{0}) < 0.2\GeV$, and could exclude up to $\Delta m(\tilde{\chi}_{1}^{\pm}, \tilde{\chi}_{1}^{0}) = 0.4\GeV$ for $m(\tilde{\chi}_{1}^{\pm}) = 100\GeV$. The blue curve presents the expected exclusion limits from the dilepton search, which could exclude up to $350\GeV$ in $m(\tilde{\chi}_{1}^{\pm})$, and for a light chargino mass of $100\GeV$ would exclude mass differences between $2$ and $15\GeV$. Improvements that are expected with the upgraded detector, and search technique improvements may further enhance the sensitivity to these models. For example the sensitivity of the disappearing tracks search can be enhanced by optimising the tracking algorithms used for the upgraded ATLAS detector allowing for an increase in tracklet efficiency, the possibility of shorter tracklets produced requiring 3 or 4 hits, and further suppression of the fake tracklet component. The dilepton search sensitivity would be expected to improve by increasing the reconstruction efficiency for low $\pt$ leptons. The addition of the electron channel would also further enhance the search sensitivity.  

\subsubsection{\theory{Complementarities between LHeC and HL-LHC for disappearing track searches}}
\label{sec:theorydisappearingtracksLHeC}
\contributors{K. Deshpande, O. Fischer, J. Zurita}
%from~\cite{Curtin:2017bxr} with , D. Curtin}
%{\bf Author(s): Oliver Fischer, Jose Zurita, D. Curtin and K. Deshpande} 

In higgsino-like SUSY models, the Higgsinos' tiny mass splittings give rise to finite lifetimes for the charginos, which is enhanced by the significant boost of the \com~system and can be used to suppress SM backgrounds ~\cite{Curtin:2017bxr}. The small mass splittings allow the Higgsinos to decay into \sloppy\mbox{$\pi^\pm, e^\pm, \mu^\pm$ + invisible} particles, with the single visible charged particle having transverse momenta in the $\mathcal{O}(0.1)\GeV$ range. 
In the clean environment (\iee{} low pile up) of the $e^- p$ collider, such single low-energy charged tracks can be reliably reconstructed, if the minimum displacement between primary and secondary vertex is at least $40\mumeters$, and the minimum $p_T$ of the charged SM particle is at least $100\MeV$.
It was shown in \citeref{Curtin:2017bxr} that the results do not crucially depend on the exact choice of these parameters. The associated DIS jet with $p_T > 20\GeV$ ensures that the event is recorded and determines the position of the primary vertex.  The charginos' decay into a neutral Higgsino and a number of SM particles with small $p_T$ defines the secondary vertex.

Tau leptons with their proper lifetime of $\sim 0.1~\mathrm{mm}$ constitute an important and irreducible background.
VBF can single- ($\tau^+ \nu_\tau$) and pair produce taus ($\tau^+ \tau^-$) together with a jet with \sloppy\mbox{$p_T > 20\GeV,  |\eta| < 4.7$} at LHeC with cross sections of $\sim0.6$ and $\sim0.3\pb$, respectively. 
Kinematically, the $\tau$ decay products can be suppressed to $10^{-3}$ (keeping ${\cal O}(1)$ of the signal) by requiring $|\eta| > 1$ (in the proton beam direction),  $\slashed{E}_{T} \gtrsim 30\GeV$) and the LLP final state energy to be very low ($\lesssim 1.5 \Delta m$ for a given chargino lifetime).
Furthermore, in the space of possible final states and decay lengths, the $\tau$'s will populate very different regions than the chargino signal, such that further suppression is possible.

\begin{figure}[t]
\centering
\begin{minipage}[t][\minipageheightdp][t]{\minipagewidthdp}
\centering
\includegraphics[width=\figuremaxwidthdp,height=\figuremaxheightdp,keepaspectratio]{\main/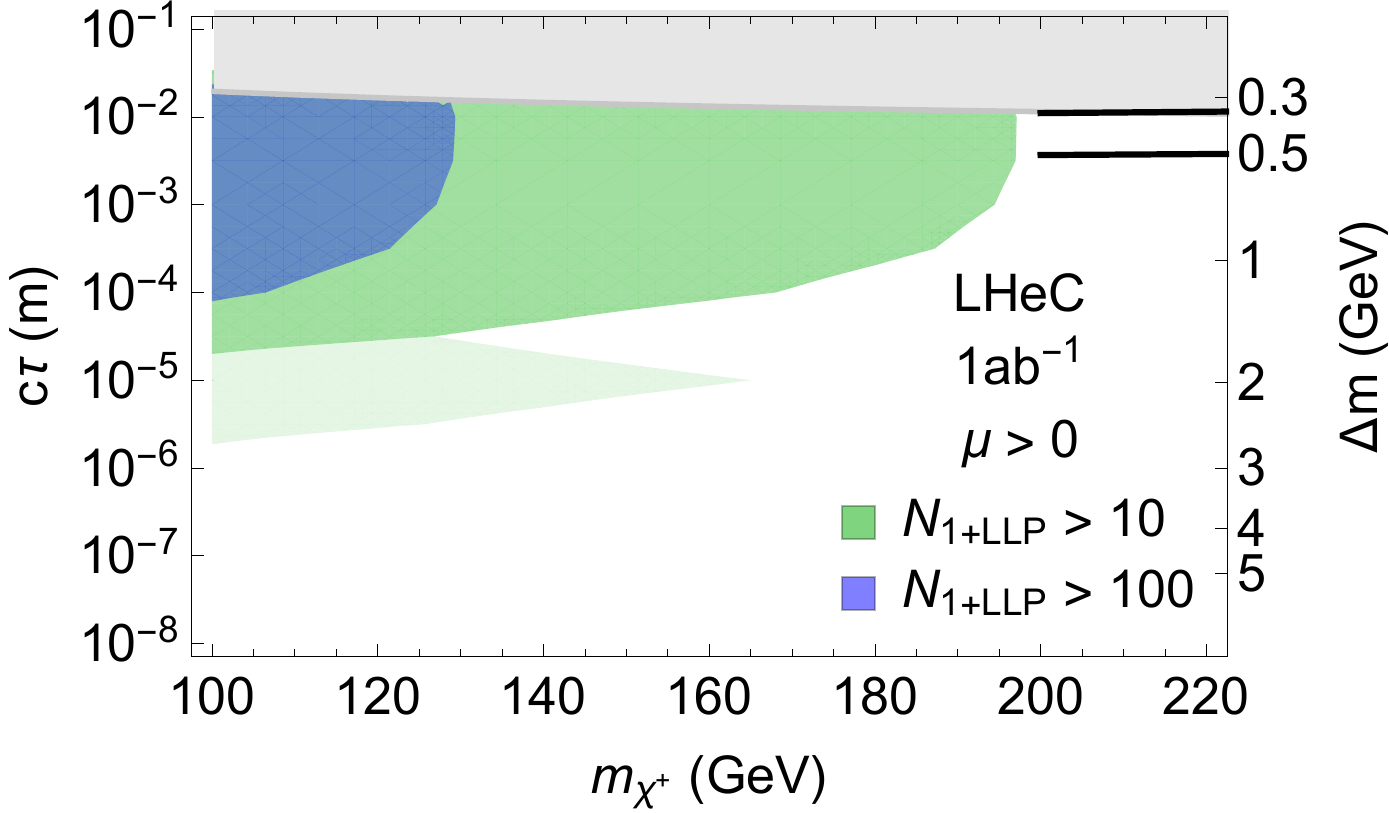}
\end{minipage}
\begin{minipage}[t][\minipageheightdp][t]{\minipagewidthdp}
\centering
\includegraphics[width=\figuremaxwidthdp,height=\figuremaxheightdp,keepaspectratio]{\main/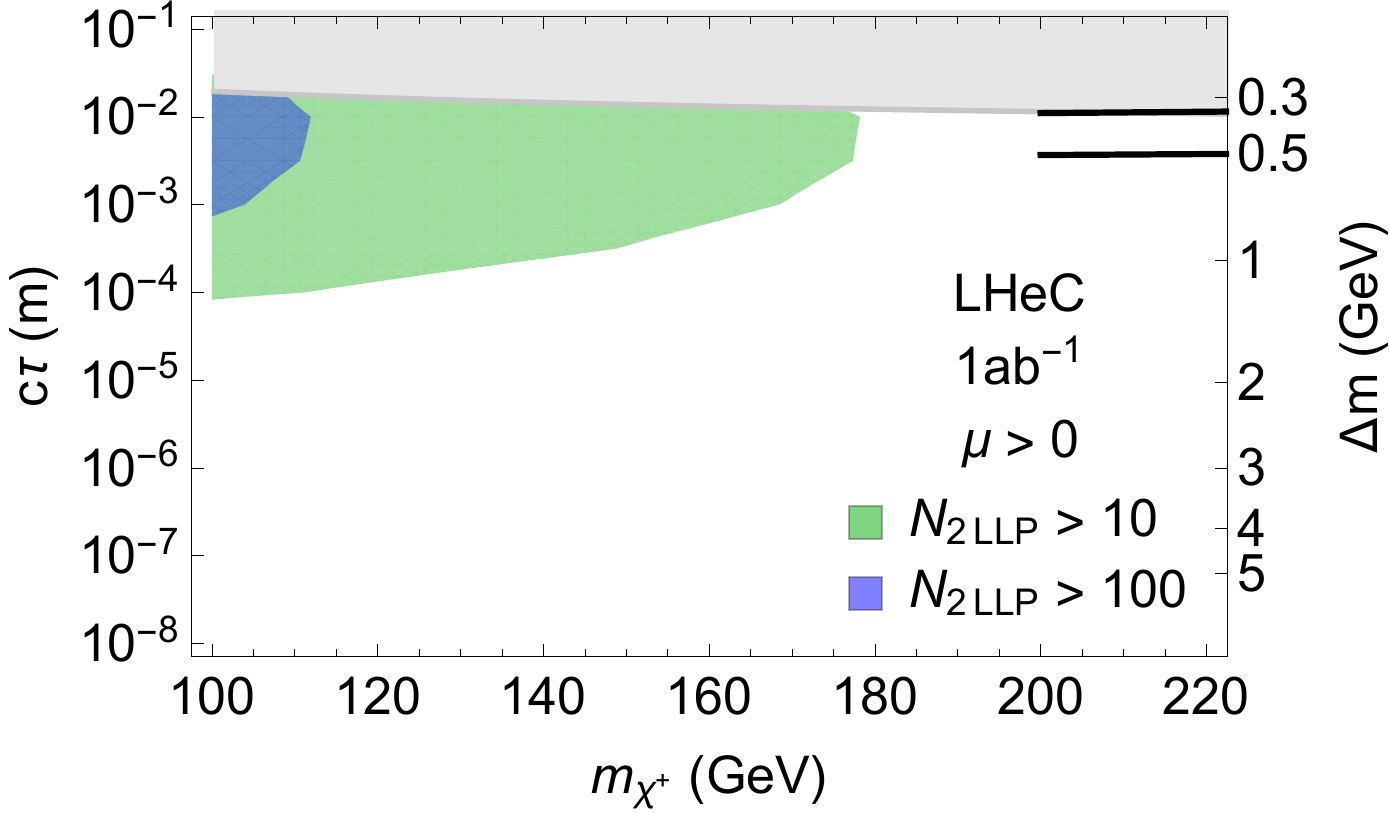}
\end{minipage}
\caption{
Regions in the $(m_{\chi^\pm}, c \tau)$ Higgsino parameter plane where more than 10 or 100 events with at least one (left) or two (right) LLPs are observed at the LHeC. Light shading indicates the uncertainty in the predicted number of events due to different hadronisation and LLP reconstruction assumptions. 
Approximately 10 signal events should be discernible against the $\tau$-background at $2\sigma$, in particular for 2 LLPs, so the green shaded region represents an estimate of the exclusion sensitivity. 
For comparison, the black curves are the optimistic and pessimistic projected bounds from HL-LHC disappearing track searches from \protect\citeref{Mahbubani:2017gjh}. The figure is from \protect\citeref{Curtin:2017bxr}.}
\label{fig:higgsinoLHeC}
\end{figure}

The probability of detecting a chargino is computed by choosing the charged particle momentum from the appropriate phase space distribution in the chargino rest frame, then computing the minimum distance the chargino must travel for the displacement of the resulting charged track to be visible.
The sensitivities of detecting at least one ($N_\mathrm{1+LLP}$), or two displaced vertices ($N_\mathrm{2LLP}$) are shown by the contours in \fig{fig:higgsinoLHeC} for $\mu > 0$. 
The darker (lighter) shading represents the contour with the lowest (highest) estimate of event yield, obtained by minimising (maximising) with respect to the two different hadronisation scenarios, and $P_{jet}$ reconstruction assumptions. 
The difference between the light and dark shaded regions can be interpreted as a range of uncertainty in projected reach.
%
%%%%%

This sensitivity for Higgsinos via LHeC searches is competitive in mass reach to the monojet projections for the HL-LHC, being sensitive to masses around $200\GeV$ for the longest theoretically motivated lifetimes (see also~\secc{sec:theoryewinosdisappear}). The LHeC search has the crucial advantage of actually observing the charged Higgsino parent of the invisible final state. Disappearing track searches at the HL-LHC presented in this report probe higher masses for the longest lifetimes, but lose sensitivity at shorter lifetimes. 
By comparison, the LHeC search is sensitive to lifetimes as short as microseconds. 
It is important to note how the robustness of the mass reach of $e^- p$ colliders arise also from the fact that results %than the disappearing track projections, since the former 
are not exponentially sensitive to uncertainties in the Higgsino velocity distribution. 

\subsubsection{\theory{Searching for Electroweakinos with disappearing tracks analysis at HL- and HE-LHC}}\label{sec:theoryewinosdisappear}
%%%% NOTE - contribution split in two parts! MD 

\contributors{T. Han, S. Mukhopadhyay, X. Wang}
	
Prospects for a disappearing charged track search are finally presented for three different scenarios of collider energy and integrated luminosity: HL-LHC, HE-LHC, and FCC-hh/SppC ($100\TeV$, $30\abinv$). The studies are documented in ~\citeref{Han:2018wus} and are complementary to the monojet prospects reported in \secc{sec:theoryewinosmono} for higgsino-like SUSY scenarios. 
%	\begin{eqnarray}
%	&{\rm HL\text{-}LHC:} \quad &14{\rm~TeV},\quad \intLumi, \nonumber \\
%	&{\rm HE\text{-}LHC:} \quad &27{\rm~TeV},\quad \intlumihelhc, \nonumber \\
%	&{\rm FCC\text{-}hh/SppC:} \quad &100{\rm~TeV},\quad 30\abinv.
%	\end{eqnarray}

As in \secc{sec:theoryewinosmono}, the significance is defined as $S/\sqrt{B+(\Delta_B B)^2 + (\Delta_S S)^2}$ where $S$ and $B$ are the total number of signal and background events, and $\Delta_S, \Delta_B$ refer to the corresponding percentage systematic uncertainties, respectively.
		
\begin{figure}[t]
\centering
\begin{minipage}[t][\minipageheightdp][t]{\minipagewidthdp}
\centering
\includegraphics[width=\figuremaxwidthdp,height=\figuremaxheightdp,keepaspectratio]{\main/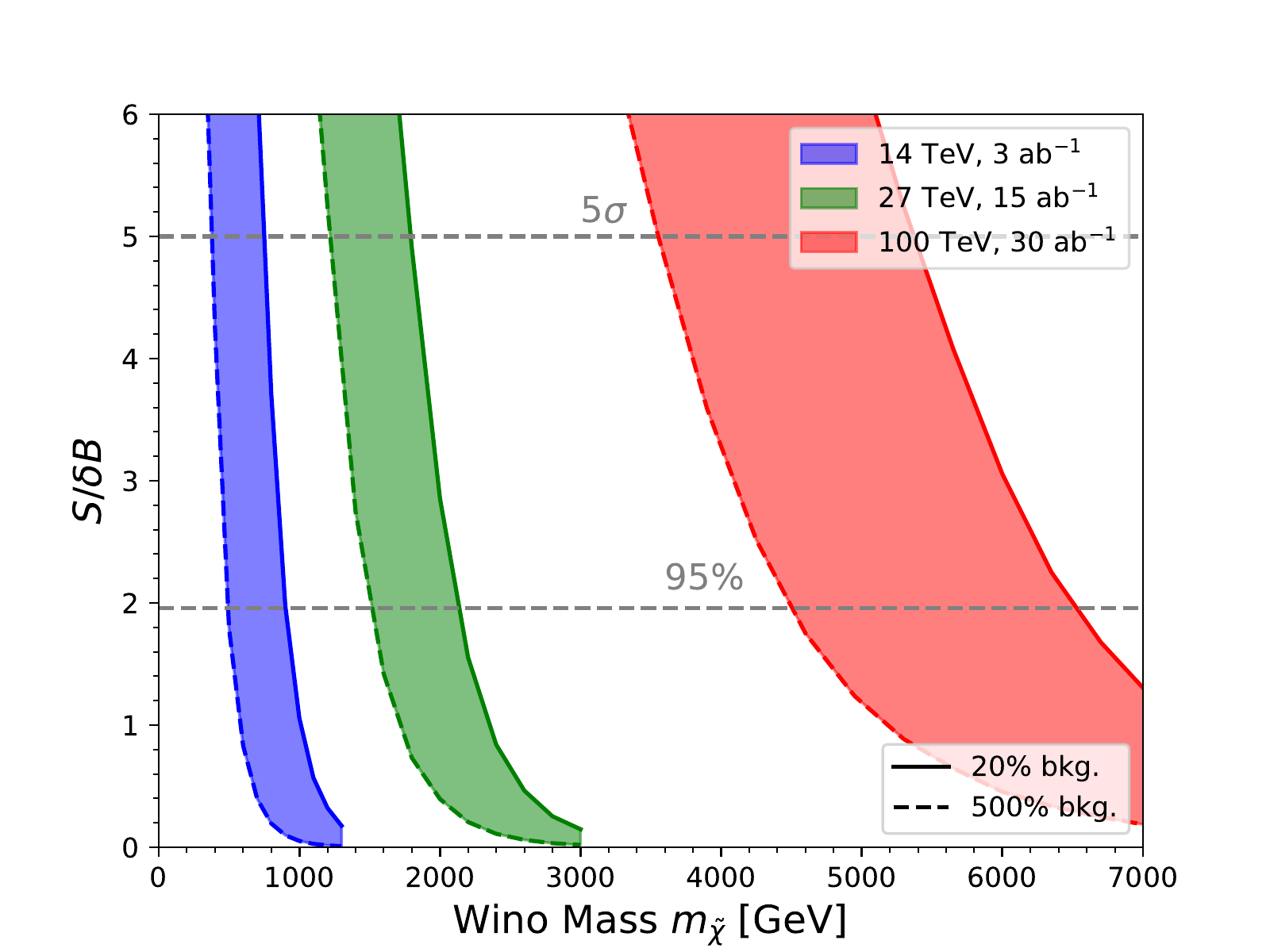}
\end{minipage}
\begin{minipage}[t][\minipageheightdp][t]{\minipagewidthdp}
\centering
\includegraphics[width=\figuremaxwidthdp,height=\figuremaxheightdp,keepaspectratio]{\main/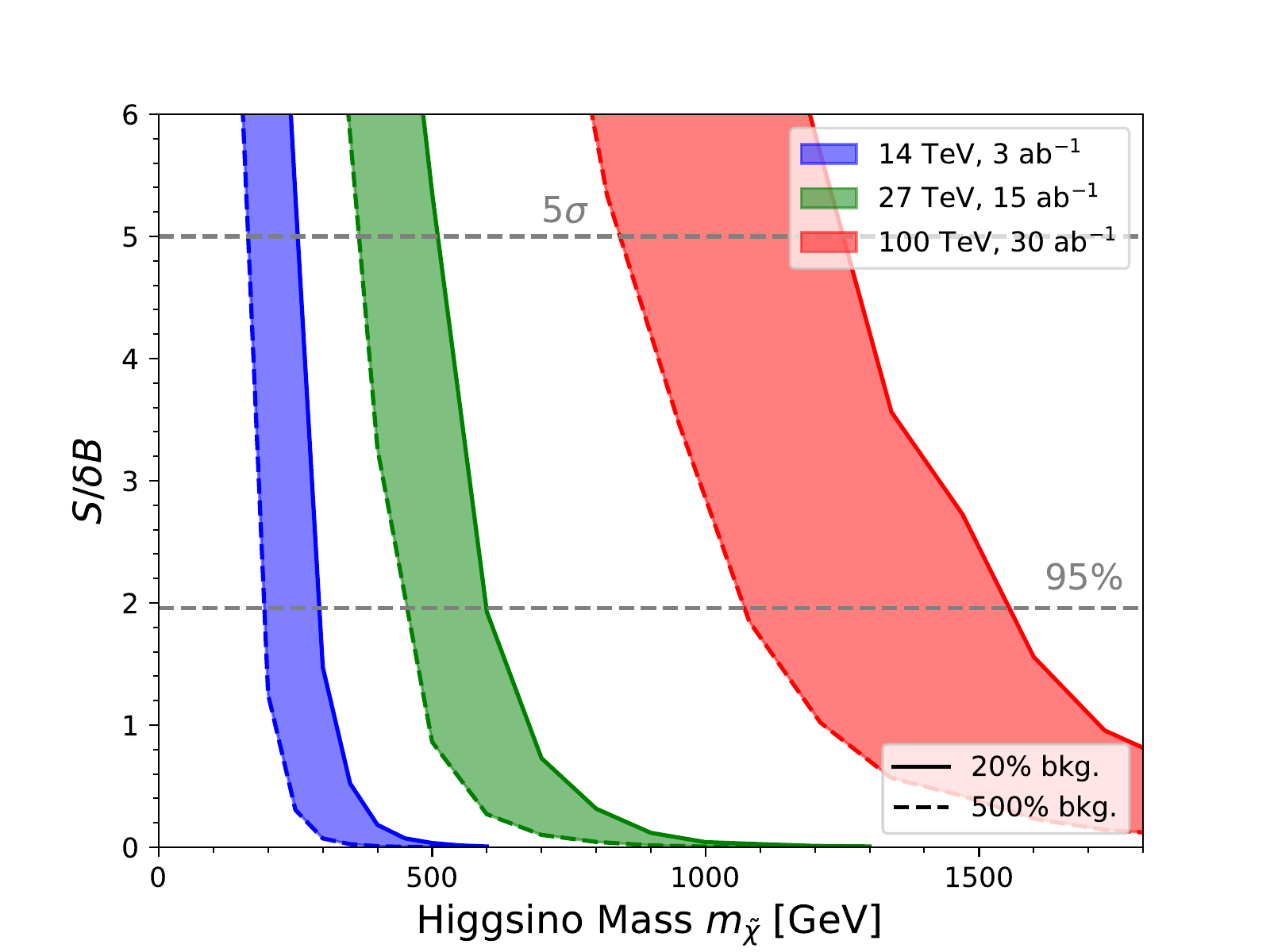}
\end{minipage}
\caption{\small Comparative reach of the HL-LHC, HE-LHC and FCC-hh/SppC options in the disappearing charged track analysis for wino-like (left) and Higgsino-like (right) DM search. The solid and dashed lines correspond to modifying the central value of the background estimate by a factor of five.}
\label{fig:sec344_reach_track}
\end{figure}

    Background and signal systematic uncertainties are assumed as $\Delta_B=20\%$ and $\Delta_S=10\%$ respectively. In \fig{fig:sec344_reach_track} we compare the reach of the HL-LHC, HE-LHC and FCC-hh/SppC options in the disappearing charged track analysis for wino-like (left) and Higgsino-like (right) DM search. The solid and dashed lines correspond to modifying the central value of the background estimate\footnote{Background is estimated by extrapolating ATLAS Run-2 analysis~\cite{Aaboud:2017mpt}. See~\cite{Han:2018wus} for details.} by a factor of five. With the optimistic estimation of the background, wino-like DM can be probed at the $95\%\cl$ up to $900$, $2100$, and $6500\GeV$, at the $14$, $27$, and $100\TeV$ colliders respectively. For the Higgsino-like scenario, these numbers are reduced to $300$, $600$, and $1550\GeV$, primarily due to the its shorter lifetime and the reduced production rate. 
    For the conservative estimation of the background, the mass reach for the wino-like states are modified to $500$, $1500$, and $4500\GeV$, respectively, at the three collider energies. Similarly, for the Higgsino-like scenario, the reach becomes $200$, $450$, and $1070\GeV$. Results for HL-LHC are also in reasonable agreement with experimental prospect studies.   The signal significance in the disappearing track search is rather sensitive to the wino and Higgsino mass values (thus making the $2\sigma$ and $5\sigma$ reach very close in mass), due to the fact that the signal event rate decreases exponentially as the chargino lifetime in the lab frame becomes shorter for heavier masses. 
	
	The improvements in going from the HL-LHC to the HE-LHC, and further from the HE-LHC to the FCC-hh/SppC are very similar to those obtained for the monojet analysis, namely, around a factor of two and three, respectively. Results for both analyses are summarised in \tabb{tab:sec344_summary}.
	
	\begin{table}[t]
		\def\arraystretch{1.0}\centering
		\small\begin{tabular}{|c|c|c|c|c|}
			\hline
			95\%\cl & Wino  & Wino & Higgsino & Higgsino \\		
			 &  Monojet & Disappearing Track &  Monojet & Disappearing  Track \\
			\hline
			$14\TeV$ &  $280\GeV$ & $900\GeV$ & $200\GeV$  & $300\GeV$\\
			\hline
			$27\TeV$ & $700\GeV$ & $2.1\TeV$ & $490\GeV$ & $600\GeV$ \\
			\hline
			$100\TeV$ &  $2\TeV$ &  $6.5\TeV$ & $1.4\TeV$ &  $1.6\TeV$ \\
			\hline
		\end{tabular}\normalsize
		\caption{\small Summary of DM mass reach at $95\%\cl$ for an EW triplet (wino-like) and a doublet (Higgsino-like) representation, at the HL-LHC, HE-LHC and the FCC-hh/SppC colliders, in optimistic scenarios for the background systematics.}
		\label{tab:sec344_summary}
	\end{table}
%    	\begin{table}
%		\def\arraystretch{1.0}\centering
%		\begin{tabular}{|c|c|c|}
%			\hline
%			95\%~\cl &  Wino &  Higgsino \\		
%			\cl &   Disappearing Track &  Disappearing  Track \\
%			\hline
%			14 TeV &   900 GeV &   300 GeV\\
%			\hline
%			27 TeV &  2.1 TeV &  600 GeV\\
%			\hline
%			100 TeV &   6.5 TeV &  1.6 TeV \\
%			\hline
%		\end{tabular}
%		\caption{\small Summary of DM mass reach at 95$\%$ \cl for an EW triplet (wino-like) and a doublet (Higgsino-like) representation, at the HL-LHC, HE-LHC and the FCC-hh/SppC colliders, in optimistic scenarios for the background systematics.}
%		\label{tab:sec344_summary}
%	\end{table}

%%%%%%%%%%%%%%%%%%%%%%%%%%%%%%
\subsection{Displaced Vertices}
Many models of new physics predict long-lived particles which decay within the detector but at an observable distance from the proton-proton interaction point (displaced signatures). 
% A growing class of new physics models predict long-lived particles potentially leading to displaced signatures. 
If the decay products of the long-lived particle include multiple  particles reconstructed as tracks or jets, the decay can produce a distinctive signature of an event containing at least one displaced vertex (DV). 
In the following sections, a number of prospects studies from ATLAS, CMS and LHCb are presented. Results are interpreted in the context of supersymmetric or higgs-portal scenarios but are applicable to any new physics model predicting one or more DVs, since the analyses are not driven by strict model assumptions. 

\subsubsection{\atlas{LLP decaying to a Displaced Vertex and $\met$ at HL-LHC}}
\label{sec:atlastllpjetmet}
\contributors{E. Frangipane, L. Jeanty, L. Lee Jr, H. Oide, S. Pagan Griso, ATLAS}
%{\bf Author(s): L. Jeanty et al. ATLAS}

%% Introduction and motivation
%One of the most intriguing scenarios of new physics to look for at the Large Hadron Collider (LHC) is new particles with measurably long lifetimes. Whether in the Standard Model (SM) or in new physics models, particles can acquire discernible lifetimes from a variety of mechanisms, including weak effective couplings to the final state due to heavy intermediate particles, which can arise in models with mass hierarchies, or due to small coupling constants, which can arise from conserved or nearly-conserved symmetries. Long lifetimes can also manifest due to decays that are phase-space suppressed or from new particles which interact only weakly with the SM particles via mediators.

%used for intro: Many models of new physics predict long-lived particles which decay within the detector but at an observable distance from the proton-proton interaction point. If the decay products of the long-lived particle include multiple charged particles reconstructed as tracks, the decay can produce a distinctive signature of an event containing at least one displaced vertex (DV). 

There are several recent papers at the LHC which have searched for displaced vertices, including \citerefs{Aaij:2017rft,Aaboud:2017iio,Sirunyan:2017jdo,Sirunyan:2018vlw}.
The projection presented here~\cite{ATL-PHYS-PUB-2018-033} requires at least one displaced vertex reconstructed within the \mbox{ATLAS} ITk, and events are required to have at least moderate missing transverse momentum ($\met$), which serves as a discriminant against background as well as an object on which to trigger. The analysis sensitivity is projected for a benchmark SUSY model of pair production of long-lived gluinos, which can naturally arise in models such as Split SUSY~\cite{Giudice:2004tc}. Each gluino hadronises into an $R$-hadron and decays through a heavy virtual squark into a pair of SM quarks and a stable neutralino with a mass of $100\GeV$.

%%% MC samples
%<<<<<<< HEAD
%This study makes use of Monte Carlo simulation samples to obtain the kinematic properties of signal events, which are then used to estimate the efficiency for selecting signal events. The pair production of gluinos from proton-proton collisions at $\sqrt{s} = 13\TeV$ was simulated in \pythia{ 6.428}~\cite{Sjostrand:2006za} at leading order with the AUET2B~\cite{ATLAS:2011krm} set of tuned parameters for the underlying event and the CTEQ6L1 parton distribution function (PDF) set. %The gluino mass, $m_{\tilde{g}}$, ranges from 2.0 to 3.8~\TeV. 
%=======
This study makes use of Monte Carlo simulation samples to obtain the kinematic properties of signal events, which are then used to estimate the efficiency for selecting signal events. The pair production of gluinos from proton-proton collisions at $\sqrt{s} = 13\TeV$ was simulated in \pythia{ 6.428}~\cite{Sjostrand:2006za} at leading order with the AUET2B~\cite{ATLAS:2011krm} set of tuned parameters for the underlying event and the CTEQ6L1 parton distribution function (PDF) set \cite{Pumplin:2002vw}. %The gluino mass, $m_{\tilde{g}}$, ranges from 2.0 to 3.8~\TeV. 
After production, the gluino hadronises into an $R$-hadron and is propagated through the ATLAS detector by \geant{4}~\cite{Agostinelli:2002hh,Aad:2010ah} until it decays. \pythia{ 6.428} is called to decay the gluino into a pair of SM quarks and a neutralino and models the three-body decay of the gluino, fragmentation of the remnants of the light-quark system, and hadronisation of the decay products. The gluino lifetime ranges from $0.1\nsec$ to $10\nsec$, and the neutralino mass is fixed to $100\GeV$. %In this study, only particle-level information about the $R$-hadron's decay products is used.
To normalise the expected number of signal events in the full HL-LHC dataset, the cross-sections for pair production of gluinos are calculated at next-to-leading order at $\sqrt{s} = 14\TeV$ and resummation of soft-gluon emission is taken into account at next-to-leading-logarithm accuracy (NLO+NLL) following the procedure outlined in \citeref{xsec}. 

\begin{figure}[t]
\centering
\begin{minipage}[t][\minipageheightdp][t]{\minipagewidthdp}
\centering
\includegraphics[width=\figuremaxwidthdp,height=\figuremaxheightdp,keepaspectratio]{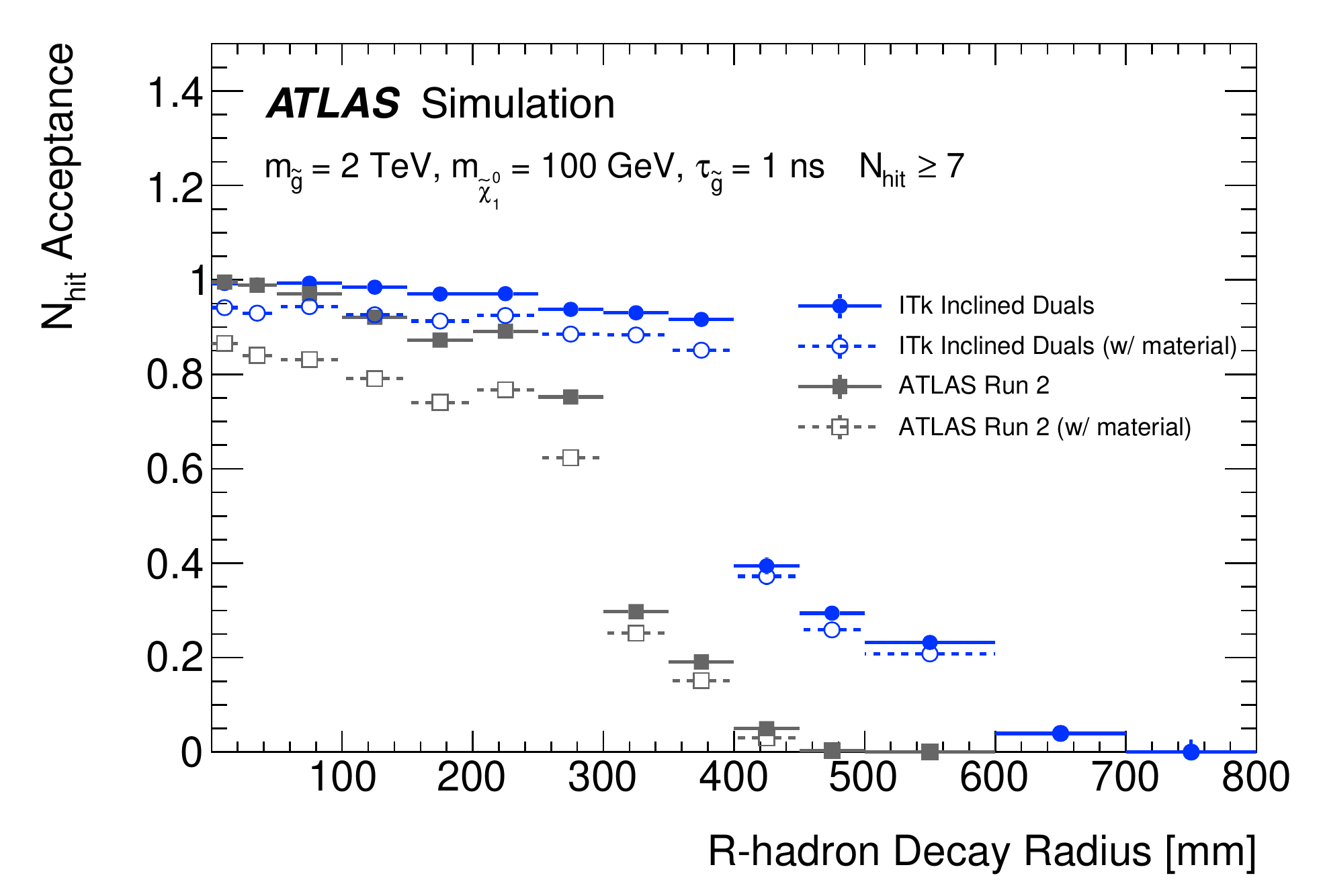}
\end{minipage}
\begin{minipage}[t][\minipageheightdp][t]{\minipagewidthdp}
\centering
\includegraphics[width=\figuremaxwidthdp,height=\figuremaxheightdp,keepaspectratio]{\main/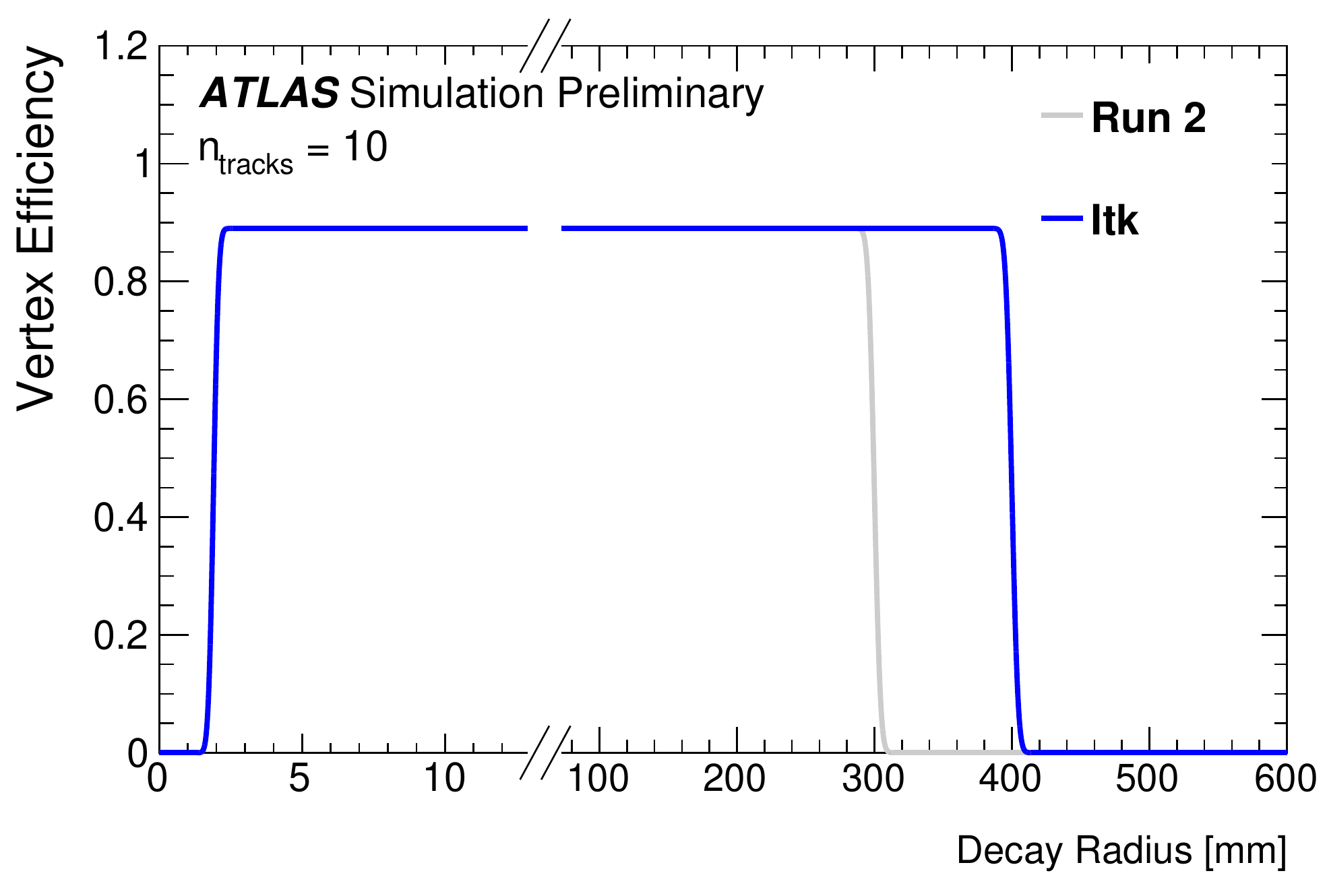}
\end{minipage}
  \caption{Left: probability that a charged particle, with $\pt >
  1\GeV$produced in the decay of a $2.0\TeV$ $R$-hadron with a lifetime of $1\nsec$,
passes through at least seven silicon layers, as a function of the decay radius of the $R$-hadron, for both the Run-2 and ITk detector layouts~\cite{Collaboration:2285585}. The probability is shown with and without
the simulated effect of material producing hadronic interactions.
 Right: parametrised efficiency for reconstructing a displaced
    vertex with $n_{\text{tracks}} = 10$, as a
    function of the decay radius of the parent particle, as measured
    in Run-2 simulation and extrapolated to the ITk geometry. }
  \label{fig:DVMET:TrkVtx}
\end{figure}

%% Tracking and vertexing performance projections
Particle-level Monte Carlo events are used to obtain kinematic distributions for the signal. The expected track reconstruction performance is estimated by factorising it into an acceptance and an efficiency term, and assuming that the efficiency performance of the Run-2 algorithm, currently close to $100\%$, will be reproduced for ITk for particles which pass the acceptance. The tracking acceptance is based on the number of hits left by a charged particle traversing the silicon sensors; at least seven hits are required for both the current ID and the future ITk. To calculate the ITk acceptance for the tracks of interest, a full simulation of the ITk geometry is used. Only charged decay products with $\pt > 1~\GeV$ are considered and material interactions with the active and passive material of the detector are taken into account. Figure~\ref{fig:DVMET:TrkVtx} (left) shows the acceptance as function of the production transverse position (radius) of the particle.
The steep drop off in efficiency in the present ID at around
$300~\mathrm{mm}$ corresponds to the farthest radial extent of the first layer of the SCT, after which it is
unlikely that a typical particle would traverse seven strip layers. In
the ITk, the equivalent drop-off does not occur until after $400~\mathrm{mm}$ due
to the larger spacing between the silicon layers.

The current displaced vertexing performance is parametrised as a function of the transverse decay position ($r_{\text{DV}}$)
and number of reconstructed tracks ($n_{\text{tracks}}$) coming from the long-lived particle decay.
To extrapolate from the Run-2 efficiency to the expected performance in ITk, the same fit values are used for each bin of
$n_{\text{tracks}}$, while the radial distance at which the vertexing efficiency starts to drop is moved from $300~\mathrm{mm}$ to $400~\mathrm{mm}$ to reflect the
change in the location of the inner silicon strip layer, as shown for one particular example in \fig{fig:DVMET:TrkVtx} (right).

%% Selections
\begin{figure}[t]
\centering
\begin{minipage}[t][\minipageheightdp][t]{\minipagewidthdp}
\centering
\includegraphics[width=\figuremaxwidthdp,height=\figuremaxheightdp,keepaspectratio]{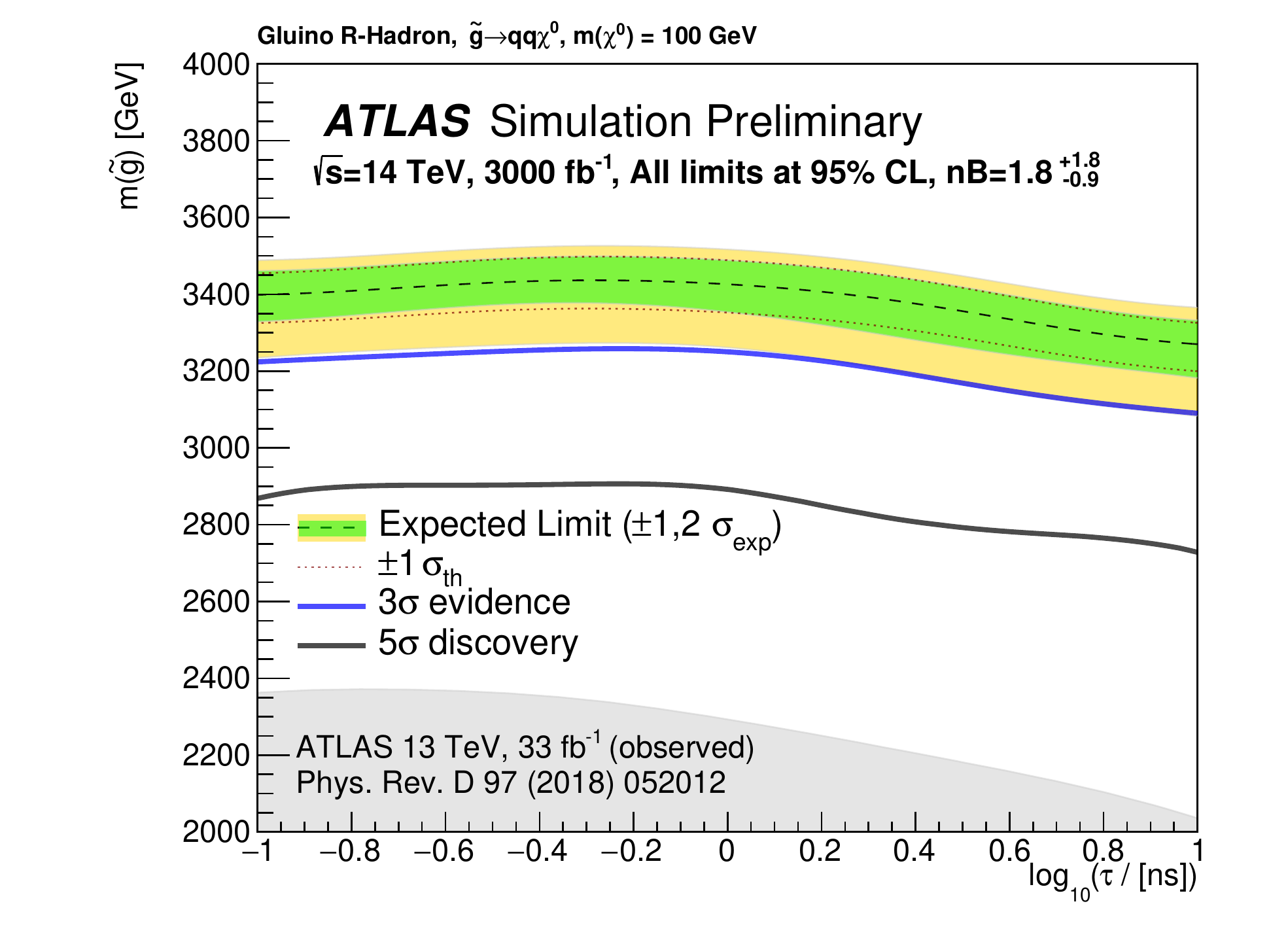}
\end{minipage}
\begin{minipage}[t][\minipageheightdp][t]{\minipagewidthdp}
\centering
\includegraphics[width=\figuremaxwidthdp,height=\figuremaxheightdp,keepaspectratio]{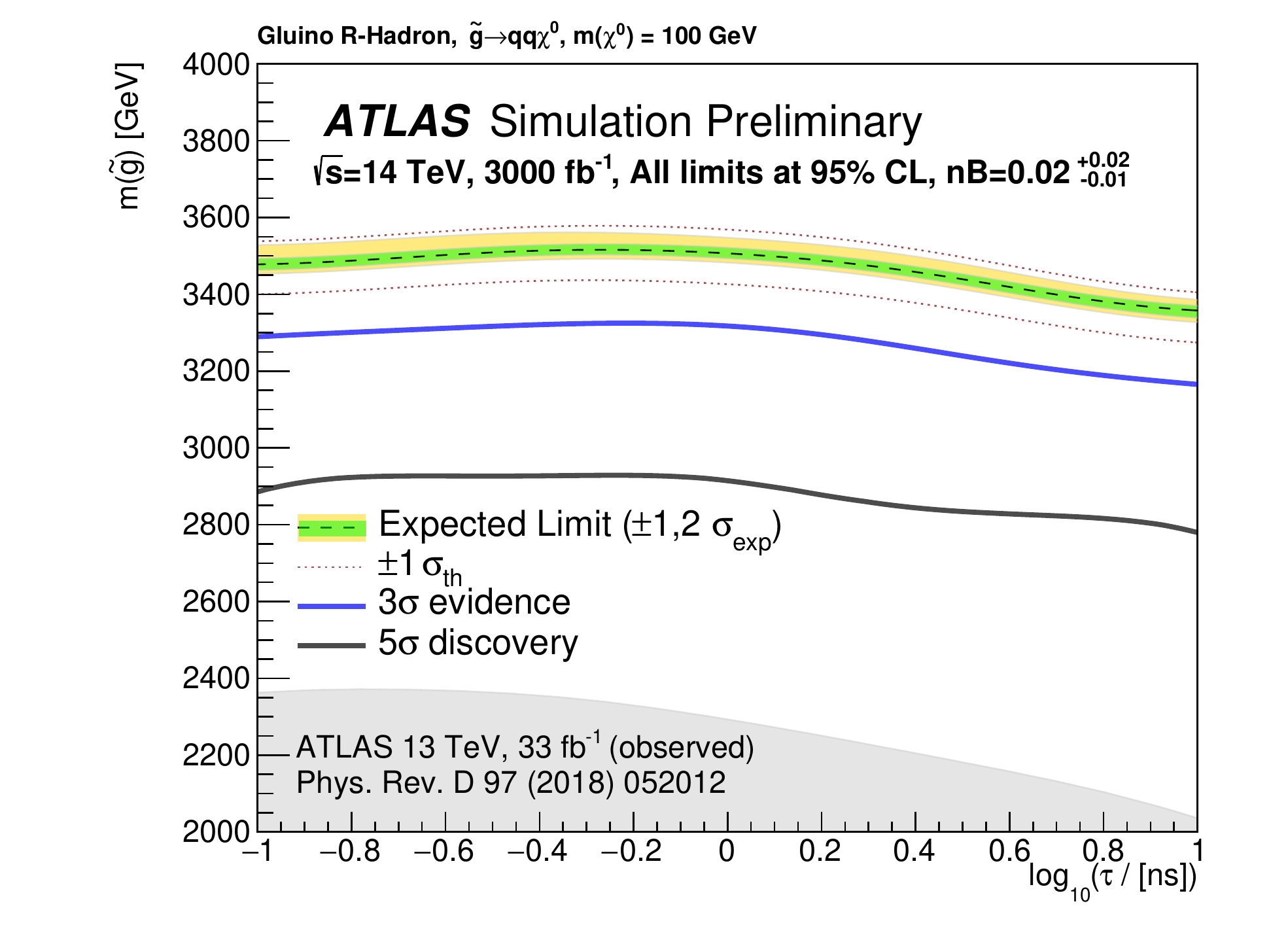}
\end{minipage}
  \caption{Projected sensitivity for the upper limit on the mass of a gluino $R$-hadron that can be observed with $3\sigma$ and $5\sigma$ confidence or excluded at $95\%\cl$, as a function of the gluino lifetime, for a background of $1.8^{+1.8}_{-0.9}$ events (left) and a background of $0.02^{+0.02}_{-0.01}$ events (right). These results are valid for a gluino which decays to SM quarks and a stable neutralino with a mass of $100\GeV$. Results assume \intLumi of collisions at $\sqrt{s} = 14\TeV$ collected with the upgraded ATLAS detector, and are compared to the observed ATLAS exclusion limits for a dataset of $33\fbinv$ at $\sqrt{s} = 13\TeV$.}
  \label{fig:exclusion_100GeVneutralino}
\end{figure}

The event selection closely follows the requirements in the recent Run-2 search for a DV and MET~\cite{Aaboud:2017iio}.
Events are required to have at least one DV within the ITk volume and at least five tracks from the gluino decay must be reconstructed.
The tracks and vertices are reconstructed with a probability given by the procedures described above;
only charged decay products with $\pt > 1\GeV$, $|\eta| < 5$, and with $6~\mathrm{mm} < r_{\text{prod}} < 400~\mathrm{mm}$ are considered.
To exclude hadronic interactions of SM particles, the vertex must not be located within a region of the detector filled with solid materials,
and the invariant mass of the reconstructed vertex must be larger than $10\GeV$.
The event must pass the MET trigger and offline requirements of the Run-2 search, \iee~MET$ > 250\GeV$;
the efficiency of passing the MET trigger and offline MET
requirements is taken from the Run-2 analysis, as parametrised in
\citeref{Hepdata} as a function of the generator-level MET and the $R$-hadron decay positions. %The generator-level MET is calculated as
%the magnitude of the transverse component of the vector sum of the momenta of the stable, weakly-interacting particles in the final
%state.

%% Background estimation and systematics
The background for this search is entirely instrumental in nature. %Therefore, it would not be reliable to perform a simulation-based projection of the expected background for a different detector and different reconstruction algorithms. For the purpose of 
For this projection, two different extrapolations of the size of current background are performed. The default extrapolation assumes that the background and its uncertainty will scale linearly with the size of the dataset, resulting in an expected background of $1.8^{+1.8}_{-0.9}$ events. However, several handles could be tightened in the analysis selection to continue to reject background without introducing appreciable signal efficiency loss. For example, additional requirements on the vertex goodness-of-fit or the compatibility of each track with the vertex could be imposed to further reduce backgrounds from low-mass vertices which are merged or crossed by an unrelated track. Therefore, a more optimistic scenario is also considered in which the total background and uncertainty are kept to the current level of $0.02^{+0.02}_{-0.01}$ events.

The signal selection uncertainties are taken to have the same relative size as in the existing Run-2 analysis. Uncertainties on the signal cross-section prediction are taken by varying the choice of PDF set and factorisation and renormalisation scales, with a reduction of $50\%$ applied to the uncertainties to account for improvements by the time the analysis will be performed. %Finally, the uncertainties on the number of expected background events are taken to be the same relative amount as in the existing analysis, for both scenarios of the background size.

%% Results
Using the number of expected signal and background events with their respective uncertainties, the expected exclusion limit at $95\%\cl$ on the gluino mass, as a function of lifetime, is calculated assuming no signal presence. In the case that signal is present, the $3\sigma$ and $5\sigma$ observation reaches are also calculated. 
The results are shown in \fig{fig:exclusion_100GeVneutralino} for both background scenarios.

The significant increase in sensitivity relative to the ATLAS result with $33\fbinv$ at $\sqrt{s} = 13\TeV$ comes in part from the increase in collision energy and integrated luminosity. For longer lifetimes, a significant gain in selection efficiency and therefore reach is also due to the larger volume of the silicon tracker, which allows displaced tracks and displaced vertices to be reconstructed at larger radii. This pushes the radius at which tracks from long-lived particles can be efficiently reconstructed from $300$ to $400~\mathrm{mm}$, with corresponding gain in acceptance for lifetimes of $10\nsec$ and greater. While the results presented here were studied only for a fixed neutralino mass of $100\GeV$, based on the results in \citeref{Aaboud:2017iio}, comparable sensitivity is expected over a large range of neutralino masses. As the neutralino mass increases for a fixed gluino mass, the multiplicity and momentum of the visible SM particles decreases, which in turn decreases the efficiency of the requirements on the track multiplicity, vertexing reconstruction, and vertex invariant mass as the difference between the neutralino mass and the gluino mass, $m_{DV}$, falls below $400\GeV$.

%\subsubsection{\cms{Prospects for triggering on displaced jets at level 1}}
%{\bf Author(s): R. Patel, Y. Gershtein, K. Ulmer et al. CMS}
% ended up in Higgs chapter

\subsubsection{\cms{Displaced muons at HL-LHC}}
\label{sec:cmsdisplacedmuons}
\contributors{K.~Hoepfner, H.~Keller, CMS}
%{\bf Author(s): K.~Hoepfner, H.~Keller, CMS}

\begin{figure}[t]
\centering
\begin{minipage}[t][\minipageheighttp][t]{\minipagewidthtp-4mm}
\centering
\raisebox{1cm}{\includegraphics[width=\figuremaxwidthtp-7mm,height=\figuremaxheighttp-7mm,keepaspectratio]{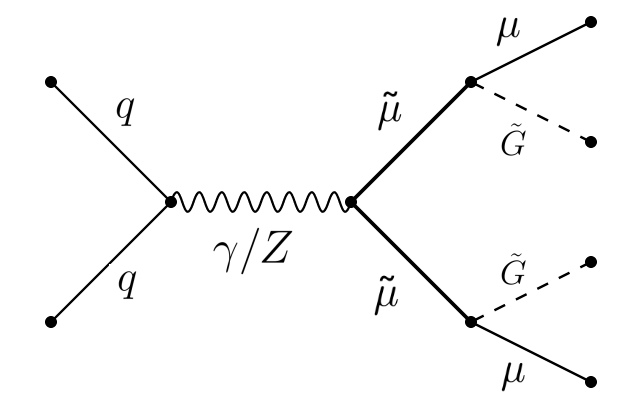}}
\end{minipage}
\begin{minipage}[t][\minipageheighttp][t]{\minipagewidthtp+2mm}
\centering
\includegraphics[width=\figuremaxwidthtp,height=\figuremaxheighttp,keepaspectratio]{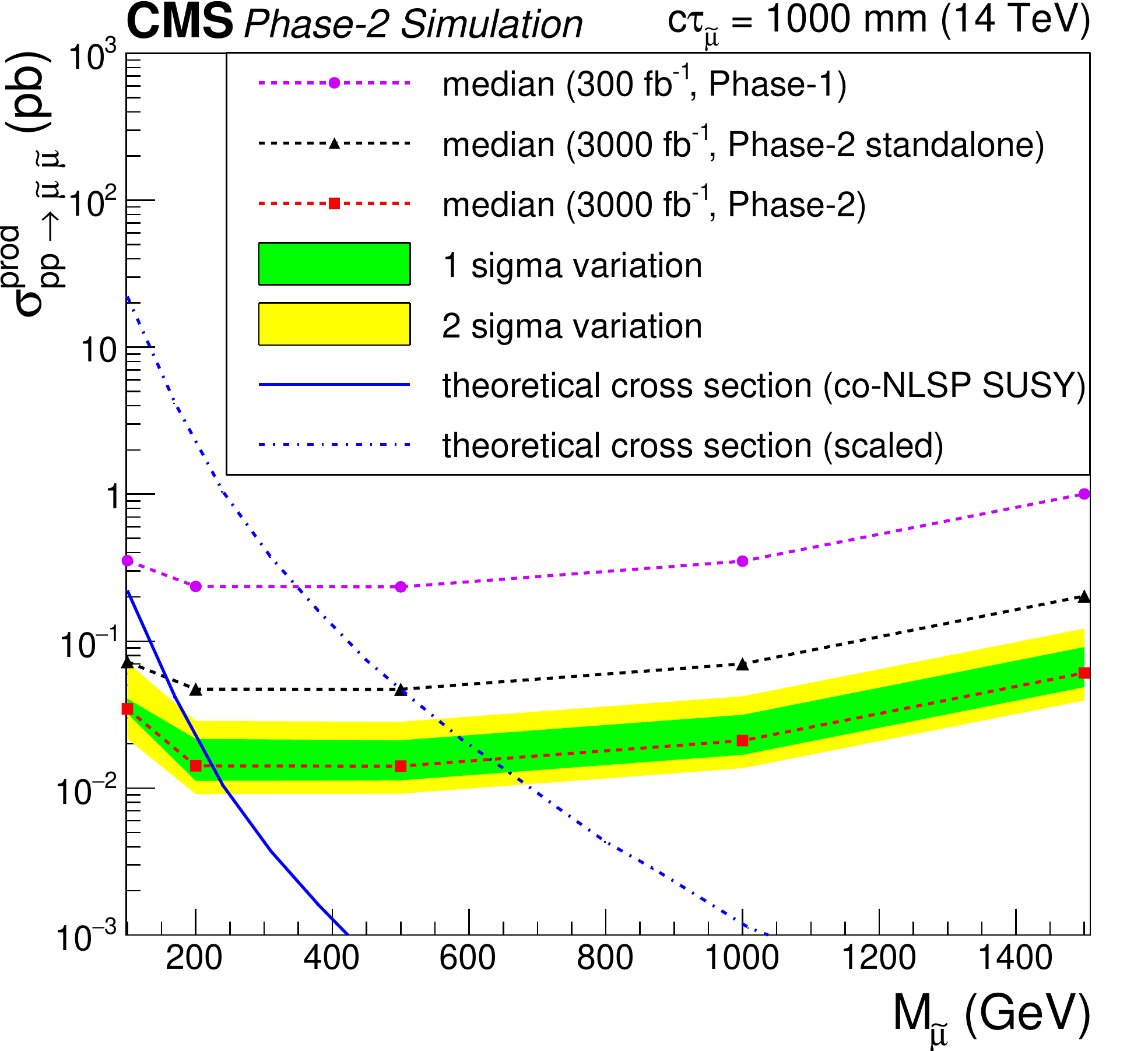}
\end{minipage}
\begin{minipage}[t][\minipageheighttp][t]{\minipagewidthtp+2mm}
\centering
\includegraphics[width=\figuremaxwidthtp,height=\figuremaxheighttp,keepaspectratio]{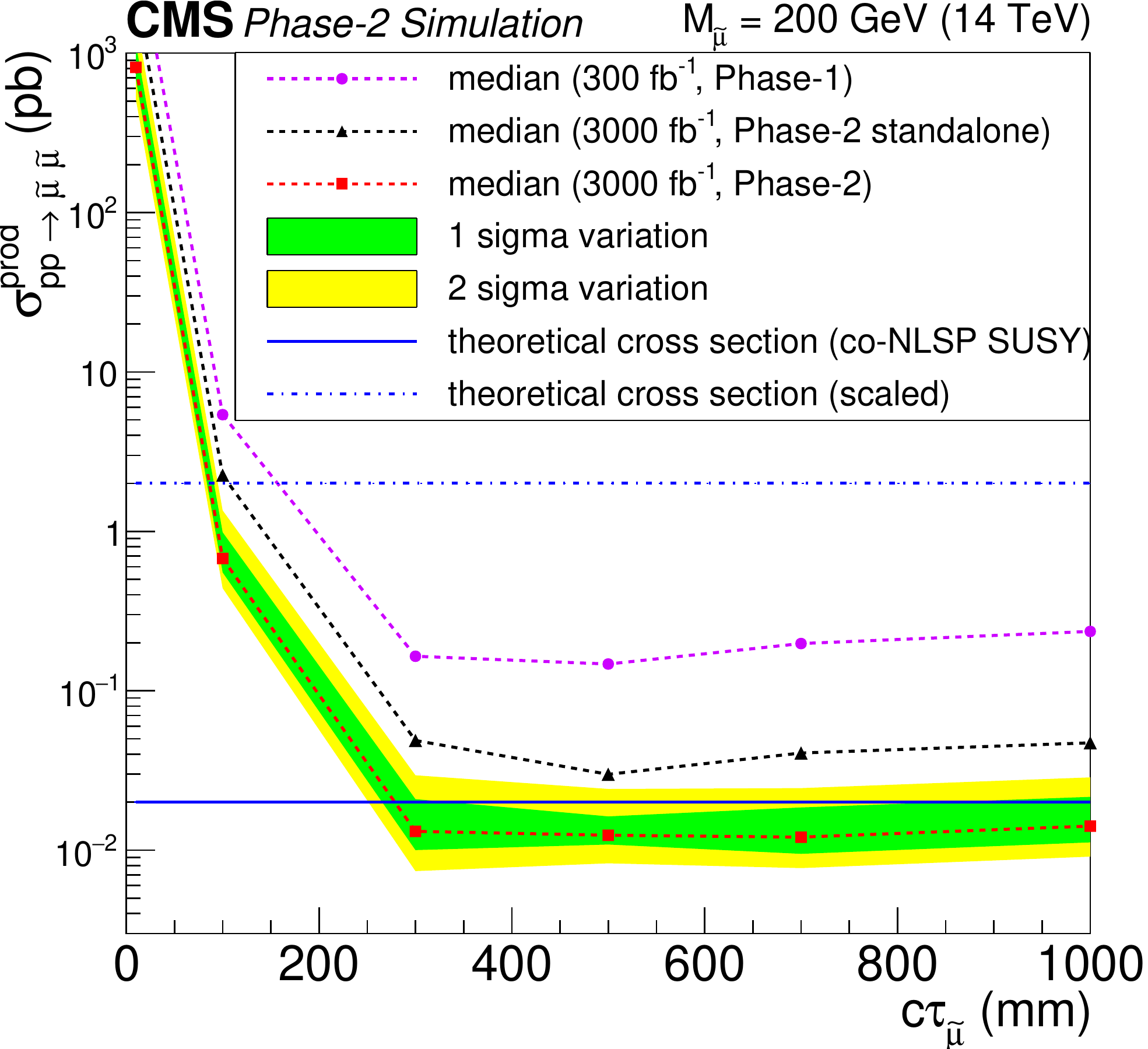}
\end{minipage}
%\begin{center}
%\begin{minipage}{0.245\textwidth}
%\centering
%\includegraphics[width=0.99\textwidth]{}
%\end{minipage}
%\begin{minipage}{0.745\textwidth}
%\centering
%\includegraphics[width=0.49\textwidth]{}
%\includegraphics[width=0.49\textwidth]{}
%\end{minipage}
\caption{Left: Feynman diagram for smuon production.
Middle and right: expected $95\%\cl$ upper limits on long-lived smuons for various mass hypotheses and $c\tau = 1$ m. In both panels, the theoretical cross section for the specific model is represented by the blue solid line. For different SUSY breaking scales, $\tan\beta$ or otherwise modified parameters, the cross sections may be 100 times larger, reflected by the blue dash-dotted line. Green (yellow) shaded bands show the one (two) sigma range of variation of the expected $95\%\cl$ limits. Phase-2 results with an average 200 pileup events and an integrated luminosity of \intLumi are compared to results obtained with $300\fbinv$. The black line shows the sensitivity without the DSA algorithm, which reduces the reconstruction efficiency by a factor three. The panel in the middle shows the limit as a function of the smuon mass and the right panel as a function of the decay length.} 
\label{fig:Limit_DisplacedMuons}
%\end{center}
\end{figure}

%Future searches at the HL-LHC should cover a large phase space and, ideally, should not be driven by strict model assumptions. 
A growing class of new physics models predict long-lived particles potentially leading to displaced signatures. 
In this study from CMS we discuss the potential for a SUSY GMSB model with heavy smuons decaying to a SM muon and a gravitino (yielding MET)~\cite{Collaboration:2283189,CMS:2018lqx}. Figure~\ref{fig:Limit_DisplacedMuons} (left) shows the model under study. In this model the smuon is produced in pairs, and  is degenerate in mass yielding long lifetimes. In such scenarios the smuon may decay after $\mathcal{O}(1\meters)$ or more such that the only detectable hits are in the muon system. Consequently the analysis uses a dedicated reconstruction algorithm for stand-alone muons (DSA) without a constraint on the vertex position. 

\begin{figure}[t]
\centering
\begin{minipage}[t][\minipageheightdp][t]{\minipagewidthdp}
\centering
\includegraphics[width=\figuremaxwidthdp,height=\figuremaxheightdp,keepaspectratio]{\main/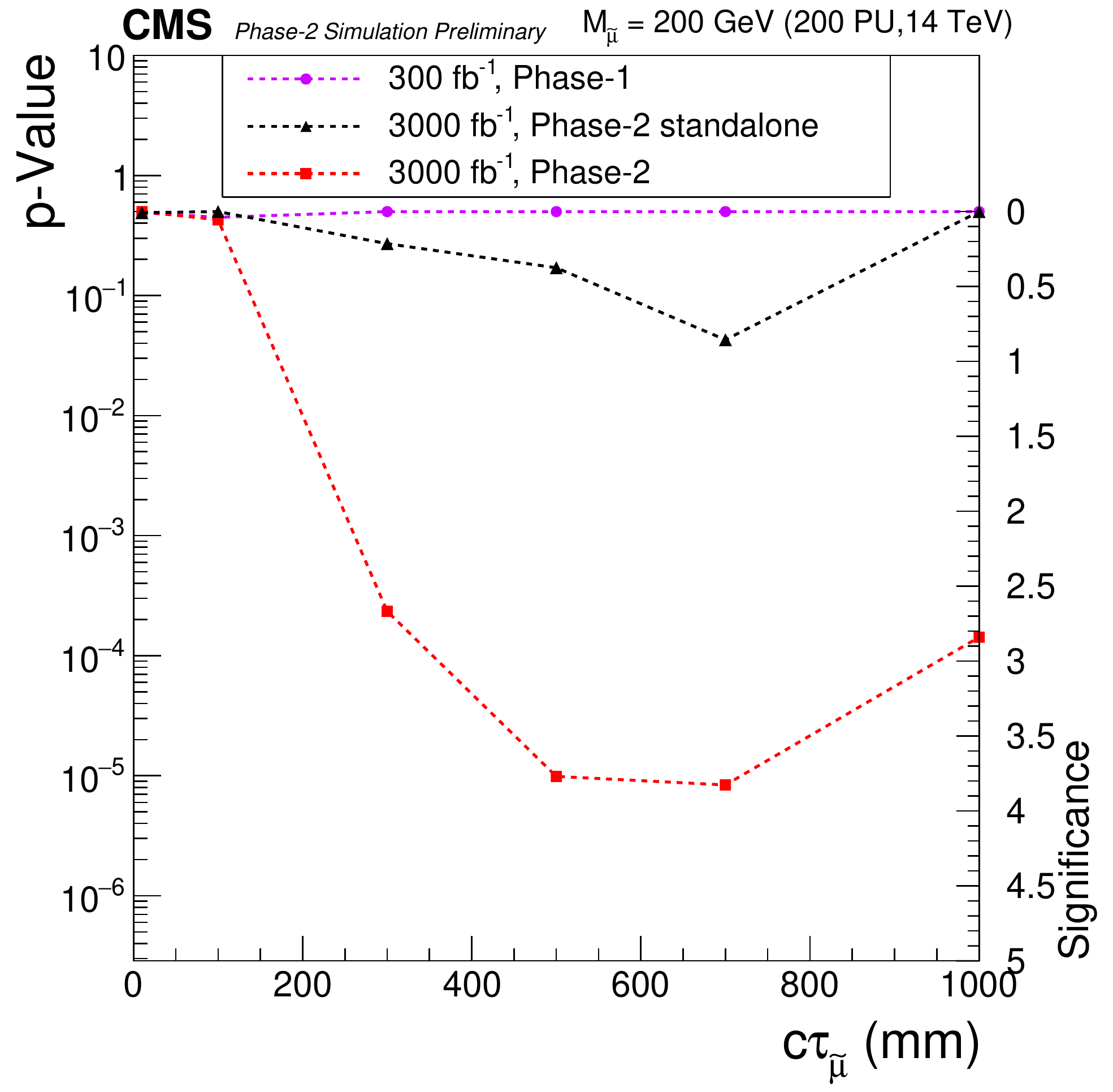}
\end{minipage}
\begin{minipage}[t][\minipageheightdp][t]{\minipagewidthdp}
\centering
\includegraphics[width=\figuremaxwidthdp,height=\figuremaxheightdp,keepaspectratio]{\main/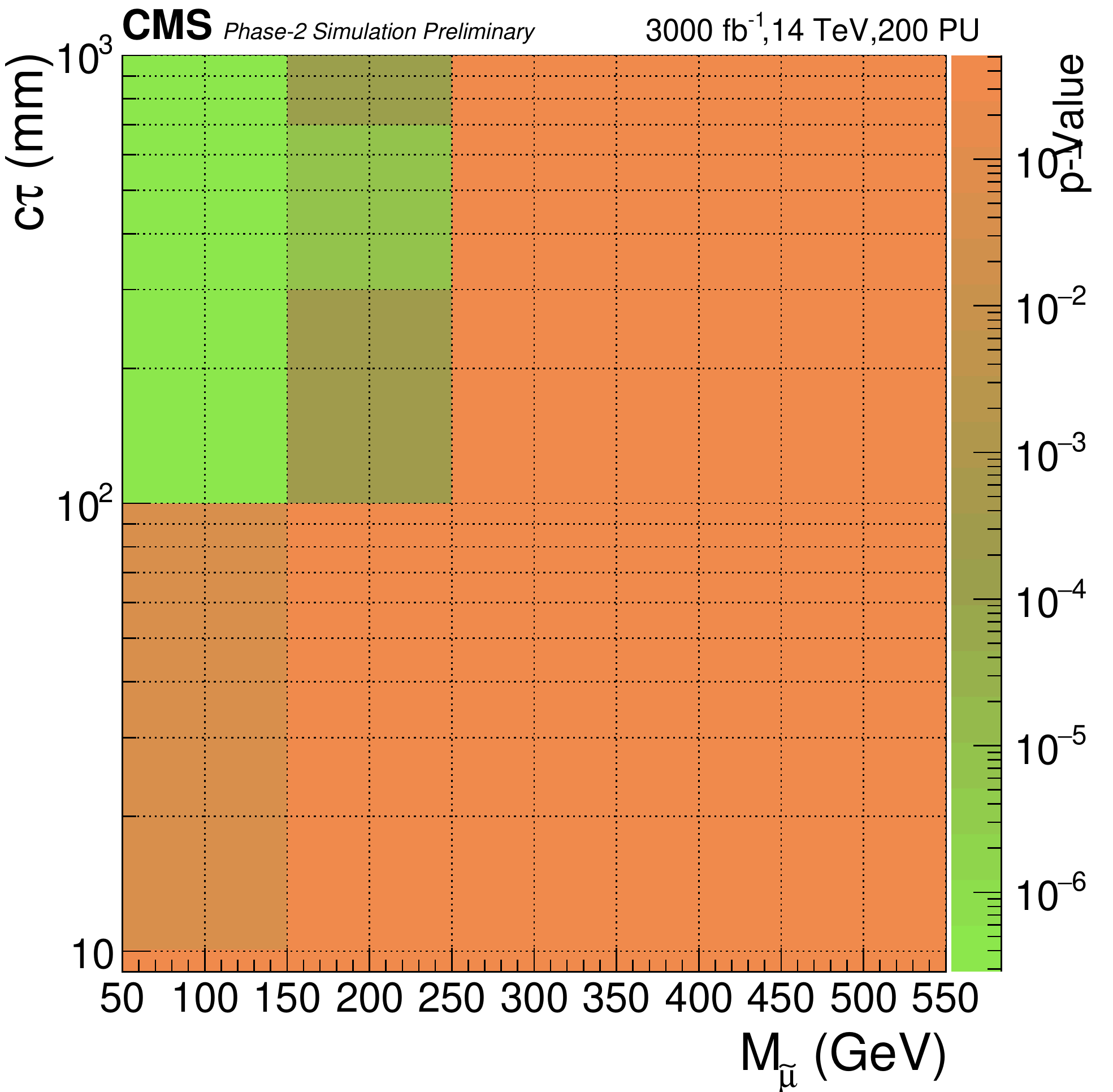}
\end{minipage}
\caption{Left: discovery significance and p-value for a fixed smuon mass of $M\tilde{\mu}= 200\GeV$. The displaced significance is compared to the algorithm with a beamspot constraint (``Phase-2 standalone''). Right: discovery sensitivity in the 2D parameter space of mass and decay length. } 
\label{fig:Discovery_DisplacedMuons}
\end{figure}

It is both challenging to trigger and to reconstruct displaced muons, especially if the displacements are large. Triggers and reconstruction algorithms, generally including the primary vertex position, 
will not be very efficient in reconstructing tracks with large impact parameters.
If the particle is sufficiently boosted, the transverse impact parameter is small(er) but the decay may occur well outside the tracker volume. In both cases, the stand-alone capabilities of the muon system constitute the only possibility for detection.

The main background for this search comes from multi-jet production (QCD), $t\bar{t}$ production, and $Z$/DY events if large impact parameters are (mis)reconstructed. Cosmic ray muons have been studied in Run-2 and are independent of the instantaneous luminosity. In the barrel they are efficiently rejected by the timing of the hits in the upper leg. Cosmic ray muons do not originate at the vertex and therefore pass the upper barrel sectors in reverse direction from outside in. The fraction of cosmic ray muons in the endcaps is negligible. Given the very low cross section of the signal process, it is essential to reduce the background efficiently.
The best background discriminator is the impact parameter significance $d_0/\sigma(d_0) \geq 10$. The muons should move in roughly opposite directions and MET should be larger than $50\GeV$ to account for the two gravitinos. After this selection the signal efficiency is about $4-5\%$ for $c\tau = 1000~\mathrm{mm}$, nearly independent of the smuon mass, and $10^{-5} - 10^{-4}$ for QCD, $t\bar{t}$, and DY backgrounds.

Figure~\ref{fig:Limit_DisplacedMuons} shows expected exclusion limits for the gauge-mediated SUSY breaking model with the smuon being a (co-)NLSP for the predicted cross section as well as for a factor 100 larger cross section. The exclusion limits are shown as functions of smuon mass in \fig{fig:Limit_DisplacedMuons} (middle) and decay length in \fig{fig:Limit_DisplacedMuons} (right). The sensitivity also depends on $c\tau$ because shorter decay lengths shift the signal closer to the background. The expected exclusion limit is around $200\GeV$ for $c\tau = 1000~\mathrm{mm}$ with \intLumi. For the same mass, a discovery sensitivity of $3\sigma$ significance can be reached, as shown in~\fig{fig:Discovery_DisplacedMuons}. This also illustrates the importance
of the lepton trigger thresholds to be kept at a few times 10 GeV, even in the environment of 200 pileup interactions.
Figure~\ref{fig:Discovery_DisplacedMuons} also shows the
discovery sensitivity in the 2-dimensional parameter space of smuon mass and decay length.

%\vspace{1cm}

\subsubsection{\lhcbC{LLPs decaying into muons and jets at the HL-LHC}}
\label{sec:lhcbllpmuons}
\contributors{A. Bay, X. Cid Vidal, E. Michielin, L. Sestini and C. V\'{a}zquez Sierra, LHCb}

The LHCb experiment has proved to be highly competent with regard to direct searches for LLPs, being able to complement ATLAS and CMS in certain parameter space regions \cite{Lee:2018pag}. 
In this section, we provide prospects in the search for R-Parity Violating (RPV) supersymmetric neutralinos decaying semileptonically into a high-$\pt$ muon and two jets. The results are taken from \citeref{LHCb:2018hiv}
which extrapolates the analysis in \citeref{Aaij:2016xmb}. The neutralinos are assumed to be produced through an exotic decay of the SM Higgs boson. Prospects are shown for the expected datasets after both planned LHCb \mbox{Upgrade~I} and \mbox{Upgrade~II} (Run-3--Run-4 and Run-5 onwards, respectively).

The trigger efficiency for this analysis is conservatively assumed to remain unchanged with respect to the published result. However, assuming a $100\%$ efficient first level trigger, after the removal of the hardware trigger level in Run-3, the overall trigger efficiency could improve by a factor of \sloppy\mbox{$2-3$}. 
Regarding pile-up effects, a moderate penalty factor is applied to account for the increased pile-up expected in Runs 3-5 at LHCb. In order to expand the projections, the results are interpolated for different masses and lifetimes that are not considered in simulation. The interpolation is linear and two-dimensional.

The results are obtained from a preliminary, unoptimised analysis of a subset of data collected for pp collisions at \com~energy of $13 \TeV$. To account for a possible deterioration in the background rejection due to multiple primary interactions at high luminosity, a penalty factor of two has been applied to the background yield. The signal efficiency is obtained from the full simulation of the Higgs boson produced via gluon-gluon fusion at $13 \TeV$. As explained, no other change in the signal and background efficiencies due to the upgrade of the detector is considered. The difference between $13$ and $14 \TeV$ energies is assumed to be negligible. Some examples of the efficiencies and background yields assumed for these calculations can be found in \tabb{tab:ul-h125}.

\begin{table}[t]
\begin{center}
{\small
\begin{tabular}{|c|c|c|c||c|}
\hline
{\bf {$m_{\boldsymbol{\tilde{\chi}^{0}_{1}}}~(\mathrm{GeV}/c^2)$}} & {\bf$\boldsymbol{c\tau_{\tilde{\boldsymbol\chi}^{0}_{1}}}~(\mathrm{mm})$} & \text{Acceptance} ($\%$) & \text{Total} ($\%$) & Background yield ($1.7\fbinv$) \\
\hline
      &  {\bf{$3$}} &  $28.0$ &   $0.27$  &  $ 2$ \\
{\bf{$20$}} & {\bf{$15$}} &  $28.1$ &   $0.30$  &   $1$ \\
     &  {\bf{$30$}} &  $27.8$ &   $0.24$  &  $ 3$ \\\hline
%%%%%%%%%%%%%%%%%%%%%%%%%%%%%%%%%%%
      &  {\bf{$3$}} &  $28.7$ &  $ 0.78$  &  $ 4$ \\
{\bf{30}} & {\bf{$15$}} &  $28.4$ &   $1.21$  &   $4$ \\
      & {\bf{$30$}}  &  $28.7$ &   $0.75$  &  $ 2$ \\\hline
%%%%%%%%%%%%%%%%%%%%%%%%%%%%%%%%%%%
{\bf{50}} & {\bf{$15$}} &  $31.5$ &   $2.33$  &  $ 2$ \\
     & {\bf{$30$}} &  $31.7$ &   $1.58$  &   $1 $\\\hline
%\\\hline
%%%%%%%%%%%%%%%%%%%%%%%%%%%%%%%%%%%
      &  {\bf{$10$}} &  $35.2$ &   $1.38$  &  $ 1$ \\
{\bf{$60$}} & {\bf{$50$}} &  $35.5$ &   $2.84$  &  $ 2$ \\
     & {\bf{$100$}} &  $35.2$ &   $2.63$  &  $ 3$ \\
 \hline
 \end{tabular} }
 \caption {\small Examples of the acceptance and total efficiencies assumed to detect a $\tilde{\chi}^{0}_{1}$ decaying semileptonically at a $pp$ collision energy of $\sqrt{s}=13\TeV$ at the LHCb detector. Reference background yields at an integrated luminosity of $1.7\fbinv$ are also presented. Differences in these yields are due to the effect of a multivariate classifier which is trained differently for each mass-lifetime case.}
\label{tab:ul-h125}
 \end{center}
 \end{table}

With the updated signal and background yields, the sensitivity projections are computed. The upper limits on the branching fraction of the Higgs boson decay to a pair of neutralinos are calculated for different 
assumptions of neutralino masses and lifetimes and for different values of integrated luminosity. The Higgs boson production cross section is assumed to be that of the SM~\cite{deFlorian:2016spz}. The actual limit is computed by comparing the $14 \TeV$ efficiencies and background yields to Run-1, and by extrapolating the results published in \citeref{Aaij:2016xmb}. The systematic uncertainties, which are sub-dominant for this result in the published analysis, are assumed to be the same as those in Run-1.

\begin{figure}[t]
\centering
\begin{minipage}[t][2\minipageheighttp][t]{\minipagewidthtp+6mm}
\centering
\includegraphics[width=\figuremaxwidthtp+3mm,height=\figuremaxheighttp,keepaspectratio]{\main/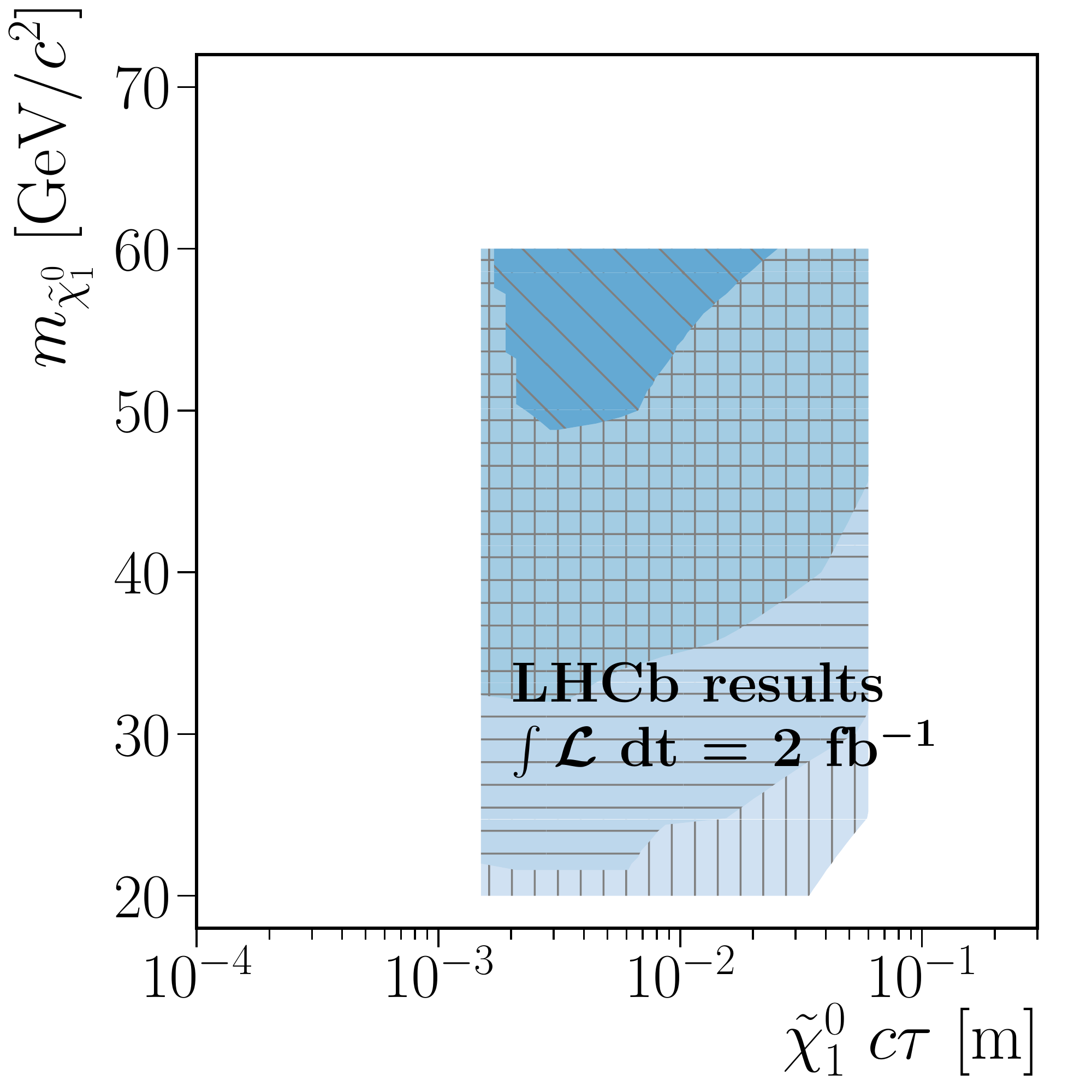}
\includegraphics[width=\figuremaxwidthtp+3mm,height=\figuremaxheighttp,keepaspectratio]{\main/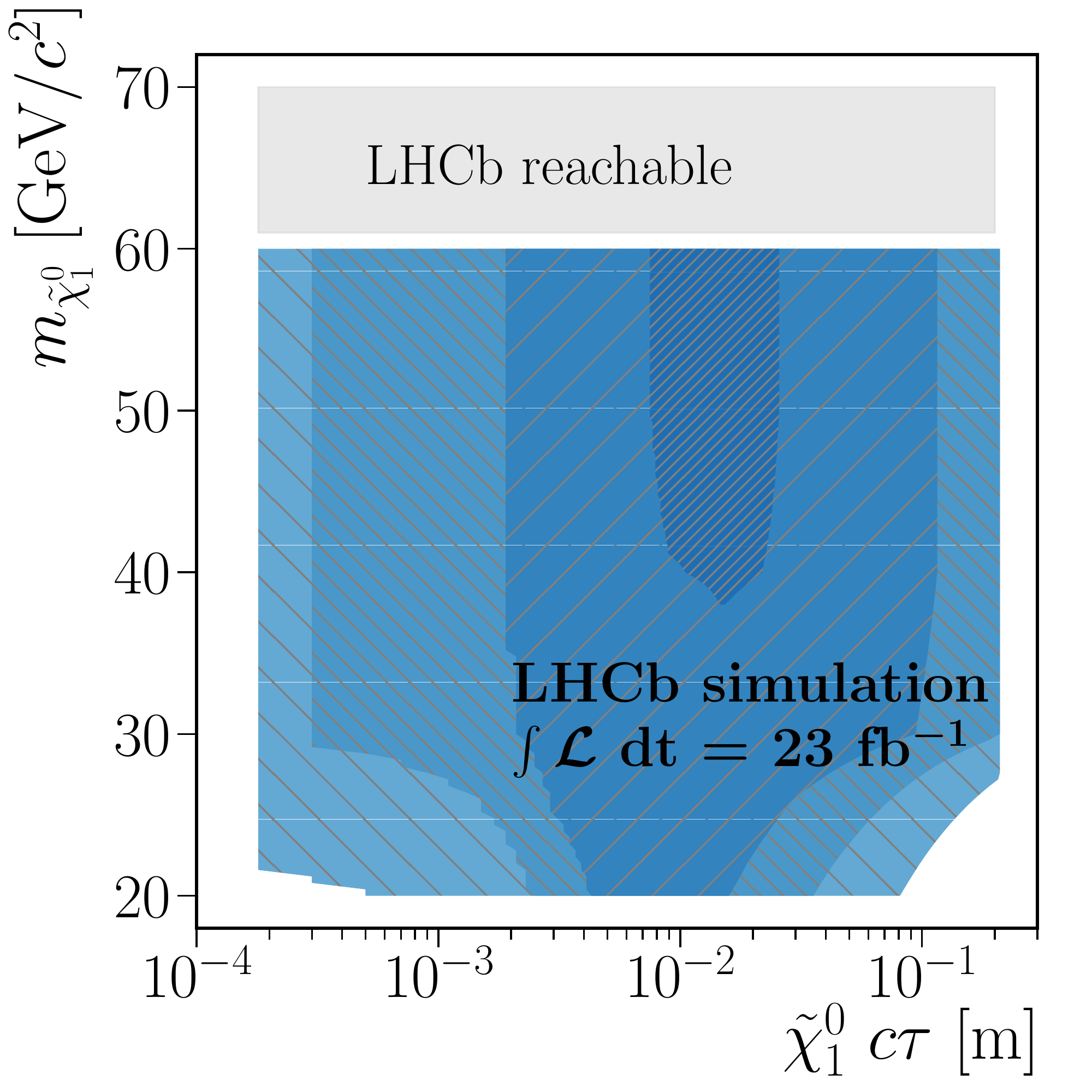}
\end{minipage}
\begin{minipage}[t][2\minipageheighttp][t]{\minipagewidthtp+6mm}
\centering
\includegraphics[width=\figuremaxwidthtp+3mm,height=\figuremaxheighttp,keepaspectratio]{\main/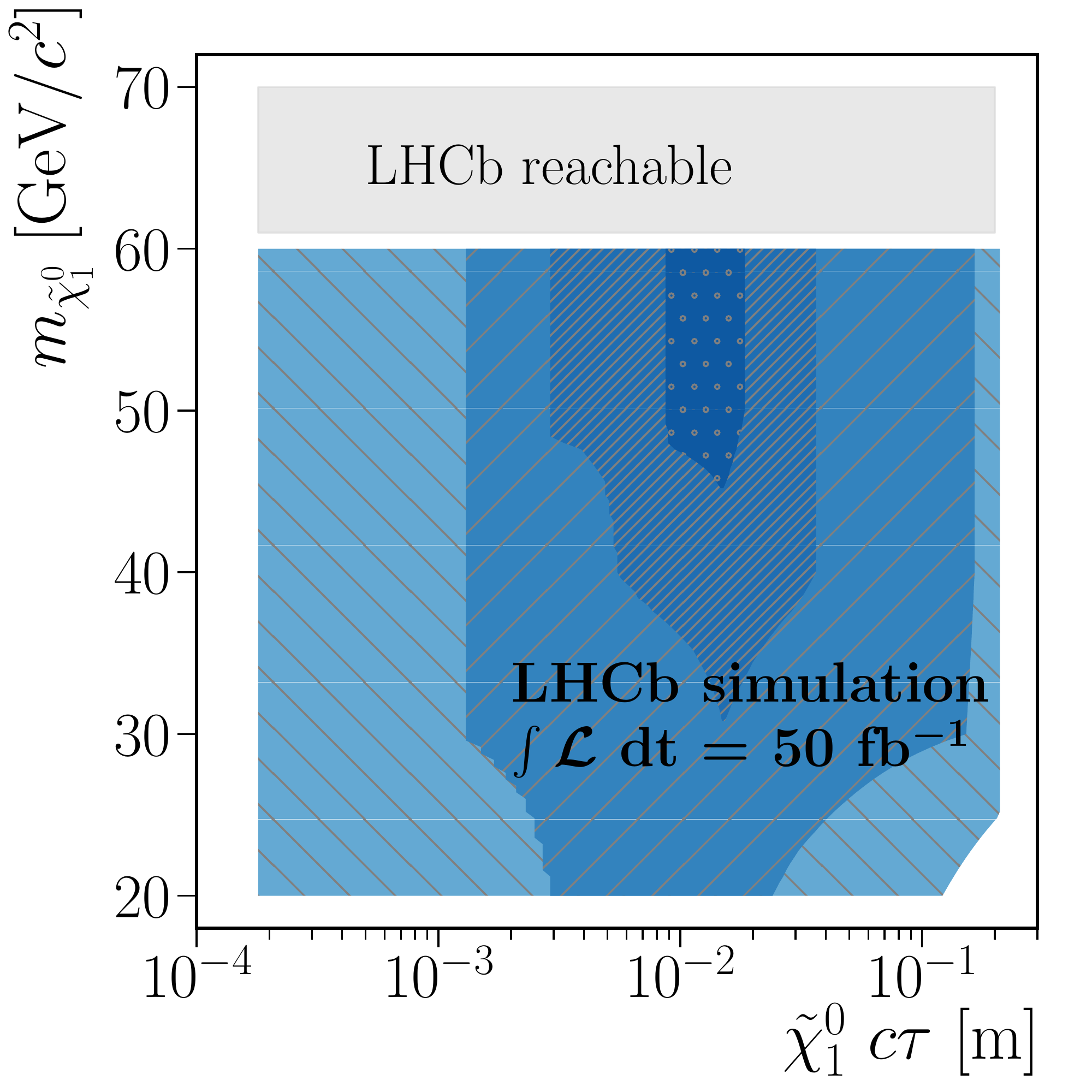}
\includegraphics[width=\figuremaxwidthtp+3mm,height=\figuremaxheighttp,keepaspectratio]{\main/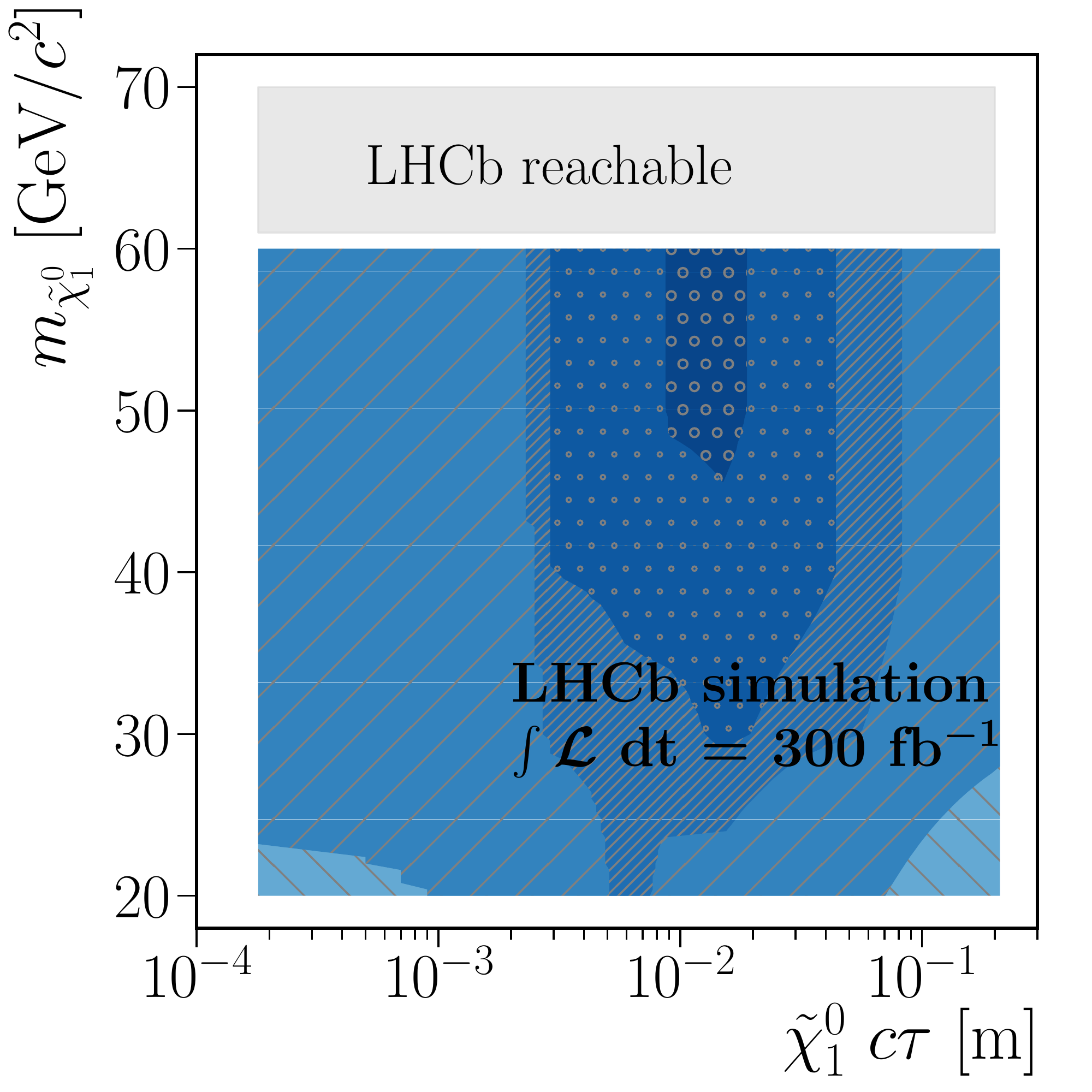}
\end{minipage}
\begin{minipage}[t][2\minipageheighttp][t]{\minipagewidthtp-12mm}
\centering
\raisebox{-3.2cm}{\includegraphics[width=\figuremaxwidthtp,height=\figuremaxheighttp+7mm,keepaspectratio]{\main/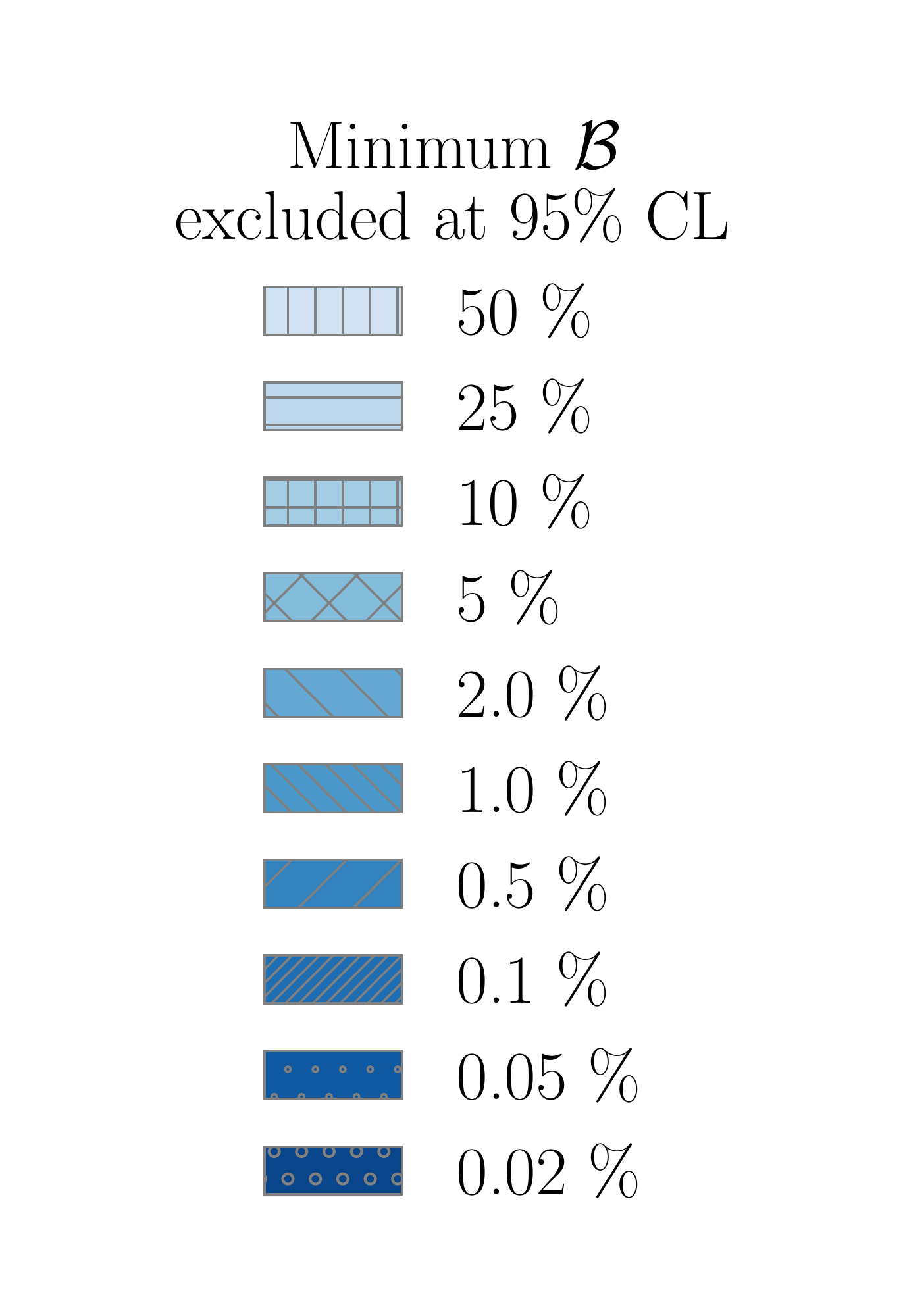}}
\end{minipage}
\vspace{-13.6cm}
%\begin{minipage}[b]{0.77\textwidth}
%  \includegraphics[width=0.49\textwidth]{\main/section4LongLivedParticles/img/LLP_compare_RPV_OBS.pdf}
%  \includegraphics[width=0.49\textwidth]{\main/section4LongLivedParticles/img/LLP_compare_RPV_UP0.pdf}\\
%   \includegraphics[width=0.49\textwidth]{\main/section4LongLivedParticles/img/LLP_compare_RPV_UP.pdf}
%  \includegraphics[width=0.49\textwidth]{\main/section4LongLivedParticles/img/LLP_compare_RPV_HL.pdf}\\
% \end{minipage}
%\begin{minipage}[t]{0.22\textwidth}
%  \vspace{-10.5cm}
% \includegraphics[width=1.5\textwidth]{\main/section4LongLivedParticles/img/LLP_compare_legend_RPV.pdf}
%\end{minipage}
  \caption{Projected sensitivities of the search for RPV supersymmetric neutralinos decaying semileptonically and produced through a Higgs boson exotic decay. The results are extrapolated from Run-1 results (top left), for luminosities of $23\fbinv$ (top right), $50\fbinv$ (bottom left) and $300\fbinv$ (bottom right). The results are presented in terms of the excluded parameter space of the neutralinos for different upper limits at $95\%\cl$ on the branching fractions of the Higgs boson decay.}
  \label{fig:RPVsearches_hllhc}
\end{figure}

The results are shown in \fig{fig:RPVsearches_hllhc}, 
for different integrated luminosities. These plots display the RPV neutralino mass and lifetime ranges excluded at $95\%\cl$. The ranges are shown for different assumed integrated luminosities and branching fractions of the Higgs boson decay to a pair of RPV neutralinos. The region for which the mass of the neutralino is above $60 \GeV$ is not shown in these projections, since no simulation was available. It is worth to notice that the lifetime range covered ($0.2<c\tau<200\mathrm{mm}$) is constrained by the physical length of the VELO detector, since the LLP is required to decay within the VELO region in order to be able to reconstruct it. For the HL-LHC, most of the LHCb accessible neutralino phase space can be excluded for a branching fraction of the $H\rightarrow \tilde{\chi}^{0}_{1} \tilde{\chi}^{0}_{1} $ decay larger than $0.5\%$.

\subsubsection{\lhcbC{LLPs decaying into dijets at the HL-LHC}}
\label{sec:lhcbllpdijets}
\contributors{X. Cid Vidal, E. Michielin, L. Sestini and C. V\'{a}zquez Sierra, LHCb}

In this section, prospects are obtained for Hidden Valley (HV)~\cite{Strassler:2006im, Pierce:2017taw} pions ($\pi_v$) decaying hadronically into a pair of jets at LHCb. The $\pi_v$, which can be long-lived, are assumed to be produced through an exotic decay of the SM Higgs boson. The prospects in this chapter are taken from \citeref{LHCb:2018hiv}, whose analysis is based on a projection of the results published in \citeref{Aaij:2017mic}. 

The simulation of the HV pions through the Higgs portal is fully specified by the mass and the lifetime of the $\pi_v$ particles, allowed to decay exclusively as $\pi_v\to\bbbar$ since this decay mode is generally preferred in this model.

The assumptions made concerning signal efficiencies and background yields are similar to those discussed in \secc{sec:lhcbllpmuons}. However, in this case, no penalty for the pile-up is applied. Signal and background yields are obtained taking into account the increase of cross sections (from $\sqrt{s}=8\TeV$ to $\sqrt{s}=14\TeV$) and of the integrated luminosities. The scaling of the signal includes both the increase in the cross section of the Higgs boson production and that of the amount of signal falling in the acceptance of the LHCb detector. As an example, \tabb{tab:dvsearches_eff} shows some of the acceptance and total efficiencies assumed for this extrapolation for different masses and lifetimes of the HV pion.
The background is scaled by a factor obtained using simulated $b\bar{b}$ events, which are expected to be the dominant contribution. The assumed yields, extrapolated from \citeref{Aaij:2017mic}, can be found in \tabb{tab:dvsearches_yield} for an integrated luminosity of $23\fbinv$. Following the same reference, the yields are divided in bins of the radial coordinate of the HV pion decay vertex position.

\begin{table}[t]
\centering
\small\begin{tabular}{|c|l|c|c|c|c|}
\hline
\multirow{2}{*}{\bf{$c\tau_{\pi_v}~(\mathrm{mm})$}} & \multirow{2}{*}{\textbf{Efficiency ($\%$)}} & \multicolumn{4}{c|}{$\boldsymbol{ m_{\pi_v}~{\rm GeV/c^{2}}}$}% 
\\ \cline{3-6} 
                                      &                                  & \textbf{$25$}                       & \textbf{$35$}                       & \textbf{$43$}                       & \textbf{$50$}                       \\ \hline
\multirow{2}{*}{\textbf{$3$}}                    & \text{Acceptance}                       & $26.8$                     & $21.2    $                 & $17.4$                     & $14.6  $                   \\ \cline{2-6} 
                                      & \text{Total}                            & \multicolumn{1}{c|}{0.9} & \multicolumn{1}{c|}{1.7} & \multicolumn{1}{c|}{1.5} & \multicolumn{1}{c|}{1.1} \\ \hline
\multirow{2}{*}{\textbf{$30$}}                   & \text{Acceptance}                       & $16.1$                     & $15.1 $                   & $13.7   $                 & $12.3  $                   \\ \cline{2-6} 
                                      & \text{Total}                            & \multicolumn{1}{c|}{0.2} & \multicolumn{1}{c|}{0.4} & \multicolumn{1}{c|}{0.4} & \multicolumn{1}{c|}{0.3} \\ \hline
\end{tabular}\normalsize
\caption{Examples of the acceptance and total efficiencies assumed to detect a $\pi_v$ particle decaying to a pair of jets at a $pp$ collision energy $\sqrt{s}=14\TeV$ at the LHCb detector. The main inefficiencies arise from the requirements to have $\pi_v$ particle in the VELO and to have the decay products in the LHCb acceptance and from the reconstruction of the secondary vertex.}
\label{tab:dvsearches_eff}
\end{table}

\begin{table}[t]
\centering
\small\begin{tabular}{|c|c|c|c|c|c|c|}
\hline
$\boldsymbol{R_{xy}~( {\rm mm})}$ & $0.4-1$           & $1-1.5 $          & $1.5-2$           &$ 2-3  $           & $3-5  $           & $5-50  $          \\ \hline
\textbf{Background yield ($\boldsymbol{23\fbinv}$)}    & $1.1\times10^5$ & $5.4\times10^5$ & $3.3\times10^5$ & $9.8\times10^5$ & $2.1\times10^6$ & $3.3\times10^5$ \\ \hline
\end{tabular}\normalsize
\caption{Background yields assumed for the HV pion analysis at an integrated luminosity of $23\fbinv$. The yields are divided in bins of $R_{xy}=\sqrt{x^2+y^2}$, where $x,y$ are the coordinates of the $\pi_v$ particle decay vertex position.}
\label{tab:dvsearches_yield}
\end{table}

\begin{figure}[t]
\centering
\begin{minipage}[t][2\minipageheighttp][t]{\minipagewidthtp+6mm}
\centering
\includegraphics[width=\figuremaxwidthtp+3mm,height=\figuremaxheighttp,keepaspectratio]{\main/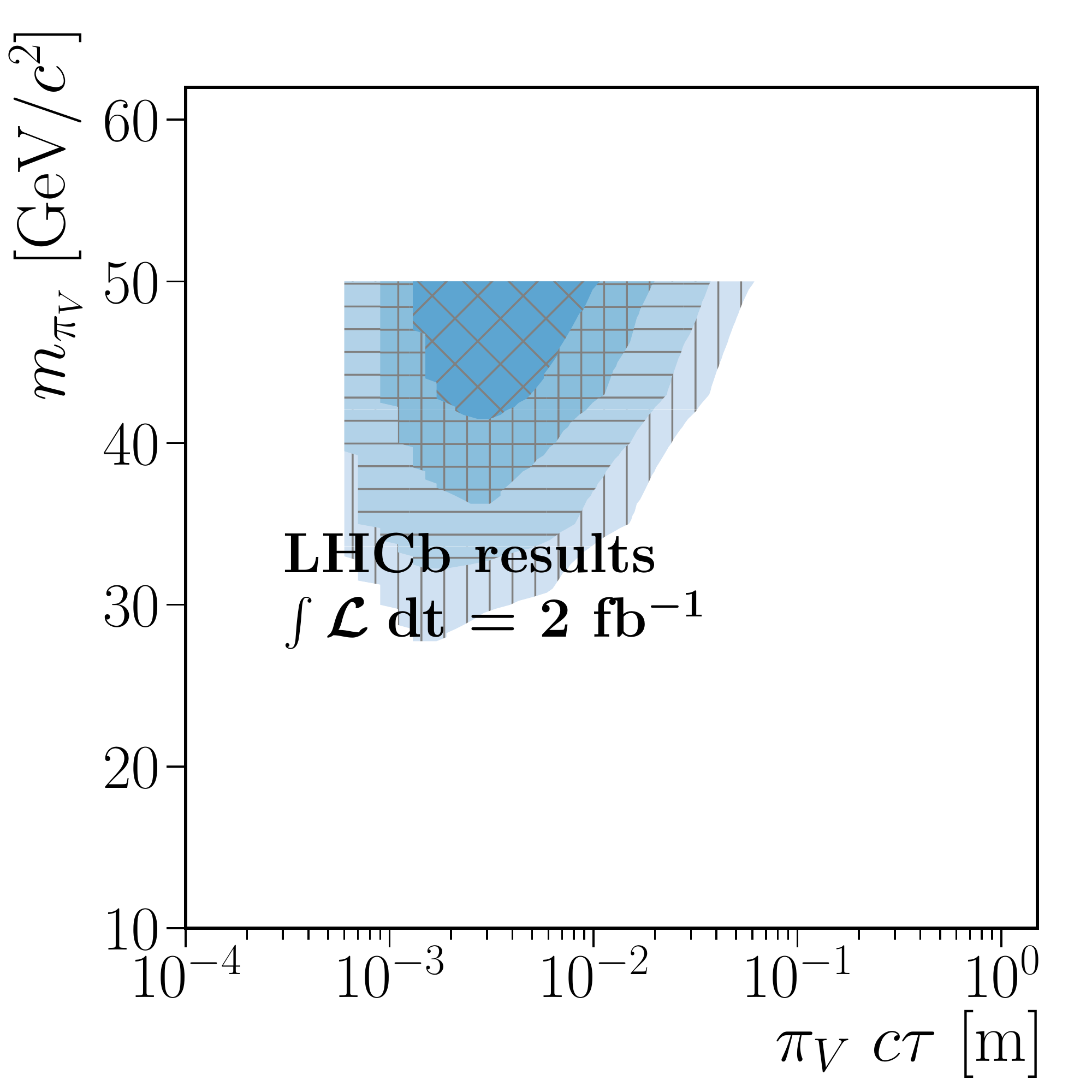}
\includegraphics[width=\figuremaxwidthtp+3mm,height=\figuremaxheighttp,keepaspectratio]{\main/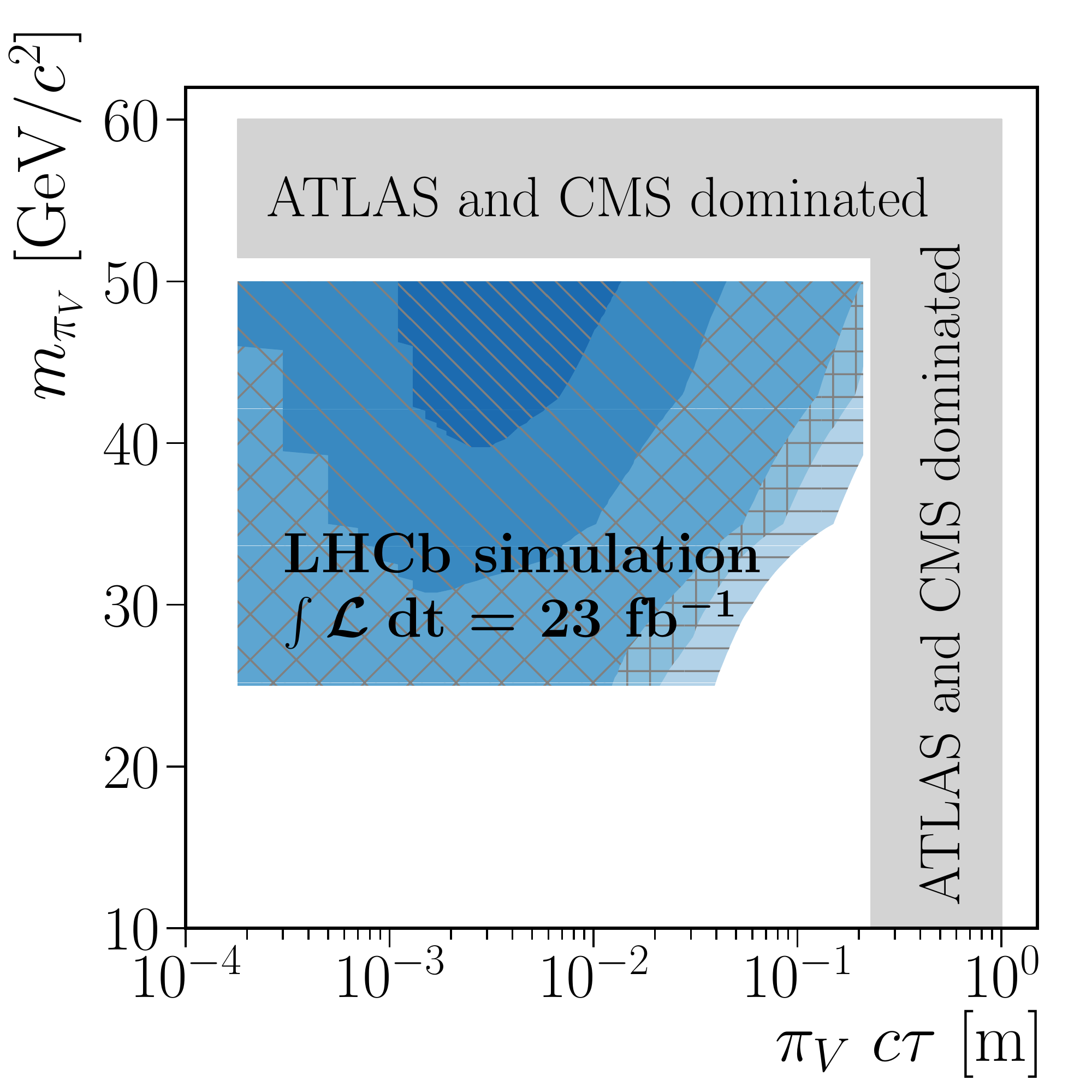}
\end{minipage}
\begin{minipage}[t][2\minipageheighttp][t]{\minipagewidthtp+6mm}
\centering
\includegraphics[width=\figuremaxwidthtp+3mm,height=\figuremaxheighttp,keepaspectratio]{\main/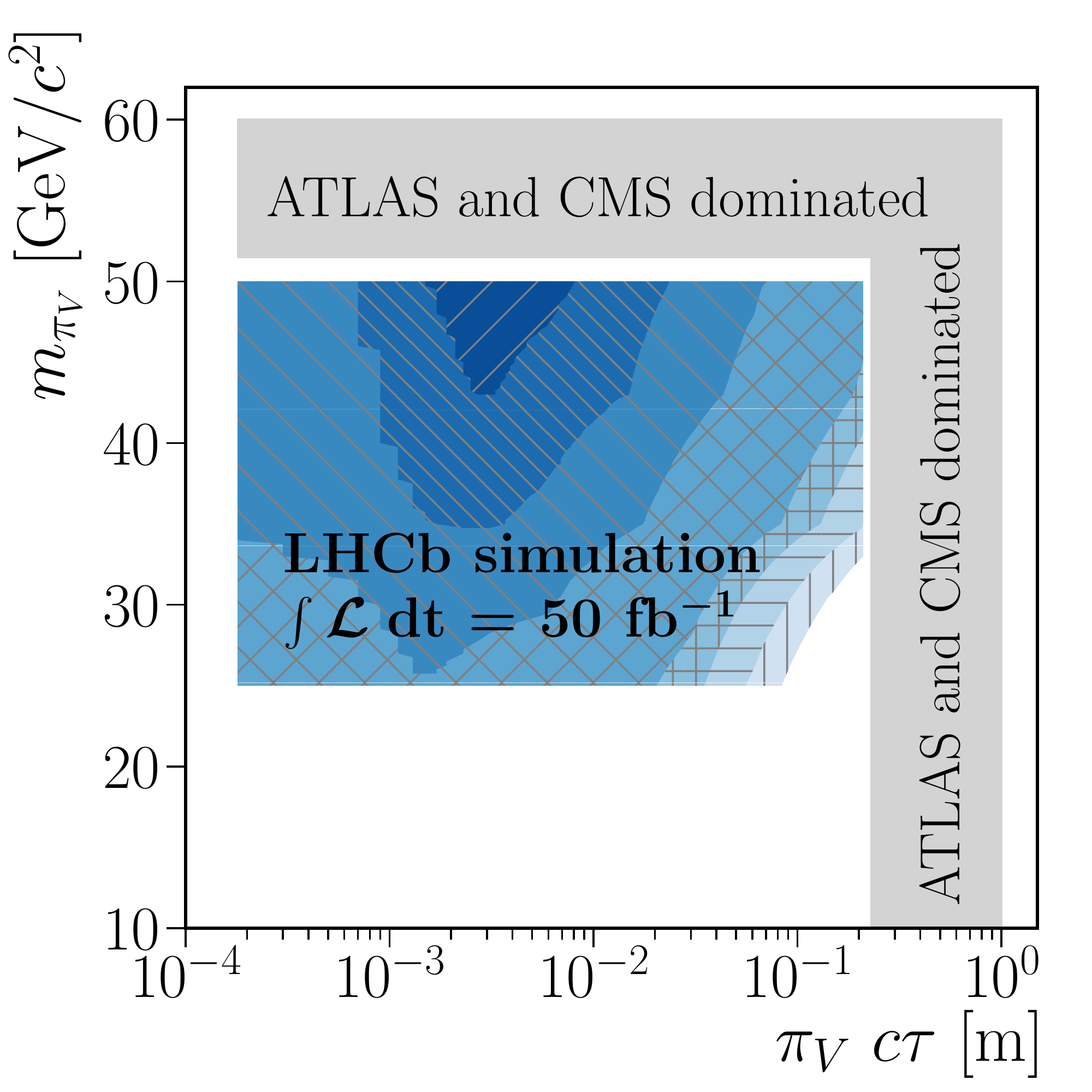}
\includegraphics[width=\figuremaxwidthtp+3mm,height=\figuremaxheighttp,keepaspectratio]{\main/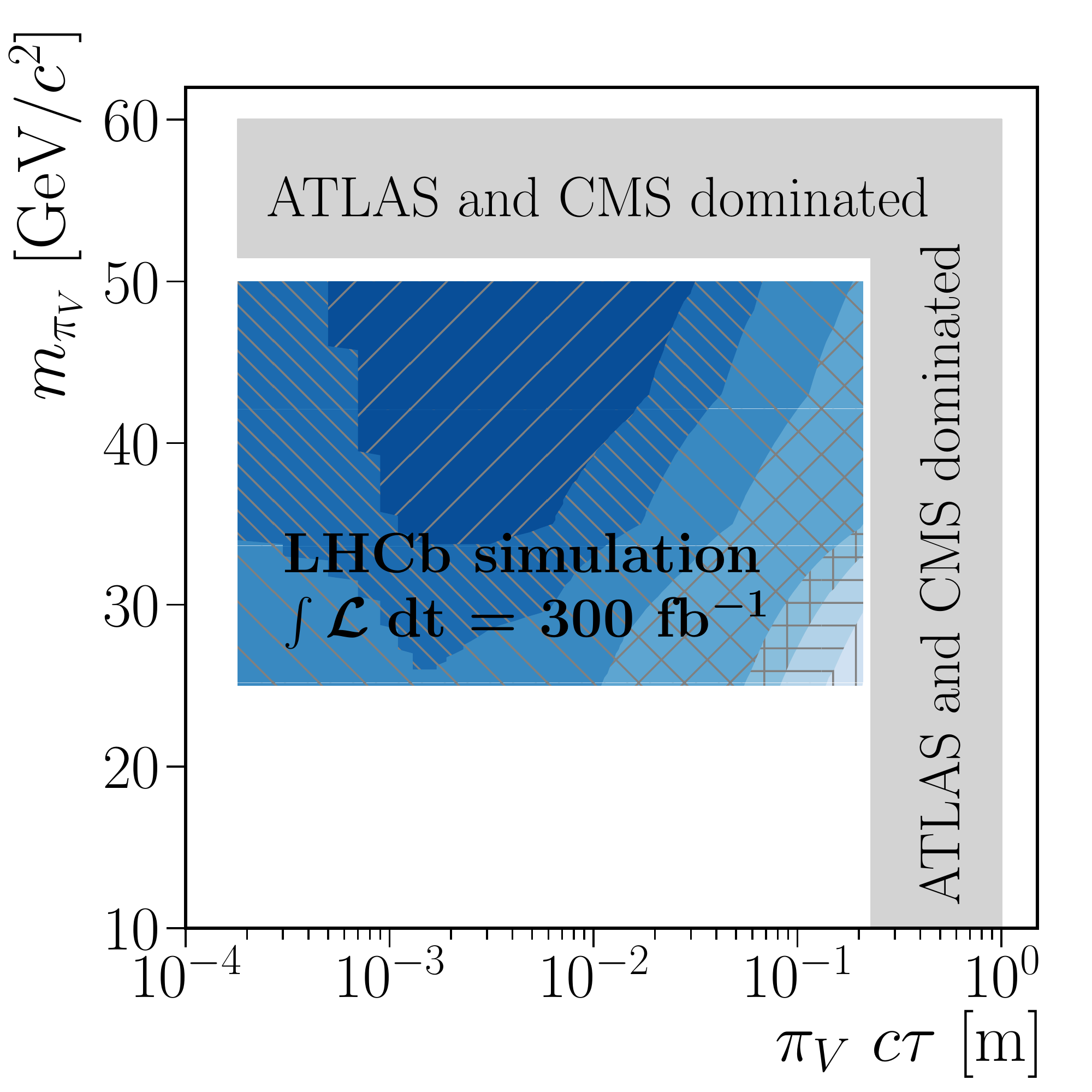}
\end{minipage}
\begin{minipage}[t][2\minipageheighttp][t]{\minipagewidthtp-12mm}
\centering
\raisebox{-2.5cm}{\includegraphics[width=\figuremaxwidthtp,height=\figuremaxheighttp-9mm,keepaspectratio]{\main/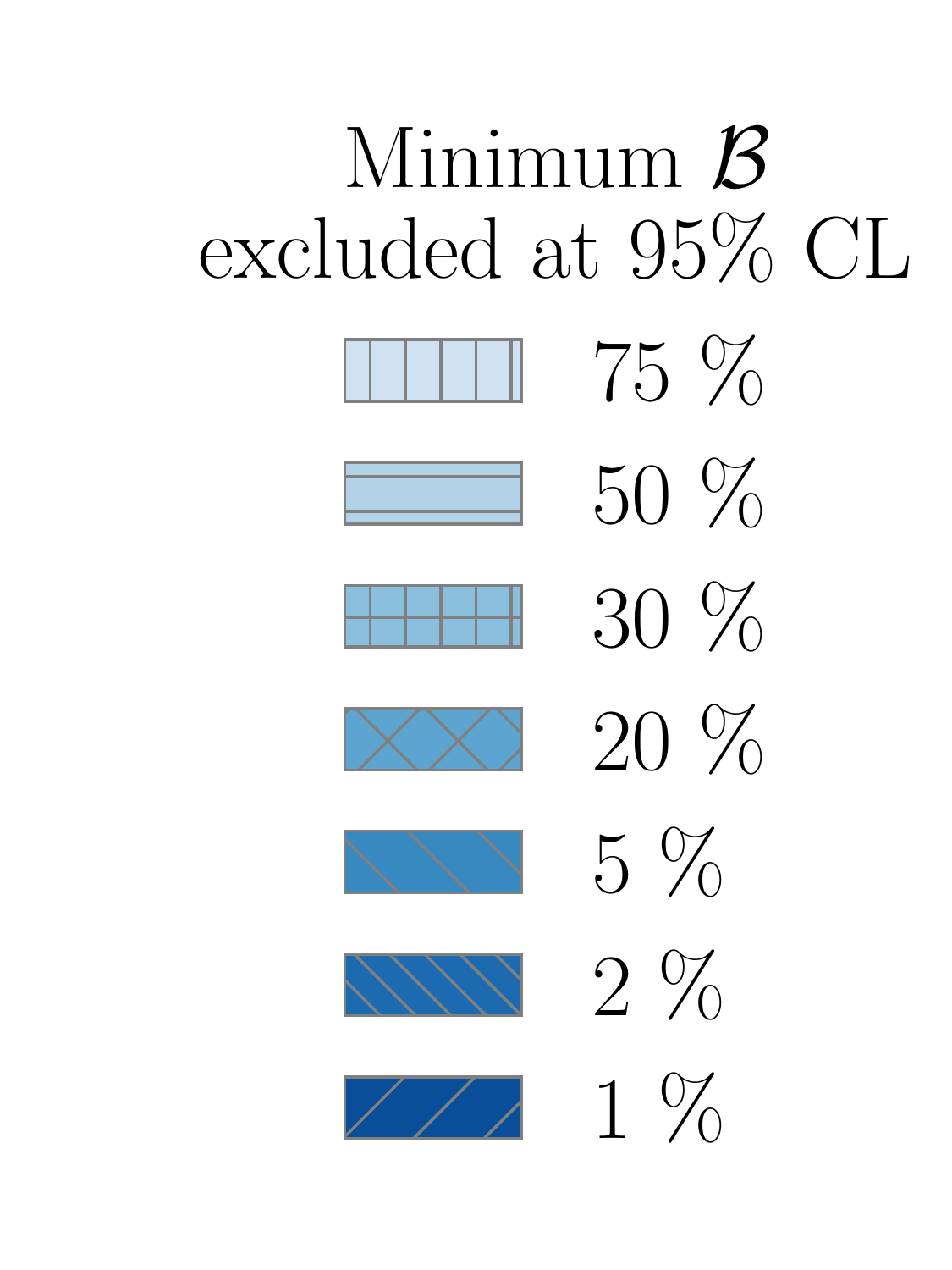}}
\end{minipage}
\vspace{-14.5cm}
%\begin{minipage}[b]{0.77\textwidth}
%  \includegraphics[width=0.49\textwidth]{}
%  \includegraphics[width=0.49\textwidth]{}\\
%   \includegraphics[width=0.49\textwidth]{}
%  \includegraphics[width=0.49\textwidth]{}\\
%\end{minipage}
%\begin{minipage}[t]{0.22\textwidth}
%  \vspace{-10cm}
%\includegraphics[width=1.3\textwidth]{}
%\end{minipage}
  \caption{Projected sensitivities of the search for HV pions decaying hadronically and produced through a Higgs boson exotic decay. The results are extrapolated from Run-1 results (top left), for luminosities of $23\fbinv$ (top right), $50\fbinv$ (bottom left) and $300\fbinv$ (bottom right). The results are presented in terms of the excluded parameter space of the HV pions for different upper limits at $95\%\cl$ on the branching fractions of the Higgs boson decay.}
  \label{fig:dvsearches_hllhc}
\end{figure}

With the updated backgrounds and expected signal yields, the CL$\rm _S$ method \cite{Read:2002hq} is used to compute the expected upper limits for different assumptions in the integrated luminosity and of the Higgs decay branching fraction. The Higgs boson production cross section is assumed to be that of the SM \cite{deFlorian:2016spz}. The systematic uncertainties, which are not dominant for the computation of these limits in the published analysis, are considered to be the same as in Run-1, and added as a correction factor to the limits obtained using just statistical uncertainties. With all these assumptions, the HV pion masses and lifetimes excluded at $95\%\cl$ are obtained. The results are shown in \fig{fig:dvsearches_hllhc}. The plots display, for different assumed integrated luminosities and branching fractions of the Higgs boson decay to a pair of HV pions, the ranges excluded at $95\%\cl$. These ranges are shown as a function of the HV pion mass and lifetime. As in \secc{sec:lhcbllpmuons}, the lifetime range covered is constrained by the physical length of the VELO detector. LHCb expects to exclude the existence of $\pi_V$ with masses between $35$ and $50\GeV/c^{2}$ and lifetimes in the $c\tau$ range $0.1-1\cmeters$, pair-produced through the decay of the Higgs boson, for branching fractions of such decay above $1\%$. The mass region below $\sim25 \GeV/c^{2}$ is expected to be accessible studying the substructure of merged jets~\cite{Bokan:1753426}.

%%%%%%%%%%%%%%%%%%%%%%%%%%%%%%
\subsection{Heavy Stable Charged Particles at HL-LHC}
Several extensions of the SM predict the existence of new
heavy particles with long lifetimes. If their lifetime exceeds a few nanoseconds, such particles can travel through the majority of the detector before decaying and therefore appear as stable. In the following, two dedicated studies performed using the upgraded CMS detector at the HL-LHC are presented for particles with non-zero electric charge and for particles with anomalously high energy loss through ionisation in the silicon sensors. Emphasis is given to detector requirements necessary to perform such specialised searches. 

\subsubsection{\cms{Heavy stable charged particle search with time of flight measurements}}
\label{sec:cmsHSCPTOF}
\contributors{C. Carrillo, J. Goh, M. Gouzevitch, G. Ramirez-Sanchez, CMS}

%Moved to the intro: 
%Several extensions of the standard model predict the existence of new heavy particles with long lifetimes. If their lifetime exceeds a few nanoseconds, such particles can travel through the majority of the detector before decaying and therefore appear as stable. 
In this section, we consider particles with non-zero electric charge which are referred to as heavy stable charged particles (HSCPs).
We concentrate on the performance in terms of specific
HSCP parameters in a model-independent way rather than providing an interpretation in a dedicated model. 
Given the wide range of new models,
it is important to stay sensitive to a wide range of unusual signatures such
as very slowly moving particles. The results presented here are from the CMS Collaboration based on~\citeref{Collaboration:2283189}.

\begin{figure}[t]
\centering
\begin{minipage}[t][\minipageheightsp][t]{\minipagewidthsp}
\centering
\includegraphics[width=\figuremaxwidthsp,height=\figuremaxheightsp,keepaspectratio]{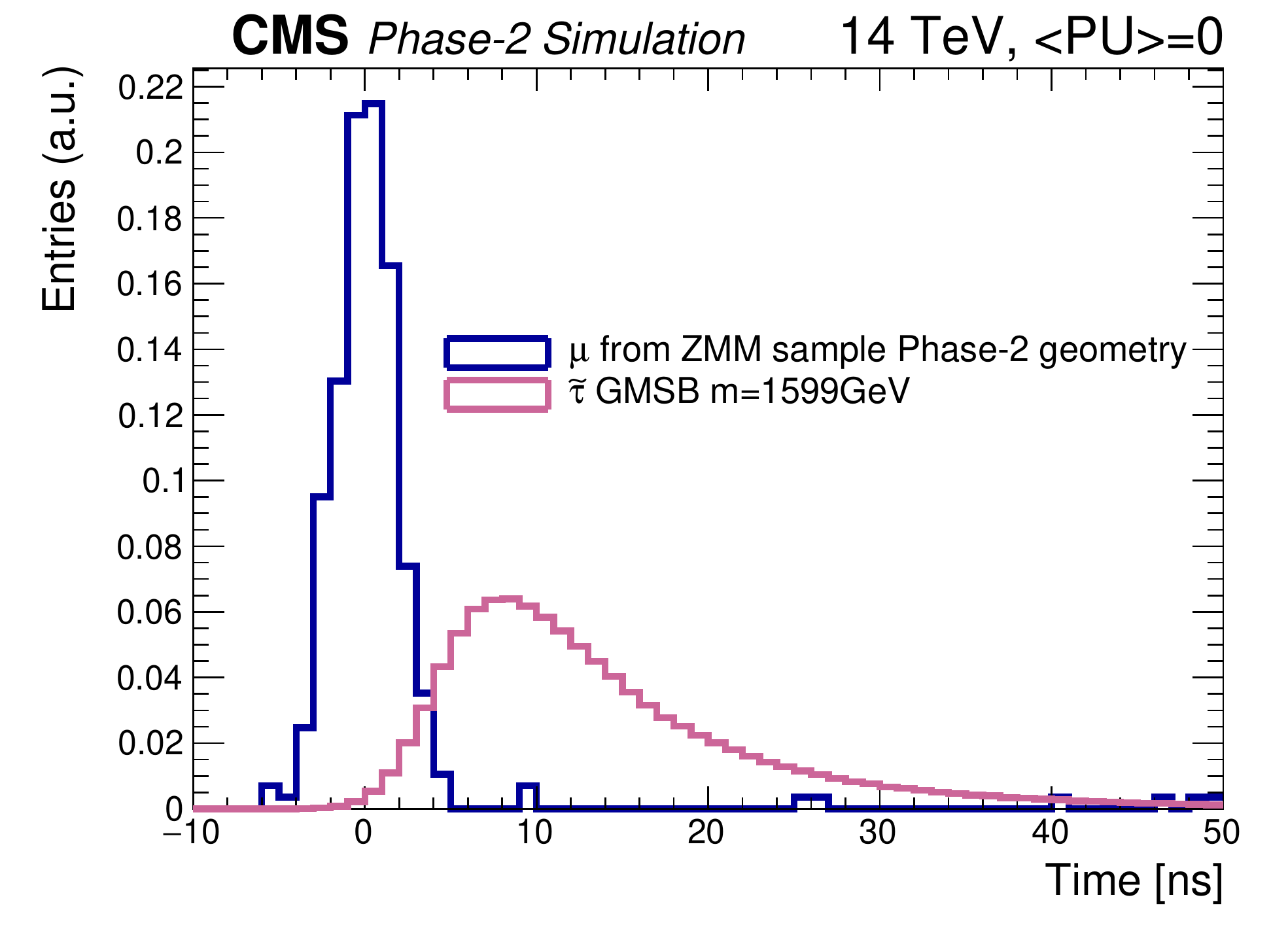}
\end{minipage}
\caption{Example of an RPC hit time measurement distribution for muons from the SM process $Z \to \mu\mu$ in comparison to events from semi-stable staus with a mass of about $1600\GeV$, produced in $pp \to \tilde{\tau} \tilde{\tau}$ processes. The relativistic muons pass through the detector at the speed of light, hence their time of arrival is centred around zero. Decay products from the slowly moving staus arrive much later, for the given mass on average by $10\nsec$.
} 
\label{fig:time}
\end{figure}

HSCPs will leave a direct signal in the tracker and muon systems of CMS and can be reconstructed similarly to muons. Depending on their mass, HSCPs can potentially move much more slowly than muons, which are typically travelling nearly at the speed of light ($\beta \approx 1$).
Therefore, HSCPs can be identified using their time-of-flight (TOF) from the
centre of CMS to the muon systems.  This is illustrated in \fig{fig:time}
for a slowly moving HSCP in comparison to relativistic muons, here from
the decay of Z bosons.  The latter are centred around zero time with respect to
their uniquely identified bunch crossing.  
This study builds on the improvements from the upgrade of the RPC link boards
in the CMS muon barrel and endcaps~\cite{Collaboration:2283189}.
While the time resolution of the present RPC system is around $25\nsec$,
the upgraded link board systems are expected to reach resolutions near
$1\nsec$ for the entire RPC system.  This upgrade enables new analysis
strategies at both the trigger and offline levels.

\begin{figure}[t]
\centering
\begin{minipage}[t][\minipageheightdp][t]{\minipagewidthdp}
\centering
\includegraphics[width=\figuremaxwidthdp,height=\figuremaxheightdp,keepaspectratio]{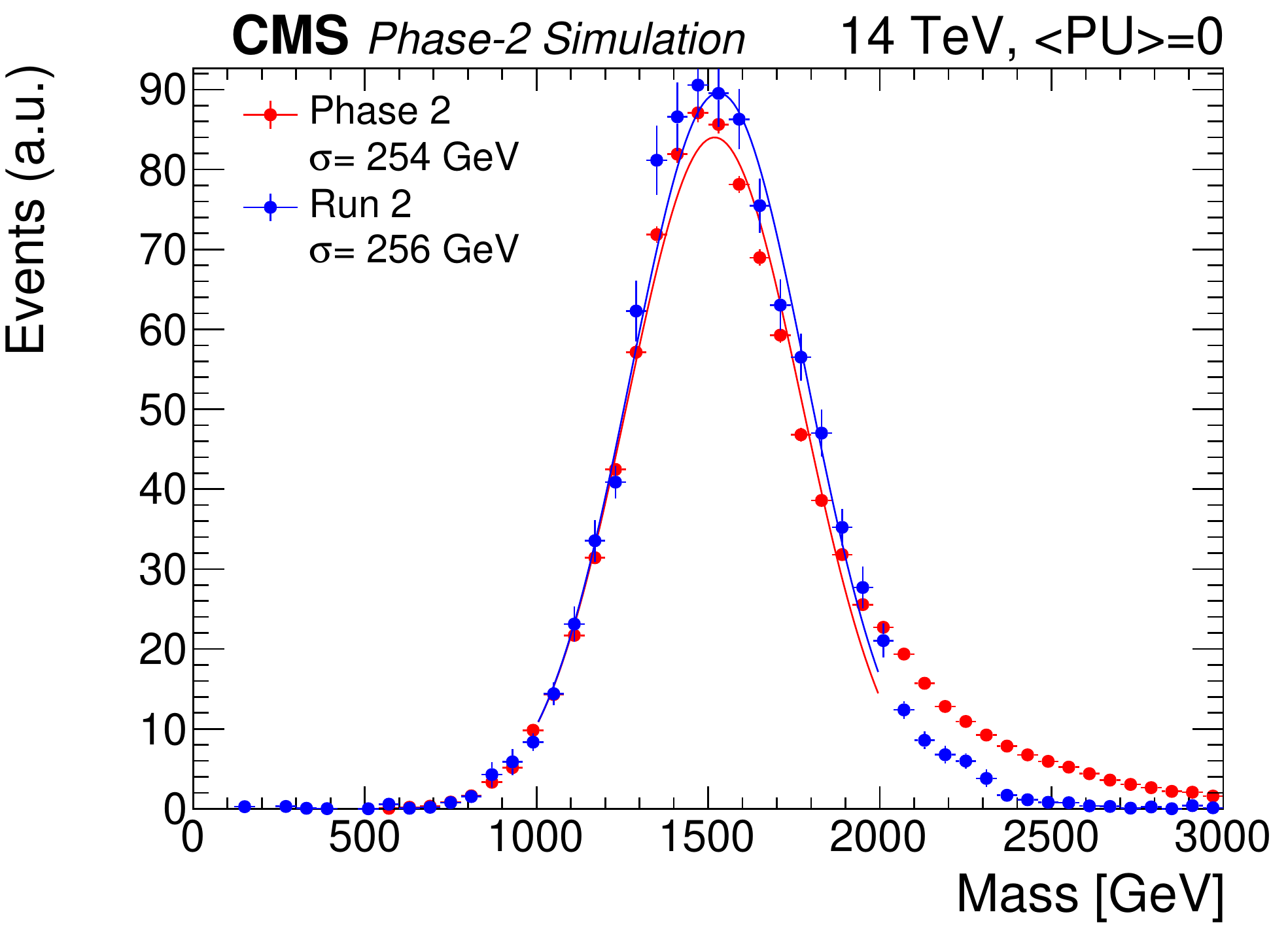}
\end{minipage}
\begin{minipage}[t][\minipageheightdp][t]{\minipagewidthdp}
\centering
\includegraphics[width=\figuremaxwidthdp,height=\figuremaxheightdp,keepaspectratio]{\main/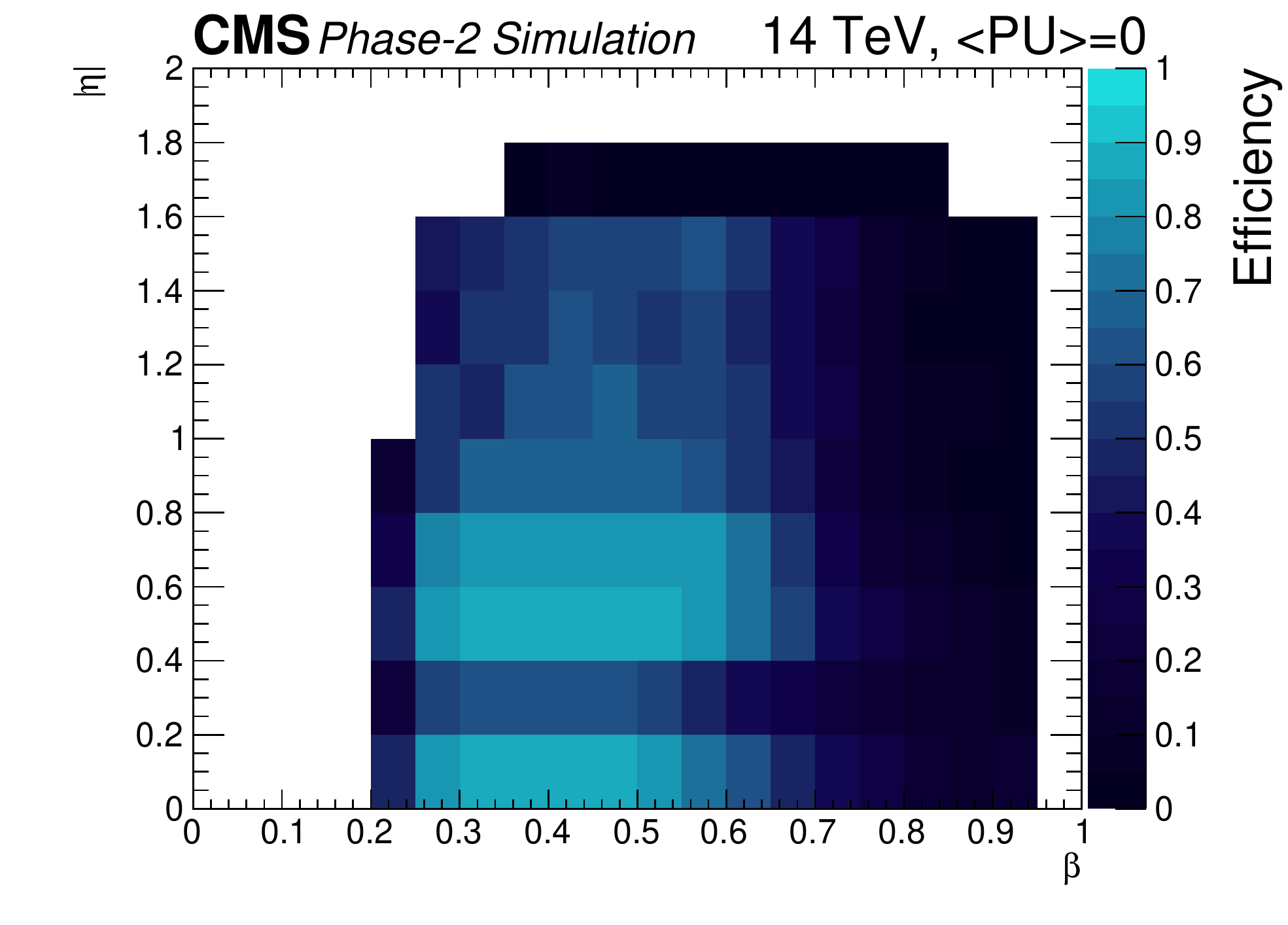}
\end{minipage}
\caption{Left: comparison of the mass resolution for a $1.6\TeV$ stau. In Run-2 the plotted resolution can only be achieved offline, while the upgraded RPC link-boards in Phase-2 provide a similar mass resolution already at the trigger level.  
Right: reconstruction efficiency for HSCPas a function of $\beta$ and $\eta$ given by the colour code of the z-axis. With the Phase-2 upgrade, events with $\beta < 0.5$ can be triggered with nearly $90\%$ efficiency for $|\eta| < 0.8$.} 
\label{fig:HSCP_beta_resolution}
\end{figure}

An HSCP trajectory is reconstructed as a slowly moving muon introducing the
parameter $\beta$ quantifying the (non)-relativistic velocity of the particle.
The velocity may be computed by measuring the time of
flight in the muon detectors at large distances from the collision point.
Particles moving slowly through the muon systems leave hits with a linear
pattern in hit-position versus time. The hits can be spread across several bunch crossings.
Therefore, muon detectors with precise timing can provide important information
for the HSCP signal searches.

Figure~\ref{fig:HSCP_beta_resolution} (left) shows the achievable mass resolution
for a supersymmetric $\tilde{\tau}$ lepton of $1.6\TeV$ mass.
The resolution for  the HSCP mass obtained for
Phase-2 at the trigger level is comparable to that realised in Run-2 studies
based on offline time-of-flight information from other muon detectors in CMS.
The information provided by the RPC trigger can be used as an independent
cross check of the reconstructed mass.
Figure~\ref{fig:HSCP_beta_resolution} (right) illustrates the expected reconstruction
efficiency as a function of $\eta$ and $\beta$. For $|\eta < 1.5|$, an efficiency of up to
$90\%$ can be reached for values of $\beta > 0.25$. In Run-2, the trigger
is highly efficient between $0.6<\beta<1$, but only about $20\%$ efficient 
for $\beta <0.5$~\cite{Khachatryan:2015lla, CMS:2016ybj}.
The large gain in efficiency for very slowly moving particles in Phase-2 enabled by the upgrade of the RPC trigger can be exploited in a model independent HSCP search.

\subsubsection{\cms{Heavy stable charged particle search with energy loss}}
\label{sec:cmsHSCPdEdx}
\contributors{J. Pazzini, J. Zobec,  CMS}
%{\bf Author(s): J. Pazzini, J. Zobec,  CMS}

It may happen that the only signs of new physics are rather exotic signatures that cannot be detected with conventional analyses. An example for such a 
signature is the production of heavy stable charged particles with long lifetimes that move slowly through the detector, heavily ionising the sensor 
material as they pass through. The supersymmetric particles stau ($\tilde{\tau}$) and gluino ($\tilde{g}$) are possible examples. Often, the cross 
section for such processes is expected to be very small and hence the HL-LHC provides a good environment for searching for such particles. Depending 
on their mass and charge, we can expect anomalously high energy loss through ionisation ($\mathrm{d}E/\mathrm{d}x$) in the silicon sensors with respect to the typical 
energy loss for SM particles ($\approx 3\MeV/\mathrm{cm}$ for minimum ionising particles (MIPs) with \sloppy\mbox{$10-1000\GeV$} momentum). 

\begin{figure}[t]
\centering
\begin{minipage}[t][\minipageheightdp][t]{\minipagewidthdp}
\centering
\includegraphics[width=\figuremaxwidthdp,height=\figuremaxheightdp,keepaspectratio]{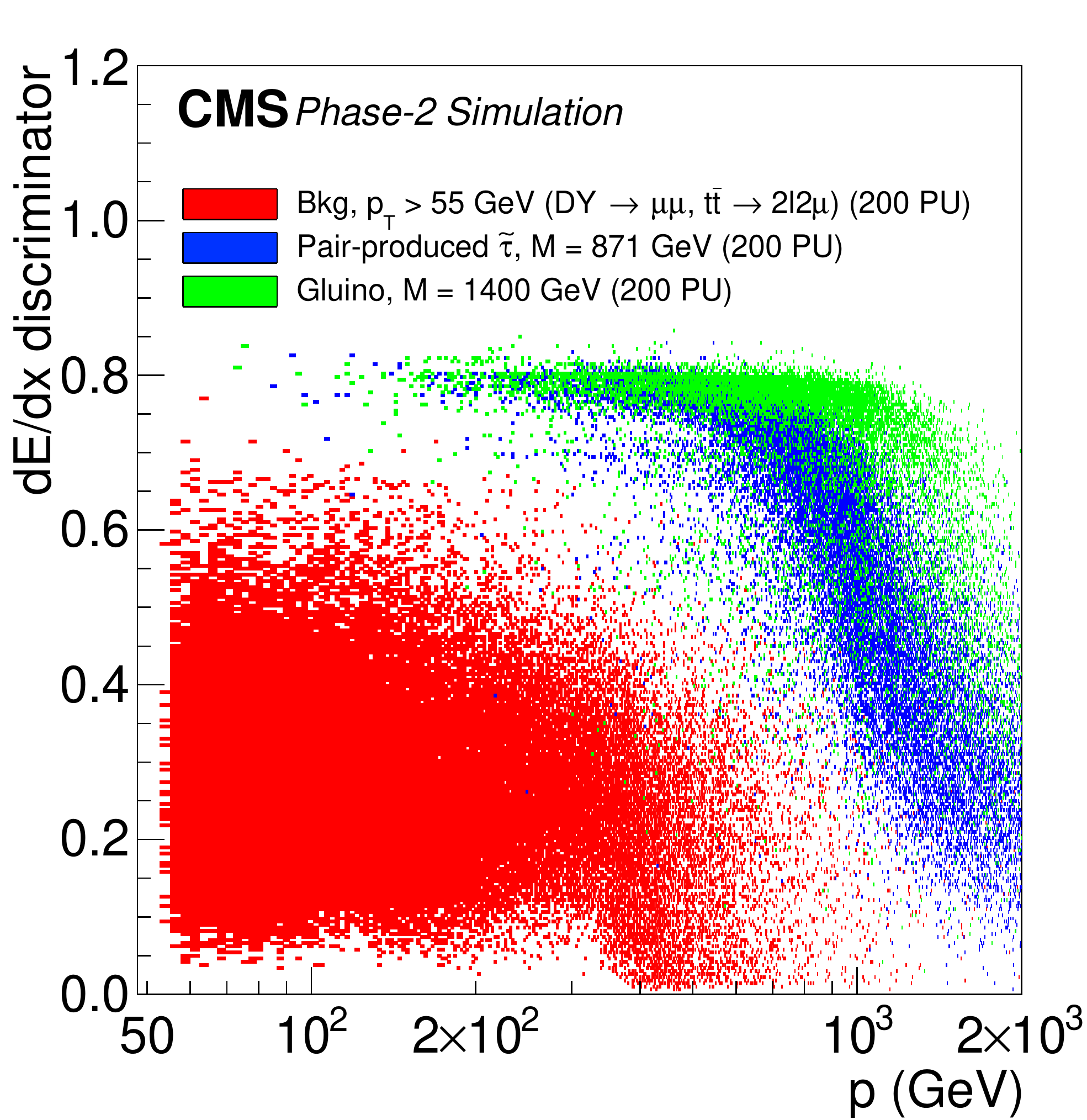}
\end{minipage}
\begin{minipage}[t][\minipageheightdp][t]{\minipagewidthdp}
\centering
\includegraphics[width=\figuremaxwidthdp,height=\figuremaxheightdp,keepaspectratio]{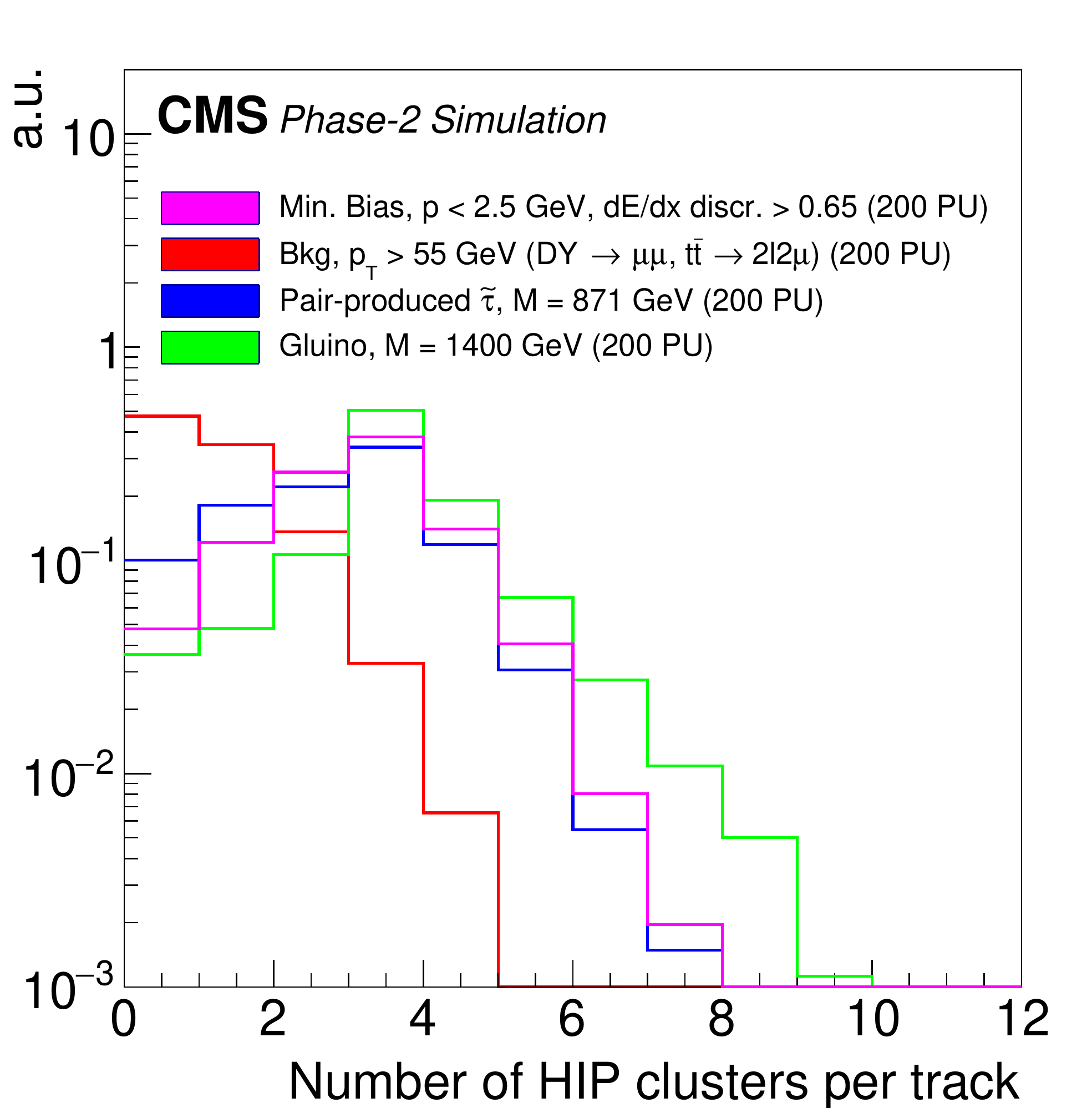}
\end{minipage}
  \caption{Left: distribution of the $\mathrm{d}E/\mathrm{d}x$ discriminator versus track momentum ($p$) for tracks with high momentum ($\pt > 55\GeV$) in background 
  events (red) and candidate signal particles. Pair produced  $\tilde{\tau}$s  with a mass of $871\GeV$ (blue), and a gluino with a mass of $1400\GeV$ 
  (green), are shown. Right: the distribution of the number of high threshold clusters with HIP flag per track for the HSCP signals, gluinos (green) 
  and  $\tilde{\tau}$s (blue), highly ionising and low-momentum protons and kaons (magenta), and tracks with high momentum ($\pt> 55\GeV$) in background 
  events (red).}
\label{fig:HSCP_variables}
\end{figure}

The present strip tracker in the CMS detector features analogue readout, and the pixel detector featured analogue readout at Phase-0 and features 
digital readout at Phase-1, 
allowing for excellent $\mathrm{d}E/\mathrm{d}x$ measurements. The Phase-2 CMS Inner Tracker will continue providing $\mathrm{d}E/\mathrm{d}x$ measurements, enabled by its Time over Threshold readout, 
while the Outer Tracker cannot provide such information, given that the readout is binary~\cite{CMS-TDR-17-001}. 
To increase the sensitivity for signatures with anomalously high ionisation loss, a second, programmable, threshold has been implemented in the readout 
electronics of some modules of the Outer Tracker, and a dedicated readout bit signals if a hit is above this second threshold~\cite{CMS-TDR-17-001}. 
Searches for heavy stable (or quasistable) 
charged particles (HSCPs) can thus be performed by measuring the energy loss in the Inner Tracker and by discriminating HSCPs from minimum ionising particles based 
on the ``HIP flag'' in the Outer Tracker. A threshold corresponding to the charge of $1.4$ MIPs is used in the simulation, and the gain in sensitivity obtained by 
using the HIP flag is studied~\cite{CMS-TDR-17-001}.

An estimator of the degree of compatibility of the track with the MIP hypothesis is defined to separate candidate HSCPs from tracks from SM background 
sources.  The high resolution $\mathrm{d}E/\mathrm{d}x$ measurements provided by the Inner Tracker modules are used for the computation of the $\mathrm{d}E/\mathrm{d}x$ 
discriminator. In \fig{fig:HSCP_variables} (left) the distribution of $\mathrm{d}E/\mathrm{d}x$ versus track momentum ($p$) for high momentum 
tracks ($\pt > 55\GeV$) selected in background events and candidate signal particles is shown. Two HSCP signals, pair produced  $\tilde{\tau}$s  with 
a mass of 871\,GeV and a gluino with a mass of 1400\,GeV, are compared to tracks from SM processes. In \fig{fig:HSCP_variables} (right) the 
distribution of the number of high threshold clusters with HIP flag per track is shown for the HSCP signals (gluinos and $\tilde{\tau}$s) compared to 
signal-like highly ionising and low-momentum protons and kaons in simulated minimum bias samples and to tracks with high momentum ($\pt> 55\GeV$) in 
simulated background events. The tracks in background events have a low number of high threshold clusters with HIP flag compared to those observed 
for tracks in HSCP signal events and slow moving protons and kaons in minimum bias events.

\begin{figure}[t]
\centering
\begin{minipage}[t][\minipageheightdp][t]{\minipagewidthdp}
\centering
\includegraphics[width=\figuremaxwidthdp,height=\figuremaxheightdp,keepaspectratio]{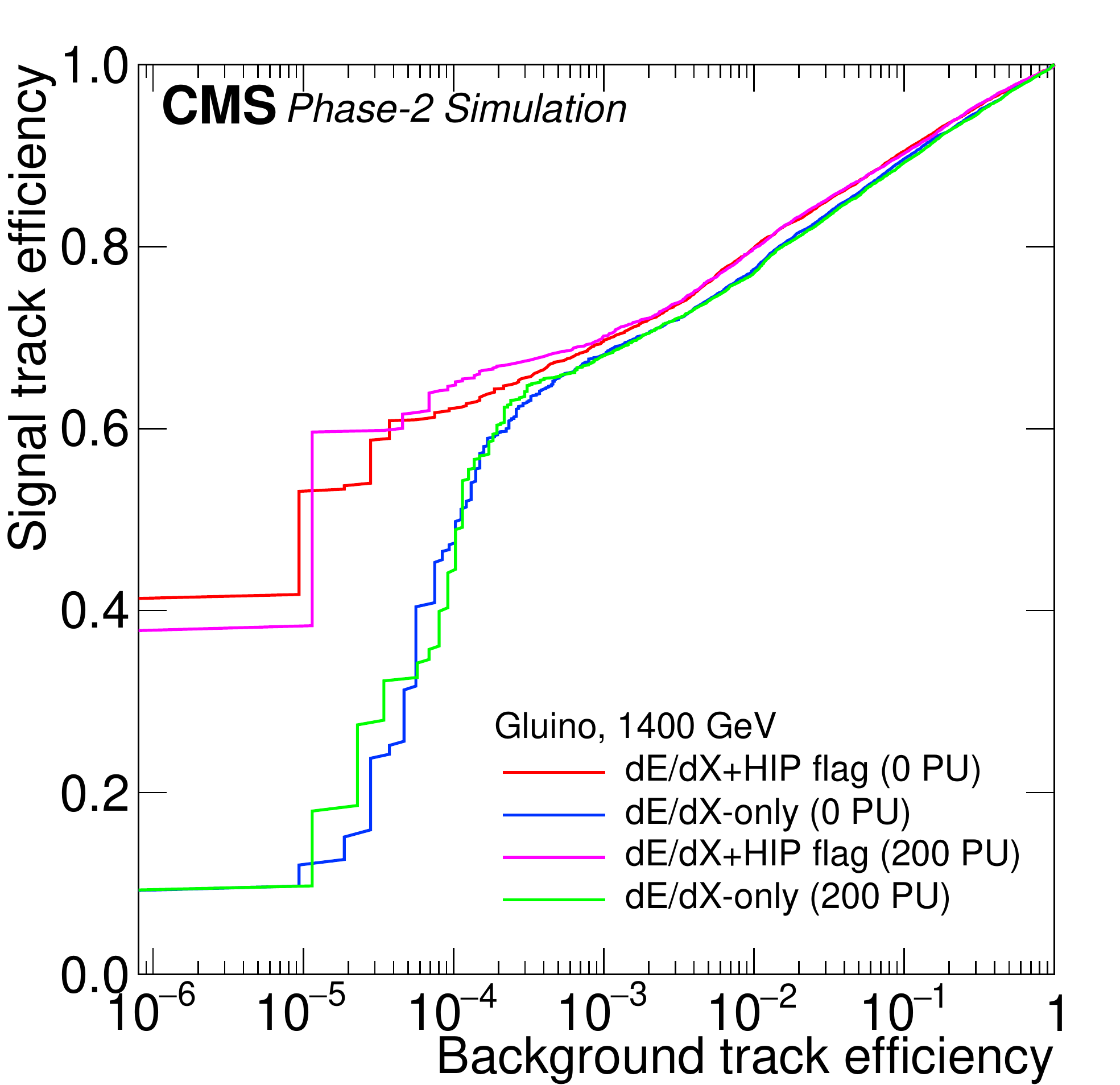}
\end{minipage}
\begin{minipage}[t][\minipageheightdp][t]{\minipagewidthdp}
\centering
\includegraphics[width=\figuremaxwidthdp,height=\figuremaxheightdp,keepaspectratio]{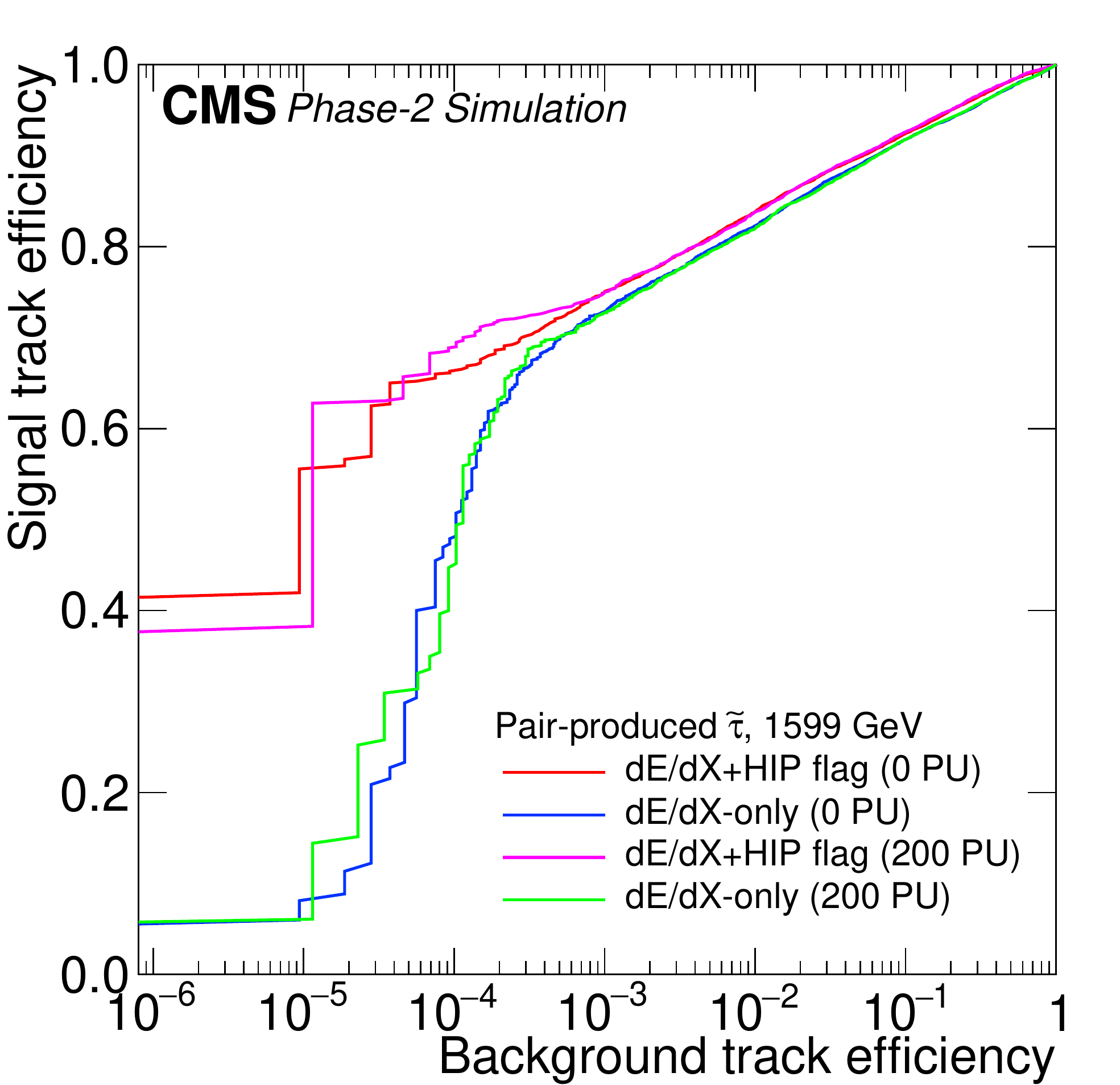}
\end{minipage}
  \caption{Performance of the $\mathrm{d}E/\mathrm{d}x$ discriminator for selecting pair produced $\tilde{\tau}$s (left) and gluinos (right) in events with $0$ PU and 
  $200$ PU. The signal versus background efficiency performance curves for a discriminator making use of both the pixel information and the Outer Tracker HIP flag 
  (red and magenta) demonstrate a better performance compared to a discriminator trained to exploit only the $\mathrm{d}E/\mathrm{d}x$ information from the pixel modules 
  (blue and green), for a background rejection of $10^{-6}$.}
\label{fig:HSCP_results}
\end{figure}

Figure~\ref{fig:HSCP_results} shows the performance of the discriminator by evaluating the signal versus background efficiency curves to identify tracks 
from signal events and reject those originating from backgrounds. The performance curves are evaluated for two different strategies for the discriminator: 
the original $\mathrm{d}E/\mathrm{d}x$ discriminator, which relies solely on the Inner Tracker modules (``$\mathrm{d}E/\mathrm{d}x$-only''), and a recomputed discriminator which includes the HIP 
flags from Outer Tracker modules (``$\mathrm{d}E/\mathrm{d}x$+HIP bit''). The signal versus background efficiency performance curves in \fig{fig:HSCP_results} demonstrate 
that for a background efficiency of $10^{-6}$, analogous to the Phase-1 analysis performance, the $\mathrm{d}E/\mathrm{d}x$+HIP-based discriminator leads to an expected 
signal efficiency of 40\%, around 4~to 8~times better than the $\mathrm{d}E/\mathrm{d}x$-only discriminator. In the $\mathrm{d}E/\mathrm{d}x$-only scenario, the efficiency for the HSCP 
signal is about 8~times smaller than that obtained in Phase-1~\cite{CMS:2016ybj} and about 64 times the Phase-1 luminosity would be required to 
reach the Phase-1 sensitivity, making this search almost untenable. The inclusion of the HIP flag for the Outer Tracker restores much of the 
efficiency, so that the same sensitivity as in Phase-1 will be realised with about four times the luminosity of Phase-1. The Phase-1 sensitivity will be 
surpassed with the full expected integrated luminosity of the HL-LHC. This study demonstrates the critical impact of the HIP flag in restoring the 
sensitivity of the CMS tracker for searches for highly ionising particles.

%fast timing is not HSCP. Moved to it's own subsection -ku
\subsection{Additional examples of specialised techniques for LLP at HL-LHC} 
\label{sec:techniques} 
Two examples of specialised techniques relevant for LLPs are presented in this section. First, CMS illustrates the importance of precise timing detectors providing efficient measure the time of flight of LLPs between primary and secondary vertices. Second, ATLAS shows how jets arising from neutral LLPs decaying within the hadronic calorimeter can be characterised to efficiently reduce pile-up dependencies and therefore improve the sensitivity to new physics of this kind. 

\subsubsection{\cmsC{Fast timing signatures for long-lived particles}}
\label{sec:cmsfasttiming}
\contributors{D. del Re, A. Ledovskoy, C. Rogan, L. Soffi, CMS}
%{\bf Author(s): A. Ledovskoy, C. Rogan, L. Soffi, CMS}

A precision MIP timing detector (MTD) allows one to assign
timing for each reconstructed vertex and to measure the time of flight of
LLPs between primary and secondary vertices. 
This section presents studies from the CMS Collaboration from~\citeref{CMS-TDR-17-006} exploring the potential of such techniques at the HL-LHC.

Using the measured displacement between primary and secondary
vertices in space and time, the velocity of LLPs in the lab frame
$\vec{\beta}^{LAB}_{P}$ (and $\gamma_{P}$) can be calculated. In such
scenarios, the LLP can decay to fully-visible or partially-invisible
systems. Using the measured energy and momentum of the visible portion
of the decay, $E^{LAB}_{P}$ and $\vec{P}^{LAB}_{P}$, one can calculate
its energy in the LLP rest frame as

\begin{equation}
  E^P_V = \gamma_P\left(E^{LAB}_V-\vec{P}^{LAB}_V\cdot\vec{\beta}^{LAB}_V\right)=\frac{m_P^2-m_I^2+m_V^2}{2 m_P},
  \label{eq:llp1}
\end{equation}
where $m_P$, $m_V$, and $m_I$ are the masses of the LLP, the visible and the invisible
systems, respectively. Assuming the mass of the invisible system is known,
the subsequent mass of the LLP can be reconstructed as

\begin{equation}
 m_P = E^P_V + \sqrt{ {E_V^P}^2 + m_I^2 - m_V^2}.
 \label{eq:llp2}
\end{equation}

The reconstruction of the decay vertex for neutral
LLPs decaying to visible or partially-invisible decay products is
enabled, thus offering unprecedented sensitivity in these searches at the
LHC. The benefits of precision timing on such LLP searches is illustrated in two representative SUSY examples.

\begin{figure}[t]
\centering
\begin{minipage}[t][\minipageheightdp][t]{\minipagewidthdp}
\centering
\raisebox{1cm}{\includegraphics[width=0.7\textwidth,keepaspectratio]{\main/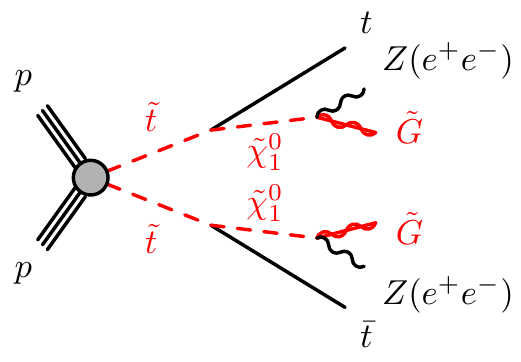}}
\end{minipage}
\begin{minipage}[t][\minipageheightdp][t]{\minipagewidthdp}
\centering
\includegraphics[width=\figuremaxwidthdp,height=\figuremaxheightdp,keepaspectratio]{\main/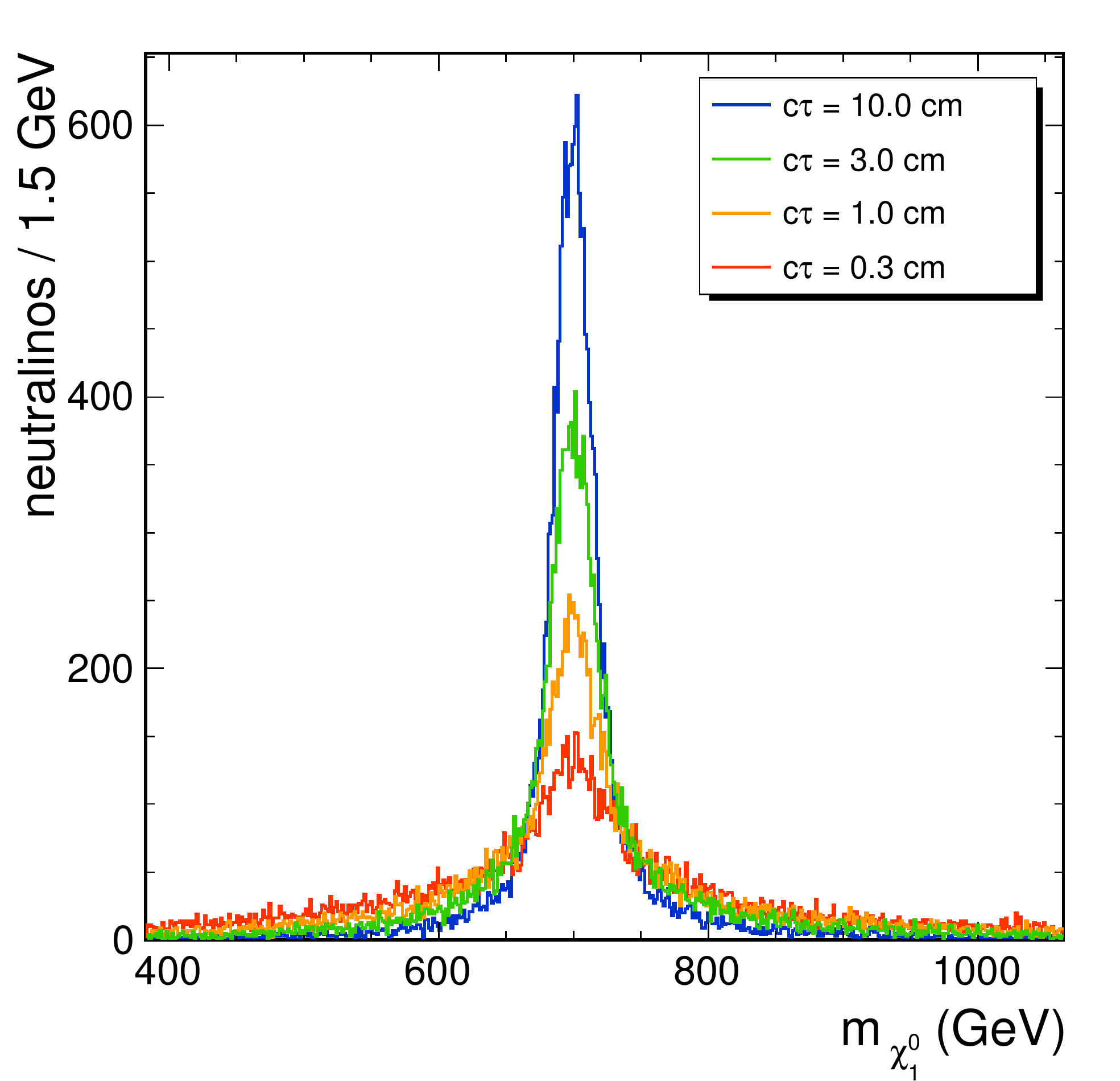}
\end{minipage}
\vspace{-0.9cm}
% \centering
% \raisebox{-0.5\height}{\includegraphics[width=0.33\textwidth]{}}
% \raisebox{-0.5\height}{\includegraphics[width=0.49\textwidth]{}}
 \caption{Diagram for top-squark pair production and decay (left) and reconstructed mass of the $\tilde{\chi}^0_1$ (right) for decays with $M(\tilde{t})=1000\GeV$ and
   $M(\tilde{\chi}^0_1)=700\GeV$. Mass distributions are shown for
   various values of $c\tau$ of $\tilde{\chi}^0_1$.
 \label{fig:LLP_stops}}
\end{figure}

The first example is a gauge mediated SUSY breaking (GMSB) scenario
where the $\tilde{\chi}^0_1$ couples to the gravitino $\tilde{G}$ via
higher-dimension operators sensitive to the SUSY breaking scale. In
such scenarios, the $\tilde{\chi}^0_1$ may have a long lifetime~\cite{Meade:2010ji}. It is
produced in top-squark pair production with $\tilde{t}\to
t+\tilde{\chi}^0_1$, $\tilde{\chi}^0_1\to Z + \tilde{G}$, and $Z\to
e^+e^-$. The decay diagram is shown in
\fig{fig:LLP_stops}~(left).

Events were generated with \pythia{ 8}~\cite{Sjostrand:2014zea}. The masses of
the top-squark and neutralino were set to $1000\GeV$ and $700\GeV$,
respectively. Generator-level quantaties were smeared according to the
expected experimental resolutions. A position resolution of $12\mumeters$
in each of three directions was assumed for the primary vertex~\cite{Chatrchyan:2014fea}.  The
secondary vertex position for the $e^+ e^-$ pair was reconstructed
assuming $30\mumeters$ track resolution in the transverse direction~\cite{Chatrchyan:2014fea}. The
momentum resolution for electrons was assumed to be $2\%$. And finally,
the time resolution of a charged track at the vertex was assumed to be
$30\psec$.

\begin{figure}[t]
\centering
\begin{minipage}[t][\minipageheightdp][t]{\minipagewidthdp}
\centering
\includegraphics[width=\figuremaxwidthdp,height=\figuremaxheightdp,keepaspectratio]{\main/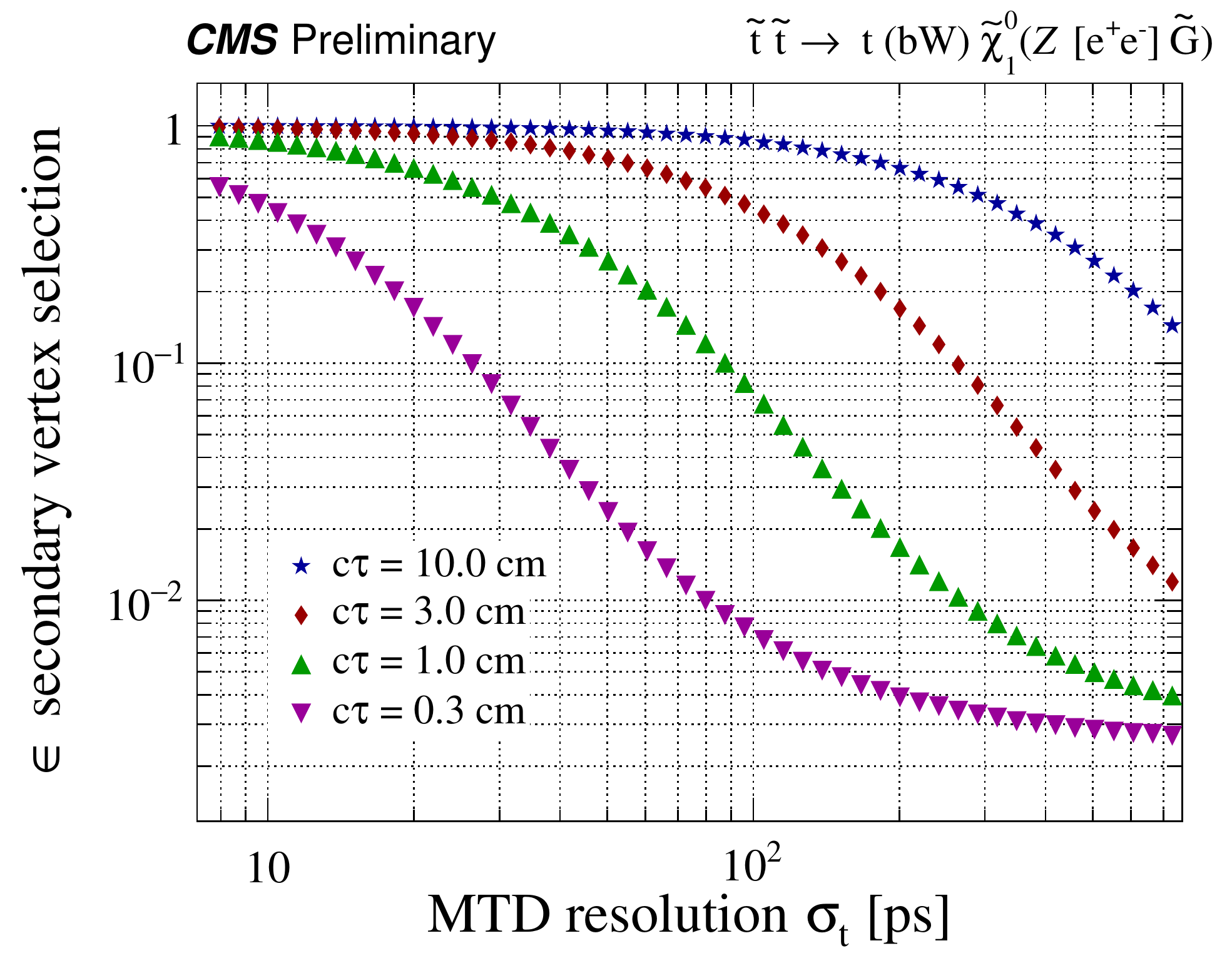}
\end{minipage}
\begin{minipage}[t][\minipageheightdp][t]{\minipagewidthdp}
\centering
\includegraphics[width=\figuremaxwidthdp,height=\figuremaxheightdp,keepaspectratio]{\main/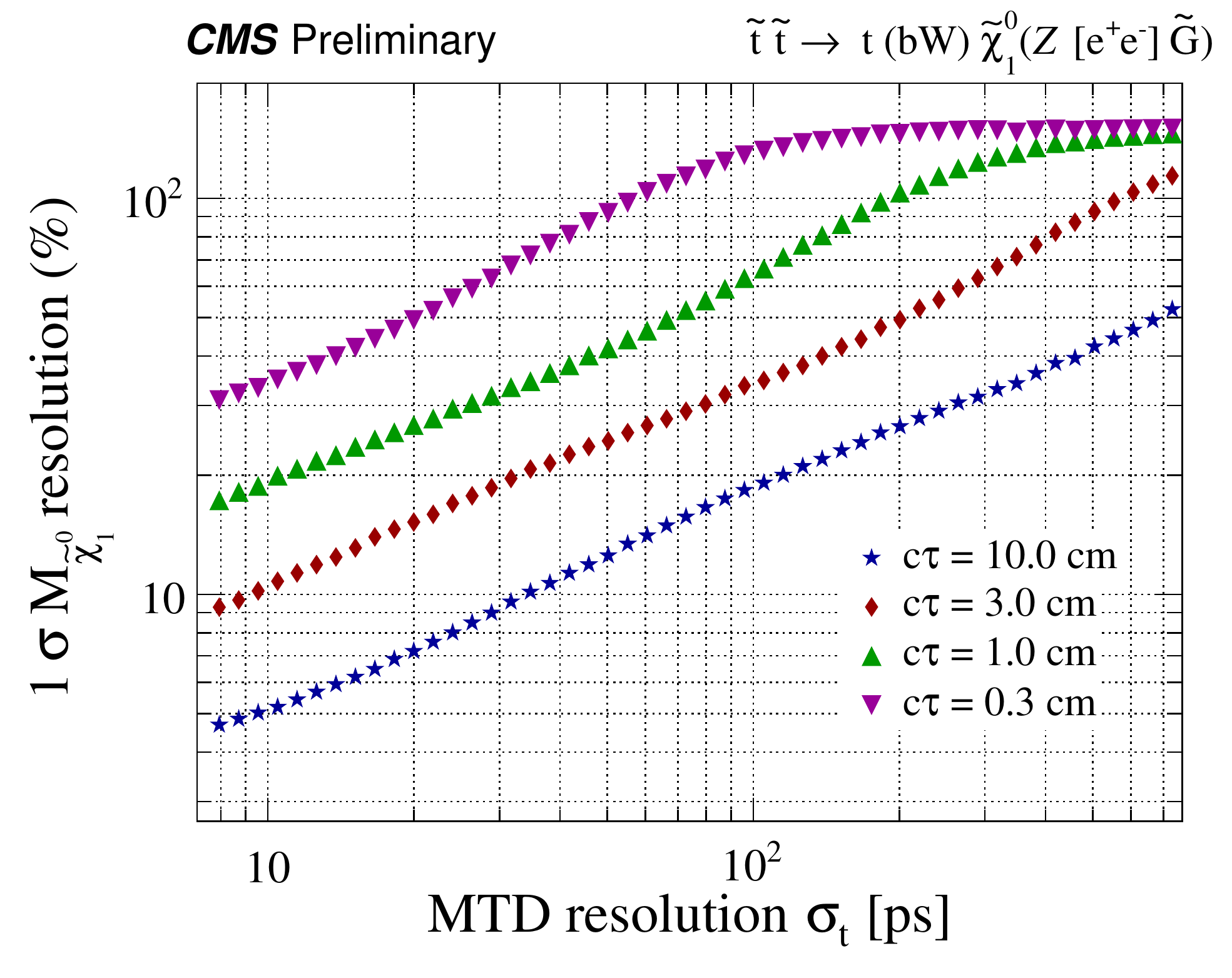}
\end{minipage}
 \caption{Efficiency (left) and mass resolution (right) as a function
   of timing resolution of MTD for reconstruction of
   $\tilde{\chi}^0_1$ mass in SUSY GMSB example of
   $\tilde{\chi}^0_1\to\tilde{G}+e^+e^-$ with
   $M(\tilde{\chi}^0_1)=700\GeV$ considering events with separation of
   primary and seconday vertices more than $3\sigma$ in both space
   and time.
 \label{fig:171025_52}}
\end{figure}

The mass of the LLP was reconstructed with \eqs{eq:llp1} and~\eqref{eq:llp2}
assuming the gravitino is massless by setting
$m_I=0$. Figure~\ref{fig:LLP_stops}~(right) shows the distribution of
the reconstructed mass of the neutralino for various $c\tau$ values of the LLP.
The fraction of events with a separation between primary and secondary
vertices of more than $3\sigma$ in both space and time as a function of
MTD resolution is show in \fig{fig:171025_52}~(left). The mass
resolution, defined as half of the shortest mass interval that
contains $68\%$ of events with $3\sigma$ displacement, as a function of
MTD resolution is shown in \fig{fig:171025_52}~(right)

\begin{figure}[t]
\centering
\begin{minipage}[t][\minipageheightdp][t]{\minipagewidthdp}
\centering
\raisebox{1.2cm}{\includegraphics[width=0.8\textwidth,keepaspectratio]{\main/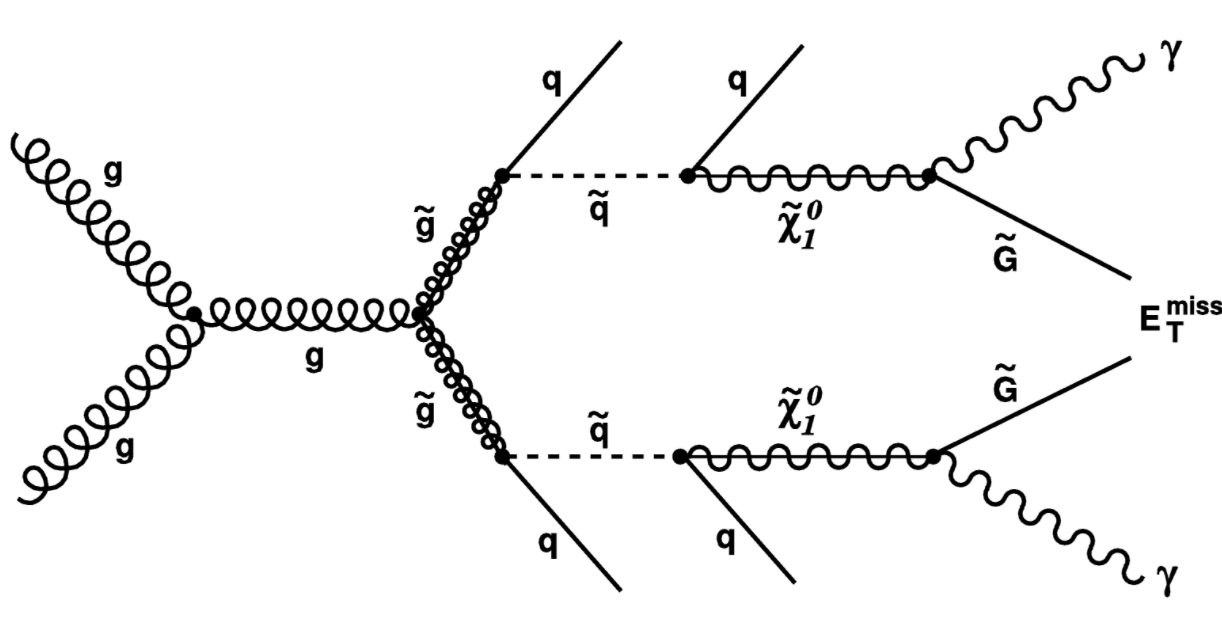}}
\end{minipage}
\begin{minipage}[t][\minipageheightdp][t]{\minipagewidthdp}
\centering
\includegraphics[width=\figuremaxwidthdp,height=\figuremaxheightdp,keepaspectratio]{\main/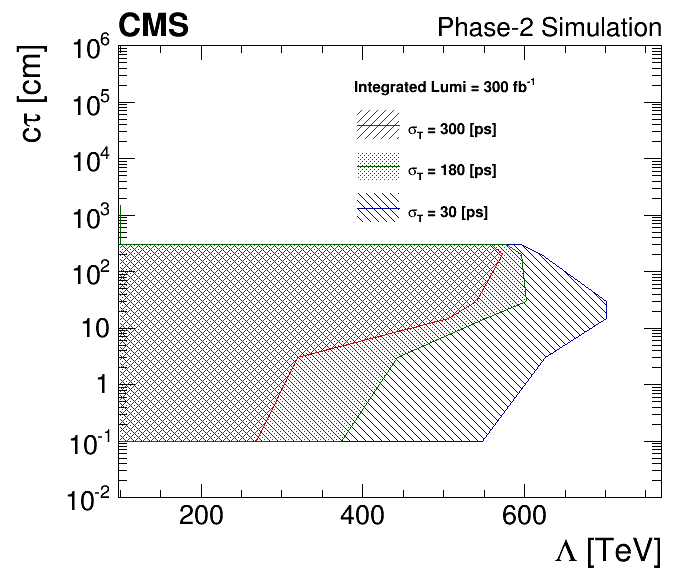}
\end{minipage}
\vspace{-1cm}
% \centering
% \raisebox{-0.5\height}{\includegraphics[width=0.4\textwidth]{}}
%% \raisebox{-0.5\height}{\includegraphics[width=0.32\textwidth]{\main/section4LongLivedParticles/img/Limits_excl_2D_Res300ps_ComparingLumi.png}}
% \raisebox{-0.5\height}{\includegraphics[width=0.50\textwidth]{}}
 \caption{Left: diagram for SUSY process that results in a diphoton
   final state through gluino production at the LHC. Right: sensitivity to GMSB
   $\tilde{\chi}^0_1\to\tilde{G}+\gamma$ signals expressed in terms of
   neutralino lifetimes and masses assuming a timing detector with
   different values of resolution and an integrated luminosity of $300
 \fbinv$.}
 \label{fig:llp-cornell}
\end{figure}

The second SUSY example is a GMSB
benchmark scenario~\cite{Strassler:2006im} where
the lightest neutralino ($\tilde{\chi}^0_1$) is the next-to-lightest
supersymmetric particle, and can be long-lived and decay to a photon and
a gravitino ($\tilde{G}$), which is the LSP.
Figure~\ref{fig:llp-cornell}(left) shows a diagram of a possible gluino
pair-production process that results in a diphoton final state.

For a long-lived neutralino, the photon from the
$\tilde{\chi}^0_1\to\tilde{G}+\gamma$ decay is produced at the
$\tilde{\chi}^0_1$ decay vertex, at some distance from the beam line,
and reaches the detector at a later time than the prompt, relativistic
particles produced at the interaction point.  The time of arrival of
the photon at the detector can be used to discriminate signal from
background.
The time of flight of the photon inside the detector is the sum of the
time of flight of the neutralino before its decay and the time of
flight of the photon itself until it reaches the detector. Since the
neutralino is a massive particle, the latter is clearly negligible with
respect to the former. It becomes clear in this sense that in order to
be sensitive to short neutralino lifetimes ($\mathcal{O}(\text{cm})$), the measurement of the photon time of flight is a crucial
ingredient of the analysis.  The excellent resolution of the MTD
detector ($\mathcal{O}(30\psec)$) can therefore be exploited to determine with high
accuracy the time of flight of the neutralino, and therefore of the
photon, also in case of a short lifetime.

A simple analysis has been performed at generator level in order to
evaluate the sensitivity of a search for displaced photons at
CMS in the scenario where a 30 ps timing resolution is available
from the MTD.  Events were generated with
\pythia{ 8}. The values of the $\Lambda$ scale
parameter were considered in the range $100-500\TeV$, and the neutralino
lifetimes ($c\tau$) explored in the range $0.1-300\cmeters$.  After requiring
the neutralino decaying within the CMS ECAL acceptance and the photon
energy being above a ``trigger-like'' threshold, the generator-level
photon time of flight was smeared according to the expected
experimental resolutions.  A cut at photon time greater than 3$\sigma$
of the considered time resolution is applied and the assumption of
background being zero in this ``signal region'' is made.  The signal
efficiency of such a requirement is computed and translated, assuming
the theoretical cross-sections provided in \citeref{Strassler:2006im}, in an
upper limit at $95\%\cl$ of \cl on the production cross-section of the
$\tilde{\chi}^0_1\to\tilde{G}+\gamma$ process.

Assuming a timing resolution of the order of $300\psec$, (thus requiring
photon time greater than $1\nsec$) close to the Run1 CMS
performance~\cite{Chatrchyan:2012jwg}, the analysis sensitivity in
terms of neutralino mass and lifetime is computed and shown in
\fig{fig:llp-cornell} (right) for a reference luminosity of 300
\fbinv, along with comparisons with improved timing resolution. 
For the
hypothesis of $\sigma_T=180$ ps a timing cut is applied at $450\psec$ and
for the $\sigma_T=30$ ps the timing is required to be larger than $100\psec$ at selection level.  As shown in the figure,
the increase of the signal efficiency at small lifetime, made possible
with the precise MTD, allows to extend the sensitivity region in the
explored phase space of short lifetime and large masses of the
neutralino.

\subsubsection{\atlas{Jets reconstruction techniques for neutral LLPs}}
\contributors{S. Pagan Griso, R. Rosten, ATLAS}

Traditional methods may fail to reconstruct, or may improperly reconstruct, objects associated with LLP decays. Searches for LLPs that are neutral under the SM gauge group might be targeted exploiting hadronic calorimeters. The techniques developed are described in the following, for more details see~\citeref{CERN-LHCC-2017-018}. 

\begin{figure}[t]
\centering
\begin{minipage}[t][\minipageheightdp][t]{\minipagewidthdp}
\centering
\includegraphics[width=\figuremaxwidthdp,height=\figuremaxheightdp,keepaspectratio]{\main/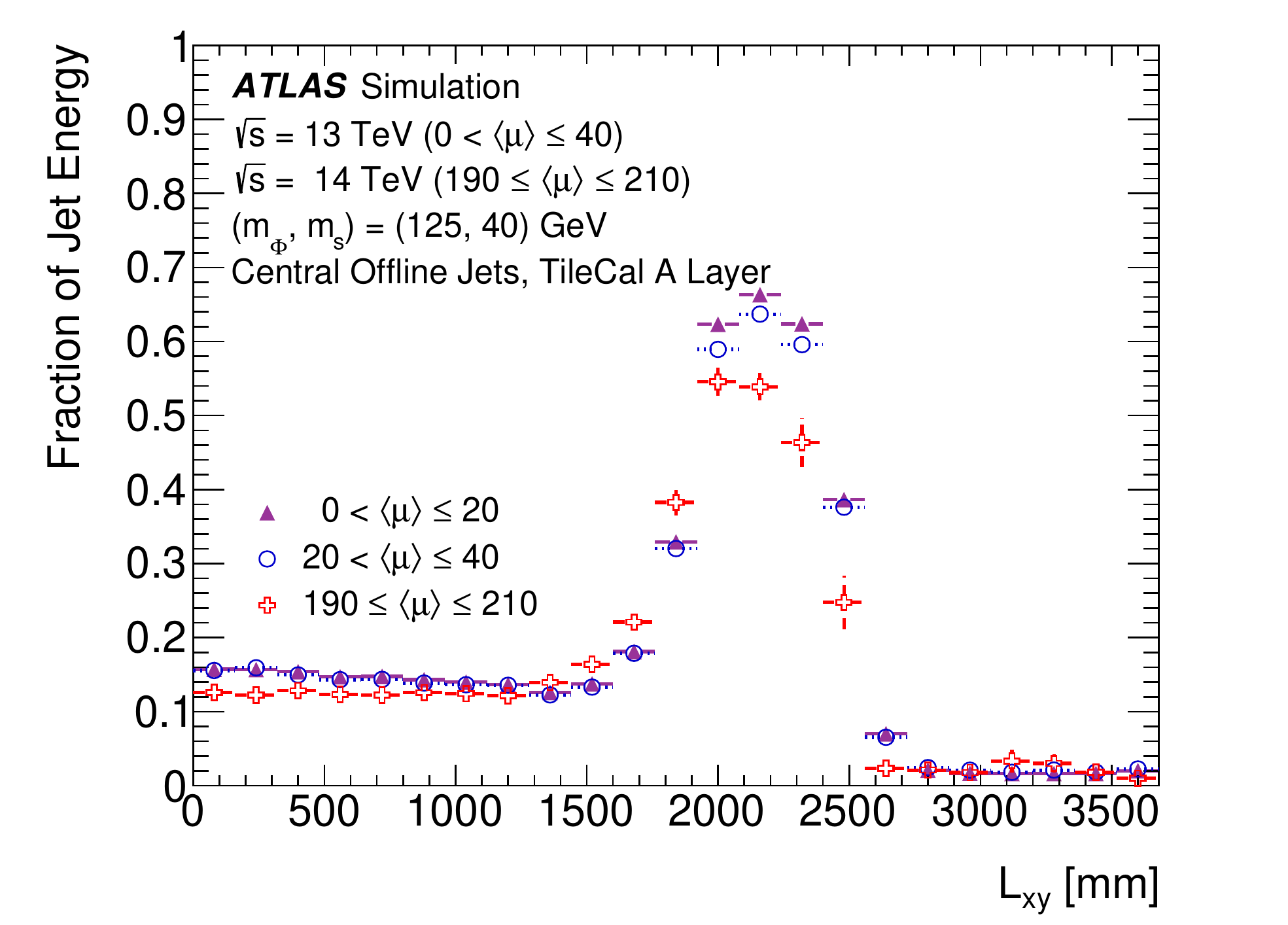}
\end{minipage}
\begin{minipage}[t][\minipageheightdp][t]{\minipagewidthdp}
\centering
\includegraphics[width=\figuremaxwidthdp,height=\figuremaxheightdp,keepaspectratio]{\main/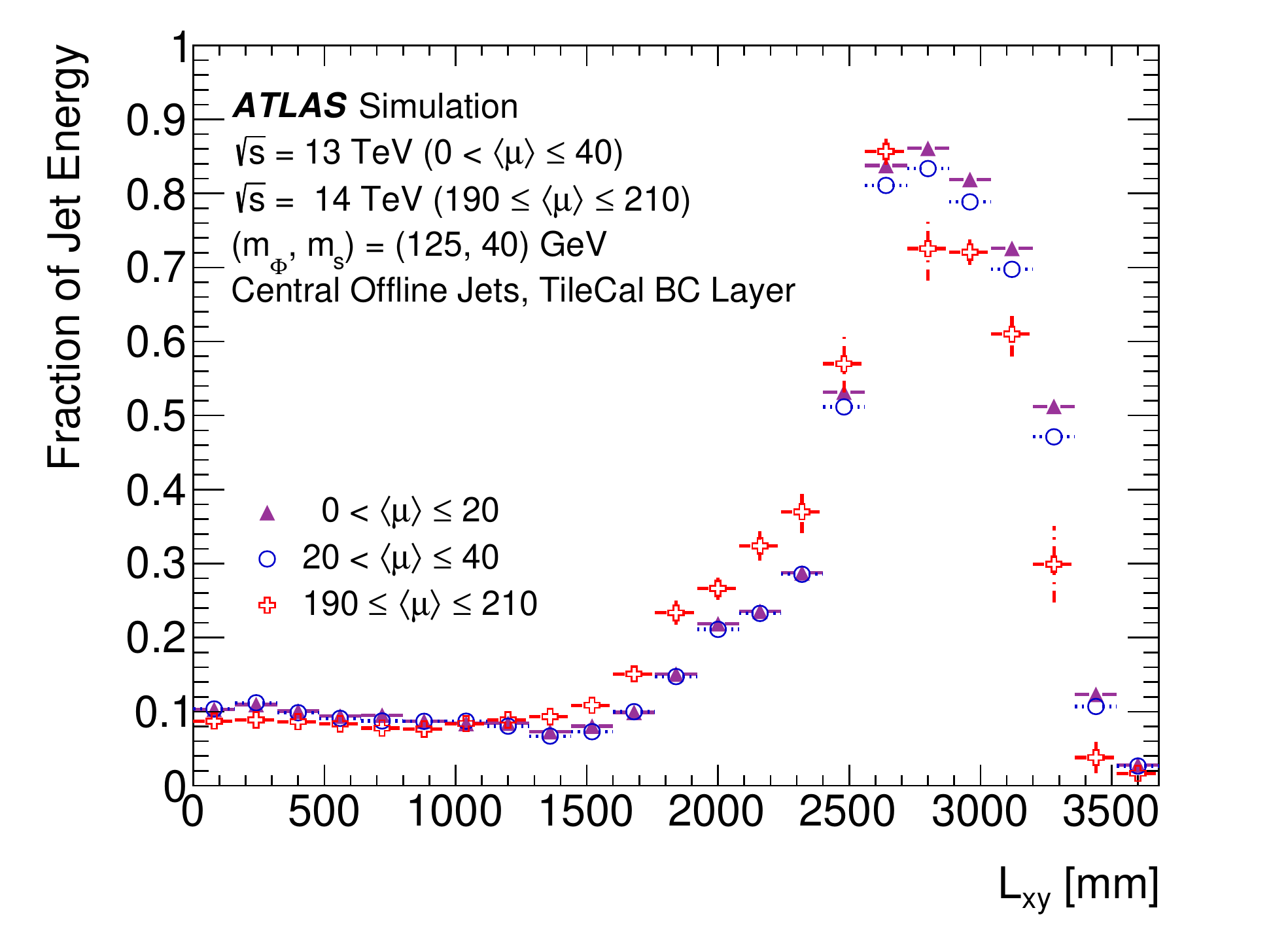}
\end{minipage}\\
\begin{minipage}[t][\minipageheightdp][t]{\minipagewidthdp}
\centering
\includegraphics[width=\figuremaxwidthdp,height=\figuremaxheightdp,keepaspectratio]{\main/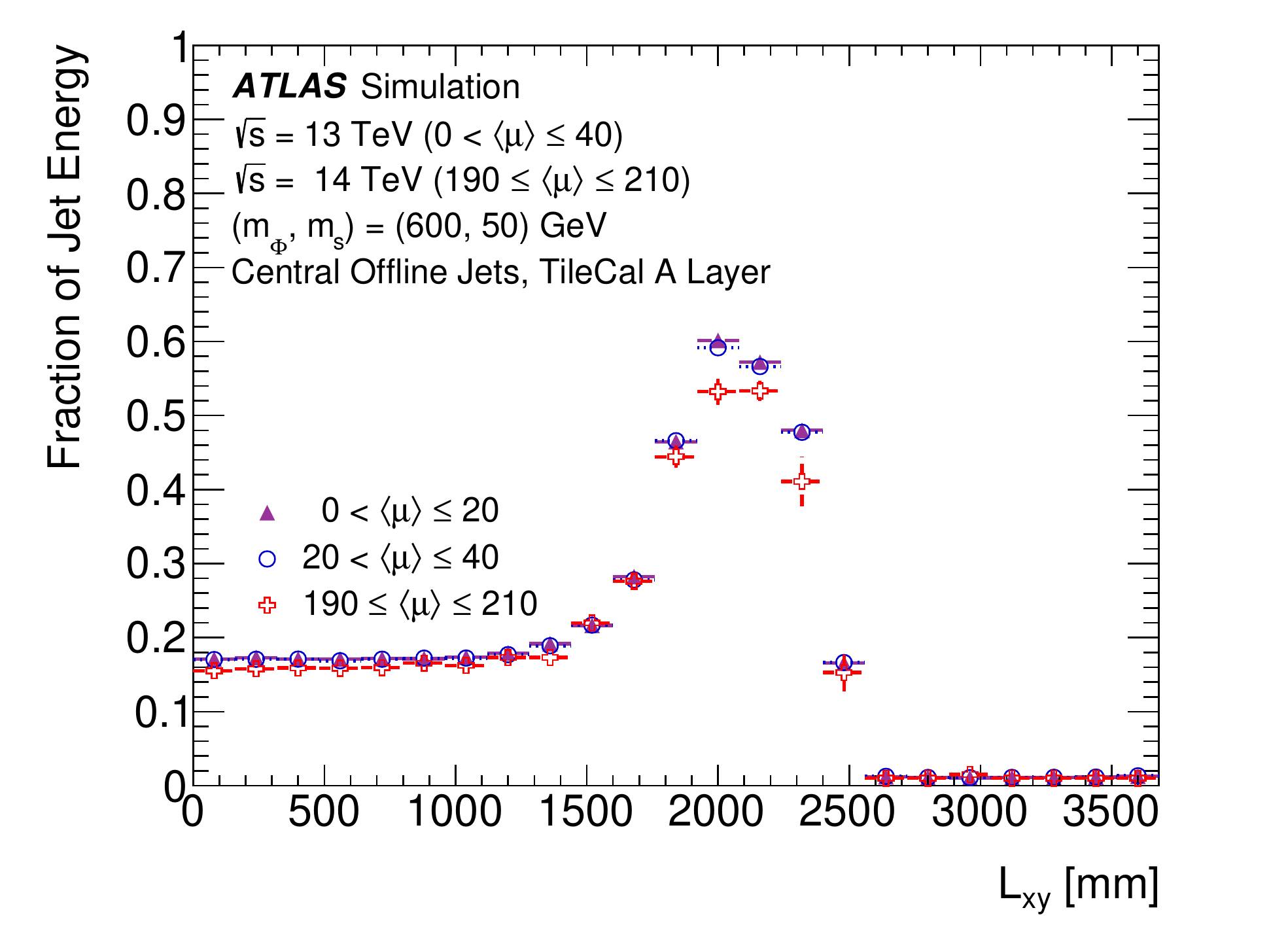}
\end{minipage}
\begin{minipage}[t][\minipageheightdp][t]{\minipagewidthdp}
\centering
\includegraphics[width=\figuremaxwidthdp,height=\figuremaxheightdp,keepaspectratio]{\main/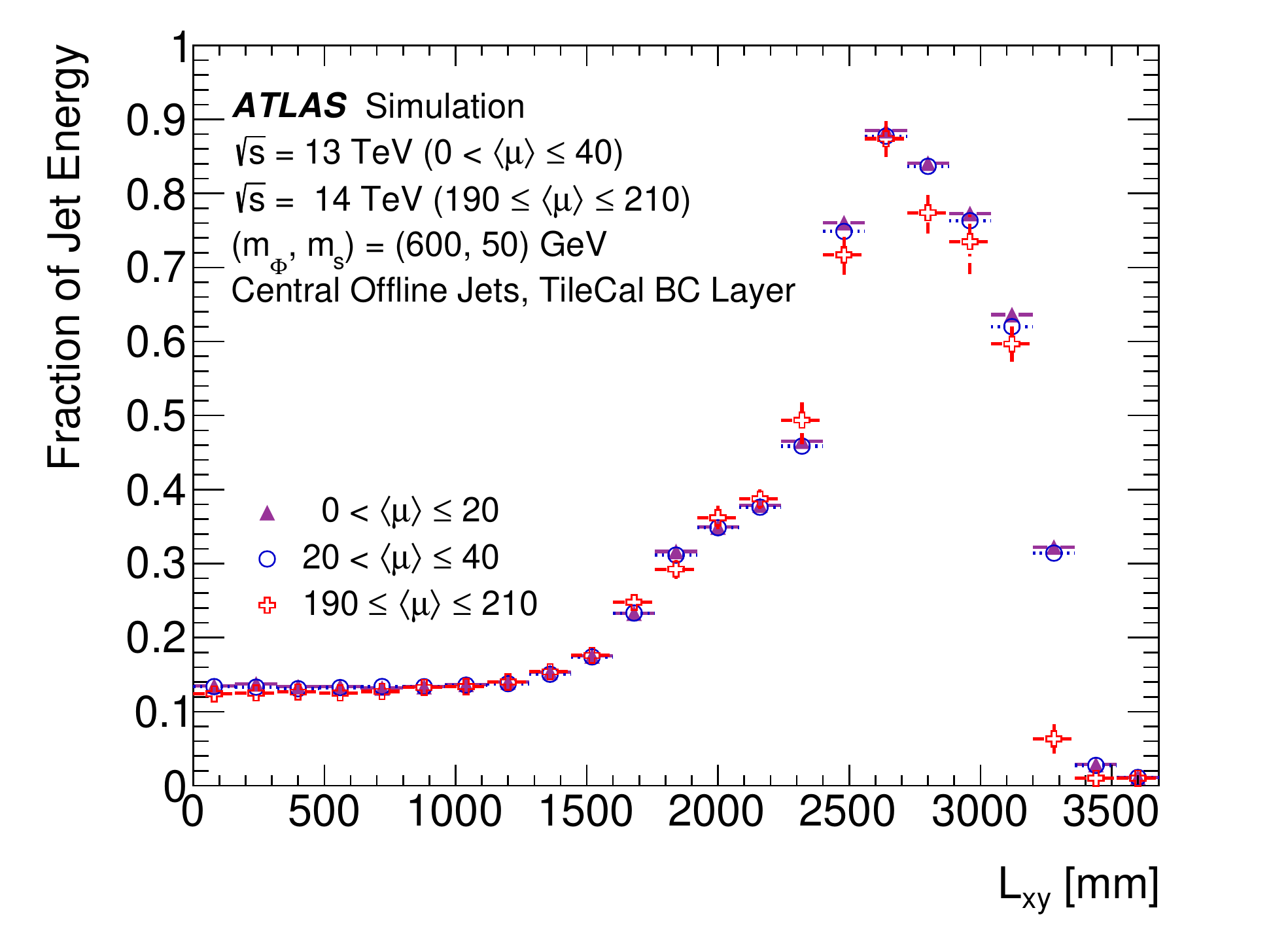}
\end{minipage}
\caption{Top: fraction of the jet energy deposited in A-layer (left) and BC-layer (right) of the Tile as a function of the transverse decay position of the LLP in events with a $125\GeV$ Higgs boson decaying to two $25\GeV$ LLPs. Bottom: same for events with a $600\GeV$ boson decaying to two $150\GeV$ LLPs. \label{fig:TileCalATLAS}}
\end{figure}

Jets resulting from neutral LLPs decaying within the hadronic calorimeter have several properties that are uncommon in jets originating at the interaction point. Within the inner detector, they naturally lack associated tracks. They likewise lack associated energy deposits in the electromagnetic calorimeter. Furthermore, the reconstructed jets are narrower than for a similar shower originating at the interaction point (IP) due to large displacement of the decay vertex. These properties, as well as those of the jet's constituents, can be used to discriminate between jets from displaced decays and those originating from the IP. On the other hand, the reconstructed jets are similar to those associated with non-collision backgrounds (NCB). Standard jet cleaning tools tend to reject jets resembling those associated with NCB, which is a primary reason the vast majority of non-LLP-dedicated jet searches will miss this signature, while the searches for an LLP would require a dedicated jet quality selection. 

The unusual signature of neutral LLP decays within the hadronic calorimeter encourages the use of dedicated triggers. At Level-1, the narrowness of the jets allows tau-candidates to be used to keep the energy threshold low while avoiding prescaling. In the higher level trigger, the low electromagnetic fraction and lack of pointing tracks can be further used to reject most jets. However, rates due to NCB, particularly beam-induced background (BIB), necessitate the use of a dedicated BIB-removal algorithm to keep rates acceptably low.

At Level-1, these dedicated triggers have benefited from the use of the Level-1 topological triggers. These have allowed for a cut on a rough estimate of the electromagnetic fraction to be applied at Level-1, allowing for the energy threshold to remain lower even as the lowest energy unprescaled Level-1 tau trigger gets pushed to higher and higher thresholds. Keeping this rate down at higher pile-up will be crucial to gathering high-statistics, high-purity samples for offline analysis. The increased longitudinal Level-1 granularity in Phase-II is especially promising for such a trigger. It may allow for a quick assessment of the energy deposited per layer in a jet, which has already been found to be a good discriminator offline for LLP jets.

Pile-up presents challenges for LLP searches at the analysis level as well. Soft energy deposits in the electromagnetic calorimeter in line with energy deposits from a neutral LLP result in jets with a higher than allowed fraction of their energy in the electromagnetic calorimeter. An alternative to using this coarse fraction is to consider the energy deposited per layer.

The model used in the generation of LLP signatures is a simplified hidden-sector toy model with a sector, containing particles neutral under the SM gauge group, weakly coupled to the SM sector. Interactions between sectors may occur via a communicator particle. 
Due to the weak coupling between sectors, the lifetime of these particles can be long. The process here is one in which a scalar boson, $\phi$, which is also the communicator, is produced during the $pp$ collision in ATLAS and decays to a pair of hidden sector particles $s$. Each LLP $s$, in turn, decays with long lifetimes via the communicator to heavy SM states. Heavy states are preferred due to the Yukawa coupling to the $\phi$ boson.

Figures~\ref{fig:TileCalATLAS} show some examples of the fraction of total jet energy at the EM-scale deposited by the LLPs produced within the $|\eta|<0.7$ rapidity range in the given layer for different slices of the average $\mu$ as a function of the LLP transverse decay position $L_{xy}$. For lighter pairs of LLPs, $s$ and their parent particle $\phi$, the effects of increasing pile-up are small, but possibly not negligible. This may motivate the introduction of multivariate analysis (MVA) technique for identifying jets consistent with displaced LLP decays. Such an MVA can be in the form of a regression that attempts to identify the decay position of the jet-initiation particle. It can be made directly pile-up-aware by including pile-up as one of its training variables. However, as long as the MVA is given the fraction of energy in the electromagnetic calorimeter, where most of the pile-up energy will be deposited, it is expected to be able to distinguish between jets initiated by decays at the same position under different pile-up conditions. 

%\subsection{Various interpretations}
%\subsubsection{\theory{Searching for Confining Hidden Valleys at the LHCb and ATLAS/CMS}}
%\contributors{A. Pierce et al.}
%{\bf Author(s): A. Pierce et al.}

\clearpage
\resetcounters

\section{High-$p_{T}$ signatures from flavour}\label{sec:flavor}

Flavour physics is often considered one of the most sensitive probes of new physics, which, depending on observable and model, can constrain new physics at scales of $1-10^5\TeV$.  Many of the observables are associated with precision measurements at the intensity frontier. This is well documented in the flavour chapter of this report, \citeref{MartinCamalich:2650175}. Any small deviations in precision low energy measurements, could signal the presence of new particles in the TeV scale, whose effect, once integrated out, is to alter precision low energy observables. Thus, in tandem with low energy flavour observables, it is important to identify the high-$p_{T}$ probes of flavour physics, relating direct tests with the indirect flavour probes.

Similar arguments hold for neutrino physics, where the generation of neutrino masses through the seesaw mechanism may generate a large variety of collider signatures, such as new resonances and cascade decays involving multiple leptons. 
In this section we first discuss the implication of neutrino mass models involving new heavy gauge bosons and sterile neutrinos, see \secc{sec:heavyneut}.  We then focus in \secc{sec:zprimeLQ} on constraints on models of $\Zp$ and leptoquarks related to $B$-decay flavour anomalies, and leptoquarks decaying to top and tau.  Finally, in \secc{sec:highpTflavanomalies}, we present a summary of the implications of these high-$p_{T}$ prospects on the parameter space of various UV models addressing the aforementioned anomalies. Notice that part of the material included in this section is also contained in the flavour chapter of this report \cite{MartinCamalich:2650175}.

%In this section 
%cite WG4 whole chapter. Introduce heavy neutrinos and sterile neutrinos. 
%Talk about complementarities (this is another case where it is relevant). 
%Say something about the possibility of explaining anomalies with Z' and LQ. 

\subsection{Neutrino masses}
\label{sec:heavyneut}
The potential Majorana nature of neutrinos as well as the origin of their tiny masses and large mixing angles remain some of the most pressing open issues in particle physics today. Models that address these mysteries, collective known as Seesaw models, hypothesise the presence of new particles that couple to SM fields via mixing/Yukawa couplings, SM gauge currents, and/or new gauge symmetries. The new predicted particles, often right-handed sterile neutrinos, can explain the generation of neutrino masses via a low-scale seesaw mechanism. Mixing between the active and sterile neutrinos is strongly constrained by precision measurements~\cite{Alonso:2012ji}. % to $|\theta|^2 < 10^{-5}$ and $|\theta|^2 <10^{-3}$ for $M < m_Z$ and $M> m_Z$, respectively. 
At the LHC, searches in purely leptonic final states have started probing masses below $m_Z$ \cite{Khachatryan:2016olu}. On the other hand, a plethora of rich physics can be studied in considerable detail at hadron colliders, and greatly complement low energy and oscillation probes of neutrinos~\cite{Atre:2009rg,Cai:2017mow}. Other complementary and promising searches for the low-scale seesaw neutrinos could be achievable also at future $ep$ colliders studying lepton-flavour violating processes, \sloppy\mbox{\egg~$e^- p \to \mu^- W+ j$}. %In this case, the models can be tested for $|\theta_e|^2 \sim 10^{-6}$ for $M > m_Z$, almost up to the threshold of $M\sim \sqrt{s}$. 
In this case, the leading production of heavy neutrinos depends on the mixing with the electron flavour, in contrast to production in $pp$ collisions, where it is proportional to the total mixing. This allows us to infer the relative strength of the mixings, especially if a hierarchy is present.
%Also searches for displaced vertices are very promising, which can test $|\theta|^2 > 10^{-8}$ for $M < m_W$ and have very little background. 
For more details, including comparison with $ee$ and $pp$ colliders, see~\citeref{Antusch:2016ejd}. 

In the following, a comprehensive summary of the discovery potential of Seesaw models at hadron colliders with collision energies of $\sqrt{s} = 14$ and $27\TeV$ is presented in~\secc{sec:neutrino551}. Heavy composite Majorana neutrinos are studied in~\secc{sec:neutrino552} using same-sign leptons signatures, and using dilepton and jets signatures in~\secc{sec:neutrino553}. 

\subsubsection{\theoryC{Neutrino mass models at the HL- and HE-LHC}}
\label{sec:neutrino551}
\contributors{T. Han, T. Li, X. Marcano, S. Pascoli, R. Ruiz, C. Weiland}

We summarise the discovery potential of seesaw models at hadron colliders with collision energies of $\sqrt{s} = 14$ and $27\TeV$. In particular, we will discuss models featuring heavy neutrinos, both pseudo-Dirac and Majorana as well as those with new gauge interaction, and models featuring scalar and fermion EW triplets.
%In Sec.~\ref{sec:lhcSeesawtype1}, models featuring heavy neutrinos, both pseudo-Dirac and Majorana as well as those with new gauge interactions, are discussed.
%Similarly, in Sec.~\ref{sec:lhcSeesawtype2} and Sec.~\ref{sec:lhcSeesawtype3}, models featuring scalar and fermion EW triplets are assessed, respectively.
For a more comprehensive reviews on the sensitivity of colliders to neutrino mass models, see \citerefs{Atre:2009rg,Cai:2017mow,Weiland:2013wha,Ruiz:2015gsa,Marcano:2017ucg} and references therein.

%%%%%%%%%%%%%%%%%%%%%%%%%%%%%%%%%%%%%%%%%%%%%%%%%%%%%%%%%%%%%%%%%%
%\subsubsection{The Type I Seesaw and Variants Discovery Potential at the HL- and HE-LHC}\label{sec:lhcSeesawtype1}
\paragraph*{The Type I seesaw and Variants Discovery Potential at the HL- and HE-LHC}\label{sec:lhcSeesawtype1}
%%%%%%%%%%%%%%%%%%%%%%%%%%%%%%%%%%%%%%%%%%%%%%%%%%%%%%%%%%%%%%%%
The Type I seesaw hypothesises the existence of fermionic gauge singlets with a Majorana mass term whose couplings 
to SM leptons generate light neutrino masses and mixing. 
In \citerefs{Kersten:2007vk,Moffat:2017feq}, it was proved that requiring all three light neutrinos to be massless is equivalent 
to the conservation of lepton number at all orders in perturbation theory. 
In other words, for low-scale seesaw models with only fermionic singlets, 
lepton number has to be nearly conserved and light neutrino masses are proportional to small lepton number violation (LNV) parameters in these models. 
For high-scale seesaws  with only fermionic singlets, light neutrino masses are inversely proportional to large LNV mass scales,
and again leads to approximate lepton number conservation at low energies.
This in turn leads to the expectation that LNV processes should be suppressed in variants of the type I seesaw unless additional particles,
whether they be fermions or scalars charged under the SM gauge couplings or 
new gauge interactions,
are introduced to decouple the light neutrino mass generation from the heavy neutrino production.
The updated discovery potential of heavy, SM singlet neutrinos at $pp$ colliders is now summarised.

%%%%%%%%%%%%%%%%%%%%%%%%%%%%%%%%%%%%%%%%%%%%%%%%%%%%%%%%%%%%%%%%%%
\paragraph*{Heavy Neutrino Production through EW Bosons at Hadron Colliders}

%-----------------------------------------
\begin{figure}[t]
\centering
\begin{minipage}[t][\minipageheightsp/2][t]{\minipagewidthsp}
\centering
\includegraphics[width=\figuremaxwidthsp*4/3,height=\figuremaxheightsp,keepaspectratio]{\main/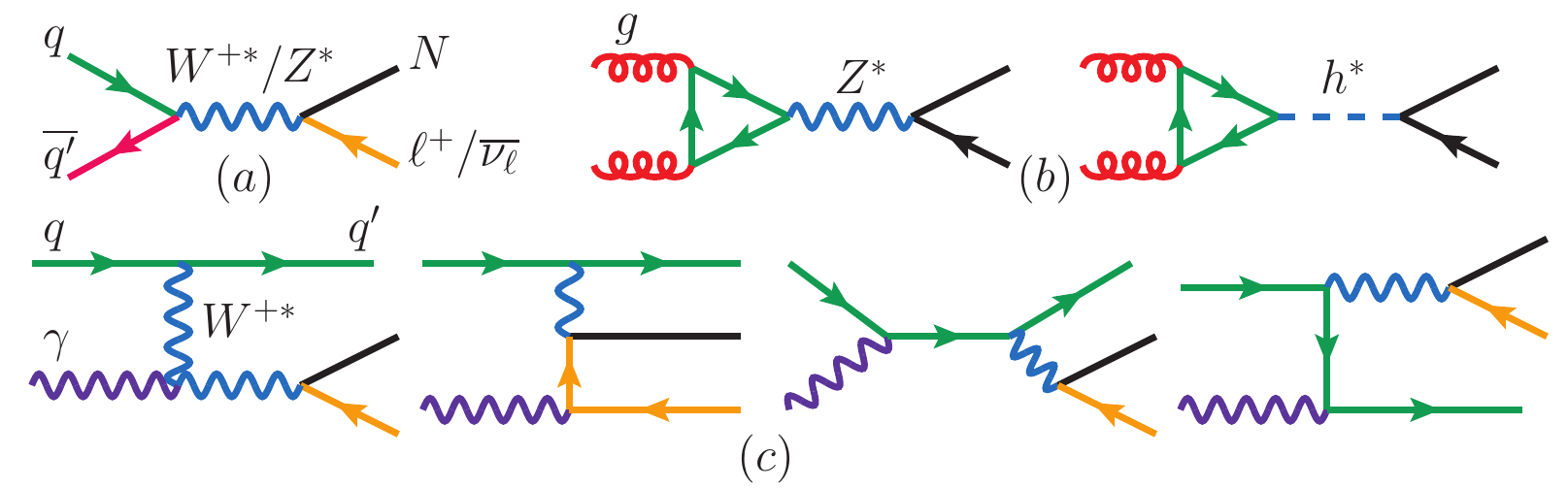}
\end{minipage}\\
\begin{minipage}[t][\minipageheightdp][t]{\minipagewidthdp}
\centering
\includegraphics[width=\figuremaxwidthdp,height=\figuremaxheightdp,keepaspectratio]{\main/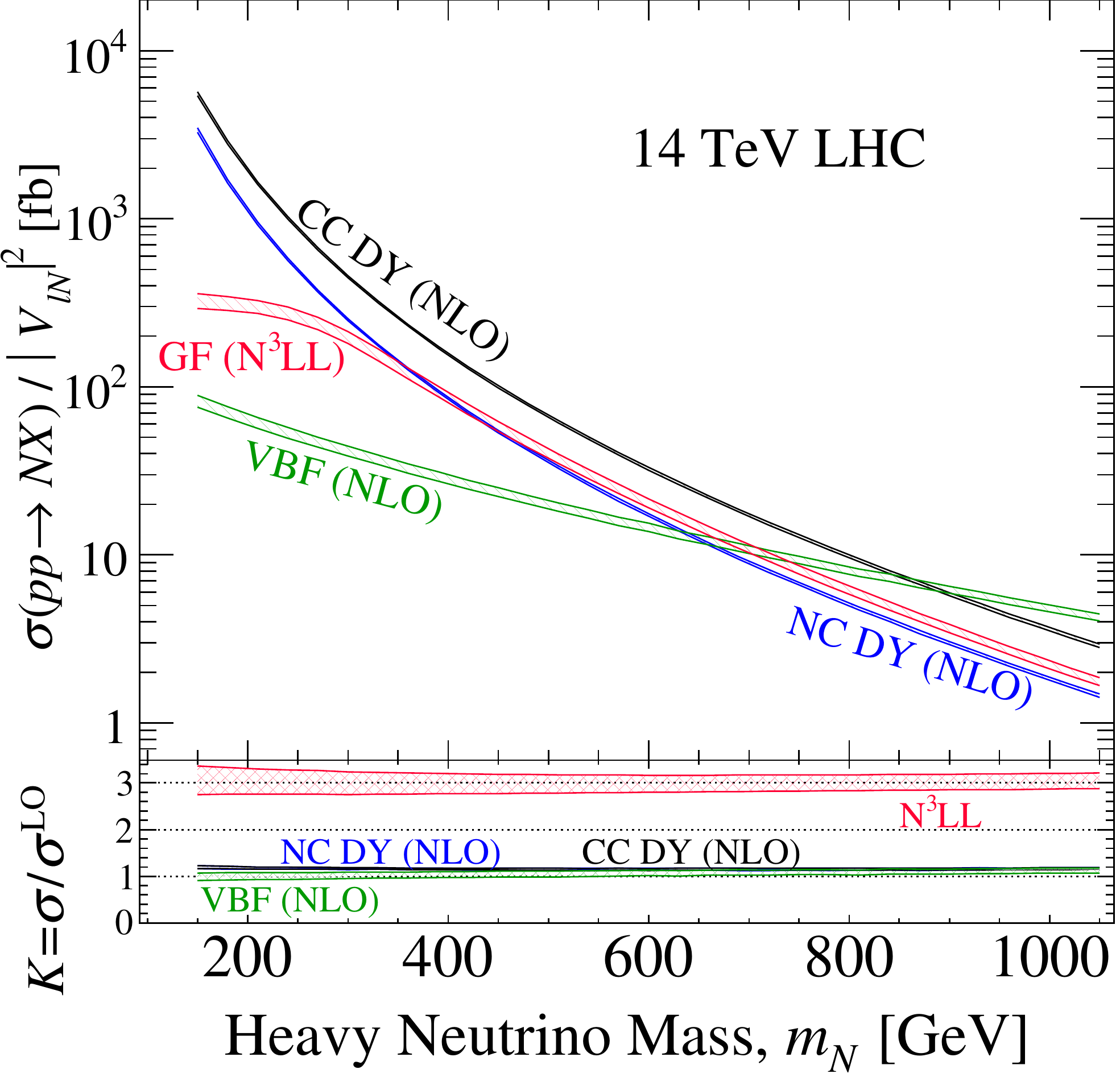}
\end{minipage}
\begin{minipage}[t][\minipageheightdp][t]{\minipagewidthdp}
\centering
\includegraphics[width=\figuremaxwidthdp,height=\figuremaxheightdp,keepaspectratio]{\main/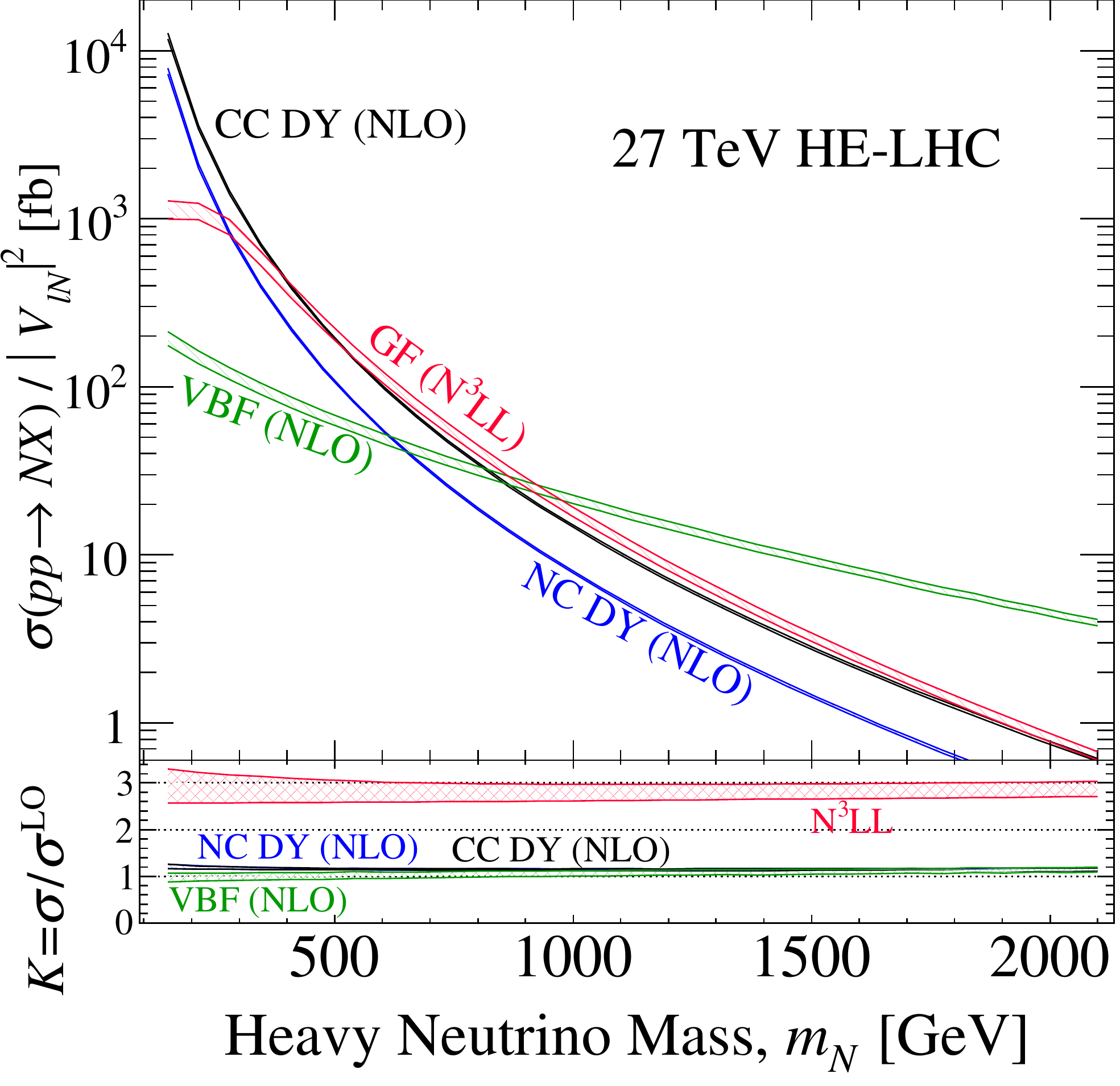}
\end{minipage}
\caption{
Top:
Born-level diagrams for heavy neutrino $N$ production via (a) Drell-Yan, (b) gluon fusion, and (c) vector boson fusion; 
from \citeref{Ruiz:2017yyf}.
Bottom:
production cross sections, divided by active-heavy mixing $\vert V_{\ell N}\vert^2$, as a function of the heavy neutrino mass
at $\sqrt{s}=14\TeV$ (left) and $27\TeV$ (right) \cite{Degrande:2016aje,Ruiz:2017yyf}.
}
\label{fig:lhcSeesaw_HeavyNProd}
\end{figure}
%-------------------------------------------------------------------

If kinematically accessible, heavy neutrinos $N$ can generically be produced in hadron collisions through both neutral current and charged current processes,
as shown in \fig{fig:lhcSeesaw_HeavyNProd} (top).
Following the prescriptions of \citerefs{Degrande:2016aje,Ruiz:2017yyf}, 
for $m_N >M_W$ and at various accuracies,
the corresponding $\sqrt{s}=14\TeV$ heavy neutrino production cross sections are presented in \fig{fig:lhcSeesaw_HeavyNProd} (bottom left).
While the Drell-Yan (DY) process dominates at low masses, 
$W\gamma$ boson fusion (VBF) dominates for $m_N\gtrsim 900-1000\GeV$~\cite{Alva:2014gxa,Ruiz:2017yyf},
with gluon fusion (GF) remaining a sub-leading channel throughout~\cite{Hessler:2014ssa,Ruiz:2017yyf}.
The situation is quite different at $27\TeV$ as shown in \fig{fig:lhcSeesaw_HeavyNProd} (bottom right). 
Indeed, GF is the leading production mode for $m_N\gtrsim 450\GeV$ until $m_N\approx 940\GeV$ where VBF takes over.
For $m_N\approx 1\TeV$, the GF, DY, and VBF mechanisms all possess fiducial cross sections in excess of $10\fb$.

%%%%%%%%%%%%%%%%%%%%%%%%%%%%%%%%%%%%%%%%%%%%%%%%%%%%%%%%%%%%%%%%%%
\paragraph*{Discovery Potential of Heavy Pseudo-Dirac Neutrinos in Low Scale Seesaws}

%-------------------------------------------------------------------
\begin{figure}[t]
\centering
\begin{minipage}[t][\minipageheightdp][t]{\minipagewidthdp}
\centering
\includegraphics[width=\figuremaxwidthdp,height=\figuremaxheightdp,keepaspectratio]{\main/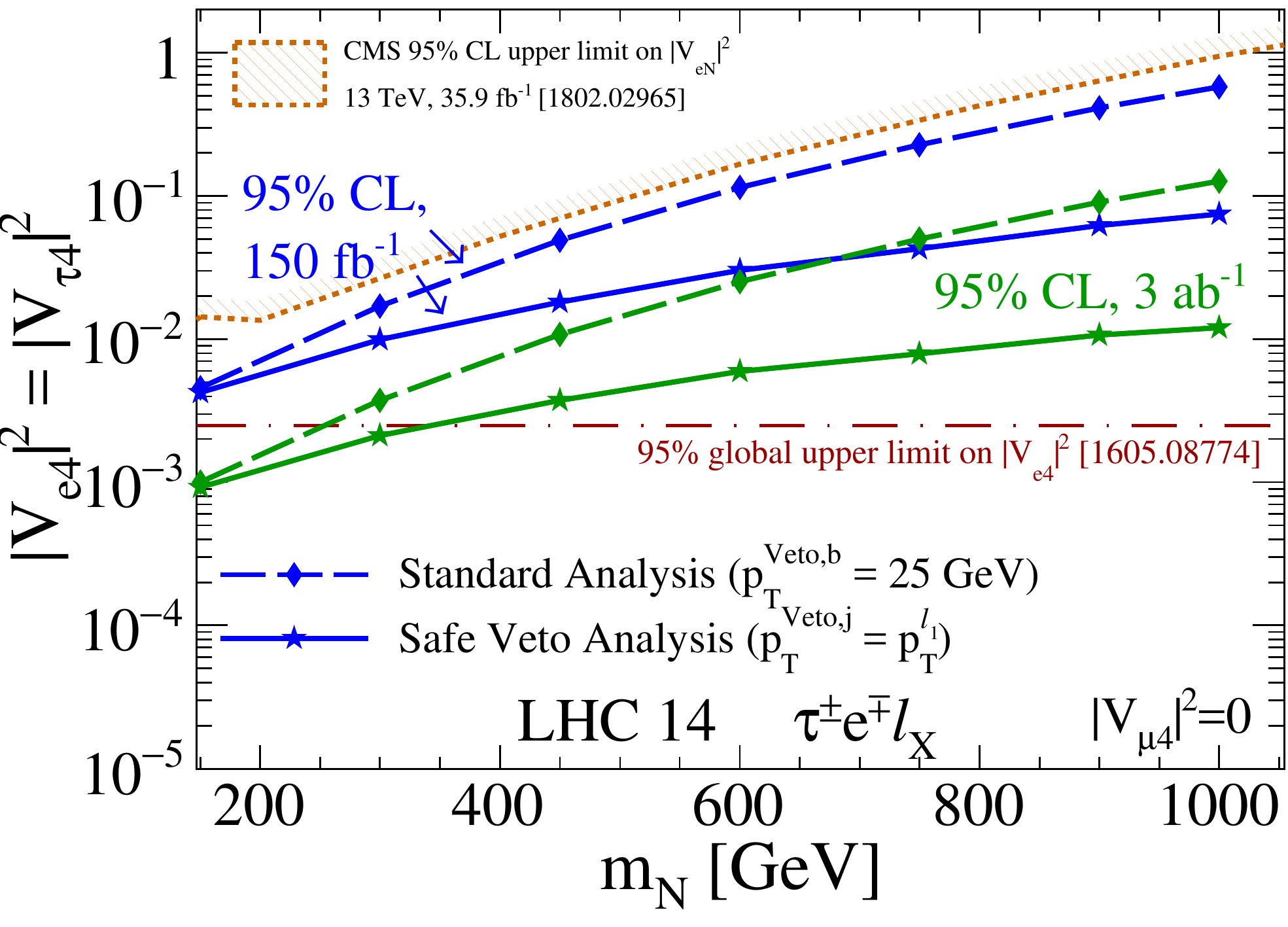}
\end{minipage}
\begin{minipage}[t][\minipageheightdp][t]{\minipagewidthdp}
\centering
\includegraphics[width=\figuremaxwidthdp,height=\figuremaxheightdp,keepaspectratio]{\main/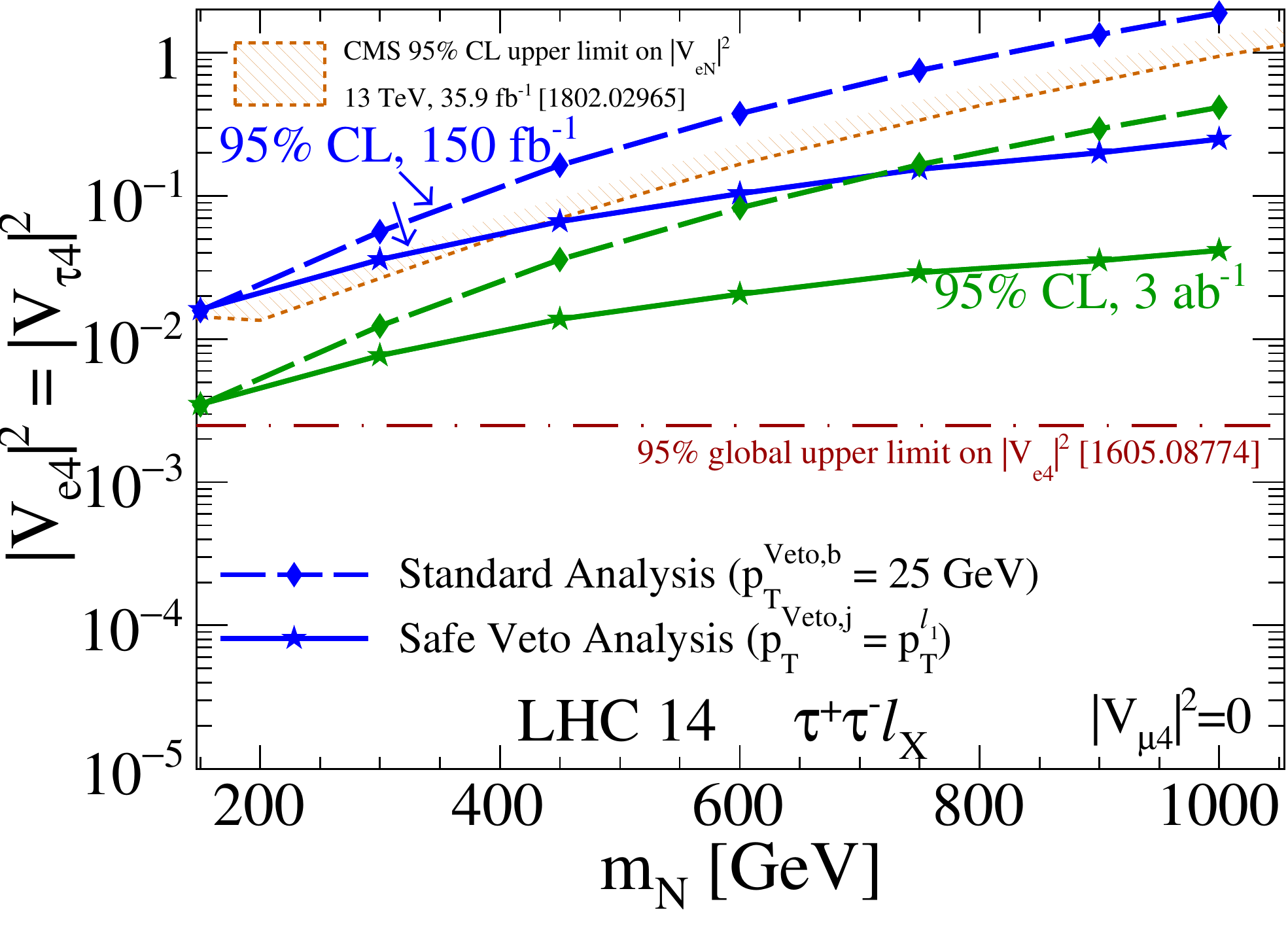}
\end{minipage}\\
\begin{minipage}[t][\minipageheightdp][t]{\minipagewidthdp}
\centering
\includegraphics[width=\figuremaxwidthdp,height=\figuremaxheightdp,keepaspectratio]{\main/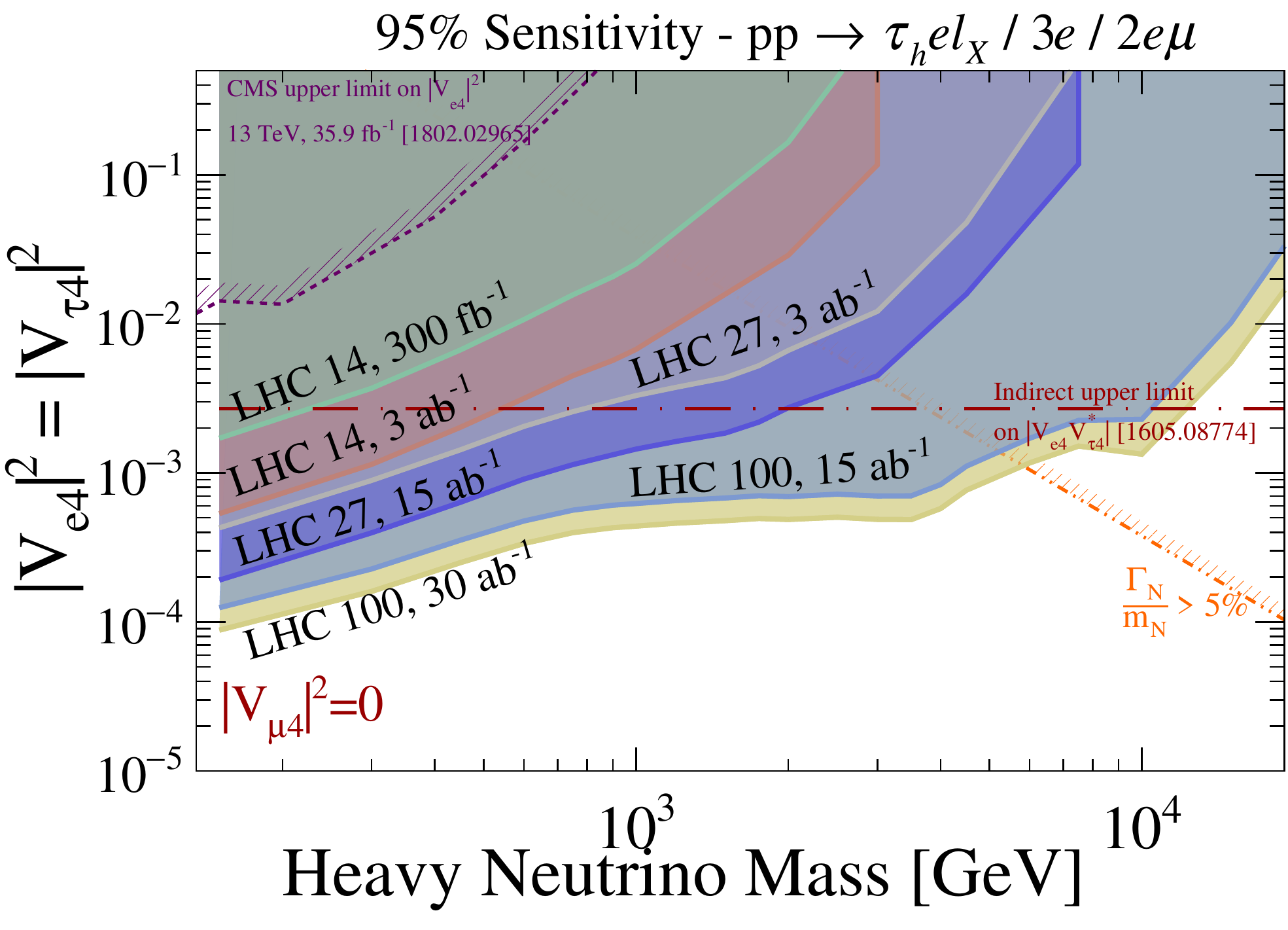}
\end{minipage}
\begin{minipage}[t][\minipageheightdp][t]{\minipagewidthdp}
\centering
\includegraphics[width=\figuremaxwidthdp,height=\figuremaxheightdp,keepaspectratio]{\main/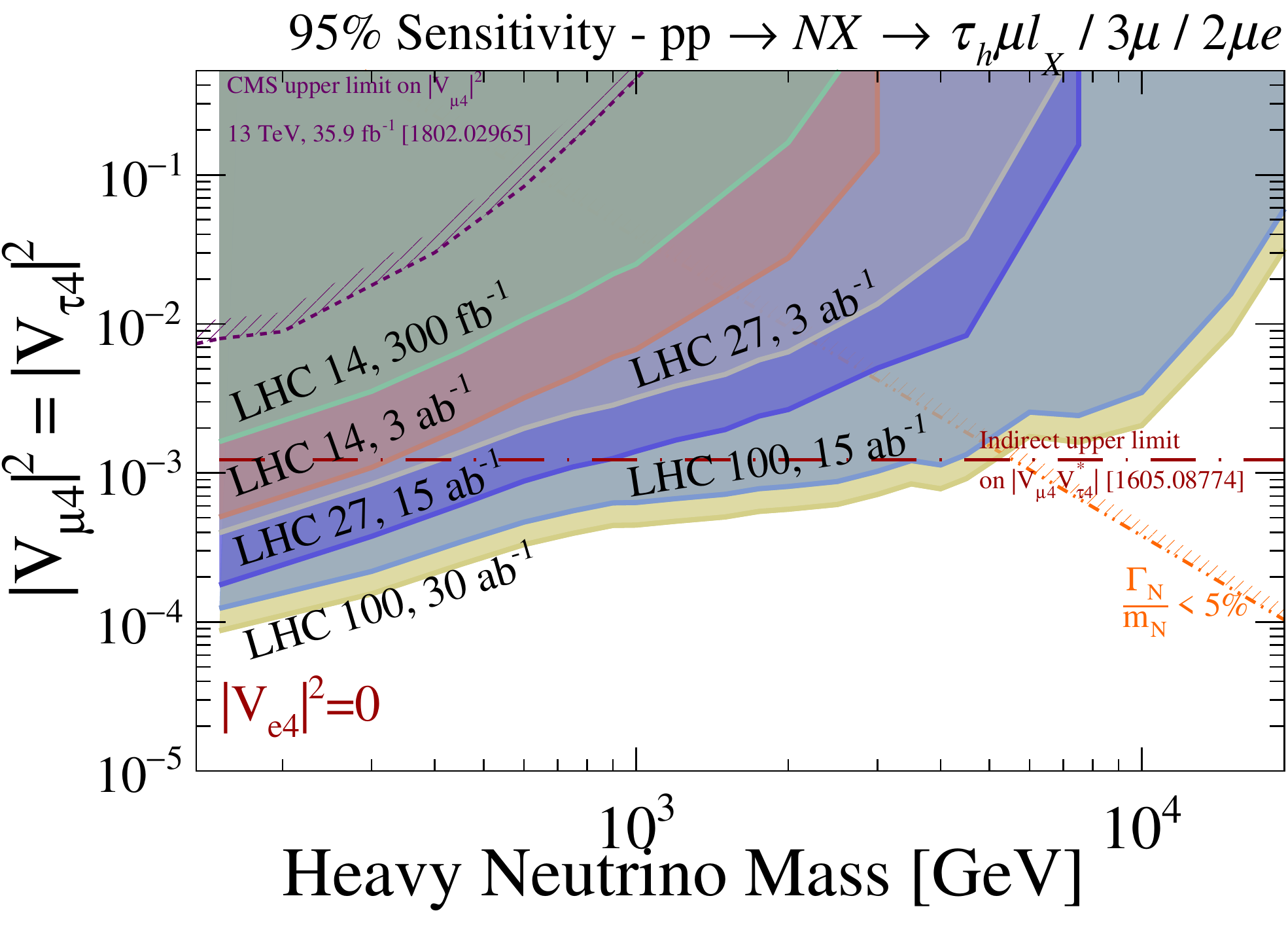}
\end{minipage}
%
%\includegraphics[width=0.49\textwidth]{}	\label{fig:jetvetotael}
%\includegraphics[width=0.49\textwidth]{}	\label{fig:jetvetotata}
%\\
%\includegraphics[width=0.49\textwidth]{}	\label{fig:MuTaujj_14TeV}
%\includegraphics[width=0.49\textwidth]{}	\label{fig:MuTaujj_27TeV}
%\end{center}
\caption{
Top:
sensitivity to the active-heavy mixing $\vert V_{\ell N}\vert^2$ as a function of the heavy neutrino mass $m_N$ in the 
trilepton final states 
(left) $\tau_h^\pm e^\mp \ell_X + \textrm{MET}$ and 
(right) $\tau_h^+ \tau_h^- \ell_X + \textrm{MET}$, 
assuming $\vert V_{e4}\vert^2 = \vert V_{\tau4}\vert^2$ and $\vert V_{\mu4}\vert^2=0$, at the $\sqrt{s}=14\TeV$ LHC.
The dash-diamond line corresponds to the standard analysis with a b-jet veto  while the solid-star line is the jet veto-based analysis~\cite{Pascoli:2018rsg,Pascoli:2018heg}.
Bottom:
%Number of $\mu^\pm \tau^\mp jj$ events as a function of the seesaw scale $M_R$ for representative scaling factors f in the ISS 
%(L) at the HL-LHC with $3\,\mathrm{ab}^{-1}$
%and 
%(R) at the HE-LHC with $3\,\mathrm{ab}^{-1}$.
%Triangles respect constraints from LFV radiative decays while crosses do not respect the $\tau \rightarrow \mu \gamma$ BaBar upper limit~\cite{Aubert:2009ag}.
for the benchmark mixing hypotheses
(left) $\vert V_{e4}\vert = \vert V_{\tau4} \vert$ with $\vert V_{\mu4}\vert = 0$ and
(right) $\vert V_{\mu4}\vert = \vert V_{\tau4} \vert$ with $\vert V_{e4}\vert = 0$,
the projected sensitivity at $\sqrt{s}=27\TeV$ and $100\TeV$ using the trilepton analysis of \citeref{Pascoli:2018rsg}.
}
\label{fig:lhcSeesaw_ISS}
\end{figure}
%-------------------------------------------------------------------

The expected suppression of LNV processes in models that contain only fermionic gauge singlets
motivates the study of lepton number conserving (LNC) processes. 
A first possibility to consider is the trilepton final state $\ell^\pm_i \ell^\mp_j \ell^\pm_k + \textrm{MET}$.
The heavy neutrino $N$ here is produced via both charged-current DY and VBF, and subsequently decays to only leptons, \iee
\begin{equation}
 p p \to \ell_N N +X \to \ell_N \ell_W W +X \to \ell_N \ell_W \ell_\nu \nu +X\,.
 \label{eq:pp3lX}
\end{equation}
Notably, a new search strategy was recently proposed in \citeref{Pascoli:2018rsg,Pascoli:2018heg} based on a dynamical jet veto selection cut
and found an increased sensitivity to active-heavy neutrino mixing by approximately an order of magnitude over the LHC's life.
This is shown in \fig{fig:lhcSeesaw_ISS} (top) specifically for the final states 
(L) $\tau_h^\pm e^\mp \ell_X + \textrm{MET}$ and
(R) $\tau_h^+ \tau_h^- \ell_X + \textrm{MET}$,
where $\tau_h$ represents a hadronically decaying $\tau$ and $\ell_X\in\{e,\mu,\tau_h\}$.
With $3\,\mathrm{ab}^{-1}$ and after taking into account global constraints on the active-heavy mixing~\cite{Fernandez-Martinez:2016lgt} 
(dot-dashed line in \fig{fig:lhcSeesaw_ISS} (top)), the HL-LHC is able to probe heavy neutrino masses up to $350\GeV$ 
and mixing down to $\vert V_{\ell N}\vert^2\simeq 10^{-3}$ could be probed.  For the benchmark mixing hypotheses 
(L) $\vert V_{e4}\vert = \vert V_{\tau4} \vert$ with $\vert V_{\mu4}\vert = 0$ and
(R) $\vert V_{\mu4}\vert = \vert V_{\tau4} \vert$ with $\vert V_{e4}\vert = 0$,
~\fig{fig:lhcSeesaw_ISS} (bottom) shows the projected sensitivity of this analysis at $\sqrt{s}=27\TeV$ and $100\TeV$ using the trilepton analysis of \citeref{Pascoli:2018rsg}.

Another possibility is to search for lepton flavour violating (LFV) final states such as
\begin{equation}
 q ~\overline{q'} \rightarrow  ~N ~\ell^\pm_1 \rightarrow ~\ell^\pm_1 ~\ell^\mp_2 ~ W^\mp  \rightarrow ~\ell^\pm_1 ~\ell^\mp_2 ~j ~j\,,
 \label{eq:LFVfinalstate}
\end{equation}
which was for example studied in the inverse seesaw (ISS), a low-scale variant of the type I, in \citeref{Arganda:2015ija}. 
Due to the strong experimental limits on $\mu\to e\gamma$ by
MEG~\cite{Adam:2013mnn}, the event rates involving taus are more promising
than those for $e^\pm \mu^\mp jj$.
Following~\citeref{Arganda:2015ija}, the number of $\tau^\pm \mu^\mp jj$ can
be estimated using the $\mu_X$-parametrisation~\cite{Arganda:2014dta}
with the neutrino Yukawa coupling
\begin{equation}
 Y_{\nu}=f \left(\begin{array}{ccc}
-1&1&0\\
1&1&0.9\\
1&1&1
\end{array}\right)\,.
\label{eq:Ytmmax1}
\end{equation}
as a representative example, and considering that only the lightest
pseudo-Dirac pair is kinematically available.
%The number of events for this channel 
%is shown in \fig{fig:lhcSeesaw_ISS}
%(L) and (R) for $\sqrt{s} = 14$ and 27~TeV, respectively, although similar
%numbers would be expected for $\tau^\pm e^\mp jj$.
%After  $\mathcal L=3~\rm{ab}^{-1}$ of data taking, more than 100 LFV
%events of this type could be produced for neutrino masses below 700 (1000)
%GeV for $pp$ collisions at 14 (27)~TeV. \pf{Why was lumi for HE-LHC only \intLumi, not \intlumihelhc?}
After $\mathcal{L} = 3(15)\abinv$, more than 100(500) LFV $\tau^\pm e^\mp j j$ events can be produced for heavy neutrino masses below $700$ ($1000$)\GeV, for $pp$ collisions at $\sqrt{s}=14(27)\TeV$.

%%%%%%%%%%%%%%%%%%%%%%%%%%%%%%%%%%%%%%%%%%%%%%%%%%%%%%%%%%%%%%%%%%
\paragraph*{Discovery Potential of Heavy Majorana Neutrinos in Phenomenological Type I Seesaw}

%-------------------------------------------------------------------
\begin{figure}[t]
\centering
\begin{minipage}[t][\minipageheightdp][t]{\minipagewidthdp}
\centering
\includegraphics[width=\figuremaxwidthdp,height=\figuremaxheightdp,keepaspectratio]{\main/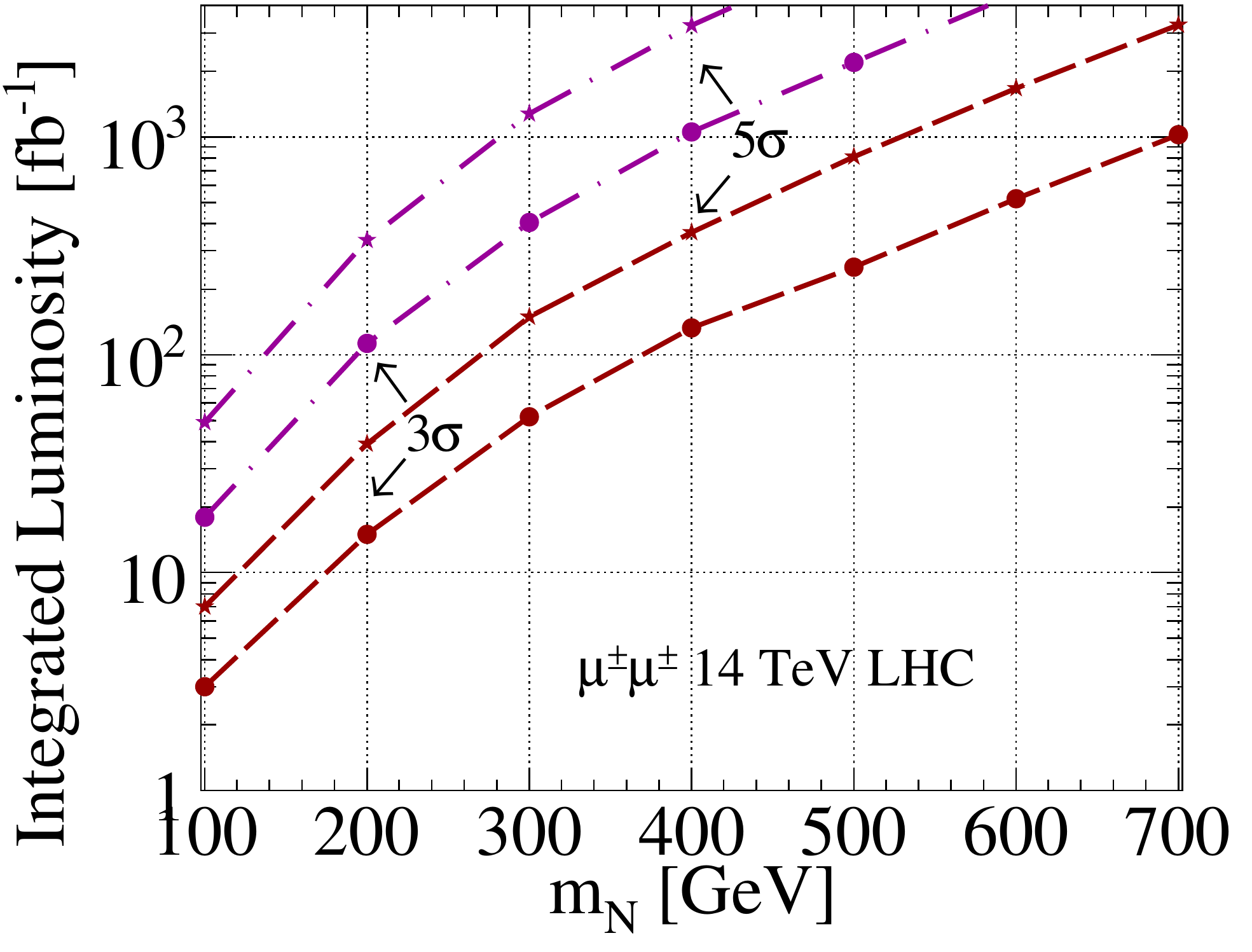}
\end{minipage}
\begin{minipage}[t][\minipageheightdp][t]{\minipagewidthdp}
\centering
\includegraphics[width=\figuremaxwidthdp,height=\figuremaxheightdp,keepaspectratio]{\main/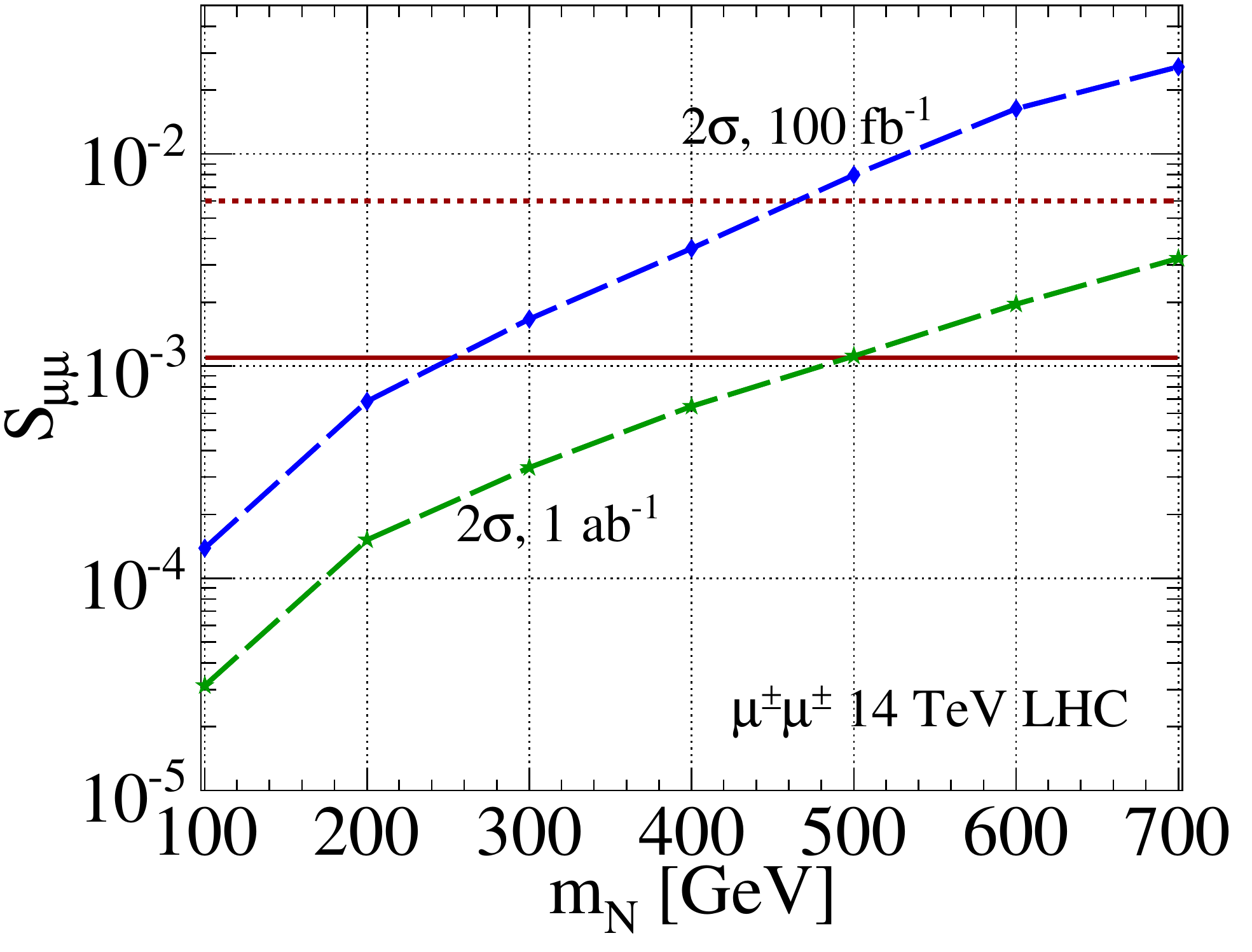}
\end{minipage}
\caption{
Left: using the LNV final state $\mu^\pm \mu^\pm jj$, required luminosity for $3(5) \sigma$ evidence (discovery) as a function of heavy neutrino mass $m_N$ assuming optimistic (brown) 
and pessimistic (purple) mixing scenarios~\cite{Alva:2014gxa}.  Right: sensitivity to $N-\mu$ mixing~\cite{Alva:2014gxa} with the optimistic (pessimistic) mixing scenario is given by the horizontal dashed (full) line.
}
\label{fig:SeesawIDiscoveryPotential}
\end{figure}
%-------------------------------------------------------------------

In the presence of additional particles that can decouple the heavy neutrino production 
from the light neutrino mass generation, \egg, new but far off-shell gauge bosons~\cite{Ruiz:2017nip},
the Majorana nature of the heavy neutrinos can leads to striking LNV collider signatures,
such as the well-studied same-sign dilepton and jets process~\cite{Keung:1983uu}
\begin{equation}
 pp \rightarrow  ~N ~\ell^\pm_1 \rightarrow ~\ell^\pm_1 ~\ell^\pm_2 ~ W^\mp  \rightarrow ~\ell^\pm_1 ~\ell^\pm_2 +nj.
 \label{eq:LNVfinalstate}
\end{equation}
Assuming that a low-scale type I seesaw is responsible for the heavy neutrino production, 
\fig{fig:SeesawIDiscoveryPotential} displays the discovery potential and active-heavy
mixing sensitivity of the $\mu^\pm\mu^\pm$ channel~\cite{Alva:2014gxa}.
Assuming the (pessimistic/conservative) mixing scenario of $S_{\mu\mu} = 1.1\times10^{-3}$~\cite{Alva:2014gxa} the HL-LHC with $3\abinv$ 
is able to discover a heavy neutrino with a mass of $m_N\simeq400\GeV$ 
and is sensitive to masses up to $550\GeV$ at $3\sigma$. 
Using only $1\abinv$, the HL-LHC can improve on the preexisting mixing constraints summarised in the pessimistic scenario 
for neutrino masses up to $500\GeV$.

%%%%%%%%%%%%%%%%%%%%%%%%%%%%%%%%%%%%%%%%%%%%%%%%%%%%%%%%%%%%%%%%%%
\paragraph*{Heavy Neutrinos and the Left-Right Symmetric Model at Colliders}\label{sec:lhcSeesawlrsmCollider}

%-------------------------------------------------------------------
\begin{figure}[!t]
\centering
\begin{minipage}[t][\minipageheightdp][t]{\minipagewidthdp}
\centering
\includegraphics[width=\figuremaxwidthdp,height=\figuremaxheightdp,keepaspectratio]{\main/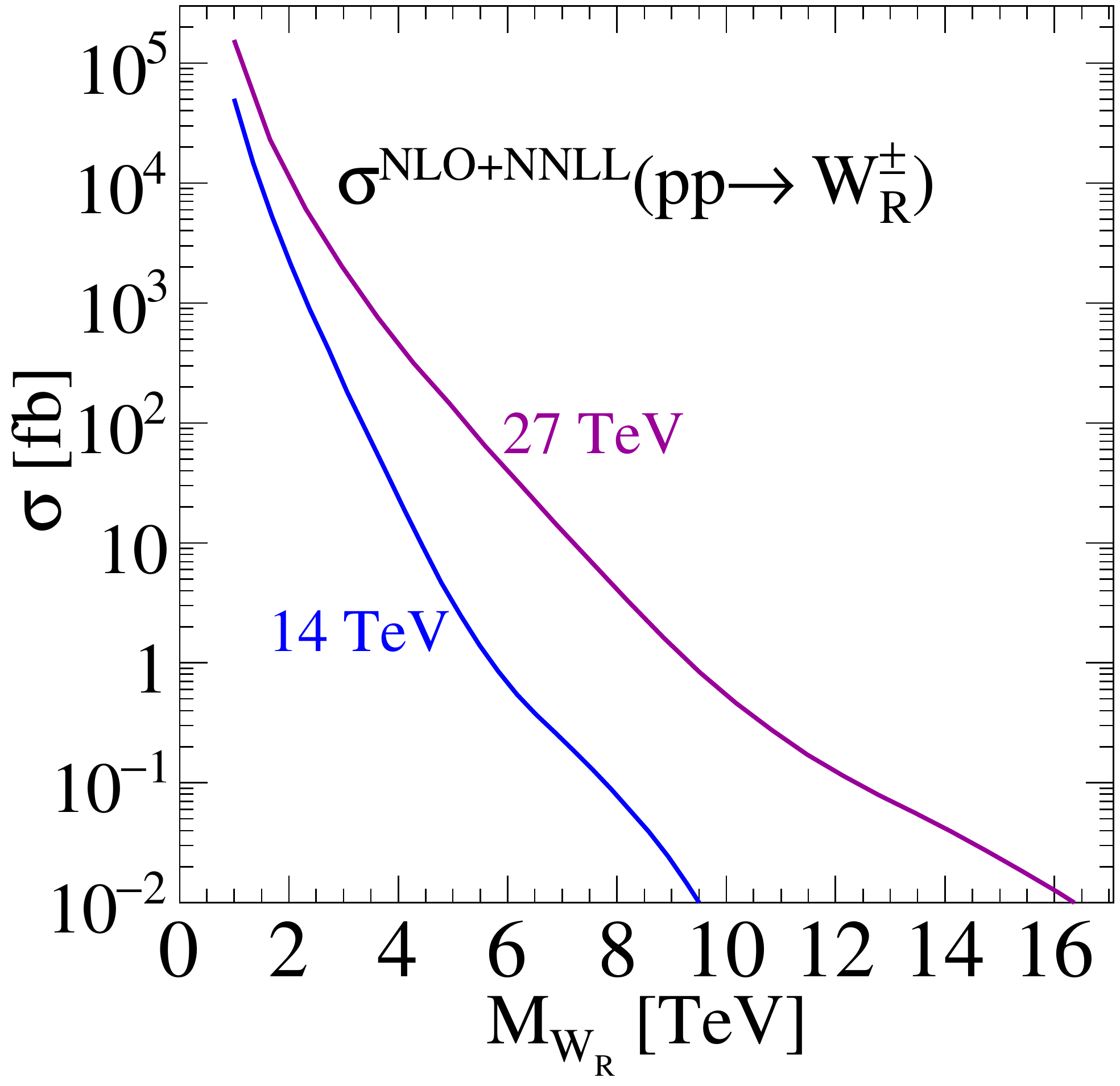}
\end{minipage}
\begin{minipage}[t][\minipageheightdp][t]{\minipagewidthdp}
\centering
\includegraphics[width=\figuremaxwidthdp,height=\figuremaxheightdp,keepaspectratio]{\main/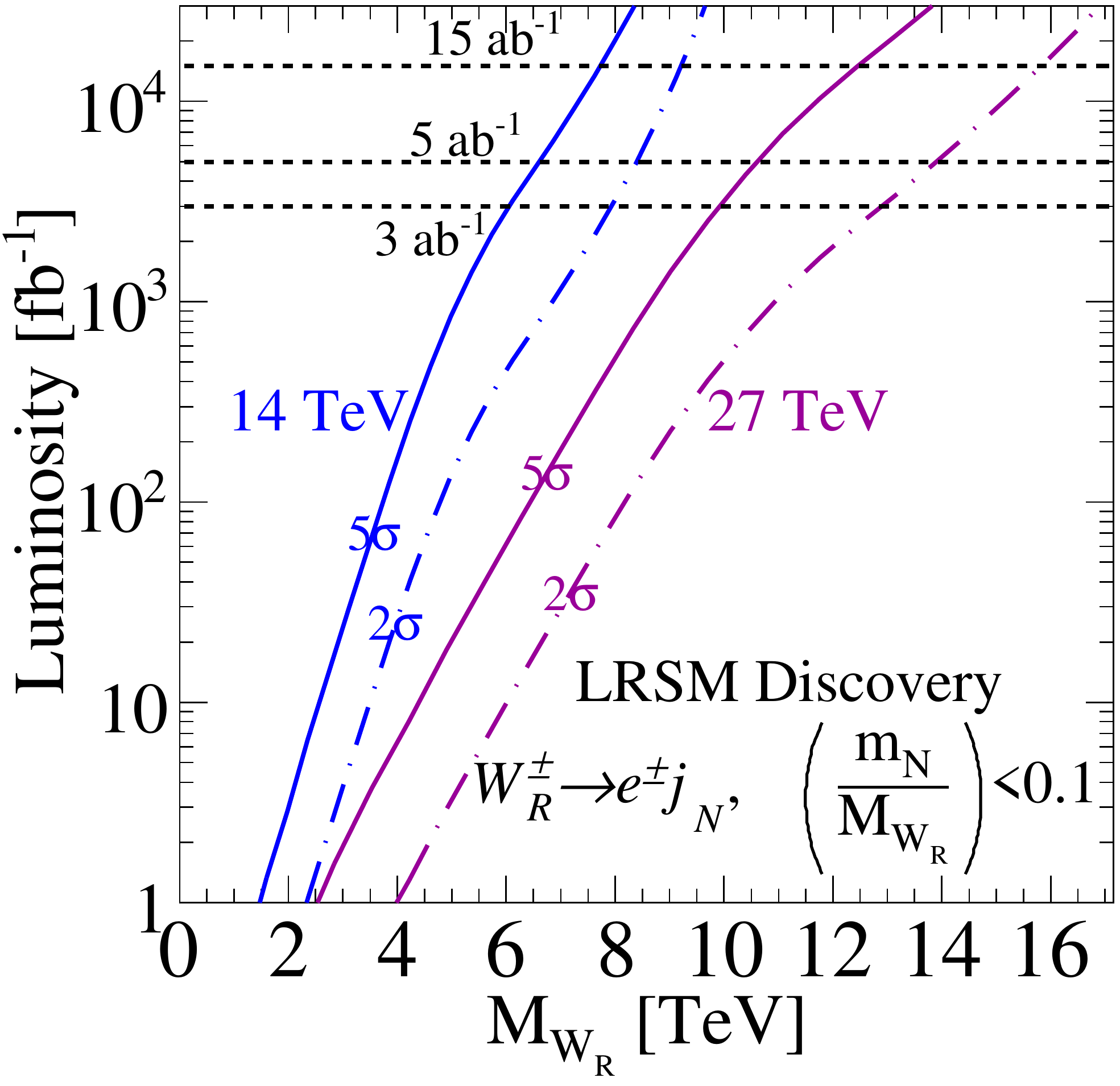}
\end{minipage}\\
\begin{minipage}[t][\minipageheightdp][t]{\minipagewidthdp}
\centering
\includegraphics[width=\figuremaxwidthdp,height=\figuremaxheightdp,keepaspectratio]{\main/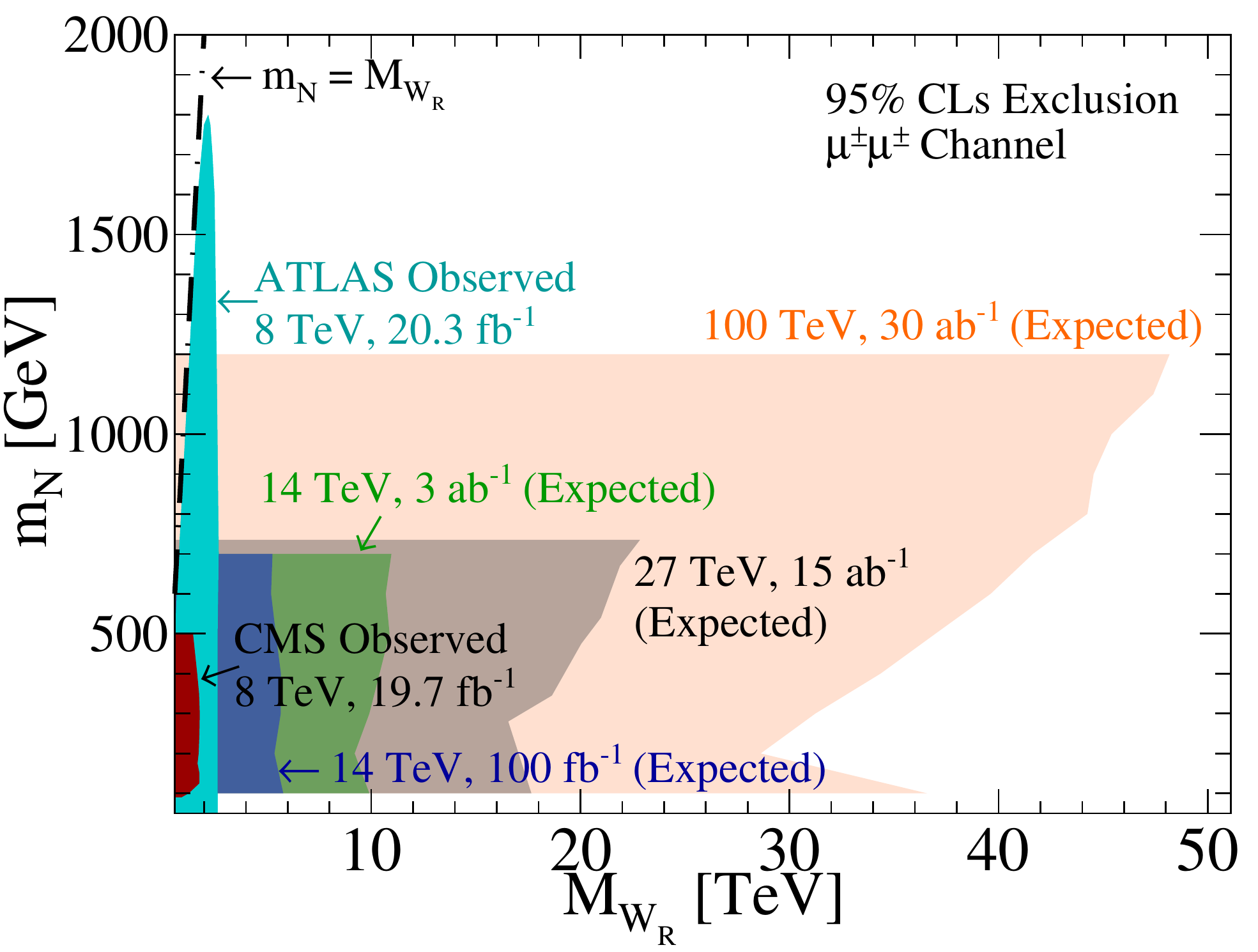}
\end{minipage}
\begin{minipage}[t][\minipageheightdp][t]{\minipagewidthdp}
\centering
\includegraphics[width=\figuremaxwidthdp,height=\figuremaxheightdp,keepaspectratio]{\main/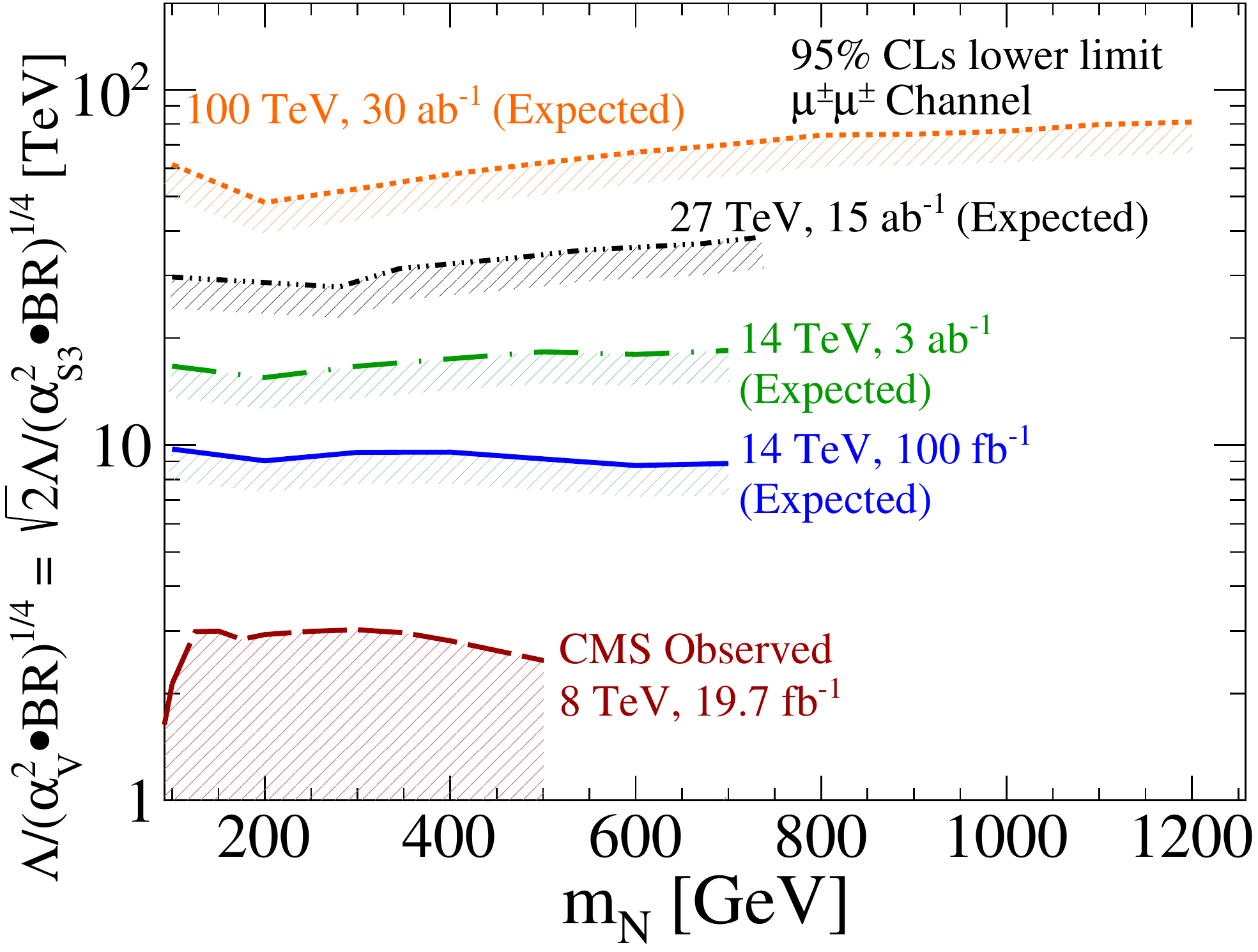}
\end{minipage}
%\begin{center}
%\includegraphics[scale=1,width=.49\textwidth]{}  \label{fig:lrsm_HLvsHL_LHC_WRXSec_vs_Mass}		
%\includegraphics[scale=1,width=.49\textwidth]{} \label{fig:lrsm_HLvsHL_LHC_WRXDisc_vs_Mass}		
%\\
%\includegraphics[scale=1,width=.49\textwidth]{}	\label{fig:lnvWithoutVR_WR_Excl_1703_04669}
%\includegraphics[scale=1,width=.49\textwidth]{} 	\label{fig:neftLimits_vs_mN_1703_04669}	
%\end{center}
\caption{
Top:
the total $pp\to W_R$  cross section at NLO+NNLL(Threshold) (left)
and
$5(2)\sigma$ discovery potential (sensitivity) via $W_R$ decays to an electron and neutrino jet $(j_N)$ (right), as a function of $W_R$ mass and at $\sqrt{s} = 14$ and $27\TeV$~\cite{Mitra:2016kov}.
Bottom:
observed and expected sensitivity to heavy Majorana neutrinos through the process $pp\to \mu^\pm N \to 2\mu^\pm+2j$ and produced via non-resonant $W_R$ (left) as well as dimension-six NEFT operators~\cite{Ruiz:2017nip} (right).
}
\label{fig:lrsm_HLvsHE_LHC}
\end{figure}
%-------------------------------------------------------------------

The Left-Right Symmetric Model (LRSM) remains one of the best motivated high-energy completions of the SM.
It addresses the origin of both tiny neutrino masses via a Type I+II seesaw hybrid mechanism 
as well as the SM's $V-A$ chiral structure through the spontaneous breakdown of an $SU(2)_L\otimes SU(2)_R$ symmetry, 
amongst other low-energy phenomena.
At collider scales, the model predicts the presence of new heavy gauge bosons that are closely aligned with their gauge states $(W_R^\pm,~Z_R')$,
heavy Majorana neutrinos $(N)$, and a plethora of neutral and electrically charged scalars $(H_i^0,~H_j^\pm,~H_k^{\pm\pm})$.
Unlike $U(1)_{BL}$ neutrino mass models, the LRSM gauge couplings are fixed to the SM Weak coupling constant, up to (small) RG-running corrections.
As a result, the Drell-Yan production mechanisms for $W_R$ and $Z_R$ result in large rates at hadron colliders.
Following the procedure of \citeref{Mitra:2016kov}, the $pp\to W_R^\pm$ cross section at NLO+NNLL(Threshold) 
is shown in \fig{fig:lrsm_HLvsHE_LHC} (top left) as a function of mass $M_{W_R}$ at $\sqrt{s} = 14$ and $27\TeV$.
At $\sqrt{s}=14(27)\TeV$, one sees that production cross sections for masses as large as $M_{W_R}\approx 5.5(9)\TeV$ are in excess of $1\fb$.
For $M_{W_R}\approx 7.5(12.5)\TeV$, rates exceed $100\ab$, 
and indicate $\mathcal{O}(10^2-10^3)$ events can be collected with $\mathcal{L}=1-15\abinv$ of data.
Such computations up to NLO in QCD with parton shower matching, including for more generic coupling,
are also publicly available following \citerefs{Mattelaer:2016ynf,Fuks:2017vtl}.

Of the many collider predictions the LRSM, one of its most promising discovery channels is the production of heavy Majorana neutrinos
from resonant $W_R$ currents with $N$ decaying via a lepton number-violating final state. At the partonic level, this is given by~\cite{Keung:1983uu}
\begin{equation}
 q_1 \overline{q_2} \to W_R \to N ~\ell^\pm_i \to \ell_i^\pm \ell_j^\pm W_R^{\mp *} \to \ell_i^\pm \ell_j^\pm q_1' \overline{q_2'}
 \label{eq:seesawLRSMDY}
\end{equation}
and has been extensively studied throughout the literature. For details; see \citeref{Cai:2017mow} and references therein.
Due to the ability to fully reconstruct \eq{eq:seesawLRSMDY}, many properties of $W_R$ and $N$ can be extracted,
including a complete determination of $W_R$ chiral couplings to quarks independent of leptons~\cite{Han:2012vk}.
Beyond the canonical $pp \to W_R \to N\ell \to 2\ell + 2j$ channel, it may be the case that the heavy neutrino is hierarchically lighter than the 
right-handed (RH) gauge bosons. 
Notably, for $(m_N/M_{W_R})\lesssim0.1$, $N$ is sufficiently Lorentz boosted that its decay products, particularly the charged lepton, 
are too collimated to be resolved experimentally~\cite{Ferrari:2000sp,Mitra:2016kov}.
Instead, one can consider the $(\ell_j^\pm q_1' \overline{q_2'})$-system as a single object, a \textit{neutrino jet}~\cite{Mitra:2016kov,Mattelaer:2016ynf}.
The hadronic process is then
\begin{equation}
 p p \to W_R \to N\ell_i^\pm \to j_N ~\ell_i^\pm,
 \label{eq:seesawLRSMnujet}
\end{equation}
and inherits much of the desired properties of \eq{eq:seesawLRSMDY}, 
such as the simultaneous presence of high-$p_T$ charged leptons and lack of MET~\cite{Mitra:2016kov,Mattelaer:2016ynf},
resulting in a very strong discovery potential.
Assuming conservative detector efficiency $(\varepsilon)$ and selection acceptance $(\mathcal{A})$ rates
of $(\varepsilon,\mathcal{A})\approx(0.33,0.64)$ based on the realistic analysis of \citeref{Mitra:2016kov},
and a branching fraction of $\BR(W_R \to Ne \to eeq\overline{q'})\approx 10\%$ for $(m_N/M_{W_R})<0.1$. 
Figure \ref{fig:lrsm_HLvsHE_LHC} (top right) shows the requisite integrated luminosity for $5(2)\sigma$ discovery of 
\eq{eq:seesawLRSMnujet} at $\sqrt{s}=14$ and $27\TeV$.
With $\mathcal{L}=3(5)\abinv$, $W_R$ as heavy as $6(6.5)\TeV$ and $10(10.5)\TeV$, respectively, can be discovered at $\sqrt{s}=14$ ($27$)\TeV.
With $\mathcal{L}=15\abinv$, mass scales as heavy $16\TeV$ can be probed at the $2\sigma$ level at $\sqrt{s}=27\TeV$.

For such heavy $W_R$ and $Z_R$ that may be kinematically outside the reach of the $\sqrt{s}=14\TeV$ LHC,
one can still produce EW- and sub-TeV scale via off-shell $W_R$ and $Z_R$ bosons~\cite{Ruiz:2017nip}.
As a result, the $pp \to W_R^* \to N\ell \to 2\ell + 2j$ process occurs instead at a hard scale $Q\sim m_N$ and 
cannot be distinguished from the phenomenological Type I seesaw without a 
detailed analysis of the heavy neutrino's chiral couplings~\cite{Han:2012vk,Ruiz:2017nip}.
However, this also means that searches for heavy $N$ in the context of the phenomenological Type I can be recast/reinterpreted in the context of the LRSM.
Subsequently, as shown in \fig{fig:lrsm_HLvsHE_LHC} (bottom left), 
$W_R$ as heavy as $8-9\TeV$ can be probed indirectly with $\mathcal{L}=1\abinv$ at $\sqrt{s}=14\TeV$~\cite{Ruiz:2017nip}.
A similar argument can be applied to heavy neutrinos produced through dimension-six
Heavy Neutrino Effective Field Theory (NEFT) operators, revealing sensitivity to mass scales up to $\Lambda \sim \mathcal{O}(10)\TeV$
over the $\sqrt{s}=14\TeV$ LHC's lifetime~\cite{Ruiz:2017nip},
as shown in \fig{fig:lrsm_HLvsHE_LHC} (bottom right).

%%%%%%%%%%%%%%%%%%%%%%%%%%%%%%%%%%%%%%%%%%%%%%%%%%%%%%%%%%%%%%%%%%
%\subsubsection{Type II Scalars Discovery Potential at the HL- and HE-LHC}\label{sec:lhcSeesawtype2}
\paragraph*{Type II Scalars Discovery Potential at the HL- and HE-LHC}\label{sec:lhcSeesawtype2}
%%%%%%%%%%%%%%%%%%%%%%%%%%%%%%%%%%%%%%%%%%%%%%%%%%%%%%%%%%%%%%%%

%-------------------------------------------------------------------
\begin{figure}[!t]
\centering
\begin{minipage}[t][\minipageheightsp][t]{\minipagewidthsp}
\centering
\includegraphics[width=\figuremaxwidthsp,height=\figuremaxheightsp,keepaspectratio]{\main/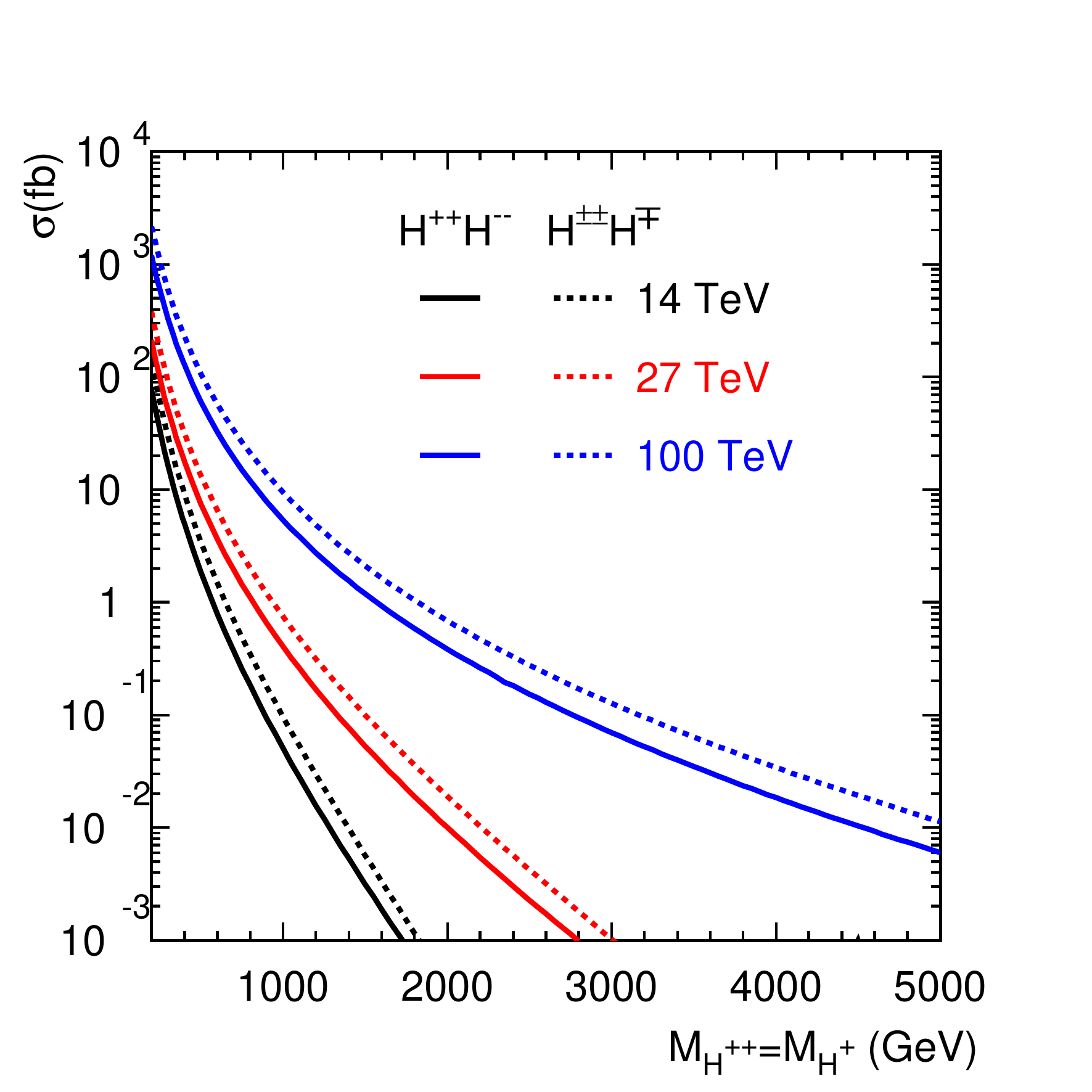}
\end{minipage}\\
\centering
\begin{minipage}[t][\minipageheightdp][t]{\minipagewidthdp}
\centering
\includegraphics[width=\figuremaxwidthdp,height=\figuremaxheightdp,keepaspectratio]{\main/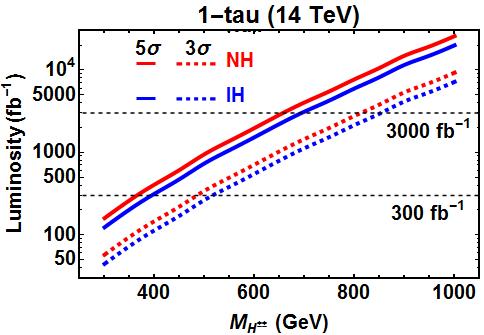}
\end{minipage}
\begin{minipage}[t][\minipageheightdp][t]{\minipagewidthdp}
\centering
\includegraphics[width=\figuremaxwidthdp,height=\figuremaxheightdp,keepaspectratio]{\main/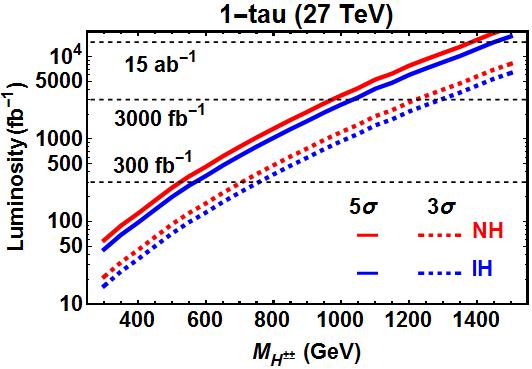}
\end{minipage}
%\begin{center}
%\qquad\quad\includegraphics[scale=1,width=.49\textwidth]{\main/section5FlavorRelatedStudies/img/lhcSeesaw_Type2_Triplet_XSec_vs_Mass.pdf} 		\label{fig:typeii-xsec}
%\\
%\includegraphics[scale=1,width=.49\textwidth]{\main/section5FlavorRelatedStudies/img/lhcSeesaw_Type2_L-MH-1tau-LHCX14_1802_0094.jpg} 	\label{fig:LMtau-typeii-LHCX14}
%\includegraphics[scale=1,width=.49\textwidth]{\main/section5FlavorRelatedStudies/img/lhcSeesaw_Type2_L-MH-1tau-LHCX27_1802_0094.jpg}	\label{fig:LMtau-typeii-LHCX27}
%%\includegraphics[scale=1,width=.45\textwidth]{\main/section5FlavorRelatedStudies/img/lhcSeesaw_Type2_L-MH-1tau-LHC100_1802_0094.jpg}	\label{fig:LMtau-typeii-LHC100}
%\end{center}
\caption{
Top: the total cross section for $pp\to H^{++} H^{--} \mathrm{and}\, H^{\pm\pm}H^\mp$ at $\sqrt{s} = 14$, $27$, and $100\TeV$.
Bottom:
requisite luminosity versus $M_{H^{\pm\pm}}$ for $5(3)\sigma$ discovery (evidence)
for the process $pp\to H^{++}H^{--}\to \tau_h \ell^\pm \ell^\mp \ell^\mp$,
where $\tau^\pm\to \pi^\pm \nu $, 
for the NH and IH at $\sqrt{s}=14$ and $27\TeV$.
}
\label{fig:lhcSeesawtypeiiScalars}
\end{figure}
%-------------------------------------------------------------------

The Type II seesaw hypothesises the existence of a new scalar SU$(2)_L$ triplet that couples to SM leptons in order to 
reproduce the light neutrino mass spectrum and oscillation data.
This is done by the spontaneous generation of a left handed Majorana mass for the light neutrinos.
Moreover, the Type II scenario is notable for the absence of sterile neutrinos, 
demonstrating that light neutrino masses themselves do not imply the existence of additional fermions.
The most appealing production mechanisms at hadron colliders of triplet Higgs bosons are the pair production
of doubly charged Higgs and the associated production of doubly charged Higgs and singly charged Higgs,
\begin{eqnarray}
pp\to Z^\ast/\gamma^\ast \to H^{++} H^{--}, \ \ \ pp\to W^\ast\to H^{\pm\pm}H^\mp.
\end{eqnarray}
followed, in the most general situation, by lepton flavour- and lepton number-violating decays to SM charged leptons.
In \fig{fig:lhcSeesawtypeiiScalars} (top left), we show the total cross section of the 
$pp\to H^{++} H^{--}$ and $pp\to  H^{\pm\pm}H^\mp$ processes as a function of triplet mass (in the degenerate limit with $M_{H^\pm}=M_{H^{\pm\pm}})$,
in collisions at $\sqrt{s} = 14$, $27$, and $100\TeV$.

In Type II scenarios, $H^{\pm\pm}$ decays to $\tau^\pm \tau^\pm$ and $\mu^\pm \mu^\pm$ pairs are comparable or greater than the
$e^\pm e^\pm$ channel by two orders of magnitude. 
Moreover, the $\tau \mu$ channel is typically dominant in decays involving different lepton flavours~\cite{Perez:2008ha,Li:2018jns}.
If such a seesaw is realised in nature, tau polarisations can help to determine the chiral property of triplet scalars:
One can discriminate between different heavy scalar mediated neutrino mass mechanisms, such as those found in the Type II seesaw and Zee-Babu model, 
by studying the distributions of tau leptons' decay products~\cite{Sugiyama:2012yw,Li:2018jns}. 
Due to the low $\tau_h$ identification efficiencies, 
future colliders with high energy and/or luminosity enables one to investigate and search for doubly charged Higgs decaying to $\tau_h$ pairs.
Accounting for constraints from neutrino oscillation data on the doubly charged Higgs BRs,
as well as  tau polarisation effects~\cite{Li:2018jns},
\fig{fig:lhcSeesawtypeiiScalars} (right) displays
the $3\sigma$ and $5\sigma$ significance in the plane of integrated luminosity versus doubly charged Higgs mass 
for $pp\to H^{++}H^{--}\to \tau^\pm\ell^\pm\ell^\mp\ell^\mp$ at 
$\sqrt{s} = 14$, $27$, and $100\TeV$, respectively.
For the one $\tau$ channel with $\tau^\pm\to \pi^\pm \overset{(-)}{\nu_\tau}$, 
the sensitivity to doubly charged Higgs mass at HL-LHC can reach $655\GeV$ and $695\GeV$ for NH and IH respectively with a luminosity of $3\abinv$. 
Higher masses, $1380-1930\GeV$ for NH and $1450-2070\GeV$ for IH, can be probed at $27\TeV$ with \intlumihelhc and $100\TeV$ with $3\abinv$.

%%%%%%%%%%%%%%%%%%%%%%%%%%%%%%%%%%%%%%%%%%%%%%%%%%%%%%%%%%%%%%%%%%
%\subsubsection{Type III Leptons Discovery Potential at HL- and HE-LHC}\label{sec:lhcSeesawtype3}
\paragraph*{Type III Leptons Discovery Potential at HL- and HE-LHC}\label{sec:lhcSeesawtype3}
%%%%%%%%%%%%%%%%%%%%%%%%%%%%%%%%%%%%%%%%%%%%%%%%%%%%%%%%%%%%%%%%

\begin{figure}[t]
\centering
\begin{minipage}[t][\minipageheightsp/2][t]{\minipagewidthsp}
\centering
\includegraphics[width=\figuremaxwidthsp*4/3,height=\figuremaxheightsp,keepaspectratio]{\main/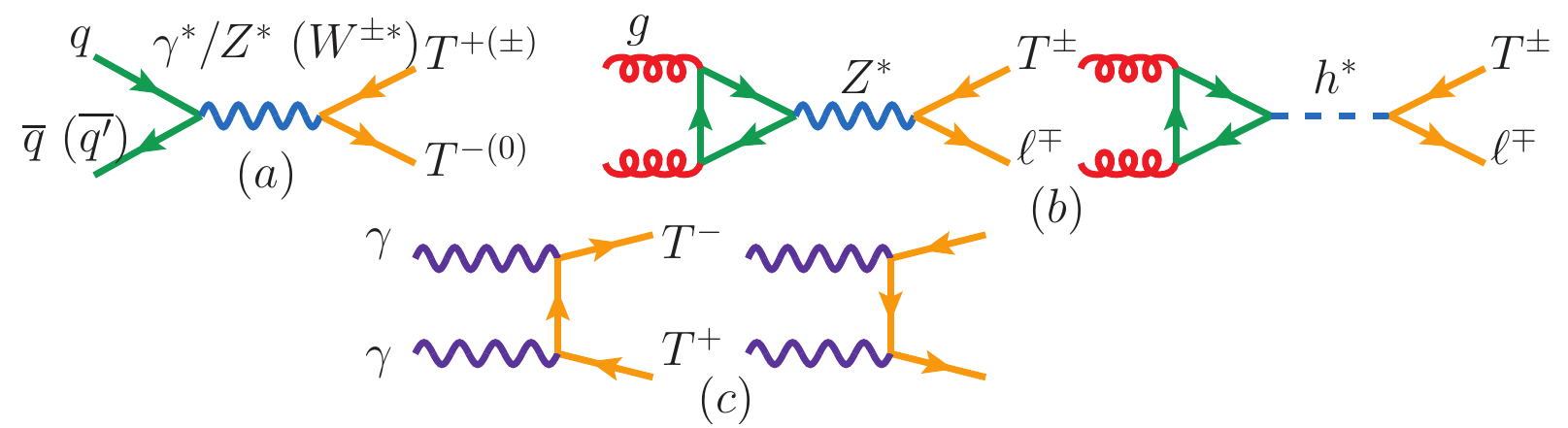}
\end{minipage}\\
\begin{minipage}[t][\minipageheightdp][t]{\minipagewidthdp}
\centering
\includegraphics[width=\figuremaxwidthdp,height=\figuremaxheightdp,keepaspectratio]{\main/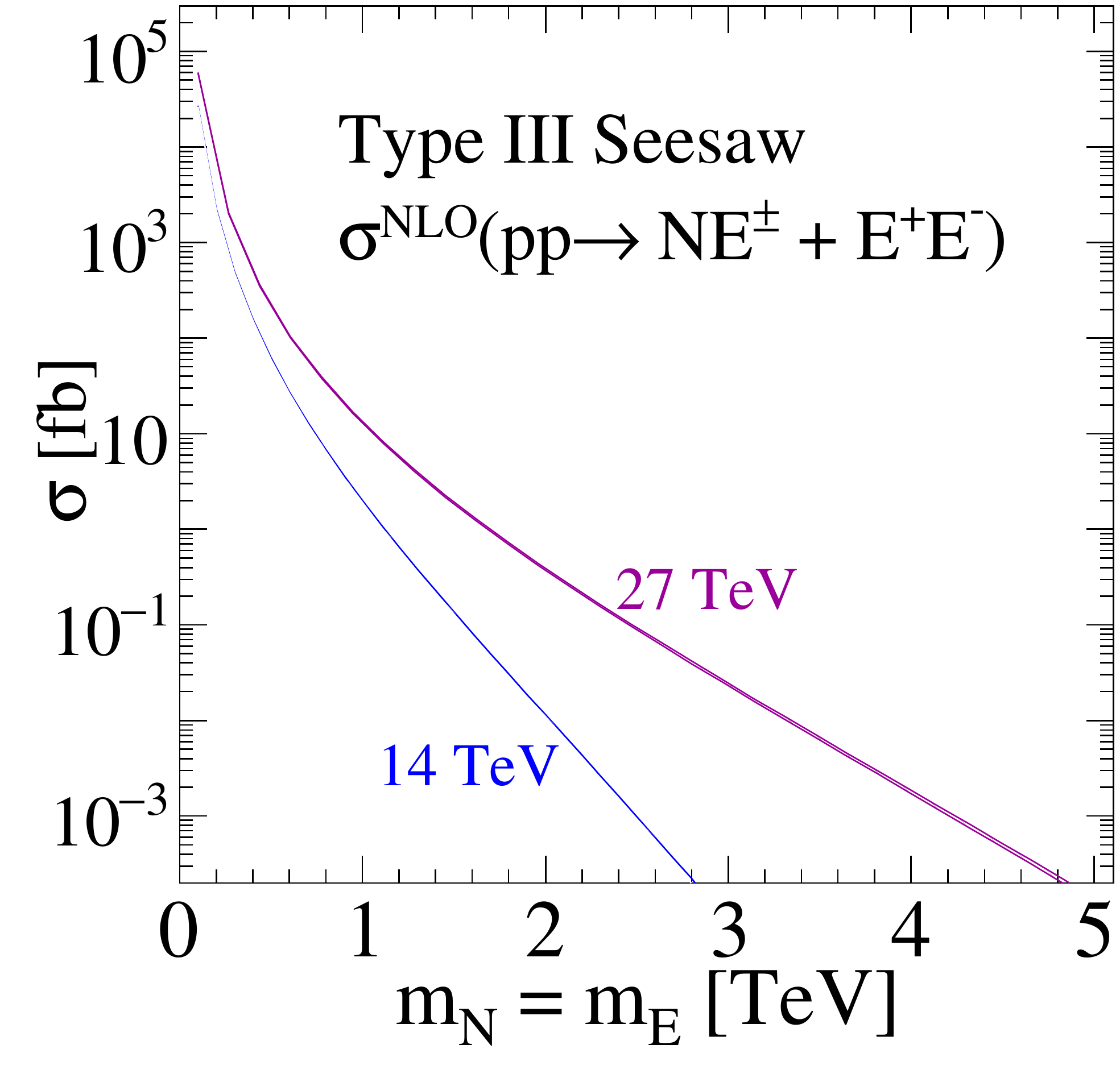}
\end{minipage}
\begin{minipage}[t][\minipageheightdp][t]{\minipagewidthdp}
\centering
\includegraphics[width=\figuremaxwidthdp,height=\figuremaxheightdp,keepaspectratio]{\main/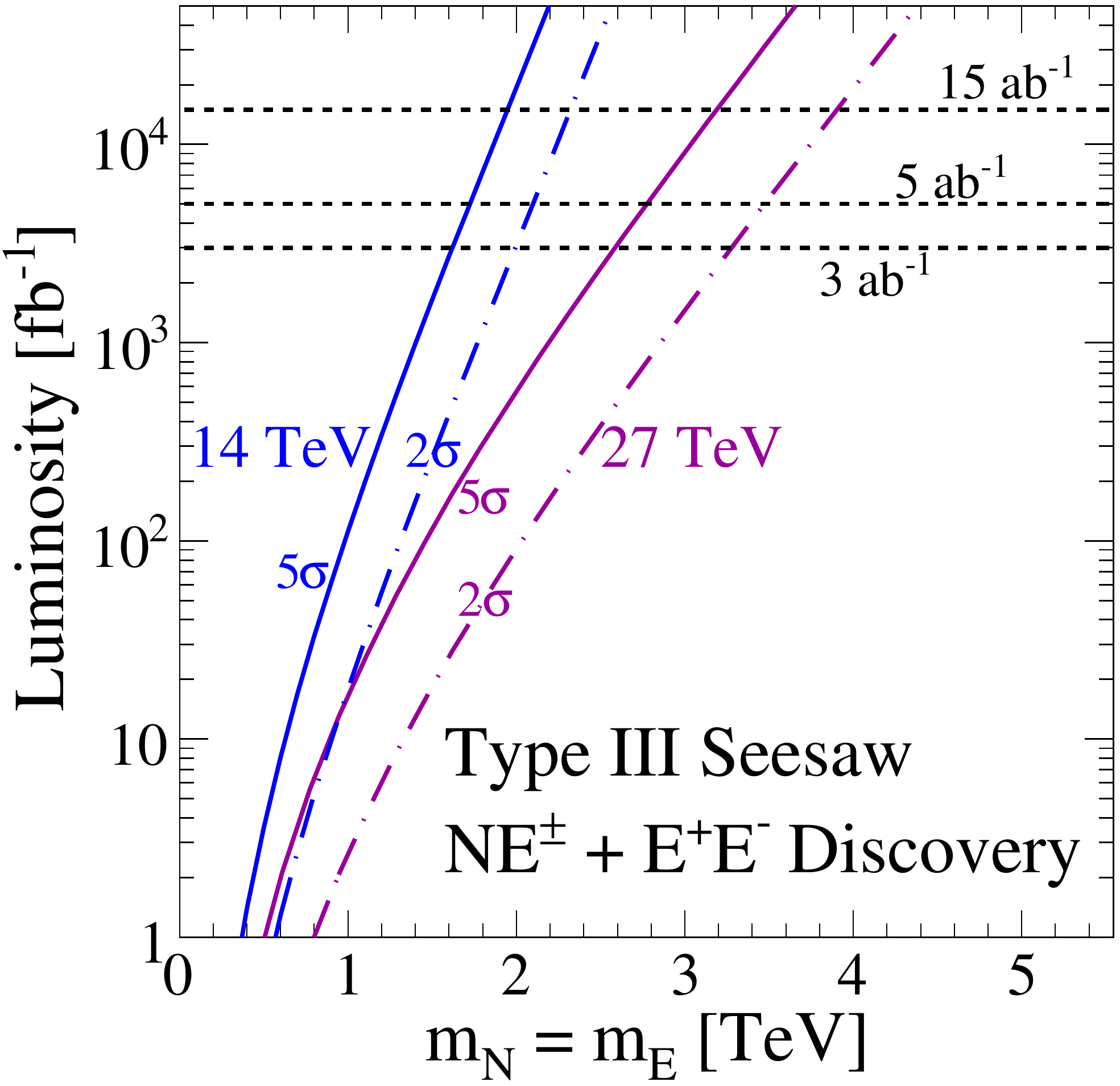}
\end{minipage}
%\begin{center}
% \includegraphics[width=0.9\textwidth]{\main/section5FlavorRelatedStudies/img/lhcSeesaw_Feynman_TypeIIIMultiProd_1711_02180.pdf}	\label{fig:type3Seesaw}
% \\
%  \includegraphics[width=.49\textwidth]{\main/section5FlavorRelatedStudies/img/lhcSeesaw_Type3_XSec_vs_Mass_1509_05416.pdf} \label{fig:type3_HLvsHL_LHC_XSec_vs_Mass}
%  \includegraphics[width=.49\textwidth]{\main/section5FlavorRelatedStudies/img/lhcSeesaw_Type3_Disc_vs_Mass_1509_05416.pdf} \label{fig:lhcSeesawtype3_HLvsHL_LHC_Disc_vs_Mass}
%\end{center}
\caption{
Top:
Born level production of Type III leptons via (a) Drell-Yan, (b) gluon fusion, and (c) photon fusion; from \citeref{Cai:2017mow}.
Bottom: at $\sqrt{s}=14$ and $27\TeV$ and as a function of heavy triplet lepton mass,
the summed inclusive production cross section [fb] of $pp\to NE^\pm + E^+E^-$, at NLO in QCD~\cite{Ruiz:2015zca} (left),
 and the requisite integrated luminosity for $5(2)\sigma$ discovery (sensitivity) to $N E^\pm + E^+E^-$ 
 based on the analyses of~\cite{Arhrib:2009mz,Li:2009mw} (right).
}
\label{fig:lhcSeesawtype3}
\end{figure}

Low-scale Type III Seesaws hypothesises the existence of heavy electrically charged $(E^\pm)$ and neutral $(N)$ leptons, 
which form a triplet under SU$(2)_L$,
that couple to both SM charged and neutral leptons through mixing/Yukawa couplings.
Due to their gauge charges, triplet leptons also couple to EW gauge bosons appreciably 
and do not feature suppressed production cross section typical of seesaw scenarios with gauge singlet fermions.
The presence of EW gauge couplings also implies that once a collider energy and mass are stipulated, 
the triplet lepton 
pair production cross section can be computed, up to small (and potentially negligible) mixing effects.
Production mechanisms commonly found in the literature, and shown in \fig{fig:lhcSeesawtype3} (top),
include charged current and neutral current Drell-Yan, photon fusion, and gluon fusion if one considers heavy-light charged lepton associated production.
A recent assessment of triplet production modes found~\cite{Cai:2017mow} that despite the sizeable luminosities
afforded to gluon fusion $(gg\to E^\pm\ell^\mp)$, including its large QCD corrections~\cite{Ruiz:2017yyf},
and photon fusion $(\gamma\gamma\to E^+E^-)$, the \sloppy\mbox{Drell-Yan} processes remain the dominant production channel of triplet leptons when kinematically accessible.
In light of this, in \fig{fig:lhcSeesawtype3} (bottom left), 
the summed cross sections for the Drell-Yan processes,
\begin{equation}
 pp\to \gamma^*/Z^* \to E^+E^- \quad\text{and}\quad 
 pp\to W^{\pm *} \to E^\pm N,
\end{equation}
are shown at NLO in QCD, following \citeref{Ruiz:2015zca}, as a function of triplet masses (assuming \sloppy\mbox{$m_N = m_E$}), at $\sqrt{s}=14$ and $27\TeV$.
For $m_{N},m_{E} \approx 1.2(1.8)\TeV$,  the production rate reaches \sloppy\mbox{$\sigma(pp\to NE+EE)\approx 1\fb$} at $\sqrt{s}=14(27)\TeV$; and for heavier leptons with \sloppy\mbox{$m_{N},m_{E} \approx 2.5(4.2)\TeV$}, one sees that $\sigma(pp\to NE+EE)\approx 1\ab$.

Another consequence of the triplet leptons coupling to all EW bosons is the adherence to the Goldstone Equivalence Theorem.
This implies that triplet leptons with masses well above the EW scale
will preferentially decay to longitudinal polarised $W$ and $Z$ bosons as well as to the Higgs bosons. 
For decays of EW boson to jets or charged lepton pairs, triplet lepton can be full reconstructed from their final-state 
enabling their properties to be studied in detail.
For fully reconstructible final-states,
\begin{eqnarray}
 N E^\pm	\rightarrow ~\ell\ell' + WZ/Wh		&\rightarrow& ~\ell\ell' ~+~ nj+mb,
 \\
 E^+ E^- 	\rightarrow ~\ell\ell' + ZZ/Zh/hh 	&\rightarrow& ~\ell\ell' ~+~ nj+mb,
\end{eqnarray}
which correspond approximately to the branching fractions $\BR(NE)\approx0.115$ and $\BR(EE)\approx0.116$,
search strategies such as those considered in \citerefs{Arhrib:2009mz,Li:2009mw} can be enacted.
Assuming a fixed detector acceptance and efficiency of $\mathcal{A}=0.75$,
which is in line to those obtained by \citerefs{Arhrib:2009mz,Li:2009mw},
\fig{fig:lhcSeesawtype3} (bottom right) shows as a function of triplet mass 
the requisite luminosity for a $5\sigma$ discovery (solid) and $2\sigma$ evidence (dash-dot)
of triplet leptons at $\sqrt{s}=14$ and $27\TeV$.
With $\mathcal{L}=3-5\abinv$, the $14\TeV$ HL-LHC can discover states as heavy as $m_{N},m_{E}=1.6-1.8\TeV$.
For the same amount of data, the $27\TeV$ HE-LHC can discover heavy leptons  $m_{N},m_{E}=2.6-2.8\TeV$;
with $\mathcal{L}=15\abinv$, one can discover~(probe) roughly $m_{N},m_{E}=3.2(3.8)\TeV$.

\subsubsection{\theoryC{Like-sign dileptons with mirror type composite neutrinos at the HL-LHC}}
\label{sec:neutrino552}
\contributors{M.~Presilla, O.~Panella, R.~Leonardi}
%{\bf Author(s): M.~Presilla~$^1$, O.~Panella~$^2$, R.~Leonardi~$^2$}\\
%$^1$ Dipartimento di Fisica e Astronomia "Galileo Galielei", Universit\`{a} degli Studi di Padova, Via Marzolo, I-35131, Padova, Italy, email: {\tt matteo.presilla@gmail.com}\\
%$^2$ INFN, Istituto Nazionale di Fisica Nucleare, Sezione di Perugia, Via A. Pascoli, 06123, Perugia, email: {\tt orlando.panella@pg.infn.it}

%\subsubsection{Introduction}
%\paragraph*{Introduction}
A composite scenario~\cite{Eichten:1983hw,Peskin:1985fm,Cabibbo:1983bk,Baur:1989kv}, where at a sufficiently high energy  scale $\Lambda$ (compositeness scale) the SM leptons and quarks show the effects of an internal substructure, has triggered  considerable recent interest both from the theoretical~\cite{Panella:2017spx,Biondini:2012ny,Leonardi:2014aa,Biondini:2017fut} and experimental~\cite{Aad:2013ab,Khachatryan:2016ac} point of view.  In particular recent studies~\cite{Leonardi:2015qna} have concentrated in searching for heavy composite Majorana neutrinos at the LHC. The scenario discussed in such studies is one in which the excited neutrino ($\nu^*$) is a Majorana particle. A recent CMS study has searched for a  heavy composite Majorana neutrino ($N$).
Using $2.6\fbinv$ data of the 2015 Run II  at $\sqrt{s}=13\TeV$, heavy composite neutrino masses are excluded, at $95\%\cl$, up to $m_N=4.35\TeV$ and $4.60$ ($4.70$)\TeV for a value of $\Lambda=M_N$,  from the $eeqq$ channel and the $\mu\mu qq$ channel, respectively~\cite{CMS:2016blm,Sirunyan:2017xnz}.
Here we focus on  a mirror type assignment for the excited composite fermions. The neutrino mass term is  built up from a Dirac mass, $m_*$,  the mass of the charged lepton component of the $SU(2)$, right-handed doublet, and $m_L$ the  Majorana mass of  the left-handed component (singlet) of the excited neutrino.
The mass matrix is diagonalised leading to two Majorana mass eigenstates. The active neutrino field $\nu^*_R$ is thus a superposition of the two mass eigenstates with mixing coefficients which depend on the ratio $m_L/m_*$.
We discuss the prospects of discovery of these physical states at the HL-LHC as compared with the previous searches of composite Majorana neutrinos at the LHC based on sequential type Majorana neutrinos.

%\subsubsection{Model}
%\paragraph*{Model}
We discuss a variant of the model analysed in \citeref{Leonardi:2015qna,Sirunyan:2017xnz} taking up the scenario in which the excited fermions are organised with a mirror $SU(2)$ structure relative to the SM fermions, \iee~the right handed components form an $SU(2)$ doublet while the left handed components are singlets~\cite{Patrignani:2016xqp}. %\pf{no such inspire record exists, please fix}. 
We construct a general Dirac-Majorana mass term and discuss its mass spectrum along with prospects of observing the resulting lepton number violating signatures at the HL-LHC~\cite{Presilla:2018ryu}.

In analogy with the usual procedure adopted in see-saw type extensions of the SM we may give a (lepton number violating) Majorana mass term  to the left handed excited neutrino ($\nu_L^*$) which is a singlet and does not actively participate to the gauge interactions in \eq{eq:mirror} --\emph{sterile neutrino}--,  while a (lepton number conserving) Dirac mass term $m^*$ is associated to the right-handed component $\nu_R^*$  which belongs to the $SU(2)$ doublet and does participate in the gauge interactions -- \emph{active neutrino}-.
%We also assume that the Dirac mass term $m_*$ arises from a Higgs mechanism via Yukawa couplings similarly to the standard model fermions so as to preserve local $SU(2)$ gauge invariance.
We can thus write down the excited neutrino Dirac-Majorana mass term appearing in the Lagrangian density of the model as:
\begin{equation}\label{eq:mass_term1}{\cal L}^{\text{D+M}} = -\frac{1}{2} m_L \,\bar{\nu}^*_L ({\nu}^*_L)^c
- m_*\, \bar{\nu}^*_L \nu^*_R\,  +\text{h.c.}
\end{equation} 

The Lagrangian mass term is easily diagonalised following standard procedures, obtaining two Majorana mass eigenstates, $\nu_{1,2}$, with (positive) mass eigenvalues given by: 
\begin{equation}
\label{eq:masseigenstates}
m_{1,2}=\sqrt{m_*^2 +\left(\frac{m_L}{2}\right)^2} \mp \frac{m_L}{2}
\end{equation}
Interacting states can be written as a mixing of the mass eigenstate according to the relation:
\begin{subequations}
\label{eq:inverted2}
\begin{align}
\label{eq:inverted2a}
(\nu_{L}^*)^c &= \,- i\cos\theta\, \nu_{1R} + \sin\theta \,\nu_{2R}\\ 
\label{eq:inverted2b}\nu_{R}^* &=\, \phantom{-}i\sin\theta\, \nu_{1R} + \cos\theta \,\nu_{2R}
\end{align}
\end{subequations} 
where the mixing angle $\theta$ is given in terms of the Dirac and Majorana masses: 
\sloppy\mbox{$\theta = -\frac{1}{2} \arctan \frac{2m_*}{m_L} $}.
%\begin{equation}
%\label{eq:mixingangle}
%\theta = -\frac{1}{2} \arctan \left( \frac{2m_*}{m_L}\right)
%\end{equation}

\begin{figure*}[t]
\centering
\begin{minipage}[t][\minipageheightdp][t]{\minipagewidthdp}
\centering
\includegraphics[width=\figuremaxwidthdp+2mm,height=\figuremaxheightdp,keepaspectratio]{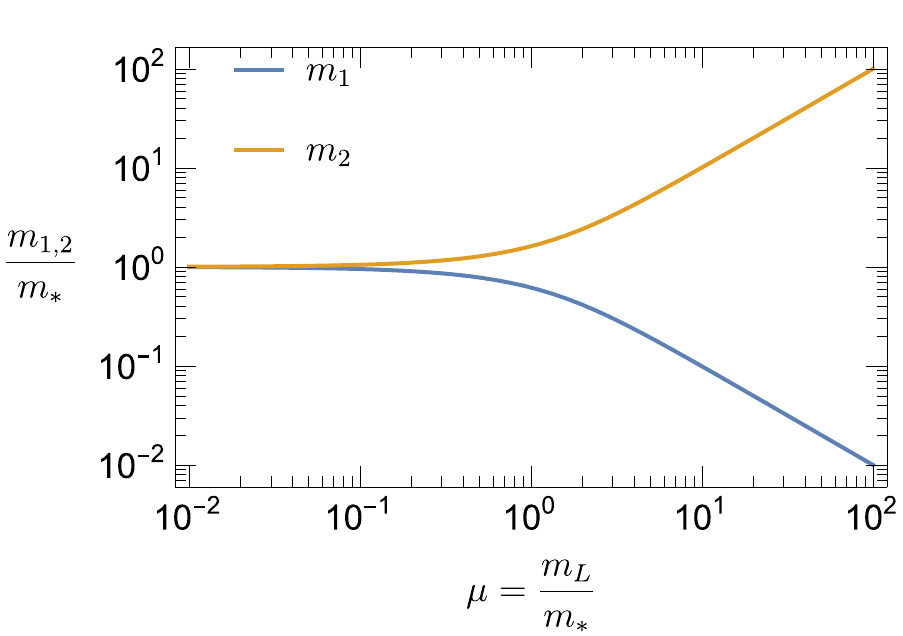}
\end{minipage}
\begin{minipage}[t][\minipageheightdp][t]{\minipagewidthdp}
\centering
\includegraphics[width=\figuremaxwidthdp,height=\figuremaxheightdp,keepaspectratio]{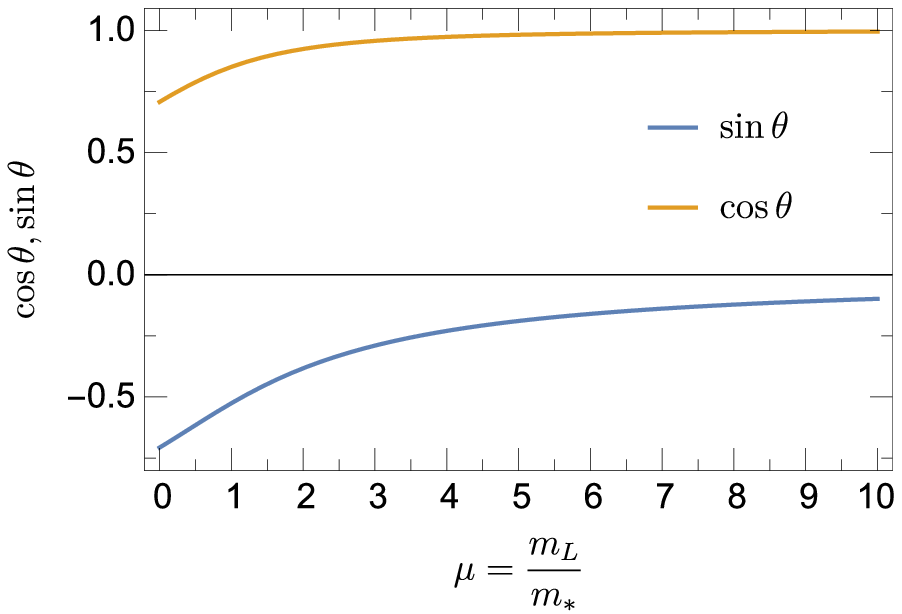}
\end{minipage}
\caption{Left: mass eigenstates $m_{1,2}$ (in units of $m_*$) from \protect\eq{eq:masseigenstates} as a function of the parameter \sloppy\mbox{$\mu=m_L/m_*$}. Right: mixing coefficients $\cos\theta$ and $\sin\theta$ from \eqs{eq:inverted2} as a function of the parameter $\mu=m_L/m_*$}
\end{figure*}
%%%%interazioni del modello

Now we take into account all the relevant effective couplings of these particles. In the  mirror type model, we consider the first lepton family and assume that the excited neutrino and the excited electron are grouped into left handed singlets and  a right-handed $SU(2)$ doublet~\cite{Patrignani:2016xqp}.
The corresponding gauge mediated Lagrangian between the left-handed SM doublet $L$ and the right-handed excited doublet $R$ via the $SU(2)_L\times U(1)_Y$ gauge fields~\cite{Takasugi:1995bb,Olive:2014kda}, which should be of the magnetic type to warrant  current conservation, can be written down:
\begin{equation}
\label{eq:mirror}
{\cal L} =\frac{1}{2 \Lambda} \, \bar{R}^* \sigma^{\mu\nu}\left( gf \frac{\bm{\tau}}{2}\cdot \bm{W}_{\mu\nu} +g'f' Y B_{\mu\nu}\right) L +h.c. \, ,
\end{equation}
where $L^T =({\nu}_L, \ell_L)$ is the ordinary 
%\pf{$L_R$ and $L_L$ need to be more clearly defined}
$SU(2)_L$ lepton doublet, and $R^T=(\nu^*_R, \ell^*_R)$ is the excited right-handed doublet; $g$ and $g'$ are the $SU(2)_L$ and $U(1)_Y$ gauge couplings and $\bm{W}_{\mu\nu}$, $B_{\mu\nu}$ are the field strength for the $SU(2)_L$ and $U(1)_Y$ gauge fields; $f$ and $f'$ are dimensionless couplings.
The relevant charged current (gauge) interaction of the excited (active) Majorana neutrino $\nu_R^*$ is easily derived from \eq{eq:mirror}
%\begin{equation}
%\label{eq:lagGI}
%\mathcal{L}_{\text{G}}=\frac{gf}{\sqrt{2}\Lambda}\, \overline{\nu}^*_R \, \sigma^{\mu \lambda} \, \ell_L \,  \partial_{\mu}  W_{\lambda} \,  + h. c.
%\end{equation}
and can be written out explicitly in terms of the Majorana mass eigenstates through \eq{eq:inverted2b}:
\begin{equation}
\label{eq:lagGImass}
\mathcal{L}_{\text{G}}=\frac{gf}{\sqrt{2}\Lambda}\,\left( -i\sin\theta\, \overline{\nu}_{1} +\cos\theta\, \overline{\nu}_{2}\right) \, \sigma^{\mu \lambda} \, \ell_L \,\,  \partial_{\mu} \, W_{\lambda} \,  + h. c.
\end{equation}

Contact interactions between ordinary fermions may arise by constituent exchange, if the fermions have common constituents, and/or by exchange of the binding quanta of the new unknown interaction whenever such binding quanta couple to the constituents of both particles~\cite{Peskin:1985fm,Olive:2014kda}. The dominant effect is expected to be given by the dimension 6 four-fermion interactions which scale with the inverse square of the compositeness scale $\Lambda$: 
\begin{subequations}
\label{eq:contact}
\begin{align}
\label{eq:Lcontact}
\mathcal{L}_{\text{CI}}&=\frac{g_\ast^2}{\Lambda^2}\frac{1}{2}j^\mu j_\mu\\
\label{eq:Jcontact}
j_\mu&=\eta_L\bar{f_L}\gamma_\mu f_L+\eta\prime_L\bar{f^\ast_L}\gamma_\mu f^\ast_L+\eta\prime\prime_L\bar{f^\ast_L}\gamma_\mu f_L + h.c.\nonumber\\&\phantom{=} +(L\rightarrow R)
\end{align}
\end{subequations}
where $g_*^2 = 4\pi$ and the $\eta$ factors are usually set equal to unity. In this work the right-handed currents will be neglected for simplicity.

The single production $q\bar{q}' \to \nu^*\ell$ proceeds through flavour conserving but non-diagonal terms, in particular with currents like  the third term in \eq{eq:Jcontact} which couple excited states with ordinary fermions and
%\begin{equation}
%\label{eq:lagCI}
%\mathcal{L}_{\text{CI}}= \frac{g_\ast^2}{\Lambda^2} \, \bar{q}_L\gamma^\mu  q_L' \, \overline{\nu^*_L}\gamma_\mu \ell_L \, .
%\end{equation} 
the contact interactions in %\eq{eq:lagCI} 
can be written out explicitly in terms of the Majorana mass eigenstates $\nu_i$ using \eq{eq:inverted2b}: 
\begin{equation}
\label{eq:lagCImass}
\mathcal{L}_{\text{CI}}= \frac{g_\ast^2}{\Lambda^2} \, \bar{q}_L\gamma^\mu  q_L' \, \left( - i\cos\theta\, \overline{\nu}_{1} + \sin\theta \,\overline{\nu}_{2} \right)\,\gamma_\mu \ell_L \, .
\end{equation}

\begin{figure*}[t!]
\centering
\begin{minipage}[t][\minipageheightdp][t]{\minipagewidthdp}
\centering
\includegraphics[width=\figuremaxwidthdp,height=\figuremaxheightdp,keepaspectratio]{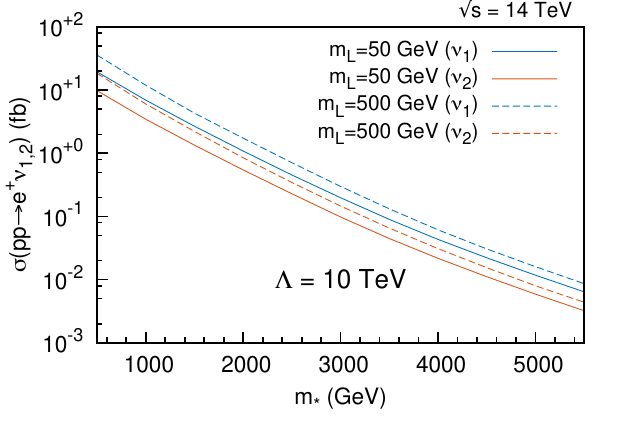}
\end{minipage}
\begin{minipage}[t][\minipageheightdp][t]{\minipagewidthdp}
\centering
\includegraphics[width=\figuremaxwidthdp,height=\figuremaxheightdp,keepaspectratio]{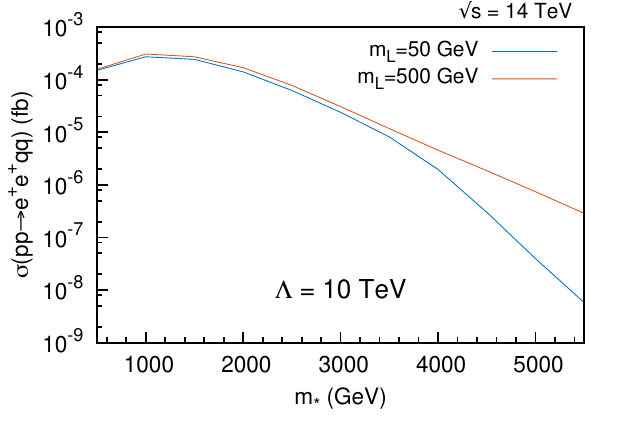}
\end{minipage}
\caption{Left: production cross section, at $\sqrt{s}=14\TeV$, for the two mass eigen-states $pp\rightarrow e^{+}\nu_{1,2}$ for two different Majorana mass values, $m_L=50, 500\GeV$. Right: cross section for the like-sign dileptons signature, $pp \to e^{+}e^{+} qq$ at the HL-LHC ($\sqrt{s}=14\TeV$).}\label{fig:xsec}
\end{figure*}

%%%conseguenze fenomenologiche del modello

The gauge interactions in \eq{eq:lagGImass} and the contact interactions in \eq{eq:lagCImass} have to be implemented in a MC generator. 
Here we have mostly used \madgraph{}~\cite{Alwall:2014hca}, complementing the results with a validation obtained with the tree-level simulator \calchep{}~\cite{Belyaev:2012qa}.
%We got rid of the four-fermion contact interaction by means of a point-like propagator as if the effective coupling is mediated by a very heavy unknown boson.

In \fig{fig:xsec} is shown the behaviour of production cross section for the two mass eigenstates and for the whole process for different model scenarios.
It is worth to notice that in the limit $m_{L} \rightarrow 0$ the cross-section of the like-sign dilepton process goes to zero. %\pf{define SS}

\begin{figure*}[t!]
\centering
\begin{minipage}[t][\minipageheightdp][t]{\minipagewidthdp}
\centering
\includegraphics[width=\figuremaxwidthdp,height=\figuremaxheightdp,keepaspectratio]{\main/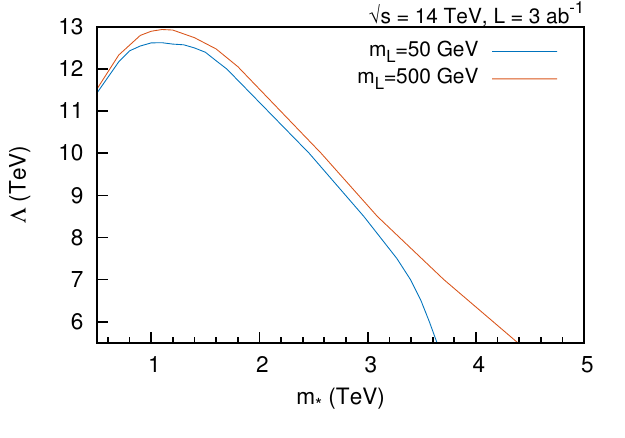}
\end{minipage}
\begin{minipage}[t][\minipageheightdp][t]{\minipagewidthdp}
\centering
\includegraphics[width=\figuremaxwidthdp,height=\figuremaxheightdp,keepaspectratio]{\main/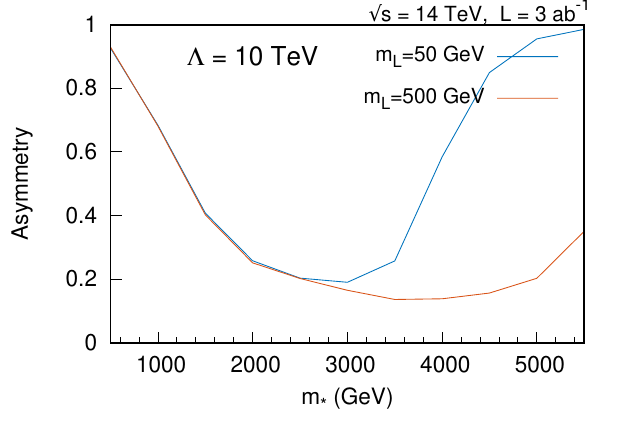}
\end{minipage}
\caption{\label{fig:results}Left: the $5\sigma$ contour levels curves, at $\sqrt{s}=14\TeV$ and ${\cal L}=3\abinv$, of the statistical significance ($S=5$) in the parameter plane $[m^*,\Lambda]$, for two different values of the Majorana mass $m_L=50, 500\GeV$. Right: the expected charge asymmetry at $\sqrt{s}=14\TeV$ and ${\cal L}=3\abinv$ of the same-sign and opposite sign channels for two different values of $m_L=50,500\GeV$ as function of $m_*$.}
\end{figure*}

We briefly discuss the potential for discovery at HL-LHC in the three-dimensional parameter space $(\Lambda, m_{L}, m_{*})$ using the fast simulation framework \delphes~\cite{deFavereau:2013fsa} for studying the hypothetical response of the CMS Phase-2 detector.
Standard model processes that could mimic the detection of a signal with lepton number violation in this rather clean signature are mainly the triple W boson production, $pp \rightarrow W^+ W^+ W^-$, and the top quark pair production $pp \rightarrow t \bar{t}$, the former being the dominant background source. %for cross-section magnitude.
Since the kinematic features of the final state reconstructed objects are similar to those of \citeref{Leonardi:2015qna}, we lower the background contribution by imposing two cuts on the leading lepton ($p_T (e_1) > 110\GeV$) and on the second-leading lepton ($p_T (e_2) > 35\GeV$).  
This particular signal region allows an efficiency in selecting signal around the $80 \%$, while beating the background sources with efficiency of $0.00044 \%$ for the $t\bar{t}$ and of $0.0034 \%$ for the  $W^+ W^+ W^-$. \\
We  compute, at a \com energy of $\sqrt{s}=14\TeV$ and with an integrated luminosity of ${\cal L}= 3\abinv$, the statistical significance, as defined by the relation $ S={\mathcal{L} \sigma_{\text{sig}} \epsilon_{\text{sig}}}/{\sqrt{\mathcal{L} \sigma_{\text{bkg}} \epsilon_{\text{bkg}}}}$, where $\epsilon_{\text{sig}},\epsilon_{\text{bkg}}$ are respectively the cumulative efficiencies of signal and background due to the event reconstruction. In \fig{fig:results} (left) we show the potential for discovery, in the plane $(m_*,\Lambda)$, of the CMS Phase-2 detector for two different values of the Majorana mass $m_L=50, 500\GeV$.

Another interesting feature of the present model is that the presence of two heavy neutrino mass eigenstates, which have also opposite \cp eigenvalues,  can lead to a charge asymmetry, \iee~to an asymmetry in the production rate of same-sign dilepton signal versus the  opposite-sign dilepton signal~\cite{Dev:2015pga}. 
In particular, the charge asymmetry ${\cal A} =( {\sigma_{e^+ e^- j j}-\sigma_{e^+ e^+ jj}})/({\sigma_{e^+ e^- j j}+\sigma_{e^+ e^+ j j}})$  could be used to test the relative strength between the Dirac and Majorana nature of the
heavy neutrinos at the HL-LHC. \\
We note that while the same-sign dilepton signature $e^+e^+jj$ is effectively background free (see above), the opposite sign dilepton signature $e^+e^-jj$ is expected to have a substantially larger SM background which however could be subtracted in order to extract the asymmetry ${\cal A}$ peculiar of the model features.
Figure \ref{fig:results} (right) shows the charge asymmetry ${\cal A}$. In line with expectations that ${\cal A} \to 1$ when $\sigma_{e^+ e^+ jj}$ is suppressed we observe that as a function of $m_*$ after the initial drop in the asymmetry, at fixed $m_L$,  then at larger values of $m_*$ the mass difference of the two eigenstates diminishes and so does the same-sign dilepton yield giving larger asymmetries close to 1 again.

%\subsubsection{Conclusions}

In summary, we have presented a mechanism of lepton  number violation within a mirror type compositeness scenario. The mirror model is realised when the left components of the excited states are singlets while the right components are active doublets which do participate to gauge transition interactions with SM fermions. We therefore introduce a \emph{left}-handed Majorana neutrino singlet of mass $m_L$ while the right-handed component belongs to a doublet and has a Dirac mass $m_*$.  This situation is exactly specular to the one encountered in typical see-saw models where the sterile neutrino is the right handed component and the active one is the left handed component.
diagonalisation of the mass matrix gives two Majorana mass eigenstates whose phenomenology at the HL-LHC is presented, providing the $5\sigma$ contour level curves of the statistical significance for a  possible discover by a CMS Phase-2 like detector. We have also shown that the charge asymmetry ${\cal A}$ of the like sign dilepton signature is a useful observable for studying the model parameter space.

%\paragraph*{Acknowledgments}

\subsubsection{\cms{Search for heavy composite Majorana neutrinos at the HL- and the HE-LHC}}
\label{sec:neutrino553}
\contributors{P. Azzi, C. Cecchi, L. Fan\'{o}, A. Gurrola, W. Johns, R. Leonardi, E. Manoni, M. Narain, O. Panella, M. Presilla, F. Romeo, S. Sagir, P. Sheldon, F. Simonetto, E. Usai, W. Zhang, CMS}

Compositeness of ordinary fermions is one possible BSM scenario that may lead to a solution of the hierarchy problem or to the explanation of the proliferation of ordinary fermions. Elementary particles are thought to be bound states of some as--yet--unobserved fundamental constituents generically referred to as \emph{preons}. Composite models also predict the existence of excited states of quarks and leptons, with masses lower than or equal to the compositeness scale $\Lambda$ which interact with SM fermions via both magnetic type gauge couplings and contact interactions.

The heavy composite Majorana neutrino $N_{\ell}$ would be a particular case of such excited states. This can be produced in association with a lepton, in $pp$ collisions, via quark--antiquark annihilation ($q\bar{q}^\prime\rightarrow \ell N_{\ell}$). 
The production and decay processes can occur via both gauge and contact interactions.

The Lagrangian density for gauge mediated interactions is
\begin{equation}
\mathcal{L}_{G} =\frac{1}{2 \Lambda} \, \overline{L^\ast_\text{R}} \sigma^{\mu\nu}\left(gf \frac{\overrightarrow{\tau}}{2}\cdot \overrightarrow{W}_{\mu\nu} +g'f' Y B_{\mu\nu}\right) L_\text{L} +h.c.,
\end{equation}
where $L_R^*$ and $L_L$ are, respectively, the right-handed doublet of the excited fermions and the left-handed doublet of the SM, $g$ and $g^\prime$ are the $SU(2)_L$ and $U(1)_Y$ gauge couplings, and $f$ and $f^\prime$ are dimensionless couplings, which are expected to be of order unity~\cite{Baur:1989kv} and henceforth simply assumed to be $1$.  
The corresponding Lagrangian describing the four-fermion contact interactions by a dimension-6 effective operator can written as
\begin{equation}
\mathcal{L}_{C} =\frac{g_\ast^2}{\Lambda^2}\frac{1}{2}j^\mu j_\mu,
\end{equation}
with
\begin{equation}
\label{Contact}
j_\mu =\eta_L\bar{\psi}_L\gamma_\mu \psi_L+\eta\prime_L\bar{\psi}^\ast_L\gamma_\mu \psi^\ast_L+\eta\prime\prime_L\bar{\psi}^\ast_L\gamma_\mu \psi_L + h.c. +(L\rightarrow R),
\end{equation}
where $g_\ast^2=4\pi$, the $\eta$ factors that define the chiral structure are usually set equal to $1$, and $\psi$ and $\psi^\ast$ are the SM and excited fermion fields~\cite{Baur:1989kv}.

The production process is dominated by the contact interaction mechanism for all values of the compositeness scale $\Lambda$ and of the mass of the neutrino $M(N_{\ell})$ relevant in this analysis, while for the decay the dominant interaction changes depending on $\Lambda$ and $M(N_{\ell})$~\cite{Leonardi:2015qna}.

This study from CMS focuses on the final state signature $\ell\ell q\bar{q}^\prime$, where $\ell$ is either an electron or a muon~\cite{CMS:2018zbo}. For the HL-LHC sensitivity study, we use MC samples for the signal and the SM backgrounds. The MC samples for the signal are generated with \calchep{ v3.6}~\cite{Belyaev:2012qa} using the NNPDF3.0 LO parton distribution functions~\cite{Ball:2014uwa}.
The background samples considered are top quark pair production (t$\bar{\mbox t}$), single top quark production (tW), Drell-Yan (DY) process, W+jets and diboson production (WW, WZ, ZZ), and are generated with \madgraph{5\_aMC@NLO}~\cite{Alwall:2014hca} using the CTEQ6L1 PDF set~\cite{Pumplin:2002vw}. 
For all of the MC samples the hadronisation of partons is simulated with \pythia{ 8}~\cite{Sjostrand:2014zea} and the expected response of the upgraded CMS detector is performed with the fast-simulation package \delphes{}~\cite{deFavereau:2013fsa}.
The contribution from additional pileup events has been included in the simulation as well.

In order to reduce the contamination from misreconstructed events, a kinematics-based selection is applied. The $p_{T}$ of the leading lepton is required to be greater than $110\GeV$, while the $p_{T}$ of the subleading lepton must greater than 40 GeV. All lepton candidates are required to be in the pseudorapidity range $|\eta|<2.4$. Restricting to the high-mass region given by $M(\ell,\ell)>300\GeV$, where $M(\ell,\ell)$ is the dilepton invariant mass, allows reducing the DY background without affecting the signal acceptance.
The large-radius jets are analysed using the PUPPI algorithm~\cite{Bertolini:2014bba}.
They are reconstructed with a distance parameter of $R=0.8$ and they are required to have a minimum $p_{T}$ of $200\GeV$, to be within the region $|\eta|<2.4$ and to be separated from leptons by a distance $\Delta R = \sqrt{(\Delta \eta)^{2} + (\Delta \phi)^{2}} > 0.8$. Requiring one or more large-radius jets is suitable regardless of whether $N_\ell$ decays through gauge or contact interactions. In fact, for gauge mediated decays of the heavy composite neutrino, the two quarks are expected to overlap and thus form a large-radius jet, while in the case of contact-mediated decays, the two quarks are well separated, but form two large-radius jets because of the overlap with final state radiation. The signal region is therefore defined by requiring two same-flavour isolated leptons (electrons or muons) and at least one large-radius jet. 

\begin{figure}[t]
\centering
\begin{minipage}[t][\minipageheightsp][t]{\minipagewidthsp}
\centering
\includegraphics[width=\figuremaxwidthsp,height=\figuremaxheightsp,keepaspectratio]{\main/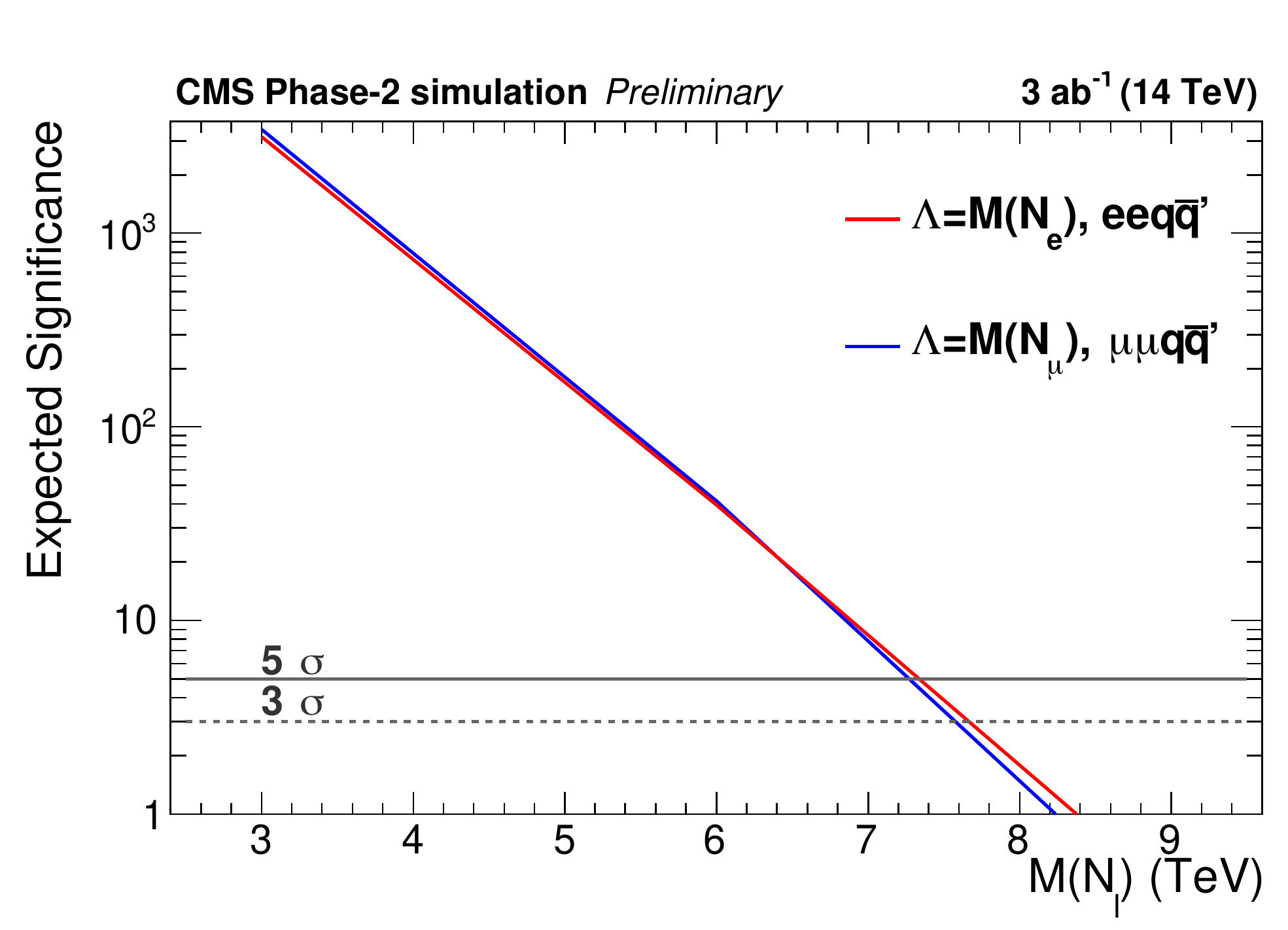}
\end{minipage}
\caption{Expected statistical significance for the HL-LHC projection of the $eeq\bar{q}^\prime$ (red line) and $\mu\mu q\bar{q}^\prime$ (blue line) channel for the case $\Lambda=M(N_\ell)$. The grey solid (dotted) line represents $5(3)\sigma$, respectively.}
\label{fig:stat-significance}
\end{figure}

A shape-based analysis is performed looking at the invariant mass distribution of the two leptons and the leading large-radius jet, $M(\ell \ell J)$.
The expected discovery sensitivity of a heavy composite Majorana neutrino, produced in association with a lepton, and decaying into a same-flavour lepton and two jets, is shown in  \fig{fig:stat-significance}. The CMS Phase-2 detector will be able to find evidence for a composite neutrino with mass below $M(N_\ell)=7.6\TeV$. 
%%%%%%%%stat signif
%%%%%%%%%%%%%%
The $M(\ell \ell J)$ distributions of signal and SM backgrounds are also used as input in the computation of an upper limit at the $95\%\cl$ on the cross section of the heavy composite Majorana neutrino produced in association with a lepton times its branching fraction to a same-flavour lepton and two quarks, $\sigma(pp \rightarrow \ell N_{\ell}) \times {\cal B}(N_{\ell} \rightarrow \ell q\bar{q}^\prime)$. A $CL_s$ criterion~\cite{Read:2002hq} is used to set upper limits. Systematic uncertainties are included on the integrated luminosity ($1\%$), pileup ($2\%$), electron ID ($0.5\%$), electron scale ($0.5\%$), muon ID ($0.5\%$), muon scale ($0.5\%$), jet energy scale ($1\%$), jet energy resolution ($1\%$), the background prediction ($0.3\%$), and Drell-Yan theory ($4\%$), and are evaluated in accordance with the most recent recommendations described in \citeref{Syst_uncert_UPSG}.

\begin{figure}[t]
\centering
\begin{minipage}[t][\minipageheightdp][t]{\minipagewidthdp}
\centering
\includegraphics[width=\figuremaxwidthdp,height=\figuremaxheightdp,keepaspectratio]{\main/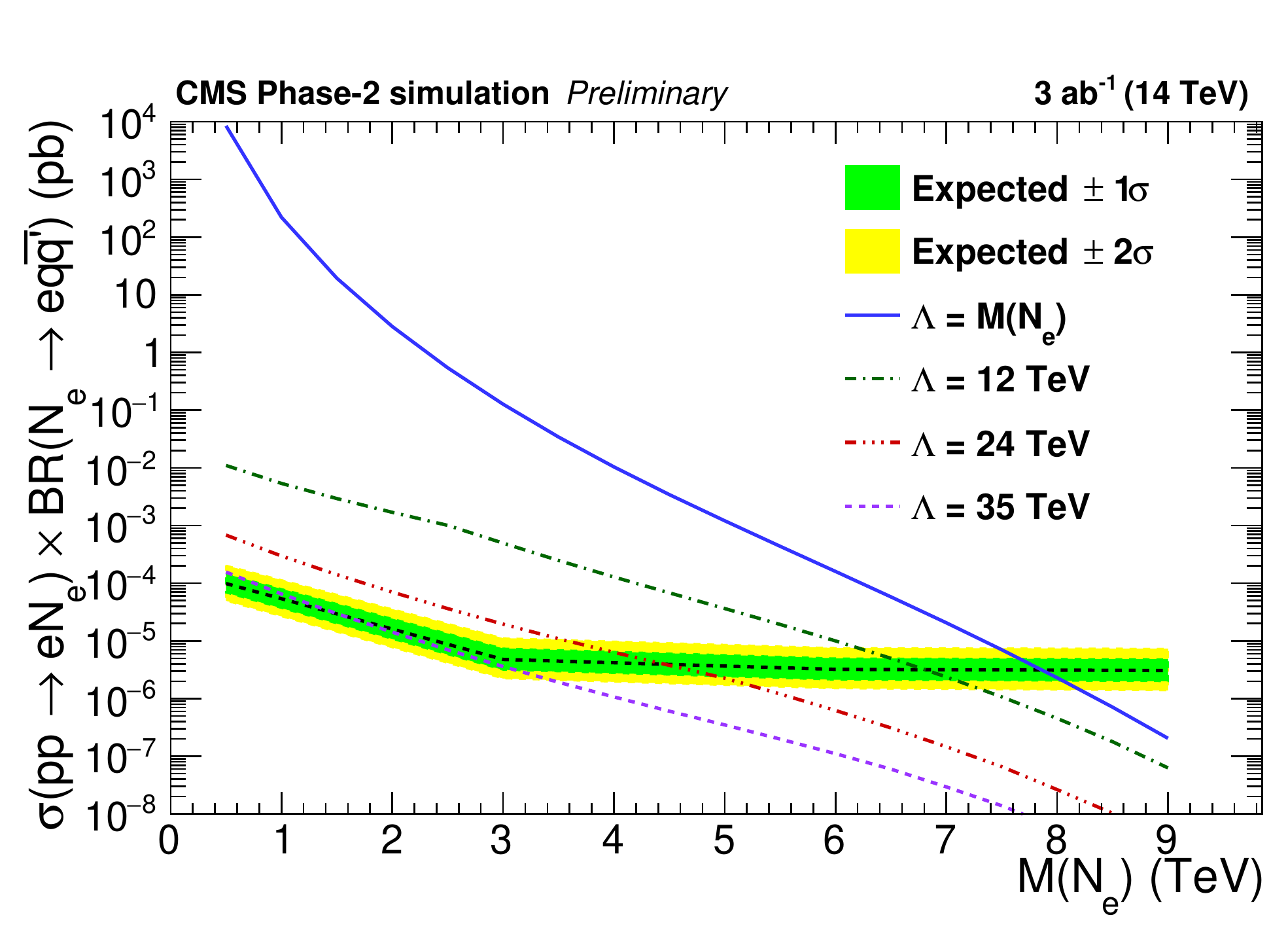}
\end{minipage}
\begin{minipage}[t][\minipageheightdp][t]{\minipagewidthdp}
\centering
\includegraphics[width=\figuremaxwidthdp,height=\figuremaxheightdp,keepaspectratio]{\main/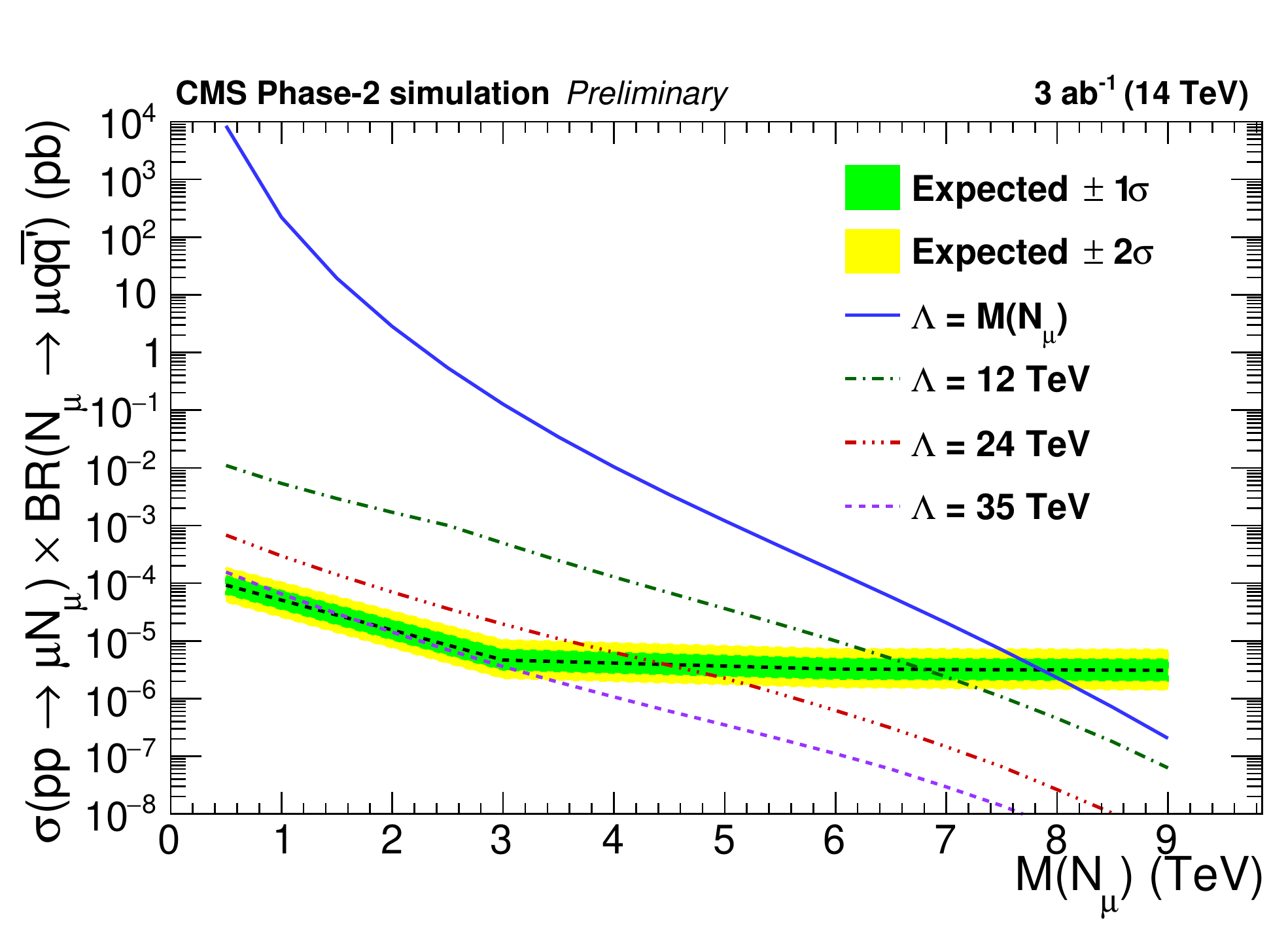}
\end{minipage}
\caption{Expected $95\%\cl$ upper limits for the HL-LHC projection (black dotted lines) on \sloppy\mbox{$\sigma(pp \rightarrow \ell N_{\ell}) \times {\cal B}(N_{\ell} \rightarrow \ell q\bar{q}^\prime)$}, obtained in the analysis of the $ee q\bar{q}^\prime$ (left) and the $\mu\mu q\bar{q}^\prime$ (right) final states, as a function of the mass of the heavy composite Majorana neutrino. The corresponding green and yellow bands represent the expected variation of the limit to one and two standard deviation(s). The solid blue curve indicates the theoretical prediction of $\Lambda=M(N_{l})$. The textured curves give the theoretical predictions for $\Lambda$ values ranging from $12$ to $35\TeV$.}
\label{fig:Limit_1D_YR}
\end{figure}

\begin{figure}[t]
\centering
\begin{minipage}[t][\minipageheightdp][t]{\minipagewidthdp}
\centering
\includegraphics[width=\figuremaxwidthdp,height=\figuremaxheightdp,keepaspectratio]{\main/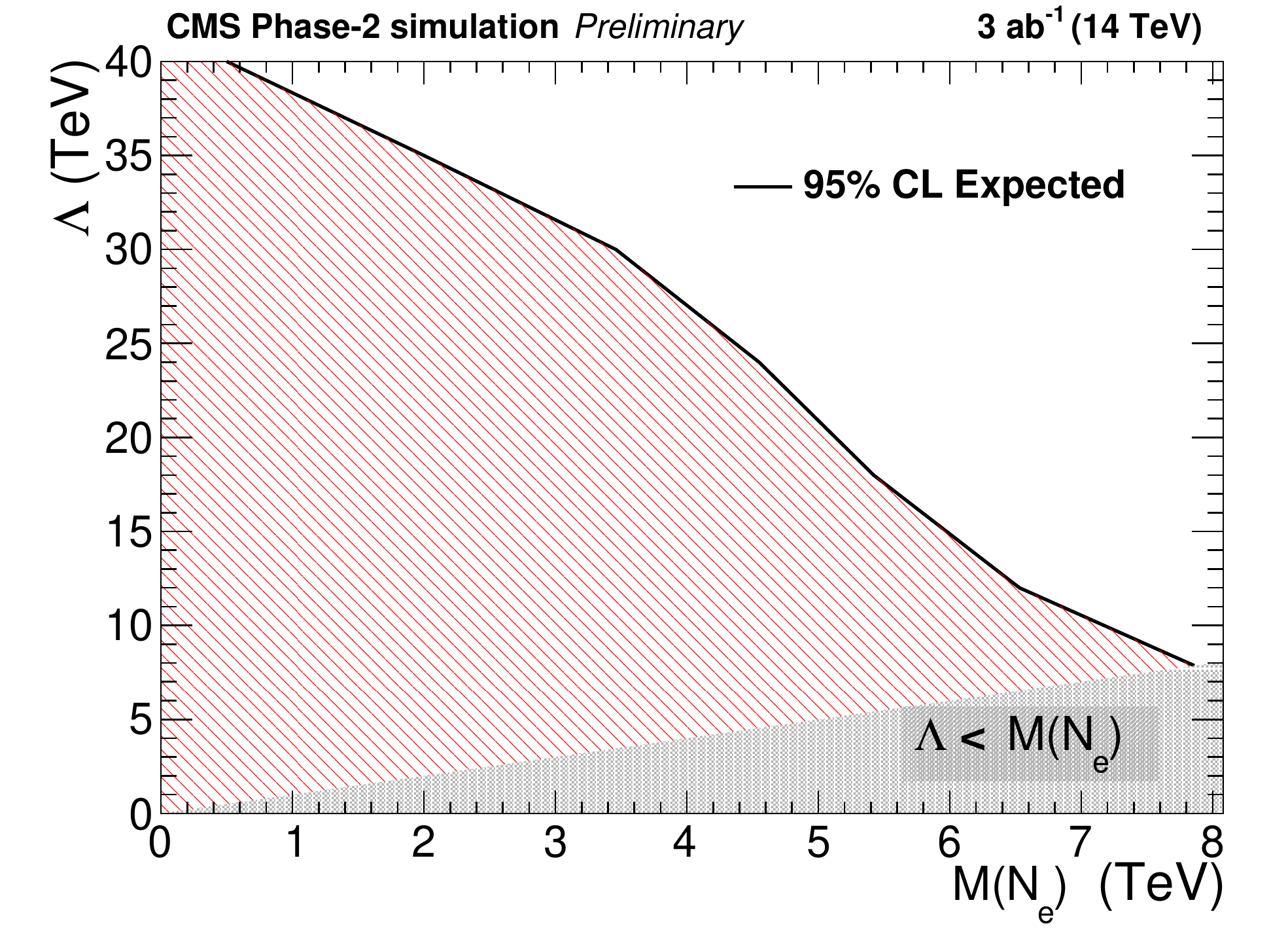}
\end{minipage}
\begin{minipage}[t][\minipageheightdp][t]{\minipagewidthdp}
\centering
\includegraphics[width=\figuremaxwidthdp,height=\figuremaxheightdp,keepaspectratio]{\main/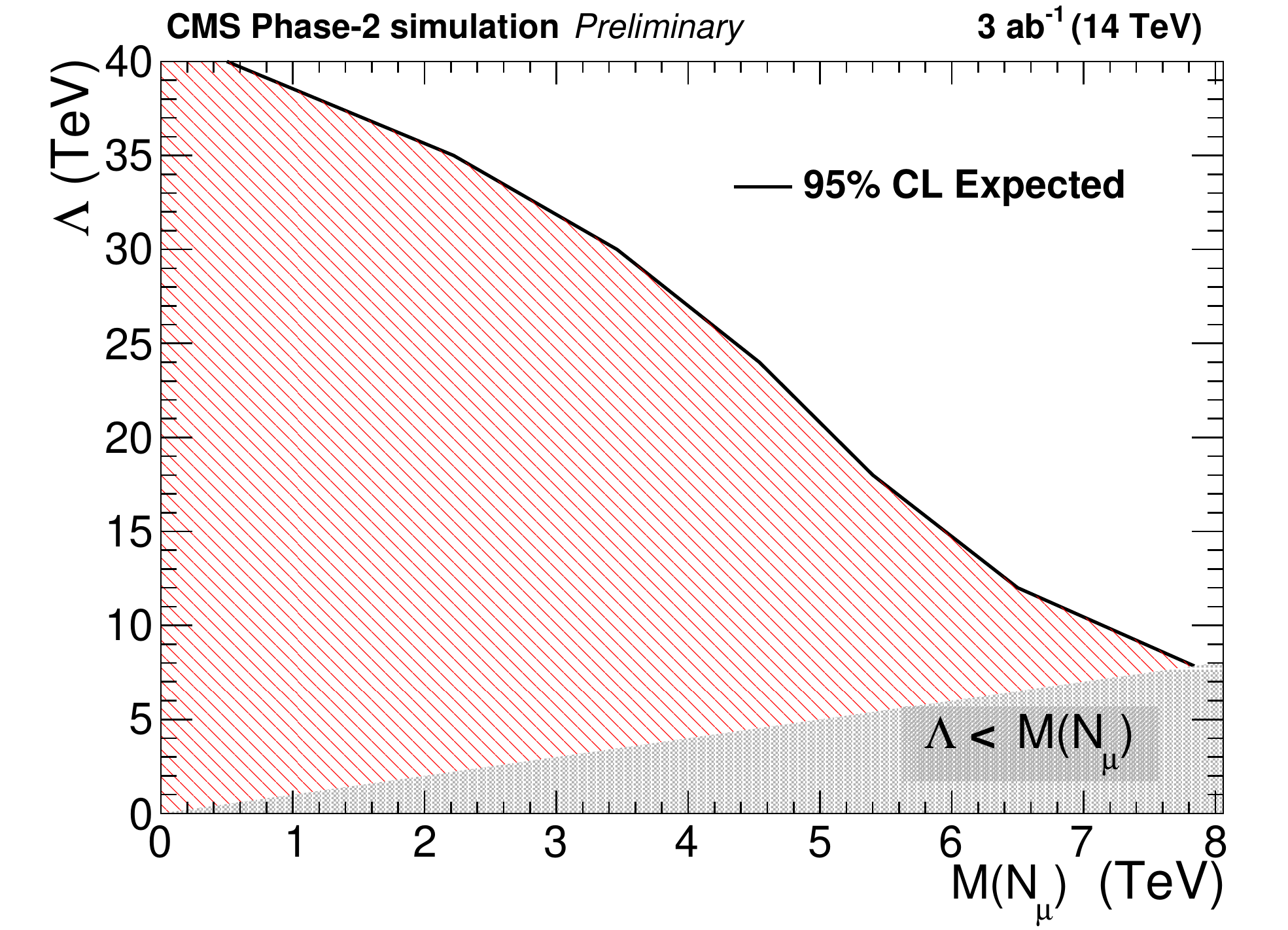}
\end{minipage}
\caption{Expected $95 \%$ \cl lower limits (black lines) on the compositeness scale $\Lambda$, obtained in the analysis of the $eeq\bar{q}^\prime$ (left) and the $\mu\mu q\bar{q}^\prime$ (right) final states, as a function of the mass of the heavy composite Majorana neutrino. The gray zone corresponds to the phase space $\Lambda < M(N_{\ell})$ not allowed by the model.}
\label{fig:Limit_2D_YR}
\end{figure}

The results are shown in \fig{fig:Limit_1D_YR} for the $ee q\bar{q}^\prime$ and $\mu\mu q\bar{q}^\prime$ channels.
Figure~\ref{fig:Limit_2D_YR} displays the corresponding upper limits on the $(\Lambda, M(N_{\ell}))$ plane.
The HL-LHC running conditions and Phase-2 detector will significantly extend the region of parameter space that can be probed. While in Run-2 the mass of the heavy composite Majorana neutrino could be excluded up to $4.60$ ($4.70$)\TeV in the \sloppy\mbox{$ee q\bar{q}^\prime$ ($\mu\mu q\bar{q}^\prime$)} channel~\cite{Sirunyan:2017xnz}, for the case of $\Lambda=M(N_{\ell})$, for the HL-LHC we could potentially exclude a composite neutrino up to a mass of $8\TeV$ with the same assumption on the compositeness scale.

%%%%%1D limilts
%%%%%%2D limits
%%%%%%%%%%%%%%

\begin{figure}[t]
\centering
\begin{minipage}[t][\minipageheightsp][t]{\minipagewidthsp}
\centering
\includegraphics[width=\figuremaxwidthsp,height=\figuremaxheightsp,keepaspectratio]{\main/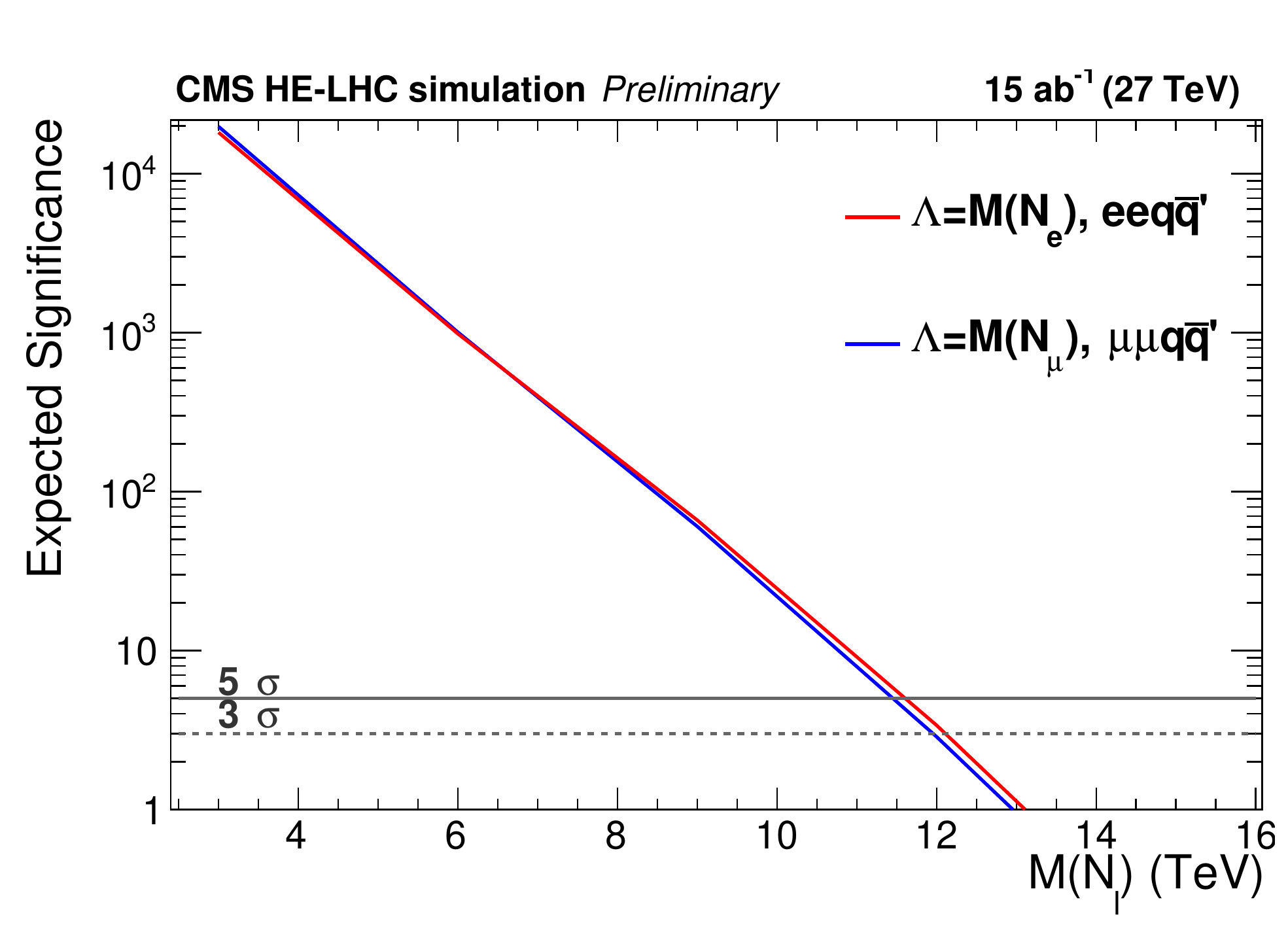}
\end{minipage}
\caption{Expected statistical significance for the HE-LHC projection of the $eeq\bar{q}^\prime$ (red line) and the $\mu\mu q\bar{q}^\prime$ (blue line) channel for the case $\Lambda=M(N_\ell)$. The gray solid (dotted) line represents $5(3)\sigma$, respectively.}
\label{fig:stat-significance-HE}
\end{figure}

\begin{figure}[t]
\centering
\begin{minipage}[t][\minipageheightdp][t]{\minipagewidthdp}
\centering
\includegraphics[width=\figuremaxwidthdp,height=\figuremaxheightdp,keepaspectratio]{\main/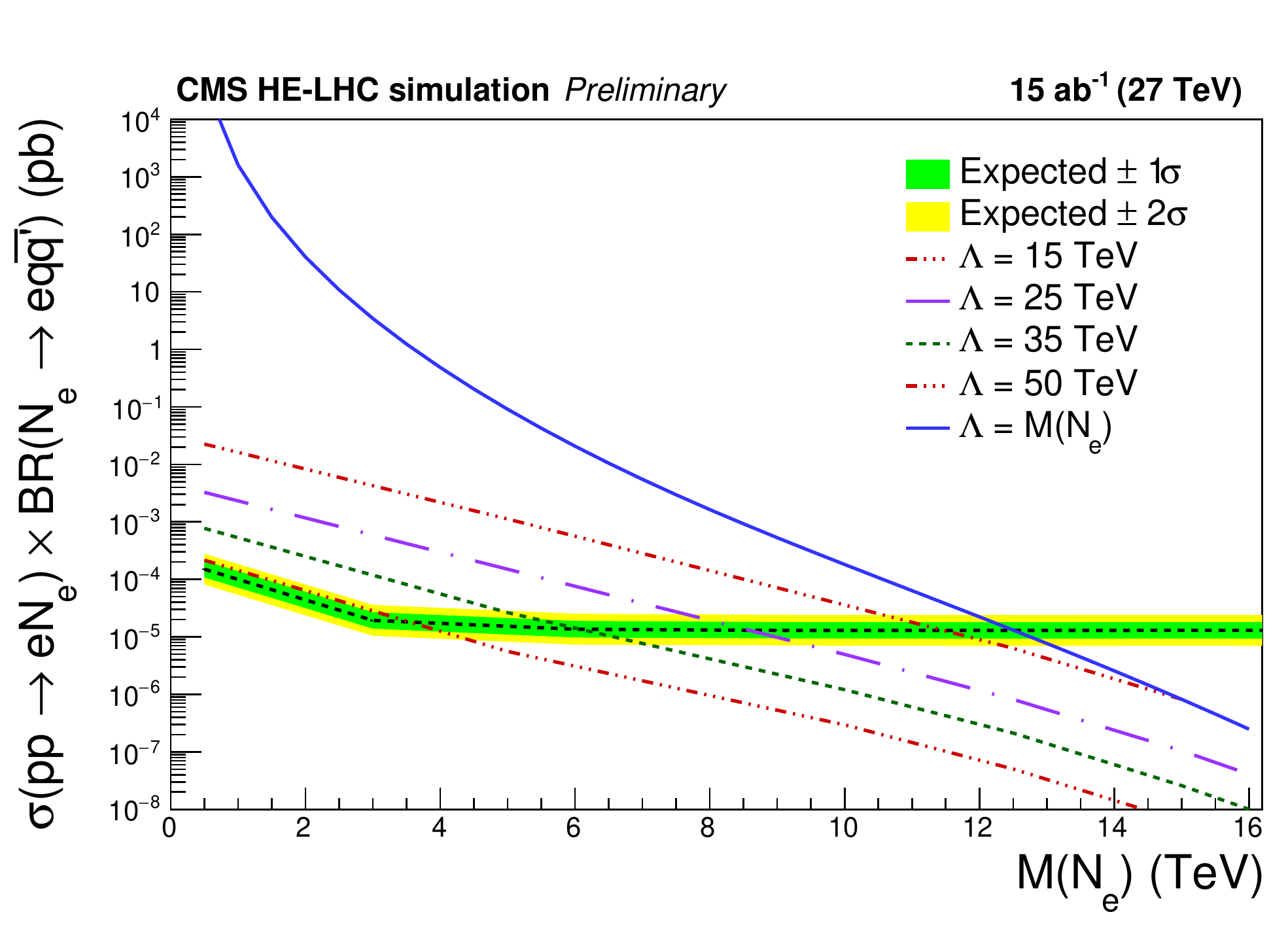}
\end{minipage}
\begin{minipage}[t][\minipageheightdp][t]{\minipagewidthdp}
\centering
\includegraphics[width=\figuremaxwidthdp,height=\figuremaxheightdp,keepaspectratio]{\main/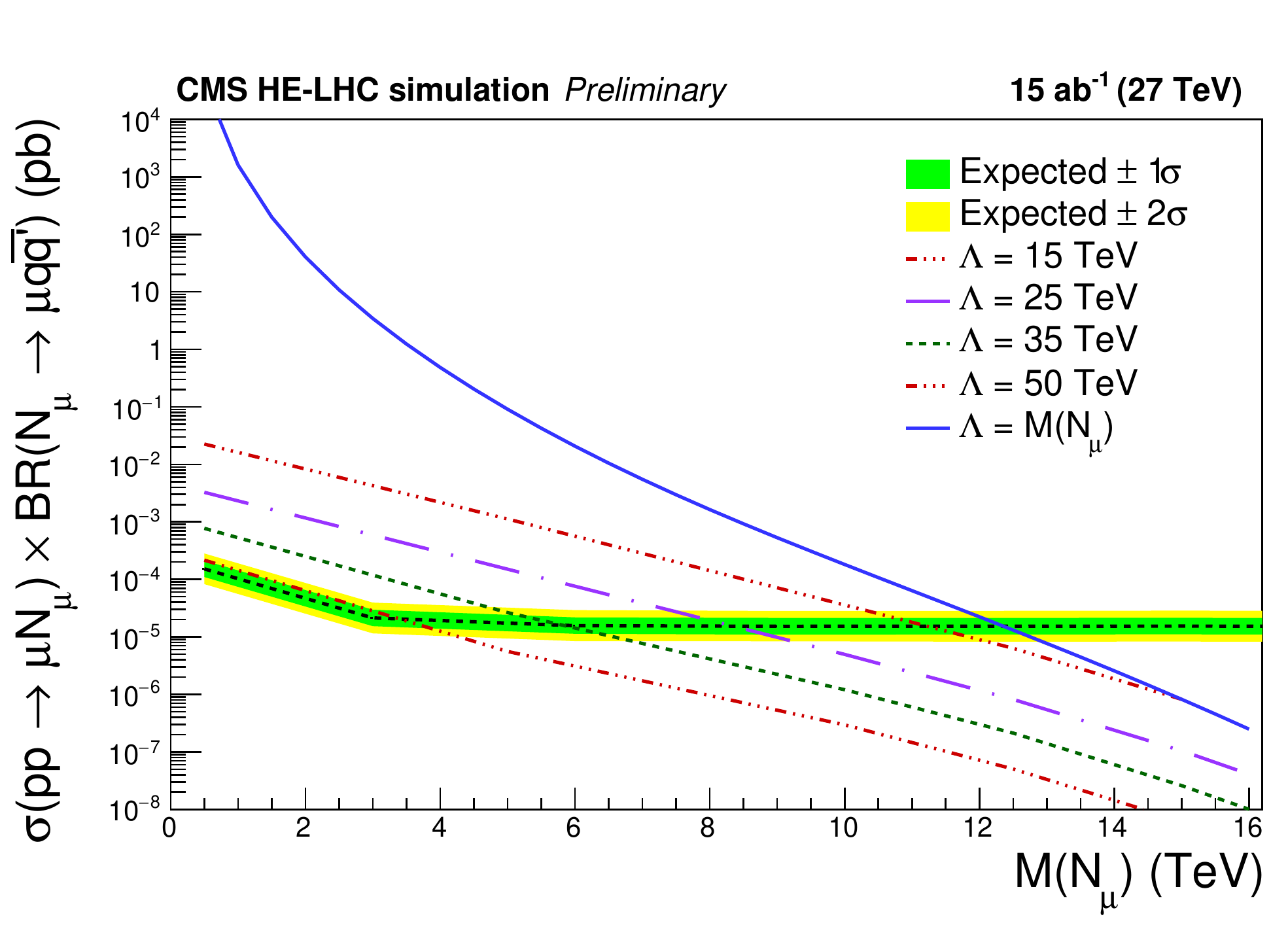}
\end{minipage}
  \caption{Expected $95\%\cl$ upper limits for the HE-LHC projection of the $ee q\bar{q}^\prime$ channel (left) and the $\mu\mu q\bar{q}^\prime$ channel (right).}
\label{fig:HCMN_limit}
\end{figure}

\begin{figure}[t]
\centering
\begin{minipage}[t][\minipageheightdp][t]{\minipagewidthdp}
\centering
\includegraphics[width=\figuremaxwidthdp,height=\figuremaxheightdp,keepaspectratio]{\main/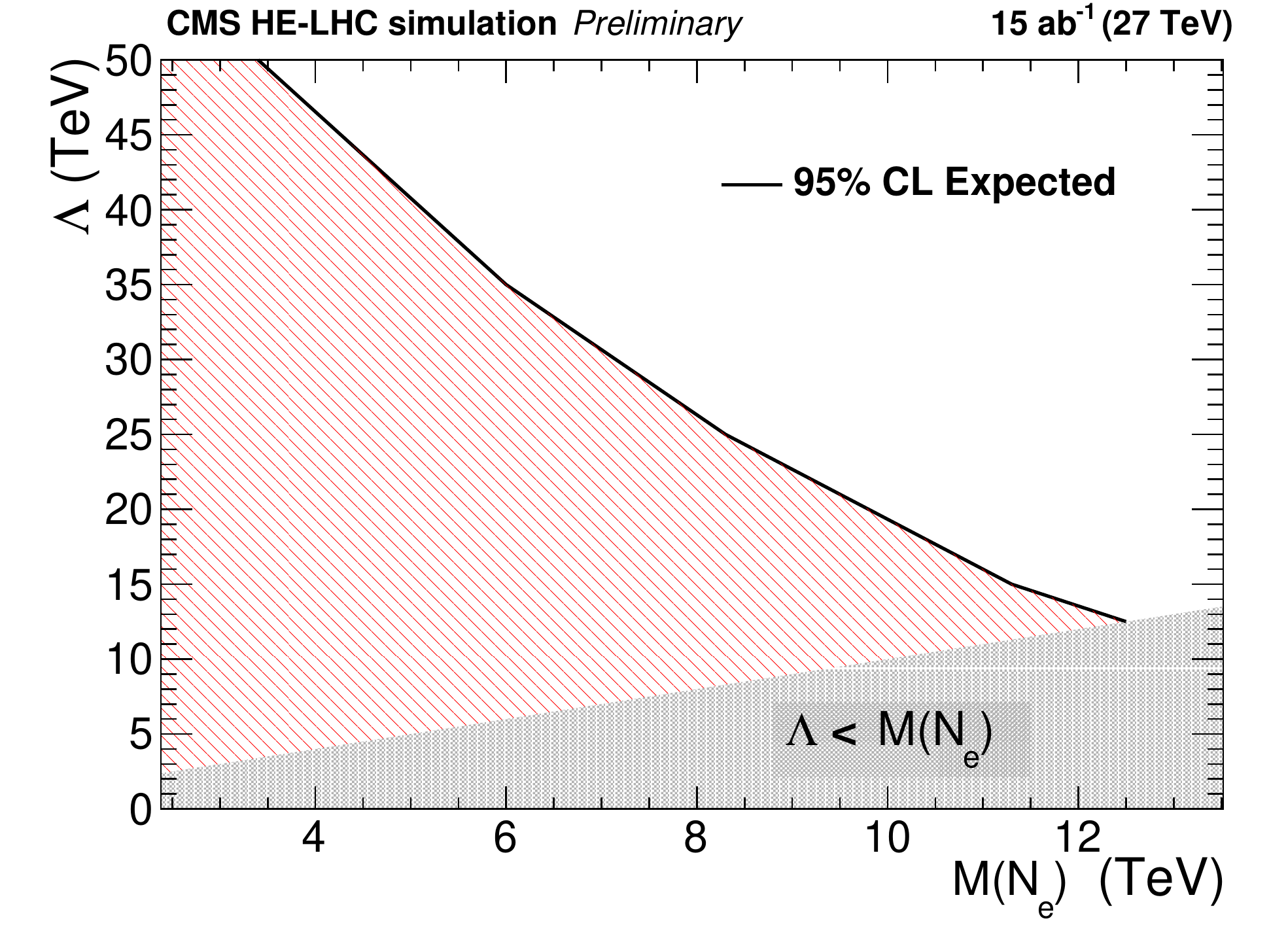}
\end{minipage}
\begin{minipage}[t][\minipageheightdp][t]{\minipagewidthdp}
\centering
\includegraphics[width=\figuremaxwidthdp,height=\figuremaxheightdp,keepaspectratio]{\main/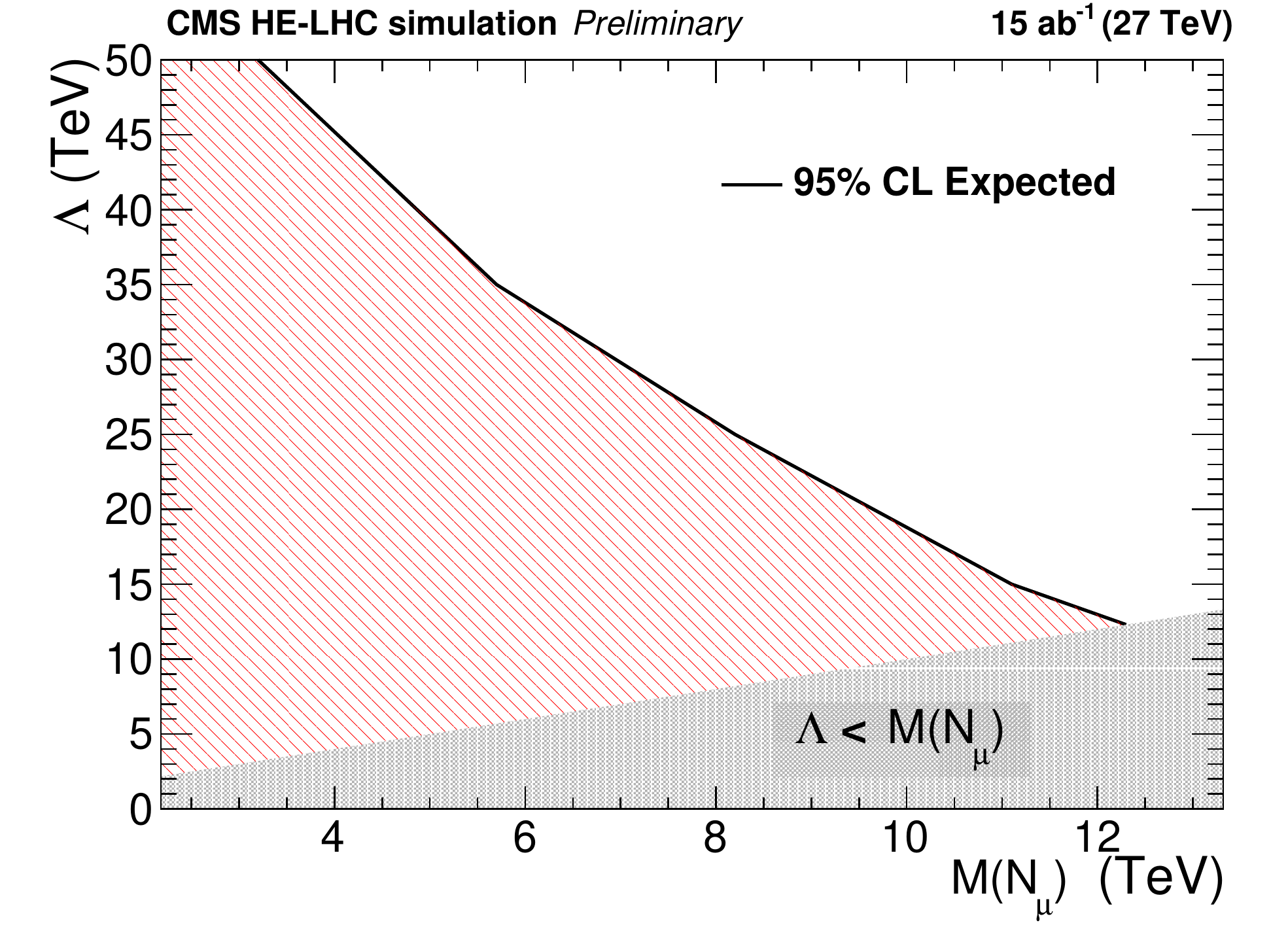}
\end{minipage}
\caption{Expected $95\%\cl$ lower limits (black lines) on the compositeness scale $\Lambda$, obtained in the analysis of the $eeq\bar{q}^\prime$ (left) and the $\mu\mu q\bar{q}^\prime$ (right) final states, as a function of the mass of the heavy composite Majorana neutrino for the HE-LHC projection. The grey zones are not allowed by the model.}
\label{fig:Limit_2D_HE}
\end{figure}

The sensitivity of the search is also considered for the High-Energy LHC at $\sqrt{s}=27\TeV$. Figure~\ref{fig:stat-significance-HE} shows that with the HE-LHC we could find evidence for a composite Majorana neutrino with mass below $M(N_{\ell}) = 12\TeV$, for $\Lambda = M(N_{\ell})$. Figure~\ref{fig:HCMN_limit} shows the results for several values of $\Lambda$.
The projection of the exclusion limits is also presented in \fig{fig:Limit_2D_HE} for the $(\Lambda, M(N_\ell))$ plane.
We conclude that, given the model condition $\Lambda = M(N_\ell)$, the HE-LHC could exclude a heavy composite Majorana neutrino with mass up to $12.5\TeV$ in both $ee q\bar{q}^\prime$ and $\mu\mu q\bar{q}^\prime$ channels.

%%%%%%%% 1D HE
%%%%%%2D limit HE

\subsection{Leptoquarks and $Z'$}
\label{sec:zprimeLQ}

Leptoquarks are hypothetical particles that carry both baryon and lepton quantum numbers. They are colour-triplets and carry fractional electric charge. The spin of a LQ state is either 0 (scalar LQ) or 1 (vector LQ). At the LHC, the pair-production of LQs is possible via gluon-gluon fusion and quark-antiquark annihilation and the production cross section only depends on the mass of the LQ. For scalar LQs, it is known at NLO in perturbative QCD~\cite{Kramer:2004df}.  The LQ may also be singly produced, in association with a lepton, but the cross section is model dependent.   
New massive vector bosons, $Z'$, are a common feature of NP models.  Typically they are assumed to couple in a flavour independent fashion.  However, it is possible to build models where these couplings are generation dependent and, after moving to the mass basis, become inter-generational.  Constraints on such couplings are weaker for the third generation. Such models have been invoked to explain several $B$-physics anomalies.

In this section the reach of the HL-LHC for LQs in the $t+\tau$ and $t+\mu$ channel is discussed in \secc{secc:cmsLQttautmu}. The reach of HL- and HE-LHC for models capable of explaining the $B$-physics anomalies is addressed in \secc{sec:theoryzprimeLQBanomalies}.  HL-LHC searches for LQs in $b+\tau$ final states is discussed in \secc{sec:leptoqbtau}, while the HE-LHC capability is considered in \secc{sec:LQHELHC}.

\subsubsection{\cms{Leptoquark searches in $t$+$\tau$ and $t$+$\mu$ decays at HL-LHC}}
\label{secc:cmsLQttautmu}
\contributors{J. Haller, R. Kogler, A. Reimers, CMS}
%{\bf Author(s): A. Reimers$^a$, R. Kogler$^a$, J. Haller$^a$, CMS}
%\noindent \textit{a) Institut f\"{u}r Experimentalphysik, Universit\"{a}t Hamburg, Germany}

%Leptoquarks (LQs) are hypothetical particles that carry both baryon and lepton quantum numbers. They are colour-triplets and carry fractional electric charge. The spin of a LQ state is either 0 (scalar LQ) or 1 (vector LQ). At the LHC, the pair-production of LQs is possible via gluon-gluon fusion and quark-antiquark annihilation. The pair production cross section depends on the mass of the LQ. For scalar LQs, it is known at NLO in perturbative QCD~\cite{Kramer:2004df}.

The reach of searches for pair production of LQs with decays to $t+\mu$ and $t+\tau$ at CMS is studied for the HL-LHC with target integrated luminosities of 
$\mathcal{L}_{\text{int}}^{\text{target}} = 300\fbinv$ and \intLumi~\cite{CMS:2018yke}. The studies are based on projecting signal and background event yields to HL-LHC conditions from published CMS results of the $t$+$\mu$~\cite{Sirunyan:2018ruf} and $t$+$\tau$~\cite{Sirunyan:2018nkj} LQ decay channels which use data from proton-proton collisions at $\sqrt{s}=13\TeV$ corresponding to $\mathcal{L}_{\text{int}} = 35.9\fbinv$ recorded in 2016. While the analysis strategies are kept unchanged with respect to the ones in \citerefs{Sirunyan:2018nkj, Sirunyan:2018ruf}, different total integrated luminosities, the higher \com~energy of $14\TeV$, and different scenarios of systematic uncertainties are considered. In the first scenario (denoted ``w/ YR18 syst. uncert.''), the relative experimental systematic uncertainties are scaled by a factor of $1 / \sqrt{f}$, with 
$f=\mathcal{L}_{\text{int}}^{\text{target}}/35.9\fbinv$, 
until they reach a defined lower limit based on estimates of the achievable accuracy with the upgraded detector~\cite{Collaboration:2650976} as described in~\secc{sec:methods:syst}. The relative theoretical systematic uncertainties are halved. In the second scenario (denoted ``w/ stat. uncert. only''), no systematic uncertainties are considered. The relative statistical uncertainties in both scenarios are scaled by $1 / \sqrt{f}$.

\begin{figure}[t]
\centering
\begin{minipage}[t][\minipageheightdp][t]{\minipagewidthdp}
\centering
\includegraphics[width=\figuremaxwidthdp,height=\figuremaxheightdp,keepaspectratio]{\main/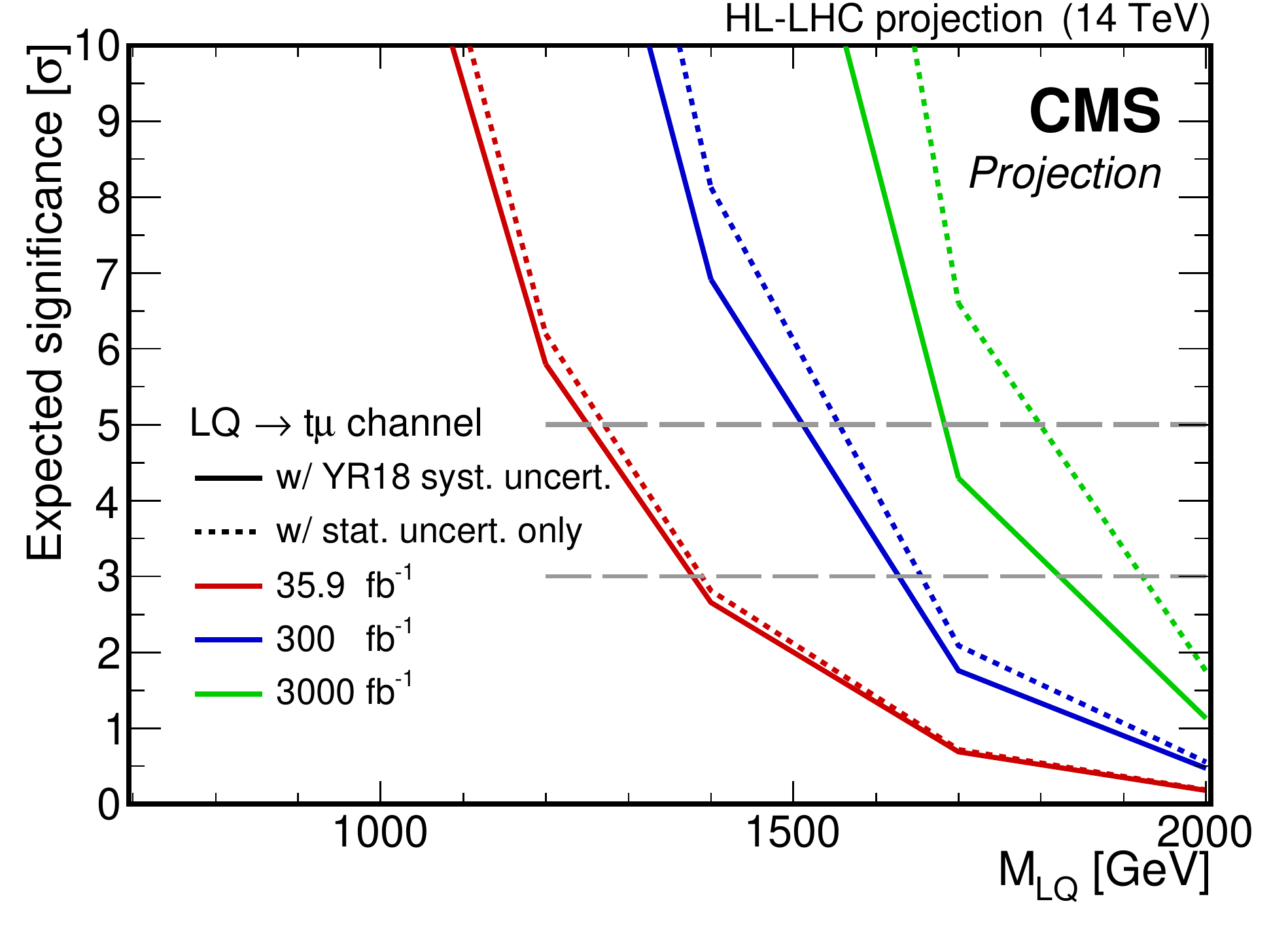}
\end{minipage}
\begin{minipage}[t][\minipageheightdp][t]{\minipagewidthdp}
\centering
\includegraphics[width=\figuremaxwidthdp,height=\figuremaxheightdp,keepaspectratio]{\main/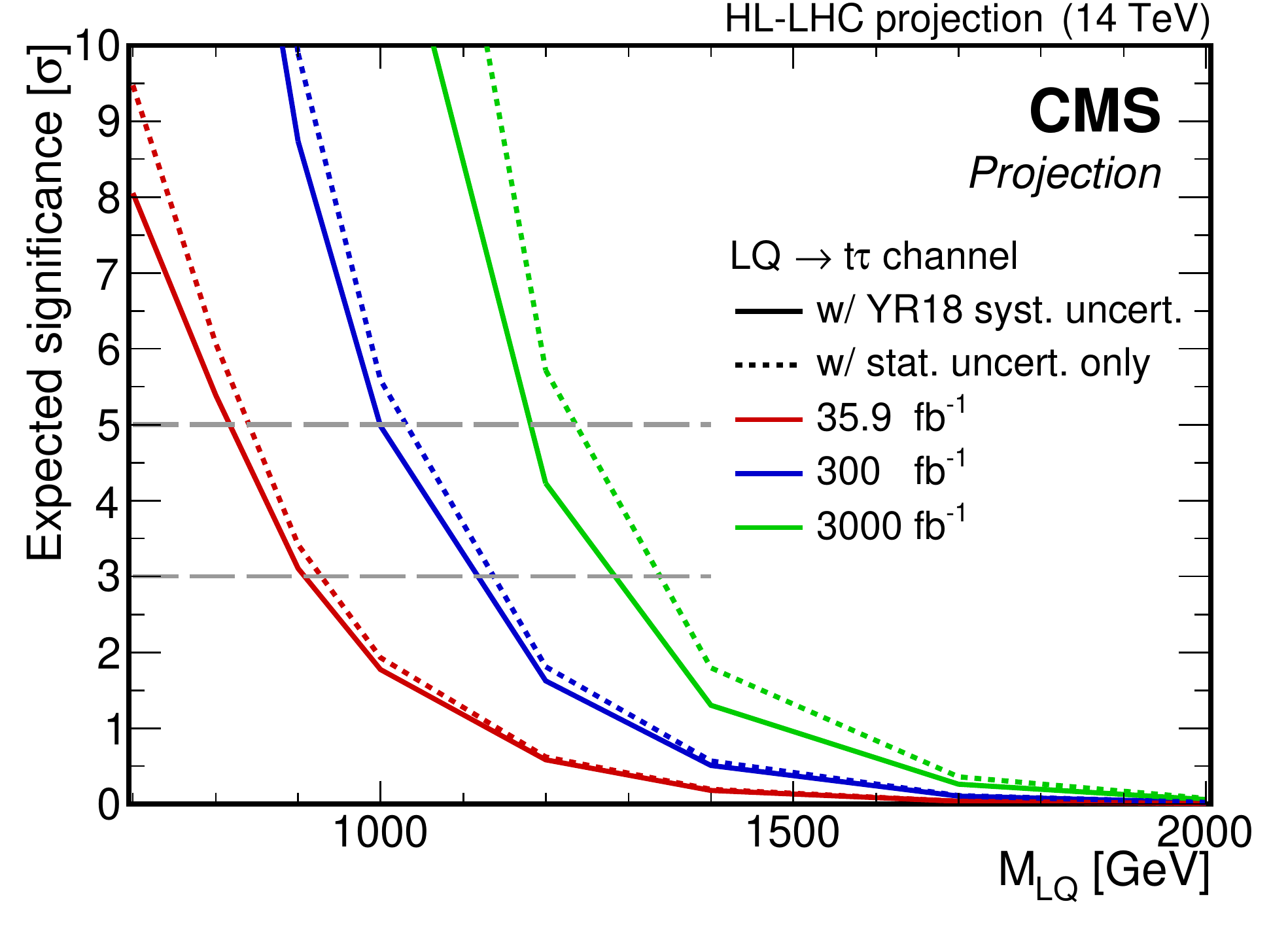}
\end{minipage}
\caption{Expected significances for an LQ decaying exclusively to top quarks and muons (left) or $\tau$ leptons (right).}
\label{fig:LQpairs:significances}
\end{figure}

Figure~\ref{fig:LQpairs:significances} presents the expected signal significances of the analyses as a function of the LQ mass for different assumed integrated luminosities in the ``w/ YR18 syst. uncert.'' and ``w/ stat. uncert. only'' scenarios. Increasing the target integrated luminosity to $\mathcal{L}_{\text{int}}^{\text{target}} = 3\abinv$ greatly increases the discovery potential of both analyses. The LQ mass corresponding to a discovery at $5\sigma$ significance with a dataset corresponding to \intLumi increases by more than 500\,GeV compared to the situation at $\mathcal{L}_{\text{int}}^{\text{target}} = 35.9\fbinv$, from about $1200\GeV$ to roughly $1700\GeV$, in the $\text{LQ}\rightarrow t\mu$ decay channel. For LQs decaying exclusively to top quarks and $\tau$ leptons, a gain of $400\GeV$ is expected, pushing the LQ mass in reach for a $5\sigma$ discovery from $800\GeV$ to $1200\GeV$. 

\begin{figure}[t]
\centering
\begin{minipage}[t][\minipageheightdp][t]{\minipagewidthdp}
\centering
\includegraphics[width=\figuremaxwidthdp,height=\figuremaxheightdp,keepaspectratio]{\main/section5FlavorRelatedStudies/img/LQpairs_Sign_tmu.pdf}
\end{minipage}
\begin{minipage}[t][\minipageheightdp][t]{\minipagewidthdp}
\centering
\includegraphics[width=\figuremaxwidthdp,height=\figuremaxheightdp,keepaspectratio]{\main/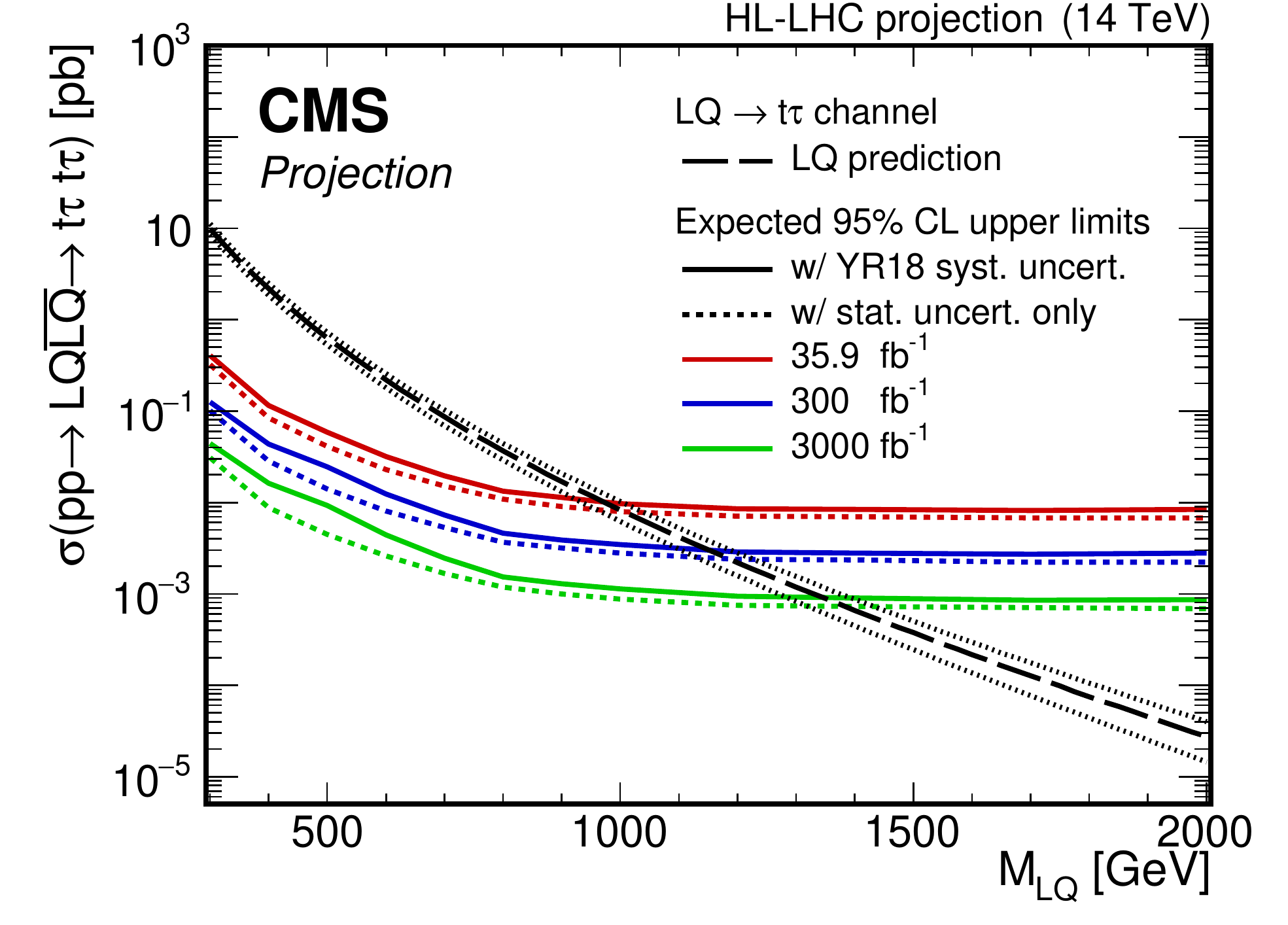}
\end{minipage}
\caption{Expected upper limits on the LQ pair production cross section at the $95\%\cl$ for an LQ decaying exclusively to top quarks and muons (left) or $\tau$ leptons (right).}
\label{fig:LQpairs:limits1d}
\end{figure}

In \fig{fig:LQpairs:limits1d}, the expected projected exclusion limits on the LQ pair production cross section are shown. Leptoquarks decaying only to top quarks and muons are expected to be excluded below masses of $1900\GeV$ for \intLumi, which is a gain of $500\GeV$ compared to the limit of $1420\GeV$ obtained in the published analysis of the 2016 dataset~\cite{Sirunyan:2018ruf}. The mass exclusion limit for LQs decaying exclusively to top quarks and $\tau$ leptons are expected to be increased by $500\GeV$, from $900\GeV$ to approximately $1400\GeV$.

\begin{figure}[t]
\centering
\begin{minipage}[t][\minipageheightdp][t]{\minipagewidthdp}
\centering
\includegraphics[width=\figuremaxwidthdp,height=\figuremaxheightdp,keepaspectratio]{\main/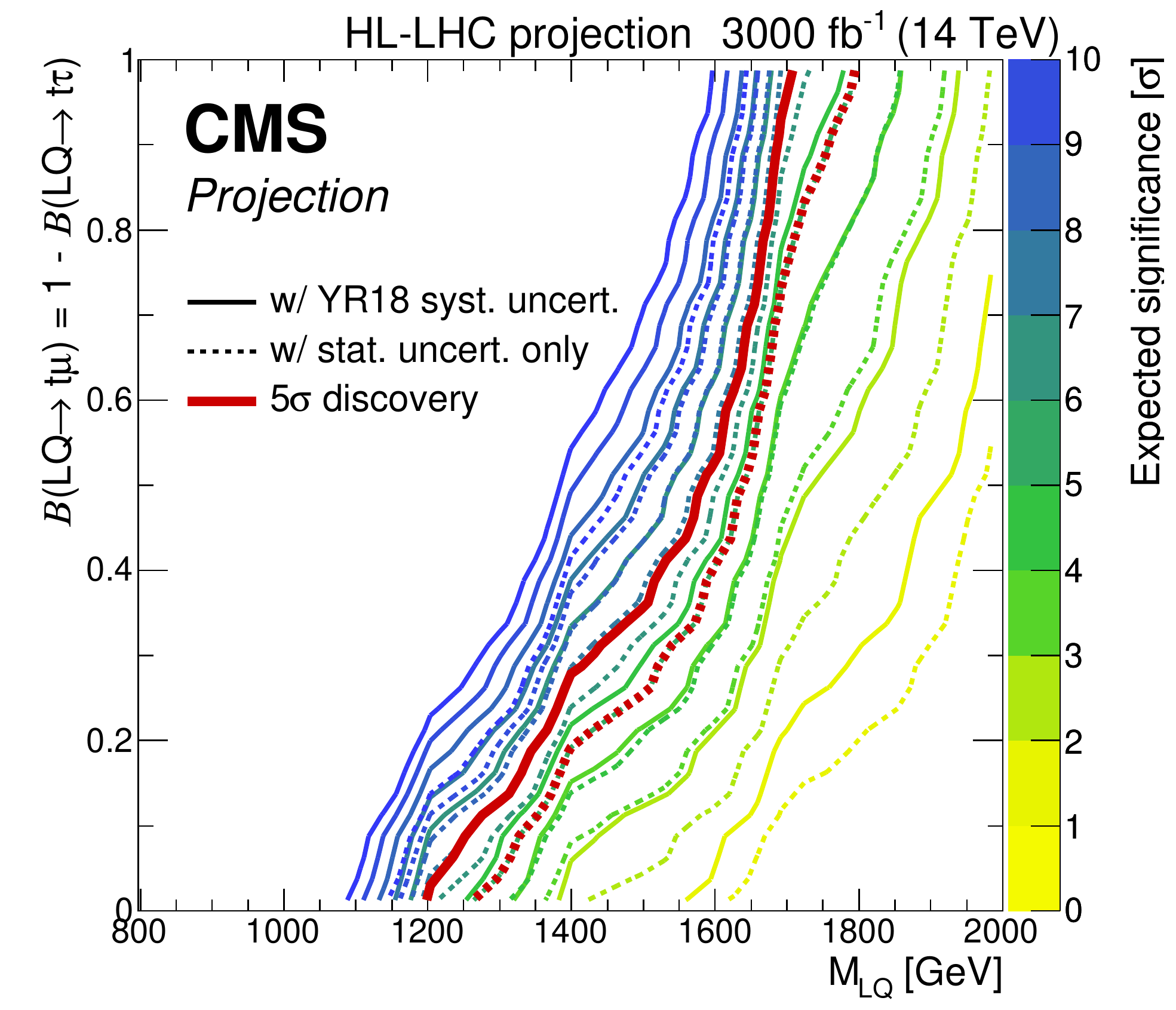}
\end{minipage}
\begin{minipage}[t][\minipageheightdp][t]{\minipagewidthdp}
\centering
\includegraphics[width=\figuremaxwidthdp,height=\figuremaxheightdp,keepaspectratio]{\main/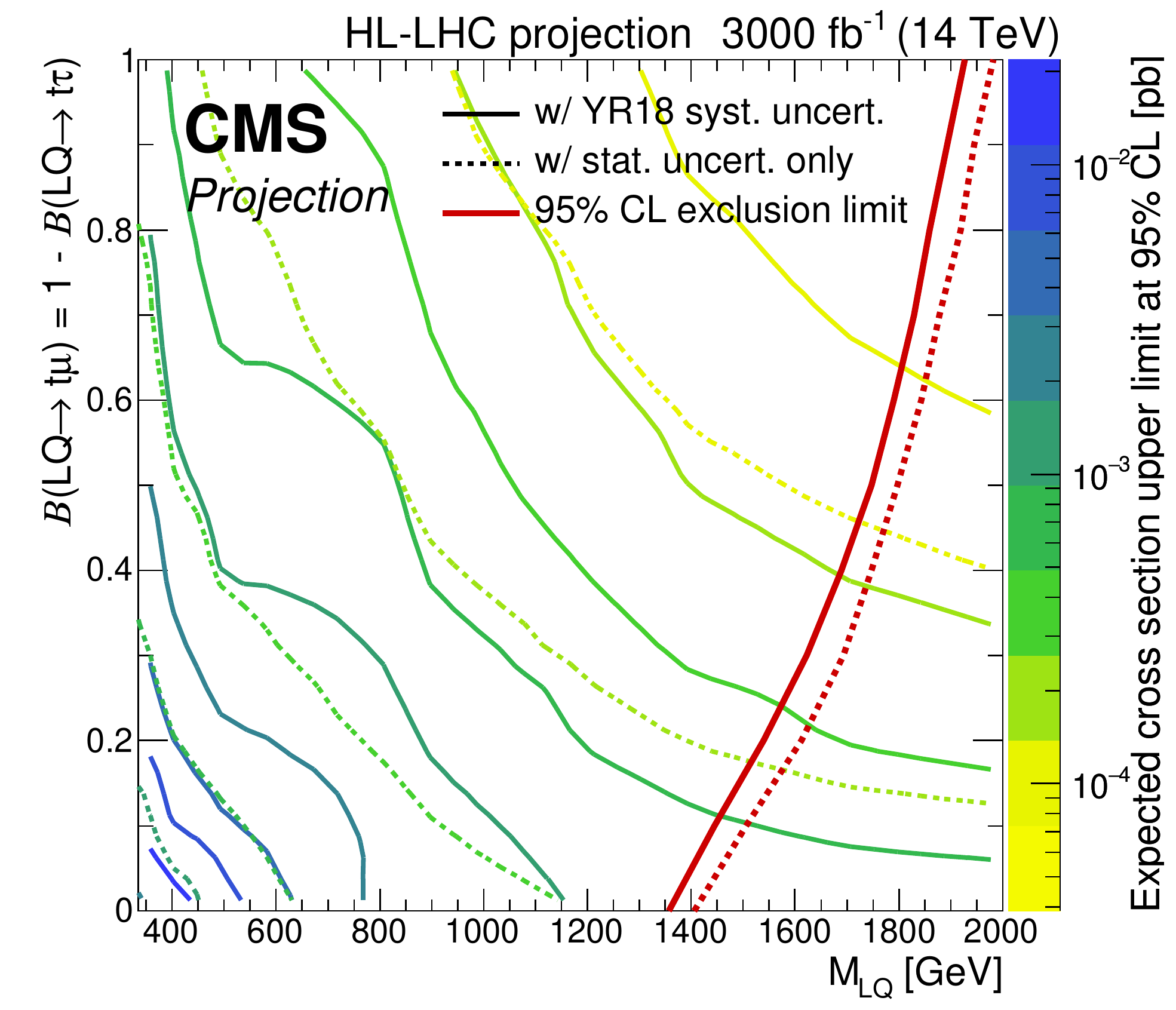}
\end{minipage}
\caption{Expected significances (left) and expected upper limits on the LQ pair-production cross section at the $95\%\cl$ (right) as a function of the LQ mass and the branching fraction. Colour-coded lines represent lines of a constant expected significance or cross section limit, respectively. The red lines indicate the $5\sigma$ discovery level (left) and the mass exclusion limit (right).}
\label{fig:LQpairs:limits2d}
\end{figure}

Figure~\ref{fig:LQpairs:limits2d} shows the expected signal significances and upper exclusion limits on the pair production cross section of scalar LQs allowed to decay to top quarks and muons or $\tau$ leptons at the $95\%\cl$ as a function of the LQ mass and a variable branching fraction $\mathcal{B}(\mathrm{LQ}\rightarrow t\mu) = 1 - \BR(\mathrm{LQ}\rightarrow t\tau)$ for an integrated luminosity of \intLumi in the two different scenarios. For all values of $\mathcal{B}$, LQ masses up to approximately $1200\GeV$ and $1400\GeV$ are expected to be in reach for a discovery at the $5\sigma$ level and a $95\%\cl$ exclusion, respectively.

\subsubsection{\theoryC{$Z'$ and leptoquarks for $B$ decay anomalies at HL- and HE-LHC}}
\label{sec:theoryzprimeLQBanomalies}
\contributors{B. Allanach, T. Tevong You}
%{\bf Author(s): B. Allanach$^1$, B. Gripaios$^2$, T. Tevong You$^{3}$}\\
%$^1$ DAMTP, University of Cambridge, CMS, Wilberforce Road, Cambridge, CB3 0WA, United Kingdom, email: {\tt B.C.Allanach@damtp.cam.ac.uk}\\
%$^2$ Cavendish Laboratory, JJ Thomson Ave, University of Cambridge, CB3 0HE, United Kingdom, email: {\tt gripaios@hep.phy.cam.ac.uk}\\
%$^3$ Gonville and Caius College, University of Cambridge, Trinity Street, CB2 1TA, United Kingdom, email: {\tt tevong@hep.phy.cam.ac.uk}
  
\begin{figure}[t]
\begin{center}
\includegraphics[width=0.60 \textwidth]{\main/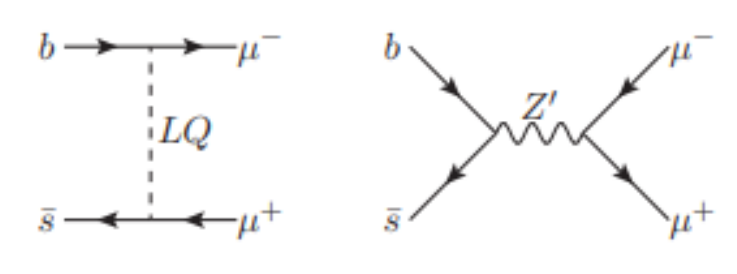}
\caption{Feynman diagrams of two simplified models for mediating
  an effective operator 
that explains discrepancies in $B \to K^{(*)} \mu^+ \mu^-$ decays as compared
to SM predictions. The
diagram on the left hand side shows mediation by a scalar LQ, whereas the
right-hand side shows mediation by a flavour
dependent $\Zp$. \label{fig:feyn}}
\end{center}
\end{figure}

Recent measurements of $R_K^{\ast}$ and other $b$ observables indicate that the $\bar b P_L s \bar \mu P_L \mu$ vertex may be receiving BSM corrections.  Here, we examine simplified models that may predict such corrections at the tree-level: $\Zp$ models and leptoquark models depicted in \fig{fig:feyn}.  

The `na\"{i}ve' $\Zp$ model contains the Lagrangian pieces
\begin{equation}
\mathcal{L}_{\Zp}^\text{min.} \supset \left( g_L^{sb} \Zp_\rho \bar{s}\gamma^\rho P_L b  + \text{h.c.} \right) + g_L^{\mu\mu} \Zp_\rho \bar{\mu} \gamma^\rho P_L \mu \, ,
\label{eq:minimalZprime}
\end{equation}
whereas the more realistic `$33\mu\mu$' $\Zp$ model contains more couplings ($SU(2)_L$ and flavour copies):
\begin{align}
\mathcal{L}_{\Zp}^{33\mu\mu} &\supset g^q_L \Zp_\rho \left[ \bar{t}\gamma^\rho P_L t + |V_{tb}|^2 \bar{b} \gamma^\rho P_L b + |V_{td}|^2\bar{d}\gamma^\rho P_L d + |V_{ts}|^2\bar{s}\gamma^\rho P_L s \right. \nonumber \\
 &\left. + \left( V_{tb} V_{ts}^* \bar{b}\gamma^\rho P_L s + V_{ts}^* V_{td}
   \bar{d}\gamma^\rho P_L s + V_{tb} V_{td}^* \bar b \gamma^\rho P_L d+
   \text{h.c.} \right) \right.\nonumber \\ &  \left.
+
   g_L^{\mu\mu}\left(\bar{\mu}\gamma^\rho P_L \mu +  
   \sum_{i,j} \bar{\nu}_i U_{i\mu} \gamma^\rho P_L U_{\mu j}^*\nu_j \right) \right],
\label{eq:3rdgenZprimeLagrangian}
\end{align}
where $U$ denotes the PMNS matrix involved in lepton mixing. 
In a fit to `clean' $b-$observables including $R_{K^{(\ast)}}$ \citeref{DAmico:2017mtc}, we have that, in the na\"{i}ve $\Zp$ model, 
\begin{equation}
|g_L^{sb} g_L^{\mu \mu}|=
(1.0 \pm 0.25) \left(\frac{M_{\Zp}}{31\TeV}\right)^2, \label{eq:cup}
\end{equation}
which we use to constrain the couplings and masses of the $\Zp$,  (the $33\mu\mu$ model's $\Zp$ couplings to $\bar s b$ and $\mu^+\mu^-$ can also be matched to \eq{eq:minimalZprime}).

There are 
particular combinations of quantum numbers allowed for the
LQs~\cite{DAmico:2017mtc, Hiller:2017bzc, Capdevila:2017bsm}, depending upon their spin. For 
the scalar case this is the triplet LQ $S_3$, with quantum numbers
$(\bar{3}, 3, \frac{1}{3})$ under $SU(3)_c \times SU(2)_L \times U(1)_Y$ and mass $M$,
whose Yukawa couplings to the third family left-handed quark and second family left-handed lepton doublets $Q_3$ and $L_2$ are of the form
\begin{equation}
y_{b \mu} Q_3L_2 S_3 + y_{s \mu} Q_2 L_2 S_3 \textrm{h.c.} \, .
\end{equation}
Its effect on the clean $b-$observables result in a constraint 
\begin{equation}
|y_{sb} y_{b \mu}|=
(1.00 \pm 0.25)\left(\frac{M}{31\TeV}\right)^2, \label{eq:cup2}
\end{equation}
from the fit~\cite{DAmico:2017mtc}.

\begin{figure}[t]
\centering
\begin{minipage}[t][\minipageheightdp][t]{\minipagewidthdp}
\centering
\includegraphics[width=\figuremaxwidthdp,height=\figuremaxheightdp,keepaspectratio]{\main/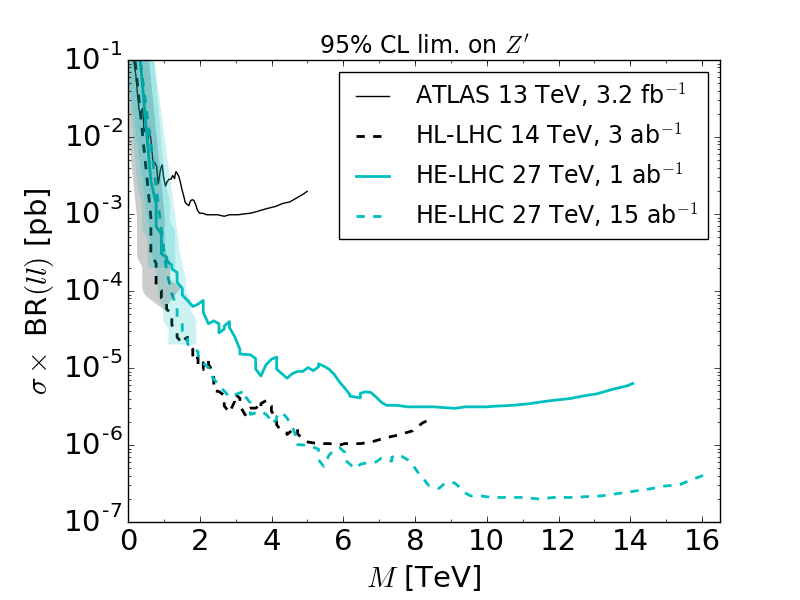}
\end{minipage}
\begin{minipage}[t][\minipageheightdp][t]{\minipagewidthdp}
\centering
\includegraphics[width=\figuremaxwidthdp,height=\figuremaxheightdp,keepaspectratio]{\main/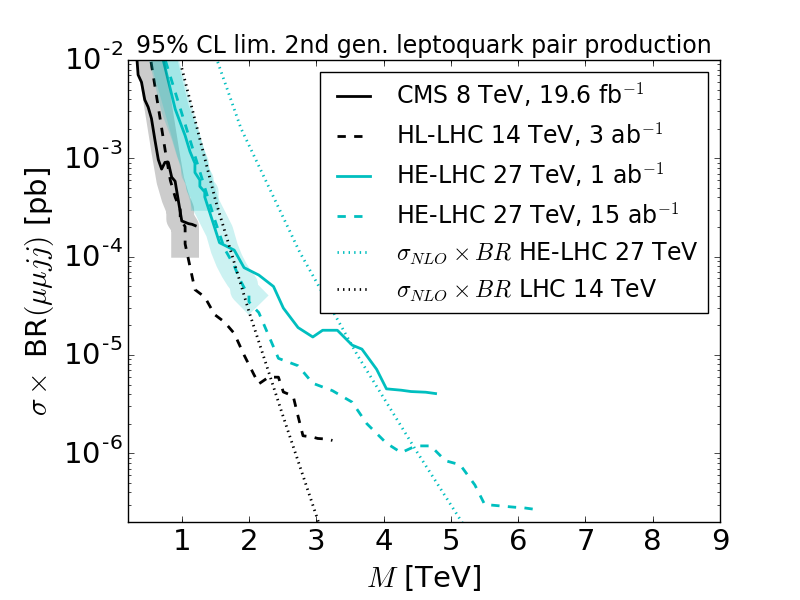}
\end{minipage}
\caption{Current bounds and projected sensitivities to the na\"{i}ve $\Zp$
model. Left: bounds from a $3.2\fbinv$ ATLAS
$\Zp$ search in the $\mu^+\mu^-$ channel~\protect\cite{Aaboud:2016cth}.
Right: bounds from a $19.6\fbinv$ search for the process \sloppy\mbox{$gg \rightarrow \bar S_3 S_3 \rightarrow (\mu^- j)(\mu^+j)$}, these
bounds were then used~\cite{Allanach:2017bta} to extrapolate to the HE-LHC and HL-LHC, and the predicted NLO production cross-sections 
for the production of the $S_3 \bar S_3$ pairs.
\label{fig:boundsSens}}
\end{figure}
\begin{figure}[t]
\begin{center}
\includegraphics[width=0.6 \textwidth]{\main/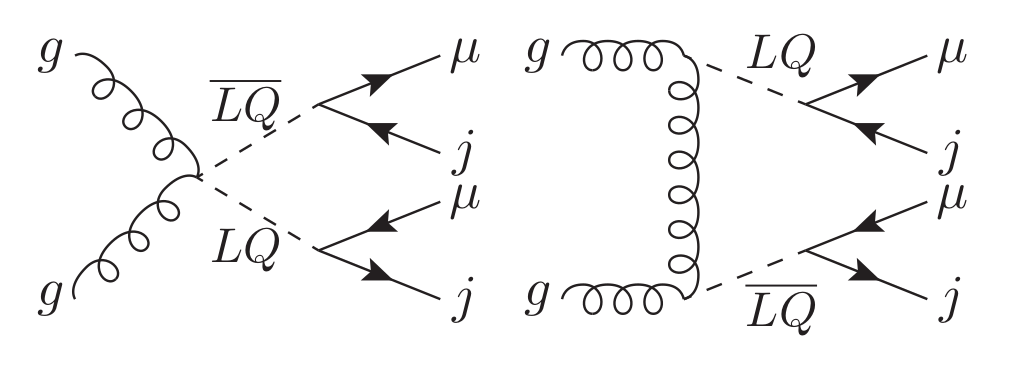}
\caption{Example processes contributing to $S_3 \bar S_3$ production and subsequent decay at the LHC. \label{fig:LQpair}}
\end{center}
\end{figure}

In \fig{fig:boundsSens}, we display the projected sensitivities on the na\"{i}ve $\Zp$ and scalar leptoquark models of the HE-LHC and HL-LHC, extrapolated from a $3.2\fbinv$ ATLAS $\Zp$ resonance search in the $\mu^+ \mu^-$ model as a function of $\Zp$ mass. 
The extrapolation is performed under the following approximations: changes in
 efficiency with respect to changing $\sqrt{s}$ or other changes to the
 operating environment 
are neglected, and the {\em narrow width} approximation is used. At strong
coupling, the narrow width approximation becomes bad. 
In the right-hand plot, the sensitivity coming from the process $gg \rightarrow \bar S_3 S_3 \rightarrow \mu^+j \mu^-j$ depicted in \fig{fig:LQpair} is shown. The sensitivities are phrased in terms of production cross-section times BRs of final states on the vertical axis. 
In the LQ model, the production cross-section depends only upon the mass $M$ of the LQ: its coupling is given by the QCD gauge coupling. We see that HL-LHC(HE-LHC) is sensitive to LQs of mass up to $2.5$ ($4.2$)\TeV for this topology.
The $B_s - \bar B_s$ mixing constraint implies $\left|y_{b\mu}y_{s\mu}^\ast\right|< M/(26\TeV)$ for the $S_3$ LQ case. Combining this with \eq{eq:cup2} yields a bound $M<40\TeV$\footnote{One can also consider singlet or triplet vector LQs whose joint constraints imply $M<20,40\TeV$, respectively, but whose sensitivities are equal to those of the $S_3$.}~\cite{DAmico:2017mtc}. Thus, the HL-LHC and HE-LHC could probe a non-negligible fraction of the viable LQ parameter space.

\begin{figure}[t]
\centering
\begin{minipage}[t][\minipageheightdp][t]{\minipagewidthdp}
\centering
\includegraphics[width=\figuremaxwidthdp,height=\figuremaxheightdp,keepaspectratio]{\main/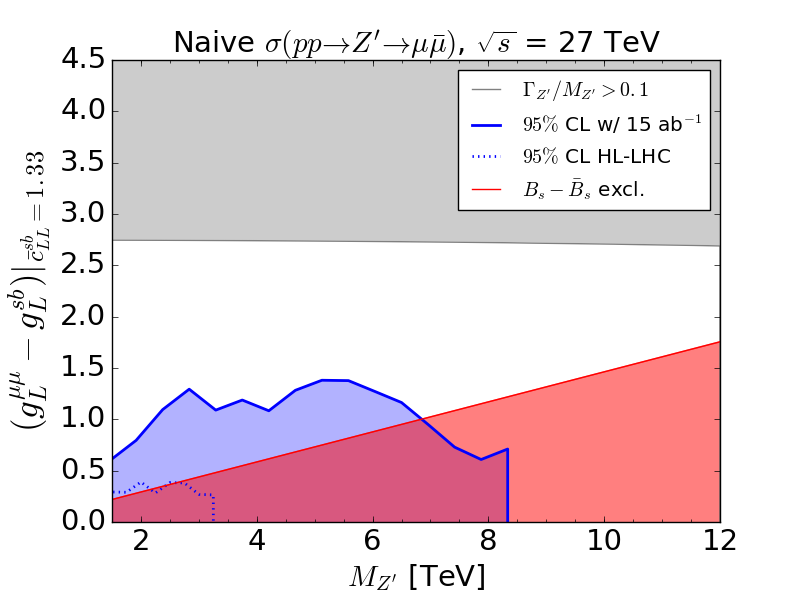}
\end{minipage}
\begin{minipage}[t][\minipageheightdp][t]{\minipagewidthdp}
\centering
\includegraphics[width=\figuremaxwidthdp,height=\figuremaxheightdp,keepaspectratio]{\main/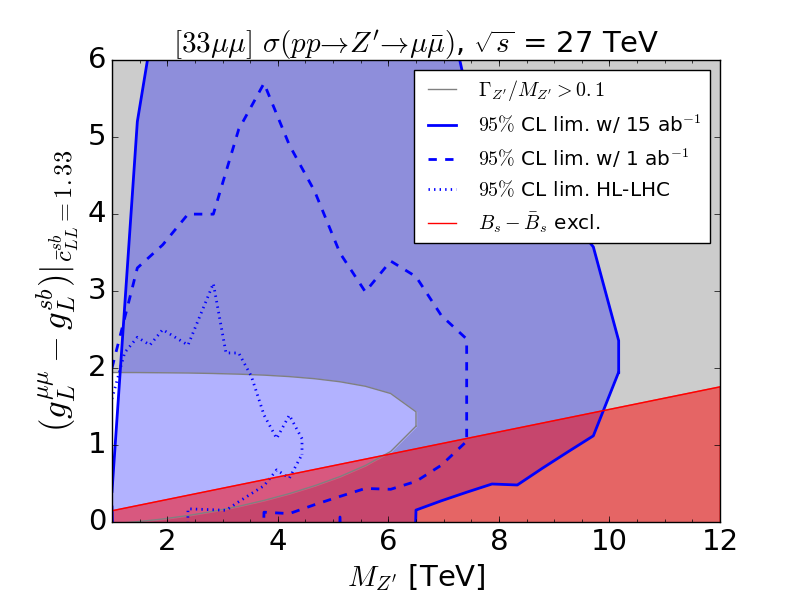}
\end{minipage}
\caption{Current bounds and projected sensitivities to $\Zp$ models explaining $R_{K^{(\ast)}}$. Each point in the plane has had the couplings adjusted to be consistent with the central value of \eq{eq:cup}. The red region is excluded from $B_s - \bar B_s$ mixing measurements, the blue region shows the expected $95\%\cl$ sensitive region and the greyed region is at width $\Gamma > 0.1 M_{\Zp}$, meaning that the extrapolation used to calculate sensitivities (which uses the narrow width approximation) is inaccurate. The blue region shows the region of sensitivity. The vertical axis shows the difference between the muonic and the quark $\Zp$ coupling, initially intended to show when one is large compared to the other. However, the $B_s-\bar B_s$ mixing constraint implies that $g_L^{sb}\ll g_L^{\mu\mu}$ and so the vertical axis is equal to $g_L^{\mu\mu}$, to a good approximation.
\label{fig:boundsCom}}
\end{figure}
For the $\Zp$ models, the production cross-section depends sensitively upon the coupling $g^{bs}_L$. If the coupling is large, $\Zp$s with masses up to $18\TeV$ can explain $R_{K}^{(\ast)}$ and still be compatible with bounds originating from $B_s - \bar B_s$ mixing measurements ($|g^{sb}_L| \leq  M_{\Zp}/148\TeV$)~\cite{DAmico:2017mtc}. We display the relevant sensitivities and bounds upon flavourful $\Zp$ models in \fig{fig:boundsCom} as a function of coupling and mass. From the left hand panel, we see that the HL-LHC only probes a small fraction of the viable na\"{i}ve $\Zp$ parameter space. However, when we examine more realistic model (the $33\mu\mu$ model) in the right-hand panel, we see that a large fraction of parameter space where the narrow width approximation applies is covered. These conclusions become stronger when one examines the sensitivity of the HE-LHC, as the solid lines display: the viable region of the $33\mu\mu$ model with narrow width approximation is completely covered, for example. 

We note that the results presented herein represent only a  rough estimation and further more detailed studies are desirable. In particular, the narrow width approximation in the case of the LQs is likely to be a rough approximation because one is really producing a pair of LQs. Our approximations effectively assume that they are produced at threshold. At higher luminosities or energies, the efficiency for identifying isolated muons will change, as well as the efficiencies for identifying jets. This could be better estimated by performing Monte-Carlo event generation studies together with a detector simulation rather than extrapolating current limits from the LHC. 

%\subsubsection{Ackowledgements}
%\paragraph*{Ackowledgements}
%This work has been partially supported by STFC consolidated grants ST/L000385/1 and ST/P000681/1.

\subsubsection{\cms{Search for leptoquarks decaying to $\tau$ and $b$ at HL-LHC}}\label{sec:leptoqbtau}
\contributors{Y. Takahashi, P. Matorras, CMS}
%{\bf Author(s): Y. Takahashi et al., CMS}

Third-generation scalar LQs have recently received considerable interest from the theory community, as the existence of leptoquarks with large couplings can explain the anomaly in the $\overline{\PB} \to \PD \tau \overline{\nu}$ and $\overline{\PB} \to \PD^{\ast} \tau \overline{\nu}$ decay rates reported by the
BaBar~\cite{Lees:2012xj, Lees:2013uzd}, Belle~\cite{Matyja:2007kt, Bozek:2010xy, Huschle:2015rga, Sato:2016svk, Hirose:2016wfn, Hirose:2017dxl},
and LHCb~\cite{Aaij:2015yra} Collaborations. 

\begin{figure}[t]
\centering
\includegraphics[width=0.3\textwidth]{\main/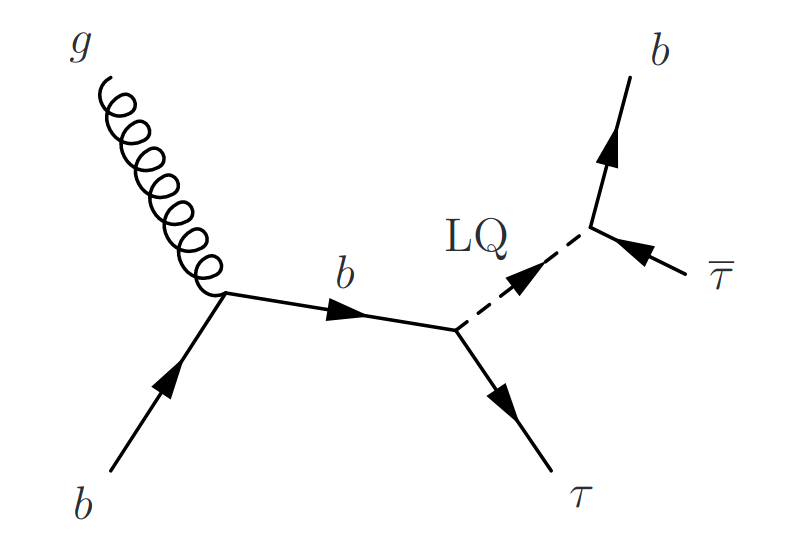}
\hspace{0.5cm}
\includegraphics[width=0.3\textwidth]{\main/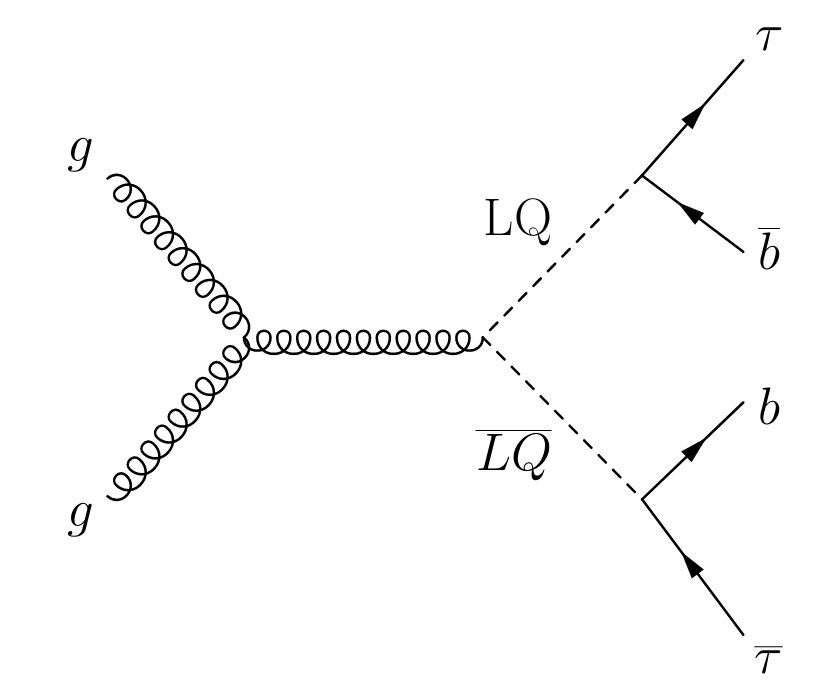}
\hspace{0.5cm}
\includegraphics[width=0.3\textwidth]{\main/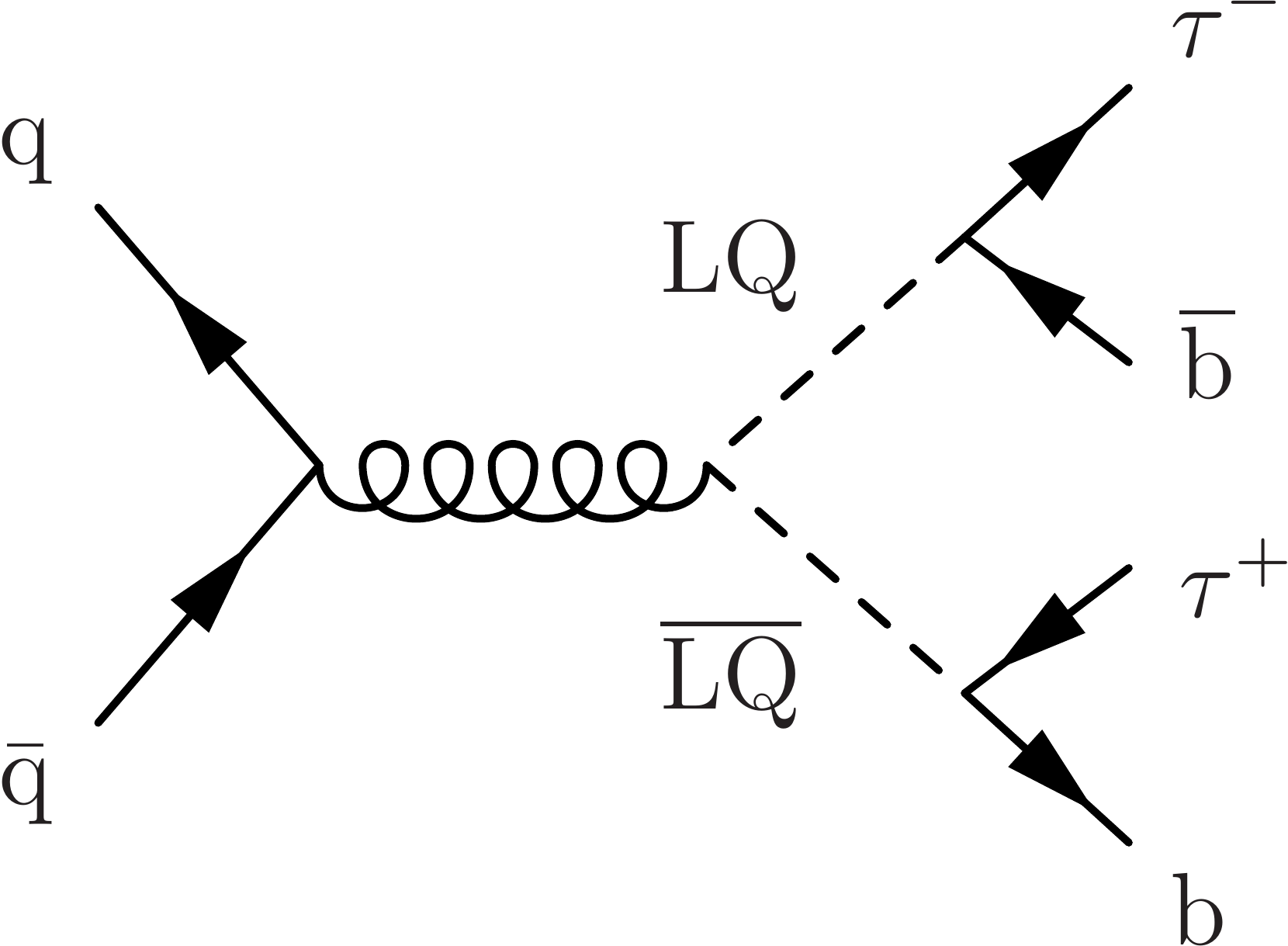}
\caption{Leading order Feynman diagrams for the production of a third-generation LQ in the single production s-channel (left) 
and the pair production channel via gluon fusion (centre) and quark fusion (right).}
\label{LQ3-diagram}
\end{figure}

This analysis from CMS~\cite{CMS-PAS-FTR-18-028} presents future discovery and exclusion prospects for singly and pair produced 
third-generation scalar LQs, each decaying to $\tauh$ and a bottom quark. Here, $\tauh$ denotes a hadronically decaying 
$\tau$ lepton. 
The relevant Feynman diagrams of the signal processes at leading order (LO) are shown in \fig{LQ3-diagram}.

\vspace{-5pt}

The analysis uses \delphes{}~\cite{deFavereau:2013fsa} event samples of simulated pp collisions at a \com~energy of $14\TeV$, corresponding to integrated luminosities of $300\fbinv$and $3\abinv$. %Additional pp interactions (pileup or PU) within the same or adjacent bunch crossings are included, with an average of 200 interactions per event. 
The matrix elements of LQ signals for both single and pair LQ production are generated at LO using version 2.6.0 of
\MGvATNLO~\cite{Alwall:2014hca} for $m_{LQ}=500$, $1000$, $1500$, and $2000\GeV$. The branching fraction $\beta$ of the LQ to a charged
lepton and a quark, in this case LQ$\to \tau b$, is assumed to be $\beta=1$. The unknown Yukawa coupling $\lambda$ of the LQ to a
$\tau$ lepton and a bottom quark is set to $\lambda=1$. The width $\Gamma$ is calculated using $\Gamma = m_{LQ} \lambda^2 /
(16\pi)$~\cite{Plehn:1997az}, and is less than $10\%$ of the LQ mass for most of the considered search range. The signal samples are
normalised to the cross section calculated at LO, multiplied by a $K$ factor to account for higher order contributions~\cite{Dorsner:2018ynv}.

Similar event selections are used in both the singly and pair produced LQ searches, except for the requirement on the number of jets. 
In both channels, two reconstructed $\tauh$ with opposite-sign charge are required, each with transverse momentum $p_{T,\tau}>50\GeV$ and a maximum pseudorapidity  $\abs{\eta_{\tau}}<2.3$. 
In the search for single production, the presence of at least one reconstructed jet with $p_T > 50\GeV$ is required, while at least
two are required in the search for pair production. Jets are reconstructed with FASTJET~\cite{Cacciari:2011ma}, using the anti-\kt algorithm~\cite{Cacciari:2008gp}, with a distance parameter of $0.4$. 

To reduce background due to Drell-Yan (particularly Z$\to\tau\tau$) events, the invariant mass of the two selected $\tauh$,
$m_{\tau\tau}$, is required to be $>95\GeV$.
In addition, at least one of the previously selected jets is required to be b-tagged to reduce QCD multijet backgrounds. 
Finally, an event is rejected if it contains an identified and isolated electron (muon), with $p_T>10\GeV$, $|\eta|<2.4$ ($2.5$).
The acceptance of the signal events is $4.9\%$ ($11\%$) for single (pair) production, where the branching ratio of two $\tau$ leptons decaying hadronically is included in the numerator of the acceptance.

\begin{figure}[t]
\centering
\begin{minipage}[t][\minipageheightdp][t]{\minipagewidthdp}
\centering
\includegraphics[width=\figuremaxwidthdp,height=\figuremaxheightdp,keepaspectratio]{\main/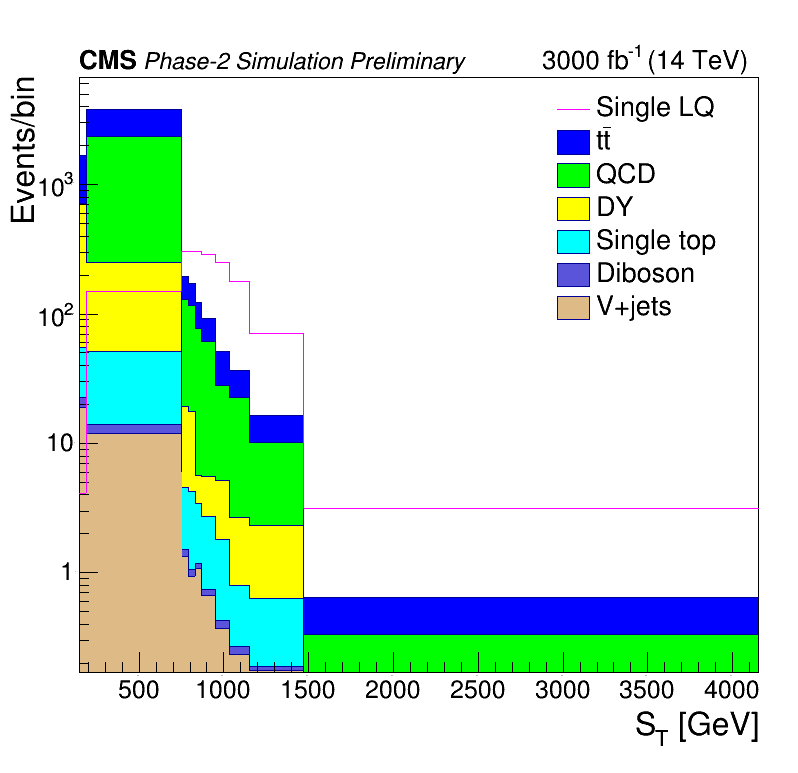}
\end{minipage}
\begin{minipage}[t][\minipageheightdp][t]{\minipagewidthdp}
\centering
\includegraphics[width=\figuremaxwidthdp,height=\figuremaxheightdp,keepaspectratio]{\main/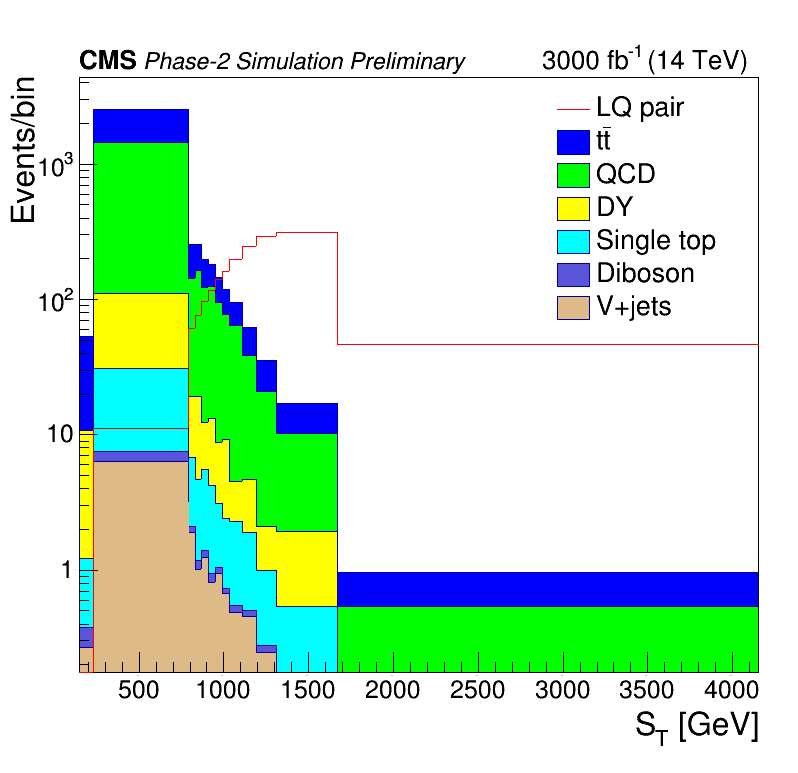}
\end{minipage}
\caption{Left: scalar sum of the \pt\ of the two selected $\tau$ leptons and the highest-$p_T$ jet in the single LQ selection region. Right: scalar sum of the $p_T$ of the two selected $\tau$ leptons and the two highest-$p_T$ jets in the LQ pair search region. The considered backgrounds are shown as stacked histograms, while empty histograms for signals for the single LQ and LQ pair channels (for $m_{LQ}=1000\GeV$) are overlaid to illustrate the sensitivity. Both signal and background are normalised to a luminosity of $3\abinv$.} %Signal is normalised to $\lambda=1$, $\beta=1$ and $\sigma=1$ pb.}
\label{fig:scalarptsums}
\end{figure}

Signal extraction is based on a binned maximum likelihood fit to the distribution of the scalar $p_T$ sum S$_{\text{T}}$, which 
is defined as the sum of the transverse momenta of the two $\tauh$ and either the highest-$p_T$ jet in the 
case of single LQ production, or the two highest-$p_T$ jets in the case of LQ pair production. These distributions are shown 
in \fig{fig:scalarptsums} for the HL-LHC $3\abinv$ scenario.

Systematic uncertainties are 
calculated by scaling the current experimental uncertainties. For uncertainties limited by statistics, including the uncertainty 
on the DY ($3.3\%$) and QCD ($3.3\%$) cross sections, a scale factor of $1/\sqrt{\mathcal{L}}$ is applied, for an integrated luminosity ratio $\mathcal{L}$. 
For uncertainties 
coming from theoretical calculations, a scale factor of $1/2$ is applied with respect to current uncertainties, as is the case for 
the uncertainties on the cross section for top ($2.8\%$) or diboson ($3\%$) events. Other experimental systematic uncertainties are 
scaled by the 
square root of the integrated luminosity ratio until the uncertainty reaches a minimum value, including uncertainties on the integrated 
luminosity ($1\%$), 
$\tau$ identification ($5\%$) and b-tagging/misidentification ($1\%/5\%$).

\begin{figure}[t]
\centering
\begin{minipage}[t][\minipageheightdp][t]{\minipagewidthdp}
\centering
\includegraphics[width=\figuremaxwidthdp,height=\figuremaxheightdp,keepaspectratio]{\main/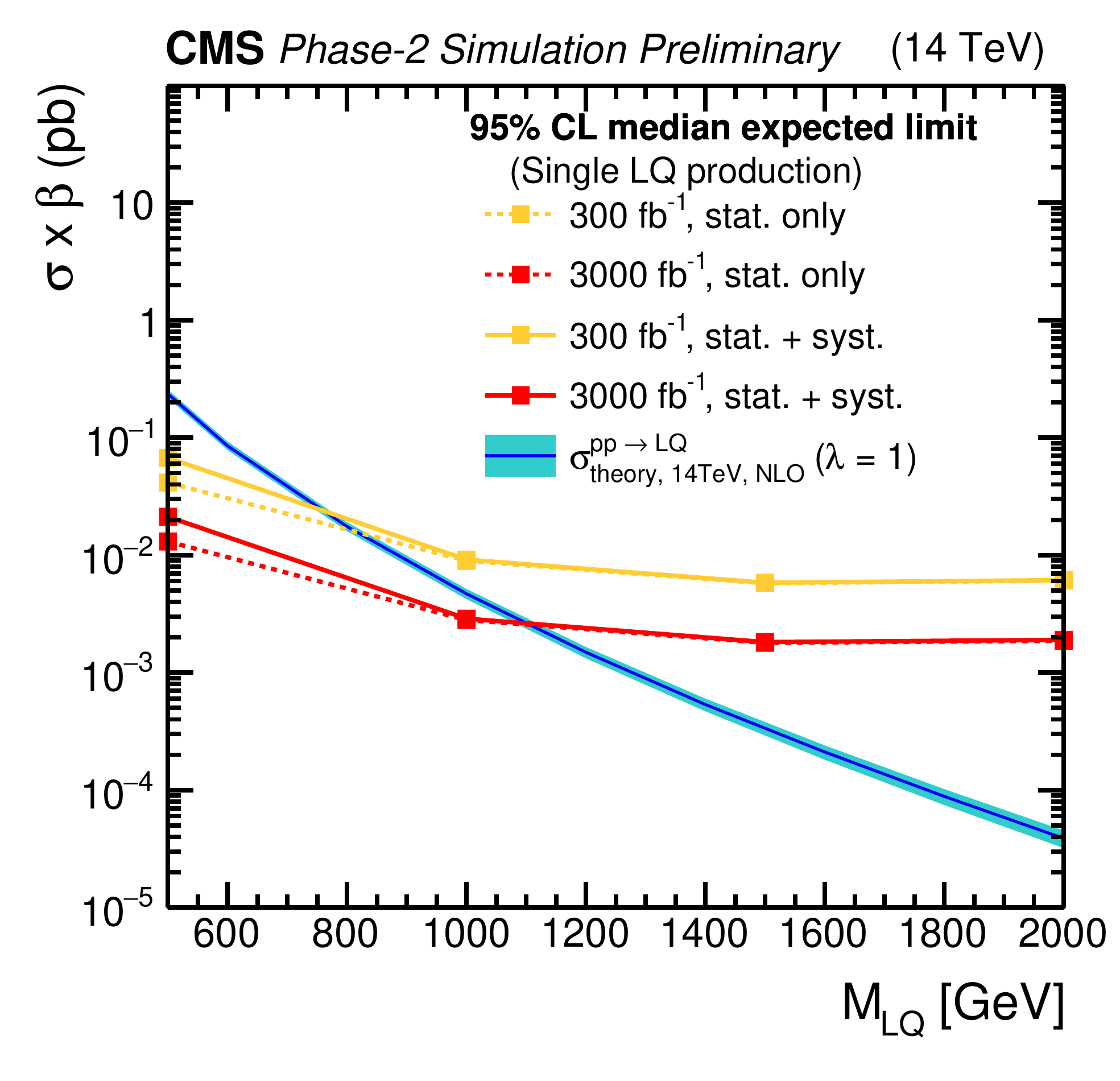}
\end{minipage}
\begin{minipage}[t][\minipageheightdp][t]{\minipagewidthdp}
\centering
\includegraphics[width=\figuremaxwidthdp,height=\figuremaxheightdp,keepaspectratio]{\main/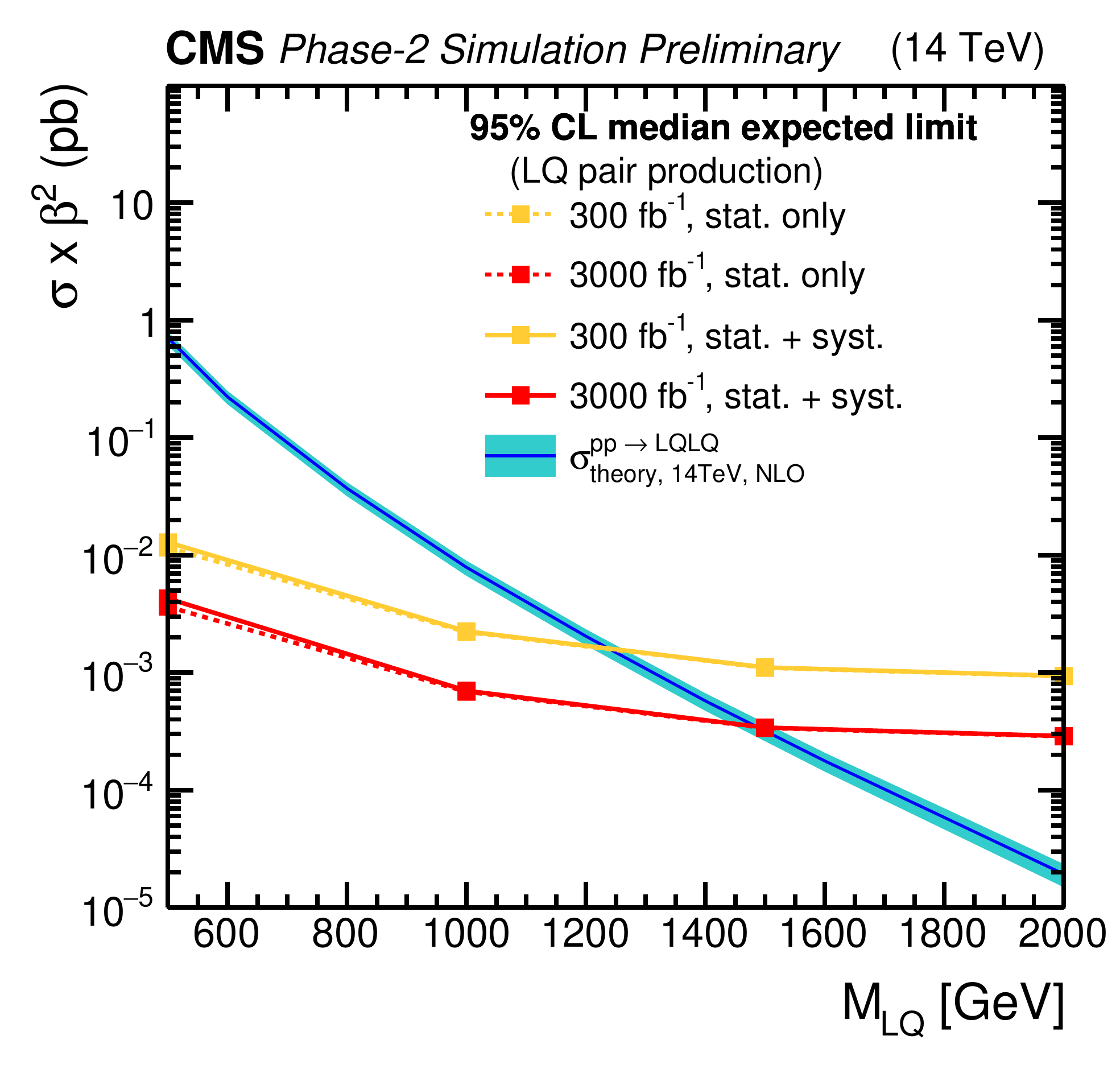}
\end{minipage}
\caption{Expected limits at $95\%\cl$ on the product of the cross section $\sigma$ and the branching fraction $\beta$, as a function
of the LQ mass, for the two high luminosity projections, $300\fbinv$ (red) and $3\abinv$ (orange), for both the stat. only  (dashed
lines) and the stat.+syst. scenarios (solid lines). This is shown in conjunction with the theoretical predictions at NLO~\cite{Dorsner:2018ynv} in cyan. Projections are calculated for both the single LQ (left) and LQ pair production (right).}
\label{fig:LQ3-projection}
\end{figure}

Figure~\ref{fig:LQ3-projection} shows an upper limit at $95\%\cl$ on the cross section times branching fraction $\beta$ as a 
function of $m_{LQ}$ by using the asymptotic CLs modified frequentist 
criterion~\cite{Cowan:2010js, Junk:1999kv,Read:2002hq, ATLAS:2011tau}.
Upper limits are calculated considering 
two different scenarios. The first one, hereafter abbreviated as "stat. only" considers only statistical uncertainties, to 
observe how the results are affected by the increase of the integrated luminosity. The second scenario, hereafter 
abbreviated as "stat.+syst.," also includes the estimate of the systematic uncertainties at the HL-LHC. 
For the single LQ production search, the theoretical prediction for the cross section assumes $\lambda = 1$ and $\beta = 1$. 

Comparing the limits with theoretical predictions assuming unit Yukawa coupling $\lambda = 1$, third-generation scalar 
leptoquarks are expected to be excluded at $95\%\cl$ for LQ masses below \sloppy\mbox{$730$ ($1250$)\GeV} for a luminosity 
of $300\fbinv$, and below $1130$ ($1520$)\GeV for $3\abinv$ in the single (pair) production channel, considering both 
statistical and systematic uncertainties. 

\begin{figure}[t]
\centering
\begin{minipage}[t][\minipageheightsp][t]{\minipagewidthsp}
\centering
\includegraphics[width=\figuremaxwidthsp,height=\figuremaxheightsp,keepaspectratio]{\main/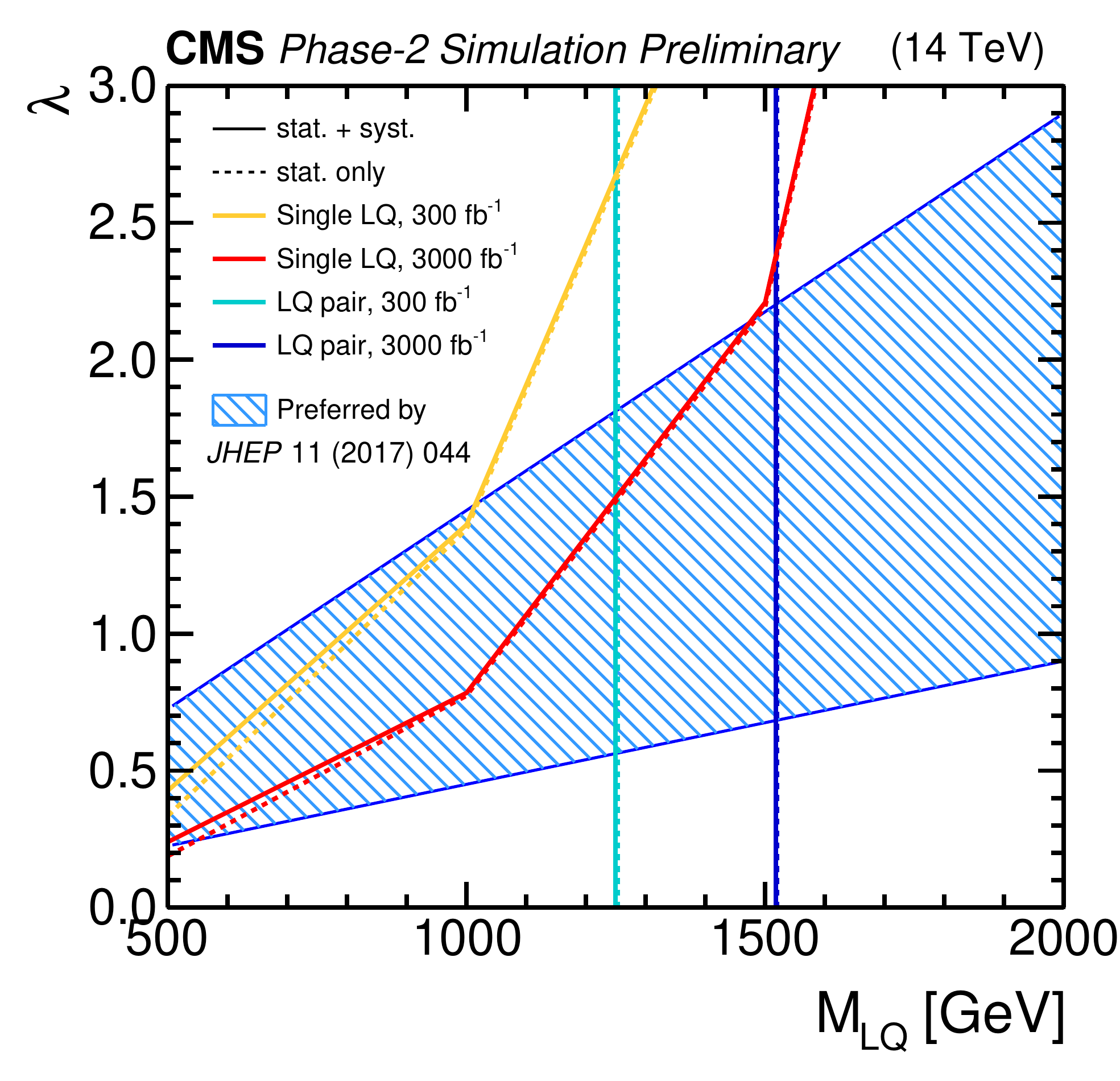}
\end{minipage}
\caption{Expected exclusion limits at $95\%\cl$ on the Yukawa coupling $\lambda$ at the LQ-lepton-quark vertex, as a function of the
LQ mass. A unit branching fraction $\beta$ of the LQ to a $\tau$ lepton and a bottom quark is assumed. Future projections for $300\fbinv$
and $3\abinv$ are shown for both the stat. only and stat.+syst. scenarios, shown as dashed and
filled lines respectively, and for both the single LQ and LQ pair production, where the latter corresponds to the vertical line
(since it does not depend on $\lambda$). The left hand side of the lines  represents the exclusion region for each of the
projections, whereas the region with diagonal blue hatching shows the parameter space preferred by one of the models proposed to
explain anomalies observed in $\PB$ physics~\cite{Buttazzo:2017ixm}.}
\label{fig:lambda_mass}
\end{figure}

Since the single-LQ signal cross section scales with $\lambda^2$, it is straightforward to recast the results presented in 
\fig{fig:LQ3-projection} in terms of expected upper limits on $m_{LQ}$  as a function of $\lambda$, as shown in 
\fig{fig:lambda_mass}. 
The blue band shows the parameter space ($95\%\cl$) for the scalar LQ preferred by the $\PB$ physics 
anomalies: $\lambda = (0.95 \pm 0.50) m_{LQ} (\mathrm{TeV})$~\cite{Buttazzo:2017ixm}. 
For the $300$ ($3000$)\fbinv luminosity scenario, the leptoquark pair production channel is more sensitive if  
$\lambda<2.7$ ($2.3$), while the single leptoquark production is dominant otherwise.
 
\begin{figure}[t]
\centering
\begin{minipage}[t][\minipageheightdp][t]{\minipagewidthdp}
\centering
\includegraphics[width=\figuremaxwidthdp,height=\figuremaxheightdp,keepaspectratio]{\main/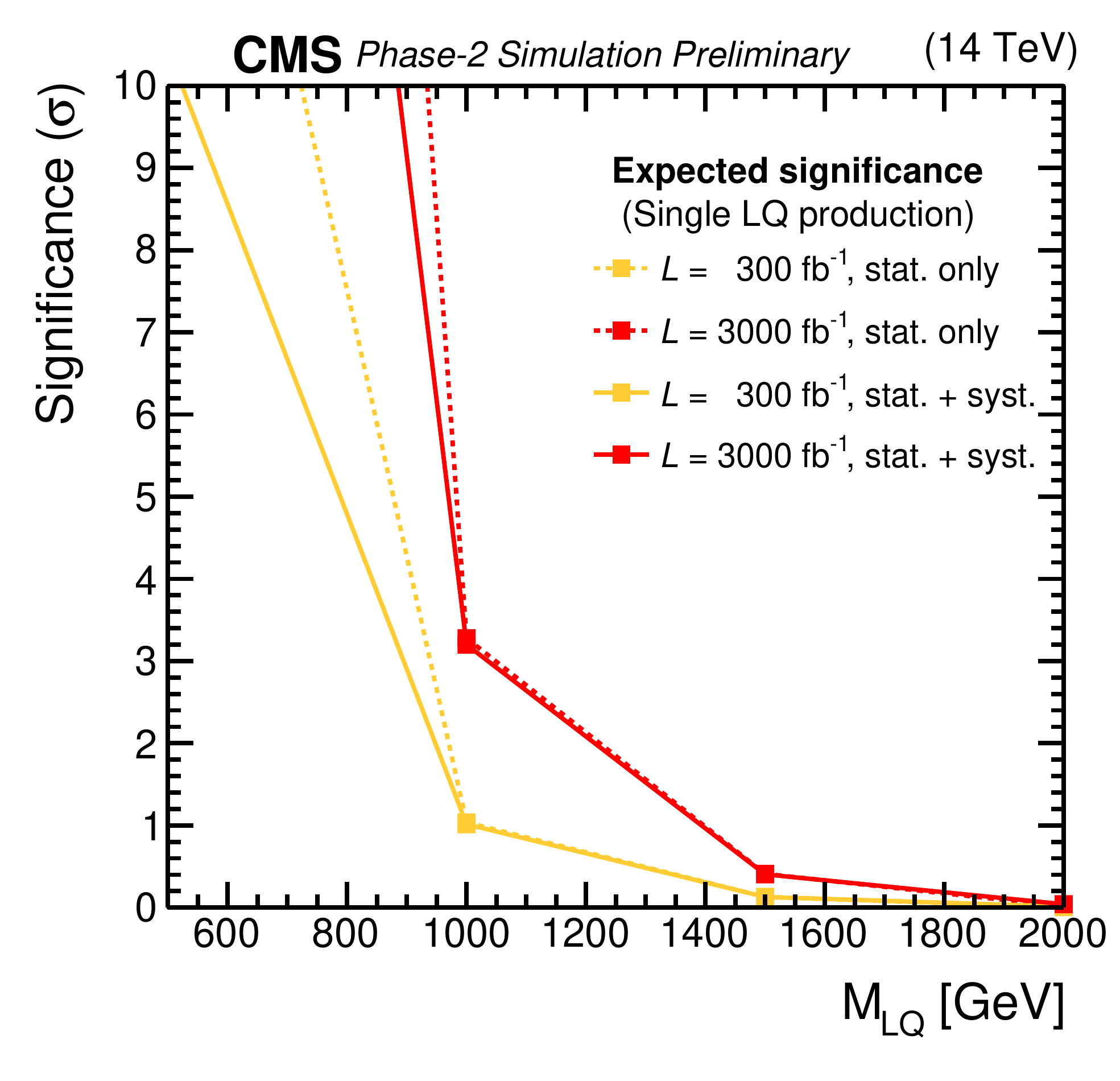}
\end{minipage}
\begin{minipage}[t][\minipageheightdp][t]{\minipagewidthdp}
\centering
\includegraphics[width=\figuremaxwidthdp,height=\figuremaxheightdp,keepaspectratio]{\main/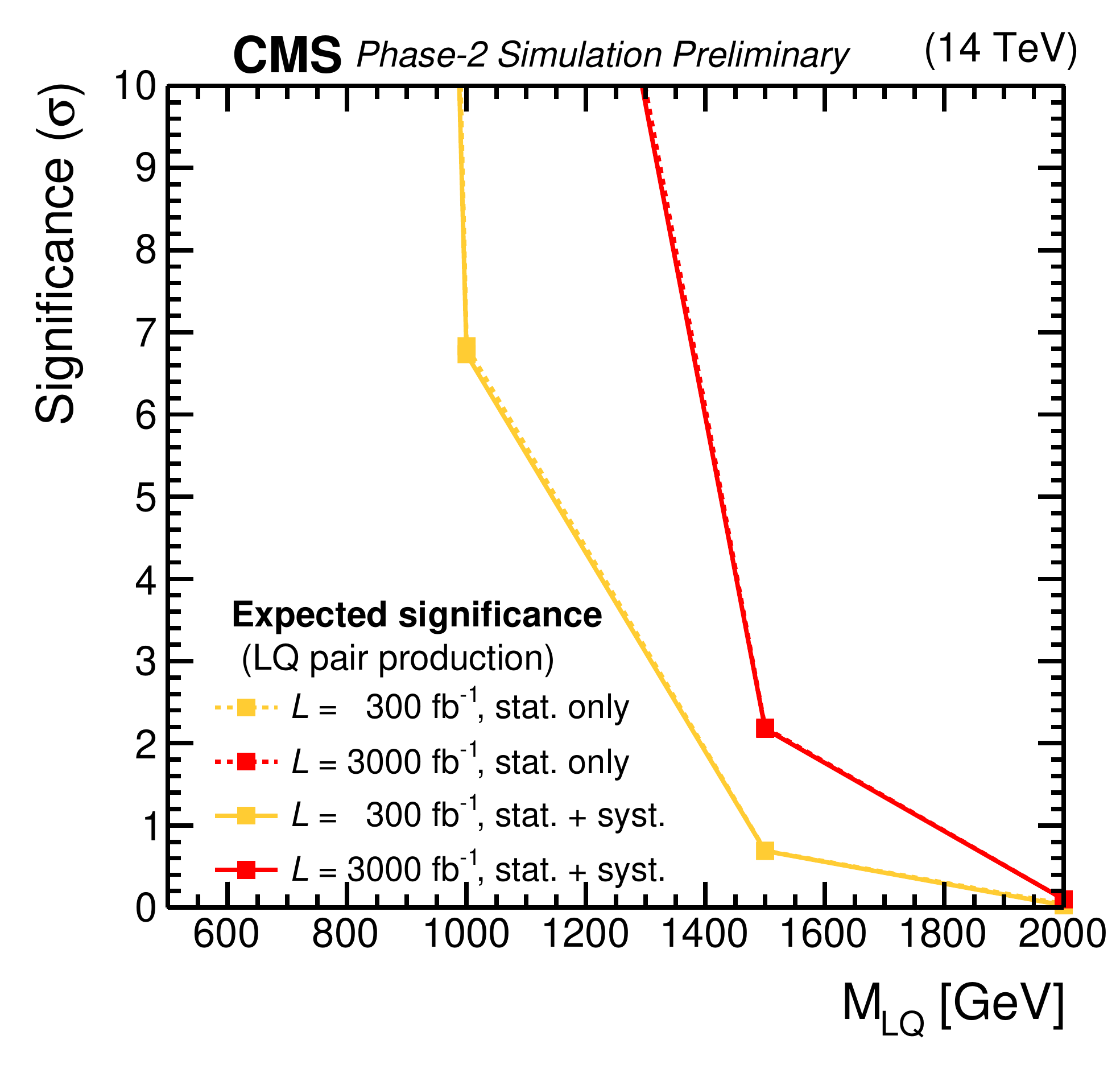}
\end{minipage}
\caption{Expected local significance of a signal-like excess as a function of the LQ mass, for the two high luminosity projections,
$300\fbinv$ (red) and $3\abinv$ (orange), assuming the theoretical prediction for the LQ cross section at NLO~\cite{Dorsner:2018ynv}, calculated with $\lambda = 1$ and $\beta = 1$. Projections are calculated for both single LQ (left) and LQ pair production (right).}
\label{fig:LQ3-discovery}
\end{figure}

Using the predicted cross section~\cite{Dorsner:2018ynv} of the signal, it is also possible to estimate the maximal LQ mass expected to be in reach for a $5\sigma$ discovery. 
Figure~\ref{fig:LQ3-discovery} shows the expected local significance of a signal-like excess as a function of the LQ mass hypothesis.

In summary, this study shows that future LQ searches under higher luminosity conditions are promising, as they are expected 
to greatly increase the reach of the search. They also show that the pair production channel is  expected to be the most 
sensitive. A significance of $5\sigma$ is within reach for LQ masses below $800$ ($1200$)\GeV for the single (pair) production 
channels in the $300\fbinv$ scenario and $1000$ ($1500$)\GeV for the $3\abinv$ scenario.

\subsubsection{\theory{HE-LHC sensitivity study for leptoquarks decaying to $\tau + b$}}\label{sec:LQHELHC}
\contributors{A. Greljo and L. Mittnacht}

We analysed the sensitivity of the $27\TeV$ $pp$ collider with $15\abinv$ of integrated luminosity to probe pair production of the scalar and vector leptoquarks decaying to $(b \tau)$ final state. We investigated events containing either one electron or muon, one hadronically decaying tau lepton, and  at least two jets. The signal events and the dominant background events ($t\bar{t}$) were generated with \MGvATNLO at leading-order. \pythia{ 6} was used to shower and hadronise events and \delphes{ 3} was used to simulate the detector response. The scalar leptoquark ($r_{23}$) and the vector leptoquark ($U_1$) UFO model files were taken from~\citeref{Dorsner:2018ynv}.

To verify the procedure, we simulated the $t\bar{t}$ background and the scalar leptoquark signal at $13\TeV$ and compared to the predicted shapes in the $S_T$ distribution from the existing CMS analysis~\cite{Sirunyan:2017yrk}. After we verified the $13\TeV$ analysis, we simulated the signal and the dominant background events at $27\TeV$. From these samples, we selected all events satisfying the particle content requirements and applied the lower cut in the $S_T$ variable. The cut threshold was chosen to maximise $s/\sqrt{b}$ while requiring at least 2 expected signal events at an integrated luminosity of $15\abinv$. In the case of the vector leptoquark we considered the Yang-Mills ($\kappa = 1$) and the minimal coupling ($\kappa = 0$) scenarios~\cite{Dorsner:2018ynv}. From the simulations of the scalar and vector leptoquark events, we found the ratio of the cross-sections, and assuming similar kinematics, we estimated the sensitivity also for the vector leptoquark $U_1$.

\begin{figure}[t]
\centering
\begin{minipage}[t][\minipageheightsp][t]{\minipagewidthsp}
\centering
\includegraphics[width=\figuremaxwidthsp,height=\figuremaxheightsp,keepaspectratio]{\main/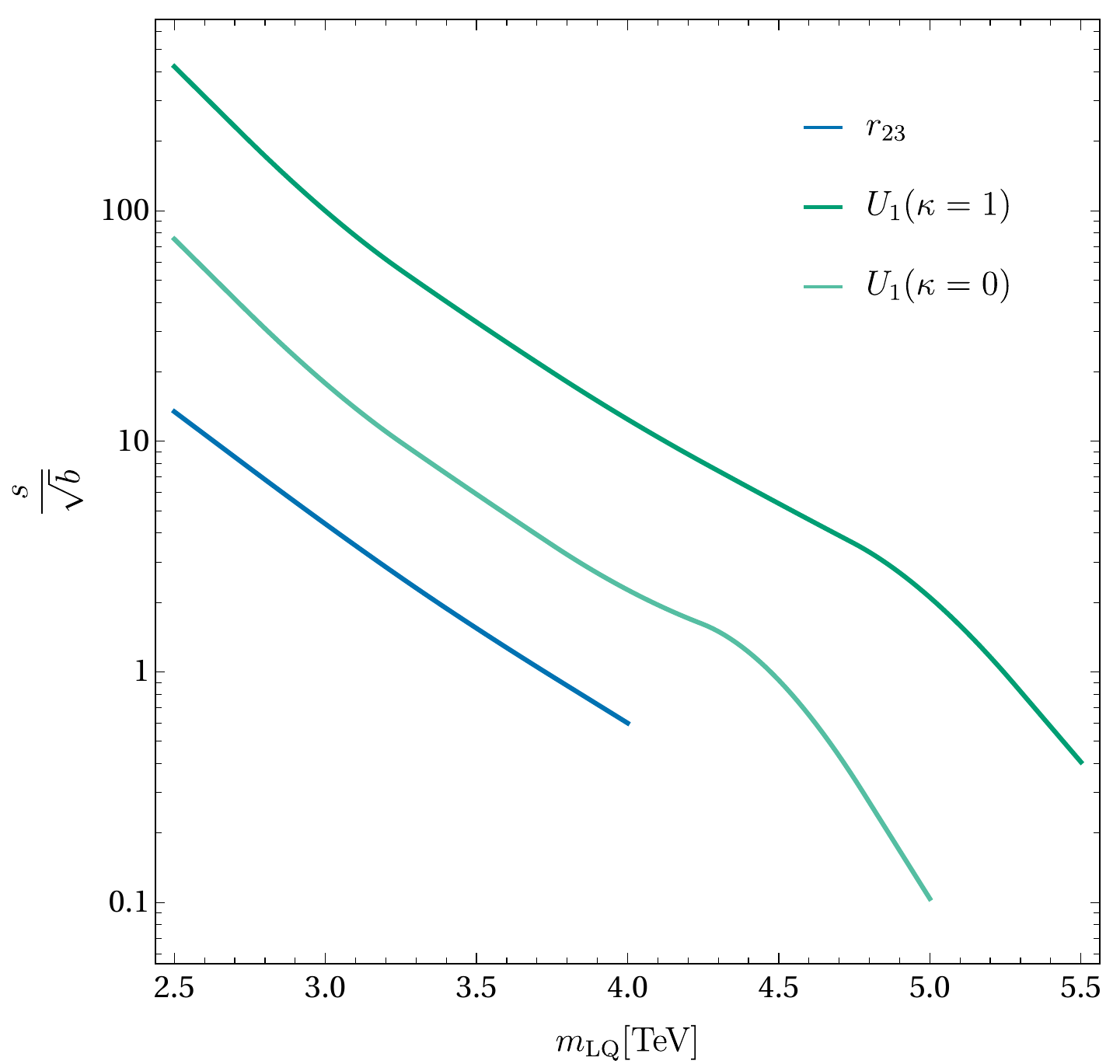}
\end{minipage}
\caption{Expected sensitivity for pair production of scalar ($r_{23}$) and vector ($U_1$) leptoquark at $27\TeV$ $p p$ collider with an integrated luminosity of $15\abinv$.} \label{fig:lukas}
\end{figure}

As shown in \fig{fig:lukas}, the HE-LHC collider will be able to probe pair produced third generation scalar leptoquark (decaying exclusively to $b \tau$ final state) up to mass of $\sim 4\TeV$ and vector leptoquark up to $\sim 4.5\TeV$ and $\sim 5.2\TeV$ for the minimal coupling and Yang-Mills scenarios, respectively. While this result is obtained by a rather crude analysis, it shows the impressive reach of the future high-energy $pp$-collider. In particular, the HE-LHC will cut deep into the relevant perturbative parameter space for $b \to c \tau \nu$ anomaly. As a final comment, this is a rather conservative estimate of the sensitivity to leptoquark models solving $R(D^{*})$ anomaly, since a dedicated single leptoquark production search is expected to yield even stronger bounds~\cite{Buttazzo:2017ixm,Dorsner:2018ynv}.
%%%%%%%%%%%

\subsection{\theory{High $p_T$ implications of flavour anomalies}}\label{sec:highpTflavanomalies}
%\contributors{Alejandro Celis, Admir Greljo, Lukas Mittnacht, Marco Nardecchia, Tevong You}
\contributors{A. Celis, A. Greljo, L. Mittnacht, M. Nardecchia, T. You}
%{\bf Author(s): Y. Takahashi et al., CMS}

%in collaboration with WG3

Precision measurements of flavour transitions at low energies, such as flavour changing $B$, $D$ and $K$ decays, are sensitive probes of hypothetical dynamics at high energy scales.  These can provide the first evidence of new BSM phenomena, even before direct discovery of new particles at high energy colliders. Indeed, the current anomalies observed in $B$-meson decays, the charge current one in $b\to c\tau \nu$ transitions, and neutral current one in $b\to s\ell^+\ell^-$,  may be the first hint of new dynamics which is still waiting to be discovered at high-$p_T$. When considering models that can accommodate the anomalies, it is crucial to analyse the constraints derived from high-$p_T$ searches at the LHC, since these can often rule out significant regions of model parameter space. Below we review these constraints, and assess the impact of the HL- and HE-LHC upgrades.
% (see also the discussion in Sections \ref{sec:7:bsll:NP},\ref{sec7:bsll:model:indep:fits}, and \ref{sec7:ModelsNP:bctaunu}).
%\td{where do we review anomalies, cite here}
% in this context.

\subsubsection{EFT analysis} 

If the dominant NP effects give rise to dimension-six SMEFT operators, the low-energy flavour measurements are sensitive to 
%the NP through combinations of the form 
$C/\Lambda^2$, with $C$ the dimensionless NP Wilson coefficient and $\Lambda$ the NP scale.    The size of the Wilson coefficient is model dependent, and thus so is the NP scale required to explain the $R_{D^{(*)}}$and $R_{K^{(*)}}$ anomalies. Perturbative unitarity sets an upper bound on the energy scale below which new dynamics need to appear~\cite{DiLuzio:2017chi}.    
%When the NP operators responsible for explaining the $R_{D^{(*)}}$and $R_{K^{(*)}}$ anomalies are aligned in the direction of the relevant flavour eigenstates one derives a 
The conservative bounds on the scale of unitarity violation are $\Lambda_U = 9.2\TeV$ and $84\TeV$ for $R_{D^{(*)}}$and $R_{K^{(*)}}$, respectively, obtained when the flavour structure of NP operators is exactly aligned with what is needed to explain the anomalies. More realistic frameworks for flavour structure, such as MFV, $U(2)$ flavour models, or partial compositeness, give rise to NP effective operators with largest effects for the third generation. This results in stronger unitarity bounds,
%  In the case of third generation alignment in the quark sector one derives 
$\Lambda_U = 1.9\TeV$ and $17\TeV$ for $R_{D^{(*)}}$ and $R_{K^{(*)}}$, respectively.  
%Two lessons can be drawn from these results:  
These results mean that: \textit{(i)} the mediators responsible for the $b\to c \tau \nu$ charged current anomalies are expected to be in the energy range of the LHC, \textit{(ii)}  the mediators responsible for the $b\to s \ell \ell$ neutral current anomalies could well be above the energy range of the LHC. However, in realistic flavour models also these mediators typically fall within the (HE-)LHC reach.       

%The first step towards explicit mediator models is to estimate the sensitivity of measuring the effects of $SU(2)_L$ gauge invariant dimension-6 operators on the tails of di-leptons produced at HL/HE LHC, \iee~$p p \to e^+ e^-$, $\mu^+ \mu^-$, and $\tau^+ \tau^-$. Assuming the mediator mass to be kinematically inaccessible for an on-shell production, this complementary test of $B$-anomalies is rather general, as it directly compares semi-leptonic four-fermion operators affecting the high-$p_T$ di-lepton production to those involved in $B$-decays. Such a comparison is possible thanks to the well-known properties of the effective theory (RGE, \etcc). Apart from the flavour structure, it is not required to specify the details of an underlining UV completion.

%In this context, it is instructive to start with an argument put forward in Ref.~\cite{Faroughy:2016osc}. 
If the neutrinos in $b \to c \tau \nu$ are part of a left-handed doublet, the NP responsible for $R_{D^{(*)}}$ anomaly
generically implies 
%Quite generically, $SU(2)_L$ gauge symmetry, with neutrinos being a part of the left-handed lepton doublet, implies that the new physics responsible for $R_{D^{(*)}}$anomaly (\iee~$b \to c \tau \nu$ transitions) is also expected to induce 
a sizeable signal in $p p \to \tau^+ \tau^-$ production at high-$p_T$. For realistic flavour structures, in which $b \to c$ transition is $\mathcal{O}(V_{cb})$ suppressed compared to $b \to b$, one expects rather large $b b \to \tau \tau$ NP amplitude. 
Schematically, $\Delta R_{D^{(*)}} \sim C_{b b \tau \tau} (1 + \lambda_{bs} / {V_{cb}})$, where $C_{b b \tau \tau}$ is the size of effective dim-6 interactions controlling $b b \to \tau \tau$, and $\lambda_{bs}$ is a dimensionless parameter controlling the size of flavour violation. Recasting ATLAS $13\TeV$, $3.2\fbinv$ search for $\tau^+ \tau^-$ \cite{Aaboud:2016cre}, \citeref{Faroughy:2016osc} showed that $\lambda_{bs} = 0$ scenario is already in slight tension with data. For $\lambda_{bs} \sim 5$, which is moderately large, but still compatible with FCNC constraints, HL (or even HE) upgrade of the LHC would be needed to cover the relevant parameter space implied by the anomaly (see 
%Fig.~[5] in~
\citeref{Buttazzo:2017ixm}). 
%However, as shown in Ref.~\cite{Buttazzo:2017ixm},  (
For large $\lambda_{bs}$ 
the limits from 
%diminishes the dominance of 
$p p \to \tau^+ \tau^-$ become comparable with 
%respect to 
direct 
the limits on $p p \to \tau \nu$ from the bottom-charm fusion. The limits on the EFT coefficients from $p p \to \tau \nu$ were derived in \citeref{Greljo:2018tzh}, and the future LHC projections are  promising. The main virtue of this channel is that the same four-fermion interaction is compared in $b \to c \tau \nu$ at low energies and $b c \to \tau \nu$ at high-$p_T$. 
%As a final comment, s
Since the effective NP scale in $R(D^{(*)})$ anomaly is low, the above EFT analyses are only indicative. 
%an indication of sensitivity. 
For more quantitative statements we review below bounds on 
%In fact, 
explicit models.
% should be considered for more quantitative statements (see discussion below).

The hints of NP in 
% new physics in the neutral current anomaly (\iee~
$R_{K^{(*)}}$ require  
%materialises as 
a $(b s) (\ell \ell)$ 
%point 
interaction.
% at low-energies. 
Correlated effects in high-$p_T$ tails of $p p \to \mu^+ \mu^- (e^+ e^-)$ distributions are expected, if the numerators (denominators) of LFU ratios $R_{K^{(*)}}$ are affected. 
%This connection has been investigated in 
Reference~\cite{Greljo:2017vvb} recast the  $13\TeV$ $36.1\fbinv$ ATLAS search \cite{ATLAS:2017wce} (see also \citeref{Aaboud:2017buh}),
%with $36.1$\fbinv of data,  
to set limits on a number of semi-leptonic four-fermion operators, and derive projections for HL-LHC (see 
%Table 1 in 
\citeref{Greljo:2017vvb}). These show that direct limits on the $(b s) (\ell \ell)$ operator from the tails of distributions will never be competitive with those implied by the rare $B$-decays~\cite{Greljo:2017vvb,Afik:2018nlr}. On the other hand, flavour conserving operators, $(q q) (\ell \ell)$, are efficiently constrained by the high $p_T$ tails of the distributions. The flavour structure of an underlining NP could thus be probed by constraining ratios $\lambda^q_{b s} = C_{b s} / C_{qq}$ with $C_{b s}$  fixed by the $R_{K^{(*)}}$ anomaly. For example, in models with MFV flavour structure, so that $\lambda^{u,d}_{b s} \sim V_{c b}$, the present high-$p_T$ dilepton data is already in slight tension with the anomaly~\cite{Greljo:2017vvb}. Instead, if  couplings to valence quarks are suppressed, \egg, if NP dominantly couples to the 3rd family SM fermions, then $\lambda^b_{b s} \sim V_{c b}$. Such NP will hardly be probed even at the HL-LHC, and  it is possible that NP responsible for the neutral current anomaly might stay undetected in the high-$p_T$ tails at HL-LHC and even at HE-LHC. Future data will cover a significant part of viable parameter space, though not completely, so that discovery is possible, but not guaranteed.

\subsubsection{Constraints on simplified models for $b\to c \tau \nu$}

Since the $b\to c \tau \nu$ decay is a tree-level process in the SM that receives no drastic suppression, models that can explain these anomalies necessarily require a mediator that contributes at tree-level:      

\noindent  $\bullet$ \, \textit{SM-like $W^{\prime}$}:  A SM-like $W^{\prime}$ boson, coupling to left-handed fermions, would explain the approximately equal enhancements observed in $R(D)$ and $R(D^{*})$.  A possible realisation is a colour-neutral real $SU(2)_L$ triplet of massive vector bosons~\cite{Greljo:2015mma}.  However, typical models encounter problems with current LHC data since they result in large contributions to $pp \to \tau^+ \tau^-$ cross-section, mediated by the neutral partner of the $W^{\prime}$~\cite{Greljo:2015mma,Boucenna:2016qad,Faroughy:2016osc}.    For $M_{W^{\prime}} \gtrsim 500\GeV$, solving the $R(D^{(*)})$ anomaly within the vector triplet model while being consistent with $\tau^+ \tau^-$ resonance searches at the LHC is only possible if the related $Z^{\prime}$ has a large total decay width~\cite{Faroughy:2016osc}. Focusing on the $W^{\prime}$, \citeref{Abdullah:2018ets} analysed the production of this mediator via $gg$ and $gc$ fusion, decaying to $\tau \nu_{\tau}$. Reference \cite{Abdullah:2018ets} concluded that a dedicated search using that a $b$-jet is present in the final state would be effective in reducing the SM background compared to an inclusive analysis that relies on $\tau$-tagging and $E_T^{\rm{miss}}$. Nonetheless, relevant limits will be set by an inclusive search in the future~\cite{Greljo:2018tzh}.

\vspace{0.2cm}

\noindent  $\bullet$  \,  \textit{Right-handed $W^{\prime}$:} \citerefs{Greljo:2018ogz,Asadi:2018wea} recently proposed that $W^{\prime}$ could mediate a right-handed interaction, with a light sterile right-handed neutrino carrying the missing energy in the $B$ decay. In this case, it is possible to completely uncorrelate FCNC constraints from $R(D^{(*)})$. The most constraining process in this case is instead $p p \to \tau \nu$. Reference \cite{Greljo:2018tzh} performed a recast of the latest ATLAS and CMS searches at $13\TeV$ and about $36\fbinv$ to constrain most of the relevant parameter space for the anomaly.

\vspace{0.2cm}

\noindent  $\bullet$\, \textit{Charged Higgs $H^{\pm}$:}  Models that introduce a charged Higgs, for instance a two-Higgs-doublet model, also contain additional neutral scalars.  Their masses are constrained by EW precision measurements to be close to that of the charged Higgs.  Accommodating the $R(D^{(*)})$ anomalies with a charged Higgs typically implies large new physics contributions to $pp \to \tau^+ \tau^-$ via the neutral scalar exchanges, so that current LHC data can challenge this option~\cite{Faroughy:2016osc}.     Note that a charged Higgs also presents an important tension between the current measurement of $R(D^*)$ and the measured lifetime of the $B_c$ \sloppy\mbox{meson~\cite{Li:2016vvp,Alonso:2016oyd,Celis:2016azn,Akeroyd:2017mhr}}.

\begin{figure}[t]
\centering
\begin{minipage}[t][\minipageheightsp][t]{\minipagewidthsp}
\centering
\includegraphics[width=\figuremaxwidthsp,height=\figuremaxheightsp,keepaspectratio]{\main/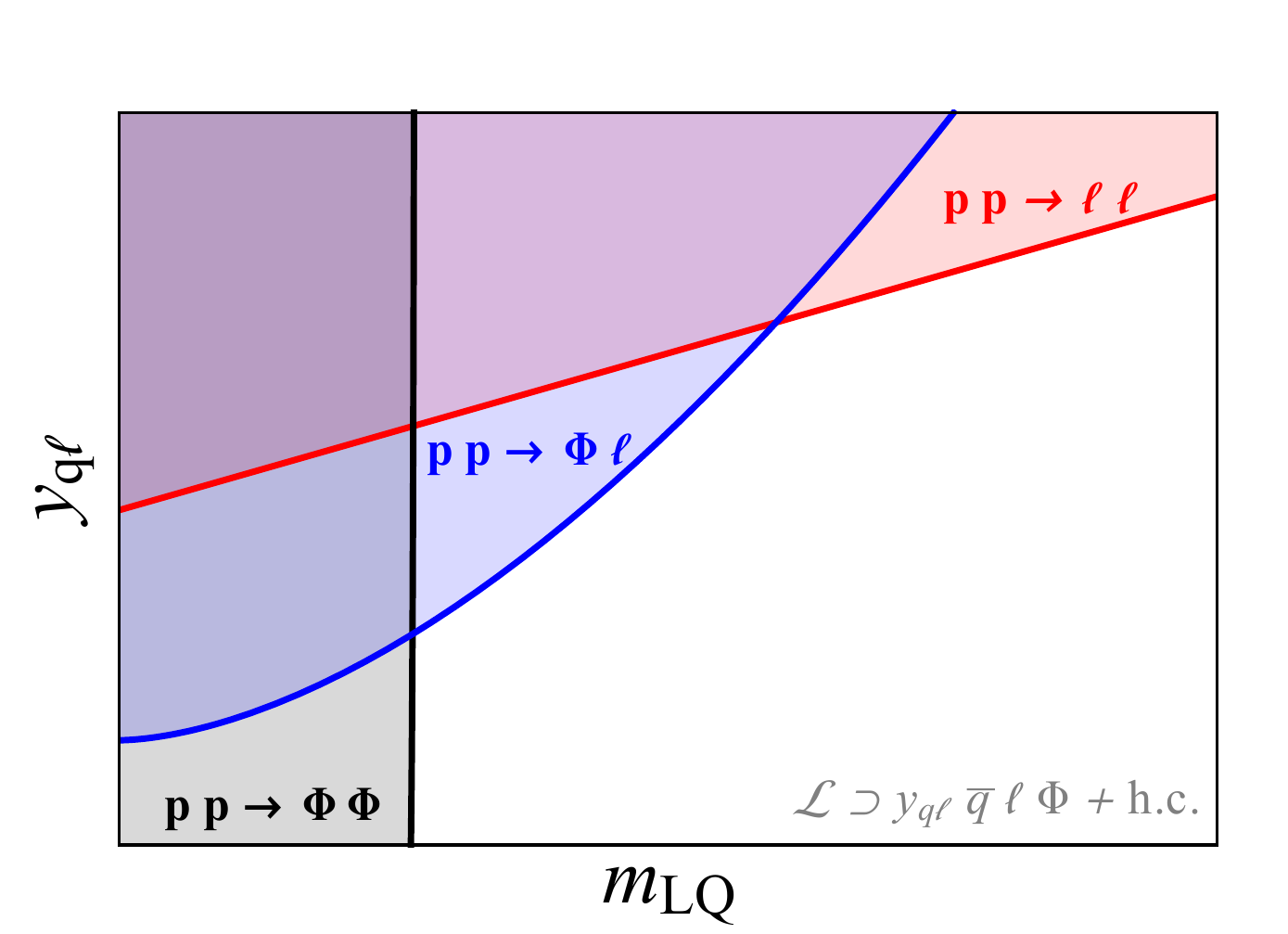}
\end{minipage}
\caption{Schematic of the LHC bounds on  LQ showing complementarity  in constraining the  $(m_{LQ},y_{q_{\ell}})$ parameters. The three cases are:
 pair production $\sigma \propto y_{q_{\ell}}^0 $, single production $\sigma \propto y_{q_{\ell}}^2 $ and Drell-Yan $\sigma \propto y_{q_{\ell}}^4$ (from \citeref{Dorsner:2018ynv}).}
\label{fig:LQstrategy}
\end{figure}

\begin{figure}[t]
\centering
\begin{minipage}[t][\minipageheightdp][t]{\minipagewidthdp}
\centering
\includegraphics[width=\figuremaxwidthdp,height=\figuremaxheightdp,keepaspectratio]{\main/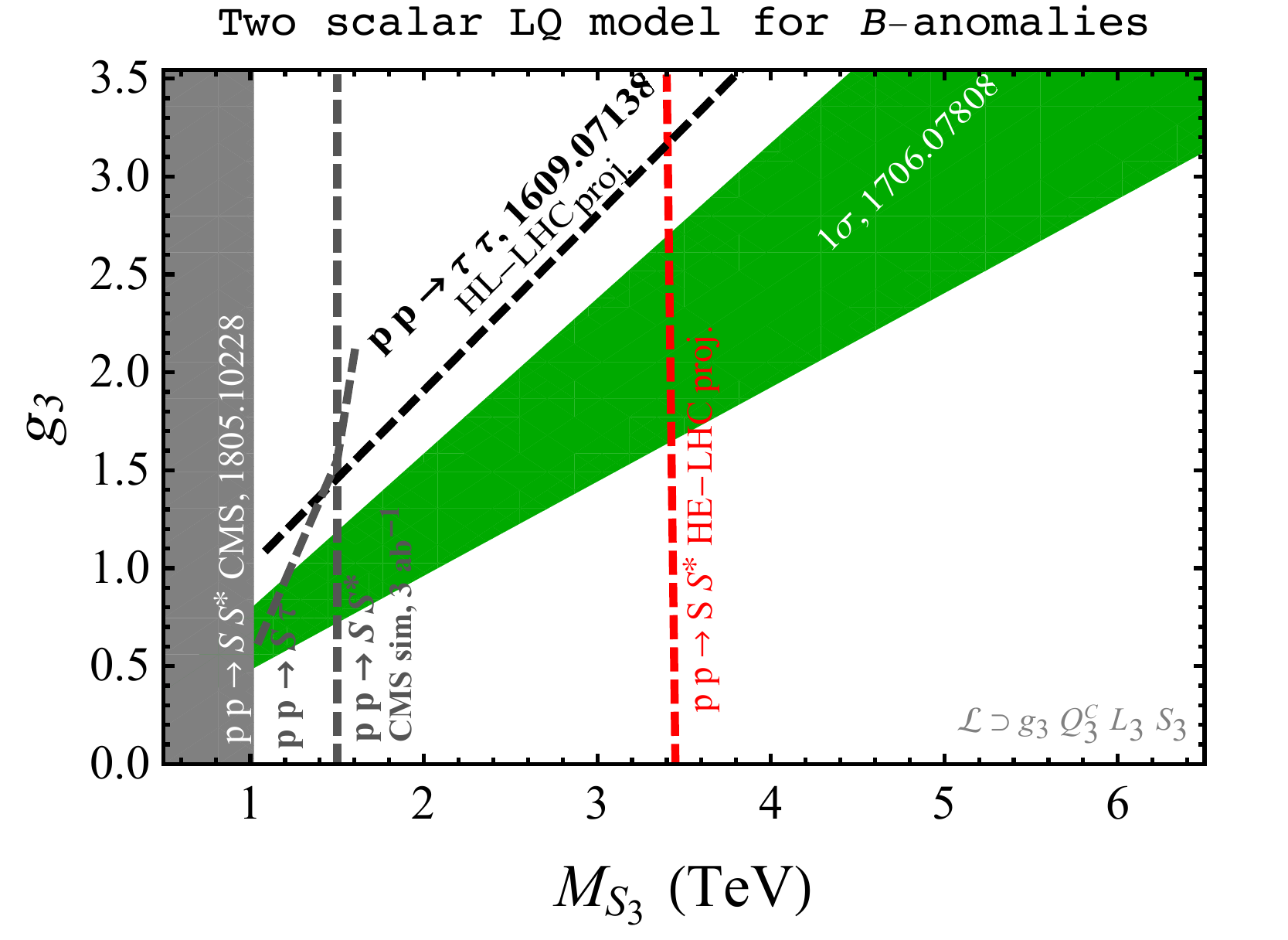}
\end{minipage}
\begin{minipage}[t][\minipageheightdp][t]{\minipagewidthdp}
\centering
\includegraphics[width=\figuremaxwidthdp,height=\figuremaxheightdp,keepaspectratio]{\main/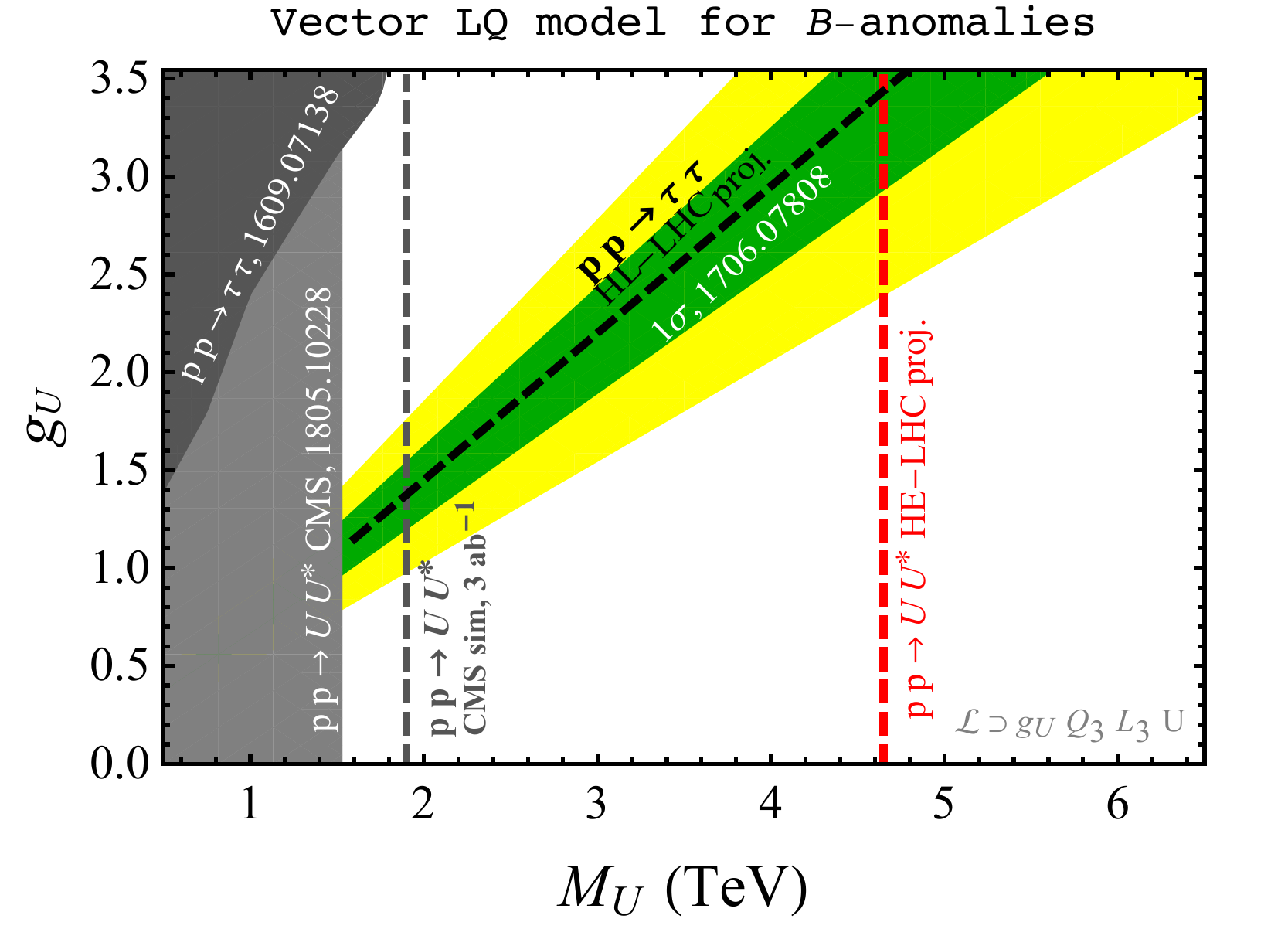}
\end{minipage}
\caption{Present constraints and HE and HL-LHC reach in the LQ mass versus coupling  plane for the scalar leptoquark $S_3$ (left), and vector leptoquark $U_1$ (right). The grey and dark grey solid regions are the current exclusions. The grey and black dashed lines are the projected reach for HL-LHC (pair production prospects are based on \secc{sec:leptoqbtau}). The red dashed lines are the projected reach at HL-LHC (see \secc{sec:LQHELHC}). The green and yellow bands are the $1\sigma$ and $2\sigma$ preferred regions from the fit to $B$ physics anomalies. The second coupling required to fit the anomaly does not enter in the leading high-\pt\ diagrams but is relevant for fixing the preferred region shown in green, for more details see \citeref{Buttazzo:2017ixm}. 
}
\label{fig:LQmediators}
\end{figure}

\vspace{0.2cm}

\noindent  $\bullet$ \, \textit{Leptoquarks:} 
The observed anomalies in charged and neutral currents appear in semileptonic decays of the $B$-mesons. This implies that the putative NP has to couple to both quarks and leptons at the fundamental level. A natural BSM option is to consider mediators that couple simultaneously quarks and leptons at the tree level. Such states are commonly referred as leptoquarks. Decay and production mechanisms of the LQ are directly linked to the physics required to explain the anomalous data.
\begin{itemize}
\item {\bf Leptoquark decays:} the fit to the $R(D^{*})$ observables suggest a rather light leptoquark (at the TeV scale) that couples predominately to the third generation fermions of the SM. A series of constraints from flavour physics, in particular the absence of BSM effects in kaon and charm mixing observables, reinforces this picture. 

\item {\bf Leptoquark production mechanism:}  The size of the couplings required to explain the anomaly is typically very large, roughly $y_{q\ell} \approx m_{LQ}$/ (1 TeV). Depending of the actual sizes of the leptoquark couplings and its mass we can distinguish three regimes that are relevant for the phenomenology at the LHC:
 	\begin{enumerate}
	\item LQ pair production due to strong interactions,
	\item Single LQ production plus lepton via a single insertion of the LQ coupling, and
	\item Non-resonant production of di-lepton through $t$-channel exchange of the leptoquark.
	\end{enumerate}
Interestingly all three regimes provide complementary bounds in the  $(m_{LQ},y_{q_{\ell}})$ plane, see \fig{fig:LQstrategy}.

\end{itemize}

Several simplified models with a leptoquark as a mediator were shown to be consistent with the low-energy data.
% In addition, the high-$p_T$ phenomenology of these models including projections was discussed.
 A vector leptoquark with $SU(3)_c\times SU(2)_L\times U(1)_Y$ SM quantum numbers $U_\mu \sim ({\bf 3}, {\bf 1}, 2/3)$ was identified as the only single mediator model which can simultaneously fit the two anomalies (see \egg~\citeref{Buttazzo:2017ixm} for a recent fit including leading RGE effects). 
 %The LHC limits from complementary processes are shown in Fig.~4 of~\cite{Buttazzo:2017ixm}. 
 In order to substantially cover the relevant parameter space, one needs future HL- (HE-) LHC, see \fig{fig:LQmediators} (right) (see also 
% Fig.~5 of~
 \citeref{Buttazzo:2017ixm} for details on the present LHC constraints). A similar statement applies to an alternative model featuring two scalar leptoquarks, $S_1, S_3$ \cite{Crivellin:2017zlb}, with the projected reach at CLIC and HL-LHC  shown in \fig{fig:LQmediators} (left) (see 
% Fig.~2 in~
 \citeref{Marzocca:2018wcf} shows details on present LHC constraints). 
 %A comparison between the reach of CLIC and LHC is shown in \fig{fig:LQmediators}.
 
Leptoquarks states are emerging as the most convincing mediators for the explanations of the flavour anomalies. It is then important to explore all the possible signatures at the the HL- and HE-LHC.
 The experimental program should focus not only on final states containing quarks and leptons of the third generation, but also on the whole list of decay channels including the off-diagonal ones ($b \mu$, $s \tau, \dots$). The completeness of this approach would shed light on the flavour structure of the putative New Physics. 
 
 Another aspect to be emphasised concerning leptoquark models for the anomalies is the fact of the possible presence of extra fields required to complete the UV Lagrangian. The accompanying particles would leave more important signatures at high $p_T$ than the leptoquark, this is particularly true for vector leptoquark extensions (see, for example, \citeref{DiLuzio:2017vat,DiLuzio:2018zxy}).  
 
As a final remark, it is important to remember that the anomalies are not yet experimentally established. Among others, this also means that the statements on whether or not  the high \pt\ LHC constraints rule out certain $R(D^{(*)})$ explanations assumes that the actual values of $R(D^{(*)})$ are given by their current global averages. If future measurements decrease the global average, the high \pt constraints can in some cases be greatly relaxed and HL- and/or HE-LHC may be essential for these, at present tightly constrained, cases.

\subsubsection{Constraints on simplified models for $b\to s ll$}

The $b\to s ll$ transition is  both loop and CKM suppressed in the SM. The explanations of the $b\to s ll$ anomalies can thus have both tree level and loop level mediators.
% enter at tree-level or at the loop level. 
Loop-level explanations typically involve lighter particles.
% that enter at a lower UV cut-off scale. 
Tree-level mediators can also be light, if sufficiently weakly coupled. However, they can also be much heavier---possibly beyond the reach of the LHC. 

%\begin{figure}[t]
%\centering
%\includegraphics[width=0.45 %\textwidth]{\main/section5FlavorRelatedStudies/img/feynmandiagrams_LQZprime.PNG}
%\caption{Tree level Feynman diagrams for leptoquarks (left) and $Z^\prime$ (right)  mediating $b \to s \mu^+ \mu^-$ transition (Fig. taken from ~\cite{Allanach:2017bta}).}
%\label{fig:feynmanLQZprime}
%\end{figure}

\noindent $\bullet$ \textit{Tree-level mediators:}\\
%For the four-fermion operators involved in 
For $b \to sll$ anomaly there are two possible tree-level UV-completions, the $Z'$ vector boson and leptoquarks, either scalar or vector.
% (see \fig{fig:bsll_diagrams} in Sec.~\ref{sec:7:bsll:NP}).
% whose Feynman diagrams are depicted in~\fig{fig:feynmanLQZprime}. The $Z^\prime$ vector boson mediates an interaction between the $b$ and $s$ quark at one vertex and two muons at the other. A leptoquark is defined as having quantum numbers such that it may interact with a quark and lepton at one vertex, and can be either scalar or vector. In this case the minimal couplings required of the leptoquark must contain a $b$ or $s$ together with a muon at a vertex. 
%Ref.~\cite{Allanach:2017bta} 
%%initiated the study of implications of the anomalies for direct discovery at future colliders. They 
%estimated the sensitivity to simplified models of $Z^\prime$ and leptoquarks at the HL- and HE-LHC (as well as the FCC-hh) in the expected parameter space where the new particles may explain the neutral current flavour anomalies. A more detailed study for the $Z^\prime$ case was then made by Ref.~\cite{Allanach:2018odd}, extending in particular to wider widths. We summarise their results here. 
For leptoquarks, \fig{fig:boundsSens} (right) shows the current $95\%\cl$ limits from $8\TeV$ CMS with $19.6\fbinv$ in the $\mu\mu jj$ final state (solid black line), as well as the HL-LHC (dashed black line) and $1$ ($10$)\abinv HE-LHC extrapolated limits (solid (dashed) cyan line). 
%is denoted by a dashed black line for HL-LHC and by a solid (dashed) cyan line for HE-LHC with 1 (10)\abinv. 
Dotted lines give the cross-sections times BR at the corresponding collider energy for pair production of scalar leptoquarks, calculated at NLO using the code of \citeref{Kramer:2004df}. We see that the sensitivity to a leptoquark with only the minimal $b-\mu$ and $s-\mu$ couplings reaches around $2.5$ and $4.5\TeV$ at the HL-LHC and HE-LHC, respectively. This pessimistic estimate is a lower bound that will typically be improved in realistic models with additional flavour couplings. Moreover, the reach can be extended by single production searches~\cite{Allanach:2017bta}, albeit in a more model-dependent way than pair production. The cross section predictions for vector leptoquark are more model-dependent and are not shown in \fig{fig:boundsSens} (right). For $\mathcal{O}(1)$ couplings the corresponding limits are typically stronger than for scalar leptoquarks. 

\begin{figure}[t]
\centering
\begin{minipage}[t][\minipageheightsp][t]{\minipagewidthsp}
\centering
\includegraphics[width=\figuremaxwidthsp,height=\figuremaxheightsp,keepaspectratio]{\main/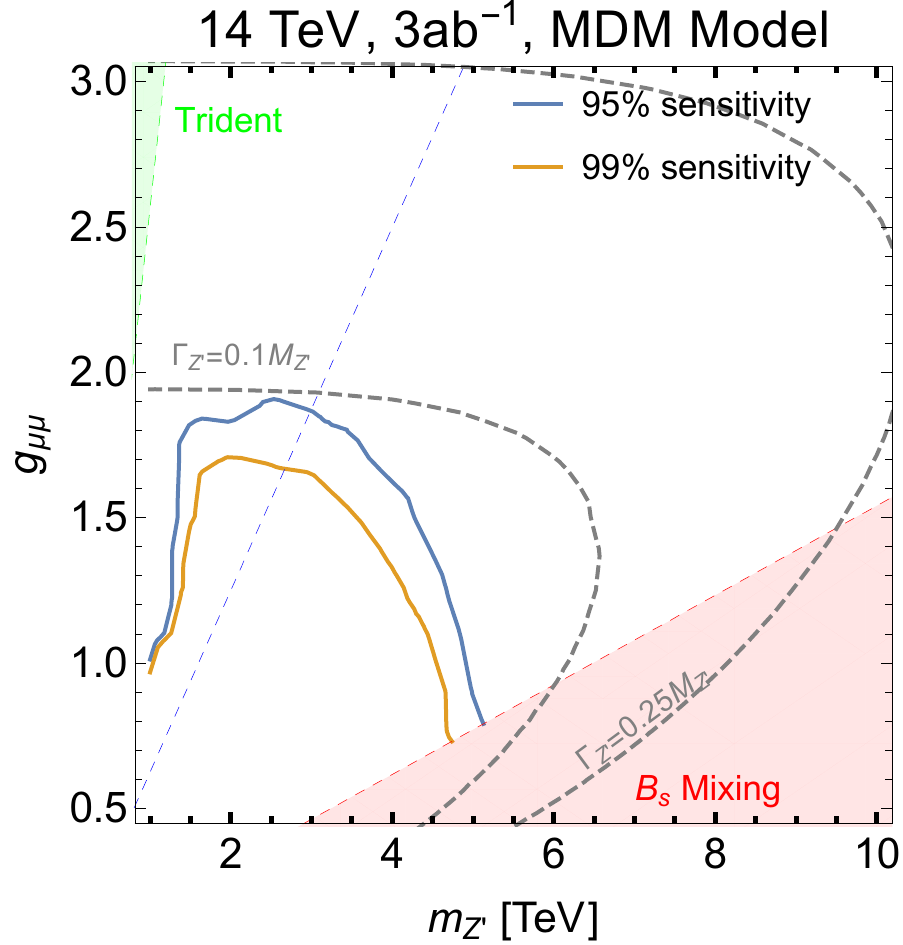}
\end{minipage}
\caption{ HL-LHC $95\%$ (blue) and $99\%$ (orange)\cl sensitivity contours to $Z^\prime$ in the ``mixed-down'' model for $g_{\mu\mu}$ vs $Z^\prime$ mass in TeV. The dashed grey contours give the $Z'$ width as a fraction of mass. The green and red regions are excluded by trident neutrino production and $B_s$ mixing, respectively. The dashed blue line is the stronger $B_s$ mixing constraint from \citeref{Bazavov:2016nty}. See also \secc{sec:theoryzprimeLQBanomalies}.}
\label{fig:Zprime14TeV}
\end{figure}

For the $Z^\prime$ mediator the minimal couplings in the mass eigenstate basis are obtained by unitary transformations from the gauge eigenstate basis, which necessarily induces other couplings.  Reference \cite{Allanach:2018odd} defined the ``mixed-up'' model (MUM) and ``mixed-down'' model (MDM) such that the minimal couplings are obtained via CKM rotations in either the up or down sectors respectively. For MUM there is no sensitivity at the HL-LHC.  The predicted sensitivity at the HL-LHC for the MDM is shown in \fig{fig:Zprime14TeV} as functions of $Z^\prime$ muon coupling $g_{\mu\mu}$ and the $Z^\prime$ mass, setting the $Z'$ coupling to $b$ and $s$ quarks such that it solves the $b\to s\ell\ell$ anomaly. The solid blue (orange) contours give the $95\%$ and $99\%\cl$  sensitivity. The red and green regions are excluded by $B_s$ mixing~\cite{Arnan:2016cpy} and neutrino trident production~\cite{Falkowski:2017pss,Falkowski:2018dsl}, respectively. The more stringent $B_s$ mixing constraint from \citeref{Bazavov:2016nty} is denoted by the dashed blue line; see, however, \citeref{Kumar:2018kmr} for further discussion regarding the implications of this bound. The dashed grey contours denote the width as a fraction of the mass. We see that the HL-LHC will only be sensitive to $Z^\prime$' with narrow width, up to masses of $5\TeV$. 

\begin{figure}[t]
\centering
\begin{minipage}[t][\minipageheightdp][t]{\minipagewidthdp}
\centering
\includegraphics[width=\figuremaxwidthdp,height=\figuremaxheightdp,keepaspectratio]{\main/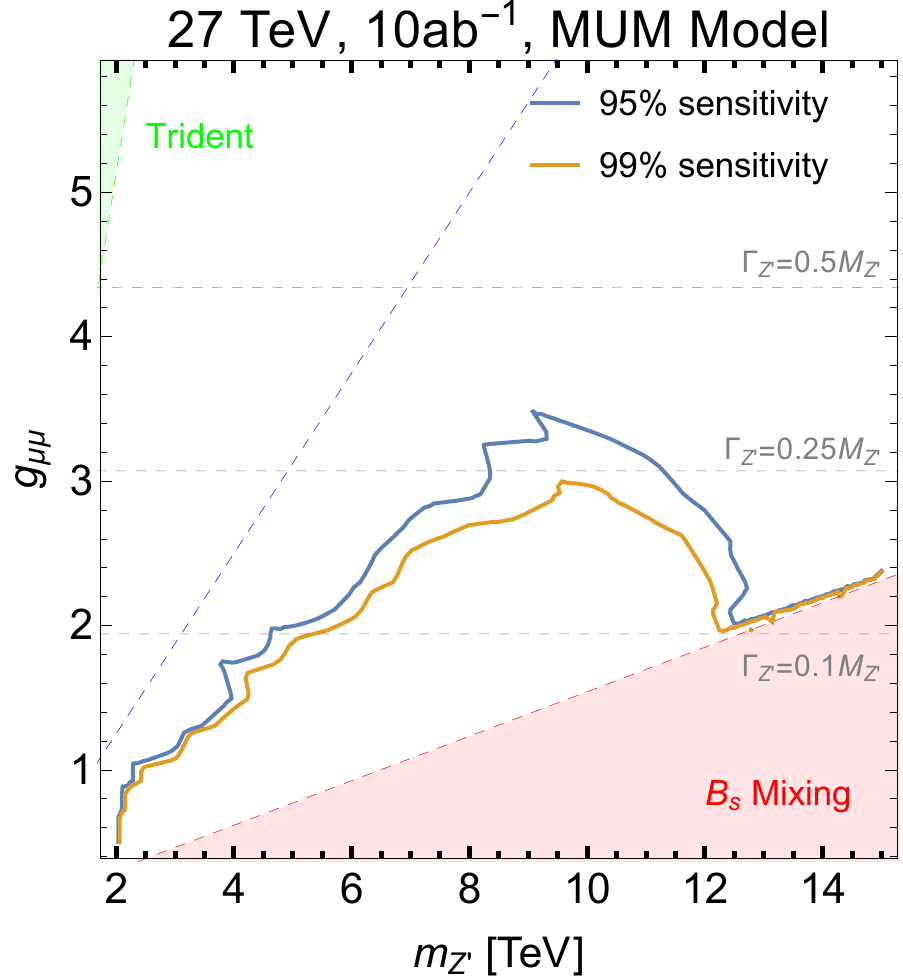}
\end{minipage}
\begin{minipage}[t][\minipageheightdp][t]{\minipagewidthdp}
\centering
\includegraphics[width=\figuremaxwidthdp,height=\figuremaxheightdp,keepaspectratio]{\main/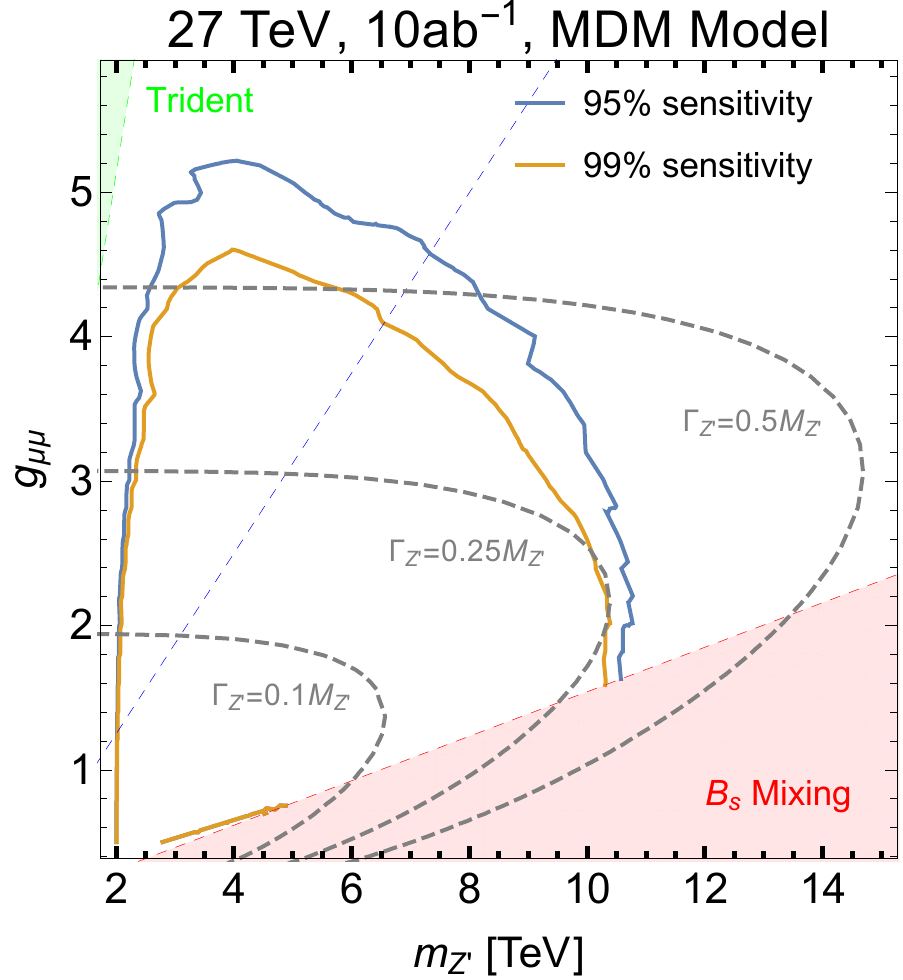}
\end{minipage}
\caption{ HE-LHC $95\%$ (blue) and $99\%$ (orange)\cl sensitivity contours to $Z^\prime$ in the ``mixed-up'' (left) and ``mixed-down'' (right) model in the parameter space of $g_{\mu\mu}$ vs $Z^\prime$ mass in TeV. The dashed grey contours are the width as a fraction of mass. The green and red regions are excluded by trident neutrino production and $B_s$ mixing. The dashed blue line is the stronger $B_s$ mixing constraint from \citeref{Bazavov:2016nty}. See also \secc{sec:theoryzprimeLQBanomalies}.}
\label{fig:Zprime27TeV}
\end{figure}

At the HE-LHC, the reach for $10\abinv$ is shown in \fig{fig:Zprime27TeV} for the MUM and MDM on the left and right, respectively. In this case the sensitivity may reach a $Z^\prime$ with wider widths up to $0.25$ and $0.5$ of its mass, while the mass extends out to $10$ to $12\TeV$. We stress that this is a pessimistic estimate of the projected sensitivity, particular to the two minimal models; more realistic scenarios will typically be easier to discover.

\vspace{0.2cm}

\noindent  $\bullet$ \textit{Explanations at the one-loop level:}\\
It is possible to accommodate the $b \to s \ell^+ \ell^-$ anomalies even if mediators only enter at one-loop.  One possibility are the mediators coupling to right-handed top quarks and to muons~\cite{Celis:2017doq,Becirevic:2017jtw,Kamenik:2017tnu,Fox:2018ldq,Camargo-Molina:2018cwu}. Given the loop and CKM  suppression of the NP contribution to the $b \to s \ell^+ \ell^-$ amplitude, these models can explain the $b \to s \ell^+ \ell^-$ anomalies for a light mediator, with mass around $\mathcal{O}(1)\TeV$ or lighter.  Constraints from the LHC and future projections for the HL-LHC were derived in \citeref{Camargo-Molina:2018cwu} by recasting di-muon resonance, $pp\to t\bar tt \bar t$ and SUSY searches.  Two scenarios were considered: \textit{i)} a scalar LQ $R_2(3, 2, 7/6)$ combined with a vector LQ $\widetilde U_{1 \alpha}(3, 1, 5/3)$, \textit{ii)} a vector boson $Z^{\prime}$ in the singlet representation of the SM gauge group. Reference \cite{Fox:2018ldq} also analysed the HL-LHC projections for the $Z^{\prime}$.  The constraints from the LHC already rule out part of the relevant parameter space and the HL-LHC will be able to cover much of the remaining regions.  Dedicated searches in the $pp\to t\bar tt \bar t$ channel and a dedicated search for $t \mu$ resonances in $t \bar t \mu^+ \mu^-$ final state can improve the sensitivity to these models~\cite{Camargo-Molina:2018cwu}.

\clearpage
\resetcounters

\section{Other BSM signatures}\label{sec:otherBSM}
New physics models aimed at extending the SM either to overcome some of its theoretical puzzles, like the EW, flavour and strong CP ``hierarchies'', or by the need of explaining new phenomena, such as neutrino masses, Dark Matter, and baryogenesis, often predict signatures that are very different from the common SUSY ones, due to the absence of large missing transverse energy. These signatures are instead usually characterised by on-shell resonances, singly or doubly produced, depending on their quantum numbers, that decay into visible SM particles. 

The lack so far of hints of new physics at the LHC, has typically required an increased level of complication in models addressing the aforementioned puzzles.  Many of these complicated, though well motivated scenarios, are not always the best reference to clearly and simply evaluate the potential of a future collider. Often, as in the case of SUSY, considering simplified models inspired by more motivated, but involved constructions, constitutes the best choice to provide target experimental signatures.  These targets are perfectly suited to evaluate the potential of future collider upgrades or new future colliders, and to compare them. Moreover, one can also try to answer questions such as what is the sensitivity to discriminate models given a discovery. 

Usual standard scenarios include singly-produced resonances, with integer spin, decaying to two SM fermions or bosons, and pair-produced heavy fermion resonances, decaying to SM bosons and fermions. Most of these cases will be covered in this section, where the main ``metric'' to evaluate collider performances is the reach in the mass of these resonances, or the reach in covering a mass vs.\ coupling parameter space.
The experimental signatures can be rather clean, like in the case of dilepton resonances, where the reach is usually limited by statistical uncertainties, or they can instead be affected by large SM backgrounds, especially in the hadronic channels, where the reach is typically limited by systematic uncertainties. 

This section presents several results considering a broad variety of signatures and new physics scenarios. Of course the list is far from complete in terms of its coverage of possible BSM scenarios. However, it provides a clear and wide enough picture of the potential of the HL- and HE-LHC in terms of reach on heavy states, and coverage of their parameter space. 

%Finally, notwithstanding that the presented results contain a very detailed study of future prospects, that should serve as a reference for future studies, the global message of this section is that the HL-LHC will be able to extend the mass reach on heavy objects typically by $\sim-\%$, while the reach on the coupling of weakly-coupled new physics will more strongly benefit from the larger statistics, with improvements on the sensitivity to the typical coupling of the order of $-\%$. Concerning the HE-LHC, the conclusion is often a doubled mass reach on heavy objects and further $-\%$ improvements in the coupling reach. \rt{This last paragraph aims at summarising the well known fact. Let's agree on the numbers to quote.}

Finally, notwithstanding that the presented results contain a very detailed study of future prospects, that should serve as a reference for future studies, the global message of this section is that the HL-LHC will be able to extend the present LHC mass reach on heavy objects typically by $\sim 20-50\%$.
%, while the reach on the coupling of weakly-coupled new physics will more strongly benefit from the larger statistics, with improvements on the sensitivity to the %typical coupling of the order of $-\%$. 
Furthermore, HL-LHC will also be able to constrain, and potentially discover, new physics that is presently unconstrained.
Concerning the HE-LHC, the conclusion is often a doubled mass reach, beyond HL-LHC, on heavy objects.
% and further $-\%$ improvements in the coupling reach.

%There are several ways in which one could organise the results presented in this section. Given the material we had at our disposal, we decided to split most of the analysis in terms of the spin of the studied resonances, which usually reflects in the final states considered. We therefore present three sections containing spin 0 and 2, spin 1 and spin 1/2 resonances. A few more analysis that are more signature based and could enter in more than one of the first three sections, have been collected in the last subsection.
The results are presented considering a categorisation in terms of the spin of the studied resonances, which usually reflects in the final states considered. We therefore present sections containing spin 0 and 2 (\secc{sec:BSMspin02}), spin 1 (\secc{sec:BSMspin1}) and spin 1/2 resonances (\secc{sec:BSMspinhalf}). Additional prospect analyses that are more signature-based and could enter in more than one of the first three sections, have been collected in the last \secc{sec:BSMsignaturebased}. 

\vspace{1cm}

\subsection{Spin 0 and 2 resonances}\label{sec:BSMspin02}
We present here several results concerning the production and decay of spin 0 and 2 particles decaying into several different SM final states. These range from resonant double Higgs production through a spin-2 KK graviton, to heavy scalar singlets that could mix with the Higgs, making the interplay with Higgs coupling measurements a crucial ingredient, from light pseudoscalar and axion-like particles to colour octet scalars.
\subsubsection{\atlas{Resonant double Higgs production in the $4 b$ final state at the HL-LHC}}
\contributors{S. Willocq and A. Miller, ATLAS}
%{\bf Author(s): S. Willocq and A. Miller, ATLAS}

The projection study in \citeref{ATL-PHYS-PUB-2018-028}, summarised here, uses the search for high-mass spin-2 KK gravitons decaying into two Higgs bosons, $HH$, as a benchmark, with each of the Higgs bosons decaying to $b\bar{b}$, thereby yielding 
a final state with two highly boosted $b\bar{b}$ systems, which are reconstructed as two large-radius jets.
The following strategy is followed to obtain sensitivity estimates at the HL-LHC: 
(i) signal and background mass distributions for the pair of candidate Higgs bosons in the event 
are taken from the most recent ATLAS data analysis at $\sqrt{s} = 13\TeV$~\cite{Aaboud:2018knk} and scaled to \intLumi;
\sloppy\mbox{(ii) simulated} signal and background mass distributions are used to 
derive mass-dependent scaling functions to extrapolate the distributions
from $\sqrt{s} = 13\TeV$ to $14\TeV$;
(iii) simulated signal and background mass distributions are used for further scaling
of the distributions to reproduce the impact of additional selection criteria
not included in Run-2 searches. 

The dominant background for the $13\TeV$ data analysis stems from multijet production. 
This background source is estimated directly from data in that analysis and represents about 
$80\%$, $90\%$, and $95\%$ of the total background in the signal region for events with
2, 3, and 4 $b$-tags (the classification of events based on the number of $b$-tags is described below).
The remaining source of background originates almost completely from $\ttbar$ production.
The shape of the dijet mass distribution for $\ttbar$ events is taken from MC samples. 
The normalisation of the $\ttbar$ background in \citeref{Aaboud:2018knk} 
is extracted from a fit to the leading large-$R$ jet mass distribution in the $13\TeV$ data.
Samples of simulated multijet background events are generated to derive scaling functions
to be applied to the background predictions from the $13\TeV$ data analysis.
These samples are generated at both $\sqrt{s} = 13\TeV$ and $\sqrt{s} = 14\TeV$.
Two different sets of MC samples are used to study the impact of differences in jet
flavour composition on the multijet background scaling functions.
The first set of events corresponds to the $2 \to 2$ processes $pp \to jj$ (with $j = g$ or $q$)
generated with \pythia{ 8} with truth jet $\pt$ in the range between $400$ and $2500\GeV$.
The second set of events corresponds to the $2 \to 4$ processes $pp \to bbbb$ generated
with \MGvATNLO  requiring $b$-quarks to have $\pt$ above $100\GeV$ and the events are required to have at least one $b$-quark 
with $\pt$ above $200\GeV$.

%The truth-level events generated to derive scaling functions are processed with 
%a reconstruction that builds a series of different jet collections.
Large-radius jets are used in the analysis. They are built from generated particles with the $\antikt$ algorithm
operating with a radius parameter $R = 1.0$. Previous studies indicate that the trimming 
effectively removes the impact of pileup up to $\mu = 300$~\cite{JetSubstructureECFA2014}.
Small-radius jets are built from generated charged particles
with the $\antikt$ algorithm and a radius of $R=0.2$. Only charged particles with $\pt > 0.5\GeV$
are used in the clustering to emulate the track jets used in the $13\TeV$ data analysis.

The event selection applied to the truth-level analysis
proceeds similarly to that used for the $13\TeV$ data analysis.
Three regions in the plane formed by the leading large-$R$ jet mass and the
subleading large-$R$ jet mass are used in the analysis.
The signal region is defined by the requirement $X_{HH} < 1.6$, with $X_{HH}$ defined as
\begin{equation}
X_{HH}\,=\,\sqrt{\left(\frac{\mleadJ\,-\,125\,\GeV{}}{0.1\,\mleadJ}\right)^2 + \left(\frac{\msublJ\,-\,120\,\GeV{}}{0.1\,\msublJ}\right)^2}~,
\label{eq:XHH}
\end{equation}
%The control region is defined by a requirement on 
% $R_{HH} = \sqrt{\left(\mleadJ\,-\,125\,\GeV{}\right)^2 + \left(\msublJ\,-\,120\,\GeV{}\right)^2} < 33$ GeV and $X_{HH} \geq 1.6$.
%The sideband region is defined by the requirements
%$R_{HH} \geq 33\GeV$and $\sqrt{\left(\mleadJ\,-\,135\,\GeV{}\right)^2 + \left(\msublJ\,-\,130\,\GeV{}\right)^2} < 58\GeV$.
where $m_{J}$ is the large-$R$ jet mass.
The choices made to define the control and sideband regions are driven by the
need to select events that are kinematically similar to those in the signal region
while providing sufficiently large samples to derive the background estimate from
the sideband region and validate it in the control region.

For the $13\TeV$ data analysis, the dominant multijet background is estimated with
a data-driven approach which utilises events with a smaller number of $b$-tags in the sideband region.
The events used for this estimation are required to have the same track-jet topology 
as in the event categories in which they are used to model the background. 
Events with 1~$b$-tag are used to model the background in the 2-tag category. 
Likewise, events with 2~$b$-tags are used to model the background in the 3- and 4-tag categories.
Various checks performed show that either multijet simulation can be used reliably to predict the shape
of the dijet mass distributions. However, differences in the flavour content of the large-$R$ jets 
in the dijet and $4b$ multijet MC samples do affect the predicted background yields in
the following study and the two samples are used to extract a range of projections at the HL-LHC.

As mentioned above, the projection for the HL-LHC proceeds in three steps. At first,  the dijet mass distributions for signal and both multijet and $\ttbar$ background events from the $13\TeV$ data analysis in Run-2 are scaled from $36.1\fbinv$ to \intLumi. The background distributions are further scaled with mass-dependent functions to extrapolate from $\sqrt{s} = 13\TeV$ to $14\TeV$. These functions take into account both increases in cross section and changes in detector performance from the Run-2 ATLAS detector to the future upgraded detector at the HL-LHC. Further mass-dependent scaling is applied to all signal and background distributions to reflect improvements in the reconstruction of highly boosted jets, as obtained by using 
variable-radius track jets~\cite{ATL-PHYS-PUB-2016-013} instead of fixed-radius ($R=0.2$) track jets, or in the background suppression by applying a requirement on
the maximum number of charged particles associated with each large-$R$ jet. 

The first improvement relative to the $13\TeV$ data analysis arises from the use of variable-radius
track jets. This circumvents the problem that $R=0.2$ track jets from the
$H \to b\bar{b}$ decay start merging for Higgs boson $\pt$ values larger than
approximately $2 m_H/R = 1250\GeV$.
The second improvement relative to the $13\TeV$ data analysis is the requirement of a maximum
number of charged particles associated with large-$R$ jets to exploit differences between
quark- and gluon-initiated jets, the latter being an important component of the multijet
background. The impact of pileup at $\mu=200$ has been studied for charged particle tracks with $\pt > 1\GeV$
associated with the primary vertex.
In the case of $\ttbar$ events, the average number of tracks associated with the primary vertex
increases by about $15\%$ due to pileup. Further pileup
suppression is possible with additional requirements on the longitudinal impact parameter or
track-vertex association probability. 
Both leading and subleading large-$R$ jets are required to have fewer than 20 charged particles with $\pt > 1\GeV$
and $\Delta R < 0.6$ with respect to the jet axis.

\begin{figure}[t!]
\centering
\begin{minipage}[t][\minipageheightdp][t]{\minipagewidthdp}
\centering
\includegraphics[width=\figuremaxwidthdp,height=\figuremaxheightdp,keepaspectratio]{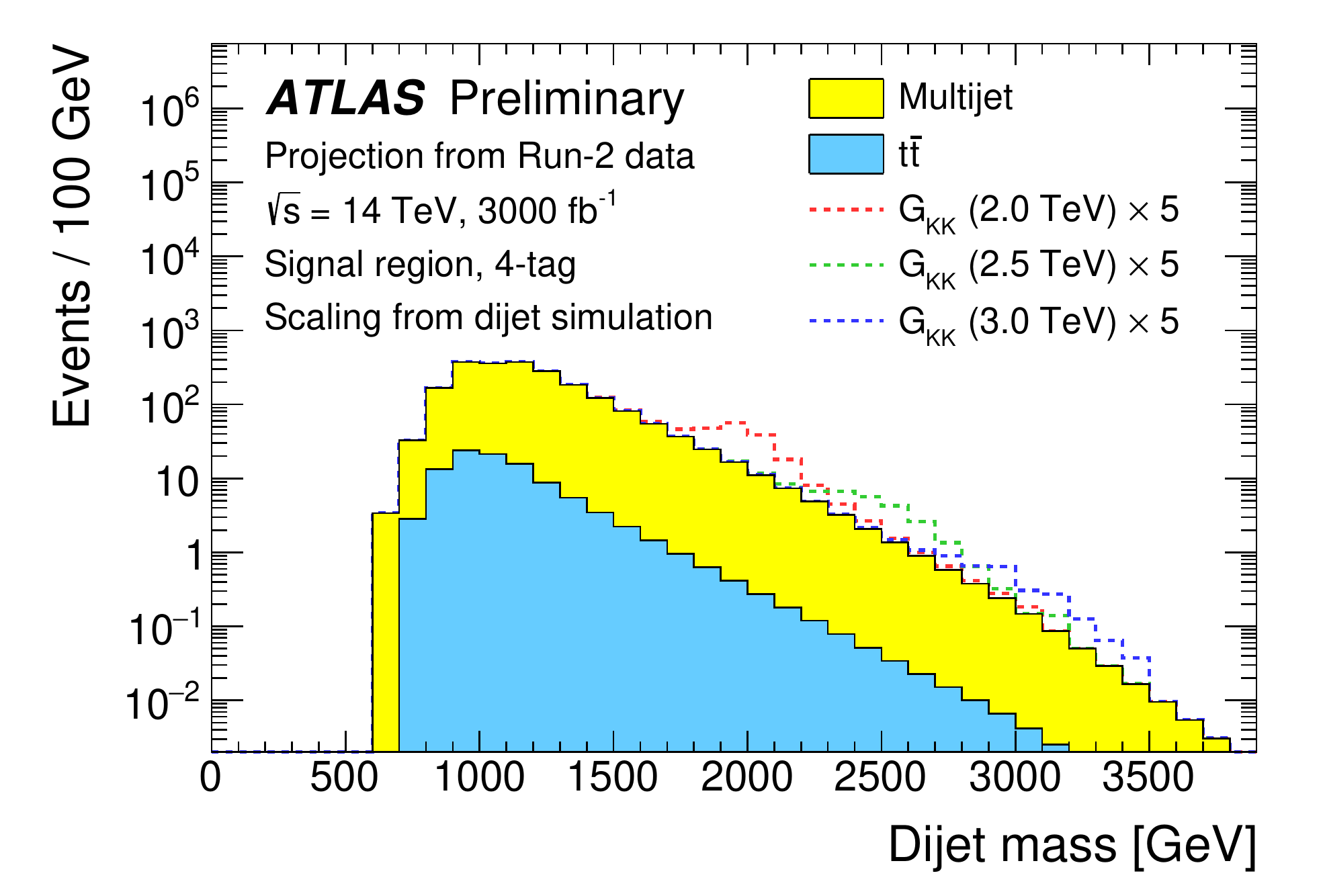}
\end{minipage}
\begin{minipage}[t][\minipageheightdp][t]{\minipagewidthdp}
\centering
\includegraphics[width=\figuremaxwidthdp,height=\figuremaxheightdp,keepaspectratio]{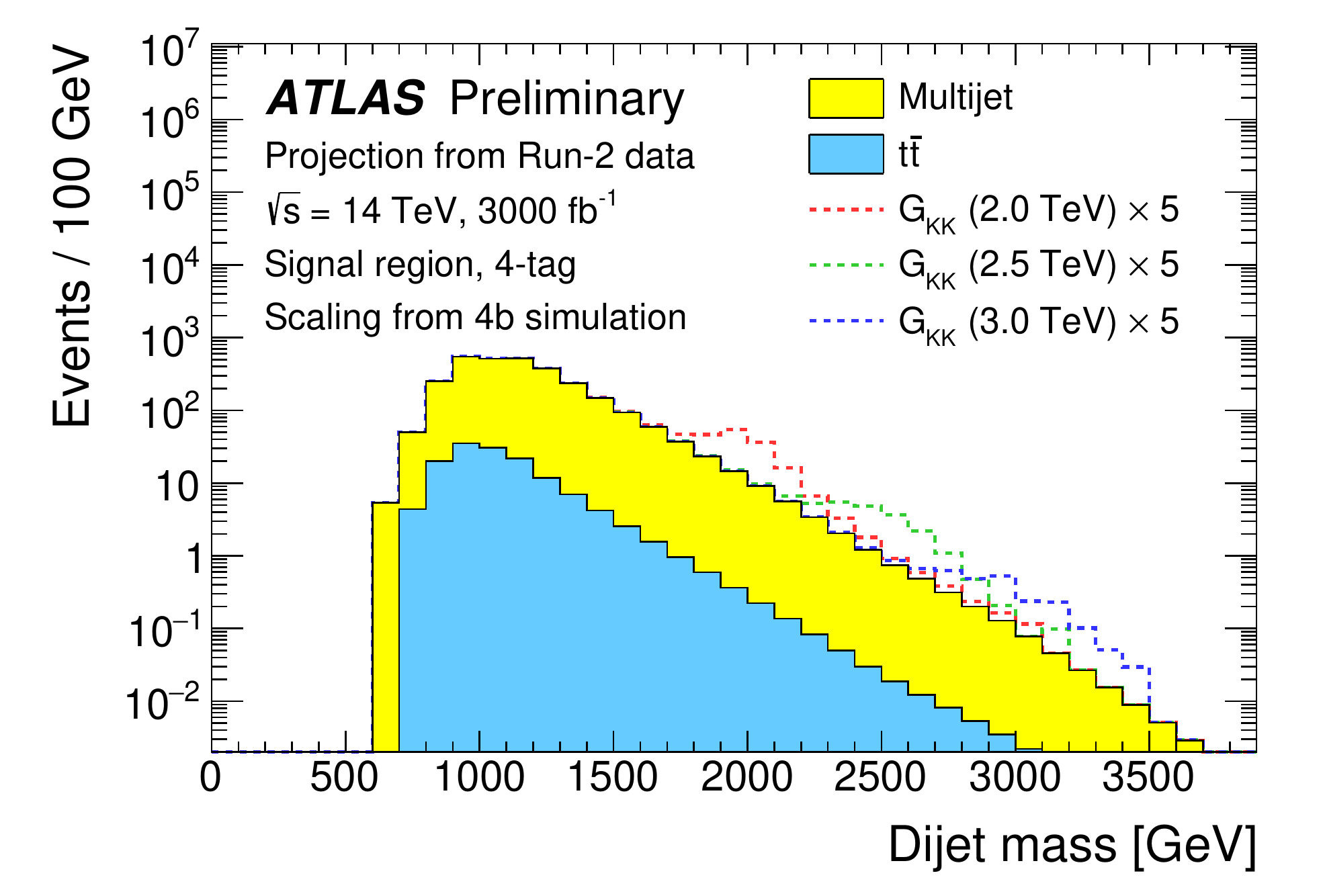}
\end{minipage}
\caption{Dijet mass distributions from the truth-level analysis for 4-tag events in the signal region
 for the expected background and signals at the HL-LHC. The multijet background is scaled using
 either the dijet (left) or the $4b$ multijet (right) MC samples. The event yields for signal events
 at $G_\mathrm{KK}$ masses of $2.0$, $2.5$, and $3.0\TeV$ are scaled up for visibility.}
\label{fig:mJJ_SR_14TeV_HHatlas}
\end{figure}

The dijet mass distributions at $\sqrt{s} = 14\TeV$ with \intLumi resulting from the scaling procedure described above, including variable-radius track jets and
the requirement on the maximum number of charged particles per large-$R$ jet, are shown in 
\fig{fig:mJJ_SR_14TeV_HHatlas} for 4-tag events in the signal region 
using either the dijet (left) or the $4b$ multijet (right) MC samples.
Systematic uncertainties are scaled down by factors of two or more (where applicable)
relative to the values from the $13\TeV$ data analysis to account for the increased precision available
with \intLumi at the HL-LHC.

%The full set of systematic uncertainties from the $13\TeV$ data analysis is
%included in the statistical analysis for the signal models. These comprise theoretical uncertainties
%in the signal acceptance as well as experimental uncertainties, which are dominant, affecting the large-$R$
%jet reconstruction (both scale and resolution uncertainties in the jet energy and mass) 
%and the $b$-tag efficiencies. Also included are uncertainties in the shape and normalisation
%of the multijet and $\ttbar$ backgrounds.
%For the background, the HL-LHC extrapolation assumes that
%the background estimate uncertainties are fully driven by the statistical uncertainties 
%in the samples used in the estimate and thus scale according to $1 / \sqrt{N}$.

%The expected $95\%\cl$ cross section upper limits for the HL-LHC projection at $\sqrt{s} = 14$~TeV
%with \intLumi are shown in \fig{fig:xslimits} either the dijet (left) or the $4b$ multijet (right) 
%MC samples. 

Upper limits on $\sigma \times \cal{B}$ at $\sqrt{s} = 14\TeV$ range from $1.44\fb$ ($1.82\fb$)
at a mass of $1.0\TeV$ to $0.025\fb$ ($0.040\fb$) at a mass of $3.0\TeV$ when dijet ($4b$) scaling and
the variable-radius track jets with a maximum requirement on the number of charged particles are applied.
The benefit from the use of variable-radius track jets becomes significant at the highest resonance
masses considered here, with an improvement in the upper limits of at least $24\%$ (depending on the
choice of scaling) at $3.0\TeV$.
The additional requirement on the maximum number of charged particles further improves the upper limits 
by factors of about $20\%$ and $45\%$ at masses of $2.0$ and $3.0\TeV$, respectively.
Systematic uncertainties have a modest impact on the limits with an effect of at most $20\%$ at
$1.0\TeV$ and decreasing to $\sim 5\%$ at high mass.
%A hypothetical relative improvement of 10\% in the $b$-tagging efficiency for true $b$-jets with 
%unchanged charm- and light-jet rejection yields improvements in the limits
%of at most 12\% when $4b$ scaling is assumed, with more modest gains when dijet scaling is assumed.
The lower mass limits on KK gravitons are summarised in \tabb{tab:MassLimits_HH4b_ATLAS}, evaluated at the $95\%\cl$, 
 using either the dijet MC samples or the $4b$ multijet MC samples to model the changes in the 
multijet background relative to the background predictions from the $13\TeV$ data analysis. 

\begin{table}[t]
\begin{center}
\begin{tabular}{lccc}
\hline
Model\T & $\sqrt{s} = 13\TeV$, $36.1\fbinv$  & \multicolumn{2}{c}{$\sqrt{s} = 14\TeV$, \intLumi} \\
        & as in \citeref{Aaboud:2018knk} &  dijet scaling & $4b$ scaling \\  % & Syst. Unc. scenario 2 \\
\hline
$\kMPl = 0.5$ \T\B &  no limit   & $2.15\TeV$  & $2.00\TeV$ \\
$\kMPl = 1.0$ \T\B &  $1.36\TeV$  & $2.95\TeV$  & $2.75\TeV$ \\
\hline
\end{tabular}
\label{tab:MassLimits_HH4b_ATLAS}
\end{center}
\caption{Expected $95\%\cl$ lower limits on $G_\mathrm{KK}$ mass for the $13\TeV$ data analysis and the extrapolation to the HL-LHC for $\kMPl = 0.5$ and $1.0$ in the bulk RS model.
 Different extrapolations are provided based on modelling of the changes in multijet background using
 either the dijet or the $4b$ multijet MC samples.}
\end{table}

\subsubsection{\cms{VBF production of resonances decaying to HH in the 4b final state at HL-LHC}}
\contributors{A. Carvalho, J. Komaragiri, D. Majumder, L. Panwar, CMS}

Several BSM scenarios predict the existence of resonances decaying to a pair of Higgs bosons, $ H $~\cite{Aad:2012tfa,Chatrchyan:2012ufa,Chatrchyan:2013lba}, such as warped extra dimensional (WED) models~\cite{Randall:1999ee}, which have a spin-0 
\sloppy\mbox{radion~\cite{Goldberger:1999uk,Csaki:1999mp,Csaki:2000zn}} and a spin-2 first Kaluza--Klein (KK) excitation of the graviton~\cite{Davoudiasl:1999jd,DeWolfe:1999cp, Agashe:2007zd}. 
Others, such as the two-Higgs doublet models~\cite{Branco:2011iw} 
(particularly, the MSSM~\cite{Djouadi:2005gj}) and the Georgi-Machacek model~\cite{Georgi:1985nv} also contain spin-0 
resonances. 
These resonances may have a sizeable branching fraction to a Higgs pair.

Searches for a new particle $X$ in the $HH$ decay channel have been performed by the
\sloppy\mbox{ATLAS~\cite{Aad:2014yja, Aad:2015uka, Aad:2015xja}} and
CMS~\cite{Khachatryan:2014jya, Khachatryan:2015year, Khachatryan:2015tha,Khachatryan:2016sey,Khachatryan:2016cfa}
Collaborations in proton-proton ($ p  p $) collisions at $\sqrt{s} =  7$ and $8\TeV$.
The ATLAS Collaboration has set limits on the production of a KK bulk graviton decaying to $HH$ in the final state with a pair of $ b $ quark and antiquark, $ b\bar{b}  b\bar{b} $, using $ p  p $ collision data at $\sqrt{s} = 13\TeV$~\cite{Aaboud:2016xco,ATLAS:2016ixk,Aaboud:2018knk}.
The CMS Collaboration has also set limits on the production of a KK bulk graviton and a radion, decaying to $HH$, in the $ b\bar{b}  b\bar{b} $ final state, using $ p  p $ collision data at $\sqrt{s} = 13\TeV$, corresponding to an integrated luminosity of 35.9\fbinv~\cite{Sirunyan:2017isc,Sirunyan:2018qca}. 
At present, the searches from ATLAS and CMS set a limit on the production cross sections and the branching fractions $\sigma( p  p  \to  X ) \BR( X  \to  H  H  \to  b\bar{b}   b\bar{b} )$ for masses of $X$, $m_{X}$ up to $3\TeV$.

The searches were confined to the $s$-channel production of a narrow resonance $X$ from the SM quark-antiquark or gluon-gluon interactions. 
The WED models that were used in the interpretations of the results have extra spatial dimension compactified between two branes (called the bulk) via an exponential metric ${\kappa l}$, where $\kappa$ is the warp factor and $l$ the coordinate of the extra spatial dimension~\cite{Giudice:2000av}. 
The reduced Planck scale, $\overline{\Mpl} \equiv \Mpl/8\pi$, \Mpl being the Planck scale and the ultraviolet cutoff of the theory $\LambdaR \equiv \sqrt{6} e^{-\kappa l} \overline{\Mpl}$~\cite{Goldberger:1999uk} are fundamental scales in these models.
A radion of mass below $1.4\TeV$ is excluded, assuming $\LambdaR = 3\TeV$, while the cross section limit for a bulk graviton decaying to $ H  H  \to  b\bar{b}   b\bar{b} $ is between $1.4$ and $4\fb$ for masses between $1.4$ and $3.0\TeV$, at $95\%\cl$~\cite{Sirunyan:2017isc,Sirunyan:2018qca}.

\begin{figure}[t]
  \begin{center}
    \includegraphics[width=0.4\textwidth]{\main/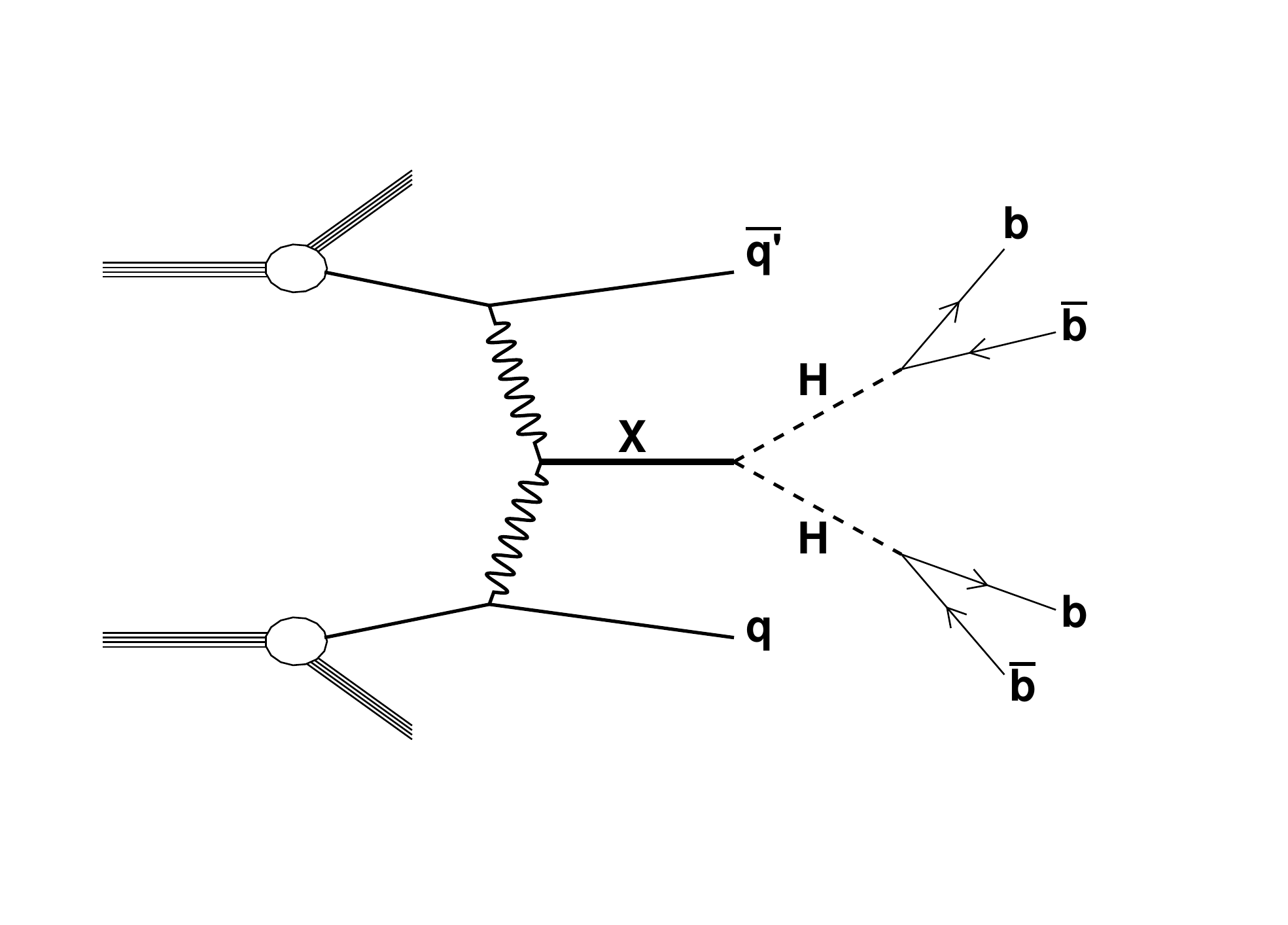}
    \caption{Diagrammatic representation of the vector boson fusion production of a resonance $X$ decaying to a pair of Higgs bosons $ H $, with both Higgs bosons decaying to $ b\bar{b} $ pairs.}
    \label{fig_tdr:feyn}
  \end{center}
\end{figure}

The search for these resonances in other production modes, such as VBF, as depicted in \fig{fig_tdr:feyn}, has not yet been explored.
While the $s$-channel production cross section of a bulk graviton, assuming $\kappa/\overline{\Mpl} = 0.5$, is in the range $0.05-5\fb$ for masses between $1.5$ and $3\TeV$, the VBF production mode is expected to have a cross section an order of magnitude smaller~\cite{Oliveira:2014kla}.
The absence of a signal from the $s$-channel process may point to highly suppressed couplings of $X$ with the SM quarks and gluons, making VBF the dominant production process in $ p  p $ collisions.
 
%The VBF production process $ p  p \to X \Pq\Paq$ should be observable at HL-LHC.

Here we explore the prospects for the search for a massive resonance produced through VBF and decaying to $HH$ at the HL-LHC with the upgraded CMS detector. For a very massive resonance, highly Lorentz-boosted Higgs bosons are more efficiently reconstructed as a single large-area jet (Higgs jet). In addition, a signal event will also have two energetic jets at large pseudorapidity $\eta$. This study from the CMS Collaboration is reported in detail in \citeref{CMS:2018zkz}

A simulation of the upgraded Phase 2 CMS detector was used for this study. 
Signal events for bulk gravitons were simulated at leading order using {\MGvATNLO  2.4.2}~\cite{Alwall:2014hca} for 
masses in the range $1.5$ to $3$\TeV\ and for a fixed width of 1\% of the mass. The {NNPDF3.0} leading order PDFs~\cite{Ball:2014uwa}, taken from the {LHAPDF6} PDF set~\cite{Harland-Lang:2014zoa,Buckley:2014ana,Carrazza:2015hva,Butterworth:2015oua}, with the four-flavour scheme, were used. The main background is given by multijet events, and has been simulated using \PYTHIA~8.212~\cite{Sjostrand:2014zea}, for events containing two hard partons, with the invariant mass of the two partons required to be greater than $1\TeV$.

For both the signal and the background processes, the showering and hadronisation of partons was simulated with \pythia{ 8}. 
The pileup events contribute to the overall event activity in the detector, the effect of which was included in the simulations assuming a pileup distribution averaging to $200$.
All generated samples were processed through a \GEANTfour-based~\cite{Agostinelli:2002hh,Allison:2006ve} simulation of the upgraded CMS detector.

The two leading-$\pt$ large-radius anti-\kt jets with a distance parameter of $0.8$ (AK8) in the event, $J_{1}$ and $J_{2}$, are required to have $\pt > 300\GeV$ and  $\abs{\eta} < 3.0$. To identify the two leading-$\pt$ AK8 jets with the boosted $H\to b\bar{b}$ candidates from the $ X \to H  H $ decay ($ H $ tagging), these jets are groomed~\cite{Salam:2009jx} to remove soft and wide-angle radiation using the modified mass drop algorithm~\cite{Butterworth:2008iy,Dasgupta:2013ihk}, with the soft radiation fraction parameter $z$ set to $0.1$ and the angular exponent parameter $\beta$ set to 0, also known as the soft-drop algorithm~\cite{Larkoski:2014wba,CMS:2017wyc}.
By undoing the last stage of the jet clustering, one gets two subjets each for $J_{1}$ and $J_{2}$.
The invariant mass of the two subjets is the soft-drop mass of each AK8 jet, which has a distribution with a peak near the Higgs boson mass $m_{H} = 125\GeV$~\cite{Aad:2015zhl,Sirunyan:2017exp}, and a width of about $10\%$.
The soft-drop mass window selection was optimised using a figure of merit of $S/\sqrt{B}$ and required to be in the range $90-140\GeV$ for both leading jets.

The $N$-subjettiness ratio $\nsub \equiv \tau_2/\tau_1$~\cite{Thaler:2011gf} has a value much smaller than unity for a jet with two subjets. For signal selection, $J_{1}$ and $J_{2}$ are required to have $\nsub < 0.6$ following an optimisation of the above figure of merit.

\begin{figure}[t]
\centering
\begin{minipage}[t][\minipageheightdp][t]{\minipagewidthdp}
\centering
\includegraphics[width=\figuremaxwidthdp,height=\figuremaxheightdp,keepaspectratio]{\main/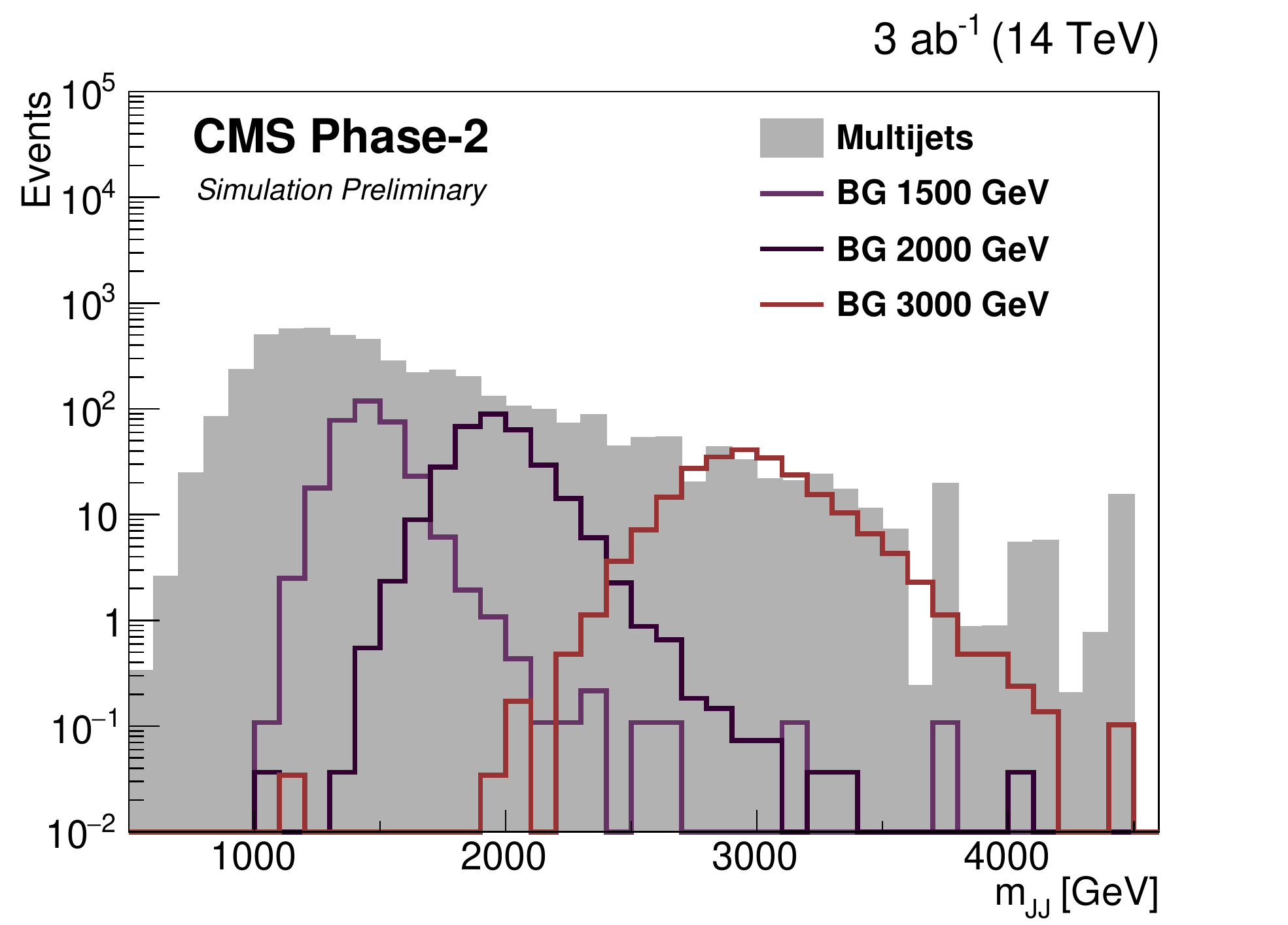}
\end{minipage}
\begin{minipage}[t][\minipageheightdp][t]{\minipagewidthdp}
\centering
\includegraphics[width=\figuremaxwidthdp,height=\figuremaxheightdp,keepaspectratio]{\main/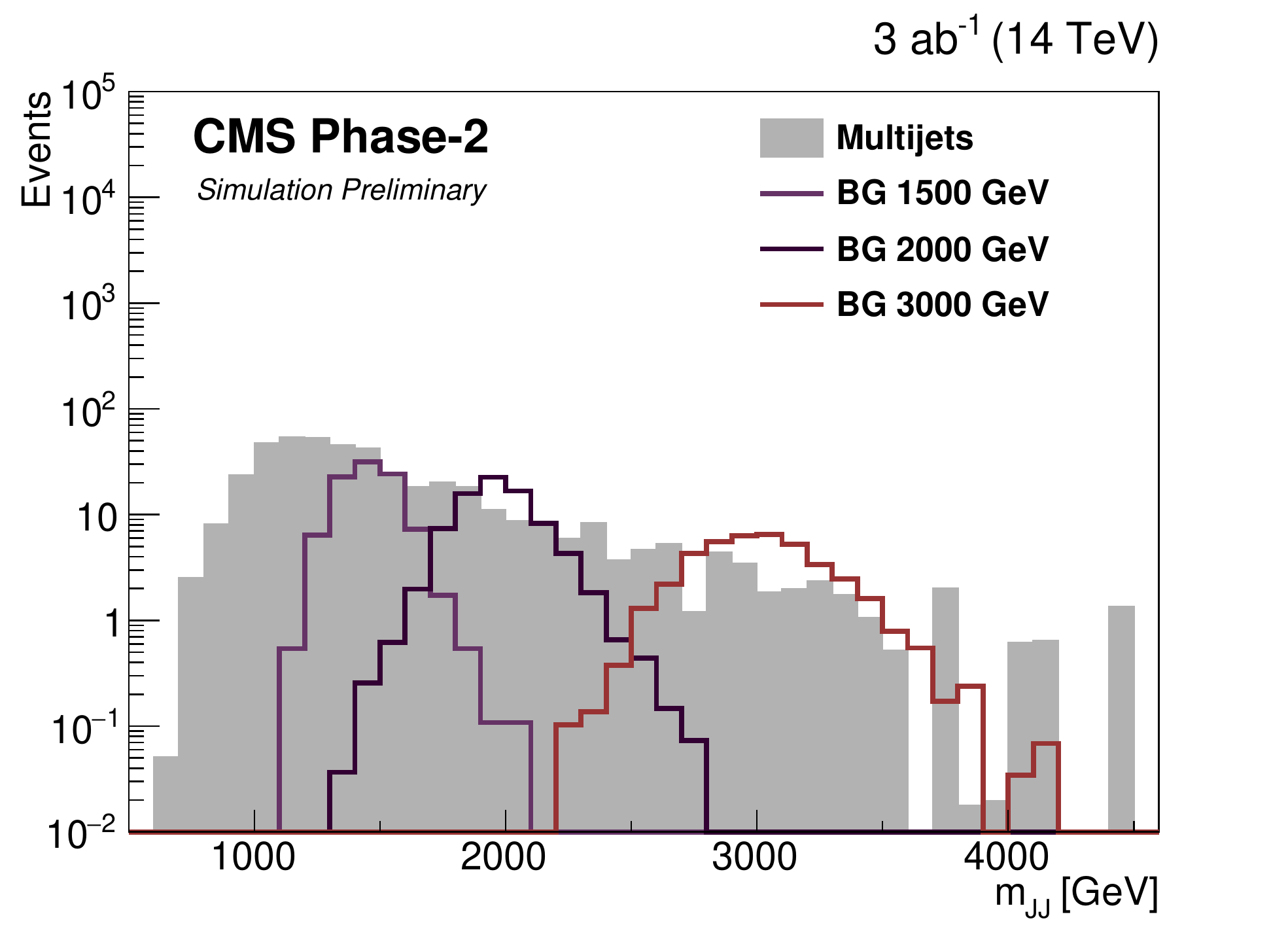}
\end{minipage}
    \caption{Estimated multijet background and the signal $m_{JJ}$ distributions for bulk gravitons (BG) of masses $1.5$, $2$, and $3\TeV$, assuming a signal cross section of $1\fb$. The distributions on the left are for the $3 b $ and those on the right are for the $4 b $ subjet $ b $-tagged categories and for an average pileup of $200$.}
    \label{fig:MJJ}
\end{figure}

\begin{table}[t]
  \begin{center}
    \small\begin{tabular}{ccccc}
      \hline\hline
      Process & \multicolumn{2}{c}{$3 b $ category} & \multicolumn{2}{c}{$4 b $ category} \\ \hline
                                  & Events & Efficiency ($\%$)     & Events & Efficiency ($\%$) \\ \hline 
      Multijets                   & $4755$   & $1.6\times 10^{-3}$ & $438$    & $1.5\times 10^{-4}$ \\ 
      BG (\mx=1.5\TeV) & $326$    & $11$                  & $95.2$   & $3.2$\\ 
      BG (\mx=2\TeV) & $316$    & $11$                  & $81.2$   & $2.7$\\ 
      BG (\mx=3\TeV) & $231$    & $7.7$                 & $41.4$   & $1.4$\\ 
      \hline\hline
    \end{tabular}\normalsize
  \end{center}
   \caption{Event yields and efficiencies for the signal and multijet background for an average pileup of $200$. The product of the cross sections and branching fractions of the signals $\sigma( p  p  \to  X \text{jj} \to  H  H  \text{jj})$ is assumed to be $1\fb$. Owing to the large sample sizes of the simulated events, the statistical uncertainties are small.}\label{tab:evtyields}
\end{table}

The $ H $ tagging of $J_{1}$ and $J_{2}$ further requires identifying their subjet pairs to be $ b $ tagged with a probability of about $49\%$ to contain at least one $ b $ hadron, and a corresponding probability of about $1\%$ of having no $ b $ or $\Pqc$ hadrons.
Events are classified into two categories: those having exactly three out of the four $ b $-tagged subjets ($3 b $ category), and those that have all four subjets $ b $-tagged ($4 b $ category).

Events are required to have at least two AK4 jets $j_{1}$ and $j_{2}$, which are separated from the $ H $ jets by $\Delta R> 1.2$, with $\pt > 50\GeV$ and $\abs{\eta}< 5$.
To pass the VBF selections, these jets must lie in opposite $\eta$ regions of the detector, and a pseudorapidity difference $\abs{\Delta\eta(j_{1}, j_{2})} > 5$.
The invariant mass $m_{jj}$ reconstructed using these AK4 jets is required to pass $m_{jj} > 300\GeV$.

The bulk graviton invariant mass $m_{JJ}$ is reconstructed from the 4-momenta of the two Higgs jets in events passing the above mentioned full selection criteria. The main multijet background is smoothly falling, above which the signal is searched for as a localised excess of events for a narrow resonance $X$. 

It is expected that the multijet background component in a true search at the HL-LHC will rely on the data for a precise estimate. Methods such as those described in \citeref{Sirunyan:2017isc} are known to provide an accurate prediction of the multijet background $m_{JJ}$ shape as well as the yield. 
%Here, the expected background yields based on simulations are described. 

The simulated multijet background sample consists of $\sim\!4$ million events none of which survive the full selection. To estimate the background, the subjet $ b $-tagging efficiency is determined using a loose set of selection criteria which require events to have $J_{1}$ and $J_{2}$ passing only the soft-drop mass and $\nsub$ requirements. The $ b $-tagging efficiency is obtained for the different subjet flavours and as a function of $\pt$ and $\eta$. 

Multijet events passing all selection criteria except the subjet $ b $ tagging are then reweighted according to the subjet efficiencies to obtain the probability of the event to pass the three of four subjet $ b $-tagging categories. The $m_{JJ}$ distributions for the multijet background after the full selection are then obtained from the weighted events in these two categories. 

From the analysis of current LHC data at $\sqrt{s} = 13\TeV$, it was found that the multijet backgrounds measured in data are a factor of $0.7$ smaller than estimated in simulation. Accordingly, the multijet background yield from simulation has been corrected by this factor, assuming this to hold also for the simulations of the multijet processes at $\sqrt{s} = 14\TeV$ .
The $m_{JJ}$ of the backgrounds thus obtained and the signals are shown in \fig{fig:MJJ}, while the event yields after full selection are given in \tabb{tab:evtyields}. 

The efficiency of events to pass the VBF jet selection depends strongly on pileup due to the combinatorial backgrounds from pileup jets, which affects both the signal and the background selection. Moreover, the VBF selection efficiency for multijets grows faster than the signal efficiency with pileup, since the latter has true VBF jets which already pass the selection in the absence of pileup. Hence, in the present search, the requirement of additional VBF jets does not result in any appreciable gain in the signal sensitivity. It is anticipated that developments in the rejection of pileup jets in the high $\eta$ region will eventually help suppress the multijets background and improve the signal sensitivity further.

The expected significance of the signal, assuming a production cross section of $1\fb$ is estimated. Several systematic uncertainties are considered.
The uncertainty in the jet energy scale amounts to $1\%$. The uncertainty in the subjet $ b $-tagging efficiency difference between the data and simulations is taken to be $1\%$. An uncertainty of $1\%$ is assigned to the integrated luminosity measurement. These uncertainties are based on the projected values for the full data set at the HL-LHC.

In addition, several measurement uncertainties are considered based on the 2016 search for a resonance decaying to a pair of boosted Higgs bosons~\cite{Sirunyan:2017isc}, which are scaled down by a factor of $0.5$ for the HL-LHC study. 
The $ H $ jet selection uncertainties include the uncertainties in the $ H $ jet mass scale and resolution ($1\%$), the uncertainty in the data to simulation difference in the selection on \nsub ($13\%$), and the uncertainty in the showering and hadronisation model for the $ H $ jet ($3.5\%$). 
The uncertainties in the signal acceptance because of the parton distribution functions ($1\%$) and the simulation of the pileup ($1\%$) are also taken into account

The expected signal significance of a bulk graviton of mass $2000\GeV$, produced through vector boson fusion, with an assumed production cross section of $1\fb$ and decaying into a pair of Higgs bosons, each of which decays to a  $b\bar{b}$ pair is found to be $2.6\sigma$
%given in \fig{fig:sigSig} 
for an integrated luminosity of \intLumi.
%
%\begin{figure}[htbp]
%  \begin{center}
%    \includegraphics[width=0.5\textwidth]{\main/section6OtherSignatures/img/Significance_WPM.pdf}
%    \caption{Expected signal significance for Bulk Gravitons of masses $1.5$, $2$, and $3$\TeV, assuming a production cross section of $1\fb$. The data set corresponds to an integrated luminosity of \intLumi and with a pileup of 200.}
%    \label{fig:sigSig}
%  \end{center}
%\end{figure}

\subsubsection{\theoryC{Heavy Higgs bosons in models with vector-like 
%Sensitivity to heavy Higgs bosons in models with vector-like 
%quarks and leptons 
fermions at the HL- and HE-LHC}}
\contributors{R. Dermisek, E. Lunghi, S. Shin}
%{\bf Author(s): Radovan Dermisek, Enrico Lunghi, Seodong Shin}
%
%
%Radovan Dermisek: Physics Department, Indiana University, Bloomington, IN 47405, USA \\
%dermisek@indiana.edu \\
%Enrico Lunghi: Physics Department, Indiana University, Bloomington, IN 47405, USA \\
%elunghi@indiana.edu \\
%Seodong Shin: Enrico Fermi Institute, University of Chicago, Chicago, IL 60637, USA \\
%Department of Physics and IPAP, Yonsei University, Seoul 03722, Korea \\
%shinseod@indiana.edu, seodongshin@yonsei.ac.kr
%

%\subsubsection{Introduction}
%\paragraph*{Introduction}

Among the simplest extensions of the SM are models with extended Higgs sector and models with additional vectorlike matter. 
Although many search strategies for individual new particles were designed, there are large regions of the parameter space that HL- or HE-LHC will not be sensitive to as a result of either small production rates or large SM backgrounds.  However, combined signatures of both extra sectors can lead to many new opportunities to search for heavy Higgs bosons and vectorlike matter simultaneously~\cite{Dermisek:2015hue}. 

For example, in a type-II two Higgs doublet model, the production cross section of the $1\TeV$ heavy \cpeven Higgs boson can be as sizeable as $\sim 2\pb$ ($10\pb$) at $13\TeV$ ($27\TeV$) \com~energy, depending on the values of $\tan\beta$.
However, in the decoupling limit, where $H \to ZZ,\; WW$ are suppressed or not present, the dominant decay modes are $t\bar t$, $hh$, $b\bar b$ and $\tau^+\tau^-$ which suffer from large SM backgrounds or small branching fractions in some regions of $\tan \beta$. On the other hand, vectorlike leptons typically have very clean signatures but their production rates are very small. Nevertheless, since vectorlike leptons can appear in decay chains of  heavy Higgs bosons, the combined signature can feature both the sizeable production rate and clean final states. 

%We consider the extension of the type-II two Higgs doublet model by vectorlike pairs of quarks and leptons introduced in \citeref{Dermisek:2015oja}. 
%In the following sections we briefly outline the model, summarise possible heavy Higgs cascade decays through vectorlike quarks and leptons and discuss  the sensitivity of the HL- and HE-LHC for selected decay modes.

%\subsubsection{Model Framework}
%\paragraph*{Model Framework}

We consider a type-II two Higgs doublet model augmented by vectorlike pairs of new quarks ($SU(2)$ doublets $Q_{L,R}$ and SU(2) singlets $U_{L,R}$ and $D_{L,R}$)
and vectorlike pairs of new leptons ($SU(2)$ doublets $L_{L,R}$,  $SU(2)$ singlets $E_{L,R}$ and  SM singlets $N_{L,R}$)~ \cite{Dermisek:2015oja}. The  $Q_{L}$, $U_{R}$,  $D_{R}$, $L_L$ and $E_R$ have the same  hypercharges as quarks and leptons in the SM, as summarised in \tabb{table:fieldcontents}.
We  further assume that the new leptons mix only with one family of SM leptons in order to simply avoid strong constraints from the experiments searching for lepton flavour violation and we consider the mixing with the second family as an example~\cite{Kannike:2011ng,Dermisek:2013gta,Falkowski:2013jya,Falkowski:2014ffa,Dermisek:2014cia,Dermisek:2014qca,Dermisek:2015vra,Dermisek:2015oja,Dermisek:2015hue,Dermisek:2016via} (the discovery potential in the case of mixing with the first family would be comparable~\cite{Dermisek:2014qca} and in the case of mixing with the third family it would be significantly weaker~\cite{Dermisek:2014qca,Kumar:2015tna}). This can be achieved by requiring that the individual lepton number is an approximate symmetry (violated only by light neutrino masses). Similarly,  we assume that the new quarks mix only with the third family of SM quarks. We consider the most general renormalisable Lagrangian consistent with these assumptions.

%%%-------------------------
\begin{table}[t]
\begin{center}
\resizebox{\columnwidth}{!}{
\small\begin{tabular}{c c c c c c c c c c c c c c}
\hline
\hline
 & ~~$\ell^i_L$ & ~~$e^i_R$ & ~~$q^i_L$ & ~~$u^i_R$ & ~~$d^i_R$ & ~~$L_{L,R}$ & ~~$E_{L,R}$ & ~~$N_{L,R}$ & ~~$Q_{L,R}$ & ~~$T_{L,R}$ & ~~$B_{L,R}$ & ~~$H_d$ & ~~ $H_u$\\
\hline
$SU(2_{L}$ & ~~\bf $2$ & ~~\bf $1$ & ~~\bf $2$ & ~~\bf $1$ & ~~\bf $1$ & ~~\bf $2$ & ~~\bf $1$ & ~~\bf $1$ & ~~\bf $2$ & ~~\bf $1$ & ~~\bf $1$ & ~~\bf $2$ & ~~\bf $2$ \\
$U(1)_{Y}$ & ~~$-\frac12$ & ~~$-1$ & ~~$\frac16$ & ~~$\frac23$ & ~~-$\frac13$ & ~~$-\frac12$ & ~~$-1$ & ~~$0$  & ~~$\frac16$ & ~~$\frac23$ & ~~-$\frac13$ & ~~$\frac12$ & ~~-$\frac12$ \\
$Z_2$ & ~~$+$ & ~~$-$ & ~~$+$ & ~~$+$ & ~~$-$ & ~~$+$ & ~~$-$ & ~~$+$ & ~~$+$ & ~~$+$ & ~~$-$ & ~~$-$ & ~~$+$ \\
\hline
\hline
\end{tabular}\normalsize}
\end{center}
\caption{Quantum numbers of SM leptons and quarks ($\ell^i_L$, $e^i_R$, $q^i_L, u^i_R, d^i_R$ for $i=1,2,3$), extra vectorlike leptons and quarks and the two Higgs doublets. The electric charge is given by $Q = T_3 +Y$, where $T_3$ is the weak isospin, which is $+1/2$ for the first component of a doublet and $-1/2$ for the second component.}
\label{table:fieldcontents}
\end{table}
%%%-----------------------

Details of the lepton sector of the model were worked out in \citeref{Dermisek:2015oja}.
After spontaneous symmetry breaking, $\left< H_u^0 \right> = v_u$ and $\left< H_d^0 \right> = v_d$ with $\sqrt{v_u^2 + v_d^2} = v = 174\GeV$ (we also define $\tan \beta \equiv v_u / v_d$), the model can be summarised by mass matrices in the charged lepton sector, with left-handed fields $( \bar \mu_{L}, \bar L^-_L, \bar E_L )$ on the left and right-handed fields $( \mu_{R} , L^-_R, 
 E_R)^T$ on the right~\cite{Dermisek:2013gta},
\begin{eqnarray}
M_e \;  
&=& 
\begin{pmatrix}
 y_\mu v_d & 0 &  \lambda_E v_d\\
  \lambda_L v_d & M_L &  \lambda v_d\\
 0 & \bar \lambda v_d & M_E 
\end{pmatrix},
\end{eqnarray}
and in the neutral lepton sector, with left-handed fields $( \bar \nu_\mu , \bar{L}_L^0 , \bar N_L)$ on the left and right-handed fields $( \nu_R = 0 ,
L_R^0 ,
N_R)^T$ on the right~\cite{Dermisek:2015oja},
\begin{eqnarray}
M_\nu  &=& 
\left( 
\begin{array}{ccc}
0 & 0 & \kappa_N v_u \\
0 & M_L & \kappa v_u \\
0 & \bar \kappa v_u & M_N \\
\end{array}
\right)~.
\label{eq:mm}
\end{eqnarray}
The superscripts on vectorlike fields represent the charged and the neutral components (we  inserted $\nu_R = 0$ for the  right-handed neutrino which is absent in our framework in order to keep  the mass matrix $3\times3$ in complete analogy with the charged sector).  The usual SM Yukawa coupling of the muon is denoted by $y_\mu$, the Yukawa couplings to $H_d$ are denoted by various $\lambda$s, the Yukawa couplings to $H_u$ are denoted by various $\kappa$s, and finally the explicit mass terms for vectorlike  leptons are given by $M_{L,E,N}$. Note that explicit mass terms between SM and vectorlike fields (\iee~$\bar \mu_L L_R$ and $\bar E_L \mu_R$) can be rotated away. These mass matrices can be diagonalised by bi-unitary transformations and we label the two new charged and neutral mass eigenstates by $e_4, \, e_5$ and $\nu_4, \, \nu_5$ respectively:
\begin{eqnarray}
U_L^\dagger M_e U_R &=& {\rm diag}
\left(
 m_\mu,  m_{e_4} ,  m_{e_5}
\right), \\
V_L^\dagger M_\nu V_R &= &  {\rm diag} 
\left(0, m_{\nu_4} , m_{\nu_5}
\right).
\label{eq:unitary}
\end{eqnarray}

Since $SU(2)$ singlets mix with $SU(2)$ doublets, the couplings of all involved particles to the $Z$, $W$ and Higgs bosons are in general modified. The flavour conserving couplings receive corrections and flavour changing couplings between the muon (or muon neutrino)  and heavy leptons are generated. The relevant formulas for these couplings in terms of diagonalisation matrices defined above can be found in \citerefs{Dermisek:2015oja, Dermisek:2013gta}. In the limit of small mixing, approximate analytic expressions  for diagonalisation matrices can be obtained which are often useful for the understanding of numerical results. These are also given in \citerefs{Dermisek:2015oja, Dermisek:2013gta}.
Details of the quark sector of the model can be found in \citeref{future:Dermisek}.
The mass matrix in the up and down quark sectors closely follow those for the neutrino and the charged leptons above. 
\begin{figure}[t]
\begin{center}
\includegraphics[width=0.65\linewidth]{\main/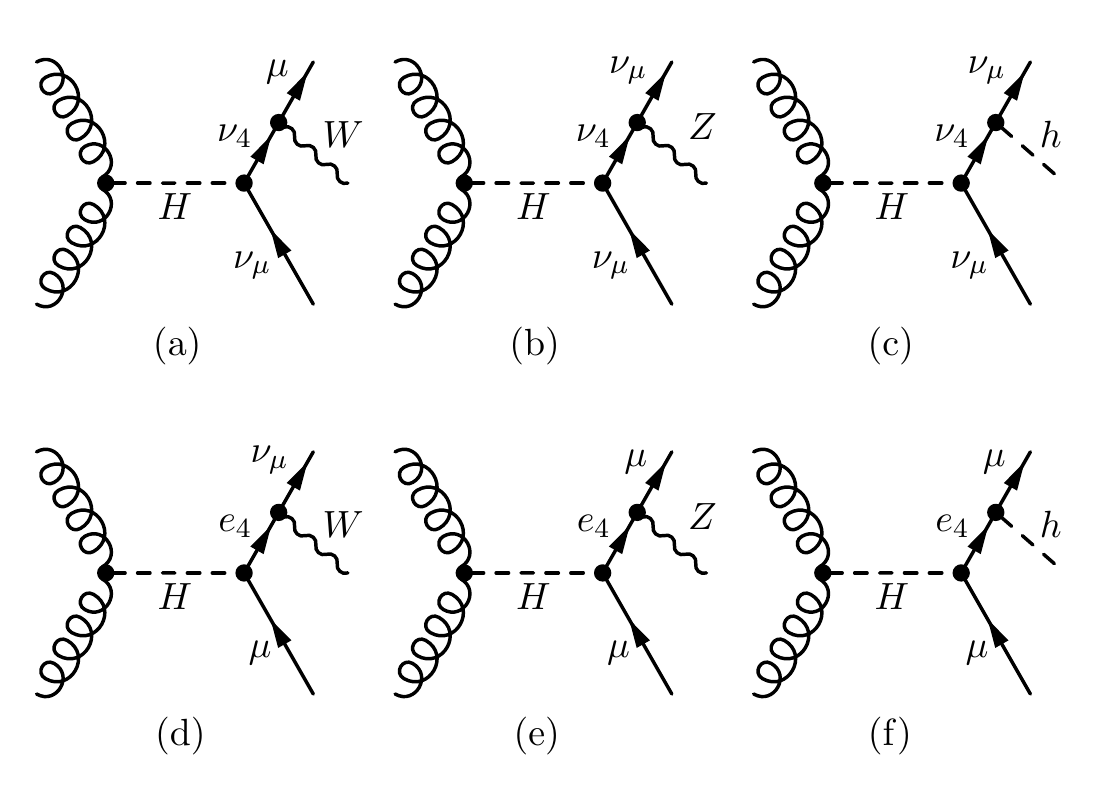}
\caption{Heavy Higgs boson cascade decays through vectorlike leptons.
\label{fig:topologies}
}
\end{center}
\end{figure}
%

%\subsubsection{Heavy Higgs cascade decays through vectorlike leptons}
%\paragraph*{Heavy Higgs cascade decays through vectorlike leptons}
%
%
%\paragraph*{Signatures and search strategies}

The generated flavour changing couplings between heavy and light leptons lead to new decay modes of heavy \cpeven (or \cpodd) Higgs boson: $H \to \nu_4 \nu_\mu$ and $H \to e_4 \mu$, where $e_4$ and $\nu_4$ are the  lightest new charged and neutral leptons. The BRs for these decay modes can be very large when the mass of  the heavy Higgs boson is below the $t \bar t$ threshold and the light Higgs boson ($h$) is SM-like so that $H \to ZZ,\; WW$ are suppressed or not present. In this case, flavour changing decays $H \to \nu_4 \nu_\mu$ or $H \to e_4 \mu$ compete only with $H\to b \bar b$,  and for sufficiently heavy $H$, also with $H \to hh$. 
Subsequent decay modes of $e_4$ and $\nu_4$, $e_4 \to W  \nu_\mu$, $e_4 \to Z \mu$,  $e_4 \to h \mu$ and $\nu_4 \to W \mu$, $\nu_4 \to Z \nu_\mu$, $\nu_4 \to h \nu_\mu$ lead to the following 6 decay chains of the heavy Higgs boson:
\begin{align}
H &\; \to \; \nu_4 \nu_\mu \; \to \; W\mu \nu_\mu, \; Z\nu_\mu \nu_\mu,\; h \nu_\mu \nu_\mu~, \label{eq:Hn4} \\  
H &\; \to \; e_4 \mu \; \to \; W\nu_\mu \mu, \; Z\mu\mu,\; h \mu\mu~, \label{eq:He4}
\end{align}
 which are also depicted in \fig{fig:topologies}. In addition, $H$ could also decay into pairs of vectorlike leptons. This is however  limited to smaller ranges for masses in which these decays are kinematically open. Moreover, the final states are the same as in pair production of vectorlike leptons. We will not consider these possibilities here. Finally, although we  focus on the second family of SM leptons in final states, the modification for a different family of leptons or quarks is straightforward.

In \citeref{Dermisek:2015hue}, it was found that in a large range of the parameter space BRs for the decay modes \eqref{eq:Hn4} and \eqref{eq:He4} can be sizeable  or even dominant while satisfying constraints from searches for heavy Higgs bosons,  pair production of  vectorlike leptons~\cite{Dermisek:2014qca} obtained from searches for anomalous production of multilepton events, and  constraints from precision EW observables~\cite{Patrignani:2016xqp}.
Since the Higgs production cross section can be very large, for example the cross section for a $200\GeV$ Higgs boson at $13\TeV$ ($27\TeV$)  LHC for $\tan \beta =1$ is $18\pb$ ($67\pb$)~\cite{Heinemeyer:2013tqa,Harlander:2012pb,Harlander:2016hcx}, the final states above can be produced in large numbers. Thus searching for these processes could  lead to the simultaneous discovery of a new Higgs boson and a new lepton. Some of the decay modes in \fig{fig:topologies} also allow for full reconstruction of the masses of both new particles in the decay chain.

The final states of the processes \eqref{eq:Hn4} and \eqref{eq:He4} are the same as final states of \sloppy\mbox{$pp \to WW, ZZ, Zh$} production or $H \to WW, ZZ$ decays with one of the gauge bosons decaying into second generation of leptons.  Since searching for leptons in final states is typically advantageous,  our processes contribute to a variety of existing searches.  Even searches for processes with fairly large cross sections can be significantly affected. For example, the contribution of  $pp \to H \to \nu_4 \nu_\mu \to W \mu \nu_\mu$  to $pp \to WW$ can be close to current limits while satisfying the constraints from  $H \to WW$. This has been studied in \citeref{Dermisek:2015oja} in the two Higgs doublet model we consider here, and also in a more model independent way in \citeref{Dermisek:2015vra}. 

The discussion of the main features of each of the heavy Higgs decay modes, of existing experimental searches to which these new process contribute, and of possible new searches can be found in \citeref{Dermisek:2015hue}. 
The processes with small SM backgrounds are the best place to look for this scenario. Examples include $H \; \to \; h \nu_\mu \nu_\mu$ and $H \; \to \; h \mu\mu$ with $h\to \gamma \gamma$. 

%--
\begin{figure}[t]
\centering
\begin{minipage}[t][\minipageheightdp][t]{\minipagewidthdp}
\centering
\includegraphics[width=\figuremaxwidthdp,height=\figuremaxheightdp,keepaspectratio]{\main/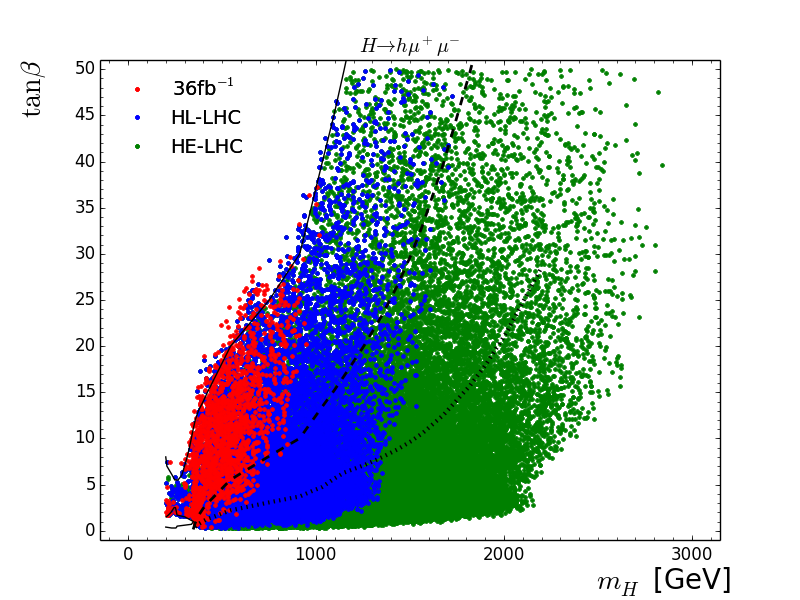}
\end{minipage}
\begin{minipage}[t][\minipageheightdp][t]{\minipagewidthdp}
\centering
\includegraphics[width=\figuremaxwidthdp,height=\figuremaxheightdp,keepaspectratio]{\main/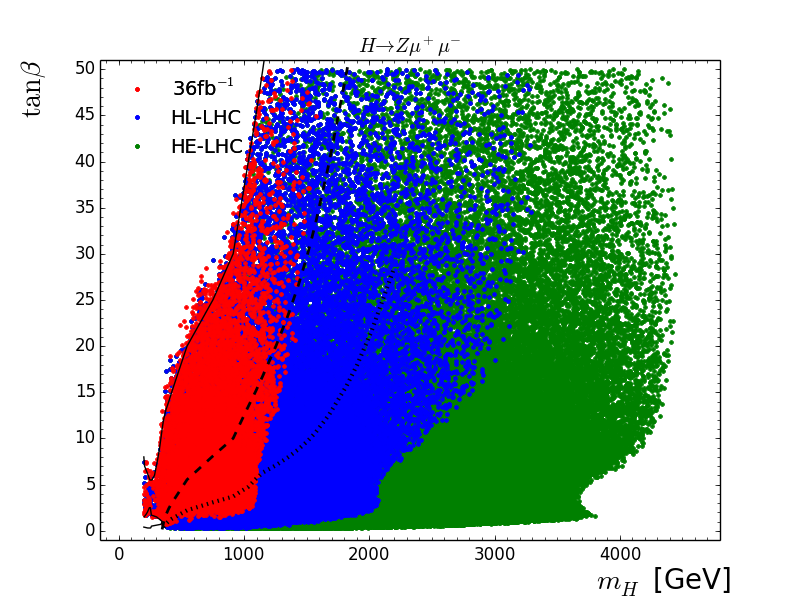}
\end{minipage}
\caption{
Experimental sensitivities for $H \to e_4^\pm \mu^\mp \to h \mu^+ \mu^-$ (left) $H \to e_4^\pm \mu^\mp \to Z \mu^+ \mu^-$ (right). Red, blue and green points are the regions of the $[m_H,\tan\beta]$ plane which are accessible with LHC with $36\fbinv$, at the HL-LHC, and at the HE-LHC, respectively. Black solid, dashed and dotted lines represent the corresponding reaches of $H\to \tau\tau$ searches.}
\label{fig:mHtanbe}
\end{figure}
%--

%\paragraph*{Reach of the HL- and HE-LHC}

We now discuss the parameter regions of heavy Higgses and vectorlike leptons which can be accessed by dedicated searches for the modes we propose with current LHC ($36\fbinv$), HL- and \sloppy\mbox{HE-LHC}. Among the processes depicted in \fig{fig:topologies}, we choose $H \to e_4^\pm \mu^\mp \to h \mu^+ \mu^-$  and $H \to e_4^\pm \mu^\mp \to Z \mu^+ \mu^-$ as representative examples. These are interesting because they result in additional resonances. The first process was analysed in \citeref{Dermisek:2016via} for $m_H \le 340\GeV$. The key selection criterion is an {\it off-$Z$} cut ($|m_{\mu^+ \mu^-} - M_Z| > 15\GeV$) which exploits the fact that the invariant mass of the two muons in the final state is distributed mostly outside of the $Z$ boson resonance region. This cut removes to a large extent background events from $Z$ + (heavy flavoured) jets with $Z \to \mu^+ \mu^-$ and raises enormously the signal significance~\cite{Dermisek:2016via}. In order to maximise the signal cross section, we focus on $h \to b \bar{b}$ (which we studied in detail in \citeref{Dermisek:2016via}) and $Z \to b \bar{b}$.

To obtain the expected sensitivities, we estimate the number of background events for the HL-LHC by rescaling the results of \citeref{Dermisek:2016via} (which are based on the search for a heavy resonance decaying to $hZ$ at ATLAS~\cite{TheATLAScollaboration:2016loc}) by the ratio of luminosities. We further estimate that the backgrounds at the HE-LHC increase by a factor $3.6 \times 5$ with respect to those at the HL-LHC, where $3.6$ is the ratio of the background cross sections calculated with \madgraph{ 5} and $5$ comes from the ratio of integrated luminosities ($\intlumihelhc / 3\abinv$).

We take a flat signal acceptance of about $30\%$ for $m_H \le 800\GeV$ based on the analysis presented in \citeref{Dermisek:2016via} for $m_H = 450, 550, 650, 750\GeV$ and $m_{e_4} = 250\GeV$. For $m_H \ge 800\GeV$ the dimuon invariant mass is much harder and concentrated above the $Z$ resonance; in this region we assume an acceptance of about $50\%$. Furthermore, we consider a flat $50\%$ detector efficiency throughout this analysis.

These estimates of background events, signal acceptances, and detector efficiencies, allow us to calculate the corresponding experimental sensitivities using the condition:
\begin{align}
N_s^{95} \le \sigma_H \cdot \BR(H \to e_4 \mu) \cdot \BR(e_4 \to h \mu,\, Z \mu) \cdot A \cdot \epsilon \cdot \mathcal L\,,
\end{align}
where $\sigma_H$ is the $pp\to H$ production cross section, $A$ is the signal acceptance, $\epsilon$ is the experimental efficiency, and $\mathcal L$ is the integrated luminosity. $N_s^{95}$ is the $95\%\cl$ upper limit on the expected number of events calculated from the background estimations described above using a modified frequentist construction (CLs)~\cite{Read:2002hq} based on a Poisson distribution. 

%--
\begin{figure}[t]
\centering
\begin{minipage}[t][\minipageheightdp][t]{\minipagewidthdp}
\centering
\includegraphics[width=\figuremaxwidthdp,height=\figuremaxheightdp,keepaspectratio]{\main/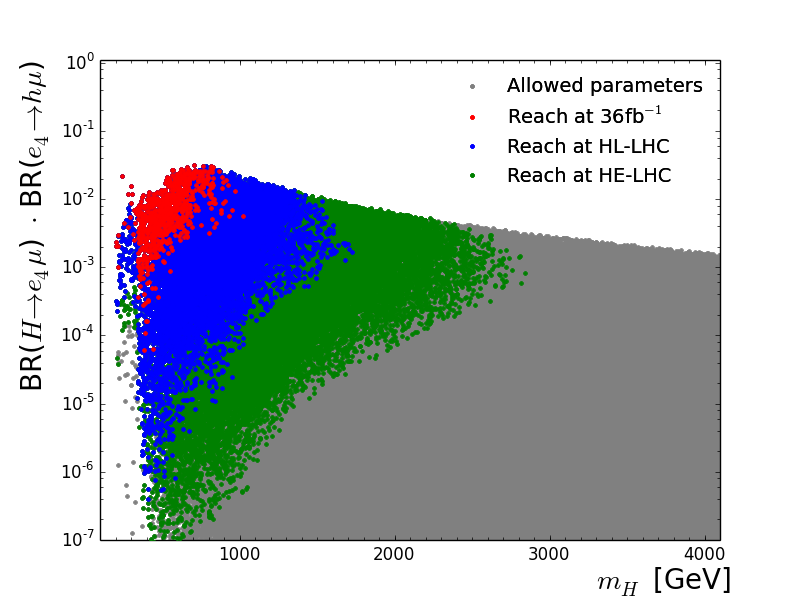}
\end{minipage}
\begin{minipage}[t][\minipageheightdp][t]{\minipagewidthdp}
\centering
\includegraphics[width=\figuremaxwidthdp,height=\figuremaxheightdp,keepaspectratio]{\main/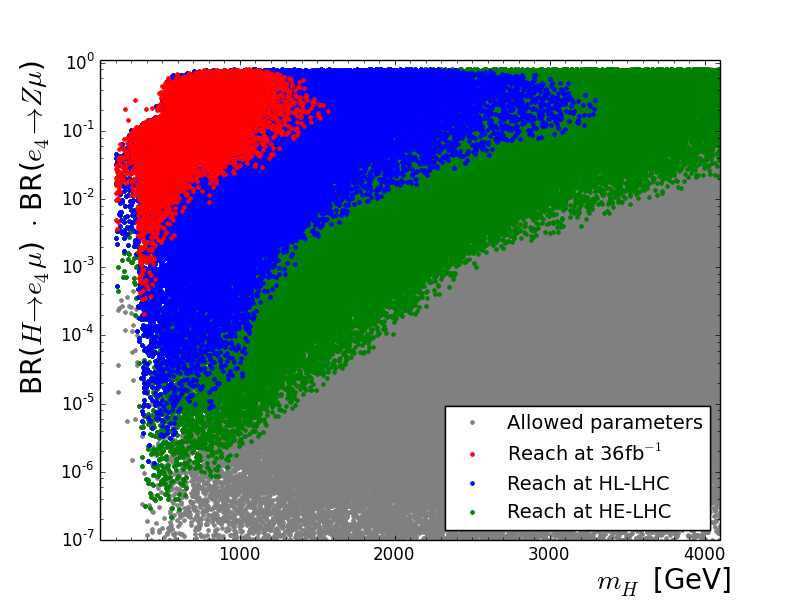}
\end{minipage}
\caption{
Possible values of the BRs $\BR(H \to e_4 \mu) \times \BR(e_4 \to h \mu)$ (left) and $\BR(H \to e_4 \mu) \times \BR(e_4 \to Z \mu)$ (right) as a function of $m_H$. Note that $\BR(H \to e_4 \mu)$ includes both $H \to e_4^+ \mu$ and $H \to e_4^- \mu^+$ decay modes. See the caption in \fig{fig:mHtanbe} for further details.
}
\label{fig:scancombinedbr}
\end{figure}
%--

Among the conventional searches of neutral heavy Higgs bosons, $H \to \tau^+ \tau^-$~\cite{Aaboud:2017sjh} currently provides the strongest constraints on our model, especially for large $\tan\beta$ region; we expect that future analysis improvements may boost the impact of $H \to b \bar b$ and $H \to t \bar t$. Searches for $H \to WW$ and $H\to ZZ$ are not constraining because these decays are heavily suppressed in the alignment limit in which tree--level couplings of heavy Higgs and weak gauge bosons vanish. Nevertheless, the processes $H \to e_4 \ell (\nu_4 \nu_\ell) \to \nu_\ell \ell W$ and $H \to e_4 \ell \to \ell^+ \ell^- Z$ are indirectly constrained by searches for \sloppy\mbox{$H \to WW$} and $H\to ZZ$ albeit with different kinematic topologies~\cite{Dermisek:2015vra, Dermisek:2015oja, Dermisek:2015hue}. 
%We leave a more detailed analyses of the $H \to WW, ZZ$ constraints at the HL- and HE-LHC to future work. 
The decay $H \to \gamma \gamma$ currently constraints our model at small $\tan\beta \lesssim 1.5$ (at the HL-LHC values of $\tan\beta$ for which we expect constraints raises to about $3.5$).  Here we focus on $H \to \tau^+ \tau^-$ as a competitive search avenue for a heavy neutral Higgs boson. 

Let us comment on the extraction of the experimental sensitivities to $H\to \tau\tau$ mentioned above. We assume that there is no change in cut acceptances and detector efficiencies for $H\to\tau\tau$ (for a given $m_H$) for the HL-LHC and the HE-LHC,\footnote{This is reasonable because the kinematics of $H$ decay products depends almost exclusively on $m_H$.} implying that $S / \sqrt{B}$ controls to a good approximation the change in sensitivity. With this assumptions, the sensitivity at the HL-LHC increases simply by the square-root of the ratio of the integrated luminosities, \iee, $\sqrt{3000./36.} = 9.13$. For the HE-LHC, we can obtain the sensitivities by further assuming that the background production cross sections ($\sigma_{\rm bkg.}$) increase by the ratio of Higgs production cross sections for $\tan\beta = 1$, \iee, 
\begin{equation}
\sigma_{\rm bkg.}^{13\TeV} / \sigma_{\rm bkg.}^{27\TeV} = \sigma (p p \to H)^{27\TeV}_{\tan\beta = 1} / \sigma (p p \to H)^{13\TeV}_{\tan\beta = 1}\,,
\end{equation}
and that the background cut acceptances remain constant. Then, the sensitivity increases as follows:
%--
\begin{align}
\frac{{\rm S}^{{\rm 27\,TeV,~15\,ab}^{-1}}}{\sqrt{{\rm B}^{{\rm 27\,TeV,~15\,ab}^{-1}}}} 
&= 
\frac{{\rm S}^{{\rm 13\,TeV,~36\,fb}^{-1}}}{\sqrt{{\rm B}^{{\rm 13\,TeV,~36\,fb}^{-1}}}} \cdot \frac{\sigma(p p \to H \to \tau^+ \tau^-)^{\rm 27\,TeV}}{\sigma(p p \to H \to \tau^+ \tau^-)^{\rm 13\,TeV}} \cdot \sqrt{\frac{\sigma_{\rm bkg.}^{\rm 13\,TeV}}{\sigma_{\rm bkg.}^{\rm 27\,TeV}} \cdot \frac{\mathcal{L}^{{\rm 15\,ab}^{-1}}}{\mathcal{L}^{{\rm 36\,fb}^{-1}}}}  
\nonumber \\
& \hskip -1cm = 
\frac{{\rm S}^{{\rm 13\,TeV,~36\,fb}^{-1}}}{\sqrt{{\rm B}^{{\rm 13\,TeV,~36\,fb}^{-1}}}} \cdot \frac{\sigma(p p \to H \to \tau^+ \tau^-)^{\rm 27\,TeV}}{\sigma(p p \to H \to \tau^+ \tau^-)^{\rm 13\,TeV}} \cdot 20.4 \cdot \sqrt{\frac{\sigma (p p \to H)^{\rm 13\,TeV}_{\tan\beta = 1}}{\sigma (p p \to H)^{\rm 27\,TeV}_{\tan\beta = 1}}} 
\,,
\label{eq:VLFsensitivity}
\end{align}
%--
where ${\rm S}^{{\rm 27\,TeV,~15\,ab}^{-1}}$ (${\rm S}^{{\rm 13\,TeV,~36\,fb}^{-1}}$) and ${\rm B}^{{\rm 27\,TeV,~15\,ab}^{-1}}$ (${\rm B}^{{\rm 13\,TeV,~36\,fb}^{-1}}$) are the number of signal and background events for $H \to \tau^+ \tau^-$ at the HE-LHC (at the LHC with $36\fbinv$). Note that the estimated sensitivity in \eq{eq:VLFsensitivity} does not include experimental systematic uncertainties, that would somewhat weaken the reach of $H\to \tau\tau$.

%--
\begin{figure}[t]
\centering
\begin{minipage}[t][\minipageheightdp][t]{\minipagewidthdp}
\centering
\includegraphics[width=\figuremaxwidthdp,height=\figuremaxheightdp,keepaspectratio]{\main/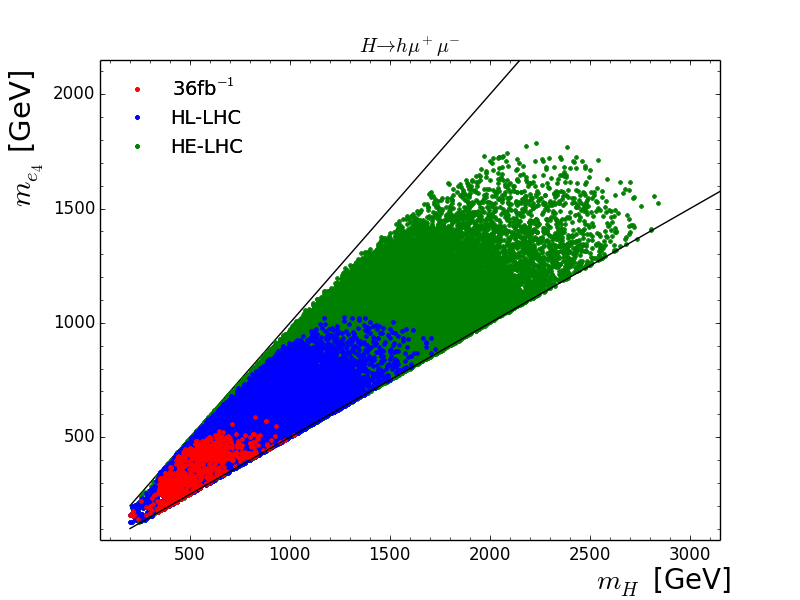}
\end{minipage}
\begin{minipage}[t][\minipageheightdp][t]{\minipagewidthdp}
\centering
\includegraphics[width=\figuremaxwidthdp,height=\figuremaxheightdp,keepaspectratio]{\main/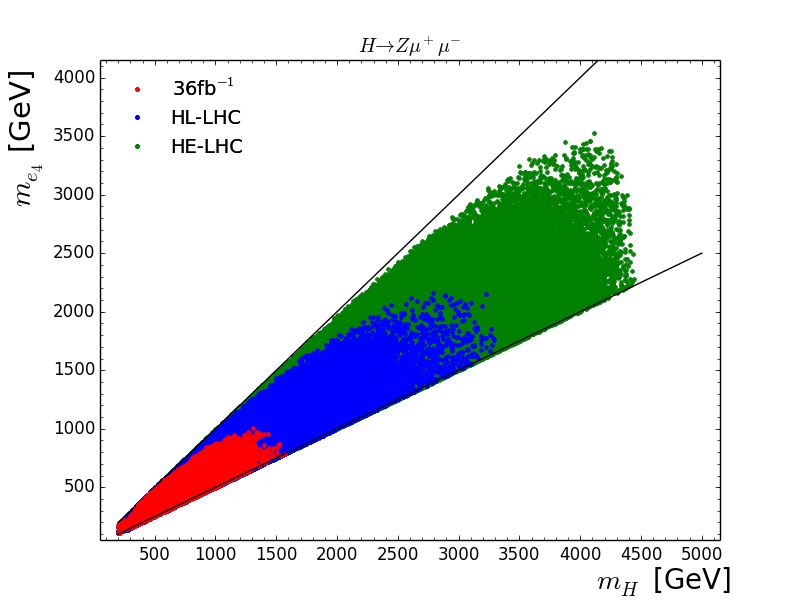}
\end{minipage}
\caption{
Experimental sensitivities for $H \to e_4^\pm \mu^\mp \to h \mu^+ \mu^-$ (left) and $H \to e_4^\pm \mu^\mp \to Z \mu^+ \mu^-$ (right) in the $[m_H,m_{e_4}]$ plane. Note that we require $m_H>m_{e_4}$ to allow the decay channel, and that we loose sensitivity for $m_H > 2 m_{e_4}$, where the $H\to e_4 e_4$ BR dominates. See the caption in \fig{fig:mHtanbe} for further details. 
}
\label{fig:mHme4}
\end{figure}
%--

Our main results are presented in \figs{fig:mHtanbe}, \ref{fig:scancombinedbr} and \ref{fig:mHme4}, where we show the experimental sensitivities for $H \to e_4^\pm \mu^\mp \to h \mu^+ \mu^-$ (left) $H \to e_4^\pm \mu^\mp \to Z \mu^+ \mu^-$ (right) obtained from reanalysing current data with integrated luminosity $36\fbinv$ (red), our estimates for the HL-LHC (blue), and \sloppy\mbox{HE-LHC (green)}. For comparison, in \fig{fig:mHtanbe}, the reach of $H \to \tau^+ \tau^-$ searches in conventional type-II two Higgs doublet model with the current data, at the HL-LHC, and the HE-LHC are shown as black solid, dashed, and dotted lines, respectively. The scatter points satisfy constraints from EWPT, Drell-Yan pair production of vectorlike leptons~\cite{Dermisek:2014qca}, heavy Higgs searches in the $H \to \tau^+ \tau^-$~\cite{Aaboud:2017sjh} and $H \to \gamma \gamma$~\cite{Khachatryan:2016yec, Aaboud:2017yyg} channels. The range of parameters that we scan over is:
\begin{align}
M_{L,N} &\in [100, 5000]\GeV \; , \\
\lambda_L, \lambda_E, \lambda, \bar \lambda &\in [-1, 1] \; , \\
\kappa_N = \kappa = \bar \kappa &= 0 \; , \\
\tan\beta & \in [0.3, 50] \; .
\label{eq:scan}
\end{align}

From the inspections of \fig{fig:mHtanbe}, we see that searches for heavy Higgs cascade decays into vectorlike leptons considerably extend the reach beyond that of $H\to \tau^+\tau^-$. Searches for $H\to h\mu^+\mu^-$ are sensitive to $m_H$ up to $1.7\TeV$ ($2.9\TeV$) and $m_{e_4}$ up to $1\TeV$ ($1.8\TeV$) at the HL-LHC (HE-LHC). The decay mode $H\to Z\mu^+\mu^-$ is even more promising and extends the sensitivity to $m_H$ up to $3.3\TeV$ ($4.5\TeV$) and $m_{e_4}$ up to $2.2\TeV$ ($3.5\TeV$) at the HL-LHC (HE-LHC).

\subsubsection{\theory{Heavy singlet scalars at HL- and HE-LHC}}
\contributors{D. Buttazzo, F. Sala, A. Tesi}
%{\bf Author(s): Dario Buttazzo$^{a,1}$, Filippo Sala$^{b,2}$, Andrea Tesi$^{c,3}$}\\
%
%{\it \small $^a$ INFN, sezione di Pisa, Largo Pontecorvo 3, I-56127 Pisa, Italy}\\
%{\it \small $^b$ DESY, Notkestrasse 85, D-22607 Hamburg, Germany}\\
%{\it \small $^c$ INFN, sezione di Firenze, Via G. Sansone 1, I-59100 Sesto F.no, Italy}\\
%{\small $^1$ dario.buttazzo@pi.infn.it, $^2$ filippo.sala@desy.de $^3$ andrea.tesi@fi.infn.it}\\
%Ackn: The work of F.S. is partly supported by a PIER Seed Project funding (Project ID PIF-2017-72).

%\subsubsection{Motivation}
%\paragraph*{Motivation}

The existence of extended Higgs sectors is predicted in several motivated extensions of the SM.
In particular, extra Higgses that are singlets under the SM gauge group arise in some of the most natural BSM constructions, like the next-to-MSSM (NMSSM, see~\citeref{Ellwanger:2009dp} for a review), as well as in Twin \cite{Chacko:2005pe} and Composite~\cite{Contino:2010rs,Panico:2015jxa} Higgs models (TH and CH models).
Independently of the hierarchy problem of the Fermi scale, extra singlets constitute a minimal possibility to realise a first-order EW phase transition~\cite{Pietroni:1992in,Curtin:2014jma,Craig:2014lda}, which is a necessary condition to achieve EW baryogenesis.
These considerations constitute a strong case for the experimental hunt of extra singlet-like scalar particles.
It is the purpose of this Section, which summarises and updates the work of \citerefs{Buttazzo:2015bka,Buttazzo:2018qqp}, to review the experimental status of the searches for such scalars, and to determine the reach of the HL- and HE-LHC. To keep this contribution brief, we focus on the case where the extra singlet is heavier than the Higgs boson.

%\subsubsection{Current exclusions and future reaches}
%\paragraph*{Current exclusions and future reaches}
\label{sec:singlets_general}

\underline{Framework}. We add to the SM a real scalar field $\phi$, so that the most general renormalisable Lagrangian reads \begin{equation}
\mathcal{L} = \mathcal{L}_{\rm SM} + \frac{1}{2}(\partial S)^2  - \mu_S^2 S^2 - a_{HS} |H|^2S- \lambda_{HS} |H|^2S^2- a_S S^3 - \lambda_S S^4\,,
\end{equation}
where $H$ is the SM Higgs doublet. 
Unless a $Z_2$ symmetry is enforced ($a_{HS} = a_S = 0$), and is not spontaneously broken, the singlet mixes with the SM Higgs as
\begin{equation}
\phi = -s_\gamma\,h_0 + c_\gamma\,s_0, \qquad h = s_\gamma\,s_0 + c_\gamma\,h_0\,,
\end{equation}
where $h_0$ and $s_0$ are the neutral \cpeven degrees of freedom contained in $H$ and $S$, $h$ and $\phi$ are the resulting mass eigenstates, and $s_\gamma, c_\gamma$ are the sine and cosine of their mixing angle $\gamma$. The signal strengths $\mu$ of $h$ and $\phi$ into SM particles, defined as cross-section times BR, read
\begin{align}
\mu_h &= c_\gamma^2 \, \mu_{\rm SM}, &
\mu_{\phi \to VV, ff} &= s_\gamma^2 \,\mu_{\rm SM} (m_\phi)\cdot (1 -\BR(\phi \to hh)), &
\mu_{\phi \to hh} &= s_\gamma^2 \, \sigma_{\rm SM} (m_\phi)\cdot \BR(\phi \to hh),
\end{align}
where $\sigma_{\rm SM}$ is the production cross-section of a SM Higgs boson, and $\mu_{\rm SM}$ is its signal strength into the pair of SM particles of interest.

\begin{figure}[t]
\centering
\begin{minipage}[t][\minipageheightdp][t]{\minipagewidthdp}
\centering
\includegraphics[width=\figuremaxwidthdp,height=\figuremaxheightdp,keepaspectratio]{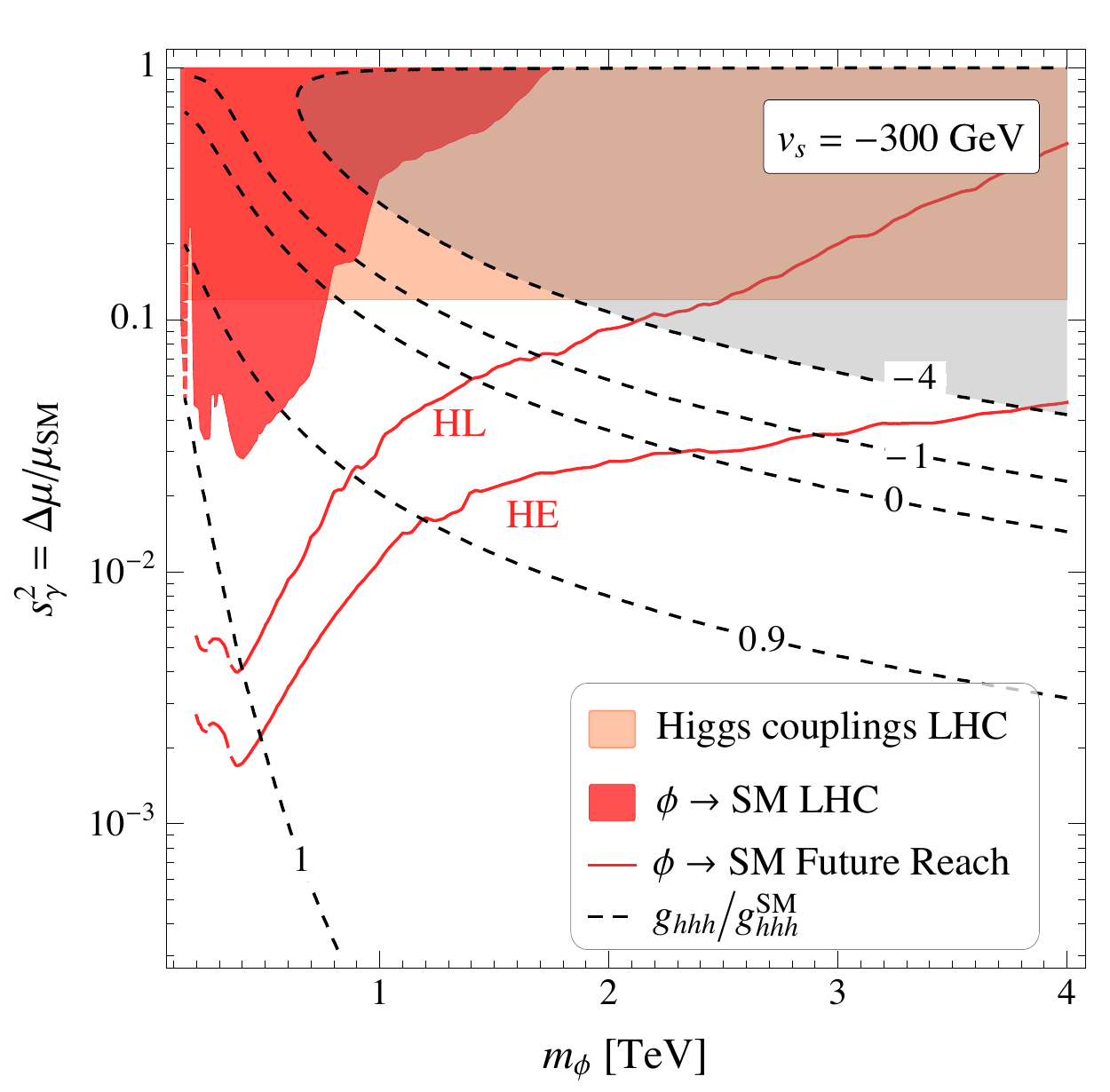}
\end{minipage}
\begin{minipage}[t][\minipageheightdp][t]{\minipagewidthdp}
\centering
\includegraphics[width=\figuremaxwidthdp,height=\figuremaxheightdp,keepaspectratio]{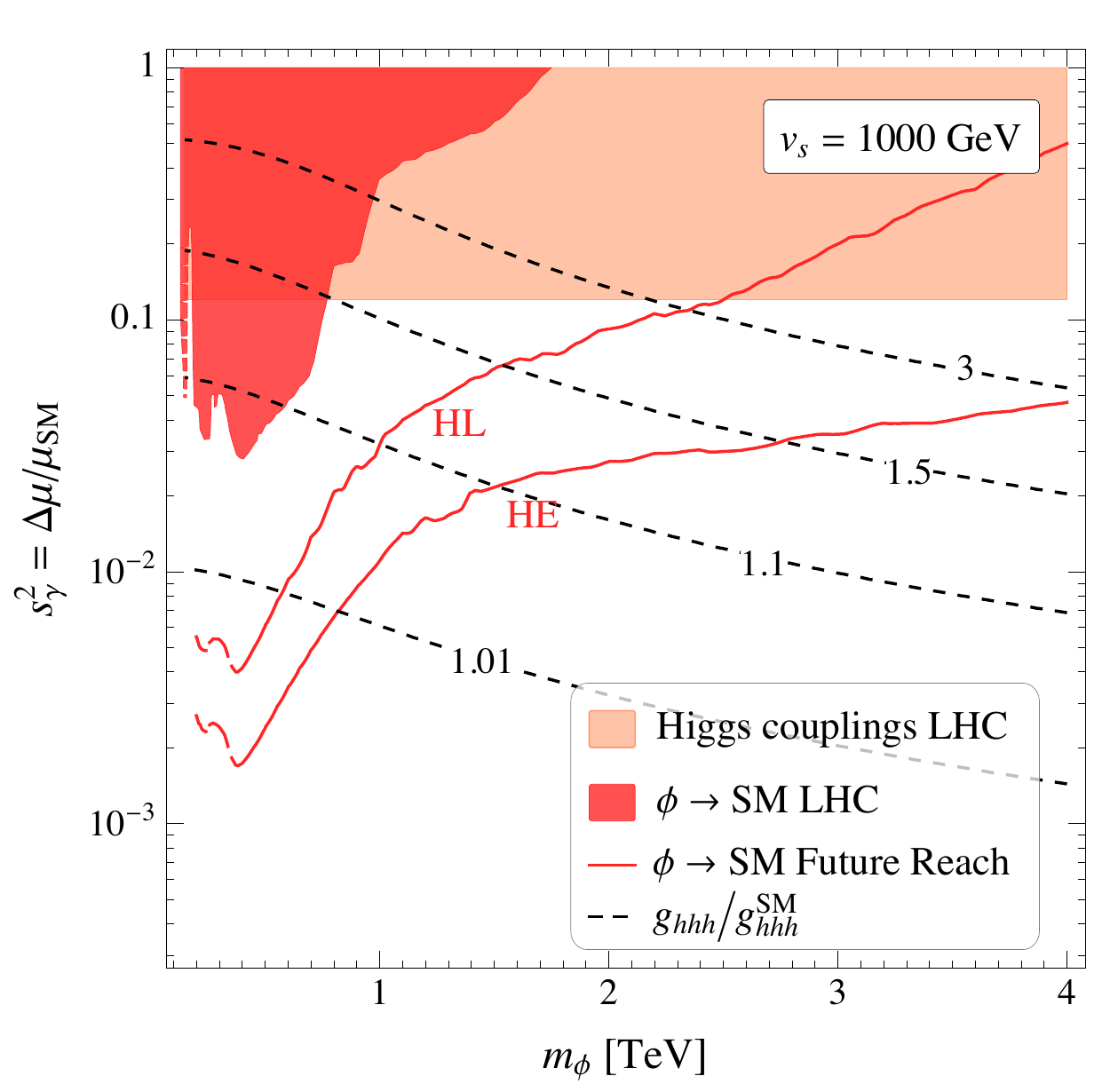}
\end{minipage}
\caption{\label{fig:ModelIndep} Shaded: LHC exclusions from resonance searches (dark red), Higgs coupling measurements (light red) and double Higgs production (gray). Dashed black lines are contours of constant ratio between the trilinear Higgs coupling and its SM value, continuous red lines are expected sensitivities from resonance searches at the HL- and HE-LHC. The singlet \vev $v_s$ is fixed to $-300\GeV$ (left) and to $1000\GeV$ (right).}
\end{figure}

\underline{Constraints from Higgs couplings}. The couplings of the SM-like Higgs boson $h$ to other SM particles are all reduced by the same amount $c_\gamma$, independently of $m_\phi$.
A combined ATLAS and CMS fit to Higgs coupling measurements from 8~TeV data yields the $2\sigma$ limit~\cite{Khachatryan:2016vau} $s_\gamma^2 \lesssim 0.12$, and the sensitivity reached with 36\fbinv of data at $13\TeV$ is comparable~\cite{CMS:2018lkl,ATLAS:2018gcr}. The HL- and HE-LHC are expected to probe values of $s_\gamma^2$ at, and possibly slightly below, the $5\%$ level~\cite{Gori:2650162}. The current exclusion and future reach from Higgs coupling measurements is shown in \fig{fig:ModelIndep}.
Constraints on the mixing angle from EW precision tests are subdominant with respect to those from Higgs coupling measurements, see \egg~\citeref{Carena:2018vpt}.

\underline{Constraints from Trilinear Higgs coupling}. The trilinear Higgs coupling $g_{hhh}$ depends on $\gamma$, $m_\phi$, and on the singlet \vev $v_s$, while its dependence on all other parameters is very mild~\cite{Buttazzo:2015bka} (we fix for definiteness $\lambda_S = \lambda_{HS} = 0.5$).
We show its ratio with respect to its SM value in \fig{fig:ModelIndep}, for two representative values of $v_s$. The gray shaded region comes from a rough bound on $g_{hhh}$ that we extract from \citeref{Aaboud:2018knk}, using the prediction of \citeref{Baglio:2012np}.\footnote{We assume that the only deviation in double Higgs production comes from deviations in $g_{hhh}$, the contribution from $pp \to \phi \to hh$ is negligibly small due to the large $m_\phi$ in the excluded region.}
Deviations of order one and larger are allowed by all current and near-future constraints, motivating in particular the HE stage of the LHC, due to the increase of the sensitivity to $g_{hhh}$ with energy~\cite{Gori:2650162}.

\underline{Constraints from direct searches of the extra singlet}.
The collider phenomenology of $\phi$ is fully controlled by only 3 parameters, $m_\phi$, $\gamma$ and $\BR(\phi \to hh)$. Analogously to the case of the triple Higgs coupling $g_{hhh}$, $\BR(\phi \to hh)$ depends dominantly on the model parameters $v_s$, $\gamma$, and $m_\phi$~\cite{Buttazzo:2015bka}. Moreover, because of the Goldstone boson equivalence theorem, it reaches the asymptotic value \sloppy\mbox{$\BR(\phi \to hh) = \BR(\phi \to ZZ) = 25\%$} for $m_\phi \gg m_h$, further reducing the number of parameters relevant for the phenomenology of $\phi$. %In what follows, for simplicity we fix ${\rm BR}_{\phi \to hh}$ to its asymptotic value.
Current resonance searches at the LHC exclude the red shaded area in \fig{fig:ModelIndep}, and are dominated by the CMS combined $ZZ$ search in \citeref{Sirunyan:2018qlb} at $13\TeV$ with $36\fbinv$ of data. We rescale the expected sensitivity at $13\TeV$~\cite{Sirunyan:2018qlb} at higher energies and luminosities using quark parton luminosities, with a procedure analogous to the one presented in \citeref{Buttazzo:2015bka}. Our results for the expected sensitivities at the HL ($14\TeV$, \intLumi) and HE ($27\TeV$, \intlumihelhc) stages of the LHC are also shown in \fig{fig:ModelIndep}.
Direct searches for the new scalar constitute the strongest probe of the parameter space of these models for $m_\phi$ below about a TeV, while larger masses are (and will be) probed more efficiently by deviations in Higgs couplings, thus making these two strategies complementary in the exploration of these  models.

\underline{Implications for the NMSSM}. The NMSSM adds to the MSSM particle content a singlet $S$, so that the superpotential reads $W = W_{\rm \tiny MSSM} + \lambda\,S H_u H_d + f(S)$. 
%\subsubsection{Interpretation in the NMSSM and in Twin and Composite Higgs}
%\paragraph*{Interpretation in the NMSSM and in Twin and Composite Higgs}
%
\begin{figure}[t]
\centering
\begin{minipage}[t][\minipageheightdp][t]{\minipagewidthdp}
\centering
\includegraphics[width=\figuremaxwidthdp,height=\figuremaxheightdp,keepaspectratio]{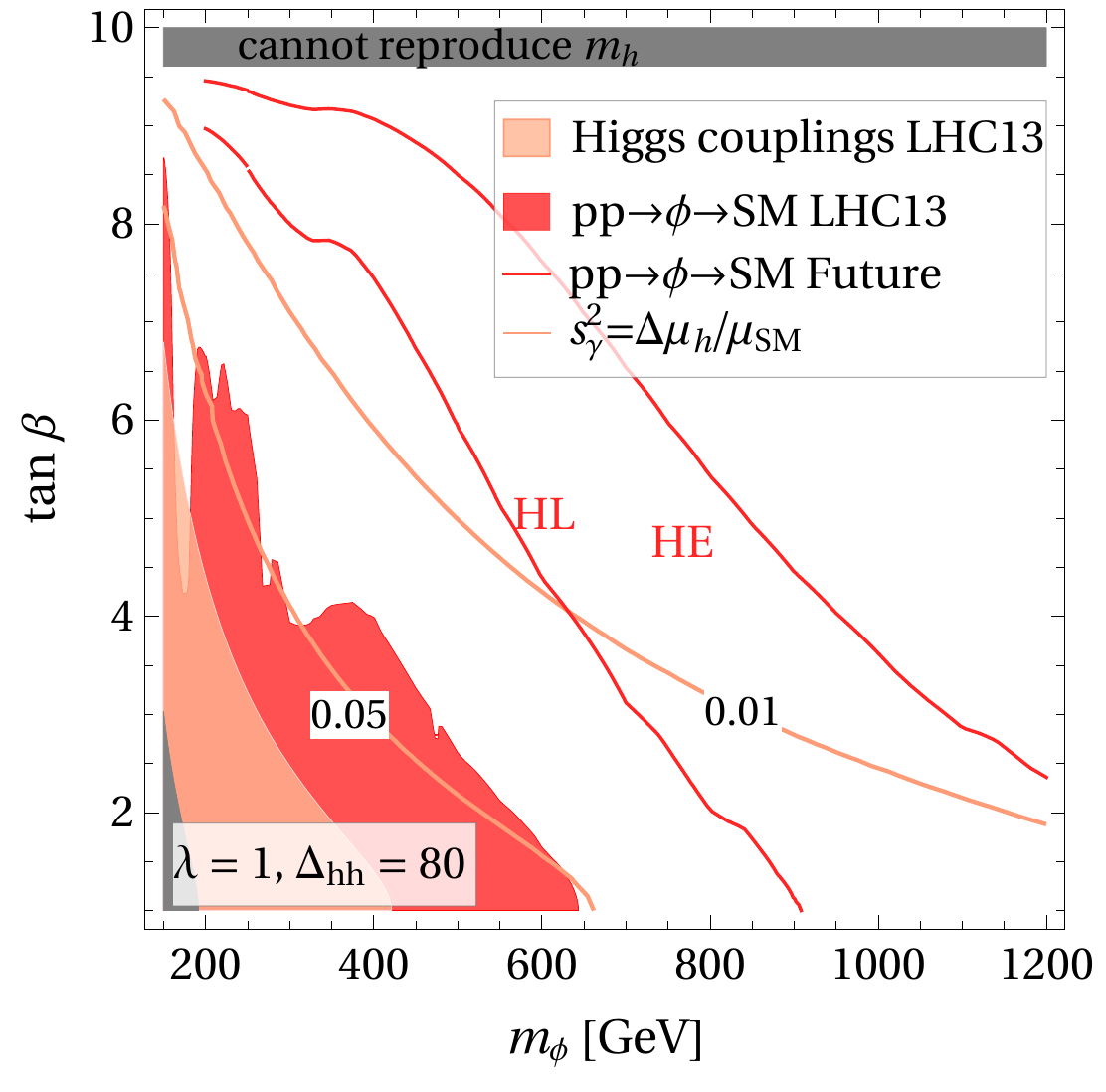}
\end{minipage}
\begin{minipage}[t][\minipageheightdp][t]{\minipagewidthdp}
\centering
\includegraphics[width=\figuremaxwidthdp,height=\figuremaxheightdp,keepaspectratio]{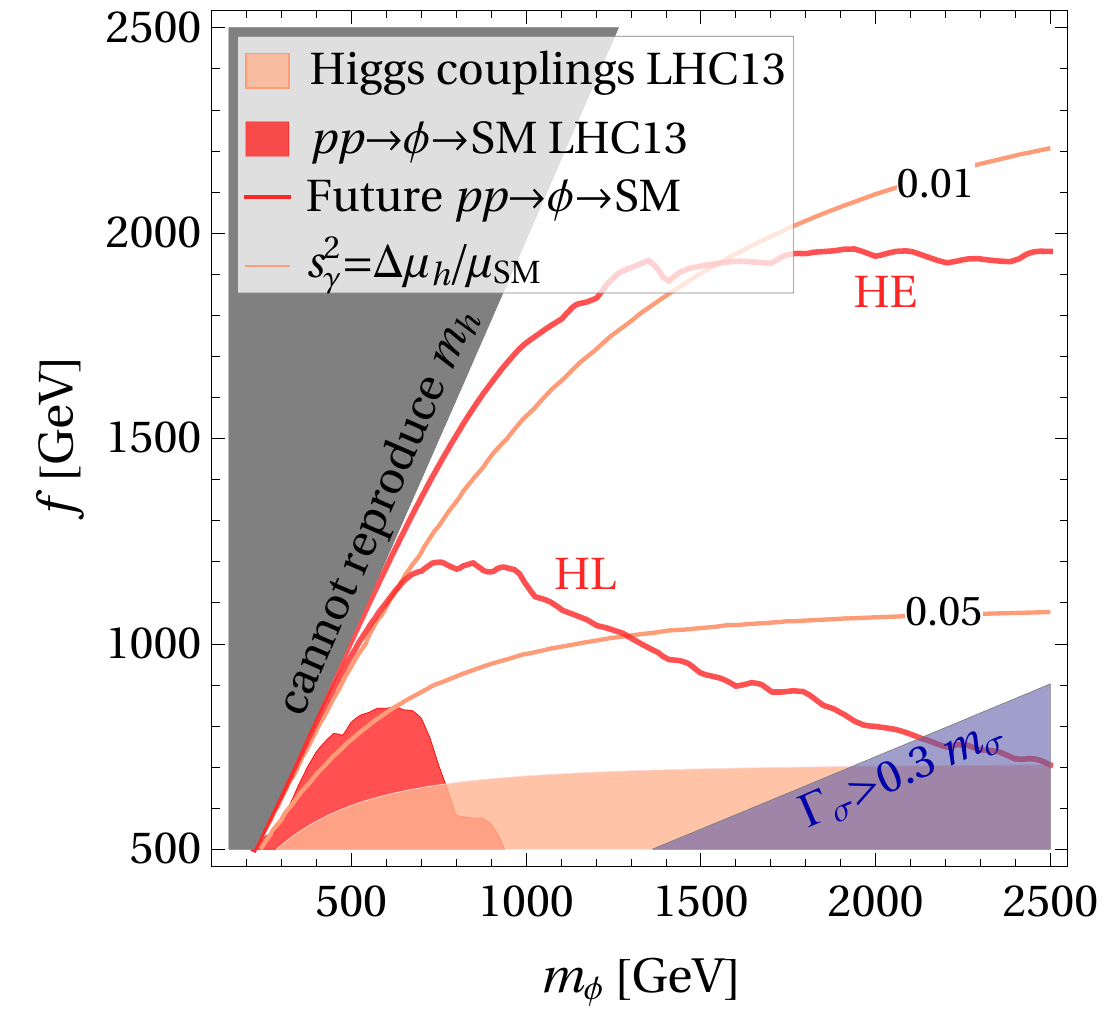}
\end{minipage}
\caption{\label{fig:NMSSM_TH}
Shaded: LHC exclusions from resonance searches (dark red) and Higgs coupling measurements (light red). %Dark gray shaded areas are unphysical.
Left: NMSSM with couplings $\lambda = 1$ and with $\Delta_{hh} = 80\GeV$. Right: Twin and Composite Higgs models. %, where in the shaded area in the bottom-right corner one has $\Gamma_\sigma > m_\sigma$.
See text for more details.
}
\end{figure}
The fine-tuning needed to reproduce the EW scale is parametrically alleviated, with respect to the MSSM, and for a given value of the stop and gluino masses, by a factor $\lambda/g$, see \egg~\citerefs{Barbieri:2006bg,Hall:2011aa,Agashe:2012zq,Gherghetta:2012gb}.
Naturalness arguments thus favour a large $\lambda$, that is however bounded from above by perturbativity.
Assuming the masses of the extra Higgs bosons in the TeV range, a model with $\lambda \simeq 2$ becomes strongly coupled at scales of order $10\TeV$, and to have a perturbative coupling up to the GUT scale one needs $\lambda \lesssim 0.7$~\cite{Espinosa:1991gr}.\footnote{It is conceivable that a strong sector exists at the scale where $\lambda$ becomes non-perturbative, and without affecting the success of GUT in the NMSSM, see \egg~the model in \citeref{Barbieri:2013hxa} and references therein.}

Here we employ the economical parametrisation of the NMSSM scalar sector put forward in \citerefs{Barbieri:2013hxa,Barbieri:2013nka}.
We then assume the extra Higgs doublet to be (slightly) decoupled, and we study the phenomenology of the Higgs-singlet scalar sector, which can be described by 4 free parameters
\begin{equation}
m_\phi, \quad \lambda,\quad t_\beta, \quad \Delta_{hh},
\label{eq:singlets_NMSSMpars}
\end{equation}
where $t_\beta$ is the ratio of the up and down Higgs \vevs, and $\Delta_{hh}$ encodes the radiative contribution to the SM-like Higgs mass $m_h^2 \lesssim m_Z^2 \,c^2_{2\beta} + \lambda^2 \,v^2\,s^2_{2\beta}/2 + \Delta_{hh}^2$. %, $\Delta_{hh} \simeq $.
The phenomenology discussed in the previous section is displayed, for the NMSSM, in the left-hand panel of \fig{fig:NMSSM_TH}.
We have fixed $\Delta_{hh} = 80\GeV$, a value obtainable for stop masses and mixing in the range of $1-2\TeV$.
The precise value of $\Delta_{hh}$ does not affect the Higgs sector phenomenology as long as it is within order $10\%$ of $80\GeV$. We have also fixed $\lambda =1$, smaller values would make all the exclusions and sensitivities weaker, while larger values would make them stronger.

As seen in \fig{fig:NMSSM_TH} left, direct searches for the extra singlet are expected to constitute the most promising probe of the $\phi - h$ parameter space. %, the reason being the fast decrease of the mixing angle $\gamma$ with increasing $m_\phi$ in the NMSSM.
Higgs coupling measurements will give a reach at most comparable to the one of resonance searches, and they will constitute a crucial complementary access to the parameter space of NMSSM scalars. Deviations in the trilinear Higgs coupling depend on more parameters than those in \eq{eq:singlets_NMSSMpars}, and can reach around $50\%$ or more if $\lambda \gtrsim 1$, see \citeref{Sala:2015lza} for a precise quantification.

\underline{Twin and Composite Higgs}.
In both TH and CH models, the SM-like Higgs boson is the pseudo-Goldstone boson associated to the spontaneous breaking of a global symmetry at a scale $f$. The EW fine-tuning of CH models is comparable to the one of typical SUSY constructions, while TH models can achieve a tuning as good as the irreducible factor $v^2/f^2$ (mainly because the top partners are neutral under colour, and can thus be light).
The radial scalar mode $\sigma$ associated to the pseudo-Goldstone Higgs has a mass $m_\sigma \sim g_* f$, where the size of $g_*$ corresponds to the typical one of the coupling of the UV completion.
TH models with weakly coupled UV completions
feature an extra scalar singlet that can be light, see \egg~\citerefs{Katz:2016wtw,Badziak:2017syq}.
The hunt for the extra scalar is a crucial test of TH models, because of the small couplings to the SM of the rest of the ``Twin'' states.

The scalar potential of TH and CH models has less free parameters, so that all the quantities relevant for the scalar phenomenology (including $g_{hhh}$ and $\BR(\phi \to hh)$) are a function of two free parameters only, that we choose to be $f$ and $m_\sigma$.
One for example obtains $s_\gamma \approx v/f$ and  $g_{hhh} \approx 1$ up to terms suppressed by the small ratio $v^2/f^2$.
In TH, an extra freedom is given by the decay channels of $\phi$ into the Twin sector, whose BR in the limit $m_\sigma \gg g f$ is fixed, through the Goldstone boson equivalence theorem, by the symmetries of the model ($3/7$ in the cases of $SO(8)/SO(7) \simeq SU(4)/SU(3)$). For simplicity, and because such decay is absent in CH models, we do not include it in our study.

The constraints and sensitivities discussed 
%in Section~\ref{sec:singlets_general} 
above
are displayed in the right-hand panel of \fig{fig:NMSSM_TH}.
Given our considerations above, in the case where these models are not too strongly coupled (we show in \fig{fig:NMSSM_TH} the region where $\Gamma_\sigma > 0.3 m_\sigma$), they are expected to manifest themselves first via new diboson resonances. On the contrary, their strong-coupling regime is expected to show up first in deviations in the Higgs couplings.

\subsubsection{\theory{Relaxion at the HL-LHC}}
\contributors{E. Fuchs, M. Schlaffer}

%\paragraph*{The relaxion framework}
The relaxion mechanism~\cite{Graham:2015cka} addresses the hierarchy problem 
differently than conventional symmetry-based solutions. 
In this framework, the Higgs mass is stabilised dynamically by the relaxion, a pseudo-Nambu-Goldstone boson.
The Higgs mass is scanned by the evolution of the relaxion field. Eventually, the relaxion stops at the field value
where the Higgs mass is much smaller than the theory's cutoff, hence addressing the fine tuning problem.  
Relaxion models do not require top, gauge or Higgs partners at the TeV scale. The possible mass range for the relaxion ranges from sub-eV to tens of GeV.  Hence this framework can lead to signatures relevant for cosmology, for the low-energy precision frontier, for the intensity frontier, and for the high energy collider frontier. For detailed studies see \citerefs{Flacke:2016szy,Choi:2016luu,Frugiuele:2018coc}.

\begin{figure}[t!]
\centering
\begin{minipage}[t][\minipageheightsp][t]{\minipagewidthsp}
\centering
\includegraphics[width=\figuremaxwidthsp,height=\figuremaxheightsp,keepaspectratio]{\main/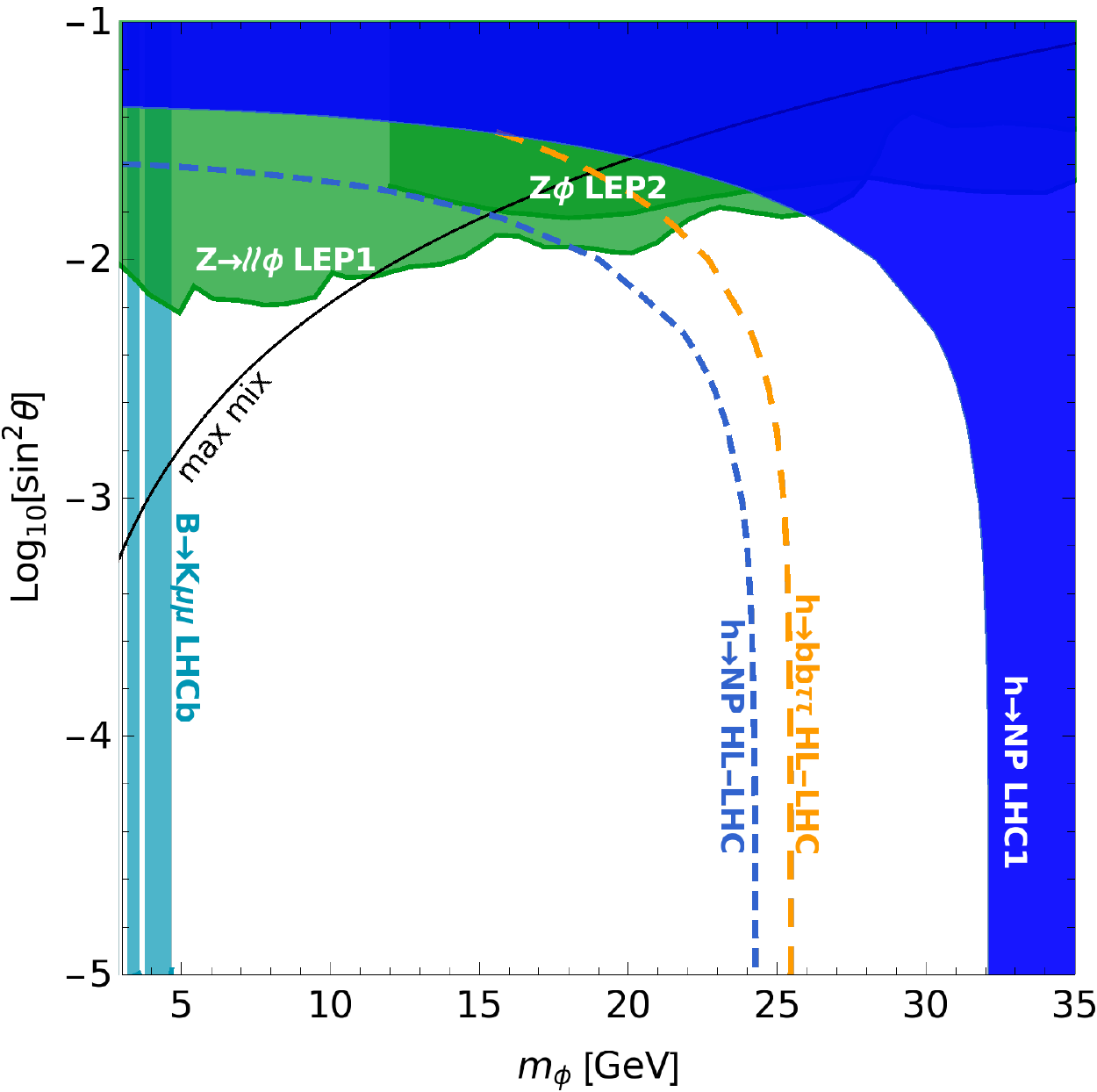}
\end{minipage}
    \caption{ Bounds and projections at the $95\%\cl$ for processes at hadron and
    lepton colliders.  \sloppy\mbox{$Z\to Z^*\phi\to \ell\bar \ell \phi$} at LEP1 with
    $\sqrt{s}=M_Z$~\cite{Acciarri:1996um} and $e^+e^-\to Z\phi$ at LEP2 with
    $\sqrt{s}=192-202\GeV$~\cite{Schael:2006cr}. 
    Bound from $B^+\to K^+ \mu^+ \mu^-$ at LHCb~\cite{Aaij:2012vr,Aaij:2015tna}. 
    Direct searches for exotic Higgs decays at the HL-LHC in
    the $bb\tau\tau$ channel inferred from \citeref{CMS:2018lqr} (orange, dashed).
    Upper bound on $\BR(h\to\textrm{NP})$ from Higgs coupling fits at the LHC Run-1 (blue area) and projection for the HL-LHC with $3\abinv$ (blue, dashed). 
    The maximal mixing
    according to \eq{eq:maxmixNatural} is indicated by the black line.  }
  \label{fig:DirIndirHLLHC}
\end{figure}

The aspects of the relaxion mechanism that are relevant for the phenomenology at colliders are summarised in the following.
The effective scalar potential of
the theory depends both on the Higgs doublet $H$ and the relaxion $\phi$,
\begin{align}
  V(H,\phi)&=\mu^2(\phi) H^\dagger H+\lambda (H^ \dagger H)^2 + V_{\rm sr}(\phi)+ V_\text{br}(h,\phi)\, ,\label{eq:pot}\\
  \mu^2(\phi)&=-\Lambda^2+g \Lambda \phi + \dots   \, ,
\label{eq:mu2}
\end{align}
where $\Lambda$ is the cutoff scale of a Higgs loop.
The relaxion scans $\mu^2$ via the slow-roll potential
\begin{equation}
  V_{\rm sr}(\phi)=r g\Lambda^3 \phi\,,
\label{eq:roll}
\end{equation}
where $g$ is a small dimension-less coupling and $r > 1/ (16 \pi^2)$ due to naturalness
requirements. Once $\mu^2(\phi)$ becomes negative, the Higgs gets a \vev
$v^2(\phi)=-\frac{\mu^2(\phi)}{\lambda}$. The non-zero \vev activates a periodic (model-dependent)
backreaction potential $V_\text{br}$ associated with the backreaction scale $\Lambda_{\textrm{br} }$ that eventually
stops the rolling of the relaxion at a value $\phi_0$, where $v(\phi_0)=246\GeV$.  Generically, the
relaxion mechanism leads to \cp violation and as a result, the relaxion $\phi$ mixes with the
Higgs $h$ and inherits its couplings to SM fields~\cite{Flacke:2016szy,Choi:2016luu}.  The relaxion
mass $m_{\phi}$ and the mixing angle $\sin\theta$ can be expressed as
\begin{align}
m_{\phi} &\simeq \frac{ \Lambda_{\textrm{br} }^2}{f} \sqrt{c_0 }\,,
\label{eq:mphiApprox}\\
\sin{\theta}
 &\simeq 8   \frac{\Lambda_{\textrm{br} }^4}{v^3 f} s_0 \,,
\label{eq:sinth}
\end{align}
where $s_0\equiv \sin\phi_0, ~c_0\equiv \cos\phi_0$, and $f$ is the scale where the shift symmetry of the backreaction potential
is broken. Combining \eqs{eq:mphiApprox} and~\eqref{eq:sinth} with
$4 \Lambda_{\textrm{br} }^2 s_0 < v^2\sqrt{c_0}$, which is fulfilled due to the suppressed value of $s_0$ at the endpoint of
the rolling~\cite{Choi:2016luu}, the mixing angle as a function of the relaxion mass is
approximately bounded by
\begin{equation}
 \sin\theta \leq 2\frac{m_{\phi}}{v}\,.
 \label{eq:maxmixNatural}
\end{equation}
This upper bound on the relaxion-Higgs mixing is indicated by the black line in \fig{fig:DirIndirHLLHC}.

Moreover, in the broken phase a trilinear relaxion-relaxion-Higgs coupling $c_{\phi\phi h}$ is
\sloppy\mbox{generated~\cite{Flacke:2016szy,Frugiuele:2018coc}},
\begin{align}
 c_{\phi\phi h} &= \frac{\Lambda_{\textrm{br} }^4}{v f^2} c_0 c_{\theta}^3
 - \frac{2 \Lambda_{\textrm{br} }^4  }{v^2 f} s_0 c_{\theta}^2 s_{\theta}
- \frac{\Lambda_{\textrm{br} }^4}{2f^3} s_0 c_{\theta}^2 s_{\theta}
- \frac{2\Lambda_{\textrm{br} }^4}{v f^2} c_0 c_{\theta} s_{\theta}^2
+ 3v \lambda c_{\theta} s_{\theta}^2
                  + \frac{\Lambda_{\textrm{br} }^4 }{v^2 f} s_0 s_{\theta}^3\,,
                  \label{eq:phiphih}
\end{align}
where $s_\theta\equiv \sin\theta,~c_\theta\equiv \cos\theta$.
Thus a bound on $c_{\phi\phi h}$ constrains the $(m_\phi,\sin\theta)$ parameter space.
In the limit of small mixing, \eq{eq:phiphih} reduces to
$ c_{\phi\phi h}  \simeq m_{\phi}^2/v$\,,
therefore directly constraining the mass. 

%\paragraph*{Exotic Higgs decay $\boldsymbol{h\to\phi\phi}$} 
There are two complementary ways to constrain $c_{\phi\phi h}$ via the exotic Higgs decay $h\to\phi\phi$: searching directly for the decay products of the $\phi$ pair, or bounding this BR by a global fit of Higgs couplings.
 
%\paragraph*{General bound on $\BR(\boldmath{h}\to \textrm{NP})$ } 
Concerning the Higgs phenomenology, the relaxion can be viewed as a singlet extension of the SM with a mixing of $\phi$ and $h$. The Higgs couplings to SM particles are reduced by the universal coupling modifier $\kappa\equiv \cos\theta$, and the total Higgs width 
$\Gamma_h^{\rm tot} = \cos^2\theta\, \Gamma_h^{\rm tot,SM} + \Gamma_h^{\rm NP}$
contains the NP contribution $\Gamma_h^{\rm NP} = \Gamma(h\to\phi\phi)$.
For the LHC, $\BR(h\to\textrm{NP})$ has been constrained via a Higgs coupling fit to Run-1 data to be at most $20\%$ at the $95\%~\cl$
while the HL-LHC has the potential to bound this BR to approximately $8.6\%$ at the $95\%\cl$~\citeref{Bechtle:2014ewa,Gori:2650162}.
% \textbf{Elina updated the above, replacing the 68\%~\cl by the approximate $95\%\cl$ bound.}
The resulting bounds in the $(m_\phi, \sin\theta)$ plane are shown in \fig{fig:DirIndirHLLHC}.

\begin{figure}[t!]
\centering
\begin{minipage}[t][\minipageheightdp][t]{\minipagewidthdp}
\centering
\includegraphics[width=\figuremaxwidthdp,height=\figuremaxheightdp,keepaspectratio]{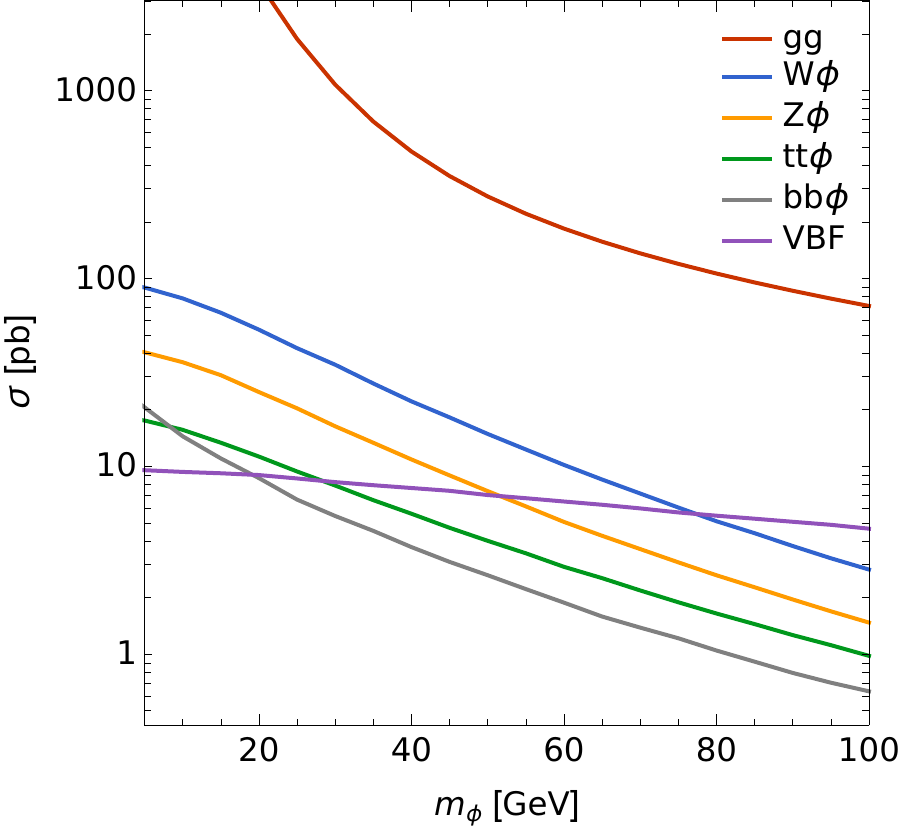}
\end{minipage}
\begin{minipage}[t][\minipageheightdp][t]{\minipagewidthdp}
\centering
\includegraphics[width=\figuremaxwidthdp,height=\figuremaxheightdp,keepaspectratio]{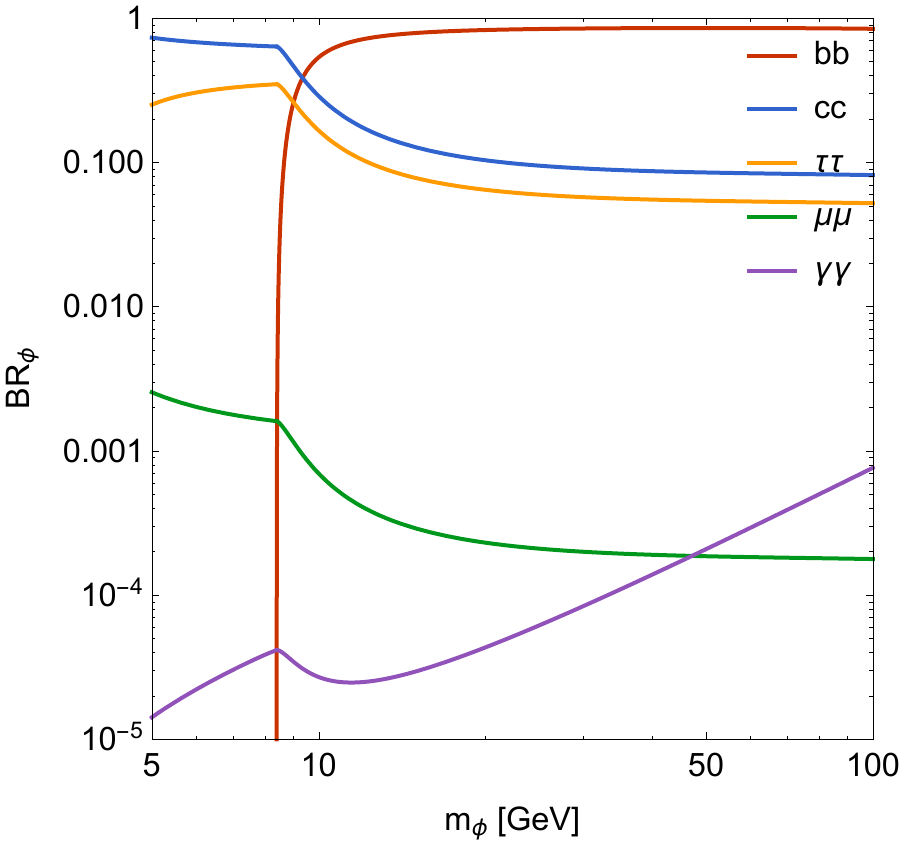}
\end{minipage}
  \caption{Production and decay of $\phi$ for $\sin^2\theta=1$.
    Left: hadronic cross sections, $\sigma(pp\to X)$ at $\sqrt{s}=14\TeV$
    for $X=\phi$ (via gluon fusion), $W\phi$, $Z\phi$, $t\bar t\phi$, $b\bar b\phi$, and $\phi jj$
    (via VBF).  The $\sigma(pp\to\phi)$ via gluon fusion is calculated using \texttt{ggHiggs
      v3.5}~\cite{Ball:2013bra,Bonvini:2014jma,Bonvini:2016frm,Ahmed:2016otz} at N$^3$LO including
    N$^3$LL resummation without a $p_T$-cut.  The remaining hadronic cross sections are obtained
    from \MGvATNLO \cite{Alwall:2014hca} at NLO with $p_T(\phi)>20\GeV$.
    Right: BRs
    ${\rm BR}(\phi\to b\bar b,\, c\bar c,\, \tau^+\tau^-, \mu^+\mu^-, \gamma\gamma)$.
   }
   \label{fig:prod_decay}
\end{figure}

The parameter space probed by the HL-LHC enters the region below the upper
 bound of the mixing and is therefore relevant for the model. The HL-LHC may exclude a
 relaxion mass above $24\GeV$ for vanishing $\sin\theta$.
 In comparison, direct relaxion production via $B\to K\phi$,
 $\phi\to\mu\mu$ at LHCb excludes $2m_\mu \leq m_\phi \lesssim 5\GeV$ also for $\sin\theta$ smaller
 than shown in \fig{fig:DirIndirHLLHC}. In contrast, the bounds set by LEP1 via
 the 3-body $Z$-decay into $f\bar f\phi$ and by LEP2 via relaxion strahlung in $Z\phi$ production
 are sensitive only to mixing angles of order  $\sin^2\theta \gtrsim 10^{-2}$ and therefore constrain mostly
 parameter space above the theoretically motivated maximal mixing.
Furthermore, the stronger bound reachable at the HL-LHC, compared to the Run-1 one, is highly beneficial for constraining the relaxion mass space down to small mixing angles.
 
% \paragraph*{Searches for the relaxion decay products}
Regarding direct searches for the relaxion decay products,
each relaxion from the Higgs decay further decays into a pair of SM particles, resulting in a four-particle final state $F$. The ATLAS and CMS
searches for such signatures yield $m_{\phi}$-dependent bounds on
$(\sigma_h/\sigma_h^{\rm SM}) \times {\rm BR}\left(h\to\phi\phi \to F\right)$.
However, none of the current
searches~\cite{CMS:2018lqr,Khachatryan:2017mnf,Aaboud:2018esj,CMS:2018wii,CMS:2016tgd,Aaboud:2018fvk,Aaboud:2018iil,Aaboud:2018gmx} is sensitive enough to probe parts of the relaxion parameter space displayed in
\fig{fig:DirIndirHLLHC}, \iee{} $5\GeV \leq m_{\phi} \leq 35\GeV$
and \sloppy\mbox{$10^{-5}\leq \sin^2\theta \leq 10^{-1}$}. 
In contrast, the HL-LHC can probe parts of the relevant parameter space via these channels.
We estimate the potential reach of the
HL-LHC with $3\abinv$ by rescaling the current limits by the ratio of luminosities and
by the ratio of the Higgs production cross sections in the dominant channels at
$8$ or $13\TeV$ with respect to $14\TeV$ 
for the search results from Run-1 or Run-2, respectively~\cite{xsecratio}.
% in the case
% of Run-1 limits at $\sqrt{s}=8\TeV$,
% additionally by the ratio of Higgs production cross sections at
% $8$ and $13\TeV$ 
According to this scaling, we find that the strongest bound at the HL-LHC is expected in the $bb\tau\tau$ channel (inferred from the CMS search at Run-2~\cite{CMS:2018lqr}), excluding $m_{\phi}>26\GeV$ 
at $95\%\cl$
as shown in \fig{fig:DirIndirHLLHC}. 
%Elina has changed this to the approx. 95%CL.

% \paragraph*{Production modes in proton collisions}

Concerning the relaxion direct production at colliders, similarly to the Higgs, the dominant production modes for the relaxion are gluon fusion ($ p  p \to \phi $), relaxion strahlung ($ p  p \to Z \phi,~W\phi$),  \sloppy\mbox{$\left\lbrace t\bar t,~b\bar b \right\rbrace$-associated} production, and vector boson fusion (VBF, $p p \to \phi j j$).
The cross sections of these processes are shown for $\sin\theta=1$ in the left panel of \fig{fig:prod_decay}. 
For the HL-LHC, a promising channel is relaxion strahlung due to the relatively large cross sections in combination with the presence of a massive gauge boson that can be tagged. Various decay modes can be considered, see the right panel of \fig{fig:prod_decay}.
So far the $Z\phi$ and $W\phi$ searches do not cover the challenging low-mass range. With high luminosity, extending these searches to lower masses could yield complementary bounds.
\subsubsection{\theoryC{The HL-LHC and HE-LHC scope for testing compositeness of 2HDMs}}
\contributors{S. De Curtis, L. Delle Rose, S. Moretti, A. Tesi, K. Yagyu}
%{\bf Author(s): S. De Curtis$^a$, L. Delle Rose$^{a}$, S. Moretti$^{b,c}$, A. Tesi$^{a}$, K. Yagyu$^{a,d}$}\\
%
%{\it \small $^a$INFN, Sezione di Firenze, and Department of Physics and Astronomy, University of Florence,} \\
%{\it \small Via G. Sansone 1, 50019 Sesto Fiorentino, Italy}\\
%{\it \small $^b$School of Physics and Astronomy, University of Southampton,} \\
%{\it \small  Highfield, Southampton SO17 1BJ, United Kingdom}\\
%{\it \small  $^c$Particle Physics Department, Rutherford Appleton Laboratory,} \\
%{\it \small  Chilton, Didcot, Oxon OX11 0QX, United Kingdom}\\
%{\it \small  $^d$ Seikei University, Musashino, Tokyo 180-8633, Japan} \\
%Ackn: SM is funded in part through the NExT Institute and the STFC CG ST/P000711/1.

%\subsubsection{Introduction}
%\paragraph*{Introduction}

Much has been written about the ability of TeV scale compositeness to naturally remedy the hierarchy problem of the SM, in particular through the pseudo-Nambu Goldstone  boson (pNGB) nature of the Higgs state. This idea is borrowed from QCD: the discovered Higgs boson is the analogue of the pion.  Just like there are $\pi$, $\eta$, \etcc{} mesons predicted by QCD, though, there could be several Higgs states predicted by compositeness beyond the one discovered. In this respect, a natural setting ~\cite{Mrazek:2011iu}   is the Composite 2-Higgs Doublet Model (C2HDM)~\cite{DeCurtis:2016scv,DeCurtis:2016tsm,DeCurtis:2017gzi}. It is built upon the experimentally established existence of a doublet structure triggering Electro-Weak Symmetry Breaking (EWSB),  generating the $W^\pm$, $Z$ and SM-like Higgs state $h$, yet it surpasses it by providing one more composite doublet of Higgs states that can be searched for at the LHC, alongside additional composite gauge bosons (the equivalent of the $\rho$, $\omega$, \etcc{} of QCD) and 
composite fermions. We concentrate here on the Higgs sector of the C2HDM (see \citeref{DeCurtis:2018iqd} for a comparative study between the 2HDM based composite solution to the hierarchy problem and the one driven instead by Supersymmetry) based on $SO(6)\to SO(4)\times SO(2)$,  in presence of so-called {\sl partial compositeness}~\cite{Kaplan:1991dc} realised through the third generation of the SM fermions.  The scalar potential and, thus, the masses of the Higgs bosons are generated at one-loop level by the linear mixing between the (elementary) SM and the (composite) strong sector fields. This implies that masses and couplings are not free parameters, unlike in the elementary realisations, but they depend upon the strong sector dynamics  and are strongly correlated.
The compositeness scale $f$ is  within the energy domain of the LHC and  the composite nature of the SM-like Higgs boson in the C2HDM can  be revealed through corrections of ${\cal O}(\xi)$, where $\xi=v^2/f^2$, with $v$ being the \vev of the SM Higgs field. 
As current lower limits on $f$ are of order $750\GeV$, this implies that such effects enter experimental observables at the $5-10\%$ level.
%Hence, it may reasonably be argued that the LHC might have problems in establishing them in the Higgs sector. \rt{Why? LHC has at least 10\% sensitivity in the Higgs sector no?}
%Indeed the ${\cal O}(\xi)$ corrections,  involving Yukawa interactions with top- and bottom-quarks or tau-leptons,  are notoriously difficult to measure at the LHC, as these lead to hadronic processes which are masked by the significant  QCD background present at the LHC. 
In order to test these effects, the best strategy is to probe gauge interactions of the SM-like Higgs boson, universally affected by ${\cal O}(\xi)$ corrections. In particular, these effects are expected to be larger than the corresponding ones in Elementary realisations of a 2HDM (E2HDM)~\cite{Branco:2011iw}, and therefore accessible at the HL-LHC.  
However, if these were to be found consistent with those of the E2HDM, the last resort in the quest to disentangle the C2HDM from the E2HDM hypothesis would be to exploit the correlation among observable processes involving extra Higgs bosons.
Hence, under the above circumstances, it becomes mandatory to assess the scope of the HL- and HE-LHC, the latter being  necessary for processes involving Higgs boson self-interactions, which have  rather small cross-sections  at the current LHC, in exploring the structure of extended Higgs sectors.
We shall do so by studying under which LHC machine conditions one could access the processes $gg \to H \to hh \to b\bar b\gamma\gamma$ and $gg \to H\to t\bar t$ (followed by semi-leptonic top decays), as these will enable one to extract crucial C2HDM parameters, wherein $H$ is the heaviest of the two \cpeven (neutral) Higgs states of the C2HDM, the lightest ($h$) being the SM-like one.  

%\subsubsection{Numerical results}
%\paragraph*{Numerical results}

The construction of the model and the fundamental parameters of the C2HDM are described in \citeref{DeCurtis:2018iqd,DeCurtis:2018zvh}. 
These correspond to the scale of compositeness $f$, the coupling of the spin-1 resonances, the masses of the heavy top partners, and the mixing between the latter and the elementary top quark (which represents the leading contribution to the effective scalar potential). 
In order to have phenomenologically acceptable configurations with EW parameters consistent with data, we require: 
\sloppy\mbox{(i) the vanishing} of the two tadpoles of the \cpeven Higgs bosons, (ii) the value of the top quark mass to match the measured one, 
and (iii) the value of the Higgs boson mass to match the measured one. 
Under these constraints, we explore the parameter space by scanning the scale of compositeness in the range $(750, 3000)$ GeV and all the other parameters in the range $(-10,10)f$.
As outputs, we obtain the masses of the charged Higgs boson $(m_{H^\pm})$, the \cpodd Higgs boson ($m_A$), and the heavier \cpeven Higgs boson ($m_H$), 
the mixing angle $\theta$ between the two \cpeven Higgs boson states ($h,H$), as well as their couplings to fermions and bosons. 
These quantities are then combined in physics observables and tested against experimental measurements through HiggsBounds~\cite{Bechtle:2013wla} and HiggsSignals~\cite{Bechtle:2013xfa},
which include {current} results from void Higgs boson searches and parameter determinations from the discovered Higgs state,
respectively. Further, we extrapolated the latter (at present counting on about 30\fbinv of accumulated luminosity after Run-1 and in Run-2) to $\sqrt{s}=14\TeV$ with $300\fbinv$(end of Run-3) and \intLumi (\sloppy\mbox{HL-LHC}), by adopting the expected experimental accuracies given in \citeref{CMS:2013xfa} (scenario 2 therein). These are listed against the so-called 
$\kappa$'s (or `coupling modifiers')  of \citeref{LHCHiggsCrossSectionWorkingGroup:2012nn}, among which those interesting us primarily are $\kappa_{VV}^h$ ($V=W^\pm,Z$), $\kappa_{\gamma\gamma}^h$ and $\kappa_{gg}^h$, generally the most constraining ones.

\begin{figure}[t]
\centering
\begin{minipage}[t][\minipageheightdp][t]{\minipagewidthdp}
\centering
\includegraphics[width=\figuremaxwidthdp,height=\figuremaxheightdp,keepaspectratio]{\main/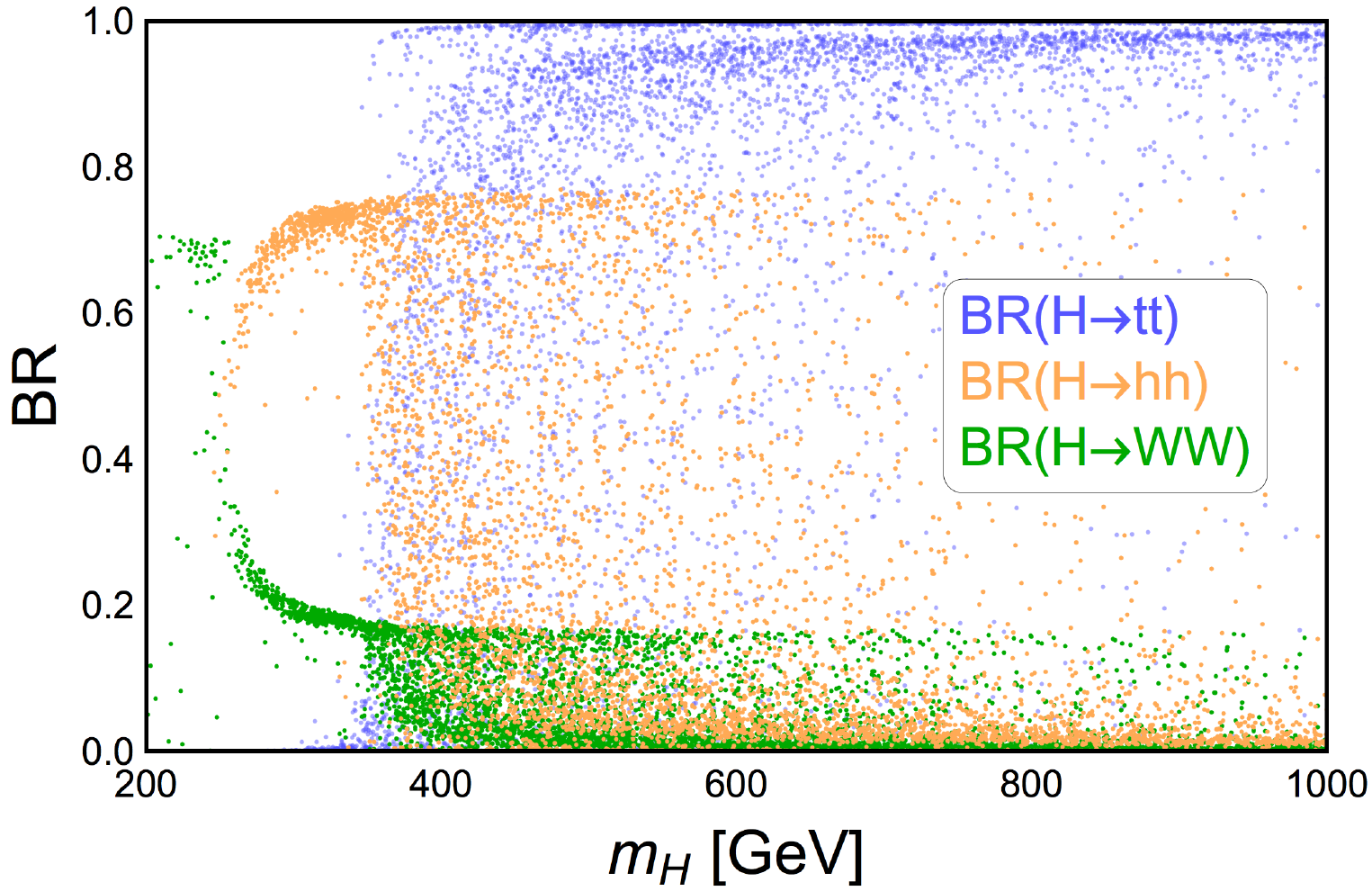}
\end{minipage}
\begin{minipage}[t][\minipageheightdp][t]{\minipagewidthdp}
\centering
\includegraphics[width=\figuremaxwidthdp,height=\figuremaxheightdp,keepaspectratio]{\main/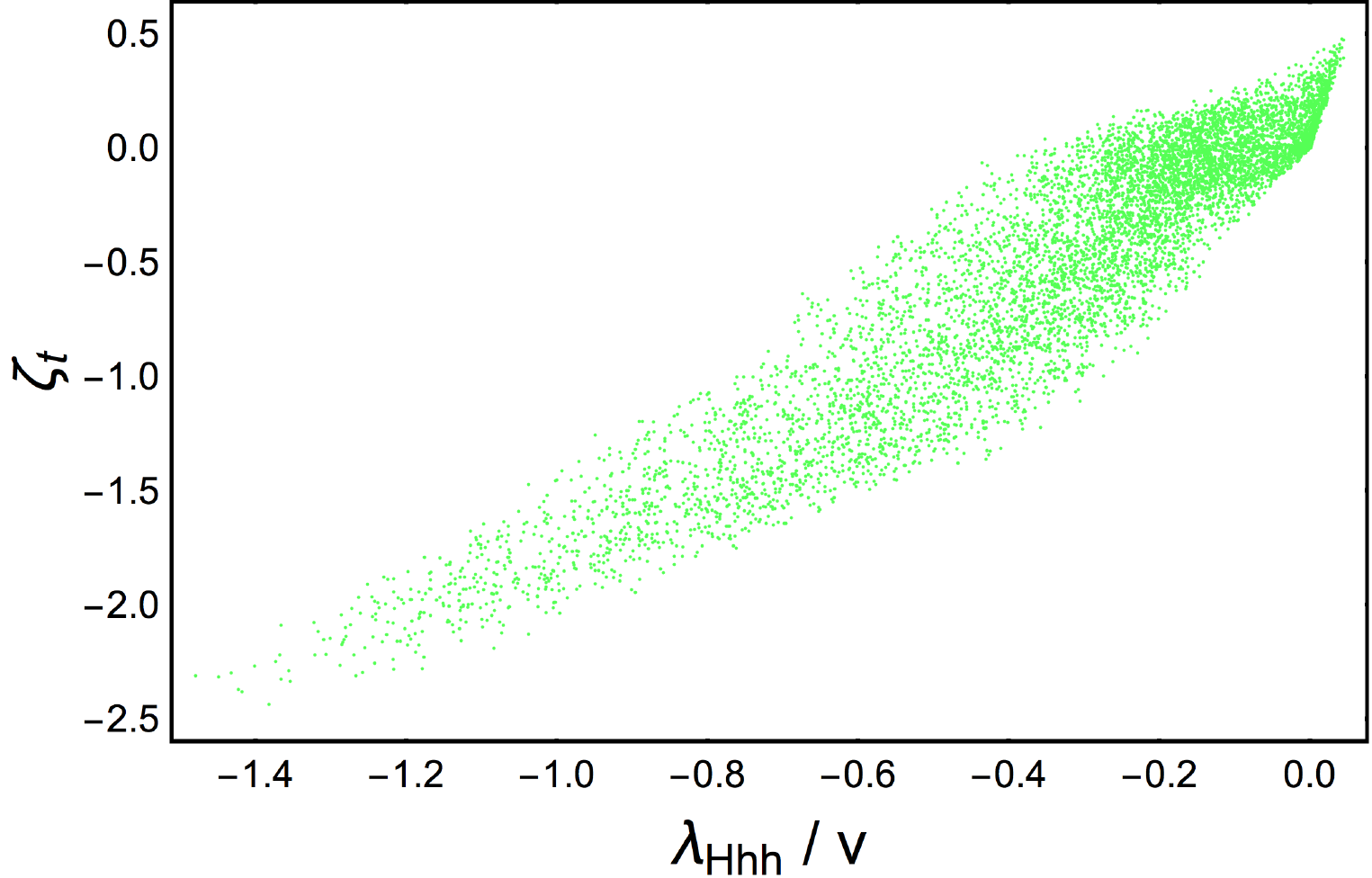}
\end{minipage}
\caption{Branching ratio of the $H$ state of the C2HDM in the $hh$ (orange) and $t\bar t$ (blue) channels (left) and correlation between the $\zeta_t$ and $\lambda_{Hhh}$ couplings obtained upon imposing present HiggsBounds and HiggsSignals constraints at $13\TeV$ (right). 
\label{fig:BRs}}
\end{figure}

Before proceeding with presenting our results, it is worth mentioning that a generic 2HDM Lagrangian 
introduces, in general, Flavour Changing Neutral Currents (FCNCs) at tree level via Higgs boson exchanges.
To avoid them, we assume here an alignment (in flavour space) between the Yukawa matrices like in the elementary Aligned 2HDM (A2HDM)~\cite{Pich:2009sp}. 
In this scenario, the coupling of the heavy Higgs $H$ to the SM top quark is controlled (modulo small corrections induced by the mixing angle $\theta \sim  \xi$) by
\begin{align}\label{eq:A2HDM}
\zeta_t=\frac{{\bar\zeta}_t-\tan\beta}{1+{\bar\zeta}_t\tan\beta},
\end{align}
where ${\bar\zeta}_t$ and $\tan\beta$ are predicted, and correlated to each other, in terms of the aforementioned fundamental parameters of the C2HDM.
Thus, being interested in the phenomenology of the $H$ state, henceforth we will map the results of our scan in terms of $m_H$ and $\zeta_t$ and restrict the parameter space to the region $m_{H,A,H^\pm} >2 m_h$.
The $\zeta_t$ parameter and the Higgs trilinear coupling $\lambda_{Hhh}$ set the hierarchy among the decay modes of the heavy state $H$. In particular, $H \rightarrow t \bar t$, when kinematically allowed, represents the main decay channel. Below the $t \bar t$ threshold, the diHiggs $H \rightarrow hh$ decay mode can reach, approximately, $80\%$, with the remaining decay space saturated by $H \rightarrow VV$. 
The corresponding BR observables are shown in \fig{fig:BRs} (left). Both of these can be notably different in the C2HDM with respect to the E2HDM, since the $Hhh$ and $Ht\bar t$ couplings can carry the imprint of compositeness (see \fig{fig:BRs} (right) for their correlation).
The hierarchy discussed above highlights the key role of the $H \rightarrow hh$ and $H \rightarrow t \bar t$ channels in the discovery and characterisation of the composite heavy Higgs boson.

\begin{figure}[t]
\centering
\begin{minipage}[t][\minipageheightdp][t]{\minipagewidthdp}
\centering
\includegraphics[width=\figuremaxwidthdp,height=\figuremaxheightdp,keepaspectratio]{\main/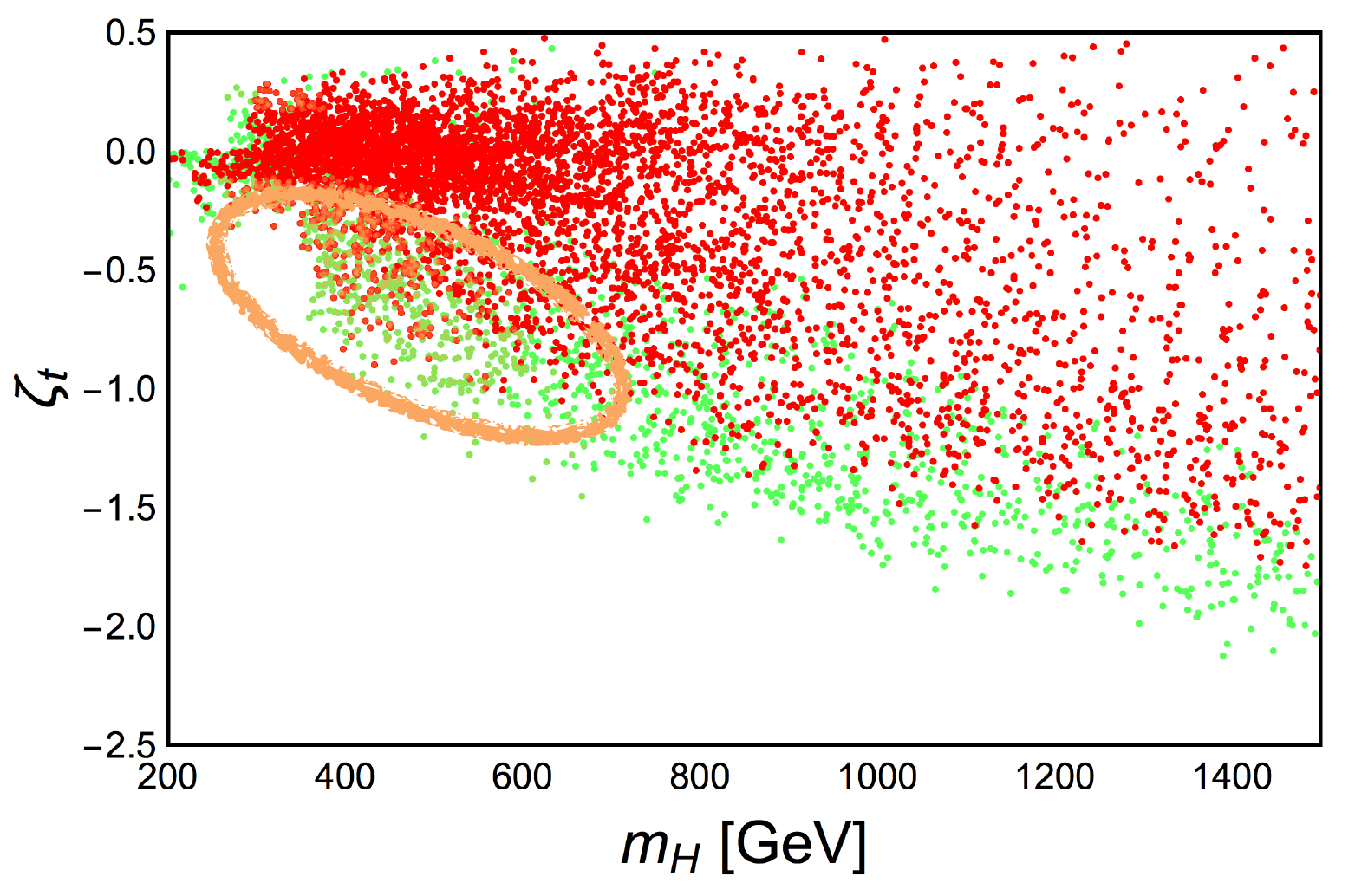}
\end{minipage}
\begin{minipage}[t][\minipageheightdp][t]{\minipagewidthdp}
\centering
\includegraphics[width=\figuremaxwidthdp,height=\figuremaxheightdp,keepaspectratio]{\main/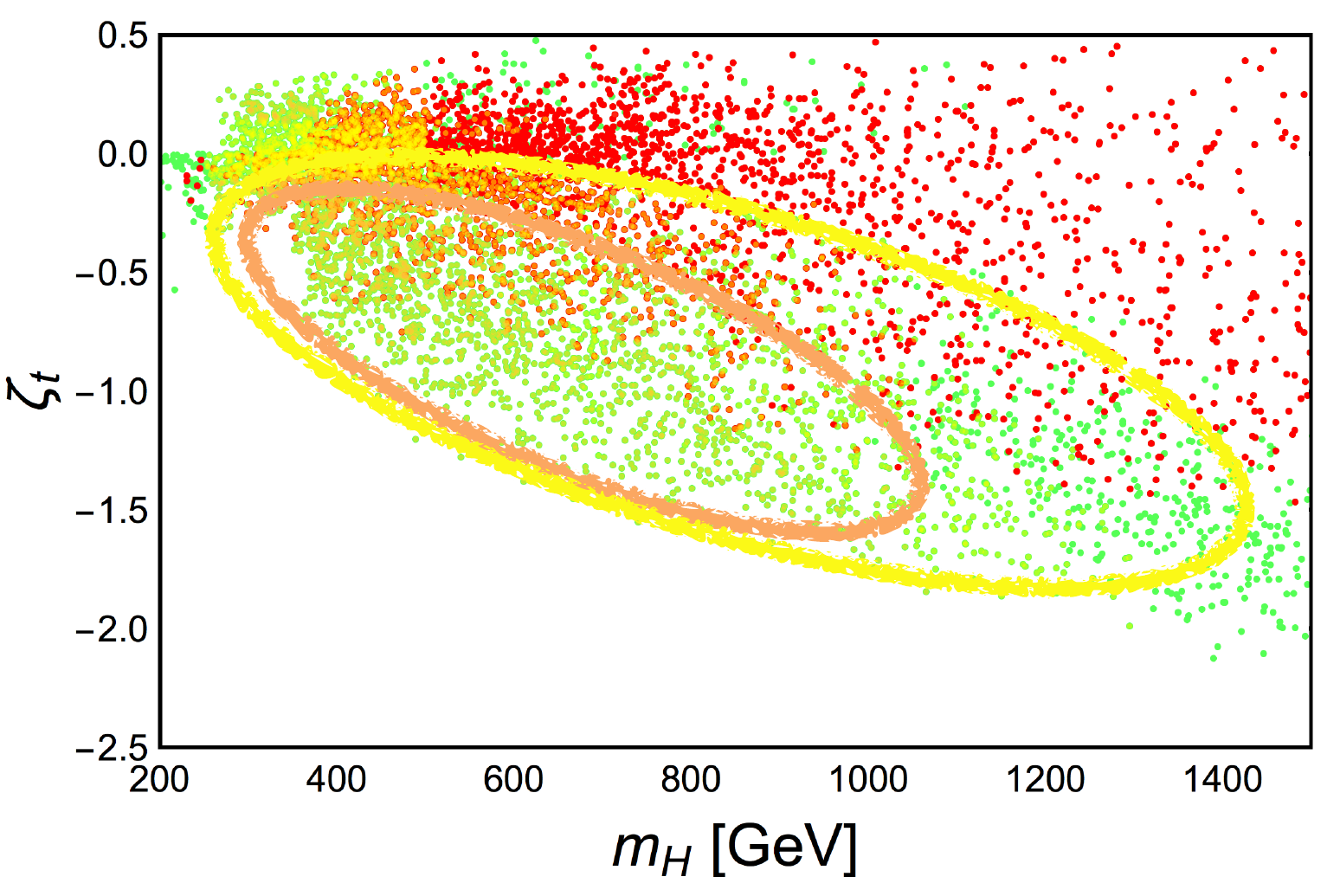}
\end{minipage}
\caption{Results of the C2HDM scan described in the text. Colour coding is as follows.
Green: all points that pass present constraints at $13\TeV$. 
Red: points that, in addition to the above, have $\kappa_{VV}^h$, $\kappa_{\gamma\gamma}^h$ and $\kappa_{gg}^h$ within the $95\%\cl$ projected uncertainty at $14\TeV$ with ${\cal L}=300\fbinv$ (left) and ${\cal L}=3\abinv$ (right).
Orange: points that, in addition to the above,  are $95\%\cl$ excluded  by the direct search $gg\to H\to hh\to b\bar b\gamma\gamma$, at $14\TeV$ with ${\cal L}=300\fbinv$ (left) and ${\cal L}=3\abinv$ (right). In the right plot the yellow points are $95\%\cl$ excluded by the same search  at the HE-LHC with ${\cal L}=15\abinv$. The orange and yellow elliptical shapes highlight the regions in which the points of the corresponding colour accumulate.
\label{fig:bbgaga}}
\end{figure}

\begin{figure}[t]
\centering
\begin{minipage}[t][\minipageheightdp][t]{\minipagewidthdp}
\centering
\includegraphics[width=\figuremaxwidthdp,height=\figuremaxheightdp,keepaspectratio]{\main/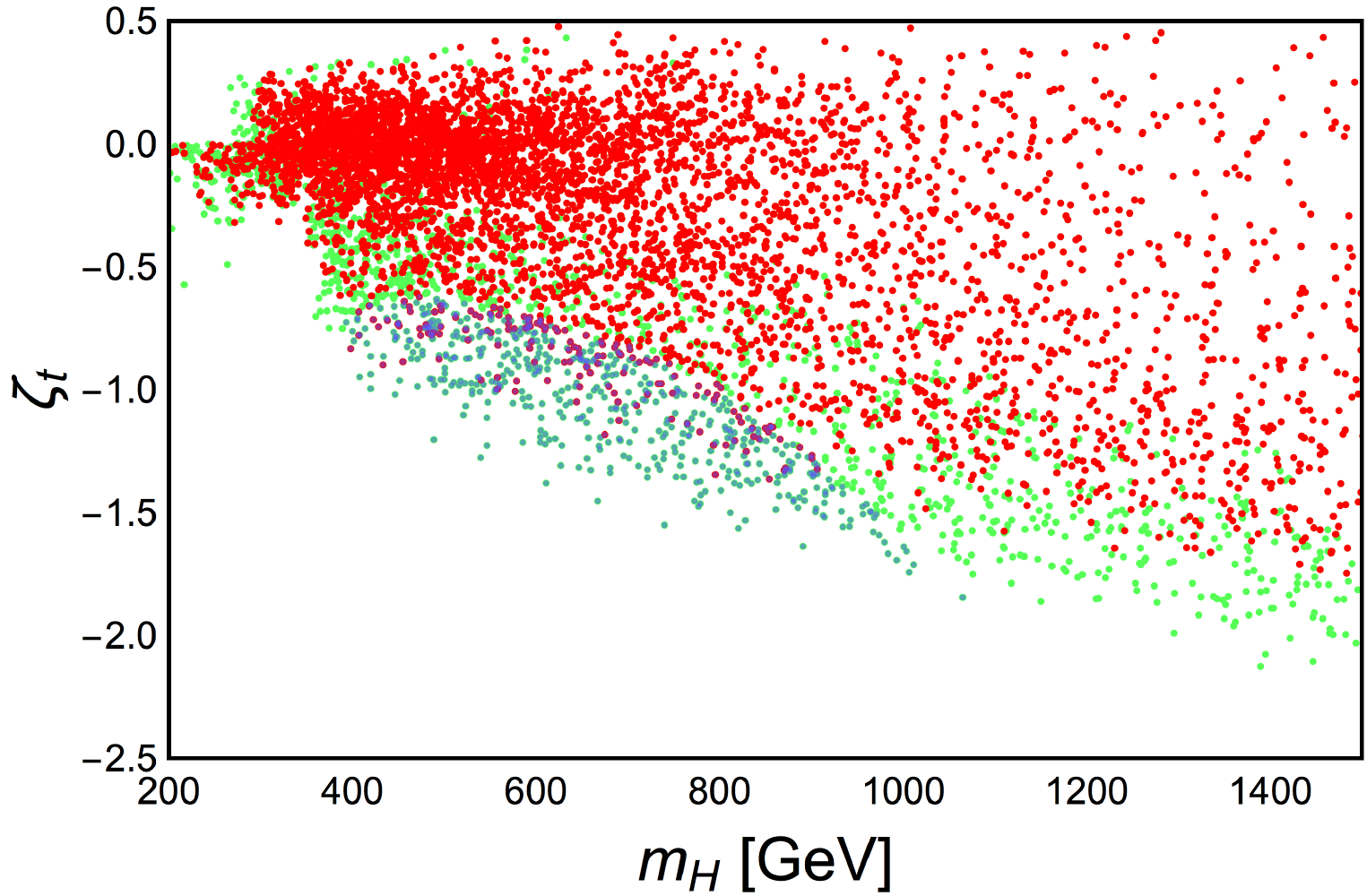}
\end{minipage}
\begin{minipage}[t][\minipageheightdp][t]{\minipagewidthdp}
\centering
\includegraphics[width=\figuremaxwidthdp,height=\figuremaxheightdp,keepaspectratio]{\main/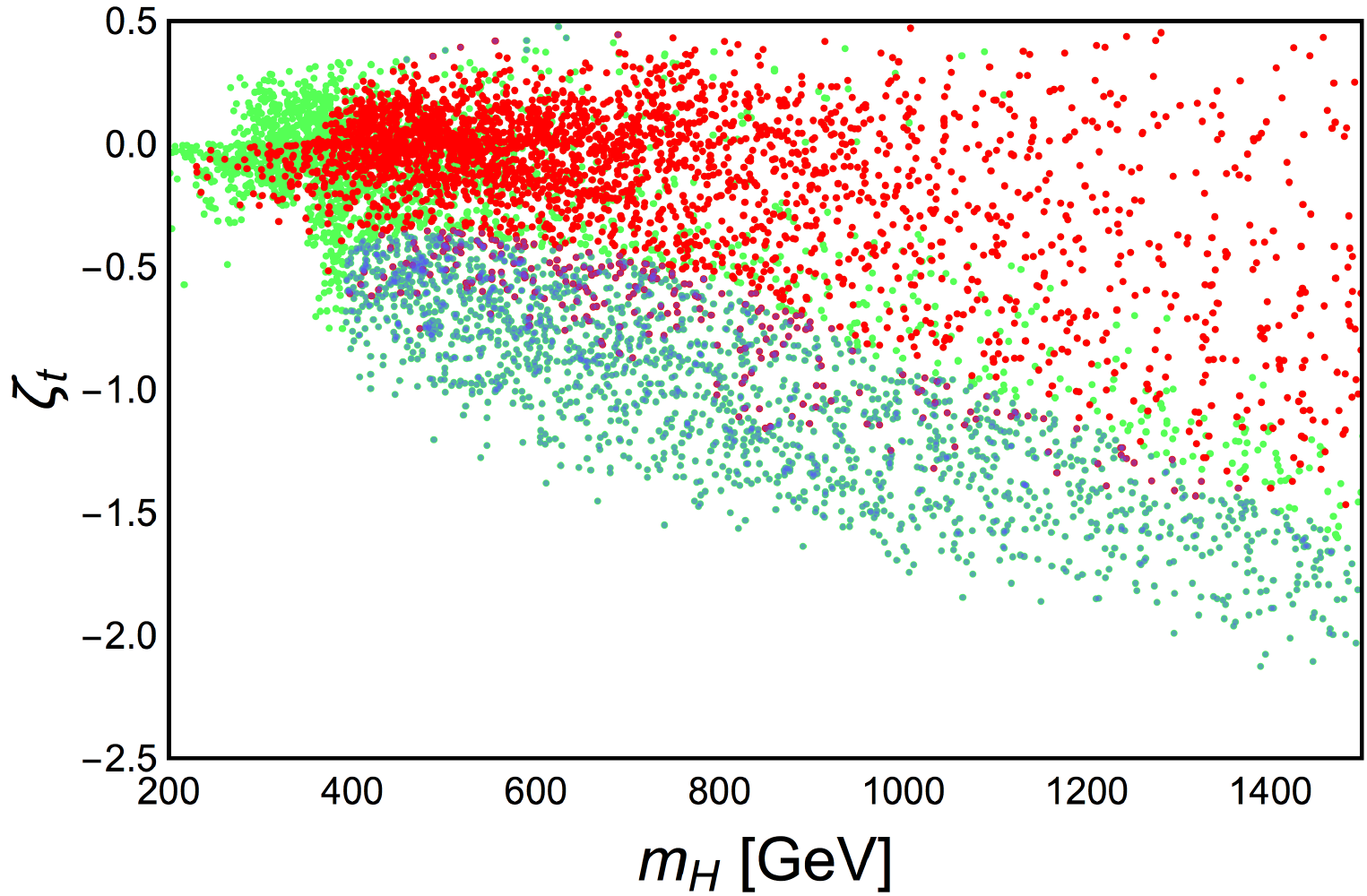}
\end{minipage}
\caption{Green an red points are as in the previous plot.
Blue: points that, in addition to the above,  are  $95\%\cl$ excluded by  the direct search $gg\to H\to t\bar t$, at $14\TeV$ with ${\cal L}=300\fbinv$ (left) and ${\cal L}=3\abinv$ (right).
\label{fig:tt}}
\end{figure}
Figures \ref{fig:bbgaga} and \ref{fig:tt} illustrate the interplay between direct and indirect searches and
the ability of the HL- and HE-LHC  to discover both the $gg\to H\to hh\to b\bar b\gamma\gamma$ and the $gg \to H\to t\bar t$ (followed by semi-leptonic top decays) signals, respectively, over regions of the C2HDM parameter space mapped onto the $(m_H,\zeta_t)$ plane, even when no deviations are visible in the aforementioned $\kappa$'s of the SM-like Higgs state $h$ (red points) at the LHC at $\sqrt{s}=14\TeV$ with ${\cal L}=300\fbinv$ and ${\cal L}=3\abinv$. 
Notice that $95\%\cl$ exclusion limits are extracted by  adopting the sensitivity projections of \citerefs{Aaboud:2018mjh,CMS:2017ihs} while compliance with the coupling modifiers is here achieved by asking that $|1 - k^h_i|$ is less than the percentage uncertainty declared in \citeref{CMS:2013xfa}, where $i=VV,\gamma\gamma$ and $gg$. Of some relevance, while tensioning the scope of the two search channels to one another, is to note that the orange  points have a large overlap with the red ones for small $| \zeta_t| $ values while 
the corresponding overlap of the blue region is smaller but it reaches larger $H$ masses. Hence, the first channel enables one to cover a larger C2HDM parameter space while the second one higher $H$ masses. Clearly, the combination of the two allows one to combine the benefits of either.  The  HE-LHC,  assuming $\sqrt s=27\TeV$ and $ {\cal L}=15\abinv$, will improve the reach in the $H$ high mass region up to $1.2\TeV$ by studying the process $gg\to H\to hh\to b\bar b\gamma\gamma$. For the $gg \to H\to t\bar t$  channel,  the naive extrapolation of the sensitivity with the parton luminosities is unreliable because it is affected by the SM $t \bar t$ threshold effects.

For a proper phenomenological analysis of the $t\bar t$ process, one should eventually account for interference effects with $gg$-induced (irreducible) background. For example, it is 
well-known that interference effects
between the $t,u$-channel $gg$-induced Leading Order (LO) QCD diagrams and the one due to a Higgs boson  
in $s$-channel via $gg$-fusion generate a characteristic peak-dip structure of the $M_{t\bar t}$ spectrum. However, this contribution is only one of the many entering at the same order in perturbation theory. In fact, \citeref{Moretti:2012mq} 
investigated the effect of all one-loop corrections of 
${\cal O}(\alpha_S^2\alpha_W)$ onto the $t\bar t$ invariant mass spectrum at the LHC
in presence of both resonant and non-resonant Higgs boson effects. It was shown therein that 
corrections of ${\cal O}(\alpha_S^2\alpha_W)$ involving a non-resonant Higgs boson
are comparable to or even larger than those involving interference with
the $s$-channel resonant Higgs boson amplitude
and that both of these are subleading with respect to all other (non-Higgs) diagrams through that
order.  Hence, a complete ${\cal O}(\alpha_S^2\alpha_W)$ calculation would be required to phenomenologically assess the relevance of such signal versus background effects in the C2HDM. 

In summary, both the HL- and HE-LHC display clear potential in accessing production and decay channels of the heavy \cpeven Higgs state of the C2HDM which can give direct access to key interactions that carry the hallmark of 
compositeness.

%\section*{Acknowledgements}
%\noindent  
%SM is funded in part through the NExT Institute and the STFC CG ST/P000711/1.

\subsubsection{\theoryC{Axion-like particles at the HL- and HE-LHC}}
\contributors{M.~Low, A.~Mariotti, D.~Redigolo, F.~Sala, K.~Tobioka}
%\textbf{Matthew Low$^a$, Alberto Mariotti$^b$, Diego Redigolo$^c$, Filippo Sala$^d$, Kohsaku Tobioka$^e$}~\\ ~\\
%\textit{\small $^a$ School of Natural Sciences, Institute for Advanced Study, Einstein Drive, Princeton, NJ 08540, USA,\\
%$^b$Theoretische Natuurkunde and IIHE/ELEM, Vrije Universiteit Brussel, and International,\\
%Solvay Institutes, Pleinlaan 2, B-1050 Brussels, Belgium\\
%$^c$School of Natural Sciences, Institute for Advanced Study, Einstein Drive, Princeton, NJ 08540, USA,\\
%Raymond and Beverly Sackler School of Physics and Astronomy, Tel-Aviv University, Tel-Aviv 69978, Israel,\\
%Department of Particle Physics and Astrophysics, Weizmann Institute of Science, Rehovot 7610001,Israel\\
%$^d$ DESY, Notkestra{\ss}e 85, D-22607 Hamburg, Germany\\
%$^e$C.N. Yang Institute for Theoretical Physics, Stony Brook University, Stony Brook, NY 11794-3800\\}

%\subsubsection{Setup and Motivation}
%\paragraph*{Setup and Motivation}

\begin{figure*}[t]
\centering
\begin{minipage}[t][\minipageheightsp][t]{\minipagewidthsp}
\centering
\includegraphics[width=\figuremaxwidthsp,height=\figuremaxheightsp,keepaspectratio]{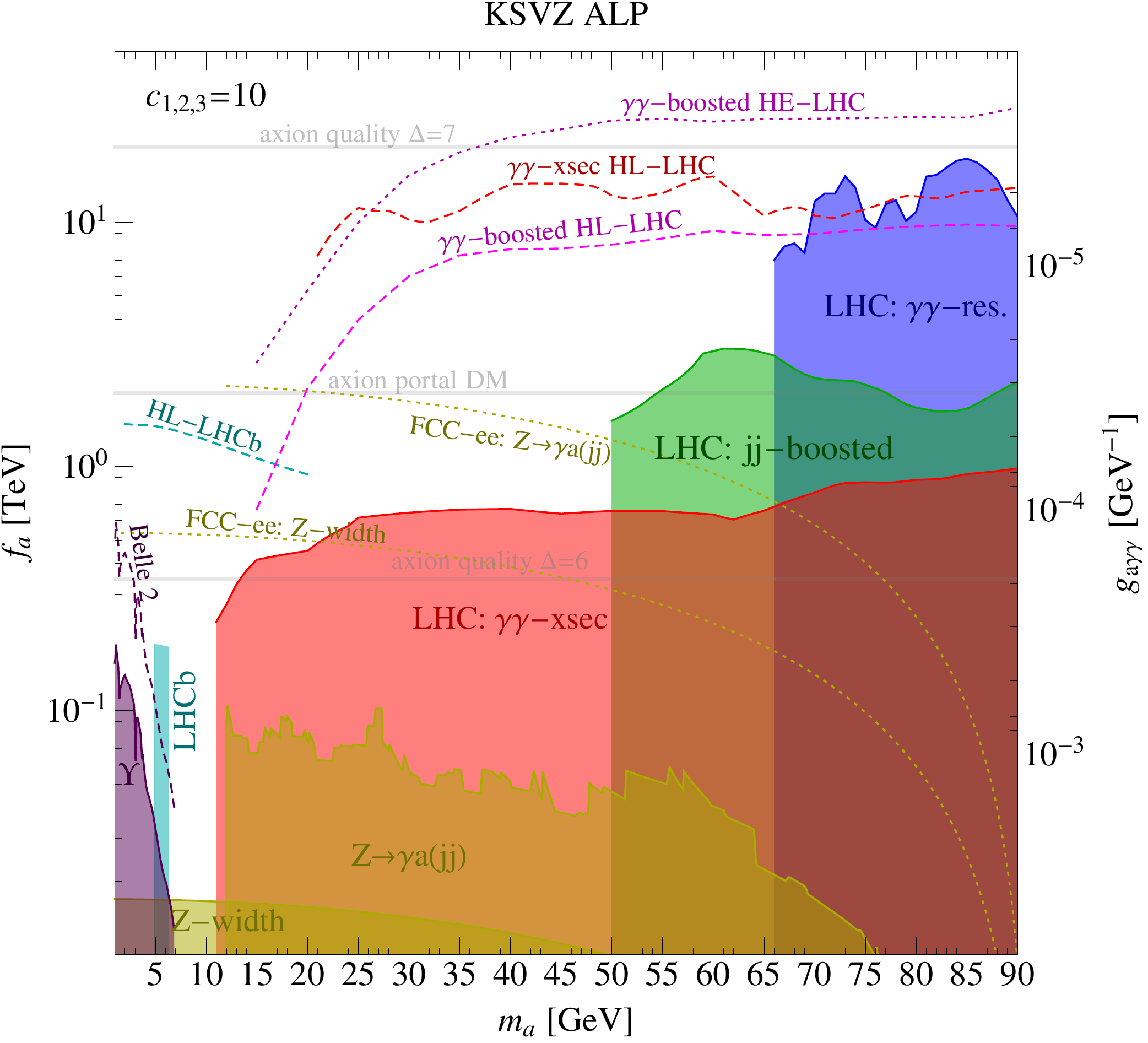}
\end{minipage}
\caption{
Status of the current best experimental constraints on the KSVZ-ALP with ABJ anomalies $c_{1,2,3}=10$. We include the BABAR bound on $\Upsilon\to \gamma a(jj)$ \cite{Lees:2011wb} {\bf(purple)} and its rescaling at Belle II \cite{deSangro:2017daz} {\bf(purple dotted)}. LHCb bound derived in \citeref{CidVidal:2018blh} from diphoton measurement around the $B_s$ mass \cite{Benson:2314368} {\bf(cyan)}. The projection for HL is also shown {\bf(dashed cyan)}. We also include LEP searches on $Z\to \gamma a(jj)$ \cite{Adriani:1992zm} and constraints from the $Z$-width \cite{Patrignani:2016xqp} {\bf(yellow)} along with both of their rescalings at FCC-ee \cite{Koratzinos:2015ywz} {\bf(yellow dotted)}. 
Constraints from inclusive cross section measurements at the Tevatron \cite{Aaltonen:2012jd} and the LHC \cite{Aad:2012tba,Chatrchyan:2014fsa,Aaboud:2017vol} derived in \citeref{Mariotti:2017vtv} {\bf(red)} and the rescaled sensitivities of the $8\TeV$ cross section measurement \cite{Aaboud:2017vol} at the \sloppy\mbox{HL-LHC} {\bf(dashed red)} are shown.  Finally, LHC bounds on boosted dijet resonances \cite{Sirunyan:2017nvi} reinterpreted for an ALP in \citeref{Mariotti:2017vtv} {\bf(green)}, LHC searches for diphoton resonances \cite{Aad:2014ioa,CMS:2015ocq,CMS:2017yta} {\bf(blue)}, and the sensitivity of the boosted diphoton resonance search based on the monojet trigger at the HL-LHC ($3\abinv$) and the HE-LHC ($15\abinv$) \cite{toappearRedigolo} {\bf(dashed/dotted magenta)} are plotted.
We also display {\bf(gray)} two theory benchmarks motivated by freeze-out of ALP-mediated Dark Matter and by the QCD axion quality problem, see \citeref{CidVidal:2018blh}.
On the r.h.s. $y$-axis we show $ g_{a\gamma\gamma}\equiv\frac{\alpha_{\text{em}}}{\pi f_a}c_\gamma$ to make contact with the QCD axion notation.} 
\label{fig:status}
\end{figure*}

The main focus of this contribution is future search strategies for axion-like particles (ALPs) in the mass range between $1$ and $90\GeV$~\cite{Mariotti:2017vtv,CidVidal:2018blh,toappearRedigolo}.  For simplicity we consider an ALP, $a$, that couples only to gauge bosons, including a non-zero coupling to gluons.
We call this type of ALP a ``KSVZ-ALP'' because it is inspired by the simplest QCD axion model~\cite{Peccei:1977hh,Kim:1979if,Shifman:1979if}.
Such an ALP is perhaps the most theoretically compelling case and also the natural target for hadron colliders, such as the HL-LHC and the HE-LHC.  The effective Lagrangian for the KSVZ-ALP, below the $Z$ mass, is 
\begin{align}
\mathcal{L}_{\text{int}}&= \frac{a}{4\pi f_a}\left[ \alpha_s c_3 G^a\tilde{G}^a+ \alpha_2 c_2 W^i\tilde{W}^i+ \alpha_1 c_1 B\tilde{B}\right],
\label{eq:LaFFdual}\\
&= \frac{a}{4\pi f_a}\left[ \alpha_s c_3 G^a\tilde{G}^a\!+ \alpha_{\text{em}} c_{\gamma} F\tilde{F}\!+\!2\alpha_2 c_W W^{-}\tilde{W}^{+}\!+\!\frac{2\alpha_{\text{em}}}{t_w} c_{Z\gamma} Z\tilde{F}\!+\alpha_2 c_w^2 c_{Z} Z\tilde{Z}\right]\,,\label{eq:La123}
\end{align}
where $\tilde{F}^{\mu\nu}=(1/2)\,\epsilon^{\mu\nu\rho\sigma}F_{\rho\sigma}$ for any field strength, $\alpha_1=\, \alpha'$ is the GUT-normalised $U(1)_Y$ coupling constant, and $t_w=s_w/c_w$ where $c_w^2=1-s_w^2=m_W^2/m_Z^2$.  The coefficients $c_i$ encode the Adler-Bell-Jackiw (ABJ) anomalies of the global $U(1)$ symmetry (of which the ALP is the pseudo-Goldstone boson) with $SU(3)$ and $SU(2)\times U(1)_Y$.
After EWSB, one can write 
\begin{equation}
c_{\gamma}=c_2+\frac{5}{3}c_1,\quad c_W=c_2, \quad c_Z=c_2+t_w^4\frac{5}{3} c_1, \quad c_{Z\gamma}=c_2-t_w^2\frac{5}{3} c_1  .
\end{equation}
For $m_a \lesssim m_Z$, the relevant two-body decays of $a$ are to two photons and to two jets, with widths
\begin{equation} \label{eq:widths}
\Gamma_{gg}=\frac{K_g\alpha_s^2c_3^2}{8 \pi^3}  \frac{m_a^3}{f_a^2},
\qquad \Gamma_{\gamma\gamma}=\frac{\alpha_{\text{em}}^2c_\gamma^2}{64 \pi^3}  \frac{m_a^3}{f_a^2},
\end{equation}
where
%both $\alpha_s$ and $\alpha_{\text{em}}$ are computed at the mass of $m_a$ and we include a k-factor
$K_g$ depends on the ALP mass and includes higher-order QCD corrections (see Appendix A in \citeref{CidVidal:2018blh}).
The couplings in \eq{eq:LaFFdual} can be generated by heavy vector-like fermions with a mass at $g_* f_a$, where $g_*$ can be as large as $4\pi$.
Explicit realisations include: i) KSVZ ``heavy axion'' models where the axion potential is UV-dominated, the axion mass is heavier than expected from QCD contributions alone, and the decay constant, $f_a$, can be as low as a TeV solving the axion quality problem (see \citerefs{Berezhiani:2000gh,Hook:2014cda,Fukuda:2015ana,Dimopoulos:2016lvn,Agrawal:2017ksf} for heavy axion models and  \citeref{Kamionkowski:1992mf} for a discussion of the axion quality problem); 
ii) ALPs arising in standard paradigms addressing the EW hierarchy problem, such as supersymmetry, where spontaneous SUSY-breaking below $M_{\text{Pl}}$ predicts, on general grounds, the existence of an $R$-axion~\cite{Nelson:1993nf,Bellazzini:2017neg}. iii) Axion portal Dark Matter scenarios where the DM is freezing out through its annihilation into gluons pairs~\cite{Mardon:2009gw,Fan:2010is}. In \fig{fig:status} we show the expected decay constant for an accidental Peccei-Quinn symmetry broken by dimension 6 operators at $M_{\text{GUT}}=10^{15}\GeV$, and the one for an axion portal which account for all of the DM relic abundance, see \citeref{CidVidal:2018blh} for more details. 

%\subsubsection{Present bounds}
%\paragraph*{Present bounds}

Barring a huge hierarchy among the anomaly coefficients, $c_{1,2} \gtrsim 10^2 c_3$, the width into gluons dominates over the one into photons, $\Gamma_{\gamma}/\Gamma_{gg}=\alpha_{\text{em}}^2/(8K_g\alpha_3^2) \sim 10^{-4}$.
ALPs that couple to gluons decay promptly in any kinematical configuration and mass range of interest for the LHC.
When accounting for the gluon coupling, searches based on rare decays of SM particles (Higgs, $Z$, $\Upsilon$, and $B$-mesons) into ALPs have a weaker bound on $f_a$ because the dominant BR of the ALP is now into hadronic final states,
while the searches look for final states with muons or photons. The only two relevant decay processes for our parameter space are $Z\to\gamma a$ and $\Upsilon_X\to \gamma a$, and the corresponding decay widths are given by
\begin{align}
&\Gamma(Z\to\gamma a)= \frac{\alpha_{\text{em}}^2c_{Z\gamma}^2}{96\pi^2t_w^2}\frac{m_Z^3}{f_a^2}\left(1-\frac{m_a^2}{m_Z^2}\right)^3 \ , \label{eq:Z}\\
&\Gamma(\Upsilon_X\to \gamma a)= \frac{\alpha_{\text{em}}c_{\gamma\gamma}^2}{\pi}\left(\frac{m_{\Upsilon_X}}{4\pi f_a}\right)^2\left(1-\frac{m_a^2}{m_{\Upsilon_X}^2}\right)^3\Gamma(\Upsilon_X\to ll)\ ,\label{eq:upsilon}
\end{align} 
where $\Gamma(\Upsilon_X\to ll)=\Gamma_{\Upsilon_X}\cdot \BR(\Upsilon_X\to ll)$, $\BR(\Upsilon_{2S,3S}\to ll)\simeq3.84\,,4.36\%$ and $\Gamma_{\Upsilon_{2S,3S}}\simeq20,32\KeV$.
These constraints, together with those from the LHC, are shown in \fig{fig:status}.

The bound from $Z$ decays, in \eq{eq:Z}, is based on the LEP search in \citeref{Adriani:1992zm} which has sensitivity down to $12\GeV$.  In this range, this search is more powerful than the inclusive bound from the total $Z$-width~\cite{Patrignani:2016xqp}.  The bound from $\Upsilon$ decays, in \eq{eq:upsilon}, is based on BABAR data~\cite{Lees:2011wb} corresponding to $1.21\cdot 10^8$ $\Upsilon_{3S}$ and $0.98\cdot 10^8$  $\Upsilon_{2S}$. 
``Standard'' inclusive diphoton resonance searches at the LHC do not probe masses below $65\GeV$~\cite{Aad:2014ioa,CMS:2015ocq,CMS:2017yta}.

A first example of what can be done to improve the low mass reach is the CMS search for a dijet resonance recoiling against a hard jet \cite{Sirunyan:2017nvi} that we rescale here for an ALP produced in gluon fusion (see \citeref{Mariotti:2017vtv} for more details). As we see in \fig{fig:status}, this probes ALPs down to $50\GeV$.
A second example is the bound from inclusive cross section measurements, derived in \citeref{Mariotti:2017vtv}, that reach masses of $10\GeV$.
References \cite{Aaltonen:2012jd,Aad:2012tba,Aaboud:2017vol,Chatrchyan:2014fsa} provide tables of the measured differential diphoton cross section per invariant mass bin, $\text{d}\sigma_{\gamma\gamma}/\text{d} m_{\gamma\gamma}$, with the relative statistical ($\Delta_{\rm stat}$) and systematical ($\Delta_{\rm sys}$) uncertainties.
A conservative bound was derived in \citeref{Mariotti:2017vtv} assuming zero knowledge of the background and requiring 
\begin{equation}
\sigma^{\text{th}}_{\gamma\gamma}(m_a)<\left[m_{\gamma\gamma}^{\rm Bin}\cdot \frac{\text{d}\sigma_{\gamma\gamma}}{\text{d} m_{\gamma\gamma}}\cdot(1+ 2\Delta_{\rm tot})\right] \cdot \frac{1}{\epsilon_S (m_a)}.
\label{eq:sensitivity}
\end{equation}
where $\Delta_{\text{tot}}=\sqrt{\Delta_{\rm stat}^2+\Delta_{\rm sys}^2}$.
The signal efficiency $\epsilon_S (m_a)$ (see \citeref{Mariotti:2017vtv} for its computation) does not go to zero below  $p_{T_{\gamma_1}}^{\text{min}}+p_{T_{\gamma_2}}^{\text{min}}$  because the ALP can still pass the cuts recoiling against unvetoed jet activity in the diphoton cross section measurements.
A lower limit on the invariant mass that can be measured, and thus on the testable $m_a$, is set~by 
\begin{equation}
m_{\gamma\gamma}> \Delta R_{\gamma\gamma}^{\text{iso}}\sqrt{p_{T_{\gamma_1}}^{\text{min}}p_{T_{\gamma_2}}^{\text{min}}}
\end{equation} 
where $p_{T_{\gamma_{1,2}}}^{\text{min}}$ are the minimal cuts on the photon transverse momenta, and $\Delta R_{\gamma\gamma}^{\text{iso}}=0.4$ is the standard isolation requirement between the photons. A small mass range around the $B_s$ mass is constrained by diphoton measurements at LHCb \cite{Benson:2314368} as first derived in \citeref{CidVidal:2018blh}. Projections for future stages of LHCb have also been derived in \citeref{CidVidal:2018blh}. 

%\subsubsection{Prospects for the HL-LHC and the HE-LHC}
%\paragraph*{Prospects for the HL-LHC and the HE-LHC}

Here we discuss new LHC search strategies 
for the HL-LHC and the HE-LHC
to improve the coverage in the $f_a$-$m_a$ plane of KSVZ-like ALPs.
The reaches of future experiments are summarised as dashed magenta lines named in \fig{fig:status},
where we also display, for comparison, LHCb, the Belle II and FCC-ee projections.\footnote{The Belle II and FCC-ee lines are obtained by rescaling of the present bounds from BABAR and LEP with the future luminosities of these experiments: 100 times more $\Upsilon_{2S}$s and $\Upsilon_{3S}$s are assumed for Belle II and $10^{12}$ $Z$s for FCC-ee.}

We show the reach of the HL-LHC from 
inclusive cross section measurements 
with the strategy outlined in the previous section, assuming the same diphoton trigger as in the $8\TeV$ run. Our signal cross section includes matching up to 2 jets and a K-factor accounting for NLO corrections computed with ggHiggs v3.5 \cite{Ball:2013bra,Bonvini:2014jma,Bonvini:2016frm,Ahmed:2016otz} which includes full NNLO and approximate $\text{N}^3\text{LO}$ corrections plus threshold resummation at $\text{N}^3\text{LL}'$.  The error on the sensitivity is assumed to be dominated by the error on the measurements.
In other words, we assume the MC uncertainties will be reduced below $\Delta_{\rm tot}$ (note that, at the present moment, MC uncertainties on the low-mass bins are at the level of 40\% using \sloppy\mbox{\sherpa~\cite{Aaboud:2017vol, Gleisberg:2008ta,Catani:2018krb}}). An unexplored direction in diphoton resonance searches is to reduce the photon isolation requirements and pass the trigger making the resonance recoiling against a hard jet.  In \fig{fig:status} we show the sensitivity at the HL-LHC and the HE-LHC of a search based on this idea. 
The signal events are required to pass the existing monojet trigger (the leading jet with $p_{T_{j_1}}>500\GeV$ and $|\eta_{j_1}|<2.5$). As a consequence the two photons produced from the boosted ALP are collimated: $\Delta R_{\gamma\gamma}\simeq 2 m_a /p_{T_{j_1}} \lesssim 0.2\left(\frac{m_a}{{\rm 50\GeV}}\right)$ where $p_{T_{\gamma}}\sim p_{T_{j_1}}/2$. To improve $S/\sqrt{B}$ we then require $p_{T_{\gamma_{1,2}}}>120\GeV$ and $\Delta R_{\gamma\gamma}<0.8$ and bin the events in invariant mass bins with a constant width of $10\GeV$. Since two photons fall into one isolation cone, $\Delta R_{\gamma\gamma}^{\text{iso}}=0.4$, standard photon isolation vetos the ALP signal. In order to access lower invariant masses, we modify the standard isolation requirement by simply subtracting the hardest photon ($\gamma_1$) in the isolation cone of every test photon,  ($\gamma_{\rm test}\neq \gamma_1$). We require\footnote{$E_{T}^{\rm iso}$ is usually calculated using calorimeter cells, while here we also include information of charged tracks. Objects with $E_{T_i}<0.5\GeV$are not included in the sum \cite{deFavereau:2013fsa}.} 
\begin{equation}
E_{T}^{\rm iso}-E_{T_{\gamma_1}}<{\rm 10\GeV}
\quad {\rm  where}\quad
E_{T}^{\rm iso}\equiv \sum_{i\neq \gamma_{\rm test}}^{\Delta R_{i,\gamma_{\rm test}}<0.4} E_{T_i} \ . \label{eq:newiso}
\end{equation}
To validate our analysis we first checked that the standard isolation with $E_{T}^{\rm iso}<10\GeV$ in \delphes{} can reproduce the photon fake rate from multijets and the real photon acceptance from the SM Higgs decay, given by \citeref{ATL-PHYS-PUB-2016-026}. We find that the isolation in \eq{eq:newiso} gives a similar fake rate and real photon acceptance to the standard ones while the photons from boosted ALP decay now pass the isolation. The modified isolation requirement allows us to go down to two-photon angular separation of $\Delta R\sim0.1$.\footnote{As minimal angular resolution we take for reference the square towers in the Layer 2 region of ATLAS ECAL~\cite{Aaboud:2016yuq}.} Below this value the two photon showers will start to overlap and the photon identification will have to be modified \cite{Ellis:2012sd,Ellis:2012zp}. A similar strategy was employed by ATLAS in $Z\to \gamma a(\gamma\gamma)$ search (see \citeref{Aad:2015bua}). 

To estimate the sensitivities in \fig{fig:status} we simulated the SM background from $2\gamma+n j$, matched for $n=1,2$, and from $j\gamma+j$ with a jet faking a photon both at $14$ and $27\TeV$. We expect the background from  $jj+j$ with two jets faking two photons to be subdominant. Finally, notice that the drop of sensitivity in the low mass range of our analysis is expected to ameliorate including the region of phase space with maximally asymmetric photon momenta $p_{T_{\gamma_{2}}}\ll p_{T_{\gamma_{1}}}$. Indeed $m_{\gamma\gamma}^{\text{min}}=\sqrt{(p_{T_{j_1}}-p_{T_{\gamma_{2}}})p_{T_{\gamma_{2}}}}\Delta R^{\text{iso}}$ after momentum conservation is imposed, so that minimising the momentum of the second photon becomes beneficial at low invariant masses (see \citeref{toappearRedigolo}). 

%\subsubsection{Conclusions}
%\paragraph*{Conclusions}
In conclusion, we showed how the sensitivity to KSVZ ALPs can be greatly improved at the HL-LHC and the HE-LHC. Both experiments have sensitivity that exceeds the one of FCC-ee. The type of searches discussed here will probe the parameter space of ``heavy axion'' models solving the strong CP-problem of the SM and probe SUSY scales, $g_{\ast} f_a$, as high as $100\TeV$ \emph{independently} of any particular assumption on the structure of the SUSY spectrum, it will also probe well motivated scenarios of heavy Dark Matter freeze-in.  

Our results have a broader application than the ones discussed here.  For example, they can still be the strongest probe in ALP scenarios where the ALP also couples to fermions and the SM Higgs, such as in composite Higgs models~\cite{Ferretti:2013kya,Cacciapaglia:2017iws} where the couplings to photons and gluons are, in fact, generated by top loops. Last but not least, the invariant mass sensitivity can likely be further extended to lower masses by studying to what extent the two collimated photons themselves can trip the monophoton trigger \cite{toappearRedigolo}, developing techniques to perform bump hunts with L1 triggers~\cite{Collaboration:2283192} and exploring the advantages of converted photons events in the inner tracker. 

\vspace{-0.3cm}

\subsubsection{\theoryC{Search for light pseudoscalar with taus at HL-LHC}}
\contributors{G. Cacciapaglia, G. Ferretti, T. Flacke, H. Serodio}

%\textbf{Authors: Giacomo Cacciapaglia$^a$, Gabriele Ferretti$^b$, Thomas Flacke$^c$ and Hugo Serodio$^d$}\\
%\textit{a) Universit\'{e} de Lyon, France; Universit\'{e} Lyon 1, Villeurbanne Cedex, France,
%b) Department of Physics, Chalmers University of Technology, Fysikg\aa rden, 41296 G\"oteborg, Sweden,
%c) Center for Theoretical Physics of the Universe, Institute for Basic Science (IBS), Daejeon, 34126, Korea,
%d) Department of Astronomy and Theoretical Physics, Lund University, SE-223 62 Lund, Sweden. }\\
%\texttt{g.cacciapaglia@ipnl.in2p3.fr, ferretti@chalmers.se, \\flacke@ibs.re.kr, hugo.serodio@thep.lu.se}

The discovered Higgs boson may be accompanied by additional light (pseudo-)scalars in models with an extended Higgs sector. We consider a pseudoscalar singlets $\phi$, generically coupling as:
\begin{equation}
 {\mathcal{L}} \supset - \sum_\psi \frac{i C^\phi_\psi m_\psi \phi}{f_\phi} \bar \psi \gamma^5 \psi
+ \frac{\phi}{16\pi^2 f_\phi}\bigg(K^\phi_{G}\, G\tilde G + K^\phi_W\,  W \tilde{W} +K^\phi_B\, B \tilde B\bigg)\,, \label{eq:lagrTCP}
\end{equation}
where $f_\phi$ is a ``decay constant'' and $\psi$ labels the SM fermions.
We focus on models of composite Higgs with partial compositeness described in terms of confining hyper-quarks~\cite{Ferretti:2013kya}. A universal feature of all these models is the presence of two such pseudoscalars ($\phi= a, \,\eta'$), whose couplings  in \eq{eq:lagrTCP} can be computed from the underlying model.  Additional couplings, including the higher dimensional couplings to the Higgs boson, not included in \eq{eq:lagrTCP}, are also determined by the underlying theory (see \citeref{Belyaev:2016ftv} for details). In \citeref{Belyaev:2016ftv} we narrowed down this class of models to a total of twelve, providing possible benchmarks in the search of new physics. 

\begin{figure*}[t] 
\centering
\begin{minipage}[t][\minipageheightsp][t]{\minipagewidthsp}
\centering
\includegraphics[height=\figuremaxheightsp,keepaspectratio]{\main/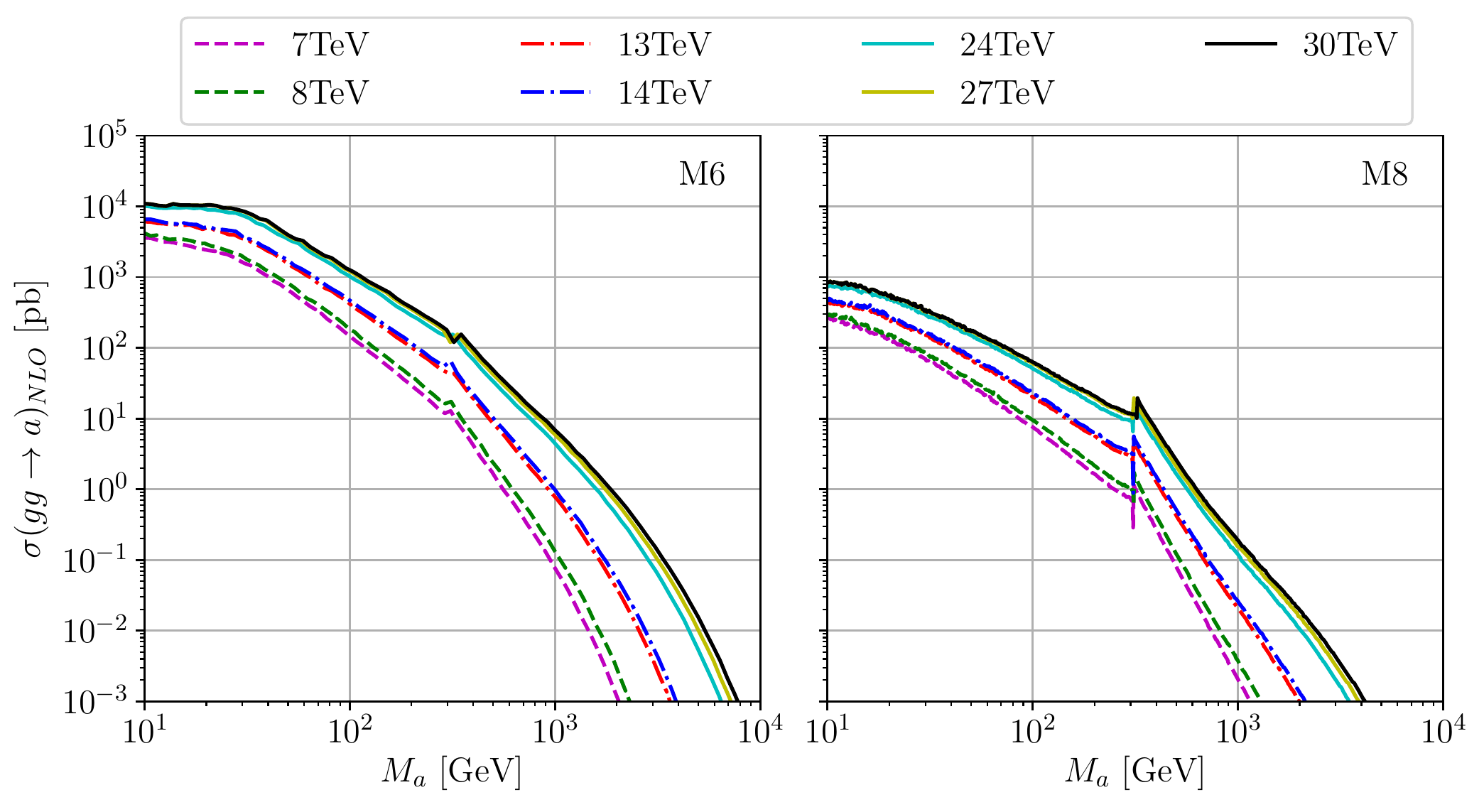}
\end{minipage}
\caption{\label{fig:Fig1TPNGB} Production cross section (dominated by gluon fusion) for the lightest pseudoscalar for the two benchmark models M6 (left) and M8 (right).}
\end{figure*}

We present results for two of the twelve models~\cite{Ferretti:2014qta,Barnard:2013zea}, denoted respectively M6 and M8 in \citeref{Belyaev:2016ftv}, which are those being studied on the lattice~\cite{Bennett:2017kga,Ayyar:2018zuk}. 
The masses of the pseudoscalars can be considered as free parameters, while their decay constants $f_\phi$ in \eq{eq:lagrTCP} are related to the composite Higgs decay constant $f$, defined by $m_W = (g/2) f \sin \theta$, with $\theta \to \pi/2$ being the Technicolour limit~\cite{Belyaev:2016ftv,Cacciapaglia:2017iws}.
For small underlying hyper-quark masses, the lighter pseudoscalar $a$ is nearly aligned with a spontaneously broken $U(1)$ symmetry and thus can be very light. Its total production cross-section is shown in \fig{fig:Fig1TPNGB} for fixed $f=1\TeV$. The second pseudoscalar $\eta'$ is related to an anomalous $U(1)$ (hence the name) and thus receives a larger mass from the strong dynamics.

\begin{figure*}[t] 
\centering
\begin{minipage}[t][\minipageheightsp][t]{\minipagewidthsp}
\centering
\includegraphics[height=\figuremaxheightsp,keepaspectratio]{\main/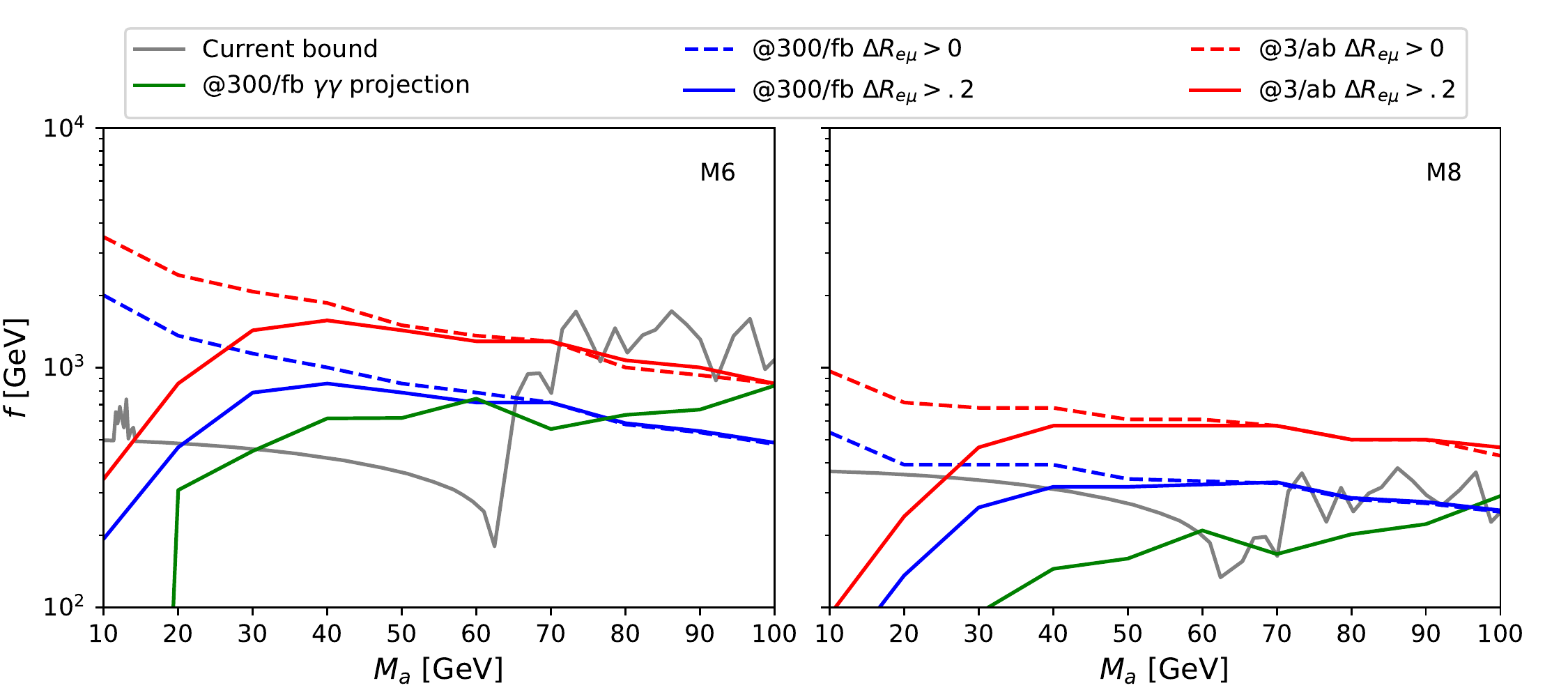}
\end{minipage}
\caption{\label{fig:Fig2TPNGB} Reach of the ditau search for the two
models M6 (left) and M8 (right), compared to the existing bounds (gray
lines). The existing bounds indicate the strongest exclusion amongst those
arising from dimuon searches~\cite{Chatrchyan:2012am}, diphoton searches~\cite{Aad:2014ioa,CMS:2017yta}, and BSM decay width
of the Higgs~\cite{Khachatryan:2016vau}. (The bounds from \citeref{Aaij:2017rft}, that can be obtained
adapting the analysis of \citeref{Haisch:2018kqx} to these models,
may also turn out to be competitive.) We have also indicated the current bounds obtained by adapting the results in \citeref{Mariotti:2017vtv} for diphotons (green).
The projected reach is computed at $14\TeV$ using the HL-LHC detector simulation, for a luminosity of $300\fbinv$(blue) and $3\abinv$ (red), and two distinct cuts on $\Delta R_{\mu e}$.}

\end{figure*}

We observe that the production cross-section of the pseudoscalars is rather large, in contrast to that of other light scalars arising in this class of models that only couple via EW interactions. Nevertheless, as their main decay channels suffer from large backgrounds, they are still fairly unconstrained in the low mass region, particularly between $14$ and $65\GeV$. 
In \citeref{Cacciapaglia:2017iws} we proposed a boosted search for the lighter pseudoscalar $a$ in the (fully leptonic, opposite flavour) ditau channel between $10$ and $100\GeV$.  Figure~\ref{fig:Fig2TPNGB} shows the reach in the $M_a/f$ plane for the two models above. A complementary proposed search in the diphoton channel has been discussed in the previous section, based on \citeref{Mariotti:2017vtv}.

A crucial discriminating variable in such search, particularly for the low mass region, is the angular separation $\Delta R_{e\mu}$ between the electron and the muon. We present the reach estimated from a cut-and-count simulation with the conservative choice $\Delta R_{e\mu}>0.2$ included or removed. In the plot we have not taken into account the systematic error in the background, but it is important to remark that, in order to take full advantage of the HL-LHC run, it should be kept below $2\%$. The additional cuts, discussed in \citeref{Cacciapaglia:2017iws}, are: $p_{T\mu}>50\GeV$, $p_{Te}>10\GeV$, $\Delta R_{\mu j}>0.5$, $\Delta R_{e j}>0.5$, $p_{Tj}>200\GeV$,  $\Delta R_{\mu e}<1$,  $m_{\mu e}<100\GeV$. Note that we also impose an upper bound on $\Delta R_{\mu e}$ to reduce the (mostly flat) $t \bar t$ background.

\begin{figure*}[t] 
\centering
\begin{minipage}[t][\minipageheightsp][t]{\minipagewidthsp}
\centering
\includegraphics[height=\figuremaxheightsp,keepaspectratio]{\main/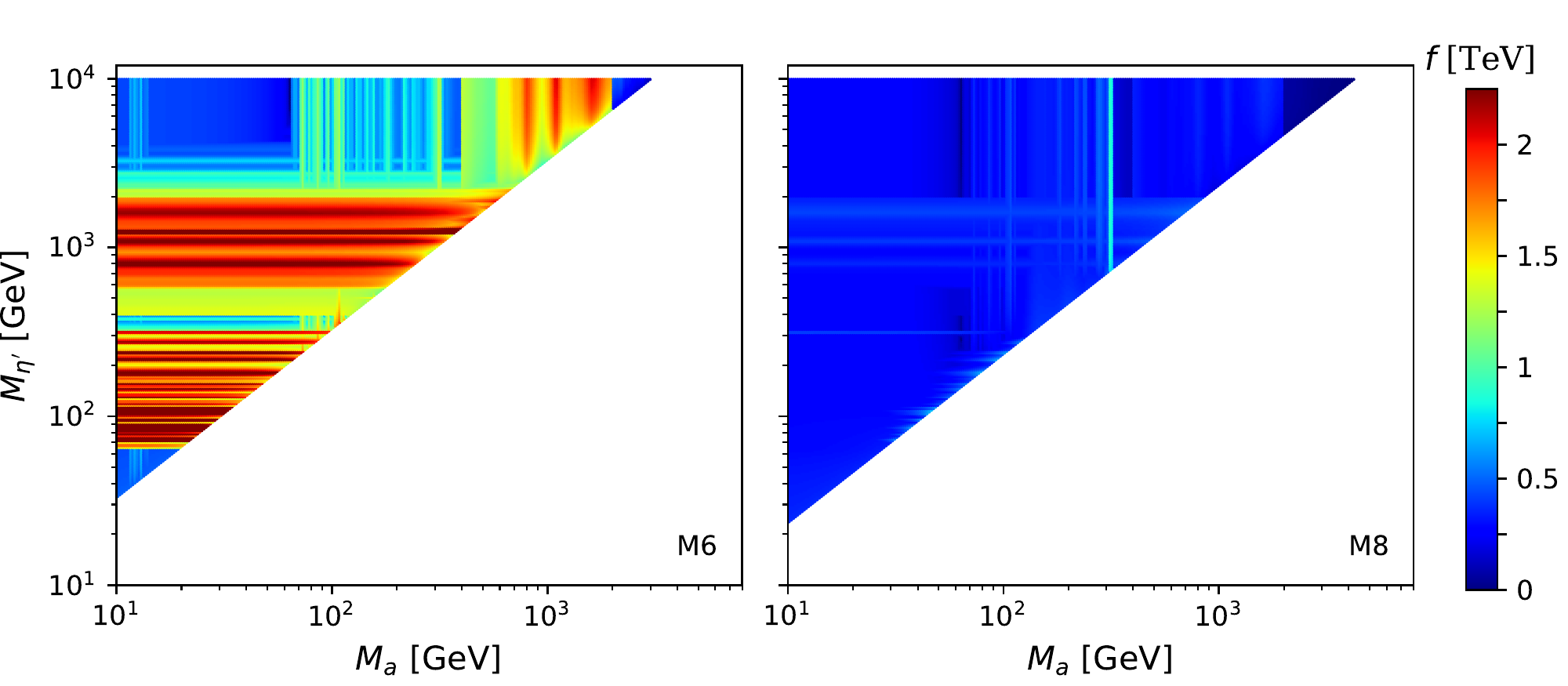}
\end{minipage}
\caption{\label{fig:Fig3TPNGB} Lower bound the Higgs decay constant $f$ for the two benchmark models M6 (left) and M8 (right) in the presence of both pseudoscalars $a$ and $\eta'$.}
\end{figure*}

The heavier pseudoscalar $\eta'$, not being a true Goldstone boson, could have a mass in the multi TeV range. Nevertheless, it may give observable signals at the LHC because it decays into final states such as $\gamma \gamma$, $Z\gamma$, $ZZ$, $WW$, $t\bar{t}$ and $Z h$ (the last one via top loops). In \fig{fig:Fig3TPNGB} we present the lower bounds on $f$ for the two models, in the $(M_a,\, M_{\eta'})$ mass plane. The white region corresponds to masses incompatible with the models~\cite{Belyaev:2016ftv}. 
The vertical band with a strong bound for $M_a \sim 215\GeV$corresponds to $Zh$ searches, which were not included in \citeref{Belyaev:2016ftv}. 

Figure~\ref{fig:Fig3TPNGB} takes into account the relevant searches performed with the 2016 data of about $36\fbinv$ of integrated luminosity at $13\TeV$.
Similar analyses can of course be performed for the remaining models, but the two presented in this work are amongst the least constrained by current data.

In conclusion, the HL and HE phases of the LHC present us with newer possibilities to search for BSM physics. The models discussed here provide concrete examples where new physics could arise both in the high and low mass regime, benefiting from both improvements.

%\textbf{Acknowledgments:} We wish to thank M. Selvaggi for help with the HL detector simulation in Delphes. Support: \textbf{G.C.:} Institut Franco-Suedois (project T{\"o}r) and  
%the Labex-LIO (Lyon Institute of Origins) under grant ANR-10-LABX-66, FRAMA (FR3127,
%F\'ed\'eration de Recherche ``Andr\'e Marie Amp\`ere''), \textbf{G.F.:} The Knut and Alice Wallenberg Foundation and the Lars Hierta Memorial Foundation \textbf{T.F.:} IBS under the project code, IBS-R018-D1, \textbf{H.S.:} ERC under the European Union's Horizon 2020 research and innovation programme (grant agreement No 668679).
\subsubsection{\theoryC{Colour octet scalar into gluons and photons at HL-LHC}}
\contributors{G. Cacciapaglia, A. Deandrea, A.M. Iyer}
%{\bf Author(s): G.Cacciapaglia$^{a,1}$,A.Deandrea$^{a,2}$, A.M. Iyer$^{b,3}$}\\
%
%{\it \small $^a$ Universit\'e de Lyon, France; Universit\'e Lyon 1, CNRS/IN2P3, UMR5822 IPNL, F-69622 Villeurbanne Cedex, France}\\ 
%{\it \small $^b$ INFN-Sezione di Napoli, Via Cintia, 80126 Napoli, Italia}

%\subsubsection{Model and effective Lagrangian description}
%\paragraph*{Model and effective Lagrangian description}
We consider a colour octet scalar $\Phi$ which is present in various extensions of the SM, and in particular in composite models for the EW sector, where such a state can be a composite object made of fundamental fermions. The colour octet $\Phi$ can be produced by single and pair production at the LHC by QCD processes.
Due to its nature and quantum numbers, in composite models, it can couple strongly to top quarks and give rise, at one loop in the 
fundamental theory, to a coupling to gluons and photons via a top loop and the topological anomaly. In particular it also gives rise to an effective vertex with a 
photon and a gluon which is highly suppressed in SM processes, giving rise to a distinctive mode for its search at the LHC. 
We therefore consider the decay mode $\Phi\rightarrow \gamma g$, thereby making it an exclusive signature for such a state. 
The effective Lagrangian for this interaction can be written as:
\begin{eqnarray}
\mathcal{L}=a_1~f_{abc}G^a_{\mu\nu}G^{b\mu\nu}\Phi^c+a_2~f_{abc}f_{ade}~\Phi^b~\Phi^d~G^c_{\mu\nu}~
G^e_{\mu\nu}+c~G^a_{\mu\nu}\Phi^aB^{\mu\nu}\,,
\end{eqnarray}
where $a_i$ are proportional to the strong coupling constant $\alpha_s$  while the ratio $c/a$ depends on the model 
under consideration. The colour octet can decay into $t\bar{t}$, $gg$, $g\gamma$, or $gZ$. The $gZ$ final state is subdominant in comparison to $g\gamma$ as it is suppressed by the Weinberg angle ($\Phi g\gamma$ vertex is derived from the $\Phi G^a_{\mu\nu}B^{\mu\nu}$ term in Lagrangian.). The corresponding comparison of rates  of $gg$  with respect to the $t\bar t$ final state is parameter dependent and is not considered here. It can however   be taken into account by the corresponding reduction of the production rate $\sigma_{eff} = \sigma \times (1-\mathcal{B}(t\bar t))^2$.
In this note we are interested in  exploring the possibility of the $g\gamma$ decay mode as a possible discovery prospect for the colour octet scalar.
For scenarios considered in \citeref{Belyaev:2016ftv}, where the colour octet arises as a bound state of colour-triplet fermions $\chi$ with hyper-charge $1/3$ or $2/3$, the branching fractions amongst the bosonic final states are fixed and given in  \tabb{tab:octetBR}.

\begin{table}[htb!]
	\begin{center}
		\small\begin{tabular}{|c|c|c|}
			\hline
			&$Y_\chi=  1/3$  & $Y_\chi= 2/3 $   \\
			\hline\hline
			$ \frac{ \BR(\Phi\to g\gamma)}{\BR(\Phi\to gg)}  $	    &  $0.048$   & $0.19$  \\
			
			\hline
		\end{tabular}\normalsize
	\end{center}
	\caption{Values of ratios of BRs in di-bosons for the pseudoscalar octet for a mass of $1\TeV$. The mass
		fixes the dependence due to the running of the strong gauge coupling,
		$\alpha_s(1\TeV) = 0.0881$ is used for this evaluation.}
	\label{tab:octetBR}
\end{table}

%\subsubsection{Pair Production}
%\paragraph*{Pair Production} 
We consider the following benchmarks for the mass of the coloured scalar $\Phi$: \sloppy\mbox{$m_\Phi=1,1.5,2,2.5\TeV$}. There are existing searches for the pair production of the scalar in the multi jet channel. For the branching fractions outlined in \tabb{tab:octetBR} we compare the efficiency for the multi jet final state corresponding to the existing searches with the following two
final states corresponding to the signal:
one single photon and three gluons; and two photons and two gluons.
The parton level events are simulated at $14\TeV$ \com~energy using \madgraph~\cite{Alwall:2014hca} and showering is done by \pythia{ 8}~\cite{Sjostrand:2007gs}. We use \delphes\,3~\cite{deFavereau:2013fsa} for the detector simulation.
The jets are reconstructed using \fastjet~\cite{Alwall:2014hca} with the standard AK4 jet reconstruction algorithm with $R=0.4$ and $p_T=60\GeV$.
The strategy for the multi-jet final state in this contribution is similar to the analysis in CMS searching in the multi-jet channel for pair produced scalars \cite{Sirunyan:2018rlj}:
\begin{itemize}
	\item The jets are reconstructed using the anti-kt algorithm with $R=0.4$ and $p_T=60\GeV$. Minimum of 4 jets are required for each event.
	\item Each jet is required to have a $p_T$ of $80\GeV$.
	\item In order to select the two best di-jet systems compatible with the signal, the four leading jets
	ordered in $p_T$ are combined to create three unique combinations of di-jet pairs per event. Out of
	the three combinations, the di-jet configuration with the smallest $\Delta R_{dijet}=\sum_{i=1,2}|\Delta R_i-0.8|$ is chosen where $\Delta R^i$ is the distance in the $\eta-\phi$ plane between the two jets in the $i^{th}$ di-jet pair.
	\item Asymmetry parameter: Once a configurations is selected, two asymmetry parameters are constructed:
	\begin{eqnarray}
	M_{asymm}=\frac{|m_{jj1} - m_{jj2} |}{m_{jj1} + m_{jj2}}\;\;\;,\;\;\;\Delta\eta_{asymm}=|\eta_{jj1}-\eta_{jj2}|.
	\end{eqnarray}
	where $m_{jjk}$ and $\eta_{jjk}$ is the dijet mass and pseudorapidity combination of the $k^{th}$ di-jet pair. Both these quantities are set $<0.1$.
	
\end{itemize}

For the 3 jets and 1 photon final state there is no existing search undertaken thus far. Thus we adapt a similar criteria described above.
\tabb{tab:eff619} gives the signal and the corresponding background efficiencies ($\epsilon $) for the two topologies and are simply $\epsilon =N_{cut}/N_{gen}$, where $N_{cut}$ is the number of events which pass the cut and $N_{gen}$ are the number of events generated.
To facilitate the collider comparison between the $gg$ and the $g\gamma$ decay for the colour octet scalar we define the following ratio:
\begin{equation}
\delta=\frac{S_{ggg\gamma}/\sqrt{B_{jjj\gamma}}}{S_{gggg}/\sqrt{B_{jjjj}}},
\label{eq:ratio}
\end{equation}
which is computed for the two hypercharge assignments in  \tabb{tab:octetBR}. 
Here the number of signal or background events ($S$ or $B$) at a given luminosity $\mathcal{L}$,  is simply $\epsilon\times \sigma\times\mathcal{L}$. The ratio however, eliminates the dependence on the luminosity \footnote{Note that for the $jjj\gamma$ signal $\sigma = \sigma_{prod}2~ \BR(\Phi\to g\gamma)$ }. 
Using the numbers from \tabb{tab:eff619}, the results for the $jjj\gamma$  are given in the first two columns  of  \tabb{tab:sig1619}.
\begin{table}[t]
	\begin{center}
		\small\begin{tabular}{|c|c|c|c|} 
			\hline
			
			&4$j$  & 3$j$+1$\gamma$& 2$j$+ 2$\gamma$\\ \hline
			\hline
			
			 Signal efficiency $\epsilon_S$ ($m_\Phi=1\TeV$)        &0.0067&0.0016&0.0593\\ \hline   
			Background Efficiency $\epsilon_B$ &0.00125&0.00035&0.000114\\\hline
			Background cross section ($\mathrm{fb}$)&$6800\times 10^{3}$&$18\times 10^{3}$&$18\times 10^{3}$\\\hline
		\end{tabular} \normalsize
	\end{center}
	\caption{ Pair production efficiencies for signal and background. The background cross-section is estimated by requiring the scalar sum of the parton $p_T>1100\GeV$. The computed efficiencies are simply the ratio of events which pass the cuts to the total number of events generated. The background for $g\gamma g\gamma$ signal is still dominated by QCD $jjja$ background in comparison to the $\gamma\gamma$ + jets. The different efficiencies are a consequence of different selection criteria.}
	\label{tab:eff619}
\end{table}

The ratio in  \eq{eq:ratio} eliminates the dependence on the production cross section and facilitates a transparent comparison of different decay modes.  Given the large backgrounds in \tabb{tab:eff619}, the signal with the cuts used is not significant, thereby requiring a more detailed analysis. We now discuss the more optimistic 2 gluons and 2 photons final state.
\begin{table}[t]
	\begin{center}
		\small\begin{tabular}{|c|c|c||c|c|} 
			\hline
			&\multicolumn{2}{c||}{$S_{ggg\gamma}$}&\multicolumn{2}{c|}{$S_{gg\gamma\gamma}$}\\\hline
			& $Y_\chi=1/3$  & $Y_\chi=2/3$&$Y_\chi=1/3$ & $Y_\chi=2/3$\\ \hline
			\hline
			
			$\delta$ &0.84&3.33&1.36& 20.56 \\
			\hline 
		\end{tabular}\normalsize
	\end{center}
	\caption{Comparison of the signal significances ratio $\delta$ (defined in \eq{eq:ratio}) between the $ggg\gamma$ and $gggg$ channels for $m_{\Phi}=1\TeV$. Similar ratio is computed for the $gg\gamma\gamma$ final state.} 
	\label{tab:sig1619}
\end{table}

%\begin{table}[t]
%	\begin{center}
%		\small\begin{tabular}{|c|c|c|} 
%			\hline
%			& $Y_\chi=1/3$  & $Y_\chi=2/3$\\ \hline
%			\hline
%			
%			$\delta$ &1.38&20.78 \\
%			\hline 
%		\end{tabular}\normalsize
%	\end{center}
%	\caption{Comparison of signal significances ratio $\delta$ (defined in \eq{eq:ratio}) between the $gg\gamma\gamma$ and $gggg$ channels.} 
%	\label{tab:sig2619}
%\end{table}

%	\subsection{2 jets+2 photons}
In the case of 2 jets + 2 photons, there are only two combinatorial possibilities for the invariant mass reconstruction. 
Let the isolated photons be denoted as $\gamma_{1,2}$ and the jets as $j_{1,2}.$
It is to be noted that the gluon jet and the photon from a given coloured scalar are fairly collimated. Thus in order to identify the correct pair of final 
states we take the hardest photon $\gamma_1$ and compute, $\Delta R_{\gamma_1 j_1}$ and $\Delta R_{\gamma_1 j_2}$ and extract the 
following:
\begin{equation}
\Delta R_{min}=min(\Delta R_{\gamma_1 j_1},\Delta R_{\gamma_1 j_2}).
\label{eq:deltaR}
\end{equation}

\begin{figure}[t!]
	\centering
	\begin{minipage}[t][\minipageheightdp][t]{\minipagewidthdp}
		\centering
		\includegraphics[width=\figuremaxwidthdp,height=\figuremaxheightdp]{\main/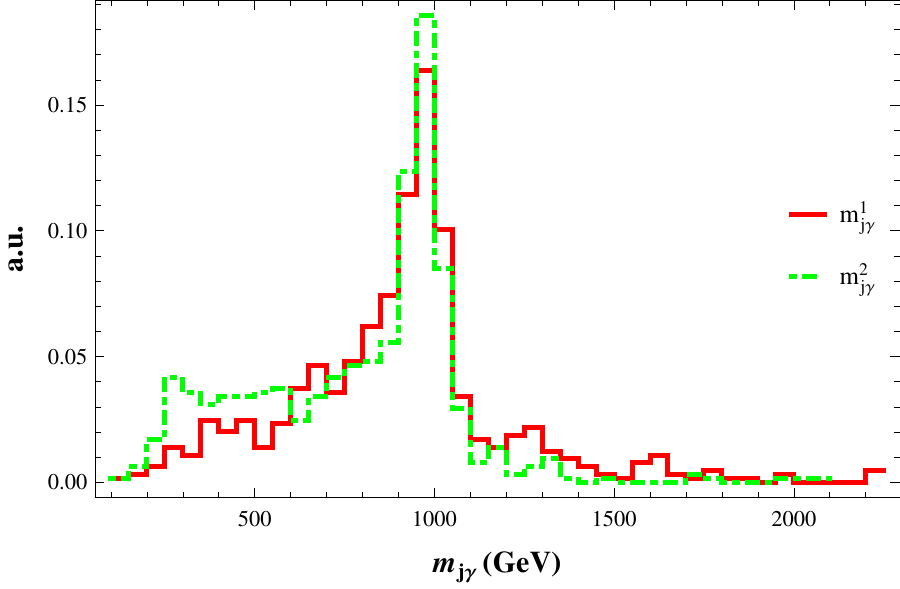} %,keepaspectratio
	\end{minipage}
	\begin{minipage}[t][\minipageheightdp][t]{\minipagewidthdp}
		\centering
		\includegraphics[width=\figuremaxwidthdp,height=\figuremaxheightdp]{\main/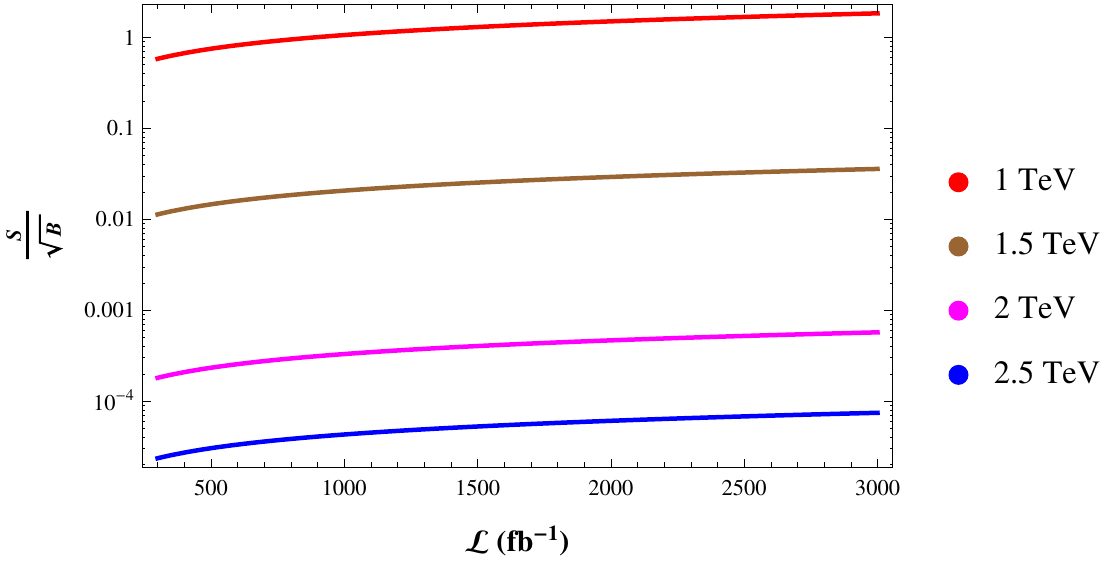} %,keepaspectratio
	\end{minipage}
	\caption{Left: distribution of the two reconstructed invariant masses $m_{j\gamma}$ for the $1\TeV$ benchmark using the criterion in \eq{eq:deltaR}. $a.u.$ on the Y-axis denotes arbitrary units. Right:  Sensitivity for the different signal benchmarks at $14\TeV$ COM~energy for luminosities up to $3\abinv$.}
	\protect\label{fig:pair}
\end{figure}

The presence of two photons  greatly limits the QCD fake rate. We further impose a hard cut of $180\GeV$ on the transverse momentum of each photon. The large signal sensitivity in this case is an artefact that a simple $p_T$ cut on the second photon reduces background drastically without affecting the signal significantly.
The last two columns of  \tabb{tab:sig1619} show the comparison of signal sensitivity between the $gg\gamma\gamma$ and $gggg$ channels, implying that they are both similar even for the pessimistic case of smaller branching fraction corresponding to $Y_\chi=1/3$. The left plot of \fig{fig:pair} shows the reconstructed invariant mass distribution and the right plot gives the signal significance as a function of luminosity for the optimistic case of $Y_\chi=2/3$. Using the computation for production cross-sections for $14\TeV$ in \cite{Gresham:2007ri}, we estimate the sensitivity for different signal benchmarks from $0.3$ to $3\abinv$. With a high luminosity run one can reach a sensitivity  close to $\sim 2.0~\sigma$ for $m_\Phi =1\TeV$. This, as well the sensitivity for the other benchmarks can be further improved  by using information of signal kinematics and invariant mass cuts.
%\begin{figure}[t!]
%	\centering
%	%	\begin{minipage}[t][\minipageheightdp][t]{\minipagewidthdp}
%	\centering
%	\includegraphics[width=5cm,height=5cm,keepaspectratio]{pair1a.pdf}
%	%	\end{minipage}
%	%	\begin{minipage}[t][\minipageheightdp][t]{\minipagewidthdp}
%	\centering
%	\includegraphics[width=6cm,height=6cm,keepaspectratio]{pair1ef.pdf}
%	%	\end{minipage}
%	\caption{Left: distribution of the two reconstructed invariant masses $m_{j\gamma}$ for the 1 TeV benchmark using the criterion in \eq{eq:deltaR}. $a.u.$ on the Y-axis denotes arbitrary units. Right:  Sensitivity for the different signal benchmarks at $14\TeV$ COM~energy for luminosities up to 3~\abinv. }
%	\protect\label{fig:pair}
%\end{figure}
%\subsubsection{Conclusion}
%\paragraph*{Conclusion}
We have performed a simplified preliminary study of the potential for the study of pair of a colour octet scalar decaying to gluon 
and photon at the LHC. This opens two additional  possibilities for their searches  corresponding to the $ggg\gamma$ and $gg\gamma\gamma$ final states. These final states not only serves as an unambiguous signature for colour octets but are also relatively cleaner  than the conventional multi jet searches.
We demonstrate that for the $gg\gamma\gamma$ channel, we can get a preliminary hint of the existence of the colour octet state at HL-LHC thereby strongly motivating the HE option of the LHC.	 
 %\bibliography{biblio}

\subsection{Spin 1 resonances}\label{sec:BSMspin1}
This section is devoted to the study of the prospects for vector   resonances. These are neutral~$\Zp$ and charged~$\Wp$, which are among the most standard benchmarks usually considered in studying the potential of future colliders, as well as RS gluons and other resonances arising from 2HDMs models. 

\subsubsection{\theory{Precision predictions for new dilepton and $t\bar t$ resonances at HL- and HE-LHC}}
\contributors{M. Altakach, J. Fiaschi, T.~Je\v{z}o, M. Klasen, I. Schienbein}
%{\bf Author(s): M. Altakach$^1$, J. Fiaschi$^2$, T.~Je\v{z}o$^3$, M. Klasen$^4$,
%I. Schienbein$^5$}
%
%\vspace*{1ex}
%
%\noindent
%$^1$ Laboratoire de Physique Subatomique et de Cosmologie, Universit\'e Grenoble Alpes, CNRS/IN2P3, 
%53 Avenue des Martyrs, F-38026 Grenoble, France
%\\
%altakach@lpsc.in2p3.fr
%
%\noindent
%$^2$ Institut f\"ur Theoretische Physik, Westf\"alische Wilhelms-Universit\"at M\"unster,
%Wilhelm-Klemm-Stra{\ss}e 9, D-48149 M\"unster, Germany
%\\
%fiaschi@uni-muenster.de
%
%\noindent
%$^3$ Physics Institute, Universit\"at Z\"urich, Z\"urich, Switzerland
%\\
%tomas.jezo@physik.uzh.ch
%
%\noindent
%$^4$ Institut f\"ur Theoretische Physik, Westf\"alische Wilhelms-Universit\"at M\"unster,
%Wilhelm-Klemm-Stra{\ss}e 9, D-48149 M\"unster, Germany
%\\
%michael.klasen@uni-muenster.de
%
%\noindent
%$^5$ Laboratoire de Physique Subatomique et de Cosmologie, Universit\'e Grenoble Alpes, CNRS/IN2P3, 
%53 Avenue des Martyrs, F-38026 Grenoble, France
%\\
%ingo.schienbein@lpsc.in2p3.fr

%\subsubsection{Introduction} 
%\paragraph*{Introduction} 

%\noindent 
We present higher order predictions for spin-1 resonance searches in two classes of observables, top-quark-pair production and dilepton production.
In the former case, we use the {\tt PBZp} code~\cite{Bonciani:2015hgv} which includes the NLO
QCD corrections to the EW production of top-antitop pairs in the presence of a new neutral gauge boson 
implemented in the parton shower Monte Carlo program \powheg{}~\cite{Frixione:2002ik,Frixione:2007vw,Alioli:2010xd}.
The dilepton cross sections are calculated using the NLO+NLL  code \RESUMMINO~\cite{Jezo:2014wra} which matches
a soft-gluon resummation at NLL accuracy to a fixed order NLO calculation.

We consider four models: the Un-Unified (UU)~\cite{Georgi:1989ic,Georgi:1989xz} and the Non-Universal (NU)~\cite{Malkawi:1996fs,Li:1981nk}
models, a leptophobic topcolour model (TC) (model IV in \citeref{Harris:1999ya}), and the SSM~\cite{Altarelli:1989ff}.
The UU and NU models belong to the general class of G(221)$=SU(2)_1 \times SU(2)_2 \times U(1)_X$ gauge theories 
with an extra $SU(2)$ gauge symmetry.
In the UU model the quarks and leptons  belong to different representations of the two $SU(2)$ gauge factors whereas in the NU model
the first two generations transform differently than the third generation. Both models take two input parameters, the mixing angle 
of the first stage symmetry breaking  $t=\tan \phi = g_2/g_1$ and the mass of the heavy resonance $M_{Z'}$ (or $M_{W'}$).
Exclusion limits on the parameters space for the G(221) models have been derived in \citeref{Hsieh:2010zr} by performing 
a global analysis of low-energy precision data. Improved limits for the $W'$ and $Z'$ masses were found in \citeref{Jezo:2014wra} using LHC 
data at $\sqrt{s}=7$ and $8\TeV$.
The TC model has three free parameters in addition to the resonance mass $M_{Z'}$: 
the width $\Gamma_{Z'}$, the relative strength ($f_1$) of the $Z'$ coupling 
to right-handed up-type quarks w.r.t. left-hand up-type quarks, and similarly the relative strength ($f_2$) of the $Z'$ coupling to right- and left-hand down-type quarks. 
Finally, in the SSM the only free parameters are the masses $M_{W'}$ and $M_{Z'}$.

\begin{figure}[t]
\centering
\begin{minipage}[t][\minipageheightsp][t]{\minipagewidthsp}
\centering
\includegraphics[width=\figuremaxwidthsp*11/6,keepaspectratio]{\main/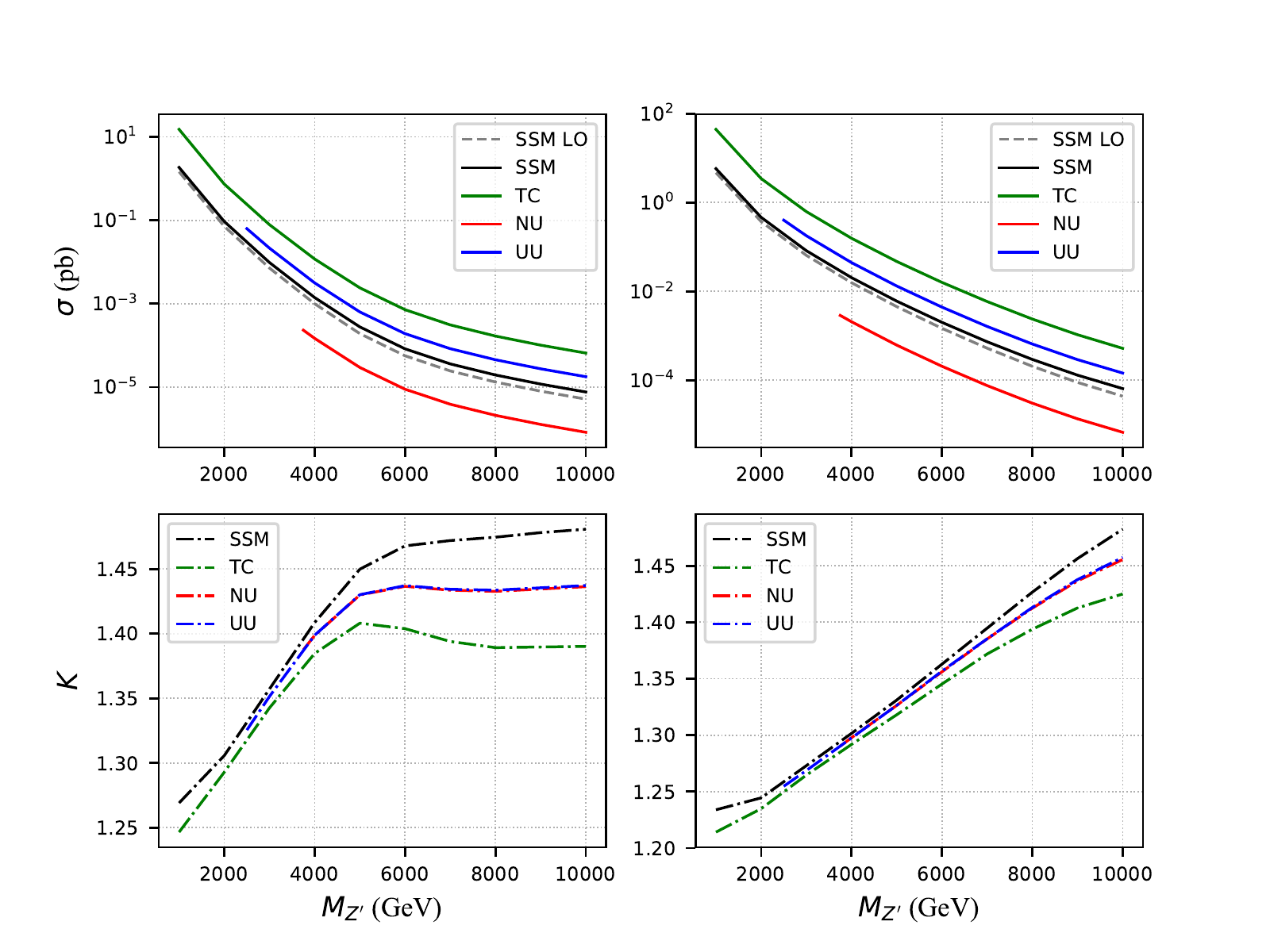}
\end{minipage}
%{\includegraphics[scale=1]{\main/section6OtherSignatures/img/POWHEG_UU_NU_TC_14_27.pdf}}
\caption{ Left: LO and NLO total cross section predictions in picobarns for $q \bar q \to Z' \to t \bar t [+g]$ (top) 
 and the NLO/LO K-factors (bottom) at a \com~energy $\sqrt{s}= 14\TeV$. Right: same as the left for a \com~energy $\sqrt{s}= 27\TeV$. 
\label{fig:POWHEG_uu_nu_tc}}
\end{figure}

We have chosen benchmark points such that the width $\Gamma_{Z'}$ in all models is the same as in the SSM.
We have calculated the width in the SSM ($\Gamma_{Z'}^{\rm SSM}$) at leading order using \pythia{ 6}~\cite{Sjostrand:2014zea} with a running 
electro-magnetic coupling, $\alpha(M_{Z'})$, such that $\Gamma_{Z'}^{\rm SSM}/M_{Z'}$ slightly increases from $3.48\%$ at $M_{Z'} =1\TeV$
to $3.61\%$ at $M_{Z'} =10\TeV$.
This is achieved by setting the parameter $t=1$ in the UU and NU models. As a consequence, the $W'$ couplings to the SM fermions
are the same in the SSM and NU cases, whereas the $Z'$ couplings are different.
For the TC model we set  $f_1 = 1$ and  $f_2=0$ which maximises the fraction of $Z'$ bosons decaying into $t \bar t$ pairs.
For the parton distribution functions (PDFs), in the $t \bar t$ case, we use a NLO PDF4LHC set for Monte Carlo studies (ISET = 90000 in LHAPDF6)  and the renormalisation and factorisation scales $\mu_R$ and $\mu_F$ are identified with the invariant mass of the system.
On the other hand, in the dilepton case, we use a NLO CT14 (ISET = 13100 in LHAPDF6) and the renormalisation and factorisation scales $\mu_R$ and $\mu_F$ are identified with the invariant mass of the system.
%\subsubsection{Results for $t\bar t$ resonances with {\tt PBZp}}
%\paragraph*{Results for $t\bar t$ resonances with {\tt PBZp}}

  \begin{figure}[t]
\centering
\begin{minipage}[t][\minipageheightsp][t]{\minipagewidthsp}
\centering
\includegraphics[width=\figuremaxwidthsp*11/6,keepaspectratio]{\main/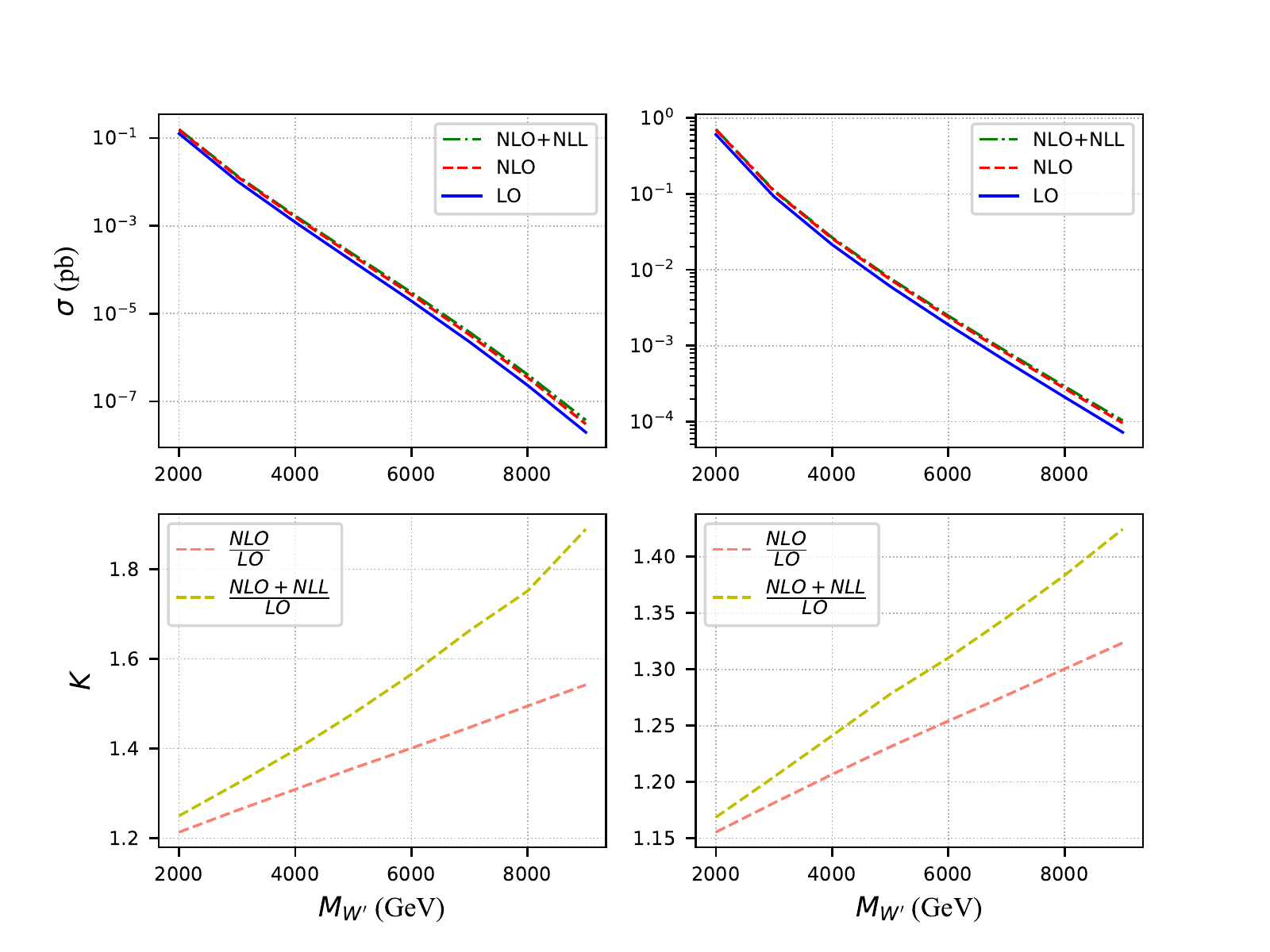}
\end{minipage}
\caption{ Left: LO, NLO, and NLO+NLL total cross section predictions in picobarn for $q \bar q \to W|W' \to e  \nu$ (top). The NLO/LO and the NLO+NLL/LO K-factors at a \com~energy $\sqrt{s}= 14\TeV$ (bottom). Right: same as the left for a \com~energy $\sqrt{s}= 27\TeV$. 
\label{fig:dilepton_ssm}}
\end{figure}

In \fig{fig:POWHEG_uu_nu_tc}, we show the total cross section for the EW production of $t \bar t$ pairs 
($q \bar q \to Z' \to t \bar t [+g]$) in picobarn at a \com~energy $\sqrt{s} = 14\TeV$ (left) and $27\TeV$ (right). 
The results are given for LO (only for SSM) and NLO cross sections together with the NLO/LO K-factors (bottom)
for the SSM, UU, NU, and TC in dependence of the $Z'$ mass.
No cut on the invariant mass of the $t \bar t$ pair has been applied.
%\subsubsection{Results for dilepton resonances with {\tt Resummino}}
%\paragraph*{Results for dilepton resonances with {\tt Resummino}}

In \fig{fig:dilepton_ssm} we show the  $W'$ production cross sections  at a \com~energy $\sqrt{s} = 14\TeV$ at NLO and NLO+NLL in the SSM as a function of the heavy gauge boson mass (top left). The ratios of the total cross sections at LHC14 at NLO and NLO+NLL over the LO cross section as a function of the $W'$ mass is also presented (bottom left).
Similarly, in the right side of \fig{fig:dilepton_ssm}  we show the same for a \com~energy $\sqrt{s} = 27\TeV$.
 Interference terms between $W$ and $W'$ gauge bosons are included. The
 invariant mass of the lepton pair is restricted to $m_{ll}>3M_{W'}/4$. Increasing the mass the threshold effects become more and more important leading to almost $16\%$ ($6\%$) increase of the cross section at $M_{W'} = 8\TeV$ for $\sqrt{s} = 14$ ($27$)\TeV.
 
\vspace{1cm}
  %\caption{NLO and NLO+NLL total cross section predictions in attobarns for %$q \bar q \to W|W' \to e  \nu$ 
 %and the NLO/LO K-factors at a \com~energy $\sqrt{S}= 14$ TeV.}
% Interference terms between $W$ and $W'$ gauge bosons are included. The
% invariant mass of the lepton pair is restricted to $Q>3M_{\Wp}/4$.}
%%%%%%%%%%%%%% Start of Table  %%%%%%%%%%%%%%%%%%%%%%%%%%%%%%%%%

%\paragraph*{Acknowledgements}
%The work of M.K.~has been supported by the BMBF under contract 05H15PMCCA and the DFG through the Research Training Network 2149
%``Strong and weak interactions - from hadrons to dark matter''.
%T.J.~was supported by the Swiss National Science Foundation~(SNF) under contracts BSCGI0-157722 and CRSII2-160814.

\subsubsection{\cmsC{Searching for a RS gluon resonance in $\ttbar$ at the HL- and HE-LHC}%Prospects for $\ttbar$ resonances}
}\label{sec:cmsKKgravitonditop}
\contributors{M. Narain, K. Pedro, S. Sagir, E. Usai, W. Zhang, CMS}
%{\bf Author(s): M. Narain$^a$, K. Pedro$^b$, S. Sagir$^{a,c}$, E. Usai$^a$, W. Zhang$^a$, CMS}
%
%\noindent \textit{a) Brown University, Providence, USA}
%\noindent \textit{b) Fermi National Accelerator Lab., Batavia, USA}
%\noindent \textit{c) Karamano\u{g}lu Mehmetbey University, Karaman, Turkey}

Many models of new physics predict heavy resonances with enhanced couplings to the third generation of the SM~\cite{Dimopoulos:1981zb,Susskind:1978ms,Hill:1993hs,Chivukula:1998wd,ArkaniHamed:2001nc,ArkaniHamed:1998rs,Randall:1999ee,Randall:1999vf}.
Thus, the study of the top quark can give important insight into the validity of such models.
This analysis from CMS~\cite{CMS:2018ohu} presents projections for a heavy resonance, in particular a Randall--Sundrum Kaluza--Klein gluon (RSG)~\cite{Randall:1999ee}, decaying into a $\ttbar$ pair using the upgraded CMS Phase-2 detector design at HL-LHC, with a \com~energy of $14\TeV$. We also present projections for $\ttbar$ resonances at a \com~energy of $27\TeV$, accessible by the HE-LHC.
Two distinct final states with either a single lepton or no leptons are considered.
The topology where the hadronic decay products of the top quark are fully merged into a single jet is studied. 
For top quarks with a large boost (transverse momentum, $\pt$, greater than $400\GeV$), an identification algorithm based on the soft-drop~\cite{Larkoski:2014wba} jet grooming algorithm is used in combination with $N$-subjettiness~\cite{Thaler:2011gf} and subjet b-tagging algorithms to identify the decay of the top quark with no leptons. No lepton isolation is imposed because leptons are not expected to be well separated from other objects in final states.

The RSG signal processes are generated using \pythia{}{ 8.212}~\cite{Sjostrand:2014zea} at leading order (LO), assuming a decay width of $17\%$, and a RS parameter $k$ value of $0.01\times m_{\textrm{Planck}}$.
The \sloppy\mbox{\powheg{ 2.0}~\cite{Nason:2004rx,Frixione:2007vw,Alioli:2010xd,Frixione:2007nw}} event generator is used to generate $\ttbar$ and single top quark events in the $t$-channel and $t\PW$ channel to NLO accuracy.
The single top quark events in the $s$-channel, $\PZ$+jets, and $\PW$+jets are simulated using \MGvATNLO  2.2.2~\cite{Alwall:2014hca}. 
The \pythia{} event generator is used to simulate the QCD multijet and $\PW\PW$ events at NLO. 
Parton showering, hadronisation, and the underlying event are simulated with \pythia{}, using the NNPDF 3.0 parton distribution functions (PDFs) and the CUETP8M1~\cite{Khachatryan:2015pea,Skands:2014pea} tune for all Monte Carlo processes, except for the $\ttbar$ sample, which is produced with the CUETP8M2T4~\cite{CMS:2016kle} tune.
The CMS Phase-2 detector simulation and the reconstruction of physics-level objects are simulated with the \delphes{} software package~\cite{deFavereau:2013fsa}. The same signal and background processes are also considered for the HE-LHC projections in both final states at $\sqrt{s} = 27\TeV$, assuming the same number of pileup interactions as the HL-LHC.
The reconstruction of physics-level objects for the HE-LHC is simulated with the \delphes{} software package with the CMS Phase-2 detector design.

The particle flow (PF) algorithm~\cite{Sirunyan:2017ulk} is used together with the pileup per particle identification (PUPPI)~\cite{Bertolini:2014bba} method to reconstruct the final state objects such as electrons, 
muons, jets, and missing transverse momentum (\ptmiss). In both final states, large-radius anti-\kt jets with a distance parameter of $0.8$ (AK8) are used.
The AK8 jets are required to have $\pt>400\GeV$, $|\eta|<4$, soft-drop mass between $105$ and $220\GeV$, and N-subjettiness ratio $\tauTT<0.65$. AK8 jets passing these requirements are referred to as $t$-tagged jets. 
In the single-lepton final state, the AK8 jets are selected if they are isolated from lepton by $\Delta R$(lepton, AK8 jet) $>$ 0.8 and events with more than one such AK8 jets are vetoed to be orthogonal to the fully hadronic final state.
In the fully hadronic final state, the sum of the $p_T$ of the two AK8 jets, $\HT$, is additionally required to be $> 1.2 \TeV$ and the angle between the two AK8 jets, $\Delta \phi$, is required to be $>2.1$.
We require single-lepton events to have exactly one electron with $\pt >80\GeV$ and $|\eta|<3$ or one muon with $\pt>55\GeV$ and $|\eta|<3$.
In order to limit the background contribution from QCD multijet events, we require $\ptmiss > 120$ ($50$)\GeV in the electron (muon) channel, where $\ptmiss$ is the magnitude of the missing transverse momentum defined as the the negative of the vector $\pt$ sum of all reconstructed PF candidates.

Additionally, the single-muon events are required to have $H_{T}^{\text{lep}}>150\GeV$, where \sloppy\mbox{$H_{T}^{\text{lep}}= \ptmiss+\pt^{\text{lep}}$}.
The single-lepton events are further required to have at least two AK4 jets with $\pt>30\GeV$ and $|\eta|<4$. 
The leading and subleading jets are required to have a $\pt$ greater than \sloppy\mbox{$185$ ($150$) and $50$ ($50$)\GeV}, respectively, in the electron (muon) channel.
Because no isolation requirement is imposed on leptons, we require that the AK4 jet that is closest to the lepton is either separated by $\Delta R > 0.4$, or the magnitude of the lepton momentum that is transverse to the jet axis is greater than $25\GeV$.

\begin{figure}[t]
\centering
\begin{minipage}[t][\minipageheightdp][t]{\minipagewidthdp}
\centering
\includegraphics[width=\figuremaxwidthdp,height=\figuremaxheightdp,keepaspectratio]{\main/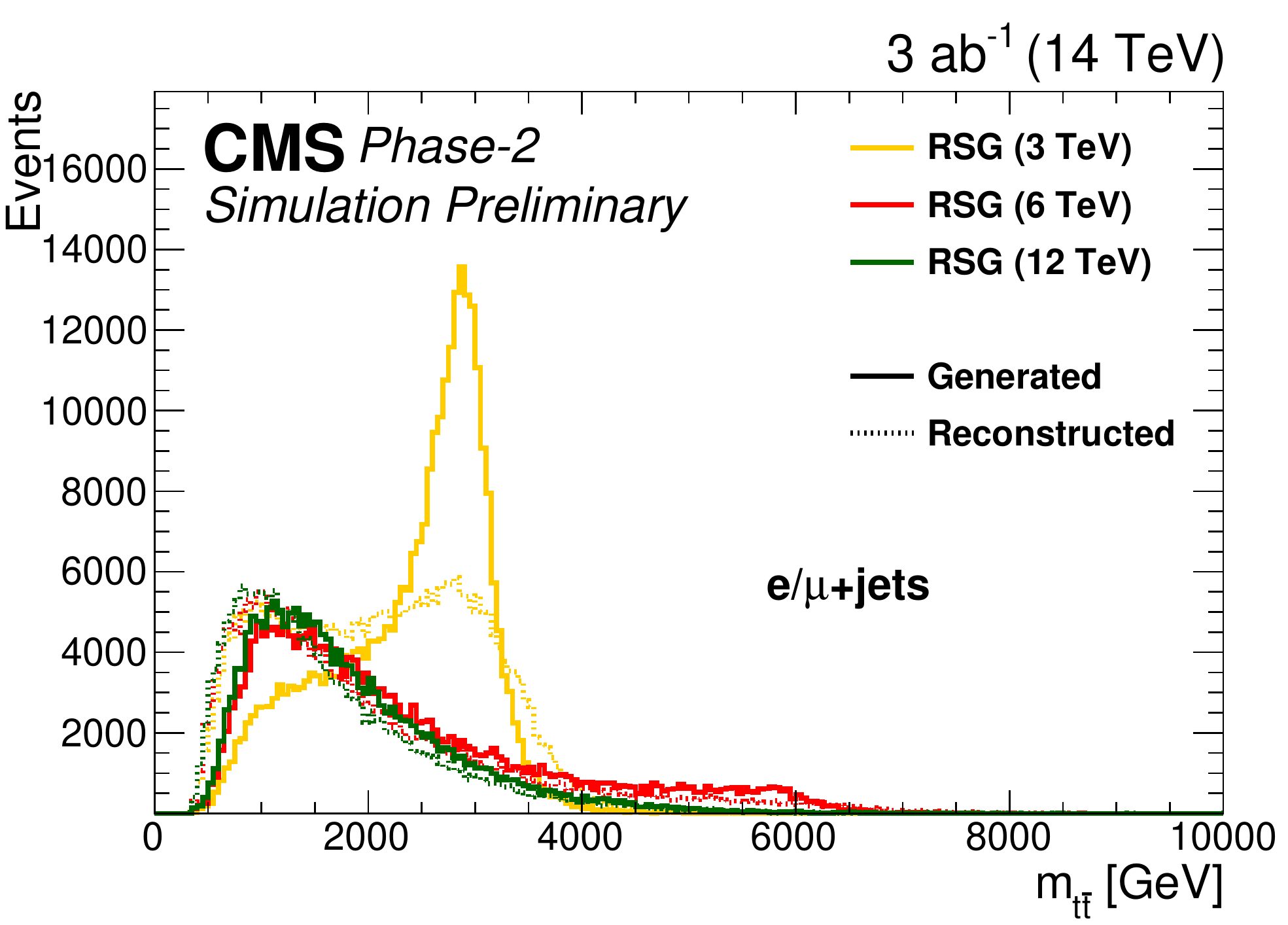}
\end{minipage}
\begin{minipage}[t][\minipageheightdp][t]{\minipagewidthdp}
\centering
\includegraphics[width=\figuremaxwidthdp,height=\figuremaxheightdp,keepaspectratio]{\main/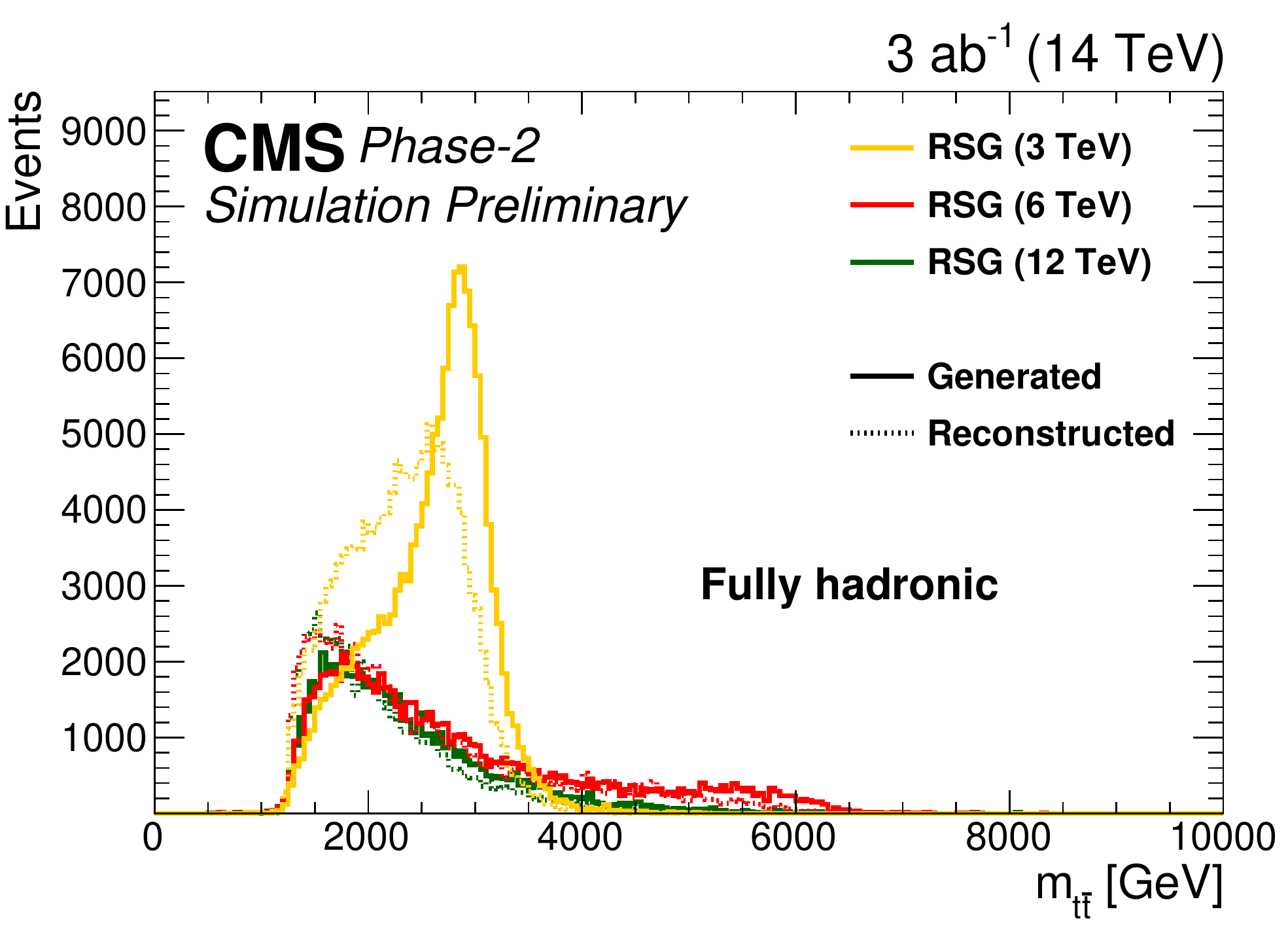}
\end{minipage}
\caption{Generated and reconstructed RSG mass distributions for the single-lepton (left) and fully hadronic (right) final states. The distributions are shown after full event selection in each final state. The signals are scaled to $1\pb$.}
\label{fig:FTR18009signals}
\end{figure}

\begin{figure}[t]
\centering
\begin{minipage}[t][\minipageheightdp][t]{\minipagewidthdp}
\centering
\includegraphics[width=\figuremaxwidthdp,height=\figuremaxheightdp,keepaspectratio]{\main/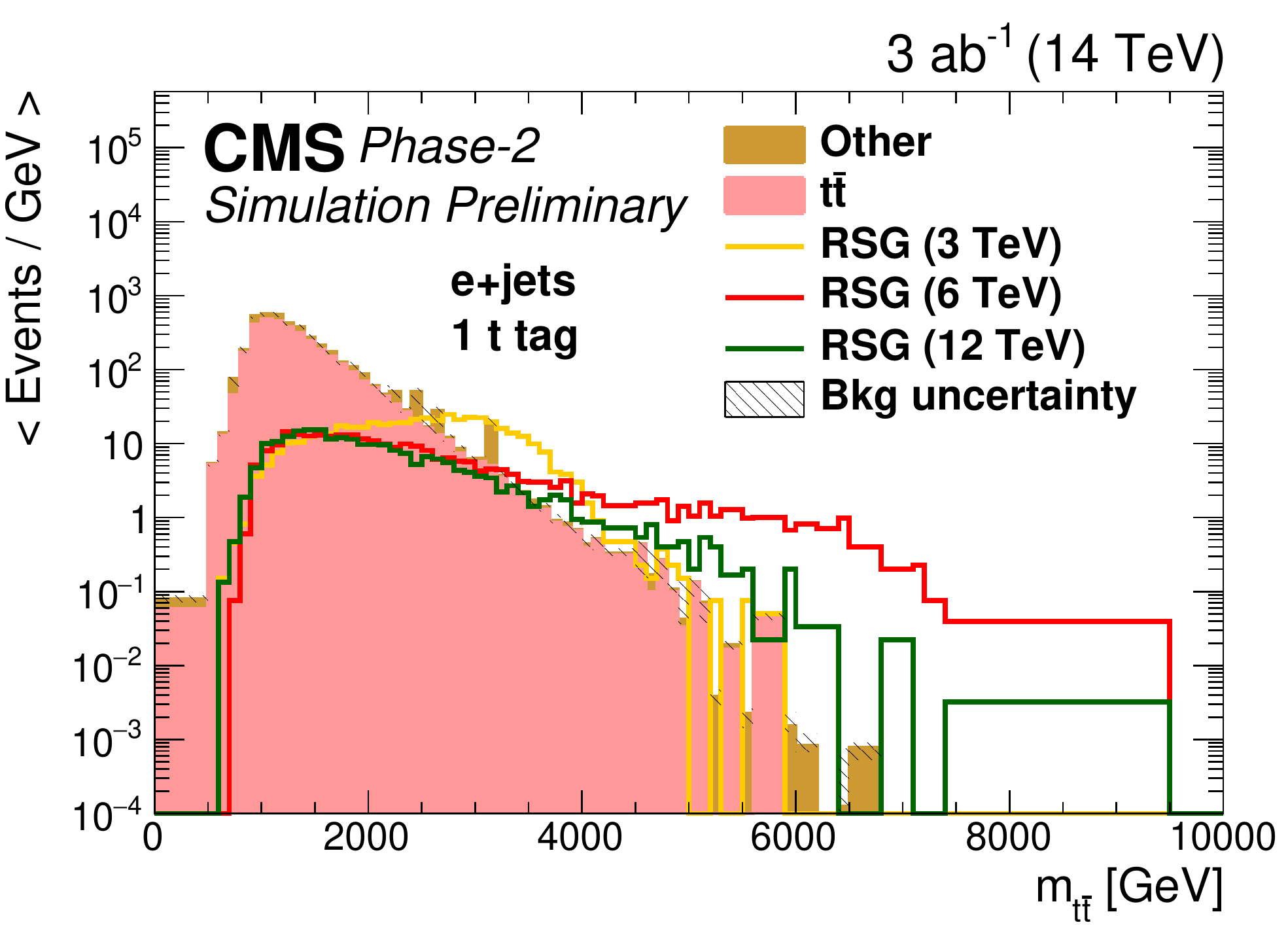}
\end{minipage}
\begin{minipage}[t][\minipageheightdp][t]{\minipagewidthdp}
\centering
\includegraphics[width=\figuremaxwidthdp,height=\figuremaxheightdp,keepaspectratio]{\main/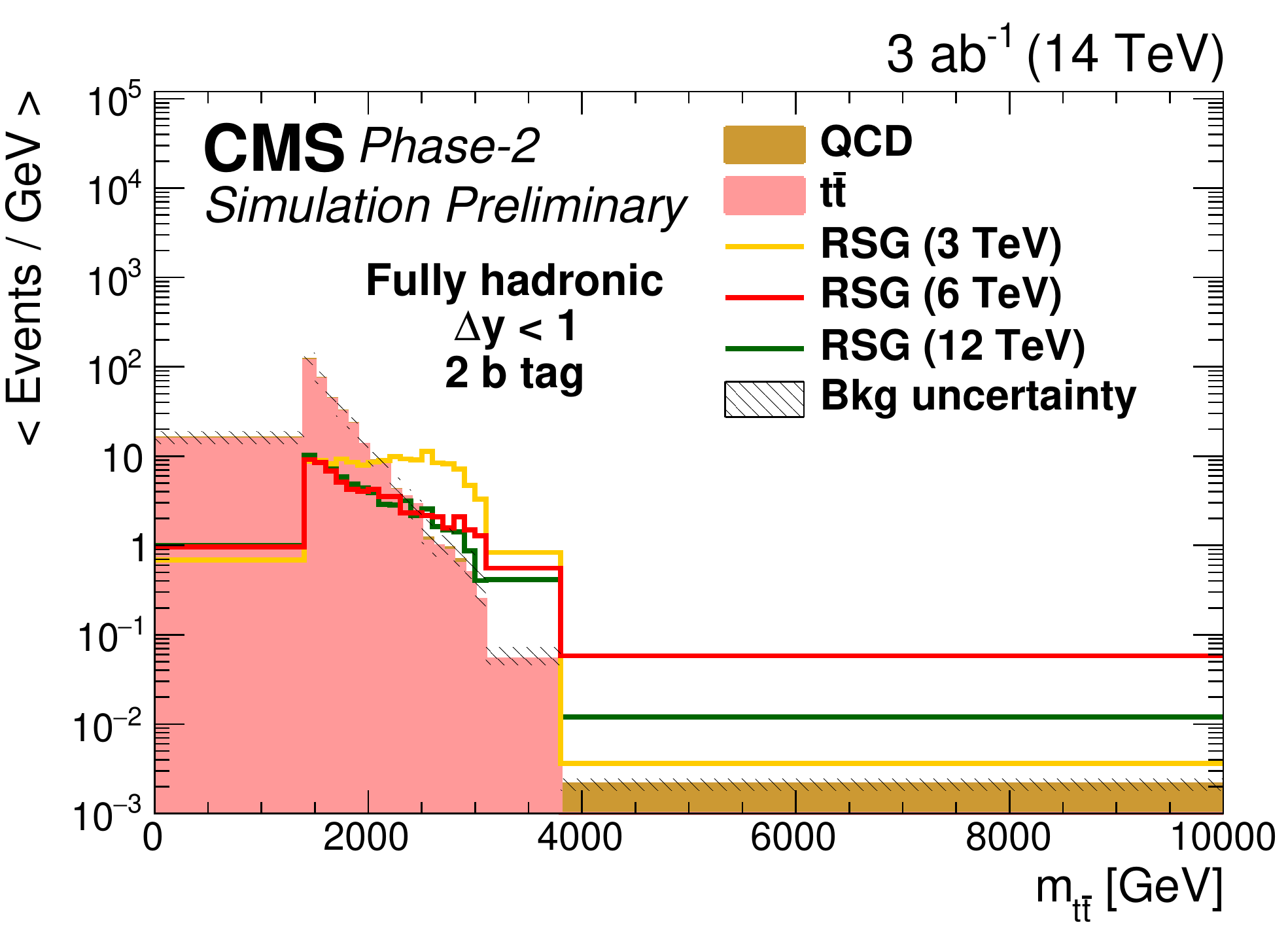}
\end{minipage}\\
\begin{minipage}[t][\minipageheightdp][t]{\minipagewidthdp}
\centering
\includegraphics[width=\figuremaxwidthdp,height=\figuremaxheightdp,keepaspectratio]{\main/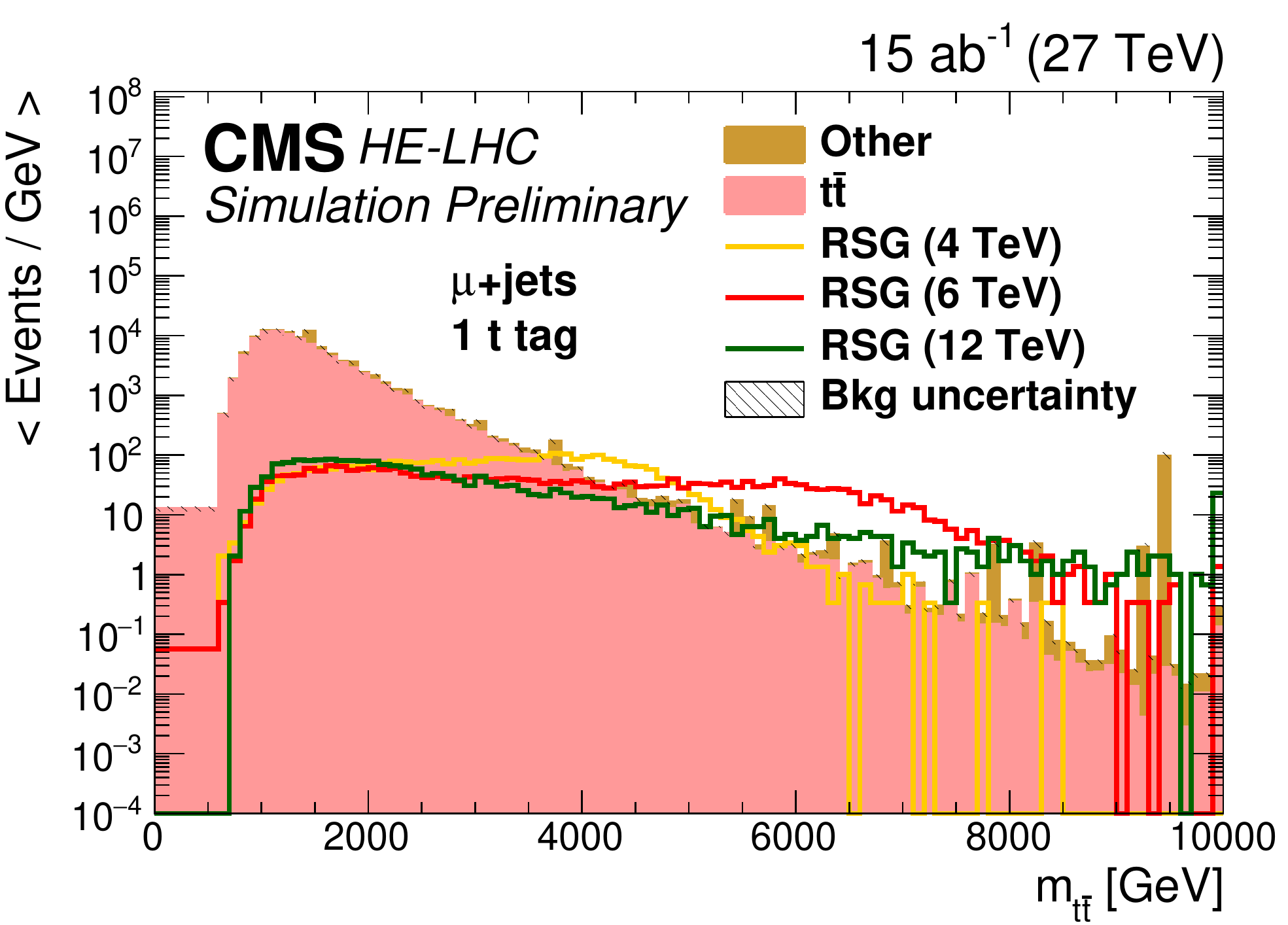}
\end{minipage}
\begin{minipage}[t][\minipageheightdp][t]{\minipagewidthdp}
\centering
\includegraphics[width=\figuremaxwidthdp,height=\figuremaxheightdp,keepaspectratio]{\main/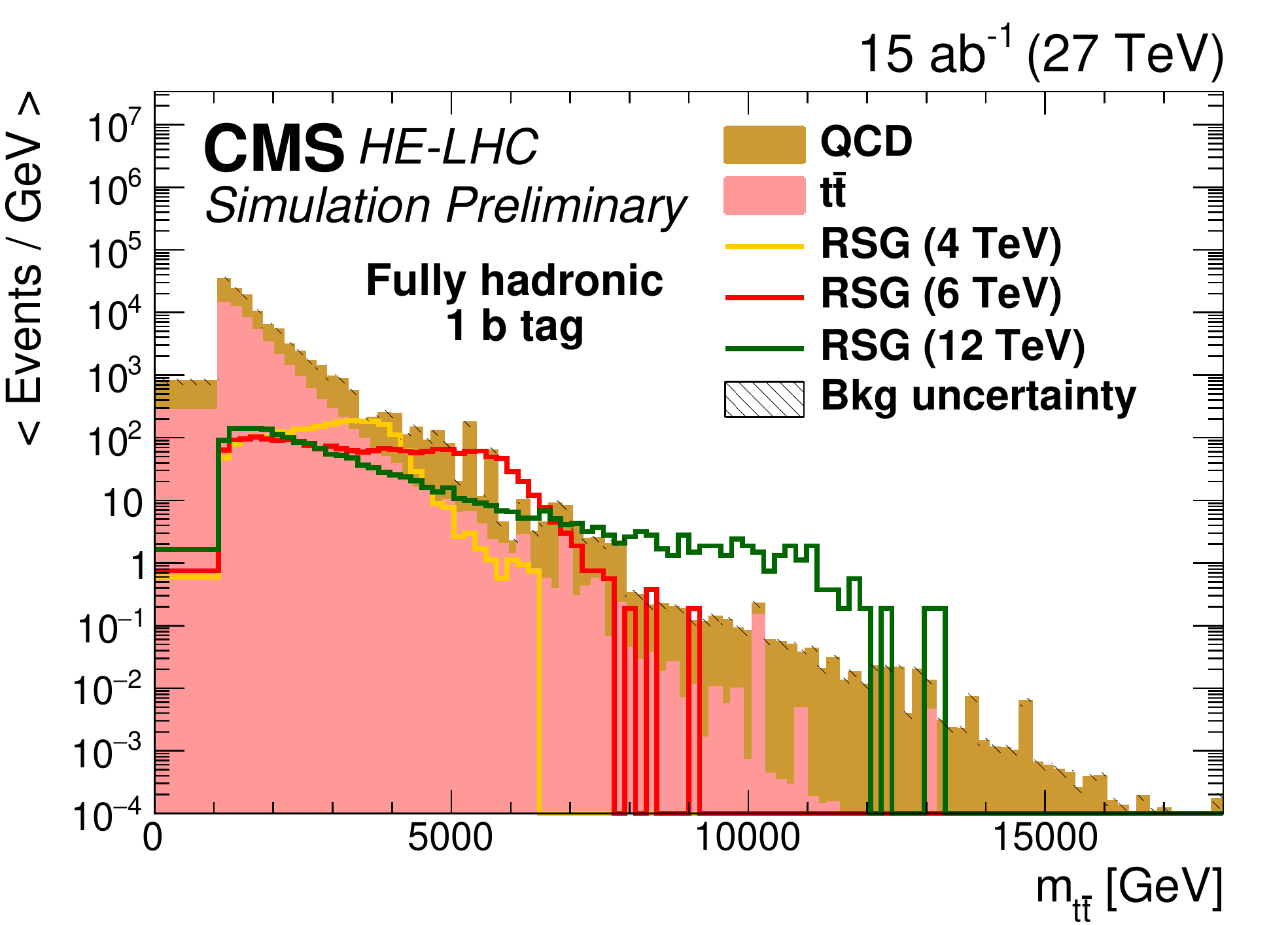}
\end{minipage}
  \caption{Distributions of $m_{\ttbar}$ in events with single-electron and one $t$-tagged jet (top left) or zero lepton, $\Delta y <1$ and two $b$ tags (top right) for 3\abinv at 14\TeV. Distributions of $m_{\ttbar}$ in events with a single-muon and one $t$-tagged jet (bottom left) or zero lepton and one $b$-tag (bottom right) for \intlumihelhc at 27\TeV.}
\label{fig:rsg:templates}
\end{figure}

We use the Theta package~\cite{theta} to derive the expected cross section limits at $95\%\cl$ on the production of a RSG decaying to $\ttbar$. 
The limits are computed using the asymptotic CLs approach. 
A binned likelihood fit on the distributions of reconstructed $\ttbar$ mass ($m_{\ttbar}$), shown in \fig{fig:FTR18009signals}, is performed in both single-lepton and fully hadronic final states.
To improve the sensitivity, the events are categorised based the number of subjet $b$ tags (0, 1, or 2) and the rapidity difference ($\abs{\Delta y(jet_1, jet_2)}$ $<$1 or $>$1) in the fully hadronic final state.
Similarly, in the single-lepton final state, the categorisation is performed using the number of $t$-tagged jets (0 or 1) and the flavour of the lepton (\Pe,~$\mu$).
Example $m_{\ttbar}$ distributions of background and signal samples are shown in \fig{fig:rsg:templates}.
Systematic uncertainties, following the recommendations in \citeref{Syst_uncert_UPSG}, are included in the fit as nuisance parameters with log-normal prior for both HL-LHC and HE-LHC. The results are limited by the statistical uncertainties in the background estimates. These uncertainties are scaled down by the projected integrated luminosity and are treated using the Barlow--Beeston light method~\cite{Barlow:1993dm,Conway:2011in}. 

\begin{figure}[t]
\centering
\begin{minipage}[t][\minipageheightdp][t]{\minipagewidthdp}
\centering
\includegraphics[width=\figuremaxwidthdp,height=\figuremaxheightdp,keepaspectratio]{\main/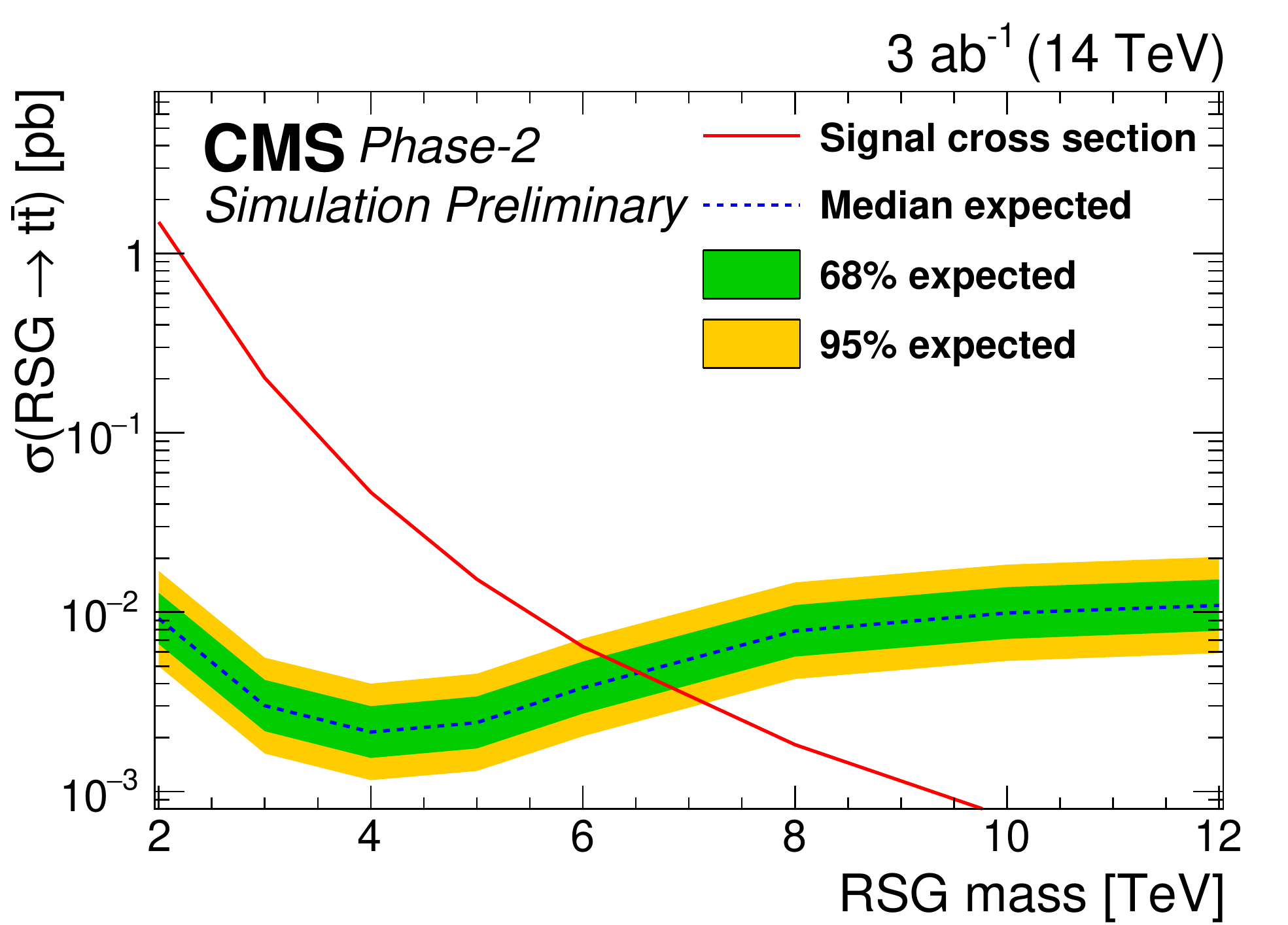}
\end{minipage}
\begin{minipage}[t][\minipageheightdp][t]{\minipagewidthdp}
\centering
\includegraphics[width=\figuremaxwidthdp,height=\figuremaxheightdp,keepaspectratio]{\main/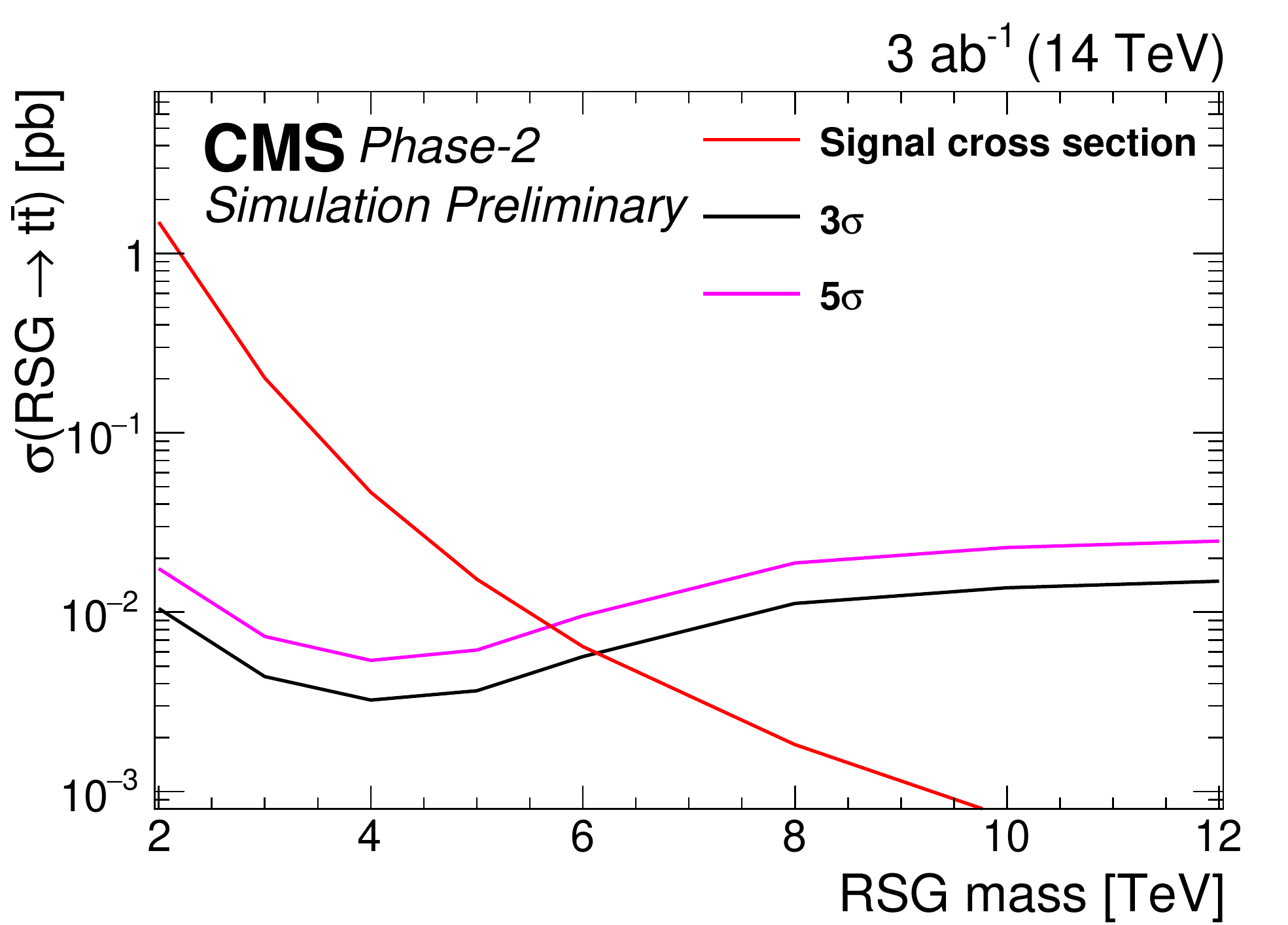}
\end{minipage}\\
\begin{minipage}[t][\minipageheightdp][t]{\minipagewidthdp}
\centering
\includegraphics[width=\figuremaxwidthdp,height=\figuremaxheightdp,keepaspectratio]{\main/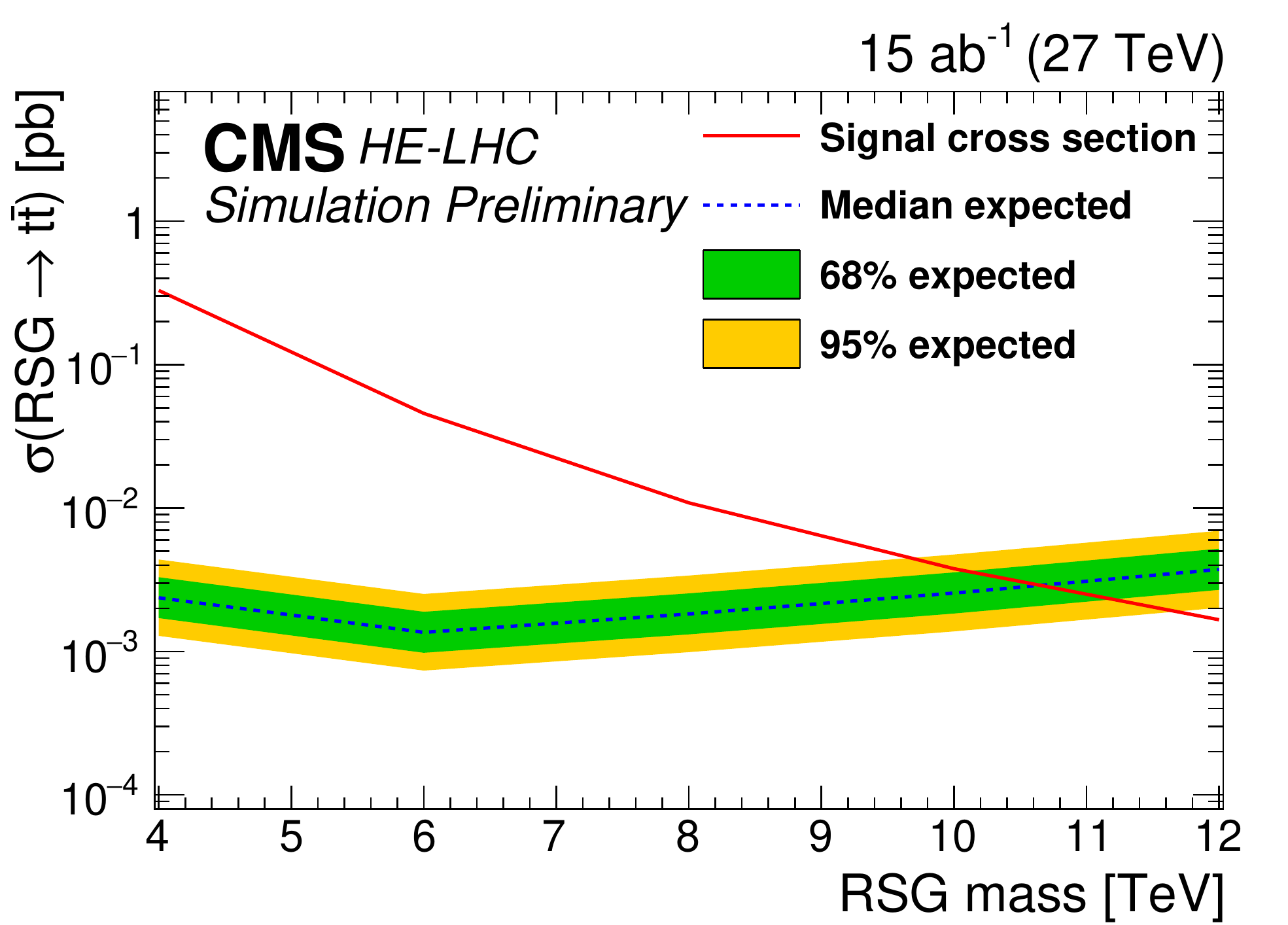}
\end{minipage}
\begin{minipage}[t][\minipageheightdp][t]{\minipagewidthdp}
\centering
\includegraphics[width=\figuremaxwidthdp,height=\figuremaxheightdp,keepaspectratio]{\main/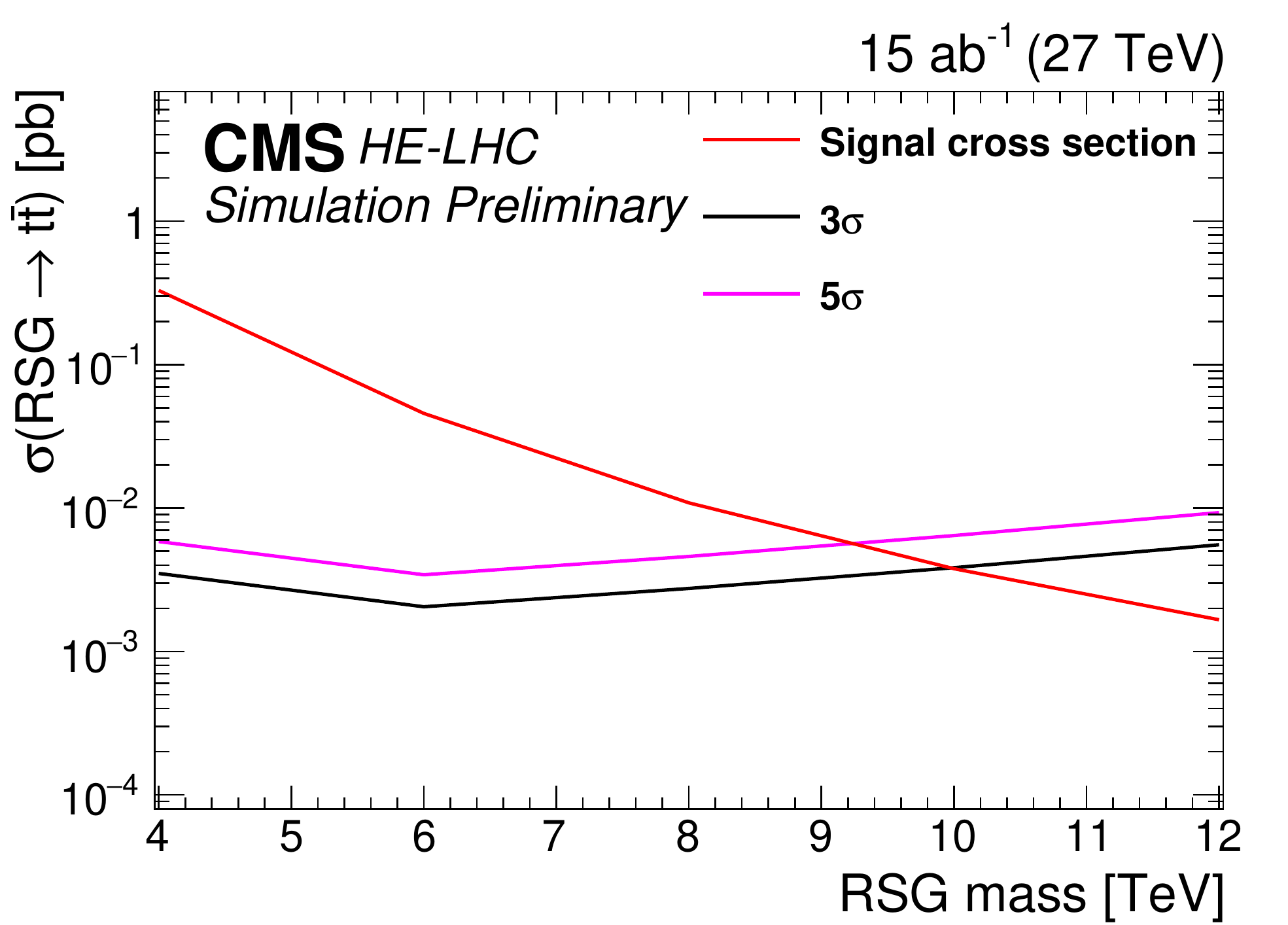}
\end{minipage}
    \caption{95\%~\cl expected upper limits (left) and $3\sigma$ and $5\sigma$ discovery reaches (right) for a RSG decaying to $\ttbar$ at 3\abinv at $14\TeV$ (top) and \intlumihelhc at 27 TeV (bottom) for the combined single-lepton and fully hadronic final states.}
    \label{fig:rsg:results}
\end{figure}

The expected limits at $95\%\cl$ and discovery reaches at 3 and $5\sigma$ for the combined single-lepton and fully hadronic final states are shown in \fig{fig:rsg:results}. The RSG with masses up to 6.6 (10.7)~{\TeV} are excluded at $95\%\cl$ for a projected integrated luminosity of 3 (15)~{\abinv} at the HL-LHC (HE-LHC). This extends the current Run-2 limits of 4.5 \TeV\ based on $36$\fbinv~\cite{Sirunyan:2018ryr}.
The discovery reach for an RSG is computed to be 5.7 (9.4)~{\TeV} at $5\sigma$ at the HL-LHC (HE-LHC).

\subsubsection{\atlas{$\Zp\rightarrow t\bar{t}$ searches at HL-LHC
%Prospects for $t\bar{t}$ resonance searches at HL-LHC
}}
\contributors{A. Duncan, ATLAS}
%{\bf Author(s): A. Duncan, M. Wielers, G. Lee, M. Marjanovic et al. ATLAS}

HL-LHC prospects at for $\Zp$ bosons in the $t\bar{t}$ final state were presented in \citeref{ATL-PHYS-PUB-2017-002}
based on the event selection and systematic uncertainties from the Run-1 analysis with $20.3\fbinv$
collected at $\sqrt{s}=8\TeV$ in \citeref{Aad:2015fna}.
These results have been updated~\cite{ATL-PHYS-PUB-2018-044} using a more recent parameterisation of the $b$-tagging efficiencies and
misidentification rates as shown in \citeref{CERN-LHCC-2017-018} and are summarised below.

\begin{figure}[t]
\centering
\begin{minipage}[t][\minipageheightsp][t]{\minipagewidthsp}
\centering
\includegraphics[width=\figuremaxwidthsp,height=\figuremaxheightsp,keepaspectratio]{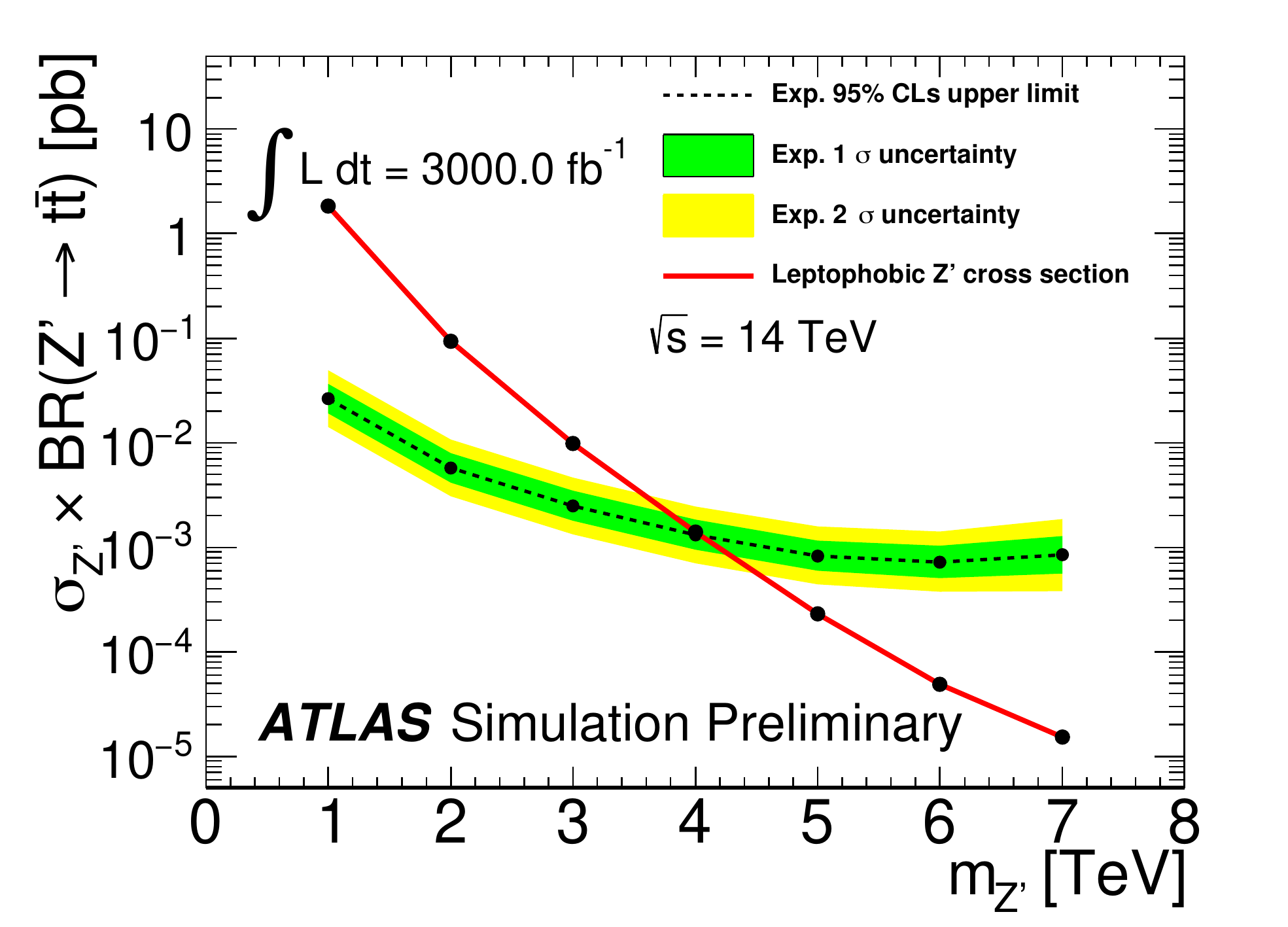}
\end{minipage}
\caption{Expected upper limits set on the cross section $\times$ BR of the Topcolour Z$^\prime$ boson for masses $1-7\TeV$, with $3\abinv$ of simulated $14\TeV$ $pp$ collisions compared to the  theoretical signal cross section.}
\label{fig:atlas_Ztt}
\end{figure}

The analysis looks for a narrow width $\Zp$ boson
in a final state in which one of the $W$ bosons from the top quark decays to two jets and the other decays to a lepton (electron or muon) and a neutrino ($t\bar{t}\rightarrow WbWb \rightarrow \ell \nu b q q^{\prime}b$). Events are required to contain exactly one lepton, several jets and at least a moderate amount of missing transverse momentum must be present. Events are separated into boosted and resolved channels with most of the signal events falling in the former category. In the resolved channel the decay products of the hadronic top-quark decay are reconstructed as three separate jets and in total events must contain at least four jets. In the boosted channel, the hadronic top-quark decay products are highly boosted and end up in one broad large-radius jet. Events are selected if at least one large-radius jet and one jet (from the other top-quark decay) is present.
Subsequently $m_{t\bar{t}}$ is reconstructed based on the reconstruction of the $W$ bosons and $b$-jets in the event. Using $m_{t\bar{t}}$ as discriminant, upper limits are set on the signal cross section times BR as a function of the $\Zp$ boson mass.
Using as benchmark a Topcolour-assisted Technicolour $\Zp_{\mathrm{TC2}}$ boson with a narrow width of $1.2\%$,
$\Zp_{\mathrm{TC2}}$ bosons can be excluded up to masses of $\simeq 4\TeV$ with \intLumi of $pp$ collisions as shown in~\fig{fig:atlas_Ztt}. 
This mass limit is conservative due to the use of systematic uncertainties from the Run-1 analysis~\cite{Aad:2015fna}. 
These uncertainties are already smaller in the Run-2 analysis~\cite{Aaboud:2018juj} and will be further reduced at the time of the HL-LHC. In particular the systematic uncertainty in the boosted channel is now reduced due to the significant improvements of the performance of boosted jets in Run-2 (in particular using more tracking information to look for sub-jets within the large-radius jets). This gain in performance also improves the signal over background ratio. In addition, the usage of the top-tagger algorithm will help to further reject background. 

\vspace{1cm}

\subsubsection{\theoryC{High mass dilepton ($ee, \mu\mu, \tau\tau$) searches at the HE-LHC}}
\label{subsubsec:hr_lep}
\contributors{C. Helsens, D. Jamin, M. Selvaggi}

%{\bf Authors: C. Helsens$^1$, D. Jamin$^2$, M. Selvaggi$^3$}\\
%\newline
%$^1$CERN EP-Departement, CH-1211 Geneva 23, Switzerland email: {\tt B.C. clement.helsens@cern.ch}\\
%$^2$Academia Sinica, Institute of  Physics, Taipei, Taiwan\\

%\newcommand*{\Zp}{\ensuremath{Z^{\prime}}}
%\newcommand*{\ZpSSM}{\ensuremath{Z^{\prime}_{\mathrm{SSM}}}}
%\newcommand*{\Z}{\ensuremath{Z}}
%\newcommand*{\Zptata}{\ensuremath{Z^{\prime}\rightarrow \tau\tau}}
%\newcommand*{\Zpee}{\ensuremath{Z^{\prime}\rightarrow e e}}
%\newcommand*{\Zpmumu}{\ensuremath{Z^{\prime}\rightarrow \mu\mu}}
%\newcommand*{\Zpll}{\ensuremath{Z^{\prime}\rightarrow \ell\ell}}
%\newcommand*{\Zptt}{\ensuremath{Z^{\prime}\rightarrow \ttbar}}
%\newcommand*{\ptZp}{\ensuremath{p_{\text{T}}^{\ensuremath{Z^{\prime}}}}}
%\newcommand*{\intlumihelhc}{\ensuremath{\mathcal{L}=15\text{ ab}^{-1}}}

%%%%%%%%%%%%%%%%%%%%%%%%%%%%%%%%%%%%%%%%%%%%%%%%%%%%%%%%%%%%%%%%%%%%%%%%%%%%%%%%%%%%%%%%%%%%
%\subsubsection{Introduction}
%\paragraph*{Introduction}
Models with extended gauge groups often feature additional $U(1)$ symmetries with corresponding heavy spin-1 bosons. These bosons, generally referred to as $\Zp$, would manifest themselves as a narrow resonance in the dilepton invariant mass spectrum. Among these models are those inspired by Grand Unified Theories, motivated by gauge unification, or a restoration of the left-right symmetry violated by the weak interaction. Examples include the $\Zp$ bosons of the $E_{6}$ motivated theories~\cite{London:1986jz,Joglekar:2016yap,Langacker:2008yv} and Minimal models~\cite{Salvioni:2009mt}. The SSM \cite{Langacker:2008yv} posits a $\ZpSSM$ boson with couplings to fermions that are identical to those of the SM $\Z$ boson.

The decay products of heavy resonances are in the multi-TeV regime and the capability to reconstruct their momentum imposes stringent requirement on the detector design. In particular, reconstructing the track curvature of multi-TeV muons requires excellent position resolution and a large lever arm. In this section, the expected sensitivity is presented for a \Zpll\ (where $\ell=\mathrm{e},\mu$) and \Zptata\ separately.

%%%%%%%%%%%%%%%%%%%%%%%%%%%%%%%%%%%%%%%%%%%%%%%%%%%%%%%%%%%%%%%%%%%%%%%%%%%%%%%%%%%%%%%%%%%%
%\subsubsection{Monte Carlo Samples}
%\paragraph*{Monte Carlo Samples}
Monte Carlo simulated event samples were used to simulate the response of the future detector to signal and backgrounds. Signals are generated with \pythia{ 8.230}~\cite{Sjostrand:2014zea} using the leading order cross-section from the generator.
All lepton flavour decays of the $\ZpSSM$ are generated assuming universality of the couplings.
The Drell-Yan background has been generated using \MGvATNLO 2.5.2~\cite{Alwall:2014hca} at leading order only. A conservative overall k-factor of $2$ has been applied to all the background processes to account for possibly large higher order QCD corrections.

%%%%%%%%%%%%%%%%%%%%%%%%%%%%%%%%%%%%%%%%%%%%%%%%%%%%%%%%%%%%%%%%%%%%%%%%%%%%%%%%%%%%%%%%%%%%
%\subsubsection{Event Selection}
%\paragraph*{Event Selection}
For the $\ell\ell$ final-states events are required to contain two isolated leptons with $\pt > 500\GeV$and $|\eta|<4$. For the $\tau\tau$ final state we focus solely on the fully hadronic decay mode which is expected to drive the sensitivity. The $\tau\tau$ event selection requires the presence of two reconstructed jets with $\pt > 500\GeV$and $|\eta|<2.5$ identified as hadronic $\tau$'s. To ensure orthogonality between the $\ell$ and $\tau$ final states, jets overlapping with isolated leptons are vetoed. Additional mass dependent selection criteria on the azimuthal angle between the two reconstructed $\tau$'s are applied to further improve the QCD background rejection (see \tabb{tab:leptonicresonances:selectiontautau}).

\begin{figure}[t]
\centering
\begin{minipage}[t][\minipageheighttp][t]{\minipagewidthtp}
\centering
\includegraphics[width=\figuremaxwidthtp,height=\figuremaxheighttp,keepaspectratio]{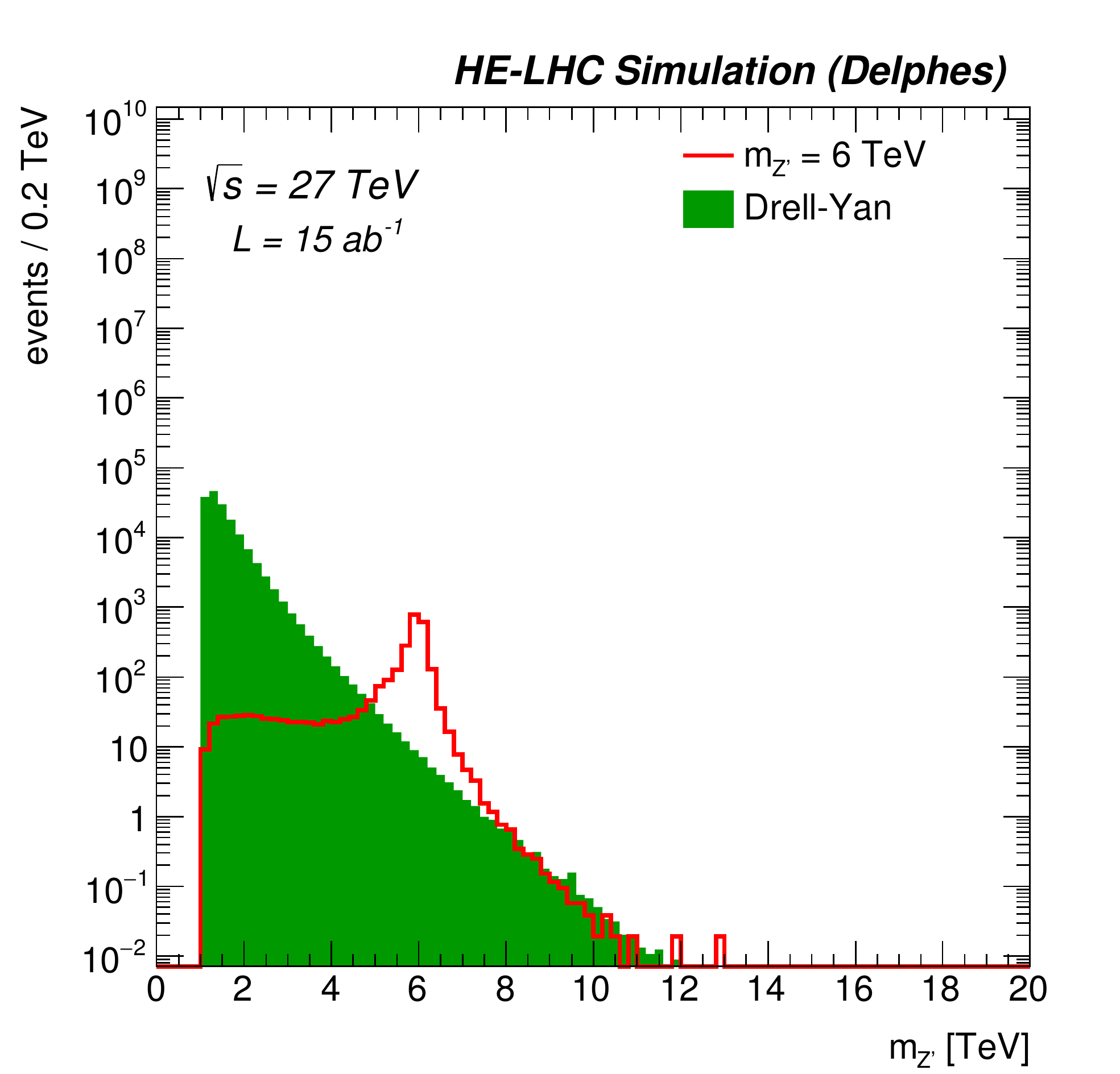}
\end{minipage}
\begin{minipage}[t][\minipageheighttp][t]{\minipagewidthtp}
\centering
\includegraphics[width=\figuremaxwidthtp,height=\figuremaxheighttp,keepaspectratio]{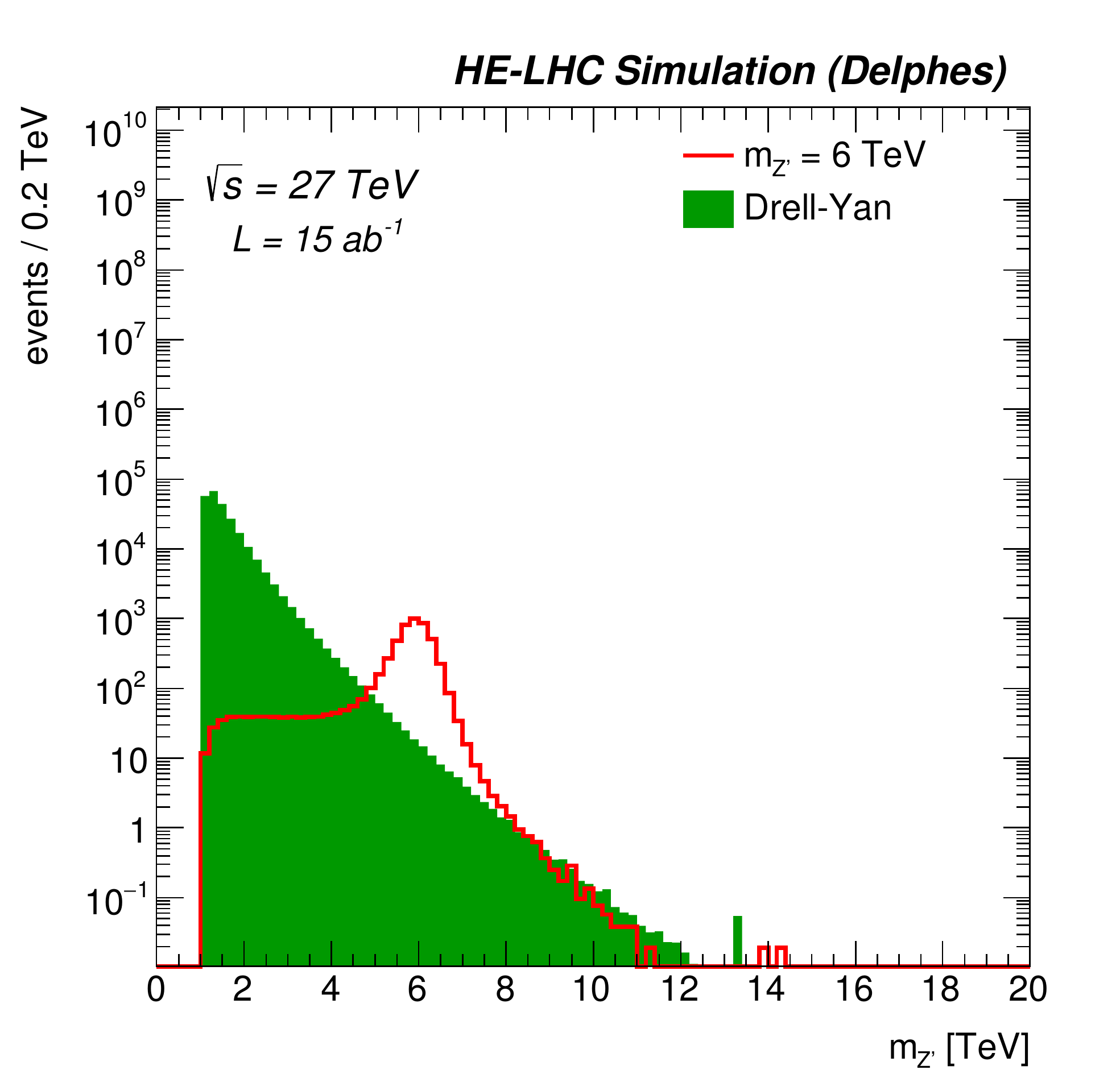}
\end{minipage}
\begin{minipage}[t][\minipageheighttp][t]{\minipagewidthtp}
\centering
\includegraphics[width=\figuremaxwidthtp,height=\figuremaxheighttp,keepaspectratio]{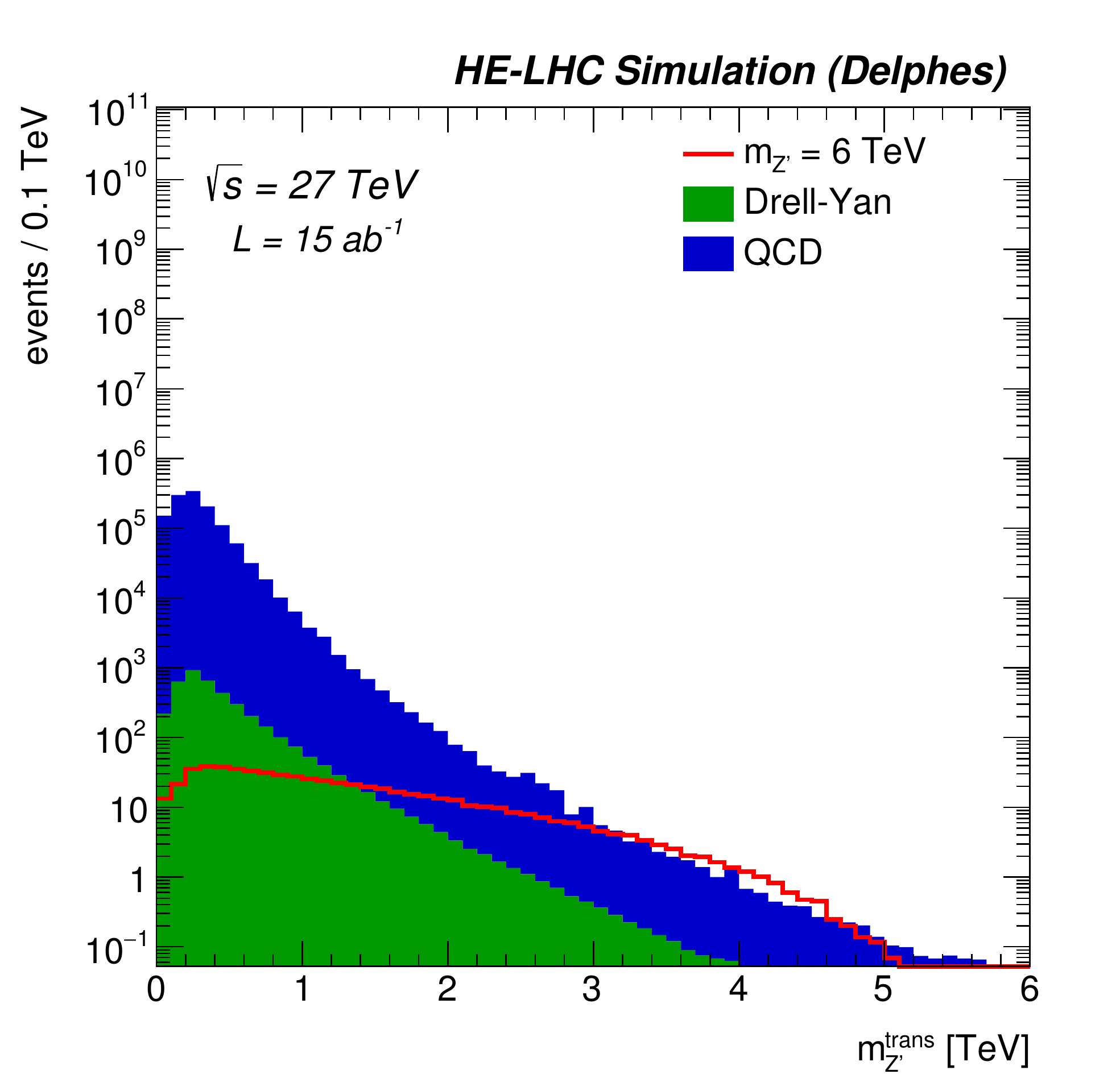}
\end{minipage}
  \caption{Left, centre: Invariant mass for a $6\TeV$ signal after full event selection for $ee$ channel (left) and $\mu\mu$ channel (centre). Right: Transverse mass for a $6\TeV$ signal after full event selection for the $\tau\tau$ channel. }
  \label{fig:leptonicresonances:masses}
\end{figure}

The left and central panels of \fig{fig:leptonicresonances:masses} show the invariant mass distribution for a $6\TeV$ $\ZpSSM$ in the $ee$ and $\mu\mu$ channels. The mass resolution is better for the $ee$ channel, as expected. The right panel of \fig{fig:leptonicresonances:masses} shows the transverse mass~\footnote{The transverse mass is defined as $m_{T}  =  \sqrt{2\ptZp*\met*(1-\cos\Delta\phi(\Zp,\met))}$.}
of a $6\TeV$ signal for the $\tau\tau$ channel. Because of the presence of neutrinos in $\tau$ decays, the true resonance mass cannot be reconstructed. Several arbitrary  choices are possible to approximate the $Z'$ mass. The transverse mass provided the best sensitivity and was therefore used to set limits and determine the discovery reach in $\tau\tau$ decay mode.

\begin{table}[t]
   \centering
\small\begin{tabular}{l|l|c|r}
   $\Zp$ mass [TeV] &  $\Delta \phi(\tau_1, \tau_2)$&  $\Delta R(\tau_1, \tau_2)$ & $\met$\\
  \hline
   $2$ & > 2.4 & > 2.4 and < 3.9 & > 80 GeV\\
   $4$ & > 2.4 & > 2.7 and < 4.4 & > 80 GeV\\
   $6$ & > 2.4 & > 2.9 and < 4.4 & > 80 GeV\\
   $8$ & > 2.6 & > 2.9 and < 4.6 & > 80 GeV\\
  $10$ & > 2.8 & > 2.9 and < 4.1 & > 60 GeV\\
  $12$ & > 2.8 & > 3.0 and < 3.6 & > 60 GeV\\
  $14$ & > 3.0 & > 3.0 and < 3.3 & > 60 GeV\\
  \end{tabular}\normalsize
  \caption{List of mass dependent cuts optimised to maximise the sensitivity for the \Zptata\ search.}
  \label{tab:leptonicresonances:selectiontautau}
\end{table}

%\MS{need editing / style}

%%%%%%%%%%%%%%%%%%%%%%%%%%%%%%%%%%%%%%%%%%%%%%%%%%%%%%%%%%%%%%%%%%%%%%%%%%%%%%%%%%%%%%%%%%%%
%\subsubsection{Results and discussion}
%\paragraph*{Results and discussion}
Hypothesis testing is performed using a modified frequentist method based on a profile likelihood that takes into account the systematic uncertainties as nuisance parameters that are fitted to the expected background predicted from Monte Carlo. For the $ee$ and $\mu\mu$ analyses, the dilepton invariant mass is used as the discriminant, while for the $\tau\tau$ channel the transverse mass is used. A $50\%$ uncertainty on the background normalisation is assumed.

\begin{figure}[t]
\centering
\begin{minipage}[t][\minipageheightdp][t]{\minipagewidthdp}
\centering
\includegraphics[width=\figuremaxwidthdp,height=\figuremaxheightdp,keepaspectratio]{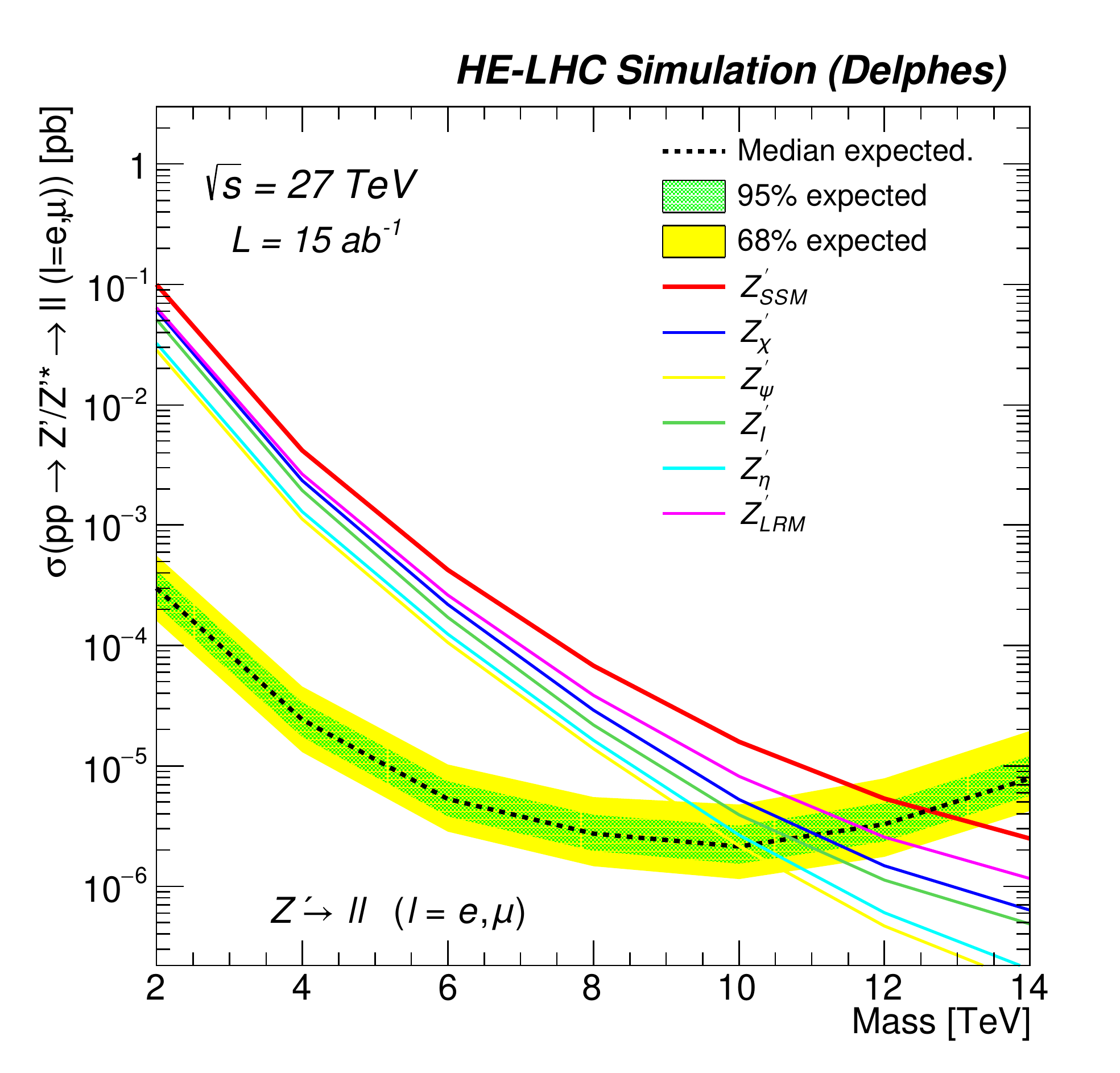}
\end{minipage}
\begin{minipage}[t][\minipageheightdp][t]{\minipagewidthdp}
\centering
\includegraphics[width=\figuremaxwidthdp,height=\figuremaxheightdp,keepaspectratio]{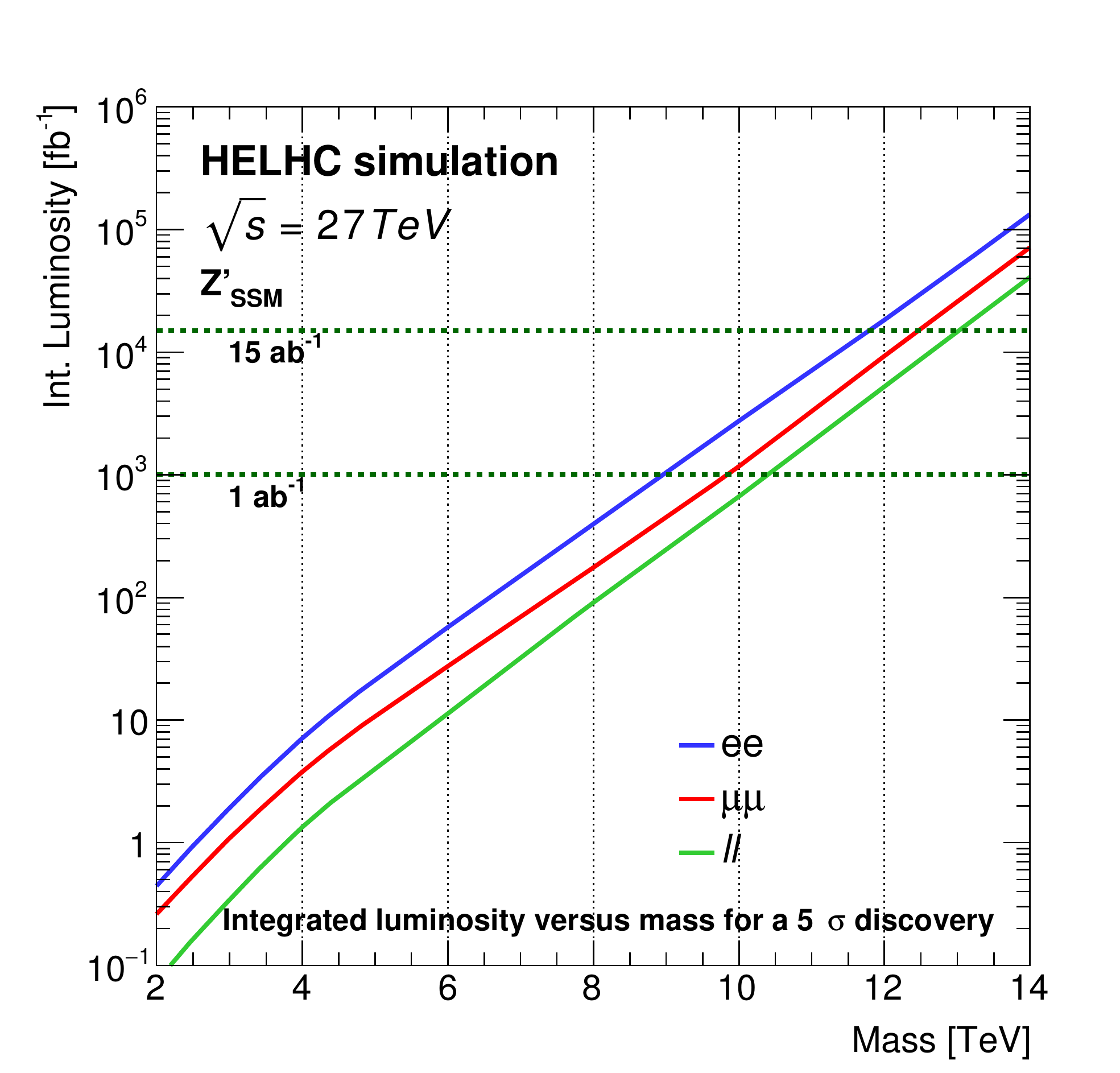}
\end{minipage}
  \caption{Exclusion limit versus mass for the dilepton (ee,$\mu\mu$) channel (left) and luminosity for a $5\sigma$ discovery (right) comparing $ee$, $\mu\mu$ and combined channels. }
  \label{fig:leptonicresonances:resultsll}
\end{figure}

\begin{figure}[t]
\centering
\begin{minipage}[t][\minipageheightdp][t]{\minipagewidthdp}
\centering
\includegraphics[width=\figuremaxwidthdp,height=\figuremaxheightdp,keepaspectratio]{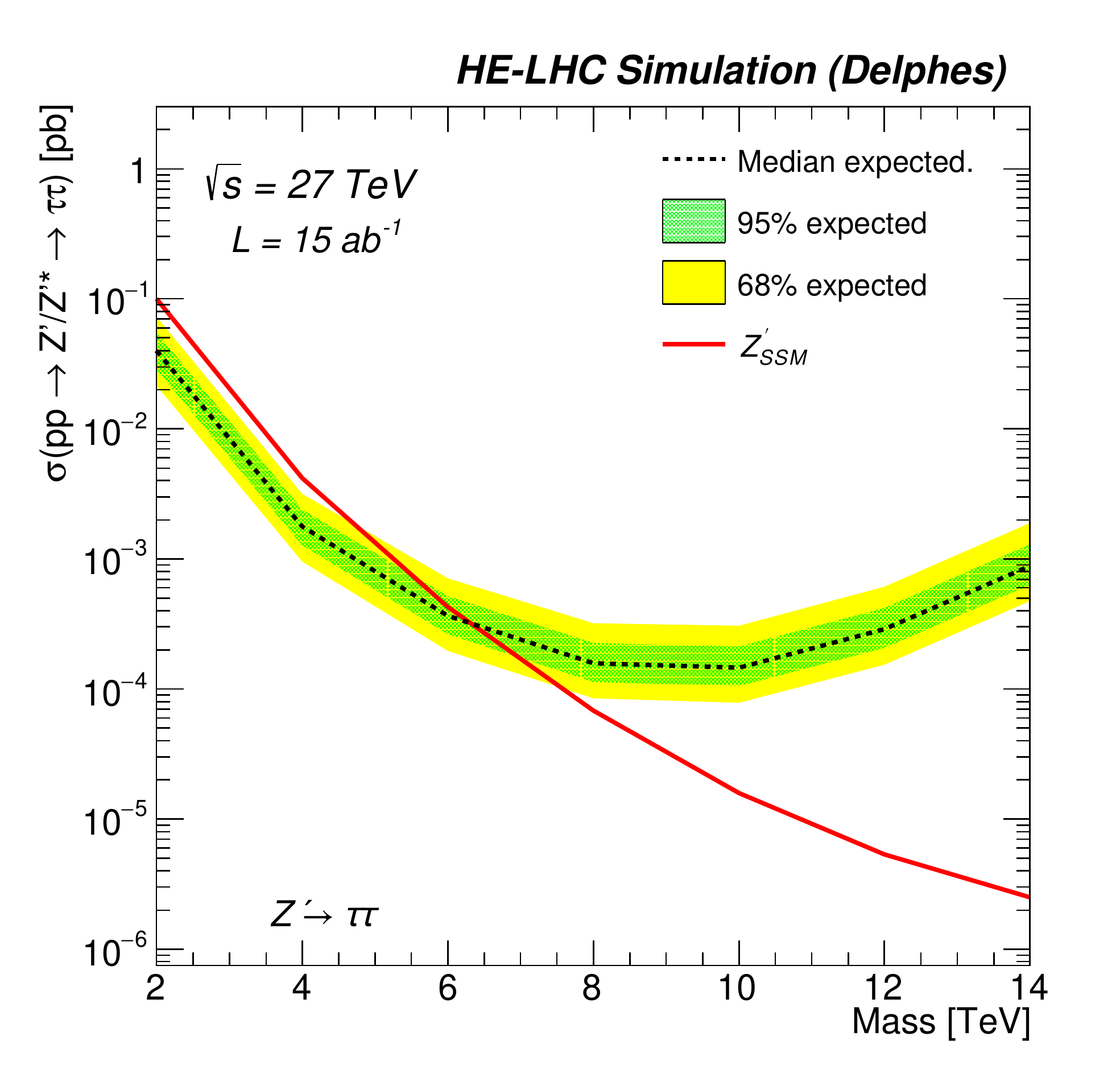}
\end{minipage}
\begin{minipage}[t][\minipageheightdp][t]{\minipagewidthdp}
\centering
\includegraphics[width=\figuremaxwidthdp,height=\figuremaxheightdp,keepaspectratio]{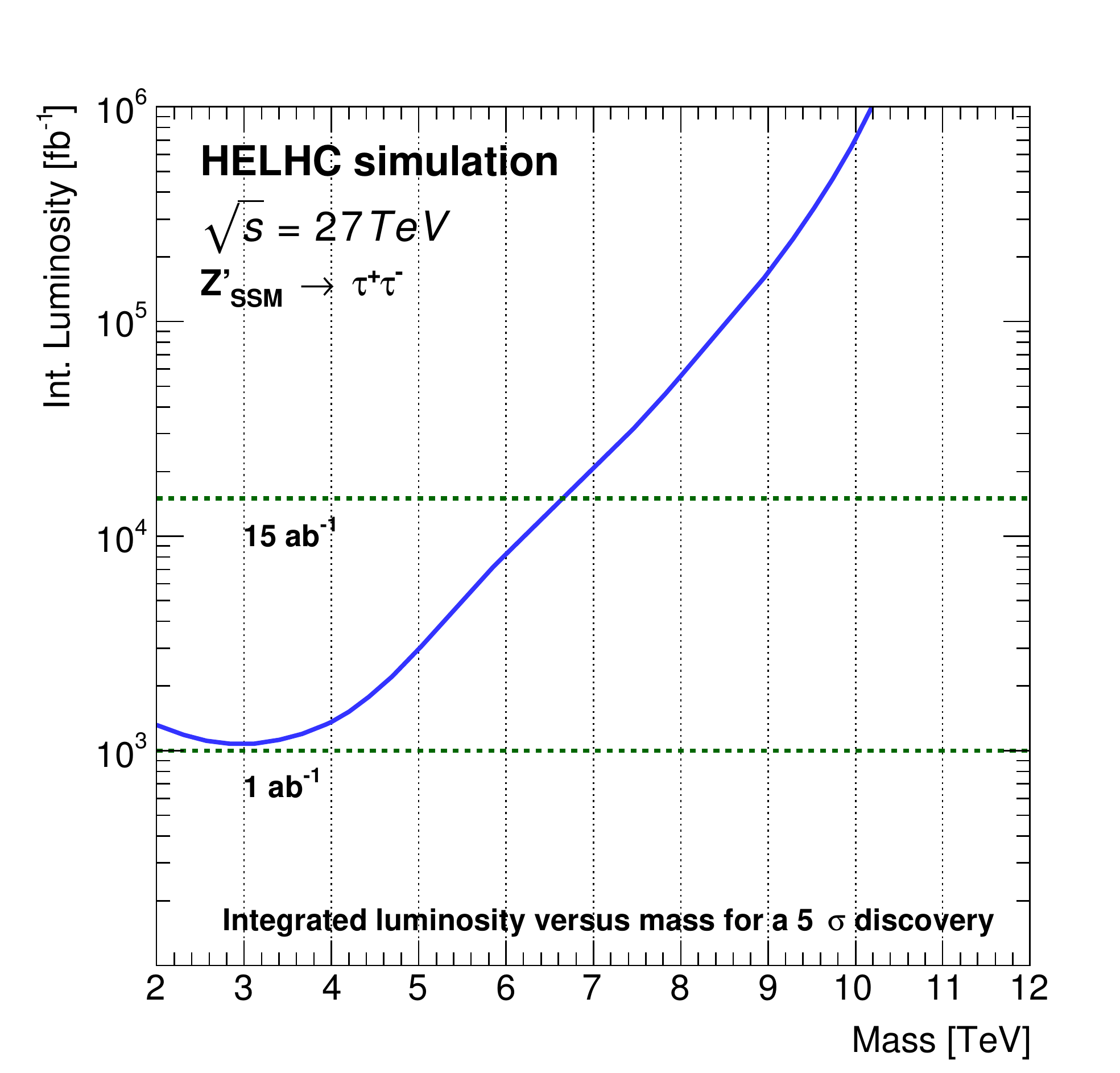}
\end{minipage}
  \caption{Exclusion limit versus mass for the ditau channel (left) and luminosity for a $5\sigma$ discovery (right). }
  \label{fig:leptonicresonances:resultstautau}
\end{figure}

The $95\%\cl$ exclusion limit obtained using \intlumihelhc of data for the combination of the ee and $\mu\mu$ channels is shown in \fig{fig:leptonicresonances:resultsll} (left) for a list of 6 different $Z'$ models. A detailed discussion on model discrimination at HE-LHC following the observation of an excess at the HL-LHC can be found in \secc{sec:Zpmodeldiscr}. We simply note here that it is possible to exclude a $Z'$ with $m_{Z'}\lesssim 10-13\TeV$ (depending on the model) at $\sqrt{s}=27\TeV$ with \intlumihelhc. Figure~\ref{fig:leptonicresonances:resultsll} (right) shows the integrated luminosity required to reach a $5\sigma$ discovery for a $\ZpSSM$ decaying leptonically as a function of the mass of the heavy resonance. Despite a worse di-lepton invariant mass resolution for the $\mu\mu$ final state, the \Zpee\ and \Zpmumu\ channel display very similar performances, due to the low background rates and a higher muon reconstruction efficiency. Acceptance for electron could have been recovered by optimising the isolation criteria, but was not done in this study. With the full dataset \intlumihelhc, a $\ZpSSM$ up to $m_{Z'}\approx 13\TeV$ can be discovered.  Figure~\ref{fig:leptonicresonances:resultstautau} shows the exclusion limits for $15\abinv$ of data (left) and the required integrated luminosity
versus mass to reach a $5\sigma$ discovery (right) for the $\tau\tau$ resonances. We find that a $\ZpSSM$ with $m_{Z'}\approx 6.5\TeV$ can be discovered or excluded. As expected, the \Zptata~final-state yields to a worse discovery potential compared to the $\ell\ell$ final states because of the presence of a much larger background contribution as well as the absence of narrow mass peak.

\subsubsection{\atlas{Prospects for $Z'\rightarrow e^+e^-,\mu^+\mu^-$ searches at the HL- and HE-LHC}}\label{sec:atlasZprimedilepton}
\contributors{M. Bugge, D. Hayden, S. Kubota, G. Lee, J-P Ochoa, J-C Rivera Vergara, M. Wielers, ATLAS}
%{\bf Author(s): M. Wielers, G. Lee, M. Marjanovic et al., ATLAS}

%Searches for Z' (SSM and other models) are performed considering the dilepton (ee, mumu) final state events. Prospects for discovery and exclusion considering 13, 14 and 15 TeV datasets as well as HE-LHC are presented. 

The sensitivity to narrow $\Zp$ bosons decaying into the $e^+e^-$ or $\mu^+\mu^-$ final state is studied for $pp$ collisions at several \com~energies: $\sqrt{s} = 13$, $14$, and $15\TeV$ with \intLumi,
as well as $\sqrt{s} = 27\TeV$ with \intlumihelhc. Results are based on studies documented in~\citeref{ATL-PHYS-PUB-2018-044}.
The latter is only studied in the $e^+e^-$ channel since the work presented here is based on the latest
layout of the upgraded ATLAS detector for the HL-LHC which is not optimised for extremely high
muon momentum measurements.
The study supersedes that from \citeref{ATL-PHYS-PUB-2014-019} since it uses the latest detector
layout and higher pileup conditions. In addition, it was found that the signal cross-sections used in the previous analysis were too high.

The projection study relies on MC simulation for the signal based on \pythia{ 8}~\cite{Sjostrand:2007gs}, the \sloppy\mbox{\textsc{NNPDF23LO}} PDF set~\cite{Ball:2012cx}, and the A14 set of tuned parameters~\cite{ATL-PHYS-PUB-2014-021} for the parton shower, hadronisation, and the underlying event. The dominant Drell-Yan background source is generated with \powhegbox~\cite{Alioli:2010xd,Frixione:2007nw} and the \textsc{CTEQ6L1} PDF set~\cite{Pumplin:2002vw} interfaced with \pythia{ 8} for the parton shower, hadronisation and the underlying event using the \textsc{AZNLO} set of tuned parameters~\cite{Aad:2014xaa}. Generated samples also include off-shell production.

\begin{figure}[t]
\centering
\begin{minipage}[t][\minipageheighttp][t]{\minipagewidthtp}
\centering
\includegraphics[width=\figuremaxwidthtp,height=\figuremaxheighttp,keepaspectratio]{\main/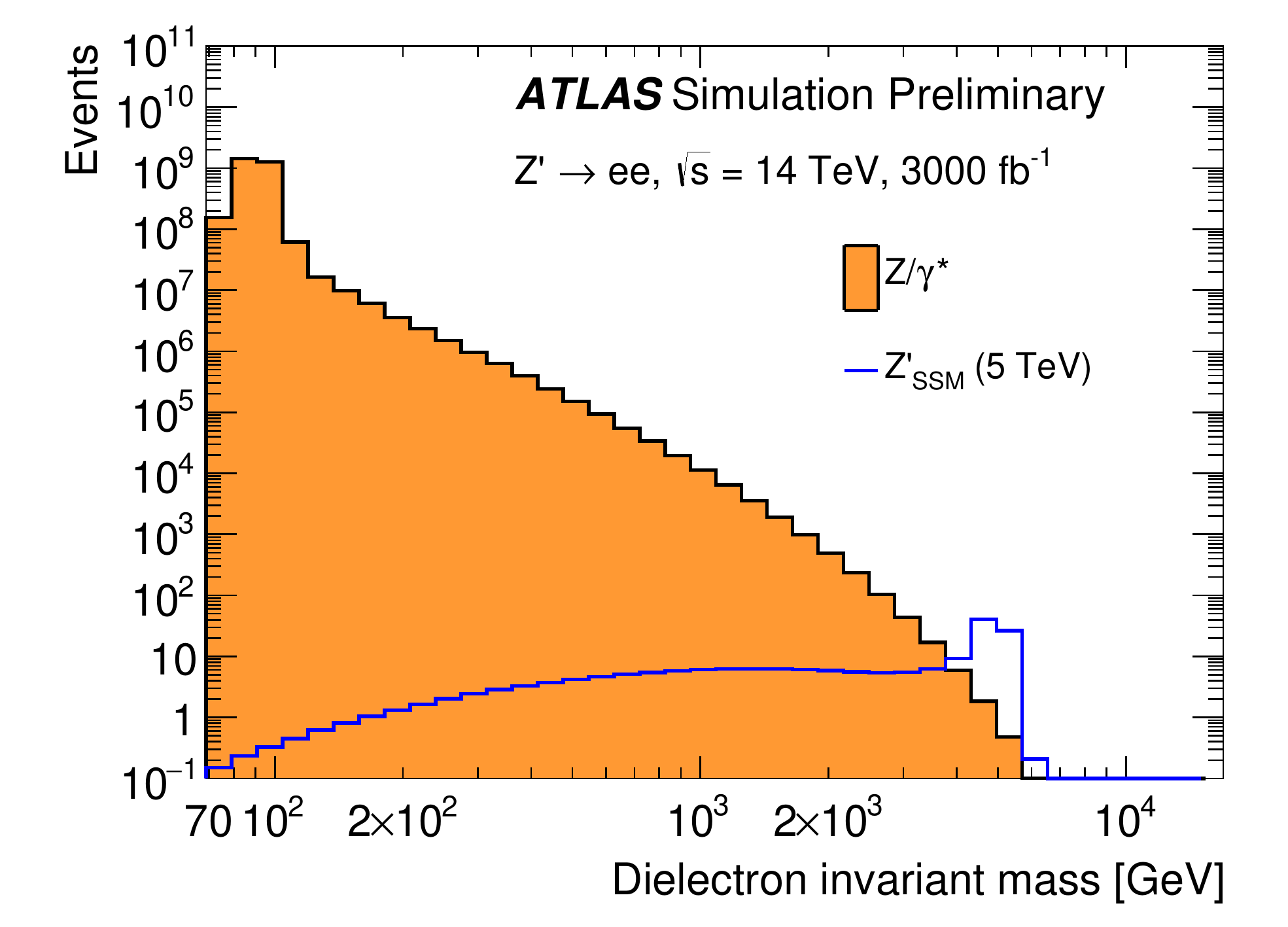}
\end{minipage}
\begin{minipage}[t][\minipageheighttp][t]{\minipagewidthtp}
\centering
\includegraphics[width=\figuremaxwidthtp,height=\figuremaxheighttp,keepaspectratio]{\main/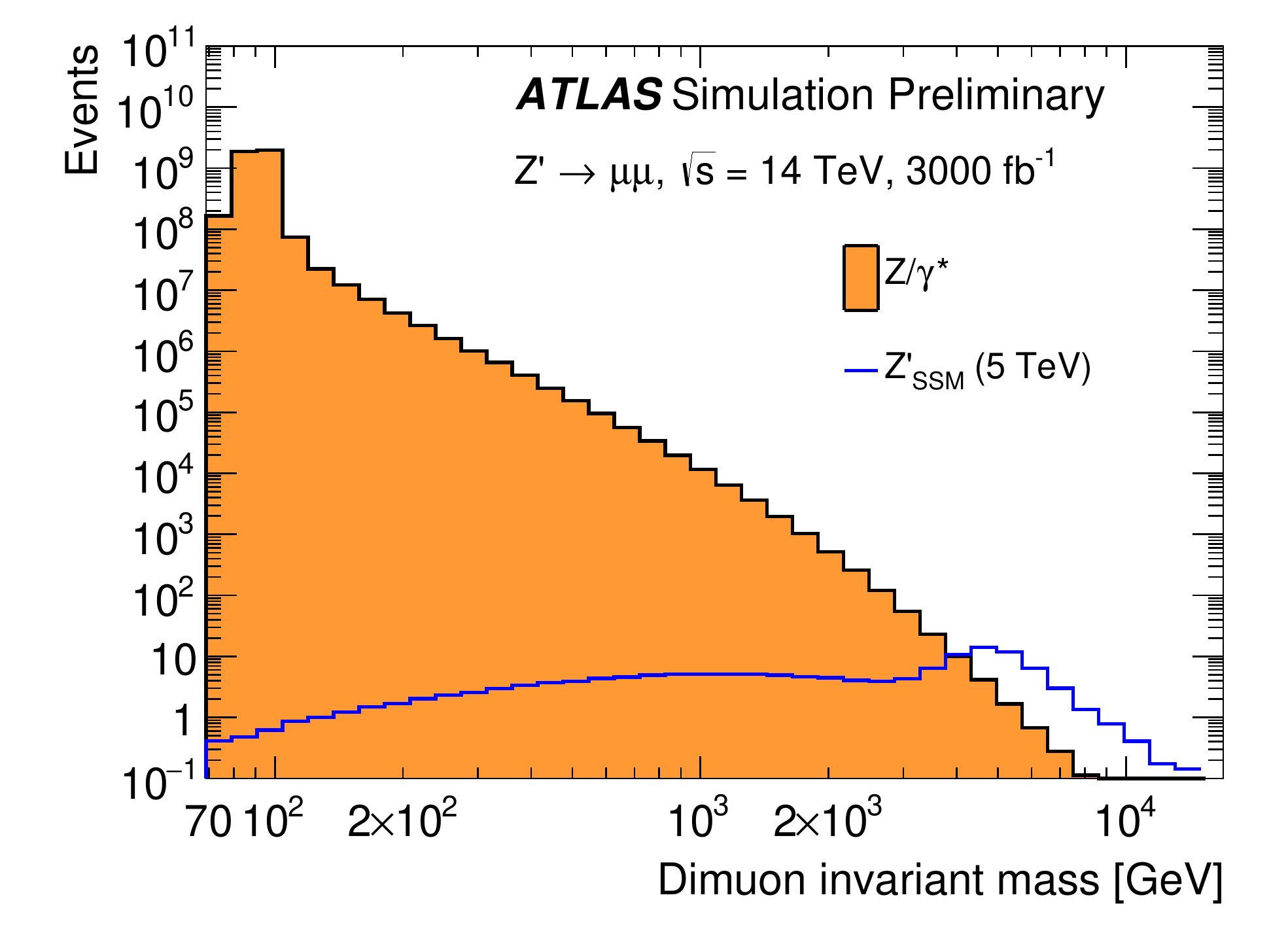}
\end{minipage}
\begin{minipage}[t][\minipageheighttp][t]{\minipagewidthtp}
\centering
\includegraphics[width=\figuremaxwidthtp,height=\figuremaxheighttp,keepaspectratio]{\main/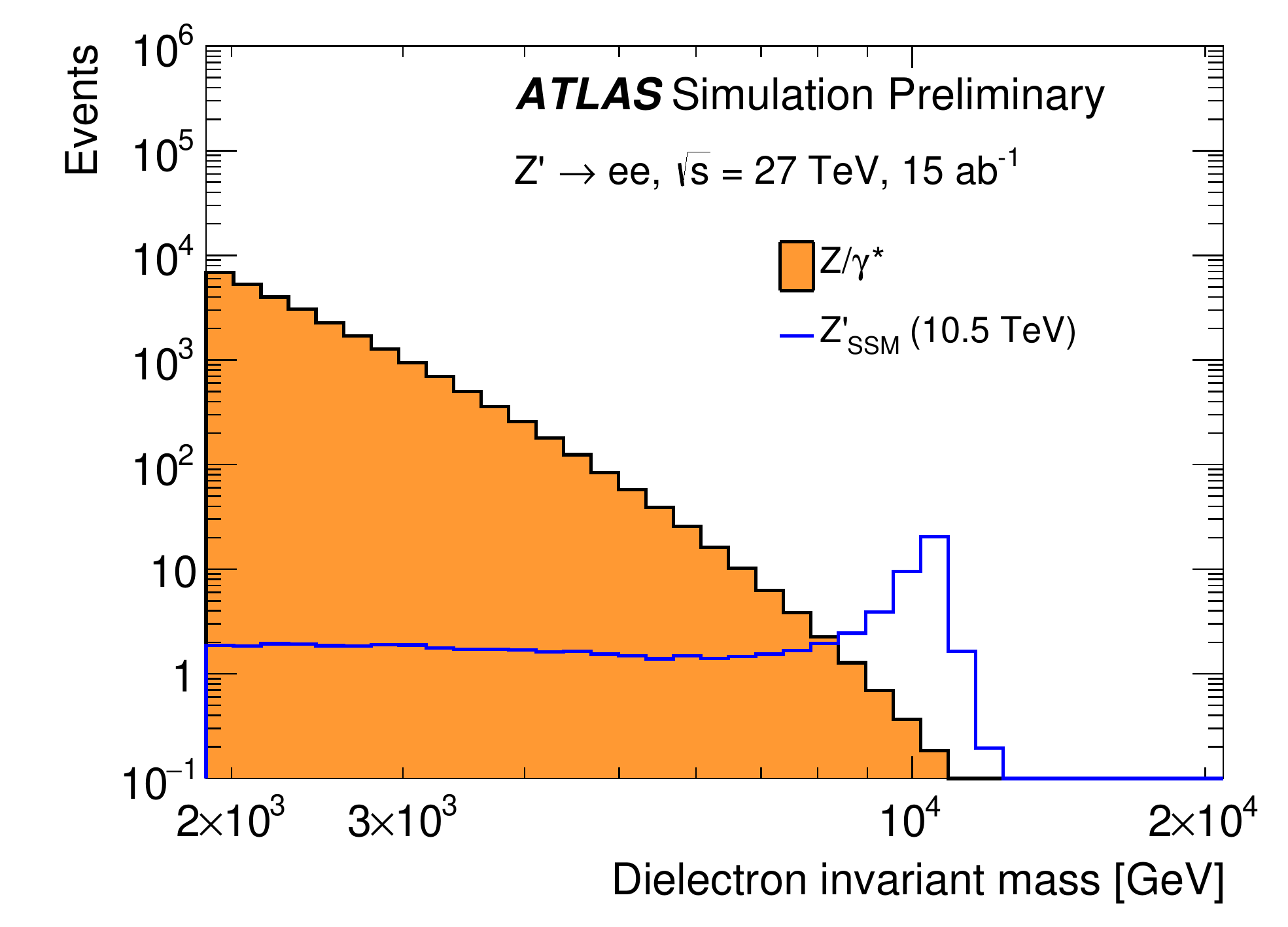}
\end{minipage}
  \caption{
    Invariant mass distributions for events satisfying all selection criteria in the dielectron and dimuon channels at $\sqrt{s}=14\TeV$ and in the dielectron channel at $\sqrt{s}=27\TeV$.
    Distributions of the Drell-Yan background and SSM $\Zp$ signal with a mass of $5.0$ ($10.5$)\TeV
    are shown for $\sqrt{s} = 14$ ($27$)\TeV.
}
\label{fig:ATLAS_Zprimell_mll}
\end{figure}

The event selection proceeds in a way similar to the analysis of the $13\TeV$ data with
36.1\fbinv~\cite{Aaboud:2017buh}.
Events must pass either the single-electron trigger requirements with $p_\mathrm{T} > 22\GeV$and
$|\eta| < 2.5$ or the single-muon trigger requirements with $p_\mathrm{T} > 20\GeV$and $|\eta| < 2.65$.
Triggered events are further required to contain exactly two electrons with $p_\mathrm{T} > 25\GeV$
and $|\eta|<2.47$ (excluding the barrel-endcap calorimeter transition region $1.37 < |\eta| < 1.52$)
or two muons with $p_\mathrm{T} > 25\GeV$and $|\eta| < 2.65$.
The electrons and muons have to satisfy the \textit{tight} and
\textit{high-$p_\mathrm{T}$}~identification requirements, respectively.
%The corresponding signal acceptance times efficiency reaches a maximum of 78\% (55\%) for $\Zp$ masses between 5 and 11~TeV in the dielectron (dimuon) channel.
%These values do not change appreciably at the different \com~energies examined here.
Invariant mass distributions of the reconstructed dielectron and dimuon candidates are shown
in \fig{fig:ATLAS_Zprimell_mll} for the $\ZpSSM$ signal and the dominant
Drell-Yan background.
Background from diboson ($WW$, $WZ$, $ZZ$) and top-quark production is not considered as their
contribution to the overall SM background is negligible for dilepton invariant masses ($m_{\ell\ell}$)
exceeding $2\TeV$. This background is more pronounced at lower masses and was found to amount to
around $10\%$ ($20\%$) of the total background for an invariant mass of $1\TeV$ ($300\GeV$),
see \citeref{Aaboud:2017buh}.
In the dielectron channel, additional background arises from $W+$jets and multijet events
in which at most one real electron is produced and one or more jets satisfy the electron selection
criteria. While this background is negligible in the dimuon channel, it amounts to approximately
$15\%$ of the total background for $m_{\ell\ell} > 1\TeV$~\cite{Aaboud:2017buh} in the dielectron channel.
This source of background is neglected in the analysis below but accounted for in the systematic
uncertainties.
The differences in the shape of the reconstructed $\Zp$ mass distributions in the dielectron and dimuon
channels arise from differences in momentum resolution for electron and muon reconstruction.
The differences in the shape of the dielectron mass distributions at $\sqrt{s}=14\TeV$ and $27\TeV$
arise from differences in the rapidity distributions.

The experimental and theoretical uncertainties assumed in this analysis are estimated from the
Run-2 results~\cite{Aaboud:2017buh} but scaled down to account for the increased statistical
precision available at the \sloppy\mbox{HL-LHC} following the recommendations in \citeref{ATLAS_PERF_Note}.
Only the largest sources of uncertainties are considered. As the uncertainties vary
with $m_{\ell\ell}$, the uncertainties are expressed relative to the value of $m_{\ell\ell}$ given in TeV.

The experimental systematic uncertainties due to the reconstruction, identification, and isolation
of electrons are negligible while those for muons add up to approximately $2.5\% \times m_{\ell\ell}~[\mathrm{TeV}]$. 
Systematic uncertainties due to the energy resolution and scale are set to $1.5\% \times m_{\ell\ell}~[\mathrm{TeV}]$. 
The uncertainties due to the resolution and reconstruction of the leptons are added in quadrature
to the dominant sources of theoretical uncertainty due to the PDFs. The uncertainties due to the choice of PDF set are taken to be 
$2.5\% \times m_{\ell\ell}~[\mathrm{TeV}]$ and the uncertainties in the parameters of the nominal PDF set are assumed
to be $5\% \times m_{\ell\ell}~[\mathrm{TeV}]$. 
Overall these uncertainties add up to $~6.5\% \times m_{\ell\ell}~[\mathrm{TeV}]$.
As the search looks for an excess in the high $m_{\ell\ell}$ tail, the sensitivity is primarily
limited by the statistical uncertainties.

\begin{figure}[t]
\centering
\begin{minipage}[t][\minipageheightsp][t]{\minipagewidthsp}
\centering
\includegraphics[width=\figuremaxwidthsp,height=\figuremaxheightsp,keepaspectratio]{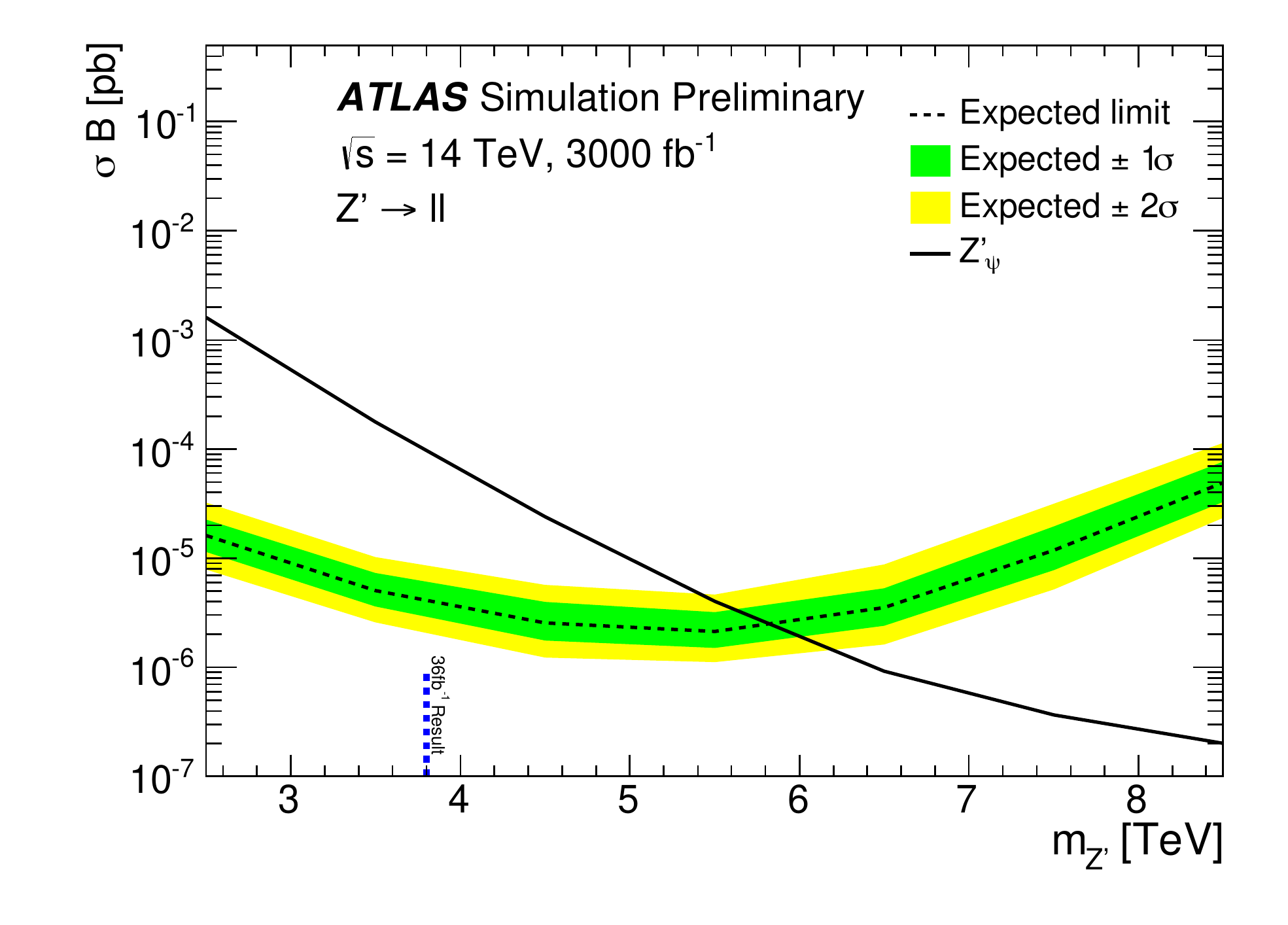}
\end{minipage}
  \caption{
    Expected (dashed black line) upper limit on cross-section times branching fraction $\sigma \times \cal{B}$  as a function of the
    $\Zp$ boson mass in the combined dielectron and dimuon channels for $\sqrt{s} = 14\TeV$ collisions
    and an integrated luminosity value of \intLumi. The $1\sigma$ (green) and $2\sigma$ (yellow)
    expected limit bands are also shown. The predicted $\sigma \times \cal{B}$ for $\Zp_\psi$ production
    is shown as a black line. These limits are based on a NNLO cross-section calculation including off-shell production ($pp \rightarrow Z'/Z^{\prime *} \rightarrow \ell \ell$). The blue marker shows the current limit obtained with the Run-2 analysis
    based on $36\fbinv$ of data.
}
  \label{fig:ATLAS_Zpll_limits}
\end{figure}

The statistical analysis relies on the Bayesian approach~\cite{Caldwell:2008fw}
used in~\citeref{Aaboud:2017buh}.
The same statistical model implementation is used in the following for both the calculation of the exclusion
limits and the discovery reach, the latter being based on a profile likelihood ratio test assuming
an asymptotic test statistic distribution.
In the absence of a signal, $95\%\cl$ upper limits are placed on the production cross section of
a $\Zp$ boson times its branching fraction $\sigma \times \cal{B}$ to a single lepton generation,
assuming lepton universality.
These limits are extracted using $\Zp$ templates binned in $m_{\ell\ell}$ for a series of $\Zp$ masses in the range between $2.5\TeV$ and $11.5\TeV$.
The interpretation of results is performed in the context of the SSM and the $E_6~\psi$ model. Exclusion limits are shown in~\fig{fig:ATLAS_Zpll_limits} assuming the $E_6~\psi$ model as a benchmark. These limits, as well as the ones presented below, are based on a NNLO cross-section calculation including off-shell production. 

Lower mass limits and the discovery reach for the different models and $\sqrt{s}$ values
at the HL-LHC are summarised in \tabb{tab:ATLAS_zplllimits}. %and visualised in Figure~\ref{fig:zplllimits}.
The projected exclusion limits extend the current $\ZpSSM$ ($\Zp_\psi$) lower mass limit of $4.5$ ($3.8$)\TeV
obtained using $36.1\fbinv$ of data taken at $\sqrt{s} = 13\TeV$ to $6.5$ ($5.8$)\TeV for $\sqrt{s} = 14\TeV$. 
Higher limits are obtained in the dielectron channel due to the superior energy resolution of
the calorimeter as compared with the momentum resolution for muons in the muon spectrometer.
Assuming similar detector performance at the LHC and HL-LHC, a corresponding lower mass limit of $5.4$ ($4.8$)\TeV are expected with $300\fbinv$at the end of Run-3. 
The $95\%\cl$ limits and discovery reach are close due to the absence of background at very high $m_{\ell\ell}$.
Compared to the results presented in \citeref{CERN-LHCC-2017-018} the discovery reach reported
here is higher due to a change in how the reach is calculated. The analysis described here
is based on the shape of the signal and background $m_{\ell\ell}$ distributions while the
$5\sigma$ significance was calculated in a mass range between $m(\Zp)/2$ to infinity in
\citeref{CERN-LHCC-2017-018}.
\begin{table}[t]
  \centering
  \small\begin{tabular}{c|cc|cc|cc}
    \hline
    \hline
    &  \multicolumn{2}{c|}{$\sqrt{s} = 13\TeV$} &  \multicolumn{2}{c|}{$\sqrt{s} = 14\TeV$} &  \multicolumn{2}{c}{$\sqrt{s} = 15\TeV$} \\
    Decay     &  Exclusion & Discovery&  Exclusion& Discovery &  Exclusion & Discovery \\
    \hline
    $\ZpSSM \to ee$  & $6.0\TeV$& $5.9\TeV$ & $6.4\TeV$ & $6.3\TeV$ & $6.7\TeV$& $6.6\TeV$\\
    $\ZpSSM \to \mu\mu$& $5.5\TeV$& $5.4\TeV$& $5.8\TeV$& $5.7\TeV$& $6.0\TeV$& $5.9\TeV$\\ \hline
    $\ZpSSM \to \ell\ell$  & $6.1\TeV$& $6.1\TeV$&$6.5\TeV$& $6.4\TeV$& $6.7\TeV$& $6.7\TeV$\\
    \hline
    \hline
    $\Zp_\psi \to ee$  & $5.3\TeV$& $5.3\TeV$& $5.7\TeV$& $5.6\TeV$ & $6.1\TeV$& $6.0\TeV$\\
    $\Zp_\psi \to \mu\mu$& $4.9\TeV$ & $4.6\TeV$& $5.2\TeV$& $5.0\TeV$& $5.5\TeV$& $5.2\TeV$\\ \hline
    $\Zp_\psi \to \ell\ell$  & $5.4\TeV$& $5.4\TeV$&$5.8\TeV$& $5.7\TeV$& $6.1\TeV$& $6.1\TeV$\\
    \hline
    \hline
  \end{tabular}\normalsize
  \caption{Expected $95\%\cl$ lower limit on the $\Zp$ mass in TeV in the dielectron and dimuon channels and their combination for two benchmark $\Zp$ models for different centre of mass energies assuming \intLumi of data to be taken at the HL-LHC. In addition, the discovery reach for finding such new heavy particles is shown.}
  \label{tab:ATLAS_zplllimits}
\end{table}

The discovery reach and lower exclusion limits at $95\%\cl$ in mass are also calculated for a detector at the HE-LHC in the dielectron channel. This is done assuming the same physics performance as for the ATLAS detector at the HL-LHC.
The exclusion limits and the discovery reach are summarised in \tabb{tab:ATLAS_zplllimits_27}.

\begin{table}[t]
  \centering
  \small\begin{tabular}{c|cc}
    \hline
    \hline
    Decay     &  Exclusion [$\mathrm{TeV}$]& Discovery [$\mathrm{TeV}$]\\
    \hline
    $\ZpSSM \to ee$ & $12.8$ & $12.8$\\
    $\Zp_\psi \to ee$ & $11.4$ & $11.2$\\ 
    \hline
    \hline
  \end{tabular}\normalsize
    \caption{Lower limits at $95\%\cl$ and discovery reach on the $\ZpSSM$ and $\Zp_\psi$ boson mass in the dielectron channel assuming \intlumihelhc of $pp$ data to be taken at the HE-LHC with $\sqrt{s}=27\TeV$.}
  \label{tab:ATLAS_zplllimits_27}
\end{table}
At the HE-LHC, $\ZpSSM$ and $\Zp_\psi$ bosons can be discovered up to $12.8\TeV$ and $11.2\TeV$, respectively,
thus increasing their discovery reach by $6.5\TeV$ compared to the HL-LHC, \iee~an increase in the
discovery potential by a factor of two.
In case $\Zp$ bosons are not discovered yet, the HE-LHC will be able to further rule out
$\ZpSSM$ and $\Zp_\psi$ bosons up to $12.8\TeV$ and $11.4\TeV$, respectively.

\subsubsection{\atlas{$W'\rightarrow e\nu, \mu\nu$ or $tb, t\rightarrow b\ell\nu$ searches at HL-LHC
%Prospects for $W'$ searches at HL-LHC
}}
\contributors{M. Bugge, J. Donini, D. Hayden, G. Lee, K. Lin, M. Marjanovic, L. Vaslin, M. Wielers, ATLAS}
%{\bf Author(s): M. Wielers, G. Lee, M. Marjanovic et al., ATLAS}

%Searches for W' are performed in lepton + missing ET final states and in top+bottom final states considering the HL-LHC dataset. 

\paragraph*{Resonances decaying into a lepton and missing transverse momentum}

The sensitivity to $W^\prime$ resonances decaying into an electron or a muon and a neutrino is studied
for \sloppy\mbox{$\sqrt{s} = 14\TeV$} $pp$ collisions at the HL-LHC~\cite{ATL-PHYS-PUB-2018-044}.
Such resonances would manifest themselves as an excess of events above the SM background at high
transverse mass $m_\mathrm{T}$.
The SM background mainly arises from processes with at least one prompt final-state electron or muon,
with the largest source being off-shell charged-current Drell-Yan (DY),
leading to a final state with an electron or a muon and a neutrino.
Other non-negligible contributions are from top-quark pair and single-top-quark production,
neutral-current DY process, diboson production, and from events in which one
final-state jet or photon satisfies the lepton selection criteria. This last component of the background,
referred to in the following as the multijet background, receives contributions from
multijet, heavy-flavour quarks and $\gamma$ + jet production; it is one of the smallest backgrounds in
this analysis. It is evaluated in a data-driven way in the Run-2 analysis and cannot be yet reliably
estimated from MC samples and is therefore not considered here.
It was found to be negligible in the muon channel at $m_\mathrm{T} > 3\TeV$ in the
Run-2 analysis based on $79.8\fbinv$ of $pp$ collisions~\cite{ATLAS:2018lcz}.
In the electron channel, the contribution constitutes around $10\%$ of the total background
at $m_\mathrm{T} \approx 3\TeV$ and mainly arises from jets misidentified as electrons. 

The projection study relies on MC simulation with the SSM $W^\prime$ signal generated using \pythia{ 8} in the same setup as for the SSM $\Zp$ signal described in~\secc{sec:atlasZprimedilepton}. This also includes off-shell production.  
The charged and neutral Drell-Yan background is also generated in the same way. Background from $t\bar{t}$ events is produced with \powhegbox{} and the \textsc{NNPDFL30NNLO} PDF set interfaced with \pythia{ 6} using the A14 tune. Diboson events are generated with \sherpa~\cite{Gleisberg:2008ta} and the \textsc{CT10} PDF set~\cite{Lai:2010vv}.

The event selection proceeds similarly to the Run-2 analysis described in \citeref{ATLAS:2018lcz}.
Events are required to satisfy the single-election or single-muon triggers.
The single electron trigger selects events containing at least one electron with $p_\mathrm{T} > 22\GeV$
and $|\eta| < 2.5$, while the single muon trigger requires a muon with $p_\mathrm{T} > 20\GeV$ and
$|\eta| < 2.65$.
Events are required to contain exactly one lepton which can be either an electron or a muon.
Muons must have $p_\mathrm{T} > 55\GeV$ and $|\eta| < 2.65$ as well as satisfy
the \sloppy\mbox{\textit{high-$p_\mathrm{T}$}} identification criteria~\cite{ATLAS_PERF_Note}. 
Electrons must have $p_\mathrm{T} > 55\GeV$ and $|\eta| < 1.37$ or $1.52 <|\eta| < 2.47$, as well as satisfy
the {\textit{tight}} identification criteria.
These $p_\mathrm{T}$ thresholds are the same as in the Run-2 analysis and are motivated by the triggers
which select events containing leptons with loose identification criteria and without isolation requirements.
Though not applied in this analysis, such events will be needed for the data-driven background subtraction
methods, as employed in Run-2, to work. The $p_\mathrm{T}$ thresholds for these ``looser'' triggers are not
yet available and therefore in the following it is assumed that the thresholds will be similar to those
used in Run-2. The magnitude of the missing transverse momentum ($E_\mathrm{T}^\mathrm{miss}$) must exceed $55\GeV$ ($65\GeV$) in
the electron (muon) channel. Events in both channels are vetoed if they contain additional leptons satisfying
loosened selection criteria, namely electrons with $p_\mathrm{T} > 20\GeV$satisfying the \textit{medium}
identification criteria or muons with \sloppy\mbox{$p_\mathrm{T} > 20\GeV$} passing the \textit{loose} muon selection. 

\begin{figure}[t]
\centering
\begin{minipage}[t][\minipageheightdp][t]{\minipagewidthdp}
\centering
\includegraphics[width=\figuremaxwidthdp,height=\figuremaxheightdp,keepaspectratio]{\main/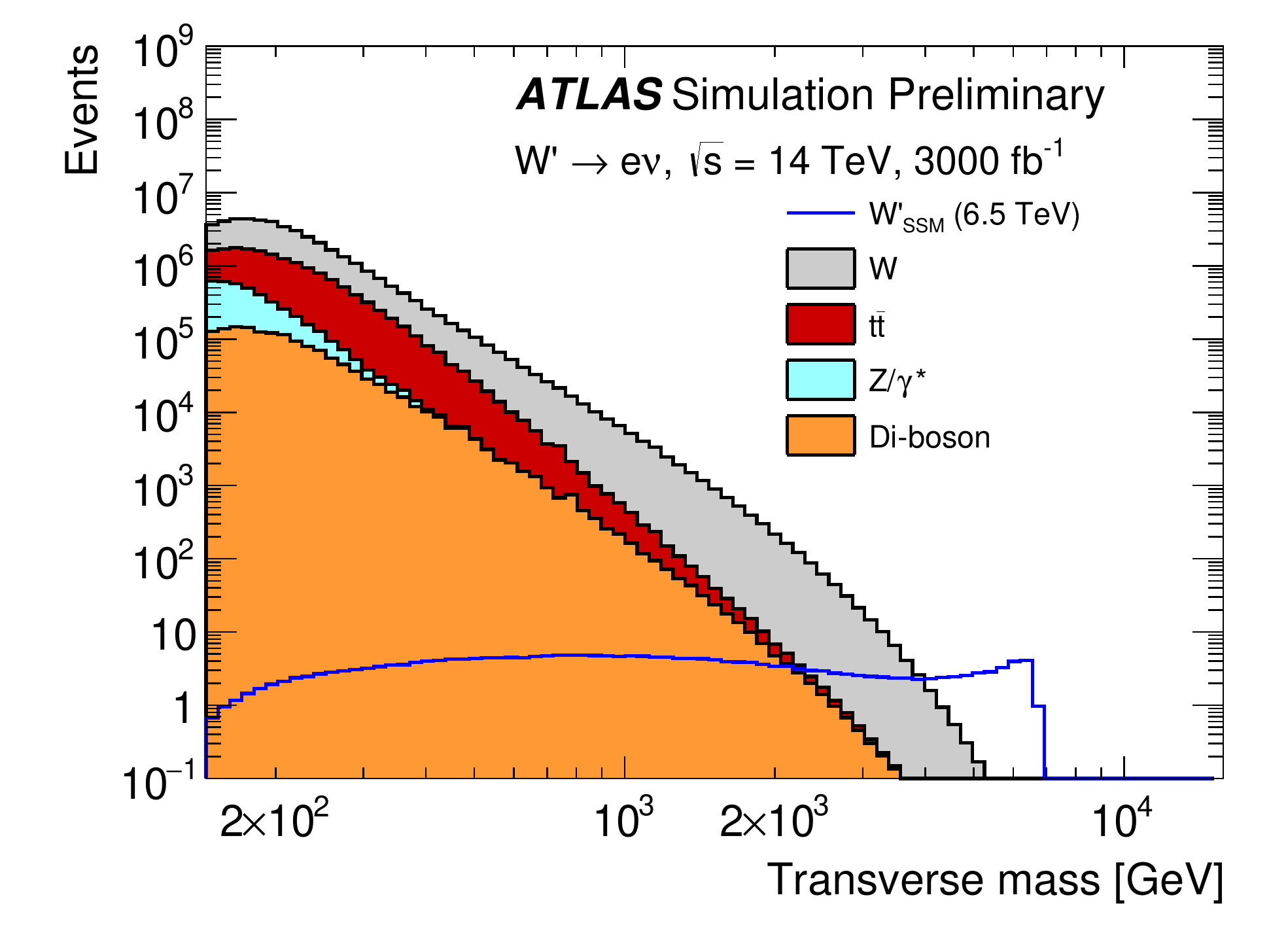}
\end{minipage}
\begin{minipage}[t][\minipageheightdp][t]{\minipagewidthdp}
\centering
\includegraphics[width=\figuremaxwidthdp,height=\figuremaxheightdp,keepaspectratio]{\main/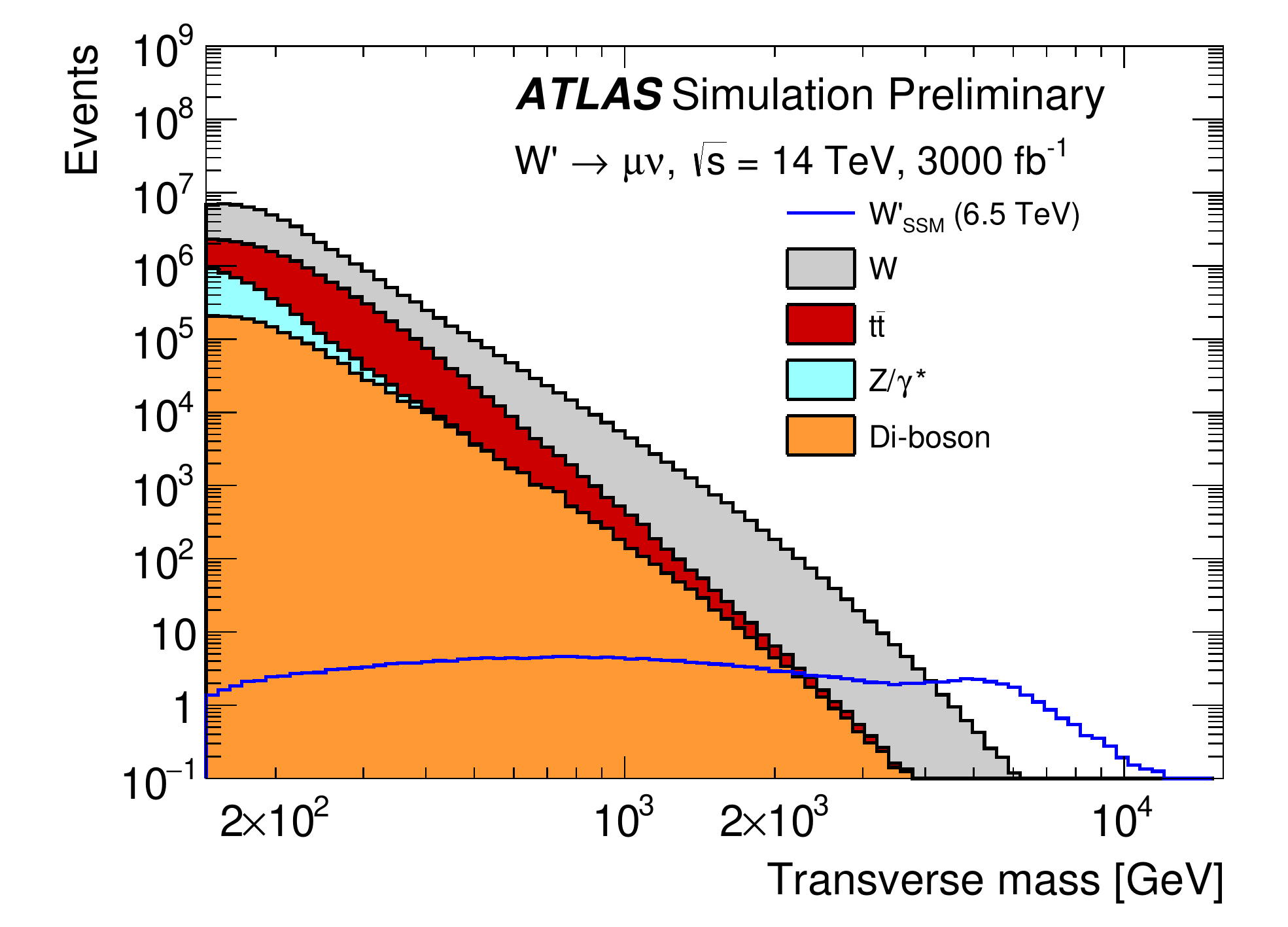}
\end{minipage}
\caption{
  Transverse mass distributions for events satisfying all selection criteria in the electron and muon
  channels of the $W^\prime\to \ell \nu$ search. The different background contributions are shown
  as a stacked sum and the expected signal distributions for a $W^\prime$ boson with a mass of $6.5\TeV$
  is shown. The bin width is constant in $\log m_\mathrm{T}$. 
}
  \label{fig:ATLAS_Wprimelv_mT}
\end{figure}

The total acceptance times efficiency in the electron (muon) channel decreases from
a value of $\sim 85\%$ ($70\%$) at a $W^\prime$ mass of $1\TeV$ to $\sim 65\%$ ($60\%$) for masses between
$5$ and $9\TeV$.
The resulting $m_\mathrm{T}$ distributions are shown in \fig{fig:ATLAS_Wprimelv_mT}
for both the expected background and the $W^\prime$ signal with a mass of $6.5\TeV$. 

Systematic uncertainties arise from both experimental and theoretical sources.
Since the uncertainties from the Run-2 analysis are found to increase as a function of $m_\mathrm{T}$
these are parametrised as a percentage of the $m_\mathrm{T}$ value expressed in units of TeV.
The uncertainties are then scaled down to account for the increased statistical power at the HL-LHC
according to recommendations in \citeref{ATLAS_PERF_Note}.
The experimental systematic uncertainties due to the reconstruction, identification, and isolation of
muons result in a value of $2.5\% \times m_{\mathrm T}~[\mathrm{TeV}]$, while these uncertainties are negligible
for electrons. 
Systematic uncertainties due to the energy resolution and scale are set to
$2.5\% \times m_{\mathrm T}~[\mathrm{TeV}]$.
The main systematic uncertainties in the $E_\mathrm{T}^\mathrm{miss}$ calculation and on the jet energy scale are found
to be negligible in Run-2 and are therefore not considered in this analysis.
Theoretical uncertainties are related to the production cross sections estimated from MC simulation.
The effects when propagated to the total background estimate are significant for charged and neutral current DY, and to some extent for top-quark production, but are negligible for diboson production.
No theoretical uncertainties are considered for the $W^\prime$ boson signal in the statistical analysis.
The largest uncertainties arise from the PDF uncertainty in the DY background.
The uncertainties due to the choice of PDF set are taken to be 
$5\% \times m_{\ell\ell}~[\mathrm{TeV}]$ and the uncertainties in the parameters of the nominal PDF set are assumed
to be $2.5\% \times m_{\ell\ell}~[\mathrm{TeV}]$. 
The uncertainty in the multijet background in the electron channel is assumed to be
$2.5\% \times m_{\mathrm T}~[\mathrm{TeV}]$.
Overall these uncertainties in the background event yield add up to $\sim 7\% \times m_\mathrm{T}~[\mathrm{TeV}]$.
As the search looks for an excess in the high $m_\mathrm{T}$ tail, the sensitivity is primarily
limited by the statistical uncertainties.

\begin{figure}[t]
\centering
\begin{minipage}[t][\minipageheightsp][t]{\minipagewidthsp}
\centering
\includegraphics[width=\figuremaxwidthsp,height=\figuremaxheightsp,keepaspectratio]{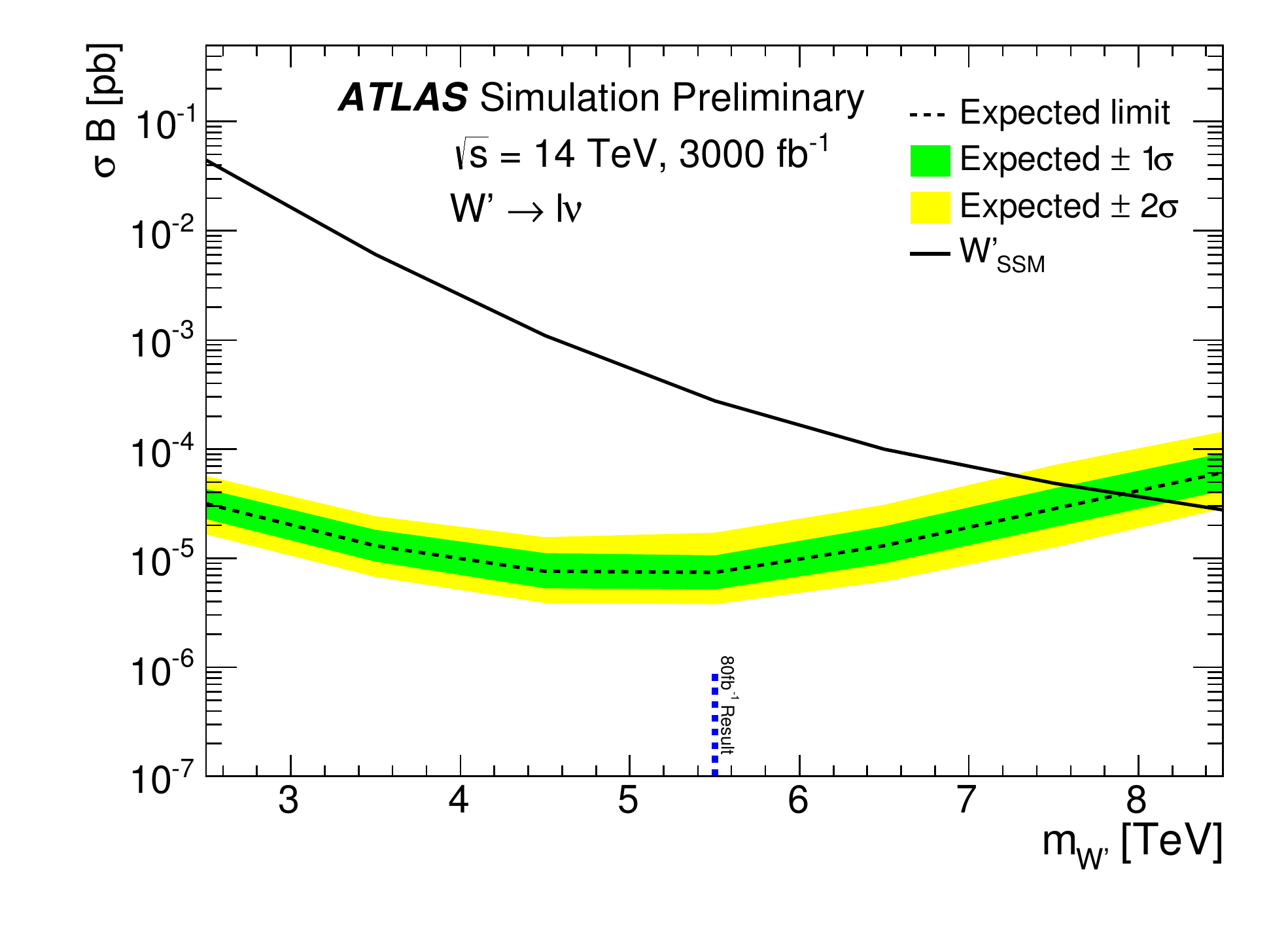}
\end{minipage}
\caption{
    Expected (dashed black line) upper limit on cross section times branching fraction ($\sigma\times\cal{B}$)
    as a function of the $W^\prime$ boson mass in the electron, muon, and combined electron and muon channels
    of the $W^\prime \to \ell \nu$ search assuming \intLumi of data. The $1\sigma$ (green) and $2\sigma$
    (yellow) expected limit bands are also shown.
    The predicted $\sigma\times\cal{B}$ for $W^\prime$ production in the SSM is shown as a black line. These limits are based on a NNLO cross-section calculation including off-shell production ($pp \rightarrow W^{'}/W^{'*} \rightarrow \ell \nu$). 
    The blue marker shows the current limits obtained with the latest Run-2 analysis based on $79.8\fbinv$
    of data.
}
  \label{fig:ATLAS_wplnulimits}
\end{figure}

The statistical analysis relies on a Bayesian approach to set cross section times branching fraction
upper limits and a profile likelihood approach to derive the discovery reach as
for the $\Zp \to \ell\ell$ search described above. The branching fraction corresponds to that
for decays into a single lepton generation, assumed to be universal in the combination of the two channels.
The $95\%\cl$ upper limit on $\sigma\times\cal{B}$ as a function of $W^\prime$ mass is shown
in \fig{fig:ATLAS_wplnulimits} for an integrated luminosity of \intLumi after
combination of the electron and muon channels.
The upper limits on $\sigma\times\cal{B}$ for $W^\prime$ bosons start to weaken above a pole mass
of $\sim 5\TeV$, which is mainly caused by the combined effect of a rapidly falling signal cross section
towards the kinematic limit and the increasing proportion of the signal being produced off-shell in
the low-$m_\mathrm{T}$ tail of the signal distribution.
The $W^\prime$ bosons in the SSM can be excluded up to masses of $7.6$ ($7.3$)\TeV in the electron (muon) channel. These limits are based on a NNLO cross-section calculation including off-shell production for the signal. 
The limits in the electron channel are stronger due to the superior energy resolution of the calorimeter
for high-momentum electrons as compared to that of the muon spectrometer for high-momentum muons.
The combination of the two channels increases the limits to just over 7.9~TeV. This is an improvement
of more than $2\TeV$ with respect to the current exclusion limits using $79.8\fbinv$ of $\sqrt{s}=13\TeV$ data.
For comparison, assuming the performance of the upgraded ATLAS detector and a luminosity of $300\fbinv$,
$W^\prime$ masses up to $6.7\TeV$ can be excluded for the combined electron and muon channels. Though the detector resolutions for the upgraded detector at the HL-LHC are applied, this is a good approximation of the reach with the current detector at the end of LHC Run-3. 

The discovery reach is based on a $5\sigma$ significance. In the context of the SSM, $W^\prime$ bosons
can be discovered up to masses of $7.7\TeV$. The discovery reach is shown in \tabb{tab:ATLAS_wplnulimits}
together with the exclusion limits discussed above.
As can be seen, the discovery reach typically is only few hundred GeV lower than the mass limits obtained
with a background-only hypothesis.
The similarity of the values for the discovery reach and the exclusion limit is expected,
as in the high-$m_\mathrm{T}$ tail the background contribution approaches zero,
while the number of signal events is about three.
The expected reach with $300\fbinv$of data will be $1.2\TeV$ lower assuming the same detector performance. 
\begin{table}[t]
  \centering
  \small\begin{tabular}{c|cc}
    \hline
    \hline
    Decay     &  Exclusion [$\mathrm{TeV}$]& Discovery [$\mathrm{TeV}$]\\
    \hline
    $W^\prime_\mathrm{SSM} \to e\nu$   & $7.6$ & $7.5$\\
    $W^\prime_\mathrm{SSM} \to \mu\nu$ & $7.3$ & $7.1$\\ \hline
    $W^\prime_\mathrm{SSM} \to \ell\nu$  & $7.9$ & $7.7$\\
    \hline
    \hline
  \end{tabular}\normalsize
  \caption{Expected $95\%\cl$ lower limit on the $W^\prime$ mass in the electron and muon channels
    as well as their combination in the context of the SSM assuming \intLumi of data.
    In addition, the discovery reach for finding such new heavy particles is shown. These limits are based on a NNLO cross-section calculation including off-shell production ($pp \rightarrow W^{'}/W^{'*} \rightarrow \ell \nu$).}
  \label{tab:ATLAS_wplnulimits}
\end{table}

\paragraph*{Resonances decaying into a top quark and a bottom quark}

The search for $W^\prime$ bosons in the lepton plus neutrino channel is sensitive to large
mass scales but it is not sensitive to right-handed $W^\prime$ bosons.
This can be alleviated by searching for $W^\prime_{\mathrm{R}} \to t\,\bar{b}$ decays
with subsequent decays $t \to W b$ and $W \to \ell \nu$. 
The final-state signature consists of two $b$-quarks, one charged lepton (electron or muon) and 
$E_{\mathrm{T}}^{\mathrm{miss}}$ from the escaping neutrino. 

Events are required to pass one of the single-lepton triggers: at least one electron
with \sloppy\mbox{$p_\mathrm{T} > 22\GeV$} and $|\eta|<2.5$ or
at least one muon with $p_\mathrm{T} > 20\GeV$ and $|\eta|<2.65$. 
Electrons must satisfy the \textit{tight} identification requirements~\cite{Aad:2014fxa} 
requirements and have $p_\mathrm{T} > 25\GeV$ and $|\eta| < 2.47$ but outside the barrel--endcap
transition region, $1.37 < |\eta| < 1.52$.
Similarly, muon candidates must meet the \textit{tight} identification criteria~\cite{Aad:2016jkr} 
and have $p_\mathrm{T} >25\GeV$ and $|\eta| <2.65$. 

The projection study relies on MC simulation for the $W^\prime$ signal
based on \sloppy\mbox{\MGvATNLO}  with the \textsc{NNPDF23LO} PDF set interfaced to \pythia{ 8} and the A14 tune for the parton shower, hadronisation, and the underlying event. Background for the various top-quark production mechanisms is generated by \powhegbox. In the case of the dominant $t\bar{t}$ background, events are produced with the \textsc{CTEQ6L1} PDF set and interfaced to \pythia{ 6} using the \textsc{Perugia2012} tune~\cite{Skands:2010ak}. $W$+jets events are produced with \MGvATNLO with the \textsc{NNPDF23NLO} PDF set interfaced to \pythia{ 8} and the A14 tune, whereas $Z$+jets events are produced with \powhegbox and the \textsc{CT10} PDF set interfaced to \textsc{Pythia8} and the AU2 tune~\cite{ATLAS:2012uec}. Diboson events are generated as for the $W^\prime \to \ell\nu$ search above.

The dominant background processes are the production of $t\bar{t}$ pairs and $W+$jets. 
Smaller contributions are also expected from single top quarks ($t$-channel, $Wt$ and $s$-channel), 
$Z+$jets and diboson ($WW$, $WZ$, and $ZZ$) production. 
All background processes are modelled with MC simulation. 
Instrumental background coming from misidentified electrons, referred to as the multijet 
background, is also present but it is very small and further suppressed by applying dedicated 
selection criteria, and it is neglected in the following.
Events are required to satisfy $E_{\mathrm{T}}^{\mathrm{miss}}>80$ ($30$)\GeV in the electron (muon)
channel as well as $m_{\textrm{T}}^W$ + $E_{\mathrm{T}}^{\mathrm{miss}} > 100\GeV$.

The $W^\prime$ candidates are built from $W$ boson and top-quark candidates.
The $W$ bosons are reconstructed from the lepton--$E_{\mathrm{T}}^{\mathrm{miss}}$ system with
the longitudinal momentum component of the neutrino from the $W$ decay extracted by imposing a
$W$-boson mass constraint.
This $W$ boson candidate is then combined with all selected jets in the event to reconstruct
a top-quark candidate as the $W$+jet combination that has a mass closest to the top-quark mass.
The jet used to form the top-quark candidate is referred to as ``$b_{\mathrm{top}}$''.
Finally, the candidate $W^\prime$ boson is reconstructed by combining the top-quark candidate with
the  highest-$p_{\mathrm{T}}$ remaining jet (referred to as ``$b_1$''). 
The invariant mass of the reconstructed $W^\prime \rightarrow t\bar{b}$ system ($m_{t\bar{b}}$) 
is the discriminating variable of this search. 
An event selection common to all signal regions is defined as: lepton 
$p_{\textrm{T}}>50\GeV$, $p_{\textrm{T}}(b_1)> 200\GeV$, and
$p_{\textrm{T}}({\textrm{top}})> 200\GeV$. As the signal events are expected to be boosted, 
the angular separation between the lepton and $b_{\text{top}}$ is required to satisfy 
$\Delta R(\ell, b_{\text{top}})<1.0$. 

The phase space is divided into eight signal regions (SR) defined by the number of jets and 
\sloppy\mbox{$b$-tagged} jets, and are labelled as ``$X$-jet $Y$-tag'' where $X=2, 3$ and $Y=1, 2$, 
that are further separated into electron and muon channels.
The signal selection acceptance times efficiency rises from $\sim 4.4\%$ ($7.7\%$) at
a $W^\prime_{\mathrm{R}}$ mass of $1\TeV$ to $4.6\%$ ($11.0\%$) at $2\TeV$, then decreasing to
$2.6\%$ ($8.2\%$) at $5\TeV$ in the electron (muon) channel. This decrease is due to the
$b$-tagging performance and the higher boost at higher mass.
The muon channel outperforms the electron channel due to overlap removal requirements, 
as they are relaxed by using a variable $\Delta R$ cone size. The variable $\Delta R$ cone size is not used 
for electrons because of the possible double counting of the energies of electron and jet.

Systematic uncertainties are evaluated following the analysis of $36.1\fbinv$ of
$\sqrt{s}=13\TeV$ $pp$ data in~\citeref{Aaboud:2018jux}
and then scaled according to the recommendations in \citeref{ATLAS_PERF_Note}. 
The uncertainty in the luminosity ($1\%$) and in the theory cross sections ($5\%$ for diboson, $10\%$ for 
$Z+$jets, and $3\%$ for single top) are included in the expected limits 
and significance calculation. The $b$-tagging and the modelling uncertainties (which are the dominant 
uncertainties in the shape of the discriminating variable from the previous analysis) are also included.

The presence of a massive resonance is tested by simultaneously fitting the $m_{tb}$
templates of the signal and background simulated event samples using a binned 
maximum--likelihood approach (ML).
%based on the \textsc{RooStats} framework~\cite{Moneta:1289965,Verkerke:2003ir,Baak:2014wma}.
Each signal region is 
treated as an independent search channel with correlated systematic uncertainties.

The normalisations of the $t\bar{t}$ and $W+$jets backgrounds were found to be different than those 
in the analysis of $36.1\fbinv$, therefore they are free parameters in the fit. They are 
constrained by Asimov dataset to one by construction. 
The other background normalisations are assigned Gaussian priors based on their 
respective normalisation uncertainties. 
The signal normalisation is a free parameter in the fit.

\begin{figure}[t]
\centering
\begin{minipage}[t][\minipageheightdp][t]{\minipagewidthdp}
\centering
\includegraphics[width=\figuremaxwidthdp,height=\figuremaxheightdp,keepaspectratio]{\main/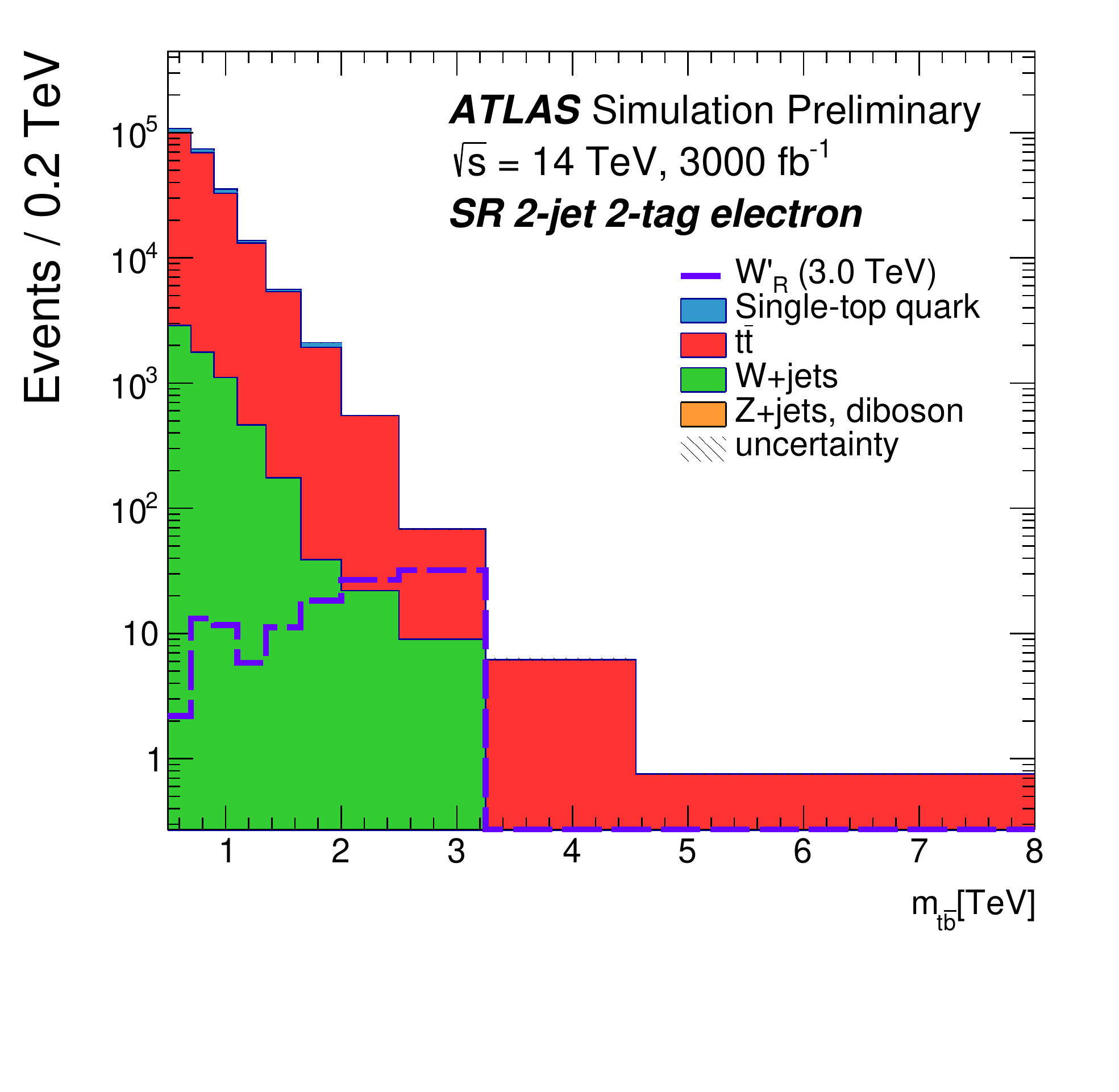}
\end{minipage}
\begin{minipage}[t][\minipageheightdp][t]{\minipagewidthdp}
\centering
\includegraphics[width=\figuremaxwidthdp,height=\figuremaxheightdp,keepaspectratio]{\main/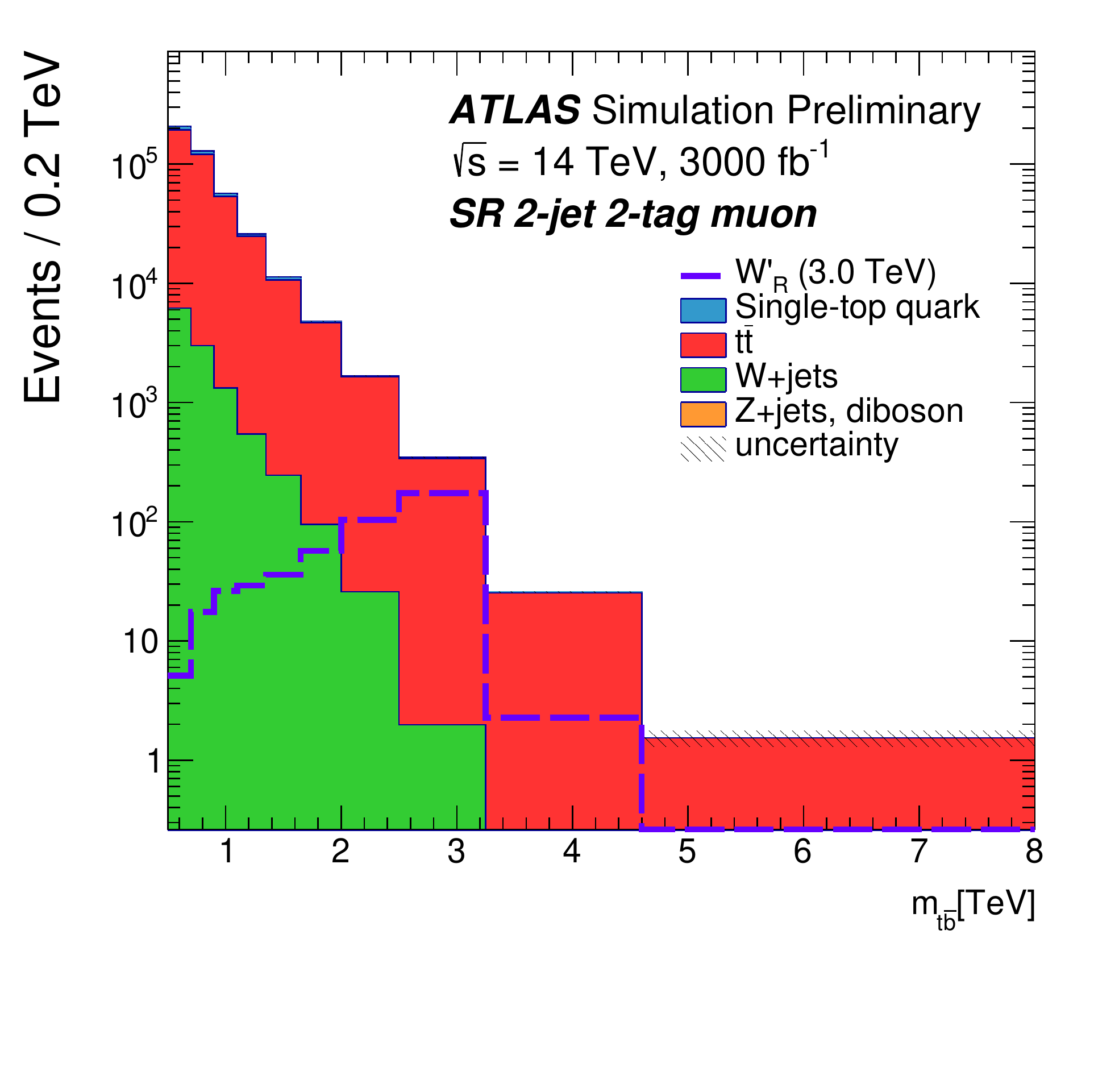}
\end{minipage}
\caption{Post-fit distributions of the reconstructed mass of the $W^\prime_{\mathrm{R}}$ boson candidate
  in the 2-jet 2-tag signal region for the electron (left) and muon (right) channels. 
  An expected signal contribution corresponding to a $W^\prime_{\mathrm{R}}$ boson mass of $3\TeV$ is shown.
  Uncertainty bands include all systematic uncertainties.}
\label{fig:SRmtb2jet}
\end{figure}

As an example, the $m_{tb}$ distributions for two of the eight signal regions after the ML fit are shown
in \fig{fig:SRmtb2jet} for the expected background and signal contribution corresponding to a
$W^\prime_{\mathrm{R}}$ boson with a mass of $3\TeV$.
The binning of the $m_{tb}$ distribution is chosen to optimise the search sensitivity while minimising statistical fluctuations. 
%Requirements are imposed on the expected number of background events per bin, and the bin width is adapted to a resolution function that represents the 
%width of the reconstructed mass peak for each studied \WpR\ boson signal sample. This results 
%in different number of bins in each region.

\begin{figure}[t]
\centering
\begin{minipage}[t][\minipageheightsp][t]{\minipagewidthsp}
\centering
\includegraphics[width=\figuremaxwidthsp,height=\figuremaxheightsp,keepaspectratio]{\main/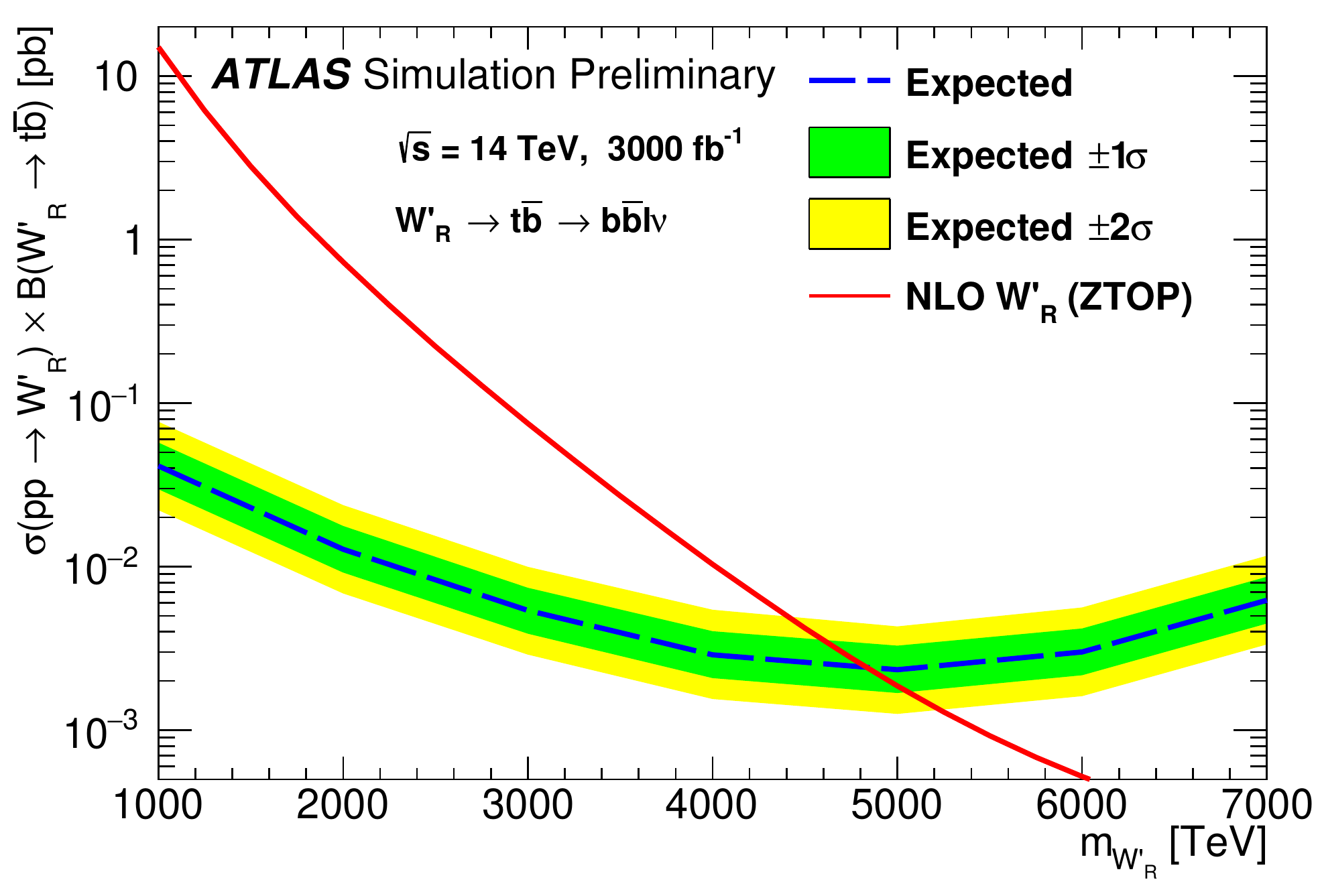}
\end{minipage}
  \caption{Upper limits at the $95\%\cl$ on the $W^\prime_{\mathrm{R}}$ production cross section times
    branching fraction as a function of resonance mass. The dashed curve and shaded bands correspond
    to the limit expected in the absence of signal and the regions enclosing one/two standard deviation
    (s.d.) fluctuations of the expected limit. The theory prediction is also shown. 
}
\label{fig:ATLAS_wpLimit}
\end{figure}

The limits are evaluated assuming the modified frequentist CL$_\text{s}$ method~\cite{Read:2002hq}
with a \sloppy\mbox{profile-likelihood-ratio} test statistic~\cite{Cowan:2010js} and using the asymptotic approximation. 
The $95\%\cl$ upper limits on the production cross section multiplied by the 
branching fraction for $W^\prime_{\mathrm{R}} \rightarrow t\bar{b}$ are shown in \fig{fig:ATLAS_wpLimit}
as a function of the resonance mass for \intLumi.
The expected exclusion limits range between $0.02\pb$ and $6\times 10^{-3}\pb$ for
$W^\prime_{\mathrm{R}}$ boson masses from $1\TeV$ to $7\TeV$.
The existence of $W^\prime_{\mathrm{R}}$ bosons with masses below $4.9\TeV$ is expected to be excluded,
assuming that the $W^\prime_{\mathrm{R}}$ coupling $g^\prime$ is equal to the SM weak coupling constant $g$. 
This would increase the limit obtained with $36.1\fbinv$~\cite{Aaboud:2018jux} by $1.8\TeV$.

The expected discovery significance is calculated using the profile likelihood test statistic for
different mass hypotheses for a luminosity of \intLumi with the asymptotic approximation.
Based on $5\sigma$ significance, it is found that $W^\prime_{\mathrm{R}}$ with masses up to $4.3\TeV$
can be discovered at the HL-LHC.

\subsubsection{\cms{Searches for $W^\prime \to \tau + \met$}}
\contributors{K.~Hoepfner, C.~Schuler, CMS}

New $\PWpr$ heavy gauge bosons %are predicted by various SM extensions. The charged version of such heavy gauge bosons is generally referred to as \PWpr. %Recently decays to third generation fermions have gained interest as possible explanations for some of the observed flavor anomalies. 
might decay as $\PWpr \rightarrow \tau \nu$. This yields to final states characterised by a single hadronically decaying tau
($\tau_h$) as the only detectable object, and missing energy due to the neutrinos.
Hadronically decaying tau leptons are selected since the corresponding branching fraction,
about $60\%$, is the largest among all $\tau$ decays.
Tau-jets are experimentally distinctive because of their low charged hadron
multiplicity, unlike QCD multi-jets,
which have high charged hadron multiplicity, or other leptonic \PWpr boson decays, which yield no jet. 
This Phase-2 study~\cite{CMS-PAS-FTR-18-030} follows closely the recently published Run-2 result~\cite{Sirunyan:2018lbg},
using hadronically decaying tau leptons.

\begin{figure}[htp]
\begin{center}
\includegraphics[width=.35\textwidth]{\main/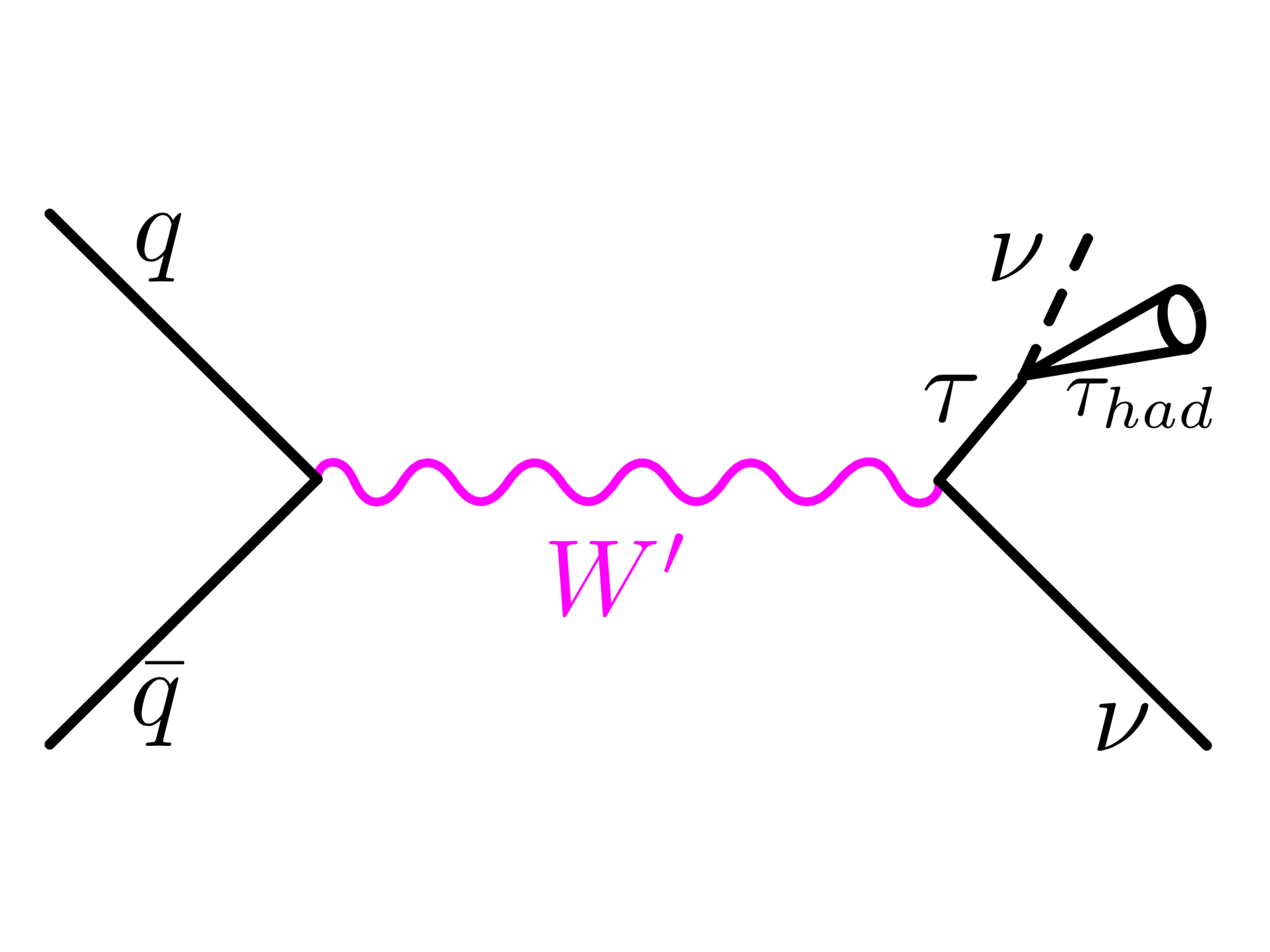}
\includegraphics[width=.49\textwidth]{\main/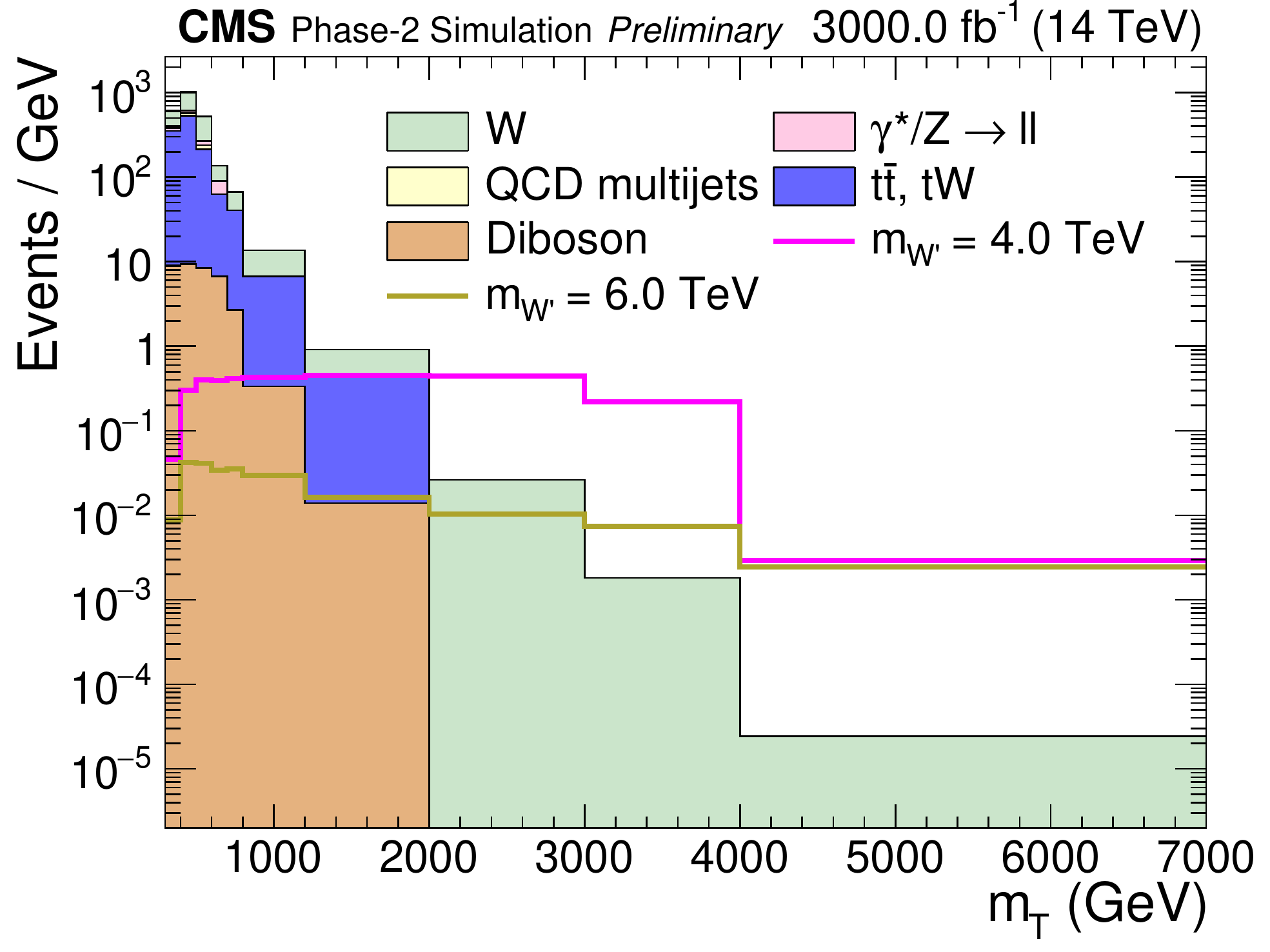}
\caption{Left: Illustration of the studied channel $W^\prime \rightarrow \tau \nu$ with the
subsequent hadronic decay of the tau ($\tau_h$).
Right: The discriminating variable, $M_T$, after all selection criteria for the HL-LHC conditions of
\intLumi and 200~PU.
The relevant SM backgrounds are shown according to the labels in
the legend. Signal examples for \PWpr boson masses of $4\TeV$ and $6\TeV$ are scaled to their SSM
LO cross section and \intLumi.}
\label{fig:feynman}
\end{center}
\end{figure}

The signature of a \PWpr boson (see \fig{fig:feynman}), is considered
similar to a high-mass W boson. 
It could be observed in the distribution of the
transverse mass ($M_T$) of the transverse momentum of the $\tau$ ($p_T^{\tau}$)
and the missing transverse momentum:
$M_T = \sqrt{2 p_T^{\tau}\met(1-cos\Delta\phi(\tau,\met))}$.
Unlike the leptonic search channels, the signal shape of $W^\prime$ bosons with hadronically
decaying tau leptons does
not show a Jacobian peak structure because of the presence of two neutrinos in the final
state. Despite the multi-particle final state,
the decay appears as a typical two-body decay; the axis of the
hadronic tau jet is back to
back with $\met$ and the magnitude of both is comparable such that their ratio is about unity.

The results are interpreted in the context of the sequential standard model in terms of \PWpr mass and coupling strength. 
A model-independent cross section limit allows interpretations in other models.
The signal is simulated at LO and the detector performance simulated with \delphes{}.
The \PWpr boson coupling strength, $g_{\PWpr}$, is given in terms of the SM weak coupling strength $g_{\PW} = e/\sin^2\theta_W\approx 0.65$.
Here, $\theta_{\PW}$ is the weak mixing angle.
If the \PWpr boson is a heavier copy of the SM \PW\ boson, their coupling ratio is $g_{\PWpr}/g_{\PW}=1$ and the 
SSM \PWpr boson theoretical cross sections, signal shapes, and widths apply.
However, different couplings are possible. Because of the dependence of the width of a particle on its couplings
the consequent effect on the transverse mass distribution, a limit can also be set on the coupling strength.

The dominant background appears in the high mass tail of the $M_T$ distribution of SM \PW\ boson events.
Subleading background contributions arise from $t\bar{t}$ and QCD multijet events. %The number of background events is reduced by the event selection.
These backgrounds primarily arise as a consequence of jets misidentified as $\tau_h$ candidates and populate the lower transverse masses while the signal exhibits an excess of events at high $M_T$.
Events with one hadronically decaying $\tau$
and $\met$ are selected if the ratio of $\pt^{\Pgt}$ to $\met$ satisfies $0.7 < \pt^{\Pgt}/\met <
1.3$ and the
angle $\Delta\phi(\vec{p_T}^{\Pgt}, \met)$ is greater than $2.4$ radians.

\begin{figure}[hbtp]\centering
    \includegraphics[width=0.49\textwidth]{\main/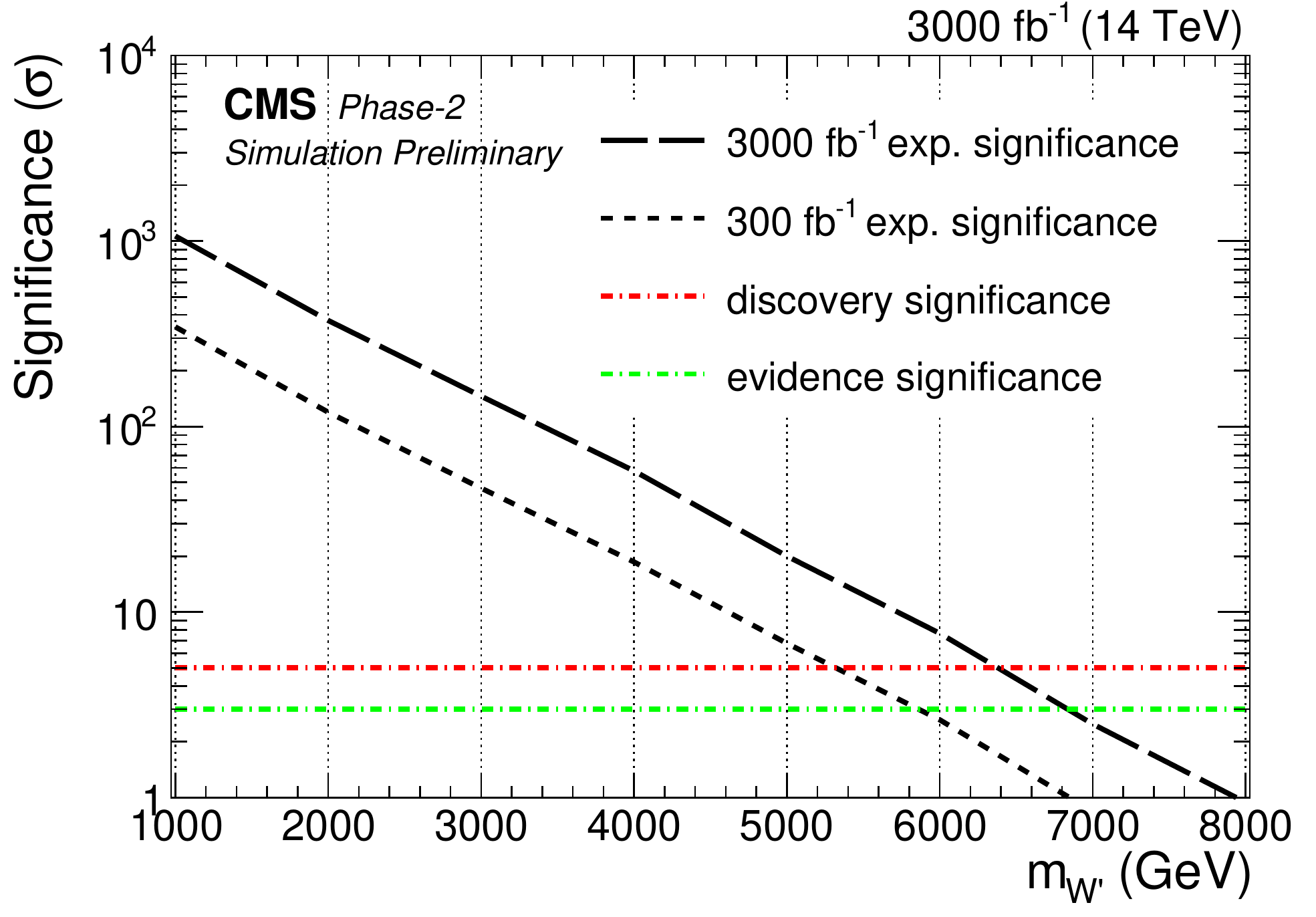}
    \includegraphics[width=0.49\textwidth]{\main/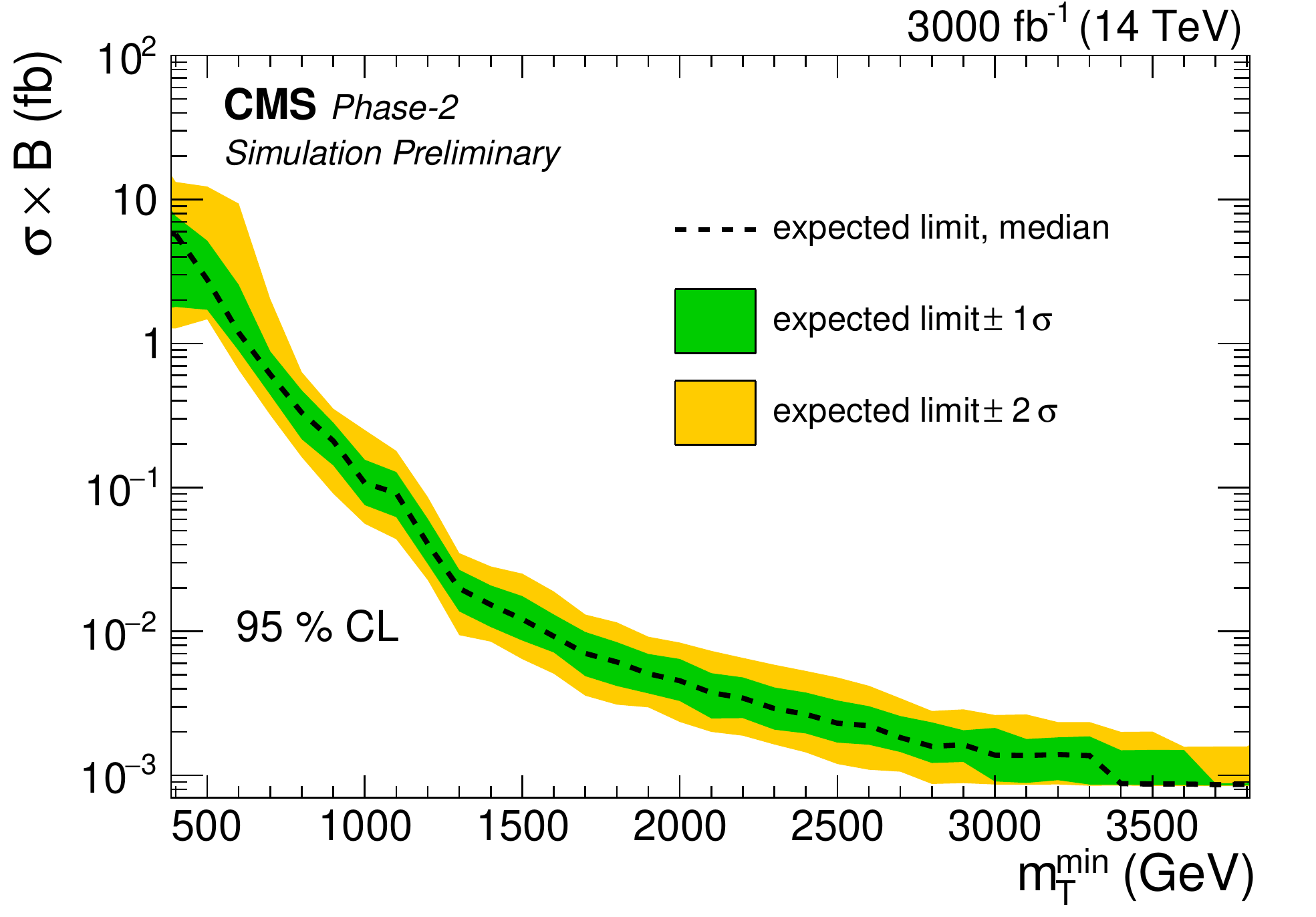}
\caption{Left: Discovery significance for SSM \PWpr to tau leptons.
Right: Model-independent cross section limit. For this, a single-bin limit is
calculated for increasing $M_T^{\rm min}$
while keeping the signal yield constant in order to avoid including any signal shape
information on this limit calculation.}
    \label{fig:WprimeResults}
\end{figure}

The physics sensitivity is studied based on the $M_T$ distribution in \fig{fig:feynman} (right).
Signal events are expected to be particularly prominent at the upper end of the $M_T$
distribution, where the expected SM background is low.
So far, there are no indications for the existence of a SSM \PWpr\ boson~\cite{Sirunyan:2018lbg}. With the
high luminosity during Phase-2, the \PWpr mass reach for potential observation increases
to $6.9\TeV$ and $6.4\TeV$
for $3\sigma$ evidence and $5\sigma$ discovery, respectively, as shown in \fig{fig:WprimeResults} (left). Alternatively, in case of no observation, one can exclude SSM \PWpr boson masses up to $7.0\TeV$
with \intLumi.
These are multi-bin limits taking into account the full $M_T$ shape.

While the SSM model assumes SM-like couplings of the fermions, the couplings could well be weaker if
further decays occur. The HL-LHC has good sensitivity to study these couplings.
The sensitivity to weaker couplings extends significantly.
%%as seen in \fig{fig:ssmlimit}-left. 
A model-independent cross section limit for new physics with $\tau$+$\met$
in the final state is depicted in \fig{fig:WprimeResults} (right), calculated as a single-bin limit
by counting the number of events above a sliding threshold $M_T^{\rm min}$.

\subsubsection{\theoryC{HL- and HE-LHC sensitivity to 2HDMs with $U(1)_X$ Gauge Symmetries}}
\contributors{D. A. Camargo, L. Delle Rose, S. Moretti, F. S. Queiroz}
Extended Higgs sectors belonging to various BSM scenarios 
%are also able to encompass all the SM Higgs boson properties. In addition, though, they 
offer the possibility to 
solve some of the open problems of the SM. Such frameworks with extended Higgs sectors have recently come together with new $U(1)_X$ gauge symmetries, offering a natural 
solutions to the DM and the neutrino mass problems.
These scenarios \cite{Ko:2013zsa,Ko:2014uka,Berlin:2014cfa,Huang:2015wts,DelleRose:2017xil,Campos:2017dgc} predict
%For instance, assuming a 2HDM sector~, 
%alongside that of
%the ensuing additional Higgs bosons,   
a rich 
new phenomenology due to the presence of a massive $\Zp$ gauge 
boson arising after the $U(1)_X$ spontaneous symmetry breaking.
 
Our goal is to explore the potential of the HL- and HE-LHC to study such a scenario. To do so, we fist use the latest available dilepton data  
from the LHC~\cite{Aaboud:2017buh, Sirunyan:2018exx} to constrain the mass and couplings of such $\Zp$ gauge bosons and, consequently, the viable parameter space of the underlying model. Then we use this result to asses the capabilities of the HL- and HE-LHC to test the existence of such heavy neutral vector bosons. An extended version of this contribution can be found in \citeref{Camargo:2018klg}.
 
The relevant part of the Lagrangian of the model we consider is
\begin{eqnarray}
\mathcal{L_{\rm NC}} \supset & - \left(\frac{g_Z}{2} J_{NC}^\mu \cos \xi \right) Z_\mu- \left(\frac{g_Z}{2} J_{NC}^\mu \sin \xi \right) \Zp_\mu \\
& + \frac{1}{4} g_X \sin \xi \left[ \left( Q_{Xf} ^R + Q_{Xf} ^L \right) \bar{\psi} _f \gamma ^\mu \psi _f + \left( Q_{Xf} ^R - Q_{Xf} ^L \right) \bar{\psi} _f \gamma ^\mu \gamma _5 \psi _f \right] Z_\mu \\
& - \frac{1}{4} g_X \cos \xi \left[ \left( Q_{Xf} ^R + Q_{Xf} ^L \right) \bar{\psi} _f \gamma ^\mu \psi _f - \left( Q_{Xf} ^L - Q_{Xf} ^R \right) \bar{\psi} _f \gamma ^\mu \gamma _5 \psi _f \right] Z' _\mu,
\label{eq:LNC1}
\end{eqnarray}
where $\xi$ represents the $Z-\Zp$ mixing parameter, $g_X$ the gauge coupling of the new abelian symmetry while 
$Q^L_X$($Q^R_X$) are the left(right)-handed fermion charges under $U(1)_X$ 
defined according to \citeref{Camargo:2018klg}. This interaction Lagrangian
represents the key information for the collider phenomenology we are going to tackle, 
because it dictates $\Zp$ production rates at the LHC as well as its most prominent decays to be searched for.

%\subsubsection{LHC Bounds}
%\paragraph*{LHC Bounds}

Present LHC bounds are obtained here by simulating at $13\TeV$ of \com~energy the process
\begin{equation}
pp\rightarrow l^+l^- +X,
\end{equation}
where $l=e,\mu$, leading to dilepton signals, and $X$ represents the surrounding hadronic activity. (The dijet signal case was studied in \citeref{Camargo:2018klg} and found to be less sensitive.) 
Since this channel is mediated by a heavy $\Zp$, alongside $\gamma$ and $Z$, a peak 
around the $\Zp$ mass would appear at large values of the invariant mass of the dilepton final state. 
We describe our results adopting both the Narrow Width Approximation (NWA) and also introducing Finite Width (FW) effects. 
In this respect, we have implemented in \feynrules~\cite{Alloul:2013bka} the models shown in \citeref{Camargo:2018klg}, where the corresponding $U(1)_X$ charge assignment is explicitly shown,
and we have simulated the partonic events with \MGvATNLO~\cite{Alwall:2011uj}. For hadronisation and detector effects we used \pythia{ 8}~\cite{Sjostrand:2007gs} and \delphes~\cite{deFavereau:2013fsa}, respectively.

%\paragraph*{HL-LHC and HE-LHC Sensitivity}\label{par:sec_HL-HE-LHC}
Assuming, for the sake of definiteness, the NWA with $g_X=0.1, 0.2$ and $0.3$, we project the current experimental limits to the HL- and HE-LHC. In order to extrapolate the current bound to a new collider configuration, one would need to scale the relevant backgrounds and find the new point giving the same number of background events, which, assuming same efficiencies and acceptances, would lead to the same excluded signal cross section, as outlined in \citeref{Thamm:2015zwa}. However, in the case of the dilepton final state, the signal and the background scale equally with energy/luminosity, so that one can employ a simpler strategy, based on a direct rescaling of the bound on the number of signal events. Therefore, in order to find the future sensitivities we just solve an equation for $M_{\rm new}$ 
(\iee, the new limit on $m_{Z'}$), knowing the current bound $M$, as follows:
\begin{equation}
\frac{ N_{\rm signal\, events} (M_{\rm new}^2, E_{\rm new},\mathcal{L}_{\rm new})}
     { N_{\rm signal\, events} (M^2, 13\,{\rm TeV},36\fbinv)}=1,
\end{equation}
with obvious meaning of the subscripts.

The results from this iteration are summarised in \tabbs{tab:LHCforcast}, \ref{tab:LHCforcast2} and \ref{tab:LHCforcast3}. 
The lower mass bounds found presently compared to the expected ones at the 
HL-LHC and/or HE-LHC clearly show how important is any LHC upgrade to test 
new physics models including a $Z'$. For some models and benchmark points, such as, \egg, $U(1)_A$ with $g_X =0.1$, 
the HE-LHC will potentially probe $\Zp$ masses up to $7\TeV$, while for others, such as, \egg,  $U(1)_F$ with $g_X =0.3$, it will potentially exclude masses up to $12\TeV$. In short, both LHC upgrades can generally extend the current reach in $m_{Z'}$ by a factor $2$ to $3$. The results are summarised in the tables. 

\begin{table*}[!t]
\centering
%\begin{tabular*}{\columnwidth}{@{\extracolsep{\fill}}lllll@{}}
\small\begin{tabular}{ccccccccc}
\hline 
Model  & $13\TeV$, $300\fbinv$&  $14\TeV$, $3\abinv$ & $27$ TeV, $300\fbinv$ & $27\TeV$, $3\abinv$ & $27\TeV$, $15\abinv$\\ \hline 
$U(1)_A$ & $3.07\TeV$ & $4.3\TeV$ & $5.02\TeV$ & $7.03\TeV$ & $8.51\TeV$ \\
$U(1)_B$ &  $3.07\TeV$ & $4.3\TeV$ & $5.02\TeV$ & $7.03\TeV$ & $8.51\TeV$\\
$U(1)_C$ & $2.37\TeV$ & $3.52\TeV$ & $3.73\TeV$ & $5.54\TeV$ & $6.96\TeV$\\
$U(1)_D$ & $4.45\TeV$ & $5.81\TeV$ & $7.76\TeV$ & $9.89\TeV$ & $11.34\TeV$\\
$U(1)_E$ &  $3.18\TeV$ & $4.45\TeV$ & $5.24\TeV$ & $7.27\TeV$ & $8.75\TeV$\\
$U(1)_F$ &  $4.55\TeV$ & $5.91\TeV$ & $7.97\TeV$ & $10.09\TeV$ & $11.54\TeV$ \\
$U(1)_G$ &  $1.73\TeV$ & $2.73\TeV$ & $2.62\TeV$ & $4.16\TeV$ & $5.45\TeV$\\
$U(1)_{B-L}$ & $2.84\TeV$ & $4.07\TeV$ & $4.60\TeV$  & $6.55\TeV$ & $8.02\TeV$\\
\hline
\end{tabular}\normalsize
\caption{HL-LHC and HE-LHC projected sensitivities for all $U(1)_X$ models studied in this work
using dilepton data at $13\TeV$, $14\TeV$ and $27\TeV$ of CM energy and for $\mathcal{L}=300\fbinv$ and $\mathcal{L}=3$ and $15\abinv$. Here, $g_X =0.1$.}
\label{tab:LHCforcast}
\end{table*}

\small\begin{table*}[!t]
\centering
%\begin{tabular*}{\columnwidth}{@{\extracolsep{\fill}}lllll@{}}
\small\begin{tabular}{ccccccccc}
\hline 
Model & $13\TeV$, $300\fbinv$& $14\TeV$, $3\abinv$  & $27$ TeV, $300\fbinv$ & $27\TeV$, $3\abinv$ & $27\TeV$, $15\abinv$\\ \hline 
$U(1)_A$ & $4.14\TeV$ & $5.49\TeV$ & $7.14\TeV$ & $9.26\TeV$ & $10.73\TeV$ \\
$U(1)_B$ & $4.14\TeV$ & $5.49\TeV$ & $7.17\TeV$ & $9.26\TeV$& $10.73\TeV$  \\
$U(1)_C$ & $3.62\TeV$ & $4.93\TeV$ & $6.09\TeV$ & $8.18\TeV$ & $9.66\TeV$ \\
$U(1)_D$ & $5\TeV$ &$6.43\TeV$ & $9\TeV$ & $11.1\TeV$ & $12.52\TeV$ \\
$U(1)_E$ & $5\TeV$ & $6.43\TeV$ & $9\TeV$ & $11.1\TeV$ & $12.52\TeV$ \\
$U(1)_F$ & $5.53\TeV$ & $6.94\TeV$ & $10.02\TeV$ & $12.09\TeV$ & $13.511\TeV$ \\
$U(1)_G$ & $2.37\TeV$ &$3.52\TeV$  & $3.73\TeV$ & $5.54\TeV$& $6.96\TeV$  \\
$U(1)_{B-L}$ & $3.83\TeV$ &$5.16\TeV$  & $6.5\TeV$  & $8.62\TeV$ & $10.10\TeV$ \\
\hline
\end{tabular}\normalsize
\caption{HL-LHC and HE-LHC projected sensitivities for all $U(1)_X$ models studied in this work
using dilepton data at $13\TeV$, $14\TeV$ and $27\TeV$ of CM energy and for $\mathcal{L}=300\fbinv$ and $\mathcal{L}=3$ and $15\abinv$. Here, $g_X =0.2$.}
\label{tab:LHCforcast2}
\end{table*}\normalsize

\small\begin{table*}[!t]
\centering
%\begin{tabular*}{\columnwidth}{@{\extracolsep{\fill}}lllll@{}}
\small\begin{tabular}{ccccccccc}
\hline 
Model & $13\TeV$, $300\fbinv$&  $14\TeV$, $3\abinv$ & $27$ TeV, $300\fbinv$ & $27\TeV$, $3\abinv$ & $27\TeV$, $15\abinv$\\ \hline 
$U(1)_A$ & $4.75\TeV$ & $6.12\TeV$ & $8.38\TeV$ & $10.5\TeV$ & $11.94\TeV$\\
$U(1)_B$ & $4.75\TeV$ & $6.12\TeV$ & $8.38\TeV$ & $10.5\TeV$ & $11.94\TeV$\\
$U(1)_C$ & $4\TeV$ & $5.38\TeV$ & $6.93\TeV$ & $9.05\TeV$& $10.52\TeV$ \\
$U(1)_D$ & $5.72\TeV$ & $7.14\TeV$ & $10.4\TeV$ & $12.48\TeV$& $13.90\TeV$ \\
$U(1)_E$ & $5.14\TeV$ & $6.53\TeV$ & $9.2\TeV$ & $11.3\TeV$ & $12.72\TeV$\\
$U(1)_F$ & $5.91\TeV$ & $7.34\TeV$ & $10.84\TeV$ & $12.87\TeV$ & $14.28\TeV$\\
$U(1)_G$ & $4\TeV$ & $5.38\TeV$ & $6.93\TeV$ & $9.05\TeV$ & $10.52\TeV$\\
$U(1)_{B-L}$ & $4.35\TeV$ & $5.70\TeV$ & $7.55\TeV$  & $9.68\TeV$ & $11.14\TeV$\\
\hline
\end{tabular}\normalsize
\caption{HL-LHC and HE-LHC projected sensitivities for all $U(1)_X$ models studied in this work
using dilepton data at $13\TeV$, $14\TeV$ and $27\TeV$ of CM energy and for $\mathcal{L}=300\fbinv$ and $\mathcal{L}=3$ and $15\abinv$. Here, $g_X =0.3$.}
\label{tab:LHCforcast3}
\end{table*}\normalsize

\vspace{1cm}

%\subsubsection{Conclusions}
%\paragraph*{Conclusions}
%
%
%We have presented the HL-LHC and HE-LHC sensitivities in the search for new heavy neutral vector bosons in several $U(1)_X$ scenarios with a 2HDM structure in their (pseudo)scalar sectors. 
%As a result, we found that the discuss upgrades of the LHC are capable of probing $\Zp$ masses well into the 10 TeV
%domain.

%\subsubsubsection*{Acknowledgments}
%\noindent
%The authors thank Werner Rodejohann and Pyung-won Ko
%for  discussions  and  comments.   DC  and  FSQ  acknowledge
%financial support from MEC and UFRN. FSQ also acknowledges the ICTP-SAIFR FAPESP grant 2016/01343-7 for additional  financial  support.   SM  is  supported  in  part  by  the
%NExT  Institute  and  acknowledges  partial  financial  support
%from  the  STFC  Consolidated  Grant  ST/L000296/1  and  the
%H2020-MSCA-RISE-2014 grant no.   645722 (NonMinimal-
%Higgs).

%\newcommand{\CH}[1]{\textbf{\textcolor{green}{CH - #1}}}
\subsubsection{\theoryC{$Z'$ discrimination at HE-LHC in case of an evidence/discovery after the HL-LHC}}\label{sec:Zpmodeldiscr}
\contributors{C. Helsens, D. Jamin, M. L. Mangano, T. Rizzo, M. Selvaggi}
%{\bf Authors: C. Helsens$^1$, D. Jamin$^2$, M. L. Mangano$^3$, T. Rizzo$^4$, M. Selvaggi$^1$}\\
%\newline
%$^1$CERN EP-Departement, CH-1211 Geneva 23, Switzerland email: {\tt B.C. clement.helsens@cern.ch}\\
%$^2$Academia Sinica, Institute of  Physics, Taipei, Taiwan\\
%$^3$CERN TH-Departement, CH-1211 Geneva 23, Switzerland\\
%$^4$SLAC National Accelerator Laboratory 2575 Sand Hill Rd., Menlo Park, CA, 94025 USA\\

%%%%%%%%%%%%%%%%%%%%%%%%%%%%%%%%%%%%%%%%%%%%%%%%%%%%%
%\subsection{Context of the study}
\paragraph*{Context of the study and HL-LHC bounds}
It is still legitimate to assume that a heavy resonance could be seen at the end of HL-LHC. If that is the case a new collider with higher energy
in the \com~is needed to study its properties as too few events will be available at $\sqrt{s}=14\TeV$. In this section we present the discrimination potential between six $Z'$ models of a HE-LHC with an assumed \com~energy of 27~TeV and an integrated luminosity of $\mathcal{L}=15\abinv$. Under the assumption that these $Z'$'s decay only to SM particles, we show that there are sufficient observables to perform this model differentiation in most cases.

\begin{figure}[t]
\centering
\begin{minipage}[t][\minipageheightdp][t]{\minipagewidthdp}
\centering
\includegraphics[width=\figuremaxwidthdp,height=\figuremaxheightdp,keepaspectratio,trim={2cm 2cm 2cm 2cm},clip]{\main/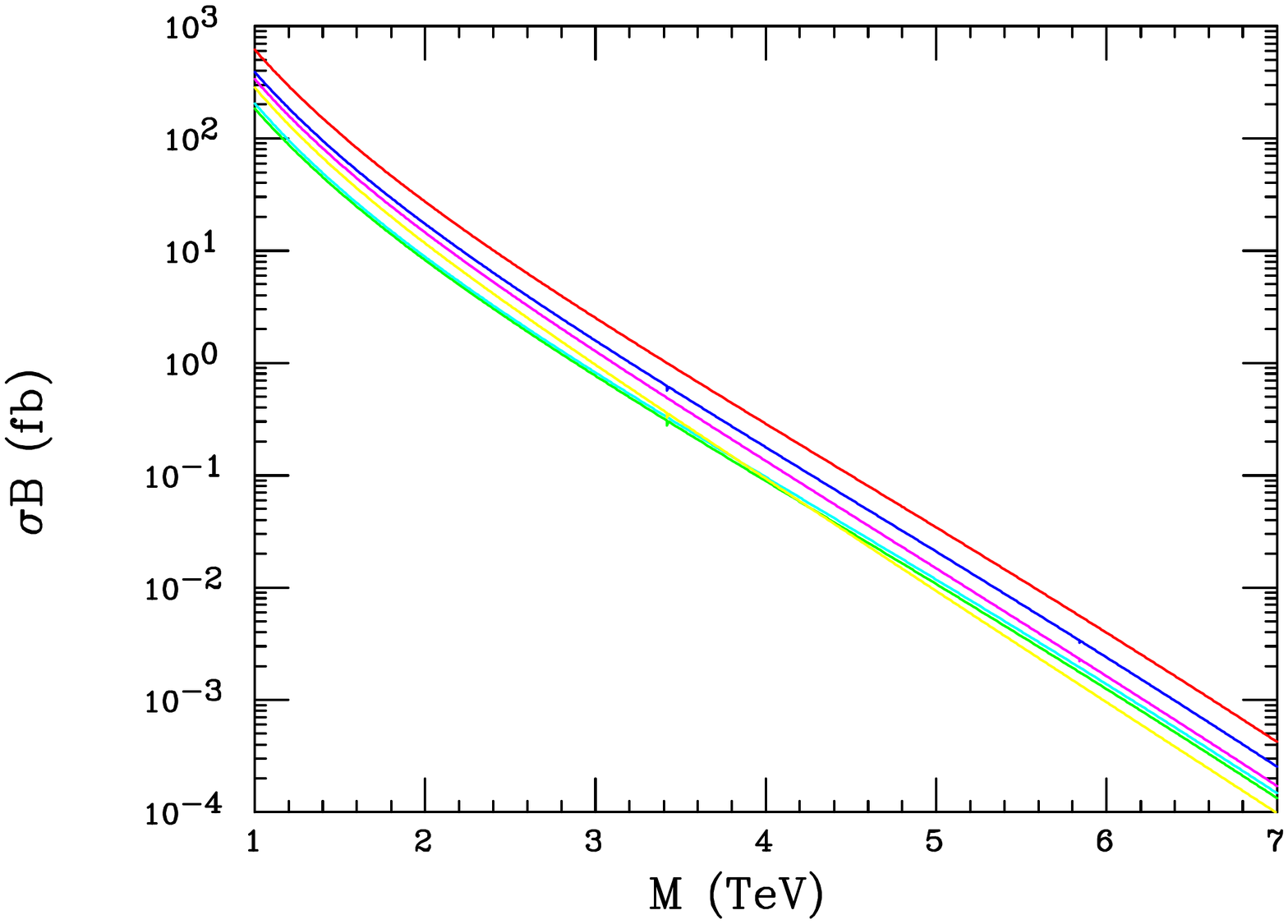}
\end{minipage}
\begin{minipage}[t][\minipageheightdp][t]{\minipagewidthdp}
\centering
\includegraphics[width=\figuremaxwidthdp,height=\figuremaxheightdp,keepaspectratio,trim={2cm 2cm 2cm 2cm},clip]{\main/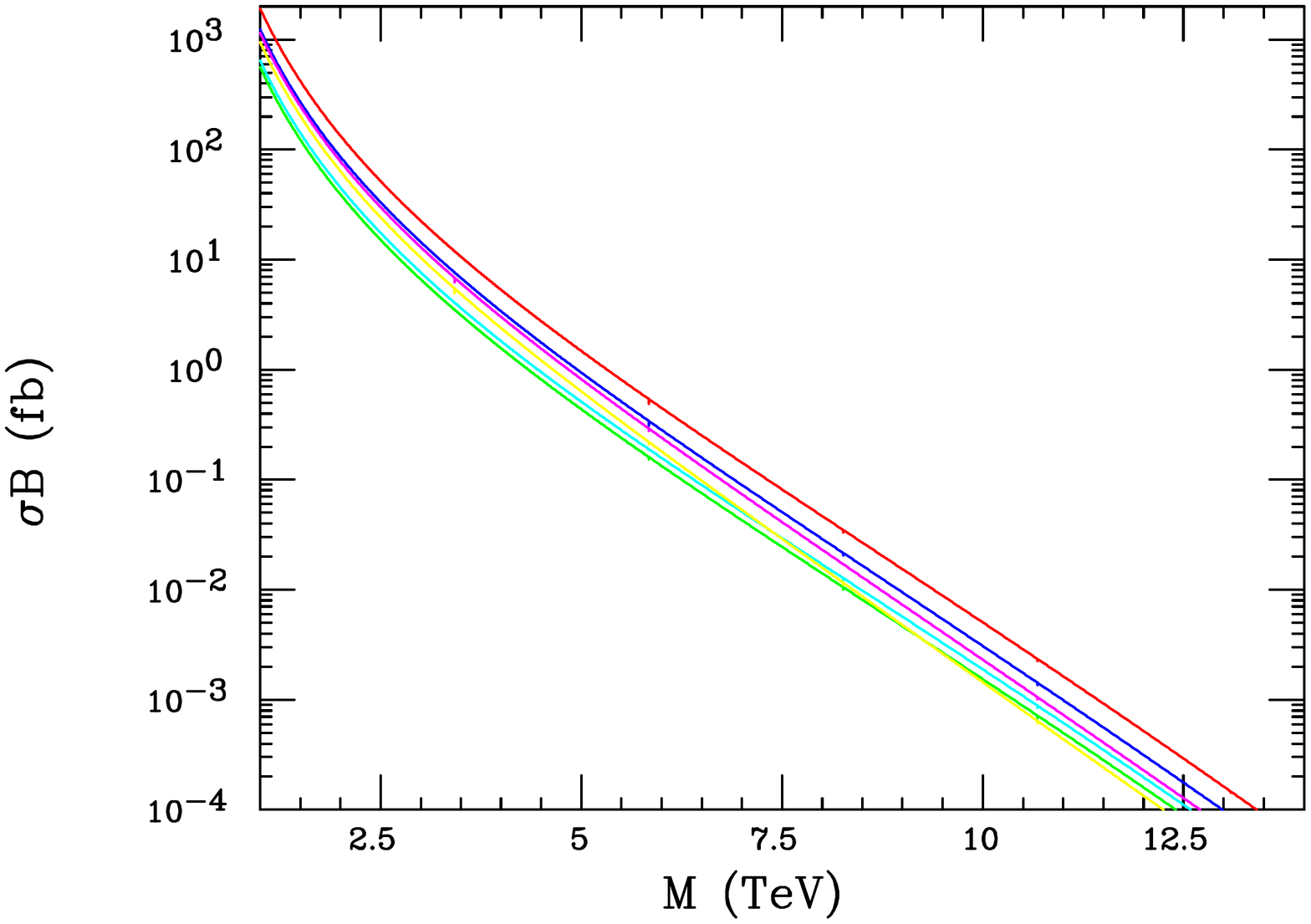}
\end{minipage}
    \caption{Left: $\sigma B_l$ in the NWA for the $Z'$ production at the $\sqrt s=14\TeV$ LHC as functions of the $Z'$ mass: SSM(red), LRM (blue), $\psi$(green), $\chi$(magenta),
$\eta$(cyan), I(yellow). Right: $\sigma B_l$ of $Z'$ in models described in (left) at $\sqrt s=27\TeV$.}
\label{fig:pheno:toy}
\end{figure}

%%%%%%%%%%%%%%%%%%%%%%%%%%%%%%%%%%%%%%%%%%%%%%%%%%%%%
%\subsection{Bounds from HL-LHC}
%\paragraph*{Bounds from HL-LHC}
As a starting point it is needed to estimate what are, for $\sqrt s=14$ TeV, the typical exclusion/discovery reaches for standard reference $Z'$ models assuming $\mathcal{L}=3\abinv$ employing only the $e^+e^-$ and $\mu^+\mu^-$ channels. %To address this and the other questions below we will use the same set of $Z'$ models as employed in Ref.~\cite{Rizzo:2014xma} and mostly in Ref.~\cite{Han:2013mra}, both of which we will refer to frequently. We employ the MMHT2014 NNLO PDF set~\cite{Harland-Lang:2014zoa} throughout with an appropriate constant $K$-factor (=1.27) to account for higher order QCD corrections. 
The production cross section times leptonic branching fraction is shown in \fig{fig:pheno:toy} (left) for these models at $\sqrt{s}=14\TeV$ in the narrow width approximation (NWA). It has been and will be assumed here that these $Z'$ states only decay to SM particles.

%Using the present ATLAS and CMS results at 13 TeV,~\cite{Aaboud:2017buh} and~\cite{Sirunyan:2018exx}, it is straightforward to estimate by extrapolation the exclusion reach at $\sqrt{s}=14\TeV$\ using the combined $ee+\mu\mu$ final states. This is given in the first column of Table~\ref{tab:pheno:spec}. 
Studies presented in this report on prospects for searches of $Z'$ by ATLAS (see~\secc{sec:atlasZprimedilepton}) shows that discovery and exclusion reaches are between $5$ and $6.5\TeV$ in $M_{Z'}$ depending on the model assumption. 
%For discovery, only the $ee$ channel is used due to poor $\mu\mu$-pair invariant mass resolution near $M_{Z'}=6$ TeV. Estimates of the $3\sigma$ evidence and $5\sigma$ discovery limits are also given in Table~\ref{tab:pheno:spec}. 
Based on these results, we will assume in our study below that we are dealing with a $Z'$ of mass $6\TeV$. Figure~\ref{fig:pheno:toy} (right) shows the NWA cross sections for the same set of models but now at $\sqrt{s}=27\TeV$ with $\mathcal{L}=15\abinv$. We note that very large statistical samples will be available for the case of $M_{Z'}=6\TeV$
for each dilepton channel.

%
%\begin{table}[t]
%\centering
%\small\begin{tabular}{|l|c|c|c|} \hline\hline
%  Model &   95$\%$ \cl  &  $3\sigma$  & $5\sigma$   \\
%\hline
%SSM    &     6.62     &  6.09    &  5.62     \\
%LRM    &   6.39     & 5.85     & 5.39  \\
%$\psi$    &  6.10   & 5.55   & 5.07  \\
%$\chi$   &  6.22    & 5.68    & 5.26   \\
%$\eta$   &  6.15     &  5.59  &  5.16   \\
%~I        & 5.98   &  5.45   &  5.05  \\
%\hline\hline
%\end{tabular}\normalsize
%\caption{ Mass reach for several $Z'$ models at $\sqrt{s}=14\TeV$\ with $\mathcal{L}=3\abinv$. }
%\label{tab:pheno:spec}
%\end{table}
%
\paragraph*{Definition of the discriminating variables}
\label{par:vardef}

The various $Z'$ models can be disentangled with the help of 3 inclusive observables: the production cross section times leptonic branching fraction $\sigma B_l$, the forward-backward asymmetry $A_{FB}$ and the rapidity ratio $r_y$. The variable $A_{FB}$ can be seen as an estimate of the charge asymmetry
\begin{equation}
A_{FB} = A_C =  \frac{\sigma(\Delta|y| > 0) - \sigma(\Delta|y| < 0)}{\sigma(\Delta|y| > 0) + \sigma(\Delta|y| < 0)},
\end{equation}
where $\Delta|y| = |y_l| - |y_{\bar{l}}|$. It has been checked that this definition is equivalent to defining
\begin{equation}
A_{FB} = \frac{\sigma_F - \sigma_B}{\sigma_F + \sigma_B},
\end{equation}
with $\sigma_F = \sigma (cos\theta^{*}_{cs})>0$ and $\sigma_B = \sigma (cos\theta^{*}_{cs})<0$ where $\theta^*_{cs}$ is the Collins-Soper frame angle. The variable $r_y$ is defined as the ratio of central over forward events:
\begin{equation}
r_y = \frac{\sigma(|y_{Z'}| < y_1)}{\sigma(y_1 < |y_{Z'}| <y_2)},
\end{equation}
where $y_1=0.5$ and $y_2=2.5$.

%%%%%%%%%%%%%%%%%%%%%%%%%%%%%%%%%%%%%%%%%%%%%%%%%%%%%
\paragraph*{Model discrimination}

The model discrimination presented in this section has been performed assuming the HE-LHC detector parametrisation~\cite{hlhelhc_web} in \delphes~\cite{deFavereau:2013fsa}. In such a detector, muons at $\eta \approx 0$ are assumed to be reconstructed with a resolution $\sigma(p)/p \approx 7\%$ for $\pt =3 \TeV$.

\subparagraph*{Leptonic final states}
\label{par:lepana}

The potential for discriminating various $Z'$ models is first investigated using the leptonic $ee$ and $\mu\mu$ final states only. The signal samples for the 6 models and the \sloppy\mbox{Drell-Yan} backgrounds have been generated with \pythia{ 8.230}~\cite{Sjostrand:2014zea} including the interference between the signal and background. The $Z'$ decays assume lepton flavour universality. For a description of the event selection and a discussion of the discovery potential in leptonic final states for the list of $Z'$ models being discussed here, the reader should refer to \secc{subsubsec:hr_lep}. We simply point out here that with $\mathcal{L}=15\abinv$, all $Z'$ models with $m_{Z'}\lesssim10\TeV$ can be excluded at $\sqrt{s}=27\TeV$.

\begin{figure}[t]
\centering
\begin{minipage}[t][\minipageheighttp][t]{\minipagewidthtp}
\centering
\includegraphics[width=\figuremaxwidthtp,height=\figuremaxheighttp,keepaspectratio]{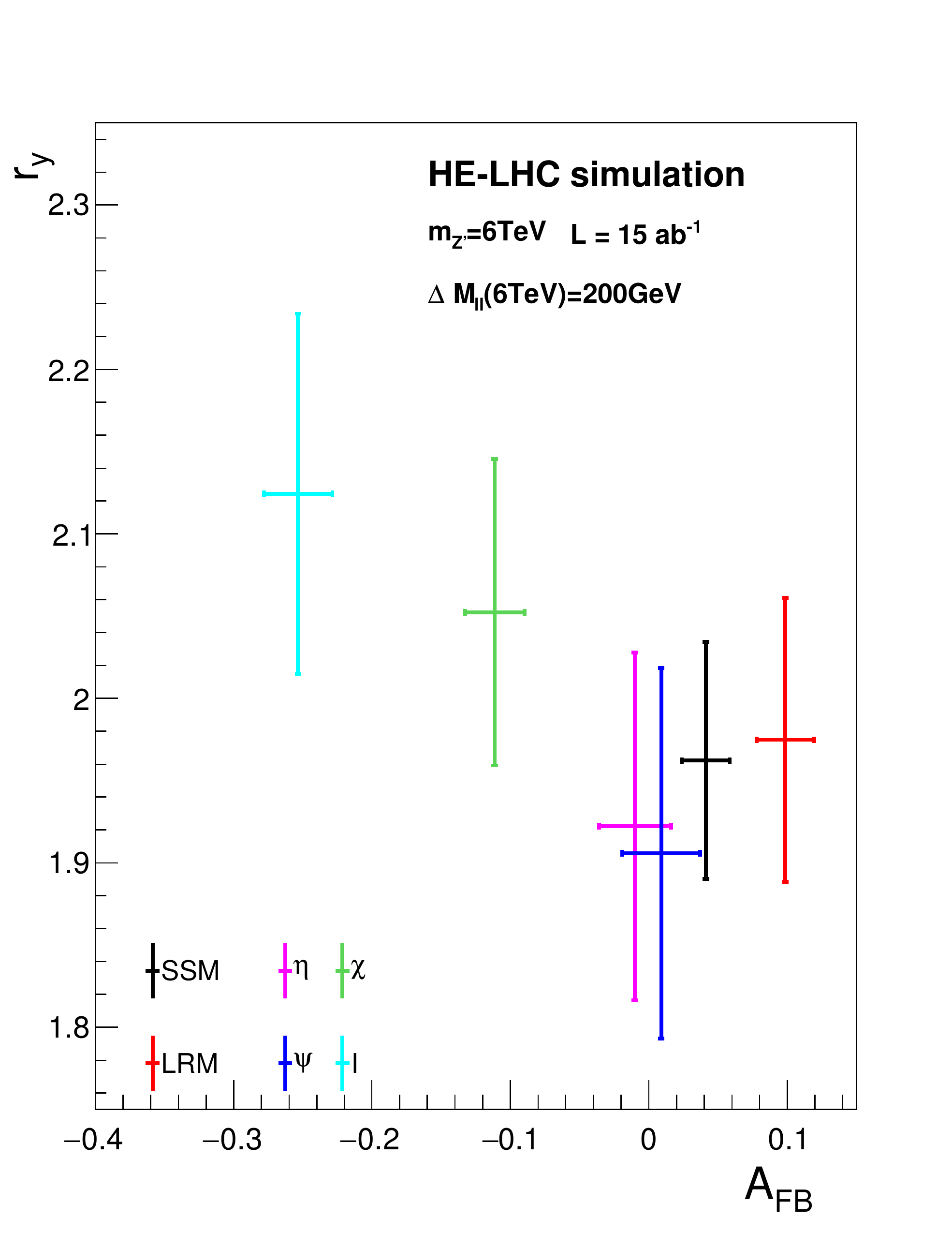}
\end{minipage}
\begin{minipage}[t][\minipageheighttp][t]{\minipagewidthtp}
\centering
\includegraphics[width=\figuremaxwidthtp,height=\figuremaxheighttp,keepaspectratio]{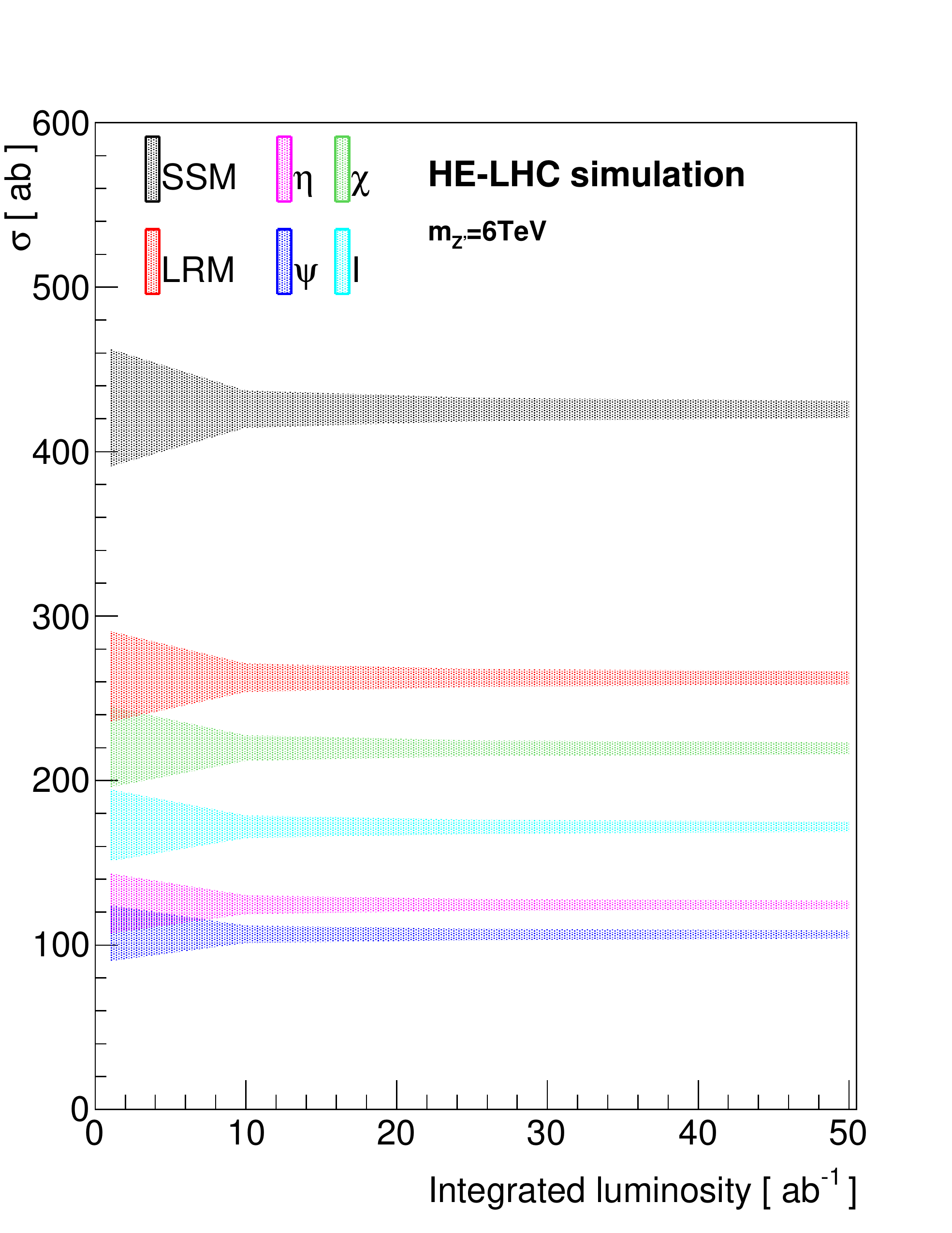}
\end{minipage}
\begin{minipage}[t][\minipageheighttp][t]{\minipagewidthtp}
\centering
\includegraphics[width=\figuremaxwidthtp,height=\figuremaxheighttp,keepaspectratio]{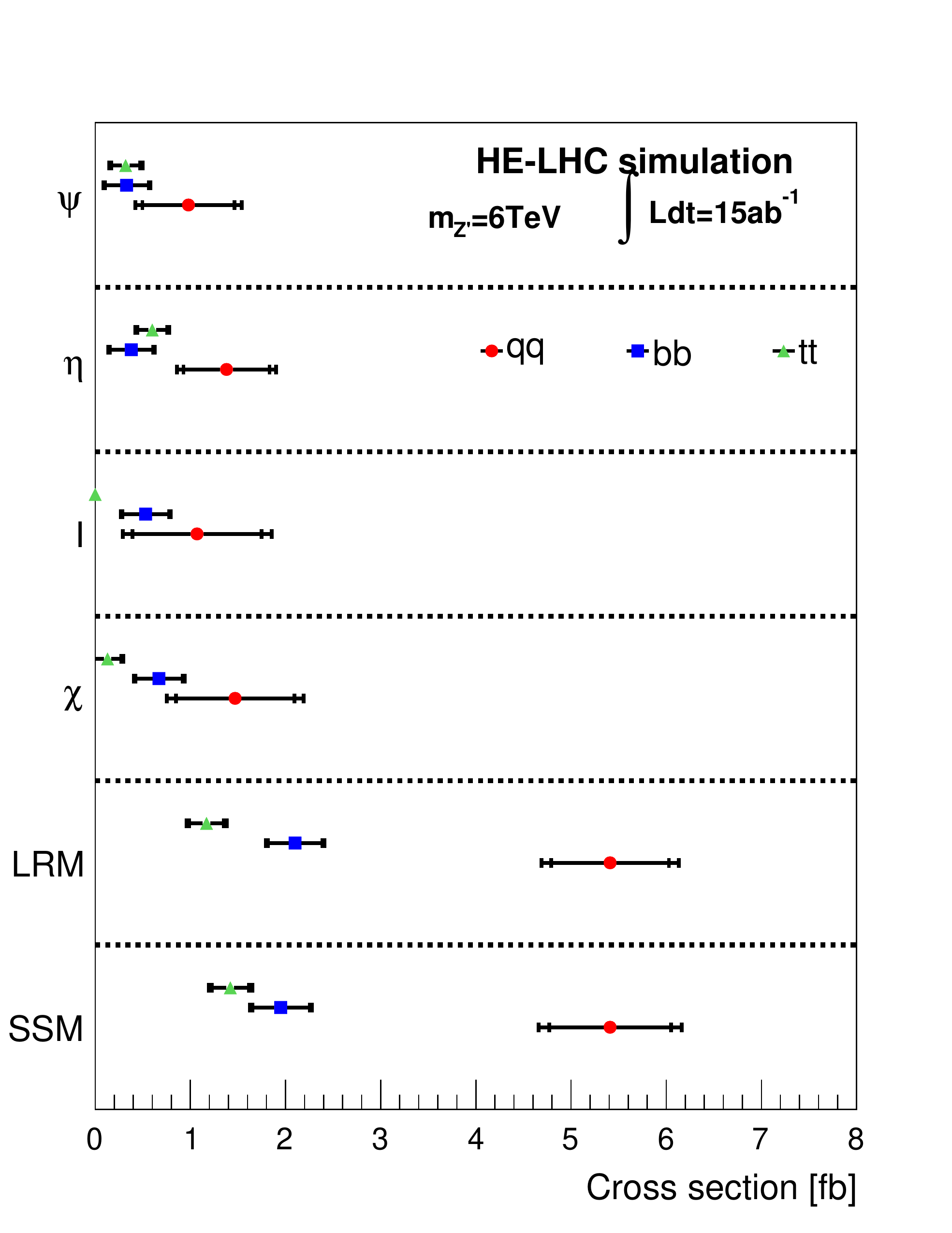}
\end{minipage}
  \caption{Left: scatter plot of $r_y$ versus $A_{FB}$ with $200\GeV$ and mass window. The full interference is included. Centre: Fitted signal cross-section together with its corresponding error versus integrated luminosity. Right: fitted cross-section of the three hadronic analyses. Statistical and full uncertainties are shown on each point.}
  \label{fig:ana:res}
\end{figure}

Figure~\ref{fig:ana:res} (left) shows the correlated predictions for the $A_{FB}$ and the rapidity ratio $r_y$ observables defined previously for these six models given the above assumptions. Although the interference with the SM background was included in the simulation, its effect is unimportant due to the narrowness of the mass window around the resonance that was employed. Furthermore, the influence of the background uncertainty on the results has been found to have little to no impact on the model discrimination potential. Therefore the displayed errors on $A_{FB}$ and $r_y$ are of statistical origin only. The results show that apart from a possible near degeneracy in models $\psi$ and $\eta$, a reasonable $Z'$ model separation can indeed be achieved.

Using a profile likelihood technique, the signal strength $\mu$, or equivalently, $\sigma B_l$, can be fitted together with its corresponding error using the the di-lepton invariant mass shape. The quantity $\sigma B_l$ and its total estimated uncertainty is shown in \fig{fig:ana:res} (centre) as a function of the integrated luminosity. The $\sigma B_l$ measurement seems to be able to resolve the degeneracy between the $\psi$ and $\eta$ models with $\mathcal{L}=15\abinv$. It should be noted however that since the cross-section can easily be modified by an overall rescaling of the couplings, further handles will be needed for a convincing discrimination.

%%%%%%%%%%%%%%%%%%%%%%%%%%%%%%%%%%%%%%%%%%%%%%%%%%%%%
%\subsubsection{Hadronic analysis}
\subparagraph*{Hadronic final states}
\label{par:hadana}

Model discrimination can be improved by including an analysis involving three $Z'$ addition hadronic final states: $t\bar{t}$, $b\bar{b}$ and $q\bar{q}$, where $q=u,d,c,s$. The sample production and event selection for the $t\bar{t}$, $q\bar{q}$ final states will be described to some extent in \secc{subsubsec:hr_had}. We simply remind the reader that the analysis involves requiring the presence of two central high $\pt$ jets. In order to ensure complete orthogonality between the various final states, jets are required to be tagged as follows. In the $Z' \rightarrow t\bar{t}$ analysis both jets should be \emph{top-tagged}. For the $Z' \rightarrow b\bar{b}$ final state both jets are required to be \emph{b-tagged} and we veto events containing at least one top-tagged jet. Finally, in the $Z' \rightarrow q\bar{q}$ analysis, we veto events that contain at least one b-tagged or top-tagged jet.

Figure~\ref{fig:ana:res} (right) summarises the discrimination potential in terms of fitted cross-section of the different models considering the three aforementioned hadronic decays, $t\bar{t}$,  $b\bar{b}$ and $q\bar{q}$. An good overall discrimination among the various models can be achieved using all possible final states. For example, the SSM and $\psi$ models, which have very close predictions for $r_y$ and $A_{FB}$, have measurably different fractions of $t\bar{t}$ or $b\bar{b}$ final states. We note however that the degeneracy between $\eta$ and $\psi$ can only be partially resolved resolved at $\approx 1 \sigma$ by exploiting the difference in $t\bar{t}$ yield.

%\paragraph*{Conclusion}
In summary, in this section we studied the discrimination potential of six $Z'$ models at HE-LHC with an assumed \com~energy of $27\TeV$ and an integrated luminosity of $\mathcal{L}=15\abinv$. The exercise has been performed assuming the evidence of an excess observed at $\sqrt{s}=14\TeV$ at a mass $m_{Z'}\approx 6\TeV$. Overall it was found that it is possible to distinguish among most models. Finally, it should be noted that further studies, perhaps employing 3-body decay modes or associated Z' production will be clearly needed to be pursued in case of discovery to further characterise the resonance properties.

\subsection{Spin $1/2$ resonances}\label{sec:BSMspinhalf}
In this section, prospect studies for spin-1/2 resonances are presented, targeting excited leptons and heavy vector-like quarks. Resonances coupled to leptons and quarks or gauge bosons and quarks are considered.

\subsubsection{\cms{Search for excited leptons at HL-LHC}}
\contributors{S. Ha, B. Kim, M. S. Kim, K. Nam, S. W. Lee, H. D. Yoo, CMS}
%{\bf Author(s): S. Ha, B. Kim, M. Kim, K. Nam, S. Lee, H. Yoo et al., CMS}

\begin{figure}[t]
\begin{center}
\includegraphics[width=0.3\textwidth]{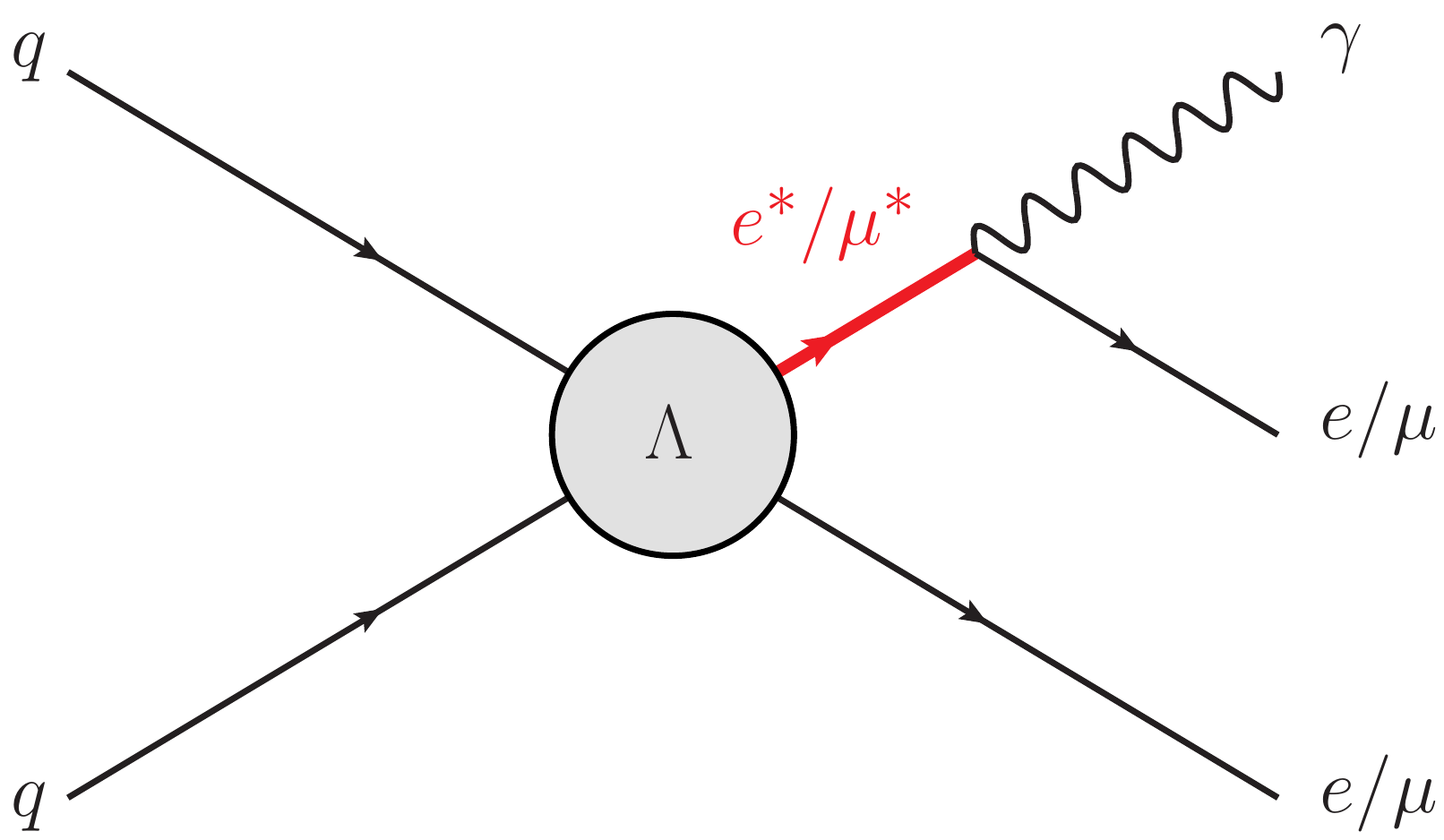}
\end{center}
\caption{\label{fig:Feynman-lstar} Feynman diagram of the production of excited leptons in $\ell\ell\gamma$ final states.}
\end{figure}

A search for excited leptons (electrons and muons) is studied at the HL-LHC with the upgraded CMS detector using simulation~\cite{CMS-PAS-FTR-18-029}.
Excited leptons are predicted by many BSM theories where quarks and leptons are not elementary but instead are themselves composite objects.
The HL-LHC environment (a \com energy of $14\TeV$ and an integrated luminosity of 3\abinv)
allows to extend the discovery potential of excited leptons.
This analysis presents a search for excited leptons ($\lstar=$ \estar, \mustar) in \sloppy\mbox{$\ell\ell\gamma$ ($\ell=$ \Pe, \Pgm)}
final states where the excited lepton decays to a SM lepton and a photon ($\lstar\rightarrow\ell\gamma$).
An illustration of the production decay mode is shown in~\fig{fig:Feynman-lstar}.

In this search, a clear signature of an opposite-sign same-flavour
lepton pair and a photon allows highly efficient signal selection to assess the CMS upgrade physics reach.
However, an ambiguity between the lepton from the excited lepton decay and the lepton from the contact interaction makes it challenging
to identify the reconstructed mass of the excited lepton due to two possible pairings of a lepton and the photon.
For this search, information from both invariant mass combinations is used to discriminate the excited lepton signal from SM background processes.
We consider a benchmark model based on the formalism described in~\citeref{Baur:1989kv}.

The signal samples are generated with PYTHIA~8.205 \cite{Sjostrand:2006za} at $\Lambda = 10\TeV$ for \lstar
masses ranging from $3.5\TeV$ to $6.5\TeV$ in steps of $250\GeV$,
where $\Lambda$ is the compositeness scale.
The simulated signal samples are generated at leading order (LO) in perturbative quantum chromodynamics.
The main background is the SM $\PZ\gamma$ process, which is generated at NLO using \MGvATNLO 2.3.3 ~\cite{Alwall:2011uj, Alwall:2014hca}.
The generated signal and background samples are interfaced to \\delphes~\cite{deFavereau:2013fsa},
which features a parametric simulation of the CMS Phase 2 detector at the particle level.

% event selection
%The signature of the signal event in this search has a same-flavour (SF) lepton pair and a photon.
%The leptons and photon from the signal event are produced centrally for $\mlstar > 3.5\TeV$
%and therefore they are separated from the PU which is more significant in the high $\eta$ region.
We select events having two isolated electrons or muons and a photon with requirements as follows.
Electron and photon candidates are required to have 
pseudorapidity $\abs{\eta} < 2.5$ and transverse momentum $\pt > 35\GeV$,
and they are excluded in the electromagnetic calorimeter barrel-endcap transition region ($1.44 < \abs{\eta} < 1.57$).
Muon candidates should be isolated with $\abs{\eta} < 2.4$ and $\pt > 35\GeV$.
The leptons are required to have opposite charge and the selected electrons 
and muons must be separated from the photon by $\Delta R = \sqrt{\Delta \eta ^2 + \Delta \phi ^2} > 0.7$.
In addition, the invariant mass of the two same flavour leptons ${m}_{\ell\ell}$ is required to be larger than
$116\GeV$ in order to suppress the dominant background contribution from real $\PZ$ boson production 
($\PZ$ resonance veto criteria).

\begin{figure}[t]
\centering
\begin{minipage}[t][\minipageheightdp][t]{\minipagewidthdp}
\centering
\includegraphics[width=\figuremaxwidthdp,height=\figuremaxheightdp,keepaspectratio]{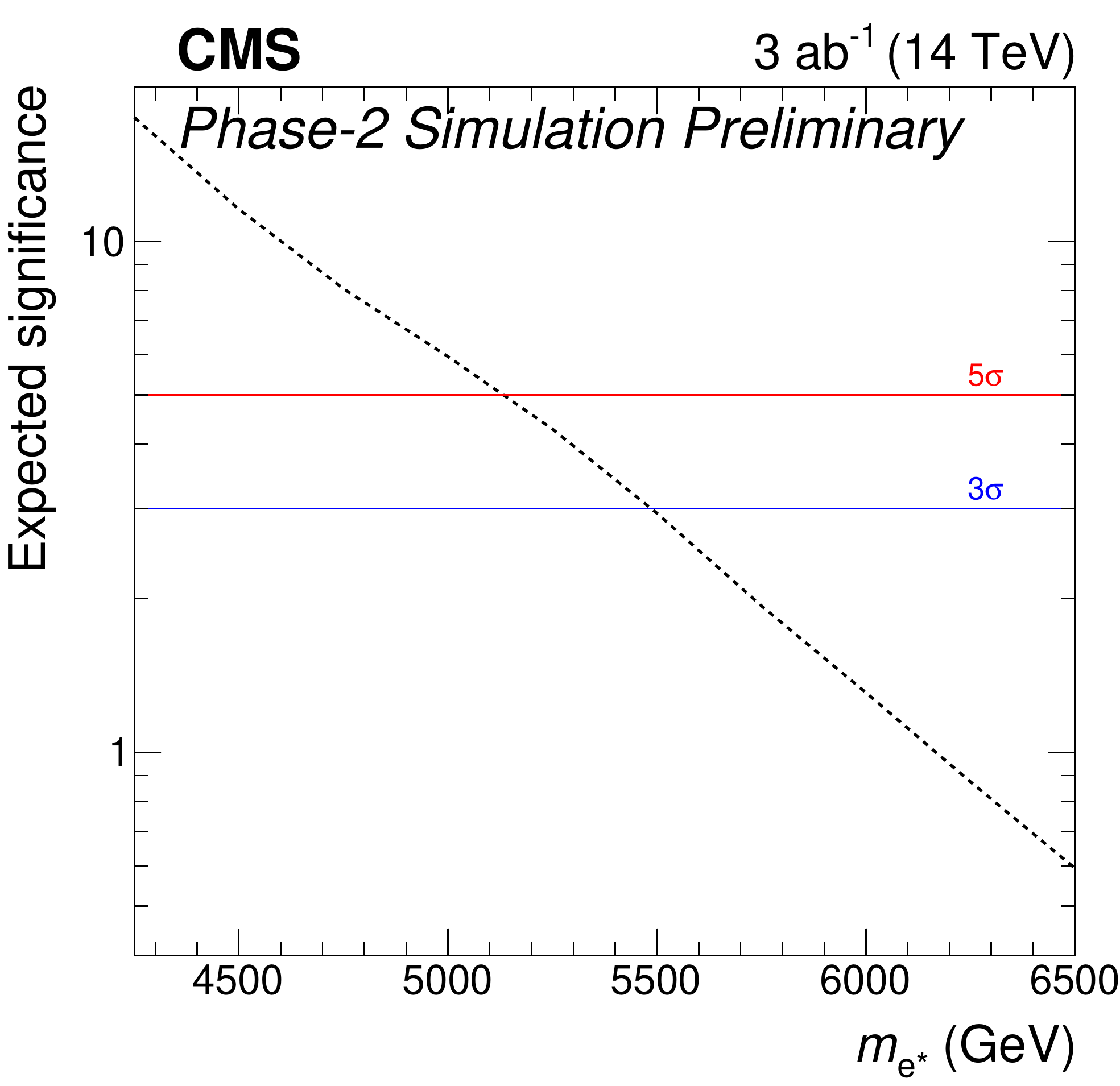}
\end{minipage}
\begin{minipage}[t][\minipageheightdp][t]{\minipagewidthdp}
\centering
\includegraphics[width=\figuremaxwidthdp,height=\figuremaxheightdp,keepaspectratio]{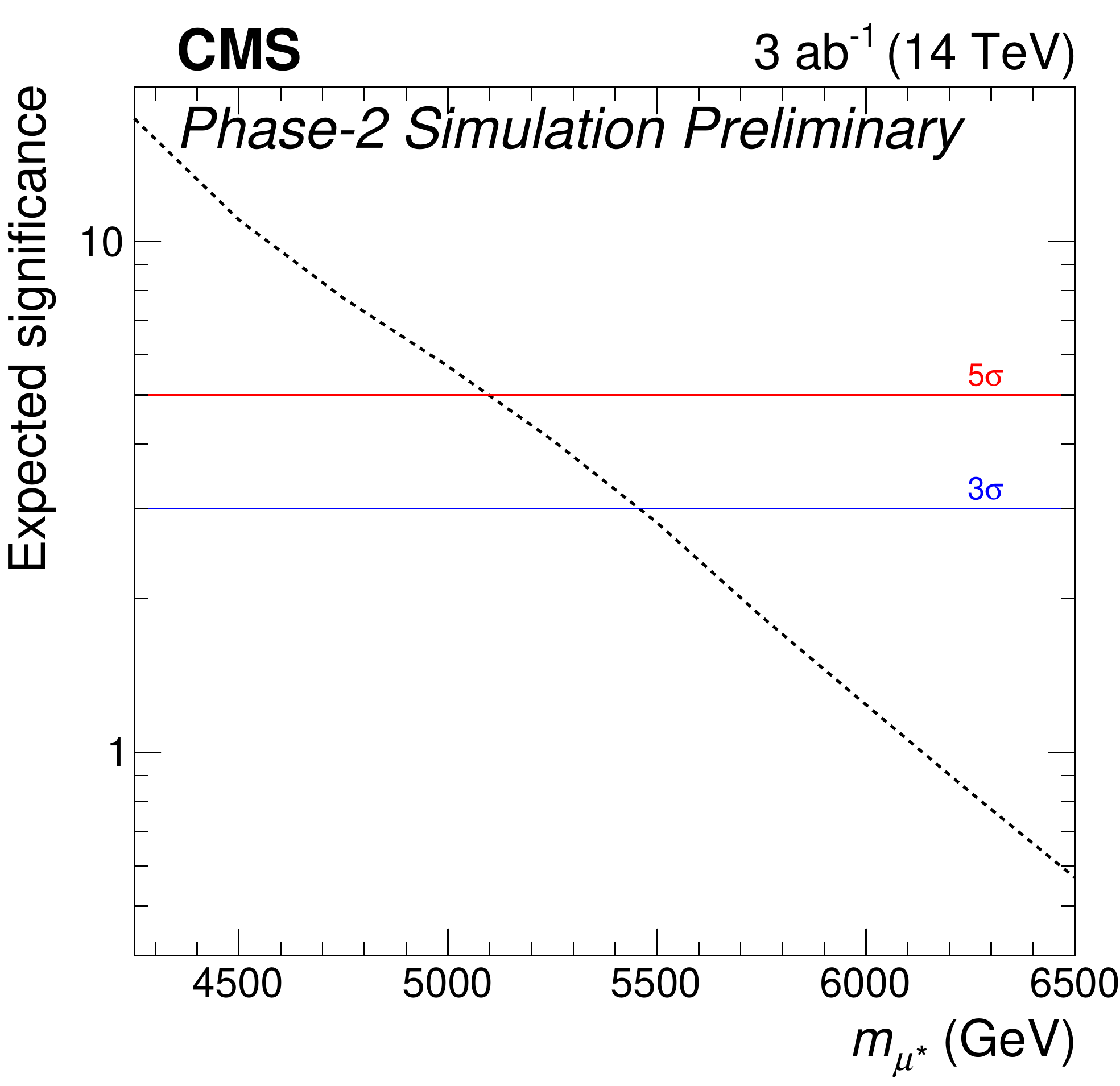}
\end{minipage}
\caption{Discovery significance for excited electrons (left) and muons (right)
with 3\abinv at the HL-LHC.
}
\label{fig:significance}
\end{figure}

% background discussion
The main SM background after the event selection is Drell-Yan production associated with a photon ($\PZ\gamma$),
%which has same signature in the final state, in case of Z boson decaying into two leptons.  
which has the same signature as the final state of the signal, when the $\PZ$ boson decays into two leptons.
This background is significantly suppressed by the $\PZ$ boson veto requirement.
Contributions of other SM processes like diboson and top quark pair production in association with a photon ($\ttbar+\gamma$)
are relatively small, in particular in the signal search region of excited lepton masses above $2\TeV$.
Simulated $\ttbar+\gamma$ events are studied in this analysis, however the background events
are imperfectly estimated due to the insufficient sample size.
Hence, we only consider the dominant $\PZ\gamma$ background in this search, and
additional background contributions are considered as systematic uncertainties on the total
background estimate.
%As any contribution from $\ttbar+\gamma$ \delphes{} samples is obtained in this analysis, this contribution is missing in the estimated background yield. To take its contribution into account, 20\% systematic uncertainty is assigned in the estimated background yield. 
The photon misidentification rate under the HL-LHC conditions is studied in~\citeref{CMS:2018rig} using PHASE-2
\delphes{} samples.
The photon misidentification rate is expected to be about $1\%$ when the photon $\pt$ is on the order of $100\GeV$, which is compatible with the 2016 result.
In the previous CMS Run-2 \lstar search~\cite{CMS-PAS-EXO-18-004}, 
we observed $20\%$ and $5\%$ contribution from $\ttbar+\gamma$ and misidentified photon backgrounds, 
respectively, at $\mlstar > 1\TeV$; therefore a $25\%$ systematic uncertainty 
is assigned for the missing background contributions.

\begin{figure}[t]
\centering
\begin{minipage}[t][\minipageheightdp][t]{\minipagewidthdp}
\centering
\includegraphics[width=\figuremaxwidthdp,height=\figuremaxheightdp,keepaspectratio]{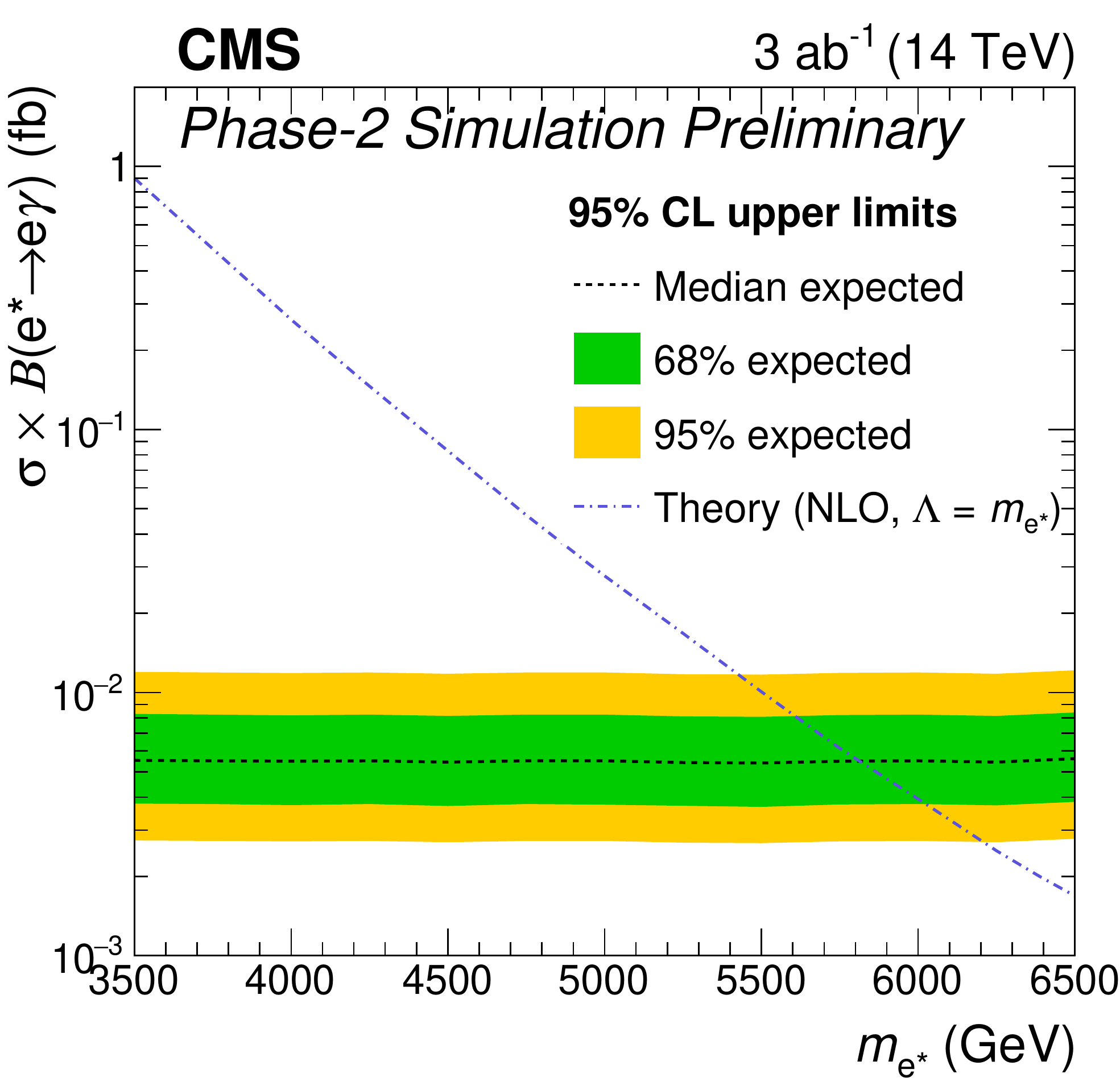}
\end{minipage}
\begin{minipage}[t][\minipageheightdp][t]{\minipagewidthdp}
\centering
\includegraphics[width=\figuremaxwidthdp,height=\figuremaxheightdp,keepaspectratio]{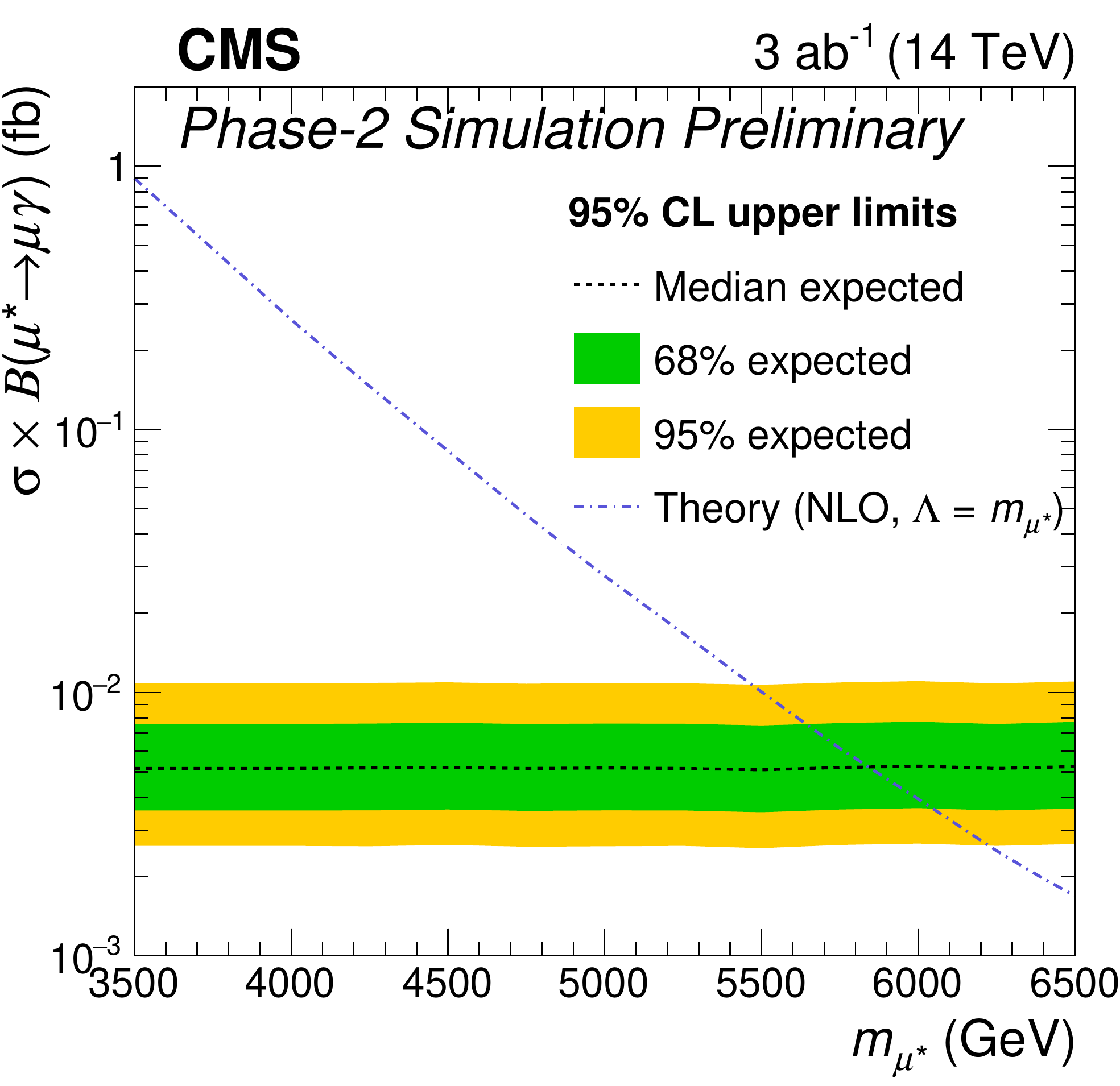}
\end{minipage}
\caption{Exclusion limits for excited electrons (left) and muons (right)
on the product of cross section and branching fraction.
}
\label{fig:limits}
\end{figure}

To distinguish between signal and background events, a two-dimensional distribution of the
two invariant masses \Mmin and \Mmax is used.
A search window is set in the two-dimensional distribution of \Mmax versus \Mmin.
For \lstar events, either \Mmin or \Mmax corresponds to the reconstructed invariant mass of \lstar.
Therefore, the mass resonance of the signal is concentrated in an ``L'' shape~\cite{cms-limit-7TeV, cms-limit-8TeV}.
On the other hand, background events have no such correlation in \Mmin and \Mmax and are scattered around at low masses below about $2\TeV$.
This distinction between signal and background events in the distribution of \Mmax versus \Mmin is
used to set the search window.
We set the lower \Mmax bound at $2\TeV$ in the two-dimensional distribution
as the search window in order to maximise the signal yields.
The reason why the L-shaped search window is not applied in this analysis is due to the insufficient MC statistics.
Therefore the results of the limits presented here are likely to be improved upon with an actual search.
The product of signal acceptance and efficiency ($\mathrm{A} \times \varepsilon_{\mathrm{sig}}$) is 
obtained using the simulated \delphes{} signal samples and the results are $58\%$ (\mustar) 
and $45\%$ (\estar) with negligible \mlstar dependence.

% systematic uncertainty
Systematic uncertainties for the performance of the lepton ($0.5\%$) and photon ($2.0\%$) reconstruction and identification,
and the integrated luminosity ($1.0\%$) follow the recommendation for upgrade analyses~\cite{CMS:2017cwx}.
The theoretical systematic uncertainty is reduced by a factor of 1/2 with respect to the 2016 result.
The statistical uncertainty in the entire signal region is dominant in this analysis.
The missing background contribution is considered to be the main systematic
uncertainty in the background estimation.

% result
The upper limit of the excited electrons and muons is determined 
under the the HL-LHC scenario, based on an integrated luminosity of $3\abinv$. 
We set $95\%\cl$ upper limits on the production cross sections, which are computed with the modified
frequentist CL$_\mathrm{s}$ method~\cite{Junk:1999kv, Read:2002hq}, with a likelihood ratio used as a test statistic.
The systematic uncertainties are treated as nuisance parameters with log-normal priors.

The discovery potential as a function of excited lepton mass shown in~\fig{fig:significance}
indicates that $3\sigma$ evidence ($5\sigma$ discovery) is possible for both excited electrons and excited muons with masses
up to $5.5$ ($5.1$)\TeV. 
Figure~\ref{fig:limits} shows the expected upper limits for \estar (left) and \mustar (right).
The expected exclusion of the excited leptons is 
$\mlstar < 5.8 \TeV$ for both \estar and \mustar
in the case where $\mlstar = \Lambda$.
While the electron channel has a lower signal yield than the muon channel, it also has lower background, and the net result is that the excluded cross sections differ only by about $10\%$, producing a similar exclusion limit on the excited lepton mass.

\subsubsection{\theoryC{VLQs at HL- and HE-LHC: discovery and characterisation}}
\contributors{D. Barducci, L. Panizzi}
%{\bf Author(s): D. Barducci$^a$, L. Panizzi$^b$\\
%$^a$ SISSA and INFN, Sezione di Trieste, via Bonomea 265, 34136 Trieste, Italy\\
%$^b$ Dipartimento di Fisica, Universit\`a di Pisa and INFN, Sezione di Pisa, Largo Pontecorvo 3, I-56127 Pisa, Italy
%}

%\subsubsection{Motivations}
%\paragraph*{Motivations}

Vector Like Quarks (VLQs) are hypothetical heavy quarks whose left- and right-handed chiral components transform under the same representation of the SM gauge group. 
In minimal extensions of the SM, VLQs couple to SM quarks via Yukawa-type interactions and gauge invariant renormalisable operators can be written only for the singlet, doublet and triplet representations of $SU(2)$~\cite{delAguila:2000rc}. It can be shown that the couplings of the VLQs with the SM bosons and quarks always have a dominant chiral component~\cite{delAguila:2000rc,Buchkremer:2013bha} and this depends only on whether their weak isospin is integer or half-integer, the other component being suppressed by a factor proportional to $m_q^{\rm SM}/m_{\text{VLQ}}$, with $m_q^{\rm SM}$ the mass of the SM quark with which the VLQ mixes.

This property affects the polarisation of the gauge bosons and quarks arising from the VLQs decay. While the gauge bosons tend to have a dominant longitudinal polarisation, the polarisation of the final state quarks allows one to extract useful information.
In particular if the VLQ decays into a top quark, its polarisation properties will affect the kinematic distributions of the final decay products. This can slightly affect the reach of new physics searches but, more importantly, in the fortunate event of a signal excess being observed, this difference can be used to probe the structure of the interactions between the VLQs and the SM sector.
\begin{figure}[t]
\centering
\begin{minipage}[t][\minipageheightsp][t]{\minipagewidthsp}
\centering
\includegraphics[width=\figuremaxwidthsp,height=\figuremaxheightsp,keepaspectratio]{\main/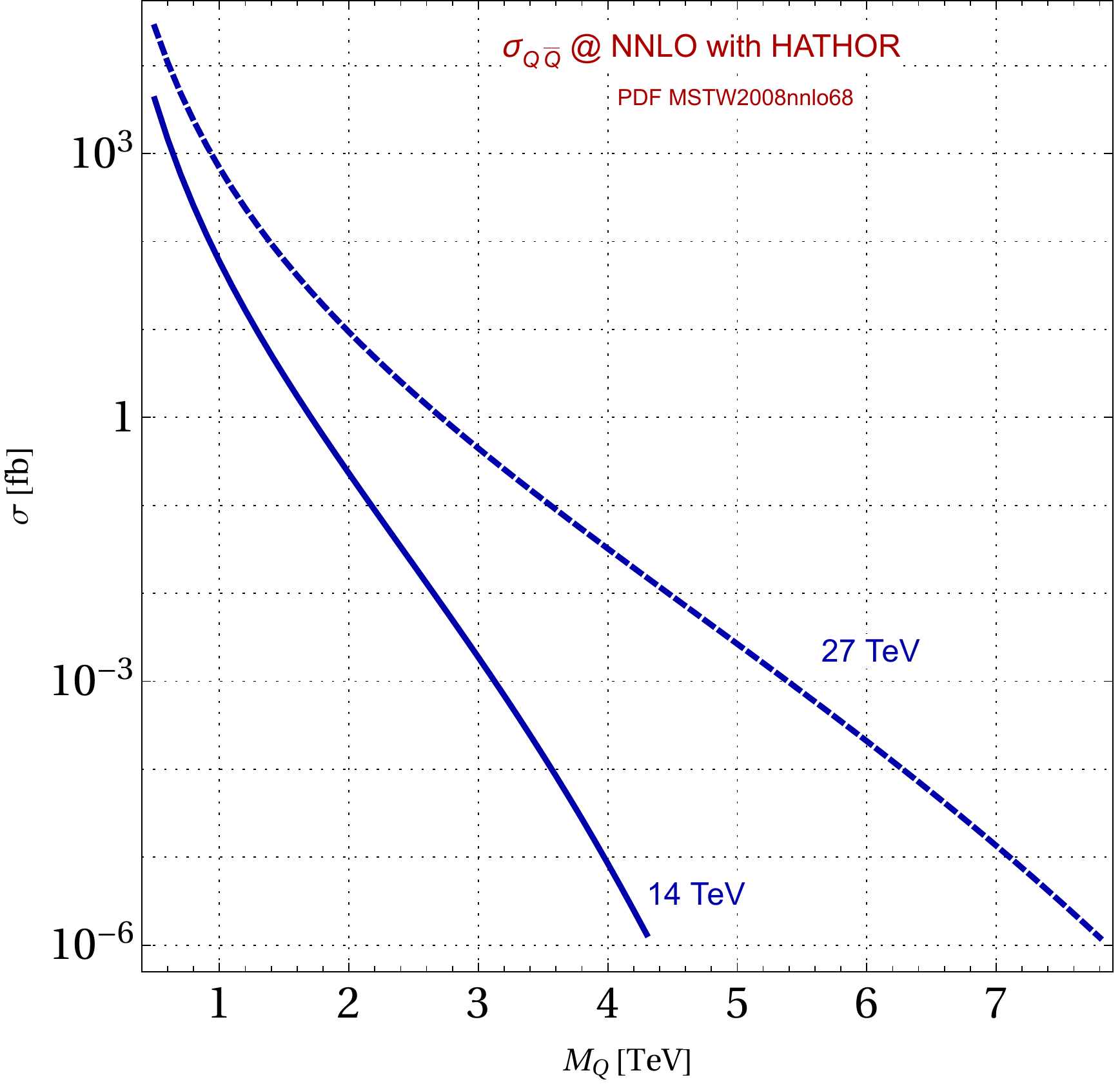}
\end{minipage}
\caption{VLQ pair production cross section for the $14$ and $27\TeV$ LHC. \label{fig:pair-prod}
}
\end{figure}

In this contribution, based on the results of~\citeref{Barducci:2017xtw}, we analyse the possibility of discriminating the chiral structure of VLQ couplings at the HL- and HE-LHC, under the hypothesis that the VLQ decays to the SM top quark. This eventually allows one to discriminate its representations under the SM $SU(2)$ gauge group.

%\subsubsection{Top quark polarisation}
%\paragraph*{Top quark polarisation}

The polar angle distribution of the top quark decay product $f$ in the top rest frame is described by
\begin{equation}
\frac{1}{\Gamma_l} \frac{{\rm d} \Gamma_l}{{\rm d}\cos \theta_{f,{\rm rest}}}=\frac{1}{2}(1+ P_t \cos\theta_{f,{\rm rest}})
\label{eq:pol-angle}
\end{equation}
where $\Gamma_l$ is the partial width, $\theta_{f,{\rm rest}}$ is the angle between the momentum of the decay product $f$ and the top spin vector and $P_t$ is the polarisation of the top. From \eq{eq:pol-angle} one sees that for positive (negative) polarised top quarks most of the decay products come in the forward direction, that is the directions of the would-be momentum of the top quark in the laboratory frame. 
In the same frame the $\theta_{f}$ distribution is now described by \eq{eq:pol-angle} combined with a boost from the top rest frame to the laboratory frame. This implies that positive polarised top quarks will produce harder decay products. 

%\subsubsection{Discovery and discrimination at HL-LHC}
%\paragraph*{Discovery and discrimination at HL-LHC}

To show how the polarisation information can be used to disentangle a VLQ chiral structure we focus on a VLQ with charge $2/3$ interacting exclusively with the top quark and the $Z$ boson. We recast a search for pair produced VLQs performed by the ATLAS collaborations in the single lepton channel with an integrated luminosity of $36.1\fbinv$ at $\sqrt{s}=13\TeV$~\cite{Aaboud:2017qpr}. 
Throughout this work the VLQ pair production cross sections have been computed with {\textsc HATHOR}~\cite{Aliev:2010zk} and are reported in \fig{fig:pair-prod}.

We then project the discovery and exclusion reach of the ATLAS search for the case of the $14\TeV$ LHC assuming the same signal acceptances than for the $13\TeV$ case and rescaling the backgrounds by the relevant parton luminosities ratios. We do these projections for higher values of integrated luminosities
and different values of the systematic uncertainties on the background determination, $\epsilon_{\rm{syst}}^{\rm{bkg}}$, 
 and then perform a $\chi^2$ fit on the leading lepton $p_T$ between the left- and right-handed coupling scenario considering the number of bins of the distribution as degrees of freedom, and under the simplifying assumption that the background (conservatively assumed to be distributed as the signal) can be subtracted with a certain efficiency, ranging from the extreme scenarios of $0\%$ and $100\%$: the relation we use is therefore $\chi^2=\sum_{i=1}^{\rm{bins}} (L_i-R_i)^2/\max[\{L_i,R_i\}+\epsilon_{\rm{cont}}^{\rm{bkg}}(B+(\epsilon_{\rm{syst}} B)^2)]$, where $L_i$ and $R_i$ are the number of events in the left- and right-handed coupling scenario, $\epsilon_{\rm{cont}}^{\rm{bkg}}$ represents the contamination percentage of background events considered for the discrimination, and we consider Poissonian uncertainties for the signal.

\begin{figure}[t]
\centering
\begin{minipage}[t][\minipageheightdp][t]{\minipagewidthdp}
\centering
\includegraphics[width=\figuremaxwidthdp,height=\figuremaxheightdp,keepaspectratio]{\main/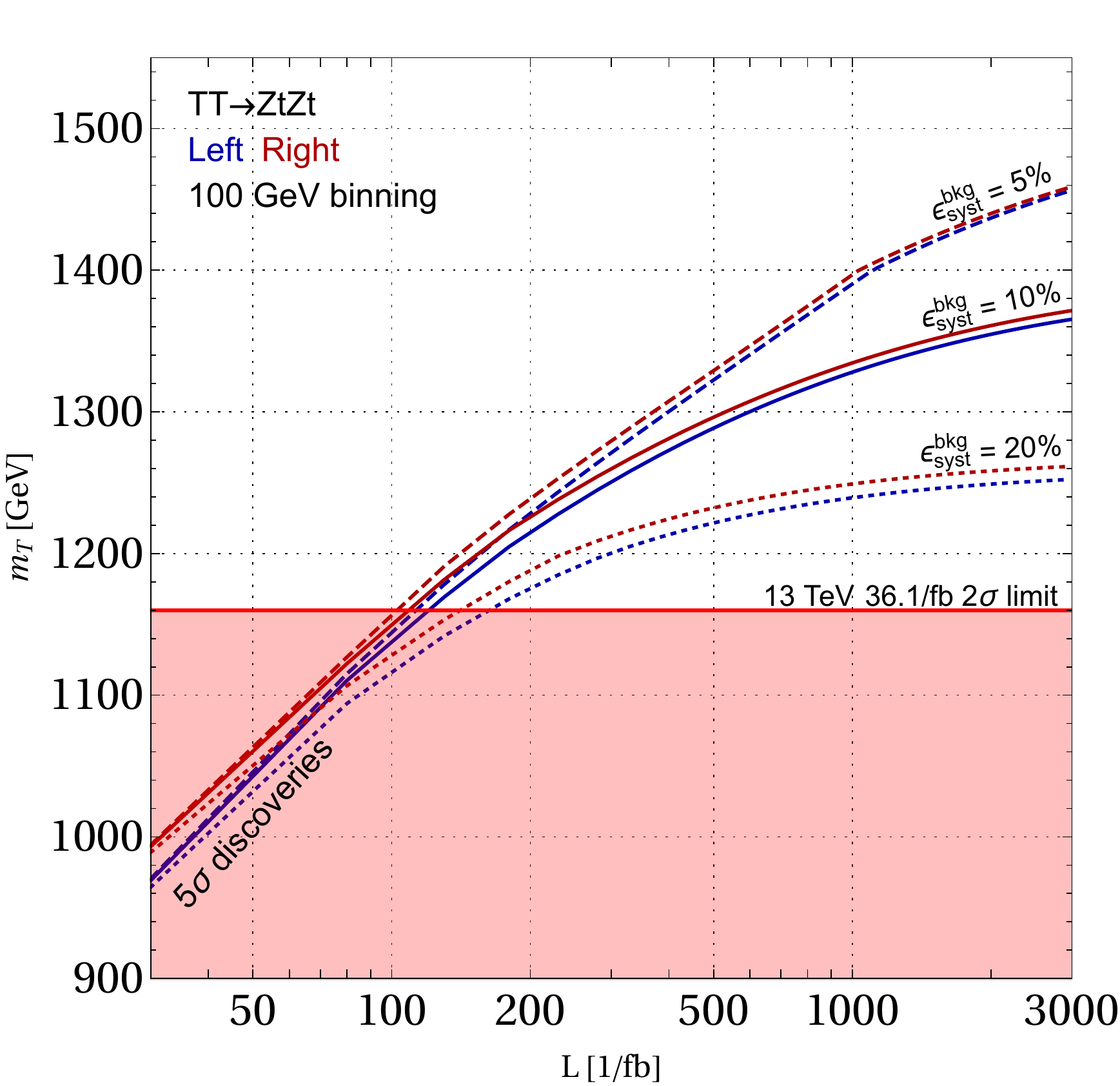}
\end{minipage}
\begin{minipage}[t][\minipageheightdp][t]{\minipagewidthdp}
\centering
\includegraphics[width=\figuremaxwidthdp,height=\figuremaxheightdp,keepaspectratio]{\main/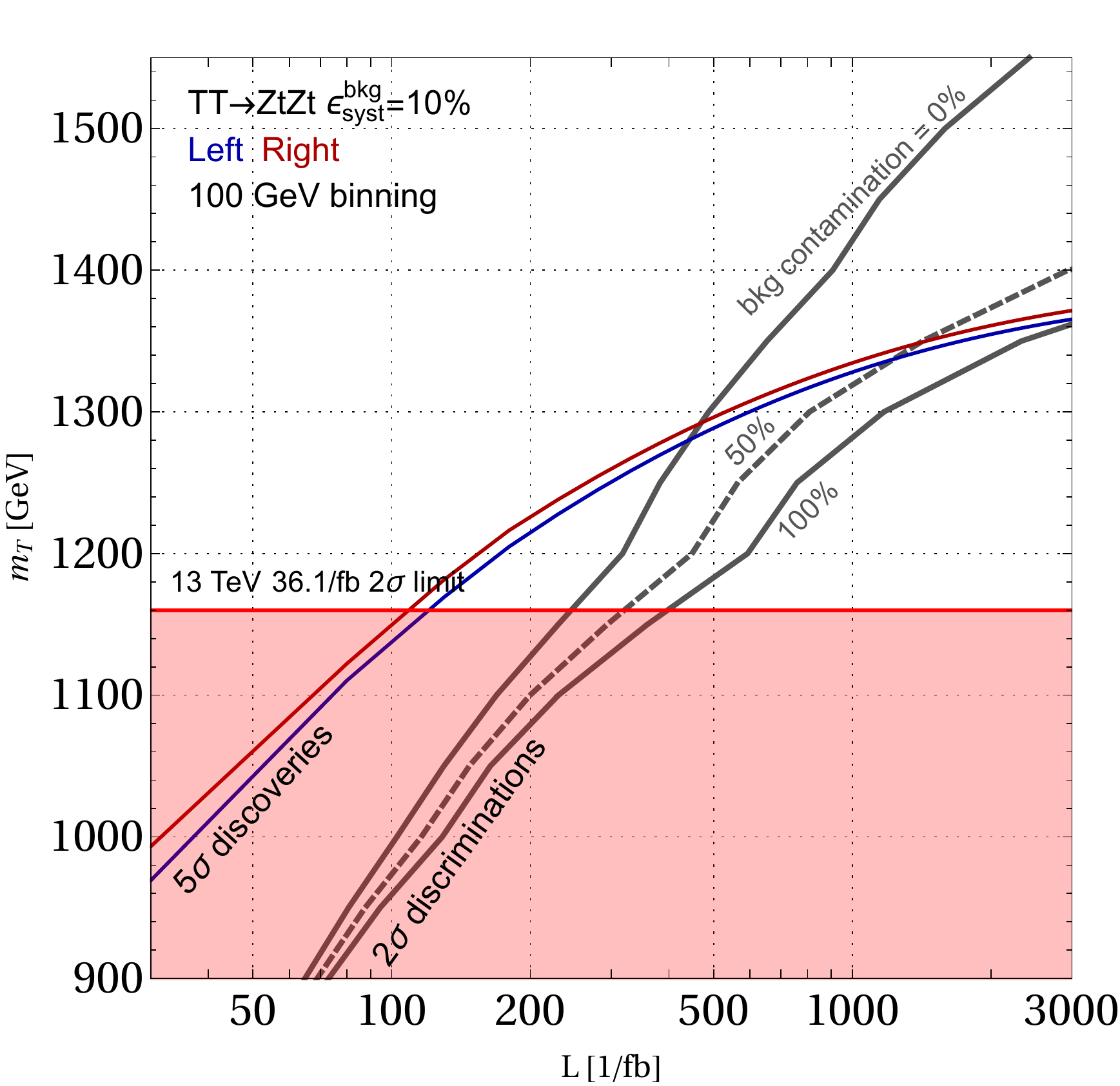}
\end{minipage}
\caption{Left: projected $5\sigma$ discovery (blue and red) for different assumptions for the systematic uncertainties on the background determination of the ATLAS single lepton search~\cite{Aaboud:2017qpr} for a $T$ VLQ decaying with $100\%$ BR into the $Zt$ final state extrapolated for the $14\TeV$ LHC (see main text for details). The blue and red lines correspond to the discovery reach for the left-handed and right-handed coupling structure. The red shaded area correspond to the experimental limit from~\cite{Aaboud:2017qpr} for the $13\TeV$ LHC from for a VLQ decaying with $100\%$ BR into the $Zt$ final state, namely $1160\GeV$. Right: projected $2\sigma$ discrimination (gray) reaches assuming $\epsilon_{\rm{syst}}^{\rm{bkg}}=10\;$\%.\label{fig:13-discr}
}
\end{figure}

The results are shown in \fig{fig:13-discr}, where we illustrate the $5\sigma$ discovery reaches for different 
$\epsilon_{\rm{syst}}^{\rm{bkg}}$ values (left panel) and the $2\sigma$ discrimination contours from the $\chi^2$ fit assuming $\epsilon_{\rm{syst}}^{\rm{bkg}}=10\%$ (right panel).
We see that, if it is possible to perform a discrimination $\chi^2$ test after removing all background events, should a VLQ with a mass lighter than around $1300\GeV$ be discovered, a mild increase in integrated luminosity will be needed to disentangle the two hypotheses, while the collected dataset will already be enough for the discrimination if a VLQ heavier than $1300\GeV$ is found with a $5\sigma$ significance. If background events cannot be removed the discrimination becomes more difficult, as the differences in the shapes become less relevant. The intersection between the discovery reach and the discrimination reach moves therefore towards higher luminosities, and in the limit of $100\%$ background contamination we find that the discovery reach corresponds to the discrimination power at the nominal luminosity of $3\abinv$.

%\subsubsection{Discovery and discrimination at HE-LHC}
%\paragraph*{Discovery and discrimination at HE-LHC}

The results obtained for the $14\TeV$ LHC show that with the signal region currently used in the ATLAS search we have considered, the LHC discovery reach will mildly increase when further data will be collected (unless one assumes a strong reduction of the current systematic uncertainties on the background determination), due to the fact that at hadron colliders the contribution of the PDFs drops when the transferred momentum of the process approaches the kinematic limit $\sqrt{s}/2$.\footnote{Clearly an optimisation of the signal regions can be performed to increase the sensitivity already at the current LHC energy.}
We then estimate here what is the mass reach of the high energy upgrade of the LHC for discovering pair-produced VLQs and discriminate their coupling structure.

\begin{figure}[t]
\centering
\begin{minipage}[t][\minipageheightdp][t]{\minipagewidthdp}
\centering
\includegraphics[width=\figuremaxwidthdp,height=\figuremaxheightdp,keepaspectratio]{\main/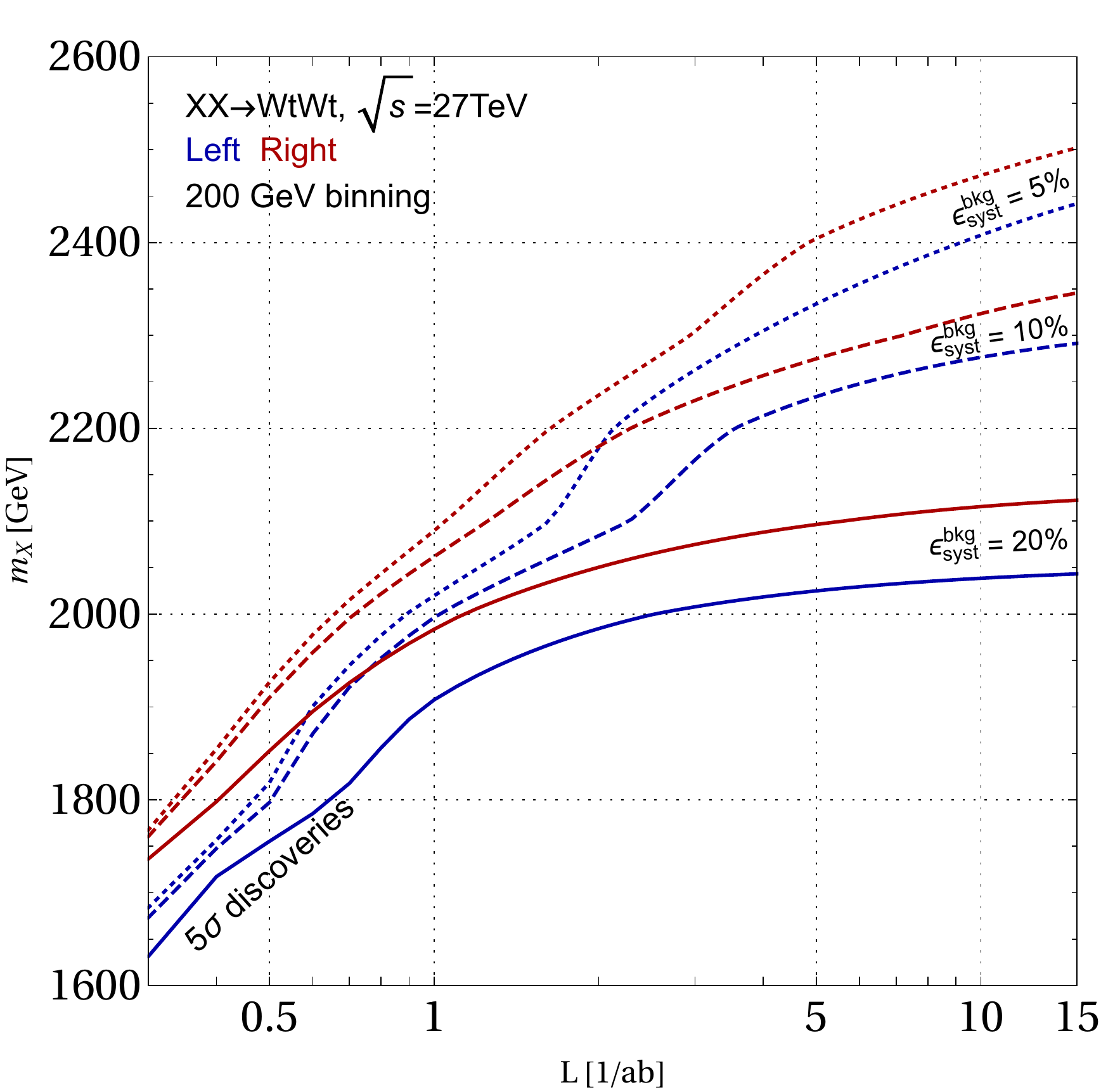}
\end{minipage}
\begin{minipage}[t][\minipageheightdp][t]{\minipagewidthdp}
\centering
\includegraphics[width=\figuremaxwidthdp,height=\figuremaxheightdp,keepaspectratio]{\main/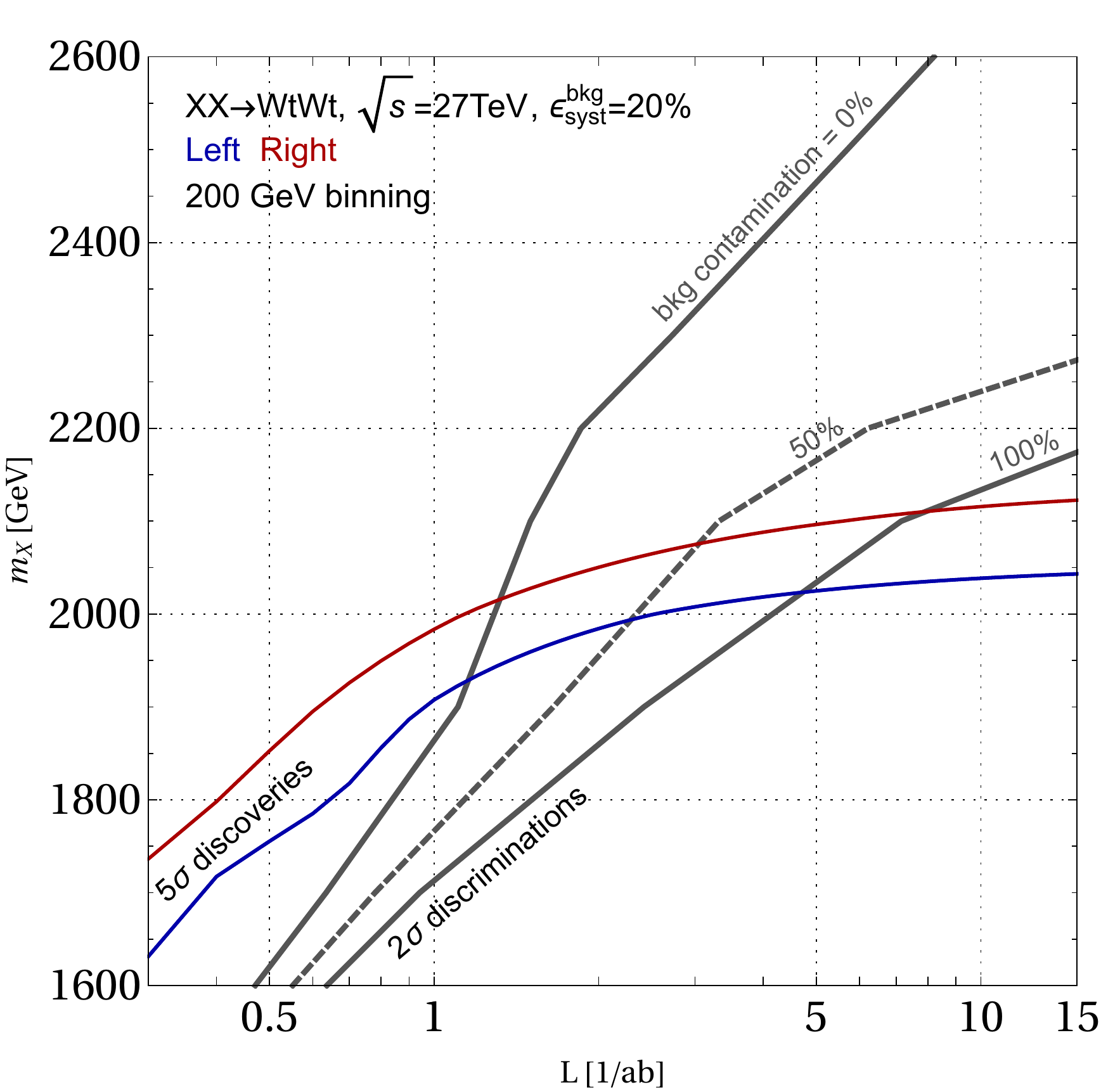}
\end{minipage}
\caption{Same as \fig{fig:13-discr} for the 2SSL search for a $X$ VLQ decaying with $100\%$ BR into the $Wt$ final state at $\sqrt{s}=27\TeV$. We use the $p_T$ distribution of the sub-leading lepton with a binning of $200\GeV$ and assume an uncertainty on the background determination of $20\%$ (right).\label{fig:27Lumi_2}}
\end{figure}

We focus on the case of a VLQ with charge $5/3$ decaying into a $W$ boson and a top quark and closely follow the search strategy defined in \citeref{Avetisyan:2013rca}. In this case we consider for discrimination the $p_T$ distribution of the sub-leading lepton, which is almost always coming from the SM top decay. With the same statistical procedure adopted above, and rescaling the backgrounds yields given in \citeref{Avetisyan:2013rca} for $\sqrt{s}=27\TeV$ we show the results in \fig{fig:27Lumi_2}. Our results show that a discrimination among the left- and right-handed coupling hypotheses is possible in all the discovery range accessible at the HE-LHC for $\epsilon_{\rm{syst}}^{\rm{bkg}}=20\%$, regardless of the background contamination. In particular, if the background can be entirely subtracted, should a VLQ with mass greater than $\sim 2\TeV$ be discovered, the collected data set will already be sufficient to exclude one of the two coupling structure, while if the background is entirely considered, discrimination is possible at the same time as discovery for a VLQ with mass greater then $\sim 2100\GeV$. 

%\subsubsection{Conclusion}
%\paragraph*{Conclusion}

%In conclusion, our study shows that if a VLQ interacting with the SM top quark is observed in the high-luminosity phase of the LHC or in high-energy future upgrades, it will be possible to exploit the kinematic properties of its decay products for discriminating the dominant chirality of its couplings. 
%We have performed a $\chi^2$ analysis to discriminate the $p_T$ distributions of the leading lepton for the pair production process of a left- or right-handed VLQ with charge 2/3 decaying to $Z$ and top, considering a single-lepton signal region used in a search from ATLAS at $13\TeV$ and projecting the results for the HL-LHC phase. Subsequently, we have used the same analysis technique considering the $p_T$ distributions of the sub-leading lepton, for a VLQ with charge 5/3 decaying to $W$ and top observed in a HE-LHC upgrade at 33 TeV considering a same-sign dilepton signal region. In both cases we find that a 2$\sigma$ discrimination is possible within the discovery range of the machines.

\subsection{Signature based analyses}\label{sec:BSMsignaturebased}
Several contributions that are constructed around experimental signatures rather than specific theoretical models are presented in this section. This includes analyses of dijets, diphotons, dibosons and ditops final state events at HL- and HE-LHC. 

\subsubsection{\theory{Coloured Resonance Signals at the HL- and HE-LHC}}
\contributors{T. Han, I. Lewis, Z. Liu}
%\textbf{Authors: Tao Han$^{a}$, Ian Lewis$^{b}$, Zhen Liu$^{c,d}$}\\
%\textit{a) PITT-PACC, Department of Physics and Astronomy, University of Pittsburgh, PA 15260, USA\\
%b) Department of Physics and Astronomy, University of Kansas, Lawrence, KS 66045, USA\\
%c) Theoretical Physics Department, Fermi National Accelerator Laboratory, Batavia, IL, 60510\\
%d) Maryland Center for Fundamental Physics, Department of Physics, University of Maryland, College Park, MD 20742, USA
%}\\
%\texttt{than@pitt.edu, ian.lewis@ku.edu, zliu2@fnal.gov}

\newcommand\rep\mathbf

% %%%%%%%%%%%%%%%%%%%%%%%%%%%%%%%%%%%%%%%%%%%%%%
%% \subsubsection{Introduction}
%\paragraph*{Introduction}
% \label{sec:intro}
% %%%%%%%%%%%%%%%%%%%%%%%%%%%%%%%%%%%%%%%%%%%%%%
%With the CERN Large Hadron Collider (LHC) accumulating an unprecedented amount of data at the energy frontier, the field of high energy physics is entering an exciting new era. 
%The HL-LHC will increase our knowledge of Nature and has great potential to discover new physics.  Additionally, the proposed HE-LHC at 27 TeV promises to substantially increase the chances of observation of high scale new physics.  
While much of the attention for new physics discovery has centred on precision measurements of the Higgs and EW sectors, the LHC is a QCD machine and most initial states are composed of coloured particles.  Hence, new coloured dijet resonances that couple to partons will be produced with favourable rates at the HL-LHC and HE-LHC.

In \citeref{Han:2010rf} we classified the possible coloured resonances at the LHC according to their spin, electric charge, and colour representation.  We now update those results for the HE-LHC. This involves comparing the cross sections of the various resonances at the LHC and HE-LHC. 
%
%The short note is organised as follows.  In Section~\ref{par:models} we review our classifications of the different possible resonances.  Section~\ref{par:2j} we update the cross sections of dijet resonances for the HE-LHC.  Finally, in Section~\ref{par:conc} we will summarise our results and provide an outlook.

%%%%%%%%%%%%%%%%%%%%%%%%%%%%%%%%%%%%%%%%%%%%%%
%\subsubsection*{Models}
%\paragraph*{Models}
%\label{par:models}
%%%%%%%%%%%%%%%%%%%%%%%%%%%%%%%%%%%%%%%%%%%%%%
Most initial states at the LHC are composed of coloured particles, \iee~quarks and gluons. We now review the possible interactions of coloured resonances with SM partons.  
%The resonances are classified according to their spin, electric charge, and $SU(3)_C$ colour representation.  
A more detailed discussion, including examples of specific realisations of the various resonances in the existing literature, is given in \citeref{Han:2010rf}.  All interactions are after EWSB.

Quark-quark annihilation can produce colour antitriplet or sextet scalars and vectors, so-called ``diquarks''.  Please note that the diquarks under discussion here are fundamental particles and not composite.  The possible scalar diquark are denoted as  $E_{N_D},~U_{N_D}$, and $D_{N_D}$ with electric charges $4/3,2/3,-1/3$ respectively.  The subscript $N_D=3,~6$ for the $\overline{\rep3}$ and $\rep{\mathbf6}$ colour representations, respectively.  Vector diquarks are represented with an additional Lorentz index $\mu$.  The interaction Lagrangian between quarks and diquarks is then
\begin{equation}\label{eq:qq}
\begin{array}{lll}
\displaystyle \mathcal{L}_{qqD} &= & \displaystyle K^j_{ab} \left[ \lambda^{E}_{\alpha\beta}
E^j_{N_D}\ \overline{u^C_{\alpha a}}P_\tau u_{\beta b}
+\lambda^{U}_{\alpha\beta} U^j_{N_D}\ \overline{d^{C}_{\alpha a}}P_\tau d_{\beta b}
+\lambda^{D}_{\alpha\beta} D^j_{N_D}\ \overline{d^C_{\alpha b}}P_\tau u_{\alpha a} \right. \\\vspace{2mm}
&+&\displaystyle \lambda^{E'}_{\alpha\beta} E^{j\mu}_{N_D}\  \overline{u^C_{\alpha a}}\gamma_\mu P_R u_{\beta b}
+ \lambda^{U'}_{\alpha\beta}  U^{j\mu}_{N_D}\ \overline{d^C_{\alpha a}}\gamma_\mu P_R d_{\beta b} 
%&&
\left.
+\lambda^{D'}_{\alpha\beta}\ D^{j\mu}_{N_D} \overline{u^C_{\alpha a}}\gamma_\mu P_\tau d_{\beta b}
\right]
+\rm{h.c.},
\end{array}
\end{equation}
where $P_{\tau}={1\over2}(1\pm\gamma_5)$ with $\tau=R,L$ for the right- and left-chirality projection operators, $a,b$ are colour indices for the $SU(3)_C$ fundamental representation, $j$ are colour indices of the $N_D$ representation of $SU(3)_C$, $\alpha,\beta$ are flavour indices,  and $K^j_{ab}$ are $SU(3)_{C}$ Clebsch-Gordan (CG) coefficients.

Quarks and gluons annihilate into colour triplet or antisextet fermions with $1/2$ or $3/2$ spin.  It is possible to produce a $\rep{15}$ colour representation, but the existence of such a fermion would spoil asymptotic freedom~\cite{Chivukula:1990di}.  The spin $1/2$ ($3/2$) fermion states are denoted by $d^*_{N_D},u^*_{N_D}$ ($d^{*\mu}_{N_D},u^{*\mu}_{N_D}$) with electric charged $-1/3$ and $2/3$, respectively. The lowest order gauge invariant interactions between a gluon, quark, and heavy fermion is dimension five:  
\begin{equation}\label{eq:qstar}
\begin{array}{lll}
\displaystyle\mathcal{L}_{qgF} &=&  \displaystyle\frac{g_{s}}{\Lambda}F^{A,\rho\sigma}
 \left[\bar u {\bar{K}_{N_D,A}} (\lambda^U_LP_L+\lambda^U_RP_R) \sigma_{\rho\sigma} u_{N_D}^{*} + \bar d {\bar{K}_{N_D,A}} (\lambda^D_LP_L+\lambda^D_RP_R) \sigma_{\rho\sigma} d_{N_D}^{*} ~~~~\right. \\\vspace{2mm}
&&  + \left.\bar u {\bar{K}_{N_D,A}} (\lambda^U_LP_L+\lambda^U_RP_R) \sigma_{\rho\sigma} \gamma_{\mu} u_{N_D}^{*\mu}  + \bar d {\bar{K}_{N_D,A}} (\lambda^D_LP_L+\lambda^D_RP_R) \sigma_{\rho\sigma} \gamma_{\mu} d_{N_D}^{*\mu} \right ] + \rm{h.c.}\,,
\end{array}
\end{equation}
where $A$ is the adjoint colour index, $F^{A,\rho\sigma}$ is the gluon field strength tensor, $\sigma_{\rho\sigma}=\frac{i}{2}[\gamma_\rho,\gamma_\sigma]$, $\Lambda$ is the scale of new physics, and $K_{N_D,A}$ are $3\times N_D$ CG coefficient matrices.  

Gluon-gluon annihilation can result in many different representations, that unlike the $\rep{15}$ fermion do not spoil asymptotic freedom.  A complete list of the possible resonances from gluon-gluon annihilation can be found in Table 1 of \citeref{Han:2010rf}.  We will focus on the theoretically motivated colour octet resonances.  Two possible resonances that can result from gluon-gluon annihilation are colour octet scalars, $S_8$, and tensors, $T_8^{\mu\nu}$.  These interactions can be described in a gauge invariant way by dimension five operators:

\begin{eqnarray}
\mathcal{L}_{gg8}=g_sd^{ABC}\bigg{(}\frac{\kappa_S}{\Lambda_S} S_{8}^{A} F^B_{\mu\nu}F^{C,\mu\nu}+\frac{\kappa_{T}}{\Lambda_T}({T^{A,\mu\sigma}_8}F^{B}_{\mu\nu}{F^{C}_{\sigma}}^{~\nu}+
f {T_{8\ \rho}^{A,\rho}} \
F^{B,\mu\nu}F^C_{\mu\nu})\bigg{)},
\label{eq:tensscal}
\end{eqnarray}
where $\Lambda_{S,T}$ are the new physics scales, and the relative coupling factor $f$ is expected to be order one.  

Finally, quark-antiquark annihilation can produce colour octet or singlet scalars and vectors with zero or unit charge.  The neutral vector-octet is denoted by $V^0_8$ and the  charged vector octet states $V^\pm_8$.  The interaction Lagrangian is then
\begin{equation}\label{eq:qqV}
\begin{array}{lll}
\displaystyle\mathcal{L}_{q\bar{q}V}&=&\displaystyle g_s\left[ {V_8}^{0,A,\mu}\ \bar{u} T^A \gamma_\mu (g^U_L P_L+g^U_{R}P_R)u+
{V_8}^{0,A,\mu}\ \bar{d} T^A \gamma_\mu (g^D_L P_L+g^D_{R}P_R)d \right .\ \\\vspace{2mm}
&&\displaystyle \left .+\left(V_8^{+,A,\mu}\ \bar{u} T^A \gamma_\mu (C_L V^{\text{CKM}}_LP_L+C_R V^{\text{CKM}}_R P_{R})d+\rm{h.c.}\right)\right ] ,
\end{array}
\end{equation}
where $V^{\text{CKM}}_{L,R}$ are the left- and right-handed Cabibbo-Kobayashi-Maskawa matrices, respectively.  To avoid constraints from flavour physics it is assumed that the CKM matrices align with the SM CKM matrices and that there are no tree level flavour changing neutral currents, \iee, $g^{U,D}_{L,R}$ and $C_{L,R}$ are flavour-diagonal.  To obtain the interactions with the colour singlet bosons it is sufficient to replace the representation matrices ${T^{A,a}}_b$ with the Kronecker delta ${\delta^a}_b$.  The couplings between octet and singlet scalar and light quarks is constrained to be small by minimal flavour violation~\cite{Manohar:2006ga}.  It is possible for a new scalar to have substantial couplings to the third generation quarks consistent with minimal flavour violation if it has a non-trivial representation under the SM flavour groups~\cite{Arnold:2009ay}.  We will only consider particles that couple to the light quarks and can be resonantly produced at the LHC.  Hence, we ignore scalar singlet and octet contributions to $s$-channel resonances with quark/anti-quark initial states.

\begin{table}[t]
\begin{center}
\small\begin{tabular}{|c|c|c|c|c|c|}  \hline
 Particle Names & $J$  & $SU(3)_{C}$  & $|Q_{e}|$ & $B$ & Related models \\
 (leading coupling) &  &  &  &  &  \\ \hline
$E_{3,6}^{\mu}\ (uu)$       &$0, 1$& ${\rep3},\ \overline{\rep6}$ &
$4\over3$ & $ -{2\over3} $ & scalar/vector diquarks \\ \hline
$D_{3,6}^{\mu}\ (ud)$       &$0, 1$& ${\rep3}, \ \overline{\rep6}$ &
$1\over3$ & $ -{2\over3} $ & scalar/vector diquarks; ${\tilde d}$ \\
\hline $U_{3,6}^{\mu}\ (dd)$       &$0, 1$& ${\rep3},\
\overline{\rep6}$ & $2\over3$ & $ -{2\over3} $ & scalar/vector
diquarks; $\tilde u$ \\  \hline\hline
 $u^{*}_{3,6}\ (ug)$ &$1\over2$, $3\over2$& $\rep3,\ \bar{\rep6}$ & ${2\over 3}$ & $ {1\over3} $ & excited $u$;
 quixes; stringy \\ \hline
  $d^{*}_{3,6}\ (dg)$ &$1\over2$, $3\over2$& $\rep3,\ \bar{\rep6}$ & ${1\over 3}$ & $ {1\over3} $ & excited $d$;
 quixes; stringy \\  \hline\hline
 $S_{8}\ (gg)$   &0& $\rep8_{S}$ &  $0$ & $0$ & $\pi_{TC},\ \eta_{TC} $ \\ \hline
  $T_{8}\ (gg)$   &2& $\rep8_{S}$ &  $0$ & $0$ & stringy \\ \hline \hline
  $V^{0}_{8}\ (u\bar u,\ d\bar d)$      &1& $\rep8$          &  $0$   & $ 0 $  & axigluon; $g^{}_{KK},\ \rho_{TC}$; coloron \\ \hline
  $V_8^{\pm}\ (u\bar d)$  &1& $\rep8$          &  $1$   & $ 0 $  & $\rho^{\pm}_{TC}$; coloron \\ \hline
\end{tabular}\normalsize
\end{center}
\caption{Summary for resonant particle names, their quantum numbers,
and possible underlying models.}
\label{tab:qnum}
\end{table}

In \tabb{tab:qnum} we summarise the different coloured resonances discussed in this section. We list our notation for the different states along with the leading couplings to SM partons and spin, colour representation, and electric charge of each state.  The subscript $S$ on $\rep8_S$ indicates that this colour octet representation is the symmetric combination of two other octets, as shown in \eq{eq:tensscal}.

%%%%%%%%%%%%%%%%%%%%%%%%%%%%%%%%%%%%%%%%%%%%%%%
%\subsubsection*{Dijet Resonance Production}
%\paragraph*{Dijet Resonance Production}
\label{par:2j}
%%%%%%%%%%%%%%%%%%%%%%%%%%%%%%%%%%%%%%%%%%%%%%
Since all of the resonances listed in \tabb{tab:qnum} couple to SM partons, they can decay back into SM partons.  Hence, they can be observed as dijet resonances.  We assume that the resonances decay exclusively back into (slim) dijets, \iee~not new particles or fat jets from top quarks.  If there are additional decay channels, our results can be simply rescaled by a BR.
 
The cross section for resonance $R$ production via quark and/or gluon initial states at the LHC is
\begin{eqnarray}
\sigma =\mathcal{L}_{12}(\tau_0)\frac{4\pi^2}{S_H}\frac{N_D(2J_R+1)}{N_1\,N_2}\frac{\Gamma(R\rightarrow X_1 X_2)}{M_R}(1+\delta_{X_1X_2}),
\end{eqnarray}
where $X_1,X_2$ are the initial state particles, $M_R$ is the resonance mass, $S_H=14,27\TeV$ is the \com~energy, and $J_R$ is the spin of the resonance.  The dimension of the colour representation of the resonance and initial state particles are denoted by $N_D$ and $N_1,N_2$, respectively.  The parton luminosity is
\begin{eqnarray}
\mathcal{L}_{12}(\tau_0)=\frac{1}{1+\delta_{X_1X_2}}\int^1_{\tau_0} \frac{dx}{x}\left(f_1(x)f_2(\tau_0/x)+f_2(x)f_1(\tau_0/x)\right),
\end{eqnarray}
where $\tau_0=M_R^2/S_H$, $f_1$ is the PDF of $X_1$, and $f_2$ is the PDF of $X_2$.

\begin{figure}[t]
\centering
%\subfigure[]{
\includegraphics[width=0.245\textwidth]{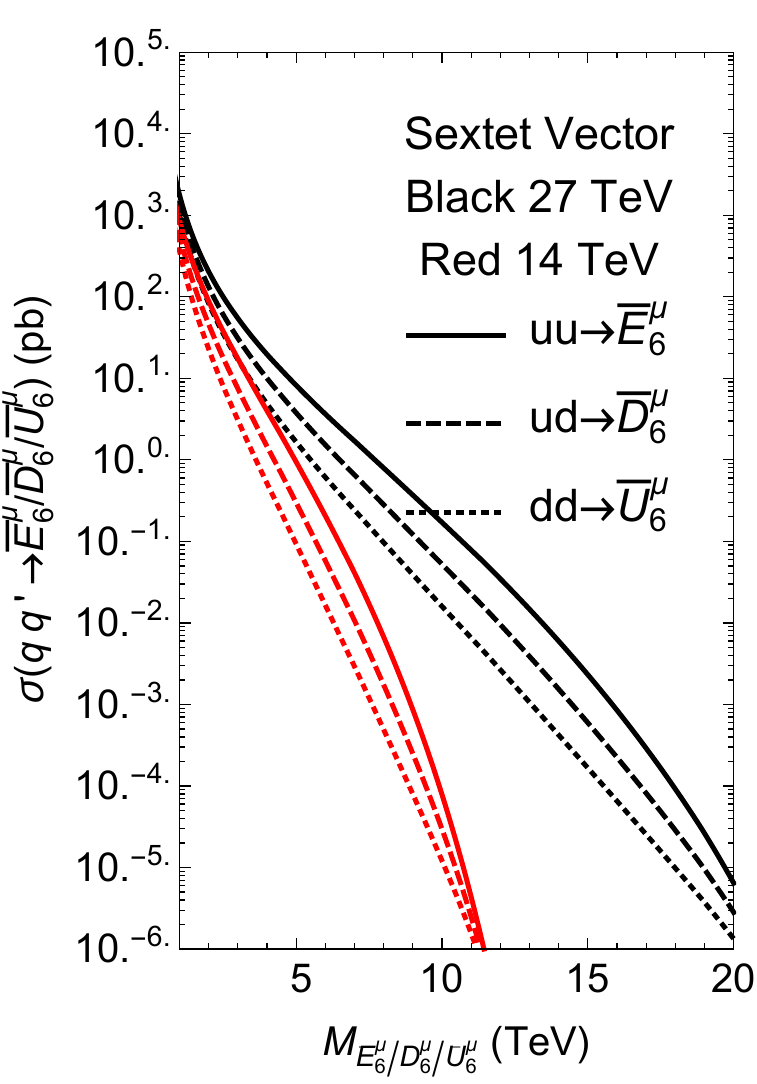}
%}
\includegraphics[width=0.245\textwidth]{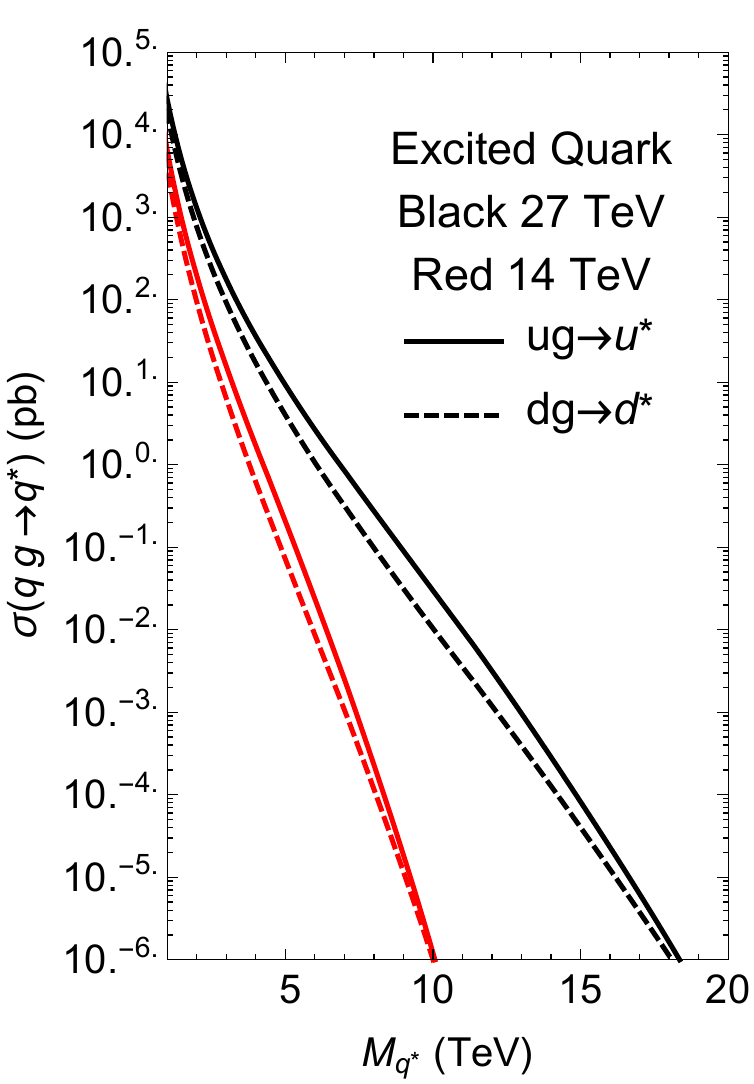}
\includegraphics[width=0.245\textwidth]{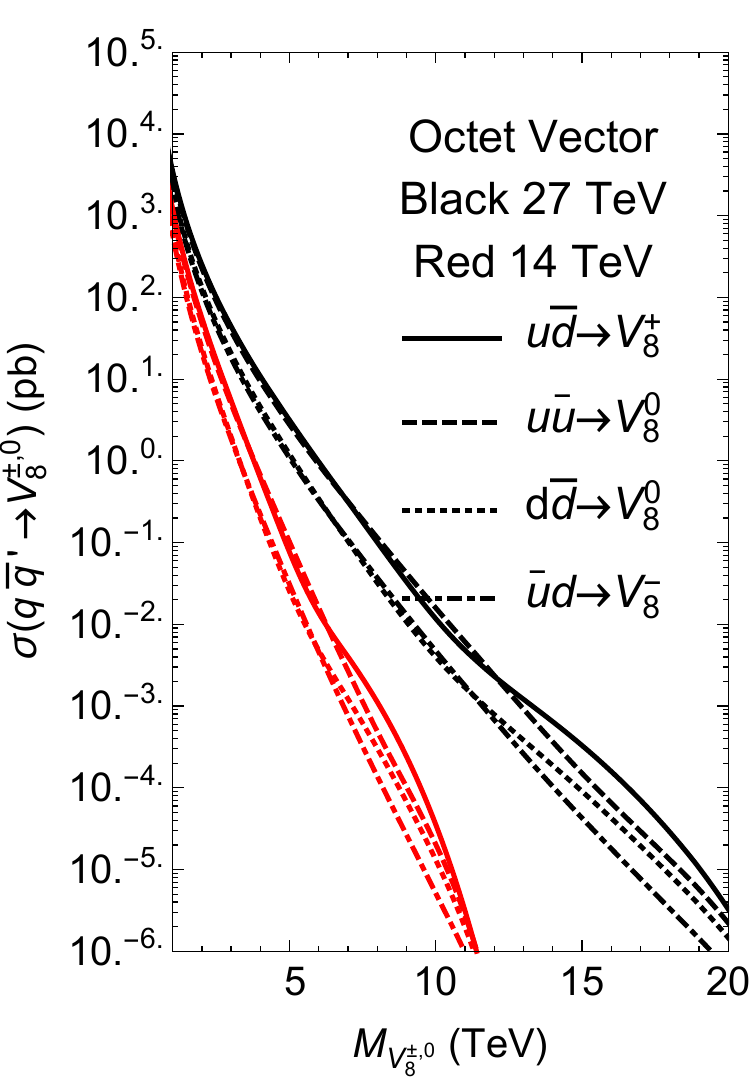}
\includegraphics[width=0.245\textwidth]{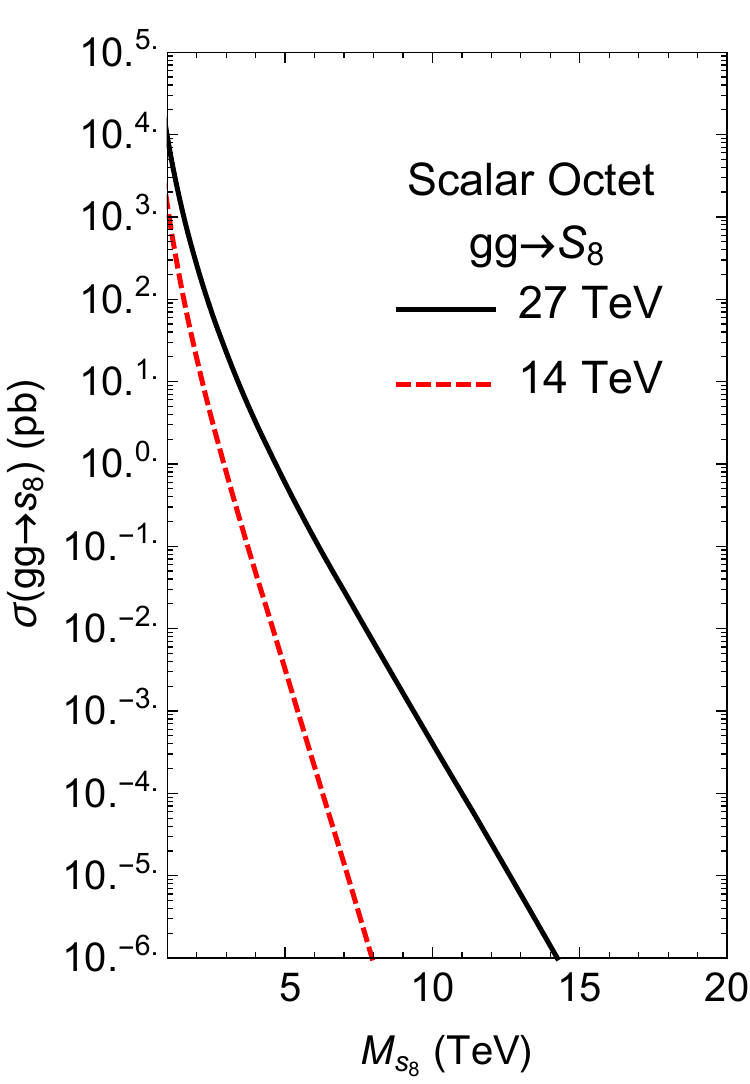}
\caption{Cross sections at HL- and HE-LHC shown in red and black curves, respectively, for (a) sextet diquarks, (b) excited quark, (c) octet vectors, and (d) $S_8$.}\label{fig:res}
\end{figure}

In \fig{fig:res} we show the cross sections of (a) sextet diquarks, (b) excited quarks, (c) octet vectors, and (d) $S_8$ for both the (red) HL- and (black) HE-LHC.  These cross sections are calculated using the NNPDF~\cite{Ball:2014uwa} PDF set, the coupling constants $\lambda^{E,U,D}$ and $\kappa_S$ are set to one, and the new physics scales $\Lambda=\Lambda_S=M_R$.    The sextet scalar cross sections are the largest for large resonance mass due to an enhancement from having two valence quarks in the initial state.  Somewhat surprisingly, despite the LHC having a reputation as having a large gluon pdf, at very high masses the vector octets produced from quark/anti-quark initial states have larger cross sections than resonances from gluon initial states.  This is because the gluon PDF drops precipitously at high momentum fraction.  In fact, at the highest resonance masses, $S_8$ produced from gluon fusion has the lowest cross sections.  Although the precise value of these rates depends on the choices of couplings and new physics scale, the gluon pdf suppression is clearly seen by how quickly the excited quark and $S_8$ cross sections decrease at high resonance masses as compared to the vector octets.

As can be clearly seen, the cross sections greatly increase at the HE-LHC.  For a resonance mass around $10-11\TeV$, we might expect $\mathcal{O}(1)$ events at the HL-LHC with $3\abinv$ of data for diquarks, excited quarks, and vector octets. For these resonances and resonance mass, at the HE-LHC with $3\abinv$ of data we may expect $\mathcal{O}(10^{3}-10^{5})$ events.  For $S_8$, for a resonance mass of $8\TeV$ for $3\abinv$ we expect $\mathcal{O}(1)$ events at the HL-LHC and $\mathcal{O}(10^{4})$ events at the HE-LHC.   This is a factor of $\mathcal{O}(10^{3}-10^{5})$ increase in the cross sections at the HE-LHC.

Additionally, at the HE-LHC, the mass reach is considerably extended.  The baseline for the HE-LHC is \intlumihelhc by its end run. For \intlumihelhc we expect $\mathcal{O}(10)$ events for $21\TeV$ diquarks ($E_6^u$), $19\TeV$ excited quark ($u^*$), $21\TeV$ octet vectors ($V_8^+$), and $15\TeV$ octet scalar $S_8$.  Comparing this to the masses for which we expect $\mathcal{O}(10)$ events by the end of the HL-LHC, we see that we may expect the mass reach of the HE-LHC to be twice that of the HL-LHC.

%%%%%%%%%%%%%%%%%%%%%%%%%%%%%%%%%%%%%%%%%%%%%%
%\subsubsection*{Conclusions}
%\paragraph*{Conclusions}
%\label{par:conc}
%
%In conclusion, the LHC is a QCD machine and coloured resonances that couple to partons are expected to be produced at favorable rates. These resonances can be observed as dijet resonances and considering the benchmark scenarios of 3\abinv at $14\TeV$ for the HL-LHC and \intlumihelhc at 27 TeV for the HE-LHC, we find that that the HE-LHC is sensitive to resonance masses up to a factor of two larger than the HL-LHC.
%%%%%%%%%%%%%%%%%%%%%%%%%%%%%%%%%%%%%%%%%%%%%%

%\bibliography{ref}
%\bibliographystyle{JHEP}
%\bibliography{\main//section6OtherSignatures/bib/coloreddijet}

\subsubsection{\theoryC{Precision searches in dijets at the HL- and HE-LHC}}
\contributors{S. V. Chekanov, J. T. Childers, J. Proudfoot, R.Wang, D. Frizzell}
%{\bf Author(s):} S. V. Chekanov$^1$, J. T. Childers$^2$, J. Proudfoot$^3$, R.Wang$^4$, D. Frizzell$^5$
%
%\vspace{1cm}
%
%$^1$ chekanov@anl.gov,  HEP Division, Argonne National Laboratory, 9700 S.~Cass Avenue, \\ Argonne, IL 60439, USA.
%
%$^2$ jchilders@anl.gov, HEP Division, Argonne National Laboratory, 9700 S.~Cass Avenue, \\ Argonne, IL 60439, USA.
%
%$^3$ proudfoot@anl.gov, HEP Division, Argonne National Laboratory, 9700 S.~Cass Avenue, \\ Argonne, IL 60439, USA.
%
%$^4$ Rui.Wang@cern.ch, HEP Division, Argonne National Laboratory, 9700 S.~Cass Avenue, \\ Argonne, IL 60439, USA.
%
%$^5$ dylan.frizzell@ou.edu, Homer L. Dodge, Department of Physics and Astronomy, \\ University of Oklahoma, Norman, OK, USA.
%
%\vspace{1cm}

Model-independent searches for deviations in dijet invariant mass distributions ($M_{jj}$) predicted by the SM is one of the central studies at the LHC.
The main goal of such searches is to find small deviations from background distributions, under the assumption that a new resonant state decaying to partons that form two jets may introduce an excess in dijet masses localised around the resonance mass.
In recent years, such searches for new physics at the LHC have been performed  by both ATLAS and CMS collaborations (see the recent studies in \citeref{Aad:2015xis,Aaboud:2016nbq, Aaboud:2017yvp, Khachatryan:2010jd, Chatrchyan:2013qha, Sirunyan:2016iap,Aaboud:2017yvp,Aaboud:2018fzt}), but no statistically significant deviations from background expectations were found. Such searches are usually performed for resonances with a width of up to $15\%$ of the resonance mass. Searches for broader resonances are usually more 
difficult since the available tools to determine the background 
hypothesis have a number of limitations that do not allow for 
reliable estimates of background shapes in the presence of broad resonances.  
Despite the generality of dijet searches, such studies can exclude a number of BSM models, such as models with  quantum black holes, excited quarks, and $Z'$ bosons and so on.
Currently, the LHC run II data provide the $95\%\cl$ exclusion limits on BSM resonances up to $6.5\TeV$ in masses.

\begin{figure}[t]
\centering
\begin{minipage}[t][\minipageheightdp][t]{\minipagewidthdp}
\centering
\includegraphics[width=\figuremaxwidthdp,height=\figuremaxheightdp,keepaspectratio]{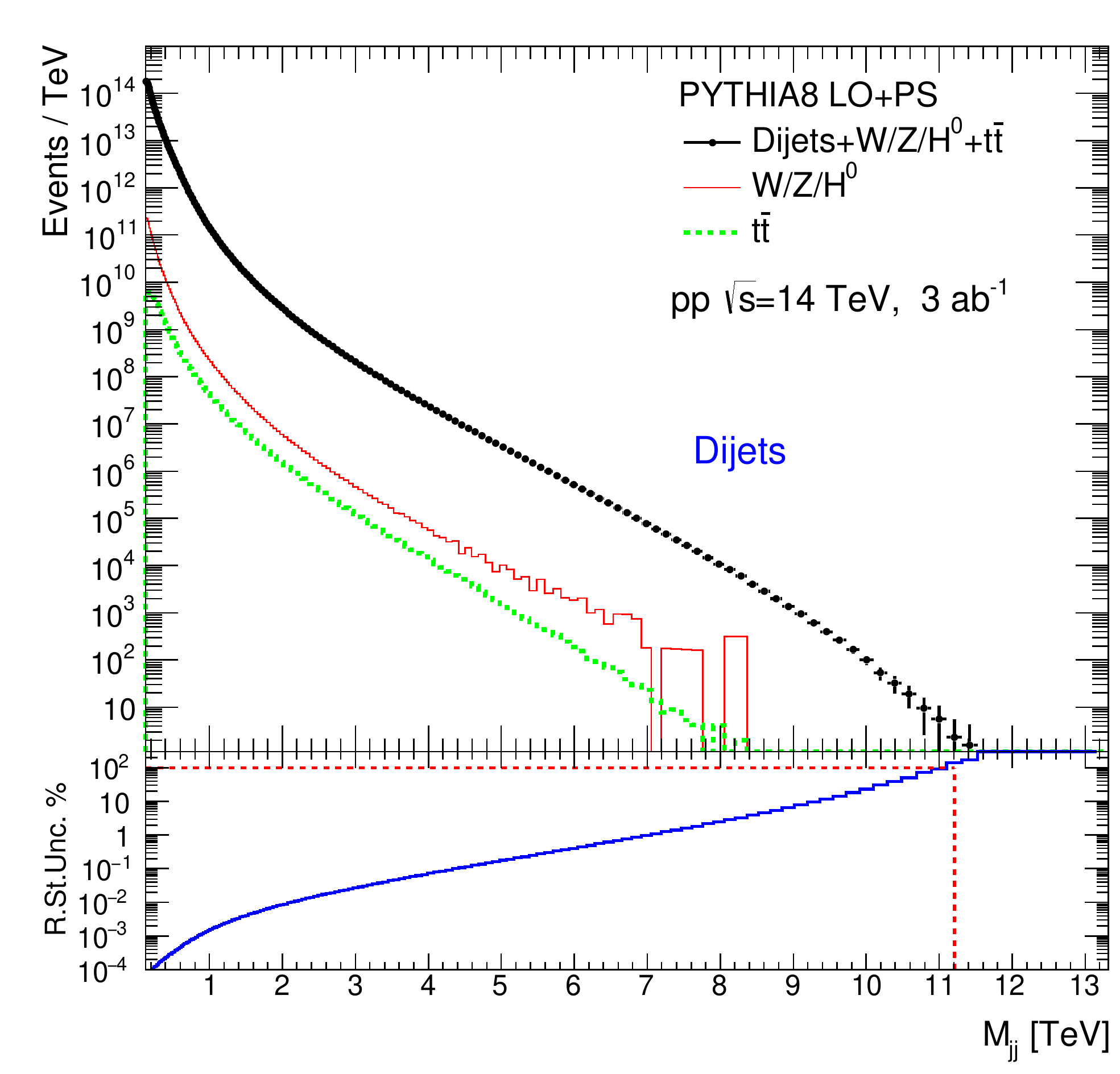}
\end{minipage}
\begin{minipage}[t][\minipageheightdp][t]{\minipagewidthdp}
\centering
\includegraphics[width=\figuremaxwidthdp,height=\figuremaxheightdp,keepaspectratio]{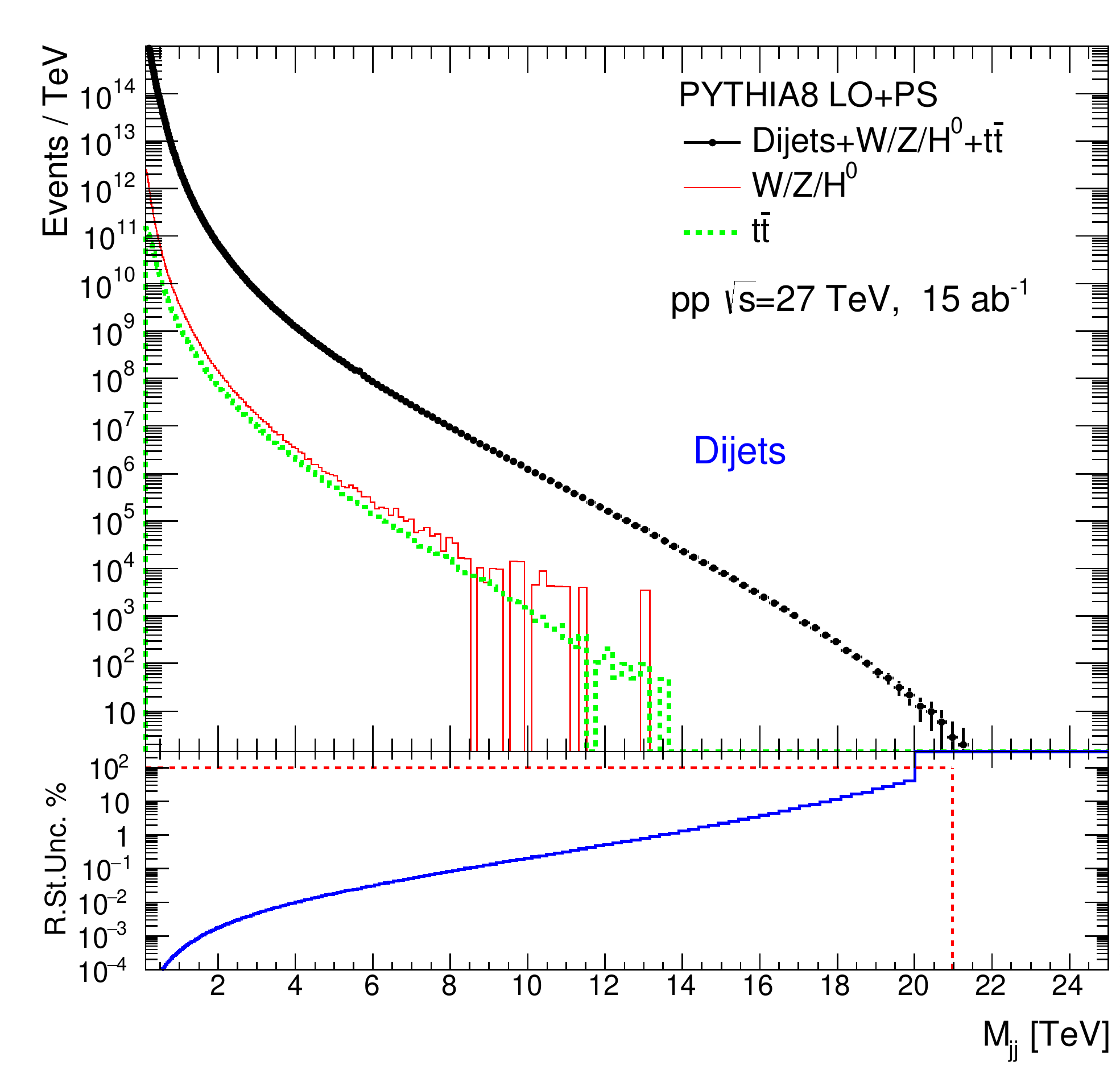}
\end{minipage}
\caption{Expectations for the dijet invariant mass distribution for \intLumi at HL-LHC and
$15\abinv$ at the \sloppy\mbox{HE-LHC} using the Pythia generator. Contributions
from  $W/Z/H^0$ -boson processes and top-quark processes are shown separately (without stacking the histograms). The bottom plots show the relative  statistical uncertainties in each bin, together with the line indicating the mass point at which the uncertainty is $100\%$. The figures are taken from~\cite{Chekanov:2017pnx}.}
\label{fig:14tev_JetJetMass_2jet}
\end{figure}

\begin{figure}[t]
\centering
\begin{minipage}[t][\minipageheightdp][t]{\minipagewidthdp}
\centering
\includegraphics[width=\figuremaxwidthdp,height=\figuremaxheightdp,keepaspectratio]{\main/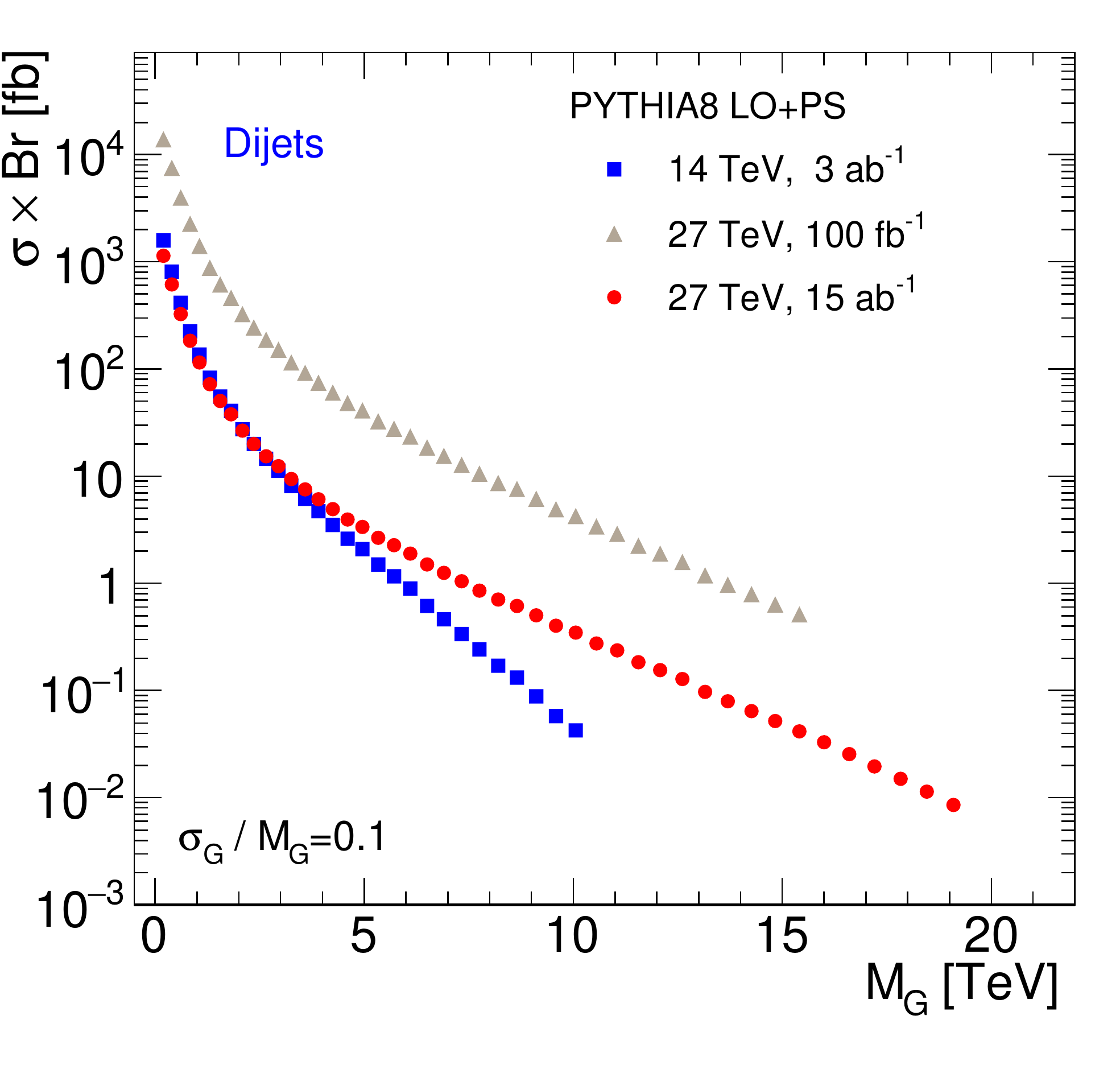}
\end{minipage}
\begin{minipage}[t][\minipageheightdp][t]{\minipagewidthdp}
\centering
\includegraphics[width=\figuremaxwidthdp,height=\figuremaxheightdp,keepaspectratio]{\main/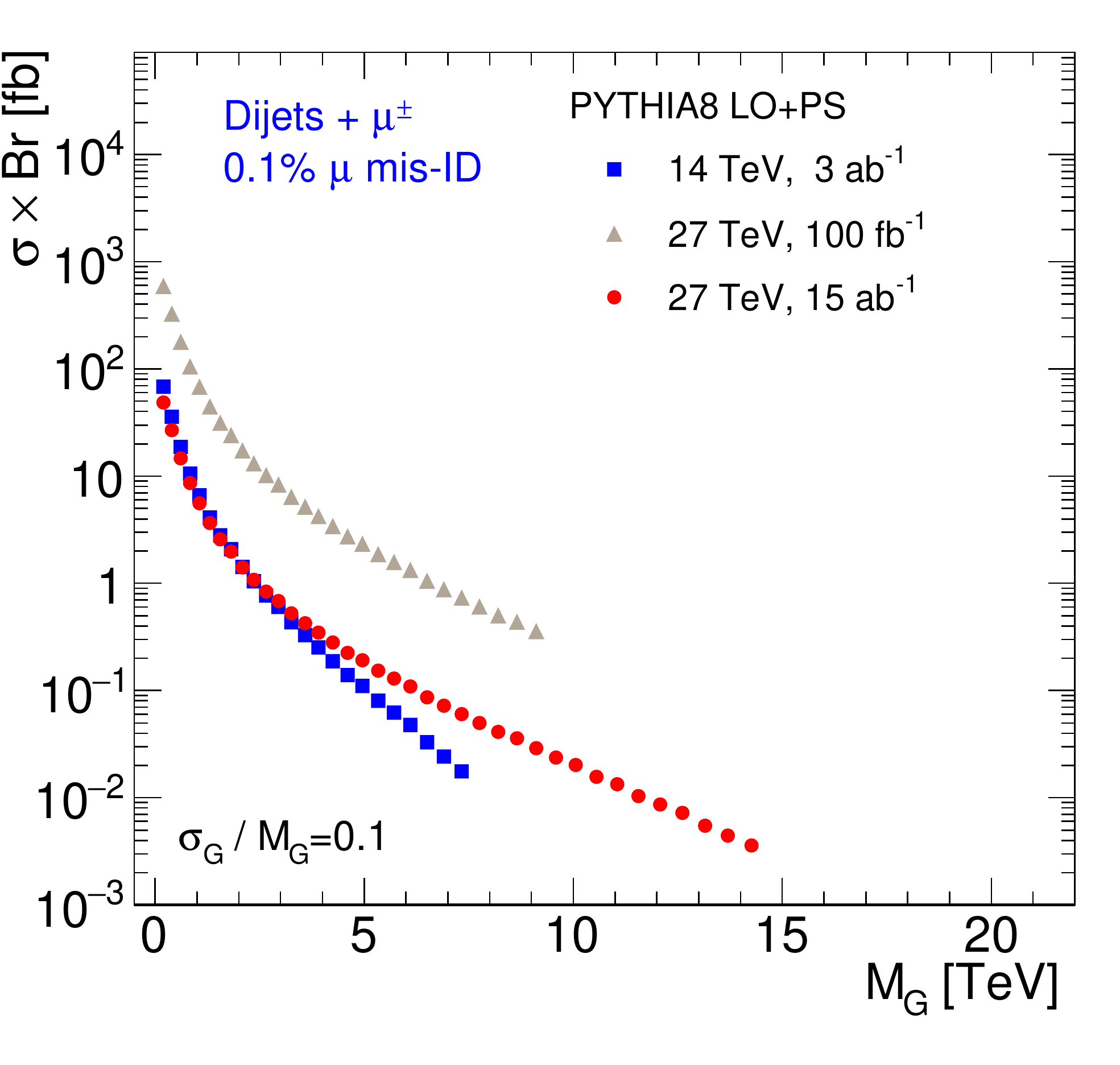}
\end{minipage}
\caption{Expected $95\%\cl$ upper limits obtained from the
$M_{jj}$  distribution on fiducial cross-section times the BR to two jets
for a hypothetical BSM signal approximated by a Gaussian contribution to the dijet mass spectrum.  The limits are obtained for the HL-LHC and HE-LHC energies. The figure shows the limits for inclusive dijet production (left) and for events with at least one isolated muon with $p_T>60\GeV$.}
\label{fig:JetJetMass_jet}
\end{figure}

Shapes of the background $M_{jj}$ distributions can be affected by several instrumental factors, making
such studies difficult for low-mass regions where statistics are large.
In the case of inclusive jet production, the rate of dijets is reduced  due to triggers with low acceptance rate, which leads to complicated shapes of in the region of $M_{jj}$ below $1\TeV$. 
This leads to difficulties in interpretation of the shapes
of the $M_{jj}$ distributions using QCD-motivated fit functions.
Recently, such searches have been extended to less inclusive events, such as events with $b-$jets and  associated leptons~\cite{Aaboud:2018tqo,ATLAS-COM-CONF-2018-020}. These studies rely on less inclusive triggers with low thresholds, thus accessing regions with small dijet masses. 

With the increase of the LHC luminosity, searches for new physics beyond the SM in the invariant mass of two jets become increasingly important since the large event rate allows one to explore a large $M_{jj}$ phase space with improved statistics.  The proposed HL-LHC and HE-LHC experiments will open a new chapter in model independent searches, both in  terms of the physics reach and the complexity of derivations of data-driven backgrounds. 
Here we will give an overview of the physics potential for model-independent searches~\cite{Chekanov:2017pnx}  in dijets for the HE-LHC and HL-LHC experiments, as well as describe some technical problems in understanding of "signal-like" feature on a smoothly falling dijet mass distributions. 
We will cover searches in inclusive jets, as well as more exclusive searches in events with $b-$jets and leptons.

The presented studies use MC event generation
with representative for the HL-LHC experiment event statistics. This  was achieved
using high-performance computers at NERSC. The \pythia{ 8}~\cite{Sjostrand:2007gs} generator with the default parameter settings and the ATLAS A14 tune~\cite{ATL-PHYS-PUB-2014-021} was used. The \com~ collision energy of $pp$ collisions
was set to $14\TeV$ and $27\TeV$ for the HL-LHC and HE-LHC respectively.
About 100 billion MC events with the multi-jet QCD, $t\bar{t}$ and $W+jet$ categories of processes
were simulated using the HepSim software~\cite{Chekanov:2014fga} 
deployed on supercomputers at NERSC.
Such a large number of events is required in order to obtain $M_{jj}$ distributions 
which are sufficiently smooth for calculations of limits. In addition, a phase-space re-weighting was used for  $2\rightarrow 2$ processes to increase the statistics in the tail of the $M_{jj}$ distribution as discussed in \citeref{Sjostrand:2007gs}. 
The jets were reconstructed with the anti-$k_T$ algorithm~\cite{Cacciari:2008gp}, as 
implemented in the FastJet package~\cite{Cacciari:2011ma}, using  a distance parameter of $R=0.4$.
The minimum transverse momenta of jets was $40\GeV$, and the pseudorapidity range  was $|\eta|<2.4$. Dijet invariant masses  were reconstructed by combining the two leading jets
having the highest transverse momentum. The $b$-jets are selected by requiring  a distance, defined in pseudorapidity and azimuthal angle,  between the $b$-quark and jet to be less than $0.4$ and
the $b$-quark $p_{T}$ to be at least $50\%$ of the jet $p_{T}$. 
A constant $10\%$ mis-tag rate was  assumed, which is sufficiently realistic~\cite{Aad:2015ydr} for large $p_T(jet)$.

In addition to jets, muons were also used for such studies. They are  required to be isolated using
a cone of the size $0.2$ in the azimuthal angle and pseudorapidity is
defined around the true direction of the lepton.
A lepton is considered to be isolated of it carries more than $90\%$ of the cone energy.
We also simulated a misidentification rate of muons (or ``fake'' rate) assuming that a muon 
can be mis-identified with a rate of $0.1\%$~\cite{Aad:2009wy}.

\begin{figure}[t]
\centering
\begin{minipage}[t][\minipageheightsp][t]{\minipagewidthsp}
\centering
\includegraphics[width=\figuremaxwidthsp,height=\figuremaxheightsp,keepaspectratio]{\main/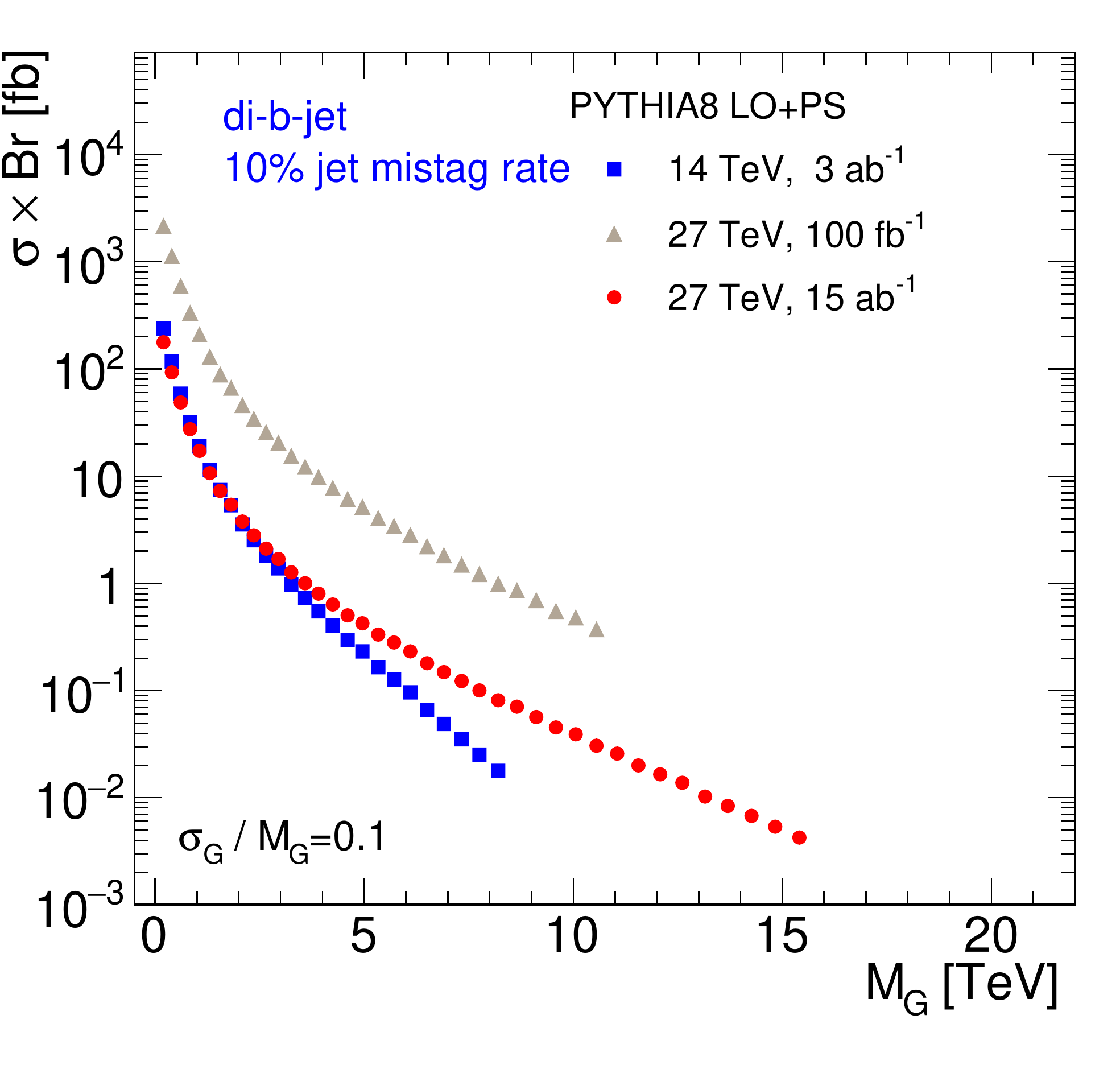}
\end{minipage}
\caption{Expected $95\%\cl$ upper limits obtained from the
$M_{jj}$  distribution for jets identified as $b$-jets.
The limits are set on fiducial cross-section times the  BR to two jets
for a hypothetical BSM signal approximated by a Gaussian contribution to the dijet mass spectrum.  The limits are obtained for the HL-LHC and HE-LHC energies.}
\label{fig:JetJetMass_bjet}
\end{figure}

Figure~\ref{fig:14tev_JetJetMass_2jet} shows two representative 
$M_{jj}$ distributions using the simulations discussed above. The results are shown for the HL-LHC and HE-LHC colliders using different integrated luminosities,  \intLumi (for HL-LHC) and $15\abinv$ (for HE-LHC). 
The figure also shows contributions to the total event rate 
from $W/Z/H^0$-boson processes combined  and
top-quark processes from the hard interactions (shown separately).
The lower panel shows the relative statistical uncertainty on the
data points, \iee~$\Delta d_i/d_i$, where $d_i$ is the number of
the events in the bins, and $\Delta d_i$ its statistical uncertainty (which is $\sqrt{d_i}$ in the case of counting statistics). For a quantitative characterisation of the dijet mass reach,
the dash lines on the lower panel show the $M_{jj}$ point at which
statistical uncertainty in a bin is $100\%$ (or $\Delta d_i/d_i=1$). We have chosen this point 
to define the statistical reach for the measurements using the $M_{jj}$ distributions.
The $M_{jj}$  mass reach at the HL-LHC is $11.2\TeV$, while the mass reach at the \com~of $27\TeV$ is close to
$21\TeV$  for the the nominal luminosity of $15\abinv$. 

\begin{figure}[t]
\centering
\begin{minipage}[t][\minipageheightdp][t]{\minipagewidthdp}
\centering
\includegraphics[width=\figuremaxwidthdp,height=\figuremaxheightdp,keepaspectratio]{\main/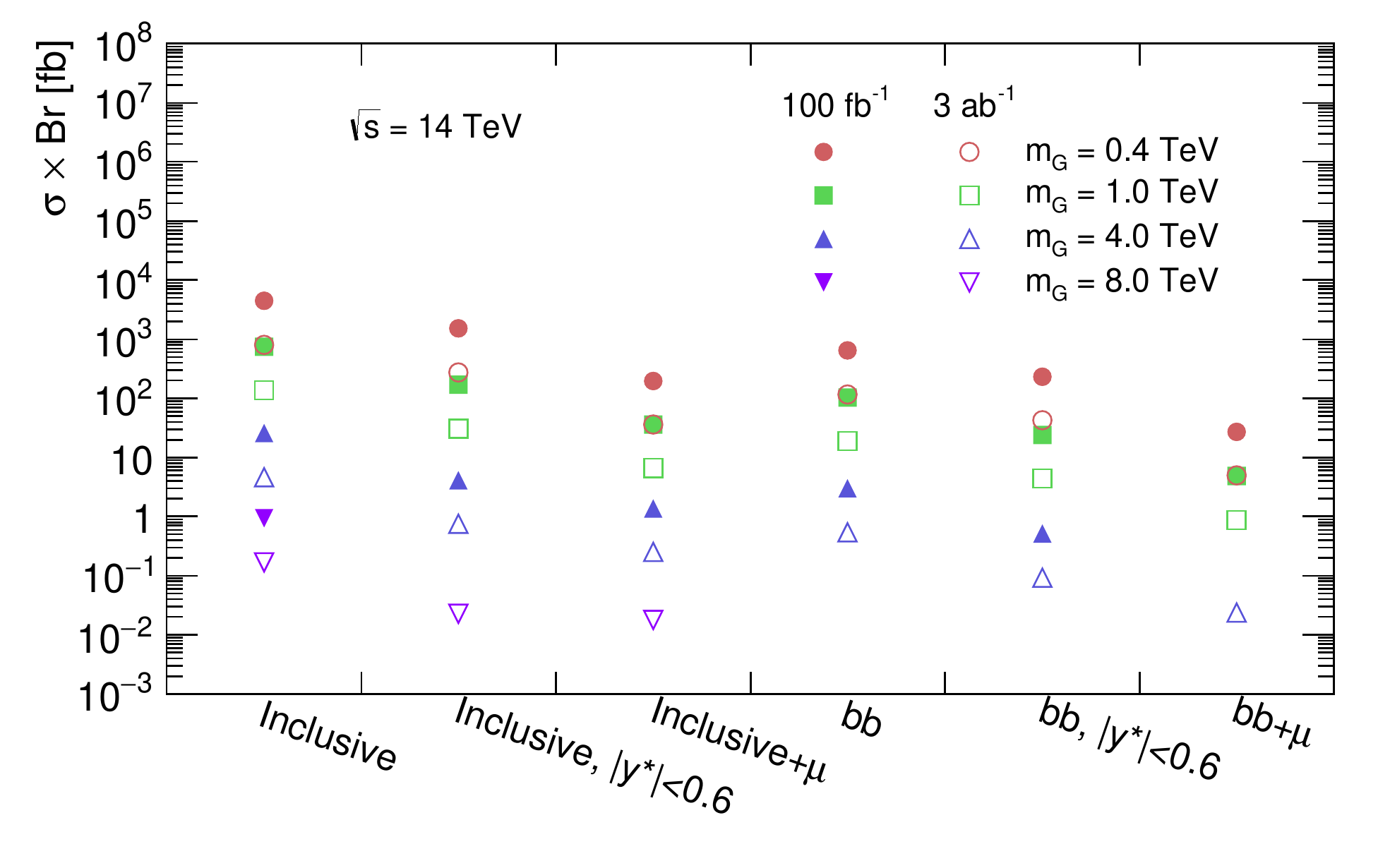}
\end{minipage}
\begin{minipage}[t][\minipageheightdp][t]{\minipagewidthdp}
\centering
\includegraphics[width=\figuremaxwidthdp,height=\figuremaxheightdp,keepaspectratio]{\main/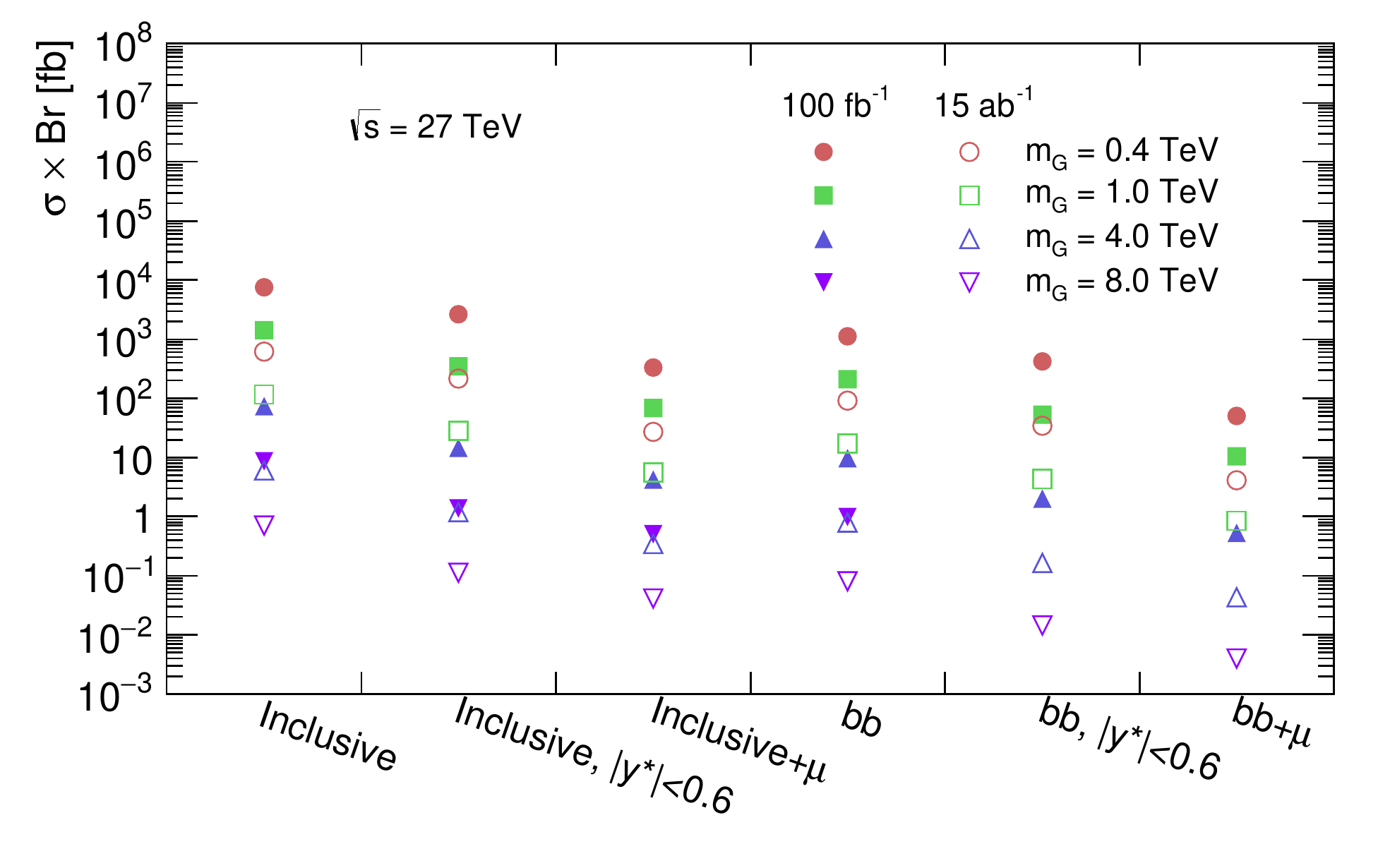}
\end{minipage}
\caption{Comparisons of the  $95\%\cl$ upper limits obtained from the
$M_{jj}$  distribution for the HL-LHC at $14\TeV$ (left) and the HE-LHC at $27\TeV$ (right). 
The limits for the masses $0.4\TeV$, $1\TeV$, $4\TeV$ and $8\TeV$ are shown for $100\fbinv$,  $3\abinv$ and $15\abinv$  for 
different channels with different selections.}
\label{fig:JetJetMass_comp}
\end{figure}

The dijet distributions discussed above were used to set  the $95\%\cl$ 
upper limit on fiducial cross-section times the 
BR for a generic Gaussian signal with
the width ($\sigma_G$) being $10\%$ of the Gaussian peak position.
Figure~\ref{fig:JetJetMass_jet} shows the  obtained limits.    
In addition to inclusive dijet events, this figure shows the upper limits
for events with at least one isolated muon with transverse momentum above $60\GeV$.
The results show that even the HE-LHC experiment with an integrated luminosity of $100\fbinv$ has advantages
over the HL-LHC (with the nominal luminosity of \intLumi) in terms of statistical sensitivity to high-mass states. 
When using the nominal integrated luminosity of $15\abinv$ for HE-LHC, 
the HE-LHC provides a factor of two 
larger reach for dijet masses compared to the HL-LHC with the nominal luminosity.

Figure~\ref{fig:JetJetMass_bjet} shows the $95\%\cl$ upper limits obtained from the $M_{jj}$ distribution for events where both leading jets were identified as $b-$jets.
As for the previous simulation based on inclusive jets, the physics reach of the HE-LHC with the nominal luminosity is a factor two 
larger than that of the HL-LHC.

Figure~\ref{fig:JetJetMass_comp} shows the comparisons of the limits for dijet and muon associated dijet channels with inclusive  or $b$-tagging selections for the $14\TeV$ and $27\TeV$ collision energies. 
In addition to the inclusive jet case, we also calculated the upper limits after applying the 
rapidity difference requirement $|y^*|<0.6$ between two  jets~\cite{ATLAS:2015nsi}  in order to enhance the sensitivity to heavy
BSM particles decaying to jets.

The studies discussed ~\cite{Chekanov:2017pnx} have also shown that searches for signals 
in dijets invariant masses require well-understood estimates for the $M_{jj}$ background shape. 
An example of such challenging background shape is shown in \fig{fig:14tev_JetJetMass_2jet}.
A data-driven determination of the shape background at the HL-LHC and  HE-LHC
should be performed with the relative statistical precision
of  $0.01\%$ per data point for $M_{jj}<1\TeV$. 
Currently, it is difficult to achieve 
such precision using the available tools used for the LHC run 2 data.

In conclusion, it was illustrated
that the HE-LHC provides substantial improvements for searches of new physics in dijet invariant masses,  
compared to searches  at the HL-LHC.
Even for a rather modest $100\fbinv$ luminosity of the HE-LHC 
project, the simulations show that the mass reach for dijet searches is 
about $50\%$ larger than that from the HL-LHC with the nominal luminosity of \intLumi.
It was also shown that the HE-LHC project with the nominal luminosity of $15\abinv$  will extend 
the HL-LHC mass reach by  a factor two. 
The reported limits can be used for exclusions of BSM resonances decaying that
decay to form two jets in inclusive dijet events, events with di-$b$-jets, and events with associated muons. Note that the actual exclusion ranges significantly depend on expected  cross-sections and BRs of BSM models for the HL-LHC and HE-LHC collision energies.

\subsubsection{\theoryC{Dissecting heavy diphoton resonances at HL- and HE-LHC}}
\contributors{B. Allanach, D. Bhatia, A. Iyer}
%{\bf Author(s): B. Allanach$^1$, D. Bhatia$^2$ and A. Iyer$^3$}\\
%$^1$ DAMTP, University of Cambridge, CMS, Wilberforce Road, Cambridge, CB3 0WA, United Kingdom email: {\tt B.C.Allanach@damtp.cam.ac.uk}\\
%$^2$ Department of Theoretical Physics, Tata Institute of Fundamental Research,
%Homi Bhabha Road, Colaba, Mumbai 400 005, India, email: {\tt disha@theory.tifr.res.in}\\
%$^3$ INFN-Sezione di Napoli, Via Cintia, 80126 Napoli, Italia, email: {\tt iyera@na.infn.it} 

We examine the phenomenology of the production of a heavy resonance $X$, 
which decays via other new on-shell particles $n$ into multi- (\iee{} three
or more) photon final states.  In the limit that $n$ has a much smaller mass than $X$, 
the multi-photon final state may dominantly appear as a two photon
final state because the $\gamma$s from the $n$ decay are highly
collinear and remain unresolved.
We discuss how to discriminate this scenario
from $X \rightarrow \gamma \gamma$: rather than discarding non-isolated
photons, it is better instead to relax the isolation criterion and instead
form photon 
jet substructure variables. 
The spins of $X$ and $n$ leave their imprint upon
the distribution of pseudorapidity gap $\Delta \eta$ between the apparent two
photon 
states. In some models, the heavy resonance $X$ may decay into $nn$ or $n \gamma$, where
$n$ is an additional light
particle,
may further decay into photons leading to a multi-photon\footnote{In the
  present paper, whenever we refer to multi-photon final states, we refer to
  three or more photons.} final state. 
Examples of such models include hidden
valley models~\cite{Strassler:2006im,Strassler:2006ri}, the
NMSSM~\cite{Ellwanger:2009dp} or Higgs portal
scenarios~\cite{Schabinger:2005ei}. 
Describing angles in terms of the pseudorapidity $\eta$ and the azimuthal
angle around the beam $\phi$, the angular separation between two photons may be 
quantified by $\Delta R=\sqrt{(\Delta \eta)^2+(\Delta \phi)^2}$. 
Neglecting its mass, 
the opening angle between the two photons coming from a highly boosted
on-shell $n$ is 
\begin{equation}\Delta  R=\frac{m_n}{\sqrt{z(1-z)}p_T(n)},
\end{equation} purely from kinematics (this was
  calculated already in the context of boosted Higgs to $b \bar b$
  decays~\cite{Butterworth:2008iy}),
  where $z$ and 
 $(1-z)$ are  the momentum fractions of the photons\footnote{The decay is
   strongly peaked towards the minimum opening angle $\Delta R=2
   m_n/p_T$~\protect\cite{Chala:2015cev}.}. 
 Thus,
\begin{equation}
\Delta R = \frac{m_n}{M_X}\frac{2 \cosh \eta(n)}{\sqrt{z(1-z)}}. \label{eq:dRest}
\end{equation}
In the limit $m_n/M_X\rightarrow 0$, $\Delta
R\rightarrow 0$ and the two photons from $n$ are 
collinear, appearing as one photon; thus several possible interpretations
can be ascribed to an apparent diphoton signal.

We assume that any couplings of new particles such as the $X$ (and the $n$, to
be introduced later) to Higgs
fields or $W^\pm, Z^0$ bosons are negligible. 
\eq{eq:Lagrangian1} gives an effective field theoretic interaction
Lagrangian for the 
coupling of $X$ to a pair of photons, when $X$ is a scalar (first line) or  a
graviton (second line). 
 \begin{eqnarray}
	\mathcal{L}_{X={\rm spin~0}}^{int} &=&    - \eta_{GX}\frac{1}{4} G_{\mu \nu}^a
                                   G^{\mu \nu a} X  - \eta_{\gamma X} \frac{1}{4}
                                   F_{\mu \nu} F^{\mu \nu} X, \nonumber\\
	\mathcal{L}_{X={\rm spin~2}}^{int} &=&    -\eta_{T\psi X}T_{fermion}^{\alpha\beta}
                                   X_{\alpha\beta}-\eta_{TGX} T_{gluon}^{\alpha\beta}
                                   X_{\alpha\beta}  - \eta_{T\gamma X} 
T_{photon}^{\alpha\beta} X_{\alpha\beta}.
	\label{eq:Lagrangian1}
\end{eqnarray}
where  $T^{\alpha\beta}_i$ is the stress-energy tensor for the  field $i$
and the $\eta_j$ are effective couplings of mass dimension -1.
$F_{\mu \nu}$ is the field strength tensor of the photon (this may be obtained
in a SM invariant way from a coupling involving the field strength tensor of
the hypercharge gauge boson),
whereas $G_{\mu 
  \nu}^a$ is the field strength tensor of a gluon of adjoint colour index $a
\in \{ 1,\ \ldots , 8 \}$.
As noted earlier, the direct decay of a vector boson into two photons is
forbidden by the 
Landau-Yang theorem~\cite{Landau:1948kw,Yang:1950rg}. 

\begin{figure}[t]
\centering
\begin{minipage}[t][\minipageheightdp][t]{\minipagewidthdp}
\centering
\includegraphics[width=\figuremaxwidthdp+6mm,height=\figuremaxheightdp-5mm,keepaspectratio]{\main/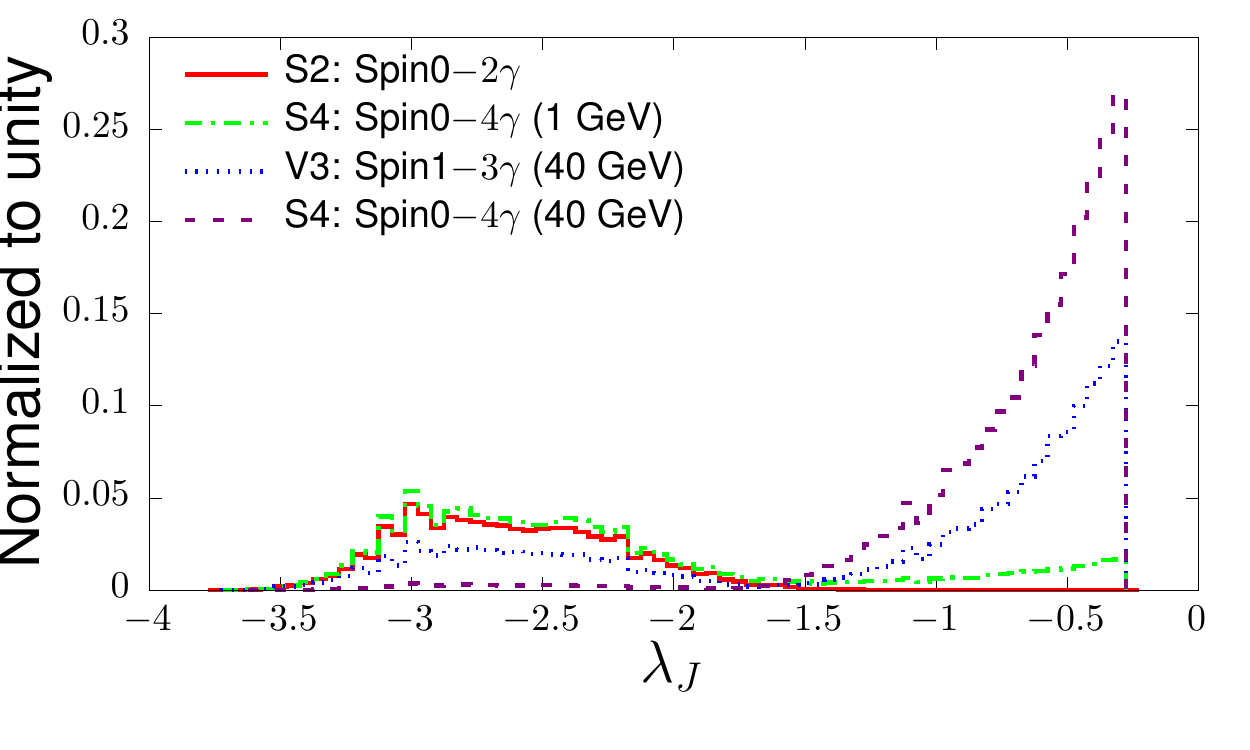}
\end{minipage}
\begin{minipage}[t][\minipageheightdp][t]{\minipagewidthdp}
\centering
\raisebox{4mm}{\includegraphics[width=\figuremaxwidthdp-8.5mm,height=\figuremaxheightdp-5mm,keepaspectratio]{\main/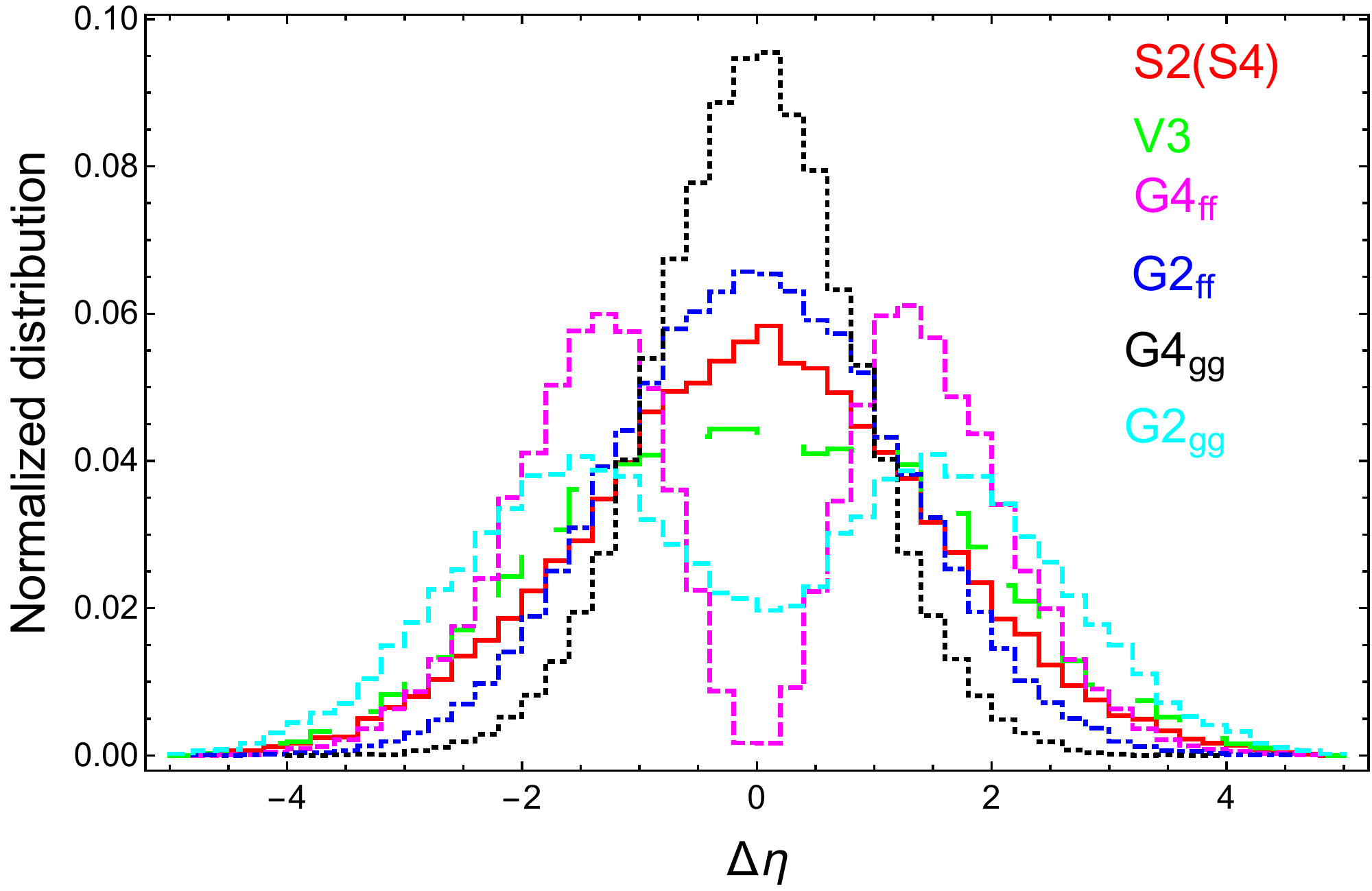}}
\end{minipage}
\vspace{-0.8cm}
\caption{Left: distribution of $\lambda_J$ variable for various signal hypotheses. $m_n$ is noted in parentheses in the legend for the relevant cases. Right: signal $\Delta \eta$ distribution for various different spin combinations at $\sqrt{s}=27\TeV$, $m_X=2.5\TeV$. The cases S2 and S4 are practically indistinguishable by eye and so we only plot one histogram for them. \label{fig:deta}\label{fig:lambdaj}}
\end{figure}

Although we assume that $n$ is electrically neutral, it may decay to two
photons through a loop-level process (as is the case for the SM
Higgs boson, for instance). 
Alternatively, if $X$ is a spin 1 particle, it could be produced by
quarks in the proton and then decay into $n \gamma$. 
The Lagrangian terms would be
\begin{equation}
\mathcal{L}^{int}_{X\,={\rm spin~1},n}= -
                                     (\lambda_{\bar q X q} \bar q_R \gamma_\mu
                                     X^\mu q_R
 + \lambda_{\bar Q X Q} \bar Q_L \gamma_\mu X^\mu Q_L + H.c.)
-\frac{1}{4}\eta_{nX\gamma}
                                     n{\tilde X}_{\mu\nu}F^{\mu\nu},
                                     \label{eq:wouldbe}
\end{equation}
where  $\lambda_i$ are  dimensionless
couplings, $q_R$ is a right-handed quark, $Q_L$ is a left-handed quark doublet
and ${\tilde X}_{\mu \nu}=\partial _\mu
X_\nu - \partial_\mu X_\nu$. 
The decay $X_{spin=1}\rightarrow n \gamma$ would
have to be a loop-level process, as explicitly exemplified in
\citeref{Chala:2015cev}, since electromagnetic gauge invariance 
forbids it at tree level. 
The possible different spins involved in multi-photon production processes, along with our nomenclature for them, are listed in \tabb{tab:cases}. 
\begin{table}[t]
\begin{center}
\small\begin{tabular}{|c|c|}\hline
Model & Process \\ \hline
$S2$ & $pp \rightarrow S \rightarrow \gamma \gamma$ \\
$S4$ & $pp \rightarrow S \rightarrow nn \rightarrow \gamma
       \gamma+\gamma\gamma$ \\ 
$V3$ & $pp \rightarrow Z' \rightarrow n \gamma \rightarrow \gamma + \gamma
       \gamma$ \\
$G2_{ff}$ & $q \bar q \rightarrow G \rightarrow \gamma \gamma$ \\ 
$G4_{gg}$ & $gg \rightarrow G \rightarrow nn \rightarrow \gamma \gamma+\gamma
            \gamma$ \\
$G4_{ff}$ & $\bar q q \rightarrow G \rightarrow nn \rightarrow \gamma \gamma+\gamma
            \gamma$ \\
\hline
\end{tabular}\normalsize
\end{center}
\caption{\label{tab:cases} Cases to discriminate with a scalar $n$ and a heavy
resonance which is: scalar ($S$), spin 1 ($Z'$) or spin 2 ($G$). We have
listed the main signal processes to discriminate between in the second column,
ignoring any proton remnants. The notation used for a given model is $Xk$:
$X=S,V,G$ labels the spin of the resonance and $k$ denotes the number of
signal photons at the parton level in the final state.} 
\end{table}
The first tool for model discrimination is the photon sub-jet variable $\lambda_J=\log(1-p_{T_L}/p_{T_J})$, where $P_{T_L}$ denotes the transverse momentum of the leading photon sub-jet whilst $p_{T_J}$ is the transverse momentum of the whole photon jet. Double pronged photon jets have a peak at $\lambda_J=-0.3$. This may be observed when the signal is multi-photon and $m_n$ is not too small, as \fig{fig:lambdaj} (left) shows.

Here, we neglect backgrounds and only focus on the signal. This is a good approximation for heavy resonances where backgrounds die off exponentially with invariant mass of the apparent diphoton pair, provided that the signal cross-section is large enough. The signal cross-section is of course set by the size of the couplings in \eqs{eq:Lagrangian1} and \eqref{eq:wouldbe}, which may be adjusted as required. Here, we consider one of the cases in \tabb{tab:cases} at a time, with cross-sections of all other cases set to zero. 

If there is a sizeable peak at $\lambda_J=-0.3$, the model possibilities are $S4,V3,G4_{gg}$ and $G4_{ff}$. Further to this, $V3$ may be further discriminated owing to its double peak structure: at $\sim-3$ and at $-0.3$. We call this case A.
If there is no sizeable peak, we can either have $S2$ or $m_n$ very small. This latter case we call case B.

After categorisation into case A or B, we can use the fact that
each case in \tabb{tab:cases} corresponds to a different distribution in $\Delta \eta$, the difference in pseudorapidity between the two initially identified photons.
We plot the $\Delta \eta$ distributions for the different spin cases in \fig{fig:deta} (right). 
We calculate $N_R$, the expected number of signal events required to disfavour a hypothesis $H_S$ over another hypothesis $H_T$ to an odds factor of $R=20$ from the $\Delta \eta$ distributions in the 
discretised Kullback-Leibler 
method proposed in \citeref{Allanach:2017qbs}.
\begin{table}[t]
	\begin{center}
		\small\begin{tabular}{|c|c|c|c|c|} 
			\hline
			
			$N_R$  & $\sqrt{s}$         & $S4$   & $G4_{gg}$ &  $G4_{ff}$ \\ \hline
			\hline
			
			$S4$   & $14\TeV$       &$\infty$&$23$&$12$\\    
			& $27\TeV$ &$\infty$&$17$&$8$ \\ \hline \hline
			
			$G4_{gg}$  & $14\TeV$   &$34$&$\infty$&$5$\\  
	&$27\TeV$			 &$24$&$\infty$&$3$\\ \hline \hline 
			$G4_{ff}$ & $14\TeV$    &$18$&$5$&$\infty$ \\  
			& $27\TeV$ &$11$&$4$&$\infty$ \\ \hline  \hline    
		\end{tabular}\normalsize
		\caption{\label{tab:spindiscrimination1lambda} Spin discrimination in case A: $N_R = {\mathcal L}
			\sigma_{tot}^{(X)}$, the expected
			number of total signal events required to be produced 
			to discriminate against the `true' row
			model versus a column model by a factor of $20$
			for $m_n=100\GeV$.} 
	\end{center}
\end{table}
\begin{table}[t]
	\begin{center}
		\small\begin{tabular}{|c||c|c|c|c|c|c|c|c|} 
			\hline
			$N_R$  	& $\sqrt{s}$    		& $S2$    		& $S4$ 		& $V3$ 	& $G2_{gg}$ 	& $G4_{gg}$ 	& $G2_{ff}$ 	& $G4_{ff}$ 	\\ \hline
			\hline
			$S2$  	& $14\TeV$ 	& $\infty$		& $19270$     	& $198$	& $25$		& $13$		& $85$		& $12$  		\\  
			      		& $27\TeV$ 	& $\infty$ 		& $8676$ 		& $269$ 	& $26$ 		& $16$ 		& $92$ 		& $14$  		\\ \hline\hline  
			$S4$  	& $14\TeV$     	& $19256$   	& $\infty$ 		& $202$	& $25$		& $13$		& $81$		& $13$		\\   
					& $27\TeV$	& $8713$   	& $\infty$  	& $311$ 	& $28$ 		& $15$ 		& $83$		& $14$		\\ \hline\hline  
			$V3$  	& $14\TeV$     	& $186$  		& $190$  		& $\infty$ 	& $55$ 		& $7$		& $30$ 		& $19$ 		\\  
					& $27\TeV$	& $261$   		& $299$ 		& $\infty$ 	& $51$ 		& $10$ 		& $40$		& $20$ 		\\ \hline\hline  
			$G2_{gg}$& $14\TeV$  	& $28$    		& $28$		& $66$	& $\infty$ 		& $4$		& $11$		& $35$      	\\  
					& $27\TeV$	& $31$  		& $33$ 		& $63$ 	& $\infty$ 		& $5$		& $13$ 		& $41$      	\\ \hline\hline  
			$G4_{gg}$& $14\TeV$  	& $21$   		& $21$		& $12$	& $6$		& $\infty$ 		& $47$		& $4$		\\  
					& $27\TeV$ 	& $26$  		& $25$ 		& $16$ 	& $7$  		& $\infty$ 		& $55$ 		& $5$		\\ \hline\hline  
			$G2_{ff}$	& $14\TeV$  	& $101$   		& $97$		& $37$	& $11$		& $33$		& $\infty$ 		& $8$  	 	\\  
					& $27\TeV$ 	& $103$  		& $93$ 		& $46$ 	& $12$ 		& $39$ 		& $\infty$ 		& $9$   		\\ \hline\hline  
			$G4_{ff}$	& $14\TeV$  	& $18$  		& $18$		& $26$	& $36$		& $4$		& $12$		& $\infty$ 		\\     
					& $27\TeV$ 	& $19$   		& $20$ 		& $29$ 	& $41$ 		& $5$ 		& $13$ 		& $\infty$ 		\\ \hline\hline  
		\end{tabular}\normalsize
		\label{tab:spindiscrimination}
	\end{center}
	\caption{Spin discrimination for case B: $N_R = {\mathcal L}
		\sigma_{tot}^{(X)}$, the expected
		number of total signal events required to be produced 
		to discriminate against the `true' row
		model versus a column model by a factor of $20$ at the $14\TeV$
		LHC for $m_n=10\GeV$. \label{tab:two}} 
\end{table}
From \tabbs{tab:spindiscrimination1lambda} and \ref{tab:two}, we see that our estimate of the required number of signal events to discriminate two signal hypotheses does not change much between the HL-LHC and the HE-LHC. For identical input parameters, we would expect the number of signal events to be higher at the higher energies, meaning that spins can be more effectively discriminated to smaller production cross-sections at the HE-LHC.

\vspace{1cm}
%\paragraph*{Acknowledgements}
%This work has been partially supported by STFC consolidated grants ST/L000385/1, ST/P000681/1.

\subsubsection{\atlasC{Prospects for diboson resonances at the HL- and HE-LHC}}
\contributors{R. Les, V. Cavaliere, T. Nitta, K. Terashi, ATLAS}
%{\bf Author(s): R. Les, V. Cavaliere, T. Nitta, K. Terashi et al. ATLAS}

Prospects are presented~\cite{ATL-PHYS-PUB-2018-022} for the search for resonances decaying to diboson ($WW$ or $WZ$, collectively called $VV$ where $V=W$ or $Z$) in the semileptonic channel where one $W$-boson decays leptonically and the other $W$ or $Z$-boson decays to quarks 
($\ell\nu qq$ channel). The results include sensitivity for such new resonances based on an integrated luminosity of $300\fbinv$or \intLumi  
of $pp$ collisions at \sloppy\mbox{$\sqrt{s}=14\TeV$} using the ATLAS upgraded detector.
Searches in other semileptonic and fully hadronic decay channels are expected to have similar sensitivities at high masses, 
as observed in the ATLAS searches with Run-2 data.
The analysis is based on event selection and classification similar to those used in the Run-1 and Run-2 ATLAS searches. 

The prospect for resonance searches presented in this article are interpreted in the context of three different models: a heavy vector triplet (HVT) \cite{Pappadopulo:2014qza}, a RS model and a narrow heavy scalar resonance.
The parameters of these models are chosen such that along the whole generated mass range, the resonance widths are less than $6\%$ of the mass value, which is smaller than the detector resolution.
The main background sources are $W$ bosons produced in association with jets ($W$+jets), with significant contributions from top-quark production (both  $t\bar{t}$ pair and single-top), non-resonant vector-boson pair production ($ZZ$, $WZ$ and $WW$) and $Z$ bosons produced in association with jets ($Z$+jets). Background originating from multi-jet processes are expected to be negligible due to the event selection requirements. 

Small- and Large-$R$ jets~(denoted by $j$ and $J$) are used in the analysis, reconstructed with the $\antikt$~algorithm and radius $0.4$ and $1.0$ respectively. It is assumed that the performance of a future $W/Z$-boson tagger at the HL-LHC conditions will have similar, if not better, 
performance as existing boson taggers. 

Events are required to have exactly one lepton satisfying the selection criteria. It is assumed that the effect of trigger thresholds is negligible for the selected leptons with $p_{T}$ studied in this note.
Events are further required to contain a hadronically-decaying $W/Z$ candidate, reconstructed either from two small-$R$ 
jets, defined as the resolved channel, or from one large-$R$ jet, designated the boosted channel.
The missing transverse energy $\met$ has to be greater than $60\GeV$,
which suppresses multijet background to a negligible level.
By constraining the $\met$ + lepton system to be consistent with the $W$ mass, the $z$ component of the neutrino  momentum
can be reconstructed by solving a quadratic equation. The smallest solution is chosen and in the case where the 
solution is imaginary, only the real part is taken. \\

\begin{figure}[t]
\centering
\begin{minipage}[t][\minipageheightdp][t]{\minipagewidthdp}
\centering
\includegraphics[width=\figuremaxwidthdp,height=\figuremaxheightdp,keepaspectratio]{\main/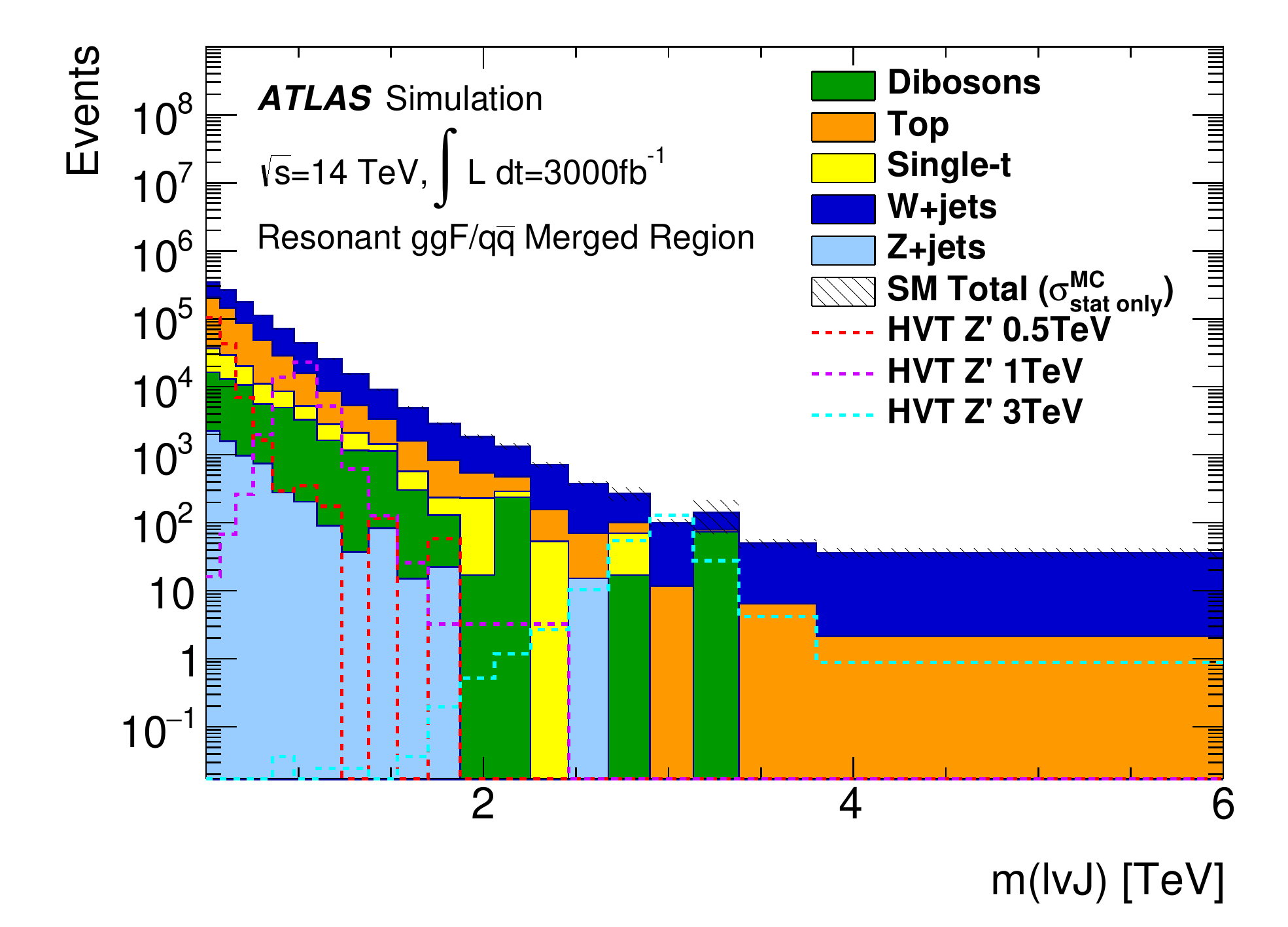}
\end{minipage}
\begin{minipage}[t][\minipageheightdp][t]{\minipagewidthdp}
\centering
\includegraphics[width=\figuremaxwidthdp,height=\figuremaxheightdp,keepaspectratio]{\main/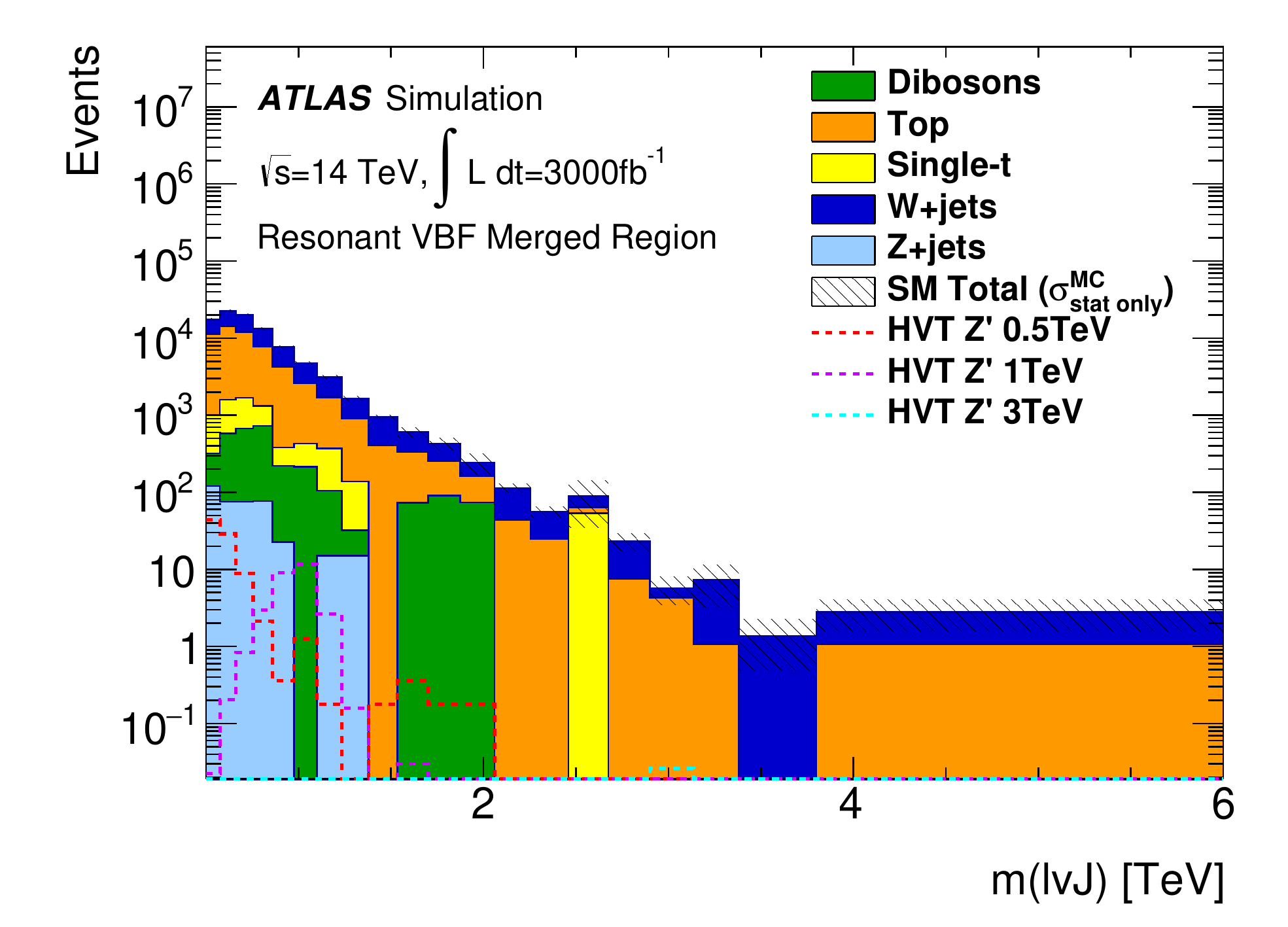}
\end{minipage}
\caption{Left: final $m(\ell\nu J)$ distribution in the merged signal region for the $\mathrm{ggF/q\bar{q}}$ search. Right: final $m(\ell\nu J)$ distributions in the merged signal region for the VBF search. Background distributions are separated into production type. HVT signals are superimposed as dashed curves where appropriate.  }
    \label{fig:distributions_VVatlas}
\end{figure}

The presence of narrow resonances is searched for in the distribution of reconstructed diboson mass using the signal shapes extracted from simulation of benchmark models.
The invariant mass of the diboson system ($m(WV)$) is reconstructed from the leptonic $W$ candidate and hadronic $W/Z$ candidate, the latter of which is obtained from two small-$R$ jets in the resolved channel ($m(\ell\nu jj)$) or large-$R$ jet in the boosted channel ($m(\ell\nu J)$). 
The background shape and normalisation are obtained from MC simulation with dedicated control regions to constrain systematic uncertainties of the background modelling and normalisation.
The search is divided into two orthogonal categories to identify the $\mathrm{ggF/q\bar{q}}$ and VBF production modes 
by identifying additional forward jets. If an event passes the VBF category selection for the additional forward jets (defined below) it is categorised as a VBF candidate event, otherwise as a $\mathrm{ggF/q\bar{q}}$ candidate event.
Events are then processed by a merged-jet selection then a two resolved-jet selection if they fail the merged selection. 
This prioritisation strategy provides the optimum signal sensitivity as it favours the merged selection which contains less background contributions.
Examples of the final distributions can be seen in \fig{fig:distributions_VVatlas}
for the merged $\mathrm{ggF/q\bar{q}}$ and VBF signal region. 

The results are extracted by performing a simultaneous binned maximum-likelihood fit to the $m(WV)$ distributions 
in the signal regions and the $W$+jets and $t \bar{t}$ control regions. 

The fit includes five background contributions, corresponding to $W$+jets, $t \bar{t}$, single-top, $Z$+jets, and diboson. 
Systematic uncertainties are taken into account as constrained nuisance parameters with Gaussian or log-normal distributions. 
For each source of systematic uncertainty, the correlations across bins of $m(WV)$ distributions and between different kinematic regions, 
as well as those between signal and background, are taken into account. 
The main background modelling systematics, namely the $W$+jets and $t\bar{t}$ shape uncertainties, are constrained by the corresponding 
control regions and are treated as uncorrelated among the resolved and merged signal regions.            

The expected upper limits are set on the signal cross section times BR as a function of the signal mass.  
For the HVT $W'$ and $Z'$ the limits are estimated to be $4.3\TeV$ with $\mathcal{L} =300\fbinv$ and $4.9\TeV$ with $\mathcal{L} =\intLumi$ of $pp$ collisions, 
using the same detector configuration and pileup conditions. For the Bulk graviton the expected limits are estimated as 2.8 and 3.3~TeV 
at $\mathcal{L} =300\fbinv$ and $\mathcal{L} =\intLumi$.
The values at $\mathcal{L} =\intLumi$ show an expected increase to the sensitivity of the search to the benchmark signals by $\sim 1\TeV$ with 
respect to existing limits in this channel in Run-2. 

\begin{figure}[t]
\centering
\begin{minipage}[t][\minipageheightdp][t]{\minipagewidthdp}
\centering
\includegraphics[width=\figuremaxwidthdp,height=\figuremaxheightdp,keepaspectratio]{\main/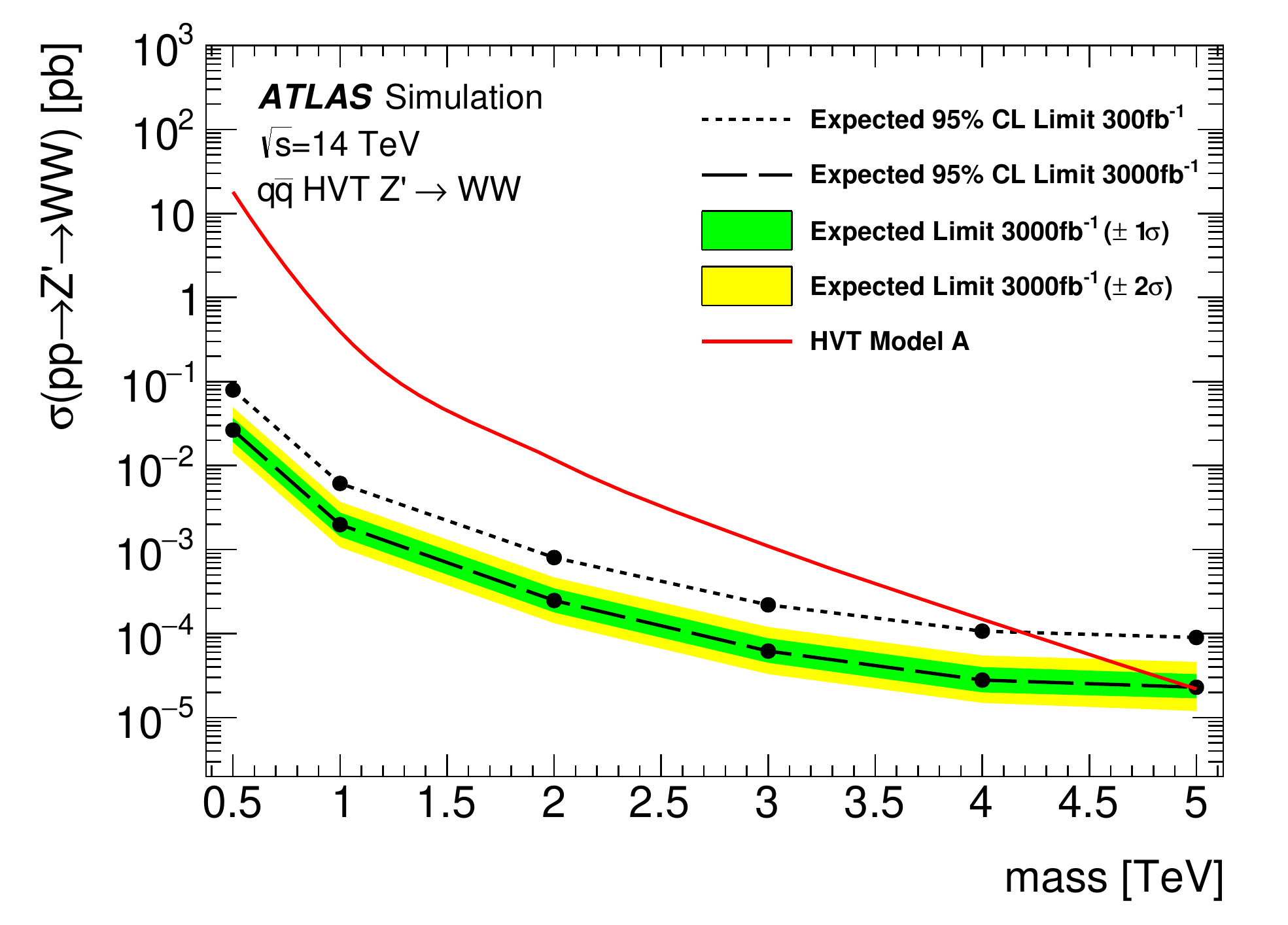}
\end{minipage}
\begin{minipage}[t][\minipageheightdp][t]{\minipagewidthdp}
\centering
\includegraphics[width=\figuremaxwidthdp,height=\figuremaxheightdp,keepaspectratio]{\main/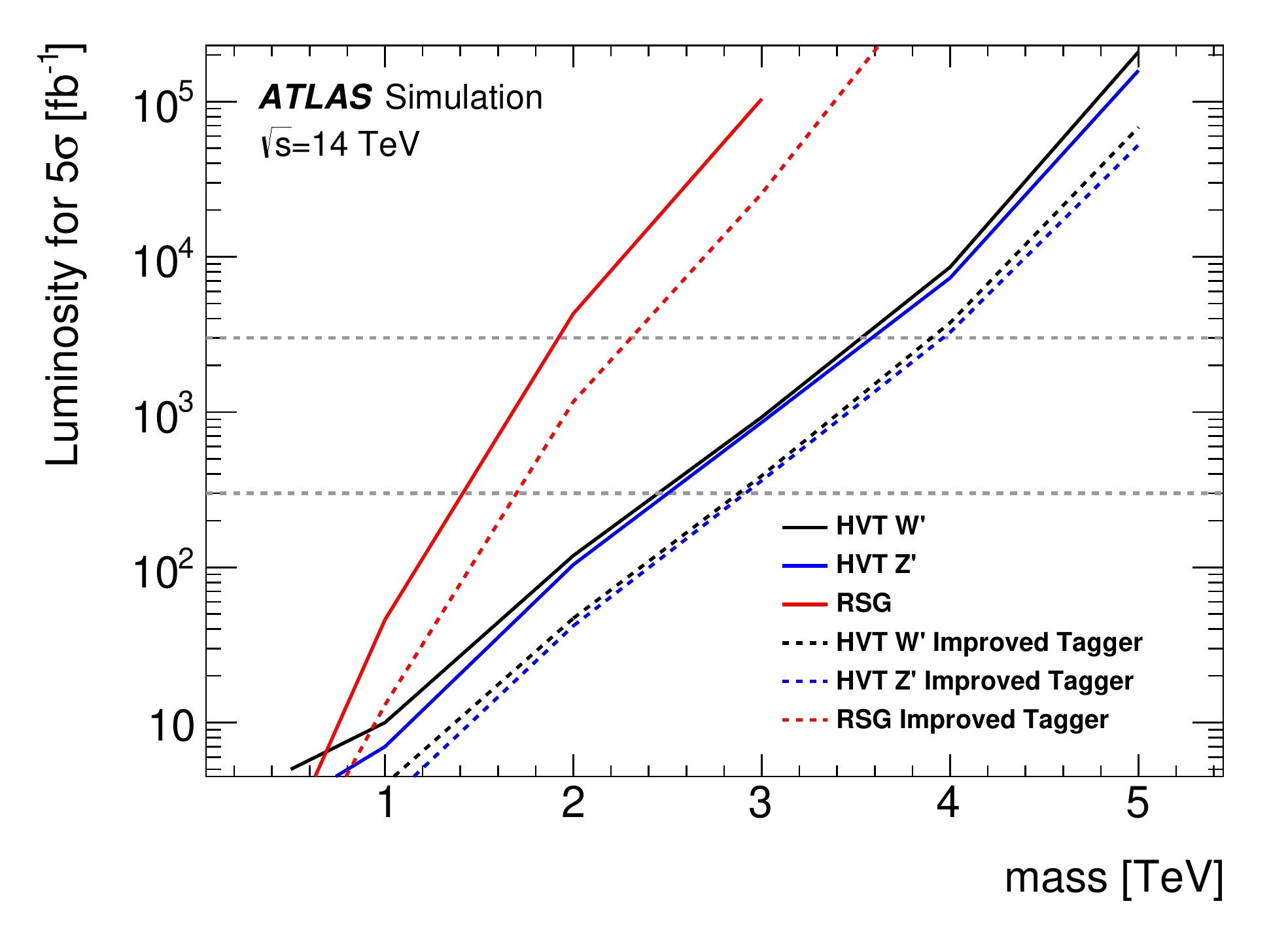}
\end{minipage}
    \caption{Left: $95\%\cl$ upper limit for the HVT $Z'$ via $\mathrm{ggF/q\bar{q}}$ production. Right: expected luminosity required to observe a $5\sigma$ signal significance for the HVT $W'$ (black), HVT $Z'$ (blue) and $G_{RS}$ (red). The solid curves shows the sensitivity using the current $W/Z$-tagger and the dashed curves for a future tagger that has a $50\%$ increased signal efficiency and a factor two increased rejection of background.}
    \label{fig:Limits_VVatlas}
\end{figure}

Figure~\ref{fig:Limits_VVatlas} (left) shows one of the upper limit plots for the $\mathrm{ggF/q\bar{q}}$ category at $\mathcal{L} =\intLumi$.   
A line showing the theoretical cross section for the HVT $Z'$ decaying into $WW$ via $\mathrm{ggF/q\bar{q}}$ 
production at each mass is superimposed and indicates the mass reach of the search. In the circumstance that HL-LHC sees an excess, the expected sensitivity can also be characterised.  
The discovery significance is defined as the luminosity required to see a $5\sigma$ effect of the signal. 
Figure~\ref{fig:Limits_VVatlas} (right) shows the expected discovery significance for the resonant search. 
The signal significance is the quadratic sum of $s/\sqrt{s+b}$, for each bin of the final discriminant distribution at that luminosity, 
$s(b)$ representing the number of signal(background) events in the bin.
In addition to the expected values, dashed curves shows the expected values for a future $W/Z$-tagger which has a $50\%$ increase 
in signal efficiency and a further factor of two in background rejection. 
These values are representative of improvements seen in a recent diboson resonance search in the fully-hadronic $VV \rightarrow qqqq$ 
analysis by using track-calo clusters as opposed to locally-calibrated topologically-clustered calorimeter jets.
Other possible improvements in $W/Z$-tagging in the HL-LHC era can originate from usage of more advanced machine-learning techniques to discriminate 
against the background contribution and better understanding of jet substructure variables with measurements at higher integrated luminosities.

\paragraph*{\textbf{Resonance search at HE-LHC}}

The prospect analysis~\cite{Cavaliere:2018zcf} at HE-LHC  mimics the analysis at HL-LHC but the \delphes{} simulation is used. 
Results are interpreted in the context of the  heavy vector triplet (HVT) model~\cite{Pappadopulo:2014qza}.
The major backgrounds $W$+jets and $t\bar{t}$ production are simulated with \madgraph{} and \textsc{aMC@NLO} respectively, interfaced with Pythia. $Z$+jets, single top and diboson contribution are not simulated and are expected to contribute at most $10\%$ to the total background.

The analysis sensitivities to the resonance signals are extracted by performing a simultaneous binned maximum-likelihood fit to the $m_{WW}$ distributions as in the HL-LHC in the signal regions and the $W$+jets and $t \bar{t}$ control regions. The signal region and control regions are defined in the same way.

\begin{figure}[t]
\centering
\begin{minipage}[t][\minipageheightdp][t]{\minipagewidthdp}
\centering
\includegraphics[width=\figuremaxwidthdp,height=\figuremaxheightdp,keepaspectratio]{\main/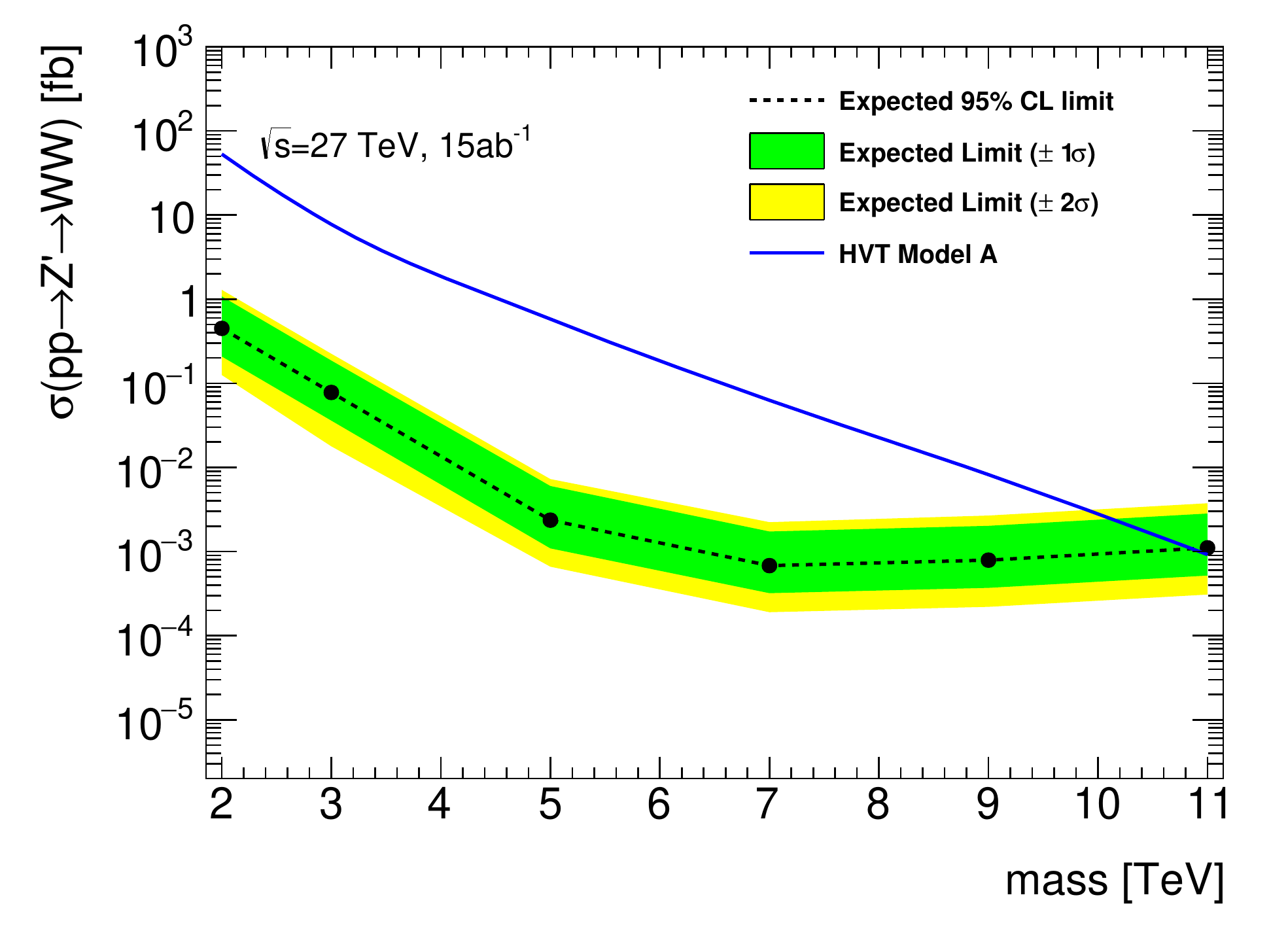}
\end{minipage}
\begin{minipage}[t][\minipageheightdp][t]{\minipagewidthdp}
\centering
\includegraphics[width=\figuremaxwidthdp,height=\figuremaxheightdp,keepaspectratio]{\main/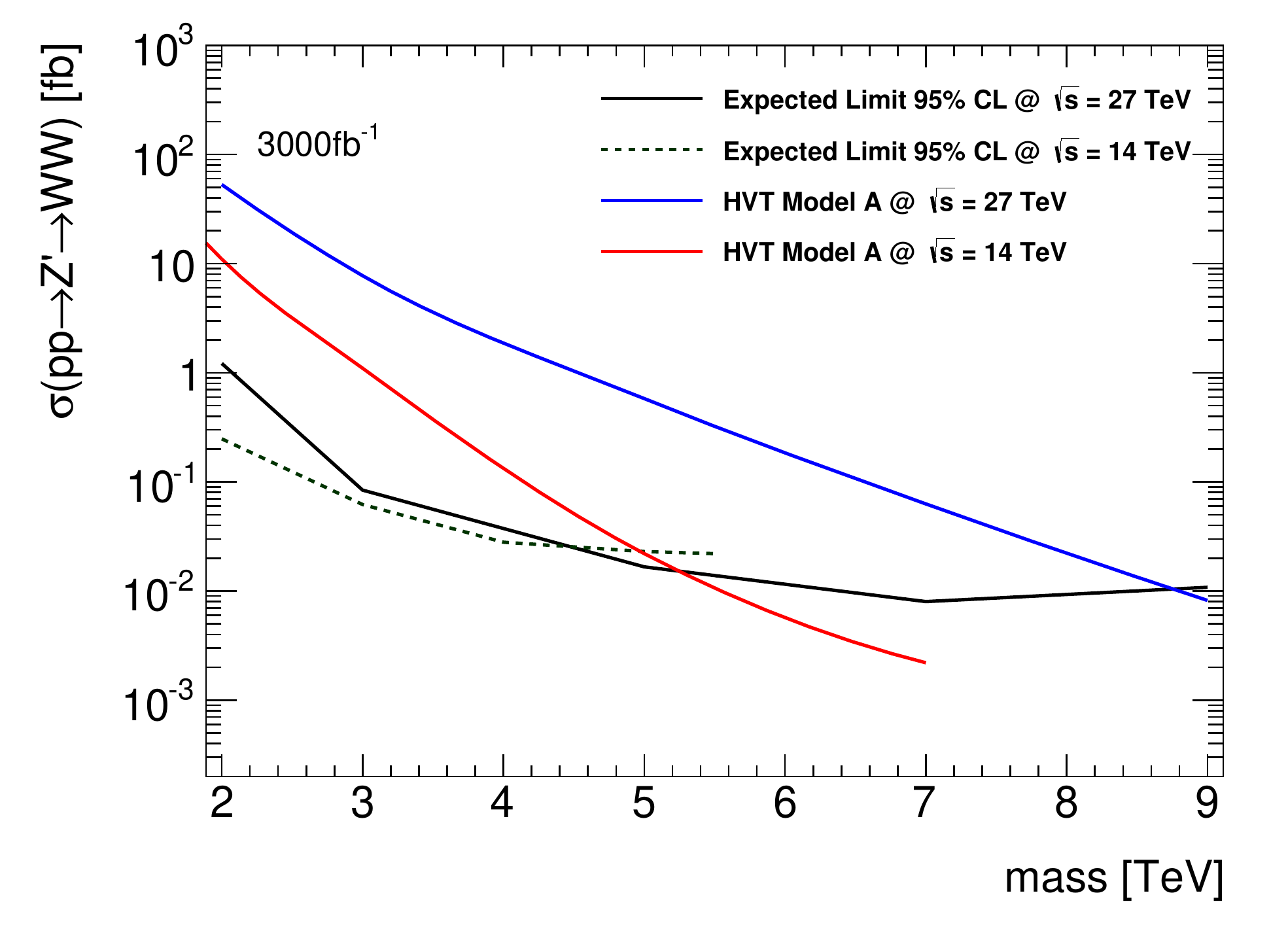}
\end{minipage}
\caption{$95\%\cl$ upper limit for the HVT $Z'$  via $\mathrm{ggF/q\bar{q}}$ production for the HE-LHC (left) and HL-LHC (right).}
\label{fig:Limits}
\end{figure}

The expected upper limits set on the signal cross section times BR as a function of the signal mass are shown in \fig{fig:Limits} at $27\TeV$ and compared to the limits obtained for HL-LHC . 
Based on the $Z'$ production cross section from the HVT signal model the exclusion mass reach is extracted to be $9$ and $11\TeV$ for integrated luminosities of $3$ and \intlumihelhc at $27\TeV$, using \delphes{} simulation of a potential detector configuration under no pileup condition.

\subsubsection{\theoryC{Prospects for Boosted Object Tagging with Timing Layers at HL-LHC}}
\label{sec:theoryboostedtiming}
\contributors{M. D. Klimek}
%\textbf{Authors: Matthew D. Klimek$^{a,b}$}\\
%\textit{a) Laboratory for Elementary Particle Physics, Cornell University, Ithaca, NY 14853, USA
%b) Department of Physics, Korea University, Seoul 02841, Republic of Korea
%}\\
%\texttt{klimek@cornell.edu}

Both CMS and ATLAS are studying new timing detectors to be installed for the high luminosity phase of the LHC. 
The primary motivation for these new detectors is to aid in mitigating the increased level of pileup that comes along with the increased luminosity.
The $\mathcal O(30\psec)$ timing resolution will permit individual proton-proton interactions within a single bunch crossing to be temporally resolved.
Beyond pileup mitigation, these detectors may be useful for novel kinds of searches.
For example, it has been proposed in \citeref{Liu:2018wte} that they could be used to set limits on long-lived particles.
In this section, we point out that this level of timing precision provides the capability of temporally resolving jets for the first time.
Currently, jet substructure techniques play a major role in the search for BSM physics.
These exploit correlations in momentum and angular distribution of jet constituents to distinguish standard QCD jets from other objects.
The capabilities of the new timing layers open up the possibility of extending jet substructure techniques to the time domain.
In this section, we will introduce the relevant concepts and suggest that time domain jet substructure can play a complementary role to existing techniques in the search of boosted resonances at the HL-LHC.

Of the various physics objects that are reconstructed by the LHC experiments, jets are unique in that they are collections of particles.
The individual jet constituents have some spread in velocity and therefore arrive at the detector over some finite span of time.
On dimensional grounds, we can estimate that the typical scale of Lorentz boosts of the jet constituents is $\gamma = E/m \sim E_j/n\Lambda_\mathrm{QCD}$, where $E_j$ is the jet energy, $n$ is the hadron multiplicity of the jet, and $\Lambda_\mathrm{QCD}$ sets the typical hadron mass.
The corresponding scale of the spread in arrival times at a detector a distance $R$ from the interaction point is then of order $\delta t\sim R\delta v\sim R\gamma^{-2}$. 
For $R\sim 1\meters$, $E_j\sim100\GeV$, $n\sim10$, and $\Lambda_\mathrm{QCD}\sim1\GeV$, we have $\delta t\sim100\psec$.
Thus we see that at typical LHC energies, the proposed timing detectors with $\mathcal O(30 \psec)$ resolution should be sensitive to the temporal structure of jets.

We note that this effect is entirely due to the hadronisation process.
Any time differences inherent to the perturbative parton shower should be of order $\Lambda_\mathrm{QCD}^{-1}$ and will be far below the resolution of the timing detectors.
(In practice, the distance from the interaction point to the timing detector will have some dependence on pseudorapidity $\eta$.
Because the showering process produces radiation in a cone around the original parton direction, the various jet constituents will then have slightly different distances to travel to the detector.
However, for reasonably central jets, this will be a small effect.)
If, in the hadronisation process, jet constituents are produced with rapidities $y$ distributed according to $dN/dy = f(y)$, then it is easy to verify that at a detector a distance $R$ in any direction from the interaction point, they will arrive distributed in time according to 
\begin{equation}
	\frac{dN}{dt} = \frac{ cR f(y(t)) }{ (ct)^2-R^2},
\end{equation}
with $ct>R$, where $c$ is the speed of light and $y(t)=\mathrm{arctanh}(R/ct)$.
Although it is not possible to calculate $f(y)$ from first principles, we note that for any reasonably well-behaved $f(y)$ the arrival time distribution will indicate a burst of jet constituents arriving promptly with a tail extending out to later times.

\begin{figure}[t]
\centering
\begin{minipage}[t][\minipageheightsp][t]{\minipagewidthsp}
\centering
\includegraphics[width=\figuremaxwidthsp,height=\figuremaxheightsp,keepaspectratio]{\main/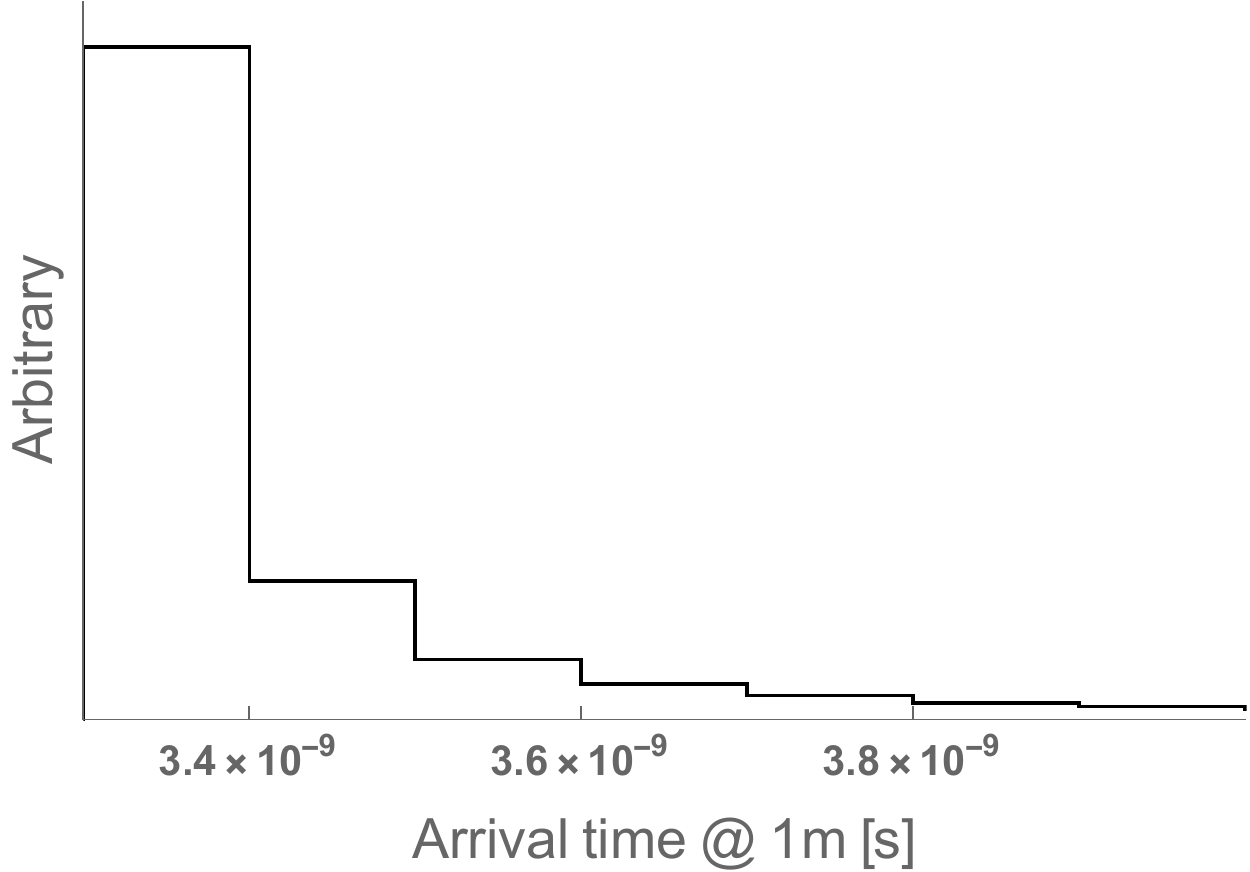}
\end{minipage}
\caption{Average arrival time profile at $1\meters$ for charged hadrons in $50\GeV$ quark jets from $e^+e^-$ annihilation as simulated by \pythia{ 8}.}
\label{fig:pythia_arrival_profile} 
\end{figure}

The hadronisation process is treated phenomenologically by a number of models, and we can verify that this behaviour is borne out.
For example, the simplest form of the Lund string hadronisation model~\cite{Andersson:1983ia} predicts $f(y)=$~constant.
PYTHIA implements a more complete version of the Lund model, and we use it to verify the predictions of the preceding paragraphs.
We generated a sample of $50\GeV$ quark jets from $e^+e^-$ annihilation in \pythia{ 8}, and computed the arrival times of all charged hadrons in each jet at a distance of $1\meters$ from the interaction point.
A histogram with $100\psec$ bin size of the averaged arrival time profile is shown in \fig{fig:pythia_arrival_profile}.
We can see that the PYTHIA output displays the expected behaviour, and we verify that the typical width of the profile is indeed $\mathcal{O}(100\psec)$.

A study of the rapidity distribution of jet constituents from arrival time profile of QCD jets could provide additional data with which to tune our models of hadronisation.
However, we will now argue that this information also provides a novel method for distinguishing normal QCD jets from jets produced by boosted objects.
As described in the preceding paragraphs, in a typical jet we expect most hadrons to arrive together and a few to arrive at later times.
Note that under a boost, the ordering of the jet constituents in velocity or rapidity is unchanged.
Consider a massive particle that decays in its rest frame producing a jet pointing along the negative $x$ axis.
The particles in the tail of its velocity distribution have velocities along the $x$ direction that are less negative than the rest of the jet.
Now consider the same massive particle but with a large boost in the positive $x$ direction so that this jet is boosted forward into the positive direction as well.
Despite the change in direction, the ordering of the velocities in the jet is the same.
The velocities of all jet constituents are now positive, and the tail of the distribution is more positive.
We see that if a jet is sufficiently boosted, the tail of its hadron distribution can arrive at the detector first.

In order to quantify this effect, we make use of a simple diagnostic that can distinguish the boosted from the unboosted jets based on their arrival time profiles.
Recall that for an unboosted jet, we expect the majority of the charged hadrons to arrive in close temporal proximity with a few arriving later, whereas for a sufficiently boosted jet, one or more charged hadrons may arrive before most of the others.
Let $t_i$ and $p_{i}$ be the arrival times and momenta of the charged hadrons in a jet and $\bar t$ be the median charged hadron arrival time.
We define the diagnostic function
\begin{equation}
	D_\tau = \frac{ \sum p_{i} \Theta(t_i-\bar t-\tau) }{ \sum p_{i} }.
\end{equation}
The parameter $\tau$ should be a value of order the time resolution of the detector.
For an unboosted jet, the median arrival time will be in the prompt burst of hadrons.
For an appropriate value of $\tau$, none of these hadrons nor any that arrive later will satisfy the theta function, and we will obtain $D_\tau=0$.
However, for boosted jets, some hadrons from the boosted tail may arrive before the main burst. Because the median will still be in the main burst, the early hadrons can satisfy the theta function, in which case we will find $D_\tau>0$.

To verify this behaviour, we generate two samples of jets in \pythia{ 8}.
The first is a sample of $500\GeV$ quark jets from $e^+e^-$ annihilation.
For comparison, we also generate a similar sample of $50\GeV$ jets, but then boost them by $v=0.98$.
This is comparable to jets that would be observed from $Z$ decay, where the $Z$ has been produced in the decay of a $1\TeV$ diboson resonance.
We choose a conservative value of $\tau=200\psec$, which is several times the expected timing detector resolution.
In the first sample, we find that our diagnostic is zero in all but $1\%$ of events.
However, for the boosted jet sample, we obtain non-zero values in $25\%$ of events.

The actual operating environment of the HL-LHC will present additional challenges coming from the noisy hadronic environment.
Jet contamination from pileup and underlying event radiation could mimic the early arrival of the tail of a boosted jet.
The information from the timing layers will be used to mitigate pileup and jet grooming techniques can be used to remove stray radiation from the jets before their temporal characteristic are analysed.
The timing substructure information can then be combined with existing jet substructure techniques to increase the efficiency for tagging boosted objects, in turn improving the reach of BSM searches.
During an HE-LHC phase, even higher boosts can be expected, demanding effective techniques for boosted object tagging.
Further details are presented in \citeref{Klimek:inprep}.

\subsubsection{\theoryC{High mass resonance searches at HE-LHC using hadronic final states}}
\label{subsubsec:hr_had}
\contributors{C. Helsens, D. Jamin, M. Selvaggi}
%%%%%%%%%%%%%%%%%%%%%%%%%%%%%%%%%%%%%%%%%%%%%%%%%%%%%%%%%%%%%%%%%%%%%%%%%%%%%%%%%%%%%%%%%%%%

\renewcommand{\pt}{\ensuremath{p_\text{T}}}
\newcommand*{\ptSup}[1]{\ensuremath{p_{\text{T}}^{\ensuremath{#1}}}}
\newcommand*{\eflow}{\ensuremath{E_\text{F}(\text{n},\alpha)}}

The presence of new resonant states~\cite{Harris:2011bh,Boelaert:2009jm,Lee:1973iz,Branco:2011iw,Hill:1994hp,Kaplan:1983sm,Bellazzini:2014yua,Randall:1999ee,Pomarol:1999ad} decaying to two highly boosted particles decaying hadronically could be observed as an excess in the invariant mass spectrum of two jets over the large SM background.
In this section we present the reach at the HE-LHC for three distinct hadronic signatures: \zptt\ , \rsg\ and \qjj. For the \zptt\, decay mode the SSM \ZpSSM~\cite{Langacker:2008yv} and a leptophobic $Z'_{\text{TC2}}$~\cite{Harris:1999ya,Harris:2011ez} have been considered as benchmarks \Zp models. For the \rsg\ and \qjj\, decay modes, a Randall-Sundrum graviton~\cite{Randall:1999ee} and excited heavy quarks~\cite{Baur:1987ga,Baur:1989kv} have been taken as a benchmarks respectively.

The decay products of the heavy resonances are typically in the multi-TeV regime and their reconstruction imposes stringent requirements on the detector design. Precise jet energy resolution requires full longitudinal shower containment. Highly boosted $W$ bosons and top quarks decay into highly collimated jets that need to be disentangled from standard QCD jets by characterising their substructure. Thus, in order to achieve high sensitivity excellent granularity is needed both in the tracking detectors and in the calorimeters.

\begin{figure}[t]
\centering
\begin{minipage}[t][\minipageheighttp][t]{\minipagewidthtp}
\centering
\includegraphics[width=\figuremaxwidthtp,height=\figuremaxheighttp,keepaspectratio]{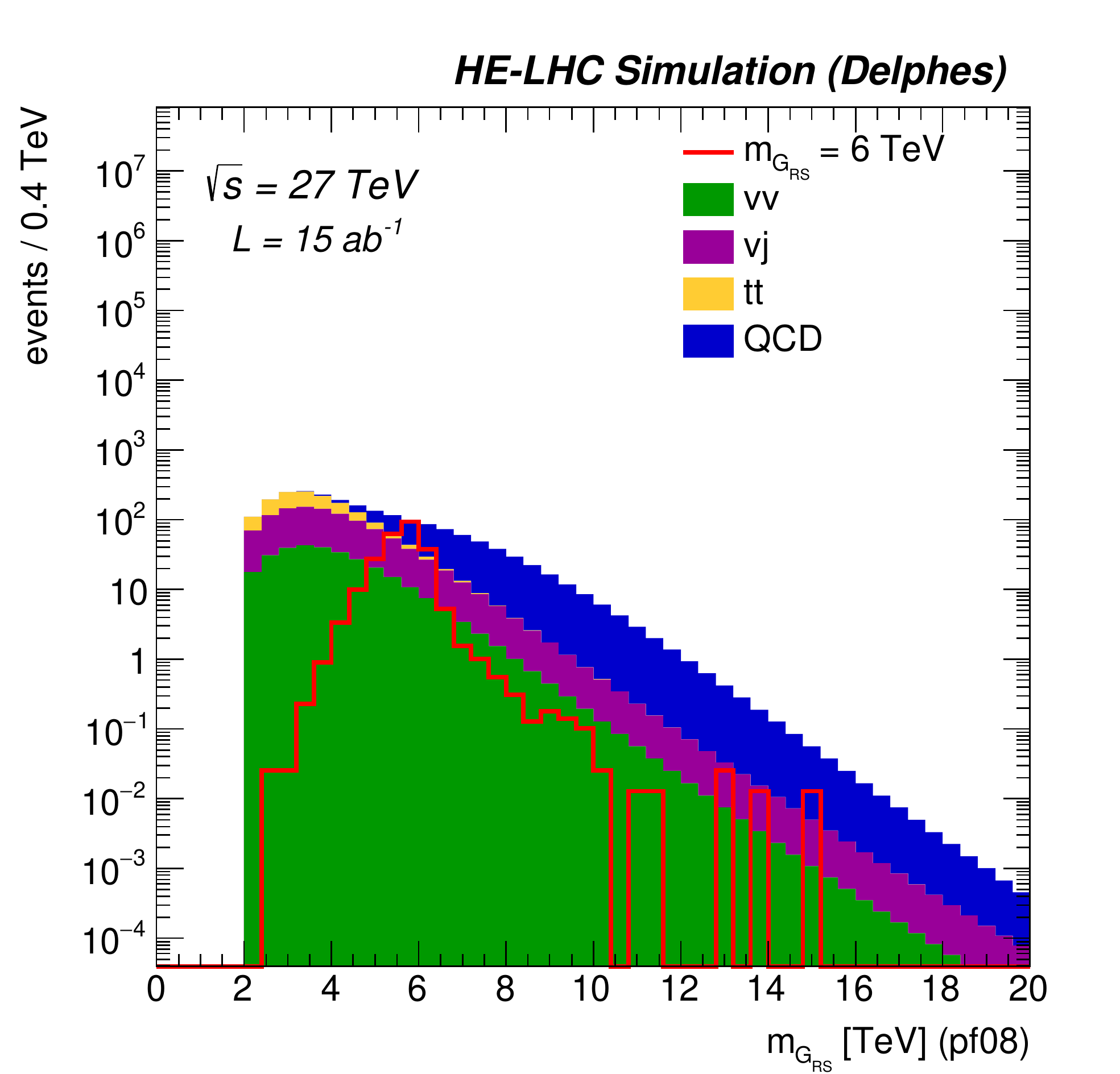}
\end{minipage}
\begin{minipage}[t][\minipageheighttp][t]{\minipagewidthtp}
\centering
\includegraphics[width=\figuremaxwidthtp,height=\figuremaxheighttp,keepaspectratio]{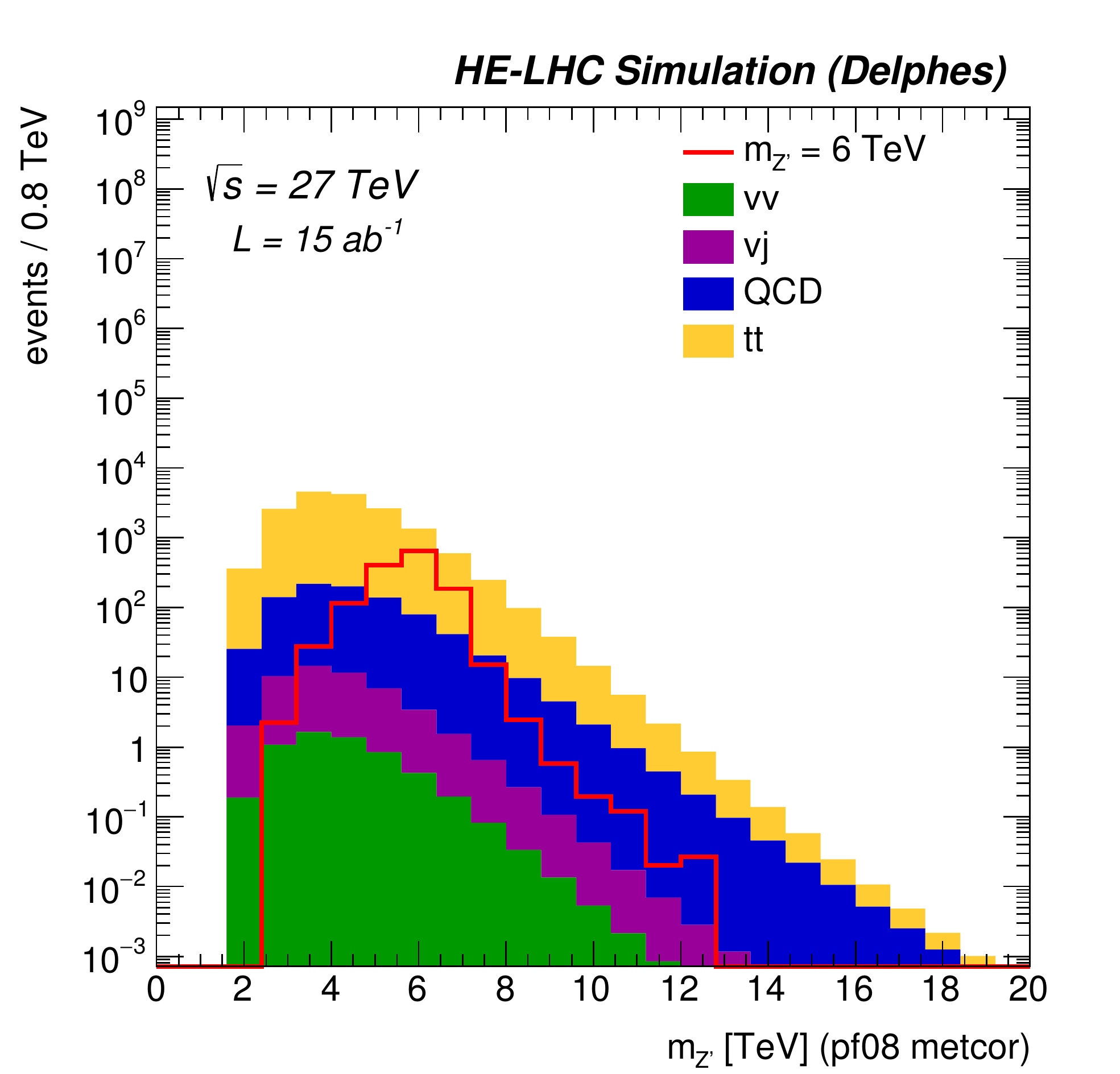}
\end{minipage}
\begin{minipage}[t][\minipageheighttp][t]{\minipagewidthtp}
\centering
\includegraphics[width=\figuremaxwidthtp,height=\figuremaxheighttp,keepaspectratio]{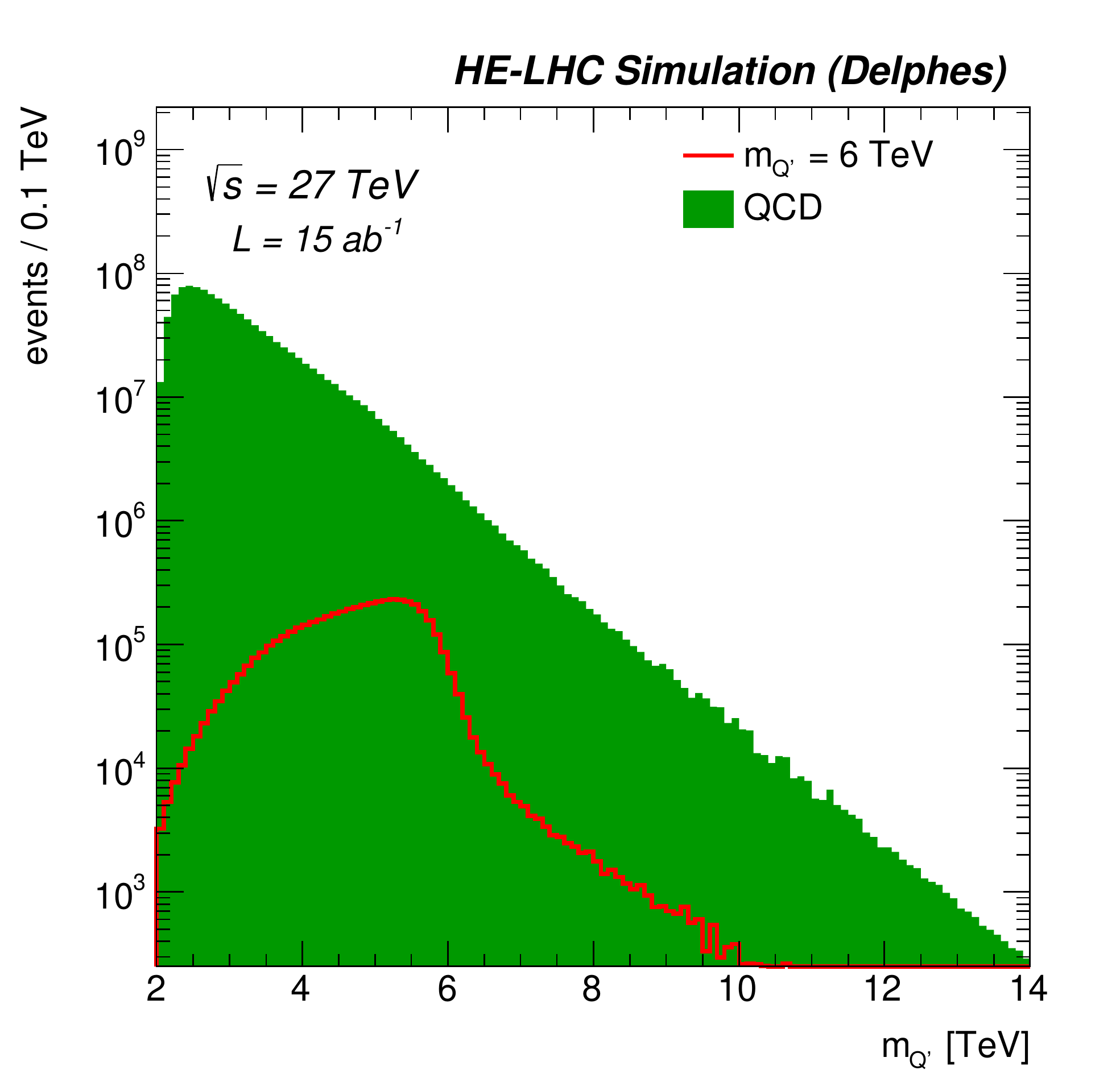}
\end{minipage}
 \caption{Invariant mass distribution of the two selected jets for the full selection for a $6\TeV$ signal for the three benchmark analyses  \rsg\ (left),  \zptt\ (centre) and \qjj\ (right)}
  \label{fig:hadres_invmass}
\end{figure}

%%%%%%%%%%%%%%%%%%%%%%%%%%%%%%%%%%%%%%%%%%%%%%%%%%%%%%%%%%%%%%%%%%%%%%%%%%%%%%%%%%%%%%%%%%%%
%\subsubsection{Monte Carlo Samples}
%\paragraph*{Monte Carlo Samples}
Signal events were generated at LO with \pythia{ 8.230}~\cite{Sjostrand:2014zea}. The considered SM backgrounds are dijet (QCD), top pairs ($\ttbar$), $VV$ and $V+\text{ jets}$ where $V=W/Z$, and were generated at LO using \MGvATNLO{}~\cite{Alwall:2014hca}. A conservative constant k-factor of 2 is applied to all the background processes to account for possibly large higher order corrections. The detector simulation was performed with \delphes~\cite{deFavereau:2013fsa} assuming an HE-LHC generic detector~\cite{hlhelhc_web}.

\begin{figure}[t]
\centering
\begin{minipage}[t][\minipageheighttp][t]{\minipagewidthtp}
\centering
\includegraphics[width=\figuremaxwidthtp,height=\figuremaxheighttp,keepaspectratio]{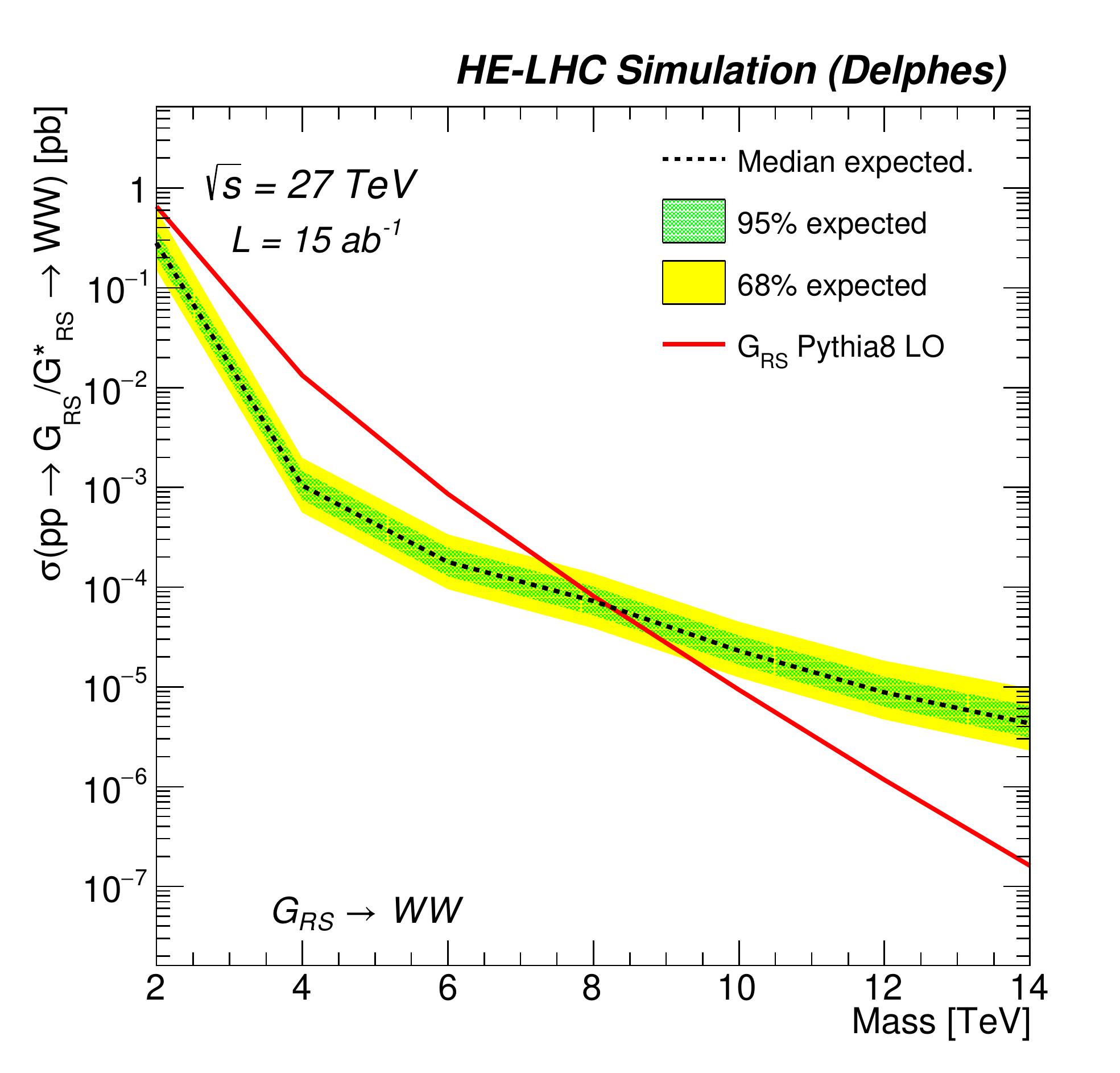}
\end{minipage}
\begin{minipage}[t][\minipageheighttp][t]{\minipagewidthtp}
\centering
\includegraphics[width=\figuremaxwidthtp,height=\figuremaxheighttp,keepaspectratio]{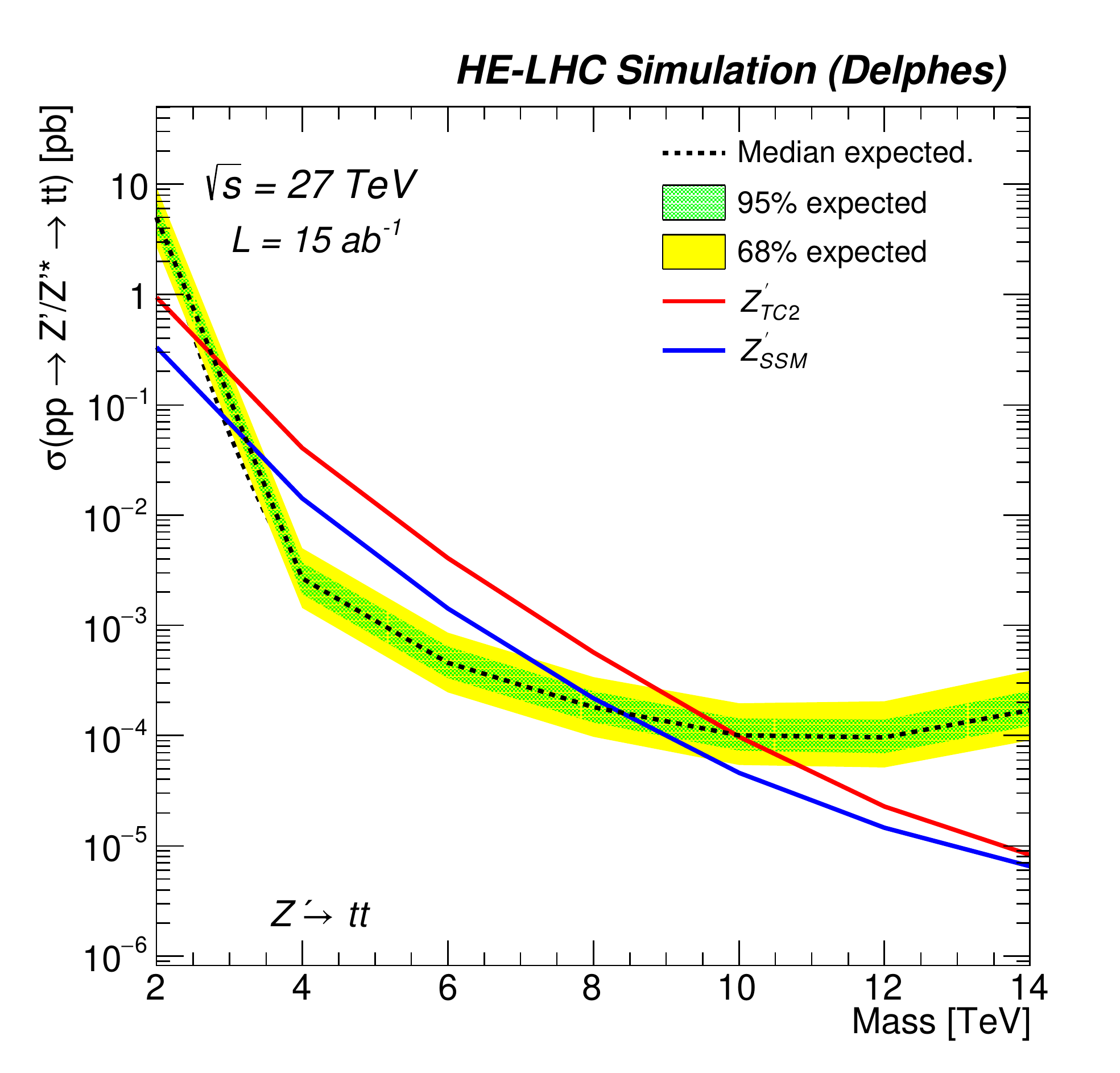}
\end{minipage}
\begin{minipage}[t][\minipageheighttp][t]{\minipagewidthtp}
\centering
\includegraphics[width=\figuremaxwidthtp,height=\figuremaxheighttp,keepaspectratio]{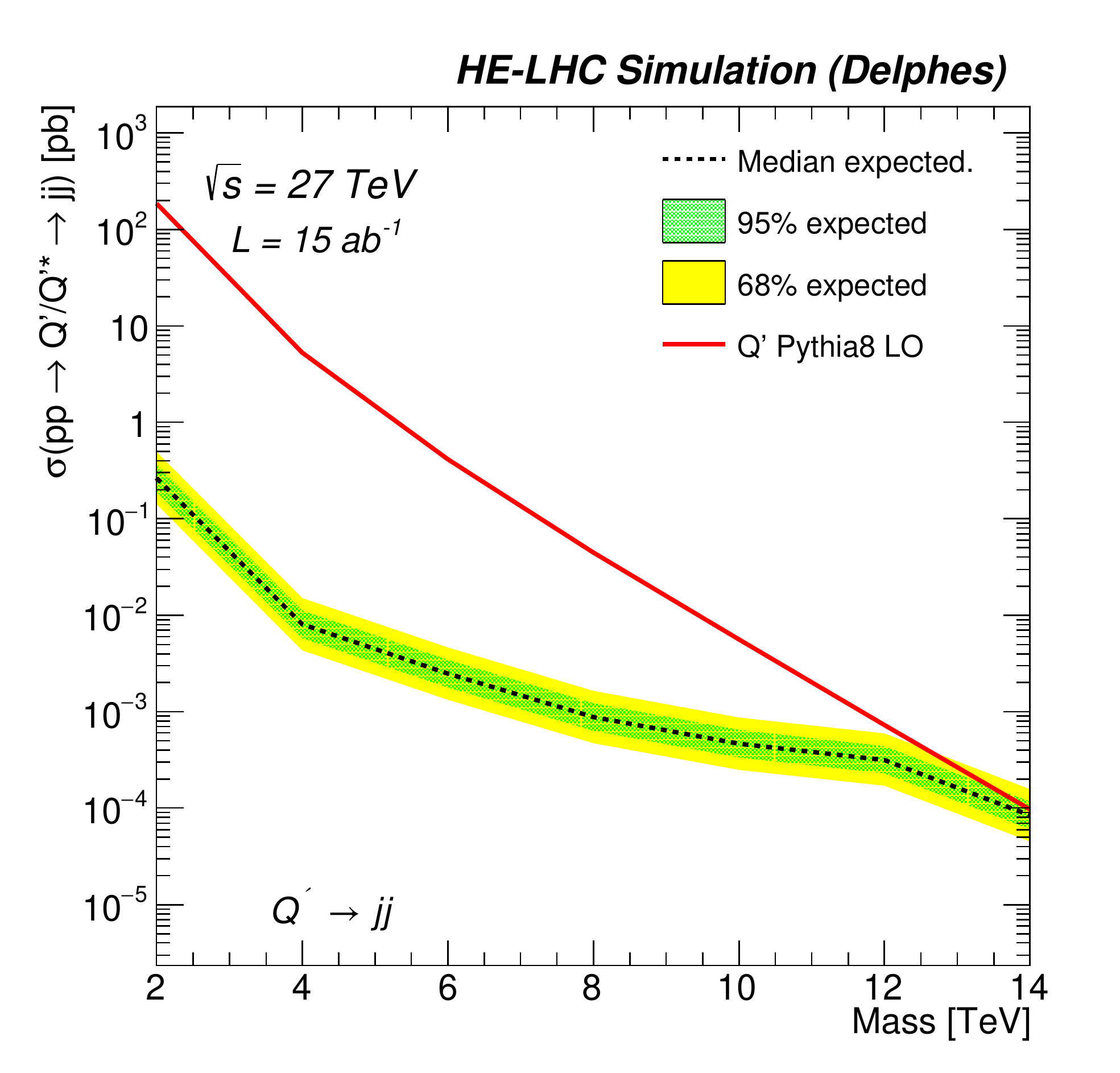}
\end{minipage}\\
\centering
\begin{minipage}[t][\minipageheighttp][t]{\minipagewidthtp}
\centering
\includegraphics[width=\figuremaxwidthtp,height=\figuremaxheighttp,keepaspectratio]{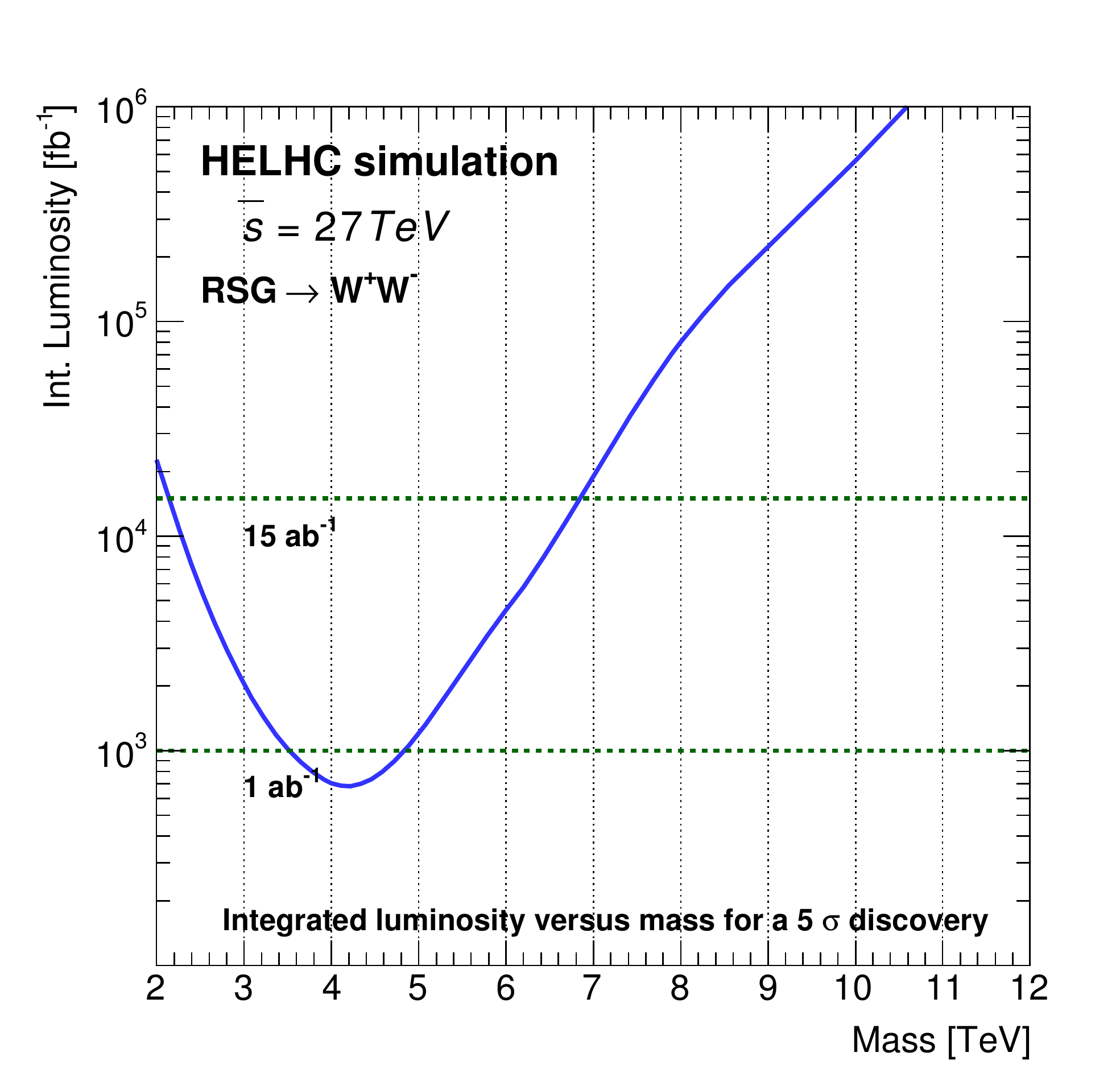}
\end{minipage}
\begin{minipage}[t][\minipageheighttp][t]{\minipagewidthtp}
\centering
\includegraphics[width=\figuremaxwidthtp,height=\figuremaxheighttp,keepaspectratio]{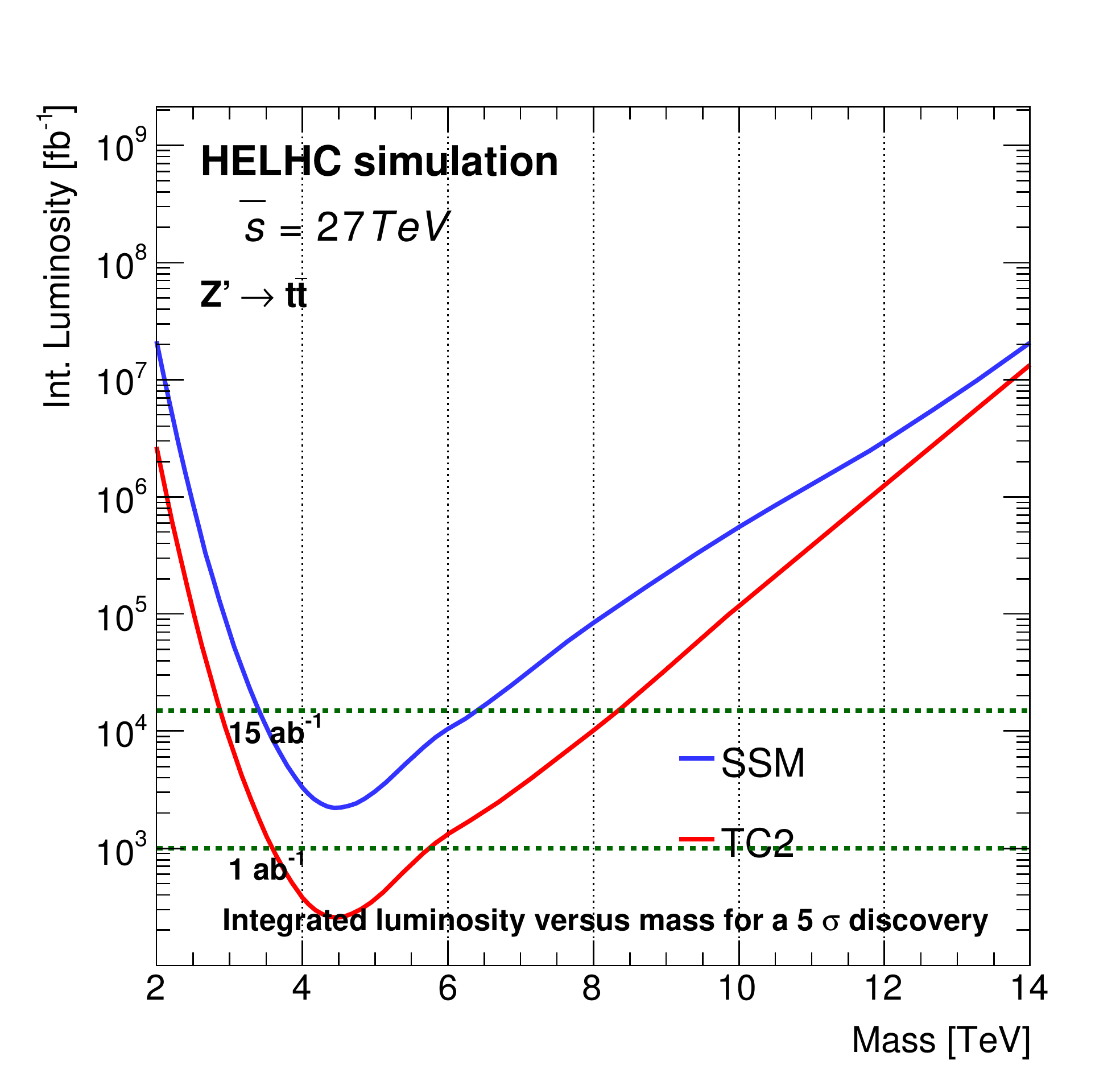}
\end{minipage}
\begin{minipage}[t][\minipageheighttp][t]{\minipagewidthtp}
\centering
\includegraphics[width=\figuremaxwidthtp,height=\figuremaxheighttp,keepaspectratio]{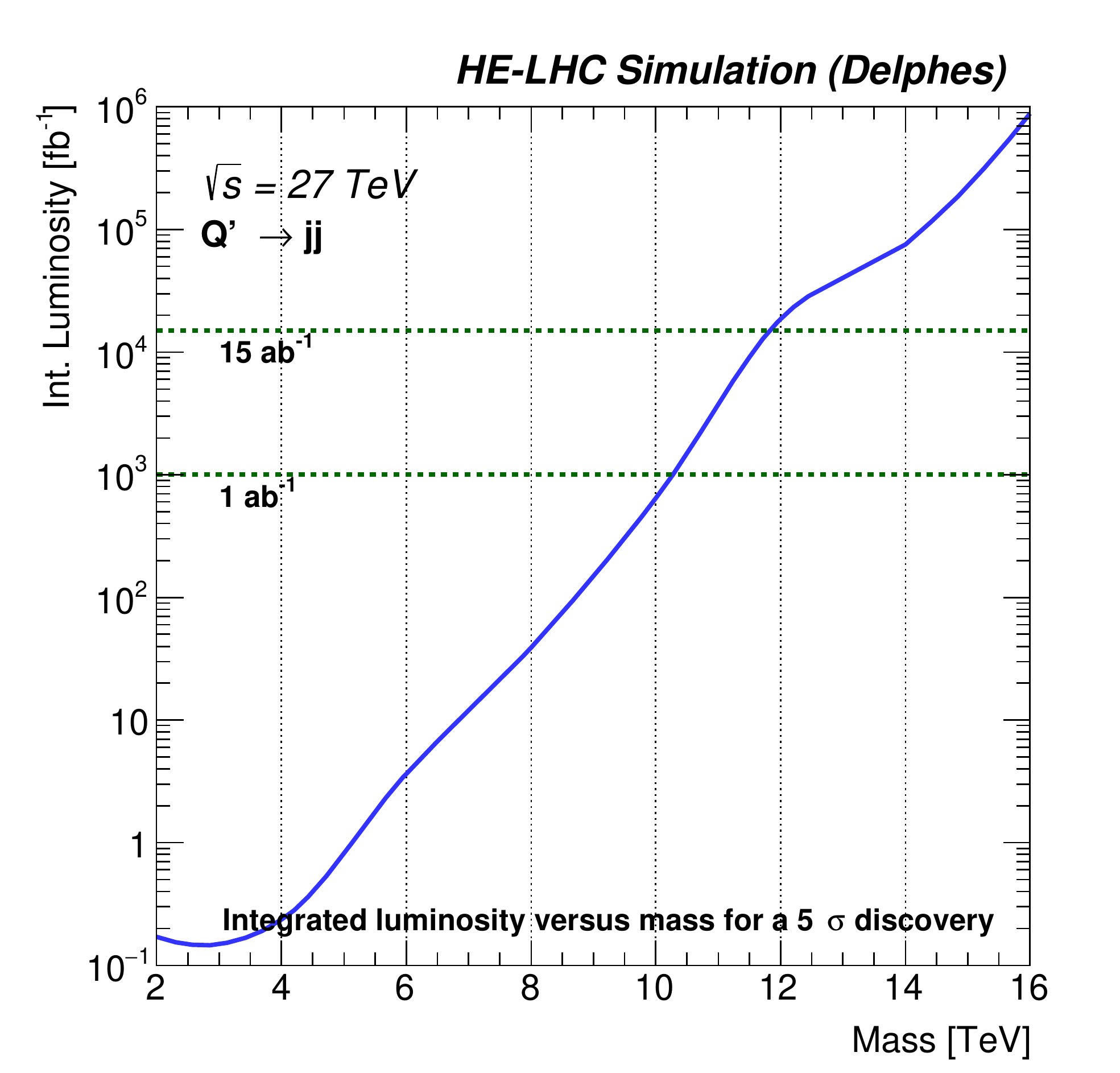}
\end{minipage}
  \caption{Top: exclusion limit at $95\%\cl$ versus heavy resonance mass for the three benchmark models: \rsg\ (left), \zptt\ (middle), and \qjj\ (right). Bottom: integrated luminosity needed for a $5\sigma$ discovery as a function of the heavy resonance mass for the three benchmark models.}
  \label{fig:hadres_limits}
\end{figure}

%%%%%%%%%%%%%%%%%%%%%%%%%%%%%%%%%%%%%%%%%%%%%%%%%%%%%%%%%%%%%%%%%%%%%%%%%%%%%%%%%%%%%%%%%%%%
%\subsubsection{Multivariate tagger}
%\paragraph*{Multivariate tagger}
\label{sec:mvatagger2}
An important ingredient of the \zptt\ and \rsg\ searches is the identification of hadronically decaying boosted tops and $W$ bosons. To this end, a tagger using jet substructure observables was developed to discriminate $W$ and top jets against QCD jets. It was found that jets using tracking only information (\emph{track-jets}) feature better angular resolution compared to pure calorimeter based jets. Therefore, track-jets are the optimal choice to build jet substructure observables.
The boosted top tagger is built from the following jet substructure observables: the soft-dropped jet mass~\cite{Larkoski:2014wba} and N-subjettiness~\cite{Thaler:2010tr} variables $\tau_{1,2,3}$ and their ratios $\tau_{2}/\tau_{1}$ and $\tau_{3}/\tau_{2}$. In addition, the $W$-jet versus QCD-jet tagger also uses an ``isolation-like'' variable that exploits the absence of high \pt\ final state-radiation (FSR) in the vicinity of the $W$ decay products. Following the strategy defined in Ref.~\cite{Mangano:2016jyj}, we call these variables \eflow\ and define them as:

\begin{equation}
\eflow =  \frac{\sum \limits_{\frac{n-1}{5}\alpha < \Delta R(k,jet)< \frac{n}{5}\alpha} \ptSup{(k)}}{\sum \limits_{\Delta R(k,jet)< \alpha} \ptSup{(k)}}
\end{equation}

We choose $\alpha=0.05$, construct 5 variables \eflow with $n=1..5$ and use them as input to the multivariate tagger. The $W$ tagging performance has significantly better performance than the top-tagging due to the use of the energy-flow variables. We choose our working points with a top and $W$ tagging efficiencies of $\epsilon_S^{\text{top}}=60\%$ and $\epsilon_S^{\text{W}}=90\%$ corresponding respectively to a background efficiency of $\epsilon_B^{\text{top}}=\epsilon_B^{\text{W}}=10\%$.

%%%%%%%%%%%%%%%%%%%%%%%%%%%%%%%%%%%%%%%%%%%%%%%%%%%%%%%%%%%%%%%%%%%%%%%%%%%%%%%%%%%%%%%%%%%%
%\subsubsection{Event Selection}
%\paragraph*{Event Selection}

The event selection proceeds as follows:  we require two jets with $\pt>1\TeV$, $|\eta|<3$ and a small rapidity gap $\Delta\eta<1.5$ between the two high \pt\ jets. For the \zptt\ and \rsg\ searches, the rapidity gap selection is relaxed to $\Delta\eta<2.4$, both jets are required to be respectively top or \sloppy\mbox{$W$-tagged}, and, to further reject background QCD jets, we require for both jets a large soft-dropped mass $m_\text{SD}>40\GeV$. Finally, for the \zptt\ search alone, we require that both selected jets must also be $b$-tagged. Since no lepton veto is applied, there is also some acceptance for leptonic decays. The sensitivity to semi-leptonic $\ttbar$ decays is enhanced by adding the $\metvec$ vector to the closest jet 4-momentum (among the two leading jets). The invariant mass of the two selected jets is used as a discriminant and is shown for the three benchmark analyses in \fig{fig:hadres_invmass}.

%%%%%%%%%%%%%%%%%%%%%%%%%%%%%%%%%%%%%%%%%%%%%%%%%%%%%%%%%%%%%%%%%%%%%%%%%%%%%%%%%%%%%%%%%%%%
%\subsubsection{Signal extraction and results}
%\paragraph*{Signal extraction and results}
Hypothesis testing is performed using a modified frequentist method based on a profile likelihood fit that takes into account the systematic uncertainties (mostly the background normalisations) as nuisance parameters. The expected exclusion limit at $95\%\cl$ and discovery reach at $5\sigma$ are shown in \fig{fig:hadres_limits} (top and bottom) for the various scenarios that have been considered. At $\sqrt{s}=27\TeV$, and with an integrated luminosity $\mathcal{L}=15\abinv$, it is possible to discover a $G_{RS}$ up to $m_{G}\approx7\TeV$, and to exclude $m_{G}\lesssim8\TeV$. For the \zptt\ search the exclusion reach is $m_{Z^{\prime}}\lesssim10(8.5)\TeV$ and it is possible to discover it up to $m_{Z^{\prime}}\approx 8(6.5)\TeV$ for the TC2 and SSM models respectively. Finally, for the excited quark model, the exclusion reach is $m_{Q^{\prime}}\lesssim14\TeV$ and the discovery reach is $m_{Q^{\prime}}\approx12\TeV$.

\subsubsection{\theoryC{On the power (spectrum) of HL/HE-LHC}}
\label{subsubsec:clockwork}
\contributors{G.~Giudice, Y.~Kats, M.~McCullough, R.~Torre, A.~Urbano}
%%%%%%%%%%%%%%%%%%%%%%%%%%%%%%%%%%%%%%%%%%%%%%%%%%%%%%%%%%%%%%%%%%%%%%%%%%%%%%%%%%%%%%%%%%%%
\begin{figure}[t!]
\centering
\begin{minipage}[t][\minipageheightdp][t]{\minipagewidthdp}
\centering
\includegraphics[width=\figuremaxwidthdp,height=\figuremaxheightdp,keepaspectratio]{\main/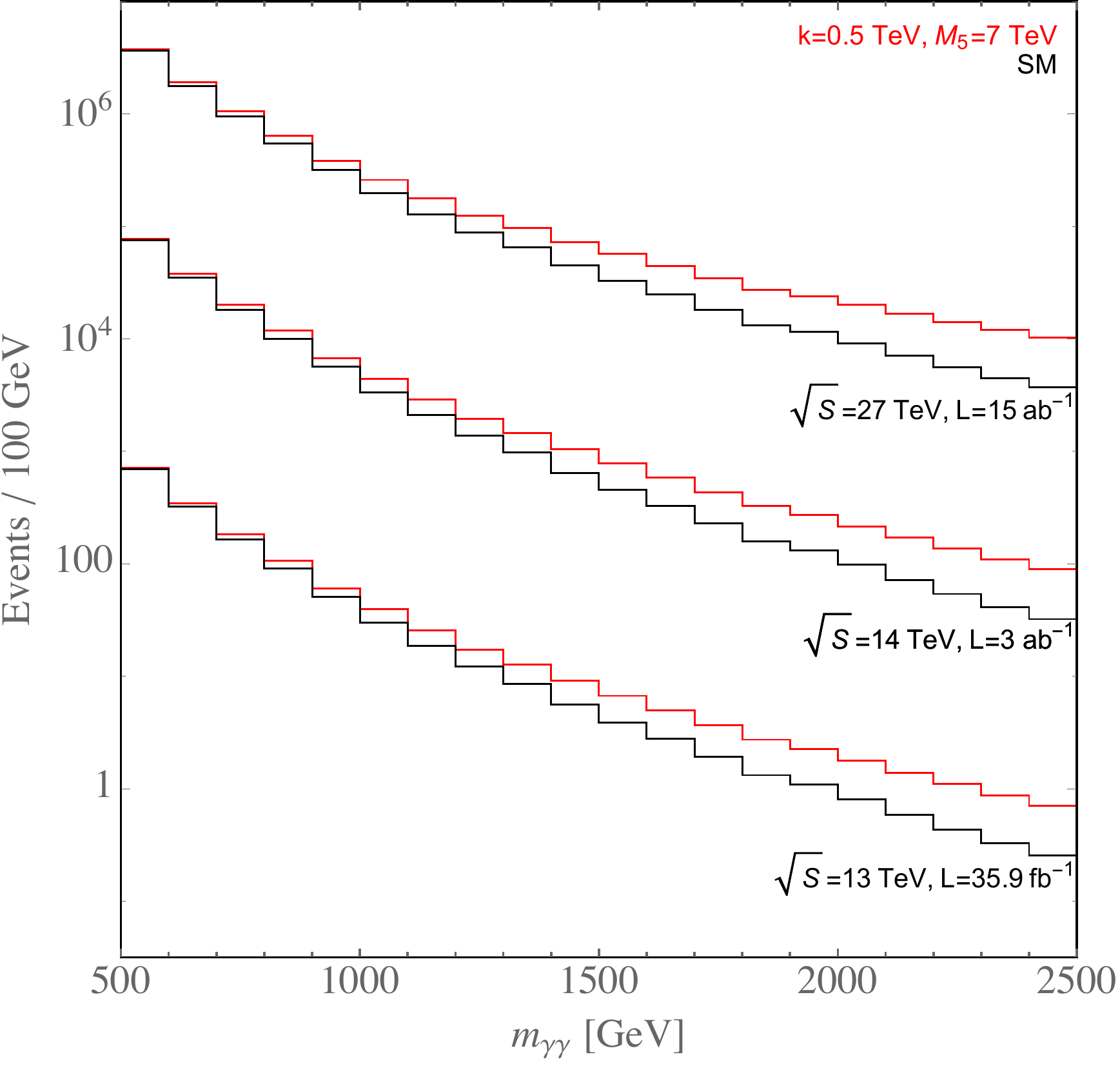}
\end{minipage}
\begin{minipage}[t][\minipageheightdp][t]{\minipagewidthdp}
\centering
\includegraphics[width=\figuremaxwidthdp,height=\figuremaxheightdp,keepaspectratio]{\main/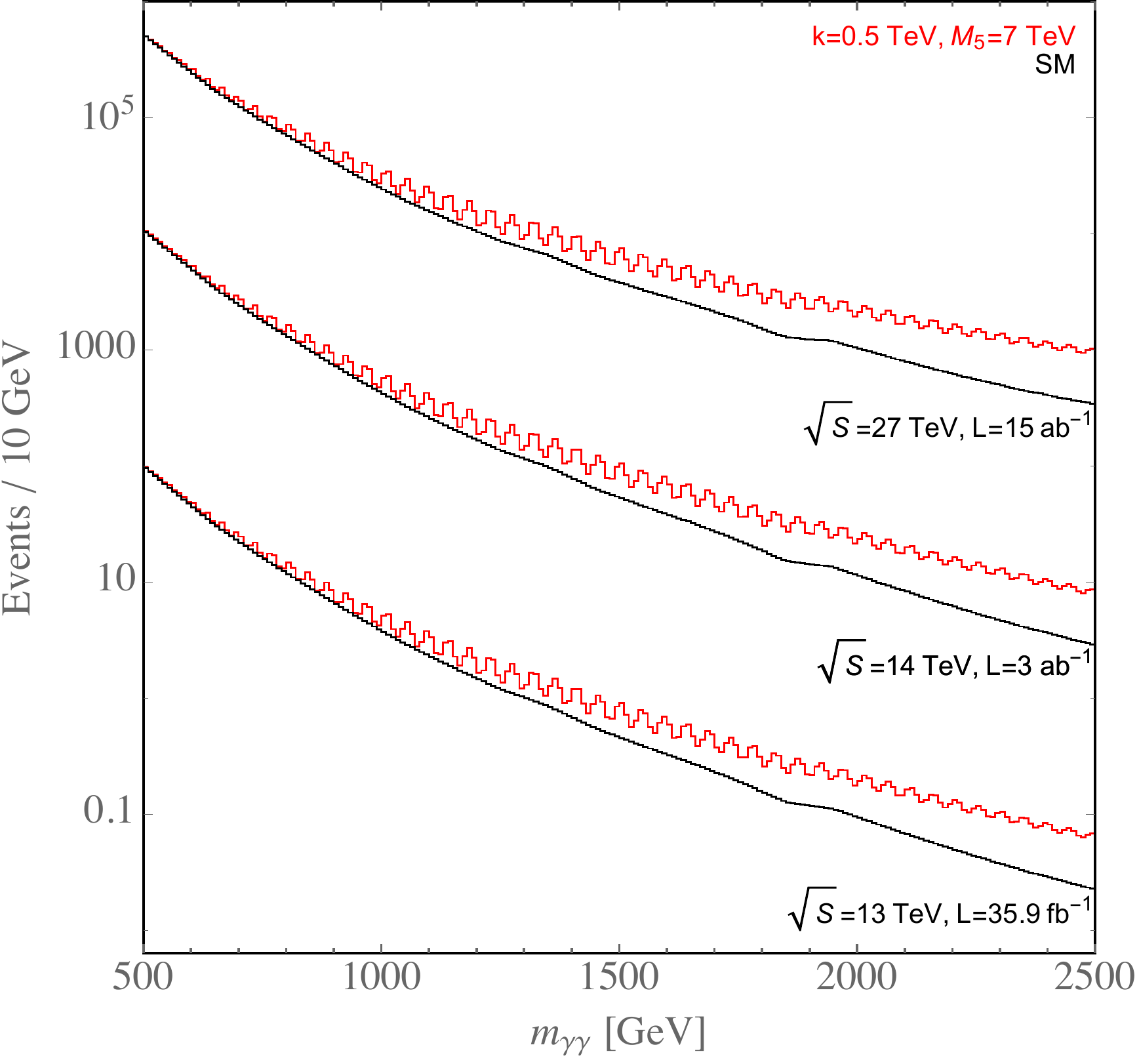}
\end{minipage}
\caption{Event rates for current integrated luminosity, HL-LHC, and HE-LHC, for the choice $k=0.5\TeV$ and $M_5 = 7\TeV$, which is a parameter point currently on the $95\%$ exclusion contour of a recent CMS analysis \cite{Sirunyan:2018wnk}.  Left: events are shown for a bin width of $100\GeV$, as in \citeref{Sirunyan:2018wnk}.  Right: events are shown for a bin width of $10\GeV$, which is close to the current experimental resolution, and reveals the oscillating nature of the signal arising due to the closely spaced resonances.}
\label{fig:spec}
\end{figure}

Typical searches for heavy new physics beyond the SM focus on particular forms of kinematic features in differential event distributions, such as isolated resonant structures, edges, and fat tails, to name a few.  However the space of theoretically consistent BSM signatures extends well beyond these classic signatures.

Recently, in \citeref{Giudice:2017fmj}, it was suggested that new physics signatures could give rise to peculiar oscillating patterns in collider data.  These signatures may be revealed as a resonance, however rather than being a resonance in the invariant mass distribution within some SM final state, this resonance would show up in Fourier space, after performing a Fourier transform on the relevant collider data.

The particular model studied in \citeref{Giudice:2017fmj}, which provides a theoretically robust example of such a signature, is the `Linear Dilaton Model' which is known to be connected to continuum limit of `Clockwork' models.  We will refer to all classes of models of this form as CW/LD.  The CW/LD models can all give rise to various forms of oscillating patterns within SM final states.  In particular, the LD model predicts closely packed resonances, starting at some threshold determined by a parameter `k' which is related to an extra-dimensional geometry.  These resonances are predicted to have mass splittings at the $\mathcal{O}(\text{few} \%)$ level.  Such mass splittings are still greater than the detector energy resolution in some high-resolution final states, such as $\gamma \gamma$ and $l^+ l^-$.

Motivated by this, in \citeref{Giudice:2017fmj} a detailed proposal for a class of collider searches in Fourier space was discussed.  In this contribution we provide a simple extension of this proposal to estimate the expected reach of the HL-LHC and HE-LHC.  The analysis is basic and has not been systematically optimised, thus the search reach based on the Fourier space analysis presented here should be considered as a preliminary estimate, however it is likely that a dedicated study could perform better.

\begin{figure}[t]
\centering
\begin{minipage}[t][\minipageheightdp][t]{\minipagewidthdp}
\centering
\includegraphics[width=\figuremaxwidthdp,height=\figuremaxheightdp,keepaspectratio]{\main/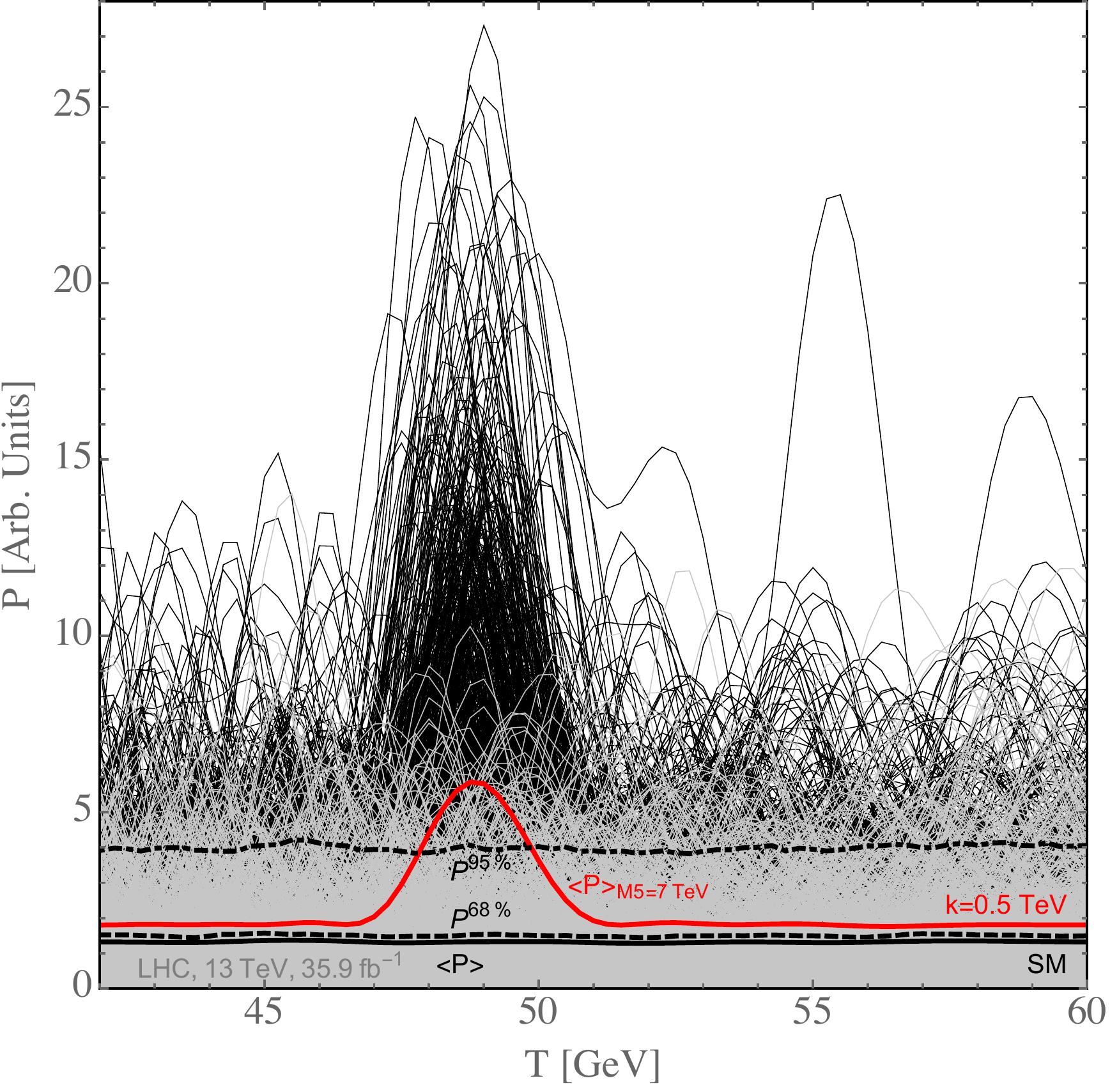}
\end{minipage}
\begin{minipage}[t][\minipageheightdp][t]{\minipagewidthdp}
\centering
\includegraphics[width=\figuremaxwidthdp,height=\figuremaxheightdp,keepaspectratio]{\main/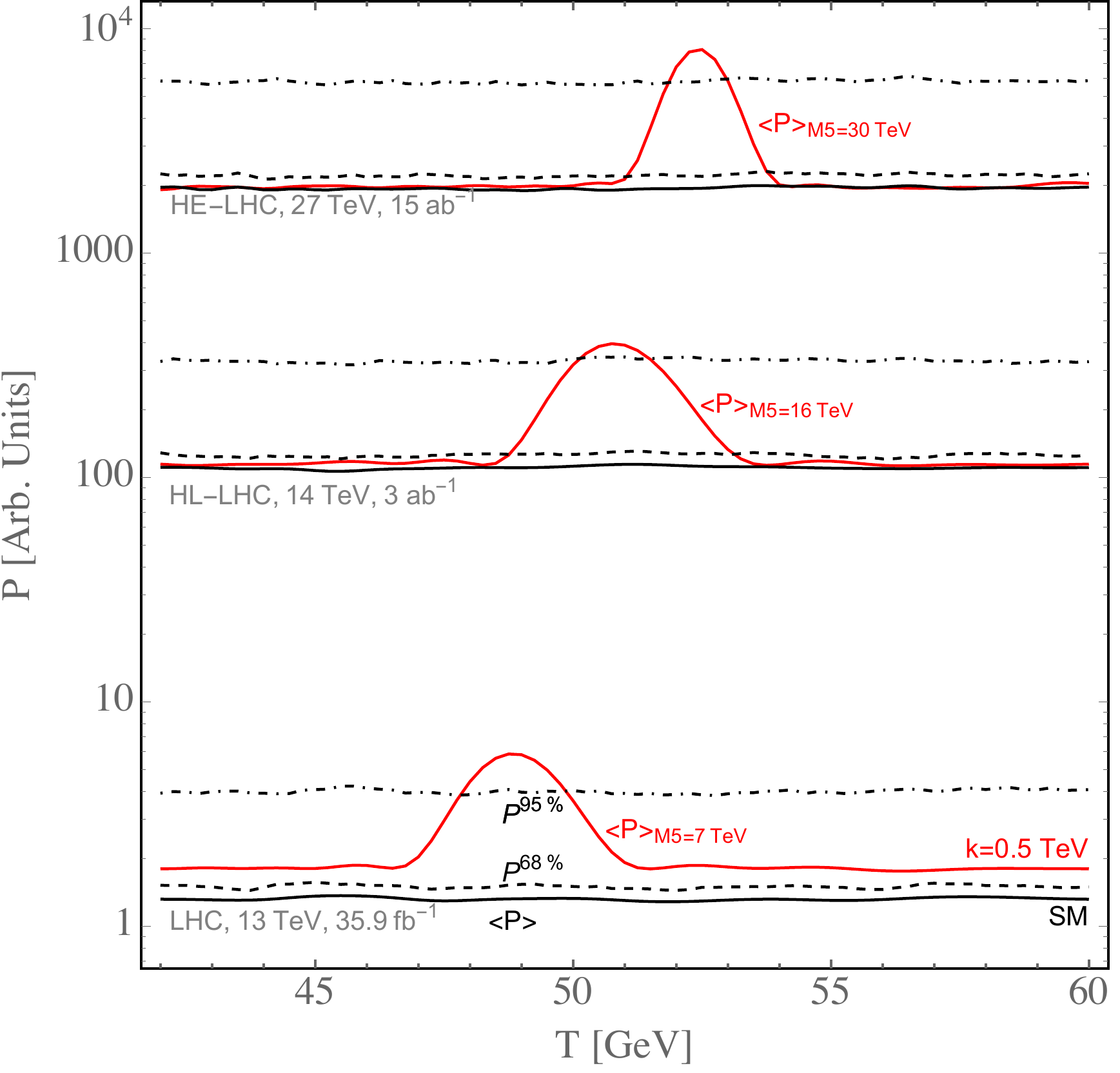}
\end{minipage}
\caption{Left: example power spectra for 1000 fake experiments at the LHC.  The individual SM lines are shown in grey, alongside the mean and confidence intervals.  The power with signal ($k=500\GeV$, $M_5=7\TeV$) included is shown in black, alongside the mean in red.
Right: the power spectrum, defined in \cite{Giudice:2017fmj}, for current integrated luminosity of $35.9\fbinv$, HL-LHC, and HE-LHC, for $k=0.5\TeV$ and a variety of values of $M_5$.  The red lines show the average power in the presence of signal and background.  The significance of the peak in the power spectrum at $T \approx 50\GeV$, which corresponds to $T=1/R$, is clearly seen for higher energy and integrated luminosity colliders.}
\label{fig:wiggleex}
\end{figure}
%\sectioneq{Analysis}
%\label{sec:analysis}

In \fig{fig:spec} we show the spectrum of events in the diphoton final state expected for two choices of binning, to reveal the oscillating nature of the signal.  The spectrum is calculated for the EBEB selections of the recent analysis \cite{Sirunyan:2018wnk}.    The SM events are taken from the NNLO prediction calculated in \citeref{Sirunyan:2018wnk}.  In \citeref{Sirunyan:2018wnk} the normalisation of the NNLO prediction is determined by fitting to the data, which is dominated by the low energy events.  To obtain the corresponding prediction at $14\TeV$ and $27\TeV$ the SM events are rescaled by the ratio of $q\overline{q}$ PDFs at the same $m_{\gamma\gamma}$, relative to $13\TeV$.  We adopt the diphoton energy resolution detailed in \citeref{Giudice:2017fmj}.

Signal events are generated and passed through the acceptance curves for the EBEB category shown in \citeref{Sirunyan:2018wnk} and we find good agreement with \citeref{Sirunyan:2018wnk} for the same signal parameters shown in Figure 5 of \citeref{Sirunyan:2018wnk}.  Similarly to the SM events, $14\TeV$ and $27\TeV$ predictions are obtained by reweighting according to the PDF ratio.

In \fig{fig:spec} the oscillatory pattern due to the multiple closely-spaced resonances is clear.  We now move to Fourier space, which is suited to searching for such oscillating patterns.  We adopt an analogous diphoton `Power' to the function detailed in \citeref{Giudice:2017fmj}, with the modification that, rather than integrating over bins, a discrete Fourier transform, which sums over bin heights at the midpoint energy, is employed.  A second modification is that rather than subtracting a fitted curve to find the residuals, we subtract the NNLO prediction detailed in \citeref{Sirunyan:2018wnk}, where the systematics related to the overall normalisation are removed by floating the distribution and fitting to the data.

The resulting power spectrum is illustrated in \fig{fig:wiggleex}. 
%\figs{fig:wiggleex} and \ref{fig:wiggle}.  
Figure \ref{fig:wiggleex} (right) was obtained by generating 5000 sets of fake experiments, where the number of events in each bin is generated from a Poisson distribution, for both the SM only and SM + signal hypotheses, before subtracting the SM prediction.  The curves for $\langle P \rangle$, $P^{68\%}$, and $P^{95\%}$ represent the mean value of $P(T)$ at a given value of $T$, over all fake experiments, and also the corresponding upper limits at a given~\cl derived from the distributions.  The average value, when signal is included, is also shown for a variety of $M_5$ values.

%\begin{figure}[t]
%\centering
%\includegraphics[width=0.5\textwidth]{\main/section6OtherSignatures/img/SummaryPlot.pdf}
%\caption{The power spectrum, defined in \cite{Giudice:2017fmj}, for current integrated luminosity of $35.9\fbinv$, HL-LHC, and HE-LHC, for $k=0.5$ TeV and a variety of values of $M_5$.  The red lines show the average power in the presence of signal and background.  The significance of the peak in the power spectrum at $T \approx 50$ GeV, which corresponds to $T=1/R$, is clearly seen for higher energy and integrated luminosity colliders.}
%\label{fig:wiggle}
%\end{figure}

In each case, the peak in Fourier space is clearly visible and approaches the $P^{95\%}$ line.  We will not attempt a robust statistical analysis here, which would require a thorough treatment of systematics in a likelihood ratio test.  Rather, we will focus on the comparison between different colliders.  Clearly, if the current LHC data can exclude a parameter point in the ballpark of $k=0.5\TeV$ and $M_5 \approx 7\TeV$, then at the HL-LHC one should be sensitive to $M_5 \approx 16\TeV$, and at the HE-LHC $M_5 \approx 30\TeV$.  Furthermore, since the statistics are much higher at these colliders one would presumably benefit from widening the range of $m_{\gamma\gamma}$ considered, thus we expect the reach to exceed these values in a dedicated analysis.

%\section{Summary}
To summarise, in this contribution we have estimated the reach of the HL-LHC and HE-LHC for detecting oscillating patterns in BSM contributions to the differential diphoton spectrum.  Such searches have not yet been performed at the LHC and they provide a new target in the search for BSM physics.  This estimate is performed in the context of the linear dilaton model, as described in \citeref{Giudice:2017fmj}.

We have found that for a warping parameter of $k=500\GeV$, which sets the mass scale of the first KK graviton resonance, if current limits on the extra-dimensional Planck scale are at the level of $M_5 \approx 7\TeV$, then with HL-LHC they should extend considerably, to around $M_5 \approx 16\TeV$, and at HE-LHC further still, to $M_5 \approx 30\TeV$.

%%%%%%%%%%%%%%%%%%%%%%%%%%%%%%%%%%%%%

\clearpage
\resetcounters

\section{Conclusions and Outlook}\label{sec:conclusions}

The LHC is performing superbly and has already moved on from discovering the Higgs boson to accurately measuring its properties.  As yet there are no clear deviations from the SM, in Higgs measurements or elsewhere.  Nonetheless, with only a fraction of the $150\fbinv$ recorded data so far analysed, it is possible that surprises await, or that some of the existing small deviations grow.  It is also possible that the new physics has small production cross sections either because it is weakly coupled to the initial state at the LHC, or that it is heavy, or both.  Weakly coupled states within kinematic reach of the LHC may require increased data sets to uncover, such as the $3\abinv$ of the HL-LHC each for ATLAS and CMS, and up to $300\fbinv$ for the Upgrade II of LHCb.  Heavy states instead might require both increased centre-of-mass energy and increased luminosity, such as the $27\TeV$ and $15\abinv$ of the HE-LHC.

The HL- and HE-LHC will both offer new possibilities to test many BSM scenarios, motivated by long-standing problems such as EW Naturalness, dark matter, the flavour problem, neutrino masses, the strong CP problem, and baryogenesis. All these new physics manifestations predict the existence of new particles, which can be searched for at HL-LHC, profiting from the much larger statistics and slightly higher energy, and at the HE-LHC, profiting from the much larger statistics and much higher energy.  In both cases the searches will also benefit from detector upgrades. Careful attention has been paid to the impact of these upgrades and the assessment of the future systematic uncertainties, especially for HL-LHC where concrete proposals for the ATLAS, CMS and LHCb experiments exist. 

\begin{figure}[t]
\centering
\includegraphics[width=\textwidth]{\main/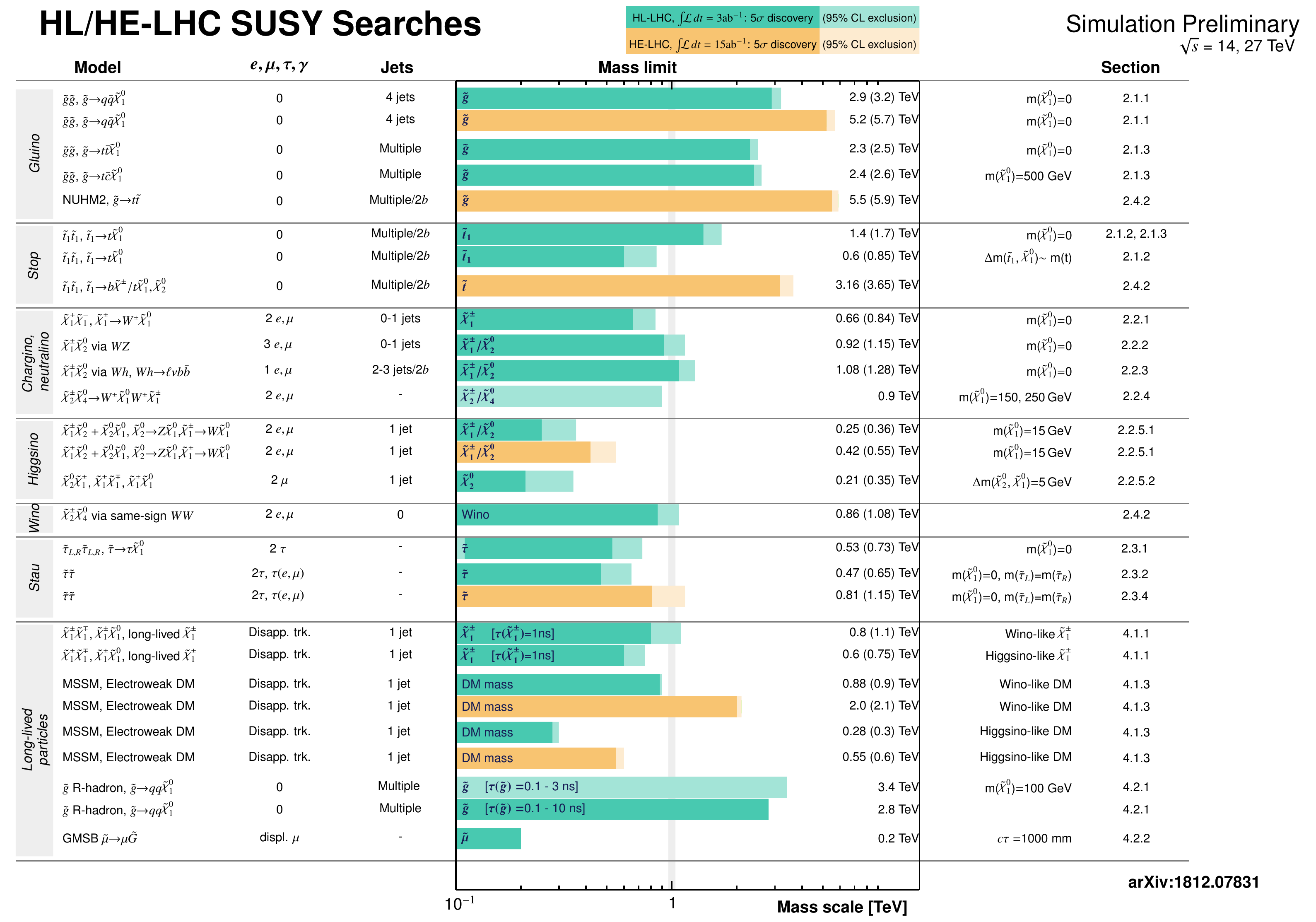}
\caption{A summary of the expected mass reach for $5\sigma$ discovery and $95\%\cl$ exclusion at the HL/HE-LHC, as presented in \secc{sec:SUSY}.}
\label{fig:SUSYsummaryplot}
\end{figure}

\begin{figure}[t]
\centering
\includegraphics[width=\textwidth]{\main/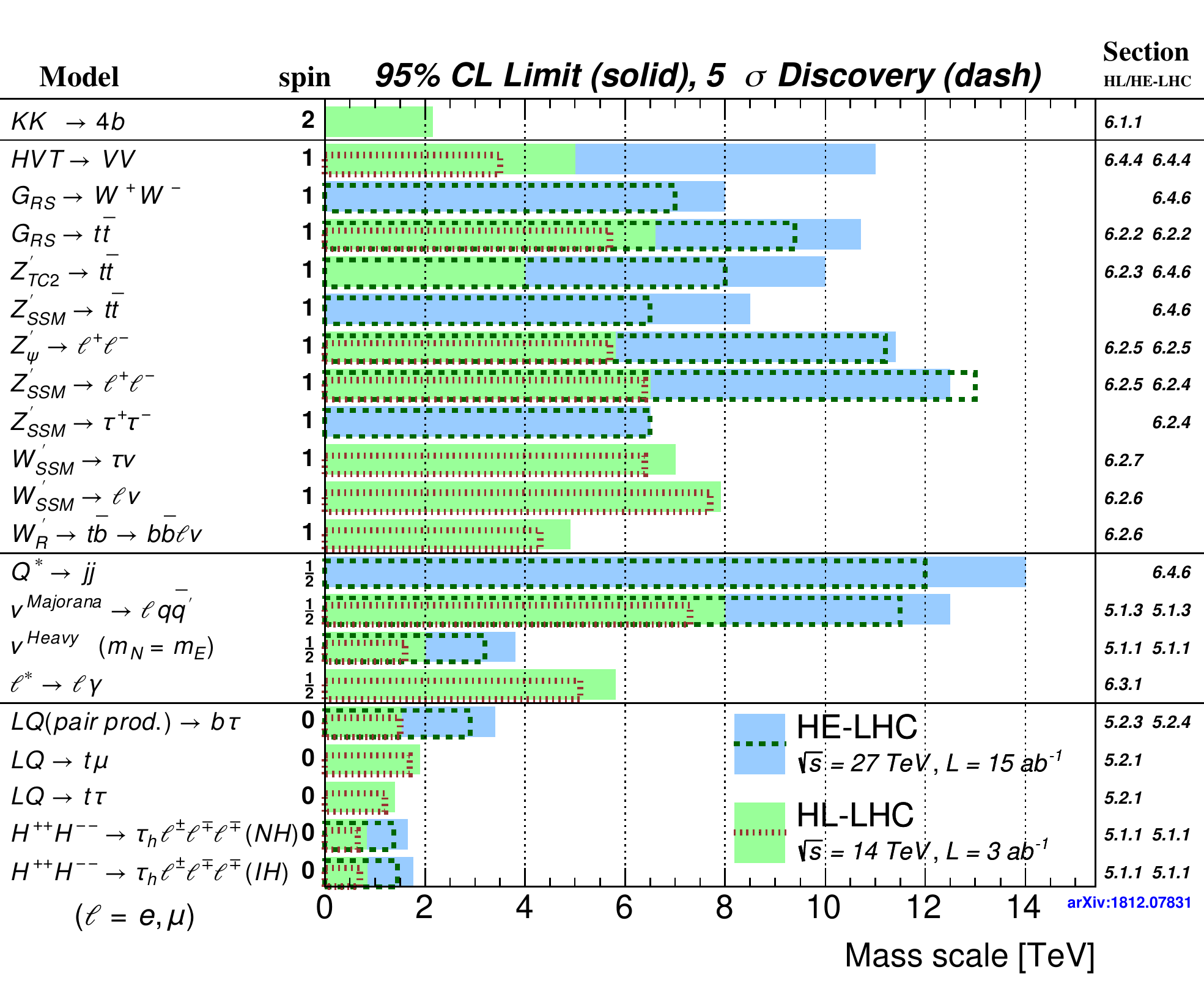}
\caption{A summary of the expected mass reach for $5\sigma$ discovery and $95\%\cl$ exclusion at the HL/HE-LHC, as presented in \seccs{sec:flavor} and \ref{sec:otherBSM}.}
\label{fig:BSMsummaryplot}
\end{figure}

This document contains numerous studies of the increased reach for many BSM scenarios that can be explored at the HL- and HE-LHC options.  As well as determining the reach in sensitivity, several studies analyse the ability for HL- and HE-LHC to characterise potential discoveries and distinguish between possible BSM explanations.  Rather than give a complete summary of all prospect analyses in this report, we instead provide representative examples of the scope of the BSM searches for the \sloppy\mbox{HL-LHC} and HE-LHC.  Broadly speaking, in most BSM scenarios, we expect the HL-LHC will increase the present reach in mass and coupling by $20-50\%$ and potentially discover new physics that is currently unconstrained.  The reach of the HE-LHC is generically more than double that of the HL-LHC. In \fig{fig:SUSYsummaryplot} we present a summary of the results of SUSY searches (both prompt and long-lived) for sparticles produced via strong- or electroweak interactions, as described in detail in \secc{sec:SUSY}. Figure \ref{fig:BSMsummaryplot} shows a summary of most of the resonance searches (both single and pair production) presented in \seccs{sec:flavor} and \ref{sec:otherBSM}.

\subsection*{HL-LHC}

%We highlight a subset of key results for prospects at the HL-LHC, selected among a large number of studies for different new physics scenarios. 
%All quoted exclusion (discovery) reaches refer to 95\% CL ($5\sigma$). 

%\paragraph*{Supersymmetry}
\textbf{Supersymmetry} 

The extension of the kinematic reach for supersymmetry searches at the HL-LHC is reflected foremost in the sensitivity to EW states, including sleptons, but also for gluinos and squarks. Studies under various hypothesis were made, including prompt and long-lived SUSY particle decays. 
Wino-like chargino pair production processes are studied considering dilepton final states. Masses up to \sloppy\mbox{$840$ ($660$)\GeV} can be excluded (discovered) for charginos decaying as $\tilde{\chi}_{1}^{\pm} \rightarrow W^{(*)} \neut$, in R-parity conserving scenarios with $\neut$ as the lightest supersymmetric particle. The results extend by about \sloppy\mbox{$500\GeV$} the mass reach obtained with $80\fbinv$ of $13\TeV$ $pp$ collisions, and extend beyond the LEP limit by almost an order of magnitude in case of massless neutralinos.
Compressed SUSY spectra are theoretically well motivated but are among the most challenging scenarios experimentally, and are barely covered by the Run-2 analyses. 
HL-LHC searches for low momentum leptons will be sensitive to $\tilde{\chi}^\pm$ masses up to $350\GeV$ for $\Delta m(\tilde{\chi}_{1}^{\pm}, \neut) \approx 5\GeV$, and to mass splittings between $0.8$ and $50\GeV$, thus bringing significant new reach to Higgsino models. Similar search techniques can also be used to target pair produced $\tilde{e}$ and $\tilde{\mu}$ in compressed scenarios. 
If $\Delta m(\tilde{\chi}_{1}^{\pm}, \neut)<1\GeV$, charginos can decay after the inner layers of the pixel detectors. 

Dedicated searches for sleptons, characterised by the presence of at least one hadronically-decaying $\tau$ and missing $E_T$, will be sensitive to \emph{currently unconstrained} pair-produced $\tilde{\tau}$: exclusion (discovery) for $m_{\tilde{\tau}}$ up to around $700$ ($500$)\GeV can be achieved under realistic assumptions of performance and systematic uncertainties.%, as shown in~\fig{fig:Disappearing} (right).
%, with a margin for further improvements using shape-fit and data-driven methods to evaluate the SM backgrounds. 

In the strong SUSY sector, HL-LHC will probe gluino masses up to 
$3.2\TeV$, with discovery reach around $3\TeV$, in R-parity conserving scenarios and under a variety of assumptions on the $\tilde{g}$ prompt decay mode. This is about $0.8-1\TeV$ above the Run-2 $\tilde{g}$ mass reach for $80\fbinv$. 
Pair-production of top squarks has been studied assuming $\tone\to t \neut$ and fully hadronic final states with large missing $E_T$. 
Top squarks can be discovered (excluded) up to masses of $1.25$ ($1.7$)\TeV for massless neutralinos, \iee{} $\Dmstop \gg m_{t}$, under realistic uncertainty assumptions. This extends by about $700\GeV$ the reach of Run-2 for $80\fbinv$. The reach in $m_{\tilde{t}}$ degrades for larger $\neut$ masses. If $\Dmstop \sim m_{t}$, the discovery (exclusion) reach is $650$ ($850$)\GeV. \\

%\paragraph*{Dark Matter and Dark Sectors}
\textbf{Dark Matter and Dark Sectors} 

Compressed SUSY scenarios, as well as other DM models, can be targeted using signatures such as mono-jet, mono-photon and vector-boson-fusion production. Mono-photon and VBF events allow targeting an EW fermionic triplet (minimal DM), equivalent to a wino-like signature in SUSY, for which there is no sensitivity in Run-2 searches with $36\fbinv$. Masses of the $\neut$ up to $310$ ($130$)\GeV can be excluded by the mono-photon (VBF) channel, with improvements possible, reducing the theoretical uncertainties. Projections for searches for a mono-$Z$ signature,
with $Z\to\ell^+\ell^{-}$ recoiling against missing $E_T$, have been interpreted in terms of models with a spin-1 mediator, and models with two Higgs doublets and an additional pseudoscalar mediator $a$ coupling to DM (2HDMa). The exclusion is expected for mediator masses up to $1.5\TeV$, and for DM and pseudoscalar masses up to $600\GeV$,  a factor of $\sim 3$ better than the $36\fbinv$ Run-2 constraints. 
The case of 2HDMa models is complemented by 4-top final states, searched in events with two same-charge leptons, or with at least three leptons. While searches using $36\fbinv$ Run-2 data have limited sensitivity considering the most favourable signal scenarios (\egg~$\tan \beta = 0.5$), HL-LHC will probe possible evidence of a signal with $\tan \beta = 1$, $m_{H}=600\GeV$and mixing angle of $\sin \theta=0.35$, assuming $m_{a}$ masses between $400\GeV$ and $1\TeV$, and will allow exclusion for all $200\GeV<m_{a}<1\TeV$.
For DM produced in association with bottom or top quarks, where a (pseudo)scalar mediator decays to a DM pair, the HL-LHC will improve the sensitivity to mediator masses by a factor of $3-8$ relative to the Run-2 searches with $36\fbinv$.

%{\color{red} %% DARK PHOTON BIT
A compelling scenario in the search for portals between the visible and dark sectors is that of the dark photon $A'$. 
%As shown in~\fig{fig:DarkPhoton}, 
Prospects for an inclusive search for dark photons decaying into muon or electron pairs indicate that the HL-LHC could cover a large fraction of the theoretically favoured $\epsilon - m_{A'}$ space, where $\epsilon$ is the kinetic mixing between the photon and the dark photon and $m_{A'}$ the dark photon mass. \\
 
%\paragraph*{Resonances} 

\textbf{Resonances} 

Several studies of resonance searches,  
in a variety of final states, have been performed and were presented here. A right-handed gauge boson with SM couplings, decaying as $W_R \to bt(\to b\ell\nu)$, can be excluded (discovered) for masses up to $4.9$ ($4.3$)\TeV, %5.7~TeV. %pf: old number? see sec 6.2.6
%and discovered at $5\sigma$ for masses up to 4.3 TeV.
$1.8\TeV$ larger than the $36\fbinv$ Run-2 result. %2.5~TeV. %pf: old again? 
For a sequential SM $W'$ boson in $\ell \nu$ final states ($\ell=e,\mu$), the mass reach improves by more than $2\TeV$ w.r.t. Run-2 ($80\fbinv$) reach, and by more than $1\TeV$ w.r.t. $300\fbinv$. 
The HL-LHC bound will be $M_{W'}=7.9\TeV$, with discovery potential up to $M_{W'}=7.7\TeV$.
Projections for searches of SSM $Z'$ bosons in the dilepton %and $t\bar{t}$ 
final state
predict exclusion (discovery) up to masses of $6.5$ ($6.4$)\TeV. The $36\fbinv$ Run-2 exclusion is $4.5\TeV$, expected to grow to $5.4\TeV$ after $300\fbinv$. %(Fig.~\ref{fig:otherbsmfig})
Using top-tagging techniques, a Randall--Sundrum Kaluza--Klein gluon decaying to $t\bar{t}$ is expected to be excluded up to $6.6\TeV$ and discovered up to $5.7\TeV$, extending by over $2\TeV$ the $36\fbinv$ bounds.

Models related to the apparent flavour anomalies in $B$ decays suggest the presence of heavy resonances, either $Z'$ or leptoquarks (LQ), coupling to second and/or third generation SM fermions.
The HL-LHC will be able to cover a significant portion of the parameter space allowed by flavour constraints, with an exclusion reach up to $4\TeV$ for the $Z'$, 
depending on the structure and size of the $Z'$ couplings. Pair produced scalar LQs coupling to $\mu$ ($\tau$) and b-quarks, on the other hand, can be excluded up to masses of $2.5$ ($1.5$)\TeV, depending on assumptions on couplings. Finally, prospect studies for third generation LQ in the $t\mu$ and $t\tau$ channels  deliver mass limits (discovery potential) increased by \sloppy\mbox{$500$ ($400$)\GeV} with respect to $36\fbinv$, with prospect discovery in the $t\mu$ channel up to $1.7\TeV$. \\

%\paragraph*{Long-lived particles}
\textbf{Long-lived particles} 

In addition to the significant expansion of expected luminosity, new detector upgrades will enable searches in the long-lived particle regime. Muons displaced from the beamline, such as found in SUSY models with $\tilde{\mu}$ lifetimes of $c\tau > 25\cmeters$, can be excluded at $95\%\cl$. New fast timing detectors will also be sensitive to displaced photon signatures arising from LLP in the $0.1 < c\tau < 300\cmeters$ range. 

Prospect studies for disappearing tracks searches using simplified models of $\tilde{\chi}^\pm$ production %which include $\tilde{\chi}_{1}^{\pm} \neut$ and $\tilde{\chi}_{1}^{\pm} \tilde{\chi}_{2}^{0}$ 
lead to exclusions of chargino masses up to $m(\tilde{\chi}_{1}^{\pm}) = 750\GeV$ ($1100\GeV$) for lifetimes of $1\nsec$ for the higgsino (wino) hypothesis. When considering the lifetime predicted by theory, masses up to $300\GeV$ and $830\GeV$ can be excluded in higgsino and wino models, respectively. This improves the $36\fbinv$ Run-2 mass reach by a factor of 2-3. The discovery reach is reduced to $160\GeV$ and $500\GeV$ respectively, due to the loss in acceptance at low lifetime ($0.2\nsec$), but sensitivity is expected to be recovered with dedicated optimisations. 
%Results are shown in~\fig{fig:Disappearing} (left).

Several studies are available also for long-lived $\tilde{g}$. As an example, we expect a $1\TeV$ extension of the $36\fbinv$ Run-2 mass reach, for models with $\tilde{g}$ lifetimes $\tau>0.1\nsec$, and an exclusion of $m_{\tilde g}$ up to $3.4-3.5\TeV$. 
Finally, the signature of long-lived dark photons decaying to displaced muons can be reconstructed with dedicated algorithms and is sensitive to very small coupling $\epsilon^2 \sim 10^{-14}$ for masses of the dark photons between $10$ and $35\GeV$. %~\cite{CMS-PAS-FTR-18-002}
%} 
Complementarities in long-lived particle searches and enhancements in sensitivity might be achieved if new proposals of detectors and experiments such as Mathusla, FASER, Codex-B, MilliQan and LHeC are realised in parallel to the HL-LHC.

\subsection*{HE-LHC}
%The HE-LHC is expected to extend the mass reach for the discovery of new particles by a factor of $\sim$2 with respect to HL-LHC. While the study of individual scenarios must account in detail for the possibly different evolution of signals and backgrounds with beam energy and include the new analysis opportunities offered by the larger statistics and kinematic reach available at 27~TeV, it is possible to provide general estimates of the improved sensitivity by extrapolating the partonic luminosities that are relevant for the production of various final states. The improvement in mass reach is rather independent of the specific type of particle(s) produced and only depends on the estimated reach at the HL-LHC. For example, new gauge bosons such as a $Z^{'}$ , whose reach at the HL-LHC is estimated to be in the range of 6~TeV, could be observed by the HE-LHC up to a mass of $\sim$12~TeV. This qualitative conclusion is verified in the more detailed studies presented in this report. In several BSM scenarios the extension of the reach by a factor of 2 relative to the HL-LHC might be regarded as an extra dent into a large range of possible masses. Nonetheless there are interesting new physics scenarios doubling the reach allow to cover an important fraction of the relevant parameter space. A subset of concrete examples are highlighted here, selected among the large number of studies presented in this document.

\textbf{Supersymmetry} 

The increase in energy from $14\TeV$ to $27\TeV$ leads to a large increase in the production cross section of heavy coloured states, a $3.5\TeV$ gluino has nearly a 400-fold increase in production cross section.  For supersymmetric spectra without compression the HE-LHC has sensitivity to gluinos  up to masses of $6\TeV$ and discovery potential of $5.5\TeV$.  The corresponding numbers for stop squarks are $3.5\TeV$ and $3\TeV$, respectively.  
These results allow to put in perspective the question to what extent classes of ``natural'' supersymmetric models are within the reach of HE-LHC and can be discovered or excluded conclusively. Examples of specific model scenarios were studied during the Workshop: while HL-LHC can only cover part of the
parameter space of the models considered, HE-LHC covers it entirely. 
The HE-LHC would allow for a $5\sigma$ discovery of most natural SUSY models via the observation of both gluinos and stops. 

If the coloured states are close in mass to the lightest supersymmetric particle (LSP), the amount of missing transverse momentum ($E_{\mathrm{T}}^{\mathrm{miss}}$) in the event is decreased. 
 The typical multijet + $E_{\mathrm{T}}^{\mathrm{miss}}$ SUSY searches are less sensitive and must be replaced with monojet-like analyses. Prospect studies show that if for example the gluino-LSP mass splitting is held at $10\GeV$, gluino masses can be excluded up to $2.6\TeV$. If the lightest coloured state is the stop, and the $\tilde{t}$-LSP splitting is such that final states include very off-shell $W$ and $b$-jets, $\tilde{t}$ masses up to about $1\TeV$ could be excluded, extending the HL-LHC reach by about a factor of two. 

The electroweakino sector of supersymmetry presents a particular challenge for hadronic machines.  If the LSP is pure higgsino or wino, there is a very small neutralino-chargino mass splitting ($\sim340\MeV$, $\sim160\MeV$ respectively) and the chargino has a correspondingly long lifetime ($c\tau \sim 5, 1\cmeters$ respectively).  The $E_{\mathrm{T}}^{\mathrm{miss}}$ is again small unless the pair produced electroweakinos recoil against an ISR jet.  Taking into account contributions from both chargino and neutralino production the monojet search will deliver a sensitivity for exclusion (discovery) of winos up to $\sim 600\GeV$ ($300\GeV$) and higgsinos up to $\sim 400\GeV$ ($150\GeV$). Taking advantage of the long lifetime of the charginos allows searches to be done for disappearing charged tracks. Considering a detector similar to the ones available for HL-LHC, winos below $\sim 1800\GeV$ ($1500\GeV$) can be excluded (discovered), while the equivalent masses for Higgsinos are $\sim500\GeV$ ($450\GeV$). 

While these results come short of covering the full range of masses for electroweakinos to be a thermal relic and account for all of DM, the mass range accessible to HE-LHC greatly extends the HL-LHC potential and can be complementary to the indirect detection probes using gamma rays from dwarf-spheroidal galaxies. \\

\textbf{Dark Matter} 

Monojet searches, as well as monophoton and vector-boson-fusion  production searches, might be sensitive to generic weakly interactive dark matter candidates beyond compressed SUSY scenarios. 
Analyses of the reach of the HE-LHC under various assumptions of the structure of the DM-SM coupling have been carried out. Models characterised by the presence of an extended Higgs sector, with Higgs doublets mixing with an additional scalar or pseudoscalar mediator that couples to DM have been studied assuming associate production of DM with a pair of top quarks.

Using leptonic decays of the tops, a fit to the distribution of the opening angle between the two leptons can help distinguish signal from SM processes, after other background-suppressing selections.  Assuming the DM is lighter than half the mediator mass, a scalar or pseudoscalar mediator can be ruled out at $95\%\cl$ up to $900\GeV$ using this technique, a factor of 2 higher in mass compared to the HL-LHC bounds.

If a dark sector exists and contains heavy coloured particles, $\mathcal{Q}$, nearly degenerate with the DM and decaying to DM and SM coloured particles, a monojet topology could be again the most sensitive. The rate and shape of the monojet distribution depend upon the mass, spin and colour representation of $\mathcal{Q}$.  Fermionic colour triplet $\mathcal{Q}$ could be discovered up to $1.1\TeV$ at the HE-LHC, a fermionic octet will be ruled out if lighter than $1.8\TeV$, and a scalar colour triplet almost degenerate in mass to the DM could be probed up to masses of $600\GeV$.  Should an excess be observed identifying the spin and colour representation, NNLO precision will be needed for the predictions of the SM backgrounds. Analyses of double-ratios of cross sections at varying $p_T$ could be utilised to partially cancel uncertainties. \\

\textbf{Resonances} 

Searches for heavy resonances will greatly benefit from the increased partonic energy of a $27\TeV$ machine.  If the resonance is narrow the search can be data driven but if the resonance is wide it will require a precise understanding of the backgrounds, which presents a formidable challenge.
The HE-LHC can expect to approximately double the mass reach on dijet resonances of the HL-LHC.  For instance, the reach for an excited quark decaying to dijets will be $14\TeV$.

Dilepton resonances, \egg~$Z'$, are present in many gauge extensions of the SM.  The exact reach in any model depends upon the coupling to light quarks and BR to charged leptons.  A $Z'$ whose couplings to SM quarks and leptons are as in the SM, a so called sequential $Z'_{SSM}$, will be discoverable in the dilepton channel up to $\sim 13\TeV$.  In the di-tau channel the reconstruction is harder, reducing the reach to $\sim 6 \TeV$.  In the di-top channel a $Z'_{SSM}$ will be discovered up to $6\TeV$ and excluded up to $8\TeV$.  It is an interesting question to ask: if a resonance is discovered can we determine the nature of the model?  Considering a $6\TeV$ $Z'$ decaying into $e, \mu, t, b, q$ final states, it is sufficient to consider three observables ($\sigma\cdot B$, the forward-backward asymmetry, and the rapidity asymmetry) to distinguish between six $Z'$ models in most cases.

Gauge extensions of the SM also often contain new heavy charged vector bosons, $W'$.  One promising search channel for these new particles is the $WZ$ final state.  More generally BSM models can contain resonances decaying to di-bosons $WZ/WW$.  Searching for a RS graviton $G_{RS}$ resonance in $WW$ final states exploiting the all-hadronic signature presents a considerable challenge and requires dedicated identification of boosted $W$-jets.  The estimated reach is $\sim 8\TeV$, almost a factor of two increase with respect to the HL-LHC.  Diboson searches in $WW\rightarrow \ell \nu qq$ final states are sensitive also to heavy vector triplet (HVT) model produced via $\mathrm{ggF/q\bar{q}}$. HVT $Z^{'}$ masses up to $11\TeV$ can be excluded by HE-LHC, extending the HL-LHC reach by about $6\TeV$. 

Searches for top partners decaying to a top and a $W$ boson in the same-sign dilepton signature can lead to discovery up to more than $2\TeV$, with the possibility to discriminate left- and right-handed couplings at the $2\sigma$ level in all the accessible discovery range. 

Models related to the generation of the observed pattern of neutrino masses and mixings offer a variety of signatures that could profit from the high energy available at HE-LHC. For instance, searches for lepton flavour violating (LFV) final states with two opposite-sign different-flavour leptons and two jets arising from the production of a heavy pseudo-Dirac neutrino and a SM lepton could test heavy neutrino masses in seesaw models up to more than $1\TeV$. 
In left-right symmetric models, a right-handed heavy $W_R$ boson decaying to a Majorana neutrinos can be discovered with $5\abinv$ up to $\sim 10.5\TeV$ and excluded up to masses of around $\sim 16\TeV$ with an integrated luminosity of $15\abinv$. This increases the reach of HL-LHC by about $10\TeV$.  Doubly charged scalars in type-II seesaw models, can also be searched for in multi-lepton signatures, with a reach in the $1.5\TeV$ region. Heavy leptons in type-III seesaw models, giving rise to final states with two leptons, jets and $b$-jets can be excluded (discovered) up to $3.8$ ($3.2$)\TeV. For comparison, the exclusion reach for the HL-LHC is about $1.6\TeV$. 

Finally, models related to observed flavour anomalies in neutral-current $B$ decays, suggest the presence of heavy resonances, either $Z'$ or leptoquarks, coupling to muons and $b$-quarks. The \sloppy\mbox{HE-LHC} will be able to cover an ample region of the parameter space allowed by flavour constraints, with an exclusion (discovery) reach up to $10$ ($7$)\TeV for the $Z'$, depending on the assumptions on the $Z'$ couplings structure and size. Pair produced leptoquarks, on the other hand, can be excluded up to masses of $4.2\TeV$, again depending on assumptions on couplings. These masses are typically a factor of $2$ to $2.5$ higher than the HL-LHC sensitivity.

\clearpage
\resetcounters
\section{Acknowledgements}
%R.~Torre and P.~Fox thank Dropbox, but not git.
We would like to thank the LHC experimental Collaborations and the WLCG for their essential support.
We are especially grateful for the efforts by the computing, generator and validation groups who were instrumental for the creation of large simulation samples. We thank the detector upgrade groups as well as the physics and performance groups for their input. Not least, we thank the many colleagues who have provided useful comments on the analyses.
The research of  A.~Aboubrahim  and P.~Nath was supported in part by NSF Grant PHY-1620575.  
The research of J.~Kalinowski and W.~Kotlarski was supported in part by the National Science Centre, Poland,  HARMONIA project under contract UMO-2015/18/M/ST2/00518 (2016-2019).
%We would like to thank Matthew Low and Lian-Tao Wang for discussions. 
The work of T.~Han, S.~Mukhopadhyay, and X.~Wangwas supported in part by the U.S.~Department of Energy under grant No.~DE-FG02- 95ER40896 and by the PITT PACC. T.~ Han also acknowledges the hospitality of the Aspen Center for Physics, which is supported by National Science Foundation grant PHY-1607611.
%S. V. Chekanov$^1$, J. T. Childers$^2$, J. Proudfoot$^3$, R.Wang$^4$, D. Frizzell
The work of S.~Chekanov, J.~Childers, J.~Proudfoot, R.~Wang, and D.~Frizzell has been created by UChicago Argonne, LLC,
Operator of Argonne National Laboratory (``Argonne'').  Argonne, a U.S. Department of Energy Office of Science laboratory,
is operated under Contract No. DE-AC02-06CH11357. It was also made possible by an allocation of computing time through the ASCR Leadership Computing Challenge (ALCC) program and used resources of the National Energy Research Scientific Computing Center, a DOE Office of Science User Facility supported by the Office of Science of the U.S. Department of Energy under Contract No. DE-AC02-05CH11231.
%
%B. Allanach$^1$, D. Bhatia$^2$ and A. Iyer
The work of B.~Allanach, D.~Bhatia,  B.~Gripaios, A.~Iyer, and T.~Tevong You has been partially supported by STFC consolidated grants ST/L000385/1, ST/P000681/1.
%B. Allanach, B. Gripaios, T. Tevong You
%This work has been partially supported by STFC consolidated grants ST/L000385/1 and ST/P000681/1.
%M. Altakach$^1$, T. Jezo$^2$, M. Klasen
The work of M.~Klasen has been supported by the BMBF under contract 05H15PMCCA and the DFG through the Research Training Network 2149
``Strong and weak interactions - from hadrons to dark matter''.
T.~Jezo was supported by the Swiss National Science Foundation~(SNF) under contracts BSCGI0-157722 and CRSII2-160814.
%S. De Curtis$^a$, L. Delle Rose$^{a}$, S. Moretti$^{b,c}$, A. Tesi$^{a}$, K. Yagyu
%S.~Moretti is funded in part through the NExT Institute and the STFC CG ST/P000711/1.  %moved to later
%Dario Buttazzo$^{a,1}$, Filippo Sala$^{b,2}$, Andrea Tesi
The work of F.~Sala is partly supported by a PIER Seed Project funding (Project ID PIF-2017-72).
%S. Biswas, E. Gabrielli, M. Heikinheimo, B. Mele
The work of M.~Heikinheimo has been supported by the Academy of Finland, Grant NO. 31013
% Giacomo Cacciapaglia, Gabriele Ferretti, Thomas Flacke, Hugo Serodio
%We wish to thank M. Selvaggi for help with the HL detector simulation in Delphes. 
G.~Cacciapaglia is supported by Institut Franco-Suedois (project T{\"o}r) and the Labex-LIO (Lyon Institute of Origins) under grant ANR-10-LABX-66, FRAMA (FR3127, F\'ed\'eration de Recherche ``Andr\'e Marie Amp\`ere'').  G.~Ferretti is supported by The Knut and Alice Wallenberg Foundation and the Lars Hierta Memorial Foundation. T.~Flacke is supported by the IBS under the project code, IBS-R018-D1.  H.~Serodio is supported by the ERC under the European Union's Horizon 2020 research and innovation programme (grant agreement No 668679).
%Daniel A. Camargo, Luigi Delle Rose, Stefano Moretti, Farinaldo S. Queiroz
%The authors thank Werner Rodejohann and Pyung-won Ko for  discussions  and  comments.   
D.~Camargo  and  F.~S.~Queiroz  acknowledge financial support from MEC and UFRN. F.~S.~Queiroz also acknowledges the ICTP-SAIFR FAPESP grant 2016/01343-7 for additional  financial  support.   
S.~Moretti  is  supported  in  part  by  the NExT  Institute  and  acknowledges  partial  financial  support from  the  STFC  Consolidated  Grant  ST/L000296/1  and STFC CG ST/P000711/1, and the H2020-MSCA-RISE-2014 grant no.~645722 (NonMinimal-Higgs).

\clearpage

% -- Add bibliography to table of contents
%\addcontentsline{toc}{chapter}{References}

% dummy reference to avoid that bibtex fails
% -- Add volume bibliography and part specific bibliographies

\bibliographystyle{report_edit}
\bibliography{\bibfiles}

\end{document}